\documentclass{ut-thesis}

\degree{Doctor of Philosophy}
\department{Computer Science}
\gradyear{2017}
\author{Trevor Brown}
\title{Techniques for Constructing Efficient Lock-free Data Structures}

\usepackage[usenames,dvipsnames,svgnames,table]{xcolor}
\definecolor{trevcolor}{RGB}{0,80,200}
\definecolor{pdfbgcolor}{RGB}{180,180,180}
\newcommand{\later}[1]{}
\newcommand{\trevorlater}[1]{}

\usepackage{hyperref}
\hypersetup{linktocpage}

\newcommand{\wcnarrow}[2]{\parbox{\namewidth}{#1} \com \mbox{#2}}

\usepackage{etoolbox}
\usepackage{environ}
\newtoggle{isthesis}
\toggletrue{isthesis}
\NewEnviron{thesisonly}{%
  \iftoggle{isthesis}
    {\BODY\xspace}
    {}%
  }
\NewEnviron{thesisnot}{%
  \iftoggle{isthesis}
    {}
    {\BODY\xspace}%
  }

\newtoggle{hidechaps}
\togglefalse{hidechaps}
\NewEnviron{hide}{%
  \iftoggle{hidechaps}
    {}
    {\BODY\xspace}%
  }

\NewEnviron{fakethm}{%
  \medskip\noindent\textbf{Theorem}\quad %
  \textit{\BODY}%
  \medskip%
}
\NewEnviron{fakeinv}{%
  \medskip\noindent\textbf{Invariant}\quad %
  \textit{\BODY}%
  \medskip%
}
\NewEnviron{fakeprop}{%
  \medskip\noindent\textbf{Proposition}\quad %
  \textit{\BODY}%
  \medskip%
}
\NewEnviron{fakelem}{%
  \medskip\noindent\textbf{Lemma}\quad %
  \textit{\BODY}%
  \medskip%
}
\NewEnviron{fakedefn}{%
  \medskip\noindent\textbf{Definition}\quad %
  \textit{\BODY}%
  \medskip%
}

\usepackage{framed}
\usepackage{tabto}
\usepackage{comment}
\excludecomment{ignore}
\usepackage{amsfonts}
\usepackage{amssymb}
\usepackage{graphicx}
\usepackage{setspace}
\usepackage{smartref} % for referencing the line numbers
\usepackage{fullpage}
\usepackage{listings}
\usepackage{ragged2e}
\usepackage{enumerate}
\usepackage{rotating}
\usepackage{verbatim}
\usepackage[normalem]{ulem}
\usepackage{paralist}
\usepackage{wrapfig}
\usepackage{paralist}
\usepackage{tabularx}
\usepackage[shortlabels]{enumitem}
\usepackage{pgf}
\newcolumntype{Y}{>{\centering\arraybackslash} X}

\newcommand{\qed}{\hfill{\rule{2mm}{2mm}}}
\newenvironment{chapscxproof}{\begin{trivlist}
\item[\hspace{\labelsep}{\bf\noindent Proof: }]
}{\qed\end{trivlist}}
\newcommand{\comnospace}{\mbox{$\triangleright$}}
\newcommand{\com}{\mbox{\comnospace\ }}

\newcommand{\func}[1]{\mbox{\sc #1}}

\newcommand{\cas}{\mbox{CAS}}

\newcommand{\true}{\textsc{True}}
\newcommand{\false}{\textsc{False}}

\newcommand{\node}{\mbox{Node}}
\newcommand{\op}{\mbox{\sct -record}}
\newcommand{\rec}{\mbox{Data-record}}

\newcommand{\listrec}{\mbox{Node}}

\newcommand{\llresults}{\info Fields}
\newcommand{\info}{\textit{info}}

\newcommand{\help}{\func{Help}}
\newcommand{\validate}{\vlt}
\newcommand{\llt}{\func{LLX}}

\newcommand{\sct}{\func{SCX}}
\newcommand{\vlt}{\func{VLX}}
\newcommand{\del}{\func{Delete}}
\newcommand{\ins}{\func{Insert}}

\newcommand{\tryins}{\func{TryInsert}}
\newcommand{\trydel}{\func{TryDelete}}
\newcommand{\tryrebalance}{\func{TryRebalance}}

\newcommand{\cleanup}{\func{Cleanup}}
\newcommand{\search}{\func{Search}}

\newcommand{\freezing}{\mbox{InProgress}}
\newcommand{\retry}{\mbox{Aborted}}
\newcommand{\done}{\mbox{Committed}}

\newcommand{\freezingdone}{allFrozen}

\newcommand{\fail}{\textsc{Fail}}
\newcommand{\finalized}{\textsc{Finalized}}

\newcommand{\fcas}{{freezing~\cas}}
\newcommand{\astep}{{abort step}}
\newcommand{\fstep}{{frozen step}}
\newcommand{\fcstep}{{frozen check step}}
\newcommand{\cstep}{{commit step}}
\newcommand{\upcas}{{update~\cas}}
\newcommand{\markstep}{{mark step}}

\newcommand{\nil}{\textsc{Nil}}

\newcommand{\presctfld}{(3)}
\newcommand{\presctinfo}{(2)}

\newcommand{\presctlinked}{1}
\newcommand{\presctabainit}{2}
\newcommand{\presctaba}{3}

\newcommand{\ritalic}[1]{r_{\mbox{\tiny\textsc{#1}}}}
\newcommand{\rx}{\ritalic x}
\newcommand{\rl}{\ritalic l}
\newcommand{\rr}{\ritalic r}
\newcommand{\rxx}{\ritalic{xx}}
\newcommand{\rxl}{\ritalic{xl}}
\newcommand{\rxr}{\ritalic{xr}}
\newcommand{\rxll}{\ritalic{xll}}
\newcommand{\rxlr}{\ritalic{xlr}}
\newcommand{\rxrl}{\ritalic{xrl}}
\newcommand{\rxrr}{\ritalic{xrr}}
\newcommand{\rxxx}{\ritalic{xxx}}
\newcommand{\rxxl}{\ritalic{xxl}}
\newcommand{\rxxr}{\ritalic{xxr}}
\newcommand{\rxxll}{\ritalic{xxll}}
\newcommand{\rxxlr}{\ritalic{xxlr}}
\newcommand{\rxxrl}{\ritalic{xxrl}}
\newcommand{\rxxrr}{\ritalic{xxrr}}

\newcommand{\rxxrll}{\ritalic{xxrll}}
\newcommand{\rxxrlr}{\ritalic{xxrlr}}
\newcommand{\rxxrrl}{\ritalic{xxrrl}}
\newcommand{\rxxrrr}{\ritalic{xxrrr}}

\newcommand{\rxxrlll}{\ritalic{xxrlll}}
\newcommand{\rxxrllr}{\ritalic{xxrllr}}
\newcommand{\rxxrlrl}{\ritalic{xxrlrl}}
\newcommand{\rxxrlrr}{\ritalic{xxrlrr}}

\newcommand{\uitalic}[1]{u_{\mbox{\tiny\textsc{#1}}}}
\newcommand{\ux}{\uitalic x}
\newcommand{\ul}{\uitalic l}

\newcommand{\uxx}{\uitalic{xx}}
\newcommand{\uxl}{\uitalic{xl}}
\newcommand{\uxr}{\uitalic{xr}}
\newcommand{\uxll}{\uitalic{xll}}
\newcommand{\uxlr}{\uitalic{xlr}}
\newcommand{\uxrl}{\uitalic{xrl}}
\newcommand{\uxrr}{\uitalic{xrr}}

\newcommand{\uxlrl}{\uitalic{xlrl}}
\newcommand{\uxlrr}{\uitalic{xlrr}}

\newcommand{\nitalic}[1]{n_{\mbox{\tiny\textsc{#1}}}}
\newcommand{\nL}{\nitalic{l}}
\newcommand{\nR}{\nitalic{r}}
\newcommand{\nLL}{\nitalic{ll}}
\newcommand{\nLLL}{\nitalic{lll}}
\newcommand{\nLR}{\nitalic{lr}}

\def\pwidth{2.5cm}
\def\indentamt{0mm}

\newcommand{\sidecom}[1]{\tabto{12.3cm} \com \mbox{#1}}
\newcommand{\medcom}[1]{\tabto{9cm} \com \mbox{#1}}
\newcommand{\dline}[2]{#1\\ \mbox{\hspace{\indentamt}#2}}
\newcommand{\tline}[3]{\dline{#1}{#2}\\ \mbox{\hspace{\indentamt}#3}}
\newcommand{\qline}[4]{\tline{#1}{#2}{#3}\\ \mbox{\hspace{\indentamt}#4}}

\lstdefinestyle{nonumbers}{numbers=none}
\newcommand{\preplisting}{\lstset{gobble=1, numbers=left, numberstyle=\tiny, numberblanklines=false, firstnumber=last, escapeinside={//}{\^^M}, breaklines=true, keywordstyle=\bfseries, morekeywords={type,subtype,break,continue,if,else,end,loop,while,do,done,exit, when,then,return,read,and,or,not,,for,each,boolean,procedure,invoke,next,iteration,until,goto}}}
\newcommand{\prepnewlisting}{\lstset{gobble=1, numbers=left, numberstyle=\tiny, numberblanklines=false, firstnumber=1, escapeinside={//}{\^^M}, breaklines=true, keywordstyle=\bfseries, morekeywords={type,subtype,break,continue,if,else,end,loop,while,do,done,exit, when,then,return,read,and,or,not,for,each,boolean,procedure,invoke,next,iteration,until,goto}}}
\newcommand{\preplistingnonumbers}{\lstset{gobble=1, numbers=none, numberstyle=\tiny, numberblanklines=false, firstnumber=last, escapeinside={//}{\^^M}, breaklines=true, keywordstyle=\bfseries, morekeywords={type,subtype,break,continue,if,else,end,loop,while,do,done,exit, when,then,return,read,and,or,not,,for,each,boolean,procedure,invoke,next,iteration,until,goto}}}

\newtheorem{thm}{Theorem}[chapter]
\newtheorem{obs}[thm]{Observation}
\newtheorem{lem}[thm]{Lemma}
\newtheorem{cor}[thm]{Corollary}
\newtheorem{con}[thm]{Constraint}
\newtheorem{defn}[thm]{Definition}
\newtheorem{prop}[thm]{Proposition}

\newcommand{\ourcomments}[2]{{\bf [[[#1--#2]]]}}

\newcommand{\eric}[1]{\ourcomments{#1}{Eric}}
\newcommand{\trevor}[1]{\ourcomments{#1}{Trevor}}
\newcommand{\faith}[1]{\ourcomments{#1}{Faith}}

\newcommand{\after}[1]{}
\newcommand{\maxslack}{overslack tree}
\newcommand{\bslack}{B-slack tree}
\newcommand{\rbslack}{relaxed \bslack}
\newcommand{\Rbslack}{Relaxed \bslack}
\newcommand{\weight}{weight}

\newcommand{\details}[1]{}

\newcommand{\iset}{A-multiset}

\usepackage[mathcal]{euscript}

\usepackage{tikz}

\usepackage[T1]{fontenc}
\usepackage[utf8]{inputenc}
\usepackage{mathptmx}
\usepackage{pifont}

\newcommand{\cmark}{\checkmark}

\lstdefinestyle{customc}{
  belowcaptionskip=1\baselineskip,
  breaklines=true,
  frame=L,
  xleftmargin=\parindent,
  language=C++,
  showstringspaces=false,
  basicstyle=\scriptsize\ttfamily,
  keywordstyle=\bfseries\color{green!40!black},
  commentstyle=\itshape\color{purple!40!black},
  identifierstyle=\color{blue},
  stringstyle=\color{orange},
  numbers=left
}

\newcommand{\record}{record}

\newcommand\leaveq{\textit{leaveQstate}}
\newcommand\enterq{\textit{enterQstate}}
\newcommand\retire{\textit{retire}}
\newcommand\isq{\textit{isQuiescent}}

\newtheorem{theorem}{Theorem}
\newtheorem{Observation}{Observation}
\newtheorem{definition}{Definition}%[section]

\usepackage{amsfonts}
\usepackage{amssymb}
\usepackage{epsfig}
\usepackage{subcaption}

\usepackage{amsmath}
\usepackage{listings}
\usepackage{array}
\usepackage{paralist}
\usepackage{url}
\usepackage{xspace}
\usepackage[final]{showlabels}
\usepackage{afterpage}

\newcommand{\fakeparagraph}[1]{\medskip\noindent\textbf{\textit{#1}.}}

\usepackage{etoolbox}
\usepackage{environ}
\newtoggle{isfullver}
\NewEnviron{fullver}{%
  \iftoggle{isfullver}
    {\BODY\xspace}
    {}%
  }
\NewEnviron{shortver}{%
  \iftoggle{isfullver}
    {}
    {\BODY\xspace}%
  }
\toggletrue{isfullver}
\usepackage{floatpag}
\floatpagestyle{empty}
\makeatletter
\AtBeginEnvironment{figure}{%
  \def\baselinestretch{1}\@currsize}%
\makeatother
\makeatletter
\AtBeginEnvironment{figure*}{%
  \def\baselinestretch{1}\@currsize}%
\makeatother

\DeclareMathOperator{\vdes}{DES}
\DeclareMathOperator{\vaddr}{ADDR}
\DeclareMathOperator{\vexp}{EXP}
\DeclareMathOperator{\vnew}{NEW}

\DeclareMathOperator{\vstate}{STATE}

\DeclareMathOperator{\vV}{V}
\DeclareMathOperator{\vR}{R}
\DeclareMathOperator{\vfld}{FLD}
\DeclareMathOperator{\vold}{OLD}
\DeclareMathOperator{\vallfrozen}{ALLFROZEN}
\DeclareMathOperator{\vnfreeze}{NFREEZE}
\DeclareMathOperator{\vnfinalize}{NFINALIZE}

\usepackage{array}
\usepackage{xcolor}
\usepackage{graphicx}
\usepackage{rotating}
\usepackage{listings}
\usepackage{paralist}
\usepackage{tikz}
\usepackage{url}
\usepackage{wrapfig}

\usepackage[T1]{fontenc}
\usepackage[utf8]{inputenc}
\usepackage{mathptmx}

\begin{document}

\begin{preliminary}
\maketitle

%% There should be NOTHING between the title page and abstract.
%% However, if your document is two-sided and you want the abstract
%% _not_ to appear on the back of the title page, then uncomment the
%% following line.
%\cleardoublepage

\begin{abstract}
%% *** Put your Abstract here. ***
%% (At most 150 words for M.Sc. or 350 words for Ph.D.)
Building a library of concurrent data structures is an essential way to simplify the difficult task of developing concurrent software.
Lock-free data structures, in which processes can \textit{help} one another to complete operations, offer the following progress guarantee: If processes take infinitely many steps, then infinitely many operations are performed.
Handcrafted lock-free data structures can be very efficient, but are notoriously difficult to implement.
We introduce numerous tools that support the development of efficient lock-free data structures, and especially trees.

We address the difficulty of using single-word compare-and-swap (CAS) to develop lock-free graph-based data structures by introducing multiword synchronization primitives called \llt\ and \sct.
These primitives fall between multi-compare-single-swap and full multiword-CAS in expressiveness, and can be implemented much more efficiently than multiword-CAS.
We use them to implement a tree update template that can be followed to produce lock-free implementations of arbitrary updates in down-trees, and use this template to produce the first lock-free implementations of several advanced trees: chromatic search trees, relaxed AVL trees, and relaxed $(a,b)$-trees.
We also introduce a new variant of a B-tree, called a \rbslack, which has significantly better worst-case space complexity, and produce a lock-free implementation using our template.

In these implementations, operations dynamically allocate memory for \textit{nodes}.
Additionally, in our implementation of \llt\ and \sct, each \sct\ operation allocates a \textit{descriptor}, which contains information that another process can use to \textit{help} the \sct\ operation complete.
These implementations must perform dynamic lock-free memory reclamation, but traditional lock-free memory reclamation algorithms are either inefficient or cannot be used with our implementations.
So, we introduce a fast epoch-based reclamation (EBR) algorithm called DEBRA+, which is the first lock-free EBR algorithm that can be used with trees.

We also devise two techniques for accelerating lock-free data structure implementations.
Using the first technique, we can efficiently eliminate dynamic allocation and reclamation of \textit{descriptors} in \sct, and a large class of other algorithms.
Using the second, we exploit hardware transactional memory support in modern processors to produce accelerated implementations of the template. %, significantly improving performance of data structures by up to 5x
%give efficient multiword synchronization primitives that can be implemented efficiently, a template that one can follow for implementing arbitrary updates in down-trees, an efficient algorithm for reclaiming memory, and techniques for
\end{abstract}

\begin{acknowledgements}
According to an old African proverb, it takes a village to raise a child. Similarly, it takes an academic community to train a researcher. Over the last few years, I have been fortunate to be surrounded by many brilliant, friendly and helpful people. I would first like to express my gratitude to my supervisor Faith Ellen. She taught me the importance of laying a rigorous theoretical foundation, completely transformed my writing, encouraged me when I would otherwise have given up on a hard problem, and continually surprised me with her keen insights. I would also like to thank Eric Ruppert, who served as a mentor and research supervisor to me during my undergraduate studies, steered me towards research, and suggested that I study under Faith.

Faith and Eric have both been excellent collaborators, as well as mentors, and they were instrumental in the development of the \llt\ and \sct\ primitives and the tree update template.
During an undergraduate research project, Ken Hoover used the tree update template to develop the lock-free relaxed AVL tree, and did some initial work on the lock-free relaxed $(a,b)$-tree, and I thank him for his help.
I would also like to acknowledge Maya Arbel-Raviv for her part in the work on weak descriptors, and thank her for our many interesting discussions.

I am indebted to the members of my supervisory committee and thesis defense committee, namely, Azadeh Farzan, Vassos Hadzilacos, Maurice Herlihy, Ryan Johnson and Sam Toueg, for the time and energy they invested. Their feedback has also been very useful, and has led to numerous new insights. Most notably, discussions with Ryan led to the crash recovery mechanism in DEBRA+. I also thank Ryan for taking the time to explore parts of the Linux kernel with me, and for helping me debug hardware transactional memory issues.

The experiments in this work would not have been possible without several large-scale systems, which were graciously provided by Oracle, the University of Rochester, the Technion (Israel Institute of Technology), and the University of Toronto. Additionally, I was able to spend much less time implementing competing algorithms thanks to researchers who published their code, as well as those who were kind enough to send me their code, namely Maya Arbel-Raviv, Alex Matveev and Shahar Timnat. Funding for my research was provided by the Natural Sciences and Engineering Research Council of Canada.

Of course, I have to mention my parents and siblings, who have been an unfailing source of joy and warmth, regardless of the challenges in their own lives. Finally, I thank the love of my life, Brittany Trueman, for taking this ride with me. She has stood by me despite all of the late nights and deadline crunch-times, and serves as a gentle reminder that there is more to life than work.
\end{acknowledgements}

\tableofcontents

%% This generates the List of Tables (on a separate page), if needed
%% (uncomment to have it appear in the document).
%\listoftables

%% This generates the List of Figures (on a separate page), if needed
%% (uncomment to have it appear in the document).
%\listoffigures
\end{preliminary}

\chapter{Introduction}
% !TEX root = paper.tex

Building a library of concurrent data structures is an essential way to simplify the difficult task of developing concurrent software.
One way to implement concurrent data structures is to use \textit{locks}.
Locks protect shared resources.
%In order to access a shared resource, a process must first acquire the lock that protects it.
Concurrent data structures can be implemented with coarse-grained locking, where a single lock protects the entire data structure, or with fine-grained locking, where there are many locks, each protecting a small part of the data structure (e.g., a tree where each node has an associated lock).
When a process acquires a lock, it obtains exclusive access to the shared resource protected by the lock.

Lock-based programming is fairly simple, and there are many concurrent lock-based data structures.
As a simple example, consider an operation for inserting a key into a binary search tree.
We describe one way this operation could be implemented using fine-grained locking.
The operation first locks the root node, and then searches for the location where a key should be inserted using hand-over-hand locking: each time it follows a child pointer, it locks the node it reaches before accessing its contents, and then subsequently unlocks the parent node.
Once the operation reaches the node where the key should be inserted, it holds locks on that node, and its parent.
The operation performs the appropriate modification to the tree, and unlocks those two nodes.
It is straightforward to argue that these insertion operations are atomic.
However, they are also \textit{inefficient}.

The locking performed by searches is extremely costly, for several reasons.
First, the cost of acquiring locks can be very high compared to the cost of reading a node's key and child pointers.
Second, locking the root, and other nodes near the top of the tree, can severely limit concurrency.
Reader-writer locks (which allow a lock to be acquired by a single writer or multiple readers) can be used to improve concurrency, but they are even more costly to acquire and release.
Third, locking nodes that are not modified by an operation essentially turns reads into writes, which negatively impacts cache performance, especially on systems with non-uniform memory architectures.

Locks also have several other downsides.
Most notably, if a process crashes while holding a lock, then it can prevent \textit{all} processes from making progress.
Additionally, locks can be susceptible to deadlock, convoying and priority inversion~\cite{Fra07}.
Deadlock occurs when two processes attempt to lock the same resources, but in different orders, and they mutually prevent one another from making progress.
Convoying occurs when many threads attempt to acquire the same lock, and processes sleep while waiting for the lock to become free.
Each time the lock becomes free, all of the processes wake up, attempt to acquire the lock, and, if unsuccessful, go back to sleep.
This corresponds to a long convoy of processes moving from the waiting queue 
(where processes reside when they are not yet ready to run) to the run queue (where they wait to be executed on a processor) and back again.
This movement of processes between scheduler queues can cause significant overhead.
Priority inversion occurs when a process that is given high priority by the scheduler attempts to acquire a lock that is held by a low priority process, but must wait for the low priority process to release the lock before it can acquire the lock and make progress.

Consequently, it is often preferable to use hardware synchronization primitives like compare-and-swap (\textsc{CAS}) instead of locks.
However, the difficulty of this task has inhibited the development of {\it lock-free} (also called \textit{non-blocking}) data structures.
These are data structures which guarantee that the system as a whole will continue to make progress % operation will eventually complete 
even if some processes crash.
In other words, they guarantee that, at all times, some operation will eventually complete.
(Naturally, these data structures cannot use locks.)
This progress guarantee is typically achieved with \textit{helping}.
When a process is blocked by an operation performed by another process, it helps the other process to make progress before continuing its own operation.

Direct implementations of lock-free data structures from CAS are notoriously complicated. %for two major reasons. %, and their proofs of correctness and progress are extremely long.
Many operations on interesting data structures access, or even modify, multiple words in memory.
With locks, one could achieve atomicity simply by acquiring locks on all of the relevant words in memory, and then performing the appropriate modification.
However, CAS operates atomically on only a \textit{single} word in memory. %, but many operations in interesting data structures access (or even modify) multiple words.
As a result, things that were straightforward with locks, such as ensuring that a node in a tree is not changed at the same time as it is deleted (or thereafter), become very difficult with CAS.
Thus, complex ad-hoc synchronization mechanisms are needed to guarantee atomicity.

We briefly discuss why these synchronization mechanisms are so complex.
In a lock-free algorithm, \textit{processes} cannot acquire exclusive access to any shared resource, or else the resource could become permanently inaccessible in the event of a process crash.
Consequently, instead of having a \textit{process} acquire exclusive access to shared resources, an \textit{operation} acquires exclusive access to shared resources, and any processes helping the operation can access these resources.
It is important to note that multiple processes can simultaneously help the same operation.
(To ensure that an operation could be helped by only one process at a time, a process would need to acquire exclusive access to a shared resource, which we have argued cannot happen.)
Therefore, the helping algorithm must be designed so that multiple processes helping the same operation do not perform conflicting changes or erroneously repeat algorithmic steps (and it must guarantee this \textit{without} having any process acquire exclusive access to any shared resource). %interfere with one another in a way that yields incorrect results.

%The synchronization mechanisms 
%
%Second, processes that help one another must carefully synchronize to avoid performing conflicting changes, or erroneously repeating algorithmic steps.
%This synchronization is often highly complicated, since \textit{processes} cannot acquire exclusive access to any shared resource in a lock-free algorithm (or else the resource could become permanently inaccessible in the event of a process crash).
%%Instead, complex lock-free data structures are typically implemented in the following way.
%Instead of having a \textit{process} acquire exclusive access to shared resources, an \textit{operation} acquires exclusive access to shared resources, and any processes helping the operation can access these resources.
%It is important to note that multiple processes can simultaneously help the same operation.
%(To ensure that an operation could be helped by only one process at a time, a process would need to acquire exclusive access to a shared resource, which we have argued cannot happen.)
%Consequently, the helping mechanism must be designed so that multiple processes helping the same operation do not interfere with one another in a way that yields incorrect results.

Since direct implementations of lock-free data structures from CAS are so complicated, it is extremely difficult to formally prove correctness and progress for them.
%The complexity of lock-free data structures implemented using CAS makes it extremely difficult to formally prove correctness and progress for them.
%%Since implementations of lock-free data structures are so complex, proving correctness and progress for them is also extremely difficult.
Moreover, when these proofs are actually produced, they are nearly impossible to check. %Not only are such proofs long and tedious, and are extremely difficult to check. %it is not unusual for a proof to be tens of pages long.
Consequently, most implementations are presented with only brief sketches of correctness and progress.
Unfortunately, these sketches often turn out to be incorrect.

Early on, researchers recognized that synchronization primitives which atomically access multiple locations make the design of lock-free data structures much easier~\cite{Barnes:1993,Israeli:1994,ST97}.
In Chapter~\ref{chap-scx}, we introduce three new primitives,
\textit{load-link-extended} (\llt), \textit{validate-extended} (\vlt) and \textit{store-conditional-extended} (\sct), which are natural generalizations of the well known \textit{load-link} (\textsc{LL}), \textit{validate} (\textsc{VL}) and \textit{store-conditional} (\textsc{SC}) primitives.
We carefully designed our primitives to strike a fine balance between ease of use and efficient implementability.
We provide a practical implementation of our primitives from \textsc{CAS}, and give complete proofs of correctness and progress.
In Chapter~\ref{chap-multiset}, we demonstrate their use by giving a simple implementation of a multiset from a singly-linked list, along with a full proof.

As we were developing our new primitives, there was a flurry of interest in lock-free trees.
In particular, numerous papers appeared containing incredibly complicated ad-hoc implementations of lock-free balanced search trees.
These implementations either had only brief correctness arguments or incredibly long proofs (with one exceeding 100 pages).
The authors of these papers never released source code for their implementations.
With the goal of providing a rigorous, provably correct alternative, 
%In order to stop this stream of papers, 
we introduced a \textit{tree update template} that can be followed to produce lock-free implementations of down-trees (trees in which all nodes except the root have in-degree one) with any kinds of update operations.
This template, which appears in Chapter~\ref{chap-template}, uses our \llt\ and \sct\ primitives.

We demonstrate the use of our template by producing lock-free implementations of several data structures that are significantly more advanced than any lock-free data structures in the literature.
Java and/or C++ code for all of our implementations is publicly available.\footnote{\url{http://implementations.tbrown.pro}}

In Chapter~\ref{chap-chromatree}, we present a lock-free implementation of a chromatic search tree, which is a generalization of a red-black tree (a kind of balanced binary search tree (BST)).
The chromatic tree is a highly advanced \textit{relaxed balance} tree.
In contrast to traditional balanced trees, which require an insertion or deletion and any necessary rebalancing to be performed as one large atomic update (potentially involving an entire root-to-leaf path), relaxed balance trees \textit{decouple} rebalancing from insertions and deletions.
Specifically, rebalancing is performed as a sequence of small atomic \textit{rebalancing steps} that can be interleaved freely with insertions, deletions, and other rebalancing steps.
This significantly improves concurrency.
However, as a consequence, the chromatic tree can transiently become slightly less balanced than a red-black tree, and there are more cases for rebalancing to handle.
In fact, there are eleven different types of rebalancing steps (each with a symmetric mirror-image version).
%Only an amortized constant number of rebalancing steps must be performed per insertion or deletion to maintain balance.
Despite the complexity of this data structure, we provide a rigorous proof of correctness and progress that is both concise (only six pages long) and easily checked.
A wide range of experiments on a large scale system showed that our implementation was significantly faster than the leading competitors.

In Chapter~\ref{chap-ravl}, we give a high level description of a lock-free relaxed AVL (RAVL) tree, which is a relaxed balance generalization of an AVL tree.
As with the chromatic tree, decoupling rebalancing so that it can be performed in small atomic rebalancing steps makes RAVL trees more concurrency friendly, but creates more cases for rebalancing.
RAVL trees 
%RAVL trees allow the properties of an AVL tree to be violated in specific ways, and
provide eleven different rebalancing steps (each with a symmetric mirror-image version). %to fix violations of the AVL tree properties.
%In the worst case, an amortized logarithmic number of rebalancing steps must be performed per insertion or deletion to maintain balance.
%Every RAVL tree can be transformed into a standard AVL tree by performing 
%Similarly to the chromatic tree, when the RAVL tree was introduced, it was assumed that rebalancing steps would be completely decoupled from insertion and deletion.
%We introduce a new rebalancing algorithm for RAVL trees that yields an upper bound of $O(c+\log n)$ on the height of the RAVL tree, where $c$ is defined as above.
Experiments show that an optimized version of RAVL trees performs as well as chromatic trees.

A B-tree of minimum degree $a \ge 2$ is a balanced tree in which all leaves have the same depth, and all nodes contain between $a$ and $2a-1$ keys.
Whenever an operation would cause a node to contain more than $2a-1$ or fewer than $a$ keys, the tree is rebalanced by \textit{splitting} the node or \textit{joining} it with a sibling, respectively (which may necessitate splitting or joining at a sequence of ancestors).
The \textit{capacity} of a node is the maximum number of keys it can contain.
The \textit{utilization} of a node is $k / c$, where $k$ is the number of keys it contains and $c$ is its capacity.
%In a B-tree, the worst case utilization of a node is $a / (2a-1)$, which tends to $1/2$ as $a$ approaches infinity.

In Chapter~\ref{chap-abtree}, we present a lock-free implementation of a relaxed $(a,b)$-tree, which is a relaxed balance generalization of a B-tree.
In a standard $(a,b)$-tree, all leaves have the same depth and contain between $a$ and $b$ keys, where $a \ge 2$ and $b \ge 2a-1$.
Observe that the worst case utilization of a node in an $(a,b)$-tree is the same as in a B-tree when $b = 2a-1$, and it decreases as $b$ grows relative to $a$.
Allowing nodes to contain fewer keys, proportional to their capacity, can reduce the number of rebalancing steps needed to maintain the structural properties of the tree.
Relaxed $(a,b)$-trees %are a relaxed balance generalization of $(a,b)$-trees that are more concurrency friendly.
%They 
allow the structural properties of $(a,b)$-trees to be violated in specific ways, and provide six different rebalancing steps to fix these violations, with the ultimate goal of transforming a relaxed $(a,b)$-tree into a standard $(a,b)$-tree.
%If $b \ge 2a$, then an amortized constant number of rebalancing steps per insertion or deletion is sufficient to maintain the $(a,b)$-tree properties.
We provide a rigorous proof of correctness and progress for our implementation in just five pages.
This in stark contrast to the lock-free B-tree of Braginsky and Petrank~\cite{BP12}, which has a 33 page proof, despite being a simpler data structure than an $(a,b)$-tree.
Experiments show that the relaxed $(a,b)$-tree has a significant performance advantage over the chromatic tree for certain workloads.

%As with the chromatic tree and RAVL tree, the relaxed $(a,b)$-tree was originally intended to have rebalancing steps that are fully decoupled from insertions and deletions.
%We introduce a new rebalancing algorithm for relaxed $(a,b)$-trees, and prove that the height of a relaxed $(a,b)$-tree is at most $O(c + \log_a n)$, where $c$ is defined as above.

When chromatic trees, RAVL trees and relaxed $(a,b)$-trees were proposed, it was assumed that rebalancing steps would be completely decoupled from insertion and deletion, so there would be no worst-case upper bound on the height of these trees.
For each tree, we introduce a simple algorithm for performing rebalancing steps that yields strong upper bounds on their heights.
We prove that the height of the chromatic tree is at most $O(c + \log n)$, where $c$ is the number of insertions and deletions currently in progress.
We sketch a proof that the RAVL tree has the same upper bound on its height.
We also prove that the height of the relaxed $(a,b)$-tree is at most $O(c + \log_a n)$.
The proofs of these upper bounds are subtle, and unlike anything in the literature on lock-free trees or relaxed balance trees.

One downside of B-trees is that half of the capacity of each node is wasted in the worst case (and even more capacity is wasted in $(a,b)$-trees with $b > 2a-1$).
%One downside of B-trees and $(a,b)$-trees is that 50\% of the capacity of each node (and even more in $(a,b)$-trees with $2a < b$) is wasted in the worst case.
The most obvious consequence of wasting space in nodes is increased tree height.
This increases search times, which are typically the dominant factor in the performance of search trees.
However, there is another, more subtle consequence in environments where memory is allocated in blocks, with a limited set of block sizes.
This is often the case in hardware, such as internet routers, where allocators are very simplistic, and typically use only a single block size to avoid fragmentation.
In this case, if we allocate a block for each node, then half of all memory is wasted in the worst case.
Fast memory is expensive, and hardware developers are very interested in building devices with less memory.
Since hardware must include sufficient resources to handle the worst case, it is \textit{not} sufficient to design a data structure with good \textit{average-case} behaviour---it must have good worst-case behaviour. %(since the hardware must include sufficient resources to handle the worst case).
%Furthermore, since hardware devices must include sufficient resources to handle the worst-case, it is not sufficient for a data structure to have good average-case behaviour to allow hardware developers to reduce the amount of memory included in their devices.

To address this problem, in Chapter~\ref{chap-bslack}, we introduce a novel data structure called a \textit{\bslack}, which is a variant of a B-tree with substantially better worst-case space complexity.
In a \bslack, all nodes contain between 0 and $b$ keys, and internal nodes contain between 2 and $b$ child pointers.
The \textit{degree} of an internal node (resp., leaf) is the number of pointers (resp., keys) it contains, and a node's \textit{slack} is $b - d$, where $d$ is its degree.
Slack represents the part of a node that is wasted.
Rather than imposing strong constraints on the degree or slack of individual nodes, the key idea is to constrain each internal node so that the sum of the slack at its children is less than $b$.
Surprisingly, this invariant is fairly straightforward to maintain, and it yields a tree with very good worst-case behaviour.
In the worst case, the average degree of nodes is very high, exceeding $b-2$ for trees of height at least three.
A rigorous and thorough mathematical analysis of the characteristics of \bslack s is presented.
The space complexity of \bslack s is significantly better than all of their competitors. % (see Appendix~\ref{sec-bslack-space-complexity}). %, even some that do not support deletion.
We also introduced relaxed \bslack s, which are a relaxed balance version of \bslack s.
\Rbslack s allow the properties of \bslack s to be violated in specific ways, and provide six different rebalancing steps to fix these violations, with the ultimate goal of transforming a \rbslack\ into a standard \bslack.
The operations on \rbslack s are relatively simple and efficient.
A Java implementation of a single-threaded \bslack\ is publicly available.

We present a provably correct lock-free implementation of a \rbslack\ using our template in Chapter~\ref{chap-lfbslack}.
This involved developing a new algorithm for determining when and where rebalancing steps should be applied, and proving that, when there are no ongoing updates, the tree is a (strict) \bslack.
Experimental results show that an optimized variant of the \rbslack\ yields significantly better performance than the chromatic tree for workloads consisting mostly of searches, and is reasonably efficient even for workloads with many updates.

In concurrent data structures that use locks, it is typically straightforward to free memory to the operating system after an object is removed from the data structure.
For example, consider a set implemented with a singly-linked list using hand-over-hand locking.
Recall that hand-over-hand locking allows a process to lock a node (other than the head node) only if it holds a lock on the node's predecessor.
To traverse from a locked node $u$ to its successor $v$, the process first locks $v$, then it unlocks $u$.
To delete a node $u$, a process first locks the head node, then performs hand-over-hand locking until it reaches $u$'s predecessor and locks it.
The process then locks $u$ (to ensure no other process holds a lock on $u$), removes it from the list, frees it to the operating system, and finally unlocks $u$'s predecessor.
It is easy to argue that no other process has a pointer to $u$ when $u$ is freed.

In contrast, memory reclamation is one of the most challenging aspects of lock-free data structure design.
The main difficulty in performing memory reclamation for a lock-free data structure is that a process can be sleeping while holding a pointer to an object that is about to be freed.
Thus, carelessly freeing an object can cause a sleeping process to access freed memory when it wakes up, crashing the program or producing subtle errors.
Since nodes are not locked, processes must coordinate to let each other know which nodes are safe to reclaim, and which might still be accessed.
(Note that reclaiming memory is similarly challenging for lock-based algorithms that have lock-free searches.)

Managed languages such as Java and C\# have automatic garbage collection, which greatly simplifies the implementation of lock-free data structures.
However, in unmanaged languages such as C and C++, programmers must manually implement lock-free memory reclamation.
Prior to this work, existing lock-free memory reclamation schemes were either inefficient, or could not easily be used with algorithms implemented from \llt\ and \sct\ (or our template).
To remedy this, we introduced DEBRA, a distributed epoch-based reclamation algorithm (Chapter~\ref{chap-debra}).
%In Chapter~\ref{chap-debra}, we introduce DEBRA, a distributed epoch-based reclamation algorithm.
Experiments show that DEBRA is highly efficient, even on systems with non-uniform memory architectures (NUMA).
However, a process that crashes in the middle of an operation can prevent all processes from reclaiming memory.
DEBRA is a good choice when process failures cannot occur, or in a lock-based data structure with lock-free searches.
We also present a fault-tolerant version of DEBRA called DEBRA+, and describe a large class of lock-free data structures that can be used with DEBRA+.

Up to this point, we have discussed tools for designing and implementing lock-free data structures.
However, there is also a need for techniques to \textit{accelerate} lock-free data structure implementations, e.g., to obtain the fastest code for data structure libraries.
We describe two techniques for doing so.

%In some simple lock-free data structures (e.g.,~\cite{Valois:1995,Harris:2001,Michael:2002,Natarajan:2014,Lea}), processes can determine how to help one another operations that block them by inspecting a small part of the data structure.
In many lock-free data structures (e.g., \cite{Ellen:2010,Howley:2012,Shafiei:2013}), each time a process performs an operation, it first creates a \textit{descriptor} that specifies the steps the process will take to perform the operation, and stores a pointer to the descriptor in the data structure so other processes can see it.
Then, whenever a process is blocked by an operation, it uses the information stored in the operation's descriptor to help it complete.
Our implementation of \llt\ and \sct\ takes this approach.
We call implementations that create a new descriptor for each operation \textit{wasteful algorithms}.
%processes publish descriptors for their operations, and helpers look at these descriptors to determine how to help.

In Chapter~\ref{chap-descriptors}, we introduce two simple \textit{descriptor} abstract data types (ADT) that attempt to capture how descriptors are used by wasteful algorithms (including our implementation of \llt\ and \sct): the immutable descriptor ADT (which represents descriptors whose fields are all immutable after they are first initialized), and the mutable descriptor ADT (which represents descriptors in which fields can be mutable).

Naturally, the descriptors used for helping must eventually be freed to the operating system, or reused.
In many applications, it is crucial that the reclamation of descriptors imposes very little runtime overhead.
Additionally, for some applications, such as embedded systems, it may be important to have a small, predictable number of descriptors in the system.
Thus, we want to reclaim or reuse descriptors in a way that minimizes time overhead and \textit{descriptor footprint}, i.e., the largest number of descriptors in the system at one time.
(Having a smaller descriptor footprint can also significantly improve the performance of processor caches, since less cache space is occupied by descriptors.)

One approach is to use a memory reclamation scheme to reclaim descriptors once they are no longer needed.
However, reclaiming descriptors for each operation can introduce significant overhead (since the memory reclamation scheme incurs overhead for each descriptor).
Additionally, it is not always easy to implement memory reclamation for descriptors, so this approach can add considerable complexity.
As a better alternative, we introduce a \textit{weak descriptor} ADT that has slightly \textit{weaker semantics} than the mutable descriptor ADT, but can be implemented \textit{without memory reclamation}.
We also identify a class of lock-free algorithms that use the descriptor ADT, and which can be \textit{transformed} to use the weak descriptor ADT.
We then present an extension to our weak descriptor ADT, and show how an even larger class of lock-free algorithms can be transformed to use this extension.
We prove correctness and progress for the extended transformation, and demonstrate its use by transforming wasteful implementations of a $k$-compare-and-swap ($k$-CAS) primitive~\cite{Harris:2002} and the \llt\ and \sct\ primitives in Chapter~\ref{chap-scx}.

We use known techniques to produce an efficient, provably correct implementation of our extended weak descriptor ADT. % in Section~\ref{sec-extended-impl}.
With this implementation, the transformed algorithms for $k$-CAS, and LLX and SCX, have some desirable properties.
In the original $k$-CAS algorithm, \textit{each operation attempt} allocates at least $k+1$ new descriptors.
In contrast, the transformed algorithm allocates only two descriptors \textit{per process, once, at the beginning of the execution}, and processes simply reuse these descriptors.
Similarly, whereas, in the original algorithm for LLX and SCX, each SCX operation creates a new descriptor, the transformed algorithm allocates only one descriptor per process, at the beginning of the execution.
Observe that this entirely eliminates dynamic allocation \textit{and} memory reclamation for descriptors (significantly reducing overhead), and results in an extremely small descriptor footprint.
Extensive experiments show that our transformed implementations perform at least as well as the original implementations, and \textit{significantly} outperform them (by up to 5x) in some workloads.

%We demonstrate the use of our transformation by applying it to lock-free implementations of: a double-compare-single-swap (DCSS) primitive, a $k$-compare-and-swap ($k$-CAS) primitive, and a binary search tree.
%We performed extensive experiments on multiple systems, over a variety of workloads, to compare our transformed implementations with implementations that allocate descriptors for each operation and use state of the art lock-free memory reclamation algorithms to reclaim descriptors.
%As expected, the results show that our transformation yields dramatic improvements in the descriptor footprint for all algorithms.
%Furthermore, our transformed implementations perform at least as well as the original implementations, and \textit{significantly} outperform them (by up to 5x) in some workloads.

Another way to accelerate lock-free data structures is to take advantage of new hardware features.
Recently, Intel introduced hardware transactional memory (HTM) in its processors.
HTM allows a programmer to run blocks of code in transactions, which either commit and take effect atomically, or abort and have no effect on shared memory.
HTM has the potential to improve the performance of handcrafted algorithms significantly.
This is because hardware transactions are extremely fast, and they can be used to optimistically avoid using other, more expensive synchronization mechanisms.
As a trivial example, a sequence of CAS instructions can be accelerated by replacing it with a transaction that performs reads, if-statements and writes.
Note that this represents a non-standard use of HTM: we are \textit{not} interested in its ease of use, but, rather, in its ability to reduce synchronization costs.

Although hardware transactions are extremely fast, it is surprisingly difficult to obtain the full performance benefit of HTM.
Intel's HTM implementation is best-effort, which means it does not guarantee that transactions will \textit{ever} be able to commit.
Even in a single threaded system, a transaction can repeatedly abort because of internal buffer overflows, page faults, interrupts, and many other events.
So, to guarantee progress, any code that uses HTM must also provide a non-transactional \textit{fallback path} to be executed if a transaction fails.
The decision of whether to allow operations on the fallback path to run concurrently with hardware transactions profoundly impacts the overall performance of algorithms implemented using HTM.
In order to support this concurrency, hardware transactions must be \textit{instrumented} with code that synchronizes with operations on the fallback path.
This can add significant overhead to hardware transactions, negating much of their benefit.
However, if operations on the fallback path are not permitted to run concurrently with hardware transactions, then numerous other performance pathologies arise.

In Chapter~\ref{chap-3path}, we explore this design space, developing three accelerated implementations of the tree update template which use two execution paths, and one which uses \textit{three} execution paths to combine the benefits of the two-path algorithms while avoiding their downsides.
We performed experiments to evaluate our new template algorithms by comparing them with the original template algorithm.
In order to compare the different template algorithms, we used each algorithm to implement two data structures: an unbalanced BST and a relaxed $(a,b)$-tree.
We then ran microbenchmarks to compare the performance (operations performed per second) of the different implementations in a variety of workloads.
The results show that our new template algorithms offer significant performance improvements.
For example, on an Intel system with 72 concurrent processes, our best implementation of the relaxed $(a,b)$-tree outperformed the implementation using the original template algorithm by an average of 410\% over all workloads.

\chapter{Model} \label{chap-model}
% !TEX root = paper.tex

We consider an asynchronous shared memory system with $n$ processes numbered one through $n$.
Processes can run at arbitrarily different speeds, and the speeds of processes are not fixed. %, so arbitrarily long delays can occur between any pair of steps by a process.
They can also experience crash failures (sometimes known as halting failures, or stop failures).
In this model, a crashed process is indistinguishable from an extremely slow process.
%One consequence is that a slow process is indistinguishable from a process that has crashed.
%
Each process has local memory that is not accessible by any other process, and there is a shared memory accessible by all processes.

\paragraph{Primitive objects}
Shared memory is divided into primitive objects, which have atomic operations that are provided directly by the hardware.
Examples include read/write registers, compare-and-swap (CAS) objects, and double-wide compare-and-swap (DWCAS) objects.

A read/write register contains a value and provides read and write operations to retrieve the current value, and replace it with a new value, respectively.
The read and write operations are atomic, and the read operation always returns the value written by the last write operation.

A CAS object is a register that also offers a CAS operation, which has two arguments: an expected value $exp$ and a new value $new$.
A CAS operation \textit{atomically} does the following.
It first reads the value in the CAS object.
If the value is equal to $exp$, then the CAS operation stores $new$ in the CAS object and returns \true.
Otherwise, the CAS operation simply returns \false.

As a theoretical object, CAS is allowed to contain arbitrarily large values.
However, in real systems, memory is organized as a finite sequence of words, which each have a finite set of possible values, and CAS operates on a \textit{single word} in memory.
For many applications, it turns out to be useful to have the ability to atomically perform CAS on \textit{two adjacent words} in memory, and DWCAS was introduced to do exactly that.
%In contrast, DWCAS atomically operates on two adjacent words.
%That is, it atomically stores two values in two adjacent words if and only if both contain their expected values.
%From a theoretical standpoint, DWCAS is not substantively different from CAS with unbounded values.
%However, DWCAS is very useful in real systems, and it is implemented on all modern Intel and AMD processors.
Although DWCAS offers no additional power over unbounded CAS in a theoretical sense, it is very useful in real systems, and is implemented on all modern Intel and AMD processors.

\paragraph{Configurations and executions}
A \textit{configuration} of the system consists of the states of all processes, and the state of shared memory.
In any given configuration, each process has a set of \textit{steps} (operations on primitive objects) that it can perform, and this set is determined by the state of the process in that configuration.
Note that a process may be able to perform a given step in one configuration, but not in another configuration.
Performing a step can change the state of the process and/or the state of shared memory.
An \textit{execution} is a (possibly infinite) sequence of alternating configurations and steps $C_0 \cdot s_0 \cdot C_1 \cdot s_1 \cdot C_2 \cdot s_2 \dots$, where each $C_i$ is a configuration, and each $s_i$ is a step that can be performed by a process in configuration $C_i$.
We say that a configuration is \textit{reachable} if it appears in some execution.

\paragraph{Records and data structures}
A \textit{\record} is a collection of primitive objects, which we refer to as \textit{fields}.
Fields can contain pointers to other \record s.
A data structure has a fixed set of \textit{entry points}, which are pointers to \record s. % and the \record s that are reachable by following pointers starting from an entry point.
A \record\ is \textit{in the data structure} if it is reachable by following pointers starting from an entry point.
A \record\ is \textit{removed from the data structure} when it changes from being in the data structure to not being in the data structure.
A \record\ is \textit{inserted into the data structure} when it changes from being not in the data structure to being in the data structure.
Any fields of a \record\ that cannot change while the \record\ is in the data structure are said to be \textit{immutable}.
Immutable fields can be arbitrarily large.
Other fields are said to be \textit{mutable}.
Each mutable field fits in a single machine word.

\paragraph{Allocating and freeing \record s}
The system has an memory \textit{allocator} that provides operations to \textit{allocate} and \textit{free} \record s.
Initially, all \record s in shared memory are \textit{unallocated}.
Accessing an unallocated record will cause program failure.
Allocating a \record\ provides the process that requested it with a pointer to it, and makes it accessible by any process that has a pointer to it.
A \record\ can also be \textit{freed}, which returns it to the \textit{unallocated} state.

%\trevor{from descriptors: memory allocation and reclamation}
%
%We study a shared memory system with $n$ threads numbered $1..n$.
%Each thread has a private memory, and there is a shared memory accessible by all threads.
%Shared memory consists of read/write registers and compare-and-swap objects.
%
%All shared memory locations are initially \textit{unallocated}, and accessing them will cause a program failure.
%The system has an \textit{allocator} that provides two operations: \textit{allocate} and \textit{free}. %threads can use to \textit{allocate} regions of memory.
%The allocate operation takes a \textit{size} argument, expressed in bytes, and returns a pointer $ptr$ to a newly allocated region of memory of the requested size.
%Once a memory region is \textit{allocated}, threads can freely access it (without causing a program failure).
%A thread can subsequently invoke \textit{free}$(ptr)$ to return that memory to the unallocated state.

\paragraph{Memory hierarchy}

The memory is organized into a hierarchy where the lowest level is \textit{main memory}.
%Although main memory is logically organized into pages, and physically organized in terms of rank, bank, device, row and column, 
Without loss of generality, we consider the \textit{cache line} granularity in main memory as the smallest unit of data.
The next levels of the hierarchy are \textit{cache} levels, which contain copies of cache lines that appear in main memory.
The cache levels are numbered, L1, L2, L3 and so on, starting with the highest level (L1) and moving down the hierarchy.
In practice, systems typically use between two and four cache levels. A cache coherence protocol ensures that processors see a consistent view of main memory despite the existence of multiple cached copies of some memory locations.
At the highest level of the memory hierarchy are \textit{registers}, special memory locations reserved in each processor for temporary computations.
Generally, operations lower in the memory hierarchy are orders of magnitude slower than operations higher in the hierarchy.

We consider a modified-exclusive-shared-invalid (MESI) cache coherence protocol.
%At all times, each cache line in a processor cache is in one of four states: modified, exclusive, shared or invalid.
%Only the last three of these states are relevant for our purposes.
Whenever a process $p$ attempts to access a cache line, it checks each of its caches one-by-one, starting from the highest level, to see whether the cache line already resides in any of its caches.
At each level of the cache hierarchy where $p$ fails to find the desired cache line, we say that it experiences a \textit{cache miss}.
The last cache level before main memory is called the \textit{last-level cache}, and a cache miss at that level is called a \textit{last-level cache miss}.
If $p$ experiences a last-level cache miss, it must retrieve a copy of the cache line from main memory.

When reading from memory, a process loads a cache line into its cache in \textit{shared} mode.
Many processes can simultaneously have the same cache line in their private caches, provided that they all loaded it in shared mode.
When writing to memory, a process loads a cache line into its cache in \textit{exclusive} mode, and makes the change to its own cached copy (gradually flushing the change downward to all lower levels of the memory hierarchy).
Loading a cache line in exclusive mode also causes any copies of that cache line in other process' caches to be \textit{invalidated}.
Whenever a process $p$ tries to access a cache line that was invalidated since $p$ last accessed it, $p$ will experience a last-level cache miss.

\paragraph{Non-uniform memory architectures (NUMAs)}

Some of the systems studied in this work have non-uniform memory architectures, in which different parts of memory can have drastically different access times for different processes.
We take a simplified view of NUMAs that captures the most important costs on many modern systems.
%Sometimes groups of processes will \textit{share the same cache} at one or more levels of the memory hierarchy.
%We take a simplified view of shared caches that captures the most important costs on many modern systems.
Processes run on one or more \textit{sockets}.
All processes on a socket share the same last-level cache (but processes on one socket cannot access the last-level cache shared by processes on another socket).
Writes by a process do \textbf{not} cause cache invalidations for any processes that share the same cache.

As an example, consider two processes $p$ and $q$ that are on the same socket (and, hence, share the same last-level cache).
Suppose $q$ loads a cache line, then $p$ writes to it, and then $q$ loads it again.
Since $p$ and $q$ share the same last-level cache, when $p$ performs its write, it simply modifies $q$'s copy that is already in the last-level cache, and does \textbf{not} invalidate it.
However, $p$'s write will still invalidate any copies of the cache line in the higher level caches, and in the last-level caches of other sockets.
Thus, the next time $q$ accesses this cache line, it will use the copy in its last-level cache.
In contrast, if $p$ and $q$ were on different sockets, then $q$'s second load would have to retrieve the cache line from main memory.

%\paragraph{Non-uniform memory architectures (NUMAs)}
%
%Some of the experimental systems studied in this work have non-uniform memory architectures, in which accessing different parts of memory can have drastically different access times for different processes.
%We take a simplified view of memory that captures the most important costs on many modern systems.
%Processes are grouped into \textit{sockets}, which are numbered starting from one.
%All processes in a socket share the same last-level cache.

\paragraph{Hardware transactional memory}

%We consider Intel's implementation of HTM.
%Arbitrary blocks of code can be executed as transactions, which either commit (and appear to take place instantaneously) or abort (and have no effect on the shared memory).
%A transaction is started by invoking \textit{txBegin}, is committed by invoking \textit{txEnd}, and can be aborted by invoking \textit{txAbort}.
%Intel's implementation of HTM is best-effort, which means that the system can force transactions to abort at any time, and no transactions are ever guaranteed to commit.
%
%Each time a transaction aborts, the hardware provides a reason why the abort occurred.
%Two reasons are of particular interest.
%\textit{Conflict} aborts occur when two processes contend on the same cache-line.
%Since a cache-line contains multiple machine words, \textit{conflict} aborts can occur even if two processes never contend on the same memory location.
%\textit{Capacity} aborts occur when a transaction exhausts some shared resources within the HTM system.
%For example, this can occur if a transaction accesses a large number of primitive objects.
%(In reality, \textit{capacity} aborts also occur for a variety of complex reasons that make it difficult to predict when they will occur.)

Transactional memory allows a programmer to execute arbitrary blocks of code atomically as transactions.
Each transaction either \textit{commits} and appears to take effect instantaneously, or \textit{aborts} and has no effect on shared memory.
The set of memory locations read (resp., written) by a transaction is called its \textit{read-set} (resp., \textit{write-set}).
The \textit{data-set} of a transaction is the union of its read-set and write-set.
If the write-set of a transaction intersects the data-set of another transaction, then the two transactions are said to \textit{conflict}.
When two transactions conflict, one of them must abort to ensure consistency.

Hardware support for transactional memory support has appeared in numerous commercially available processors, including Intel's Haswell, Broadwell and Skylake microarchitectures, and recent IBM POWER, BlueGene/Q and zEC architectures.
This support consists of instructions for starting, committing and aborting transactions, and various other platform specific offerings.
In these HTM implementations, transactions can abort, not only because of conflicts, but also for other spurious reasons.
These HTM implementations are \textit{best-effort}, which means they offer no guarantee that any transaction will ever commit.%
\footnote{Technically, some IBM architectures also offer \textit{constrained transactions}, which guarantee certain types of transactions will not fail spuriously. However, the constraints are highly restrictive. For example, a constrained transaction can contain at most 32 instructions, cannot use loops, and cannot write to more than four different cache lines. There are many additional restrictions.}
We consider Intel's implementation of HTM.

%\trevor{move this later?}
%Since transactions are never guaranteed to commit, in order to guarantee progress, alternative \textit{non-transactional} code must be provided (to be executed in the event that a transaction aborts).
%%We use \textit{fast path} to denote the transactional code path, and \textit{fallback path} to denote the non-transactional code path.
%We say a process is running on the \textit{fast path} (resp., \textit{fallback path}) if it is executing transactional code (resp., non-transactional code).
%%In some cases, the fallback code path is designed 
%Note that a process on the fast path may see inconsistent state if it runs concurrently with a process on the fallback path (since the fallback path is not atomic).
%Thus, a program must either guarantee mutual exclusion between the fast path and fallback path, or carefully design the code for these two paths so it is safe for processes on the fast path to run concurrently with processes on the fallback path.

\paragraph{Linearizability}

All of the implementations discussed in this work are \textit{linearizable}.
Linearizability is a correctness condition introduced by Herlihy and Wing \cite{HW90:toplas}.
A concurrent execution $\alpha$ is linearizable if linearization points can be selected for each completed operation, and for a subset of the operations that started but did not complete, such that the linearization point for an operation occurs during the operation, and the result of each completed operation in $\alpha$ is the same as it would be if the operations were executed atomically at their linearization points.
An algorithm is linearizable if every concurrent execution is linearizable.

\paragraph{Progress conditions}

In this work, we consider lock-free (also called non-blocking) data structures.
Lock-free data structures guarantee that, from every reachable system configuration, some operation will eventually complete even if some processes crash.
However, individual operations may starve.
Note that this rules out the use of locks, since a process that crashes while holding a lock can cause deadlock (potentially blocking all other processes).
In contrast, wait-free data structures offer a stronger guarantee that \textit{every} operation terminates after a finite number of steps (unless the process executing the operation crashes).
This stronger guarantee typically comes at the cost of reduced performance and increased complexity over lock-free algorithms.

\chapter{Lock-free synchronization primitives} \label{chap-scx}
% !TEX root = paper.tex

\begin{thesisnot}
Building a library of
concurrent data structures 
% old list mixed ADTs and data structures: like dictionaries, queues, stacks,  lists, and trees 
% we could list some data structures only:  like stacks, heaps, search trees and hash tables
% but this seems unnecessary. 
% are useful abstractions that
is an essential way to simplify the difficult task of developing concurrent software. 
% for the average programmer. [Deleted this because LLX/SCX is still intended for library designers,
% not average programmer]
There are many lock-based data structures, %in the literature, 
but locks 
%can be a difficult abstraction to work with, and 
%are susceptible to priority inversion, lock-convoys, deadlock, and a lack of fault tolerance.
are not fault-tolerant and are susceptible to %well-known 
problems such as deadlock \cite{Fra07}.
It is often preferable to use hardware synchronization primitives like compare-and-swap 
(\textsc{CAS}) instead of locks.
However, the difficulty of this task has inhibited the development of 
%practical 
{\it non-blocking} data structures.
These are data structures
% IT'S IMPT TO DEFINE THIS, BECAUSE OF THE CONFUSION OF TERMINOLOGY (NON-BLOCKING VS LOCK-FREE)
which guarantee that some operation will eventually complete even if some processes crash.
\end{thesisnot}

\begin{thesisnot}
Our goal is to facilitate the implementation of high-performance, provably correct, non-blocking data structures on any system that supports a hardware \textsc{CAS} instruction.
\end{thesisnot}
\begin{thesisonly}
The goal of this chapter is to facilitate the implementation of high-performance, provably correct, non-blocking data structures on any system that supports a hardware \textsc{CAS} instruction.
\end{thesisonly}
%To do so, 
We introduce three new operations,
\textit{load-link-extended} (\llt), \textit{validate-extended} (\vlt) and \textit{store-conditional-extended} (\sct), which are natural generalizations of the well known \textit{load-link} (\textsc{LL}), \textit{validate} (\textsc{VL}) and \textit{store-conditional} (\textsc{SC}) operations.
We provide a practical implementation of our new operations
from \textsc{CAS}, and give complete proofs of correctness and progress.
%Complete proofs of correctness
%appear in \cite{Brown:2013}.
%Our implementation is proved to be linearizable and 
%satisfy various progress properties.
%guarantees.
%\faith{Do we want to be more specific about progress?}
%\eric{No; too complicated.  I added a mention of progress properties though}
We also show how these operations make the implementation of non-blocking data structures and their proofs of correctness substantially less difficult, as compared to using \textsc{LL},
\textsc{VL}, \textsc{SC}, and \textsc{CAS} directly.
%\faith{How do we show that this? By example?}
%\eric{I think this is already explained well enough in subsequent paragraphs.  This paragraph
% is meant to be an overview of the whole paper}

\llt, \sct\ and \vlt\ operate on {\it \rec s}.
Any number of types of \rec s can be defined, each type containing a fixed number of {\it mutable} fields (which can be updated), and a fixed number of {\it immutable} fields (which cannot).
Each \rec\ can represent a natural unit of a data structure, such as a node of 
a tree or a table entry.
A successful \llt\ operation returns a snapshot of the mutable fields of one \rec. (The immutable fields can be read directly, since they never change.)
\begin{ignore}
(In actuality, an \llt$(r)$ can fail to obtain a snapshot of the mutable fields of $r$, but we defer discussion of this fact for the moment.) However, 
\end{ignore}
An \sct\ operation by a process $p$ is used to atomically store a value in one mutable field of one \rec\ {\it and} {\it finalize} a set of \rec s, meaning that those \rec s cannot undergo any further changes. 
The \sct\ succeeds only if each \rec\ in a specified set has not changed since $p$ last performed an $\llt$ on it.
A successful \vlt\ on a set of \rec s simply assures the caller that each of these \rec s has not changed since the caller last performed an \llt\ on it. 
A more formal specification of the behaviour of these operations is given in Section \ref{sec-operations}.

Early on, researchers recognized that operations accessing multiple locations atomically make the design of non-blocking data structures much easier \cite{Barnes:1993,Israeli:1994,ST97}.
Our new primitives do this in three ways.
First, they operate on \rec s, rather than individual words, to allow the data structure designer to think at a higher level of abstraction.  
Second, and more importantly, a \vlt\ or \sct\ can depend upon multiple \llt s. % instead of just one.
Finally, the effect of an \sct\ can apply to multiple \rec s, modifying one and finalizing others.

The precise specification of our operations was chosen to balance ease of use and efficient implementability.
They are more restricted than multi-word CAS \cite{Israeli:1994}, multi-word RMW \cite{AMTT97}, or transactional memory \cite{ST97}.
% I deleted the following sentence, since it broke the flow of the paragraph and doesn't really say anything interesting.
%The key challenge for implementing all of these primitives,
%including \sct, is to update multiple words simultaneously using only
%single-word CAS.
On the other hand, the ability to finalize \rec s makes \sct\ more general than $k$-compare-single-swap \cite{LMS09}, which can only change one word.
We found that atomically changing one pointer and finalizing a collection of \rec s provides 
just enough power to implement numerous pointer-based data structures in which operations replace a small portion of the data structure.
To demonstrate the usefulness of our new operations, in Chapter~\ref{chap-multiset}, we give an implementation of a simple, linearizable, non-blocking multiset based on an ordered, singly-linked list.
\begin{ignore}
In a companion paper
\cite{paper2},
%, {\em A General Technique for Non-blocking Balanced Search Trees}, 
also submitted to this conference,
we give a general scheme for implementing linearizable, non-blocking
%balanced search
trees using \llt\ and \sct,
and an experimental comparison between a balanced chromatic search tree
built using this scheme
and existing
%dictionary implementations.
data structures for dictionaries.
Proving the correctness of the chromatic tree by directly reasoning about individual
CAS steps (instead of \llt s and \sct s) would be infeasible. 
%\faith{IS THIS DESCRIPTION OF THE EXPERIMENT ACCURATE?
%We could replace it with the following, vaguer, alternative:
%"we give a general scheme for implementing linearizable, non-blocking,
%balanced search trees using \llt, \sct \ and \vlt,
%and present experiments to demonstrate that the resulting
%data structures perform well on real systems."}
%In Sec.~\ref{sec-exp}, experiments are presented to demonstrate these operations can be implemented in such a way that the resulting data structures perform well on real systems.
\end{ignore}

\begin{thesisnot}
Our implementation of \llt, \vlt, and \sct\ is designed for an asynchronous system where processes may crash.
We assume shared memory locations can be accessed by single-word CAS, read and write instructions.
%For simplicity, we 
%defer memory management to future work; instead we 
We assume a safe garbage collector
(as in the Java environment) that will not reallocate a memory location if any process can reach it by following pointers.
This allows records to be reused. 
\trevor{mention the discussion of how to reclaim memory without garbage collection}
\end{thesisnot}

Our implementation has some desirable performance properties.
A \vlt\ on $k$ \rec s only requires reading $k$ words of memory.
If \sct s being performed concurrently depend on \llt s of disjoint sets of \rec s, they all succeed.
If an \sct\ encounters no contention with any other \sct\ and finalizes $f$ \rec s, then a total of $k+1$ CAS steps and $f+2$ writes are used for the \sct\ and the $k$ \llt s on which it depends.
We also prove progress properties that suffice for building non-blocking data structures using \llt\ and \sct.

\begin{ignore}
Our implementation of \sct\ has been optimized for the contention-free case:
If a process performs an \sct\ that depends on $k$ \llt s 
and no other process is concurrently performing an \sct\ that depends on those $k$ \rec s,
then the sequence of $k$ \llt s and one \sct\ requires 
reading the $k$ \rec s and performing $k+1$ CAS steps and $f+2$ writes,
where $f$ is the number of finalized \rec s.
\end{ignore}

\begin{ignore}
Our implementations of \llt, \sct \ and \vlt\ are wait-free, but invocations can fail as a result of subtle interactions with a concurrent \sct.
For this reason, it makes sense to express progress in terms of successful operations.
If operations are performed infinitely often, then operations succeed infinitely often.
We call this \textit{non-blocking} progress.
(Some authors refer to this as \textit{lock-freedom}.)
Furthermore, invocations of \sct\ will succeed infinitely often if a process \textit{sets up} an invocation of \sct\ infinitely often, by performing a sequence of \llt s (subject to some constraints, discussed below) which, if successful, are followed by an invocation of \sct.
(This is important because a process must perform a sequence of successful \llt s before it can \textit{invoke} \sct.)
\end{ignore}

\begin{ignore}
Our implementations of \llt, \sct \ and \vlt \ are linearizable \cite{HW90:toplas}, but we do not linearize any unsuccessful operation.
Is not meaningful to linearize unsuccessful invocations of \llt, since they cannot occur in a sequential execution.
We do not linearize any unsuccessful invocation $I$ of \sct \ or \vlt, because it is possible for $I$ to be unsuccessful as a result of subtle interactions with an invocation of \sct \ that is linearized after $I$ finishes.
Thus, we linearize all successful operations, and make other arguments about unsuccessful operations.
\end{ignore}

%\eric{Can following paragraph be cut to save space?}
%The remainder of this paper is organized as follows.
%In Sec.~\ref{sec-rel}, we discuss related work.
%In Sec.~\ref{sec-operations}, we give precise specifications for
%\llt, \sct\ and \vlt\ and discuss linearizability and
%progress properties that implementations of these operations
%should have.
%The example of a multiset implemented using these operations
%is presented and proved correct in Sec.~\ref{sec-multiset}.
%describe their usage, and give a precise definition of what it means for an implementation of these operations to be correct.
%Sec.~\ref{sec-impl} provides a detailed discussion of our implementation of \llt, \sct\ and \vlt\ using \textsc{READ}, \textsc{WRITE}, and \textsc{CAS},  together with an overview of the proofs of correctness and progress.
%Full proofs appear in the Appendix.
%Experiments are presented in Sec.~\ref{sec-exp} to demonstrate
%that the resulting implementation of a multiset performs well on
%real systems.
%, and Sec.~\ref{sec-data} describes the underlying data structure.
%Sec.~\ref{sec-dotreeupdate} describes the operation \dotreeupdate.
%The pseudocode is discussed in Sec.~\ref{sec-pseudocode}.
%Correctness arguments are made in Sec.~\ref{sec-correctness} (with a full proof deferred to the Appendix).
%In Sec.~\ref{sec-exp}, experiments are presented to demonstrate these operations can be implemented in such a way that the resulting data structures perform well on real systems.
%Concluding remarks and a discussion of future work appears in Sec.~\ref{sec-conclusion}.

\section{Related work} \label{sec-rel}

Transactional memory \cite{HM93,ST97} is a
%another
general approach to simplifying the design of concurrent algorithms by
%allowing
providing atomic access to multiple objects.
It allows
a block of code
%blocks of code to be
designated as a transaction to be executed 
atomically, with respect to other transactions.
Our \llt/\vlt/\sct\ primitives may be viewed as
implementing a restricted kind of transaction, in which 
each transaction can perform any number of reads
followed by a single write and then finalize any number of words.
%\trevor{The $k$-CSS paper makes much ado about saying this very thing.  Since finalizing is essentially what makes us different, we might want to say something about that, perhaps: we can do a single write, and also mark a number of locations so that they cannot be modified by future transactions.}
It is possible to implement general transactional memory in a non-blocking manner (e.g., \cite{Fra07,ST97}).
% Original STM paper by Shavit Touitou DC 1997 is another non-blocking
% Others satisfy weaker progress guarantee (e.g. Herlihy Luchangco Moir Scherer PODC03 is obs-free)
However, at present, 
implementations of transactional memory in software incur significant overhead, and hardware transactional memory (HTM) implementations have significant limitations and are not widely available.
% and handcrafted implementations of individual data structures outperform those
% built by applying transactional memory in a straightforward way.
So, there is still a need for more specialized techniques 
for designing
%that allow the design of
shared data structures
that combine ease of use and efficiency.
%Bronson et~al. introduced transactional predication \cite{Bronson:2010:TPH:1835698.1835703}, which applies STM techniques, in a minimal capacity, to get nonblocking implementations of concurrent sets and maps.
%to extend concurrent sets and maps that are implemented using other
%techniques (e.g., locking or direct application of \cas).
%\trevor{I now wonder if this reference is worth making.}

Most shared-memory systems  provide CAS operations in hardware.
However, LL and SC operations have often been seen
as more convenient primitives for building algorithms.
Anderson and Moir gave the first
wait-free implementation of small LL/SC objects from CAS 
using $O(1)$ steps per operation~\cite{Anderson:1995}.
See \cite{JP05:opodis} for a survey of %numerous 
other implementations that use less space
or handle larger LL/SC objects.    
%Several papers have shown how to implement larger LL/SC objects, each of which can span $w$
%words of memory.
%For example, Doherty et al.\ gave a non-blocking implementation using 
%$O((n+m)w)$ space \cite{DHLM04} and Jayanti
%gave a wait-free implementation that uses $O(w)$ steps per operation
%using $O((n^2+m)w)$ space \cite{JP05:opodis}.
% Others multiword LL/SC implementations I won't mention:
% N=#procs
% Anderson and Moir WDAG95 is less space efficient than Jayanti
% Michael DISC04 has same space complexity as Jayanti, but much slower SCs.
% Jayanti Petrovic ICDCS05 achieves O(w) steps per op but has space complexity O(nmw)
% GFH09:icise gives a simple implementation using O(N+M) space which they claim is non-blocking
% but it seems to use atomic inc and dec operations on the CAS objects too (if those are implemented
% from a CAS in a loop, it's less clear that the result would be non-blocking).  Basic idea
% is to use indirection to point to multiword record.  The only non-trivial stuff is in the memory
% management, which uses the inc and dec operations to update reference counts
%However,
%none of these allow the outcome of an SC operation to depend on multiple LL operations, as in our 
%\llt\ and \sct\ operations.

Many non-blocking implementations of primitives that access multiple objects
use the {\it cooperative technique}, first described by Turek, Shasha and Prakash \cite{TSP92}
and Barnes \cite{Barnes:1993}.
%This is a general way of constructing non-blocking implementations
%of shared data structures.  
Instead of using locks that give a process exclusive access
to a part of the data structure, this approach gives exclusive access to {\it operations}.  If the
process performing an operation that holds a lock is slow, other processes can {\it help}
complete the operation and release the lock.

The cooperative technique was also used recently
for a wait-free universal construction \cite{CER10} and
to obtain non-block\-ing binary
search trees \cite{Ellen:2010} and Patricia tries \cite{Shafiei:2013}.  
The approach used here is similar.

Israeli and Rappoport \cite{Israeli:1994} used a version of the cooperative technique to implement multi-word CAS from single-word CAS (and sketched how this could be used to implement multi-word SC operations).
However, their approach applies single-word CAS to very large words.
%Their implementation assumes CAS objects that can atomically operate on a field and an array of bits (one per process).
%See %Sundell's recent paper 
%\cite{Sun11} for a survey of subsequent implementations.
The most efficient implementation of $k$-word CAS ($k$-CAS) \cite{Sun11}
%performs a $k$-word CAS by first using CAS to replace each word with a pointer to a record containing information about the operation, and then using CAS to replace these pointers with the desired new values.
first uses single-word
CAS to replace each of the $k$ words with a pointer to a \textit{descriptor} record containing information about the operation, and then uses single-word CAS to replace each of these pointers with the desired new value and update a \textit{status} field of the descriptor.
In the absence of contention, this takes $2k+1$ CAS steps.
%%The implementation in \cite{IR94} requires fewer CAS instructions in this case, but it assumes CAS objects that can atomically operate on a field and an array of bits (one per process), which introduces substantial overhead.
%In contrast, in our implementation, an \sct\ that depends on \llt s of $k$ \rec s performs $k+1$ single-word CAS steps when there is no contention, no matter how many words each record contains.
%%Because \sct\ is more restrictive than multi-word CAS,
%%we are able to implement it more efficiently.
%So, our weaker primitives can be significantly more efficient than multi-word CAS or multi-word RMW \cite{AMTT97,Attiya:2011}, which is even more general.
%%For example, deleting a node from a doubly linked  can be done with
%%4-CAS (to change two pointers that point to the node and make
%%sure that the node's two pointers haven't changed.)
%%IS THERE AN IMPLEMENTATION OF THIS TO REFERENCE (Mark, Nir, Tsigas?)

Whenever a \rec\ $r$ is removed from a data structure by a multi-word CAS, care must be taken to ensure that other processes do not concurrently update any field of $r$.
Furthermore, other processes must not update any field of $r$ \textit{after} it is removed.
%One way to prevent \textit{concurrent} updates to fields of $r$ is to have the multi-word CAS that removes $r$ depend on every mutable field of $r$.
%However, this does not prevent changes to $r$ \textit{after} it is removed.
%(Additionally, this approach can require the multi-word CAS to operate on a large number of fields, especially when nodes have many child pointers.)
The standard way to prevent these erroneous changes %prevent changes to $r$ after it is removed 
is to set a \textit{marked} bit in $r$ when it is removed, and to have each multi-word CAS depend on the \textit{marked} bits of the nodes it accesses.
%Then, each multi-word CAS operation verify that the \textit{marked} bit of each \rec\ it accesses is not set.
Consider an operation that atomically marks $k$ \rec s and changes a pointer to remove them from the data structure.
This can be done with a $(k+1)$-CAS, or with an \sct\ that depends on \llt s of $k$ \rec s.
As we saw above, if this $(k+1)$-CAS is implemented using the most efficient multi-word CAS algorithm currently available, it will perform $2k+3$ single-word CAS steps with no contention.
In contrast, in our implementation, an \sct\ will perform only $k+1$ CAS steps with no contention.
So, our weaker primitives can be significantly more efficient than multi-word CAS or multi-word RMW \cite{AMTT97,Attiya:2011}, which is even more general.
%
%
%
%If $k$ \rec s are removed from a data structure by a multi-word CAS,
%then the multi-word CAS must depend on every mutable field
%of these records to prevent another process from concurrently
%updating any of them.
%
%It is possible to use $k$-word CAS to apply to $k$ \rec s instead
%of $k$ words with indirection:
%Every \rec\ is represented by a single word containing a pointer to
%the contents of the record.
%To change any fields of the \rec, a process swings the pointer to a
%new copy of its contents containing the updated values.
%However, the extra level of indirection affects all reads,
%% also slows down reads.
%%this would  add a level of indirection for all reads,
%slowing them down considerably.

\begin{ignore}
%\faith{Trevor agreed to describe something about one such implementation.}
An \sct\ that depends on \llt s of $k$ \rec s could be directly
implemented from a multi-word CAS
%in two ways.  A multiword CAS could be
atomically applied to all words that make up the $k$ \rec s.
%but
However, this would be impractical for large \rec s.
Furthermore, even if each \rec\ contains just a single word and a bit to support finalizing, 
using the implementation described above to perform a $2k$-word CAS would take $4k+1$ CAS instructions when there is no contention.
%so that the multi-word CAS implementation described above only has to operate on $2k$ locations, it would perform $4k+1$ CAS instructions when there is no contention.
In contrast, our implementation performs an \sct\ using only $k+1$ CAS instructions in this case.
%\rec s composed of more than one word.
\faith{The following doesn't make sense, at least without additional
explanation: You would have to require that whenever a field of any
record is changed, a new copy of the entire record is made.
But this is what our implementation does.}
\trevor{In fact, even if you require what Faith suggests above, in order to change a record $x$, you will have to replace every record on the path from $x$ to its entry point.  Thus, it would cripple concurrency.  I added some more stuff, just above, that might be a reasonable replacement for what is written below.}
Alternatively,
a new copy can be made of each \rec\ that is to be
finalized or whose contents are to be updated
and the \sct\ can be performed by swinging pointers to the new copies
%the contents of a \rec\ could be updated by copying the entire record 
%and swinging a pointer to achieve the \sct\
using a $k$-word CAS. However, this would 
add a level of indirection for all reads, slowing them down considerably.
\trevor{Why indirection?}
%\eric{Does this make sense?}
In contrast, our implementation does not require extra indirection, 
and performs an \sct\ using only $k+1$ single-word CAS steps
when there is no contention.
\trevor{Isn't our real advantage that we do fewer CASs than existing multiword CAS implementations, both because we unfreeze atomically with a regular write, and because finalizing does not require CAS?  I believe existing multiword CAS implementations would use at least $2k$ CASs just to replace the $k$ \rec s, then another $2f$ to do the necessary finalizing. %  (for them, doing an \sct\ costs $k$ CASs to ``freeze,'' one to change a value, then $f$ to finalize.  for us, it costs $k$ to freeze, one to change values, and we get finalizing ``for free.''}
}
\end{ignore}

Luchangco, Moir and Shavit \cite{LMS09} defined the $k$-compare-single-swap ($k$-CSS) primitive,
%operation
which atomically tests
wheth\-er $k$ specified memory locations contain specified values and, if all tests succeed,
writes a value to one of the locations.
They provided an {\it obstruction-free} implementation of $k$-CSS,  meaning
that  a process
performing a $k$-CSS is guaranteed to terminate if it runs alone.
%Analogously, 
They implemented $k$-CSS using an obstruction-free implementation of LL/SC
from CAS.
Specifically, to try to update location $v$ using $k$-CSS, a process performs
LL($v$), followed by two collects of the other $k-1$ memory locations.
If $v$ has its specified value,
both collects return their specified values, and the contents of these
memory locations do not change between the two collects, the process
performs SC to change the value of $v$.
Unbounded version numbers are used both in their implementation
of LL/SC and to avoid the ABA problem between the two collects.

Our \llt\ and \sct\ primitives can be viewed as multi-\rec-LL and
single-\rec-SC primitives, with the additional power to finalize \rec s.
%Our primitives can be viewed as $k$-LL-single-SC (which apply to \rec s instead of words).
%The advantage of our primitives is that an \sct\ can also finalize \rec s.
% but the most 
%important difference between our work and theirs is the ability of \sct\ %operations to finalize \rec s when the \sct's update happens.  
We shall see that this extra ability
%is an extremely useful capability
is  extremely useful for implementing 
pointer-based data structures.
In addition, our implementation of \llt\ and \sct\ allows us to develop shared
data structures that satisfy the non-blocking progress condition, which is stronger
than obstruction-freedom.
%$k$-CSS: $2k$ reads, 2 \cas s, 1 write.  
%\sct\ and its linked \llt s: $6|V|+\sum_{r \in V} |r|$ reads (where $|r|$ is the number of mutable 
%fields of $r$), $|V|+1$ \cas s, $2+|V|$ writes.  \llt/\sct\ implementing $k$-CSS: $7k$ reads, 
%$k+1$ \cas s, $k+2$ writes.  \sct\ is ill suited to implement $k$-CSS for two reasons.  First, \llt\ 
%and \sct\ are more powerful than $k$-CSS (finalizing).  Second, \llt\ computes and returns a snapshot.  
%If we only want to do \sct, then we don't need this snapshot, and we can eliminate 
%$|V| + \sum_{r \in V} |r|$ reads.}

\begin{ignore}
FOLLOWING PARAGRAPH GETS TOO OFF-TOPIC, I THINK.

When developing a concurrent data structure, STM can be used to both implement the abstract data type (ADT) operations, and to compose several of these operations into larger atomic transactions.
Bronson et~al. introduced transactional predication \cite{Bronson:2010:TPH:1835698.1835703}, which applies STM techniques, in a minimal capacity, to extend concurrent set and maps that are implemented using other techniques (e.g., locking or direct application of \cas).
The goal is to allow efficient composition of their operations into small transactions, \textit{without} using STM to completely re-implement the individual operations.
Using this technique, they added functionality to several leading concurrent data structures, including the non-blocking, randomized skip-list of the Java Foundation Classes.
The modified implementations were competitive with their original counterparts for single-operation transactions, while outperforming competing data structures that were implemented purely by STM techniques by a wide margin for multi-operation transactions.
Their work underscores the value of fast implementations of concurrent data structures, and provides a way to combine a significant benefit of STM with the raw performance of hand-optimized, non-blocking structures.
\end{ignore}

\section{The primitives} \label{sec-operations}

Our primitives operate on a collection of \rec s of various user-defined types.
%, which we refer to as the data structure.
%As was described in the previous Section, an instance of the data structure is represented by a number of \rec s of various types, with each type having 
Each type of \rec\ has a fixed number of mutable fields (each fitting into a single word), and a fixed number of immutable fields (each of which can be large).  Each field is given a value when the \rec\ is created.  Fields can contain pointers that refer to other \rec s.  
%\eric{I suggest deleting next two sentences}
%We can deal with arrays by creating one \rec\ type for each possible array size.  
%Then, if we would like to create a \rec\ with an array of variable size, we can simply give the 
%\rec\ a pointer to an array.
\rec s are accessed using \llt, \sct\ and \validate,
and reads of individual mutable or immutable fields of a \rec.
Reads of mutable fields are permitted because a snapshot of a \rec's fields is sometimes excessive, 
and it is sometimes sufficient (and more efficient) to use reads instead of \llt s.
%implementations of some data structure operations can be derived by using read
%simple reads
%instead of \llt.

%\eric{Next paragraph is new.  I think it needs to be here to explain why we need to avoid the ABA problem in the multiset example.}

An implementation 
of LL and SC from \cas\ has to ensure that,
between when a process performs LL and when it next performs SC
on the same word, the value of the word has not changed.
Because the value of the word could change and then change back
to a previous value, it is not sufficient to check that the word
has the same value when the LL and the SC are performed.
This is known as the ABA problem.
It also arises for implementations of \llt\ and \sct\ from \cas.
A general technique to overcome this  problem is described
in Section~\ref{sec-impl-aba}.
However, if the data structure designer can guarantee that the 
ABA problem will not arise
(because each \sct\ never attempts to store a value into
a field that previously contained that value),
our implementation can be used in a more efficient manner.

\begin{ignore}
The ABA problem occurs when a process reads the same value in a memory location
twice and assumes that the field has not changed between the  
reads but, meanwhile, it has changed to a different value and then back again.
In Sec.~\ref{sec-impl}, we implement \llt, \sct\ and \vlt\ from \cas, for which 
the ABA problem can cause difficulties.  
Our implementation works most efficiently when the user guarantees that the ABA 
problem will not arise because \sct s on each mutable field 
never attempt to store a value that was previously there.
This is easy to do directly for some algorithms.%non-blocking data structures.
Otherwise, it can be done using the general, but less efficient, technique described in Sec.~\ref{sec-impl-aba}.
\end{ignore}

%\begin{figure}[tb]
%	\centering
%	\includegraphics[scale=0.75]{chap-scx/fig-example-rotation-sequential-draft}
%	\caption{A rotation in a binary search tree.  Numbers (and variables) appearing next to nodes are weights.}
%	\label{fig-rotation-sequential}
%\end{figure}
%
%\begin{figure}[tb]
%	\centering
%	\includegraphics[scale=0.75]{chap-scx/fig-example-rotation-implementation-draft}
%	\caption{Using \sct\ to implement the rotation in Figure~\ref{fig-example-rotation-sequential}.}
%	\label{fig-example-rotation-implementation}
%\end{figure}

Before giving the precise specifications of the behaviour of \llt\ and \sct,
we describe how to use them,
with the implementation of a multiset as a running example.
%This is followed by precise specifications of the behaviour of these
%operations.
The multiset abstract data type supports three operations: \func{Get}$(key)$,
which returns the number of occurrences of $key$ in the multiset, \func{Insert}$(key, count)$, which inserts $count$ occurrences of $key$ into the multiset, and \func{Delete}$(key, count)$, which deletes $count$ occurrences of $key$ from the multiset and returns \true, provided there are at least $count$ occurrences of $key$ in the multiset. Otherwise, it simply returns \false. 
%We assume the keys are drawn from a totally ordered universe.

Suppose we would like to implement a multiset using a sorted, singly-linked list.
%The list has one node for each distinct key in the multiset, and is sorted according to those keys.
We represent each node in the list by a \rec \ with an immutable field $key$, which contains a key in the multiset, and mutable fields: $count$, which records the number of times $key$ appears in the multiset, and $next$, which points to the next node in the list.
The first and last elements of the list are sentinel nodes
with count 0 and
with special keys $-\infty$ and $\infty$, respectively, which never occur in the multiset.

Figure~\ref{fig-example-multiset} shows how updates to the list are handled.  
Insertion behaves differently depending
on whether the key is already present.
Likewise, deletion behaves differently depending
on whether it removes all copies of the key.
%There are two types of insertions, depending
%on whether the key is already present, and two types of deletions, 
%depending on  whether the deletion removes all copies of the key. 
%%FAITH: THE FOLLOWING WAS REMOVED BECAUSE WE ALREADY
%%TOLD THE READER IN THE INTRO THAT THE DESIGN OF THIS
%%DATA STRUCTURE WOULD BE GIVEN IN SECTION 5.
%The multiset implementation will be explained in detail in Section~\ref{sec-multiset}, but for now 
For example,
consider the operation \func{Delete}$(d, 2)$ depicted in Figure~\ref{fig-example-multiset}(c).
This operation removes node $r$ by changing $p.next$ %from $r$ 
to point to a new copy of $rnext$.
A new copy is used to avoid the ABA problem, since $p.next$
may have pointed to $rnext$ in the past.
%(It will become clear why we copy $rnext$ when we discuss the ABA problem in Section \ref{?}.)
%We now describe how a process uses \llt \ and \sct\
%to perform the \func{Delete}$(d,2)$. First, the process.
To perform the \func{Delete}$(d,2)$,
a process first invokes \llt s on $p$, $r$, and $rnext$. 
% (in some order).
Second, it creates a copy $rnext'$ of $rnext$.
Finally, it performs an \sct\ that depends on these three \llt s.  This \sct\ attempts to change $p.next$ to point to $rnext'$.
This \sct \ will succeed only if none of $p$, $r$ or $rnext$ have changed since the aforementioned \llt s.
Once $r$ and $rnext$ are removed from the list, we want subsequent invocations of \llt \ and \sct \ to be able to detect this, so that we can avoid, for example, erroneously inserting a key into a deleted part of the list.
Thus, we specify in our invocation of \sct \ that $r$ and $rnext$ should be \textit{finalized} if the \sct \ succeeds.
Once a \rec\ is finalized, it can never be changed again.

\llt\ takes (a pointer to) a \rec\ $r$ as its argument.  Ordinarily, it returns either a snapshot of $r$'s mutable fields or \finalized. 
%We do allow \llt$(r)$ to fail if it is concurrent with an \sct\ involving $r$; 
%in this case the \llt\ returns \fail.
If an \llt$(r)$ is concurrent with an \sct\ involving $r$, it is also allowed to fail
and return \fail.
\sct\ takes four arguments:
%an ordered set $V$
a sequence $V$ of (pointers to) \rec s upon which the \sct\ depends, 
a subsequence $R$ of $V$ containing (pointers to) the \rec s to be finalized,
a mutable field $fld$ of 
a \rec\ in $V$ to be modified, and a value $new$ to store in this field.
%The \sct\ returns \true\ if it succeeds and \false\ otherwise.
\vlt\ takes a sequence $V$ of (pointers to) \rec s as its only argument.
Each \sct\ and \vlt\ and returns a Boolean value.
%%FAITH CHANGED
 
For example, in Figure~\ref{fig-example-multiset}(c), the \func{Delete}($d,2$) operation invokes \sct($V, R, fld, new$), where $V = \langle p, r, rnext \rangle$, $R = \langle r, rnext\rangle$, $fld$ is the next pointer of $p$, and $new$  points to the %newly created 
node $rnext'$.
%The set $R$ should be $\{r, rnext\}$ because we are removing nodes $r$ and $rnext$ from the list, and we do not want any other operations to modify them after they are removed.

%Thus, for our example in Figure~\ref{fig-example-rotation-implementation}, a process $p$ performs the rotation by invoking \llt\ on $A, B, C, D$, and $E$, in some order, creating \rec s $A', B', C'$, and $D'$, and finally invoking \sct$(V, R, fld, new)$, with arguments $V = \{A, B, C, D, E\}$, $R = \{A, B, C, D\}$, a pointer $fld$ to the child pointer of $E$ that should be changed, and a pointer $new$ to node $D'$.  The set $R$ should be $\{A, B, C, D\}$ because we are removing nodes $A, B, C$, and $D$ from the tree, and we do not want any subsequent operations to modify them after they are removed.\\

A terminating \llt\  is called {\it successful} if it returns a snapshot or \finalized, and {\it unsuccessful} if it returns \fail.
A terminating \sct\ or \vlt\ is called {\it successful} if it returns \true, and {\it unsuccessful} if it returns \false.
Our operations are wait-free, but an operation may not terminate if the process performing
it fails, in which case the operation is neither successful nor unsuccessful.
We say an invocation $I$ of \llt$(r)$ by a process $p$
 is \textit{linked to} an invocation $I'$ of \sct$(V, R, fld, new)$ or \vlt$(V)$ by process $p$ if  $r$ is in $V$,
$I$ %returns a value different from \fail\ or \finalized, 
returns a snapshot, and between $I$ and $I'$, process $p$ performs
no invocation of \llt$(r)$ or \sct$(V', R', fld', new')$
and no unsuccessful invocation of \vlt$(V')$, for any $V'$ that
contains $r$. 
Before invoking \vlt$(V)$ or \sct$(V, R, fld, new)$, a process must
perform an \llt$(r)$ linked to the invocation
%must perform an \llt$(r)$ linked to this operation
for each $r$ in $V$.

%This is called {\it setting up} the \vlt\ or \sct.
%A process $p$ \textit{sets up} an invocation of \sct$(V, R, fld, new)$ by invoking \llt$(r)$ for each 
%$r \in V$, and then invoking \sct$(V, R, fld, new)$ if these invocations of \llt$(r)$ all return values 
%different from \fail\ or \finalized.

\subsection{Correctness properties}
%%FAITH THIS TITLE WAS MOVED FROM 3 PARAGRAPHS EARLIER.
%We now give the correctness specification for \llt, \sct\ and \vlt.
An implementation of \llt, \sct\ and \vlt\ is {\it correct} if,
for every execution, there is a linearization of all successful \llt s,
all successful \sct s, a subset of the non-terminating \sct s,
all successful \vlt s, and all reads, such that the following conditions
are satisfied.
\begin{compactenum}[{\bf C\arabic{enumi}}:]
	\item 
		Each read of a field $f$ of a \rec\ $r$ returns
		the last value stored in $f$ by an \sct\ linearized before the read
		(or $f$'s initial value, if no such \sct\ has modified $f$).
	\item 
%		Each \llt$(r)$ that returns a value different from \fail \ or \finalized \
Each linearized \llt($r$) that does not return \finalized\
		returns	the last value stored in each mutable field $f$ of $r$
		by an \sct\ linearized before the \llt\ (or $f$'s initial value, if no such \sct\ has modified~$f$).
	\item 
		Each linearized \llt$(r)$ returns \finalized\ if and only if it is linearized
		after an %\textit{successful}
		\sct($V, R, fld,$ $new$) with $r$ in $R$.
	\item 
For each linearized invocation $I$ of \sct($V, R,$ $fld, new$) or \vlt$(V)$,
and for each $r$ in $V$, 
no \sct($V'$, $R'$, $fld'$, $new'$) with $r$ in $V'$ is linearized between the
\llt$(r)$ linked to $I$ and $I$.
%\faith{Is this better? If you prefer, you can replace the end of the 
%sentence following ``between the'' with
%linearization point of the \llt$(r)$ linked to $I$ and
%the linearization point of $I$.}
%No \sct$(V', R', fld', new')$ with $r \in V'$ is linearized
%between when an invocation $I$ of \sct($V, R,$ $fld, new$) or \vlt$(V)$
%is linearized and an \llt$(r)$ linked to $I$ is linearized.
%\eric{Is between confusing when it is used "backwards" like this?}
%SHOULD an \llt BE REPLACED BY the \llt?
%there is no linearized \sct$(V', R', fld', new')$ with $r \in V'$
%If an invocation $I$ of \sct$(V, R, fld, new)$ or \vlt$(V)$ is linearized then,
% for all $r \in V$, there has been no \sct$(V', R', fld', new')$
%with $r \in V'$ linearized since the \llt$(r)$ linked to $I$.
\end{compactenum}

The first three properties assert that successful reads and \llt s return
%the
correct answers.
The last property says that an invocation of \sct\ or \vlt\
does not succeed when it should not.
However, an \sct\ can fail if it is concurrent with another \sct\ 
that accesses some \rec\ in common.
LL/SC also exhibits analogous failures in real systems.
\begin{ignore}
However, {\it spurious failures} of \sct\ are allowed.
%it is not stated as an ``if and only if'' condition; in other words,
%we allow ``spurious failures'' of \sct. 
This is similar to how LL/SC behaves in real systems.
\end{ignore}
Our progress properties limit
%Below, we give a progress condition that limits
the situations in which
this
%such spurious failures
can occur.

%Before we specify our implementation's progress guarantee, we make the following definitions.  First, if a process crashes while executing an invocation of \sct, we think of that \sct\ as continuing forever.  However, each \sct\ has a bounded \textit{threat Section} wherein it can cause
%a concurrent \sct\ or \vlt\ to return \false.
%Second, we define the \llt$(r)$ \textit{linked to} an invocation $I$ of \sct\ which is performed by some process $p$ to be the last \textit{successful} \llt$(r)$ performed by $p$ prior to $I$.

\subsection{Progress properties} 
\label{progress-spec}

In our implementation, \llt, \sct\ and \vlt\ are wait-free, but they can fail spuriously.
%\trevor{Is this confusing?  Should we just say ``are wait-free, but they may fail?''}
%%FAITH NO. I THINK IT IS BETTER THE WAY IT IS.
%%THE MEANING IT CONVEYS IS SLIGHTLY DIFFERENT.
%invocations of \llt\ to return \fail, as a result of subtle interactions with a concurrent invocation of \sct.
So, to be able to build non-blocking data structures, we must state our progress properties in terms of {\it successful} \llt, \sct\ and \vlt\ operations.
The first progress property guarantees that 
%Our first (minor) property is fairly trivial to prove; it 
\llt s on finalized \rec s succeed.
\begin{compactenum}[{\bf P\arabic{enumi}}:]
	\item 
        Each terminating \llt$(r)$ returns \finalized\ if it begins after the end of a successful \sct($V, R, fld, new$) with $r$ in $R$ or after another \llt$(r)$ has returned \finalized.
\label{sct-prop-progress-llt}
\end{compactenum}

%\trevor{fix progress properties for queries and updates}
The next progress property guarantees non-blocking progress for queries built from \llt\ and \vlt, and for updates built from \llt\ and \sct.
Recall that \vlt\ and \sct\ can be invoked only after performing a sequence of \llt s that return snapshots.
Specifying a progress guarantee for \sct\ and \vlt\ is subtle, because if processes repeatedly perform \llt\ on \rec s that have been finalized, or repeatedly perform failed \llt s, then they may never be able to invoke \sct\ or \vlt.
In particular, it is \textit{not} sufficient to simply prove that \llt s return snapshots infinitely often, since \textit{all} of the \llt s in a sequence must return snapshots before a process can invoke \sct\ or \vlt.
Additionally, to ensure that changes to a data structure can continue to occur, we make a general assumption that there is always at least one non-finalized \rec.

We give two definitions which are helpful for clearly stating the progress properties for \sct\ and \vlt.
An \sct-\func{Update} algorithm performs \llt s on a sequence $V$ of \rec s and invokes \sct$(V, R, fld, new)$ if all of these \llt s return snapshots.
A successful \sct-\func{Update} is one in which the \sct\ returns \true.
Similarly, a \vlt-\func{Query} algorithm performs \llt s on a sequence $V$ of \rec s and invokes \vlt$(V)$ if all of these \llt s return snapshots.
A successful \vlt-\func{Query} is one in which the \vlt\ returns \true.
\begin{compactenum}[{\bf P\arabic{enumi}}:]
\setcounter{enumi}{1}
\item Suppose that
    (a) there is always some non-finalized \rec\ reachable by following pointers from an entry point, 
    (b) for each \rec\ $r$, each process performs finitely many invocations of \llt$(r)$ that return \finalized, and
    (c) processes perform infinitely many executions of \sct-\func{Update} and/or \vlt-\func{Query} algorithms.
    Then, infinitely many \sct\ or \vlt\ operations succeed.
\label{sct-prop-progress-sct}
%
%\item Suppose, for each \rec\ $r$, each process performs finitely many invocations of \llt$(r)$ that return \finalized. Then, if invocations of \vlt\ (resp., \sct) are \textit{set up} infinitely often, they succeed infinitely often.
\end{compactenum}
One way to prevent processes from performing an infinite number of invocations of \llt$(r)$ on a finalized \rec\ $r$ is to have each process keep track of the \rec s it knows are finalized.
However, in many natural applications, for example, the multiset implementation in Chapter~\ref{chap-multiset}, this kind of explicit bookkeeping can be avoided.

Our implementation of \llt, \sct\ and \vlt\ in Section \ref{sec-impl}
actually satisfies stronger progress properties
than the ones described above.  For example, a \vlt($V$) or
\sct$(V,R,fld,new)$
is guaranteed to succeed if there is no concurrent \sct$(V',R',fld',new')$
such that $V$ and $V'$ have one or more elements in common.
%%FAITH CHANGED
% whose $V$-set \trevor{sequence?} overlaps with $V$.
However, for the purposes of the specification of the primitives, 
we decided to give progress guarantees that are sufficient to prove that algorithms that
use the primitives are non-blocking, but weak enough that it may be possible to design
other, even more efficient implementations of the primitives.  For example, our
specification would allow some spurious failures of the type that occur in common
implementations of ordinary LL/SC operations (as long as there is some 
guarantee that not all operations can fail spuriously).

\begin{ignore}
properly set up, which means
the process must first perform invocations of \llt\ that
return values different from \fail\ or \finalized.
We make a general assumption (which we prove is satisfied by our algorithms) that there is a bound on the number of times that any process will perform an \llt$(r)$ that returns \finalized, for any \rec\ $r$.
Then, provided there still exist some \rec s that are not finalized, 
the first progress property ensures that
eventually \llt s on them will succeed, so it will always be possible to invoke \sct s.
(In a typical data structure implementation from \llt\ and \sct, 
the existence of some non-finalized \rec s will obviously be satisfied.
For example, in our multiset, no node that is reachable by following $next$ pointers
starting from the $head$ is ever finalized.)

%Because of this, we prove numerous additional progress properties.
%Notably, if \sct \ is invoked infinitely often, then invocations of \sct \ succeed infinitely often.
%Then, in order to prove that invocations of \sct \ succeed infinitely often, we need only show that processes can perform sufficient successful invocations of \llt\ to invoke \sct \ infinitely often.
%We address this by proving that, for any \rec \ $r$ and any process $p$, an invocation $S$ of \sct \ can cause at most one invocation $I$ of \llt$(r)$ performed by process $p$ to be unsuccessful, and only if $S$ is concurrent with $I$.
%
%Our implementation of \llt/\sct/\vlt\ offers the following progress guarantee.
\end{ignore}

\section{Implementation of primitives} \label{sec-impl}

%\eric{Faith said she would add some glue here}

%\subsection{Data structure and freezing}

%--Motivate freezing (show problems arise with only CAS, describe similarity to flagging and marking)?

%\newcommand{\wcnarrow}[2]{\parbox{\namewidth}{#1} \com \mbox{#2}}
\begin{figure}[tb]
\def\namewidth{18mm}
\preplisting
\begin{framed}
\begin{lstlisting}[mathescape=true,style=nonumbers]
 type $\rec$
   //\com User-defined fields 
   //\wcnarrow{$m_1, \ldots, m_y$}{mutable fields}
   //\wcnarrow{$i_1, \ldots, i_z$}{immutable fields}
   //\com Fields used by  \llt /\sct\ algorithm
   //\wcnarrow{$\info$}{pointer to an \op}
   //\wcnarrow{$marked$}{Boolean}
\end{lstlisting}
\preplisting
\begin{lstlisting}[mathescape=true,style=nonumbers]
 type $\op$
   //\wcnarrow{$V$}{sequence of \rec s}
   //\wcnarrow{$R$}{subsequence of $V$ to be finalized}
   //\wcnarrow{$fld$}{pointer to a field of a \rec\ in $V$}
   //\wcnarrow{$new$}{value to be written into the field $fld$}
   //\wcnarrow{$old$}{value previously read from the field $fld$} 
   //\wcnarrow{$state$}{one of \{\freezing, \done, \retry\}}
   //\wcnarrow{$\freezingdone$}{Boolean}
   //\wcnarrow{$\llresults$}{sequence of pointers, one read from the}
   //\wcnarrow{\mbox{ }}{\info\ field of each element of $V$}
\end{lstlisting}
\end{framed}
    \vspace{-5mm}
	\caption{Type definitions for shared objects used to implement \llt, \sct, and \vlt.}
	\label{code1}
\end{figure}

%Now that we have introduced the notion of freezing,
%We first describe 
The shared data structure used
to implement \llt, \sct\ and \vlt\ 
%, and describe the %high level behavior of the
%implementation of the operations.
%used to implement the \llt, \sct\ and \vlt\ operations 
consists of a set of \rec s and a set of \op s. (See Figure~\ref{code1}.)  
Each \rec\ contains  user-defined mutable and immutable fields.  It also contains
a 
$marked$ bit, which is used to finalize the \rec,
and an $\info$ field.
The marked bit is initially \false\ and only ever changes from
\false\ to \true.
%%FAITH CHANGED
The $\info$ field points to an \op\ that describes the last \sct\ that accessed the \rec.
Initially, it points to a {\it dummy} \op.
When an \sct\ accesses a \rec, it changes the $\info$ field of the \rec\ to point to its \op.
While this \sct\ is  active, the $\info$ field acts as a kind of lock on the \rec,
granting exclusive access to this \sct, rather than to a process.
(To avoid confusion, we call this {\it freezing}, rather than locking, a \rec.)
We ensure that an \sct\ $S$ does not change a \rec\ for its own purposes while it is 
%already 
frozen for another \sct\ $S'$.
%Before we can describe how \llt, \sct\ and \vlt\ are implemented, 
%we must first introduce the notion of freezing.
%A \rec\ can only be frozen for a single \sct\ at a time, and 
Instead, $S$
uses the information in the \op\ of $S'$ to help  $S'$ complete 
(successfully or unsuccessfully),
so that the \rec\ can be unfrozen. 
This cooperative approach is used to ensure progress.
%To implement freezing in a non-blocking manner, any invocation $S$ of \sct \ that freezes $r$ will 
%change $r.\info$ two point to an \op\ that contains all information about $S$ that a process would 
%need to help it complete (successfully or unsuccessfully).

%Each invocation $S$ of \sct$(V, R, fld, new)$ starts by freezing every $r \in V$ to prevent other 
%invocations of \sct \ from modifying these \rec s while $S$ is working with them.

\begin{figure}[tb]
%%	\begin{minipage}{0.49\textwidth}
        \centering
    	\includegraphics[scale=0.065]{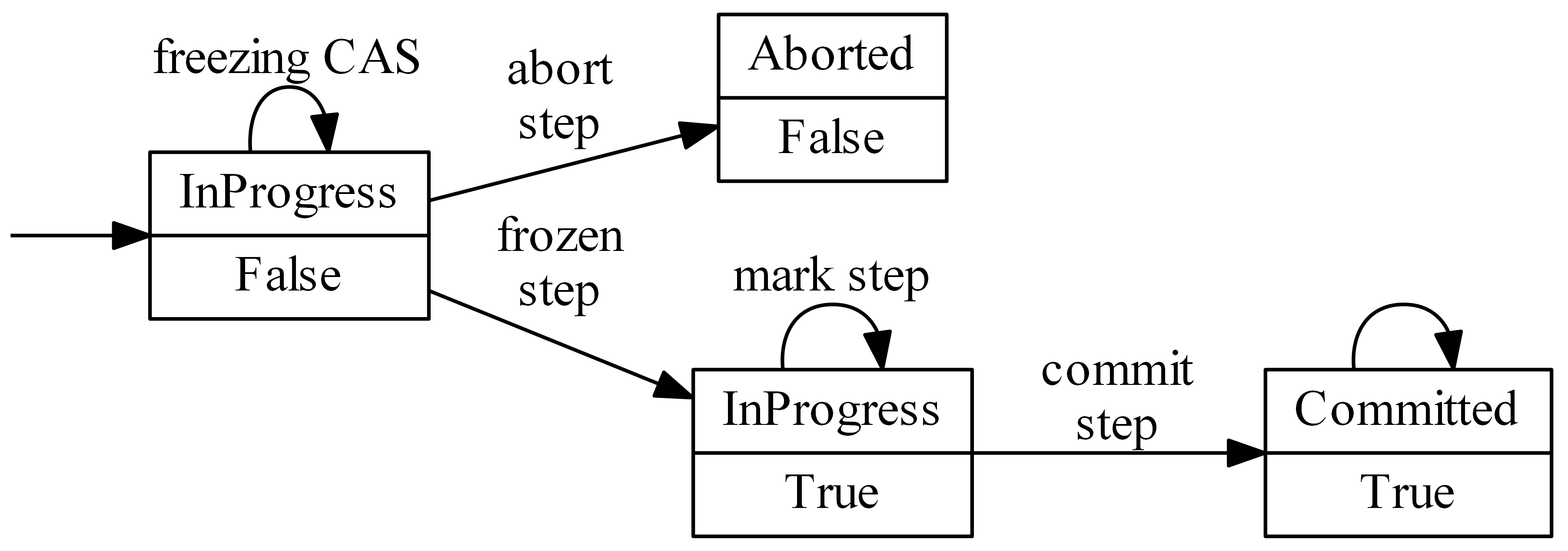} %\\
%\textbf{(a)}
%%    \end{minipage}
%    \hspace{0.02\textwidth}
%	\begin{minipage}{0.49\textwidth}
%        \centering
%    	\includegraphics[scale=0.4]{chap-scx/info_state_transitions3.png} \\
 %  	\textbf{(b)}
%    \end{minipage}
\caption{Possible 
[$state$, $\freezingdone$] field transitions
%transitions for the [$state$, $\freezingdone$] fields
of an \op.}
\label{fig-state-allfrozen-transitions}
\end{figure}

An \op\ contains enough information to allow any process to complete an \sct\ operation 
that is in progress. 
%$V$ is the sequence of \rec s upon which the \sct\ depends.  
%$R$ is the subset of $V$ representing \rec s to be finalized by the \sct.
%The mutable field of a \rec\ in $V$ is stored in $fld$, and $new$ is the new value to be written 
%there by the \sct.
$V, R, fld$ and $new$ store the arguments of the \sct\ operation that created the \op.
Recall that $R$ is a subsequence of $V$ and $fld$ points to a mutable field $f$ of some \rec\ $r'$ in $V$.
%%FAITH: old was not previously defined.
%and $old$  stores 
The value that was read from $f$ by the \llt$(r')$ linked to the \sct\ is 
stored in $old$.
The \op\ has one of three states, \freezing, \done\ or \retry,
which is stored in its $state$ field. This field is
%The $state$ field describes the current state of the \sct\ (\freezing, \done\ or \retry),
initially \freezing.
The \op\ of each \sct\ that terminates is eventually set
to \done\ or \retry, depending on whether or not it successfully
makes its desired update.
%changes.
The dummy \op\ always has $state$ = \retry.
%Its value changes in accordance with the diagram in
%Figure~\ref{fig-state-transitions}.
%Figure~\ref{fig-state-transitions2}(a).
The $\freezingdone$ bit, which is initially \false, gets 
set to \true\ after all \rec s in $V$ have been frozen for the \sct.
The values of $state$ and $\freezingdone$ change in accordance with the diagram in
Figure~\ref{fig-state-allfrozen-transitions}.
%Figure~\ref{fig-state-transitions2}(a).
The steps in the 
%pseudocode
algorithm
that cause these changes are also 
indicated.
%Unsuccessful steps, which have no effect, are not shown.
The $\llresults$ field stores, for each $r$ in $V$, the value of $r$'s \info\ field that was
read by the \llt$(r)$ linked to the \sct.

% Next sentence was already said in the specification section
%In order to prevent \sct s from interfering with one another, we require that $fld$ point 
%to a mutable field $f$ of a \rec\ $r'$ that appears in $V$ (so that an \sct \ has to 
%freeze a \rec \ before changing one of its mutable fields).

%Expanded version of Figure~\ref{fig-state-transitions} showing
%\textbf{(a)} 
%\ (initially [\freezing, \false]) as well as precisely when successful \fcas s and \markstep s occur.
%Only {\it successful} steps (i.e., that have any effect) are shown.

%\begin{figure}[tb]
%\centering
%\mbox{
%	%\hspace{-1.5cm}
%	\includegraphics[scale=0.4]{chap-scx/state_transitions4.png}
%}
%\caption{%Expanded version of Figure~\ref{fig-state-transitions} showing
%All possible transitions for the [$state$, $\freezingdone$] fields of an \op\ (initially [\freezing, \false]) as well as precisely when successful \fcas s and \markstep s occur.  Only \textbf{successful} steps (i.e., that have any effect) are shown.}
%\label{fig-state-transitions2}
%\end{figure}
%
%\begin{figure}[tb]
%\centering
%\includegraphics[scale=0.4]{chap-scx/info_state_transitions3.png}
%\caption{Possible transitions for a \rec\ $r$ between frozen and unfrozen.
%The top half of each node displays whether $r$ is marked.
%The bottom half displays the $state$ of the \op\ pointed to by $r.\info$.
%Black edges represent changes to $r.\info.state$ or $r.marked$.
%Gray edges represent changes to $r.\info$ (i.e., so that it points to a new \op).
%%Nodes with outgoing gray edges represent an unfrozen $r$.  Nodes with no outgoing gray edges represent a frozen $r$.
%Boldness indicates $r$ is frozen.
%}
%\label{fig-info-state-transitions}
%\end{figure}

\begin{figure}[tb]
%	\begin{minipage}{0.49\textwidth}
        \centering
    	\includegraphics[scale=0.065]{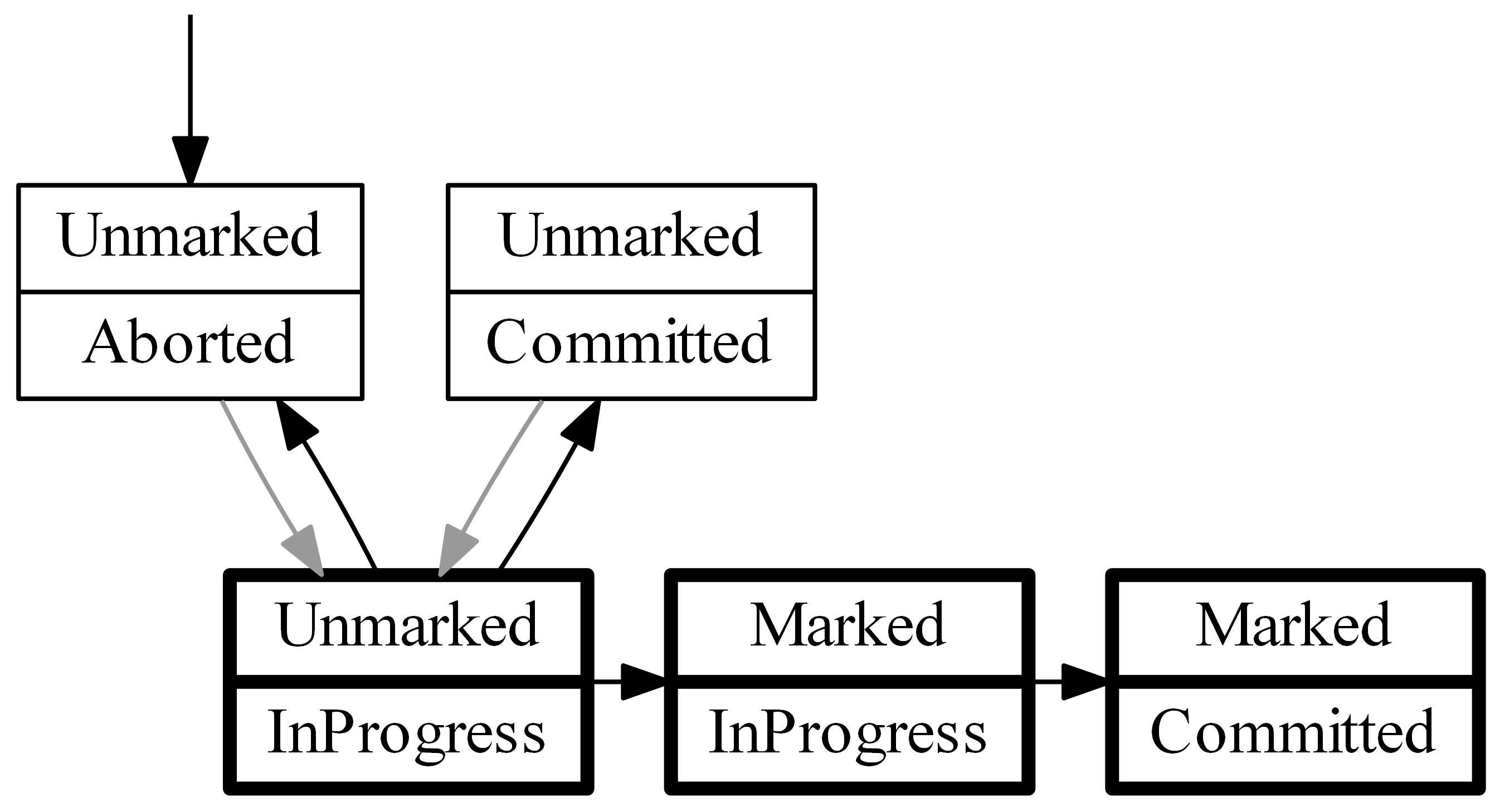} %\\
%    \end{minipage}
\caption{
%\textbf{(b)}
Possible transitions for the
$marked$ field of a \rec\ and the $state$ of the \op\ pointed to by
the {\it info} field of the \rec .}
%a \rec\ $r$ between frozen and unfrozen.
%Each node displays whether $r$ is marked (top) and the $state$ of the \op\ pointed to by $r.\info$ (bottom).
%Black edges represent changes to $r.\info.state$ or $r.marked$.
%Gray edges represent changes to $r.\info$. %(i.e., so that it points to a new \op).
%Boldness indicates $r$ is frozen.}
\label{fig-state-transitions2}
\end{figure}

%We now precisely characterize when a \rec\ is \textit{frozen} or \textit{finalized}.
We say that a \rec\ $r$ is \textit{marked} when $r.marked = \true$.
A \rec\ $r$ is \textit{frozen} for an \op\ $U$ if $r.\info$ points to $U$ and either $U.state$ is \freezing, or $U.state$ is \done\ and $r$ is marked. % (i.e., it is finalized).
%Consequently, a finalized \rec\ that points to an \op\ $U$ is forever frozen for $U$.
While a \rec\ $r$ is frozen for an \op\ $U$, a mutable field $f$ of $r$ can be changed 
only if $f$ is the field pointed to by $U.fld$ (and it can only be changed by a process 
helping the \sct\ that created $U$).
Once a \rec\ $r$ is marked and $r.\info.state$ becomes \done, $r$
 will never be modified again in any way.
%A \rec\ $r$ is \textit{finalized} when $r$ is marked and $r.$\info\ points to an \op\ whose $state$ is \done.
%\eric{This may be slightly confusing because we say earlier that the finalization of nodes happens atomically with the update to $fld$, but here we are saying that it happens at the commit step (which is a bit later).  It doesn't matter, because all records are frozen between the update and the commit anyway, so it probably isn't worth changing.}
%We show that, once a \rec\ is finalized, it is never again modified, in any way.
%(Note that the set $R$ can be characterized as the set of \rec s that should be finalized if the \sct\ succeeds.)
Figure~\ref{fig-state-transitions2} shows how 
a \rec\ can change between frozen and unfrozen.
The three bold boxes represent frozen \rec s.
The other two boxes represent \rec s that are not frozen.
A \rec\ $r$ can only become frozen when $r.\info$ is changed (to point to a new \op\ whose state is \freezing). This is represented by the grey edges.
The black edges represent changes to $r.\info.state$ or $r.marked$.
%Each node in the diagram represents whether a \rec \ $r$ is frozen.
A frozen \rec\ $r$ can only become unfrozen when $r.\info.state$ is changed.

\begin{ignore}
When a \rec\ $r$ is created, its $\info$ field points to a dummy \op\ with $state =$ \retry.
%Since all \rec s frozen by an \op\ $U$ point to $U$, we can atomically unfreeze them when aborting this \sct\ by simply setting $U.state :=$ \retry.
%Similarly, we can atomically unfreeze all unmarked \rec s in $V$, and finalize all \rec s in $R$ when this \sct\ is successful by setting $U.state :=$ \done.
%
Whether a \rec\ $r$ is frozen %(or finalized) 
depends both on whether it is marked, and on the state of the \op\ its $\info$ field points to, %Figure~\ref{fig-state-transitions} (which shows how an \op's $state$ can change) provides only part of the information needed to understand how $r$ changes between unfrozen and frozen.
%Hence,
so it is helpful to visualize in a single figure how $r.\info.state$ and $r.marked$ can change.  We
prove that these changes obey the combined transition diagram in Figure~\ref{fig-state-transitions2}.
Each node in the diagram represents whether a \rec \ $r$ is frozen. %, and contains a pair of values: $r.marked$ (on top), and $r.\info.state$.
%Each black edge represents a change in $r.marked$ or $r.\info.state$.
%Each gray edge represents a change to $r.\info$, writing a pointer to an \op\ (which we will see below has never before appeared in $r.\info$).
%Boldness indicate that $r$ is frozen.
%%A \rec\ in a frozen-state with no outgoing gray edges is frozen.  Conversely, a \rec\ in a frozen-state with an outgoing gray edge is unfrozen.
%%A \rec\ in the frozen-state $\langle T,\done \rangle$ is finalized.
Notably, %we can see from this figure that 
a frozen \rec\ $r$ can only become unfrozen when $r.\info.state$ is changed, and $r$ can only become frozen when $r.\info$ is changed (to point to a new \op\ whose state is \freezing).
%that has never been pointed to by $r.\info$).
\end{ignore}

\subsection{Constraints} 
\label{constraints}
\label{sec-impl-aba}

For the sake of efficiency, we have designed our implementation of \llt, \vlt\ and \sct\ to work
only if the primitives are used in a way that satisfies certain constraints, described in this section.
We also describe general %(but somewhat inefficient) 
ways to ensure these constraints
are satisfied.
However, there are often quite natural ways to ensure
the constraints are satisfied
without resorting to the extra work required by the general solutions.

%By definition, this problem does not arise in algorithms that are implemented with LL/SC, but we must explicitly account for it, since we use \cas\ to implement \llt/\sct.
%In particular, since we use
Since our implementation of \llt, \sct\ and \vlt\ uses helping to guarantee progress, each \cas\ of an \sct\ might be repeatedly performed by several helpers, possibly after the original invocation of \sct\ has terminated.
To avoid difficulties, we must show there is no ABA problem in the fields affected by these CAS steps.

The \info\ field of a \rec\ $r$ is modified by \cas\ steps that attempt to freeze
$r$ for an \sct.
All such steps performed by processes helping one invocation of 
\sct\ try to \cas\ the \info\ field of $r$
from the same old value to the same new value, and that new value is 
a pointer to a newly created \op.
Because the \op\ was freshly allocated for the \sct , %a location that has never been used before,
the ABA problem will not arise in the \info\ field.
%This approach is compatible with any safe memory reclamation scheme that reuses an old address only once no process has (or can obtain) a pointer to it. %can reach it by following pointers.)
% %freezes each $r \in V$ at most once.
 
We must also avoid the ABA problem in mutable fields of \rec s, which are modified using CAS steps by invocations of \sct .
Formally, it suffices to prove the following constraint is satisfied.
\begin{compactitem}
\item
{\bf Constraint}: For every invocation $S$ of \sct$(V, R, fld, new)$, $new$ is not the initial value of $fld$, and no invocation of \sct$(V', R', fld, new)$ was linearized before the $\llt(r)$ linked to $S$ was linearized, where $r$ is the \rec\ that contains $fld$.
\end{compactitem}

This constraint can be satisfied using a similar approach to the one used for \info\ fields. %way that the ABA problem was avoided for \info\ fields.
%A similar approach could be used to avoid the ABA problem in a mutable field of a \rec: the new value %given to the \sct\ could
Specifically, the new value can 
be placed inside a wrapper object that is freshly allocated.
(This is referred to as Solution 3 in~\cite{Dechev2010}, which discusses several general solutions to the ABA problem.)
However, the extra level of indirection slows down accesses to fields.
Observe that, if mutable fields are only ever changed to point to newly allocated \rec s, then the \rec s themselves serve the same function as wrapper objects.
In this case, the constraint is satisfied without the overhead of wrapper objects.
The multiset implementation in Chapter~\ref{chap-multiset} works this way.
%Other general solutions to the ABA problem are discussed in~\cite{Dechev2010}.

%Formally, to avoid the ABA problem, it suffices to prove the following constraint is satisfied.
%\begin{compactitem}
%\item
%{\bf Constraint}: For every invocation $S$
%of  \sct$(V, R$, $fld, new)$,
%$new$ is not the initial value of $fld$ and no invocation of
%\sct$(V', R', fld, new)$  was linearized before the $\llt(r)$ linked to
%$S$ was linearized,  where $r$ is the \rec\ that contains $fld$.
%\end{compactitem}
%%Although this constraint can always be satisfied using the wrapper technique,
%%in many cases, there are simple, more efficient ways to ensure that it
%%this constraint
%%is satisfied.
%%avoid the ABA problem directly,
%The multiset in Chapter~\ref{chap-multiset} provides an example of a simple, more efficient way to ensure that this constraint is always satisfied.
%%and the trees in \cite{paper2} are examples.
%%as in our multiset example in Section~\ref{sec-multiset}.
%%%FAITH CHANGED IT REPEATED MATERIAL OCCURRING 2 PARGS ABOVE

\begin{ignore}
Instead, for the sake of efficiency, we directly prove that the data structures
we implement using \llt\ and \sct\ avoid the ABA problem.
Specifically, 

We  instead put a precondition on \sct\ that will avoid the ABA problem.
Let $f$ be a mutable field of a \rec \ $r$. %and $S$ be an invocation of \sct$(V, R, fld, new)$.
An invocation $S$ of \sct$(V, R, fld, new)$ must ensure that  
$new$ is not the initial value of $fld$, and no invocation of \sct$(V', R', fld, new)$ 
was linearized before the $\llt(r)$ linked to $S$ was linearized.
This constraint can always be satisfied using the wrapper technique mentioned above, but
in many cases there are simple ways to satisfy it directly, as 
in our multiset example in Sec.~\ref{sec-multiset}.
\end{ignore}

To guarantee progress for invocations of \sct, %ensure property P2,
% that states that \sct s eventually succeed.  However,
%in order to do this
we place a constraint on the way \sct\ is used.
Our implementation of \sct$(V, R, fld, new)$ does something similar to locking each \rec\ in $V$.
Livelock could occur if different invocations of \sct \ do not process \rec s in the same order.
%(See Section ~\ref{sec-impl}.)
One way to prevent this is to define a total ordering on all \rec s (for example, ordering them by their locations in memory) and having each invocation of \sct\ sort its $V$ sequence using this ordering. %each sequence
%% $V$ If it is true for V, then it is also true for its subsequence R.
%passed to an invocation of  \sct\ could  be sorted using this ordering. %\trevor{added simply}.
%the sequences of pointers to \rec s $V$ that are passed
%to invocations of \sct\ are
However, this sorting can be expensive.
%%FAITH CHANGED From would to could. If V is short, it is not expensive.
%\trevor{Can we say something like ``this is an expense we can often avoid?''} FAITH: NO. IT REPEATS WHAT WE  SAY THREE SENTENCES LATER.
In fact, to guarantee progress, it is not necessary for {\it all} \sct s to order their $V$ sequences consistently.
%%FAITH CHANGED $V$-sets to sequences $V$
Instead, it suffices to show that, if all the \rec s stop changing, then the sequences passed to later invocations of \sct\ are all consistent with some total order.
%More precisely, 
Formally, use of our implementation of \sct\ requires adherence to the following constraint.
\begin{compactitem}
\item
{\bf Constraint}:
Consider each execution that contains
%FAITH CHANGED any TO each.
%for every execution, if there is
a configuration $C$ after which the value of no
field of any \rec\ changes.
%then
There must be a total order
on all \rec s created during this execution such that,
if \rec\ $r_1$ appears before
\rec\ $r_2$ in the sequence $V$ passed to an invocation 
of \sct\ whose linked \llt s begin after $C$,
then $r_1 < r_2$.
\end{compactitem}
\begin{ignore}
%In order to guarantee \sct s will succeed in our implementation, we must place a constraint on the $V$ sequences that are passed to invocations of \sct.
%This is because, as we shall see in Section~\ref{sec-impl}, an invocation of \sct$(V, R, fld, new)$ does something akin to acquiring locks on each \rec\ in $V$, and livelock could occur if different invocations of \sct \ do not process \rec s in the same order.
%We could easily sidestep this issue by specifying that the \rec s in each $V$ sequence must be ordered according to some total ordering on the universe of all \rec s that are ever created in an execution (e.g., by sorting \rec s by their locations in memory).
However, %this constraint is stronger than
%necessary;
% it needs to be, and we
%we instead impose
 the following weaker constraint suffices.
%can sometimes derive simpler, more efficient algorithms using our primitives
%if they need 
%that 
%only satisfy the following, weaker constraint.
\begin{compactitem}
\item
{\bf Constraint}: Suppose there is a time $T$ after which, for every field of every \rec\ $r$, 
every read or \llt\ of $r$ returns
the same value.  Then, there must exist a partial order of all \rec s such that
if $r_1$ and $r_2$ appear in the $V$-sequence of some \sct\ whose linked \llt s begin
after $T$, then $r_1<r_2$ according to the partial order.
\end{compactitem}
%\trevor{I changed the preceding constraint.  The previous one was insufficient to guarantee progress.}
%Let $S_1$ be an invocation of \sct$(V_1, R_1, fld_1, new_1)$ ending at time $t_1'$, $S_2$ be an invocation of \sct$(V_2, R_2, fld_2, new_2)$ ending at time $t_2'$, $t_1$ be when the earliest \llt \ linked to $S_1$ occurs, and $t_2$ be when the earliest \llt \ linked to $S_2$ occurs.
%If there is no invocation of \sct \ linearized during the interval $[min(t_1, t_2), max(t_1', t_2')]$, then $V_1$ and $V_2$ must be ordered consistently (i.e., any pair of elements that appear in both $V_1$ and $V_2$ must occur in the same order in both sequences).
Although this constraint is quite technical,
%this weaker condition is quite technical, we shall see that
it can simplify the design of algorithms that use our primitives and their proofs of correctness.
Intuitively, it says that the order of $V$-sequences of \sct s must be consistent
only if all the \rec s stop changing.
\end{ignore}
This property is often easy to satisfy in a natural way.
In many \textit{search} data structures (where operations first traverse the data structure and then perform an update), this constraint is satisfied simply by ordering the elements of each $V$ sequence using the order they are encountered during the traversal.
For example, if one were using \llt\ and \sct\ to implement an {\it unsorted} singly-linked list, then this constraint would be satisfied if the nodes in each sequence $V$ occur 
%$V$-sequence are simply listed
in the order they are encountered by following next pointers from the beginning of the list, {\it even if} some operations could reorder the nodes in the list.
While the list is changing, such a sequence may have repeated elements and might not be consistent with any total order.

Although this simple approach satisfies the constraint for many data structures, it does not always work.
For example, consider a doubly-linked list in which some operations traverse left-to-right and some operations traverse right-to-left.
In a static list, a left-to-right traversal can visit \rec s $A$ then $B$, and a right-to-left traversal can visit $B$ then $A$.
In this case, it is not sufficient to order $V$ sequences by the order in which \rec s are visited by traversals.
Instead, an update after a left-to-right traversal could use the order in which it visited \rec s during the traversal, and an update after a right-to-left traversal could use the \textit{reverse} of the order in which it visited \rec s.

\subsection{Detailed algorithm description and sketch of proofs}
% argument}

%\eric{The precondition of LLX (r has been initiated) is neither defined nor explained}

\begin{figure*}[p!]
%\vspace{-2mm}
\small
\def\pwidth{4cm}
\prepnewlisting
%\hrule
\vspace{-2mm}
\begin{framed}
\begin{lstlisting}[mathescape=true]
    //\llt$(r)$ by process $p$
    //\com Precondition: $r \neq \nil$.
      $marked_1 := r.marked$ // \label{ll-read-marked1} \sidecom{order of lines~\ref{ll-read-marked1}--\ref{ll-read-marked2} matters}
      $r\info := r.\info$ // \label{ll-read} 
      $state := r\info.state$ // \label{ll-read-state}
      $marked_2 := r.marked$ // \label{ll-read-marked2}
      if $state = \retry$ or $(state = \done$ and not $marked_2)$ then //  \label{ll-check-frozen} \sidecom{if $r$ was not frozen at line~\ref{ll-read-state}}
        read $r.m_1,...,r.m_y$ //and record the values in local variables $m_1,...,m_y$%
        \label{ll-collect}
        if $r.\info = r\info$ then//\label{ll-reread}\sidecom{if $r.\info$ points to the same} 
          //store $\langle r, r\info, \langle m_1, ..., m_y \rangle \rangle$ in $p$'s local table %
\sidecom{\op\ as on line~\ref{ll-read}}\label{ll-store}
          return $\langle m_1, ..., m_y \rangle$ // \label{ll-return}  \vspace{2mm}%
      
      if $state = \freezing$ then $\help(r\info)$ //\label{ll-help}
      if $marked_1$ then// \label{ll-check-finalized}
        return $\finalized$ // \label{ll-return-finalized}
      else
        return $\fail$ // \label{ll-return-fail} \vspace{2mm} \hrule \vspace{2mm}%

    //\sct$(V, R, fld, new)$ by process $p$
    //\tline{\com Preconditions: (\presctlinked) for each $r$ in $V$, $p$ has performed an invocation $I_r$ of \llt$(r)$ linked to this \sct}%
            {\hspace{19.5mm}(\presctabainit) $new$ is not the initial value of $fld$}%
            {\hspace{19.5mm}(\presctaba) for each $r$ in $V$, no $\sct(V', R', fld, new)$ was linearized before $I_r$ was linearized}
      //\dline{Let $\llresults$ be a pointer to a newly created table in shared memory containing,}%
              {for each $r$ in $V$, a copy of $r$'s \info\ value in $p$'s local table of \llt\ results}%
              \label{sct-create-llresults}
      //Let $old$ be the value for $fld$ stored in $p$'s local table of \llt\ results\label{sct-create-old}
      return $\help(\mbox{pointer to new \op} (V, R, fld, new, old, \freezing,  \false, \llresults ))$ // \label{sct-create-op}\label{sct-call-help} \vspace{2mm} \hrule \vspace{2mm}%

    //\help$(scxPtr)$ 
      //\com \mbox{Freeze all \rec s in $scxPtr.V$ to protect their mutable fields from being changed by other \sct s}
      for each $r$ in $scxPtr.V \mbox{ enumerated in order}$ do//\label{help-fcas-loop-begin}
        //Let $r$\info\ be the pointer indexed by $r$ in $scxPtr.\llresults$ \label{help-rinfo}
        if not $\cas(r.\info,r\info,scxPtr)$ then //\sidecom{\textbf{\fcas}}\label{help-fcas}
          if $r.\info \neq scxPtr$ then // \label{help-check-frozen} 
            //\com \mbox{Could not freeze $r$ because it is frozen for another \sct}
            if $scxPtr.\freezingdone = \true$ then//\sidecom{\textbf{\fcstep}}\label{help-fcstep}
              //\com the \sct\ has already completed successfully 
              return $\true$ // \label{help-return-true-loop} 
            else
              //\com Atomically unfreeze all \rec s frozen for this \sct 
              $scxPtr.state := \retry$ //\sidecom{\textbf{\astep}}\label{help-astep}
              return $\false$ // \label{help-return-false} \vspace{2mm}
      //\com Finished freezing \rec s (Assert: $state \in \{\freezing, \done\}$) 
      $scxPtr.\freezingdone := \true$//\sidecom{\textbf{\fstep}}\label{help-fstep}
      for each $r$ in $scxPtr.R$ do $r.marked := \true$ //\sidecom{\textbf{\markstep}}\label{help-markstep}
      //$\cas(scxPtr.fld, scxPtr.old, scxPtr.new)$ \sidecom{\textbf{\upcas}}\label{help-upcas} \vspace{2mm}
      //\com Finalize all $r$ in $R$, and unfreeze all $r$ in $V$ that are not in $R$ 
      $scxPtr.state := \done$//\sidecom{\textbf{\cstep}}\label{help-cstep}
      return $\true$ // \label{help-return-true} \vspace{2mm} \hrule \vspace{2mm}%

    //\validate$(V)$ by process $p$ 
    //\mbox{\com Precondition: for each \rec\ $r$ in $V$, $p$ has performed an \llt$(r)$ linked to this \vlt}
      for each $r$ in $V$ do
        //Let $r\info$ be the \info\ field for $r$ stored in $p$'s local table of \llt\ results\label{vlt-info} 
        if $r\info \neq r.\info$ then return $\false$ //\tabto{8cm}\mbox{\com $r$ changed since \llt$(r)$ read $\info$}%
        \label{vlt-reread} 
      return $\true$ //\tabto{4.5cm}\mbox{\com At some point during the loop, all $r$ in $V$ were unchanged} %
      %\vspace{2mm} \hrule %
\end{lstlisting}
\end{framed}
    \vspace{-5mm}
	\caption{Pseudocode for \llt, \sct\ and \validate.}
	\label{code-main}
\end{figure*}

Pseudocode for our implementation of \llt, \vlt\ and \sct\ appears in Figure~\ref{code-main}. 
If $x$ contains a pointer to a record, then $x.y := v$
assigns the value $v$ to field $y$ of this record,
\&$x.y$ denotes the address of this field and all other
occurrences of $x.y$ denote the value stored in this field.

\begin{fakethm}
The algorithms in Figure~\ref{code-main} satisfy properties C1 to C4 and P1 to P2 in every execution where the constraints of Section~\ref{constraints} are satisfied.
\end{fakethm}

%The detailed proof of correctness~\cite{techreport1}
%%in Appendix \ref{sec-proof}
%is quite involved, so we only sketch the main ideas here.

%INFO ABOUT HELP MOVED FROM HERE.

An \llt$(r)$ returns a snapshot, \fail, or \finalized.
At a high level, it works as follows.
If the \llt\ determines that $r$ is not frozen and $r$'s $\info$ field does not change
while the \llt\ reads the mutable fields of $r$, 
the \llt\ returns the values read as a snapshot.
Otherwise, the \llt\ helps the \sct\ that it saw froze $r$,
%has frozen $r$ (if any), %to ensure progress,
and returns \fail\ or \finalized.
If the \llt\ returns \fail, it is not linearized.
We now discuss in more detail how \llt\ operates and is linearized in the other two cases.

First, suppose the \llt($r$) returns a snapshot at line~\ref{ll-return}.
Then, the test at line~\ref{ll-check-frozen} evaluates to \true.  So,
either $state=\retry$, which means $r$ is not frozen at line~\ref{ll-read-state},
or $state=\done$ and $marked_2=\false$. This also means $r$ is not frozen at 
line~\ref{ll-read-state}, since $r.marked$ cannot change from \true\ to \false.
%It follows trivially from the definition of frozen that $r$ is {\it not} frozen 
%if and only if either $r.\info.state=\done$ and $r.marked=\false$, or $r.\info.state=\retry$.
%Thus, the test at line~\ref{ll-check-frozen} evaluates to true only if
%$r$ was not frozen at 
%line~\ref{ll-read-state} (since $r.marked$ cannot change from \true\ to \false).
The \llt\ reads $r$'s mutable fields (line~\ref{ll-collect}) and rereads
$r.\info$ at line~\ref{ll-reread}, finding it
the same as on 
%unchanged since
line~\ref{ll-read}.
%\faith{\fcas has not yet been introduced in the text, so the sentence
%"(It is fairly easy to prove that
%each \fcas\ changes $r.\info$ to a pointer to a new \op, so there is no ABA problem in $r.\info$.)" was 
%changed to:}
In Section~\ref{sec-impl-aba}, we
explained 
%\faith{I changed explain to explained, since the explanation occurs earlier.}
why this implies that
%show that this means that
%Line \ref{help-rinfo} is the only place where $r.\info$ is changed
%and it is only changed to point to a new \op. Therefore
$r.\info$ did not change between lines~\ref{ll-read} and \ref{ll-reread}.
Since $r$ is not frozen at line~\ref{ll-read-state}, we know from Figure~\ref{fig-state-transitions2} that $r$ is unfrozen at all times between line~\ref{ll-read-state} and~\ref{ll-reread}.
%The key property we prove for freezing says
We prove that mutable fields can change only while $r$ is frozen,
so the values read by line \ref{ll-collect}
constitute a snapshot of $r$'s mutable fields.  Thus, we can linearize the 
\llt\ at line~\ref{ll-reread}.

Now, suppose the \llt($r$) returns \finalized.  Then, 
the test on line~\ref{ll-check-finalized} evaluated to \true, so $r$ was already marked when line~\ref{ll-read-marked1} was performed (and it will remain marked forever).
%If {\it rinfo.state} = \freezing\ when line~\ref{ll-check-finalized}
%was performed, $\help(r\info)$ was called and returned \true.
Immediately before line~\ref{ll-check-finalized}, the \llt\ performs $\help(r\info)$ if $state = \freezing$.
Below, we argue that either $state = \done$ or this invocation of $\help(r\info)$ will perform a \cstep\ and change $r\info.state$ to \done\ before returning.
%\trevor{check that we actually argue the preceding fact in prose, or that we just mean it's argued in the proof!}
By Figure~\ref{fig-state-transitions2}(a), the $state$ of an \op\ never changes after 
it is set to \done. 
So, after line~\ref{ll-check-finalized}, 
{\it rinfo.state} = \done\ and, thus, $r$ has been finalized.
Hence, the \llt\ can be linearized at line \ref{ll-return-finalized}.
%If that line calls $\help(r\info)$, it returns \true, and we argue below that
%this happens only after the \sct\ that $r\info$ describes has been helped to complete
%successfully (and hence $r\info.state$ is
%has been set to  \done).
%So, by line~\ref{ll-return-finalized}, the \sct\ described by $r\info$ is
%committed and, moreover, it changed $r.marked$ to \true\ prior to line \ref{ll-read-marked1},
%so $r$ has been finalized by that \sct. 
%Thus, we can linearize the \llt\ at line \ref{ll-return-finalized}.

%THIS PARAGRAPH MOVED FROM ABOVE TO BE CLOSER TO THE REST OF SCX.
When a process performs an \sct, it first creates a new \op\ and then invokes \help\ (line~\ref{sct-create-op}).
The \help\ routine performs the real work of the \sct. It is also used by a process to help other processes
complete their \sct s (successfully or unsuccessfully).
%Recall that the values in an \op's $old$ and $\llresults$ fields were read by
%an earlier \llt.
The values in an \op's $old$ and $\llresults$
%For simplicity in the pseudocode, these values 
come from a table in the local memory of the process that invokes the \sct,
%this table
which stores the results of the last \llt\ it performed on each \rec.  (In
practice, the memory required for this table could be greatly reduced
%in the typical case where
when a process knows which of these values
%field values
are needed for future \sct s.
Alternatively, instead of using a process-local table of values, one could have \llt\ return the values it reads, and then simply pass the appropriate values as arguments to invocations of \sct\ and \vlt.)

Consider an invocation of \help($U$) by process $p$ to carry out the work of the invocation 
$S$ of \sct($V, R, fld, new$) that is described by the \op\ $U$.
%When a process $p$ calls \help, it
First, $p$
attempts to freeze each $r$ in $V$ by performing a {\it \fcas} to store a pointer to $U$ in $r.\info$ (line~\ref{help-fcas}).
Process $p$ uses the value read from $r.\info$ by the $\llt(r)$ linked to $S$ 
as the old value for this CAS
and, hence, it will succeed only if $r$ has not been frozen for any other \sct\ since then.
If $p$'s \fcas\ fails, it checks whether some other helper has successfully frozen
the \rec\ with a pointer to $U$ (line~\ref{help-check-frozen}).

If every $r$ in $V$ is successfully frozen, $p$ performs a {\it \fstep} to set $U.\freezingdone$ to 
\true\ (line~\ref{help-fstep}).
After this \fstep, the \sct\ is guaranteed not to fail, meaning that no process
will  perform an \astep\ while
helping this 
\sct.
%MOVED THIS COMMENT TO WHERE IT IS USED.
%(In fact, we prove that if any process helping this \sct\ ever performs a \fstep, 
%the first such \fstep\ will occur before any helper reaches line~\ref{help-$fcstep}.)
%then no helper will enter the if block at line~\ref{help-check-frozen} until after a \fstep, 
%after which every help  entering the if block will return \true\ at line~\ref{help-return-true-loop}.)
Then, for each $r$ in $R$,
$p$ performs a {\it \markstep} to set $r.marked$ to \true\ 
(line~\ref{help-markstep}) and,
from Figure~\ref{fig-state-transitions2},
$r$ remains frozen from then on.
Next, $p$ performs an {\it \upcas}, storing $new$ in the field pointed to by $fld$ (line~\ref{help-upcas}), if successful.
We prove that, among all the \upcas\ steps on $fld$ performed by the helpers of $U$,
only the first can succeed.
Finally, $p$ unfreezes all $r$ in $V$ that are not in $R$ by performing a {\it \cstep} that changes $U.state$ to \done\ (line~\ref{help-cstep}).
%Since each $r \in R$ has already had its $marked$ bit set, after this \cstep, each $r \in R$ will 
%%be finalized, and will 
%remain frozen thereafter.

Now suppose that, when $p$ performs line~\ref{help-check-frozen}, it
finds that some \rec\ $r$ in $V$ is already frozen for another invocation $S'$ of \sct.
If $U.\freezingdone$ is \false\ at line~\ref{help-fcstep},
then we can prove that
%as mentioned above, 
no helper of $S$ will ever reach line~\ref{help-fstep}, so $p$ can
abort $S$.  To do so,
it unfreezes each $r$ in $V$ that it has frozen by performing an {\it \astep}, which changes $U.state$ to \retry\ (line~\ref{help-astep}), and then returns \false\ (line~\ref{help-return-false}) to indicate that $S$ has been aborted.
% we consider two possibilities.
If $U.\freezingdone$ is \true\ at line~\ref{help-fcstep}, 
it means that each element of $V$, including $r$, was
successfully frozen by some helper of $S$ and then, later, a process
froze $r$ for $S'$.
Since $S$ cannot be aborted after $U.\freezingdone$ was set to \true,
its state must have changed from \freezing\ to \done\ before $r$
was frozen for another \op.
Therefore, $S$ was successfully completed and
\begin{ignore}
some helper of $S$ successfully froze all elements of~$V$, including $r$, and then 
a helper of $S'$ later froze $r$ for $S'$.  However, before a process could have frozen $r$ for
$S'$, the \llt($r$) linked to $S'$
must have ensured that $S$ has been helped to completion (either committing it or
aborting it).  As mentioned above, $S$ cannot be aborted if a process sets $U.\freezingdone$ to \true,
so $S$ must have been helped to a successful completion.  Thus,
\end{ignore}
$p$ can return \true\ at line~\ref{help-return-true-loop}.

\begin{ignore}
We now describe how to linearize \sct s.
In Sec.~\ref{sec-impl-aba} we explain how the implementation avoids the ABA problem on
mutable fields of a \rec s.
It follows that only the first \upcas\ on $fld$ executed by the helpers of each
\sct\ can succeed, since all of these CAS steps use the same expected value that was
read from $fld$ by the \llt\ linked to the \sct.
Moreover, we prove that if the first \upcas\ does succeed, then no \rec\ in the
\sct's $V$-sequence has been modified since the linked \llt, and no helper can return \false.
(In particular, the invocation of \help\ called by the \sct\ itself will then
return \true\ too.)
\end{ignore}
%We linearize an invocation of \sct\ if and only if the first \upcas\ performed by one of its helpers succeeds.
%\trevor{We don't actually say that we linearize the \sct\ at the first such \upcas.  Is this okay?}
%This includes all \sct s that return \true, and may include some non-terminating \sct s.
We linearize an invocation of \sct\ at the first \upcas\ performed
by one of its helpers.
% provided it succeeds. 
%Otherwise, it is not linearized.
We prove that this \upcas\ always succeeds.
%In particular, we linearize all \sct s that return \true, and may linearize
%some non-terminating \sct s.
Thus, all \sct s that return \true\ are linearized,
as well as possibly some non-terminating \sct s.
The first \upcas\ of  \sct($V,R,fld,new$) modifies the value of $fld$, so
a read($fld$) that occurs immediately after the \upcas\ will return
the value of $new$.
Hence, the linearization point of an \sct\ must occur at its first \upcas.
%%FAITH CHANGED
There is one subtle issue about this linearization point: 
If an \llt($r$)  is linearized 
between the \upcas\ and \cstep\ of an \sct\ that finalizes $r$,
it might not return \finalized, violating condition C3.
%the second correctness condition would be violated.
%\trevor{Should this be C3, rather than C2? This part seems unclear. I think the point is that, since we allow regular reads, we have to linearize the \sct\ at its first \upcas\ (or else we cannot linearize a read of this field that happens just after this \upcas). Because of this, if an \llt\ were linearized between this \upcas\ and the first \cstep\ (which finalizes the \rec), then this \llt\ would return a snapshot (when it should really return \finalized).}
However, this cannot happen.
Before the \llt\ is linearized on line \ref{ll-return-finalized},
%on line \ref{ll-check-finalized},
the \llt\ either sees $state \neq \freezing$ or helps the \sct.
Since the \llt\ has seen that the node $r$ is marked, and $r$ \textit{cannot} be marked and point to an \op\ with $state$ \retry\ (as Figure~\ref{fig-state-transitions2} shows), the \llt\ will either see $state = \done$, or help the \sct\ perform a \cstep.

\begin{ignore}
Although \rec s are finalized at 
the \upcas, \llt s detect whether to return \finalized\ by checking whether the \cstep\ 
has occurred.  
This could potentially cause a problem if an \llt($r$)  is linearized 
between the \upcas\ and \cstep\ of a \sct\ that finalizes $r$, but we show that this
cannot occur, since the \llt\ would help the \sct\ perform its \cstep\ before the \llt's linearization point.
% Moreover, the finalized record cannot change between the \upcas\ and the \cstep\ since
% it is still frozen during that time.
\end{ignore}

An invocation $I$ of \vlt$(V)$ is executed by a process $p$ after $p$ has performed an invocation of \llt$(r)$ linked to $I$, for each $r$ in $V$.
\vlt$(V)$ simply checks, for each $r$ in $V$, that the \info\ field of $r$ is the same as when it was read by $p$'s last
%successful
\llt$(r)$ and, if so, \vlt$(V)$ returns \true.
%\trevor{The word successful is unnecessary.}
In this case, we prove that each \rec\ in $V$ does not change between the linked \llt\ and the time its \info\ field is reread.  Thus, the \vlt\ can be linearized at the first time it executes  line \ref{vlt-reread}.
Otherwise, the \vlt\ returns \false\ to indicate that the \llt\ results may not constitute a snapshot.

We remark that our use of the cooperative method avoids costly recursive helping.
If, while $p$ is helping $S$, it cannot freeze all of $S$'s \rec s because one of them is already frozen for a third \sct, then $p$ will simply  perform an \astep,  which unfreezes all \rec s that $S$ has frozen.

%As a minor optimization, note that it is pointless to freeze (or finalize) any \rec\ that has no mutable fields (so \help\ can skip %expensive 
%\fcas s on such \rec s).
%\trevor{Note: this last sentence make a claim that I don't think is sound.  Sure, it's pointless to freeze something that has no mutable fields.  However, it might not be pointless to finalize it.  Someone might use whether it is finalized to determine whether it is in the tree.  (Moreover, if you don't finalize some node that you remove, you won't satisfy the constraint in appendix B.)}

\begin{figure*}[tb]
	\centering
    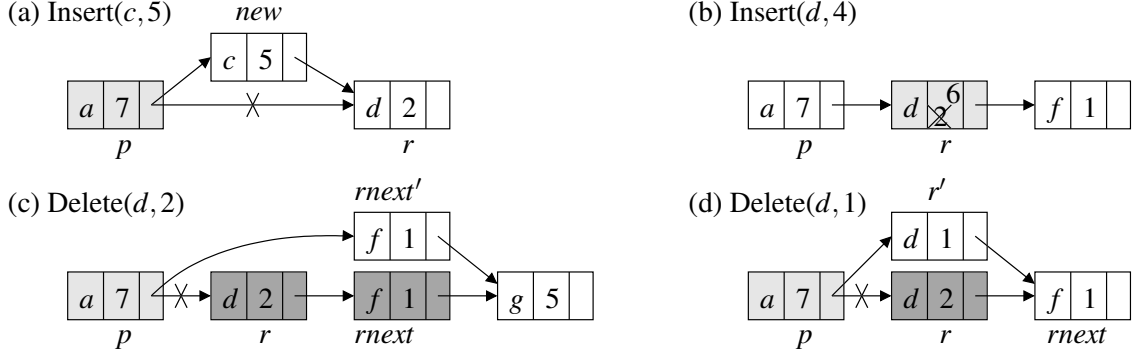
	\caption{Using \sct\ to update a multiset.  \llt s of all shaded nodes are linked to the \sct. Darkly shaded nodes are finalized by the \sct.
Where a field has changed, the old value is crossed out.}
	\label{fig-example-multiset}
\end{figure*}%\eric{If there is time, can shrink fonts in Figure \ref{fig-example-multiset}}

% PROGRESS

%\trevor{edit this proof to account for there only being P1 and P2!}
We briefly sketch why the progress properties 
described in Section~\ref{progress-spec} are satisfied.
It follows easily from the code that
an invocation of \llt($r$) returns \finalized\ 
if it begins after the end of an \sct\ that finalized $r$ or another \llt\ sees that $r$ is finalized.
This establishes P1.
%\rec\ $r$ has been finalized.
%The first progress property follows trivially from the code.%: once a \rec is finalized, 
%\llt s will be able to detect this.
%The \info\ field of a \rec\ is only modified during \help.
%To prove the progress property P2, we consider two cases.

We now sketch the proof of P2.
Consider an execution in which processes execute infinitely many \sct-\func{Update} and/or \vlt-\func{Query} algorithms, but only a finite number succeed.
In this execution, only finitely many \op s are created.
Each process calls \help($U$) if it sees that $U.state = \freezing$, which it can do at most once for each \op\ $U$.
Since every \cas\ is performed inside the \help\ routine, there is some time $t$ after which
no process performs a \cas, calls \help, or sees a \op\ whose $state$ is \freezing.
An \sct\ or \vlt\ can fail only when an \info\ field is modified by a concurrent \sct\ operation, and an \llt\ can only fail for the same reason, or when it sees a \op\ whose $state$ is \freezing.
Therefore, all \llt s, \vlt s and \sct s that begin after $t$ will succeed.
Since processes execute infinitely many \sct-\func{Update} and/or \vlt-\func{Update} algorithms, they must execute infinitely many after $t$.
The first \sct-\func{Update} or \vlt-\func{Update} algorithms to occur after $t$ will succeed, yielding a contradiction.

\begin{ignore}
If there are several \sct s running, the consistency constraint on the
orders of their $V$-sequences ensures that livelock cannot occur:
Assume \sct s stop succeeding at some point. Then, by
the constraint on the sequences passed to invocations of \sct s,
eventually they will
will attempt to freeze nodes in a consistent order, and one of them will
eventually freeze all of its nodes and complete successfully, which is a contradiction.
\end{ignore}

\begin{ignore}
\trevor{Here's a summary of how we prove progress.  I know this is a wall of text...  I just wanted to give you a summary, so you could see everything on one screen, and either strip out most of the detail, or decide we don't need it.}\\
If \sct s are set up infinitely often, then \sct s occur infinitely often.
\begin{compactitem}
\item introduce blaming (unsuccessful guys blame \sct s)
\item we prove that the in-degree of each node in the blame graph is bounded
\item suppose \sct s are set up infinitely often, but \sct s stop happening at time $t$
\item then, only a finite number of \llt s can blame \sct s (and, hence, can be unsuccessful)
\item using this, and our assumption about how many \llt s can return \finalized, we argue that, eventually, every \llt\ returns a snapshot
\item therefore \sct\ is invoked infinitely often (yielding a contradiction)
\end{compactitem}
If \sct s occur infinitely often, then \sct s are linearized infinitely often.
\begin{compactitem}
\item eventually, the data structure stops changing.
\item we first prove that, if $S_1$ blames $S_2$ for $r$, and $S_2$ blames $S_3$ for $r'$, then $r$ precedes $r'$ in the $V$ sequence of $S_2$.
\item we then show that, once there are enough \sct s whose first linked \llt\ starts after the data structure stops changing, there will be a sufficiently long path in the blame graph to imply that $r$ precedes $r$.
\item this contradicts the constraint in section 3, which implies that the $V$ sequences of these \sct s must induce a \textit{strict} partial order.
\end{compactitem}
Non-blocking progress.
\begin{compactitem}
\item suppose \sct s occur infinitely often
\item then, we have already argued they succeed infinitely often
\item otherwise, since there are only finitely many \sct s, and only \sct s can be blamed, and the in-degree of each node in the blame graph is bounded, eventually, every invocation of \llt\ or \vlt\ succeeds.
\end{compactitem}
\trevor{We should probably talk about setting up \vlt s, the same way we do \sct s...  Currently, we haven't proved that you will be able to set up \vlt s, but the proof is just like the proof that you will be able to set up \sct s.}
\end{ignore}

\subsection{Additional properties} \label{sec-additional-properties}

Our implementation of \sct\ satisfies some additional properties, which are helpful
for designing certain kinds of non-blocking data structures so that query operations
can run efficiently.
Consider a pointer-based data structure with a fixed set of \rec s called 
{\it entry points}.
An operation on the data structure starts at an entry point and follows pointers to
visit other \rec s.  (For example, in our multiset example, the head of the linked list
is the sole entry point for the data structure.)
%Thus, 
We say that a \rec\ is {\it in the data structure} if it can be reached by following
pointers from an entry point, and a \rec\ $r$ is {\it removed from the data structure}
by an \sct\ if $r$ is in the data structure immediately prior to the
linearization point of the \sct\ and is not in the data structure immediately afterwards.

If the data structure is designed so that a \rec\ is finalized when (and only when) 
it is removed from the data structure, then we have the following additional properties.
%\begin{prop}
%\label{searches-work}
%Suppose each linearized \sct$(V,R,fld$, $new)$ removes precisely the \rec s in $R$ from the data structure.
%\begin{compactitem}
%\item
%If \llt$(r)$ returns a value different from \fail\ or \finalized, $r$ is in the data structure just before the \llt\ is linearized.
%\item
%If an \sct$(V,R,fld,new)$ is linearized and $new$ is (a pointer to) a \rec, then %$new$
%this \rec\ 
%%%FAITH CHANGED
%is in the data structure just after the \sct\ is linearized.
%\item
%If an operation reaches a \rec\ $r$ by following pointers read from other \rec s, starting from an entry point, then $r$ was in the data structure at some earlier time during the operation.
%\end{compactitem}
%\end{prop}

\begin{fakeprop}
Suppose each linearized \sct$(V,R,fld$, $new)$ removes precisely the \rec s in $R$ from the data structure.
\begin{compactitem}
\item
\textit{If \llt$(r)$ returns a value different from \fail\ or \finalized, $r$ is in the data structure just before the \llt\ is linearized.}
\item
\textit{If an \sct$(V,R,fld,new)$ is linearized and $new$ is (a pointer to) a \rec, then this \rec\ is in the data structure just after the \sct\ is linearized.}
\item
\textit{If an operation reaches a \rec\ $r$ by following pointers read from other \rec s, starting from an entry point, then $r$ was in the data structure at some earlier time during the operation.}
\end{compactitem}
\end{fakeprop}

%\begin{proofsketch}
The first two properties are straightforward to prove.
The last property is proved by induction on the \rec s reached.
For the base case, entry points are always reachable.  For the induction step,
consider the time when an operation reads a pointer to $r$ from another 
\rec\ $r'$ that the operation
reached earlier.  By the induction hypothesis, there was an earlier time $t$ during the operation
when $r'$ was in the
data structure.  If $r'$ already contained a pointer to $r$ at $t$, then $r$ was also
in the data structure at that time.  Otherwise, an \sct\ wrote a pointer to $r$ in $r'$ after $t$,
and just after that update occurred, $r'$ and $r$ were in the data structure 
(by the second part of the proposition).
%\end{proofsketch}

The last property is a particularly useful one for linearizing query operations.  It means that 
operations that search through a data structure can use simple reads of pointers 
instead of the more expensive \llt\ operations.  
Even though the \rec\ that such a search operation reaches
may have been removed from the data structure by the time it is reached, the lemma guarantees
that there {\it was} a time during the search when the \rec\ was in the data structure.
For example, we use this property to linearize searches in our multiset algorithm in Chapter~\ref{chap-multiset}.

\section{Formal proof} \label{sec-proof}

\begin{figure}[p]
\includegraphics[width=\linewidth]{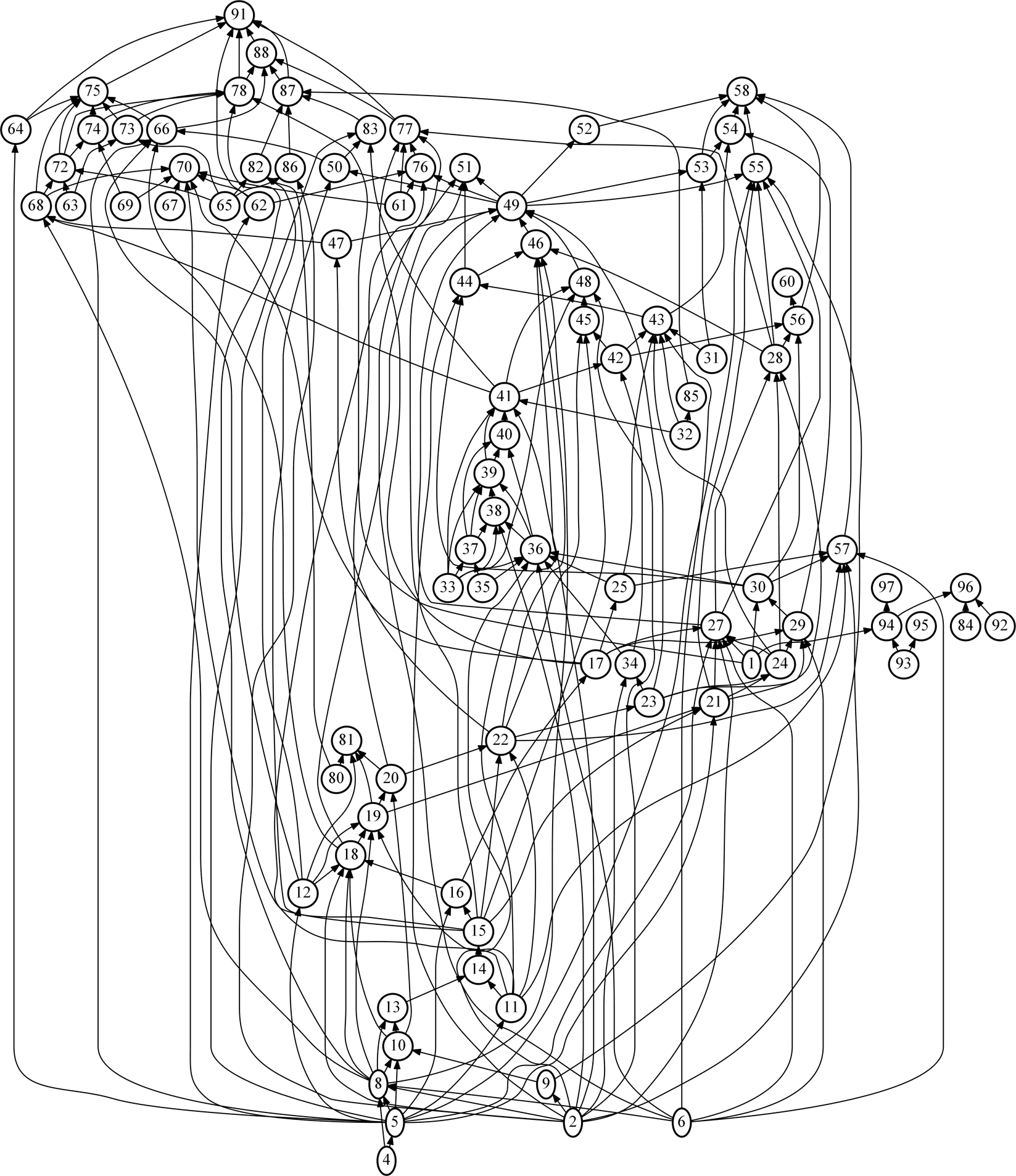}
\caption{Dependency graph for the formal correctness proof.}
\label{fig-scx-proof-dependencies}
\end{figure}

We now give a formal proof of correctness for our implementation of \llt, \sct\ and \vlt.
%The results in this proof are numbered consecutively.
A dependency graph, which appears in Figure~\ref{fig-scx-proof-dependencies}, gives a visual overview of the following results, and how they refer to one another.
An edge is drawn from $A$ to $B$ if $B$ refers to $A$.
To save space, rather than annotating nodes of the graph with ``Lemma 3.60'' and ``Corollary 3.13,'' we simply write ``60'' and ``13.''

\subsection{Basic properties}

We begin with some elementary properties that are needed to prove basic lemmas about freezing.
In particular, we show that the $\info$ field of a \rec\ cannot experience an ABA problem.
%Initially, we prove this claim only for preconditions~\prehelprecurse, \prehelpremoved\ and \prehelpinforead\ of \help.
%The proof of the remaining Precondition~\prehelpinfocontent\ is more difficult, and must be proved together with some properties of the $state$ field of an \op.

\begin{defn} \label{defn-llt-linked-to-sct}
Let $I'$ be an invocation of \sct$(V, R, fld, new)$ or \vlt$(V)$ by a process $p$, and $r$ be a \rec\ in $V$.
We say an invocation $I$ of \llt$(r)$ is \textbf{linked to} $I'$ if and only if:
\begin{enumerate}
\item $I$ returns a value different from \fail\ or \finalized, and
\label{prop-returns-value-different-from-fail-or-finalized}
\item no invocation of \llt$(r)$, \sct$(V', R', fld', new')$, or \vlt$(V')$, where $V'$ contains $r$, is performed by $p$ between $I$ and $I'$.
\label{prop-no-sct-or-vlt-between-linked-llt-and-sct-or-vlt}
\end{enumerate}
\end{defn}

\begin{obs} \label{obs-op-invariants}
An \op\ $U$ created by an invocation $S$ of \sct\ satisfies the following invariants.
\begin{enumerate}
	\item{$U.fld$ points to a mutable field $f$ of a \rec\ $r'$ in $U.V$.}
			\label{inv-fld}
	\item{The value stored in $U.old$ was read at line~\ref{ll-collect}
			from $f$ by the \llt$(r')$ linked to $S$.}
			\label{inv-old}
%	\item{$U.new$ is a value that can be written to the field pointed to by $U.fld$.}
%			\label{inv-new}
%	\item{$U.R \subseteq U.V$} %\{r \in U.R \mid r$ has at least one mutable field$\} \subseteq U.V$}
%			\label{inv-removed}
	\item{For each $r$ in $U.V$, the pointer indexed by $r$ in $U.\llresults$
			was read from $r.\info$ at line~\ref{ll-read}
			by the \llt$(r)$ linked to $S$.}
			\label{inv-llresults}
%	\item{The values stored in $U.\llresults$ were read before
%			the value stored in $U.old$.}
%			\label{inv-read-order}
	\item{For each $r$ in $U.V$,
			the \llt$(r)$ linked to $S$
			must enter the if-block
			at line~\ref{ll-check-frozen},
			and see $r.\info = r\info$
			at line~\ref{ll-reread}.}
			\label{inv-unfrozen}
%	\item{For each $r \in U.V$,
%			the \llt$(r)$ linked to $S$ terminated before the \sct\ begins.}
%			\label{inv-read-age}
\end{enumerate}
\end{obs}
%\eric{Part 4 of the above observation appears to be false.  Why is it needed anyway?  Do you just want to say that the value stored in $U.\llresults$ that corresponds to $r'$ was read before the value stored in $U.old$?  If you want it to be true as stated, you should re-read fld on line 23 instead of looking it up in the table.}
\begin{chapscxproof}
%All of these invariants rely on the fact that the fields of an \op\ do not change after they are initialized.  In the case of sets like $U.V$ and $U.R$, the contents of the sets do not change.  In the case of the table pointed to by $U.\llresults$, its contents do not change, either.
%
%Invariant~\ref{inv-fld} follows immediately from Precondition~\presctfld\ of \sct.
%Invariant~\ref{inv-new} follows immediately from Precondition~\presctnew\ of \sct.
%Invariant~\ref{inv-removed} follows immediately from Precondition~\presctremoved\ of \sct.
%Invariant~\ref{inv-llresults} follows from Precondition~\presctinfo\ of \sct, line~\ref{sct-create-llresults}, and a trivial inspection of \llt.
%Invariant~\ref{inv-old} follows from Precondition~\presctinfo\ of \sct, line~\ref{sct-create-old}, and a trivial inspection of \llt.
%Invariant~\ref{inv-read-order} follows from the same trivial inspection of \llt.
%
%Finally, we prove Invariant~\ref{inv-unfrozen}.
%By the definition of an \llt\ linked to an \sct,
%the \llt$(r)$ linked to $S$ must see
%\isreserved$(r, r\info) = \false$ at line~\ref{ll-check-frozen}
%(or else it will return \fail\ or \finalized),
%and this must occur before $S$ begins.
None of the fields of an \op\ except $state$ change after they are initialized at line~\ref{sct-create-op}.  The contents of the table pointed to by $U.\llresults$ do not change, either.  Therefore, it suffices to show that these invariants hold when $u$ is created.
The proof of these invariants follows immediately from the precondition of \sct, the pseudocode of \llt, and the definition of an \llt\ linked to an \sct.
%\qed
\end{chapscxproof}

The following two definitions ease discussion of the important steps that access shared memory.

\begin{defn} \label{defn-helping}
A process is said to be \textbf{helping} an invocation of \sct\ that created an \op\ $U$
whenever it is executing \help$(ptr)$, where $ptr$ points to $U$.
(For brevity, we sometimes say a process is ``helping $U$'' instead of ``helping the \sct\ that created $U$.'')
\end{defn}

Note that, since \help\ does not call itself directly or indirectly, a process cannot be helping two different invocations of \sct\ at the same time.

\begin{defn} \label{defn-belongs}
We say that a \fcas, \upcas, \fstep, \markstep, \astep, \cstep\ or \fcstep\ $S$ \textbf{belongs} to an \op\ $U$ when $S$ is performed by a process %executing \help$(ptr)$, where $ptr$ points to $U$.
helping $U$.
We say that a \fstep, \markstep, \astep, \cstep\ or \fcstep\ is \textbf{successful} if it changes the the field it modifies to a different value.
A \fcas\ or \upcas\ is successful if the \cas\ succeeds.
Any step is \textbf{unsuccessful} if it is not successful.
%When we use \textit{belongs} in the present tense (without any further qualifiers), as in ``if $p$, then a \fstep\ does not belong to $U$,'' we are making a statement about the entire execution.
\end{defn}

%\eric{I edited the following proof a bit}

\begin{lem} \label{lem-no-steps-belong-to-dummy-op}
No \fcas, \upcas, \fstep, \markstep, \astep, \cstep\ or \fcstep\ belongs to the dummy \op.
\end{lem}
\begin{chapscxproof}
According to Definition~\ref{defn-belongs}, we must simply show that no process ever helps the dummy \op.
An invocation $H$ of \help$(scxPtr)$ is only invoked at line~\ref{sct-call-help} or line~\ref{ll-help}.
If $H$ occurs at line~\ref{sct-call-help}, then $scxPtr$ cannot be the dummy \op, since the argument to $H$ is a newly created \op.
So, suppose $H$ occurs at line~\ref{ll-help} in \llt.
Then the \llt\ sees $state = \freezing$ just before performing $H$.
Thus, $scxPtr$ had $state = \freezing$ at some point before $H$.
Since the dummy \op\ initially has state \retry, and \freezing\ is never written into any $state$ field, $scxPtr$ cannot be the dummy \op.
\end{chapscxproof}

%\eric{I combined two lemmas into the following one}

%\begin{lem} \label{lem-fcas-old-created-before-new}
%Each \fcas\ $F$ uses an old value which points to an \op\ created strictly before the \op\ pointed to 
%by $F$'s new value.
%\end{lem}
%\begin{chapscxproof}
%When a \fcas\ attempts to change an $\info$ field $f$ from $x$ to $y$, $y.\llresults$ contains $x$ (by 
%line~\ref{help-rinfo}).  Then, since $y.\llresults$ does not change after the \op\ pointed to by $y$ is 
%created, the \op\ pointed to by $x$ was created before the \op\ pointed to by $y$.
%\qed
%\end{chapscxproof}

\begin{lem} \label{lem-no-aba-info}
%There is no ABA problem on $\info$ fields.
Every update to the $\info$ field of a \rec\ $r$ changes $r.\info$ to a value that has never previously appeared there. Hence, there is no ABA problem on $\info$ fields.
\end{lem}
\begin{chapscxproof}
We first note that $r.\info$ can only be changed by a \fcas\ at line~\ref{help-fcas}.
When a \fcas\ attempts to change an $\info$ field $f$ from $x$ to $y$, $y.\llresults$ contains $x$ (by 
line~\ref{help-rinfo}).  Then, since $y.\llresults$ does not change after the \op\ pointed to by $y$ is 
created, the \op\ pointed to by $x$ was created before the \op\ pointed to by $y$.
So, letting $a_1, a_2, ...$ be the sequence of \op s ever pointed to by $r.\info$, we know 
%by Lemma~\ref{lem-fcas-old-created-before-new} 
that $a_1, a_2, ...$ were created (at line~\ref{sct-create-op}) in that order.  Since we have assumed memory allocations always receive new addresses, $a_1, a_2, ...$ are distinct.
%\qed
\end{chapscxproof}

\begin{defn}
A \fcas\ \textbf{on} a \rec\ $r$ is one that operates on $r.\info$.
A \markstep\ \textbf{on} a \rec\ $r$ is one that writes to $r.marked$.
\end{defn}

\begin{lem} \label{lem-only-first-fcas-can-succeed}
For each \rec\ $r$ in the $V$ sequence of an \op\ $U$, only the first \fcas\ belonging to $U$ on $r$ can succeed.
\end{lem}
\begin{chapscxproof}
Let $ptr$ be a pointer to $U$, and {\it fcas} be the first \fcas\ belonging to $U$ on $r$.  Let $r\info$ be the old value used by {\it fcas}.  By Definition~\ref{defn-belongs}, the new value used by {\it fcas} is $ptr$. 
Since {\it fcas} belongs to $U$, Lemma~\ref{lem-no-steps-belong-to-dummy-op} implies that $U$ is not the dummy \op\ initially pointed to by each \info\ field.
Hence, $U$ was created by an invocation of \sct, so Observation~\ref{obs-op-invariants}.\ref{inv-llresults} implies that $r.\info$ contained $r\info$ during the $\llt(r)$ linked to $S$.
Since the $\llt(r)$ linked to $S$ terminates before the start of $S$, and $S$ creates $U$, the $\llt(r)$ linked to $S$ must terminate before any invocation of \help$(ptr)$ begins.
From the code of \help, {\it fcas} occurs in an invocation of \help$(ptr)$.
Thus, $r.\info$ contains $r\info$ at some point before {\it fcas}.
If {\it fcas} is successful, then $r.\info$ contains $r\info$ just before {\it fcas}, and $ptr$ just after.
Otherwise, $r.\info$ contains $r\info$ at some point before {\it fcas}, but contains some other value just before {\it fcas}.
In either case, Lemma~\ref{lem-no-aba-info} implies that $r.\info$ can never again contain $r\info$ after {\it fcas}.
Finally, since each \fcas\ belonging to $U$ on $r$ uses $r\info$ as its old value (by line~\ref{help-rinfo} and the fact that table $U.\llresults$ does not change after it is first created), there can be no successful \fcas\ belonging to $U$ on $r$ after {\it fcas}.
\end{chapscxproof}

\subsection{Changes to the \info\ field of a \rec\ and the \textit{state} field of an \op}%Properties of \fcas s, \fstep s, \fcstep s, \astep s and \cstep s}

We prove that freezing of nodes proceeds an orderly way.
The first lemma shows that a process cannot freeze a node that is frozen 
by a different operation that is still in progress.

\begin{lem} \label{lem-no-info-change-while-freezing}
The $\info$ field of a \rec\ $r$ cannot be changed while $r.\info$ points to an \op\ with $state$ \freezing.
\end{lem}
\begin{chapscxproof}
Suppose an $\info$ field of a \rec\ $r$ is changed while it points to an \op\ $U$ with $U.state =$ \freezing.
This change can only be performed by a successful \fcas\ {\it fcas} whose old value is a pointer to $U$ and whose new value is a pointer to $W$.
Let $S$ be the invocation of \sct\ that created $W$.
From line~\ref{help-rinfo}, we can see that the old value for {\it fcas} (a pointer to $U$) is stored in the table $W.\llresults$ and,
by Observation~\ref{obs-op-invariants}.\ref{inv-llresults},
this value was read from $r.\info$ (at line~\ref{ll-read}) by the \llt$(r)$ linked to $S$.
Hence, the \llt$(r)$ linked to $S$ reads $U.state$ at line~\ref{ll-read-state}.
By Observation~\ref{obs-op-invariants}.\ref{inv-unfrozen}, the \llt$(r)$ linked to $S$ passes the test at line~\ref{ll-check-frozen} and enters the if-block.
This implies that, when $U.state$ was read at line~\ref{ll-read-state}, either it was \done\ and $r$ was unmarked, or it was \retry.
Thus, $U.state$ must be \retry\ or \done\ prior to {\it fcas}, and the claim follows from the fact that \freezing\ is never written to $U.state$.
%\qed
\end{chapscxproof}

%\begin{cor} \label{cor-no-info-change-until-first-uass-or-bcas}
%The $\info$ field of a \rec\ $r$ that points to an \op\ $U$ created by an invocation of \sct\ (i.e., not the dummy \op) cannot be changed until the first \cstep\ or \astep\ belonging to $U$.
%\end{cor}
%\begin{chapscxproof}
%From line~\ref{sct-create-op}, we know that $U.state$ is initially \freezing.  Hence, Lemma~\ref{lem-no-info-change-while-freezing} applies, and $r.\info$ does not change until $U.state$ changes.  From the code, $U.state$ is only changed by a \cstep\ or \astep\ belonging to $U$.
%%\qed
%\end{chapscxproof}

It follows from Lemma \ref{lem-no-info-change-while-freezing} 
that if a node is frozen for an operation $O$, it remains frozen for $O$ until
$O$ is committed or aborted.

\begin{lem} \label{lem-if-succ-fcas-then-point-u-until-bcas-or-uass}
If there is a successful \fcas\ {\it fcas} belonging to an \op\ $U$ on a \rec\ $r$, and some time $t$ after the first \fcas\ belonging to $U$ on $r$ and before the first \astep\ or \cstep\ belonging to $U$, then $r.\info$ points to $U$ at $t$.
\end{lem}
\begin{chapscxproof}
%Suppose there is a successful \fcas\ $succ$ belonging to $U$ on $r$. 
Since {\it fcas} belongs to $U$, by Lemma~\ref{lem-no-steps-belong-to-dummy-op}, $U$ cannot be the dummy \op, so $U$ is created at line~\ref{sct-create-op}, where $U.state$ is initially set to \freezing.
Let $t'$ be when the first \astep\ or \cstep\ belonging to $U$ on $r$ occurs.
By Lemma~\ref{lem-only-first-fcas-can-succeed}, {\it fcas} must be the first \fcas\ belonging to $U$ on $r$.
Thus, $t$ is after {\it fcas} occurs, and before $t'$.
Immediately following {\it fcas}, $r.\info$ points to $U$.
From the code, $U.state$ can only be changed by an \astep\ or \cstep\ belonging to $U$.
Therefore, $U.state =$ \freezing\ at all times after {\it fcas} and before $t'$.
By Lemma~\ref{lem-no-info-change-while-freezing}, $r.\info$ cannot change after {\it fcas}, and before $t'$.
Hence, $r.\info$ points to $U$ at $t$.
%\qed
\end{chapscxproof}

%\begin{figure}[h!]
%\centering
%\includegraphics[scale=1]{chap-scx/lem-first-fass-before-first-fcstep-edited.png}
%\caption{diagram to accompany Lemma~\ref{lem-first-fass-before-first-fcstep}.}
%\label{fig-lem-first-fass-before-first-fcstep}
%\end{figure}

The following result proves that a \fstep\ occurs only after all \rec s are successfully frozen.

\begin{lem} \label{lem-if-fass-then-all-succ-fcas}
If a \fstep\ belongs to an \op\ $U$ then, for each $r$ in $U.V$, there is a successful \fcas\ belonging to $U$ on $r$ that occurs before the first \fstep\ belonging to $U$.
\end{lem}
\begin{chapscxproof}
Suppose a \fstep\ belongs to $U$.  Let \textit{fstep} be the first such \fstep\ and let $H$ be the invocation of \help\ that performs \textit{fstep}.
Since \textit{fstep} occurs at line~\ref{help-fstep}, for each \rec\ $r$ in $U.V$, $H$ must perform a successful \fcas\ belonging to $U$ on $r$ or see $otherPtr = scxPtr$ in the preceding loop.
If $H$ performs a successful \fcas\ belonging to $U$ on $r$, then we are done.
Otherwise, $r.\info = scxPtr$ at some point before \textit{fstep}.
Since \textit{fstep} belongs to $U$, $scxPtr$ points to $U$.
From Lemma~\ref{lem-no-steps-belong-to-dummy-op}, $U$ is not the dummy \op\ to which $r.\info$ initially points.
Hence, some process must have changed $r.\info$ to point to $U$, which can only be done by a successful \fcas.
%\qed
\end{chapscxproof}

Finally, we show that \rec s are frozen in the correct order.
(This is used below to prove that livelock cannot occur.)

%\eric{I edited the following proof a bit.}

\begin{lem} \label{lem-fcas-on-ri-only-after-succ-fcas-on-previous}
Let $U$ be an \op, and $\langle r_1, r_2, ..., r_l\rangle$ be the sequence of \rec s in $U.V$.
For $i \geq 2$, a \fcas\ belonging to $U$ on \rec\ $r_i$
can occur only after a successful \fcas\ belonging to $U$ on $r_{i-1}$.
\end{lem}
\begin{chapscxproof}
Let {\it fcas} be a \fcas\ belonging to $U$ on $r_i$, for some $i \geq 2$.
Let $H$ be the invocation of \help\ which performs {\it fcas}.
%Then, throughout $H$, $scxPtr$ points to $U$.
The loop in $H$ iterates over the sequence
$r_1, r_2, ..., r_l$, so 
$H$ performs {\it fcas} in iteration $i$ of the loop.
Since $H$ reaches iteration $i$, $H$ must perform iteration $i-1$.
Thus, by the code of \help, $H$ must perform a \fcas\ {\it fcas$'$} belonging to $U$ on $r_{i-1}$ at line~\ref{help-fcas} before {\it fcas}.
%Since no successful \fcas\ belonging to $U$ on $r_{i-1}$
%occurs before {\it fcas}, {\it fcas$'$} must fail.
If {\it fcas$'$} succeeds, then the claim is proved.
Otherwise, $H$
%must subsequently read $r_{i-1}.\info$ at line~\ref{read-otherflag} and store the value it read in local variable $otherPtr$, after which it
will check whether $r_{i-1}.\info$ is equal to $scxPtr$ at line~\ref{help-check-frozen}.
Since \help\ does not return in iteration $i-1$, $r_{i-1}.\info = scxPtr$.
This can only be true if $U$ is the dummy \op, or there has already been a successful \fcas \ belonging to $U$ on $r_{i-1}$.
Since {\it fcas} belongs to $U$, Lemma~\ref{lem-no-steps-belong-to-dummy-op} implies that $U$ cannot not be the dummy \op.
%Hence, a successful \fcas\ belonging to $U$ on $r_{i-1}$ has occurred.
%\qed
\end{chapscxproof}

\subsection{Proving $state$ and \textit{info} fields change as described in Fig.~\ref{fig-state-transitions2}}

\begin{figure}[tb]
\centering
\mbox{
	\hspace{-1.5cm}
	\includegraphics[scale=0.5]{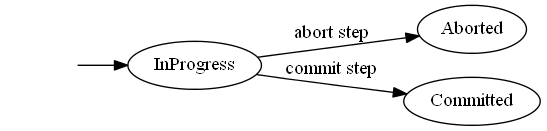}
}
\caption{Possible transitions for the $state$ field of an \op\ (initially \freezing).}
\label{fig-state-transitions}
\end{figure}

In this section, we first prove that an \op's $state$ transitions respect Figure~\ref{fig-state-transitions}, then we expand upon the information in Figure~\ref{fig-state-transitions} by showing that \fstep s, \astep s, \cstep s and successful \fcas s proceed as illustrated in Figure~\ref{fig-state-transitions2}.

We now prove an \op\ $U$'s $state$ transitions respect Figure~\ref{fig-state-transitions} by noting that a $U.state$ is never changed to \freezing, and proving that it does not change from \retry\ to \done\ or vice versa.
Since $U.state$ can only be changed by a \cstep\ or \astep\ belonging to $U$, and each \cstep\ is preceded by a \fstep\ (by the code of \help), it suffices to show that there cannot be both a \fstep\ and an \astep\ belonging to an \op.
%While proving this claim, we also prove the following supporting lemmas concerning \fcas s.
%\begin{itemize}
%	\item{If there is a successful \fcas\ belonging to $U$ on each \rec\ in $U.V$, then a \fstep\ and a 
%\cstep\ belonging to $U$ occur prior to the termination of any instance of \help$(ptr)$ with $ptr$ a 
%pointer to $U$.}
%\end{itemize}

%$\bigstar$
%The next five lemmas show that there cannot be both an \astep\ and a \cstep\ belonging to the same \op.
%% Note that a \fcstep\ belonging to an \op\ $U$ can occur even if a \fstep\ does not belong to $U$, but only if there is some $r \in U.V$ on which there is no successful \fcas\ belonging to $U$.

\begin{lem} \label{lem-if-all-succ-fcas-then-no-fcstep-until-fass}
Let $U$ be an \op.
Suppose that there is a successful \fcas\ belonging to $U$ on each \rec\ in $U.V$.
Then, a \fcstep\ belonging to $U$ cannot occur until after a \fstep\ belonging to $U$ has occurred.
\end{lem}
\begin{chapscxproof}
To derive a contradiction, suppose that the first \fcstep\ {\it fcstep} belonging to $U$ occurs before any \fstep\ belonging to $U$.
Let $H$ be the invocation of \help\ in which {\it fcstep} occurs.
%Suppose a \fcstep\ belongs to $U$.
%Let {\it fcstep} be the first such \fcstep, and let $H$ be the invocation of \help\ in which {\it fcstep} occurs.
%
%We show that a \fstep\ must precede {\it fcstep}.
Before {\it fcstep}, $H$ performs an unsuccessful \fcas\ at line~\ref{help-fcas} on one \rec\ in $U.V$.
The hypothesis of the lemma says that there is a successful \fcas, {\it fcas}, belonging to $U$ on this same \rec.
By Lemma~\ref{lem-only-first-fcas-can-succeed}, {\it fcas} must occur before $H$'s unsuccessful \fcas.
Thus, {\it fcas} occurs before {\it fcstep}, which occurs before any \fstep\ belonging to $U$.
From the code of \help, a \fstep\ belonging to $U$ precedes the first \cstep\ belonging to $U$, which implies that {\it fcas} and {\it fcstep} occur before the first \cstep\ belonging to $U$.
Further, the code of \help\ implies that no \astep\ can occur before {\it fcstep}.
Thus, {\it fcas} and {\it fcstep} occur strictly before the first \cstep\ or \astep\ belonging to $U$.
After {\it fcas}, and before {\it fcstep}, $H$ sees $r.\info \neq scxPtr$ at line~\ref{help-check-frozen}.
Since $scxPtr$ points to $U$, this implies that $r.\info$ points to some \op\ different from $U$.
However, Lemma~\ref{lem-if-succ-fcas-then-point-u-until-bcas-or-uass} implies that $r.\info$ must point to $U$ when $H$ performs line~\ref{help-check-frozen}, which is a contradiction.
%
%If no \cstep\ or \astep\ belonging to $U$ occurs before {\it fcstep} (which occurs after {\it fcas}), then Lemma~\ref{lem-if-succ-fcas-then-point-u-until-bcas-or-uass} implies that $r.\info$ points to $U$ at line~\ref{help-check-frozen}, which contradicts.
%
%From the code
%
%By Lemma~\ref{lem-if-succ-fcas-then-point-u-until-bcas-or-uass}, $r.\info$ points to $U$ from {\it fcas} until the first \cstep\ or \astep\ belonging to $U$.
%Thus, a \cstep\ or \astep\ belonging to $U$ occurs before line~\ref{help-check-frozen}.
%However, from the code, we can see that no \astep\ belonging to $U$ can occur before the first \fcstep\ belonging to $U$.
%This implies that a \cstep\ belonging to $U$ must occur before {\it fcstep}.
%Finally, a \cstep\ is preceded by a \fstep\ in the code, so a \fstep\ belonging to $U$ must occur before {\it fcstep}.
\end{chapscxproof}

%\begin{cor} \label{cor-if-succ-fcass-then-fass}
%If some invocation $H$ of \help$(ptr)$ where $ptr$ points to an \op\ $U$ terminates, and a successful \fcas\ on $r$ belongs to $U$ for each $r \in U.V$, then a \fstep\ belongs to $U$.
%\end{cor}
%\begin{chapscxproof}
%Suppose an invocation $H$ of \help$(ptr)$ terminates, and that a successful \fcas\ on $r$ belongs to $U$ for each $r \in U.V$.  $H$ can only return at line~\ref{help-return-true-loop}, \ref{help-return-false} or \ref{help-return-true}.
%If $H$ returns at line~\ref{help-return-true-loop} or \ref{help-return-false}, then a \fcstep\ belongs to $U$.  However, by Lemma~\ref{lem-if-all-succ-fcas-then-no-fcstep-until-fass}, such a \fcstep\ must be preceded by a \fstep\ belonging to $U$.
%Otherwise, $H$ returns at line~\ref{help-return-true}, after performing a \fstep\ (by the pseudocode of \help).
%%\qed
%\end{chapscxproof}

\begin{cor} \label{cor-first-fass-before-first-fcstep}
If a \fstep\ belongs to an \op\ $U$, then the first such \fstep\ must occur before any \fcstep\ belonging to $U$.
\end{cor}
\begin{chapscxproof}
Suppose a \fstep\ belongs to $U$.  By Lemma~\ref{lem-if-fass-then-all-succ-fcas}, we know there is a successful \fcas\ belonging to $U$ on $r$ for each $r$ in $U.V$.  Thus, by Lemma~\ref{lem-if-all-succ-fcas-then-no-fcstep-until-fass},
a \fcstep\ belonging to $U$ cannot occur until a \fstep\ belonging to $U$ has occurred.
%\qed
\end{chapscxproof}

\begin{lem} \label{lem-fass-then-no-bcas}
There cannot be both a \fstep\ and an \astep\ belonging to the same \op.
%If a \fstep\ belongs to an \op\ $U$, then no \astep\ belongs to $U$.
\end{lem}
\begin{chapscxproof}
Suppose a \fstep\ belongs to $U$.  By
Corollary~\ref{cor-first-fass-before-first-fcstep},
the first such \fstep\ precedes the first \fcstep,
which, 
by the pseudocode of \help,
precedes the first \astep. 
This \fstep\ sets
$U.\freezingdone$
to \true\ and
$U.\freezingdone$
is never changed from \true\ to \false.
Therefore, any process that performs a \fcstep\ belonging to $U$
will immediately return \true, without performing an \astep.
Thus, there can be no \astep\ belonging to $U$.
%\qed
\end{chapscxproof}

\begin{cor} \label{cor-no-change-from-done-retry}
An \op\ $U$'s $state$ cannot change from \done\ to \retry\ or from \retry\ to \done.
\end{cor}
\begin{chapscxproof}
Suppose $U.state =$ \done.
Then a \cstep\ belonging to $U$ must have occurred.
Since each \cstep\ is preceded by a \fstep,
Lemma~\ref{lem-fass-then-no-bcas} implies that no \astep\ belongs to $U$.
Thus, $U.state$ can never be set to \retry.

Now suppose that $U.state =$ \retry. Then either $U$ is the dummy \op\ or an \astep\ belongs to $U$.  If $U$ is the dummy \op\ then, by Lemma~\ref{lem-no-steps-belong-to-dummy-op}, no \cstep\ belongs to $U$,
so $U.state$ never changes to \done.
Otherwise, $U.state$ is initially \freezing, so there must have been an \astep\ belonging to $U$.  Hence, by Lemma~\ref{lem-fass-then-no-bcas}, no \fstep\ can belong to $U$.
If there is no \fstep\ belonging to $U$,
then there can be no \cstep\ belonging to $U$ (by the pseudocode of \help).
Therefore $U.state$ can never be set to \done.
%\qed
\end{chapscxproof}

\begin{cor} \label{cor-state-transitions-respect-figure}
The changes to the $state$ field of an \op\ respect Figure~\ref{fig-state-transitions}.
\end{cor}
\begin{chapscxproof}
Immediate from Corollary~\ref{cor-no-change-from-done-retry} and the fact that an \op's state field cannot change to \freezing\ from any other state.
%\qed
\end{chapscxproof}

The next five lemmas prove that any successful \fcas s belonging to an \op\ $U$ must occur prior to the first \fstep\ or \astep\ belonging to $U$.  This result allows us to fill in the gaps between Figure~\ref{fig-state-transitions} and Figure~\ref{fig-state-transitions2}(a).%  These lemmas will also and will eventually be used in proving progress arguments.
%\begin{itemize}
%	\item{}
%\end{itemize}

%Before we proceed to prove properties of \upcas s, we first demonstrate that every successful \fcas\ belonging to an \op\ $U$ must occur before the first \fstep\ or \astep\ belonging to $U$.  Ultimately, this will allow us to prove that any successful \fcas\ belonging to $U$ must occur during the invocation of \sct\ that created $U$.

%\eric{I edited the following proof a little (especially claim 4).}

%\eric{I think there is a slight problem with the following proof that should be fixed after Feb 10.  (See .tex file for details)}

\after{This next lemma seems to be not quite true if V seq can have duplicates.  I think the way to fix it is to define things slightly differently.  Instead of talking about $fcas on r_i$ we should talk about the $fcas of U for position i$ (meaning an fcas that occurs during the ith iteration of the loop) or some such.  This will probably have an effect on how we blame for failed SCXs when proving progress.  Eric}

\begin{lem} \label{lem-abort-fcas-flow}
Let $U$ be an \op, and let $\langle r_1, r_2, ..., r_l\rangle$
be the sequence of \rec s in $U.V$. % enumerated in order.
Suppose an \astep\ belongs to $U$ and let $astep$ be the first such \astep.
Then, there is a
%unique
%\rec\ $r_k \in \{r_1, ..., r_l\}$ such that
$k \in \{1, \ldots, l\}$ such that
\begin{enumerate}[1.]
	\item{a \fcas\ belonging to $U$ on $r_k$ occurs prior to $astep$,}
	\label{abort-fcas-flow-claim-fcas-on-rk}
	\item{there is no successful \fcas\ belonging to $U$ on $r_k$,}
\label{abort-fcas-flow-claim-no-succ-fcas-on-rk}
	\item{for each $i \in \{1, ..., {k-1}\}$,
a successful \fcas\ belonging to $U$ on $r_i$ occurs prior to $astep$,
and no successful \fcas\ belonging to $U$ on $r_i$ occurs after $astep$, and}
	\label{abort-fcas-flow-claim-succ-fcas-on-r1-etc}
	\item{$r_k.\info$ changes after the \llt$(r_k)$ linked to $S$ reads $r_k.\info$ at line~\ref{ll-reread} and before the first \fcas\ belonging to $U$ on $r_k$.
	} \label{abort-fcas-flow-claim-rk-changes}
\end{enumerate}
\end{lem}
\begin{chapscxproof}
Let $H$ be the invocation of \help\ that performs $astep$ and $k$ be the iteration of the loop in \help\ during which $H$ performs $astep$.
The loop in $H$ iterates over the sequence $r_1, r_2, ..., r_l$ of \rec s. %of $U.V$ 

%We first prove
%claim~\ref{abort-fcas-flow-claim-fcas-on-rk} and
%claim~\ref{abort-fcas-flow-claim-no-succ-fcas-on-rk}
%together.
Claim~\ref{abort-fcas-flow-claim-fcas-on-rk} follows from the definition of $k$:  before $H$ performs $astep$, it performs a \fcas\ belonging to $H$ on $r_k$ at line~\ref{help-fcas}.
%Before $H$ performs $astep$, it sees $r.\info \neq scxPtr$ at line~\ref{help-check-frozen} in iteration $k$, prior to which it performs a \fcas\ belonging to $U$ on $r_k$ at line~\ref{help-fcas}.
%This proves claim~\ref{abort-fcas-flow-claim-fcas-on-rk}.

To derive a contradiction, suppose Claim 2 is false, i.e., there is a \textit{successful} \fcas\ belonging to $U$ on $r_k$.  By Claim 1, {\it fcas} is before $astep$.
From Corollary~\ref{cor-no-change-from-done-retry} and the fact that $astep$ occurs, we know that no \cstep\ belongs to $U$.
By Lemma~\ref{lem-if-succ-fcas-then-point-u-until-bcas-or-uass}, $r_k.\info$ points to $U$ at all times between the first \fcas\ belonging to $U$ on $r_k$ and $astep$.
However, this contradicts the fact that $r_k.\info$ does not point to $U$ when $H$ performs line \ref{help-check-frozen} just before performing $astep$.

We now prove Claim~\ref{abort-fcas-flow-claim-succ-fcas-on-r1-etc}.
By Claim \ref{abort-fcas-flow-claim-fcas-on-rk}, prior to $astep$,
$H$ performs a \fcas\ belonging to $U$ on $r_k$.
By Lemma~\ref{lem-fcas-on-ri-only-after-succ-fcas-on-previous}, this can only occur after a successful \fcas\ belonging
to $U$ on $r_i$, for all $i <k$.
By Lemma~\ref{lem-only-first-fcas-can-succeed}, there is no successful
\fcas\ belonging to $U$ on $r_i$ after $astep$.

We now prove Claim~\ref{abort-fcas-flow-claim-rk-changes}.
By Claim~\ref{abort-fcas-flow-claim-fcas-on-rk} and Claim~\ref{abort-fcas-flow-claim-no-succ-fcas-on-rk}, an unsuccessful \fcas \ {\it fcas} belonging to $U$ on $r_k$ occurs prior to $astep$.
By line~\ref{help-rinfo} and Observation~\ref{obs-op-invariants}.\ref{inv-llresults}, the old value for {\it fcas} is read from $r_k.\info$ and stored in $r\info$ at line~\ref{ll-read} by the \llt$(r_k)$ linked to $S$.
By Observation~\ref{obs-op-invariants}.\ref{inv-unfrozen}, the \llt$(r_k)$ linked to $S$ again sees $r_k.\info = r\info$ at line~\ref{ll-reread}.
Thus, since {\it fcas} fails, $r_k.\info$ must change after the \llt$(r_k)$ linked to $S$ performs line \ref{ll-reread} and before {\it fcas} occurs.
%Hence, $r_k.\info$ contains $r\info$ when it is read by the \llt$(r_k)$ linked to $S$ at 
%line~\ref{ll-read}, and again when it is read at line~\ref{ll-reread}.
%By Lemma~\ref{lem-no-aba-info}, $r_k.\info$ does not change between these lines.
%Thus, $r_k.\info$ must change after it is read by the \llt$(r_k)$ linked to $S$ at line~\ref{ll-reread} 
%and before {\it fcas} occurs.
%\qed
\end{chapscxproof}

\begin{lem} \label{lem-no-succ-fcas-after-fass-or-bcas}
No \fcas\ belonging to an \op\ $U$ is successful after the first \fstep\ or \astep\ belonging to $U$.
\end{lem}
\begin{chapscxproof}
First, suppose a \fstep\ belongs to $U$, and let $fstep$ be the first such \fstep.
Then, by Lemma~\ref{lem-if-fass-then-all-succ-fcas}, there is
a successful \fcas\ belonging to $U$ on $r$ for each $r$ in $U.V$
that occurs before $fstep$.
By Lemma~\ref{lem-only-first-fcas-can-succeed}, only the first \fcas\ belonging to $U$ on $r$ can be successful.
Hence, no \fcas\ belonging to $U$ is successful after $fstep$.

Now, suppose an \astep\ belongs to $U$, and let $astep$ be the first such \astep.
Let $\langle r_1, r_2, ..., r_l\rangle$ be the sequence of \rec s in $U.V$. % enumerated in order.
%By Corollary~\ref{cor-no-change-from-done-retry}, there is no \cstep\ belonging to $U$.
By Lemma~\ref{lem-abort-fcas-flow}, there is a $k \in \{1, ..., l\}$ such that no successful \fcas\ belonging to $U$ is performed on any $r_i \in \{r_1, ..., r_{k-1}\}$ after $astep$, and no successful \fcas\ belonging to $U$ on $r_k$ ever occurs.
By Lemma~\ref{lem-fcas-on-ri-only-after-succ-fcas-on-previous},
there is no \fcas\ belonging to $U$ on any $r_i \in \{r_{k+1}, ..., r_l\}$.
%\qed
\end{chapscxproof}

\begin{cor} \label{cor-if-succ-fcas-then-point-u-until-bcas-or-uass}
If there is a successful \fcas\ {\it fcas} belonging to an \op\ $U$ on a \rec\ $r$, then {\it fcas} occurs before time $t$, when the first \astep\ or \cstep\ belonging to $U$ occurs. Moreover, $r.\info$ points to $U$ at all times after {\it fcas} occurs, and before $t$.
\end{cor}
\begin{chapscxproof}
By Lemma~\ref{lem-no-succ-fcas-after-fass-or-bcas}, {\it fcas} occurs before $t$.
The claim then follows from Lemma~\ref{lem-if-succ-fcas-then-point-u-until-bcas-or-uass}. %, $r.\info$ points to $U$ at all times after {\it fcas} occurs, and before $t$.
\end{chapscxproof}

\begin{lem} \label{lem-state-transitions2}
Changes to the $state$ and $\freezingdone$ fields of an \op, as well as \fstep s, \astep s, \cstep s, \markstep s and successful \fcas s can only occur as depicted in Figure~\ref{fig-state-allfrozen-transitions}.
\end{lem}
\begin{chapscxproof}
Initially, the dummy \op\ has $state =$ \retry\ and $\freezingdone =$ \false\
and, by Lemma \ref{lem-no-steps-belong-to-dummy-op}, they never change.

Every other \op\ $U$ initially has $state =$ \freezing\ and
$\freezingdone =$ \false.
Only \astep s, \fstep s, and \cstep s
can change $state$ or $\freezingdone$.
From the code of \help, each transition shown in Figure \ref{fig-state-transitions2}(a),
results in the indicated values for $state$ and $\freezingdone$.
A \cstep\ on line \ref{help-cstep} must be preceded by a \fstep\ on line \ref{help-fstep}.
Therefore, from [\freezing, \false], the only outgoing transitions are
to [\retry, \false] and [\freezing, \true].
By Lemma \ref{lem-fass-then-no-bcas}, there cannot be both
a \fstep\ and an \astep\ belonging to $U$.
Hence, from [\retry, \false], there cannot be a \fstep\ or \cstep\
and there cannot be an \astep\ from  [\freezing, \true] or [\done, \true]. 

By Lemma \ref{lem-no-succ-fcas-after-fass-or-bcas}, successful \fcas s can only occur when $state =$ \freezing\ and $\freezingdone =$ \false.
%By Corollary~\ref{cor-marksteps}, successful \markstep s can only occur where indicated.
From the code of \help, for each $r$ in $U.R$, the first \markstep\ belonging to $U$ on $r$ must occur after the first \fstep \ belonging to $U$ and before the first \cstep \ belonging to $U$.
Since each $r$ in $U.R$ initially has $r.marked = \false$, and $r.marked$ is only changed at line~\ref{help-markstep}, where it is set to \true, only the first \markstep\ belonging to $U$ on $r$ can be successful.
%\qed
\end{chapscxproof}

\subsection{The period of time over which a \rec\ is frozen}

%\textbf{[ Re-write this glue ]}

We now prove several lemmas which characterize the period of time over which a \rec\ is frozen for an \op.
We first use the fact that the $state$ of an \op\ cannot change from \retry\ to \done\ to extend Lemma~\ref{lem-no-info-change-while-freezing} to prove that the \info\ field of a \rec\ cannot be changed while the \rec\ is frozen for an \op.
%The remaining three lemmas prove that a finalized \rec\ is frozen forever and define intervals over which all \rec s in an \op's $V$ sequence will be frozen.
In the following, we often say ``after X and before the first time Y happens.''
In the event that Y never happens, this phrasing should be interpreted to mean simply ``after X.''

%\begin{lem} \label{lem-if-fcas-then-not-frozen-between-llt-and-fcas}
%Let $U$ be an \op\ created by an invocation $I$ of \sct.
%If a successful \fcas\ belonging to $U$ on $r$ occurs then, at all times after the \llt$(r)$ linked to $I$ executes line~\ref{ll-read-state} and before the first \fcas\ belonging to $U$ on $r$, $r$ is not frozen.
%\end{lem}
%\begin{chapscxproof}
%
%\end{chapscxproof}

\begin{figure}[tb]
	\centering
	\begin{tabular}{|l|c|c|}
		\hline
		& \hspace{1mm} $r.\info.state$ \hspace{1mm} & $r.marked$ \\
		\hline
		Frozen & \done & \true \\
		& \freezing & \{\true, \false\} \\
		\hline
		Unfrozen & \done & \false \\
		& \retry & \{\true, \false\} \\
		\hline
	\end{tabular}
	\caption{%A precise characterization of 
	When a \rec\ $r$ is frozen,
			in terms of $r.\info.state$ and $r.marked$.}
	\label{fig-freezing-table}
\end{figure}

\begin{lem} \label{lem-records-frozen}
If a \fstep\ belongs to an \op\ $U$ then,
for each $r$ in $U.V$, a \fcas\ belonging to $U$ on $r$ precedes the first \fstep\ belonging to $U$, and $r$ is frozen for $U$ at all times after the first \fcas\ belonging to $U$ on $r$
and before the first \cstep\ belonging to $U$.
\end{lem}
\begin{chapscxproof}
Fix any $r$ in $U.V $.
If a \fstep\ belongs to $U$ then,
by Lemma~\ref{lem-if-fass-then-all-succ-fcas},
it is preceded by
a successful \fcas\ belonging to $U$ on $r$.
Further, by Lemma~\ref{lem-fass-then-no-bcas},
no \astep\ belongs to $U$.
Thus, by Corollary~\ref{cor-if-succ-fcas-then-point-u-until-bcas-or-uass},
$r.\info$ points to $U$ at all points between
time $t_0$, when the first \fcas\ belonging to $U$ on $r$ occurs,
and time $t_1$, when the first \cstep\ belonging to $U$ occurs
(after the first \fstep).
Since no \astep\ belongs to $U$,
$U.state =$ \freezing\ at all times before $t_1$.
Hence, by the definition of freezing (see Figure~\ref{fig-freezing-table}),
$r$ is frozen for $U$
at all times between $t_0$ and $t_1$.
%\qed
\end{chapscxproof}

\begin{cor} \label{cor-records-frozen-from-fass-to-uass}
If a \fstep\ belongs to an \op\ $U$, then
each $r$ in $U.V$ is frozen for $U$
at all times between
the first \fstep\ belonging to $U$
and the first \cstep\ belonging to $U$.
\end{cor}
\begin{chapscxproof}
Suppose there is a \fstep\ belonging to $U$.
By Lemma~\ref{lem-records-frozen}, each $r$ in $U.V$ is frozen for $U$ at all times between the first \fcas\ belonging to $U$ on $r$
and the first \cstep\ belonging to $U$.
It then follows directly from the pseudocode of \help\ that the first \fstep\ belonging to $U$ must follow the first \fcas\ belonging to $U$ on $r$,
for each $r$ in $U.V$, and precede the first \cstep\ belonging to $U$.
%\qed
\end{chapscxproof}

\begin{cor} \label{cor-marksteps}
A successful \markstep\ belonging to $U$ can occur only while $r$ is frozen for $U$.
\end{cor}
\begin{chapscxproof}
Immediate from Lemma~\ref{lem-state-transitions2} and Corollary~\ref{cor-records-frozen-from-fass-to-uass}.
%\qed
\end{chapscxproof}

\begin{lem} \label{lem-become-frozen-only-by-info-change}
A \rec\ can only be changed from unfrozen to frozen by a change in its $\info$ field (which can only be the result of a \fcas).
\end{lem}
\begin{chapscxproof}
Let $r$ be a \rec\ whose \info\ field points to an \op\ $U$.
According to the definition of a frozen \rec\ (see Figure~\ref{fig-freezing-table}), if $r.\info$ does not change, then $r$ can only become frozen if $U.state$ changes from \done\ or \retry\ to \freezing, or from \retry\ to \done\ (provided $r$ is marked).  However, both cases are impossible by Corollary~\ref{cor-state-transitions-respect-figure}.
%\qed
\end{chapscxproof}

%\eric{Trevor:  As we discussed, we need a different word than finalized to mean $marked and committed$ (because a \rec\ gets finalized at the \upcas\ of the \sct\ that finalizes it, not
%when it becomes committed.
%I have used a macro permafrozen for this.  I tried to make sure that I got correctly
%disentangled $finalized$ and $permafrozen$}

\newcommand{\permafrozen}{permafrozen}

\begin{defn}
A \rec\ $r$ is called \textbf{\permafrozen\ for} \op\ $U$ if $r$ is marked, $r.\info$ points to $U$ and the $U.state$ is \done.  Notice that a \rec\ that is \permafrozen\ for $U$ is also frozen for $U$.
\end{defn}

\begin{lem} \label{lem-finalized-forever-frozen}
Once a \rec\ $r$ is \permafrozen\ for \op\ $U$, it remains \permafrozen\ for $U$ thereafter.
\end{lem}
\begin{chapscxproof}
By definition, when $r$ is \permafrozen\ for $U$, it is frozen for $U$, $U.state$ is \done\ and $r.marked = \true$.
Once $r.marked$ is set to \true, it can never be changed back to \false.
By Corollary~\ref{cor-state-transitions-respect-figure},
$U.state$ was never \retry,
$U.state$ will remain \done\ forever, and
$r$ will be frozen for $U$ as long as $r.\info$ points to $U$.
It remains only to prove that $r.\info$ cannot change while $r$ is \permafrozen\ for $U$.
Note that $r.\info$ can be changed only by a successful \fcas.

To obtain a contradiction,
suppose a \fcas\ {\it fcas} changes $r.\info$ from $U$ to $W$ while $r$ is \permafrozen\ for $U$.
By Lemma~\ref{lem-no-steps-belong-to-dummy-op},
$W$ is not the dummy \op.
Let $S$ be the invocation of \sct\ that created $W$.
From the code of \help, $r$ is in $W.V$. So, by the precondition of \sct, there is an invocation of \llt($r$) linked to $S$.
By Observation~\ref{obs-op-invariants}.\ref{inv-llresults} and line~\ref{help-rinfo}, the old value for {\it fcas} (a pointer to $U$) was read at line~\ref{ll-read} of the \llt$(r)$ linked to $S$.
Let $I$ be the invocation of $\llt(r)$ linked to $S$.
Since we have argued that $U.state$ is never \retry, $U.state \in \{\freezing, \done\}$ when $I$ reads $ state$ from $U.state$ at line~\ref{ll-read-state}.

If $state = \freezing$ then $I$ does not enter the if-block at line~\ref{ll-check-frozen}, and returns \fail \ or \finalized, which contradicts Definition~\ref{defn-llt-linked-to-sct}.\ref{prop-returns-value-different-from-fail-or-finalized}.

Now, consider the case where $state = \done$.
If we can argue that $r$ is marked when $I$ performs line~\ref{ll-read-marked2}, then we shall obtain the same contradiction as in the previous case.
Since $state = \done $, a \cstep \ belonging to $U$ occurs before $I$ performs line~\ref{ll-read-state}.
By Lemma~\ref{lem-state-transitions2}, any successful \markstep\ belonging to $U$ occurs prior to this \cstep.
Therefore, if $r$ is in $U.R$, then $r$ will be marked when $I$ performs line~\ref{ll-read-marked2}, and we obtain the same contradiction.
The only remaining possibility is that $r$ is not in $U.R$, and $r$ is marked by a successful \markstep\ $mstep$ belonging to some other \op\ $U'$ \textit{after} $I$ performs line~\ref{ll-read-marked2}, and before {\it fcas} occurs (which is while $r$ is \permafrozen\ for $U$).
Since $r.\info$ points to $U$ when $I$ performs line~\ref{ll-read}, and again when {\it fcas} occurs, Lemma~\ref{lem-no-aba-info} implies that $r.\info$ points to $U$ throughout this time.
However, this contradicts Corollary~\ref{cor-marksteps}, which states that $mstep$ can only occur while $r.\info$ points to $U'$.
%\qed
\end{chapscxproof}

\begin{lem} \label{lem-frozen-forever-after-markstep}
Suppose a successful \markstep\ $mstep$ belonging to an \op \ $U$ on $r$ occurs.
Then, $r$ is frozen for $U$ when $mstep$ occurs, and forever thereafter.
\end{lem}
\begin{chapscxproof}
By Corollary~\ref{cor-marksteps}, $mstep$ must occur while $r$ is frozen for $U$.
From the code of \help, a \fstep \ belonging to $U$ must precede $mstep$, and $r$ must be in $V$ (since it is marked at line~\ref{help-markstep}).
Thus, Corollary~\ref{cor-records-frozen-from-fass-to-uass} implies that $r$ is frozen for $U$ at all times between $mstep$ and the first \cstep\ belonging to $U$.
Since $r.marked$ is never changed from \true \ to \false, $mstep$ must be the first \markstep\ that ever modifies $r.marked$.
From the code of \help, $mstep$ must precede the first \cstep \ belonging to $U$.
If any \cstep\ belonging to $U$ occurs after $mstep$, immediately after the first such \cstep, $r$ will be marked, and $r.\info.state$ will be \done, so $r$ will become \permafrozen\ for $U$.
By Lemma~\ref{lem-finalized-forever-frozen}, $r$ will remain frozen for $U$, thereafter.
%\qed
\end{chapscxproof}

\begin{lem} \label{lem-r-not-frozen-in-good-llt}
Suppose $I$ is an invocation of $\llt(r)$ that returns a value different from \fail \ or \finalized.
Then, $r$ is not frozen at any time in $[t_0, t_1]$, where $t_0$ is when $I$ reads $r\info.state$ at line~\ref{ll-read-state}, and $t_1$ is when $I$ reads $r.\info$ at line~\ref{ll-reread}.
\end{lem}
\begin{chapscxproof}
We prove that $r$ is not frozen at any time between $t_0$ and $t_1$.
Since $I$ returns a value different from \fail\ or \finalized, it enters the if-block at line~\ref{ll-check-frozen}, and sees $r.\info = r\info$ at line~\ref{ll-reread}.
Therefore, it sees either $state =$ \done\ and $r.marked = \false$, or $state =$ \retry\ at line~\ref{ll-check-frozen}.
In each case, Corollary~\ref{cor-state-transitions-respect-figure} guarantees that $r\info.state$ will never change again after time $t_0$.
Thus, if $state = \retry$, then $r$ is not frozen at any time between $t_0$ and $t_1$.
Now, suppose $state = \done$.
We prove that $r.marked$ does not change between $t_0$ and $t_1$.
A pointer to an \op\ $W$ is read from $r.\info$ and stored in the local variable $r\info$ at line~\ref{ll-read}, before $t_0$.
At line~\ref{ll-reread}, $r.\info$ still contains a pointer to $W$.
By Lemma~\ref{lem-no-aba-info}, $r.\info$ must not change between line~\ref{ll-read} and line~\ref{ll-reread}.
Therefore, $r.\info$ points to $W$ at all times between $t_0$ and $t_1$.
By Corollary~\ref{cor-marksteps}, a successful \markstep\ can occur between $t_0$ and $t_1$ only if it belongs to $W$.
Since $state = \done$, a \cstep \ belonging to $W$ must have occurred before $t_0$.
By Lemma~\ref{lem-state-transitions2}, any successful \markstep\ belonging to $W$ must have occurred before $t_0$.
Therefore, $W.state =$ \done\ and $r.marked = \false$ throughout [$t_0, t_1$].
%\qed
\end{chapscxproof}

\begin{cor} \label{cor-r-not-frozen-in-linked-llt}
Let $S$ be an invocation of \sct, and $r$ be any \rec \ in the $V$ sequence of $S$. %\in$ $U.V$.
Then, $r$ is not frozen at any time in $[t_0, t_1]$, where $t_0$ is when the \llt$(r)$ linked to $S$ reads $r\info.state$ at line~\ref{ll-read-state}, and $t_1$ is when the \llt$(r)$ linked to $S$ reads $r.\info$ at line~\ref{ll-reread}.
\end{cor}
\begin{chapscxproof}
Immediate from Definition~\ref{defn-llt-linked-to-sct}.\ref{prop-returns-value-different-from-fail-or-finalized} and Lemma~\ref{lem-r-not-frozen-in-good-llt}.
%We prove that $r$ is not frozen at any time between $t_0$ and $t_1$.
%By definition, the \llt$(r)$ linked to $S$ returns a value different from \fail\ or \finalized, which implies that it enters the if-block at line~\ref{ll-check-frozen}, and sees $r.\info = r\info$ at line~\ref{ll-reread}.
%Since this \llt\ enters the if-block at line~\ref{ll-check-frozen}, it sees either $state =$ \done\ and $r.marked = \false$, or $state =$ \retry.
%In each case, Corollary~\ref{cor-state-transitions-respect-figure} guarantees that $r\info.state$ will never change again after time $t_0$.
%Thus, if $state = \retry$, then $r$ is not frozen at any time between $t_0$ and $t_1$.
%Now, suppose $state = \done$.
%We prove that $r.marked$ does not change between $t_0$ and $t_1$.
%A pointer to an \op\ $W$ is read from $r.\info$ and stored in the local variable $r\info$ at line~\ref{ll-read}, before $t_0$.
%At line~\ref{ll-reread}, $r.\info$ still contains $r\info$.
%By Lemma~\ref{lem-no-aba-info}, $r.\info$ must not change between line~\ref{ll-read} and line~\ref{ll-reread}.
%Therefore, $r.\info$ points to $W$ at all times between $t_0$ and $t_1$.
%By Corollary~\ref{cor-marksteps}, a successful \markstep\ can occur between $t_0$ and $t_1$ only if it belongs to $W$.
%Since $state = \done$, a \cstep \ belonging to $W$ must have occurred before $t_0$.
%By Lemma~\ref{lem-state-transitions2}, any successful \markstep\ belonging to $W$ must have occurred before $t_0$.
%Therefore, $W.state =$ \done\ and $r.marked = \false$ throughout [$t_0, t_1$].
%\qed
\end{chapscxproof}

\subsection{Properties of \upcas\ steps}

%We prove five results in this section.
%The other lemmas merely simplify the proofs of these five results, and are not used elsewhere in the proof.
%We discuss the five results below as they are proved (together with their supporting lemmas).
%We begin with simple observations.

%\begin{lem} \label{lem-records-frozen-for-first-upcas}
%The first \upcas\ belonging to an \op\ $U$ on a \rec\ $r$ occurs while $r$ is frozen for $U$.
%\end{lem}
%\begin{chapscxproof}
%Let $I$ be the first \upcas\ belonging to $U$ on $r$.  Clearly, $I$ occurs after the first \fstep\ belonging to $U$, since a \fstep\ precedes $I$ in the code.  Similarly, $I$ must occur before the first \cstep\ belonging to $U$, since $I$ if the first \upcas, and an \upcas\ precedes the \cstep\ in the code.
%Thus, by Corollary~\ref{cor-records-frozen-from-fass-to-uass}, all $r \in V$ are frozen for $U$ when $I$ occurs.%, and $I$ can only modify a \rec\ in $V$ (by restrictions on the arguments of \sct). 
%%\qed
%\end{chapscxproof}

\begin{obs} \label{obs-immutable-fields-do-not-change}
An immutable field of a \rec\ cannot change from its initial value.
\end{obs}
\begin{chapscxproof}
This observation follows from the facts that
\rec s can only be changed by \sct\ and 
an invocation of \sct\ can only accept a pointer
to a mutable field as its $fld$ argument (to modify).
\end{chapscxproof}

\begin{obs} \label{obs-only-upcas-modifies-records}
Each mutable field of a \rec\ can be modified only by a successful \upcas.
\end{obs}

\begin{obs} \label{obs-all-upcas-write-new}
Each \upcas\ belonging to an \op\ $U$ is of the form \cas$(U.fld, U.old, U.new)$. 
Invariant: $U.fld$ and $U.new$ contain the arguments $fld$ and $new$, respectively, that were passed to the invocation of \sct$(V, R, fld, new)$ that created $U$.
\end{obs}
\begin{chapscxproof}
An \upcas\ occurs at line~\ref{help-upcas} in an invocation of \help$(scxPtr)$, where it operates on $scxPtr.fld$, using $scxPtr.old$ as its old value, and $scxPtr.new$ as its new value. The fields of $scxPtr$ do not change after $scxPtr$ is created at line~\ref{sct-create-op}. At this line, the arguments $fld$ and $new$ that were passed to the invocation of \sct$(V, R, fld, new)$ are stored in $scxPtr.fld$ and $scxPtr.new$, respectively.
%\qed
\end{chapscxproof}

%$\bigstar$
%We now give a lemma that helps us prove the next result in this section, as well as the first result in the next section.  \eric{This is not a very useful sentence}

\begin{lem} \label{lem-first-upcas-while-frozen}
The first \upcas\ belonging to an \op\ $U$ on a \rec\ $r$ occurs while $r$ is frozen for $U$. 
\end{lem}
\begin{chapscxproof}
Let $upcas$ be the first \upcas \ belonging to $U$.
%Suppose a successful \upcas\ $upcas$ belongs to an \op\ $U$. By Lemma~\ref{lem-only-first-upcas-can-succeed}, it is the first \upcas\ belonging to $U$. 
By line~\ref{help-upcas}, such an \upcas\ will modify $U.fld$ which, by Observation~\ref{obs-op-invariants}.\ref{inv-fld}, is a mutable field of a \rec\ $r$ in $U.V$.
Since $upcas$ is preceded by a \fstep\ in the pseudocode of \help, a \fstep\ belonging to $U$ must precede $upcas$.
Hence, Corollary~\ref{cor-records-frozen-from-fass-to-uass} applies, and each $r$ in $U.V$ is frozen for $U$ at all times between the first \fstep\ belonging to $U$ and the first \cstep\ $cstep$ belonging to $U$.
From the code of \help, if $cstep$ exists, then it must occur after $upcas$.
Thus, when $upcas$ occurs, $r$ is frozen for $U$.
\end{chapscxproof}

In Section~\ref{sec-impl} we described a constraint on the use of \sct\ that allows us to implement an optimized version of \sct\ (which avoids the creation of a new \rec\ to hold each value written to a mutable field), and noted that the correctness of the unoptimized version follows trivially from the correctness of the optimized version.
In order to prove the next few lemmas, we must invoke this constraint. % (which we reproduce below for ease of reference).
In fact, we assume a weaker constraint, and are still able to prove what we would like to.
We now give this weaker constraint, and remark that it is automatically satisfied if the constraint in Section~\ref{sec-impl} is satisfied.

\begin{con} \label{con-use-of-sct}
Let $fld$ be a mutable field of a \rec \ $r$.
If an invocation $S$ of \sct$(V, R, fld, new)$ is linearized, then:
%A process \textbf{cannot} invoke \sct$(V, R, fld, new)$:
\begin{itemize}
    \item $new$ is not the initial value of $fld$, and
	\item no invocation of \sct$(V', R', fld, new)$ is linearized before the $\llt(r)$ linked to $S$ is linearized.
%	\item where $new$ is the initial value of the field pointed to by $fld$, or
%	\item{after any process starts a \textbf{successful} invocation of \sct$(V', R', fld, new)$}
%	\item{after performing a successful invocation of \sct$(V', R', fld, new)$, or}
%	\item{after another process invokes \sct$(V', R', fld, new)$.}
\end{itemize}
\end{con}

We prove the following six lemmas solely to prove that only the first \upcas\ belonging to an \op\ can succeed.
This result is eventually used to prove that exactly one successful \upcas\ belongs to any \op\ which is helped to \textit{successful} completion.

We need to know about the linearization of \sct s and linked \llt s to prove the next lemma, which uses Constraint~\ref{con-use-of-sct}. % refers to linearization points.
Let $S$ be an invocation of \sct, and $U$ be the \op \ that it creates.
As we shall see in Section~\ref{sec-corr-lin}, we linearize $S$ if and only if there is an \upcas\ 
belonging to $U$, and $S$ is linearized at its first \upcas.
Each invocation of \llt \ linked to an invocation of \sct \ is linearized at line~\ref{ll-reread}.
\after{Would it make more sense to put the formal discussion of linearization points here, since you use them in the next proof (instead of merely having a forward pointer to them).}

%\begin{lem} \label{lem-unsucc-scx-no-upcas}
%Let $U$ be an \op\ created by an invocation $S$ of \sct.
%If $S$ returns \false,
%then no \upcas\ belongs to $U$.
%\end{lem}
%\begin{chapscxproof}
%Let $ptr$ be a pointer to $U$. Let $H$ be the invocation of \help$(ptr)$ performed by $S$.
%Since $S$ returns \false, we know from line~\ref{sct-call-help} that $H$ returns \false.
%From the code of \help, $H$ returns at line~\ref{help-return-false}.
%Just before this $ H $ performs an \astep\ belonging to $U$ at line~\ref{help-astep}.
%Thus, by Lemma~\ref{lem-fass-then-no-bcas}, no \fstep\ belongs to $U$.
%However, by the code of \help, a \fstep\ belonging to $U$ must precede any \upcas \ belonging to $U$. Therefore, no \upcas\ belongs to $U$.
%%\qed
%\end{chapscxproof}

\begin{lem} \label{lem-two-scts-cannot-write-same-value}
%If an invocation $S$ of \sct$(V, R, fld, new)$ is linearized, then no invocation of \sct$(V', R', fld, new)$ is linearized.
No two \upcas s belonging to different \op s can attempt to change the same field to the same value.
\end{lem}
\begin{chapscxproof}
Suppose, to derive a contradiction, that \upcas s belonging to two different \op s $U$ and $U'$ attempt to change the same (mutable) field of some \rec\ $r$ to the same value.
Let $upcas$ and $upcas'$ be the first \upcas \ belonging to $U$ and $U'$, respectively.  Let $S$ and $S'$ be the invocation of \sct\ that created $U$ and $U'$, respectively.
From Observation~\ref{obs-all-upcas-write-new} and the fact that $upcas$ and $upcas'$ attempt to change the same field to the same value, we know that $S$ and $S'$ must have been passed the same $fld$ and $new$ arguments.
Note that $S$ and $S'$ are linearized at $upcas$ and $upcas'$, respectively.
Without loss of generality, suppose $S$ is linearized after $S'$.
By Constraint~\ref{con-use-of-sct}, $S'$ is linearized after the invocation $I$ of $\llt(r)$ linked to $S$ is linearized.

By Lemma~\ref{cor-r-not-frozen-in-linked-llt}, $r$ is not frozen when $I$ is linearized.
By Lemma~\ref{lem-first-upcas-while-frozen}, $r$ is frozen for $U'$ when $upcas'$ occurs (which is after $I$ is linearized).
By Lemma~\ref{lem-become-frozen-only-by-info-change}, $r$ can become frozen for $U'$ only by a successful \fcas \ belonging to $U'$ on $r$.
Therefore, a successful \fcas\ {\it fcas$'$} belonging to $U'$ on $r$ occurs after $I$ is linearized, and before $upcas'$.
By Lemma~\ref{lem-first-upcas-while-frozen}, $r$ is frozen for $U$ when $upcas$ occurs (which is after $upcas'$), which implies that a successful \fcas \ {\it fcas} belonging to $U$ on $r$ occurs after $upcas'$, and before $upcas$.
To recap, $I$ is linearized before {\it fcas$'$}, which is before $upcas'$, which is before {\it fcas}, which is before $upcas$.
By line~\ref{help-rinfo} and Observation~\ref{obs-op-invariants}.\ref{inv-llresults}, the old value $old$ for {\it fcas} is read from $r.\info$ and stored in $r\info$ at line~\ref{ll-read} by $I$.
Since $I$ performs line~\ref{ll-read} before it is linearized, $old$ is read from $r.\info$ before {\it fcas$'$}.
Since {\it fcas$'$} changes $r.\info$ to point to $U'$, Lemma~\ref{lem-no-aba-info} implies that $r.\info$ does not point to $U'$ at any time before {\it fcas$'$}.
Therefore, $old$ is not $U'$.
Since {\it fcas} is successful, $r.\info$ must be changed to $old$ at some point after {\it fcas$'$}, and before $upcas$.
However, this contradicts Lemma~\ref{lem-no-aba-info}, since $r.\info$ had already contained $old$ before {\it fcas$'$}.
%
%Suppose two \upcas s $upcas_0$ and $upcas_1$ belonging to \op s $U_0$ and $U_1$, respectively, attempt to change the same field to the same value.  We show that $U_0$ and $U_1$ must be the same \op.
%Consider the invocations $S_0$ and $S_1$ of \sct\ that created $U_0$ and $U_1$, respectively.
%From Observation~\ref{obs-all-upcas-write-new} and the fact that $upcas_0$ and $upcas_1$ attempt to change the same field to the same value, we know that $S_0$ and $S_1$ must have been passed the same $fld$ and $new$ arguments.
%By Constraint~\ref{con-use-of-sct}, either $S_0$ and $S_1$ are the same invocation of \sct, or at least one of $S_0$ and $S_1$ is unsuccessful.
%However, by Lemma~\ref{lem-unsucc-scx-no-upcas}, an \upcas \ cannot belong to an unsuccessful invocation of \sct.
%Hence, $S_0$ and $S_1$ must be the same invocation of \sct.
%Since an \sct \ creates only one \op \ at line~\ref{sct-create-op}, $U_0$ and $U_1$ must be the same \op.
\end{chapscxproof}

\begin{lem} \label{lem-upcas-cannot-write-initial-value}
An \upcas\ never changes a field back to its initial value.
\end{lem}
\begin{chapscxproof}
By Observation~\ref{obs-all-upcas-write-new}, each \upcas\ belonging to an \op\ $U$ attempts to change a field to the value $new$ that was passed as an argument to the invocation of \sct\ that created $U$.  Since Constraint~\ref{con-use-of-sct} implies that $new$ cannot be the initial value of the field, we know that no \upcas\ can change the field to its initial value.
%\qed
\end{chapscxproof}

\begin{lem} \label{lem-upcas-cannot-use-same-old-and-new-values}
No \upcas\ has equal $old$ and $new$ values.
\end{lem}
\begin{chapscxproof}
Let $upcas$ be an \upcas\ and let $U$ be the \op\ to which it belongs.
By Observation~\ref{obs-all-upcas-write-new}, the old value used by $upcas$ is $U.old$, and the new value used by $upcas$ is $U.new$. Let $f$ be the field of a \rec\ pointed to by $U.fld$; this is the field to which $upcas$ is applied. By Lemma~\ref{lem-upcas-cannot-write-initial-value}, $U.new$ cannot be the initial value of $f$. If $U.old$ is the initial value of $f$, then we are done. So, suppose $U.old$ is not the initial value of $f$. Since a mutable field can only be changed by a successful \upcas, there exists a successful \upcas\ $upcas'$ which changed $f$ to $U.old$ prior to $upcas$. By Observation~\ref{obs-op-invariants}.\ref{inv-old}, $U.old$ was read from $f$ prior to the start of the invocation $S$ of \sct\ that created $U$ and, therefore, prior to $upcas$. Hence, when $upcas'$ occurs, $U$ has not yet been created. Note that $upcas'$ must occur in an invocation of \help$(ptr')$ where $ptr'$ points to some \op\ $U'$ different from $U$.
However, by Lemma~\ref{lem-two-scts-cannot-write-same-value}, $upcas$ and $upcas'$ use different new values, so $U.old$ (the new value for $upcas'$) must be different from $U.new$ (the new value for $upcas$).
%
%In order to derive a contradiction, suppose an \upcas\ uses old value $old$ and new value $new$, where $old = new$.  Let $upcas$ be the first such \upcas\ occurring in the execution, $U$ be the \op\ to which it belongs, and $f$ the field (pointed to by $U.fld$) to which it writes.
%Then, by Lemma~\ref{lem-upcas-cannot-write-initial-value}, $old$ cannot be the initial value of $f$.
%Further, by Observation~\ref{obs-op-invariants}.\ref{inv-old}, $old$ was read from $f$ prior to the \sct\ that created $U$ and, in turn, prior to $upcas$.
%Thus, an \upcas\ $upcas'$ wrote $old$ to $f$ before $upcas$.
%Since $upcas$ is the first \upcas\ to use the same old and new values, $upcas'$ must have an old value $old'$ and new value $new'$ (where $new' = old$) which satisfy $old' \neq new'$.
%Then, by Observation~\ref{obs-all-upcas-write-new}, $upcas$ and $upcas'$ cannot belong to the same \op\ (since $old' \neq old$).
%However, by Lemma~\ref{lem-two-scts-cannot-write-same-value}, no two invocations of \sct\ can attempt to write the same value to a field, so there cannot be two \op s $U$ and $U'$ with $U.fld = U'.fld$ and $U.new = U'.new$, and we have a contradiction.
%\qed
\end{chapscxproof}

\begin{lem} \label{lem-only-one-succ-upcas}
At most one successful \upcas\ can belong to an \op.
\end{lem}
\begin{chapscxproof}
We prove this lemma by contradiction. Consider the earliest point in the execution when the lemma is violated.
Let $upcas'_U$ be the earliest occurring second successful \upcas\ belonging to any \op, and $U$ be the \op\ to which it belongs, and let $upcas_U$ be the preceding successful \upcas\ belonging to $U$.
Further, let $f$ be the field upon which $upcas'_U$ operates, and let $old$ and $new$ be the old and new values used by $upcas_U$, respectively.  
(By Observation~\ref{obs-all-upcas-write-new}, $upcas_U$ and $upcas'_U$ attempt to change the same field from the same old value to the same new value.)
By Lemma~\ref{lem-upcas-cannot-use-same-old-and-new-values}, we know that $old \neq new$.
Then, since $upcas'_U$ is successful, there must be a successful \upcas\ $upcas'_W$ belonging to some \op\ $W$ which changes $f$ to $old$ between $upcas_U$ and $upcas'_U$.
By Lemma~\ref{lem-upcas-cannot-write-initial-value}, $old$ is not the initial value of $f$. Hence, there must be another successful \upcas\ $upcas_W$ which changes $f$ to $old$ before $upcas_U$.
By Lemma~\ref{lem-two-scts-cannot-write-same-value}, $upcas_W$ must belong to $W$, so $upcas_W$ and $upcas'_W$ both precede $upcas'_U$.
This contradicts the definition of $upcas'_U$.
%\qed
\end{chapscxproof}

\begin{lem} \label{lem-no-aba-on-mutable-fields}
An \upcas\ never changes a field to a value that has already appeared there.  (Hence, there is no ABA problem on mutable fields.)
\end{lem}
\begin{chapscxproof}
Suppose a successful \upcas\ $upcas$ belonging to an \op\ $U$ changes a field $f$ to have value $new$.
By Lemma~\ref{lem-upcas-cannot-write-initial-value}, $new$ is not the initial value of $f$. By Lemma~\ref{lem-only-one-succ-upcas}, all successful \upcas s that change $f$ must belong to different \op s. Hence, Lemma~\ref{lem-two-scts-cannot-write-same-value} implies that no \upcas\ other than $upcas$ can change $f$ to $new$. 
%\qed
\end{chapscxproof}

Lemma \ref{lem-only-one-succ-upcas} proved that at most 
one \upcas\ of each \op\ can succeed.
Now we prove that such a successful \upcas\ must be the {\it first} one belonging to \op.

\begin{lem} \label{lem-only-first-upcas-can-succeed}
Only the first \upcas\ belonging to an \op\ $U$ can succeed.
\end{lem}
\begin{chapscxproof}
Let $upcas$ be the first \upcas\ belonging to $U$, and $f$ be the field that $upcas$ attempts to modify. %, and $old$ and $new$ be the old and new values for $upcas$, respectively.
If $upcas$ succeeds then, by Lemma~\ref{lem-only-one-succ-upcas}, there can be no other successful \upcas\ belonging to $U$.
So, suppose $upcas$ fails.
By Observation~\ref{obs-all-upcas-write-new}, each \upcas\ belonging to $U$ uses the same old value $U.old$. 
%and $U.old$ does not change after $U$ is created at line~\ref{sct-create-op}, 
By Observation~\ref{obs-op-invariants}.\ref{inv-old}, $U.old$ was read from $f$ prior to the start of the invocation $S$ of \sct\ that created $U$ and, therefore, prior to $upcas$.
Then, since $upcas$ fails, $f$ must change between when $U.old$ is read from $f$ and when $upcas$ occurs.
By Observation~\ref{obs-only-upcas-modifies-records}, $f$ can only be changed by an \upcas. By Lemma~\ref{lem-no-aba-on-mutable-fields}, each \upcas\ applied to $f$ changes it to a value that it has not previously contained. Therefore, $f$ will never again be changed to $U.old$. Hence, every subsequent \upcas\ belonging to $U$ will fail.
%\qed
\end{chapscxproof}

\subsection{Freezing works}

%\textbf{[add glue here, for the start of the section]}

In addition to being used to prove the remaining lemmas of this section, the following two results are used to prove linearizability in Section~\ref{sec-corr-lin}.
Intuitively, they allow us to determine whether a \rec \ has changed simply by looking at its \info\ field, and whether it is frozen.
%Intuitively, the next two lemmas prove that an unfrozen \rec\ whose \info\ field points to an \op\ $U$ at two different times $t_0$ and $t_1$ does not change between $t_0$ and $t_1$ (i.e., none of its fields change).

\begin{cor} \label{cor-upcas-only-modifies-frozen}
An \upcas\ belonging to an \op\ $U$ on a \rec\ $r$ can succeed only while $r$ is frozen for $U$. 
\end{cor}
\begin{chapscxproof}
Suppose a successful \upcas\ $upcas$ belongs to an \op\ $U$.
By Lemma~\ref{lem-only-first-upcas-can-succeed}, it is the first \upcas\ belonging to $U$. 
The claim follows from Lemma~\ref{lem-first-upcas-while-frozen}.
%
%By line~\ref{help-upcas}, such an \upcas\ will modify $U.fld$ which, by Observation~\ref{obs-op-invariants}.\ref{inv-fld}, points to a mutable field of a \rec\ $r \in U.V$.
%Since $upcas$ is immediately preceded by a \fstep\ in the pseudocode of \help, a \fstep\ belonging to $U$ must precede $upcas$.
%Hence, Corollary~\ref{cor-records-frozen-from-fass-to-uass} applies, and each $r \in U.V$ is frozen for $U$ at all times between the first \fstep\ belonging to $U$ and the first \cstep\ belonging to $U$ (which must follow $upcas$, by the pseudocode of \help).
%Thus, when $upcas$ occurs, $r$ is frozen for $U$.
%\qed
\end{chapscxproof}
By Observation~\ref{obs-only-upcas-modifies-records}, a mutable field of $r$ can only change while $r$ is frozen.

\begin{lem} \label{lem-read-unfrozen-info-twice-no-change}
If a \rec\ $r$ is not frozen at time $t_0$, $r.\info$ points to an \op\ $U$ at or before time $t_0$, and $r.\info$ points to $U$ at time $t_1 > t_0$, then no field of $r$ is changed during $[t_0,t_1]$.
\end{lem}
\begin{chapscxproof}
Since $r.\info$ points to $U$ at or before time $t_0$, and again at time $t_1$, Lemma~\ref{lem-no-aba-info} implies that $r.\info$ must point to $U$ at all times in $[t_0, t_1]$.
Further, from Lemma~\ref{lem-become-frozen-only-by-info-change}, $r$ can only be changed from unfrozen to frozen by a change to $r.\info$. %, which we have argued is impossible.
Therefore, at all times in $[t_0, t_1]$, $r$ is not frozen.
By Corollary~\ref{cor-upcas-only-modifies-frozen}, and Observation~\ref{obs-only-upcas-modifies-records}, each mutable field of $r$ can change only while $r$ is frozen.
By Corollary~\ref{cor-marksteps}, $r.marked$ can change only while $r$ is frozen.
Finally, by Observation~\ref{obs-immutable-fields-do-not-change}, immutable fields do not ever change.
Hence, no field of $r$ changes during $[t_0, t_1]$.
%
%Since $r.\info$ points to $U$ at time $t_0$, and again at time $t_1$, Lemma~\ref{lem-no-aba-info} implies that $r.\info$ must point to $U$ at all times in $[t_0, t_1]$.
%Further, from Lemma~\ref{lem-become-frozen-only-by-info-change}, $r$ can only be changed from unfrozen to frozen by a successful \fcas\ applied to $r.\info$. 
%Between $t_0$ and $t_1$, there is no successful \fcas\ belonging to an \op\ other than $U$ that changes $r.\info$, since it would change the value of $r.\info$. 
%
%Now we show that, after $t_0$, there is no successful \fcas\ belonging to $U$. If $U$ is the dummy \op, then this is true by Lemma~\ref{lem-no-steps-belong-to-dummy-op}. Otherwise, there must have been a successful \fcas\ belonging to $U$ prior to $t_0$, since $r.\info$ points to $U$ at $t_0$. By Lemma~\ref{lem-only-first-fcas-can-succeed}, no subsequent \fcas\ belonging to $U$ on $r$ is successful. Therefore, at all times in $[t_0, t_1]$, $r$ is not frozen. Hence, Corollary~\ref{cor-upcas-only-modifies-frozen} implies that no \upcas\ on $r$ can succeed in $[t_0, t_1]$.
%By Observation~\ref{obs-only-upcas-modifies-records}, a mutable field can only be changed by a successful \upcas.
%Finally, by Observation~\ref{obs-immutable-fields-do-not-change}, immutable fields do not change. Hence, no fields of $r$ change. 
%\qed
\end{chapscxproof}

The remaining results of this section describe intervals over which certain fields of a \rec\ do not change.
Suppose $U$ is an \op \ created by an invocation $S$ of \sct, $r$ is a \rec\ in $U.V$, and $I$ is the invocation of $\llt(r)$ linked to $S$.
Intuitively, we use the preceding lemma to prove, over the next two lemmas, that no field of $r$ changes between when $I$ last reads $r.\info$, and when $r$ becomes frozen for $U$.
We then use this result in Section~\ref{sec-corr-help} to prove that $S$ succeeds if and only if this holds for each $r$ in $V$.
The remaining results of this section are used primarily to prove that exactly one successful \upcas\ belongs to $U$ if a \fstep\ belongs to $U$ (and $S$ does not crash, or some process helps it complete).

\begin{cor} \label{cor-read-unfrozen-info-then-fcas-no-change}
Let $U$ be an \op, and $S$ be the invocation of \sct\ that created $U$.
If there is a successful \fcas\ belonging to $U$ on $r$, then no field of $r$ changes after the \llt$(r)$ linked to $S$ reads $r\info.state$ at line~\ref{ll-read-state}, and before this \fcas\ occurs.
\end{cor}
\begin{chapscxproof}
Let {\it fcas} be a successful \fcas\ belonging to $U$ on $r$.
Note that the \llt$(r)$ linked to $S$ terminates before $S$ begins.
Since $S$ creates $U$ and {\it fcas} changes $r.\info$ to point to $U$, $S$ begins before {\it fcas}.
We now check that Lemma~\ref{lem-read-unfrozen-info-twice-no-change} applies. By Corollary~\ref{cor-r-not-frozen-in-linked-llt}, $r$ is not frozen when the \llt$(r)$ linked to $S$ executes line~\ref{ll-read-state}.
From line~\ref{help-rinfo} of \help\ and Observation~\ref{obs-op-invariants}.\ref{inv-llresults}, we know the old value $r\info$ for {\it fcas} is read from $r.\info$ at line~\ref{ll-read} by the \llt$(r)$ linked to $S$.
Further, since {\it fcas} succeeds, $r.\info$ contains $r\info$ just prior to {\it fcas}.
Thus, Lemma~\ref{lem-read-unfrozen-info-twice-no-change} applies, and proves the claim.
%\qed
\end{chapscxproof}

%The next two results extend the period of time over which the mutable fields of a \rec \ do not change.
%These results are used in Section~\ref{sec-corr-help} to prove that exactly one successful \upcas\ belongs to an \op\ $U$ created by a successful invocation of \sct.

\begin{lem} \label{lem-fcas-then-upcas-no-change}
If an \upcas \ belongs to an \op\ $U$ then, for each $r$ in $U.V$, there is a successful \fcas\ belonging to $U$ on $r$, and no mutable field of $r$ changes during $[t_0(r), t_1)$, where $t_0(r)$ is when the first such \fcas\ occurs, and $t_1$ is when the first \upcas \ belonging to $U$ occurs.
\end{lem}
\begin{chapscxproof}
Suppose an \upcas\ belongs to an \op\ $U$.
Let $upcas$ be the first such \upcas.
Since each \upcas\ is preceded in the code by a \fstep, a \fstep\ also belongs to $U$.
Fix any $r$ in $U.V$.
By Lemma~\ref{lem-fass-then-no-bcas}, there is a successful \fcas\ belonging to $U$ on $r$.
By Lemma~\ref{lem-records-frozen}, $r$ is frozen for $U$ at all times in $[t_0(r), t_2)$, where $t_2$ is when the first \cstep\ belonging to $U$ occurs.
Since an \upcas\ belonging to an \op\ $W$ can modify $r$ only while $r$ is frozen for $W$ (by Corollary~\ref{cor-upcas-only-modifies-frozen}), any \upcas\ that modifies $r$ during $[t_0(r), t_2)$ must belong to $U$.
From the code of \help, $t_0(r) < t_1 < t_2$.
%By inspection of \help, one can easily see that $upcas$ (which occurs at time $t_1$) must occur after $t_0(r)$ and before $t_2$.
However, since the first \upcas\ belonging to $U$ occurs at $t_1$, no \upcas\ belonging to $U$ can occur during $[t_0(r), t_1)$.
%\qed
\end{chapscxproof}

\after{After Feb 10, may want to reword next lemma to cover the case where $S$ crashes
before any \cstep\ belonging to $U$ occurs}

\begin{lem} \label{lem-marked-changes}
Let $U$ be an \op \ created by an invocation $S$ of \sct, and $r$ be a \rec\ in $U.V$.  Let $t_0$ be when the \llt$(r)$ linked to $S$ reads $U.state$ at line~\ref{ll-read-state}, and $t_2$ be when the first \cstep \ belonging to $U$ occurs.
If an \upcas\ belongs to $U$ then, for each $r$ in $U.V$, between $t_0$ and $t_2$, $r.marked$ can be changed (from \false\ to \true, by the first \markstep\ belonging to $U$ on $r$) only if $r$ is in $U.R$.
\end{lem}
\begin{chapscxproof}
Fix any $r$ in $U.V$.
The fact that $r.marked$ can be changed only from \false\ to \true\ follows immediately from the fact that $r.marked$ is initially \false, and is only changed at line~\ref{help-markstep}.
It also follows that a successful \markstep\ on $r$ must be the first \markstep\ on $r$.
The rest of the claim is more subtle.
Suppose a successful \markstep\ $mstep$ belonging to an \op\ $W$ on $r$ occurs during $(t_0,t_2)$.
By Lemma~\ref{lem-frozen-forever-after-markstep}, $r$ is frozen for $W$ when $mstep$ occurs.
From the code of \help, a \fstep \ $fstep$ belonging to $U$ precedes the first \upcas \ belonging to $U$, and a \fcas \ belonging to $U$ on $r$ precedes $fstep$.
By Lemma~\ref{lem-if-fass-then-all-succ-fcas}, there is a successful \fcas\ belonging to $U$ on $r$.
By Lemma~\ref{lem-only-first-fcas-can-succeed}, it must be the first \fcas \ {\it fcas} belonging to $U$ on $r$.
Let $t_1$ be when {\it fcas} occurs.
Note that $t_0 < t_1 < t_2$.
By Corollary~\ref{cor-read-unfrozen-info-then-fcas-no-change}, $r$ does not change during $[t_0, t_1)$.
Thus, $mstep$ cannot occur in $[t_0, t_1)$.
By Lemma~\ref{lem-records-frozen}, $r$ is frozen for $U$ at all times during $[t_1, t_2]$.
This implies $U = W$, so $mstep$ is the first \markstep\ belonging to $U$ on $r$.
Finally, since there is a \markstep\ belonging to $U$ on $r$, we obtain from line~\ref{help-markstep} that $r$ is in $U.R$.
\end{chapscxproof}

\subsection{Correctness of \help} \label{sec-corr-help}

The following lemma shows that a helper of an \op\ cannot return \true\ until after the \op\ is \done.
We shall use this to ensure that the \sct\ does not return \true\ until after the \sct\ has taken effect.

\begin{lem} \label{lem-no-return-true-in-loop-until-uass}
%$\bigstar$ \textbf{(inline in L\ref{lem-if-fass-belongs-to-op-then}?)}
An invocation of \help$(scxPtr)$ where $scxPtr$ points to an \op\ $U$
cannot return from line~\ref{help-return-true-loop} before the first \cstep\ belonging to $U$.
\end{lem}
\begin{chapscxproof}
Suppose an invocation $H$ of \help$(scxPtr)$ returns
%\true\
at line~\ref{help-return-true-loop}.
Before returning, $H$ sees $r.\info \neq scxPtr$ at line~\ref{help-check-frozen}, which implies that $r.\info$ does not point to $U$.
Prior to this, $H$ performs a \fcas\ belonging to $U$ at line~\ref{help-fcas}.
By line~\ref{help-fcstep}, a \fstep\ belongs to $U$.
Then, since Lemma~\ref{lem-records-frozen} states that $r.\info$ points to $U$ at all times between the first \fcas\ belonging to $U$ and the first \cstep\ belonging to $U$, a \cstep\ belonging to $U$ must occur before $H$ returns.
%\qed
\end{chapscxproof}

Next, we obtain an exact characterization of the \upcas\ steps that succeed.

\begin{lem} \label{lem-first-upcas-succ}
If there is an \upcas \ belonging to an \op \ $U$, then the first \upcas \ belonging to $U$ is successful and changes the mutable field pointed to by $U.fld$ from $U.old$ to $U.new$.
No other \upcas \ belonging to $U$ is successful.
\end{lem}
\begin{chapscxproof}
Let $t_0(r)$ be when the \llt$(r)$ linked to $S$ reads $r\info.state$ at line~\ref{ll-read-state}, and $t_1$ be when the first \upcas\ belonging to $U$ occurs.
Since an \upcas\ belongs to $U$, Lemma~\ref{lem-fcas-then-upcas-no-change} implies that, for each $r$ in $U.V$, no mutable field of $r$ changes between $t_0(r)$ and $t_1$.
From Observation~\ref{obs-op-invariants}.\ref{inv-old} and the code of \llt, we see that the value stored in $U.old$ is read from the field pointed to by $U.fld$ after time $t_0(r)$.
Further, since the \llt$(r)$ linked to $S$ terminates before $S$ begins (by the definition of an \llt\ linked to an \sct) and, in turn, before $U$ is created, we know the value stored in $U.old$ was read before any \upcas\ belonging to $U$ occurred.
Then, since $U.fld$ is a mutable field of a \rec\ in $U.V$, this field does not change between $t_0(r)$ and $t_1$.
By Observation~\ref{obs-all-upcas-write-new}, any \upcas\ belonging to $U$ will attempt to change $U.fld$ from $U.old$ to $U.new$, so the first \upcas\ belonging to $U$ will succeed.
%Finally, by Lemma~\ref{lem-only-first-upcas-can-succeed}, no other \upcas\ belonging to $U$ can succeed.
Lemma~\ref{lem-only-first-upcas-can-succeed} completes the proof.
%\qed
\end{chapscxproof}

Our next lemma shows that the \sct s that are linearized have the desired effect.

\begin{lem} \label{lem-if-fass-belongs-to-op-then}
Let $U$ be an \op \ created by an invocation $S$ of \sct, and $ptr$ point to $U$.
If either 
\begin{compactitem}
\item
a \fstep\ belongs to $U$ and some invocation of \help$(ptr)$ terminates, or 
\item
$S$ or any invocation of \help$(ptr)$ returns \true,
\end{compactitem}
 then the following claims hold.
\begin{enumerate}
\item{
	Every invocation of \help$(ptr)$ that terminates returns \true.
} \label{claim-help-true-then-all-helpers-true}
\item{
	Exactly one successful \upcas\ belongs to $U$,
	and it is the first \upcas\ belonging to $U$.
	It changes the mutable field pointed to by $U.fld$
	from $U.old$ to $U.new$. %, where $U.old \neq U.new$.
} \label{claim-help-true-then-fass}
\item{
	A \fstep\ of $U$ and a \cstep\ of $U$
	occur before any invocation of \help$(ptr)$ returns.%
	% for linearization arguments as well as the next point!
}\label{claim-help-true-then-returns-after-uass}
\item{
	At all times after the first \cstep\ for $U$,
	each $r$ in $U.R$ is \permafrozen\ for $U$.
} \label{claim-help-true-then-finalized}
\end{enumerate}
\end{lem}
\begin{chapscxproof}
We first simplify the lemma's hypothesis.
If $S$ returns \true, then $S$'s invocation of \help$(ptr)$ has returned \true.
If some invocation of \help$(ptr)$ returns \true,
a \cstep\ belongs to $U$ by Lemma~\ref{lem-no-return-true-in-loop-until-uass},
and that \cstep\ is preceded by a \fstep\ of $U$.
So, for the remainder of the proof, we can assume that a \fstep\ belongs to $U$ and some
invocation of \help($ptr$) terminates.

%We first show that a \fstep \ belongs to $U$, and some invocation $H$ of \help$(ptr)$ terminates.
%Suppose $S$ returns \true.
%Then, $S$ performs an invocation $H$ of \help$(ptr)$ that returns \true.
%By Lemma~\ref{lem-no-return-true-in-loop-until-uass}, a \cstep \ belonging to $U$ occurs before $H$ 
%returns.
%From the code of \help, a \fstep\ must precede the first \cstep\ belonging to $U$.
%Now, suppose some invocation of \help$(ptr)$ returns \true.
%Then, as we argued above, there must be \fstep\ belonging to $U$.

\textbf{Proof of Claim~\ref{claim-help-true-then-all-helpers-true}:}
Since a \fstep \ belongs to $U$, Lemma~\ref{lem-fass-then-no-bcas} implies that no \astep \ belongs to $U$.
Thus, $H$ cannot return at line~\ref{help-return-false}, which implies that $H$ must return \true.

\textbf{Proof of Claim~\ref{claim-help-true-then-fass} and Claim~\ref{claim-help-true-then-returns-after-uass}:}
By Claim~\ref{claim-help-true-then-all-helpers-true}, $H$ must return \true.
If $H$ returns at line~\ref{help-return-true},
then it does so %immediately 
after performing an \upcas\ and a \cstep, each belonging to $U$.
Otherwise, $H$ returns at line~\ref{help-return-true-loop}.
However, by Lemma~\ref{lem-no-return-true-in-loop-until-uass}, no invocation of \help$(ptr)$ can return at line~\ref{help-return-true-loop} until the first \cstep\ belonging to $U$, which is necessarily preceded by an \upcas\ for $U$ (by inspection of \help).
Thus, Claim~\ref{claim-help-true-then-returns-after-uass} is proved.
Lemma~\ref{lem-first-upcas-succ} proves Claim~\ref{claim-help-true-then-fass}.

\textbf{Proof of Claim~\ref{claim-help-true-then-finalized}:}
By Corollary~\ref{cor-records-frozen-from-fass-to-uass}, every $r$ in $U.V$ is frozen for $U$ from the first \fstep\ belonging to $U$ until the first \cstep\ belonging to $U$.
From the code of \help, each $r$ in $U.R$ is marked before the first \cstep \ belonging to $U$, and Lemma~\ref{lem-marked-changes} implies that they are still marked when the first \cstep \ belonging to $U$ occurs.
Further, immediately after the first \cstep\ belonging to $U$ (which must exist by Claim~\ref{claim-help-true-then-returns-after-uass}), $U.state$ will be \done, so each $r$ that is in both $U.V$ and $U.R$ will be \permafrozen\ for $U$.
Since $R$ is a subsequence of $V$, and $U.R$ and $U.V$ do not change after they are obtained at line~\ref{sct-create-op} from $R$ and $V$, respectively, %$(U.V \cap U.R) = U.R$.
it follows from Lemma~\ref{lem-finalized-forever-frozen} that each $r$ in $U.R$ remains \permafrozen\ for $U$ forever.
%\qed
\end{chapscxproof}

Now we show that \sct s that are not linearized do not modify any mutable fields, and do not return \true.

\begin{lem} \label{lem-help-false}
Let $U$ be an \op\ created by an invocation $S$ of \sct, and $ptr$ be a pointer to $U$.
If $S$ or any invocation $H$ of \help$(ptr)$ returns \false, then the following claims hold.
\begin{enumerate}
\item{Every invocation of \help$(ptr)$ that terminates returns \false.}
\label{claim-help-false}
\item{An \astep \ belonging to $U$ occurs before any invocation of \help$(ptr)$ returns.}
\label{claim-help-false-abort}
\item{No \upcas\ belongs to $U$.}
\label{claim-help-false-no-upcas}
\end{enumerate}
\end{lem}
\begin{chapscxproof}
Note that, if $S$ returns \false, then its invocation of \help$(ptr)$ returns \false, so an invocation $H$ exists and returns \false.
By Lemma~\ref{lem-if-fass-belongs-to-op-then}.\ref{claim-help-true-then-all-helpers-true}, if any invocation of \help$(ptr)$ returned \true, then $H$ would have to return \true.
Since $H$ returns \false, every terminating invocation of \help$(ptr)$ must return \false, which proves Claim~\ref{claim-help-false}.
We now prove Claim~\ref{claim-help-false-abort} and Claim~\ref{claim-help-false-no-upcas}.
Consider the invocation $H'$ of \help$(ptr)$ that returns earliest.
By Claim~\ref{claim-help-false}, $H'$ returns \false.
Before $H'$ returns \false, an \astep\ belonging to $U$ is performed at line~\ref{help-astep}.
Thus, Lemma~\ref{lem-fass-then-no-bcas} implies that no \fstep\ belongs to $U$.
By the code of \help, each \upcas\ belonging to $U$ follows a \fstep\ belonging to $U$.
\end{chapscxproof}

Now that we have proved each invocation of \help\ that returns \true\ or \false\ has its expected effect, we must prove that the return value is always correct.  (Otherwise, for example, two invocations of \sct \ with overlapping $V$ sequences could interfere with one another, but still return \true.)

\begin{lem} \label{lem-help-true-iff-records-unchanged}
Let $U$ be an \op\ created by an invocation $S$ of \sct, and $ptr$ be a pointer to $U$.
Any invocation $H$ of \help$(ptr)$ that terminates returns \true\ if no $r$ in $U.V$ changes from when the \llt$(r)$ linked to $S$ reads $r.\info$ at line~\ref{ll-reread} at time $t_0(r)$ to when the first \fcas\ belonging to $U$ on $r$ at time $t_1(r)$.  Otherwise, $H$ returns \false.
\end{lem}
\begin{chapscxproof}
Since $H$ terminates, it must return \true\ or \false.
Hence, it suffices to prove $H$ returns \true\ if and only if no $r$ in $U.V$ changes between $t_0(r)$ and $t_1(r)$.%  We prove the ``only if'' direction in Case I.  However, we prove the contrapositive of the ``if'' direction in Case II, as this is easier to prove with the preceding lemmas.

\textbf{Case I: } Suppose $H$ returns \true.
By Lemma~\ref{lem-if-fass-belongs-to-op-then}.\ref{claim-help-true-then-returns-after-uass}, a \fstep\ belongs to $U$.  Hence, by Lemma~\ref{lem-if-fass-then-all-succ-fcas}, there is a successful \fcas\ belonging to $U$ for each $r$ in $U.V$.
Then, by Corollary~\ref{cor-read-unfrozen-info-then-fcas-no-change}, no field of $r$ changes between $t_0(r)$ and $t_1(r)$.
%% TODO: previous line requires a bit of thought, since you have to realize that the t_0(r) we are talking about here occurs after the t_0(r) talked about in the referenced corollary...  maybe this should be fixed?

\textbf{Case II: }
Suppose $H$ returns \false.  We show that some $r$ in $U.V$ changes between $t_0(r)$ and $t_1(r)$.
Since $H$ can only return \false\ at line~\ref{help-return-false}, immediately before $H$ returns, it performs an \astep\ belonging to $U$.
Thus, Lemma~\ref{lem-abort-fcas-flow}.\ref{abort-fcas-flow-claim-rk-changes} applies and the claim is proved.
%\qed
\end{chapscxproof}

%\hrule
%
%\vspace{1mm}
%\textit{(Lemmas not recently checked below here.)}

\subsection{Linearizability of \llt/\sct/\vlt} \label{sec-corr-lin}

\begin{ignore}
Linearizability is a correctness condition introduced by Herlihy and Wing \cite{HW90:toplas}.  A concurrent execution $\alpha$ is linearizable if linearization points can be selected for each linearizable operation such that the linearization point for an operation occurs during the operation, and the result of each operation in $\alpha$ is the same as it would be if the operations were executed atomically at their linearization points.  An algorithm is then linearizable if linearization points can be chosen such that every concurrent execution is linearizable.
\end{ignore}

As described in Section~\ref{sec-operations}, we linearize all reads, all invocations of \llt\ that do not return \fail, all invocations of \vlt\ that return \true, and all invocations of \sct\ that modify the sequential data structure (all that return \true, and some that do not terminate).
In our implementation, subtle interactions with a concurrent invocation of \sct\ can cause an invocation of \llt \ to return \fail.
Since this cannot occur in a linearized execution,  we do not linearize  any such invocation of \llt.
Similarly, we allow some invocations of \sct \ and \vlt \ to return \false \ because of interactions with concurrent invocations of \sct. 
%Consequently, $I$ may return \false\ because of an invocation of \sct \ that is linearized \textit{after} $I$.
Rather than distinguishing between the invocations of \sct \ or \vlt \ that return \false \ because of an earlier linearized invocation of \sct\ (which is allowed by the sequential specification), and those that return \false \ because of contention, we simply opt not to linearize any invocation of \sct \ or \vlt \ that returns \false.
Alternatively, we could have accounted for the invocations that returns \false \ because of contention by allowing spurious failures in the sequential specification of the operations.
However, this would unnecessarily complicate the sequential specification.
Intuitively, an algorithm designer using \llt, \sct \ and \vlt \ is most likely to be interested in invocations of \sct \ and \vlt \ that return \true\ since these, respectively, change the sequential data structure, and indicate that a set of \rec\ have not changed since they were last passed to successful invocations of \llt\ by this process.
Knowing whether an invocation of \sct \ or \vlt \ was unsuccessful because of a change to the sequential data structure, or because of contention, is less likely to be useful.

Before we give the linearization points, we state precisely which invocations of \sct \ we shall linearize.
Let $S$ be an invocation of \sct, and $U$ be the \op\ it creates.
We linearize $S$ if and only if there is an \upcas \ belonging to $U$.
By Lemma~\ref{lem-if-fass-belongs-to-op-then}, every successful invocation of \sct\ will be linearized.
(By Lemma~\ref{lem-help-false}, no unsuccessful invocation of \sct\ will be linearized.)

We first give the linearization points of the operations, then we prove our \llt/\sct/\vlt\ implementation respects the correctness specification given in Section~\ref{sec-operations}.
\\

\noindent\textbf{Linearization points:}
\begin{itemize}
	\item{
		An \llt$(r)$ that returns values at line~\ref{ll-return}
		is linearized at line~\ref{ll-reread}.
		We linearize an \llt$(r)$ that returns \finalized\
		at line~\ref{ll-return-finalized}.
	}
	\item{
		Let $U$ be an \op \ created by an invocation $S$ of \sct. 
		Suppose there is an \upcas \ belonging to $U$.
		We linearize $S$ at the first such \upcas\
		(which is the unique successful \upcas \ belonging to $U$,
		 by Lemma~\ref{lem-first-upcas-succ}).
	}
	\item{
		An invocation $I$ of \vlt\ that returns \true\ is linearized
		at the first execution of line~\ref{vlt-reread}.
	}
	\item{
		We assume reads are atomic.
		Hence, a read is simply linearized when it occurs.
	}
\end{itemize}

\begin{lem} \label{lem-lin-points-during-ops}
The linearization point of each linearized operation occurs during the operation.
\end{lem}
\begin{chapscxproof}
This is trivial to see for reads, and invocations of \llt\ and \vlt.
Let $U$ be an \op \ created by a linearized invocation $S$ of \sct, and $ptr$ be a pointer to $U$.
This claim is not immediately obvious for $S$, since the first \upcas \ belonging to $U$ may be performed by a process helping $U$ to complete (not the process performing $S$).
Since $S$ is linearized, there is an \upcas \ $upcas$ belonging to $U$, which can only occur in an invocation of  \help$(ptr)$.
Since $ptr$ points to $U$, which is created by $S$, $upcas$ must occur after the start of $S$.
If $S$ does not terminate, then we are done.
Otherwise, Lemma~\ref{lem-if-fass-belongs-to-op-then}.\ref{claim-help-true-then-returns-after-uass} implies that a \cstep \ belongs to $U$, and the first such \cstep \ occurs before any invocation of \help$(ptr)$ returns.
From the code of \help, $upcas$ must occur before before the first \cstep \ belonging to $U$, so $upcas$ must occur before any invocation of \help$(ptr)$ returns.
Finally, since $S$ invokes \help$(ptr)$, $upcas$ must occur before $S$ terminates.
%
%Since $I$ returns \true, its invocation of \help$(ptr)$ must return \true.
%By Lemma~\ref{lem-if-fass-belongs-to-op-then}.\ref{claim-help-true-then-returns-after-uass}, a \cstep\ belonging to $U$ occurs before any invocation of \help$(ptr)$ returns.
%From the code, any \cstep\ belonging to $U$ is preceded by an \upcas\ belonging to $U$.
%Since $I$ is linearized at the first \upcas\ belonging to $U$, $I$ must be linearized before it terminates.
%Finally, since any \upcas\ belonging to $U$ must occur in an invocation of \help$(ptr)$, and no such invocation can begin before $U$ is created by $I$, we know $I$ must be linearized after it begins.
%\qed
\end{chapscxproof}

We first show that each read returns the correct result according to its linearization point.

\after{Shorten the next proof.  It's too pedantic.}

\begin{lem} \label{lem-lin-read}
If a read $R_f$ of a field $f$ is linearized after a successful invocation of \sct$(V, R, fld, new)$, where $fld$ points to $f$, then $R_f$ returns the parameter $new$ of the last such \sct.
Otherwise, $R_f$ returns the initial value of $f$.
%%Version 3
%Consider a read $R_f$ of a field $r.f$.
%Let $S$ be the last successful invocation of \sct$(V, R, fld, new)$,
%with $fld$ a pointer to $r.f$, that is linearized before $R_f$.
%If $S$ exists, then $R_f$ returns $new$.
%%Version 1
%Consider a read $R_f$ of a field $r.f$.
%If $R_f$ is linearized after a successful invocation of
%\sct$(V, R, fld, new)$, with $fld$ a pointer to $r.f$,
%then $R_f$ returns the argument $new$ passed to the last such \sct.
%%Version 2
%If a read $R_f$ of a field $r.f$ occurs after a successful invocation of
%\sct$(V, R, fld, new)$, with $fld$ a pointer to $r.f$,
%has been linearized, then $R_f$ returns the argument $new$
%passed to the last such \sct.
%Otherwise, $R_f$ returns the initial value of $r.f$.
\end{lem}
\begin{chapscxproof}
We proceed by cases.

\textbf{Case I:} $f$ is an immutable field.
In this case, a pointer to $f$ cannot be the $fld$ parameter of an invocation of \sct.
Further, by Observation~\ref{obs-immutable-fields-do-not-change}, $f$ cannot be modified after its initialization, so $R_f$ returns the initial value of $f$.

\textbf{Case II:} $f$ is a mutable field.
Suppose there is no successful invocation of \sct$(V, R, fld, new)$, where $fld$ points to $f$, linearized before $R_f$.
Since an invocation of \sct \ is linearized at its first \upcas, there can be no \upcas \ on $f$ prior to $ R_f $.
Since $f $ can only be modified by successful \upcas, $ R_f $ must return the initial value of $f $.

Now, suppose there is a successful invocation of \sct$(V, R, fld, new)$, where $fld$ points to $f$, linearized before $R_f$.
Let $S$ be the last such invocation of \sct\ linearized before $R_f$, and $U$ be the \op \ it creates.
By Lemma~\ref{lem-if-fass-belongs-to-op-then}.\ref{claim-help-true-then-fass}, there is exactly one successful \upcas \ $upcas$ belonging to $U$, occurring at the linearization point of $S$.
Since each successful \upcas \ is the linearization point of some invocation of \sct, and no invocation of \sct$ (V', R', fld, new') $ is linearized between $S$ and $ R_f $, no successful \upcas \ occurs between $S$ and $ R_f $.
Since a mutable field can only be changed by \upcas, $ R_f $ returns the value stored by the successful \upcas \ $upcas$ belonging to $U$.
By Lemma~\ref{lem-if-fass-belongs-to-op-then}.\ref{claim-help-true-then-fass}, $upcas$ changes $f $ to $U.new$.
Finally, since $U.new$ does not change after it is obtained from $new$ at line~\ref{sct-create-op}, the claim is proved.
%\qed
\end{chapscxproof}

Next, we prove that an \llt\ that returns a snapshot does return the correct result according to its linearization point.

\begin{cor} \label{cor-lin-llt-success}
Let $r$ be a \rec \ with mutable fields $f_1, ..., f_y$, and $I$ be an invocation of $\llt(r)$ that returns a tuple of values $\langle m_1, ..., m_y \rangle$ at line~\ref{ll-return}. %, one for each mutable field of $r$.
For each mutable field $f_i$ of $r$, if $I$ is linearized after a successful invocation of \sct$(V, R, fld, new)$, where $fld$ points to $f_i$, then $m_i$ is the parameter $new$ of the last such invocation of \sct.
Otherwise, $m_i$ is the initial value of $f_i$.
%A successful \llt$(r)$ returns the last value written to each mutable field $f$ of $r$ by an \sct\ (or $f$'s initial value, if no \sct\ has modified $f$).
\end{cor}
\begin{chapscxproof}
Since $I$ returns at line~\ref{ll-return}, 
%$r\info$ is read from $r.\info$ at line~\ref{ll-read} at some time $t_0$, and $r.\info$ is reread and seen to contain $r\info$ at line~\ref{ll-reread} at some later time $t_1$.
the same value is read from $r.info$ on line~\ref{ll-read} and line~\ref{ll-reread} at times $t_0$ and $t_1$, respectively. 
By Lemma~\ref{lem-r-not-frozen-in-good-llt}, $r$ is unfrozen at line~\ref{ll-check-frozen}, which is between $t_0$ and $t_1$.
Thus, Lemma~\ref{lem-read-unfrozen-info-twice-no-change} implies that $r$ does not change during $[t_0,t_1]$.
Since the values returned by the \llt\ are read from the fields of $r$ between $t_0$ and $t_1$ (at line~\ref{ll-collect}), each read returns the same result as it would if it were executed atomically at the linearization point of the \llt\ (line~\ref{ll-reread}).
Finally, Lemma~\ref{lem-lin-read} completes the proof.
%\qed
\end{chapscxproof}

Next, we show that an \llt$(r)$ returns \finalized\ only if $r$ really has been finalized.

\begin{lem} \label{lem-lin-llt-finalized}
Let $I$ be an invocation of \llt$(r)$, and $U$ be the \op \ to which $I$ reads a pointer at line~\ref{ll-read}.
If $I$ returns \finalized, then an invocation $S$ of \sct$ (V, R, fld, new) $ that created $U$ is linearized before $I$, and $r$ is in $R$.
\end{lem}
\begin{chapscxproof}
%Let $U$ be the \op \ pointed to by $r\info$, and $S$ be the \sct\ that creates $U$.
%Let $U$ be the \op\ to which $I$ reads a pointer at line~\ref{ll-read}.
Suppose $I$ returns \finalized.
Then, $I$ is linearized at line~\ref{ll-return-finalized}.
At line~\ref{ll-help}, $I$ will either see $state \neq \freezing$ or will invoke \help$(r\info)$ (where $r\info$ is a pointer to $U$).
Thus, when $I$ is linearized, $U.state$ is either \retry\ or \done.
Since $I$ sees $marked_1 = \true$, $r$ is marked when $I$ performs line~\ref{ll-read-marked1}.
By Lemma~\ref{lem-state-transitions2}, once $r$ is marked, it can never again point to an \op\ with $state$ \retry.
Therefore, $U.state$ must be \done\ when $I$ is linearized.
%%
%%
%%
%idea:
%    either see state != inprog or help
%    thus, r\info.state is aborted or committed when I is linearized
%    since marked1 implies that r is marked, Lemma~\ref{lem-state-transitions2} implies that r cannot point to an scx record w/state aborted. so r\info.state must be committed.
%%
%%
%%
%When $I$ performs line~\ref{ll-check-finalized}, either $r\info.state$ is \done \ or $I$'s invocation of \help$(r\info)$ returns \true.
%We show that, in either case, a \cstep \ belonging to $U$ must have occurred before $I$ performs line~\ref{ll-return-finalized}.
%If $r\info.state$ is \done, then this follows immediately from the fact that no \op \ has \done\ as its initial state.
%Otherwise, Lemma~\ref{lem-if-fass-belongs-to-op-then}.\ref{claim-help-true-then-returns-after-uass} implies that a \cstep \ belonging to $U$ occurs before $I$'s invocation of \help$(r\info)$ returns.
Since $U.state = \done$, Lemma~\ref{lem-no-steps-belong-to-dummy-op} implies that $U$ is not the dummy \op, so there must be an invocation $S$ of \sct$(V, R, fld, new)$ that created $U$.
From the code of \help, an \upcas\ belonging to $U$ occurs before the first \cstep\ belonging to $U$.
Since $S$ is linearized at its first \upcas, $S$ is linearized before $I$.

It remains to show that $r$ is in $R$.
Since $U.R$ does not change after it is obtained from $R$ at line~\ref{sct-create-op}, it suffices to show $r$ is in $U.R$.
By line~\ref{ll-check-finalized}, $marked_1 = \true$, which means that $r$ is marked when $I$ reads a pointer to $U$ from $r.\info$ at line~\ref{ll-read}.
%%%%%%%%%%%%%%%%%
Let $t_0$ be when $I$ performs line~\ref{ll-read}, and $t_1$ be when the first \cstep \ belonging to $U$ occurs.
We consider two cases.
Suppose $t_1 < t_0$.
Then, when $I$ performs line~\ref{ll-read}, $r$ is marked, $r.\info$ points to $U$, and $U.state = \done$, which means that $r$ is \permafrozen\ for $U$.
By Lemma~\ref{lem-finalized-forever-frozen}, $r$ is frozen for $U$ at all times after $t_0$.
Now, suppose $t_0 < t_1$.
In this case, Lemma~\ref{lem-state-transitions2} implies $U.state = \freezing$ when $I$ performs line~\ref{ll-read}, and that $U.state$ will never be \retry.
By Lemma~\ref{lem-no-info-change-while-freezing}, $r.\info$ must point to $U$ at all times in $[t_0, t_1]$.
Thus, at $t_1$, $r$ is marked and $r.\info$ points to $U$, which means that $r$ is \permafrozen\ for $U$.
%
%From the code of \help, a \fstep\ belonging to $U$ must precede the first \cstep \ belonging to $U$.
%By Corollary~\ref{cor-records-frozen-from-fass-to-uass}, $r\info$ must point to $U$ at all times between the first \fstep \ belonging to $U$ and the first \cstep \ belonging to $U$.
%
By Lemma~\ref{lem-finalized-forever-frozen}, $r$ is frozen for $U$ at all times after $t_1$.
In each case, $r$ is frozen for $U$ at all times in some (non-empty) suffix of the execution.
Since $r$ is marked, there must be a successful \markstep\ belonging to some \op\ $W$ on $r$.
By Lemma~\ref{lem-frozen-forever-after-markstep}, $r$ is frozen for $W$ at all times after this \markstep.
Therefore, $U = W$, which means there is a \markstep\ belonging to $U$ on $r$.
Finally, line~\ref{help-markstep} implies that $r$ is in $U.R$.
%\qed
\end{chapscxproof}

\begin{lem} \label{lem-lin-llt-finalized-if-after-sct}
Let $r$ be a \rec, $I$ be an invocation of \llt$(r)$ that terminates, and $S$ be a linearized invocation of \sct$(V, R, fld, new)$ with $r$ in $R$.
$I$  returns \finalized\ if it is linearized after $S$, or begins after $S$ is linearized.
(This implies $I$ will be linearized in both cases.)
%(This implies $I$ will be linearized if it begins after $S$ is linearized.)
\end{lem}
\begin{chapscxproof}
Let $U$ be the \op\ created by $S$.

\textbf{Case I:} $I$ begins after $S$ is linearized.
In this case, $S$ is linearized at a successful \upcas\ $upcas$ belonging to $U$ that occurs before $I$ begins.
From the code of \help, a \markstep\ on $r$ belonging to $U$ must occur before $upcas$.
Consider the first \markstep\ $mstep$ on $r$.
Since $r.marked$ is initially \false, $mstep$ must be successful.
Let $W$ be the \op\ to which $mstep$ belongs.
By Lemma~\ref{lem-frozen-forever-after-markstep}, $r$ is frozen for $W$ at all times after $mstep$.
By Lemma~\ref{cor-upcas-only-modifies-frozen}, $r$ is frozen for $U$ when $upcas$ occurs (which is after $mstep$).
Since $r$ can be frozen for only one \op\ at a time, $W = U$.
Therefore, $r.\info$ points to $U$ throughout $I$.
So, when $I$ performs line~\ref{ll-check-frozen}, it will see $state = \done$ and $marked_2 = \true$.
Moreover, when it subsequently performs line~\ref{ll-check-finalized}, it will see %$r\info.state = \done$ and 
$marked_1 = \true$, so it will return \finalized.
%This implies that $I$ will not enter the if-block at line~\ref{ll-check-frozen}.
%
%\eric{I don't really understand the rest of the proof of this case, since $rinfo$ points to $W$, not $U$.}
%Suppose a \cstep \ belonging to $U$ occurs before $I$ performs line~\ref{ll-check-finalized}.
%Then, Lemma~\ref{lem-state-transitions2} implies that $r\info.state$ will be \done \ when $I$ performs line~\ref{ll-check-finalized}, and $I$ will return \finalized.
%
%Now, suppose no \cstep \ belonging to $U$ occurs before $I$ performs line~\ref{ll-check-finalized}.
%From the code of help, a \fstep \ belonging to $U$ must occur before the first \upcas \ belonging to $U$ (which precedes  line~\ref{ll-check-finalized}).
%By Lemma~\ref{lem-state-transitions2}, $r\info.state$ will be \freezing \ when $I$ performs line~\ref{ll-check-finalized}, so $I$ will invoke \help$(r\info)$.
%By Lemma~\ref{lem-if-fass-belongs-to-op-then}.\ref{claim-help-true-then-all-helpers-true}, this invocation of \help$(r\info)$ will succeed, and $I$ will return \finalized.

\textbf{Case II:} $I$ is linearized after $S$.
$I$ can either be linearized at line~\ref{ll-reread} or at line~\ref{ll-return-finalized}.
If it is linearized at line~\ref{ll-return-finalized}, then it returns \finalized, and we are done.
Suppose, in order to derive a contradiction, that $I$ is linearized at line~\ref{ll-reread}.
Then, $I$ returns at line~\ref{ll-return} and, by Corollary~\ref{cor-r-not-frozen-in-linked-llt}, $r$ is unfrozen  at all times during $[t_0, t_1]$, where $t_0$ is when $I$ performs line~\ref{ll-read-state}, and $t_1$ is when $I$ performs line~\ref{ll-reread}.
Since $I$ is linearized at time $t_1$, $S$ must be linearized at an \upcas \ $upcas$ belonging to $U$ that occurs before time $t_1$.
By Corollary~\ref{cor-upcas-only-modifies-frozen}, $upcas$ can only occur while $r$ is frozen for $U$.
Since $r$ is unfrozen at all times during $[t_0,t_1]$, $upcas$ must occur at some point before $t_0$.
From the code of \help, a \markstep\ belonging to $U$ must occur before $upcas$.
Consider the first \markstep\ $mstep$ belonging to any \op\ $W$ on $r$.
As we argued in the previous case, $r$ is frozen for $W$ at all times after $mstep$.
However, this contradicts our argument that $r$ is unfrozen at all times during $[t_0,t_1]$ (since $t_0$ is after $mstep$).
Thus, $I$ cannot be linearized at line~\ref{ll-reread}, so $I$ must return \finalized.
%
%
%
%
%
%
%
%Hence, Corollary~\ref{cor-records-frozen-from-fass-to-uass} applies, and $r$ is frozen for $U$ at all times between $upcas$ and the first \cstep \ belonging to $U$.
%Since we have argued above that $r$ is unfrozen when $I$ performs line~\ref{ll-read-state}, a \cstep \ $cstep$ belonging to $U$ must occur before line~\ref{ll-read-state}.
%
%From line~\ref{ll-check-finalized}, and the fact that $r\info.R$ does not change after it is created by $S$ at line~\ref{sct-create-op}, we know $r \in r\info.R $ immediately after $S$ performs line~\ref{sct-create-op}.
%Hence, $r \in R$.
%Thus, immediately after $cstep$, $r$ is finalized and frozen for $U$.
%Further, by Lemma~\ref{lem-finalized-forever-frozen}, $r$ is finalized and frozen for $U$ at all times after $cstep$.
%Since $cstep$ occurs before time $t_0$, $I$ must read \done\ from $r\info.state$ at time $t_0$, and see $r\in r\info.R $ at line~\ref{ll-r-not-in-R}.
%Thus, $I$ will see $isFrozen = \true$ at line~\ref{ll-check-frozen}, and will not return at line~\ref{ll-return}, which contradicts our assumption.
%\qed
\end{chapscxproof}

%\begin{lem} \label{lem-info-read-in-linked-llt}
%In the iteration for \rec\ $r$ of the loop of \vlt\ (\sct), the value stored in local variable $r\info$ was read from $r.\info$ by the \llt$(r)$ linked to the invocation $I$ of \vlt\ (\sct).
%\end{lem}
%\begin{chapscxproof}
%By Precondition~\prevltinfo\ of \vlt\ (Precondition~\presctinfo\ of \sct), there must be an \llt$(r)$ linked to $I$. 
%From line~\ref{vlt-info} (line~\ref{help-rinfo}) and Observation~\ref{obs-op-invariants}.\ref{inv-llresults}, we see that, in the iteration of the loop of \vlt\ (\sct) for $r$, the value stored in $r\info$ is read at line~\ref{ll-read} of \llt$(r)$ linked to $I$.
%%\qed
%\end{chapscxproof}

The following lemma proves that an \sct\ succeeds only when it is supposed to, according to the
correctness specification.

\begin{lem} \label{lem-lin-sct-vlt}
If an invocation $I$ of \vlt$(V)$ or \sct$(V, R, fld, new)$ is linearized then, for each $r$ in $V$, no \sct$(V', R', fld', new')$ with $r$ in $V'$ is linearized between the \llt$(r)$ linked to $I$ and $I$.
\end{lem}
\begin{chapscxproof}
Fix any $r$ in $V$. 
By the preconditions %Precondition~\prevltinfo\ 
of \sct\ and \vlt, there must be an \llt$(r)$ linked to $I$.
If $I$ is an invocation of \sct, then let $U$ be the \op \ that it creates.
Let $L$ be the $\llt(r)$ linked to $I$, $t_0$ be when $L$ performs line~\ref{ll-read}, $t_1$ be when $L$ performs line~\ref{ll-read-state} and $t_2$ be when $L$ is linearized (at line~\ref{ll-reread}).
Since $L$ is linked to $I$, $t_2$ exists.
Let $t_3$ be when $I$ is linearized.
For \sct, $t_3$ is when the first \upcas\ belonging to $U$ occurs.
For \vlt, $t_3$ is when $I$ first performs line~\ref{vlt-reread}.
Since $I$ is linearized, $t_3$ exists.
Finally, we define time $t_4$.
For \sct, $t_4$ is when the first \cstep\ belonging to $U$ occurs, or the end of the execution if there is no \cstep\ belonging to $U$ (or $\infty$ if the execution is infinite).
For \vlt, $t_4$ is when $I$ sees $r\info = r.\info$ at line~\ref{vlt-reread} (in the iteration for $r$).
If $I$ is an invocation of \vlt\ then, since $I$ is linearized, it returns \true.
This implies that $I$ must see $r\info = r.\info$ at line~\ref{vlt-reread} in the iteration for $r$, so $t_4$ exists.
Clearly, $t_4$ exists if $I$ is an invocation of \sct.
%Otherwise, if $I$ is an invocation of \sct \ then, since $I$ returns \true, Lemma~\ref{lem-if-fass-belongs-to-op-then}.\ref{claim-help-true-then-fass} implies that a \cstep \ belonging to $U$ occurs, so $t_4 $ exists.

We now prove $ t_0 <t_1 <t_2 <t_3 <t_4$.
From the code of \llt, $t_0 <t_1 <t_2$.
Suppose $I$ is an invocation of \vlt.
Since $L$ terminates before $I$ begins, $t_2 < t_3$.
Trivially, $t_3 < t_4$.
Now, suppose $I$ is an invocation of \sct.
Since each \upcas \ belonging to $U$ occurs in an invocation of \help$(ptr)$ where $ptr$ points to $U$, and $U$ is created during $I$, $t_3 $ must occur after the start of $I$.
Since $L$ terminates before $I$ begins, $t_2 < t_3$.
From the code of \help, the first \upcas \ belonging to $U$ precedes the first \cstep \ belonging to $U$ (as well as the other options for $t_4$), so $t_3 < t_4$.

Next, we prove that, at all times in $[t_1, t_4)$, $r$ is either frozen for $U$, or not frozen (i.e., $r$ is not frozen for any \op \ different from $U$ at any point during $[t_1, t_4)$).
We consider two cases.

\textbf{Case I:}
Suppose $I$ is an invocation of \vlt.
Then, $r.\info$ contains $r\info$ at $t_0$, and again at $t_4$.
By Lemma~\ref{lem-no-aba-info}, $r.\info$ must contain $r\info$ at all times in $[t_0, t_4]$.
By Corollary~\ref{cor-r-not-frozen-in-linked-llt}, $r$ is unfrozen at time $t_1$.
By Lemma~\ref{lem-become-frozen-only-by-info-change}, $r$ can only be changed from unfrozen the frozen by a change to $r.\info$.
Since $r$ does not change during $[t_0, t_4]$, $r$ is unfrozen at all times in $[t_1, t_4]$.

\textbf{Case II:}
Suppose $I$ is an invocation of \sct.
From the code of \help, a \fstep\ belonging to $U$ precedes the first \upcas \ belonging to $U$.
By Lemma~\ref{lem-if-fass-then-all-succ-fcas}, a successful \fcas \ belonging to $U$ on $r$ precedes the first \fstep \ belonging to $U$.
Let $t_2'$ be when the first successful \fcas \ {\it fcas} belonging to $U$ on $r$ occurs.
It follows that $t_2' < t_3$.
Since each  \fcas \ belonging to $U$ occurs in an invocation of \help$(ptr)$ where $ptr$ points to $U$, and $U$ is created during $I$, each \fcas \ belonging to $U$ must occur after the start of $I$.
Recall that $L$ terminates before the start of $I$.
Hence, $t_2' > t_2$.
By Observation~\ref{obs-op-invariants}.\ref{inv-llresults} and line~\ref{help-rinfo}, the old value for {\it fcas} is the value $r\info$ that was read from $r.\info$ at line~\ref{ll-read} of $L$ (at $t_0$).
Since {\it fcas} is successful, $r.\info$ must contain $r\info$ just before {\it fcas}.
Therefore, $r.\info$ contains $r\info$ at $t_0$, and again at $t_2'$.
By the same argument we made in Case I (but with $t_2'$ instead of $t_4$), Lemma~\ref{lem-no-aba-info}, Corollary~\ref{cor-r-not-frozen-in-linked-llt}, and Lemma~\ref{lem-become-frozen-only-by-info-change} imply that $r$ is unfrozen at all times in $[t_1, t_2']$.
By Lemma~\ref{lem-records-frozen}, $r$ is frozen for $U$ at all times in $(t_2',t_4)$, which proves this case.

At last, we have assembled the results needed to obtain a contradiction.
Suppose, to derive a contradiction, that an invocation $S$ of \sct$(V', R', fld', new')$ with $r$ in $V'$ is linearized between $L$ and $I$ (i.e., in $(t_2, t_3)$).
Let $W$ be the \op \ created by $S$.
$S$ is linearized at the first \upcas \ $upcas$ belonging to $W$.
From the code of \help, a \fstep \ belonging to $W$ precedes $upcas$, and $upcas$ precedes any \cstep  \ belonging to $W$.
By Lemma~\ref{lem-records-frozen}, a successful \fcas \ {\it fcas} belonging to $W$ on $r$ precedes $upcas$, and $r$ is frozen for $W$ at all times after {\it fcas}, and before the first \cstep\ belonging to $W$.
Therefore, $r$ is frozen for $W$ when $upcas$ occurs in $(t_2, t_3)$.
Since, at all times in $[t_1, t_4)$, $r$ is either frozen for $U$, or not frozen, we must have $W = U$.
This is a contradiction, since $upcas$ occurs before the \textit{first} \upcas \ belonging to $U$ occurs (at $t_3$).
Thus, $S$ cannot exist.
%\qed
\end{chapscxproof}

\begin{thm}
\label{thm-llt-sct-vlt-correct} % !!! NOTE:  2 parts of this lemma have been moved
% to lemma \ref{finalize-progress} to mirror the organization in Sec 3 (because they describe
% progress properties, so if you used the label thm-llt-sct-vlt-correctness, make sure you 
% update it to the right one.
%
%
Our implementation of \llt/\sct/\vlt\ satisfies the correctness specification discussed in Section~\ref{sec-operations}.
That is, we linearize all successful \llt s, all successful \sct s, a subset of the \sct s that never terminate, all successful \vlt s, and all reads, such that:
\begin{enumerate}
	\item
		Each read of a field $f$ of a \rec\ $r$ returns
		the last value stored in $f$ by a linearized \sct\
		(or $f$'s initial value, if no linearized \sct\ has modified $f$).
	    \label{claim-llt-sct-vlt-correctness-read}
	\item{%
  		Each linearized \llt($r$) that does not return \finalized\
		returns	the last value stored in each mutable field $f$ of $r$
		by a linearized \sct\ (or $f$'s initial value, if no linearized \sct\ has modified~$f$).}%
        \label{claim-llt-sct-vlt-correctness-llt-values}
	\item{
		Each linearized \llt$(r)$ returns \finalized\ if and only if it is linearized
		after an \sct($V, R, fld,$ $new)$ with $r$ in $R$.}
        \label{claim-llt-sct-vlt-correctness-llt-finalized-if-lin-after-sct}
	\item
	    If an invocation $I$ of \sct$(V, R, fld, new)$ or \vlt$(V)$ returns \true\ then,
	    for all $r$ in $V$, there has been no \sct$(V', R', fld', new')$
	    with $r$ in $V'$ linearized since the \llt$(r)$ linked to $I$.
	    \label{claim-llt-sct-vlt-correctness-sct-vlt}
\end{enumerate}
\end{thm}
\begin{chapscxproof}
By Lemma~\ref{lem-lin-points-during-ops}, the linearization point of each operation occurs during that operation.
Claim~\ref{claim-llt-sct-vlt-correctness-read} follows immediately from Lemma~\ref{lem-lin-read}.
Claim~\ref{claim-llt-sct-vlt-correctness-llt-values} is immediate from Lemma~\ref{cor-lin-llt-success}.
The only-if direction of Claim~\ref{claim-llt-sct-vlt-correctness-llt-finalized-if-lin-after-sct} follows from Lemma~\ref{lem-lin-llt-finalized}, and the if direction follows from Lemma~\ref{lem-lin-llt-finalized-if-after-sct}.
Claim~\ref{claim-llt-sct-vlt-correctness-sct-vlt} is immediate from Lemma~\ref{lem-lin-sct-vlt}.
\end{chapscxproof}

\subsection{Progress guarantees}

\begin{lem} \label{lem-wait-free}
\llt, \sct \ and \vlt \ are wait-free
\end{lem}
\begin{chapscxproof}
%Each invocation of \sct\ performs an invocation $H$ of \help. 
The loop in $H$ iterates over the elements of the finite sequence $V$ and performs a constant amount of work during each iteration.  If $H$ does not return from the loop, then it performs a constant amount of work after the loop and returns. The claim then follows from the code.
%\qed
\end{chapscxproof}

\begin{lem}\label{finalize-progress} Our implementation satisfies the first progress property in Section \ref{sec-operations}:  Each terminating \llt$(r)$ returns \finalized\ if it begins after the end of a successful \sct($V, R,$ $fld, new$) with $r$ in $R$ or after another \llt$(r)$ has returned \finalized. 
%\begin{enumerate}
%	\item
%	    Each \llt$(r)$ that terminates
%	    returns \finalized\ if it %is linearized after $S$, or
%	    begins after the end of an invocation of \sct$(V, R, fld, new)$ with $r$ in $R$. %is linearized.
%	    \label{claim-llt-sct-vlt-correctness-llt-finalized-if-start-after-sct}
%	\item 
%	    If invocation $I$ of $\llt(r)$ that returns \finalized\ precedes
%	    an invocation $I'$ of $\llt(r)$ that terminates,
%	    then $I'$ returns \finalized.
%	    \label{claim-llt-sct-vlt-correctness-llt-finalized-if-after-finalized}
%\end{enumerate}
\end{lem}
\begin{chapscxproof}
Consider a terminating invocation $I'$ of \llt($r$).  
%Claim~\ref{claim-llt-sct-vlt-correctness-llt-finalized-if-start-after-sct} also follows from Lemma~\ref{lem-lin-llt-finalized-if-after-sct}.
If $I'$ begins after the end of a successful \sct($V,R,fld,new$) with $r$ in $R$, the claim follows from Lemma~\ref{lem-lin-llt-finalized-if-after-sct}.
If $I'$ begins after another invocation $I$ of \llt$(r)$ has returned \finalized, then
%Claim~\ref{claim-llt-sct-vlt-correctness-llt-finalized-if-after-finalized} requires a small amount of reasoning.
%By Lemma~\ref{lem-lin-llt-finalized}, 
$I$ is linearized after an invocation $S$ of \sct$(V, R, fld, new)$ with $r$ in $R$.
Since $I$ precedes $I'$, $I'$ starts after $S$ is linearized.
By Lemma~\ref{lem-lin-llt-finalized-if-after-sct}, $I'$ returns \finalized.
\end{chapscxproof}

We now begin to prove the non-blocking progress properties.
First, we describe how we assign blame to an \sct\ for each failed invocation of \llt, \vlt\ or \sct.

\begin{defn} \label{defn-blame-llx}
Let $I$ be an invocation of \llt \ that returns \fail.
If $I$ enters the if-block at line~\ref{ll-check-frozen}, then let $U$ be the \op \ pointed to by $r.\info$ when $I$ performs line~\ref{ll-reread}.
Otherwise, let $U$ be the \op \ pointed to by $r.\info$ when $I$ performs line~\ref{ll-read}.
We say $I$ \textbf{blames} the invocation $S$ of \sct \ that created $U$.
(We prove below that $S$ exists.)
\end{defn}

\begin{lem} \label{lem-if-llt-fail-then-blame-sct}
If an invocation $I$ of \llt\ returns \fail, then it blames some invocation of \sct.
\end{lem}
\begin{chapscxproof}
Suppose $I$ enters the if-block at line~\ref{ll-check-frozen}.
Then, $I$ must see $r.\info \neq r\info $ at line~\ref{ll-reread}.
Let $U$ be the \op \ pointed to by $r.\info$ when $I$ performs line~\ref{ll-reread}.
Since $I$ reads $r\info$ from $r.\info$ at line~\ref{ll-read}, $r.\info$ must change to point to $U$ between when $I$ performs line~\ref{ll-read} and line~\ref{ll-reread}. 
Thus, there must be a successful \fcas \ belonging to $U$ on $r$ between these two times.
Since a successful \fcas \ belongs to $U$, Lemma~\ref{lem-no-steps-belong-to-dummy-op} implies that $U$ cannot be the dummy \op.
Therefore, $U$ must be created by an invocation of \sct.

Now, suppose $I$ does not enter the if-block at line~\ref{ll-check-frozen}.
Then, from the code of \llt, $r\info.state$ cannot be \retry \ when $I$ performs line~\ref{ll-read-state}, so the \op \ pointed to by $r\info$ is not the dummy \op.
Since $I$ reads the value stored in $r\info$ from $r.\info$ at  line~\ref{ll-read}, the \op \ pointed to by $r.\info$ when $I$ performs line~\ref{ll-read} must have been created by an invocation of \sct.
%\qed
\end{chapscxproof}

\begin{defn} \label{defn-blame-vlt}
Let $I$ be an invocation of \vlt$(V)$ that returns \false, and $r$ be the \rec\ in $V$ for which $I$ sees $ r.\info \neq r\info $ at line~\ref{vlt-reread}.
Consider the first successful \fcas \ on $r$ between when the \llt$(r)$ linked to $I$ reads $r.\info$ at line~\ref{ll-read}, and when $I$ sees $ r.\info \neq r\info $ at line~\ref{vlt-reread}.
Let $U$ be the \op \ to which this \fcas \ belongs, and $S$ be the invocation of \sct \ that created $U$.
(We prove below that $S$ exists.)
We say $I$ \textbf{blames} $S$ \textbf{for} $r$.
\end{defn}

\begin{lem} \label{lem-if-vlt-false-then-blame-sct}
If an invocation $I$ of \vlt \ returns \false, then it blames some invocation of \sct.
\end{lem}
\begin{chapscxproof}
Since $I$ returns \false, it sees $ r\info \neq r.\info $ at line~\ref{vlt-reread}, for some $r$ in $V$.
Let $p$ be the process that performs $I$.
By line~\ref{vlt-info}, $r\info$ is a copy of $r$'s \info\ value in $p$'s local table of \llt \ results.
By the precondition of \vlt \ and the definition of an \llt$(r)$ linked to $I$, this value is read from $r.\info$ at line~\ref{ll-read} by the \llt$(r)$ linked to $I$.
Therefore, $r.\info$ must change between when the \llt$(r)$ linked to $I$ performs line~\ref{ll-read} and when $I$ sees $ r\info \neq r.\info $ at line~\ref{vlt-reread}.
Thus, there must be a successful \fcas \ belonging to some \op \ $U$ on $r$ between these two times.
Since a \fcas \ belongs to $U$, Lemma~\ref{lem-no-steps-belong-to-dummy-op} implies that $U$ is not the dummy \op, so $U$ must have been created by an invocation of \sct.
%\qed
\end{chapscxproof}

\begin{defn} \label{defn-blame-scx}

Let $U$ be an \op \ created by an invocation $S$ of \sct \ that returns \false, and $U'$ be an \op \ created by an invocation $S'$ of \sct.
Consider the \rec s $r$ that are in both $U.V$ and $U'.V $, and for which there is no successful \fcas\ belonging to $U$ on $r$.
Let $r'$ be the \rec\ among these which occurs earliest in $U.V$. %the partial order $\mathcal{O}$.
We say $S$ \textbf{blames} $S'$ \textbf{for} $r'$ if and only if there is a successful \fcas \ on $r'$ belonging to $U'$, and this \fcas \ is the earliest successful \fcas \ on $r'$ to occur between when the \llt$(r')$ linked to $S$ reads $r'.\info$ at line~\ref{ll-read} and the first \fcas \ belonging to $U$ on $r'$.
\end{defn}

\begin{lem} \label{lem-if-abort-then-blame-different-scx}
Let $U$ be an \op\ created by an invocation $S$ of \sct.
If $S$ returns \false, then it blames some other invocation of \sct.
\end{lem}
\begin{chapscxproof}
Since $S$ returns \false, Lemma~\ref{lem-help-false}.\ref{claim-help-false-abort} implies that an \astep \ belongs to $U$.
By Lemma~\ref{lem-abort-fcas-flow}, %.\ref{abort-fcas-flow-claim-no-succ-fcas-on-rk}, 
there is a \rec \ $r_k$ in $U.V$ such that there is a \fcas\ belonging to
$U$ on $r_k$, but no successful one.
%on which there is no successful \fcas \ belonging to $U$.
Moreover,
%By Lemma~\ref{lem-abort-fcas-flow}.\ref{abort-fcas-flow-claim-rk-changes}, 
$r_k.\info$ changes after time $t_1$, when the \llt$(r_k)$ linked to $S$ reads $r_k.\info$ at line~\ref{ll-reread}, and before time $t_2$, when the first \fcas\ belonging to $U$ on $r_k$ occurs.
Since the \llt$(r_k)$ linked to $S$ terminates before $U$ is created (at line~\ref{sct-create-op} of $S$), and a \fcas\ belonging to $U$ can only occur after $U$ is created, we know $t_1 < t_2$.
Let $t_0$ be the time when the \llt$(r_k)$ performs line~\ref{ll-read}.
Note that $t_0 < t_1 < t_2$.
Since $r_k.\info$ can only be changed by a successful \fcas, there must be a successful \fcas \ on $r_k$  during $(t_1, t_2)$.
Let {\it fcas} the the earliest successful \fcas\ on $r_k$  during $(t_0, t_2)$, and let $ U' $ be the \op\ to which it belongs.
Since %$r_k.\info$ can only be changed by a successful \fcas, and 
{\it fcas} occurs before the first \fcas \ belonging to $U$ on $r_k$, we know that $U \neq U' $.
Let \[\rho = \{ r \mid r \mbox{ is in } U.V \mbox{ and } r \mbox{ is in } U'.V \mbox{ and } \nexists \mbox{ successful \fcas \ belonging to } U \mbox{ on } r \}.\]
By the code of \help, a \fcas \ belonging to $ U' $ can only modify a \rec\ in $ U'.V $. %so $ r_k \in U'.V $.
Thus, $r_k \in \rho$.

We now show $r_k$ is the element of $\rho$ that occurs earliest in $U.V$. %the partial order $\mathcal{O}$.
Suppose, to derive a contradiction, that some $r_i \in \rho$ comes before $r_k$ in $U.V$. %$\mathcal{O}$.
By Lemma~\ref{lem-abort-fcas-flow}.\ref{abort-fcas-flow-claim-fcas-on-rk} and Lemma~\ref{lem-abort-fcas-flow}.\ref{abort-fcas-flow-claim-no-succ-fcas-on-rk}, there is an unsuccessful \fcas \ belonging to $U$ on $r_k$.
By Lemma~\ref{lem-fcas-on-ri-only-after-succ-fcas-on-previous}, before this unsuccessful \fcas, there must be a successful \fcas\ belonging to $U$ on $r_i$.
However, this implies $r_i \notin \rho$, which is a contradiction.

Let $S'$ be the invocation of \sct \ that creates $U'$.
Thus far, we have shown that $S$ blames $S'$.
It remains to show that $S \neq S'$.
By Lemma~\ref{lem-no-steps-belong-to-dummy-op}, a \fcas \ or \astep \ cannot belong to the dummy \op.
Therefore, neither $U$ nor $ U' $ can be the dummy \op.
Since $U \neq U'$, $U$ and $U'$ must be created by different invocations of \sct.
%\qed
\end{chapscxproof}

We now prove that an invocation of \llt$(r)$ can return \fail \ only under certain circumstances.

\begin{defn} \label{defn-threatening-section}
Let $U$ be an \op \ created by an invocation $S$ of \sct.
The \textbf{threatening section} of $S$ begins with the first \fcas \ belonging to $U$, and ends with the first \cstep \ or \astep \ belonging to $U$.
\end{defn}

\begin{lem} \label{lem-succ-fcas-or-upcas-only-during-sct}
Let $U$ be an \op \ created by an invocation $S$ of \sct.
The threatening section of $S$ lies within $S$, and every successful \fcas\ or \upcas\ belonging to $U$ occurs during $S$'s threatening section.
\end{lem}
\begin{chapscxproof}
Let $ptr$ be a pointer to $U$, and $t_0$ and $t_1$ be the times when $S$'s threatening section begins and ends, respectively.
%By Definition~\ref{defn-threatening-section}, 
Since each \fcas, \upcas, \astep\ or \cstep\ belonging to $U$ occurs in an invocation of \help $ (ptr) $, and $S$ creates $U$, we know that these steps can only occur after $S$ begins.
Hence, $t_0$ is after $S$ begins.
Clearly every \fcas \ belonging to $U$ occurs after $t_0$.
From the code of \help, the first \upcas \ belonging to $U$ occurs between $t_0$ and $t_1$.
By Lemma~\ref{lem-only-first-upcas-can-succeed}, this is the only \upcas \ belonging to $U$ that can succeed.
By Lemma~\ref{lem-only-first-fcas-can-succeed}, no \fcas \ belonging to $U$ can succeed after the first \fstep \ or \astep \ belonging to $U$.
From the code of help, the first \fstep \ belonging to $U$ must occur before the first \cstep \ belonging to $U$.
Thus, every successful \fcas \ belonging to $U$ occurs between $t_0$ and $t_1$.
From the code of \sct, $S$ performs an invocation $H$ of \help $ (ptr) $ before it returns, and $H$ will perform either a \cstep \ or \astep \ belonging to $U$, so long as it does not return from line~\ref{help-return-true-loop}.
By Lemma~\ref{lem-no-return-true-in-loop-until-uass}, $H$ cannot return from line~\ref{help-return-true-loop} until after the first \cstep \ belonging to $U$.
Therefore, a \cstep \ or \astep \ belonging to $U$ must occur before $S$ terminates, so $t_1$ is before $S$ terminates.
%\qed
\end{chapscxproof}

\begin{obs} \label{obs-if-fcas-then-r-in-V}
Let $S$ be an invocation of \sct$(V, R, fld, new)$, and $U$ be the \op \ it creates.
If there is a \fcas \ belonging to $U$ on $r$, then $r$ is in $V$.
\end{obs}
\begin{chapscxproof}
From the code of \help, there will only be a \fcas \ belonging to $U$ on $r$ if $r$ is in $U.V$, and line~\ref{sct-create-op} implies that $r$ in $V$.
%\qed
\end{chapscxproof}

\begin{lem} \label{lem-llx-can-only-fail-if-concurrent-scx}
An invocation $I$ of \llt $ (r)$ can return \fail \ only if it overlaps the threatening section of some invocation of \sct $ (V, R, fld, new)$ with $r$ in $V$.
\end{lem}
\begin{chapscxproof}
By Lemma~\ref{lem-if-llt-fail-then-blame-sct}, $I$ blames an invocation $S$ of \sct.
Let $U$ be the \op \ created by $S$, and $ ptr $ be a pointer to $U$.
By Definition~\ref{defn-blame-llx}, $I$ reads a pointer to $U$ from $r.\info$.
Since $U$ is not the dummy \op, $r.\info$ can only point to $U$ after a successful \fcas \ belonging to $U$ on $r$.
By Observation~\ref{obs-if-fcas-then-r-in-V}, $r$ is in $V$.
We now show that $I$ overlaps the threatening section of $S$.
Consider the two cases of Definition~\ref{defn-blame-llx}.

\textbf{Case I:} $I$ enters the if-block at line~\ref{ll-check-frozen}, and reads a pointer to $U$ from $r.\info$ at line~\ref{ll-reread}.
In this case, from the code of \llt, we know that $r.\info$ changes between when $I$ performs line~\ref{ll-read} and when $I$ performs line~\ref{ll-reread}.
Since $r.\info$ can only be changed to point to $U$ by a successful \fcas\ belonging to $U$, there must be a successful \fcas \ belonging to $U$ during $I$.
By Lemma~\ref{lem-succ-fcas-or-upcas-only-during-sct}, $I$ must overlap the threatening section of $S$.

\textbf{Case II:} $I$ does not enter the if-block at line~\ref{ll-check-frozen}.
Since $I$ reads a pointer to $U$ from $r.\info$ at line~\ref{ll-read}, we know that $r\info$ is a pointer to $U$.
By the test at line~\ref{ll-check-frozen}, either $state= \done $ and $marked_2=\true $, or $state = \freezing $.

Suppose $state= \freezing$.
Then, $U.state = \freezing $ when $I$ performs line~\ref{ll-read-state}.
Since $U$ is not the dummy \op, a pointer to $U$ can appear in $r.\info$ only after a successful \fcas \ belonging to $U$.
By Corollary~\ref{cor-state-transitions-respect-figure}, $U.state $ can only be \freezing \ before the first \cstep \ or \astep \ belonging to $U$.
Therefore, Definition~\ref{defn-threatening-section} implies that $I$ performs line~\ref{ll-read-state} during the threatening section of $S$.

Now, suppose $state = \done $ and $marked_2=\true$.
By Corollary~\ref{cor-state-transitions-respect-figure}, $U.state =$ \done \ at all times after $I$ performs line~\ref{ll-read-state}.
We consider two sub-cases.
If $marked_1= \true $, then $I$ will return \finalized \ if it reaches line~\ref{ll-check-finalized}.
Since we have assumed that $I$ returns \fail, this case is impossible.
Otherwise, a \markstep\ $mstep$ belonging to some \op \ $W$ changes $r.marked $ to \true\ between line~\ref{ll-read-marked1} and line~\ref{ll-read-marked2}.
It remains only to show that $mstep$ occurs during the threatening section of the invocation of \sct \ that created $W$.
%after the first \fcas \ belonging to $W$, and before the first \cstep\ belonging to $W$
Since $r.marked $ is initially \false, and is never changed from \true \ to \false, $mstep$ must be the first \markstep\ belonging to $W$ on $r$.
From the code of \help, a \fstep \ belonging to $W$ must precede $mstep$.
Therefore, Lemma~\ref{lem-fass-then-no-bcas} implies that no \astep\ belonging to $W$ ever occurs.
From the code of \help, $mstep$ must occur after the first \fcas \ belonging to $W$, and before the first \cstep \ belonging to $W$.
By Definition~\ref{defn-threatening-section}, $mstep$ occurs during the threatening section of the invocation of \sct \ that created $W$.
%
%Hence, from the test at line~\ref{ll-check-frozen}, either $U.state=$ \done \ and $r$ is marked, or $U.state =$ \freezing.
%Suppose $U.state=$ \done \ and $r$ is marked.
%By Corollary~\ref{cor-state-transitions-respect-figure}, $U.state =$ \done \ at all times after $I$ performs line~\ref{ll-read-state}.
%Since $r.marked$ is initially \false, and is only changed at line~\ref{help-markstep} (where it is set to \true), $I$ will return \finalized\ if it reaches line~\ref{ll-check-finalized}.
%We have assumed that $I$ returns \fail, so this case is impossible.
%Now, suppose $U.state = $ \freezing \ when $I$ performs line~\ref{ll-read-state}.
%Since $U$ is not the dummy \op, a pointer to $U$ can appear in $r.\info$ only after a successful \fcas \ belonging to $U$.
%By Corollary~\ref{cor-state-transitions-respect-figure}, $U.state $ can only be \freezing \ before the first \cstep \ or \astep \ belonging to $U$.
%Therefore, Lemma~\ref{lem-succ-fcas-or-upcas-only-during-sct} implies that $I$ performs line~\ref{ll-read-state} during the threatening section of $S$.
%\qed
\end{chapscxproof}

We now prove that an invocation of \sct \ or \vlt \ can return \false \ only under certain circumstances.

\begin{defn} \label{defn-vulnerable-section}
The \textbf{vulnerable interval} of an invocation $I$ of \sct \ or \vlt \ begins at the earliest starting time of an \llt$(r)$ linked to $I$, and ends when $I$ ends.
\end{defn}

\begin{lem} \label{lem-if-sct-or-vlt-blames-then-succ-fcas}
Let $I$ be an invocation of \sct \ or \vlt, and $U$ be an \op \ created by an invocation $S$ of \sct.
If $I$ blames $S$ for a \rec\ $r$, then a successful \fcas \ belonging to $U$ on $r$ occurs during $I$'s vulnerable interval.
\end{lem}
\begin{chapscxproof}
Suppose $I$ is an invocation of \sct.
Let $U_I$ be the \op \ created by $I$.
By Definition~\ref{defn-blame-scx}, a successful \fcas \ belonging to $U$ on $r$ occurs between when the \llt$(r)$ linked to $I$ performs line~\ref{ll-read}, and the first \fcas \ {\it fcas} belonging to $U_I$ on $r$.
By Lemma~\ref{lem-succ-fcas-or-upcas-only-during-sct}, {\it fcas} occurs during $I$.
Now, suppose $I$ is an invocation of \vlt.
Then, by Definition~\ref{defn-blame-vlt}, a successful \fcas \ belonging to $U$ on $r$ occurs between when the \llt$(r)$ linked to $I$ performs line~\ref{ll-read}, and when $I$ sees $ r.\info \neq r\info $ at line~\ref{ll-reread}.
%\qed
\end{chapscxproof}

\begin{obs} \label{obs-if-sct-or-vlt-blames-for-r-then-r-in-blaming-V}
If an invocation $I$ of \sct$ (V, R, fld, new)$ or \vlt$ (V)$ blames an invocation of \sct \ for a \rec \ $r$, then $r$ is in $V$.
\end{obs}
\begin{chapscxproof}
Suppose $I$ is an invocation of \sct.
Let $U$ be the \op \ created by $I$.
By Definition~\ref{defn-blame-scx}, $r$ is in $U.V $.
Since $ U.V $ does not change after $U$ is created at line~\ref{sct-create-op} of $I$, $r$ is in $V$.
Now, suppose $I$ is an invocation of \vlt.
In this case, the claim is immediate from Definition~\ref{defn-blame-vlt}.
%\qed
\end{chapscxproof}

\begin{obs} \label{obs-if-sct-or-vlt-blames-for-r-then-r-in-blamed-V}
If an invocation $I$ of \sct\ or \vlt\ blames an invocation $S$ of \sct$(V, R, fld, new)$ for a \rec \ $r$, then $ r$ is in $V$.
\end{obs}
\begin{chapscxproof}
Let $U$ be the \op \ created by $S$.
By Lemma~\ref{lem-if-sct-or-vlt-blames-then-succ-fcas}, there is a successful \fcas \ belonging to $U$ on $r$.
The claim then follows from Observation~\ref{obs-if-fcas-then-r-in-V}.
%\qed
\end{chapscxproof}

\begin{lem} \label{lem-sct-or-vlt-false-only-if}
An invocation $I$ of \sct$(V, R, fld, new)$ or \vlt$(V)$ ending at time $t$ can return \false \ only if its vulnerable interval overlaps the threatening section of some other \sct$(V', R', fld', new')$,
% with nonempty $V \cap V'$.
where some \rec\ appears in both $V$ and $V'$.
\end{lem}
\begin{chapscxproof}
Suppose $I$ returns \false.
By Lemma~\ref{lem-if-vlt-false-then-blame-sct} and Lemma~\ref{lem-if-abort-then-blame-different-scx}, $I$ blames an invocation $S$ of \sct$(V', R', fld', new')$, where $ I \neq S $, for a \rec \ $r$.
Let $U$ be the \op \ created by $S$, and $U_I$ be the \op \ created by $I$.
By Lemma~\ref{lem-if-sct-or-vlt-blames-then-succ-fcas}, a successful \fcas \ {\it fcas} belonging to $U$ on $r$ occurs during $I$'s vulnerable interval.
By Lemma~\ref{lem-succ-fcas-or-upcas-only-during-sct}, {\it fcas} occurs during the threatening section of $S$.
Therefore, $I$'s vulnerable interval overlaps the threatening section of $S$.
By Observation~\ref{obs-if-sct-or-vlt-blames-for-r-then-r-in-blaming-V} and Observation~\ref{obs-if-sct-or-vlt-blames-for-r-then-r-in-blamed-V}, $r$ is in both $V$ and $V'$.
\end{chapscxproof}

We now prove bounds on the number of invocations of \llt, \sct \ and \vlt \ that can blame an invocation of \sct.

\begin{lem} \label{lem-uass-or-bcas-before-llx-returns-fail}
Let $I$ be an invocation of \llt$(r)$ that returns \fail, and $U$ be the \op\ created by the invocation of \sct \ that is blamed by $I$.
A \cstep \ or \astep \ belonging to $U$ occurs before $I$ returns.
\end{lem}
\begin{chapscxproof}
By Definition~\ref{defn-blame-llx}, we know that $I$ reads a pointer to $U$ from $r.\info$.
By Lemma~\ref{lem-if-llt-fail-then-blame-sct}, $U$ is not the dummy \op.
Thus, $U$ can have state \done\ or \retry\ only after a \cstep\ or \astep\ belonging to $U$ has occurred.
If $I$ returns at line~\ref{ll-return}, then it saw $state \in \{\retry, \done\}$ at line~\ref{ll-check-frozen}, and we are done.
So, suppose $I$ returns at line~\ref{ll-return-finalized} or line~\ref{ll-return-fail}.
Then, before returning, $I$ performs line~\ref{ll-help}, where it either sees $state \in \{\retry, \done\}$ or invokes \help$(r\info)$.
If $I$ sees $state \in \{\retry, \done\}$, then we are done.
So, suppose $I$ invokes \help$(r\info)$.
From the code of \help, if $I$'s invocation of \help  \ returns \false, then $I$ performs an \astep \ belonging to $U$ during its invocation of \help.
Otherwise, by Lemma~\ref{lem-if-fass-belongs-to-op-then}.\ref{claim-help-true-then-returns-after-uass}, a \cstep \ belonging to $U$ occurs before $I$'s invocation of \help\ returns.
\end{chapscxproof}

\after{Is it actually true that two llts by a proc can blame the same sct, or can we tighten the next lemma to say $at most one$ instead of $at most two$?}

\begin{lem} \label{lem-sct-only-blamed-by-one-llt-per-process}
Each invocation of \sct \ can be blamed by at most two invocations of \llt\ per process.
\end{lem}
\begin{chapscxproof}
Let $S$ be an invocation of \sct.
To derive a contradiction, suppose there is some process $p$ that blames $S$ for three failed
invocations of \llt : $I_1, I_2$ and $I_3$ (which are performed by $p$ in this order).
By Definition~\ref{defn-blame-llx}, $I_1$, $I_2$ and $I_3$ each read a pointer to $U$ from $r.\info$, either at line~\ref{ll-read} or at line~\ref{ll-reread}.
Since $r.\info$ points to $U$ at some point during $I_1$, and again at or after the time $I_2$ performs line~\ref{ll-read}, we know from Lemma~\ref{lem-no-aba-info} that $r.\info$ points to $U$ when $I_2$ performs line~\ref{ll-read}.
By the same argument, $r.\info$ points to $U$ when $I_3$ performs line~\ref{ll-read}.
Thus, in both $I_2$ and $I_3$, the local variable $r\info$ is a pointer to $U$.

We first argue that $I_2$ cannot enter the if-block at line~\ref{ll-check-frozen}.
Suppose, to obtain a contradiction, that $I_2$ enters the if-block.
Since we know that $r.\info$ points to $U$ when $I_2$ performs line~\ref{ll-read} and when $I_3$ performs line~\ref{ll-read}, Lemma~\ref{lem-no-aba-info} implies that $r.\info$ points to $U$ when $I_2$ performs line~\ref{ll-reread}.
Thus, $I_2$ will see $r.\info = r\info$ at line~\ref{ll-reread}, and will return at line~\ref{ll-return}, which contradicts our assumption that $I_2$ returns \fail\ at line~\ref{ll-return-fail}.

Since $I_2$ does not enter the if-block at line~\ref{ll-check-frozen}, either $state = \done$ and $marked_2 = \true$ or $state = \freezing$ in $I_2$.
By Lemma~\ref{lem-uass-or-bcas-before-llx-returns-fail}, a \cstep \ or \astep \ belonging to $U$ occurs prior to the termination of $I_1$, which is before the start of $I_2$.
By Corollary~\ref{cor-state-transitions-respect-figure}, $r\info.state $ does not change after it is set to \done \ or \retry \ by this \cstep \ or \astep, so $state \neq \freezing$ in $I_2$.
Therefore, $state = \done$ and $marked_2 = \true$ in $I_2$, which means that the \rec\ $r$ is marked at some point during $I_2$.
By inspection of the code, once a \rec\ is marked, it remains marked forever.
Thus, when $I_3$ performs line~\ref{ll-read-marked1}, it will see $r.marked = \true$, so $marked_1 = \true$ in $I_3$.
Since an invocation of \llt\ can return \fail\ only if $marked_1 = \false$, $I_3$ cannot return \fail, which is a contradiction.
\end{chapscxproof}

%We know prove a bound on the number of invocations of \llt, \sct \ and \vlt \ that can blame an invocation of \sct.

\begin{lem} \label{lem-sct-only-blamed-by-v-scts-or-vlts-per-process}
Each invocation of \sct$(V, R, fld, new)$ can be blamed by at most $|V|$ invocations of \sct \ or \vlt \ per process.
\end{lem}
\begin{chapscxproof}
By Observation~\ref{obs-if-sct-or-vlt-blames-for-r-then-r-in-blamed-V}, if an invocation of \sct \ or \vlt \ blames an invocation of \sct $ (V, R, fld, new) $ for $r$, then $r$ is in $V$.
Thus, it suffices to prove that an invocation $S$ of \sct$ (V, R, fld, new) $
cannot be blamed for any $r$ in $V$ by more than one invocation of \sct \ or \vlt\ performed by process $p$. %, can be blamed for $r$ by at most one invocation of \sct \ or \vlt, for each $ r \in V $, and process $p$.

Let $I$ and $I'$ be invocations of \sct \ or \vlt \ performed by process $p$, and $U$, $U'$ and $U_S$ be the \op s created by $I$, $I'$ and $S$, respectively.
Without loss of generality, let $I'$ occur after $I$.
Suppose, in order to derive a contradiction, that $I$ and $I'$ both blame %the same invocation 
$S$ %of \sct$(V, R, fld, new)$
for the same \rec \ $r$.
Let $t_0$ ($t_0'$) be the time when the \llt$(r)$ linked to $I$ ($I'$) performs line~\ref{ll-read}, and $t_1$ ($t_1'$) be the time when $I$ ($I'$) finishes.
By Lemma~\ref{lem-if-sct-or-vlt-blames-then-succ-fcas}, a successful \fcas \ belonging to $U_S$ occurs between $t_0$ and $t_1$, and a successful \fcas \ belonging to $U_S$ occurs between $t_0'$ and $t_1'$.
If we can show $ t_0 < t_1 < t_0' < t_1' $, then we shall have demonstrated that there must be two such \fcas s, which contradicts Lemma~\ref{lem-only-first-fcas-can-succeed}.

Since the \llt$(r)$ linked to $I$ ($I'$) terminates before $I$ ($I'$), we know $t_0 < t_1$ ($t_0' < t_1'$).
%It remains only to show $ t_1 < t_2 $.
By Observation~\ref{obs-if-sct-or-vlt-blames-for-r-then-r-in-blaming-V}, $r$ is in the $V$ sequences of invocations $I$ and $I'$.
Hence, Definition~\ref{defn-llt-linked-to-sct}.\ref{prop-no-sct-or-vlt-between-linked-llt-and-sct-or-vlt} implies that $t_1 \notin [t_0', t_1']$, and $t_1' \notin [t_0, t_1]$.
Since $I'$ occurs after $I$, $ t_1 < t_1'$.
Therefore, $ t_0 < t_1 < t_0' < t_1'$.
%\qed
\end{chapscxproof}

We now define the blame graph and prove a number of its properties.

\begin{defn} \label{defn-blame-graph}
We define the \textbf{blame graph} for an execution to be a directed graph whose nodes are the invocations of \llt, \vlt\ and \sct, with an edge from an invocation $I$ to another invocation $I'$ if and only if $I$ blames $I'$.  (Note that only the nodes corresponding to invocations of \sct\ can have incoming edges.)
\end{defn}

%The first property we prove is a bound on the in-degree of a node in the blame graph.
%
%\begin{lem} \label{lem-blame-graph-in-degree-bound}
%%Let $ D $ be the size of the largest $V$ set passed to any invocation of \sct \ in the execution.
%Let $ | P | $ be the number of processes that take steps.
%Each node in the blame graph corresponding to an invocation of \sct$ (V, R, fld, new) $ has in-degree at most $ | P | (1+| V |) $.
%\end{lem}
%\begin{chapscxproof}
%Immediate from Lemma~\ref{lem-sct-only-blamed-by-one-llt-per-process} and Lemma~\ref{lem-sct-only-blamed-by-v-scts-or-vlts-per-process}.
%%\qed
%\end{chapscxproof}

The next property we prove is that, for each execution, there is a bound on the length of the longest path in the blame graph.
%Recall that there are a number of \rec s which serve as entry points to the data structure.
%
As mentioned in Section \ref{sec-operations}, we require the following constraint in order to prove this bound exists.

\begin{con} \label{con-partial-order}
If there is a configuration $C$ after which the value of no
field of any \rec\ changes,
then
 there is a total order $\prec$
on all \rec s created during the execution such that,
if \rec\ $r_1$ appears before data
\rec\ $r_2$ in the sequence $V$ passed to an invocation 
of \sct\ whose linked \llt s begin after $C$,
then $r_1 \prec r_2$.
%Consider the set of invocations of \sct\ \{$S \mid $ no \sct\ is linearized at or after the start of the first $\llt(r)$ linked to $S$\}. %, and before the end of $S$\}.
%The $V$ sequences of \sct s in this set must induce a strict partial order $\mathcal{O}$ on the set of \rec s that are ever created.
%Let $U_1$ and $U_2$ be \op s created by invocations $S_1$ and $S_2$ of \sct, respectively, and $[t_0, t_1]$ be the smallest interval that completely contains the vulnerable intervals of $S_1$ and $S_2$.
%If no invocation of \sct \ is linearized during $[t_0, t_1]$, then the ordering of \rec s in $U_1.V$ must agree with that of $U_2.V$.
\end{con}

%In other words, all of the \rec s in the $V$ set of each \op \ are totally ordered, and this ordering is consistent with the ordering of the \rec s in the $V$ set of any other \op.

%\eric{This next defn doesn't really make sense when V can have duplicates.  For example if $V=\langle r,  r\rangle$, then $r$ strictly preceds $r$ in $\prec_U$.  Do we need it to make sense when V can contain duplicates, or can we just define it for SCXs whose linked LLXs come after the time that all records stop changing?  It seems like you can just define it for SCX's after the stabilization point, because you really only need this ordering in Lemma \ref{lem-if-no-succ-sct-after-time-t-then}, which only talks about such \sct s.}
%
%\begin{defn} \label{defn-partial-order-operator}
%Let $U$ be an \op, %\ created by an invocation of \sct\ whose linked \llt s ,
%and $r$ and $r'$ be \rec s in $U.V$.
%We define $r \prec_U r'$ to hold precisely when $r$ comes before $r'$ in $U.V$. %$\mathcal{O}$.
%We define $r \preceq_U r'$ to hold precisely when $r = r'$ or $r \prec_U r'$.
%%We define $r \prec r'$ to hold precisely when $r$ comes before $r'$ in $\mathcal{O}$.
%\end{defn}

%\eric{Again, next lemma is problematic if V can have duplicates. If $V=\langle r , r', r\rangle$ it says after the very first \fcas (on $r$), $r'$ will also be frozen (because $r' \prec_U r$).  Does it only have to be true for scxs whose linked llxs start after the stabilization point?}

\begin{lem} \label{lem-after-fcas-previous-recs-frozen}
Let $U$ be an \op\ created by an invocation of \sct\ whose linked \llt s begin after the configuration $C$ that is specified in Constraint~\ref{con-partial-order}.
Immediately after a successful \fcas\ belonging to $U$ on $r$, $r.\info$ points to $U$ and, for each $ r'$ in $U.V $, where $r' \prec r$, $r'.\info$ points to $U$ and a successful \fcas \ belonging to $U$ on $r'$ has occurred.
\end{lem}
\begin{chapscxproof}
Let {\it fcas} be a successful \fcas \ belonging to $U$ on $r$, and let $r'$ be any \rec\ in $U.V$ that satisfies $ r' \prec r $.
By Constraint~\ref{con-partial-order}, $r'$ must occur before $r$ in the sequence $U.V$.
By Lemma~\ref{lem-fcas-on-ri-only-after-succ-fcas-on-previous}, a successful \fcas\ {\it fcas$'$} belonging to $U$ on $r'$ occurs prior to {\it fcas}.
Thus, Corollary~\ref{cor-if-succ-fcas-then-point-u-until-bcas-or-uass} implies that $r'.\info$ points to $U$ at all times after {\it fcas$'$} and before the first \cstep \ or \astep \ belonging to $U$.
Similarly, $r.\info$ points to $U$ at all times after {\it fcas} and before the first \cstep \ or \astep \ belonging to $U$.
By Lemma~\ref{lem-no-succ-fcas-after-fass-or-bcas}, {\it fcas} must precede the first \fstep\ or \astep \ belonging to $U$.
From the code of \help, the first \fstep\ belonging to $U$ must precede the first \cstep\ belonging to $U$.
Hence, {\it fcas} and {\it fcas$'$} both precede the first \cstep \ or \astep \ belonging to $U$.
Therefore, immediately after {\it fcas}, the \info\ fields of $r$ and $r'$ both point to $U$.
%Since a \fcas \ belongs to $U$, lemma 4 implies that $U$ cannot be the dummy \op.
%Hence, $U.state$ is initially \freezing.
%Finally, since a \cstep \ or \astep \ belonging to $U$ does not occur until after {\it fcas}, $r$ and $r'$ are both frozen for $U$ immediately after {\it fcas}.
%\qed
\end{chapscxproof}

\begin{lem} \label{lem-if-blame-chain-then-recs-ordered}
Let $U_1$, $U_2$ and $U_3$ be \op s respectively created by invocations $S_1$, $S_2$ and $S_3$ of \sct\ whose linked \llt s begin after the configuration $C$ that is specified in Constraint~\ref{con-partial-order}, and $r$ and $r'$ be \rec s.
If $S_1$ blames $S_2$ for $r$, and $S_2$ blames $S_3$ for $r'$, then $r \prec r'$.
\end{lem}
\begin{chapscxproof}
Since $S_1$ blames $S_2$ for $r$, we know from Definition~\ref{defn-blame-scx} that $r$ is in $U_2.V$.
Similarly, since $S_2$ blames $S_3$ for $r'$, we know $r'$ is in $U_2.V$.
Furthermore, a successful \fcas\ belonging to $U_2$ on $r$ occurs, and no successful \fcas\ belonging to $U_2$ on $r'$ occurs.
By Lemma~\ref{lem-fcas-on-ri-only-after-succ-fcas-on-previous}, $r$ must occur before $r'$ in the sequence $U_2.V$.
Thus, Constraint~\ref{con-partial-order} implies $r \prec r'$.
%%Thus, $r \prec_{U_2} r'$. %$r$ and $r'$ must be ordered by $\mathcal{O}$.
%Suppose, to derive a contradiction, that $r = r'$ or $r' \prec_{U_2} r$.
%Since $S_1$ blames $S_2$ for $r$, we know from Definition~\ref{defn-blame-scx} that there is a successful \fcas \ {\it fcas} belonging to $U_2$ on $r$.
%By Lemma~\ref{lem-after-fcas-previous-recs-frozen}, prior to {\it fcas}, there is a successful \fcas \ {\it fcas$'$} belonging to $U_2$ on $r'$.
%However, since $S_2$ blames $S_3$ for $r'$, Definition~\ref{defn-blame-scx} implies that {\it fcas$'$} cannot exist, yielding a contradiction.
%\qed
\end{chapscxproof}

\begin{lem} \label{lem-one-to-one-correspondence-upcas-and-sct}
There can be only as many successful \upcas s as there are invocations of \sct \ that either return \true, or do not terminate.
\end{lem}
\begin{chapscxproof}
From the code, an \upcas \ can only occur in an invocation of \help$ (ptr) $, where $ptr$ points to an \op \ $U$.
Further, from the code of \help, there is at least one \fcas \ belonging to $U$ or \fstep \ belonging to $U$.
Hence, Lemma~\ref{lem-no-steps-belong-to-dummy-op} implies that $U$ is not the dummy \op.
Thus, $U$ is created by an invocation of \sct \ at line~\ref{sct-create-op}.
By Lemma~\ref{lem-only-first-upcas-can-succeed}, only the first \upcas \ belonging to an \op \ can succeed.
By Lemma~\ref{lem-help-false}.\ref{claim-help-false-no-upcas}, no \upcas \ belongs to an \op\ created by an unsuccessful invocation of \sct.
%\qed
\end{chapscxproof}

\after{I'm still a bit uncomfortable about what entry points really mean, formally (Eric).}

We think of processes as accessing such a data structure via
a fixed number of special \rec s called \textit{entry points}, each of which has a single mutable pointer to a \rec.
We assume there is always some \rec\ reachable by following pointers from an entry point that is not finalized. %We assume the $R$ set of \sct\ does not contain any pointer to an entry point. 
(This assumption that entry points cannot be finalized is not crucial, but it 
simplifies the statement of some progress guarantees.)

\begin{defn} \label{defn-rec-initiated}
A \rec \ is \textbf{initiated} at all times after it first becomes reachable by following \rec \ pointers from an entry point.
\end{defn}

\begin{obs} \label{obs-only-upcas-can-initiate}
The only step in an execution that can cause a \rec\ to become initiated is a successful \upcas.
\end{obs}
%\begin{chapscxproof}
%Follows immediately from Observation~\ref{obs-only-upcas-modifies-records}.% and the fact that entry point is a \rec.
%%a mutable field of a \rec \ can only be modified by a successful \upcas.
%%The claim then follows from the fact that, for each entry point $e$, the only \rec \ initially reachable by following \rec \ pointers from $e$ is itself.
%%\qed
%\end{chapscxproof}

\begin{lem} \label{lem-if-blame-for-r-then-r-already-initiated}
Let $S_1$ and $S_2$ be invocations of \sct, and let $r$ be a \rec.
If $S_1$ blames $S_2$ for $r$, then $r$ was initiated before the start of $S_1$, and before the start of $S_2$.
\end{lem}
\begin{chapscxproof}
Let $U_1$ and $U_2$ be the \op s created by $S_1$ and $S_2$, respectively.
By Definition~\ref{defn-blame-scx}, $r$ is in both $U.V$ and $U'.V$.
By Observation~\ref{obs-op-invariants}.\ref{inv-llresults}, there are invocations of \llt$(r)$ linked to $S_1$ and $S_2$, respectively.
By the precondition of \llt, $r$ must be initiated before the \llt$(r)$ linked to $S_1$, and before the \llt$(r)$ linked to $S_2$.
Finally, the \llt$(r)$ linked to $S_1$ must terminate before $S_1$ begins, and the \llt$(r)$ linked to $S_2$ must terminate before $S_2$ begins.
%\qed
\end{chapscxproof}

\after{Instead of defining $\sigma$, define its complement, since we mostly
talk about things {\it not} in $\sigma$}

\begin{lem} \label{lem-if-no-succ-sct-after-time-t-then}
If no \sct\ is linearized after some time $t$, then the following hold.
\begin{enumerate}
\item A finite number $N$ 
of \rec s  are ever initiated in the execution.%
\label{claim-if-no-succ-sct-after-time-t-then-finite-number-of-initiated-rec}
\item Let $\sigma$ be the set of invocations of \sct \ in the execution whose vulnerable intervals start at or before $t$.  The longest path in the blame graph consisting entirely of invocations of \llt, \sct, and \vlt \ that are not in $\sigma$ has length at most $N+2$
\label{claim-if-no-succ-sct-after-time-t-then-no-path-longer-than}
\end{enumerate}
\end{lem}
\begin{chapscxproof}
Claim~\ref{claim-if-no-succ-sct-after-time-t-then-finite-number-of-initiated-rec} follows immediately from Observation~\ref{obs-only-upcas-can-initiate} and Lemma~\ref{lem-one-to-one-correspondence-upcas-and-sct}.

We now prove claim~\ref{claim-if-no-succ-sct-after-time-t-then-no-path-longer-than}.
Suppose, in order to derive a contradiction, that there is a path of length at least $N +3$ in the blame graph consisting entirely of invocations of \llt, \sct, and \vlt \ that are not in $\sigma$.
Since only invocations of \sct \ can be blamed, at least $N+2 $ of the nodes on this path must correspond to invocations of \sct.
Let $ S_1, S_2,..., S_{N+2}$ be invocations of \sct \ corresponding to any $N+2 $ consecutive nodes on this path, and let $U_1, U_2, ..., U_{N+2}$ be the \op s they created, respectively.
For each $ i \in\{ 1, 2,...,N+1\} $, let $r_i$ be the \rec \ for which $S_i$ blames $S_{i+1}$.
Since no invocation of \sct \ is linearized after $t$, and the vulnerable sections of $ S_1, S_2,..., S_{N+2}$ all start after $t$, no invocation of \sct\ is linearized after the first $\llt(r)$ linked to any of these invocations of \sct.
Therefore, from Lemma~\ref{lem-if-blame-chain-then-recs-ordered} and the fact that, for each $i \in \{1, 2, ..., N\}$, $S_i$ blames $S_{i+1}$ for $r_i$ and $S_{i+1}$ blames $S_{i+2}$ for $r_{i+1}$, we obtain $r_i \prec r_{i+1}$.
%
%Thus, Constraint~\ref{con-partial-order} implies that $r_i \prec r_{i+1}$, for all $i \in \{1, 2, ..., N\}$.
By Lemma~\ref{lem-if-blame-for-r-then-r-already-initiated}, before any invocation of \sct \ in \{$ S_1, S_2,..., S_{N+2} $\} begins, $r_1, r_2, ..., r_{N+1}$ have all been initiated.
Therefore, some \rec \ $r$ appears twice in $\{r_1, r_2, ..., r_{N+1}\}$.
Since the $\prec$ relation is transitive, we obtain $r \prec r$, which is a contradiction.
\end{chapscxproof}

We now prove the main progress property for \sct.

\begin{lem} \label{lem-sct-progress}
If invocations of \sct \ complete infinitely often, then invocations of \sct \ succeed infinitely often.
\end{lem}
\begin{chapscxproof}
Suppose, to derive a contradiction, that after some time $t'$, invocations of \sct \ are performed infinitely often, but no invocation of \sct \ is successful.
Then, since we only linearize successful \sct s, and a subset of the non-terminating \sct s, there is a time $t \ge t'$ after which no invocation of \sct\ is linearized.
Let $\sigma$ be the set of invocations of \sct \ in the execution whose vulnerable intervals start at or before $t$. 
%Note that any invocation of \sct \ not in $\sigma$ must be unsuccessful. %Any pair of invocations of \sct \ not in $\sigma$ must be unsuccessful.%, and must have created \op s whose $V$ sets are ordered consistently.
%We first show that only finitely many invocations of \sct \ can blame invocations in $\sigma$.
By Lemma~\ref{lem-sct-only-blamed-by-one-llt-per-process} and Lemma~\ref{lem-sct-only-blamed-by-v-scts-or-vlts-per-process},  the in-degree of each node in the blame graph is bounded.
Since $\sigma$ is finite, and the in-degree of each node in $\sigma$ is bounded, only a finite number of invocations of \sct \ can blame invocations in $\sigma$.
Now, consider any maximal path $\pi$ consisting entirely of invocations of \llt, \sct, and \vlt \ that are \textit{not} in $\sigma$.
By Lemma~\ref{lem-if-no-succ-sct-after-time-t-then}.\ref{claim-if-no-succ-sct-after-time-t-then-no-path-longer-than}, $\pi$ has length at most $N+3$.
Since no invocation of \sct \ is successful after $t$, the invocation $S$ of \sct \ corresponding to the last node on path $\pi$ must be unsuccessful.
By Lemma~\ref{lem-if-abort-then-blame-different-scx}, $S$ must blame some other invocation of \sct.
Since $\pi$ is maximal, $S$ must blame an invocation of \sct \ in $\sigma$.
Thus, there can be only finitely many of these paths (of bounded length).
However, this contradicts our assumption that invocations of \sct \ occur infinitely often.
%\qed
\end{chapscxproof}

Unfortunately, since \sct\ and \vlt\ cannot be invoked unless a sequence of invocations of \llt\ (linked to the \sct\ or \vlt) return snapshots, the previous result is not strong enough unless we can guarantee that processes can invoke \sct\ and/or \vlt\ infinitely often.
We first give two definitions that help clarify the progress guarantees for \sct\ and \vlt.

\begin{defn} \label{defn-set-up-sct}
%Suppose $p$ is a process, and $V$ is a sequence of \rec s such that, for each $r' \in V$, $p$ has not performed an $\llt(r')$ that returned \finalized.
An \sct-\func{Update} algorithm performs \llt s on a sequence $V$ of \rec s and invokes \sct$(V, R,$ $fld, new)$ if all of these \llt s return snapshots.
%A process $p$ performs an \scx-\func{update} by executing an algorithm that
% an invocation of \sct$(V, R, fld, new)$ by executing any algorithm that invokes \llt$(r)$ for each $r$ in $V$, and then invokes \sct$(V, R, fld, new)$ if none of these \llt s return \fail\ or \finalized.
A \textit{successful} \sct-\func{Update} is one in which the \sct\ returns \true.
\end{defn}

\begin{defn}
A \vlt-\func{Query} algorithm performs \llt s on a sequence $V$ of \rec s and invokes \vlt$(V)$ if all of these \llt s return snapshots.
A \textit{successful} \vlt-\func{Query} is one in which the \vlt\ returns \true.
\end{defn}

\begin{thm} \label{thm-llt-sct-vlt-progress}
Our implementation of \llt/\sct/\vlt\ satisfies the following progress properties.
\begin{enumerate}
\item If operations (\llt, \sct, \vlt) are performed infinitely often, then operations succeed infinitely often.
\label{claim-progress-if-operations-io-then-succ-io}
\item Suppose that
    (a) there is always some non-finalized \rec\ reachable by following pointers from an entry point, 
    (b) for each \rec\ $r$, each process performs finitely many invocations of \llt$(r)$ that return \finalized, and
    (c) processes perform infinitely many executions of \sct-\func{Update} and/or \vlt-\func{Query} algorithms.
    Then, infinitely many \sct\ or \vlt\ operations succeed.
\label{claim-progress-if-set-up-sct-io-then-succ-io}
%\item Suppose, for each \rec\ $r$, each process performs finitely many invocations of \llt$(r)$ that return \finalized. Then, if processes attempt to \textit{set up} invocations of \sct\ infinitely often, invocations of \sct\ succeed infinitely often.
%\label{claim-progress-if-set-up-sct-io-then-succ-io}
%\item If there is always some \rec\ reachable by following pointers from an entry point that is not finalized, then invocations of \vlt\ (resp., \sct) can be set up infinitely often.
%\label{claim-progress-if-not-finalized-then-can-set-up-sct-io}
\end{enumerate}
\end{thm}
\begin{chapscxproof}
Both claims have similar proofs, by cases.
%Claim~\ref{claim-progress-if-not-finalized-then-can-set-up-sct-io} is obvious.
%The first two claims have similar proofs, by cases.
%%The proofs of these claims are quite similar, and both proceed by cases.

\textbf{Proof of Claim~\ref{claim-progress-if-operations-io-then-succ-io}.}
Suppose operations are performed infinitely often.
%We consider two cases.

\textbf{Case I:} invocations of \sct \ are performed infinitely often.
In this case, Lemma~\ref{lem-sct-progress} implies that invocations of \sct \ will succeed infinitely often, and the claim is proved.

\textbf{Case II:} after some time $t$, no invocation of \sct \ is performed.
In this case, the blame graph contains a finite number of invocations of \sct.
By Lemma~\ref{lem-sct-only-blamed-by-one-llt-per-process} and Lemma~\ref{lem-sct-only-blamed-by-v-scts-or-vlts-per-process},  the in-degree of each node in the blame graph is bounded.
By Lemma~\ref{lem-if-llt-fail-then-blame-sct} and Lemma~\ref{lem-if-vlt-false-then-blame-sct}, each unsuccessful invocation of \llt \ or \vlt \ blames an invocation of \sct.
Therefore, only finitely many invocations of \llt \ and \vlt \ can be unsuccessful.
Thus, eventually, every invocation of \llt\ or \vlt\ succeeds.
%This contradicts our assumption that unsuccessful invocations of \llt \ and \vlt \ are performed infinitely often.

\textbf{Proof of Claim~\ref{claim-progress-if-set-up-sct-io-then-succ-io}.}
Suppose the antecedent holds, and processes perform infinitely many executions of \sct-\func{Update} or \vlt-\func{Query} algorithms.

\textbf{Case I:} invocations of \sct\ are performed infinitely often.
In this case, Lemma~\ref{lem-sct-progress} implies that invocations of \sct\ will succeed infinitely often, and the claim is proved.

\textbf{Case II:} eventually, no invocation of \sct\ is performed.
Then, as we argued in Case II of the proof of Claim~\ref{claim-progress-if-operations-io-then-succ-io}, only finitely many invocations of \llt\ can be unsuccessful.
This implies that, after some time $t$, every invocation of \llt\ is successful.
Furthermore, by antecedent (b) of this claim, for each \rec\ $r$, each process will perform finitely many invocations of \llt$(r)$ that return \finalized.
By Lemma~\ref{lem-if-no-succ-sct-after-time-t-then}, there are a finite number of \rec s that are ever initiated in the execution.
Therefore, each process can perform only finitely many invocations of \llt\ that return \finalized.
Consequently, after some time $t'$, every invocation of \llt\ by any process will return a snapshot.
Any \sct-\func{Update} that starts after $t'$ will perform \llt s that return snapshots, and will invoke \sct, violating our assumption in this case.
Thus, after $t'$, processes execute infinitely many \vlt-\func{Query} algorithms, but no \sct-\func{Update} algorithms.
Any \vlt-\func{Query} algorithm that starts after $t'$ will perform \llt s that return snapshots, and will invoke \vlt.
Thus, infinitely many invocations of \vlt\ are performed after $t'$.
Since none of these invocations are concurrent with any invocation of \sct, Lemma~\ref{lem-sct-or-vlt-false-only-if} implies that all of these invocations of \vlt\ must succeed.
\end{chapscxproof}

\newpage
\section{Additional properties of \llt/\sct/\vlt} \label{sec-properties}

In this section we prove some additional properties of \llt/\sct/\vlt\ that are intended to simplify the design of certain data structures.
%In the following, we use the term \textbf{data structure} to refer to the intermediate data structure. 
%\eric{We haven't defined intermediate data structure (and we shouldn't)!}
% (wherein each \rec \ has mutable and immutable fields, but no $\info$ field or $marked$ bit).
At this level, a \textit{configuration} consists of the state of each process, and a collection of \rec s (which have only mutable and immutable fields).
A \textit{step} is either a \func{Read}, or a linearized invocation of \llt, \sct\ or \vlt.

%An execution is a (possibly infinite) sequence of alternating \textbf{configurations} and \textbf{steps}, beginning with an initial configuration.
%Configurations and steps are defined in the usual way.
%A configuration consists of the contents of shared memory and the state of each process.
%A step is a read from or write to shared memory, or a \cas.
%We sometimes refer to a time immediately before (after) a step $s$, with the understanding that we are actually referring to the configuration that immediately precedes (follows) $s$.
%\textbf{[[[[Put this somewhere else (earlier)?]]]]}

\begin{defn} \label{defn-rec-in-added-removed}
A \rec \ $r$ is \textbf{in the data structure} in some configuration $C$ if and only if $r$ is reachable by following pointers from an entry point.
We say a \rec\ $r$ is \textbf{removed (from the data structure) by} some step $s$ if and only if $r$ is in the data structure immediately before $s$, and $r$ is not in the data structure immediately after $s$.
We say a \rec \ $r$ is \textbf{added (to the data structure) by} some step $s$ if and only if $r$ is not in the data structure immediately before $s$, and $r$ is in the data structure immediately after $s$.
\end{defn}

Note that a \rec \ can be removed from or added to the data structure only by a linearized invocation of \sct.

%\trevor{[[[Rough notes: Constraint~\ref{con-mark-all-removed-recs} exists for two reasons.
%First, we use it to show that no \rec \ changes after it is removed (for correctness of \search\ and \func{Get}).
%Second, we use it to show that, if no successful invocations of \sct\ occur after some time then, eventually, no invocation of \llt\ can return \finalized\ (for progress of, e.g., multiset \ins\ and \dotreeupdate).]]]}

If the following constraint is satisfied, then the results of this section apply. %The results of this section apply if the following constraint is satisfied.

\begin{con} \label{con-mark-all-removed-recs}
%\begin{enumerate}
%\item 
For each linearized invocation $S$ of \sct$(V, R, fld, new)$, $R$ contains precisely the \rec s that are removed from the data structure by $S$. %, and $fld$ is not a field of a \rec\ in $R$.
%\label{claim-con-finalize-removed}
%\item A process never invokes \sct$(V, R, fld, new)$ with $fld$ a pointer to a field of a \rec \ $r \in R$.
%\label{claim-con-no-change-then-finalize}
%\end{enumerate}
\end{con}

\begin{lem} \label{lem-if-initiated-rec-not-in-data-structure-then-does-not-change}
If a \rec \ $r$ is removed from the data structure for the first time by step $s$, then no linearized invocation of \sct$(V, R, fld, new)$, where $fld$ is a mutable field of $r$, occurs at or after $s$.
(Hence, $r$ does not change at or after $s$.)
\end{lem}
\begin{chapscxproof}
The only step that can change $r$ is a linearized invocation of \sct.
The invocation $S'$ of \sct$(V', R', fld', new')$ that removes $r$ modifies a mutable field of some \rec\ different from $r$.
Thus, $fld'$ is not a field of $r$.
Since this is the only change to the data structure when $s$ occurs, $r$ does not change when $s$ occurs.
Suppose, to derive a contradiction, that an invocation $S$ of \sct$(V, R, fld, new)$, where $fld$ is a mutable field of $r$, occurs after $s$.
Then, since $r$ is in $V$, the precondition of \sct\ implies that an invocation $I$ of $\llt(r)$ linked to $S$ must occur before $S$.
By Constraint~\ref{con-mark-all-removed-recs}, $r$ is in $R'$.
Thus, if $I$ occurs after $S'$, then it returns \finalized, which contradicts Definition~\ref{defn-llt-linked-to-sct}.
Otherwise, $S'$ occurs between $I$ and $S$, so $S$ cannot be linearized, which contradicts our assumption.
\end{chapscxproof}

%\begin{cor} \label{cor}
%Let $fld$ be a field of a \rec\ $r$, and $S$ be a linearized invocation of \sct$(V, R, fld, new)$.
%Just before $S$ is linearized, $r$ is in the data structure.
%\end{cor}
%\begin{chapscxproof}
%From the precondition of \sct, before $S$, there must be an invocation $I$ of $\llt(r)$ linked to $S$.
%From the precondition of \llt, $r$ must be initialized before $I$.
%Thus, $r$ was in the data structure, at some time before $S$.
%Since $r$ changes when $S$ is linearized, 
%\end{chapscxproof}

%\begin{lem} \label{lem}
%No field of a \rec\ $r$ is changed from its initial value before $r$ is initiated.
%\end{lem}
%\begin{chapscxproof}
%Suppose, to derive a contradiction, that an invocation $S'$ of \sct$(V', R', fld, new')$ 
%\end{chapscxproof}

\begin{lem} \label{lem-rec-in-data-structure-just-before-llt}
If an invocation $I$ of $\llt(r)$ returns a value different from \fail\ or \finalized, then $r$ is in the data structure just before $I$ is linearized.
\end{lem}
\begin{chapscxproof}
By the precondition of \llt, $r$ is initiated and, hence, in the data structure, at some point before $I$.
Suppose, to derive a contradiction, that $r$ is not in the data structure just before $I$ is linearized.
Then, some linearized invocation of \sct$(V, R, fld, new)$ must remove $r$ before $I$ is linearized.
By Constraint~\ref{con-mark-all-removed-recs}, $r$ is in $R$.
However, this implies that $I$ must return \finalized, which is a contradiction.
\end{chapscxproof}

\begin{lem} \label{lem-rec-in-data-structure-after-linearized-sct}
If $S$ is a linearized invocation of \sct$(V, R, fld, new)$, where $new$ is a \rec, then $new$ is in the data structure just after $S$.
\end{lem}
\begin{chapscxproof}
Note that $fld$ is a mutable field of a \rec\ $r$ in $V$.
We first show that $r$ is in the data structure at some point before $S$.
By the precondition of \sct, before $S$, there is an $\llt(r)$ linked to $S$.
By the precondition of \llt, $r$ must be initiated when this linked \llt\ occurs.
Thus, Definition~\ref{defn-rec-initiated} and Definition~\ref{defn-rec-in-added-removed} imply that $r$ is in the data structure at some point before $S$.
Suppose, to derive a contradiction, that $r$ is not in the data structure just after $S$.
Then, $r$ must either be removed by $S$, or by some previous step.
However, this directly contradicts Lemma~\ref{lem-if-initiated-rec-not-in-data-structure-then-does-not-change}.
\end{chapscxproof}

Let $C_1$ and $C_2$ be configurations in the execution.
We use $C_1 < C_2$ to mean that $C_1$ precedes $C_2$ in the execution.
We say $C_1 \le C_2$ precisely when $C_1 = C_2$ or $C_1 < C_2$.
We denote by $[C_1, C_2]$ the set of configurations $\{C \mid C_1 \le C \le C_2\}$.

\after{I think the proof of the following lemma could be simplified by first proving
a version of the second part of the claim (for any field f), then taking the special case of a pointer
field to prove the first part.  May require some rewording of the statement, though.}

\begin{lem} \label{lem-if-rec-traversed-then-rec-in-data-structure}
Let $r_1,r_2,...,r_l$ be a sequence of \rec s, where $r_1$ is an entry point, and $C_1,C_2,...,C_{l-1}$ be a sequence of configurations satisfying $C_1 < C_2 < ... < C_{l-1}$.
If, for each $i \in \{1, 2, ..., l-1\}$, a field of $r_i$ points to $r_{i+1}$ in configuration $C_i$, then $r_{i+1}$ is in the data structure in some configuration in $[C_1, C_i]$.
%(And $r_1$ is always in the data structure.)
Additionally, if a mutable field $f$ of $r_l$ contains a value $v$ in some configuration $C_l$ after $C_{l-1}$ then, in some configuration in $[C_1, C_l]$, $r_l$ is in the data structure and $f$ contains $v$.
\end{lem}
\begin{chapscxproof}
We prove the first part of this result by induction on $i$.

Since each entry point is always in the data structure, and $r_1$ points to $r_2$ in configuration $C_1$, $r_2$ is in the data structure in $C_1$.
Thus, the claim holds for $i=1$.

Suppose the claim holds for $i$, $1 \le i \le l-2$.
We prove it holds for $i+1$.
If $r_i$ is in the data structure when it points to $r_{i+1}$ in $C_i$, then $r_{i+1}$ is in the data structure in $C_i$, and we are done.
Suppose $r_i$ is \textit{not} in the data structure in $C_i$.
By the inductive hypothesis, $r_i$ is in the data structure in some configuration in $[C_1, C_{i-1}]$.
Let $s$, $C_1 < s < C_i$, be the first step such that $r_i$ is removed from the data structure by $s$.
%Consider the first step $s$ occurring after $C_1$, and before $C_i$, such that $r_i$ is removed from the data structure at $s$.
In the configuration $C$ just before $s$, $r_i$ is in the data structure.
By Lemma~\ref{lem-if-initiated-rec-not-in-data-structure-then-does-not-change}, $r_i$ does not change at or after $s$. 
Thus, $r_i$ does not change after $C$.
Since $C$ occurs before $C_i$, and $r_i$ points to $r_{i+1}$ in $C_i$, $r_i$ must point to $r_{i+1}$ in $C$.
Therefore, in $C$ (which satisfies $C_1 \le C < C_i$), $r_i$ is in the data structure and points to $r_{i+1}$.

The second part of the proof is quite similar to the inductive step we just finished.
Suppose $f$ contains $v$ in $C_l$.
If $r_l$ is in the data structure in $C_l$, then we are done.
Suppose $r_l$ is not in the data structure in $C_l$.
We have shown above that $r_l$ is in the data structure in some configuration in $[C_1, C_l]$.
Let $s'$, $C_1 < s' < C_l$, be the first step such that $r_l$ is removed from the data structure by $s'$.
In the configuration $C'$ just before $s'$, $r_l$ is in the data structure.
By Lemma~\ref{lem-if-initiated-rec-not-in-data-structure-then-does-not-change}, $r_l$ does not change at or after $s'$. 
Thus, $r_l$ does not change after $C'$.
Since $C'$ occurs before $C_l$, and $f$ contains $v$ in $C_l$, $f$ must contain $v$ in $C'$.
Therefore, in $C'$ (which satisfies $C_1 \le C' < C_l$), $r_l$ is in the data structure and $f$ contains $v$.
\end{chapscxproof}

%\begin{lem} \label{lem}
%Want to say something like: function like \dotreeupdate \ that starts after an invocation of \sct$(V, R, fld, new)$ is linearized will not perform $\llt(r)$ for any $r \in R$.
%
%Let $r_1,r_2,...,r_l$ be a sequence of \rec s, where $r_1$ is an entry point, and $C_1,C_2,...,C_{l-1}$ be a sequence of configurations satisfying $C_1 < C_2 < ... < C_{l-1}$.
%If an invocation $S$ of \sct$(V, R, fld, new)$ is linearized before $C_1$, then .............
%
%\textbf{It seems like anything useful that we want to say here we either cannot yet say, or is already said in the constraint above...  gonna come back to this after starting to work on progress for the multiset...}
%%For each linearized invocation $S$ of \sct$(V, R, fld, new)$, $R$ contains precisely the \rec s that are removed from the data structure by $S$.
%\end{lem}
%\begin{chapscxproof}
%
%\end{chapscxproof}
%
%\eric{Suggestions for multiset section (which I don't want to edit while Trevor is working on it):
%
%Move pseudocode for multisets to beginning of the section.
%
%Start the section with a sentence introducing the pseudocode.
%Maybe also mention how the data structure is initialized.
%
%Put Obs \ref{obs-multiset-satisfies-con-mark-all-removed-recs} after Lemma
%\ref{lem-multiset-constraints-invariants}
%}
%
%

\section{Modifications to enable memory reclamation} \label{sec-memory}

Each invocation of \sct\ creates a new \op , which must eventually be reclaimed.
In managed languages such as Java and C\#, automatic garbage collection can be used to reclaim \op s.
In unmanaged languages, more specialized techniques must be used.
We start by describing a modification to the \sct\ algorithm that enables garbage collection in managed languages, and then describe how memory can be reclaimed in unmanaged languages.

\begin{figure}
\FrameSep3pt
\centering
\noindent
\begin{minipage}[t]{0.45\linewidth}
\begin{framed}
\includegraphics[width=\linewidth]{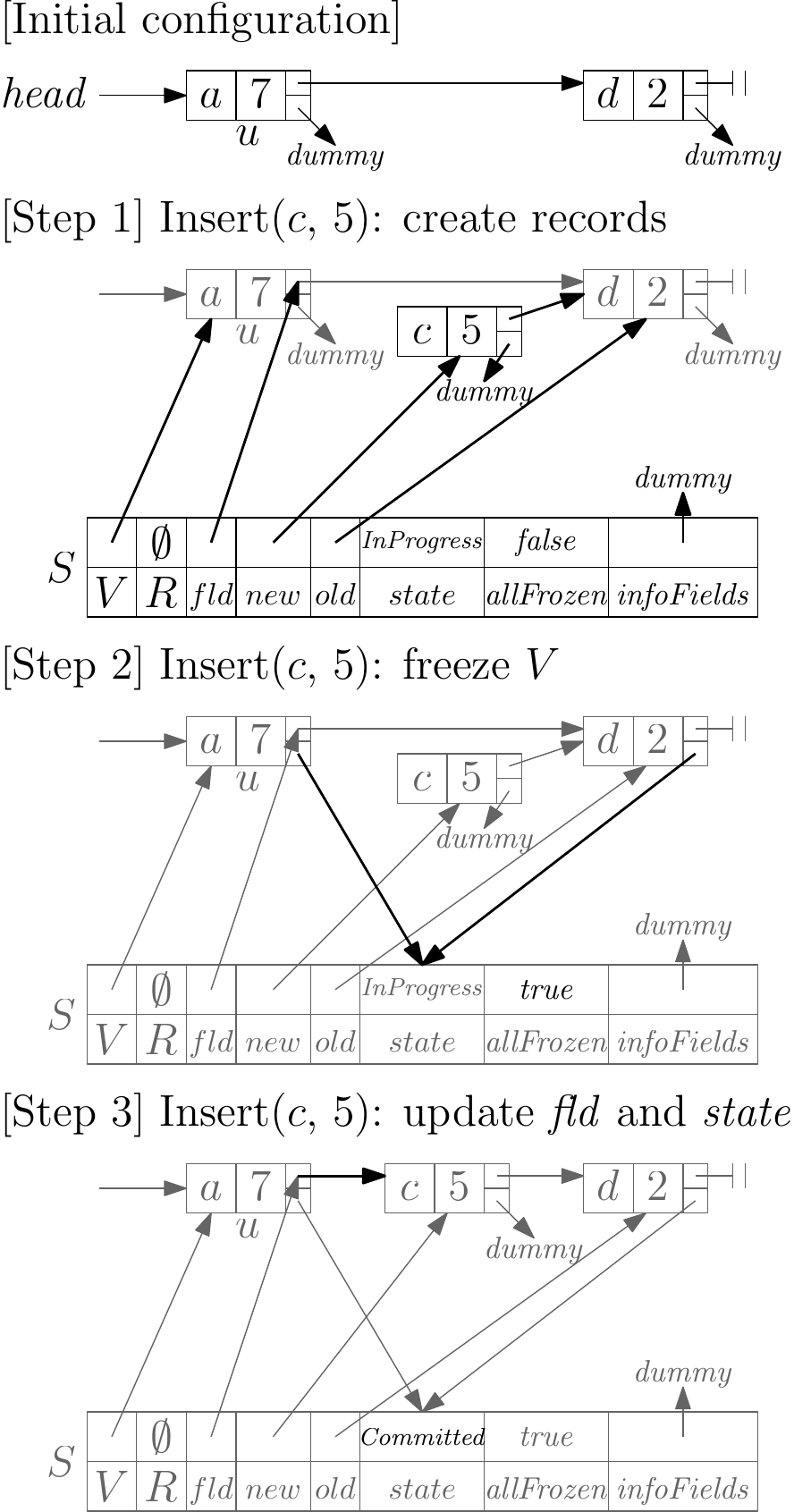}
\end{framed}
\end{minipage}
\begin{minipage}[t]{0.45\linewidth}
\begin{framed}
\includegraphics[width=\linewidth]{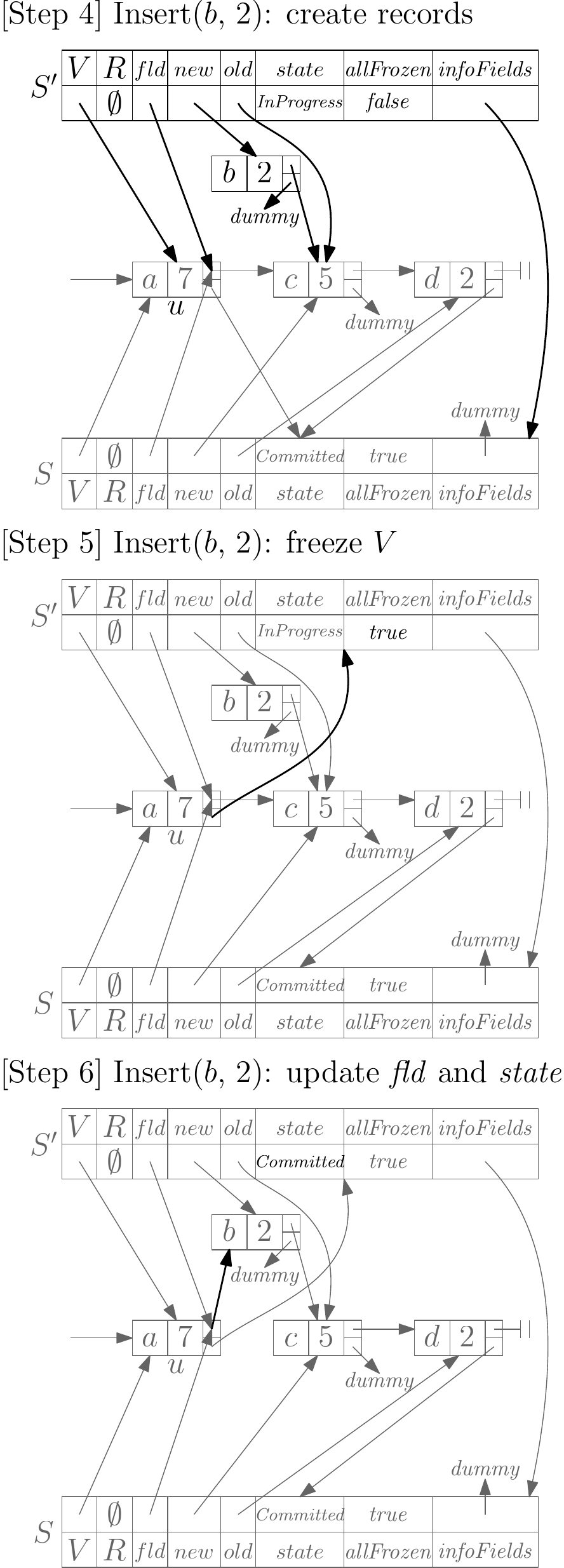}
\end{framed}
\end{minipage}
\FrameSep0pt
\caption{
    Example execution of a multiset in which \sct\ operations create reachable garbage.
}
\label{fig-garbage-example}
\end{figure}

\begin{figure}
\centering
\includegraphics[width=0.5\linewidth]{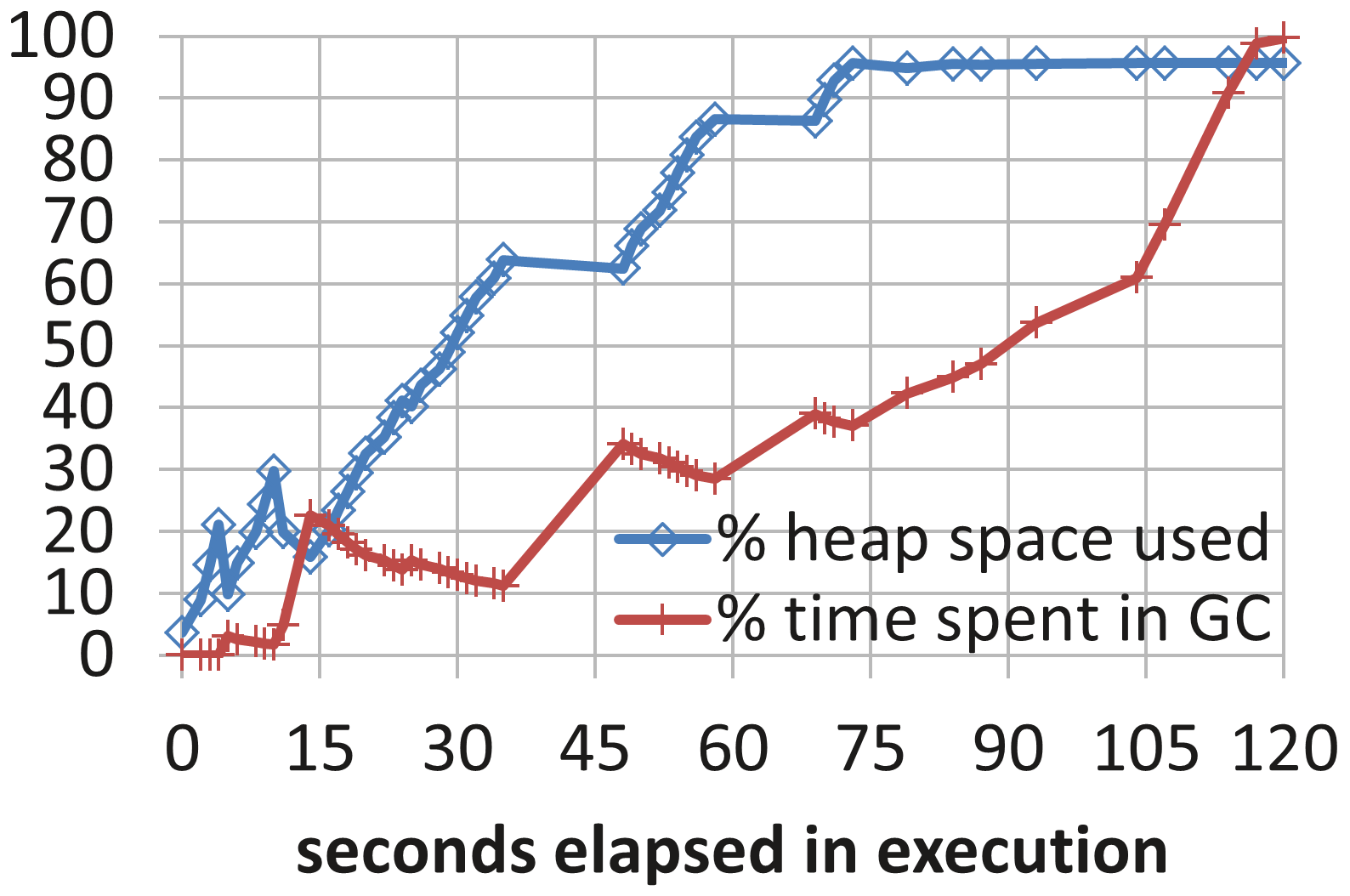}
\caption{Experiment showing total memory usage and time spent in garbage collection for a Java microbenchmark on a balanced binary search tree implemented from \llt\ and \sct .}
\label{fig-mem-leak}
\end{figure}

\subsection{Garbage collection}

At a high level, garbage collection reclaims records only once all processes can no longer access them.
Thus, as long as any process can reach a record by following pointers, %from a record in the data structure, 
the record can not be reclaimed.
The vanilla \sct\ algorithm presented above is poorly suited for automatic garbage collection, because it allows garbage to remain reachable for an arbitrarily long time.

To show this experimentally, we produced a Java implementation of a lock-free balanced binary search tree (discussed in Chapter~\ref{chap-template}), and ran a microbenchmark wherein 8 processes perform 50\% \ins\ and 50\% \del\ operations on keys drawn uniformly randomly from $[0, 10^6)$ for 120 seconds.
During this experiment, we used the NetBeans 7.2 profiler to collect statistics on the total memory usage of the Java virtual machine (JVM), and on the fraction of the execution time that was spent performing garbage collection.
We ran the experiment on an Intel i7-2600k with 4 cores and 2 hyperthreads per core and 16GB of RAM, running Windows 7, with the Java 1.7.0\_07 64-bit server VM.
The Java heap was automatically sized by the JVM.
(I.e., we did not specify explicit minimum or maximum sizes for the heap.)
Measurements showed that it was approximately 3GB.

The results of the experiment appear in Figure~\ref{fig-mem-leak}.
Garbage collection performed reasonably well until approximately 15 seconds into the experiment.
As the execution continued, more and more of the total heap space was occupied, and the fraction of execution time spent performing garbage collection increased (in response to the increased memory pressure).
Eventually, nearly 100\% of the execution time was spent in garbage collection, and insertions and deletions become virtually impossible, because garbage collection could not free enough memory.
Note that gaps in the measurements (e.g., between 35 and 48 seconds) were caused by long garbage collection delays.

\paragraph{The problem}

We explain the poor performance of garbage collection using an example, which is illustrated in Figure~\ref{fig-garbage-example}.
There, newly added or changed elements are drawn in black, and elements unchanged from the previous step are drawn in gray.
Nodes are drawn with a key, number of copies, and pointers to the next node (the upper pointer) and to an \op\ (the lower pointer).
The dummy \op\ is denoted simply with \textit{dummy}.

In this example, \ins$(c, 5)$ performs an \sct\ that creates an \op\ $S$ (Step~1), freezes the node $u$ containing key $a$ (Step~2), then updates the next pointer of $u$ to perform the insertion and sets $S.state$ to \textit{Committed} (Step~3).
Note that $u$ will continue to point to $S$ until it is next frozen by an \sct .
Next, \ins$(b, 2)$ performs an \sct\ that creates an \op\ $S'$ (Step~4), freezes $u$ (Step~5), then updates $u.next$ and $S'.state$ (Step~6).
%The second \op\ $S'$ also 
Since $u$ pointed to $S$ when \ins$(b, 2)$ performed its \sct , $S'$ has a pointer to $S$ in its \textit{infoFields}.
(Recall that \textit{infoFields} contains the old values used for freezing CAS steps.)
Thus, $S$ continues to be reachable (through $S'$). %, since it is reachable via $S$.
%Thus, $S$ is reachable for as long as $S'$ is reachable.
Note that $u$ will continue to point to $S'$ until it is next frozen by an \sct .
Furthermore, the next time $u$ is frozen, the \sct\ that freezes it will have a pointer to $S'$ in its \op\ (in the \textit{infoFields} field), so $S'$ will continue to be reachable (and, hence, so will $S$).
In this way, long chains of pointers between \op s can prevent them from being reclaimed.
Similarly, as long as an \op\ remains reachable, so do the nodes that it points to (and the \op s they each point to, etc.).

\begin{figure}
\centering
\includegraphics[width=0.5\linewidth]{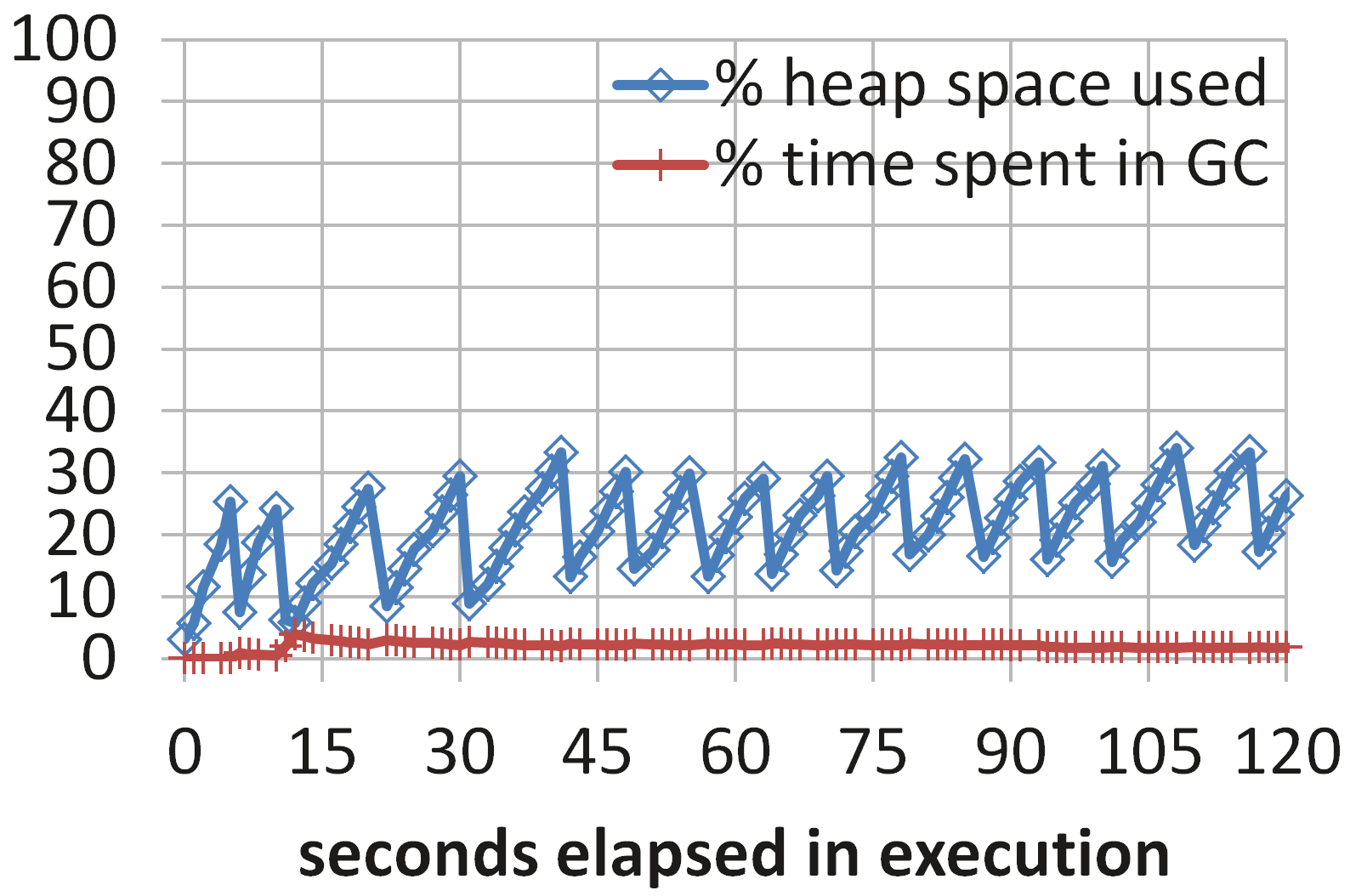}
\caption{Experiment showing the result of modifying the \help\ procedure to break chains of pointers once they are no longer necessary.}
\label{fig-mem-leak-fixed}
\end{figure}

\paragraph{A solution}

We solve this problem by modifying the \help\ procedure so that it breaks these chains of pointers once they are no longer necessary.
Specifically, after a process finishes helping an \op\ $S$ and sets \textit{S.state} to \textit{Committed} or \textit{Aborted}, it then replaces all pointers in $S.V$, $S.R$, $S.$\textit{fld}, $S.$\textit{new}, $S.$\textit{old} and $S.$\textit{infoFields} with \nil.
Note that this change makes it possible for helpers to see \nil\ in some or all of these fields.
However, these fields can contain \nil\ only after \textit{state} is \textit{Committed} or \textit{Aborted} (which means helping is no longer necessary).
Therefore, each invocation of \help$(ptr)$, where $ptr$ is a pointer to an \op\ $S$, begins by making a copy of $S$, and then checking whether $S.state$ contains \textit{Committed} or \textit{Aborted}.
If so, the invocation of \help\ simply returns, without helping.
Otherwise, it proceeds as usual, referring to the copy of $S$, instead of reading the fields of $S$ directly.
Figure~\ref{fig-mem-leak-fixed} shows the result of this fix on the experiment described above.

\subsection{Unmanaged languages}

In unmanaged languages, where automatic garbage collection is unavailable, memory must be reclaimed manually.
In order to safely reclaim an \op\ $S$, a process must know that it can no longer be accessed by any other process.
This requires knowing $S$ is not reachable (by following pointers) from any node in the data structure, or from any pointer stored in the private memory of a process.
Most state of the art lock-free memory reclamation schemes (e.g.,~\cite{Alistarh:2015,Brown:2015,Michael2004}) require an algorithm to invoke a \textit{retire} procedure when a \rec\ is no longer reachable starting from any node in the data structure.
The memory reclamation scheme then determines when \rec s are no longer reachable starting from any pointer in the private memory of a process, and subsequently frees them.

\subsection{Reachability from nodes in the data structure}

We modify \llt\ and \sct\ to use a (very) limited form of reference counting.
This allows us to determine whether $S$ is reachable from a node in the data structure.
To do this, we augment the \textit{state} field of each \op\ with a natural number $k$ (so the \textit{state} field now contains $\langle s, k \rangle$, where $s \in \{$\textit{InProgress}, \textit{Committed}, \textit{Aborted}$\}$).
\llt\ ignores the new $k$ component of the \textit{state} field and continues to use only the $s$ component.
We modify \help\ as follows.
Just before the loop at line~\ref{help-fcas-loop-begin}, we initialize a new variable $k$ to zero.
At the end of each iteration of this loop, we increment $k$.
Instead of setting $S.$\textit{state} to \textit{Aborted} at line~\ref{help-astep}, we use CAS to change $S.$\textit{state} from \textit{InProgress} to $\langle \textit{Aborted}, k \rangle$.
We call this an \textit{abort CAS}, and say it \textit{belongs to} $S$ if it attempts to modify $S.$\textit{state}.
Similarly, instead of setting $S.$\textit{state} to \textit{Committed} at line~\ref{help-cstep}, we use CAS to change $S.$\textit{state} from \textit{InProgress} to $\langle \textit{Committed}, k \rangle$.
We call this a \textit{commit CAS}, and say it \textit{belongs to} $S$ if it attempts to modify $S.$\textit{state}.
Every time a process performs a successful \fcas\ belonging to $S$ (at line~\ref{help-fcas}) that changes the \textit{info} field of a \rec\ from $old$ to $S$, the process now also performs an atomic fetch and decrement (just after the \fcas) to change $old.$\textit{state} from $\langle s, k \rangle$ to $\langle s, k-1 \rangle$ (where $s$ is \textit{Committed} or \textit{Aborted}).
We claim that, if $k-1 = 0$, then $old$ is no longer reachable from any node in the data structure.

\paragraph{Correctness}

More specifically, we claim the following.
If any process performs an abort CAS (resp., commit CAS) belonging to $S$, then the first abort CAS (resp., commit CAS) belonging to $S$ will succeed, and it will change $S.$\textit{state} to $\langle \textit{Aborted}, k \rangle$ (resp., $\langle \textit{Committed}, k \rangle$), where $k$ is the number of successful freezing CAS steps that belong to $S$.
(Observe that $k$ is then equal to the number of pointers to $S$ that were stored in \rec s by \fcas\ steps.)
Furthermore, $S.$\textit{state} will not change after this successful abort CAS (resp., commit CAS).

In the following, we refer to the original \llt\ and \sct\ algorithm presented in Section~\ref{sec-impl} as the \textit{original algorithm}, and we refer to the version that performs the reference counting described above as the \textit{modified algorithm}.

In the original algorithm, \textit{state} was changed in two places: at line~\ref{help-cstep}, where it was changed to \textit{Committed} by a write, and at line~\ref{help-astep}, where it was changed to \textit{Aborted} by a write.
Furthermore, \textit{state} could change only from \textit{InProgress} to \textit{Aborted} or \textit{Committed}.
Therefore, whenever line~\ref{help-astep} was executed, \textit{state} could contain \textit{InProgress} or \textit{Aborted}, but not \textit{Committed}.
Similarly, whenever line~\ref{help-cstep} was executed, \textit{state} could contain \textit{InProgress} or \textit{Committed}, but not \textit{Aborted}.
This implies that, if a process helping $S$ executed line~\ref{help-astep} (resp., line~\ref{help-cstep}), then no process helping $S$ could execute line~\ref{help-cstep} (resp., line~\ref{help-astep}).
It follows that, in the modified algorithm, if a process helping $S$ executes an abort CAS at line~\ref{help-astep} (resp., commit CAS at line~\ref{help-cstep}), then no process helping $S$ executes a commit CAS (resp., abort CAS).

Suppose some process performs an abort CAS belonging to $S$.
Then, by the argument above, there is no commit CAS belonging to $S$.
Moreover, \textit{state} must contain \textit{InProgress} or $\langle \textit{Aborted}, k \rangle$ (for some $k$) whenever an abort CAS is performed.
The first time a process performs an abort CAS belonging to $S$, $S.$\textit{state} must contain \textit{InProgress}, so the abort CAS will succeed and change $S.$\textit{state} to $\langle \textit{Aborted}, k \rangle$.
After this abort CAS, $S.$\textit{state} will never again contain \textit{InProgress}, so no subsequent abort CAS belonging to $S$ can succeed.
Thus, $S.$\textit{state} will not change after the successful abort CAS.
Finally, Lemma~\ref{lem-abort-fcas-flow} implies that there is a successful \fcas\ belonging to $S$ on each of the first $k$ \rec s in $V$, and there are no other successful \fcas\ steps belonging to $S$.
Consequently, $k$ is the number of successful \fcas\ steps that belong to $S$.

Now, suppose some process performs a commit CAS belonging to $S$.
Then, by the argument above, there is no abort CAS belonging to $S$.
Moreover, \textit{state} must contain \textit{InProgress} or $\langle \textit{Committed}, k \rangle$ (for some $k$) whenever a commit CAS is performed.
The first time a process performs a commit CAS belonging to $S$, $S.$\textit{state} must contain \textit{InProgress}, so the commit CAS will succeed and change $S.$\textit{state} to $\langle \textit{Committed}, k \rangle$.
After this commit CAS, $S.$\textit{state} will never again contain \textit{InProgress}, so no subsequent commit CAS belonging to $S$ can succeed.
Thus, $S.$\textit{state} will not change after the successful commit CAS.
%%%%%%%%%%%%%%%%%%%%%%%%%%%%%%%%%%%%%%%%%%%%%%%%%%%%%%%%%
Finally, we prove that $k$ is the number of successful \fcas\ steps that belong to $S$.
Any process that performs a commit CAS belonging to $S$ at line~\ref{help-cstep} must also perform a \fstep\ belonging to $S$ at line~\ref{help-fstep}.
Therefore, Lemma~\ref{lem-if-fass-then-all-succ-fcas} and Lemma~\ref{lem-only-first-fcas-can-succeed} imply that there is exactly one successful \fcas\ belonging to $S$ for each \rec\ in $V$.
In other words, there are $|V|$ successful \fcas\ steps belonging to $S$.
Since a commit CAS is executed only if a process exits the loop at line~\ref{help-fcas-loop-begin} after iterating through each \rec\ in $V$, every commit CAS attempts to change \textit{state} to $\langle \textit{Committed}, |V| \rangle$.
Thus, $k = |V|$. % is the number of successful \fcas\ steps that belong to $S$.

\subsection{Reachability from private pointers}

%We show how to perform reclamation of \op s for a large class of algorithms that use \llt\ and \sct .
Suppose all invocations of \llt\ and \sct\ are performed by \textit{operations}, and processes do not keep any pointers in private memory between operations.
(That is, each time a process finishes an operation, it ``forgets'' all of the pointers in its private memory.)
Then, we can use a distributed epoch-based reclamation scheme called DEBRA (described in Chapter~\ref{chap-debra}) to reclaim \op s once they are no longer reachable from any pointer in the private memory of a process.
%DEBRA makes an assumption that the records it it used to reclaim are accessed only by \textit{operations}, and these operations access records only by following pointers starting from nodes in the tree.
Using DEBRA entails invoking a pair of procedures \textit{leaveQstate}$()$ and \textit{enterQstate}$()$ at the beginning and end of each operation, respectively, and invoking another procedure \textit{retire}$(S)$ whenever an \op\ $S$ is no longer reachable from any node in the tree (but may be reachable from the private memory of some process).
In the absence of process failures, every \op\ is eventually reclaimed by DEBRA.
However, DEBRA is not fault tolerant.
For lock-free algorithms that satisfy some additional properties, a fault tolerant version called DEBRA+ can be used.

For lock-free algorithms that cannot use DEBRA, it may be possible to use other lock-free reclamation schemes such as Hazard Pointers~\cite{Michael2004}, Beware and Cleanup~\cite{Gidenstam2009} or ThreadScan~\cite{Alistarh:2015}.
The details of the algorithm implemented using \llt\ and \sct\ can affect whether it is possible to use each of these reclamation schemes (see Chapter~\ref{chap-debra}).

\chapter{Multiset implemented with LLX/SCX} \label{chap-multiset}
% !TEX root = paper.tex

%\chapter{An example: multiset} \label{chap-multiset}

\section{Implementation}

\begin{figure*}[ptb]
\begin{framed}
\prepnewlisting
\def\namewidth{12mm}
\begin{lstlisting}[mathescape=true,style=nonumbers]
  type// \listrec
    //\com Fields from sequential data structure
    //\wcnarrow{$key$}{key (immutable)}
    //\wcnarrow{$count$}{occurrences of $key$ (mutable)}
    //\wcnarrow{$next$}{next pointer (mutable)}
    //\com Fields defined by \llt /\sct\ algorithm
    //\wcnarrow{$\info$}{a pointer to an \op}
    //\wcnarrow{$marked$}{a Boolean value} \vspace{2mm}
  //\textbf{shared} \listrec\ $tail := \mbox{new }\listrec ( \infty, 0, \nil )$
  //\textbf{shared} \listrec\ $head := \mbox{new }\listrec ( -\infty, 0, tail )$
\end{lstlisting}%
\prepnewlisting
\hrule
%\vspace{-2mm}
\begin{lstlisting}[mathescape=true]
  //\func{Get}$(key)$
    $\langle r, - \rangle := \search(key)$
    if $key = r.key$ then return $r.count$ 
    else return $0$//\vspace{2mm}\hrule\vspace{2mm}   
  //\search$(key)$
    //\com Postcondition: $p$ and $r$ point to \listrec s with $p.key < key \le r.key$.
    $p := head$ // \label{multiset-search-p}
    $r := p.next$
    while $key > r.key$ do // \label{multiset-search-loop}
      $p := r$ // \label{multiset-search-advance-p}
      $r := r.next$ // \label{multiset-search-advance-r}
    return $\langle r, p \rangle$ //\vspace{2mm}\hrule\vspace{2mm}
  //\ins$(key, count)$ \tabto{4cm} \com Precondition: $count > 0$
    while $\true$ do
      $\langle r, p \rangle := \search(key)$ // \label{multiset-insert-search}
      if $key = r.key$ then // \label{multiset-insert-check-key}
        $localr := \llt(r)$//\label{multiset-insert-llt-r}
        if $localr \notin \{\fail, \finalized\}$ then 
          //\mbox{\textbf{if} $\sct$($\langle r \rangle, \langle \rangle, \&r.count,$ $localr.count + count$) \textbf{then return}}\label{multiset-insert-sct1}
      else
        $localp := \llt(p)$//\label{multiset-insert-llt-p}
        if $localp \notin \{\fail, \finalized\}$ and $r = localp.next$ then
          //\mbox{\textbf{if} $\sct$($\langle p \rangle, \langle \rangle, \&p.next,$ $\mbox{new }\listrec( key, count, r )$) \textbf{then return}} \label{multiset-insert-sct2} \vspace{2mm} \hrule \vspace{2mm}
  //\del$(key, count)$ \tabto{4cm} \com Precondition: $count > 0$
    while $\true$ do
      $\langle r, p \rangle := \search(key)$ // \label{multiset-delete-search}
      $localp := \llt(p)$//\label{multiset-delete-llt-p}
      $localr := \llt(r)$//\label{multiset-delete-llt-r}
      if $localp, localr \notin \{\fail, \finalized\}$ and $r = localp.next$ then
        if $key \neq r.key$ or $localr.count < count$ then return $\false$//\label{multiset-delete-return-false}
        else if $localr.count > count$ then // \label{multiset-delete-check-count-less}
          if $\sct(\langle p \rangle, \langle r \rangle, \&p.next, \mbox{new }\listrec( r.key, localr.count - count,\ localr.next))$//\mbox{\textbf{ then return} $\true$}\label{multiset-delete-sct1}
        else //\com assert: $localr.count = count$  \label{multiset-delete-check-count-else}
          if $\llt(localr.next) \notin \{\fail, \finalized\}$ then//\label{multiset-delete-llt-rnext}
            if $\sct(\langle p,r,localr.next \rangle,$ $\langle r,localr.next\rangle,$ $\&p.next, \mbox{new copy of } localr.next)$//\mbox{\textbf{ then return} $\true$}\label{multiset-delete-sct2}
\end{lstlisting}
\end{framed}
\vspace{-5mm}
	\caption{Pseudocode for a multiset, implemented with a singly linked list. % (with dummy nodes at the head and tail).
	%Data structure includes two dummy nodes for the end of the list.
	%Data structures used to implement a multiset, including two dummy nodes for the ends of the list.
	}
	\label{code-list}
\end{figure*}

We now give a detailed description of the implementation of a multiset
using \llt\ and \sct.
We assume that keys stored in the multiset are drawn from a totally ordered set and $-\infty < k < \infty$ for every key $k$ in the multiset.
As described in Section~\ref{sec-operations}, we use a singly-linked
list of nodes, sorted by key.
To avoid special cases, it always has a sentinel node, $head$,
with key $-\infty$ at its beginning and a sentinel node, $tail$,
with key $\infty$ at its end.
The definition of \listrec, the \rec\ used to represent a node,
and the pseudocode are presented in
Figure~\ref{code-list}.
\begin{ignore}
We now give a detailed description of the implementation of a multiset
using \llt\ and \sct\ that was introduced in Section \ref{sec-operations}.
We assume that keys stored in the multiset are drawn from a totally ordered set.
Each key of the multiset is stored in a \rec\ called a \listrec\ along with a count (see Figure \ref{code-list}).  The \listrec s are arranged in a singly linked list sorted by keys.
To avoid special cases, we have two sentinel nodes, $head$ and $tail$, at the beginning 
and end of the list.  These sentinel nodes have keys $-\infty$ and $\infty$, where 
$-\infty < k < \infty$ for every other possible key $k$.
Pseudocode is presented in
Figure~\ref{code-list}. 
\end{ignore}

\func{Search}$(key)$ traverses the list, starting from 
%the
$head$,
by reading $next$ pointers until reaching the first node $r$
whose key is at least $key$.
This node and the preceding node $p$ %(which may be $head$) 
are returned.
\func{Get}$(key)$ performs \func{Search}($key$), outputs $r$'s count
if $r$'s key matches $key$, and outputs 0, otherwise.

%\begin{figure}[tbp]
%        \prepnewlisting
%\begin{lstlisting}[mathescape=true,style=nonumbers]
% type// \listrec
%    //\com Fields from the sequential data structure
%    //\wc{$key$}{key (immutable)}
%    //\wc{$count$}{occurrences of $key$ (mutable)}
%    //\wc{$next$}{next pointer (mutable)}
%    //\com Fields defined by the \llt /\sct\ algorithm
%    //\wc{$\info$}{a pointer to an \op}
%    //\wc{$marked$}{a Boolean value}
%      
% //\textbf{shared} \listrec\ $tail := \mbox{new }\listrec ( \infty, 0, \nil )$
% //\textbf{shared} \listrec\ $head := \mbox{new }\listrec ( -\infty, 0, tail )$
% \end{lstlisting}%
%\caption{Data structures used to implement a multiset, including two dummy nodes for the ends of the list.}
%\label{multiset-datatypes}
%\end{figure}
	
%\vspace{-2mm}
%      //\dline{\com Postconditions: (1) $r$ points to a \listrec, and $key \le r.key$.}%
%      {\hspace{5.75mm} (2) $p$ points to a \listrec\ or $p = head$.}

%\eric{Why do we need the precondition $count > 0$ for delete?
%I took out the precondition $key\neq \infty$; I wrote in the text that infinities are not actual keys.}
%\faith{The precondition is there because it doesn't make sense to delete
%0 elements. It is a no-op.}

An invocation $I$ of \func{Insert}($key,count)$
starts by calling \func{Search}$(key)$.  Using the nodes $p$ and $r$
that are returned, it updates the data structure.
It decides whether $key$ is already in the multiset
(by checking whether $r.key = key$) and, if so,
it invokes $\llt(r)$ followed by an \sct\ linked to $r$ to
increase $r.count$ by $count$, as depicted in
Figure~\ref{fig-example-multiset}(b).
Otherwise, $I$ performs the update depicted in
Figure~\ref{fig-example-multiset}(a):
It invokes $\llt(p)$, checks that $p$ still points to $r$,
creates a node, $new$, and invokes an \sct\ linked to $p$
to insert $new$ between $p$ and $r$.
If $p$ no longer points to $r$, the \llt\ 
returns \fail\ or \finalized, or the \sct\ returns \false,
then $I$ restarts.

An invocation $I$ of \func{Delete}($key,count)$ also begins by calling
\func{Search}$(key)$.  It invokes \llt\ on the nodes $p$ and $r$
and then checks that $p$ still points to $r$. If
%$r$'s key is not $key$ or
$r$ does not contain at least $count$ copies of $key$, then $I$ returns \false.
If $r$ contains exactly $count$ copies,
then $I$ performs the update depicted in Figure~\ref{fig-example-multiset}(c)
to remove node $r$ from the list.
To do so, it invokes \llt\ on the node, $rnext$, that $r.next$ points to,
makes a copy $rnext'$ of $rnext$,
and invokes an \sct\ linked to $p, r$ and $rnext$
to change $p.next$ to point to $rnext'$.
This \sct\ also finalizes the nodes $r$ and $rnext$,
which are thereby removed from the data structure.
The node $rnext$ is replaced by a copy to
avoid the ABA problem in $p.next$.
If $r$ contains more than $count$ copies,
then $I$ replaces $r$ by a new copy $r'$ with an appropriately reduced count
using an \sct\ linked to $p$ and $r$,
as shown in Figure~\ref{fig-example-multiset}(d).
This \sct\ finalizes $r$.
If
an
%either
%CHANGED BECAUSE THERE CAN BE 3 LLXs.
\llt\ returns \fail\ or \finalized, or the \sct\ returns \false\,
then $I$ restarts.

%We say a node is \textit{in the data structure} whenever it is
%reachable by following pointers from $head$.
%We say a node is \textit{removed from the data structure} by an
%invocation $S$ of \sct\ if %and only if, 
%it is in the data structure immediately before $S$
%and is not in the data structure immediately after $S$.

\section{Correctness and progress} \label{app-multiset}

%\trevor{edit this proof preview into a sketch.}

This section gives a detailed proof of correctness for the multiset implementation.
We start with a high-level sketch.

\subsection{Proof sketch}
%\medskip\fakeparagraph{Proof sketch}
The proof begins by showing that this multiset implementation satisfies some basic invariants.
Notably, (1) $head$ always points to a node, (2) if a node has key $\infty$ then its $next$ pointer is $nil$, and (3) if a node's key is not $\infty$ then its $next$ pointer points to a node with a strictly larger key.
%Notably, we show that the following are true at all times.
%
%\begin{fakeinv}
%The following are true at all times.
%\begin{compactitem}
%\item $head$ always points to a node.
%\item If a node has key $\infty$, then its $next$ pointer is $\nil$.
%\item If a node's key is not $\infty$, then its $next$ pointer
%points to some node with a strictly larger key.
%\end{compactitem}
%\end{fakeinv}
%
%It follows that the data structure is always a sorted list.
These invariants imply that the data structure is always a sorted list.

Next, we prove that the \rec s removed from the data structure by a linearized invocation of \sct($V$, $R$, $fld$, $new$) are exactly the \rec s in $R$.
%We prove the following lemma by considering
%%each of the types of 
%%%FAITH CHANGED What is a type of SCX?
%the \sct s performed by update operations shown in Figure \ref{fig-example-multiset}.
%\begin{lem}
%\label{R-sets-correct}
%The \rec s removed from the data structure by
%a linearized invocation of \sct($V$, $R$, $fld$, $new$)
%are exactly the \rec s in $R$.
%\end{lem}
This allows us to apply the proposition in Section~\ref{sec-additional-properties} to prove that there is a time during each \search\ when the nodes $r$ and $p$ that it returns are both in the list and $p.next = r$.

Each \func{Get} and each
\func{Delete} that returns \false\ is linearized at the linearization
point of the \func{Search} it performs. 
%\trevor{I added commas.}
%%FAITH CHANGED I Took them out, but added a second each, instead.
Every other \func{Insert} or \func{Delete}
is linearized at its successful \sct. 
%\trevor{This should be linearized \sct.}
%
%The linearization point for \func{Search} is more complicated.
%This is because it is possible that when a \func{Search} first arrives at
%a node, the node has already been removed from the list.
%The following result, proved in the Appendix, provides a linearization
%point.
%Lemma 95
%
%\begin{lem}
%For each node visited during 
%an invocation of \func{Search}, there is a time during the
%invocation when that node is in the data structure.
%\end{lem}
%
%We linearize each \func{Search} at any point during its invocation when
%the node $r$ it returns is in the data structure.
%The next result, shows
Linearizability of all operations then follows from the following invariant:
%Next, we prove the following invariant,
%which allows us to argue that these linearization points are correct.
%
%\begin{lem}
At every time $t$, the multiset of keys
in the data structure is equal to the multiset of keys that would result
from the atomic execution of the sequence of operations linearized up to time $t$.
%\end{lem}
Next, we prove that all operations (\ins, \del\ and \search) return the correct values.
Finally, we prove progress by showing the number of invocations of \llt\ that return \finalized\ is bounded (a requirement for using \llt\ and \sct), and then using the progress properties of \llt\ and \sct.

\subsection{Complete proof}
%\medskip\noindent\textbf{Complete proof}\medskip

In the following, we define the \textbf{response} of a \search\ to be a step at which a value is returned.
Note that we specify Lemma~\ref{lem-multiset-constraints-invariants}.\ref{claim-multiset-finalized-before-removed}, instead of directly proving the considerably simpler statement in Constraint~\ref{con-mark-all-removed-recs}, so that we can reuse the intermediate results when proving linearizability.
%We export an intermediate result in Lemma~\ref{lem-multiset-constraints-invariants}.\ref{claim-multiset-sct-no-aba}, for the same reason.

\begin{lem} \label{lem-multiset-constraints-invariants}
The multiset algorithm satisfies the following properties.
\begin{enumerate}
%\item    Each \sct\ at line~\ref{multiset-insert-sct1} strictly increases the value of the $count$ field it modifies.
%\label{claim-multiset-sct-increases-count}
\item    Every invocation of \llt \ or \sct \ has valid arguments, and satisfies its preconditions.
\label{claim-multiset-llt-sct-preconditions}
\item    Every invocation of \search\ satisfies its postconditions.
\label{claim-multiset-search-postconditions}
%\item    Let $fld$ be a mutable field of a \rec \ $r$.
%If an invocation $S$ of \sct$(V, R, fld, new)$ is linearized, then $new$ is not the initial value of $fld$, and no invocation of \sct$(V', R', fld, new)$ is linearized before the $\llt(r)$ linked to $S$.
%         
%\label{claim-multiset-sct-no-aba}
%\item    A process never invokes \sct$(V, R, fld, new)$ where $fld$ points to a field of a \rec \ in $R$.
%\label{claim-multiset-no-change-then-finalized}
\item    Let $S$ be an invocation of \sct$(V, R, fld, new)$ performed by an invocation $I$ of \ins\ or \del, and $p$, $r$ and $rnext$ refer to the local variables of $I$.
         If $I$ performs $S$ at line~\ref{multiset-insert-sct1}, then no \rec\ is added or removed by $S$, and $R = \emptyset$.
         If $I$ performs $S$ at line~\ref{multiset-insert-sct2}, then only $new$ is added by $S$, no \rec\ is removed by $S$, and $R = \emptyset$.
         If $I$ performs $S$ at line~\ref{multiset-delete-sct1}, then only $new$ is added by $S$, only $r$ is removed by $S$, and $R = \{r\}$.
         If $I$ performs $S$ at line~\ref{multiset-delete-sct2}, then only $new$ is added by $S$, only $r$ and $rnext$ are removed by $S$, and $R = \{r, rnext\}$.
\label{claim-multiset-finalized-before-removed}
\item    The $head$ entry point always points to a \listrec, the $next$ pointer of each \listrec\ with $key \neq \infty$ points to some \listrec\ with a strictly larger key, and the $next$ pointer of each \listrec\ with $key = \infty$ is \nil. %points to some \listrec\ with $key = \infty$ (possibly itself). %and the $count$ field of each \listrec\ is positive.
\label{claim-multiset-sorted-list}
\end{enumerate}
\end{lem}
\begin{chapscxproof}
We prove these claims by induction on the sequence of steps taken in the execution.
The only steps that can affect these claims are invocations of \llt \ and \sct, and responses of \search es.
\textbf{Base case.}
Clearly, %Claim~\ref{claim-multiset-sct-increases-count}, 
Claim~\ref{claim-multiset-llt-sct-preconditions}, Claim~\ref{claim-multiset-search-postconditions} %, Claim~\ref{claim-multiset-sct-no-aba} %, Claim~\ref{claim-multiset-no-change-then-finalized} 
and Claim~\ref{claim-multiset-finalized-before-removed} hold before any such step occurs.
%By Observation~\ref{obs-only-upcas-modifies-records} and Observation~\ref{obs-immutable-fields-do-not-change}, no mutable or immutable field of a \rec \ changes before the first successful \upcas \ $upcas$ in the execution.
Before the first \sct, the data structure is in its initial configuration.
Thus, 
%From the initial configuration of the data structure, %the only \rec s that are initiated before $upcas$ are the $head$ entry point, and the \listrec\ with $key = \infty$ that is pointed to by $head.next$.
%Thus, Claim~\ref{claim-multiset-sorted-list} holds before any step occurs.
Claim~\ref{claim-multiset-sorted-list} holds before any step occurs.
%Since no \rec s can be removed from the data structure until $upcas$, Claim~\ref{claim-multiset-finalized-before-removed} is vacuously true before any step occurs.
%Now that we have shown all of these claims are initially true, we suppose they hold before some step $s$, and prove that they still hold after $s$.
\textbf{Inductive step.}
Suppose these claims hold before some step $s$.
We prove they hold after $s$.

%\textbf{Proof of Claim~\ref{claim-multiset-sct-increases-count}.}
%The only step that can affect the claim is a linearized invocation $s$ of \sct$(V, R, fld, new)$ performed at line~\ref{multiset-insert-sct1}.
%From the code, $fld$ is $r.count$.
%Since $s$ is linearized, no invocation of \sct$(V'', R'', fld'', new'')$ with $r \in V''$ is linearized between $I$ and $s$.
%Thus, $r.count$ does not change between when $I$ and $s$ are linearized.
%From the code, $new$ is $count$ plus the value read from $r.count$ by $I$.
%Therefore, $new$ is strictly larger than $r.count$ was when $I$ was linearized, which implies that $new$ is strictly larger than $r.count$ when $s$ is linearized.
%This immediately implies that $new$ is not the initial value of $fld$.
%Suppose an invocation $S'$ of \sct$(V', R', fld, new)$ is linearized before $I$, to derive a contradiction.
%If $S'$ is the last invocation of \sct\ that modifies $fld$ before $I$ is linearized, then $I$ reads $new$ from $fld$, so $s$ must change $fld$ to a value strictly larger than $new$, which is a contradiction.
%Otherwise, since we have argued each \sct\ that modifies $fld$ in between when $S'$ and $I$ are linearized strictly increases the value stored in $fld$, 

\textbf{Proof of Claim~\ref{claim-multiset-llt-sct-preconditions}.}
The only steps that can affect this claim are invocations of \llt \ and \sct.

Suppose $s$ is an invocation of \llt.
By inductive Claim~\ref{claim-multiset-finalized-before-removed} and Observation~\ref{obs-multiset-satisfies-con-mark-all-removed-recs}, Constraint~\ref{con-mark-all-removed-recs} is satisfied at all times before $s$ occurs.
The only places in the code where $s$ can occur are at lines~\ref{multiset-insert-llt-r}, \ref{multiset-insert-llt-p}, \ref{multiset-delete-llt-p}, \ref{multiset-delete-llt-r} and \ref{multiset-delete-llt-rnext}.
Suppose $s$ occurs at line~\ref{multiset-insert-llt-r}, \ref{multiset-insert-llt-p}, \ref{multiset-delete-llt-p} or \ref{multiset-delete-llt-r}.
Then, by inductive Claim~\ref{claim-multiset-search-postconditions}, argument to $s$ is non-\nil.
We can apply Lemma~\ref{lem-if-rec-traversed-then-rec-in-data-structure} to show that the argument to $s$ is in the data structure and, hence, initiated, at some point during the last \search\ before $s$.
Now, suppose $s$ occurs at line~\ref{multiset-delete-llt-rnext} (so $localr.next$ is the argument to $s$).
Then, $key = r.key$ when line~\ref{multiset-delete-return-false} is performed so, by the precondition of \del, $r.key \neq \infty$.
By inductive Claim~\ref{claim-multiset-sorted-list}, $r.next \neq \nil$ when \llt$(r)$ is performed at line~\ref{multiset-delete-llt-r}, so $localr.next \neq \nil$.
We can apply Lemma~\ref{lem-if-rec-traversed-then-rec-in-data-structure} to show that $localr.next$ is in the data structure and, hence, initiated, at some point between the start of the last \search\ before $s$ and the last \llt$(r)$ before $s$ (which reads $localr.next$ from $r.next$).

Suppose $s$ is a step that performs an invocation $S$ of \sct$(V, R, fld, new)$.
Then, the only places in the code where $s$ can occur are at lines~\ref{multiset-insert-sct1}, \ref{multiset-insert-sct2}, \ref{multiset-delete-sct1} and \ref{multiset-delete-sct2}.
It is a trivial exercise to inspect the code of \ins\ and \del, and argue that the process that performs $s$ has done an $\llt(r)$ linked to $S$ for each $r \in V$, that $R \subseteq V$, and that $fld$ points to a mutable field of a \rec \ in $V$.
It remains to prove that Precondition~\presctinfo\ and Precondition~\presctfld\ of \sct\ are satisfied.
Let $I$ be the invocation of $\llt(r)$ linked to $S$.
Suppose $s$ occurs at line~\ref{multiset-insert-sct1}.
The only step that can affect the claim is a linearized invocation $s$ of \sct$(V, R, fld, new)$ performed at line~\ref{multiset-insert-sct1}.
From the code, $fld$ is $r.count$.
Since $s$ is linearized, no invocation of \sct$(V'', R'', fld'', new'')$ with $r \in V''$ is linearized between $I$ and $s$.
Thus, $r.count$ does not change between when $I$ and $s$ are linearized.
From the code, $new$ is $count$ plus the value read from $r.count$ by $I$.
Therefore, $new$ is strictly larger than $r.count$ was when $I$ was linearized, which implies that $new$ is strictly larger than $r.count$ when $s$ is linearized.
This immediately implies Precondition~\presctinfo\ and Precondition~\presctfld\ of \sct. %that $new$ is not the initial value of $fld$.
%Suppose an invocation $S'$ of \sct$(V', R', fld, new)$ is linearized before $I$, to derive a contradiction.
%If $S'$ is the last invocation of \sct\ that modifies $fld$ before $I$ is linearized, then $I$ reads $new$ from $fld$, so $s$ must change $fld$ to a value strictly larger than $new$, which is a contradiction.
%Otherwise, since we have argued each \sct\ that modifies $fld$ in between when $S'$ and $I$ are linearized strictly increases the value stored in $fld$, 
%
Now, suppose $s$ occurs at line~\ref{multiset-insert-sct2}, \ref{multiset-delete-sct1} or \ref{multiset-delete-sct2}.
%It is easy to see that no process invokes $s$ after any process starts a successful invocation of \sct$(V', R', fld, new)$.
Then, $new$ is a pointer to a \listrec\ that was created after $I$.
Thus, no invocation of \sct$(V', R', fld, new)$ can even \textit{begin} before $I$.
We now prove that $new$ is not the initial value of the field pointed to by $fld$.
From the code, $fld$ is $p.next$.
If $p.next$ is initially \nil, then we are done.
Otherwise, $p.next$ initially points to some \listrec\ $r'$.
Clearly, $r'$ must be created before $p$.
Hence, $r'$ must be created before the invocation of \search\ followed a pointer to $p$.
Since $new$ is a pointer to a \listrec\ that is created after this invocation of \search, $new \neq r'$.

\textbf{Proof of Claim~\ref{claim-multiset-search-postconditions}.}
To affect this claim, $s$ must be the response of an invocation of \search$(key)$.
We prove a loop invariant that states $r$ is a \listrec, and either $p$ is a \listrec\ and $p.key < key$ or $p = head$.
Before the loop, $p = head$ and $r = head.next$.
By inductive Claim~\ref{claim-multiset-sorted-list} $head.next$ is always a \listrec, so the claim holds before the loop.
Suppose the claim holds at the beginning of an iteration.
Let $r$ and $p$ be the respective values of local variables $r$ and $p$ at the beginning of the iteration, and $r'$ and $p'$ be their values at the end of the iteration.
From the code, $p' = r$ and $r'$ is the value read from $r.next$ at line~\ref{multiset-search-advance-r}.
By the inductive hypothesis, $p'$ is a \listrec.
Since the loop did not exit before this iteration, $key > p'.key$.
Further, since \search$(key)$ is invoked only when $key < \infty$ (by inspection of the code and preconditions), $p'.key < \infty$.
By inductive Claim~\ref{claim-multiset-sorted-list}, $p'.next = r.next$ always points to a \listrec, so $r'$ is a \listrec, and the inductive claim holds at the end of the iteration.
Finally, the exit condition of the loop implies $key \le r'.key$, so \search\ satisfies its postcondition.

%By inductive Claim~\ref{claim-multiset-sorted-list}, at all times before $I$ returns, $head.next$ points to a \listrec, and the $next$ field of each \listrec\ points to a \listrec.
%Thus, the code of \search\ implies that $r$ is a pointer to a \listrec.
%The exit condition of the loop implies that $key \le r.key$.
%Suppose $I$ does not enter the loop at line~\ref{multiset-search-loop}.
%Then, by line~\ref{multiset-search-p}, $p = head$.
%Now, suppose $I$ enters the loop.
%Then, $I$ must perform line~\ref{multiset-search-advance-p}.
%The first time $I$ performs line~\ref{multiset-search-advance-p}, $p$ is assigned the value $r$, which was read from $head.next$, which we have argued is a pointer to a \listrec.
%Further, since $I$ passes the test at line~\ref{multiset-search-loop}, $p.key < key$.
%It is trivial to prove (with a loop invariant) that, at the end of each iteration of the loop, $p$ contains a value that was read from the $next$ pointer of a \listrec, and $p.key < key$.
%%Thus, the value $p$ returned by $I$ is a pointer to a \listrec\ which satisfies $p.key < key$.

\textbf{Proof of Claim~\ref{claim-multiset-finalized-before-removed}.}
Since a \rec \ can be removed from the data structure only by a change to a mutable field of some other \rec, this claim can be affected only by linearized invocations of \sct.
Suppose $s$ is a linearized invocation of \sct$(V, R, fld, new)$.
Then, $s$ can occur only at line~\ref{multiset-insert-sct1}, \ref{multiset-insert-sct2}, \ref{multiset-delete-sct1} or \ref{multiset-delete-sct2}.
Let $I$ be the invocation of \ins\ or \del\ in which $s$ occurs.
We proceed by cases.

Suppose $s$ occurs at line~\ref{multiset-insert-sct1}.
Then, $fld$ is a pointer to $r.count$. %, and inductive Claim~\ref{claim-multiset-sct-no-aba} implies that $new$
Thus, $s$ changes a $count$ field, \textit{not} a $next$ pointer.
Since this is the only change that is made by $s$, no \rec\ is removed by $s$, and no \rec\ is added by $s$.
Since $R = \emptyset$, the claim holds.

Suppose $s$ occurs at line~\ref{multiset-insert-sct2}.
Then, $fld$ is a pointer to $p.next$, and $new$ is a pointer to a new \listrec.
Before performing $s$, $I$ performs an invocation $L_1$ of \llt$(p)$, which returns a value different from \fail, or \finalized, at line~\ref{multiset-delete-llt-p}.
Just after performing $L_1$, $I$ sees that $localp.next = r$.
Note that $L_1$ is linked to $s$.
Since $s$ is linearized, and $p \in V$, $p.next$ does not change in between when $L_1$ and $s$ are linearized.
Therefore, $s$ changes $p.next$ from $r$ to point to a new \listrec\ whose $next$ pointer points to $r$.
Since $s$ is linearized, Lemma~\ref{lem-if-initiated-rec-not-in-data-structure-then-does-not-change} implies that $p$ must be in the data structure just before $s$ (and when its change occurs).
Since this is the only change that is made by $s$, no \rec\ is removed by $s$, and $new$ points to the only \rec\ that is added by $s$.
Since $R = \emptyset$, the claim holds.

Suppose $s$ occurs at line~\ref{multiset-delete-sct1} or line~\ref{multiset-delete-sct2}.
Then, $fld$ is a pointer to $p.next$, and $new$ is a pointer to a new \listrec.
Before performing $s$, $I$ performs invocations $L_1$ and $L_2$ of \llt$(p)$ and \llt$(r)$, respectively, which each return a value different from \fail, or \finalized.
Note that $L_1$ and $L_2$ are linked to $s$.
Just after performing $L_1$, $I$ sees that $localp.next = r$.
Since $s$ is linearized, and $p \in V$, $p.next$ does not change in between when $L_1$ and $s$ are linearized.
Similarly, since $r \in V$, $r.next$ does not change between when $L_1$ and $s$ are linearized.
Before $s$, $I$ sees $key = r.key$ at line~\ref{multiset-delete-return-false}.
By the precondition of \del, $r.key \neq \infty$.
Thus, inductive Claim~\ref{claim-multiset-sorted-list} (and the fact that keys do not change) implies that $r.next$ points to some \listrec\ $rnext = localr.next$ at all times between when $L_2$ and $s$ are linearized.
We consider two sub-cases.

\textit{Case I:}
$s$ occurs at line~\ref{multiset-delete-sct1}.
Therefore, $s$ changes $p.next$ from $r$ to point to a new \listrec\ whose $next$ pointer points to $rnext$ and, when this change occurs, $r.next$ points to $rnext$.
Since $s$ is linearized, Lemma~\ref{lem-if-initiated-rec-not-in-data-structure-then-does-not-change} implies that $p$ must be in the data structure just before $s$ (and when its change occurs).
Since this is the only change that is made by $s$, $r$ points to the only \rec\ that is removed by $s$, and $new$ points to the only \rec\ that is added by $s$.
Since $R = \{r\}$, the claim holds.

\textit{Case II:}
$s$ occurs at line~\ref{multiset-delete-sct2}.
%When $I$ copies $localr.next$ at line~\ref{multiset-delete-sct2}, inductive Claim~\ref{claim-multiset-sorted-list} implies that it reads a pointer to some \listrec\ $rnext'$ from $rnext.next$.
Since $rnext \in V$, $rnext.next$ does not change between when the \llt\ at line~\ref{multiset-delete-llt-rnext} and $s$ are linearized.
Thus, $rnext.next$ contains the same value $v$ throughout this time.
Therefore, $s$ changes $p.next$ from $r$ to point to a new \listrec\ whose $next$ pointer contains $v$ and, when this change occurs, $p.next$ points to $r$, $r.next$ points to $rnext$, and $rnext.next$ contains $v$.
Since $s$ is linearized, Lemma~\ref{lem-if-initiated-rec-not-in-data-structure-then-does-not-change} implies that $p$ must be in the data structure just before $s$ (and when its change occurs).
Since this is the only change that is made by $s$, $r$ and $rnext$ point to the only \rec s that are removed by $s$, and $new$ points to the only \rec \ that is added by $s$.
Since $R = \{r, next\}$, the claim holds.

\textbf{Proof of Claim~\ref{claim-multiset-sorted-list}.}
This claim can be affected only by a linearized invocation of \sct\ that changes a $next$ pointer.
%Since each $next$ pointer can be changed only by a successful \upcas, only successful \upcas s can affect this claim.
%Suppose $s$ is an \upcas.
Suppose $s$ is a linearized invocation of \sct$(V, R, fld, new)$.
Then, $s$ can occur only at line~\ref{multiset-insert-sct2}, \ref{multiset-delete-sct1} or \ref{multiset-delete-sct2}.
We argued in the proof of Claim~\ref{claim-multiset-finalized-before-removed} that, in each of these cases, $s$ changes $p.next$ from $r$ to point to a new \listrec, and that this is the only change that it makes.
Let $I$ be the invocation of \ins\ or \del\ in which $s$ occurs.
%We proceed by cases.

Suppose $s$ occurs at line~\ref{multiset-delete-sct1}.
We argued in the proof of Claim~\ref{claim-multiset-finalized-before-removed} that, at all times between when the \llt$(r)$ at line~\ref{multiset-delete-llt-r} and $s$ are linearized, $p.next$ points to $r$ and $r.next$ points to some \listrec\ $rnext$.
Therefore, $new.key = r.key$ and $new.next$ points to $rnext$.
We show $r$ is a \listrec\ (and not the $head$ entry point), and $r.key \neq \infty$.
Since $r.next$ points to a \listrec\ $rnext \neq \nil$, $r \neq \nil$ and $r.key \neq \infty$ (by the inductive hypothesis).
Similarly, since $p.next$ points to $r$, either $r = \nil$ or $r$ is a \listrec, so we are done.
Since $r.next$ points to $rnext$ just before $s$ is linearized, setting $new.next$ to point to $rnext$ does not violate the inductive hypothesis.
Since $p.next$ points to $r$, the inductive hypothesis implies that either $p$ is the $head$ entry point or $p.key < r.key$.
Clearly, setting $p.next$ to point to $new$ does not violate the inductive hypothesis in either case.

Suppose $s$ occurs at line~\ref{multiset-insert-sct2}.
Then, $new.key = key$ and $new.next$ points to $r$.
Before $s$, $I$ invokes \search$(key)$ at line~\ref{multiset-insert-search}, and then sees $key \neq r.key$ at line~\ref{multiset-insert-check-key}.
By inductive Claim~\ref{claim-multiset-search-postconditions}, this invocation of \search\ satisfies its postconditions, which implies that $r$ points to a \listrec\ which satisfies $key < r.key$.
Since \search$(key)$ is invoked only when $key < \infty$ (by inspection of the code and preconditions), $key < \infty$.
Thus, setting $new.next$ to point to $r$ does not violate the inductive hypothesis.
The post conditions of \search\ also imply that either $p$ is a \listrec\ and $p.key < key$ or $p = head$.
Therefore, setting $p.next$ to point to $new$ does not violate the inductive hypothesis.

Suppose $s$ occurs at line~\ref{multiset-delete-sct2}.
We argued in the proof of Claim~\ref{claim-multiset-finalized-before-removed} that, at all times between when the \llt$(r)$ at line~\ref{multiset-delete-llt-r} and $s$ are linearized, $p.next$ points to $r$, $r.next$ points to some \listrec\ $rnext$ (pointed to by $localr.next$) and $rnext.next$ points to some \listrec\ $rnext'$.
Thus, $new.key = rnext.key$ and $new.next$ points to $rnext'$.
Since $rnext.next$ points to a \listrec\ $rnext'$, $rnext.key < \infty$ (by the inductive hypothesis).
Therefore, since $rnext.next$ points to $rnext'$ just before $s$, setting $new.next$ to point to $rnext'$ does not violate the inductive hypothesis.
By the inductive hypothesis, either $p$ is the $head$ entry point, or $p.key < r.key < rnext.key = new.key < \infty$.
Clearly, setting $p.next$ to point to $new$ does not violate the inductive hypothesis in either case.
\end{chapscxproof}

\begin{cor} \label{cor-multiset-data-structure-always-sorted-list}
The $head$ entry point always points to a sorted list with strictly increasing keys. %, followed by an infinite list of \listrec s with key $\infty$.
\end{cor}
\begin{chapscxproof}
Immediate from Lemma~\ref{lem-multiset-constraints-invariants}.\ref{claim-multiset-sorted-list}.
\end{chapscxproof}

\begin{obs} \label{obs-multiset-satisfies-con-mark-all-removed-recs}
Lemma~\ref{lem-multiset-constraints-invariants}.\ref{claim-multiset-finalized-before-removed} implies Constraint~\ref{con-mark-all-removed-recs}.
%Lemma~\ref{lem-multiset-constraints-invariants}.\ref{claim-multiset-sct-no-aba} implies Constraint~\ref{con-use-of-sct}.
\end{obs}

We now argue that the multiset algorithm satisfies a constraint placed on the use of \llt\ and \sct.
This constraint is used to guarantee progress for \sct. %, is reproduced here, for ease of reference.
%Consider any execution that contains a configuration $C$ after which no field of any \rec\ changes.
%There must be a total order on all \rec s created during this execution such that, if \rec\ $r_1$ appears before \rec\ $r_2$ in the sequence $V$ passed to an invocation of \sct\ whose linked \llt s begin after $C$, then $r_1 < r_2$.

\begin{obs} \label{obs}
Consider any execution with %that contains
a configuration $C$ after which no field of any \rec\ changes.
There is a total order on all \rec s created during this execution such that, if \rec\ $r_1$ appears before \rec\ $r_2$ in the sequence $V$ passed to an invocation $S$ of \sct\ whose linked \llt s begin after $C$, then $r_1 < r_2$.
\end{obs}
\begin{chapscxproof}
Since the \llt s linked to $S$ begin after $C$, it follows immediately from the multiset code that $V$ is a subsequence of nodes in the list.
By Corollary~\ref{cor-multiset-data-structure-always-sorted-list}, they occur in order of strictly increasing keys, so $r_1$ before $r_2$ in $V$ implies $r_1.key < r_2.key$.
Thus, we take the total order on keys to be our total order.
\end{chapscxproof}

\begin{defn} \label{defn-multiset-in-data-structure}
The number of occurrences of $key \neq \infty$ \textbf{in the data structure} at time $t$ is $count$ if there is a \rec\ $r$ in the data structure at time $t$ such that $r.key = key$ and $r.count = count$, and zero, otherwise.
\end{defn}

We call an invocation of \ins\ or \del\ \textbf{effective} if it performs a linearized invocation of \sct\ (which either returns \true, or does not terminate).
From the code of \ins\ and \del, each \textbf{effective} invocation of \ins\ or \del\ performs exactly one linearized invocation of \sct, each invocation of \ins\ that returns is effective, and each invocation of \del\ that returns \true\ is effective.
We linearize each effective invocation of \ins\ or \del\ at its linearized invocation of \sct.
The linearization point for an invocation $I$ of \del$(key, count)$ that returns \false\ is subtle.
Suppose $I$ returns \false\ after seeing $r.key \neq key$.
Then, we must linearize it at a time when the nodes $p$ and $r$ returned by its invocation $I'$ of \search\ are both in the data structure and $p.next$ points to $r$.
By Observation~\ref{obs-multiset-satisfies-con-mark-all-removed-recs}, Constraint~\ref{con-mark-all-removed-recs} is satisfied.
This means we can apply Lemma~\ref{lem-if-rec-traversed-then-rec-in-data-structure} to show that there is a time during $I'$ when $p$ is in the data structure and $p.next = r$ (so $r$ is also in the data structure).
We linearize $I$ at the last such time.
Now, suppose $I$ returns \false\ after seeing $r.count < count$.
Then, we must linearize it at a time when the node $r$ returned by its invocation $I'$ of \search\ is both in the data structure, and satisfies $r.count < count$.
As in the previous case, we can apply Lemma~\ref{lem-if-rec-traversed-then-rec-in-data-structure} to show that there is a time after the start of $I'$, and at or before when $I$ reads a value $v$ from $r.count$ at line~\ref{multiset-delete-return-false}, such that $r$ is in the data structure and $r.count = v$.
We linearize $I$ at the last such time.
Similarly, we linearize each \func{Get} at the last time after the start of the \func{Search} in \func{Get}, and at or before when the \func{Get} reads a value $v$ from $r.count$, such that $r$ is in the data structure and $r.count = v$.
Clearly, each operation is linearized during that operation.

\begin{lem}
At all times $t$, the multiset $\sigma$ of keys in the data structure is equal to the multiset $\sigma_L$ of keys that would result from the atomic execution of the sequence of operations linearized up to time $t$.
\end{lem}
\begin{chapscxproof}
We prove this claim by induction on the sequence of steps taken in the execution.
Since $next$ pointers and $count$ fields can be changed only by linearized invocations of \sct \ (and $key$ fields do not change), we need only consider linearized invocations of \sct \ when reasoning about $\sigma$.
Thus, invocations of \ins\ and \del\ that are not effective cannot change the data structure.
Since invocations of \func{Get} do not invoke \sct, they cannot change the data structure.
Therefore, we need only consider effective invocations of \ins\ and \del\ when reasoning about $\sigma_L$.
Since each effective invocation of \ins\ or \del\ is linearized at its linearized invocation of \sct, the steps that can affect $\sigma$ and $\sigma_L$ are exactly the same.
\textbf{Base case.}
Before any linearized \sct\ has occurred, no $next$ pointer has been changed.
Thus, the data structure is in its initial configuration, which implies $\sigma = \emptyset$.
Since no effective invocation of \ins\ or \del\ has been linearized, $\sigma_L = \emptyset$.
\textbf{Inductive step.}
Let $s$ be a linearized invocation $S$ of \sct$(V, R, fld, new)$, $I$ be the (effective) invocation of \ins\ or \del\ that performs $S$, and $p$, $r$ and $rnext$ refer to the local variables of $I$.
Suppose $\sigma = \sigma_L$ before $s$.
Let $\sigma'$ denote $\sigma$ after $s$, and $\sigma_L'$ denote $\sigma_L$ after $s$.
We prove $\sigma' = \sigma_L'$.
%$S$ occurs at line~\ref{multiset-insert-sct1}, \ref{multiset-insert-sct2}, \ref{multiset-delete-sct1} or \ref{multiset-delete-sct2}.
%We proceed by cases.

Suppose $S$ is performed at line~\ref{multiset-insert-sct1}.
Then, $I$ is an invocation of \ins$(key, count)$, and $\sigma_L' = \sigma_L + \{count$ copies of $key\}$.
By Lemma~\ref{lem-multiset-constraints-invariants}.\ref{claim-multiset-finalized-before-removed}, no \rec \ is added or removed by $S$.
Before $I$ performs $S$, $I$ performs an invocation $L$ of $\llt(r)$ linked to $S$ at line~\ref{multiset-insert-llt-r}.
Since $S$ is linearized, no mutable field of $r$ changes between when $L$ and $S$ are linearized.
Therefore, the value $localr.count$ that $L$ reads from $r.count$ is equal to the value of $r.count$ at all times between when $L$ and $S$ are linearized, and line~\ref{multiset-insert-sct1} implies that $S$ changes $r.count$ from $localr.count$ to $localr.count + count$.
Since $S$ is linearized, Lemma~\ref{lem-if-initiated-rec-not-in-data-structure-then-does-not-change} implies that $r$ must be in the data structure just before $S$ is linearized.
By Lemma~\ref{lem-multiset-constraints-invariants}.\ref{claim-multiset-sorted-list}, $r$ is the only \listrec\ in the data structure with key $key$, so $\sigma$ contains exactly $v$ copies of $key$ just before $S$ is linearized.
Since this is the only change made by $S$, $\sigma' = \sigma + \{count$ copies of $key\}$, and the inductive hypothesis implies $\sigma' = \sigma_L'$.

Suppose $S$ is performed at line~\ref{multiset-insert-sct2}.
Then, $I$ is an invocation of \ins$(key, count)$, and $\sigma_L' = \sigma_L + \{count$ copies of $key\}$.
By Lemma~\ref{lem-multiset-constraints-invariants}.\ref{claim-multiset-finalized-before-removed}, no \rec \ is removed by $S$, and only $new$ is added by $S$.
From the code of \ins, $new.key = key$ and $new.count = count$.
Therefore, $\sigma' = \sigma + \{count$ copies of $key\}$, and the inductive hypothesis implies $\sigma' = \sigma_L'$.

Suppose $S$ is performed at line~\ref{multiset-delete-sct1}.
Then, $I$ is an invocation of \del$(key, count)$.
Before $I$ performs $S$, $I$ performs an invocation $L$ of $\llt(r)$ linked to $S$ at line~\ref{multiset-delete-llt-r}.
Since $S$ is linearized, no mutable field of $r$ changes between when $L$ and $S$ are linearized.
Thus, the value $localr.count$ that $L$ reads from $r.count$ is equal to the value of $r.count$ at all times between when $L$ and $S$ are linearized.
This implies that $I$ sees $r.key = key$ and $r.count \ge count$ at line~\ref{multiset-delete-return-false}.
By Lemma~\ref{lem-multiset-constraints-invariants}.\ref{claim-multiset-finalized-before-removed}, $r$ is the only \rec\ removed by $S$, and $new$ is the only \rec\ added by $S$.
By Definition~\ref{defn-rec-in-added-removed}, $r$ must be in the data structure just before $S$ is linearized.
By Lemma~\ref{lem-multiset-constraints-invariants}.\ref{claim-multiset-sorted-list}, $r$ is the only \listrec\ in the data structure with key $key$.
Hence, $\sigma$ contains exactly $localr.count$ copies of $key$ just before $S$ is linearized.
From the code of \del, $new.key = r.key$ and $new.count = localr.count - count$.
Therefore, $\sigma' = \sigma - \{count$ copies of $key\}$.
By the inductive hypothesis, $\sigma = \sigma_L$.
Thus, there are $localr.count \ge count$ copies of $key$ in $\sigma_L$.
Therefore, if $I$ is performed atomically at its linearization point, it will enter the if-block at line~\ref{multiset-delete-check-count-less}, so $\sigma_L' = \sigma_L - \{count$ copies of $key\} = \sigma'$.

Suppose $S$ is performed at line~\ref{multiset-delete-sct2}.
Then, $I$ is an invocation of \del$(key, count)$.
Before $I$ performs $S$, $I$ performs an invocation $L$ of $\llt(r)$ linked to $S$ at line~\ref{multiset-delete-llt-r}.
Since $S$ is linearized, no mutable field of $r$ changes between when $L$ and $S$ are linearized.
Thus, the value $localr.count$ that $L$ reads from $r.count$ is equal to the value of $r.count$ at all times between when $L$ and $S$ are linearized.
This implies that $I$ sees $r.key = key$ and $r.count \ge count$ at line~\ref{multiset-delete-return-false}, and $count \ge r.count$ at line~\ref{multiset-delete-check-count-less}.
Hence, $r.count = count$ at all times between when $L$ and $S$ are linearized.
Let $rnext$ be the \listrec\ pointed to by $I$'s local variable $localr.next$.
(We know $rnext$ is a \listrec, and not \nil, from $r.key = key < \infty$ and Lemma~\ref{lem-multiset-constraints-invariants}.\ref{claim-multiset-sorted-list}.)
After $L$, $I$ performs an invocation $L'$ of $\llt(rnext)$ linked to $S$ at line~\ref{multiset-delete-llt-rnext}.
By the same argument as for $r.count$, the value $v$ that $L'$ reads from $rnext.count$ is equal to the value of $rnext.count$ at all times between when $L'$ and $S$ are linearized.
By Lemma~\ref{lem-multiset-constraints-invariants}.\ref{claim-multiset-finalized-before-removed}, $r$ and $rnext$ are the only \rec s removed by $S$, and $new$ is the only \rec \ added by $S$.
By Definition~\ref{defn-rec-in-added-removed}, $r$ and $rnext$ must be in the data structure just before $S$.
By Lemma~\ref{lem-multiset-constraints-invariants}.\ref{claim-multiset-sorted-list}, $r$ is the only \listrec\ in the data structure with key $key$, and $rnext$ is the only \listrec\ in the data structure with its key.
Hence, $\sigma$ contains exactly $r.count = count$ copies of $key$, and exactly $v$ copies of $rnext.key$.
From the code of \del, $new.key = rnext.key$ and $new.count = rnext.count = v$.
Therefore, $\sigma' = \sigma - \{count$ copies of $key\}$
By the inductive hypothesis, $\sigma = \sigma_L$.
Thus, there are exactly $count$ copies of $key$ just before $I$ in the linearized execution.
From the code of \del, in the linearized execution, $I$ will enter the else block at line~\ref{multiset-delete-check-count-else}, so $\sigma_L' = \sigma_L - \{count$ copies of $key\} = \sigma'$.
\end{chapscxproof}

\begin{lem}
Each invocation of \func{Get}$(key)$ that terminates returns the number of occurrences of $key$ in the data structure just before it is linearized.
\end{lem}
\begin{chapscxproof}
Consider any invocation $I$ of \func{Get}$(key)$.
Let $I'$ be the invocation of \search$(key)$ performed by \func{Get}$(key)$, and $p$ and $r$ refer to the local variables  of $I'$.
By Lemma~\ref{lem-multiset-constraints-invariants}.\ref{claim-multiset-search-postconditions}, $I'$ satisfies its postcondition, which means that $key \le r.key$, and either $p.key < key$ or $p = head$.
We proceed by cases.
Suppose $key = r.key$.
Then, after $I'$, $I$ reads a value $v$ from $r.count$ and returns $v$.
By Observation~\ref{obs-multiset-satisfies-con-mark-all-removed-recs}, Constraint~\ref{con-mark-all-removed-recs} is satisfied.
By Lemma~\ref{lem-if-rec-traversed-then-rec-in-data-structure}, there is a time after the start of $I'$, and at or before when $I$ reads $r.count$, such that $r$ is in the data structure and $r.count = v$.
$I$ is linearized at the last such time.
By Corollary~\ref{cor-multiset-data-structure-always-sorted-list}, $r$ is the only \rec\ in the list that contains key $key$.
Suppose that either $key < r.key$ and $p = head$, or $key < r.key$ and $p.key < key$.
Then, $I$ returns zero.
By Lemma~\ref{lem-if-rec-traversed-then-rec-in-data-structure}, at sometime during $I'$, $p$ was in the data structure and $p.next$ pointed to $r$.
$I$ is linearized at the last such time.
By Corollary~\ref{cor-multiset-data-structure-always-sorted-list}, the data structure contains no occurrences of $key$ when $I$ is linearized.
\end{chapscxproof}

\begin{lem}
Each invocation $I$ of \del$(key, count)$ that terminates returns \true \ if the data structure contains at least $count$ occurrences of $key$ just before $I$ is linearized, and \false \ otherwise.
\end{lem}
\begin{chapscxproof}
\textbf{Case I:} $I$ returns \false.
In this case, $I$ satisfies $key \neq r.key$ or $localr.count < count$ at line~\ref{multiset-delete-return-false}.
Suppose $key \neq r.key$.
Then, by the postcondition of \search, $key < r.key$, and either $p.key < key$ or $p = head$.
By Observation~\ref{obs-multiset-satisfies-con-mark-all-removed-recs}, Constraint~\ref{con-mark-all-removed-recs} is satisfied.
By Lemma~\ref{lem-if-rec-traversed-then-rec-in-data-structure}, there is a time during the preceding invocation $I'$ of \search, when $p$ was in the data structure and $p.next$ pointed to $r$.
$I$ is linearized at the last such time.
Corollary~\ref{cor-multiset-data-structure-always-sorted-list} implies that there are no occurrences of $key$ in the data structure when $I$ is linearized.
By the precondition of \del, $count > 0$, so the claim is satisfied. %there are less than $count$ occurrences of $key$ when $I$ is linearized.

Now, suppose $localr.count < count$ at line~\ref{multiset-delete-return-false}.
By Lemma~\ref{lem-if-rec-traversed-then-rec-in-data-structure}, there is a time after the start of $I'$, and before $I$'s \llt$(r)$ reads $localr.count$ from $r.count$, such that $r$ is in the data structure and $r.count = localr.count$.
$I$ is linearized at the last such time.
By Corollary~\ref{cor-multiset-data-structure-always-sorted-list}, $r$ is the only \rec\ in the list that contains key $key$, so there are $r.count < count$ occurrences of $r.key = key$ in the data structure when $I$ is linearized.

\textbf{Case II:} $I$ returns \true.
In this case, $I$ satisfies $key = r.key$ and $localr.count \ge count$ at line~\ref{multiset-delete-return-false}, and $I$ is linearized at an invocation $S$ of \sct\ at line~\ref{multiset-delete-sct1} or \ref{multiset-delete-sct2}.
In each case, Lemma~\ref{lem-multiset-constraints-invariants}.\ref{claim-multiset-finalized-before-removed} implies that $r$ is removed by $S$, so $r$ is in the data structure just before $S$ is linearized.
Hence, $r$ is in the data structure just before $I$ is linearized.
Before $I$ performs $S$, $I$ performs an invocation $L$ of $\llt(r)$ linked to $S$ at line~\ref{multiset-delete-llt-r} that reads $localr.count$ from $r.count$.
%By Lemma~\ref{lem-rec-in-data-structure-just-before-llt}, $r$ is in the data structure just before $I$ is linearized.
Since $S$ is linearized, no mutable field of $r$ changes between when $L$ and $S$ are linearized.
Therefore, the value of $localr.count$ is equal to the value of $r.count$ at all times between when $L$ and $S$ are linearized.
Thus, just before $I$ is linearized, $r$ is in the data structure and $r.count \ge count$.
%Since $I$ is linearized at a step, rather than a configuration, this still holds \textit{when} $I$ is linearized.
Finally, Corollary~\ref{cor-multiset-data-structure-always-sorted-list} implies that $r$ is the only \rec\ in the list that contains key $key$, so the claim holds.
\end{chapscxproof}

In Section~\ref{progress-spec}, we explained that a process cannot invoke \sct$(V, R, fld, new)$ until it performs successful invocations of \llt\ on each $r \in V$, so an \llt$(r)$ that returns \finalized\ can prevent a process from invoking \sct$(V, R, fld, new)$.
Thus, if a process can repeatedly perform \llt$(r)$ on a finalized \rec\ $r$, then doing so may prevent it from performing any successful \sct\ operations.
As we discussed there, one way to prevent this from happening is to have each process explicitly keep track of all \finalized\ \rec s it has seen (so it can avoid performing \llt\ on them again).
We now prove that this kind of explicit bookkeeping is unnecessary for the multiset algorithm, because processes will not repeatedly invoke \llt$(r)$ on a finalized \rec\ $r$.

\begin{lem} \label{lem-multiset-only-one-finalized-llt}
No process performs more than one invocation of $\llt(r)$ that returns \finalized, for any \rec\ $r$.
\end{lem}
\begin{chapscxproof}
Let $r$ be a \rec.
Suppose, to derive a contradiction, that a process $p$ performs two invocations $L$ and $L'$ of $\llt(r)$ that return \finalized.
Without loss of generality, let $L$ occur before $L'$.
From the code of \ins\ and \del, $p$ must perform an invocation of \search, $L$, another invocation $I$ of \search, and then $L'$.
Since $L$ returns \finalized, it is linearized after an invocation $S$ of \sct$(V, R, fld, new)$ with $r \in R$.
By Lemma~\ref{lem-multiset-constraints-invariants}.\ref{claim-multiset-finalized-before-removed}, $r$ is removed from the data structure by $S$.
We now show that $r$ cannot be added back into the data structure by any subsequent invocation of \sct.
From the code of \ins\ and \del, each invocation of \sct$(V', R', fld', new')$ that changes a $next$ pointer is passed a newly created \listrec, that is not known to any other process, as its $new'$ argument.
This implies that $new'$ is not initiated, and cannot have previously been removed from the data structure.
Therefore, $r$ is not in the data structure at any point during $I$.
By Observation~\ref{obs-multiset-satisfies-con-mark-all-removed-recs}, Constraint~\ref{con-mark-all-removed-recs} is satisfied.
By Lemma~\ref{lem-if-rec-traversed-then-rec-in-data-structure}, $r$ is in the data structure at some point during $I$, which is a contradiction.
\end{chapscxproof}

Finally, we use the progress guarantees provided by \llt\ and \sct\ to prove that the multiset is lock-free. % all operations on the multiset are lock-free.

\begin{lem} \label{lem-multiset-progress}
If operations (\ins, \del\ and \func{Get}) are invoked infinitely often, then operations complete infinitely often.
\end{lem}
\begin{chapscxproof}
Suppose, to derive a contradiction, that operations are invoked infinitely often but, after some time $t$, no operation completes.
If \sct s are performed infinitely often, then they will succeed infinitely often and, hence, operations will succeed infinitely often.
Thus, there must be some time $t' \ge t$ after which no \sct\ is performed.
Then, after $t'$, the data structure does not change, and only a finite number of nodes with keys different from $\infty$ are ever added to the data structure.
Consider an invocation $I$ of \search$(key)$ that is executing after $t'$.
Each time $I$ performs line~\ref{multiset-search-advance-r}, it reads a \listrec\ $rnext$ from $r.next$, and $rnext.key > r.key$.
Therefore, by Corollary~\ref{cor-multiset-data-structure-always-sorted-list}, $I$ will eventually see $r.key = \infty$ at line~\ref{multiset-search-loop}.
%Since $\infty$ cannot be passed as the $key$ argument to \ins, \del\ or \func{Get}, every invocation of \search\ eventually completes.
This implies that every invocation of \func{Get} eventually completes.
Therefore, \ins\ and \del\ must be invoked infinitely often after $t'$.
From the code of \ins\ (\del), in each iteration of the while loop, a \search\ is performed, followed by a sequence of \llt s.
If these \llt s all return values different from \fail\ or \finalized, then an invocation of \sct\ is performed.
Since every invocation of \search\ eventually completes, Definition~\ref{defn-set-up-sct} implies that invocations of \sct\ are set up infinitely often.
Thus, invocations of \sct\ succeed infinitely often.
From the code of \ins\ and \del, after performing a successful invocation of \sct, an invocation of \ins\ or \del\ will immediately return.
\end{chapscxproof}

\chapter{A template for implementing trees} \label{chap-template}
% !TEX root = paper.tex

\begin{thesisonly}
The binary search tree (BST) is among the most important data structures.
Previous concurrent implementations of balanced BSTs without locks either used coarse-grained transactions, which limit concurrency, or lacked rigorous proofs of correctness.
In this Chapter, we describe a general technique for producing a lock-free implementation of \textit{any} data structure based on a down-tree (a directed acyclic graph of indegree one), with updates that modify any connected subgraph of the tree atomically.
Our approach drastically simplifies the task of proving correctness.
This makes it feasible to develop provably correct implementations of non-blocking balanced BSTs with updates that synchronize on a small constant number of nodes.
\end{thesisonly}

\begin{thesisnot}
The binary search tree (BST) is among the most important data structures.
Previous concurrent implementations of balanced BSTs without locks
%Although there are fast concurrent implementations of unbalanced BSTs without using locks,
%this has not been the case for balanced BSTs.
%Although balanced binary search trees are among the most important data structures, there has %been little success in providing fast concurrent implementations of them without using locks.
%Most previously published attempts
either used coarse-grained transactions, which limit concurrency, or lacked rigorous proofs of correctness.
In this paper, we describe a general technique for implementing 
\textit{any} data structure based on a down-tree (a directed acyclic graph 
of indegree one),
%It can be used to implement
with updates that modify any connected subgraph of the tree atomically.
% such as those in Figure~\ref{fig-tree-pictures}.
The resulting implementations are non-blocking, which means that some process is always guaranteed to make progress, even if processes crash.
Our approach drastically simplifies the task of proving correctness.
This makes it feasible to develop provably correct implementations of non-blocking balanced BSTs with fine-grained synchronization
(i.e., with updates that synchronize on a small constant number of nodes).
\end{thesisnot}

As with all concurrent implementations, the implementations obtained using our technique are more efficient if each update to the data structure involves a small number of nodes near one another.
We call such an update {\em localized}.
We use \emph{operation} to denote an operation of the abstract data type (ADT) being implemented by the data structure.
Operations that cannot modify the data structure are called \emph{queries}.
% (henceforth called an operation) either performs an atomic update, or is called a \textit{query}.
For some data structures, such as Patricia tries and leaf-oriented BSTs,
operations modify the data structure using a single localized update.
In some other data structures, operations that modify the data structure
%every atomic update %performed by an operation  is inherently localized.
%In cases where updates are not localized, sometimes they
%Some atomic updates that are not localized
can be split into several
%atomic
localized updates that can be freely interleaved.

A particularly interesting application of our technique is to implement \textit{relaxed-balance} versions of sequential data structures efficiently.
Relaxed-balance data structures decouple
updates that rebalance the data structure from operations,
%operations from updates that rebalance the data structure,
and allow
%the
updates that accomplish rebalancing to be delayed and freely interleaved with
other updates.
%updates performed by operations.
For example, a chromatic tree is a relaxed-balance version of a red-black tree (RBT) which splits up the insertion or deletion
of a key and any subsequent rotations into a sequence of 
%atomic
localized updates.
%For example, consider a red-black tree (RBT), which inserts a key and performs several rotations together as a single atomic action.
%A chromatic tree splits these steps into a sequence of atomic local actions.
%
%ADT operations that modify such data structures are typically divided into several atomic local updates.
There is a rich literature of relaxed-balance versions of sequential
data structures \cite{DBLP:journals/acta/Larsen98},
and several papers (e.g., \cite{LOS01}) have described general techniques
that can be used to easily produce
%\fchanged{large classes of them from}
them from large classes of
%relaxations of large classes of
existing sequential data structures.
The small number of nodes involved in each update makes relaxed-balance data structures perfect candidates for efficient implementation using our technique.

\subsubsection*{Our Contributions}

\begin{itemize}
\item We provide a simple template that can be filled in to obtain an implementation of any update for a data structure based on a down-tree.
We prove that any data structure that follows our template for all of its updates will automatically be linearizable and non-blocking.
The template takes care of all process coordination, so the data structure designer is able to think of updates as atomic steps.
%This makes the implementation and proofs of correctness
%% of data structures based on down-trees 
%significantly shorter and simpler.

\item  To demonstrate the use of our template, we provide a complete, provably correct, non-blocking linearizable implementation of a chromatic tree \cite{NS96}, which is a
relaxed-balanced version of a RBT. %red-black tree (RBT).
To our
%the authors'
knowledge, this is the first provably correct, non-blocking balanced BST in which updates synchronize on a small constant number of nodes.
%We prove our
Our %We don't give a proof in the paper!
chromatic trees always have height $O(c+\log n)$, where $n$ is the number of keys stored in the tree and $c$ is the number of insertions and deletions that are in progress (proved in Section~\ref{height-bound}).
%Our data structure stores an ordered dictionary, providing inserts and deletes as well as searches, successor queries, and predecessor queries.

%\trevor{possibly something from this inserted}
%We also provide mechanisms for implementing queries.
%Some queries can be performed efficiently using only reads.
%For example, in a search tree where the keys of the nodes are immutable, one can search for a key and retrieve its associated value by simply reading keys and child pointers, and the value in the node reached.
%We prove such a search is linearizable even if concurrent update operations occur along the path (see Sec.~\ref{}).
%The implementation of \llt\ and \sct\ also provides an easy way to take a snapshot of a set of nodes, which facilitates the implementation of more complex queries.

\item 
We show that sequential implementations of some queries
are linearizable,
%give correct, linearizable answers,
even though they completely ignore concurrent updates.
For example, an ordinary BST search (that works when there is no concurrency) also works in our chromatic tree.
Ignoring updates makes searches very fast.
We also describe how to perform
successor queries in our chromatic tree, which
%%some
%more complex queries
interact properly with updates that follow our template.
%for example,
%%predecessor and
%successor queries in our chromatic tree
%(Section \ref{sec-chromatic-succ}).

%\item There is a rich literature of relaxed-balance data structures waiting to be implemented using our template.
%See \cite{DBLP:journals/acta/Larsen98} for a survey.
%Several papers (e.g., \cite{LOS01}) have also described general techniques that can be used
%to easily produce relaxations of large classes of existing sequential data structures.

\item We show experimentally that our Java implementation of a chromatic tree rivals, and often significantly outperforms, known highly-tuned concurrent dictionaries,
over a variety of workloads, contention levels and thread counts.
For example, with 128 threads, our algorithm outperforms Java's non-blocking skip-list by 13\% to 156\%, the lock-based AVL tree of Bronson et~al. by 63\% to 224\%, and a RBT that uses software transactional memory (STM) by 13 to 134 times (Section~\ref{sec-chromatic-exp}).
\end{itemize}

%\begin{figure}[tb]
%\centering
%\includegraphics{chap-template/abstractions.pdf}
%\caption{
%Positioning this work in a stack of abstractions. \textbf{[[[note: these citations are hard-coded]]]}
%}
%\label{fig-abstractions}
%\end{figure}

\section{Related work} \label{sec-chromatic-related}

There are many lock-based implementations of search tree data structures.
(See \cite{AKKMT12,BCCO10:ppopp} for state-of-the-art examples.)
Here, we focus on implementations that do not use locks.
Valois \cite{Valois:1995} sketched an implementation of %approach to implementing
non-blocking
node-oriented
%\fchanged{(i.e., dictionary keys are stored at internal nodes, as well as leaves)}
BSTs from CAS.
Fraser \cite{Fraser2004} gave a non-blocking BST %implementation 
using 8-word CAS, but did not provide 
a full proof of correctness.  He also described how
%the
multi-word CAS can be implemented 
from single-word CAS instructions.
Ellen et al.~\cite{Ellen:2010} gave a
provably correct, non-blocking
implementation of leaf-oriented
BSTs directly from single-word CAS.
A similar approach was used for $k$-ary search trees \cite{BH11}
and Patricia tries \cite{Shafiei:2013}.
%Their trees were generalized to $k$-ary search trees \cite{BH11}. %\cite{BA12,BH11}.
%They
All three
used the cooperative technique originated by Turek, Shasha and 
Prakash~\cite{TSP92} and Barnes~\cite{Barnes:1993}.
Howley and Jones \cite{Howley:2012} used a similar approach to build
%a
node-oriented BSTs.  They tested
their implementation using a model checker, but did not prove it correct.
Natarajan and Mittal \cite{Natarajan:2014} give another leaf-oriented
BST implementation, together with a sketch of
correctness.
%a correctness proof.
Instead of marking nodes, it marks edges.
This enables insertions to be accomplished by a single CAS, so 
%helping is not needed.
they do not need to be helped.
It also combines deletions that would otherwise conflict.
%It allows deletions that modify overlapping parts of the tree to succeed simultaneously.}
%Natarajan and Mittal \cite{NM14-incompletecitation} give another leaf-oriented BST %implementation that is different from \cite{EFRB10:podc} in that it allows deletions that modify 
%overlapping parts of the tree to succeed simultaneously, and requires less auxiliary information %in the data structure to facilitate helping. 
All of these trees are not balanced, so the height of a tree with $n$ keys can be $\Theta(n)$.

Tsay and Li \cite{TL94} gave a general approach for implementing trees in a wait-free manner
using LL and SC operations (which can, in turn be implemented from CAS, e.g., \cite{Anderson:1995}).  However,
their technique requires every process accessing the tree (even for read-only operations such
as searches) to copy an entire path
of the tree starting from the root.
Concurrency is severely limited, since every operation
must change the root pointer.
Moreover, an extra level of indirection is required
%in
for
every child pointer.

Red-black trees  \cite{Bay72,GS78} are
well known BSTs
%a classical type of BST
that have height $\Theta(\log n)$.
Some
%previous
attempts have been made to implement RBTs without using locks.
It was observed that the approach of Tsay and Li could be used to 
implement wait-free RBTs
\cite{NSM13}
%\cite{Sav09}
and, furthermore, this could be done so
that only updates must copy a path;
searches may simply read the path.
However, the concurrency of updates is still very limited.
%\eric{I reworded the next sentence to make it consistent with
%a sentence in the intro about previously published RBTs }
Herlihy et al.~\cite{Herlihy2003} and Fraser and Harris \cite{Fra07} 
experimented on RBTs implemented using
software transactional memory (STM),
which only satisfied obstruction-freedom, a weaker progress property.
%and satisfying the weaker progress property of obstruction-freedom.
Each insertion or deletion, together with necessary rebalancing is enclosed in a single
large transaction, which can touch all nodes on a path from the root to a leaf.
\begin{ignore}
\eric{Very little detail given in those articles.
As far as I can tell from looking at p.100 of Herlihy et al and Fraser and Harris's source 
code of rb\_stm.c, they just put the sequential code for an insert, delete, lookup inside 
a giant transaction.}
\end{ignore}

Some researchers have attempted fine-grained approaches to
build non-blocking balanced search trees, but they all
use extremely complicated process coordination schemes.
Spiegel and Reynolds \cite{SR10} described a non-blocking 
data structure that 
combines elements of B-trees and skip lists.
Prior to this paper, it was the leading implementation of an ordered dictionary.
However, the authors provided only
%At most, they provide
a brief justification of correctness. %, the data structure has not been proved correct.
Braginsky and Petrank~\cite{BP12} described
%rather complex
%non-blocking implementations of other B-tree variants.
a B+tree implementation.
Although they
%provide
have posted
a correctness proof, it is very long and
complex.
%complicated.
%\trevor{This last comment will likely be obviated by an explicit comparison in the conclusion of the complexity of our proofs.}
%\begin{ignore}
%\eric{Details of Spiegel and Reynolds algorithms are omitted in paper, but some more details appear in Spiegel's thesis available from his web page, and the real details are in the 
%Java code for their implementation, available from https://github.com/mspiegel/lockfreeskiptree.
%They claim searches are wait-free.  There is a 3-page ``proof'' of correctness on pp.82-85 of the thesis, that doesn't convince me (but I didn't spend much time on it).
%}
%\eric{I deleted ``The only previous nonblocking implementations of
%search trees with detailed proofs of correctness are those in
%\cite{EFRB10:podc} and \cite{BP12}.
%%\eric{Is this fair to \cite{Kim05}, which also includes a proof?  }
%because it is not fair to \cite{Kim05} which also included a proof, and
%it suggests that we have some confidence in the proof of \cite{BP12}.
%Faith can remove this comment if she agrees with the deletion of the sentence.
%}
%
%Except for those that use software transactional memory,
%the previous implementations of balanced search trees
%use extremely complicated process coordination schemes.
%%prohibitively complex (some requiring over 100 pages to describe). 
%Furthermore, rigorous proofs of correctness have not appeared.
%\end{ignore}

\medskip

In a balanced search tree, a process is typically responsible
for restoring balance after an insertion or deletion
by performing a series of rebalancing steps along the path
from the root to the location
where the insertion or deletion occurred.
Chromatic trees, introduced by Nurmi and Soisalon-Soininen \cite{NS96},
decouple the updates that perform the insertion or deletion from the updates that perform the rebalancing steps.
Rather than treating an insertion or deletion and its associated rebalancing steps as a single, large
%atomic
update, it is broken into smaller, localized updates that can be interleaved, allowing more concurrency.
This decoupling originated in the work
of Guibas and Sedgewick \cite{GS78} and Kung and Lehman \cite{KL80}.
We use the leaf-oriented chromatic trees by
Boyar, Fagerberg and Larsen \cite{Boyar97amortizationresults}.
They provide a family of local rebalancing steps which can be executed in any
order, interspersed with insertions and deletions.
Moreover, an amortized \textit{constant} number of rebalancing steps
per \ins\ or \del\ is sufficient to restore balance for any sequence of operations.
We have also used our template to implement a non-blocking version of Larsen's
leaf-oriented relaxed AVL tree \cite{Lar00}.
In such a tree, an amortized \textit{logarithmic} number of rebalancing steps
per \ins\ or \del\ is sufficient to restore balance.

\begin{ignore}
In a balanced search tree, a process is typically responsible
for restoring balance after an insertion or deletion
by performing a series of %changes to the tree, called 
%{\em rebalancing steps},
rebalancing steps
along the path 
from the root to the location
where the update occurred.
Chromatic trees, introduced by Nurmi and Soisalon-Soininen \cite{NS96},
%were designed as a version of RBTs that relax the balance conditions
%to make them more amenable to concurrent (lock-based) implementations.
%They %completely 
decouple updates from the rebalancing steps.  
Rather than treating the update and its associated rebalancing steps as a single, large
atomic update, they are broken into smaller, localized updates that can be interleaved, allowing more concurrency.
%In our implementation of search trees, we use 
The idea of decoupling updates and rebalancing steps
in concurrent lock-based implementations of search trees 
originated in the work
of Guibas and Sedgewick \cite{GS78} and Kung and Lehman \cite{KL80}.
%on search trees with relaxed
%balanced properties for concurrent implementations.
%A great deal of research has been done on relaxed-balance trees since then.
%Chromatic trees were introduced by Nurmi and Soisalon-Soininen \cite{NS96}
%as a version of RBTs with relaxed balance properties that could be implemented using locks.
We use the leaf-oriented chromatic trees by 
Boyar, Fagerberg and Larsen \cite{Boyar97amortizationresults}.
They provide a family of local rebalancing operations which can be executed in any
order, interspersed with insertions and deletions.
Moreover, an amortized \textit{constant} number of rebalancing steps
per \ins\ or \del\ are sufficient to restore balance for any sequence of updates.
We have also used our template to implement a non-blocking version of Larsen's leaf-oriented relaxed AVL tree \cite{Lar00}.
In such a tree, an amortized \textit{logarithmic} number of rebalancing steps per \ins\ or \del\ are sufficient to restore balance.
\end{ignore}

There is also a node-oriented relaxed AVL tree by
Boug\'{e} et~al.~\cite{BGMS98}, in which
%A relaxed balance version of an AVL tree by Boug\'{e} et~al.~\cite{BGMS98} has recently been %implemented several times, using locks and lock-based STM.
%In such a tree,
an amortized \textit{linear} number of rebalancing steps per \ins\ or \del\ is sufficient to restore balance.
Bronson et~al.~\cite{BCCO10:ppopp} developed a highly optimized fine-grained locking implementation
of this data structure
using optimistic concurrency techniques to improve search performance.
Deletion of a key stored in an internal node with two children is done
by simply marking the node and a later insertion of the same key can
reuse the node by removing the mark.
If all internal nodes are marked, the tree is essentially leaf-oriented.
%equivalent to a leaf-oriented tree.}
%This implementation is a partially external tree, which means that keys are stored both in the %internal nodes and in the leaves, and it performs logical deletion when deleting a key from a %node that has two children.
Crain et~al.
gave a different implementation using lock-based STM
\cite{Crain:2012:SBS:2145816.2145837}
and locks \cite{Crain:2013},
in which \emph{all} deletions are done by marking the node containing the key.
Physical removal of nodes and rotations are performed
by one separate thread. Consequently, the tree can become very unbalanced.
%THE FOLLOWING WAS REMOVED BECAUSE IT ONLY APPEARS IN THEIR JOURNAL VERSION.
%They only show that the tree eventually becomes
%balanced when contention disappears and updates stop.
%To improve performance, they used nontransactional reads.
% implemented Boug\'{e}'s tree using lock based STM and described how to keep transactions %small by traversing the tree using non-transactional reads.
%STM can be implemented in a non-blocking way, but such implementations currently involve %too much overhead to be practical.
%Implementing their data structure using non-blocking STM would introduce too much overhead %to be practical.
%Moreover, the effort expended to keep transactions small appears to be greater than the effort %that would be required to produce an implementation using our template.
%Their C implementation is more than five times longer than our Java implementation of a %chromatic tree (which is a considerably more complex sequential data structure).
Drachsler et~al.~\cite{Drachsler2014}
give another fine-grained lock-based implementation,
in which deletion physically removes the node containing the key
and searches are non-blocking.
Each node also contains predecessor and successor pointers,
so when a search ends at an incorrect leaf, sequential search can be performed
to find the correct leaf.
%which physically removes a node during deletion
%describe a fine-grained lock-based implementation of a threaded BST in which search 
%operations can ignore locks, and describe how to use the same approach to obtain an %implementation of Boug\'{e}'s tree.
%There is no non-blocking implementation of  a node-oriented relaxed AVL tree,
A non-blocking implementation of Boug\'{e}'s tree has not appeared,
but our template would make it easy to produce one.

%\trevor{Need to cite the ``speculation friendly'' balanced BST using STM by Crain, et~al. that appeared at PPoPP'12, since it decouples updates from rebalancing steps and uses fine-grained transactions.
%It is difficult to say whether we should discuss Crain's data structure here, or in the paragraph on fine-grained approaches.}

\section{\llt, \sct\ and \vlt\ primitives} \label{sec-primitives}

\begin{thesisonly}
Our tree update template uses the \llt\ and \sct\ primitives described in Chapter~\ref{chap-scx}.
The benefit of using \llt\ and \sct\ is two-fold: the template can be described quite simply, and much of the complexity of its correctness proof is encapsulated in that of \llt\ and \sct.
Recall that the implementation of the primitives from CAS in Chapter~\ref{chap-scx} is more efficient if the user of the primitives can guarantee that the following two constraints are satisfied.
The first constraint prevents the ABA problem for the \cas\ steps that actually perform the updates, and the second prevents processes from encountering livelock due to freezing \rec s in different orders.

\begin{compactenum}[\hspace{3.4mm}{\bf Constraint} \bfseries 1:]
\item Each invocation of \sct$(V, R, fld, new)$ tries to change $fld$ to a value $new$ that it never previously contained. \label{constraint-sct-aba}
\item Consider each execution that contains a configuration $C$ after which the value of no field of any \rec\ changes. There is a total order of all \rec s created during this execution such that, for every \sct\ whose linked \llt s begin after $C$, the $V$ sequence passed to the \sct\ is sorted according to the total order. \label{constraint-total-order}
\end{compactenum}

\noindent
It is easy to satisfy these two constraints using standard approaches (e.g., by attaching a version number to each field and sorting $V$ sequences).
However, we shall see that both constraints %Constraints 1 and 2
are \textit{automatically} satisfied in a natural way when \llt\ and \sct\ are used according to our tree update template.

We assume there is a \rec\ $entry$ which acts as the entry point to the data structure and is never deleted.
This \rec\ points to the root of a down-tree.
We represent an empty down-tree by a pointer to an empty \rec.
%An empty down-tree can be represented by a \nil\ pointer or an empty \rec.
A \rec\ is {\em in the tree} if it can be reached by following pointers
from $entry$.
A \rec\ $r$ is {\em removed from the tree} by an \sct\ if $r$
is in the tree immediately prior to the linearization point of the \sct\ and
is not in the tree immediately afterwards.
Data structures produced using our template \textit{automatically} satisfy one additional constraint:
\after{note: there is something strange with the definition of ``removed from the tree'' if multiple \sct s can be linearized at the same time.}

\begin{compactenum}[\hspace{3.4mm}{\bf Constraint} \bfseries 1:]
\setcounter{enumi}{2}
\item A \rec\ is finalized when (and only when) it is removed from the tree. \label{constraint-finalized-iff-removed}
\end{compactenum}

\noindent
Recall that, under this additional constraint, the implementation of \llt\ and \sct\ in Chapter~\ref{chap-scx} also guarantees the following three properties.
These properties are useful for proving the correctness of our template.
%We talk about \textit{linearized} \sct s, rather than \textit{successful} ones, because some non-terminating \sct s can be linearized, and unsuccessful \sct s are not.
\begin{compactitem}
\item
If \llt$(r)$ returns a snapshot, %value different from \fail\ or \finalized,
then $r$ is in the tree
just before the \llt\ is linearized.
\item
If an \sct$(V,R,fld,new)$ is linearized and $new$ is (a pointer to) a \rec, then this \rec\ 
is in the tree
immediately after
%when
the \sct\ is linearized.
\item
If an operation reaches a \rec\ $r$ by following pointers read from other \rec s, starting from $entry$, then $r$ was in the tree at some
earlier time during the operation.
\end{compactitem}
\noindent
In the following, we sometimes abuse notation by treating the
sequences $V$ and $R$ as sets, in which case
we mean the set of all \rec s in the sequence.

The memory overhead introduced by the implementation of \llt\ and \sct\ is fairly low.
Each node in the tree is augmented with a pointer to a descriptor and a bit.
Every node that has had one of its child pointers changed by an \sct\ points to a descriptor.
(Other nodes have a \nil\ pointer.)
A descriptor can be implemented to use only three machine words after the update it describes has finished.
\end{thesisonly}

\begin{thesisnot}
The load-link extended (\llt), store-conditional extended (\sct) and validate-extended (\vlt) primitives are multi-word generalizations of the well-known load-link (LL), store-conditional (SC) and validate (VL) primitives, and they have been implemented from single-word \cas\ \cite{Brown:2013}.
The benefit of using \llt, \sct\ and \vlt\ to implement our template is two-fold: the template can be described quite simply, and much of the complexity of its correctness proof is encapsulated in that of \llt, \sct\ and \vlt.

Instead of operating on single words, \llt, \sct\ and \vlt\ operate on \rec s, each of which consists of a fixed number of mutable fields (which can change), and a fixed number of immutable fields (which cannot). 
\llt$(r)$ attempts to take a snapshot of the mutable fields of a \rec\ $r$.
If it is concurrent with an \sct\ involving~$r$, it may return \fail, instead.
Individual fields of a \rec\ can also be read directly.
An \sct$(V,R,fld,new)$ takes as arguments a sequence $V$ of \rec s, a subsequence $R$ of $V$, a pointer $fld$ to a mutable field of one \rec\ in~$V$, and a new value $new$ for that field.
The \sct\ tries to atomically store the value $new$ in the field that $fld$ points to and {\it finalize} each \rec\ in $R$.
Once a \rec\ is finalized, its mutable fields cannot be changed by any subsequent \sct, and any \llt\ of the \rec\ will return \finalized\ instead of a snapshot.

Before a process invokes \sct\ or \vlt($V$), it must perform an \llt$(r)$ on each \rec\ $r$ in $V$.
The last such \llt\ by the process is said to be {\it linked} to the \sct\ or \vlt, and the linked \llt\ must return a snapshot of $r$ (not \fail\ or \finalized).
An \sct($V, R, fld, new$) by a process modifies the data structure only if each \rec\ $r$ in $V$ has not been changed since its linked \llt($r$); otherwise the \sct\ fails.
Similarly, a \vlt$(V)$ returns \true\ only if each \rec\ $r$ in $V$ has not been changed since its linked \llt($r$) by the same process; otherwise the \vlt\ fails.
\vlt\ can be used to obtain a snapshot of a set of \rec s.
Although \llt, \sct\ and \vlt\ can fail, their failures are limited in such a way that we can use them to build non-blocking data structures. %(details in \cite{paper1}).
See \cite{Brown:2013} for a more formal specification of these primitives.

These new primitives were designed to balance ease of use and efficient  implementability using single-word \cas.
The implementation of the primitives from CAS in \cite{Brown:2013} is more efficient if the user of the primitives can guarantee that two constraints, which we describe next, are satisfied.
The first constraint prevents the ABA problem for the \cas\ steps that actually perform the updates.

\begin{compactenum}[\hspace{3.4mm}{\bf Constraint} \bfseries 1:]
\setcounter{enumi}{0}
\item Each invocation of \sct$(V, R, fld, new)$ tries to change $fld$ to a value $new$ that it never previously contained. \label{constraint-sct-aba}
\end{compactenum}
%For every invocation $S$ of  \sct$(V, R, fld, new)$,
%$new$ is not the initial value of $fld$ and no invocation of
%\sct$(V', R', fld, new)$  was linearized before the $\llt(r)$ linked to
%$S$ was linearized,  where $r$ is the \rec\ that contains $fld$.

The implementation of \sct\ does something akin to locking the elements of $V$ in the order they are given. 
Livelock can be easily avoided by requiring all $V$ sequences to be sorted according to some total order on \rec s. %, so the second constraint is used to avoid livelock.
%We could easily avoid livelock by requiring all $V$ sequences to be ordered by some total ordering on \rec s.
However, this ordering is necessary only to guarantee that \sct s continue to succeed.
Therefore, as long as \sct s are still succeeding in an execution, it does not matter how $V$ sequences are ordered.
This observation leads to the following constraint, which is much weaker.

\begin{compactenum}[\hspace{3.4mm}{\bf Constraint} \bfseries 1:]
\setcounter{enumi}{1}
\item Consider each execution that contains a configuration $C$ after which the value of no field of any \rec\ changes. There is a total order of all \rec s created during this execution such that, for every \sct\ whose linked \llt s begin after $C$, the $V$ sequence passed to the \sct\ is sorted according to the total order. \label{constraint-total-order}
\end{compactenum}

It is easy to satisfy these two constraints using standard approaches (e.g., by attaching a version number to each field and sorting $V$ sequences).
However, we shall see that Constraints 1 and 2 are \textit{automatically} satisfied in a natural way when \llt\ and \sct\ are used according to our tree update template.

Under these constraints, the implementation of \llt,  \sct, and \vlt\ in \cite{Brown:2013} guarantees that there is a linearization of all \sct s that modify the data structure (which may include \sct s that do not terminate because a process crashed, but \textit{not} any \sct s that fail), and all \llt s and \vlt s that return, but do not fail.
%a value different from \fail.

%\eric{We should probably also talk  about $additional properties$
%which are used to argue later that searches can be linearized.}

We assume there is a \rec\ $entry$ which acts as the entry point to the data structure and is never deleted.
This \rec\ points to the root of a down-tree.
We represent an empty down-tree by a pointer to an empty \rec.
A \rec\ is {\em in the tree} if it can be reached by following pointers from $entry$.
A \rec\ $r$ is {\em removed from the tree} by an \sct\ if $r$ is in the tree immediately prior to the linearization point of the \sct\ and is not in the tree immediately afterwards.
Data structures produced using our template \textit{automatically} satisfy one additional constraint:
\after{note: there is something strange with the definition of ``removed from the tree'' if multiple \sct s can be linearized at the same time.}

\begin{compactenum}[\hspace{3.4mm}{\bf Constraint} \bfseries 1:]
\setcounter{enumi}{2}
\item A \rec\ is finalized when (and only when) it is removed from the tree. \label{constraint-finalized-iff-removed}
\end{compactenum}

\noindent
Under this additional constraint, the implementation of \llt\ and \sct\ in \cite{Brown:2013} also guarantees the following three properties.
%We talk about \textit{linearized} \sct s, rather than \textit{successful} ones, because some non-terminating \sct s can be linearized, and unsuccessful \sct s are not.
\begin{compactitem}
\item
If \llt$(r)$ returns a snapshot, %value different from \fail\ or \finalized,
then $r$ is in the tree
just before the \llt\ is linearized.
\item
If an \sct$(V,R,fld,new)$ is linearized and $new$ is (a pointer to) a \rec, then this \rec\ 
is in the tree
immediately after
%when
the \sct\ is linearized.
\item
If an operation reaches a \rec\ $r$ by following pointers read from other \rec s, starting from $entry$, then $r$ was in the tree at some
earlier time during the operation.
\end{compactitem}
\noindent These properties are useful for proving the correctness of our template.
In the following, we sometimes abuse notation by treating the
sequences $V$ and $R$ as sets, in which case
we mean the set of all \rec s in the sequence.

The memory overhead introduced by the implementation of \llt\ and \sct\ is fairly low.
Each node in the tree is augmented with a pointer to a descriptor and a bit.
Every node that has had one of its child pointers changed by an \sct\ points to a descriptor.
(Other nodes have a \nil\ pointer.)
A descriptor can be implemented to use only three machine words after the update it describes has finished.
%The implementation of \llt\ and \sct\ in \cite{Brown:2013} assumes garbage collection, and we do the same in this work.
%This assumption can be eliminated by using, for example, the new efficient memory reclamation scheme of Aghazadeh et~al. \cite{AGW13}.
\end{thesisnot}

\section{Tree update template} \label{sec-dotreeupdate}

\begin{figure}[tb]
	\centering
	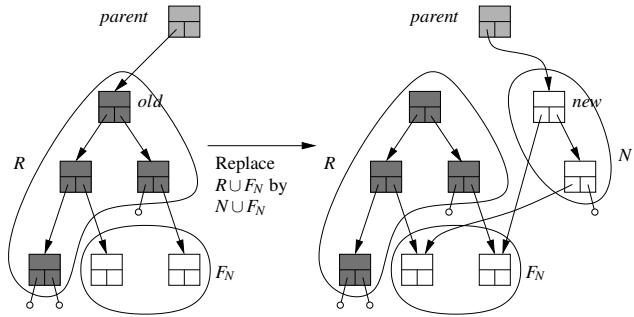
	\caption{Example of the tree update template.
			$R$ is the set of nodes to be removed, % = \{r, r_1, r_2\}$,
			$N$ is a tree of new nodes that have never before appeared in the tree, and
			$F_N$ is the set of children of $N$ (and of $R$). %(which should not be removed), % = \{f_1, f_2, f_3\}$,
			%Pointers that are not shown are not affected by the update.  
			Nodes in $F_N$ may have children.  
			The shaded nodes (and possibly others) are in the sequence $V$  of the \sct\ that performs the update.
			The darkly shaded nodes are finalized by the \sct.
			%% TODO: REWRITE THIS OUTSIDE OF CAPTION
			%Note that $N+F = New$, a non-empty tree constructed in \dotreeupdate}.
			%We say ``$R\cup F$ is replaced by $N\cup F$'' rather than ``$R$ is replaced by $N$,''
			%because $N$ may be empty (i.e., node $n$ might not exist).
			}
	\label{fig-replace-subtree}
\end{figure}

Our tree update template implements updates that atomically replace an old connected subgraph in a down-tree by a new connected subgraph.
Such an update can implement any change to the tree, such as an insertion into a BST or a 
rotation used to rebalance a RBT.
The old subgraph includes all nodes with a field (including a child pointer) to be modified.
The new subgraph may have pointers to nodes in the old tree.
Since every node in a down-tree has indegree one, the update can be performed by changing a single child pointer of some node $parent$.
(See Figure~\ref{fig-replace-subtree}.)
However, problems could arise if a concurrent operation changes the part of the tree being updated.
For example, nodes in the old subgraph, or even $parent$, could be removed from the tree before $parent$'s child pointer is changed.
Our template takes care of the process coordination required to prevent such problems.

\begin{figure}[tb]
	\centering
	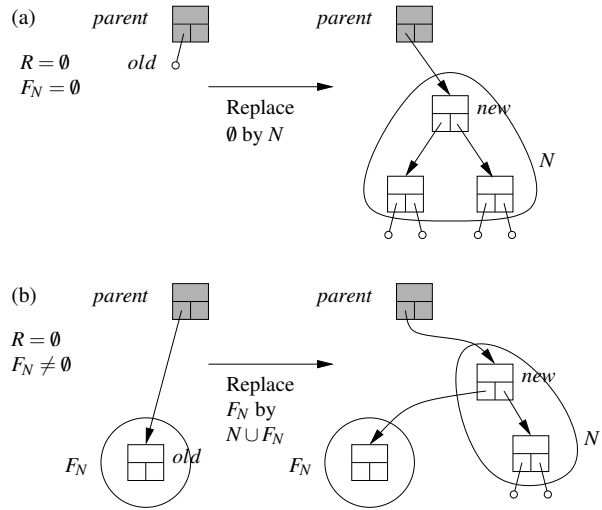
%	\vspace*{-2mm}
	\caption{Examples of two special cases of the tree update template when no nodes are removed from the tree.  
	(a) Replacing a \nil\ child pointer:  In this case, $R=F_N=\emptyset$.  
	(b) Inserting new nodes in the middle of the tree:  In this case, $R=\emptyset$ and $F_N$ consists of a single node.}
			%% TODO: REWRITE THIS OUTSIDE OF CAPTION
			%Note that $N+F = New$, a non-empty tree constructed in \dotreeupdate}.
			%We say ``$R\cup F$ is replaced by $N\cup F$'' rather than ``$R$ is replaced by $N$,''
			%because $N$ may be empty (i.e., node $n$ might not exist).
			
	\label{fig-replace-subtree2}
\end{figure}

Each tree node is represented by a \rec\ with a fixed number of child pointers as its mutable fields (but different nodes may have different numbers of child fields).  
Each child pointer points to a \rec\ or contains \nil\ (denoted by $\multimap$ in our figures).
For simplicity, we assume that any other data in the node is stored 
in immutable fields.
Thus, if an update must change some of this data, it makes a new copy of the node with
the updated data.
%A special \rec, \RootPtr, which consists of a single child pointer,
% with a single child pointer, %$c_1$ 
%(initially $\nil$),
%serves as the entry point to the tree.
%serves as an entry point to the data structure.
%\eric{Is this accurate?  Why does it have to be a data record?  Isn't root just an (immutable) pointer to the root node?}
%\faith{The tree could be empty or the root might change.}
%We represent each leaf by a \rec\ with no child pointers.
%The child pointers of a leaf contain $\bot$.

\begin{figure}[tbp]
\def\pwidth{4cm}
\prepnewlisting
%\hrule
\vspace{-2mm}
\begin{lstlisting}[mathescape=true]
    $\func{Template}(args)$
      $m := \func{SearchPhase}(args)$ // \hfill \com search in the tree starting at \textit{entry} to identify a location for the update
      return $\func{UpdatePhase}(m)$

    $\func{UpdatePhase}(m)$
      if $\func{UpdateNotNeeded}(m)$ then return $\func{Result}(m)$ //\label{code-template-return-result1}
      $i := 0$
      $n_i := \mbox{root of}\ m$
      loop
        $s_i := \llt(n_i)$
        if $s_i \in \{\fail, \finalized\}$ or $\func{Conflict}(n_i, s_i, m)$ then return $\fail$ //\label{code-template-return-fail1}
        $s_i' :=$// immutable fields of $n_i$
        exit loop when $\func{Condition}(s_0, s_0', \ldots, s_i, s_i')$ // \hfill \com \func{Condition} must eventually return $\true$
        $n_{i+1} := \func{NextNode}(s_0, s_0', \ldots, s_i, s_i')$ // \hfill \com returns a non-\nil\ child pointer from one of $s_0, \ldots, s_i$
        $i := i+1$
      end loop
      if $\func{UpdateNotNeeded}(s_0, s_0', \ldots, s_i, s_i')$ then return $\func{Result}(s_0, s_0', \ldots, s_i, s_i')$ //\label{code-template-return-result2}
      if $\sct(\func{\sct-Arguments}(s_0, s_0', \ldots, s_i, s_i'))$ then return $\func{Result}(s_0, s_0', \ldots, s_i, s_i')$ //\label{code-template-return-result3}
      else return $\fail$ //\label{code-template-return-fail2}
\end{lstlisting}
%\vspace*{-3mm}
	\caption{The tree update template. \func{Condition}, \func{NextNode}, \func{Conflict}, \func{SCX-Arguments} and \func{Result} can be filled in with any locally computable functions that satisfy their postconditions.
    \func{SearchPhase} should be filled in with a search procedure for the data structure.
%\func{SCX-Arguments} and \func{Conflict} satisfy their postconditions, and \func{Condition} eventually returns \true.} % PC1 to PC\ref{con-R-non-empty-then-GR-a-non-empty-tree}.
}
	\label{code-dotreeupdate}
\end{figure}

At a high level, an update that follows the template proceeds in two phases: the \textit{search phase} and the \textit{update phase}.
In the search phase, the update searches for a location where it should occur.
Then, in the update phase, the update performs \llt s on a connected subgraph of nodes in the tree, including $parent$ and the set $R$ of nodes to be removed from the tree. 
Next, it decides whether the tree should be modified, and, if so, performs an \sct\ that atomically changes a child pointer, as shown in Figure~\ref{fig-replace-subtree}, and finalizes the nodes in $R$.
Figure~\ref{fig-replace-subtree2} shows two special cases where $R$ is empty.
%We actually provide two versions of the template: \func{SearchUpdate} and \func{UpdatePhase}.
%In the former, the search for the location where the update should occur is included as part of the atomic update.
%The latter is a simplified version of the former, in which the search is \textit{not} part of the atomic update. %performed non-atomically, before the tree is atomically modified.
(As a minor note, search operations that do not perform any update can also be said to follow the template.)

%Our template for performing an update is fairly simple: An update first searches for a location where an update should occur, then performs \llt s on a connected subgraph of nodes in the tree, including $parent$ and the set $R$ of nodes to be removed from the tree. 
%Next, it decides whether the tree should be modified, and, if so, performs an \sct\ that atomically changes a child pointer, as shown in Figure~\ref{fig-replace-subtree}, and finalizes the nodes in $R$.
%Figure~\ref{fig-replace-subtree2} shows two special cases where $R$ is empty.
%An update that performs this sequence of steps is said to \textit{follow} the template.

\paragraph{Detailed description}
Figure~\ref{code-dotreeupdate} presents code for the template. %both versions of the template.
%An update is said to \textit{follow} \func{SearchUpdate}  (resp., \func{UpdatePhase}) if it performs the sequence of steps shown in \func{SearchUpdate}) (resp., \func{UpdatePhase}).
%We explain \func{SearchUpdate} in detail, and note that \func{UpdatePhase} is just a simplified version of it.
An update is said to \textit{follow} the template if it performs the sequence of steps in \func{Template}.

Consider an update \func{UP} that follows the template.
\func{UP} first invokes a \func{SearchPhase} procedure to find the location in the tree where the update should occur.
\func{SearchPhase} reads a sequence of child pointers starting from $entry$ to identify some node $n_0$, and returns the \textit{relevant} part $m$ of the subtree rooted at $n_0$ that it observed.
The node $n_0$ \textit{must} be in the data structure at some time during \func{SearchPhase}.
(This requirement will be used to prove progress.)
Intuitively, $m$ contains only information that: (a) is observed by the search in the subtree rooted at $n_0$, and (b) is relevant to the update.
We require $m$ to be a connected subgraph rooted at $n_0$. %, %contain $n_0$, and to be a connected subgraph, 
(Consequently, $m$ cannot be empty.)
It is not necessary for $m$ to contain a value for every field of a node.
%but it need not contain a value for every field of a node. %can contain any subset of the fields of any given node.
For instance, it may contain a value for the left child pointer of a node, but not for the right child pointer.
%(One can think of missing fields as \textit{do not care} values.)
As we will see, the update will modify the tree \textit{only} if the fields of nodes in $m$ agree with the values stored in $m$ when the update is linearized.
Next, \func{UP} invokes \func{UpdatePhase}$(m)$.

%During its traversal, \func{UP} can remember a part $m$ of the subtree rooted at $n_0$ that it observed.
In \func{UpdatePhase}, \func{UP} first determines whether it should modify the data structure by performing some deterministic local computation (denoted by \func{UpdateNotNeeded} in Figure~\ref{code-dotreeupdate}) using the contents of $m$. % of the data structure that it remembers from its traversal.
If \func{UP} decides that an update is not needed (e.g., because \func{UP} is a deletion of a key that is not in the data structure), then it returns a result calculated locally by the \func{Result} function (and the update \textit{succeeds}).
Now, suppose \func{UP} decides that an update is needed.
Then, \func{UP} performs \llt s on a sequence $\sigma = \langle n_0, n_1, \ldots\rangle$ of nodes starting with $n_0$.
This sequence must contain every node in $m$.
For maximal flexibility of the template, 
the sequence  $\sigma$ can be constructed on-the-fly, as \llt s are performed.
Thus, \func{UP} chooses a non-\nil\ child of one of the previous nodes to be the next node of $\sigma$ by performing some deterministic local computation (denoted by \func{NextNode} in Figure~\ref{code-dotreeupdate}) using any information that is available locally, namely, the contents of $m$, snapshots of mutable fields returned by \llt s on the previous elements of $\sigma$, and values read from immutable fields of previous elements of $\sigma$.
(This flexibility can be used, for example, to avoid unnecessary \llt s when deciding how to rebalance a BST.) %As an example of how this flexibility can be used, in a balanced BST, \func{UP} can decide exactly how it should rebalance the tree as it is performing \llt s.)

After each invocation of \llt, \func{UP} checks if the result of the \llt\ was \fail\ or \finalized.
If so, \func{UP} returns \fail\ to indicate that it was aborted because of a concurrent update on an overlapping portion of the tree.
Otherwise, \func{UP} performs another local computation, denoted by \func{Conflict}$(n_i, s_i, m)$, which returns \true\ if the results of the \llt$(n_i)$ disagree with $m$, and \false\ otherwise. % in a way that should cause \func{UP} to fail. %the part $m$ of the data structure that \func{UP} remembers from its traversal.
%returns \true\ if the results of the \llt\ disagree with the part $m$ of the data structure remembered during the traversal.
For example, suppose \func{UP} is a deletion of a key $k$, and $m$ contains a leaf $l$ with key $k$ and a node $p$ that points to $l$. % returns it finds a leaf $l$ containing $k$ and its parent $p$ during its traversal.
Then, if \func{UP} performs an \llt$(p)$ which shows that $p$ does not point to $l$, \func{Conflict} must return \true\ (and cause \func{UP} to return \fail). % if \func{UP} performs an \llt$(p)$ which shows that $p$ no longer points to $l$.
Note that, if \func{Conflict}$(n_i, s_i, m)$ returns \true, then a field of $n_i$ changed since the value of that field in $m$ was read in \func{SearchPhase}.
Thus, \func{Conflict} can cause \func{UP} to fail only if \func{UP} is concurrent with a successful update.
This implies that UP can return \fail\ without threatening the non-blocking progress property.
%To guarantee progress, we allow \func{Conflict}$(m, n_i, s_i)$ to return \true\ only if some field of $n_i$ has changed since the value of that field in $m$ was read.
%(With this restriction, \func{Conflict} can cause \func{UP} to fail only if it is concurrent with another update that successfully modifies the data structure.)
(Similarly, whenever one can prove that a node has changed since UP performed \llt\ on it, UP can immediately return \fail. Note that this possibility is not reflected explicitly in the template.)

Next, \func{UP} performs another local computation (denoted by \func{Condition} in Figure~\ref{code-dotreeupdate}) %using locally available information
to decide whether more \llt s should be performed.
To avoid infinite loops, this function must eventually return \true\ in any execution of \func{UP}.
(This condition is trivially satisfied if the loop in the template has a bounded number of iterations.)
%If any \llt\ in the sequence returns \fail\ or \finalized, then \func{UP} also returns \fail, to indicate that the attempted update has been aborted because of a concurrent update on an overlapping portion of the tree.
%
%\trevor{make sure this makes sense}
If all of the \llt s successfully return snapshots, then \func{UP} decides whether it should modify the tree by once again invoking \func{UpdateNotNeeded}. %another local computation (denoted by \func{UpdateNotNeeded}).
If it decides to modify the tree, it invokes \sct.
If \func{UP} invokes \sct\ and the \sct\ fails, then \func{UP} returns \fail.
In all other cases, \func{UP} \textit{succeeds} and returns a result computed locally by the \func{Result} function.

Just before \func{UP} performs an \sct, it invokes a function \func{\sct-Arguments} which uses locally available information to construct the arguments $V, R, fld$ and $new$ for the \sct.
%This function returns the arguments $V$, $R$, $fld$ and $new$.
%We now specify constraints on the output of \func{Change}.
%Consider the invocation of \sct$(V, R, fld, new)$ that is performed by \func{UP}$(top, args)$, where 
%$V, R, fld,$ and $new$ are created by \func{Change}.
%(Because of this, what follows is essentially a sequence of constraints on \func{Change}.)
The postconditions that must be satisfied by \func{\sct-Arguments} are somewhat technical, but intuitively, they are meant to ensure that the arguments produced describe an update as shown in Figure~\ref{fig-replace-subtree} or Figure~\ref{fig-replace-subtree2}.
The update must remove a connected set $R$ of nodes from the tree and replace it by a connected set $N$ of newly-created nodes that is rooted at $new$ by changing the child pointer stored in $fld$ to point to $new$.
%
%(If a data structure has updates that delete the root, then \textit{root} should simply contain a single pointer to the \textit{real root} of the data structure.
%Then, one can delete the \textit{real root} by replacing it with a new, \nil-node that represents the empty tree.)
In order for this change to occur atomically, we include $R$, the nodes in $m$, and the node containing $fld$, in $V$.
This ensures that if any of these nodes has changed since it was last accessed by one of \func{UP}'s \llt s, the \sct\ will fail.
The sequence $V$ may also include any other nodes in $\sigma$.
%Note that $N \neq \emptyset$, since $new \in N$.
%
%
Formally, we require \func{SCX-Arguments} to satisfy ten postconditions.
The first three are basic requirements of \sct.
\begin{compactenum}[\hspace{3.4mm}\bfseries PC1:]
\item $V$ is a subsequence of $\sigma$.
\label{con-llt-on-all-nodes-in-V}
\item The node $parent$ containing the mutable field $fld$ is in $V$.
\label{con-parent-in-V}%
\item $R$ is a subsequence of $V$.
\label{con-R-subsequence-of-V}
\item All nodes in $m$ are in $V$.
\label{con-m-in-V}
\end{compactenum}

The next three postconditions guarantee that the $R$ sequence of an \sct\ contains exactly the nodes that it removes from the tree.
The first two are quite simple.
Let $G_N$ be the directed graph $(N \cup F_N,E_N)$, where $E_N$ is the set of all child pointers of nodes in $N$ when they are initialized, and $F_N = \{ y : y\not\in N \mbox{ and }(x,y) \in E_N$ for some $x\in N$\}.
Let $old$ be the value read from $fld$ by the \llt\ on $parent$.
\begin{compactenum}[\hspace{4.1mm}\bfseries PC1:]
\setcounter{enumi}{4}
\item 
If $old=\nil$ then $R=\emptyset$ and $F_N = \emptyset$. 
\label{con-old-nil-then-R-empty}
\item 
If $R = \emptyset$ and $old \neq \nil$, then $F_N = \{ old \}$.
\label{con-fringe-of-GN-is-old}
\end{compactenum}

Stating the last of these three postconditions formally requires some care, since the tree may be changing while \func{UP} performs its \llt s.
%To keep correctness proofs (very) simple, one should be able to show that these postconditions are satisfied \textit{without worrying about concurrent modifications}.
%Thus, we cannot express them in terms of the shared data structure (which can change).
%Instead, we state these postconditions in terms of the static view of the data structure that \func{UP} obtains from the \llt s it performs.
If $R \neq \emptyset$, let $G_R$ be the directed graph $(R \cup F_R,E_R)$,
where $E_R$ is the union of the sets of edges representing child pointers read from each $r \in R$
when it was last accessed by one of \func{UP}'s \llt s and
$F_R = \{ y : y\not\in R \mbox{ and }(x,y) \in E_R$ for some $x\in R$\}.
$G_R$ represents \func{UP}'s view of the nodes in $R$ according to its \llt s, and $F_R$ is the \textit{fringe} of $G_R$.
%If the tree does not change during \func{UP},
If other processes do not change the tree while \func{UP} is being performed,
then $F_R$ contains the nodes
that should remain in the tree, but
%in the tree
whose parents will be removed and replaced.
Therefore, we must ensure that the nodes in $F_R$ are reachable from nodes in $N$ (so they are not accidentally removed from the tree).
%Similarly,
Let $G_\sigma$ be the directed graph $(\sigma \cup F_\sigma,E_\sigma)$,
where $E_\sigma$ is the union of the sets of edges representing child pointers read from each $r \in \sigma$ when it was last accessed by one of \func{UP}'s \llt s and $F_\sigma = \{ y : y\not\in \sigma \mbox{ and }(x,y) \in E_\sigma$ for some $x\in \sigma$\}.
Since $G_\sigma$, $G_R$ and $G_N$ are not affected by concurrent updates, the following postcondition can be proved using purely sequential reasoning, ignoring the possibility that concurrent updates could modify the tree during \func{UP}.
\begin{compactenum}[\hspace{4.1mm}\bfseries PC1:]
\setcounter{enumi}{6}
\item 
If $G_\sigma$ is a down-tree and $R \neq \emptyset$,
then $G_{R}$ is a non-empty down-tree rooted at $old$
and $F_N = F_R$.
\label{con-R-non-empty-then-GR-a-non-empty-tree}
\end{compactenum}%

The next two postconditions are used to satisfy Constraint~\ref{constraint-sct-aba}, which is used to prove there is no ABA problem.
\begin{compactenum}[\hspace{4.1mm}\bfseries PC1:]
\setcounter{enumi}{7}
\item 
$G_{N}$ is a non-empty down-tree rooted at $new$.
\label{con-GN-non-empty-tree}
\item
%\trevor{I changed this to talk about allocating memory instead of creating nodes to try to answer a reviewer's comment.}
\func{UP} allocates memory for all nodes in $N,$ including $new$.%
\label{con-new-nodes}%
\end{compactenum}

\noindent
%Postcondition PC\ref{con-new-nodes} requires $new$ to be a newly-created node, in order to satisfy Constraint~\ref{constraint-sct-aba}.
Note that there is no loss of generality in requiring $new$ to be a newly-allocated node:
If we wish to change a child $y$ of node $x$ to \nil\ (to chop off the entire subtree rooted at $y$)
or to a descendant of $y$ (to splice out a portion of the tree),
then, instead, we can replace $x$ by a new copy
of $x$ with an updated child pointer.
%\trevor{Is this redundant after the above comment about deleting the last node in the tree?}
Likewise, if we want to delete the entire tree, then $entry$ can be changed to point to a new, empty \rec.

The next postcondition is used to satisfy Constraint~\ref{constraint-total-order}, which is used to prove progress.
\begin{compactenum}[\hspace{4.1mm}\bfseries PC1:]
\setcounter{enumi}{9}
\item 
The sequences $V$ constructed by all updates %that follow the template
that take place entirely % between two consecutive modifications to the tree structure
during a period of time when no \sct s change the tree structure
must be ordered consistently according to a fixed tree traversal algorithm (for example, an in-order traversal or a breadth-first traversal).
\label{con-V-sequences-ordered-consistently}
\end{compactenum}

%We require the \func{SearchPhase} procedure invoked by \func{SearchUpdate} to satisfy the following 

%\faith{If the label con-fringe-of-GN-is-F was used, it will have
%to be updated.}
%Collectively, these properties %constraints 
%are used to prove that the data structure is always a tree, that each 
%operation that follows our template is linearizable, that each linearized \func{UP} %$(top, args)$ 
%changes $fld$
%from  $old$ to $new$
%(which replaces a connected subgraph containing nodes $R \cup F_N$ with a connected subgraph containing nodes $N \cup F_N$), and
%%(which replaces the subgraph $G_R$ with $G_N$, if $R \neq \emptyset$, and replaces $old$ by $G_N$, otherwise), and
%%\faith{Is it important that we say anything about the replacement if $R= \emptyset$?}
%%\trevor{In response to Faith's comment, I tried to succinctly describe what happens when $R \neq \emptyset$.}
%%\faith{I think your change is good, so I hid your comment.}
%%and replaces the subgraph $G_R$ with $G_N$, and
%that no node in $R$ is subsequently modified or reinserted
%into the tree. %, and that these operations are non-blocking.
%

\paragraph{Properties of updates that follow the template}
%\trevor{rewrite to account for \func{SearchPhase} and the delayed traversal property.}
%As we will see in Section~\ref{app-tree-proof}, $m$ is useful for proving that updates in which the \func{SearchPhase} procedure also satisfies
% including their linearizability of updates including
Lock-free progress is guaranteed for all updates that follow the template.
We briefly discuss the correctness properties provided by the template. %template guarantees about updates that follow it.
We consider each of the different lines where an update \func{UP} that follows the template can return from \func{UpdatePhase}.
Ad-hoc correctness arguments are needed if \func{UP} returns after an invocation of \func{UpdateNotNeeded} at line~\ref{code-template-return-result1} or line~\ref{code-template-return-result2}, because the behaviour of \func{UpdateNotNeeded} depends on the semantics of the data structure being implemented.
If \func{UP} returns \fail\ at line~\ref{code-template-return-fail1} or line~\ref{code-template-return-fail2}, then it has no effect on shared memory. % (except for any changes it makes while helping other updates during its invocations of \llt).
The last place where \func{UP} can return is after performing a successful \sct\ at line~\ref{code-template-return-result3}.
In this case, we prove that the \textit{update phase} of \func{UP} (i.e., its invocation of \func{UpdatePhase}) is atomic, and that the fields of all nodes in $m$ agree with their values in $m$ when the update phase occurs.
%By the semantics of \llt\ and \sct, and the postconditions above, any update that returns at line~\ref{code-template-return-result3} atomically replaces a connected subgraph $R$ by a new connected subgraph $N$ (which are both constructed by \func{SCX-Arguments}).
%% and the fields of all nodes in $m$ agree with their values in $m$ when the update occurs.
%Since the update reaches line~\ref{code-template-return-result3}, it must perform $s_i := \llt(n_i)$ for each $n_i$ in $m$, and then see that $n_i$'s fields agree with $m$ when \func{Conflict}$(n_i, s_i, m)$ is invoked.
%Since the \sct\ succeeds, none of $n_i$'s fields can change between the $\llt(n_i)$ and the \sct.
%Therefore, the fields of all nodes in $m$ agree with their values in $m$ when the update occurs.
Furthermore, if \func{SearchPhase} satisfies the following property, then the \textit{entire template update is atomic} (including \func{SearchPhase}).

\begin{compactenum}[\hspace{3.4mm}{\bf Delayed traversal property (DTP)}:]
\setcounter{enumi}{0}
\item Suppose an invocation of \func{SearchPhase}$(args)$ terminates and returns $m$ in configuration $C$, and an \textit{atomic} invocation $S'$ of \func{SearchPhase}$(args)$ is performed in a later configuration $C'$.
If all of the nodes in $m$ are in the tree in $C'$, and their fields agree with the values in $m$, then $S'$ returns $m$.
\end{compactenum}

\section{Correctness proof} \label{app-tree-proof}

In the following, we use \llt\ and \sct\ as atomic primitives.
Every invocation of \sct\ either succeeds or fails.
A \textit{successful} \sct\ modifies the data structure, and returns \true.
A \textit{failed} \sct\ does not modify the data structure, and returns \false.
An \llt\ either succeeds or fails.
A \textit{successful} \llt\ returns \finalized\ or a snapshot.
A \textit{failed} \llt\ returns \fail.
Recall that an invocation of \llt\ is \textit{linked to} an invocation $I'$ of \sct$(V, R, fld, new)$ or \vlt$(V)$ by process $p$ if $r$ is in $V$, $I$ returns a snapshot, and between $I$ and $I'$, process $p$ performs no invocation of \llt$(r)$ or \sct$(V', R', fld', new')$ and no unsuccessful invocation of \vlt$(V')$, for any $V'$ that contains $r$.
We use the term \textit{template operation} to refer to any operation that follows the tree update template.
We call a template operation an \textit{effective update} if it performs a successful \sct.

We now sketch the main ideas of the proof.
Consider a data structure in which all updates %are performed
%We assume that \sct s on nodes are invoked only
% following 
follow the tree update template and \func{\sct-Arguments} satisfies postconditions PC\ref{con-llt-on-all-nodes-in-V} 
to PC\ref{con-R-non-empty-then-GR-a-non-empty-tree}.
We linearize each \textit{effective} update at %(the linearization point of) 
its \sct.
We prove, by induction on the sequence of steps in an execution,
that the data structure is always a tree, each call to \llt\ and \sct\ satisfies its preconditions, Constraints~\ref{constraint-sct-aba} to \ref{constraint-finalized-iff-removed} are satisfied, and each effective update atomically replaces a connected subgraph containing nodes $R \cup F_N$ with another connected subgraph containing nodes $N \cup F_N$ (finalizing and removing the nodes in $R$ from the tree and adding the new nodes in $N$ to the tree).
Next, we prove that no node in the tree is finalized, every removed node is finalized, and removed nodes are never reinserted.
Finally, we prove that effective updates have \textit{atomic update phases}, and effective updates whose \func{SearchPhase} procedures satisfy DTP are \textit{entirely atomic} (including the search phase).
%\eric{Oops.  This is still wrong.  We wanted to say $R\cup F$ is replaced by $N\cup F$, but
%$F$ is no longer defined, except informally in the diagrams.  What is written here now is incorrect if $R$ is empty (because then $F_R$ is  empty too, but $F$ contains one node}.
%\eric{The definition of $F$ has disappeared.  Should we say that we replace $R\cup F_R$ by $N\cup F_R$?}
%\faith{I meant to change this and then I forgot. Thanks for catching it.}
%We also prove no node in the tree is finalized, every removed node is finalized, and removed nodes are never reinserted. % into the tree.

\begin{ignore}
%\trevor{I tried to write a basic rundown of the proof.}
We make a general assumption that \sct s on nodes are invoked only by operations that follow the tree update template.
Then, the tree update template is correct if the data structure is always a tree, and each operation that follows the template atomically replaces a connected sub-graph containing nodes $R \cup F$ with another connected sub-graph containing nodes $N \cup F$, finalizing and removing the nodes in $R$ from the tree, and adding the nodes in $N$ to the tree.
We first prove by induction on the sequence of steps in the execution that each operation that follows the template satisfies the preconditions of \llt\ and \sct, that the data structure is always a tree, and that a successful operation effects the sort of transformation we described.
This follows primarily from the properties of an operation that follows the template and the semantics of \llt\ and \sct.
We also prove that no node in the tree is finalized, every removed node is finalized, and removed nodes are never added back into the tree.
\trevor{fix this reference}
In the process of proving the above, we satisfy a constraint that allows us to offer the following result from \cite{paper1}, which tells us that searches can be implemented using only \func{READ}s in a data structure whose updates follow the template.
If a process $p$ follows child pointers starting from a node in the tree at time $t$, then each node reached at time $t'$ by following one of these child pointers was in the tree at some time in $[t, t']$.
Furthermore, if $p$ read $v$ from a mutable field $r.fld$ of a node in the chain at some time $t'' \ge t'$ then, at some time in $[t, t'']$, $r$ was in the tree and $r.fld$ contained $v$.1
\end{ignore}

%We linearize each \textit{effective} update \func{UP} at %(the linearization point of) 
%its \sct\ and prove the following correctness properties. % for each effective update \func{UP}.
%\begin{compactenum}[\hspace{4.1mm}\bfseries C1:]
%\setcounter{enumi}{0}
%\item If UP were performed atomically at its linearization point, then it would perform \llt s on the same nodes, and these \llt s would return the same values.
%\label{prop-corr-lin-same-llt}%
%\end{compactenum}%
%%We show that, for each such operation $O$,
%%%that follows the template, 
%%if $O$ were performed atomically at its linearization point, then it would perform \llt s on the same nodes, and these \llt s would return the same values.
%This implies that UP's \func{SCX-Arguments} and \func{Result} computations must be the same as they would be if UP were performed atomically at its linearization point, so we obtain the following.
%\begin{compactenum}[\hspace{4.1mm}\bfseries C1:]
%\setcounter{enumi}{1}
%\item If UP were performed atomically at its linearization point, then it would perform the same \sct\ (with the same arguments) and return the same value. % in the concurrent and linearized executions.
%\label{prop-corr-lin-same-sct}%
%\end{compactenum}%
%%
%%\eric{The parag below  seems like a secondary property, so I moved it here because its former location interrupted the discussion of the main linearization argument.  It's not clear what a $search$ is in an arbitrary down tree so I changed the last sentence to talk about $some query operations$}
%%\faith{good change!}
Additionally, a property was proved in Chapter~\ref{chap-scx} that allows some query operations to be performed very efficiently using only \func{read}s, for example, \func{Get} in Chapter~\ref{chap-chromatree}.
%\func{SearchPhase}
%.  (For example, searches are done this way in Section \ref{sec-chromatic}.)
\begin{compactenum}[\hspace{4.1mm}\bfseries QueryProp:]
\setcounter{enumi}{2}
\item If a process $p$ follows child pointers starting from a node in the tree at time $t$ and reaches a node $r$ at time $t' \geq t$,
then $r$ {\it was} in the tree at some time between $t$ and $t'$.
Furthermore, if $p$ reads $v$ from a mutable field of $r$ at time
$t'' \ge t'$ then, at some time between $t$ and $t''$, node $r$ was in the tree and this field contained $v$.
\label{prop-corr-queries}%
\end{compactenum}%

\paragraph{Formal proof}
We now proceed with the formal proof.
Since we refer to the preconditions of \llt\ and \sct\ in the following, we reproduce them here, for convenience.
\begin{compactitem}
\item \llt$(r)$: $r$ has been initiated (previously inserted into the tree)
\item \sct$(V, R, fld, new)$:
    \begin{compactenum}
    \item for each $r \in V$, $p$ has performed an invocation $I_r$ of $\llt(r)$ linked to this \sct
    \item $new$ is not the initial value of $fld$
    \item for each $r \in V$, no \sct$(V', R', fld, new)$ occurred before $I_r$
    \end{compactenum}
\end{compactitem}

\medskip
The following lemma establishes Constraint~\ref{constraint-finalized-iff-removed}, and some other properties that will be useful when proving linearizability.

%\eric{Has this paper said what the preconditions of \llt and \sct are?  Below they are referred to by number}

\begin{lem} \label{lem-dotreeup-constraints-invariants}
The following properties hold in any execution of template operations.
\begin{enumerate}
\item    Let $S$ be a successful invocation of \sct$(V, R, fld, new)$, 
         and $G$ be the directed graph induced by the edges read by the \llt s linked to $S$.
%and $N$ and $G$ be as described in the code of \dotreeupdate.
         $G$ is a sub-graph of the data structure at all times after the last \llt\ linked to $S$ and before $S$, and no node in the $N$ set of $S$ is in the data structure before $S$.
\label{claim-dotreeup-G-in-data-structure}
\item    Every \llt\ or \sct\ performed by a template operation has valid arguments, and satisfies its preconditions.
%         Additionally, if an invocation of \dotreeupdate\ performs a successful invocation of \sct$(V, R, fld, new)$, then the elements of $V$ are distinct.
\label{claim-dotreeup-llt-sct-preconditions}
\item    Let $S$ be a successful invocation of \sct$(V, R, fld, new)$, where $fld$ is a field of $parent$, and $old$ is the value read from $fld$ by the $\llt(parent)$ linked to $S$. %be the root of $G_R$. %, and $N$, $F$ and $top$ be as described in the code of \dotreeupdate.
         $S$ changes $fld$ from $old$ to $new$, replacing a connected subgraph containing nodes $R \cup F_N$ with another connected subgraph containing nodes $N \cup F_N$.
         Further, the \rec s added by $S$ are precisely those in $N$, and the \rec s removed by $S$ are precisely those in $R$.
\label{claim-dotreeup-finalized-before-removed}
\item    At all times, $root$ is the root of a tree of \node s.
         (We interpret $\bot$ as the empty tree.)
\label{claim-dotreeup-tree}
\end{enumerate}
\end{lem}
\begin{chapscxproof}
We prove these claims by induction on the sequence of steps taken in the execution.
%The only steps that can affect these claims are successful invocations of \llt\ and \sct.
Clearly, these claims hold initially.
%Clearly, these claims hold before any such step has occurred.
Suppose they hold before some step $s$.
We prove they hold after $s$.
Let $O$ be the operation that performs $s$.

\textbf{Proof of Claim~\ref{claim-dotreeup-G-in-data-structure}.}
To affect this claim, $s$ must be a successful invocation of \sct$(V, R, fld, new)$.
Since $s$ is successful, the semantics of \sct\ imply that, for each $r \in V $, no successful \sct$(V', R', fld', new')$ with $r \in V'$ occurs between the invocation $I$ of $\llt(r)$ linked to $s$ and $s$.
%Thus, $O$ satisfies Constraints~\conDoTreeUpRootAndR, \conDoTreeUpParent\ and \conDoTreeUpDistinct\ of \dotreeupdate.
Thus, for each $r \in V $, no mutable field of $r$ changes between $I$ and $s$.
We now show that all nodes and edges of $G$ are in the data structure at all times after the last \llt\ linked to $s$, and before $s$.
Fix any arbitrary $r \in V$.
By inductive Claim~\ref{claim-dotreeup-llt-sct-preconditions}, $I$ satisfies its precondition, so $r$ was initiated when $I$ started and, hence, was in the data structure before $I$.
By the semantics of \sct, since $I$ returns a value different from \fail\ or \finalized, no successful invocation of \sct$(V'', R'', fld'', new'')$ with $r \in R''$ occurs before $I$.
By %Observation~\ref{cor-dotreeup-satisfies-con-mark-all-removed-recs} and
inductive Claim~\ref{claim-dotreeup-finalized-before-removed}, Constraint~\ref{constraint-finalized-iff-removed} is satisfied at all times before $s$.
%Hence, we can freely apply results from Appendix~\ref{sec-properties} at all times before $s$.
Thus, $r$ is not removed before $I$.
Since $s$ is successful, no successful invocation of \sct$(V'', R'', fld'', new'')$ with $r \in R''$ occurs between $I$ and $s$.
(If such an invocation were to occur then, since $r$ would also be in $V''$, the semantics of \sct\ would imply that $s$ could not be successful.)
Since $I$ occurs before $s$, %Therefore, Constraint~\ref{constraint-finalized-iff-removed} implies that
$r$ is not removed before $s$.
When $I$ occurs, since $r$ is in the data structure, all of its children are also in the data structure.
Since no mutable field of $r$ changes between $I$ and $s$, all of $r$'s children read by $I$ are in the data structure throughout this time.
Thus, each node and edge in $G$ is in the data structure at all times after the last \llt\ linked to $s$, and before $s$.

Finally, we prove that no node in $N$ is in the data structure before $s$ is successful.
Since $O$ follows the tree update template, $s$ is its only modification to shared memory.
Since each $r' \in N$ is newly created by $O$, it is clear that $r'$ can only be in the data structure after $s$.
%Tree update template Property~\ref{con-N-not-in-data-structure} completes the proof.

\textbf{Proof of Claim~\ref{claim-dotreeup-llt-sct-preconditions}.}
Suppose $s$ is invocation of $\llt(r)$.
Then, $r \neq \nil$ (by the discussion in Sec.~\ref{sec-dotreeupdate}).
By the code in Figure~\ref{code-dotreeupdate}, either $r = top$, or $r$ was obtained from the return value of some invocation $L$ of \llt$(r')$ previously performed by $O$.
If $r$ was obtained from the return value of $L$, %then, $L$ satisfied its precondition by the inductive hypothesis, so $r$ is initiated before (and, hence, when) $s$ begins.
then Lemma~\ref{lem-rec-in-data-structure-just-before-llt} implies that $r'$ is in the data structure when $L$ occurs.
Hence, $r$ is in the data structure when $L$ is occurs, which implies that $r$ is initiated when $s$ occurs.
Now, suppose $r = top$.
By the precondition of $O$, $r$ was reached by following child pointers from $root$ since the last operation by $p$.
By %Observation~\ref{cor-dotreeup-satisfies-con-mark-all-removed-recs} and 
inductive Claim~\ref{claim-dotreeup-finalized-before-removed}, Constraint~\ref{constraint-finalized-iff-removed} is satisfied at all times before $s$.
Therefore, we can apply Lemma~\ref{lem-if-rec-traversed-then-rec-in-data-structure}, which implies that $r$ was in the data structure at some point before the start of $O$ (and, hence, before $s$).
%Therefore, the precondition of $O$ and Lemma~\ref{lem-if-rec-traversed-then-rec-in-data-structure} imply that $r$ was in the data structure at some point before the start of $O$ (and, hence, before $s$).
By Definition~\ref{defn-rec-in-added-removed}, $r$ is initiated when $s$ begins.

Suppose $s$ is an invocation of \sct$(V, R, fld, new)$.
%Let $parent$ be the node that contains the mutable (child) field $fld$ changed by $s$.
By PC\ref{con-parent-in-V}, $fld$ is a mutable (child) field of some node $parent \in V$.
By PC\ref{con-R-subsequence-of-V}, $R$ is a subsequence of $V$.
Therefore, the arguments to $s$ are valid.
By PC\ref{con-llt-on-all-nodes-in-V} and the definition of $\sigma$, for each $r \in V$, $O$ performs an invocation $I$ of $\llt(r)$ before $s$ that returns a value different from \fail\ or \finalized\ (and, hence, is linked to $s$), %.
%By Property~\ref{defn-llt-linked-to-sct}, $I$ is linked to $s$, 
so $s$ satisfies Precondition~\presctlinked\ of \sct.

We now prove that $s$ satisfies \sct\ Precondition~\presctabainit.
%We first show that $new$ is not the initial value of $fld$.
Let $parent.c_i$ be the field pointed to by $fld$.
If $parent.c_i$ initially contains $\bot$ then, by PC\ref{con-GN-non-empty-tree}, %and Property~\ref{con-fringe-of-GN-is-F},
$new$ is a \node, and we are done.  %\eric{May need modifying after Faith's change to constraints.}
Suppose $parent.c_i$ initially points to some \node\ $r$.
We argued in the previous paragraph that $O$ performs an \llt$(parent)$ linked to $s$ before $s$.
By %Observation~\ref{cor-dotreeup-satisfies-con-mark-all-removed-recs} and 
inductive Claim~\ref{claim-dotreeup-finalized-before-removed}, Constraint~\ref{constraint-finalized-iff-removed} is satisfied at all times before $s$.
Therefore, we can apply Lemma~\ref{lem-if-rec-traversed-then-rec-in-data-structure} to show that $r$ was in the data structure at some point before the start of $O$ (and, hence, before $s$).
However, by inductive Claim~\ref{claim-dotreeup-G-in-data-structure} (which we have proved for $s$), %Property~\ref{con-N-not-in-data-structure}, 
$new$ cannot be initiated before $s$, so $new \neq r$.

Finally, we show $s$ satisfies \sct\ Precondition~\presctaba.
Fix any $r' \in V$, and let $L$ be the \llt$(r')$ linked to $s$ performed by $O$.
To derive a contradiction, suppose a successful invocation $S'$ of \sct$(V', R', fld, new)$ occurs before $L$ (which occurs before $s$).
By Lemma~\ref{lem-rec-in-data-structure-after-linearized-sct} (which we can apply since Constraint~\ref{constraint-finalized-iff-removed} is satisfied at all times before $s$), $new$ would be in the data structure (and, hence, initiated) before $s$ occurs.
However, this contradicts our argument that $new$ cannot be initiated before $s$ occurs.

\textbf{Proof of Claim~\ref{claim-dotreeup-finalized-before-removed} and Claim~\ref{claim-dotreeup-tree}.}
%$\bigstar$
To affect these claims, $s$ must be a successful invocation of \sct$(V, R, fld, new)$.
%%%%%%%%%%%%%%%%% TODO: %%%%%%%%%%%%%%%%%%%%%% \eric{I think t should be replaced by s throughout this argument, although there is something weird here about whether s is an invocation or a  step or a linearization point or...  But leave it for now; we'll fix it later.}
The semantics of \sct\ and the fact that $s$ is successful imply that, for each $r \in V$, no successful \sct$(V', R', fld', new')$ with $r \in V'$ occurs after the invocation $I$ of $\llt(r)$ linked to $s$ and before $s$.
Thus, $O$ satisfies tree update template PC\ref{con-old-nil-then-R-empty}, PC\ref{con-R-non-empty-then-GR-a-non-empty-tree}, PC\ref{con-GN-non-empty-tree} and PC\ref{con-fringe-of-GN-is-old}.
%Further, for each $r \in V$, no mutable field of $r$ changes between $I$ and $s$. \trevor{needed?}
By inductive Claim~\ref{claim-dotreeup-G-in-data-structure} (which we have proved for $s$), all nodes and edges in $G$ are in the data structure just before $s$, and no node in $N$ is in the data structure before $s$.
Let $parent.c_i$ be the mutable (child) field changed by $s$, and $old$ be the value read from $parent.c_i$ by the $\llt(parent)$ linked to $s$.
%
%We prove by cases that the following claims hold just before $s$.
%\begin{compactitem}
%\item each node and edge of $G_R$ is in the data structure, and each node in $G_R$ is a descendent of $old$
%\item no $R \cap F$ is empty, and no node in $R$ is a descendent of a node in $F$
%\item every path from $root$ to a node in $\{$descendents of $old\} - R$ passes through a node in $F$
%\item $parent \notin (R \cup F \cup N)$ % is not a descendent of $old$
%\end{compactitem}

Suppose $R \neq \emptyset$ (as in Fig.~\ref{fig-replace-subtree}).
%In this case, $F = Fringe(R, G_R)$, which implies that each node in $F$ is in $G_R$.
Then, by PC\ref{con-R-non-empty-then-GR-a-non-empty-tree}, $G_R$ is a tree rooted at $old$ and $F_N = F_R$.
%Since $R$ is a subsequence of $V$, and $O$ performs an \llt$(r)$ linked to $s$ for each $r \in V$, $O$ performs an \llt$(r)$ linked to $s$ for each $r \in R$.
Since $G_R$ is a sub-graph of $G$, inductive Claim~\ref{claim-dotreeup-G-in-data-structure} implies that each node and edge of $G_R$ is in the data structure just before $s$.
%Therefore, $G_R$ is a sub-graph of $G$, which implies that each node and edge of $G_R$ is in the data structure just before $s$.
Further, since $O$ performs an \llt$(r)$ linked to $s$ for each $r \in R$, and no child pointer changes between this \llt\ and $s$, $G_R$ contains every node that was a child of a node in $R$ just before $s$.
Thus, $F_R$ contains every node $r \notin R$ that was a child of a node in $R$ just before $s$.
This implies that, just before $s$, for each node $r \notin R$ in the sub-tree rooted at $old$, $F_R$ contains $r$ or an ancestor of $r$.
By inductive Claim~\ref{claim-dotreeup-tree}, just before $s$, every path from $root$ to a descendent of $old$ passes through $old$.
Therefore, just before $s$, every path from $root$ to a node in $\{$descendents of $old\} - R$ passes through a node in $F_R$. %% this seems like something we want to export from all cases
Just before $s$, by the definition of $F_R$, and the fact that the nodes in $R$ form a tree, $R \cap F_R$ is empty and no node in $R$ is a descendent of a node in $F$.
%Since, just before $s$, the nodes in $R$ form a tree, and each node in $F$ is in $Fringe(R, G_R)$, $R \cap F$ is empty and no node in $R$ is a descendent of a node in $F$.
%%% maybe split up cases based on what is above here... actually, that seems like a bad idea %%
By PC\ref{con-GN-non-empty-tree}, $G_N$ is a non-empty tree rooted at $new$ with node set $N \cup F_N = N \cup F_R$, where $N$ contains nodes that have not been in the data structure before $s$.
Since $parent.c_i$ is the only field changed by $s$, $s$ replaces a connected sub-graph with node set $R \cup F_R$ by a connected sub-graph with node set $N \cup F_R$.
We prove that $parent$ was in the data structure just before $s$.
Since $s$ modifies $parent.c_i$, just before $s$, $parent$ must not have been finalized.
Thus, no successful \sct$(V', R', fld', new')$ with $parent \in R'$ can occur before $s$.
By inductive Claim~\ref{claim-dotreeup-finalized-before-removed}, Constraint~\ref{constraint-finalized-iff-removed} is satisfied at all times before $s$, so $parent$ cannot be removed from the data structure before $s$.
By inductive Claim~\ref{claim-dotreeup-llt-sct-preconditions}, the precondition of the \llt$(parent)$ linked to $s$ implies that $parent$ was initiated, so $parent$ was in the data structure just before $s$.
Since no node in $N$ is in the data structure before $s$, the \rec s added by $s$ are precisely those in $N$.
Since, just before $s$, no node in $R$ is in $F$, or a descendent of a node in $F$, and every $r \in \{$descendents of $old\}-R$ is reachable from a node in $F_R$, the \rec s removed by $s$ are precisely those in $R$.

Now, to prove Claim~\ref{claim-dotreeup-tree}, we need only show that $parent.c_i$ is the root of a sub-tree just after $s$.
%%argue parent is not in any subtree rooted at an element of F
We have argued that $old$ is the root of $G_R$ just before $s$.
Since $parent.c_i$ points to $old$ just before $s$, the inductive hypothesis implies that $old$ is the root of a subtree, and $parent$ is not a descendent of $old$.
%
%Since $parent.c_i$ points to $old$ just before $s$, the inductive hypothesis implies that $old$ is a child of $parent$ just before $s$.
%%%% this next step appears to rely on the particular case that we are in
%By Property~\ref{con-R-non-empty-then-GR-a-non-empty-tree}, $G_R$ is a tree rooted at $old$, and we have argued that it is a sub-graph of the data structure just before $s$.
%Thus, just before $s$, $parent$ is not a descendent of $old$. %% this sounds like what we want to export for our case
Therefore, just before $s$, $parent$ is not in any sub-tree rooted at a node in $F_R$.
This implies that no descendent of $old$ is changed by $s$.
%Therefore, for each \node\ $r \in F$, no descendent of $r$ is changed by $s$.
By inductive Claim~\ref{claim-dotreeup-tree}, each $r \in F$ is the root of a sub-tree just before $s$, so each $r \in F$ is the root of a sub-tree just after $s$.
%
%%%%%%%%%%%%%%%%%%%%%%%%%%%%%%%%%%%%%%%%%%%
%
Finally, since PC\ref{con-GN-non-empty-tree} states that $G_N$ is a non-empty down-tree rooted at $new$, and we have argued that $G_N$ has node set $N \cup F_R$, $parent$ is the root of a sub-tree just after $s$.

The two other cases, where $R = \emptyset$ and $old = \nil$ (as in Fig.~\ref{fig-replace-subtree2}a), and where $R = \emptyset$ and $old \neq \nil$ (as in Fig.~\ref{fig-replace-subtree2}b), are similar (and substantially easier to prove).
\end{chapscxproof}

\begin{cor} \label{cor-dotreeup-satisfies-con-mark-all-removed-recs}
A \rec\ is finalized when (and only when) it is removed from the tree (Constraint~\ref{constraint-finalized-iff-removed}).
%Constraint~\ref{constraint-finalized-iff-removed} is implied by Lemma~\ref{lem-dotreeup-constraints-invariants}.\ref{claim-dotreeup-finalized-before-removed}.
\end{cor}

\begin{lem} \label{lem-dotreeupdate-rec-cannot-be-added-after-removal}
After a \rec\ $r$ is removed from the data structure, it cannot be added back into the data structure.
\end{lem}
\begin{chapscxproof}
Suppose $r$ is removed from the data structure.
The only thing that can add $r$ back into the data structure is a successful invocation $S$ of \sct$(V, R, fld, new)$.
%By Constraint~\ref{con-dotreeup-exclusively-does-sct}, 
Such an \sct\ must occur in an effective update $O$.
By Lemma~\ref{lem-dotreeup-constraints-invariants}.\ref{claim-dotreeup-finalized-before-removed}, every \rec\ added by $S$ is in $O$'s $N$ set.
By Lemma~\ref{lem-dotreeup-constraints-invariants}.\ref{claim-dotreeup-G-in-data-structure}, no node in $N$ is in the data structure at any time before $S$.
Thus, $r$ cannot be added to the data structure by $S$.
\end{chapscxproof}

\begin{obs}
A template operation $O$ can modify the tree only if it is an effective update (i.e., it performs a successful invocation of \sct).
\end{obs}
%\begin{chapscxproof}
%Immediate from the template code, and the fact that $O$ does \textit{not} perform a successful \sct.
%\end{chapscxproof}

\begin{lem} \label{lem-effective-updatephase-atomic}
Consider an effective update $O$.
The invocation $I$ of \func{UpdatePhase}$(m)$ performed by $O$ behaves exactly as it would if it were performed atomically at $O$'s invocation $S$ of \sct.
(That is, $I$ returns the same value, and performs the same invocations of \llt\ and \sct\ (with the same arguments), as it would if it were performed atomically at $S$.)
\end{lem}
\begin{chapscxproof}
Let $I_L$ be the invocation of \func{UpdatePhase}$(m)$ performed by the update in the linearized execution that corresponds to $O$ (i.e., that performs $S$).
Note that $I_L$ occurs atomically at $S$.
We prove that the arguments and return values of the \llt s performed by $I$ and $I_L$ are the same.
Recall that $s_0, s_0', ..., s_i, s_i'$ consist of the return values of the \llt s performed by the operation, and the immutable fields of the nodes accessed by these \llt s.
It immediately follows that $I$ and $I_L$ have the same inputs to their local \func{SCX-Arguments}$(s_0, s_0', ..., s_i, s_i')$ and \func{Result}$(s_0, s_0', ..., s_i, s_i')$ computations.
Consequently, $I$ and $I_L$ perform invocations of \sct\ with exactly the same arguments, and have the same return value.

We start by proving that each \llt\ performed by $I$ returns the same value as it would if it were performed atomically at $S$ (which is also when $I_L$ occurs).
Since $S$ is successful, by the definition of \llt\ and \sct, for each $\llt(r)$ performed by $I$, no invocation of \sct$(V', R', fld', new')$ with $r \in V'$ occurs after this \llt\ and before $S$.
Thus, no $r \in V$ changes after $I$'s \llt$(r)$ and before $S$.
%\trevor{i don't understand why i'm talking about $parent$ here...}
%Thus, the $\llt(parent)$ performed by $I$ returns the same result that it would if it were performed atomically when $S$ occurs.

We now prove that $I$ and $I_L$ perform the same \llt s, which return the same values.
Let $I^k$ and $I_L^k$ be the $k$th \llt s by $I$ and $I_L$, respectively, $a^k$ and $a_L^k$ be the respective arguments to $I^k$ and $I_L^k$, and $v^k$ and $v_L^k$ be the respective return values of $I^k$ and $I_L^k$.
%We denote the prefix length $k$ of a sequence $s$ by $s^k$.
We prove by induction that $a^k = a_L^k$ and $v^k = v_L^k$ for all $k \ge 1$.

\textbf{Base case:}
Since $I$ and $I_L$ have the same argument $m$, they both identify the same starting node $n_0$ for their \llt s.
Hence, $a^1 = a_L^1 = n_0$.
Since each \llt\ performed by $I$ returns the same value as it would if it were performed atomically at $S$, $v^1 = v_L^1$.

\textbf{Inductive step:}
Suppose the inductive hypothesis holds for $k-1$, where $k > 1$. %$a^{k-1} = a_L^{k-1}$ and $v^{k-1} = v_L^{k-1}$ for $k > 1$.
The \func{NextNode} computation from which $I$ obtains $a^k$ depends only on $v^1, ..., v^{k-1}$ and the immutable fields of nodes $a^1, ..., a^{k-1}$.
Similarly, the \func{NextNode} computation from which $I_L$ obtains $a_L^k$ depends only on $v_L^1, ..., v_L^{k-1}$ and the immutable fields of nodes $a_L^1, ..., a_L^{k-1}$.
Thus, by the inductive hypothesis, $a^k = a_L^k$.
Since each \llt\ performed by $I$ returns the same value as it would if it were performed atomically at $S$, we have $v^k = v_L^k$.
Therefore, the inductive hypothesis holds for $k$, and the claim is proved.
\end{chapscxproof}

\begin{lem} \label{lem-effectivedtp-searchatomic}
Consider an effective update $O$ that satisfies DTP.
The invocation of \func{SearchPhase} performed by $O$ returns the same value $m$ that it would if it were performed atomically at $O$'s \sct.
%Moreover, the invocation of \func{UpdatePhase} performed by $O$ has the same argument and behaviour as it would if it were atomic
%Then, $O$ is atomic.
%(That is, $O$ behaves exactly as it would if it were executed atomically at its \sct.)
%\trevor{need to be more precise: something about invoking UpdatePhase with the same argument as it would if it were executed atomically at its linearization point, and updatephase doing all the same llxs and the same scx.}
\end{lem}
\begin{chapscxproof}
We first argue that, when $O$ performs its \sct, all of the nodes in $m$ are in the tree, and their fields match the values in $m$.
Since $O$ is an effective update, it performs a successful \sct$(V, R, fld, new)$.
By inspection of the code in Figure~\ref{code-dotreeupdate}, since $O$ performs an \sct, for each node $n_i \in V$, $O$ must perform an invocation of \llt$(n_i)$ that returns a value different from \fail\ or \finalized, and subsequently perform an invocation of \func{Conflict}$(n_i, s_i, m)$ that returns \false.
By PC\ref{con-m-in-V}, $V$ includes every node in $m$.
Therefore, by the semantics of \llt\ and \sct, for each node $n_i \in m$, at all times after $O$'s \llt$(n_i)$ and before its \sct, $n_i$ is in the tree and its fields match their values stored in $m$.
The claim then follows from the fact that $O$ satisfies DTP.
\end{chapscxproof}

\begin{thm} \label{thm-effectivedtp-atomic}
Every effective update whose \func{SearchPhase} satisfies DTP is atomic.
\end{thm}
\begin{chapscxproof}
Consider an effective update $O$ that satisfies DTP.
The argument $m$ to $O$'s invocation of \func{UpdatePhase} is returned from $O$'s invocation of \func{SearchPhase}.
By Lemma~\ref{lem-effectivedtp-searchatomic}, $O$'s invocation of \func{SearchPhase} would return the same argument $m$ if it were performed atomically at $O$'s \sct.
By Lemma~\ref{lem-effective-updatephase-atomic}, $O$'s invocation of \func{UpdatePhase}$(m)$ would have the same behaviour if it were performed atomically at $O$'s \sct.
Thus, $O$ has the same effect on the tree (and returns the same value) as it would if it were performed atomically at its \sct.
\end{chapscxproof}

\section{Progress proof}
\label{template-progress}

Recall the following definitions and progress property from Chapter~\ref{chap-scx}.
We use this progress property to prove progress for template operations.
An \sct-\func{Update} algorithm performs \llt s on a sequence $V$ of \rec s and invokes \sct$(V, R, fld, new)$ if all of these \llt s return snapshots.
A successful \sct-\func{Update} is one in which the \sct\ returns \true.
Similarly, a \vlt-\func{Query} algorithm performs \llt s on a sequence $V$ of \rec s and invokes \vlt$(V)$ if all of these \llt s return snapshots.
A successful \vlt-\func{Query} is one in which the \vlt\ returns \true.
Note that every execution of a template operation that does not return at line~\ref{code-template-return-result1} or line~\ref{code-template-return-result2} is an execution of an \sct-\func{Update} algorithm.
\begin{compactenum}[{\bf P\arabic{enumi}}:]
\setcounter{enumi}{1}
\item Suppose that
    (a) there is always some non-finalized \rec\ reachable by following pointers from an entry point, 
    (b) for each \rec\ $r$, each process performs finitely many invocations of \llt$(r)$ that return \finalized, and
    (c) processes perform infinitely many executions of \sct-\func{Update} and/or \vlt-\func{Query} algorithms.
    Then, infinitely many \sct\ or \vlt\ operations succeed.
\end{compactenum}

We briefly describe the proof at a high level.
First, to ensure that processes can perform infinitely many template operations, we argue that all template operations are lock-free (resp., wait-free) if their invocations of \func{SearchPhase} are lock-free (resp., wait-free).
Next, we show that every data structure whose operations are template operations satisfies conditions (a), (b) and (c) of P2.
Observe that, if each \sct\ and \vlt\ is performed by a template operation, then P2 implies the following: if infinitely many template operations are performed, then infinitely many template operations succeed.

\begin{lem} \label{lem-dotreeup-wait-free}
Let $\mathcal{P} \in \{$lock-freedom, wait-freedom$\}$.
If the \func{SearchPhase} of a template operation satisfies $\mathcal{P}$, and the implementation of \llt\ and \sct\ given in Chapter~\ref{chap-scx} is used, then the entire template operation satisfies $\mathcal{P}$.
\end{lem}
\begin{chapscxproof}
A template operation consists of an invocation of \func{SearchPhase}, possibly followed by a loop in which \llt, \func{Condition}, \func{Conflict} and \func{NextNode} are invoked (see Fig.~\ref{code-dotreeupdate}), possibly followed by an invocation of \sct.
All of the other steps comprise a finite local computation.
The implementation of \llt\ and \sct\ given in Chapter~\ref{chap-scx} is wait-free.
Similarly, \func{NextNode}, \func{Condition} and \func{Conflict} perform finite local computations, and \func{Condition} must eventually return \true\ in every template operation that invokes it, causing the operation to exit the loop.
Thus, apart from the \func{SearchPhase}, the template operation is wait-free.
\end{chapscxproof}

\begin{lem} \label{lem-dotreeup-only-one-finalized-llt}
No process performs more than one invocation of $\llt(r)$ that returns \finalized, for any $r$, during template operations.
\end{lem}
\begin{chapscxproof}
Fix any \rec\ $r$.
Suppose, to derive a contradiction, that a process $p$ performs two different invocations $L$ and $L'$ of $\llt(r)$ that return \finalized, during template operations.
Then, since each template operation returns \fail\ immediately after performing an $\llt(r)$ that returns \finalized, $L$ and $L'$ must occur in different template operations $O$ and $O'$.
Without loss of generality, suppose $O$ occurs before $O'$.

Since $L$ returns \finalized, it occurs after an invocation $S$ of \sct$(\sigma, R, fld, new)$ with $r \in R$ (by the semantics of \llt\ and \sct).
By Lemma~\ref{lem-dotreeup-constraints-invariants}.\ref{claim-dotreeup-finalized-before-removed}, $r$ is removed from the data structure by $S$.
By Lemma~\ref{lem-dotreeupdate-rec-cannot-be-added-after-removal}, $r$ cannot be added back into the data structure.
Thus, $r$ is not in the data structure at any time after $S$ occurs (and, hence, after $O$ occurs).
%\trevor{the following requires $n_0$ to be in the tree at some point after the search phase starts. we must constrain search phase so it satisfies this requirement.}
Recall that the first node $n_0$ on which $O'$ performs \llt\ is the root of the part $m$ of the tree returned by an invocation of \func{SearchPhase}, and that $n_0$ is in the data structure at some point during this invocation of \func{SearchPhase}.
%By inspection of the template in Figure~\ref{code-dotreeupdate}, the first node $n_0$ on which $O'$ performs \llt\ is located by following child pointers from the $root$ entry point.
Thus, $n_0$ must be in the tree at some time after $O$ but before $O'$.
Consequently, $n_0$ cannot be $r$.

By inspection of the template in Figure~\ref{code-dotreeupdate}, $O'$ first performs \llt\ on $n_0$, then performs \llt\ \textit{only} on nodes that it obtained from the results of previous \llt s, and finally performs $S$.
Since $r \neq n_0$, $L'$ (which has $r$ as its argument) must have obtained $r$ from the result of an invocation $L''$ of \llt$(r')$ by $O'$ that occurs after $O$ and before $L'$.
By Lemma~\ref{lem-rec-in-data-structure-just-before-llt}, $r'$ must be in the data structure just before $L''$.
Since $L''$ returns $r$, $r'$ also points to $r$ just before $L''$.
Therefore, $r$ is in the data structure just before $L''$, which contradicts the fact that $r$ is not in the data structure at any time after $O$.
\end{chapscxproof}

\begin{thm} \label{thm-dotreeup-progress}
%Suppose \func{SearchPhase} is lock-free.
If \func{SearchPhase} is lock-free, and template operations are performed infinitely often, then template operations succeed infinitely often.
\end{thm}
\begin{chapscxproof}
Suppose, to derive a contradiction, that template operations are performed infinitely often but, after some configuration $C$, no template operation succeeds.
We show that the conditions of P2 are satisfied.
Recall that the entry point to the data structure is never deleted (see Section~\ref{sec-primitives}).
Consequently, Constraint~\ref{constraint-finalized-iff-removed} implies that the entry point is never finalized, so condition (a) of P2 is satisfied.
Condition (b) of P2 is satisfied by Lemma~\ref{lem-dotreeup-only-one-finalized-llt}.
If a template operation performs a successful \sct, or returns at line~\ref{code-template-return-result1},~\ref{code-template-return-result2} or~\ref{code-template-return-result3}, then it is successful.
Consequently, after $C$, no successful \sct\ occurs, and no template operation returns at line~\ref{code-template-return-result1},~\ref{code-template-return-result2} or~\ref{code-template-return-result3}.
It follows that, after $C$, every template operation returns \fail\ at line~\ref{code-template-return-fail1}, after performing an \llt\ that returns \fail\ or \finalized, or at line~\ref{code-template-return-fail2}, after performing an unsuccessful \sct.
Therefore, every execution of a template operation after $C$ is an execution of an \sct-\func{Update} algorithm, and Lemma~\ref{lem-dotreeup-wait-free} implies that each such operation is lock-free.
Hence, there must be infinitely many executions of \sct-\func{Update} algorithms, so condition (c) of P2 is satisfied.
By P2, there must be infinitely many successful invocations of \sct\ or \vlt.
Since there are no invocations of \vlt, and \sct\ is invoked only by template operations, and each template operation that performs a successful \sct\ succeeds, there must be infinitely many successful template operations, which is a contradiction.
\end{chapscxproof}

\chapter{Chromatic tree implemented with the template} \label{chap-chromatree}
% !TEX root = paper.tex

\section{Chromatic trees}

Here, we show how the tree update template can be used to implement an ordered dictionary ADT using
chromatic trees.
%%Due to space restrictions, 
%Due to space restrictions, we only sketch the algorithm and its correctness proof.
%%\trevor{Obviously the following sentence has the change.  Should we refer to a full version of this paper, or might seeing the full details cause a reviewer to believe that this extended abstract omits too much?}
%All details of the implementation
%and its correctness proof are in the full version of the paper.
%Complete pseudocode and proofs appear in Appendix~\ref{chromatic}.
The ordered dictionary stores a set of keys, each with an associated value, 
where the keys are drawn from a totally ordered universe.
The dictionary supports five operations.
%the following operations. 
If $key$ is in the dictionary, \func{Get}$(key)$ returns its associated value.
Otherwise, \func{Get}$(key)$ returns $\bot$.
\func{Successor}$(key)$ returns the smallest key in the dictionary that is larger than $key$ (and its associated value), or $\bot$ if no key in the dictionary is larger than $key$.
\func{Predecessor}$(key)$ is analogous.
%queries can be handled in a symmetric way.
%returns the largest key in the dictionary that is smaller than $key$ (and its associated value), or $\bot$ if there is no such key. 
%\func{InsertIfAbsent}$(key, value)$, which associates $value$ with $key$ and returns \true\ if $key$ is not in the dictionary, and returns \false\ otherwise,
\ins$(key, value)$ replaces the value associated with $key$ by $value$ and returns the previously associated value, or $\bot$ if $key$ was not in the dictionary.
If the dictionary contains $key$, \del$(key)$ removes it and returns the value that was associated immediately beforehand.
%If there is an association between $key$ and any value,
%\del$(key)$ removes it and returns the value.
Otherwise, \del($key$) simply returns $\bot$.
%\del$(key)$ removes any association between $key$ and a value and returns this value (or $\bot$ if there was none).

A RBT is a BST in which the root and all leaves are coloured black, and every other node is coloured either red or black, subject to the constraints that no red node has a red parent, and the number of black nodes on a path from the root to a leaf is the same for all leaves.
These properties guarantee that the height of a RBT is logarithmic in the number of nodes it contains.
We consider search trees that are leaf-oriented, meaning the dictionary keys are stored in the leaves, and internal nodes store keys that are used only to direct searches towards the correct leaf.
In this context, the BST property says that, for each node $x$, all descendants of $x$'s left child have keys less than $x$'s key and all descendants of $x$'s right child have keys that are greater than {\it or equal to} $x$'s key.

To decouple rebalancing steps from insertions and deletions, so that each is localized, and rebalancing steps can be interleaved with insertions and deletions, it is necessary to relax the balance properties of RBTs.
%\trevor{Maybe the previous sentence can disappear after the explanation in the introduction.}
A {\it chromatic tree} \cite{NS96} is a relaxed-balance RBT in which colours are replaced by non-negative integer weights, where weight zero corresponds to red and weight one corresponds to black.  As in RBTs, 
the sum of the weights on each path from the root to a leaf is the same.  However, 
%a chromatic tree  allows 
RBT properties can be violated in the following two ways.
First, a red child node may have a red parent, in which case we say that a
\textit{red-red violation} occurs at this child.
Second, a node may have weight $w>1$, in which case we say that
$w-1$ \textit{overweight violations} occur at this node.
%\eric{Is my change here correct: do we count $w-1$ violations for each a node of weight $w$?}
The root always has weight one,
%(Each operation that modifies the root blindly sets its weight to one.)
so no violation can occur there. %at the root.

{\it Rebalancing steps} are localized updates to a chromatic tree that are performed at the location of a violation.
Their goal is to eventually eliminate all red-red and overweight violations, while maintaining the invariant that the tree is a chromatic tree.
If no rebalancing step can be applied to a chromatic tree (or, equivalently, the 
chromatic tree contains no violations), then it is a RBT.
We use the set of rebalancing steps of Boyar, Fagerberg and 
Larsen~\cite{Boyar97amortizationresults} (which appear in Figure~\ref{fig-chromatic-rotations}).
%(shown in Figure \ref{fig-chromatic-rotations} of Appendix \ref{chromatic}),
This set of rebalancing steps has a number of desirable properties:
No rebalancing step increases the number of violations in the tree,
%Each rebalancing step either {\it eliminates} one violation
%reduces the number of violations in the tree, or removes one violation, and creates another, closer to the root (where it must eventually be eliminated).
%or {\it moves} a violation closer to the root (i.e., removes one violation and creates another higher up the tree).
%Thus,
rebalancing steps can be performed in any order, and,
after sufficiently many rebalancing steps, the tree will always become a RBT.
Furthermore, in any sequence of insertions, deletions and rebalancing steps starting from an empty chromatic tree, the amortized number of rebalancing steps
%required per update is constant (specifically,
is at most three per insertion and one per deletion.
%(on average, at most three per \ins\ and one per \del).
%When a rebalancing step removes one violation, and creates another, we say the step \textit{moves} a violation.
%We say a rebalancing step \textit{eliminates} a violation if it removes a violation and does not create any.

\begin{figure}[tb]
\centering
\begin{minipage}{0.33\textwidth}
\centering
\includegraphics[width=\linewidth]{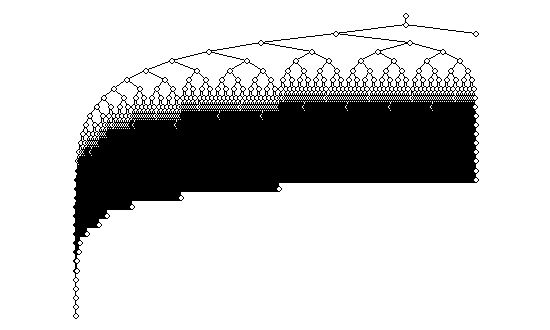}
(a)
\end{minipage}
\begin{minipage}{0.33\textwidth}
\centering
\includegraphics[width=\linewidth]{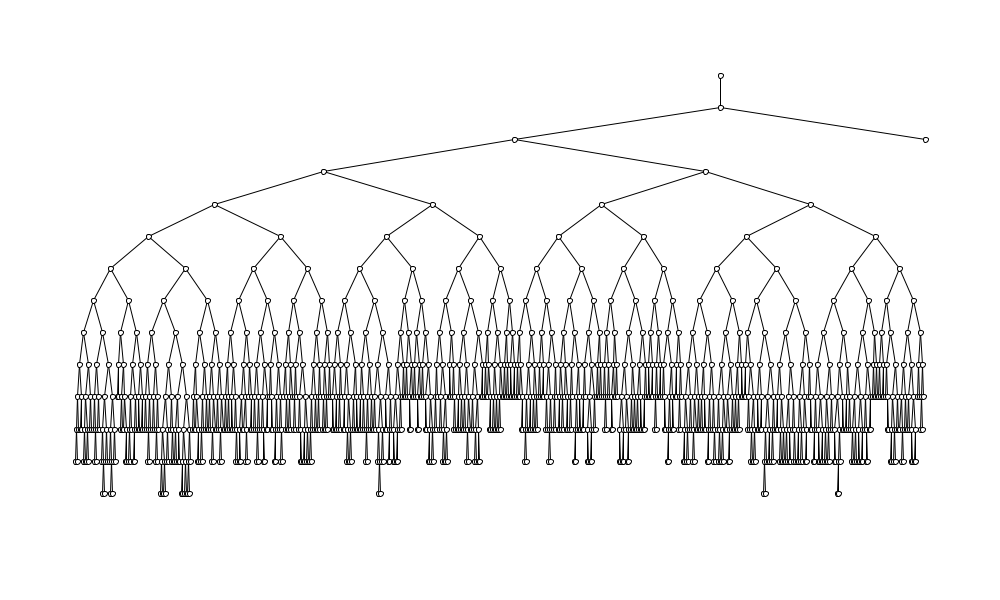}
(b)
\end{minipage}
\begin{minipage}{0.33\textwidth}
\centering
\includegraphics[width=\linewidth]{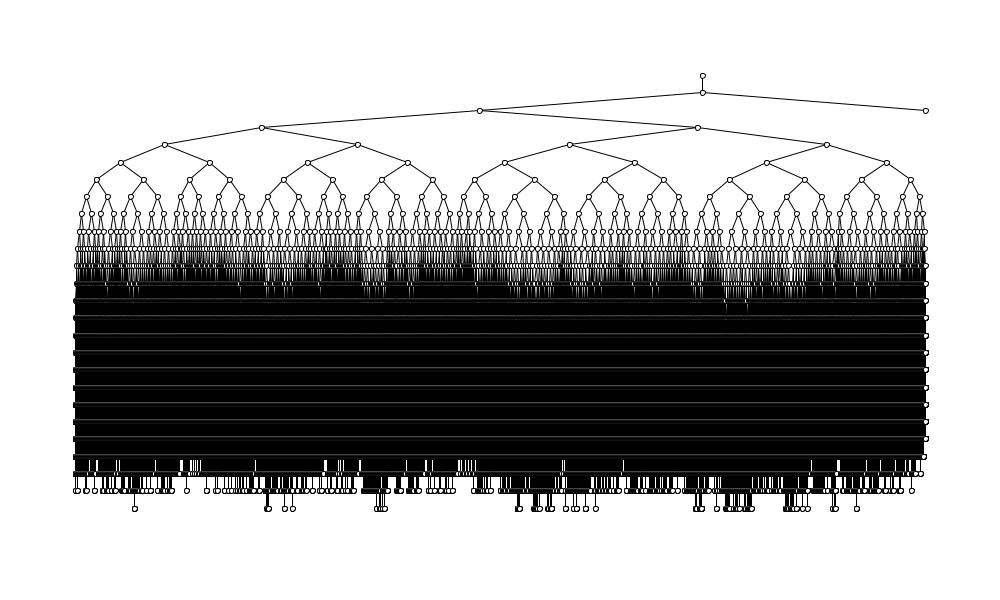}
(c)
\end{minipage}
\caption{Visualizations of chromatic trees: (a) a single process inserting 0, 1, 2, ..., 99999. (b) four processes performing millions of random insertions and deletions of keys in $[0,10^3)$. (c) four processes performing millions of random insertions and deletions of keys in $[0,10^6)$.}
\label{fig-chromatic-example-trees}
\end{figure}

Figure~\ref{fig-chromatic-example-trees} gives visualizations of three example chromatic trees.
The first is the result of a single process sequentially inserting the keys 0, 1, 2, ..., 99,999.
The depth of the tree is 34, but the average depth of leaves is approximately 19.
This demonstrates the impact of rebalancing in the canonical worst-case example.
The second tree was produced by having four processes uniformly randomly perform 50\% insertions and 50\% deletions of keys drawn uniformly from $[0, 10^3)$ for six seconds.
The resulting tree contains approximately 500 keys, and has depth 13, with an average leaf depth of 11.
The third tree was produced just like the second, but with the larger key range $[0,10^6)$.
The resulting tree contains approximately 500,000 keys, and has depth 25, with an average leaf depth of 21.

\section{Implementation} \label{sec-chromatic-impl}

\begin{figure}[tb]
%\begin{wrapfigure}{R}{0.28\textwidth}
%\begin{minipage}{0.49\textwidth}
%\vspace{-2mm}
\centering
\raisebox{10.9mm}{
\includegraphics[scale=.7]{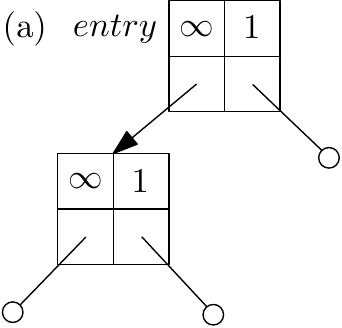}}\hspace*{3mm}
\includegraphics[scale=.7]{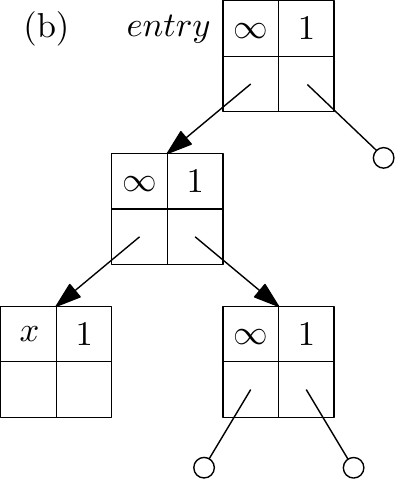}
%\end{minipage}
\caption{(a) empty tree, (b) non-empty tree.}% (weight on right).}
\label{fig-treetop}
%\end{wrapfigure}
\vspace{-2mm}
\end{figure}

We represent each node by a \rec\ with two mutable child pointers, and immutable fields $k$, $v$ and $w$ that contain the node's key, associated value, and weight, respectively.
(See Figure~\ref{code-chromatic-data}.)
%An immutable $isRoot$ bit indicates whether a node that is in the tree is the \textit{real root} of the tree, i.e., the leftmost grandchild of the {\it root} sentinel node.
%(When any update replaces a node, it uses the node's $isRoot$ and $k$ fields to determine whether it should blindly set the weight of the replacement node to one.
%More precisely, an update identifies a sentinel node by checking $k = \infty$, and the \textit{real root}, which contains a non-$\infty$ key, by checking the $isRoot$ field.)
The child pointers of a leaf are always \nil, and the value field of an internal node is always \nil.
%Leaves have no child pointers, but have an additional immutable field $v$ for the value associated with the leaf's key.
%\eric{I moved the stuff about sentinel nodes to the appendix to save space.  It seems like a detail of implementation.}
%
%
%
To avoid special cases when the chromatic tree is empty, we add sentinel nodes % with the special key $\infty$
at the top of the tree (see Figure \ref{fig-treetop}).
The entry point to the data structure, $entry$, is the topmost sentinel node.
The chromatic tree is rooted at the leftmost grandchild of $entry$.
For convenience we use $root$ to denote the current leftmost grandchild of $entry$.
The sentinel nodes have key $\infty$ to avoid special cases for \func{Get}, \ins\ and \del, and weight one to avoid special cases for rebalancing steps.
The node $root$ also always has weight one.
%Without having a special case for \ins$(x, val)$, we automatically get the sentinel nodes in Figure~\ref{fig-treetop}(b), which also eliminate special cases for \del.
%\trevor{TODO: the following sentence is confusing.}
The sum of weights is the same for all paths from the root of the chromatic tree to its leaves. %, but not for paths that include $entry$ or the sentinel nodes.
%The requirement that the sum of weights is the same for all paths applies only within the chromatic tree.
%In particular, this does not apply to $entry$ or the sentinel nodes.

Detailed pseudocode for \func{Get}, \ins\ and \del\ is given in Figure~\ref{code-chromatic-search}, \ref{code-chromatic-del} and~\ref{code-chromatic-ins}.
Note that an expression of the form $P\ ?\ A : B$ evaluates to $A$ if the predicate $P$ evaluates to true, and $B$ otherwise.
The expression $x.y$, where $x$ is a \rec, denotes field $y$ of $x$, and the expression $\&x.y$ represents a pointer to field $y$.

\begin{figure}[tbp]
\begin{framed}
%\hspace*{-7mm}
%\begin{minipage}[t]{85mm}
\def\namewidth{18mm}
\preplisting
\begin{lstlisting}[mathescape=true,style=nonumbers]
 type// \node
     //\com User-defined fields
     //\wcnarrow{$left, right$}{child pointers (mutable)}
     //\wcnarrow{$k, v, w$}{key, value, weight (immutable)}
     //\com Fields used by \llt/\sct\ algorithm
     //\wcnarrow{$\info$}{pointer to \op}
     //\wcnarrow{$marked$}{Boolean}
\end{lstlisting}
\end{framed}
%\end{minipage}
	\caption{Data definition for a node in the chromatic tree.}
	\label{code-chromatic-data}
\end{figure}

\begin{figure}
\begin{framed}
\prepnewlisting
%\vspace{-5mm}
%\hrule
%\vspace{-2mm}
\begin{lstlisting}[mathescape=true]
 //\func{Get}$(key)$
   $\langle -, -, -, l \rangle := \func{Search}(key)$
   return $(key = l.k)\ ?\ l.v : \nil$// \\ \vspace{-2mm} \hrule \vspace{1mm} %
      
 //\func{Search}$(key)$ %
   %//\com Returns the leaf found by doing a BST search for $key$, along with its parent, grandparent and great grandparent
      
   $ggp := \nil; gp := \nil; p := entry; l := entry.\mbox{\textit{left}}$
   while $l$// is internal
     $ggp := gp; gp := p; p := l$
     $l := (key < p.k)\ ?\ p.\mbox{\textit{left}} : p.\mbox{\textit{right}}$
   return $\langle ggp, gp, p, l \rangle$// %\\ \vspace{-2mm} \hrule \vspace{1mm} %
\end{lstlisting}
\end{framed}
	\caption{Peudocode for \func{Get} and \func{Search}.}
	\label{code-chromatic-search}
\end{figure}

%The tree contains \textit{sentinel} nodes (as shown in Figure~\ref{fig-treetop}), which simplify some operations by eliminating special cases.

%\eric{Faith suggested putting a brief description of GET and SUCCESSOR here.  I added a brief
%discussion of GET, since that is a good lead-in to INS/DEL.  But I figured Successor isn't
%as necessary to cover here, because it's difficult to describe it very succinctly.
%Do you like the new paragraph?  }
\func{Get}, \ins\ and \del\ each execute an auxiliary procedure, \func{Search}($key$), which appears in Figure~\ref{code-chromatic-search}.
\func{Search}$(key)$ starts at $entry$ and traverses nodes as in an ordinary BST
search, using \func{Read}s of child pointers until reaching a leaf,
which it then returns (along with the leaf's parent, grandparent and great grandparent).
%It then returns a pair of pointers to the parent and grandparent of the leaf that was reached. %(using \nil\ instead of a pointer when one of these nodes does not exist).
%It then returns $\langle p, gp \rangle$, where $p$ and $gp$ are pointers to the parent and grandparent of the leaf that was reached, or \nil s if the nodes do not exist.
%It then returns $\langle p, gp \rangle$, where $p$ (resp., $gp$) is a pointers to the parent (resp., grandparent) of the leaf that was reached, or \nil\ if the parent (resp., grandparent) does not exist.
%If either node does not exist, then \nil\ is returned instead of a pointer to it.
Because of the sentinel nodes, the leaf's parent always exists, and the grandparent %(which is needed to delete the leaf) 
exists whenever the chromatic tree is non-empty.
The grandparent is used whenever a \func{Delete} actually deletes a key (which can happen only if the tree is non-empty) and in some rebalancing steps (which are performed only in a non-empty tree).
The great grandparent is used only in some rebalancing steps (which are performed only when the great grandparent exists).
If the grandparent (or great grandparent) does not exist, then \func{Search} returns \nil\ in its place.
%If the tree is empty, \func{Search} returns \nil\ instead of the grandparent.
%(Similarly, if the great grandparent does not exist, then \func{Search} returns \nil\ in its place.)
We define the \textit{search path} for $key$ at any time to be the path that \func{Search}($key$) would follow, if it were done instantaneously.
The \func{Get}($key$) operation simply executes a \func{Search}($key$) and then returns
the value found in the leaf if the leaf's key is $key$, or $\bot$ otherwise. % (see Figure~\ref{code-chromatic0}).
% I REMOVED FOLLOWING SENTENCE, SINCE WE EXPLAIN THIS MORE LATER.
%Although the tree may change while the \func{Search} is being performed, 
%we prove that there is a time during the \func{Search} when the leaf reached {\it was} on the search path
%for $key$.
%
%\faith{What happens if the tree is empty? Shouldn't Search start at
%\RootPtr? It doesn't have a value, so should we make \RootPtr = \nil
%a special case?}

\begin{figure}[tbp]
\vspace{-6mm}
\def\figsize{1}
\centering
%\begin{tabularx}{\textwidth}{YYY}
%\includegraphics[scale=1]{chap-template/INS1.pdf} & \includegraphics[scale=1]{chap-template/INS2.pdf} & \includegraphics[scale=1]{chap-template/DEL.pdf} \\
%\func{InsertNew} & \func{InsertReplace} & \del
%\end{tabularx}
%\centering
%
%\vspace{-1.2cm}
\begin{tabular}{|m{0.42\textwidth}|m{0.45\textwidth}m{0mm}|}
\hline
\centering \includegraphics[scale=\figsize]{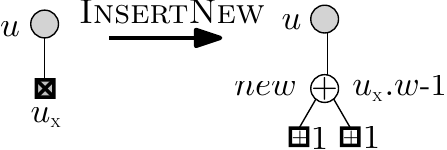} &
\centering \includegraphics[scale=\figsize]{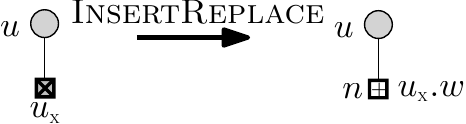} & \\
\hline
\centering \includegraphics[scale=\figsize]{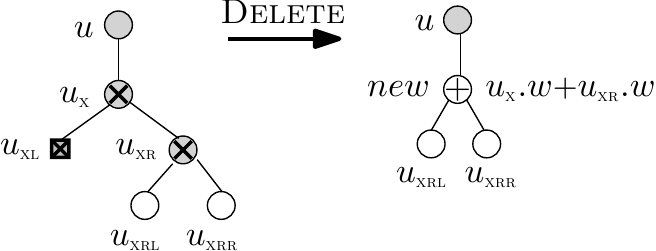} &
\centering \includegraphics[scale=\figsize]{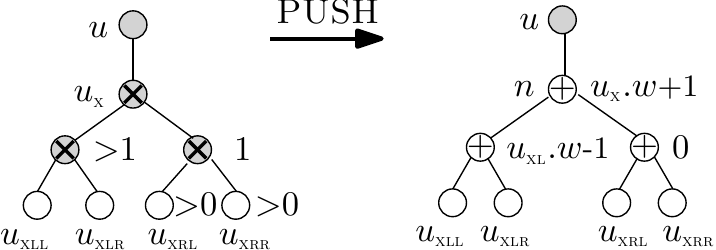} & \\
\hline
\centering \includegraphics[scale=\figsize]{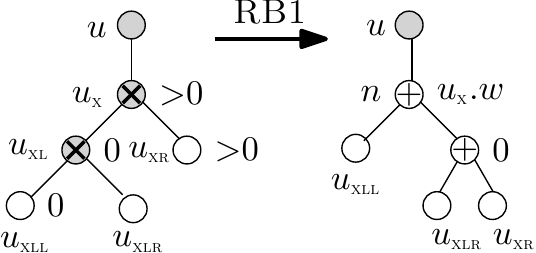} &
\centering \includegraphics[scale=\figsize]{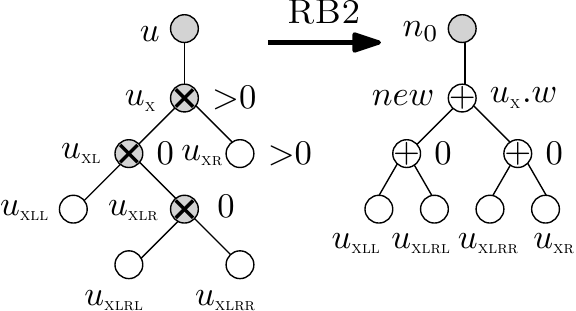} & \\
\hline
\centering \includegraphics[scale=\figsize]{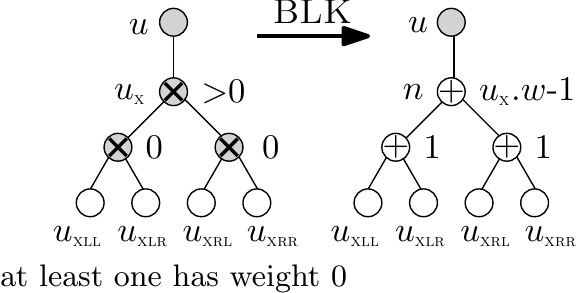} &
\centering \includegraphics[scale=\figsize]{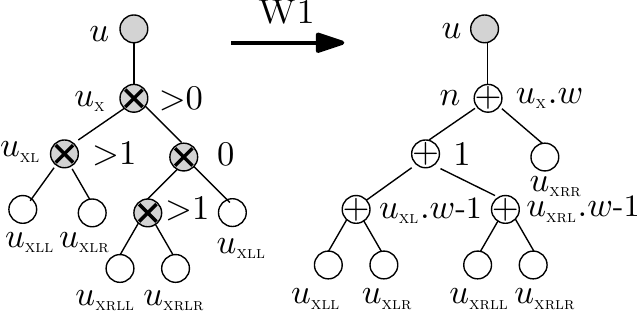} & \\
\hline
\centering \includegraphics[scale=\figsize]{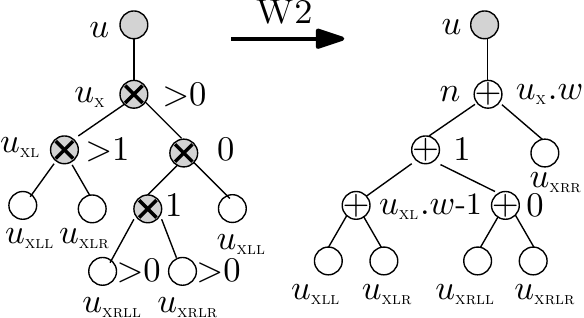} &
\centering \includegraphics[scale=\figsize]{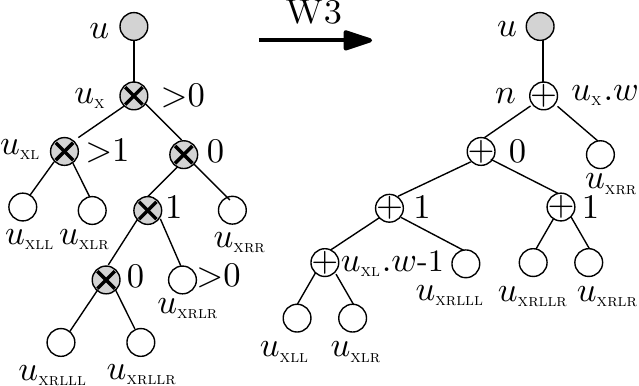} & \\
\hline
\centering \includegraphics[scale=\figsize]{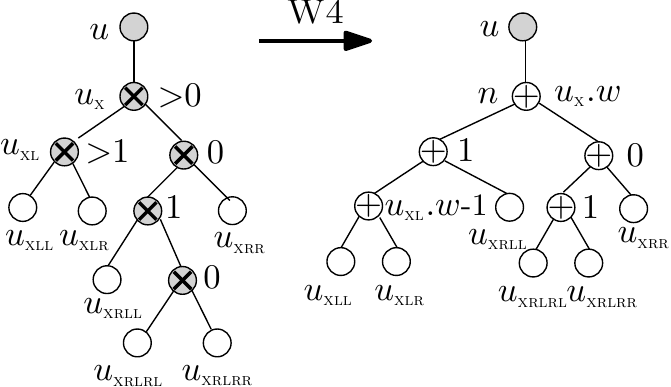} &
\centering \includegraphics[scale=\figsize]{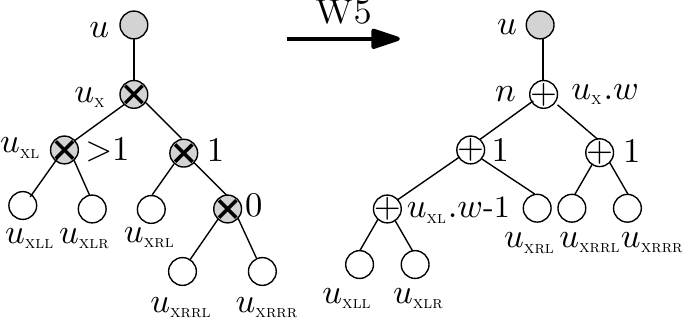} & \\
\hline
\centering \includegraphics[scale=\figsize]{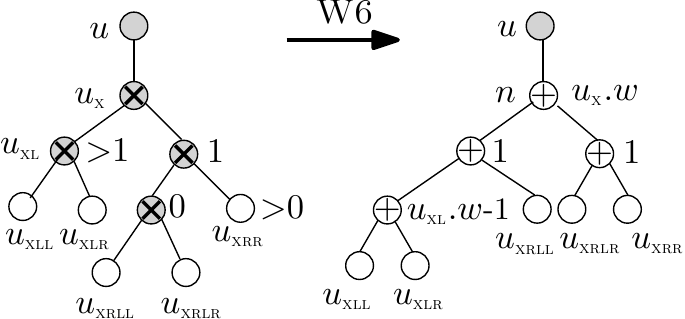} &
\centering \includegraphics[scale=\figsize]{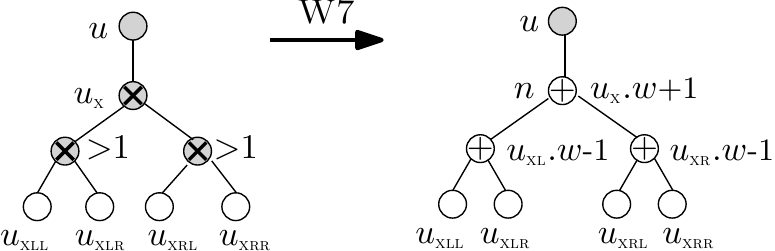} & \\
\hline
\end{tabular}

%\includegraphics[scale=0.8]{chap-template/chromatic-rotations1.png}
%\caption{}
%\label{fig-chromatic-rotations1}
%\end{figure}
%\begin{figure}[tbp]
%\centering
%\includegraphics[scale=0.8]{chap-template/chromatic-rotations2.png}
\vspace{-3mm}
\caption{Transformations for chromatic search trees.  Each transformation also has a mirror image.}
\label{fig-chromatic-rotations}
\end{figure}

Figure~\ref{fig-chromatic-rotations} shows the complete set of transformations for the chromatic tree.
%
%%\eric{I added some explanation to this paragraph.  Remove this comment if you like it (and it hasn't already been done elsewhere).}
%The rebalancing steps, which are shown in Figure~\ref{fig-chromatic-rotations}, are a slight modification of those in \cite{Boyar97amortizationresults}.\footnote{Specifically, we do not allow \func{W1}, \func{W2}, \func{W3} or \func{W4} to be applied when the node labeled $\ux$ has weight 0.  Under this restriction, this set of rebalancing steps has the desirable property that when a violation moves,  it remains on the search path to the key whose insertion or deletion originally caused the violation.  It is easy to verify that an alternative rebalancing step can always be performed when $\ux.w = 0$, so this modification does not affect the chromatic tree's convergence to a RBT.}
\func{InsertNew}, \func{InsertReplace} and \func{Delete} change the keys and values in the tree.
The other transformations are rebalancing steps.
(Each rebalancing step also has a symmetric mirror-image version, denoted by an S after the name, except \func{BLK} and \func{W7}, which are their own mirror images.)
The weight of a node appears to its right.
The name of a node appears below it or to its left.
We use a simple naming scheme for the nodes in the diagram.
Consider the node $\ux$.
We denote its left child by $\uxl$, and its right child by $\uxr$.
Similarly, we denote the left child of $\uxl$ by $\uxll$, and so on.
(The subscript {\sc x} indicates that we do not care whether a node is a left or right child.)
Each transformation shown in Figure \ref{fig-chromatic-rotations} is achieved by an \sct\ 
that swings a child pointer of $u$ and depends on  \llt s of all of the shaded nodes.
The nodes marked with $\times$ are finalized (and removed from the data structure).
The nodes marked by a $+$ are newly created nodes.
Nodes drawn as squares are leaves.
Circular nodes may be either internal nodes or leaves.
Nodes with no marking or shading, and no weight written beside them, do not necessarily have to exist for an operation to be applied.\footnote{However, sometimes these nodes are guaranteed to exist because, e.g., the tree is a full binary tree, all leaves have positive weight, and internal nodes do not change into leaves (or vice versa).}
For example, \func{Delete} can be applied even if $\uxr$ is a leaf.
In this case, the children of $\uxr$ are drawn simply to show where they will be reattached to the tree after the transformation, if they \textit{do} exist.
%
%Any nodes with no marking or shading, and with no weight written beside them, may or may not exist.
%(That is, the ``parent'' of such a node in Figure~\ref{fig-chromatic-rotations} may actually be a leaf.
%Note, however, that a node must exist if it has a sibling (since the tree is a full binary tree), 
%because there are some constraints 
%These nodes are drawn only to show where they will be reattached to the tree after the transformation, if they do exist.)
%%\eric{I know this wording is not quite right, because child pointers cannot be \nil in
%%our representation of chromatic trees.  What's a better way to say it?} 
%%Keys are stored in newly created nodes such that an %The keys stored in newly created nodes are such that an 
%%in-order traversal encounters them
%%the same as in the removed nodes (so that an in-order traversal encounters them in the same order).
%%Figure~\ref{code-chromatic-dorb2} implements
%%one of the rebalancing steps.  The others can be generated from their diagrams in a similar way.

At a high level, \ins\ and \del\ are quite similar.
\ins($key$, $value$) and \del($key$) each perform \func{Search}($key$) 
and then make the required update at the leaf reached,
in accordance with the tree update template.
\ins\ and \del\ can perform three different updates (shown in Figure~\ref{fig-chromatic-rotations}). % contains the updates that can be performed by \ins\ and \del.
\textsc{InsertNew} inserts a key that was not previously in the tree by replacing a leaf with a subtree containing a new internal (routing) node and two new leaves.
\textsc{InsertReplace} changes the value associated with a key in the tree by replacing a leaf with a new leaf.
\textsc{Delete} removes a key and its associated value from the tree by replacing a leaf, its sibling, and its parent with a new (routing) internal node.
If the modification performed by an \ins\ or \del\ fails, then the operation restarts from scratch. 
If it succeeds, it may increase the number of violations in the tree by one, 
and the new violation occurs on the search path to $key$.
If a new violation is created, then an auxiliary procedure \cleanup\ is invoked to fix it before the \ins\ or \del\ returns.

Observe that an \ins$(x, val)$ in an empty tree will perform \textsc{InsertNew}, which will yield the tree in Figure~\ref{fig-treetop}(b).
%%Thus, \ins\ does not need a special case when the tree is empty.
%Observe that, without having a special case, an \ins$(x, val)$ in an empty tree (which will perform \textsc{InsertNew}) will automatically yield the tree in Figure~\ref{fig-treetop}(b). %automatically create the sentinel nodes in Figure~\ref{fig-treetop}(b) (without explicitly handling this special case).
Similarly, deleting the last key in the tree will yield the tree in Figure~\ref{fig-treetop}(a).
%Furthermore, deleting the last key in the tree will automatically delete sentinel extra sentinel nodes eliminate special cases for \del.
Thus, the sentinel nodes are automatically maintained by insertions and deletions without adding any special cases.

\begin{figure}[p]
\begin{framed}
%\newcommand{\wcnarrow}[2]{\parbox{\namewidth}{#1} \com \mbox{#2}}
%\hspace*{-7mm}
%\begin{minipage}[t]{85mm}
\def\namewidth{18mm}
\preplisting
%\begin{lstlisting}[mathescape=true,style=nonumbers]
%    type// \node
%        //\com User-defined fields
%        //\wcnarrow{$left, right$}{child pointers (mutable)}
%        //\wcnarrow{$k, v, w$}{key, value, weight (immutable)}
%        //\com Fields used by \llt/\sct\ algorithm
%        //\wcnarrow{$\info$}{pointer to \op}
%        //\wcnarrow{$marked$}{Boolean}
%%\end{lstlisting}
%%\end{minipage}
%%\hspace*{-5mm}
%%\begin{minipage}[t]{110mm}
%\def\namewidth{17mm}
%\preplisting
%\begin{lstlisting}[mathescape=true,style=nonumbers]
%    type// \lrec
%        //\com User-defined fields
%        //\wcnarrow{$k, v, w$}{key, value, weight (immutable)}
%        //\com Fields used by \llt/\sct\ algorithm
%        //\wcnarrow{$\info$}{pointer to \op}
%        //\wcnarrow{$marked$}{Boolean}
%\end{lstlisting}
%\end{minipage}
%
%\prepnewlisting
%\hrule
%\vspace{-2mm}
\begin{lstlisting}[mathescape=true]
 //\ins$(key, value)$
   //\com Associates $value$ with $key$ in the dictionary and returns the old associated value, or $\bot$ if none existed %Replaces $\langle key, old \rangle$ with $\langle key, value \rangle$ in the dictionary, and returns $old$, or $\bot$ if $key$ was not in the dictionary
      
   do
     $result := \tryins(key, value)$
   while $result = \fail$
   $\langle createdViolation, value \rangle := result$
   if $createdViolation$ then $\cleanup(key)$
   return $value$// \vspace{2mm}\hrule\vspace{2mm} %
    
 //\tryins$(key, value)$ 
   //\tline{\com Returns $\langle \true, \bot \rangle$ if $key$ was not in the dictionary %
               and inserting it caused a violation,} %
              {$\langle \false, \bot \rangle$ if $key$ was not in the dictionary %
               and inserting it did not cause a violation,} %
              {$\langle \false, oldValue \rangle$ if $\langle key, oldValue \rangle$ was in the dictionary, and %} %
              %{
              \fail\ if we should try again}

   //\com Search for $key$ in the tree
   $\langle -, -, p, l \rangle := \func{Search}(key)$//\label{ins-search-line}
      
   //\com Template iteration 0 (parent of leaf)
   $result := \llt(p)$
   if $result \in \{\fail, \finalized\}$ then return $\fail$ else $\langle p_L, p_R \rangle := result$
   if $l \notin \{p_L, p_R\}$ then return $\fail$// \medcom{\func{Conflict}: verify $p$ still points to $l$}

   //\com Template iteration 1 (leaf)
   $result := \llt(l)$
   if $result \in \{\fail, \finalized\}$ then return $\fail$ else $\langle l_L, l_R \rangle := result$

   //\com Computing \func{SCX-Arguments} from locally stored values (and immutable fields)
   $V := \langle p, l \rangle$ //\label{ins-V}
   $R := \langle l \rangle$ //\label{ins-R}
   $fld := (l = p_L)\ ?\ \&p.\mbox{\textit{right}} : \&p.\mbox{\textit{left}}$ //\label{ins-fld}
   $newL :=$ //new leaf with weight 1, key $key$ and value $value$ //\label{ins-newL}
   if $l.k = key$ then //\label{ins-check-key-already-present}
     $oldValue := l.v$
     $new := newL$ //\label{ins-new-newL}
   else
     $oldValue := \bot$
     if $l \mbox{ is a sentinel node}$ then $w := 1$ else $w := l.w - 1$
     if $key < l.k$ then
       $new :=$ //new node with weight $w$, key $l.k$, value \nil, and children $newL$ and $l$ \label{ins-new-internal1}
     else
       $new :=$ //new node with weight $w$, key $key$, value \nil, and children $l$ and $newL$ \label{ins-new-internal2}

   if $\sct(V, R, fld, new)$ then return $\langle (new.w = p.w = 0), oldValue \rangle$
   else return $\fail$
\end{lstlisting}
\end{framed}
	\caption{Pseudocode for \func{Insert} and \tryins .} % (which follows the tree update template).}
	\label{code-chromatic-ins}
\end{figure}

\begin{figure}[p]
\vspace{-5mm}
\begin{framed}
\preplisting
\begin{lstlisting}[mathescape=true]
 //\del$(key)$ 
   //\com Deletes $key$ and returns its associated value, or returns $\bot$ if $key$ was not in the dictionary
   do
     $result := \trydel(key)$
   while $result = \fail$
   $\langle createdViolation, value \rangle := result$
   if $createdViolation$ then $\cleanup(key)$
   return $value$// \vspace{2mm}\hrule\vspace{2mm} %
    
 //\trydel$(key)$
   //\qline{\com Returns $\langle \false, \bot \rangle$ if $key$ was not in the dictionary,} %
              {$\langle \false, oldValue \rangle$ if $\langle key, oldValue \rangle$ was in the dictionary and deleting it did not create a violation,} %
              {$\langle \true, oldValue \rangle$ if $\langle key, oldValue \rangle$ was in the dictionary and deleting it created a violation, and} %
              {\fail\ if we should try again}

   //\com Search for $key$ in the tree
   $\langle -, gp, p, l \rangle := \func{Search}(key)$//\label{del-search-line}
      
   //\com \func{UpdateNotNeeded}: check if tree is empty %(\func{Condition} returns false and \func{SCX-Arguments} returns \nil)

   if $gp = \nil$ then return $\langle \false, \nil \rangle$//\label{del-no-gp}
   
   //\com \func{UpdateNotNeeded}: check if $key$ is not in the dictionary %(\func{Condition} returns false and \func{SCX-Arguments} returns \nil)

   if $l.k \neq key$ then return $\langle \false, \bot \rangle$ //\label{del-notin}
   
   //\com Template iteration 0 (grandparent of leaf)
   $result := \llt(gp)$
   if $result \in \{\fail, \finalized\}$ then return $\fail$ else $\langle gp_{L}, gp_{R} \rangle := result$
   if $p \notin \{gp_{L}, gp_{R}\}$ then return $\fail$// \medcom{\func{Conflict}: verify $gp$ still points to $p$}\label{code-chromatic-trydel-conflict1}

   //\com Template iteration 1 (parent of leaf)
   $result := \llt(p)$
   if $result \in \{\fail, \finalized\}$ then return $\fail$ else $\langle p_{L}, p_{R} \rangle := result$
   if $l \notin \{p_{L}, p_{R}\}$ then return $\fail$// \medcom{\func{Conflict}: verify $p$ still points to $l$}\label{code-chromatic-trydel-conflict2}
      
   //\com Template iteration 2 (leaf)
   $result := \llt(l)$
   if $result \in \{\fail, \finalized\}$ then return $\fail$ else $\langle l_{L}, l_{R} \rangle := result$
   $s := (key < p.k)\ ?\ p_{R} : p_{L}$//\label{del-getsibling}
      
   //\com Template iteration 3 (sibling of leaf)
   $result := \llt(s)$
   if $result \in \{\fail, \finalized\}$ then return $\fail$ else $\langle s_{L}, s_{R} \rangle := result$
    
   //\com Computing \func{\sct-Arguments} from locally stored values (and immutable fields)
   $w := (p.k = \infty$ or $gp.k = \infty)\ ?\ 1 : p.w + s.w$//\label{del-weight}
   $new :=$// new node with weight $w$, key $s.k$, value $s.v$, and children $s_L$ and $s_R$\label{del-create-new}
   $V := (key < p.k)\ ?\ \langle gp, p, l, s \rangle : \langle gp, p, s, l \rangle$
   $R := (key < p.k)\ ?\ \langle p, l, s \rangle : \langle p, s, l \rangle$
   $fld := (key < gp.k)\ ?\ \&gp.\mbox{\textit{left}} : \&gp.\mbox{\textit{right}}$
      
   if $\sct(V, R, fld, new)$ then return $\langle (w > 1), l.v \rangle$
   else return $\fail$
\end{lstlisting}
\end{framed}
\vspace*{-5mm}
	\caption{Code for \del\ and \trydel .}
	\label{code-chromatic-del}
\end{figure}

\subsection{Detailed description of insertion}

\func{Insert}$(key, value)$ invokes \tryins\ to search for a leaf containing $key$ and perform the localized update that actually associates $value$ with $key$.
In Figure~\ref{code-dotreeupdate}, the template is presented using a loop, however \tryins\ requires only two iterations, so we unroll the loop.
To make this transformation as clear as possible, we still conceptually organize the steps of \tryins\ into iterations.

\tryins\ begins by invoking \func{Search}$(key)$ to find the leaf $l$ on the search path to $key$ and its parent $p$.
In terms of the template, \func{Search} represents the \func{SearchPhase} procedure, $p$ is $n_0$, and $p$ and $l$ are in the part $m$ of the tree returned by \func{SearchPhase}$(key)$.
(Specifically, $m$ contains an edge from $p$ to $l$, and the keys of $p$ and $l$.)

In iteration 0 (where $i=0$), \tryins\ performs \llt$(p)$.
Then, it uses the result of the \llt$(p)$ to verify that $p$ still points to $l$.
This verification step is conceptually part of the \func{Conflict} procedure.
It is necessary because $p$ might be changed between the \func{Search} and \llt$(p)$ so that it no longer points to $l$.
If $p$ no longer points to $l$ (so it no longer matches $m$), then \tryins\ returns \fail , which indicates that the \ins\ should be retried.
In terms of the template, this corresponds to an invocation of \func{Conflict} returning \true.
As in the template, \tryins\ returns \fail\ if any of its invocations of \llt\ return \fail\ or \finalized .
In iteration 1, \tryins\ performs \llt$(l)$.
This is the last iteration (which corresponds to \func{Condition} returning \true\ in the template).

\tryins\ then computes \func{\sct-Arguments} over the next few lines.
\tryins\ uses locally stored values to construct the sequences $V$ and $R$ that it will use for its \sct, ordering their elements according to a breadth-first traversal, in order to satisfy PC\ref{con-V-sequences-ordered-consistently}.
Line~\ref{ins-fld} determines which child pointer $fld$ of $p$ should be changed by the insertion using the result of its \llt$(p)$.
Line~\ref{ins-newL} creates a new leaf $newL$ with weight 1, and the new key and value.
Line~\ref{ins-check-key-already-present} then checks whether $l$ contains $key$.
If so, \tryins\ takes $newL$ to be the \sct\ argument $new$ at line~\ref{ins-new-newL} (so that $newL$ will be inserted in place of $l$). %the single node $new$ that will be inserted in place of $l$.
Otherwise, \tryins\ creates an internal node $new$ (at line~\ref{ins-new-internal1} or~\ref{ins-new-internal2}) with the appropriate weight, routing key, the value \nil, and the children $newL$ and $l$, ordered according to their keys.

Finally, \tryins\ invokes \sct$(V, R, fld, new)$.
If the \sct\ succeeds, then \tryins\ returns the result of the expression $(new.w = p.w = 0)$, which indicates whether \tryins\ created a red-red violation at $new$, and $oldValue$, which contains the value that was previously associated with $key$ (or $\bot$ if key was not in the tree).
Otherwise, \tryins\ returns \fail .

An invocation of \ins\ performs at most one successful invocation of \tryins , and succeeds only if it performs a successful \tryins .
Unsuccessful invocations of \tryins\ cannot create violations.
A successful invocation of \tryins\ can create at most one red-red violation at $new$, and cannot create any other violations.
Thus, the expression $(new.w = p.w = 0)$ returned by \tryins\ indicates whether \tryins\ created any violation.
If \tryins\ created a violation, then \ins\ invokes \cleanup$(key)$ (which is described in more detail, below) to fix it before \ins\ returns.

A simple inspection of the pseudocode suffices to prove that \tryins\ follows the template, and the arguments to \sct\ satisfy the postconditions of \sct-\func{Arguments}. % postconditions PC1 to PC\ref{con-new-nodes}, and that invocations of \tryins\ follow the template.

\subsection{Detailed description of deletion}

\del($key$) invokes \trydel\ to search for a leaf containing $key$ and perform the localized update that actually deletes $key$ and its associated value.
As in the description of insertion, we unroll the template loop, and organize the steps of \trydel\ into iterations.

By inspection of the pseudocode, one can see that \trydel\ follows the template.
\trydel\ begins by invoking \func{Search}$(key)$ to find the leaf $l$ on the search path to $key$ and its parent $p$ and grandparent $gp$.
In terms of the template, \func{Search} represents the \func{SearchPhase} procedure, $gp$ is $n_0$, and $gp, p$ and $l$ are in the part $m$ of the tree returned by \func{SearchPhase}$(key)$.
(Specifically, $m$ contains an edge from $gp$ to $p$, an edge from $p$ to $l$, and the keys of $gp, p$ and $l$.)
If the grandparent does not exist, then the tree is empty (and it looks like Figure~\ref{fig-treetop}(a)), so \trydel\ returns successfully at line~\ref{del-no-gp}.
In terms of the template, this corresponds to an invocation of \func{UpdateNotNeeded}$(m)$ that returns \true.
Similarly, if $l$ does not contain $key$, we can show there is a time during the \func{Search} when the tree does not contain $key$, and \trydel\ returns successfully at line~\ref{del-notin}.
Note that, if \trydel\ does \textit{not} return at line~\ref{del-notin}, then $l$ contains $key$.

%Note that invocations of \trydel\ that return at line~\ref{del-no-gp} do not follow the template, and their correctness is argued separately.
%other invocations of \trydel\ follow the template.

In iteration 0 (where $i=0$), \trydel\ performs \llt$(gp)$, and uses the result to verify that $gp$ still points to $p$ (as in $m$).
If $gp$ no longer points to $p$, then \trydel\ returns \fail .
In terms of the template, this corresponds to an invocation of \func{Conflict} returning \true.
As in the template, \trydel\ returns \fail\ if any of its invocations of \llt\ return \fail\ or \finalized .
In iteration 1, \trydel\ performs \llt$(p)$, and uses the result to verify that $p$ still points to $l$.
If $p$ no longer points to $l$, then \trydel\ returns \fail .
In iteration 2, \trydel\ performs \llt$(l)$.
At line~\ref{del-getsibling}, \trydel\ uses the result of its previous \llt$(p)$ to obtain a pointer to the sibling, $s$, of the leaf to be deleted.
In iteration 3, \trydel\ performs \llt$(s)$.
This is the last iteration (which corresponds to \func{Condition} returning \true\ in the template).

\trydel\ then computes \func{\sct-Arguments} over the next few lines.
Line~\ref{del-weight} computes the weight of the node $new$ in the depiction of \del\ in Figure~\ref{fig-treetop}, ensuring that it has weight one if it is taking the place of a sentinel or $root$.
Line~\ref{del-create-new} creates $new$, reading the key, and value directly from $s$ (since they are immutable) and the child pointers from the result of the \llt$(s)$. % (since they are mutable).
Next, \trydel\ uses locally stored values (and the immutable key of $p$) to construct the sequences $V$ and $R$ that it will use for its \sct, ordering their elements according to a breadth-first traversal, in order to satisfy PC\ref{con-V-sequences-ordered-consistently}.
Finally, \trydel\ uses $key$ and $gp.k$ to decide which child pointer $fld$ of $gp$ should be changed by the deletion, and invokes \sct$(V, R, fld, new)$ to perform the modification.
If the \sct\ succeeds, then \trydel\ returns the immutable value stored in node $l$, and the result of the expression $w > 1$, which indicates whether \trydel\ created an overweight violation at the new node.
%\trydel\ can create an overweight violation (but not a red-red violation), and $w > 1$ holds if and only if it did.
If the \sct\ fails, then \trydel\ returns \fail .
%
%\after{The template says that we read immutable fields of a \rec\ after  we read its mutable fields, but here we are reading them in the opposite order.
%Let's try to fix this for the camera ready copy.}
%
%A simple inspection of the pseudocode suffices to prove that \func{\sct-Arguments} satisfies postconditions PC1 to PC\ref{con-new-nodes}, and that \trydel\ follows the template.
Inspection of the pseudocode suffices to prove that the arguments to \sct\ satisfy postconditions of \sct-\func{Arguments}.

An invocation of \del\ performs at most one successful invocation of \trydel , and succeeds only if it performs a successful \trydel .
Unsuccessful invocations of \trydel\ cannot create violations.
A successful invocation of \trydel\ can create overweight violations at $new$, but cannot create any other violations.
Thus, the expression $w > 1$ returned by \trydel\ indicates whether \trydel\ created any violation.
If \trydel\ creates a new violation, then \del\ invokes \cleanup$(key)$ to fix it.

\subsection{The rebalancing algorithm}
%\section{The rebalancing algorithm} \label{sec-chromatic-rebalancing-alg}

%\trevor{This paragraph is enormous, but I don't know if/how we should break it.}
Since rebalancing is decoupled from updating,
there must be
%we must design
a scheme that determines 
when processes should perform rebalancing steps to eliminate violations.
%\trevor{A reviewer was confused about why we modified the rebalancing steps in \cite{Boyar97amortizationresults}, and believed that our modifications were necessary for us to be able to implement the chromatic tree using our template.  Here, I'm trying to explain that \cite{Boyar97amortizationresults} suggests using a queue of problems to keep track of violations that need to be fixed, and that our algorithm avoid the use of a problem queue, but needs some extra knowledge of how violations will move around in the tree, so that the algorithm can keep track of violations without explicitly storing information about their movements.}
In \cite{Boyar97amortizationresults}, the authors suggest maintaining one or more \textit{problem queues} which contain pointers to nodes that contain violations, and dedicating one or more \textit{rebalancing processes} to simply perform rebalancing steps as quickly as possible.
This approach does not yield a bound on the height of the tree, since rebalancing may lag behind insertions and deletions.
It is possible to obtain a height bound with a different queue based scheme, but we present a way to bound the tree's height without the (significant) overhead of maintaining any auxiliary data structures.
%Note that this effort is orthogonal to the task of applying our template to implement chromatic trees.
The linchpin of our method is the following claim concerning violations, which is satisfied by each rebalancing step in Figure~\ref{fig-chromatic-rotations}.
%invariant.
%\begin{compactenum}[\hspace{6mm}\bfseries {I}1:]
%\begin{compactenum}[\hspace{3mm}\bfseries {INV}:]
\begin{compactenum}[\hspace{3mm}\bfseries {VIOL}:]
%\begin{compactenum}[\hspace{3mm}\bfseries {}]
 \item If a violation is on the search path to $key$ before a rebalancing step, then the violation is still on the search path to $key$ after the rebalancing step, or it has been eliminated.
\end{compactenum}
%\label{inv-moving-viol}
%\end{compactenum}

The rebalancing steps shown in Figure~\ref{fig-chromatic-rotations} are actually a slight modification of those in \cite{Boyar97amortizationresults}.
Specifically, in the original rebalancing steps, \func{W1}, \func{W2}, \func{W3} and \func{W4} (and their symmetric versions) require the node labeled $\ux$ to have non-negative weight.
In our rebalancing steps, we require $\ux$ to have positive weight.
The reason for this restriction is as follows.

While studying the original rebalancing steps, we realized that most of them satisfied VIOL.
However, if \func{W1}, \func{W2}, \func{W3} or \func{W4} (or a symmetric version) were performed when $\ux$ had weight 0, then VIOL could be violated.
In each of these cases, we found that another rebalancing step which satisfies VIOL could be applied instead.
%\trevor{give example, here, of case where VIOL is violated, and the other rebalancing step we can perform, instead?}
(We argue, below, that a rebalancing step can always be applied whenever there is a violation in the tree.)
Thus, under this restriction, all rebalancing steps satisfy VIOL (a fact that is easily verified).
Consequently, each violation created by \ins$(key, value)$ or \del$(key)$ stays on the search path to $key$ until it is eliminated.

%While studying the rebalancing steps in \cite{Boyar97amortizationresults}, we realized that most of them satisfy VIOL.
%Specifically, VIOL is violated only if \func{W1}, \func{W2}, \func{W3} or \func{W4} (or a symmetric rebalancing step) is performed when the node labeled $\ux$ has weight 0.
%Furthermore, any time a rebalancing step would violate VIOL, another rebalancing step that satisfies VIOL can be applied instead.
%(For instance, 
%Hence, we always choose to perform rebalancing steps such that each violation created by an \ins$(key)$ or \del$(key)$ stays on the search path to $key$ until it is eliminated.
%%The rebalancing steps, which are shown in Figure~\ref{fig-chromatic-rotations}, are a slight modification of those in \cite{Boyar97amortizationresults}.\footnote{Specifically, we do not allow \func{W1}, \func{W2}, \func{W3} or \func{W4} to be applied when the node labeled $\ux$ has weight 0.  Under this restriction, this set of rebalancing steps has the desirable property that when a violation moves,  it remains on the search path to the key whose insertion or deletion originally caused the violation.  It is easy to verify that an alternative rebalancing step can always be performed when $\ux.w = 0$, so this modification does not affect the chromatic tree's convergence to a RBT.}

\begin{figure}[tb]
\begin{framed}
\preplisting
\begin{lstlisting}[mathescape=true]
 //\cleanup$(key)$
   //\com Ensures the violation created by an \ins\ or \del\ of $key$ gets eliminated
   while $\true$
     $ggp := \nil$;  $gp:=\nil$;  $p:=\nil$;  $l := entry$ //\medcom Save four last nodes traversed\label{cleanup-start}
     while //\true 
       if //$l$ is a leaf then \textbf{return} \medcom Arrived at leaf without finding a violation\label{cleanup-terminate}
       if $key < l.key$ then {$ggp := gp$;  $gp := p$;  $p := l$; $l:=l.left$} //\label{move-l-left}
       else {$ggp := gp$;  $gp := p$;  $p := l$;  $l:=l.right$} //\label{move-l-right}
       if $l.w > 1$ or ($p.w = 0$ and $l.w = 0$) then //\medcom Found a violation\label{find-violation}
         $\tryrebalance(ggp,gp,p,l)$ //\medcom Try to perform a rebalancing step
         exit loop //\medcom Go back to $entry$ and traverse again\label{cleanup-end}
\end{lstlisting}
\end{framed}
	\caption{Pseudocode for \cleanup.}
	\label{code-chromatic-cleanup}
\end{figure}

In our implementation, each \func{Insert} or \func{Delete} that increases the number of violations in the tree cleans up after itself by invoking \func{Cleanup}$(key)$.
Pseudocode for \func{Cleanup} appears in Figure~\ref{code-chromatic-cleanup}.
\func{Cleanup}$(key)$ behaves like \func{Search}($key$) until it finds the first node $l$ on the search path where a violation occurs.
Then, \func{Cleanup}$(key)$ attempts to eliminate or move the violation at $l$ by %calling another procedure \tryrebalance, which applies one localized rebalancing step at $n_3$, following the tree update template.
invoking another procedure \tryrebalance\, which applies one localized rebalancing step at $l$, following the tree update template.
(\tryrebalance\ is described in more detail, below.)
%(\tryrebalance\ is similar to \del, and pseudocode is omitted, due to lack of space.)
\func{Cleanup}$(key)$ repeats these actions, searching for $key$ and invoking \tryrebalance\ to perform a rebalancing step, until the search goes all the way to a leaf without finding a violation.

In order to prove that each \ins\ or \del\ cleans up after itself, we must prove that while an invocation of \func{Cleanup}$(key)$ searches for $key$ by reading child pointers, it does not somehow miss the violation it is responsible for eliminating, even if a concurrent rebalancing step moves the violation upward in the tree, above where \func{Cleanup} is currently searching.
To see why this is true, consider any rebalancing step that occurs while \cleanup\ is searching.
The rebalancing step is implemented using the tree update template, and looks like Figure~\ref{fig-replace-subtree}.
It takes effect at the point it changes a child pointer $fld$ of some node $parent$ from a node $old$ to a node $new$.
If \func{Cleanup} reads $fld$ while searching, we argue that it does not matter whether $fld$ contains $old$ or $new$.
First, suppose the violation is at a node that is removed from the tree by the rebalancing step, or a child of such a node.
If the search passes through $old$, it will definitely reach the violation, since nodes do not change after they are removed from the tree.
If the search passes through $new$, VIOL implies that the rebalancing step either eliminated the violation, or moved it to a new node on the search path through $new$.
Finally, if the violation is further down in the tree, below the section modified by the concurrent rebalancing step, a search through either $old$ or $new$ will reach it.

\begin{figure}[tbp]
\centering
\includegraphics[scale=0.42]{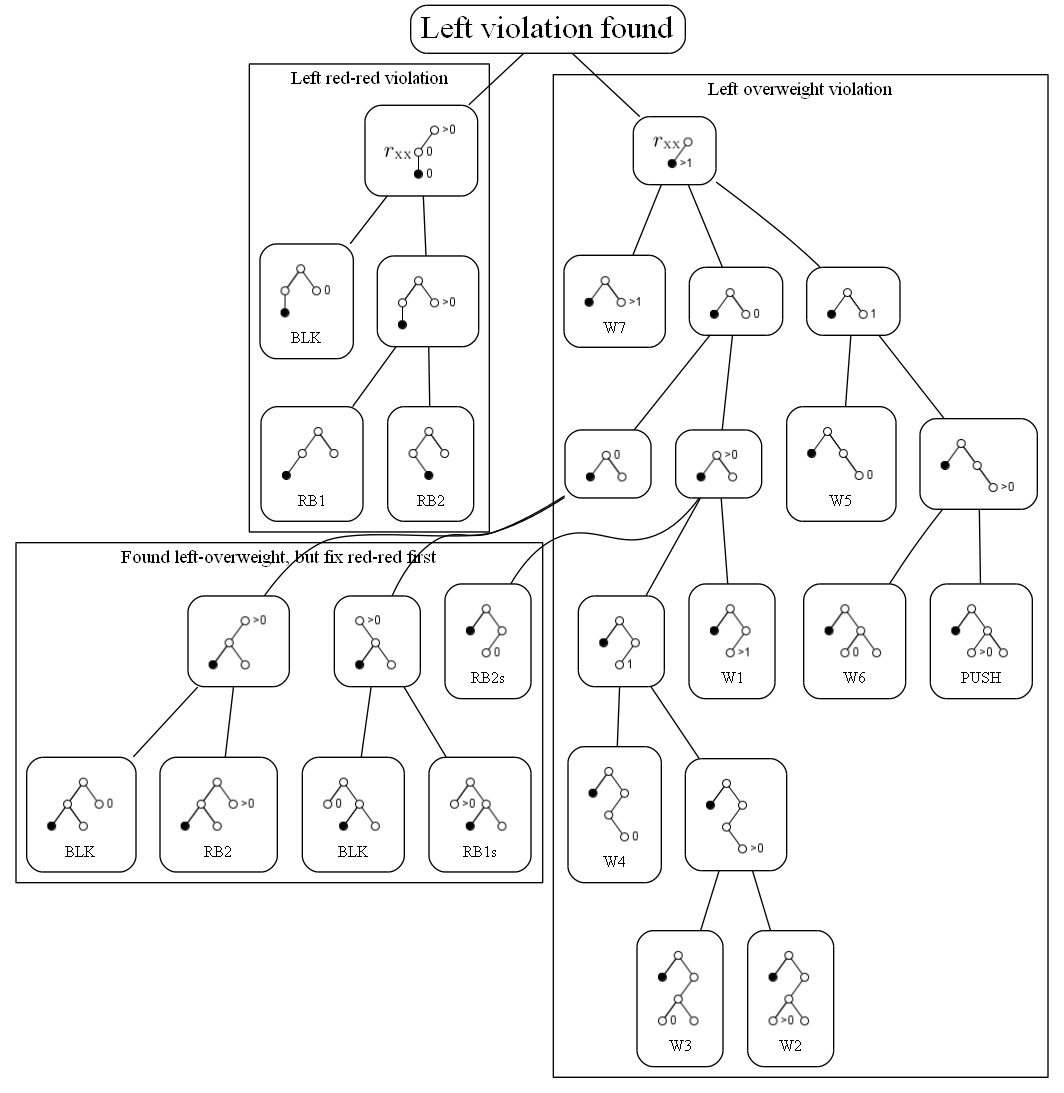}
\caption{Decision tree used by the algorithm to determine which rebalancing operation to apply when a violation is encountered at a node (shaded black). % that is a left child.  \eric{Confusing:  There are some cases in upper-left box where black dot is a right child of its parent}
The corresponding diagram to cover right violations can be obtained by: horizontally flipping each small tree diagram, changing each rebalancing step \func{DoX} to its symmetric version \func{DoXs}, and changing each symmetric version back to its original version.}
\label{fig-decision-tree}
\end{figure}

%cleanup (modification to rebalancing steps / VIOL, cleanup doesn't miss violations)
%
%tryrebalance (can always perform a rebalancing step if there is a violation)
%
%example rebalancing step

\subsection{Deciding which rebalancing step to perform}

\begin{figure}[tbp]
\begin{framed}
\preplisting
\begin{lstlisting}[mathescape=true]
 //\tryrebalance$(ggp, gp, p, l)$
 //\com Precondition: $l.w > 1$ or $l.w = p.w = 0 \neq gp.w$%, } %
         {%     $l$ has been a child of $p$, $p$ has been a child of $gp$, and %
          %     $gp$ has been a child of $ggp$}
             
   $r := ggp$
   if $(result := \llt(r)) \in \{\fail, \finalized\}$ then return else $\langle \rl, \rr \rangle := result$ //\label{tryrebalance-llt-ggp}
   if $gp \notin\{\rl, \rr\}$ then return       //\medcom{\func{Conflict}: verify $ggp$ still points to $gp$}

   $\rx := gp$
   if $(result := \llt(\rx)) \in \{\fail, \finalized\}$ then return else $\langle \rxl, \rxr \rangle := result$ //\label{tryrebalance-llt-gp}
   if $p \notin\{\rxl, \rxr\}$ then return      //\medcom{\func{Conflict}: verify $gp$ still points to $p$}

   $\rxx := p$
   if $(result := \llt(\rxx)) \in \{\fail, \finalized\}$ then return else $\langle \rxxl, \rxxr \rangle := result$ //\label{tryrebalance-llt-p}
   if $l \notin\{\rxxl, \rxxr\}$ then return    //\medcom{\func{Conflict}: verify $p$ still points to $l$}
   
   $\rxxx := l$
   if $\rxxx.w > 1$ then            //\medcom Overweight violation at $l$ \label{tryrebalance-if-overweight}
     if $\rxxx = \rxxl$ then        //\medcom Left overweight violation ($l$ is a left child) \label{tryrebalance-if-left-overweight}
       if $(result := \llt(\rxxl)) \in \{\fail, \finalized\}$ then return else $\langle \rxxll, \rxxlr \rangle := result$
       //$\func{OverweightLeft}($all $r$ variables$)$
     else //\com $\rxxx = \rxxr$      \medcom Right overweight violation ($l$ is a right child)
       if $(result := \llt(\rxxr)) \in \{\fail, \finalized\}$ then return else $\langle \rxxrl, \rxxrr \rangle := result$
       //$\func{OverweightRight}($all $r$ variables$)$
   else                             //\medcom Red-red violation at $l$ \label{tryrebalance-if-redred}
     if $\rxx = \rxl$ then          //\medcom Left red-red violation ($p$ is a left child) \label{tryrebalance-if-left-redred}
       if $\rxr.w = 0$ then
         if $(result := \llt(\rxr)) \in \{\fail, \finalized\}$ then return else $\langle \rxrl, \rxrr \rangle := result$
         //$\func{DoBLK}(\langle r,\rx,\rxx,\rxr \rangle,$ all $r$ variables$)$
       else if $\rxxx = \rxxl$ then
         //$\func{DoRB1}(\langle r,\rx,\rxx\rangle,$ all $r$ variables$)$
       else //\com $\rxxx = \rxxr$
         if $(result := \llt(\rxxr)) \in \{\fail, \finalized\}$ then return else $\langle \rxxrl, \rxxrr \rangle := result$
         //$\func{DoRB2}(\langle r,\rx,\rxx,\rxxr \rangle,$ all $r$ variables$)$
     else //\com $\rxx = \rxr $       \medcom Right red-red violation ($p$ is a right child)
       if $\rxl.w = 0$ then
         if $(result := \llt(\rxl)) \in \{\fail, \finalized\}$ then return else $\langle \rxll, \rxlr \rangle := result$
         //$\func{DoBLK}(\langle r,\rx,\rxl,\rxx \rangle,$ all $r$ variables$)$
       else if $\rxxx = \rxxr$ then
         //$\func{DoRB1s}(\langle r,\rx,\rxx \rangle,$ all $r$ variables$)$
       else //\com $\rxxx = \rxxl$
         if $(result := \llt(\rxxl)) \in \{\fail, \finalized\}$ then return else $\langle \rxxll, \rxxlr \rangle := result$
         //$\func{DoRB2s}(\langle r,\rx,\rxx,\rxxl \rangle,$ all $r$ variables$)$
\end{lstlisting}
\end{framed}
	\caption{Pseudocode for \tryrebalance .
	% if each invocation of a subroutine is replaced by the code of that subroutine (and if this is recursively applied to the invocations made by \func{OverweightLeft} and \func{OverweightRight}).
	}
	\label{code-chromatic-tryrebalance}
\end{figure}

\begin{figure}[tbp]
\begin{framed}
%   //%\dline{\com Precondition: $l \in \{p_L,p_R\rangle, p = gp_L, gp \in \{ggp_L,ggp_R\}$ and,}
%            {for each $x \in \{l,p,gp,ggp\}$, $x_L$ ($x_R$) was %
%             read from the left (right) pointer of $x$}
%\begin{minipage}{\textwidth}
\preplisting
\begin{lstlisting}[mathescape=true]
 //\func{OverweightLeft}$(r,\rx,\rxx,\rxxl,\rl,\rr,\rxl,\rxr,\rxxr)$
   if $\rxxr.w = 0 $ then
     if $\rxx.w = 0 $ then
       if $\rxx = \rxl$ then
         if $\rxr.w = 0$ then
           if $(result := \llt(\rxr)) \in \{\fail, \finalized\}$ then return else $\langle \rxrl, \rxrr \rangle := result$
           //$\func{DoBLK}(\langle r,\rx,\rxx,\rxr\rangle,$ all $r$ variables$)$
         else //\com $\rxr.w >0 $
           if $(result := \llt(\rxxr)) \in \{\fail, \finalized\}$ then return else $\langle \rxxrl, \rxxrr \rangle := result$
           //$\func{DoRB2}(\langle r,\rx,\rxx,\rxxr\rangle,$ all $r$ variables$)$
       else //\com $\rxx = \rxr $
         if $\rxl.w = 0$ then
           if $(result := \llt(\rxl)) \in \{\fail, \finalized\}$ then return else $\langle \rxll, \rxlr \rangle := result$
           //$\func{DoBLK}(\langle r,\rx,\rxl,\rxx\rangle,$ all $r$ variables$)$
         else //\com $\rxx = \rxl$
           //$\func{DoRB1s}(\langle r,\rx,\rxx\rangle,$ all $r$ variables$)$
     else //\com $\rxx.w >0 $
       if $(result := \llt(\rxxr)) \in \{\fail, \finalized\}$ then return else $\langle \rxxrl, \rxxrr \rangle := result$
       if $(result := \llt(\rxxrl)) \in \{\fail, \finalized\}$ then return else $\langle \rxxrll, \rxxrlr \rangle := result$
       if $\rxxrl.w > 1$ then
         //$\func{DoW1}(\langle\rx,\rxx,\rxxl,\rxxr,\rxxrl\rangle,$ result, all $r$ variables$)$
       else if $\rxxrl.w = 0$ then
         //$\func{DoRB2s}(\langle\rx,\rxx,\rxxr,\rxxrl\rangle,$ all $r$ variables$)$
       else //\com $\rxxrl.w = 1$
         if $\rxxrlr = \nil$ then return//\label{overweightleft-check-nil-rxxrlr} \hfill \com Special case: a node we performed \llt\ on was modified $ $
         if $\rxxrlr.w = 0$ then
           if $(res := \llt(\rxxrlr)) \in \{\fail, \finalized\}$ then return else $\langle \rxxrlrl, \rxxrlrr \rangle := res$//\label{overweightleft-bad-llt-rxxrlr}
           //$\func{DoW4}(\langle\rx,\rxx,\rxxl,\rxxr,\rxxrl,\rxxrlr\rangle,$ all $r$ variables$)$
         else //\com $\rxxrlr.w > 0$
           if $\rxxrll.w = 0$ then
             if $(res := \llt(\rxxrll)) \in \{\fail, \finalized\}$ then return else $\langle \rxxrlll, \rxxrllr \rangle := res$//\label{overweightleft-bad-llt-rxxrll}
             //$\func{DoW3}(\langle\rx,\rxx,\rxxl,\rxxr,\rxxrl,\rxxrll\rangle,$ all $r$ variables$)$
           else //\com $\rxxrll.w > 0$
             //$\func{DoW2}(\langle\rx,\rxx,\rxxl,\rxxr,\rxxrl\rangle,$ all $r$ variables$)$
   else if $\rxxr.w = 1 $ then
     if $(result := \llt(\rxxr)) \in \{\fail, \finalized\}$ then return else $\langle \rxxrl, \rxxrr \rangle := result$
     if $\rxxrr = \nil$ then return//\label{overweightleft-check-nil-rxxrr} \hfill \com Special case: a node we performed \llt\ on was modified $ $
     if $\rxxrr.w = 0 $ then
       if $(result := \llt(\rxxrr)) \in \{\fail, \finalized\}$ then return else $\langle \rxxrrl, \rxxrrr \rangle := result$//\label{overweightleft-bad-llt-rxxrr}
       //$\func{DoW5}(\langle\rx,\rxx,\rxxl,\rxxr,\rxxrr\rangle,$ all $r$ variables$)$
     else if $\rxxrl.w = 0 $ then
       if $(result := \llt(\rxxrl)) \in \{\fail, \finalized\}$ then return else $\langle \rxxrll, \rxxrlr \rangle := result$//\label{overweightleft-bad-llt-rxxrl}
       //$\func{DoW6}(\langle\rx,\rxx,\rxxl,\rxxr,\rxxrl\rangle,$ all $r$ variables$)$
     else //\com $\rxxr.w > 0$ and $\rxxrl.w > 0$
       //$\func{DoPush}(\langle\rx,\rxx,\rxxl,\rxxr\rangle,$ all $r$ variables$)$
   else //\com $\rxxr.w > 1$
     if $(result := \llt(\rxxr)) \in \{\fail, \finalized\}$ then return else $\langle \rxxrl, \rxxrr \rangle := result$
     //$\func{DoW7}(\langle\rx,\rxx,\rxxl,\rxxr\rangle,$ all $r$ variables$)$ \hrule %

 //\func{OverweightRight}$(r,\rx,\rxx,\rxxr,\rl,\rr,\rxl,\rxr,\rxxl)$
 //\com Obtained from \func{OverweightLeft} by flipping each $R$ in the subscript of an $r$ variable to an $L$ (and vice versa), and by flipping each rebalancing step \func{DoX} to its symmetric version \func{DoXs} (and vice versa).
\end{lstlisting}
%\end{minipage}
\end{framed}
    \vspace{-5mm}
	\caption{Pseudocode for \func{OverweightLeft} (and effectively also \func{OverweightRight}).}
	\label{code-chromatic-overweight-left}
\end{figure}

Whenever \cleanup\ finds a violation, it invokes \tryrebalance , and provides it with pointers to the node $l$ where the violation was found, along with the node's parent $p$, grandparent $gp$ and great grandparent $ggp$.
In terms of the template, the search in \cleanup\ corresponds to the \func{SearchPhase} procedure, $ggp$ is $n_0$, and $ggp$, $gp$, $p$ and $l$ are in the part $m$ of the tree returned by \func{SearchPhase}.
(Specifically, $m$ contains an edge from $ggp$ to $gp$, an edge from $gp$ to $p$, an edge from $p$ to $l$, and the keys of $ggp, gp, p$ and $l$.)
In the first three template iterations, \tryrebalance\ performs \llt\ on $ggp$, $gp$ and $p$ (at lines~\ref{tryrebalance-llt-ggp},~\ref{tryrebalance-llt-gp} and~\ref{tryrebalance-llt-p}), and verifies that $ggp$ still points to $gp$, $gp$ still points to $p$, and $p$ still points to $l$.
In terms of the template, these verification steps are part of the \func{Conflict} procedure.
If any of these verification steps fails, then a node no longer matches $m$ (so \func{Conflict} conceptually returns \true), and \tryrebalance\ returns \fail.
As in the template, if an \llt\ returns \fail\ or \finalized, then \tryrebalance\ returns \fail.
If \tryrebalance\ returns \fail, then \cleanup\ will restart its search for violations from $entry$.
In the remaining template iterations, \tryrebalance\ traverses the decision tree in Figure~\ref{fig-decision-tree} to decide which rebalancing step should be performed.
Finally, \tryrebalance\ invokes a procedure (e.g., \func{DoBLK} or \func{DoRB2}) to perform the specific rebalancing step according to Figure~\ref{fig-chromatic-rotations}.

We now describe the implementation of the decision tree.
Pseudocode for \tryrebalance , including the implementation of the decision tree, appears in Figure~\ref{code-chromatic-tryrebalance} and~\ref{code-chromatic-overweight-left}.
The first task in \tryrebalance\ is to identify whether the violation found by \cleanup\ is a left or right overweight violation, or a left or right red-red violation.
A left (resp., right) overweight violation is an overweight violation at a node that is a left (right) child.
A left (resp., right) red-red violation is a red-red violation at a node whose \textit{parent} is a left (right) child.
\tryrebalance\ identifies the type of violation by first determining whether the violation is an overweight violation or a red-red violation at line~\ref{tryrebalance-if-overweight}, and then further determining whether it is a left or right violation at line~\ref{tryrebalance-if-left-overweight} (if it is an overweight violation) or line~\ref{tryrebalance-if-left-redred} (if it is a red-red violation).
This information is sufficient to determine where \tryrebalance\ should start in the decision tree shown in Figure~\ref{fig-decision-tree} (or in the symmetric, mirror-image version of this decision tree).

Once the violation type has been identified, \tryrebalance\ performs a sequence of \llt s, and uses their results (and the immutable fields of nodes) to decide which path to traverse in the decision tree.
More specifically, at each node of the decision tree, \tryrebalance\ decides which child to proceed to by looking at the weight of one node, as indicated in the child (in Figure~\ref{fig-decision-tree}).
(For clarity, we factorize the \textit{left overweight} and \textit{right overweight} parts of the decision tree traversal into two other functions, \func{OverweightLeft} and \func{OverweightRight}.)
The leaves of the decision tree are labeled by the rebalancing step to apply.
%Since every leaf is labeled by a rebalancing step, it is straightforward to verify that a rebalancing step can always be applied whenever \cleanup\ finds a violation.
Note that this decision tree is a component of the sequential chromatic tree algorithm that was left to the implementer in \cite{Boyar97amortizationresults}.

Whenever \tryrebalance\ follows a pointer to a node in the chromatic tree, either we must be able to argue that the pointer is not \nil, or \tryrebalance\ must explicitly check whether the pointer is \nil .
It is easy to prove that most pointers followed by \tryrebalance\ are non-\nil\ with the help of three simple invariants (which are easy to prove if rebalancing steps are atomic).
First, each node is created with two \nil\ child pointers or two non-\nil\ child pointers, and a child pointer does not change from a non-\nil\ value to \nil, or vice versa.
(Thus, if \tryrebalance\ has followed one child pointer of a node, then the other child pointer is non-\nil .)
Second, each node with weight zero has two non-\nil\ child pointers, and node weights never change.
Third, every node that has a red-red violation always has a parent, grandparent and great grandparent.
These invariants suffice to prove that only two explicit \nil\ checks are necessary, specifically, at lines~\ref{overweightleft-check-nil-rxxrlr} and~\ref{overweightleft-check-nil-rxxrr}.

In each of these cases, \tryrebalance\ returns without performing any rebalancing step, and \cleanup\ repeats its search for violations.
This creates two potential problems.
First, we would like to argue that a rebalancing step can always be applied, whenever the chromatic tree contains a violation, but in this case we return without performing a rebalancing step.
Second, we would like to argue that returning without performing a rebalancing step will not cause an infinite loop where \cleanup\ repeatedly finds the same violation, and \tryrebalance\ repeatedly returns without fixing it.

We prove both of the above claims for the case where \tryrebalance\ returns because of the \nil\ check at line~\ref{overweightleft-check-nil-rxxrr}.
(The proof for the other case is similar.)
Consider the node labeled \textit{W5} in Figure~\ref{fig-decision-tree} and its parent.
Moving from the parent to the node labeled \textit{W5} in the decision tree corresponds to following the right child pointer of the node $\rxxr$ (where $\rxxr$ is the right child of the node labeled $\rxx$).
We prove that, in a chromatic tree, the right child pointer of $\rxxr$ must be non-\nil .
Suppose, to obtain a contradiction, that it is \nil\ (i.e., $\rxxr$ is a leaf).
Recall that the sum of node weights is always the same on all paths from the root to a leaf in a chromatic tree, and all node weights are non-negative.
Let $SW(u)$ be the sum of weights from $root$ to a node $u$.
Since $\rxxr$ is a leaf, it follows that $SW(\rxxl) \le SW(\rxxr)$ (and, if $\rxxl$ is also a leaf, then $SW(\rxxl) = SW(\rxxr)$).
By definition, $SW(\rxxl) = SW(\rxx) + \rxxl.w$, and $SW(\rxxr) = SW(\rxx) + \rxxr.w$.
Furthermore, in this case, $\rxxr.w = 1$, so $SW(\rxxr) = SW(\rxx) + 1$.
However, $\rxxl$ is overweight (indicated with shading in Figure~\ref{fig-decision-tree}), so $\rxxl.w > 1$, which implies $SW(\rxxl) > SW(\rxxr)$, which is a contradiction.
Consequently, if \tryrebalance\ sees that the right child pointer of $\rxxr$ is \nil\ at line~\ref{overweightleft-check-nil-rxxrr}, then the view of the chromatic tree that it obtained from its invocations of \llt\ was not a consistent view of the chromatic tree.
In other words, another update must have concurrently modified the tree.
Thus, the configuration of nodes observed by \tryrebalance\ is transient, and, when \cleanup\ repeats its search for violations, it will eventually see a different configuration (possibly after helping the other update to complete).

%Showing that \tryrebalance\ follows the template (i.e., by defining the procedures in Figure~\ref{code-dotreeupdate}) is complicated by the fact that it must decide which of the chromatic tree's 22 rebalancing steps to perform.
%It is more convenient to unroll the loop that performs \llt s, and write \tryrebalance\ using conditional statements.
%A helpful technique is to consider each path through the conditional statements in the code, and check that the procedures \func{Condition}, \func{NextNode}, \func{\sct-Arguments} and \func{Result} can be defined to produce this single path.
%It is sufficient to show that this can be done for each path through the code, since it is always possible to use conditional statements to combine the procedures for each path into procedures that handle all paths.

\subsection{Implementing a rebalancing step}

\begin{figure}[tb]
\begin{minipage}{0.38\textwidth}
%\begin{framed}
\vspace{-1.6cm}
\hspace{-1.5cm}
\includegraphics[scale=1]{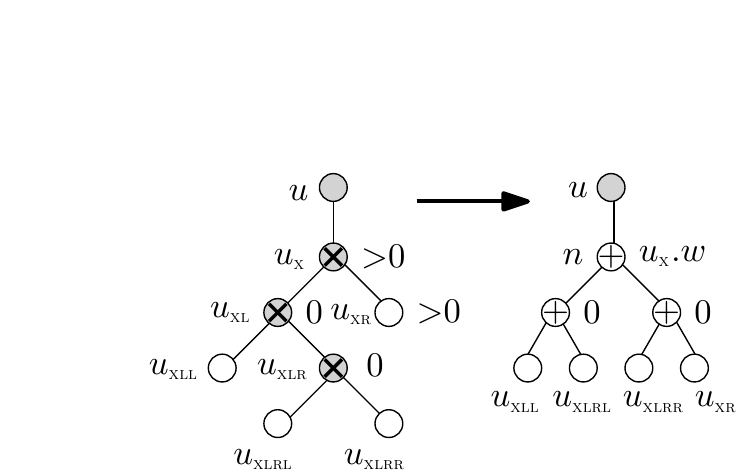}
%\end{framed}
\end{minipage}
\begin{minipage}{0.62\textwidth}
\preplistingnonumbers
\begin{framed}
\begin{lstlisting}[mathescape=true]
 //$\func{DoRB2}(u,\ux,\uxl,\uxr,\uxll,\uxlr,\uxlrl,\uxlrr)$
   //\com Create new nodes according to the right-hand diagram
   //create node $\nL$ with $k=\uxl.k$, $w=0$, {\it left} $ = \uxll$, $right=\uxlrl$
   //create node $\nR$ with $k=\ux.k$, $w=0$, {\it left} $=\uxlrr$, $right=\uxr$
   //create node $n$ with $k=\uxlr.k$, $w=\ux.w$, {\it left} $=\nL$, $right=\nR$
   //\com Perform the \sct\ to swing the child pointer of node $u$
   if $\ux = \ul$ then $ptr := \&u.left$ else $ptr := \&u.right$
   $\sct(\langle u,\ux,\uxx,\uxlr \rangle, \langle \ux,\uxl,\uxlr \rangle, ptr, n)$
\end{lstlisting}
\end{framed}
\end{minipage}
	\caption{
	Implementing rebalancing step \func{RB2}.
	Other rebalancing steps are handled similarly using the diagrams shown in Figure~\ref{fig-chromatic-rotations}.
	}
	\label{code-chromatic-dorb2}
\end{figure}

We now describe how the individual rebalancing steps in Figure~\ref{fig-chromatic-rotations} are implemented. % from \llt\ and \sct, using the tree update template.
As an example, we consider one of the rebalancing steps, named \func{RB2} (shown in Figure~\ref{code-chromatic-dorb2}), which eliminates a red-red violation at node $\uxlr$.
The other rebalancing steps are implemented similarly.

Pseudocode for \func{RB2} appears in Figure~\ref{code-chromatic-dorb2}.
This code is invoked from \tryrebalance, after performing a sequence of \llt s and deciding which rebalancing step to perform.
We first describe the steps taken in \tryrebalance\ before \func{RB2} is invoked.
%To implement \func{RB2}, a sequence of \llt s are performed in \tryrebalance, while deciding which rebalancing step to perform.
\tryrebalance\ performs \llt s on each of the shaded nodes in Figure~\ref{code-chromatic-dorb2}, starting with $u$.
\llt\ is performed on $u$ because it will be changed by the rebalancing step.
\llt\ is performed on $\ux$, $\uxl$ and $\uxlr$, because the rebalancing step will remove them from the tree.
%A pointer to $\ux$ is obtained from the result of \llt($u$), pointers to $\uxl$ and $\uxr$ are obtained from the result of \llt($\ux$), and a pointer to $\uxlr$ is obtained from the result of \llt($\uxl$).
%Since the rebalancing step would remove $\uxlr$ from the tree, an \llt\ is performed on it, too.
%As we mentioned above, if any of these \llt s return \fail\ or \finalized, then \tryrebalance\ will return \fail.
For \func{RB2} to be applicable, $\ux$ and $\uxr$ must have positive weights and $\uxl$ and $\uxlr$ must both have weight 0.
Since weight fields are immutable, \tryrebalance\ can verify constraints on weights at any time. % after the pointers to $\ux$, $\uxr$, $\uxl$, and $\uxlr$ have been obtained.
One might wonder why \llt\ is not performed on $\uxr$, since RB2 should be applied only if $\ux$ has a right child with positive weight.
Suppose \tryrebalance\ simply verifies that $\uxr.w > 0$ without performing \llt\ on $\uxr$.
Since weight fields are immutable, the only way that $\uxr.w$ can change is if the right child of $\ux$ changes.
Thus, if $\uxr.w$ changes after \tryrebalance\ performs \llt$(\ux)$, then the right child of $\ux$ will change, so the \sct\ will fail.
%However, if weights were mutable, then we would have 
%to include $f_3$ in $V$ to ensure that the \sct\ only succeeds
%if $f_3.w$ does not change.

Now we describe the behaviour of \func{RB2}.
First, $n$ and its two children are created.
$N$ consists of these three nodes.
The keys stored in the newly created nodes are the same as in the removed nodes (so that an in-order traversal encounters them in the same order).
Finally, \sct$(V, R, fld, new)$ is invoked, where $fld$ is the child pointer of $u$ that pointed to $\ux$ in the result of \llt($u$).

If the \sct\ modifies the tree, then no node $r \in V$ has changed since the update performed \llt$(r)$.
In this case, the \sct\ replaces the directed graph $G_R$ by the directed graph $G_N$ and the nodes in $R$ are finalized.
This ensures that other updates cannot erroneously modify these old nodes after they have been replaced.
The nodes in the set $F_R = F_N = \{\uxll, \uxlrl, \uxlrr, \uxr\}$ each have the same keys, weights, and child pointers before and after the rebalancing step, so they can be reused.
$V = \langle u, \ux, \uxx, \uxlr \rangle$ is the sequence of nodes on which \tryrebalance\ performs \llt s, and $R = \langle \ux, \uxl, \uxlr \rangle$ is a subsequence of $V$, so PC\ref{con-llt-on-all-nodes-in-V}, PC\ref{con-parent-in-V} and~PC\ref{con-R-subsequence-of-V} are satisfied.
Clearly, we satisfy PC\ref{con-GN-non-empty-tree} and~PC\ref{con-new-nodes} when we create $new$ and its two children.
It is easy to verify that PC\ref{con-m-in-V}, PC\ref{con-old-nil-then-R-empty}, PC\ref{con-fringe-of-GN-is-old} and~PC\ref{con-R-non-empty-then-GR-a-non-empty-tree} are satisfied.
If the tree does not change during the update, then the nodes in $V$ are ordered consistently with a breadth-first traversal of the tree, so 
%Since this is true for all updates, 
PC\ref{con-V-sequences-ordered-consistently} is satisfied.

\subsection{Successor queries}

%\section{\func{Successor} queries} \label{sec-chromatic-succ}

\begin{figure}[tb]
\begin{framed}
\preplisting
%\hrule
%\vspace{-2mm}
\begin{lstlisting}[mathescape=true]
 //$\func{Successor}(key)$
   //\com Returns the successor of $key$ and its associated value (or $\langle \bot, \bot\rangle$ if there is no such successor)
   $l := entry$
   loop until $l$// is a leaf
     if $\llt(l) \in \{\fail, \finalized\}$ then $\mbox{retry }\func{Successor}(key)\mbox{ from scratch}$
     if $key < l.key$ then
       $lastLeft := l$
       $l := l.left$
       $V := \langle lastLeft \rangle$
     else
       $l := l.right$
       //add $l$ to end of $V$ \vspace{2mm}%
      
   if $lastLeft = entry$ then return $\langle \bot, \bot \rangle$ //\medcom Dictionary is empty \label{succ-return-empty}
   else if $key < l.k$ then return $\langle l.k, l.v \rangle$ //\label{succ-return-l}
   else //\medcom Find next leaf after $l$ in in-order traversal
     $succ := lastLeft.right$
     loop until $succ$// is a leaf
       if $\llt(succ) \in \{\fail, \finalized\}$ then $\mbox{retry }\func{Successor}(key)\mbox{ from scratch}$
       //add $succ$ to end of $V$
       $succ := succ.left$
     if $succ.key = \infty$ then $result:=\langle \bot,\bot\rangle$ else $result:=\langle succ.k,succ.v\rangle$
     if $\vlt(V)$ then return $result$ //\label{succ-return-succ}
     else $\mbox{retry }\func{Successor}(key)\mbox{ from scratch}$
\end{lstlisting}
\end{framed}
%\vspace*{-3mm}
	\caption{Code for \func{Successor}.}
	\label{code-chromatic-succ}
\end{figure}

\func{Successor}($key$) runs an ordinary BST search algorithm, using \llt s to read the child fields of each node visited, until it reaches a leaf.
%begins by using \llt s to search for $key$ until
%it fails or accesses a finalized node (in which case it restarts) or
%reaching a leaf.
If the key of this leaf
is larger than $key$, it is returned and the operation is
linearized at any time during the operation
when this leaf was on the search path for $key$.
Otherwise, \func{Successor} finds the next leaf.
To do this, it remembers the last time it followed a left child pointer and, instead, follows one right child pointer, and then left child pointers until it reaches a leaf, using \llt s to read the child fields of each node visited.
%
%in an in-order traversal of the tree starting from their lowest common ancestor \trevor{Explain this better},
%using \llt s to read the child fields of each node visited.
If any \llt\ it performs
returns \fail\ or \finalized, \func{Successor} restarts.
Otherwise, it performs a validate-extended (\vlt), which returns \true\ only if all nodes on the path connecting the two leaves have
not changed.  If the \vlt\ succeeds, the key of the second leaf found is returned and
the query is linearized at the linearization point of the \vlt.
If the \vlt\ fails, \func{Successor} restarts.

%Successor notes:
%\begin{compactitem}
%\item \llt\ everything on the search path from the root to $key$.
%\item if any \llt\ returns \fail\ or \finalized, then retry the successor query from scratch.
%\item remember the last node $lastLeft$ from which we went left
%\item everything after that on the path has keys $\le key$, so their left subtrees cannot contain the successor to $key$
%\item go right from $lastLeft$ (and \llt\ it)
%\item go all the way left to a leaf, \llt ing as we go
%\item finally, do \vlt$(V)$ to check whether anything changed after we read it
%\item clearly, if nothing we read changed (or was removed), then we found the smallest element in the right subtree of $lastLeft$
%\end{compactitem}

%\trevor{Maybe describe how freezing can be used to check the preconditions of the rebalancing step, for a node that is not in $\{parent\}\cup R$ that would be good, because it would illustrate why $V$ sometimes has to be bigger than just $\{parent\}\cup R$ (originally Eric's comment)}
%
%\eric{Talk about successor or predecessor function}

\section{Correctness proof}
\label{chromatic-correctness}

\after{
(1) Proving successor's calls to LLX/VLX satisfy their preconditions
(2) Proving if an insert or delete creates a violation, then it performs cleanup
(3) Describing how to check whether a node is a sentinel (to check node r, evaluate (r.key = \infty or r = entry.left.left))
(4) Proving that the way we check whether something is a sentinel plays nicely with the rest of the algorithm
}

%\begin{obs} \label{obs-chromatic-removed-nodes-form-a-tree}
%Consider a successful operation $O$ that follows the tree update template.
%Let $x$ be the node pointed to by the child pointer that $O$ changes, just before $O$.
%The set of nodes removed from the data structure by $O$ form a tree rooted at $x$.
%\end{obs}
%\begin{chapscxproof}
%Immediate from the constraints of the tree update template.
%\end{chapscxproof}

%\begin{cor} \label{cor-chromatic-removed-nodes-pointed-to-only-by-removed-nodes}
%A node that has been removed from the data structure can only be pointed to by another node that has been removed from the data structure and, hence, only by a node that is not in the data structure.
%\end{cor}
%\begin{chapscxproof}
%Immediate from Lemma~\ref{lem-dotreeupdate-rec-cannot-be-added-after-removal} and Definition~\ref{defn-rec-in-added-removed}.
%(Technically, this should be a straightforward induction on the sequence of changes to the data structure in an execution.)
%\end{chapscxproof}

%\eric{Glue to introduce next lemma or two}

%The implementation of \llt, \sct, and \vlt\ in  \cite{paper1} are linearizable, so we may treat them as atomic operations.
%The linearizability of operations on the search tree is fairly easy to prove using the fact
%that all tree updates are performed atomically.
%\faith{this sentence seems like it is  saying A therefore A.}
We start with a high-level correctness argument for the chromatic tree.
As mentioned above, \func{Get}($key$) invokes \func{Search}$(key)$, which traverses a path from $entry$ to a leaf by reading child pointers.
Even though this search can pass through nodes that have been removed by concurrent updates,
we prove by induction that every node visited {\it was} on the search path for $key$ at some time during the search.
\func{Get} %($key$)
can thus be linearized at some time $t$ when the leaf it reaches was on the search path for $key$.
(Moreover, we prove that this leaf is the only one in the tree that could possibly contain $key$.)

We prove that each insertion, deletion and rebalancing step that performs an \sct\ follows the template (and, hence, has an atomic update phase).
Each \ins\ that terminates performs a successful \sct, and we linearize the \ins\ at this \sct.
Each \del\ that terminates performs a successful \sct, or returns at line~\ref{del-no-gp} or \ref{del-notin}.
If a \del\ performs a successful \sct, we linearize it at this \sct.
Otherwise, the \del\ behaves like a query, and is linearized in the same way as \func{Get}.
We argue that each invocations of \func{Search} performed by \ins\ or \del\ satisfies the delayed traversal property (DTP), which was introduced in Section~\ref{sec-dotreeupdate}.
This allows us to invoke Theorem~\ref{thm-effectivedtp-atomic}, which proves that \ins\ and \del\ are atomic (including the search phase).
Because no rebalancing step modifies the set of keys stored in leaves, the set of leaves always represents the set of dictionary entries.
%In order to prove that the deletions that perform successful invocations of \sct, insertions and rebalancing steps are linearizable, it suffices to prove that they follow the template.

\paragraph{Detailed proof}

We first start with two simple definitions, and then prove the major result of this section.

\begin{defn} \label{defn-searchpath}
The \textbf{search path to} $key$ \textbf{starting at} a node $r$ is the path that an ordinary BST search starting at $r$ would follow.
If $r = entry$, then we simply call this the search path to $key$.
\end{defn}

Note that this search is well-defined even if the data structure is not a BST.
Moreover, the search path starting at a node is well-defined, even if the
node has been removed from the tree.
In any case, we simply look at the path that an ordinary BST search would follow, if it were performed on the data structure.  
Additionally, observe that the \func{Search} procedure invoked by \tryins\ and \trydel\ is an ordinary BST search, and, hence, an atomic \func{Search}$(key)$ traverses exactly the search path to $key$.
%This also allows us to define the range of a node.

\begin{defn}
The \textbf{range} of a node $u$ in some configuration is the set of keys for which $u$ is on the search path.
\end{defn}

The next lemma is the main result in this section.
It establishes several results which are used to prove that searches and updates are linearizable.
The first result states that \tryins, \trydel\ and \tryrebalance\ follow the tree update template, which is a prerequisite to invoking the results in Section~\ref{app-tree-proof}.
Our proof that \tryrebalance\ follows the tree update template is complicated slightly by the fact that its subroutine, \func{OverweightLeft}, returns \fail\ at line~\ref{overweightleft-check-nil-rxxrlr} and line~\ref{overweightleft-check-nil-rxxrr} if it sees a \nil\ child pointer.
(The \func{OverweightRight} routine is similar.)
These return statements do not explicitly fit into the template, so, to guarantee non-blocking progress, we must prove that they are executed by an update only if it is concurrent with another successful update.

The second result states that the top of the tree is always as shown in Fig.~\ref{fig-treetop}.

The third and fourth claims are used to linearize searches.
Specifically, since a search only reads child pointers, and the tree may change as the search traverses the tree, we must show that it still ends up at the correct leaf.
In other words, we must show that the search is linearizable even if it traverses some nodes that are no longer in the tree.
Note that these claims are proved in a somewhat similar way to \cite{Ellen:2010}, but the proofs here must deal with the additional complication of rebalancing operations occuring while a search traverses the tree.

The fifth claim states that the search phase of each insertion (resp., deletion) satisfies the delayed traversal property.
This allows us to argue in the sixth claim that the entire insertion (resp., deletion) is atomic.
The sixth claim also states that the update phase of each rebalancing step is atomic.
(The search phase of a rebalancing step is not part of the atomic operation, but this is fine, because a chromatic tree allows rebalancing steps to be freely applied anywhere their preconditions are met, and in any order. Thus, it is \textit{not} important for a rebalancing step to be performed specifically at the location that would be found by an atomic search.)

The final claim establishes the chromatic tree structure.

%\trevor{try proving that the search phase in ins/del satisfies DTP in the same inductive proof with lemmas 6.3-6.5 (and the lemma extracted from my comment in lemma 6.3).}
%
%\trevor{does the following depend on the entire insert/delete operations being atomic? yes, in a subtle way. see below. i think it also depends on the search tree property.}

\begin{lem} \label{lem-chromatic}
Our implementation of a chromatic tree satisfies the following claims.
\begin{compactenum}
\item \tryins\ and \trydel\ follow the tree update template and satisfy all constraints specified by the template.
%\label{claim-chromatic-invariants-ins-del-follow-template}
%\item
If an invocation of \tryrebalance\ does not return at line~\ref{overweightleft-check-nil-rxxrlr} or line~\ref{overweightleft-check-nil-rxxrr}, then it follows the tree update template and satisfies all constraints specified by the template. %, otherwise it does not perform an \sct.
Otherwise, it follows the tree update template up until it returns without performing an \sct, and it satisfies all constraints specified by the template.
%\label{claim-chromatic-invariants-rebalance-follow-template}
\label{claim-chromatic-invariants-follow-template}
\item The tree rooted at $root$ always looks like Fig.~\ref{fig-treetop}(a) if it is empty, and Fig.~\ref{fig-treetop}(b) otherwise.
\label{claim-chromatic-invariants-top-of-tree}

%\item Updates to the tree do \textit{not} shrink the range of any node in the tree.
%(Formally, consider a node $u$ in the tree before and after an update \textit{UP}. The range of $u$ after \textit{UP} is a superset of the range of $u$ before \textit{UP}.)
\item If a node $v$ is in the data structure in some configuration $C$ and $v$ was on the search path for key $k$ in some earlier configuration $C'$, then $v$ is on the search path for $k$ in $C$. (Equivalently, updates to the tree do not shrink the range of any node in the tree.) \label{lem-chromatic-searchpath}
\item If an invocation of \func{Search}$(k)$ reaches a node $v$, then there was some earlier configuration during the search when $v$ was on the search path for $k$. \label{lem-chromatic-searches}
\item The \func{Search} procedure used by \tryins\ and \trydel\ satisfies DTP. \label{lem-chromatic-dtp}
\item Invocations of \tryins, \trydel\ and \tryrebalance\ that perform a successful \sct\ are atomic.\label{lem-chromatic-atomicity}
\item The tree rooted at the left child of $entry$ is always a chromatic tree. \label{lem-chromatic-searchtree}
%\item At all times, the tree rooted at the left child of $root$ is a BST. 
\end{compactenum}
\end{lem}
\begin{chapscxproof}
We prove these claims together by induction on the sequence of steps in an execution.
That is, for each claim $C$, we assume that all of the claims hold before an arbitrary step in the execution, and prove that $C$ holds after the step.

\medskip

\noindent\textbf{Claim~\ref{claim-chromatic-invariants-follow-template}:}
This claim follows almost immediately from inspection of the code.
The only subtlety is showing that no process invokes $\llt(r)$ where $r = \nil$. %, and (2) the only step that can affect this claim is an invocation of \llt.
Suppose the inductive hypothesis holds just before an invocation of $\llt(r)$.

For \func{InsertNew}, \func{InsertReplace} and \del, $r \neq \nil$ follows from inspection of the code, inductive Claim~\ref{claim-chromatic-invariants-top-of-tree}, and the fact that every key inserted or deleted from the dictionary is less than $\infty$ (so every key inserted or deleted minimally has a parent and a grandparent).

For rebalancing steps, $r \neq \nil$ follows from inspection of the code and the decision tree in Fig.~\ref{fig-decision-tree}, using a few facts about the data structure.
\tryrebalance\ performs \llt s on its arguments $ggp, gp, p, l$, and then possibly on a sequence of other nodes in the subtree rooted at $gp$, as it follows the decision tree.
From Fig.~\ref{fig-treetop}(b), it is easy to see that any node with weight $w \neq 1$ minimally has a parent, grandparent, and great-grandparent.
Thus, the arguments to \tryrebalance\ are all non-\nil.
By Claim~\ref{lem-chromatic-searchtree}, each leaf has weight $w \ge 1$, every node has zero or two children, and the child pointers of a leaf do not change.
This is enough to argue that all \llt s performed by \tryrebalance, and nearly all \llt s performed by \func{OverweightLeft} and \func{OverweightRight}, are passed non-\nil\ arguments.
Without loss of generality, we restrict our attention to \llt s performed by \func{OverweightLeft}.
The argument for \func{OverweightRight} is symmetric.
The only \llt s that require different reasoning are performed at lines~\ref{overweightleft-bad-llt-rxxrlr}, \ref{overweightleft-bad-llt-rxxrll}, \ref{overweightleft-bad-llt-rxxrr} and \ref{overweightleft-bad-llt-rxxrl}.
For lines~\ref{overweightleft-bad-llt-rxxrlr} and \ref{overweightleft-bad-llt-rxxrr}, the claim follows immediately from lines~\ref{overweightleft-check-nil-rxxrlr} and line~\ref{overweightleft-check-nil-rxxrr}, respectively.
Consider line~\ref{overweightleft-bad-llt-rxxrll}.
If $\rxxrll = \nil$ then, since every node has zero or two children, and the child pointers of a leaf do not change, $\rxxrl$ is a leaf, so $\rxxrlr = \nil$.
Therefore, \func{OverweightLeft} will return before it reaches line~\ref{overweightleft-bad-llt-rxxrll}.
By the same argument, $\rxxrl \neq \nil$ when line~\ref{overweightleft-bad-llt-rxxrl} is performed.
Thus, $r \neq \nil$ no matter where the \llt\ occurs in the code.

\medskip

\noindent\textbf{Claim~\ref{claim-chromatic-invariants-top-of-tree}:}
The only step that can modify the tree (and, hence, affect this claim) is an invocation $S$ of \sct\ performed by an invocation $I$ of \tryins, \trydel\ or \tryrebalance.
Suppose the inductive hypothesis holds just before $S$.
By Claim~\ref{claim-chromatic-invariants-follow-template}, $I$ follows the tree update template up until it performs $S$.
By Lemma~\ref{lem-effective-updatephase-atomic} and Lemma~\ref{lem-dotreeup-constraints-invariants}, the update phase of $I$ is atomic.
Thus, $S$ atomically performs one of the transformations in Fig.~\ref{fig-chromatic-rotations}.
By inspection of these transformations, when the tree is empty, \func{InsertNew} at the left child of $entry$ changes the tree from looking like Fig.~\ref{fig-treetop}(a) to looking like Fig.~\ref{fig-treetop}(b) and, otherwise, does not affect the claim.
When the tree has only one node with $key \neq \infty$, \del\ at the leftmost grandchild of $entry$ changes the tree back to looking like Fig.~\ref{fig-treetop}(a) and, otherwise, does not affect the claim.
\func{InsertReplace} does not affect the claim.

Without loss of generality, suppose $S$ performs a left rebalancing step.
(The argument for right rebalancing steps is symmetric.)
Each rebalancing step in \{\func{BLK}, \func{RB1}, \func{RB2}\} applies only if $\uxl.w = 0$, and every other rebalancing step applies only if $\uxl.w > 1$.
By the inductive hypothesis, just before $S$, all nodes that are not in the subtree rooted at the leftmost grandchild of $entry$ have weight one.
Therefore, $S$ must change a child pointer in the subtree rooted at the leftmost grandchild of $entry$.
Since the child pointer changed by $S$ was traversed while a process was searching for a key that it inserted or deleted, and every such key must be less than $\infty$, $S$ can replace only nodes with $key < \infty$ (and, hence, cannot affect the claim).

Note: using similar reasoning, it is easy to verify that, each time a process accesses a field of a node $u$, $u \neq \nil$.

\medskip

\noindent\textbf{Claim~\ref{lem-chromatic-searchpath}:}
Initially, the claim is trivially true (since the tree only contains sentinel nodes, which are on every search path).
In order for $v$ to change from being on the search path for $k$ in configuration $C'$ to no longer being on the search path for $k$ in configuration $C$, the tree must change between $C'$ and $C$.
Thus, there must be a successful \sct\ $S$ between $C'$ and $C$.
Moreover, this is the only kind of step that can affect this claim.
We show $S$ preserves the property that $v$ is on the search path for $k$.

By inductive Claim~\ref{claim-chromatic-invariants-follow-template}, $S$ is performed by a template operation.
Thus, by Lemma~\ref{lem-dotreeup-constraints-invariants}, $S$ changes a pointer of a node from $old$ to $new$, removing a connected set $R$ of nodes (rooted at $old$) from the tree, and inserting a new connected set $N$ of nodes.
If $v$ is not a descendant of $old$ immediately before $S$, then this change cannot remove $v$ from the search path for $k$.
So, suppose $v$ is a descendant of $old$ immediately prior to $S$.

Since $v$ is in the data structure in both $C'$ and $C$, it must be in the data structure at all times between $C'$ and $C$ by Lemma~\ref{lem-dotreeupdate-rec-cannot-be-added-after-removal}.
Therefore, $v$ is a descendant of $old$, but $S$ does not remove $v$ from the tree.
Recall that the fringe $F_R$ is the set of nodes that are children of nodes in $R$, but are not themselves in $R$ (see Figure~\ref{fig-replace-subtree} and Figure~\ref{fig-replace-subtree2}).
By definition, $v$ must be a descendant of a node $f \in F_R$.
Moreover, since $v$ is on the search path for $k$ just before $S$, so is $f$.
We argue, for each possible tree modification in Figure~\ref{fig-chromatic-rotations}, that if any node in $F_R$ is on the search path for $k$ prior to $S$, then it is still on the search path for $k$ after $S$.
We proceed by cases.

\textit{Case~1:} Suppose $S$ is part of an invocation of \tryins, which atomically implements the \func{InsertNew} and \func{InsertReplace} modifications in Figure~\ref{fig-chromatic-rotations}.
Since $S$ replaces a leaf with either a new leaf, or a new internal node and two new leaves, the fringe set is empty.
Thus, the claim is vacuously true.

\textit{Case~2:} Suppose $S$ is part of an invocation of \trydel, which atomically implements the \func{Delete} modification in Figure~\ref{fig-chromatic-rotations}.
(The argument for its mirror image is symmetric.)
$S$ replaces a leaf $\uxl$, its sibling $\uxr$, and their parent $\ux$, with a new copy $new$ of $\uxr$.
If $\uxr$ is a leaf, then $F_R$ is empty, and the claim is vacuously true.
Otherwise, $F_R$ consists of the children $\uxrl$ and $\uxrr$ of $\uxr$.
Since $new$ has the same key as $\uxr$, the ranges of $\uxrl$ and $\uxrr$ after $S$ are supersets of what they were before $S$.
Thus, the claim holds.

\textit{Case~3:} Suppose $S$ is part of an invocation of \tryrebalance, which atomically implements the remaining modifications in Figure~\ref{fig-chromatic-rotations} (and their mirror images).
In each rebalancing step, $|R| = |N|$ and the set of keys in $R$ is the same as the set of keys in $N$.
Furthermore, in-order traversals on $R \cup F_R$ just before $S$ and $N \cup F_R$ just after $S$ yields the same sequence of keys.
By Claim~\ref{lem-chromatic-searchtree}, this sequence is sorted.
Consequently, the nodes in $R$ partition the range of $old$ in exactly the same way that the nodes in $N$ partition the range of $new$.
Thus, the claim holds.

\medskip

\noindent\textbf{Claim~\ref{lem-chromatic-searches}:}
Consider a read $r$ of a child pointer in an invocation $I$ of \func{Search}$(k)$.
This is the only kind of step that can affect this claim.
Let $v$ be the node pointed to by the value returned by $r$.
We prove that $r$ preserves the claim.
If $r$ is the first read of a child pointer by $I$, then $v$ is $entry$, which is always on the search path for $k$, so $r$ preserves the claim.

Now, suppose there is a previous read $r'$ of a child pointer by $I$.
By the inductive hypothesis, the node $v'$ that was returned by $r'$ was on the search path for $k$ in some configuration $C'$ after the beginning of $I$ and before $r'$.
Our goal is to prove that there is a configuration $C$, after $C'$ and before $r$, when $v$ is on the search path for $k$.
If $v'$ is in the tree when $I$ performs $r$, then $C$ is the configuration just before $r$.
Otherwise, $C$ is the last configuration before $v'$ was removed from the tree.

Without loss of generality, assume $k < v'.key$.
(The argument when $k \geq v'.key$ is symmetric.)
By inductive Claim~\ref{lem-chromatic-searchtree}, the data structure is a chromatic search tree just before $r$, so $I$ must reach $v$ by following the left child pointer of $v'$.
We now prove that $v'.left$ points to $v$ in $C$.
Suppose $v'$ is in the tree when $I$ performs $r$ (so $C$ is just before $r$).
This case follows immediately from our assumption that $r$ reads a pointer to $v$ from $v'.left$.
Now, suppose $v'$ is not in the tree when $r$ occurs, so $C$ is the last configuration before $v'$ was removed.
Recall that $v'$ is in the tree in $C'$.
By Lemma~\ref{lem-dotreeupdate-rec-cannot-be-added-after-removal}, $v'$ cannot be added back into the data structure after it is removed.
Since $v'$ is in the tree in $C'$, and is not in the tree when $I$ subsequently performs $r$, $v'$ must be removed after $C'$ and before $r$ occurs.
By inductive Claim~\ref{claim-chromatic-invariants-follow-template} and template Constraint~\ref{constraint-finalized-iff-removed}, $v'$ becomes finalized precisely when it is removed.
Since $v'$ cannot change after it is finalized, and $v'.left$ points to $v$ when $r$ occurs (which is after $v'$ is removed), we can see that $v'.left$ must point to $v$ in $C$ (which is just before $v'$ is removed).

Finally, we prove that $v$ is on the search path for $k$ in $C$.
Since $v'$ was on the search path for $k$ in $C'$, and it is in the data structure in $C$, which is after $C'$ but before $r$, inductive Claim~\ref{lem-chromatic-searchpath} implies that $v'$ is on the search path for $k$ in $C$.
Since $k < v'.key$ and $v = v'.left$, $v$ is also on the search path for $k$ in $C$.

\medskip

\noindent\textbf{Claim~\ref{lem-chromatic-dtp}:}
Initially, the claim holds vacuously (since no steps have been taken).
Suppose an invocation $S$ of \func{Search}$(k)$ in \trydel\ terminates and returns $m$.
(The proof for \tryins\ is similar.)
Consider any configuration $C$, after $S$ returns $m$, in which all of the nodes in $m$ are in the tree and their fields agree with the values in $m$.
Suppose the inductive hypothesis up until $C$.
We prove that an invocation $S'$ of \func{Search}$(k)$ in \trydel\ would return $m$ if $S'$ were performed atomically just after configuration $C$.

The value $m = \langle -, gp, p, l \rangle$ returned by $S$ contains a leaf $l$, its parent $p$ and its grandparent $gp$.
By inductive Claim~\ref{lem-chromatic-searches}, $gp$, $p$ and $l$ were each on the search path at some point during $S$ (which is before $C$).
Since $gp$, $p$ and $l$ are in the tree in $C$, inductive Claim~\ref{lem-chromatic-searchpath} implies that they are all on the search path for $k$ in $C$.
Therefore, $S'$ will visit each of them.
Conceptually, $m$ also encodes the facts that $gp$ points to $p$ and $p$ points to $l$.
These facts are checked at line~\ref{code-chromatic-trydel-conflict1} and line~\ref{code-chromatic-trydel-conflict2} (as part of the \func{Conflict} procedure).
By our assumption (that the fields of the nodes in $m$ in configuration $C$ agree with their values in $m$), these facts also hold when $S'$ is performed.
Consequently, $S'$ will also return $m = \langle -, gp, p, l \rangle$.

\medskip

\noindent\textbf{Claim~\ref{lem-chromatic-atomicity}:}
By Claim~\ref{lem-chromatic-dtp} and Theorem~\ref{thm-effectivedtp-atomic}, all invocations of \tryins\ and \trydel\ that perform a successful \sct\ are atomic.
By Lemma~\ref{lem-effective-updatephase-atomic}, all invocations of \tryrebalance\ that perform a successful \sct\ are atomic.

\medskip

\noindent\textbf{Claim~\ref{lem-chromatic-searchtree}:}
Only successful invocations of \sct\ can affect this claim.
These are performed only in \tryins, \trydel\ and \tryrebalance.
The claim holds in the initial state of the tree shown in Figure~\ref{fig-treetop}(a).
We show that every successful invocation $S$ of \sct\ preserves the claim.
We proceed by cases.

\textit{Case~1:} $S$ is performed in an invocation $I$ of \tryins$(key, value)$ or \trydel$(key)$.
By Claim~\ref{lem-chromatic-atomicity}, the entirety of $I$ is atomic, including its search phase.
Thus, when $I$ occurs, its search procedure returns the unique leaf on the search path for $key$.
Consequently, $I$ atomically performs one of the transformations \func{InsertNew}, \func{InsertReplace} or \func{Delete} in Figure~\ref{fig-chromatic-rotations} to replace $l$ (and possibly some of its neighbouring nodes).
Since $I$ is entirely atomic (including its search phase), and it simply performs one of the chromatic tree updates, it is easy to verify that it preserves the claim.

\textit{Case~2:} $S$ is performed in an invocation $I$ of \tryrebalance, which performs one of the rebalancing transformations \func{Push}, \func{RB1}, \func{RB2}, \func{BLK}, \func{W1}, \func{W2}, \func{W3}, \func{W4}, \func{W5}, \func{W6}, or \func{W7}.
By Claim~\ref{lem-chromatic-atomicity}, the update phase of $I$ is atomic (but the search phase is not necessarily atomic).
Therefore, $I$ atomically performs one of the rebalancing transformations at some location in the tree (but not necessarily the same rebalancing transformation, at the same location, that it would perform if $I$'s search were also part of the atomic update).
All of the rebalancing transformations preserve the claim (regardless of where in the tree they are performed).
\end{chapscxproof}

We define the linearization points for chromatic tree operations as follows.
\begin{compactitem}
\item \func{Get}($key$) is linearized at a time during the operation when the leaf reached was on the search path for $key$.
(This time exists, by Lemma~\ref{lem-chromatic}.\ref{lem-chromatic-searches}.)
\item An \ins\ is linearized at its successful \sct\ inside \tryins\ (if such an \sct\ exists).
(Note: every \ins\ that terminates performs a successful \sct.)
\item A \del\ that returns $\bot$ is linearized at a time during the operation when the leaf returned by its last invocation of \func{Search} was on the search path for $key$.
(This time exists, by Lemma~\ref{lem-chromatic}.\ref{lem-chromatic-searches}.)
\item A \del\ that does not return $\bot$ is linearized at its successful \sct\ inside \trydel\ (if such an \sct\ exists).
(Note: every \del\ that terminates, but does not return $\bot$, performs a successful \sct.)
\item A \func{Successor} query that returns at line \ref{succ-return-empty} is linearized when it performs \llt($entry$).
\item A \func{Successor} query that returns at line \ref{succ-return-l} at the time during the operation that $l$ was on the search path for $key$.
(This time exists, by Lemma~\ref{lem-chromatic}.\ref{lem-chromatic-searches}.)
\item A \func{Successor} query that returns at line \ref{succ-return-succ} is linearized when it performs  its successful \vlt.
\end{compactitem}
It is easy to verify that every operation that terminates is assigned a linearization point during
the operation.

\begin{thm}
The chromatic search tree is a linearizable implementation of an ordered dictionary
with the operations \func{Get}, \ins, \del, \func{Successor}.
\end{thm}
\begin{chapscxproof}
Lemma~\ref{lem-chromatic}.\ref{lem-chromatic-atomicity} proves that the \sct s implement atomic changes to the tree as shown in Figure~\ref{fig-chromatic-rotations}.
By inspection of these transformations, the set of keys and associated values stored in leaves are not altered by any rebalancing steps.
Moreover, the transformations performed by each linearized \ins\ and \del\ maintain the invariant that the set of keys and associated values stored in leaves of the tree is exactly the set that should be in the dictionary.

When a \func{Get}$(key)$ is linearized, the search path for $key$ ends at the leaf returned by its invocation of \func{Search}.
If that leaf contains $key$, \func{Get} returns the associated value, which is correct.
If that leaf does not contain $key$, then, by Lemma~\ref{lem-chromatic}.\ref{lem-chromatic-searchtree}, it is nowhere else in the tree, so \func{Get} is correct to return $\bot$.

If \func{Successor}($key$) returns $\langle \bot,\bot\rangle$ at line \ref{succ-return-empty}, then at its linearization point, the left child of the $entry$ is a leaf.
By Lemma \ref{lem-chromatic}.\ref{claim-chromatic-invariants-top-of-tree}, the dictionary is empty.

If \func{Successor}($key$) returns $\langle l.k,l.v\rangle$ at line \ref{succ-return-empty}, then
at its linearization point, $l$ is the leaf on the search path for $key$.
So, $l$ contains either $key$ or its predecessor or successor at the linearization point.
Since $key<l.k$, $l$ is $key$'s successor.

Finally, suppose \func{Successor}($key$) returns $\langle succ.k,succ.v\rangle$ at line \ref{succ-return-succ}.
Then $l$ was on the search path for $key$ at some time during the search.
Since $l$ is among the nodes validated by the $\vlt$, it is not finalized, so it is still on the search path for $key$ at the linearization point, by Lemma~\ref{lem-chromatic}.\ref{lem-chromatic-searchpath}.
Since $key \geq l.k$, the successor of $key$ is the next leaf after $l$ in an in-order traversal of the tree.
Leaf $l$ is the rightmost leaf in the subtree rooted at the left child of $lastLeft$ and the key returned is the leftmost leaf in the subtree rooted at the right child of $lastLeft$.
The paths from $lastLeft$ to these two leaves are not finalized and therefore are in the tree.
Thus, the correct result is returned.
\end{chapscxproof}

\section{Progress proof}

%\trevor{One reviewer tripped up on this explanation, so we should probably give a bit more of a sketch of how this proof works.  (Not only did he not understand what we were proving w.r.t. progress, he also seems to have interpreted what we wrote below to mean that progress is only guaranteed if a finite number of \ins\ and \del\ operations occur (which, admittedly, is pretty stupid).)}
Our goal is to prove that, if processes take steps infinitely often, then chromatic tree operations succeed infinitely often.
At a high level, this follows from Theorem~\ref{thm-dotreeup-progress} (the final progress result for template operations), and the fact that at most $3i+d$ rebalancing steps can be performed after $i$ insertions and $d$ deletions have occurred (proved in \cite{Boyar97amortizationresults}). %only an amortized constant number of rebalancing steps can be performed for each \ins\ or \del. %, if \ins\ and \del\ operations stop happening, only a bounded number of rebalancing steps can be performed.
%This section provides a formal proof % of non-blocking progress, 
%using the results established in Section~\ref{app-tree-proof}.
Theorem~\ref{thm-dotreeup-progress} applies only if processes perform infinitely many template operations, so we must prove that processes will perform infinitely many template operations if they take steps infinitely often.
The subtlety is that some invocations of \tryrebalance\ might not follow the template.
One can imagine a pathology in which non-blocking progress is violated, because processes perform only finitely many invocations of \tryrebalance\ that follow the template, but perform infinitely many invocations that do \textit{not} follow the template.
We first prove that this does not happen.
Then, we prove the main result.

\begin{lem} \label{lem-chromatic-tryrebalance-follows-template-infinitely-often}
%If an execution contains infinitely many invocations of \tryrebalance, then it also contains infinitely many successful invocations of \sct.
If infinitely many invocations of \tryrebalance\ are performed, then infinitely many invocations follow the tree update template.
% invocations of \tryrebalance\ follow the tree update template.
\end{lem}
\begin{chapscxproof}
As we explained near the beginning of this section, each invocation either follows the template or returns at line~\ref{overweightleft-check-nil-rxxrlr} or line~\ref{overweightleft-check-nil-rxxrr} without invoking \sct.
We prove that each invocation $I$ of \tryrebalance\ that returns at line~\ref{overweightleft-check-nil-rxxrlr} or line~\ref{overweightleft-check-nil-rxxrr} is concurrent with a template operation that performs a successful \sct\ during $I$ (and, hence, follows the template).
Suppose not, to derive a contradiction.
Suppose $I$ returns at line~\ref{overweightleft-check-nil-rxxrlr}.
Consider the configuration immediately before $I$ returns.
By inspection of \func{OverweightLeft}, we have $\rxxrl.w = 1$ and $\rxxrlr = \nil$, so $\rxxrl$ is a leaf.
Moreover, since the tree does not change during $I$, we also have the following facts: $\rxxr$ is the parent of $\rxxrl$, $\rxxr.w = 0$, $\rxxl$ is the sibling of $\rxxr$, and $\rxxl.w > 1$.
Therefore, the sum of weights on a path from $root$ to a leaf in the sub-tree rooted at $\rxxr$ is different from the sum of weights on a path from $root$ to a leaf in the sub-tree rooted at $\rxxl$.
So, the tree is not a chromatic tree, which is a contradiction.
The proof when $I$ returns at line~\ref{overweightleft-check-nil-rxxrr} is similar, and is left as an exercise.
\end{chapscxproof}

\begin{thm}
The chromatic tree operations are non-blocking.
\end{thm}
\begin{chapscxproof}
To derive a contradiction, suppose there is some configuration $C$ after which some processes continue to take steps but no successful chromatic tree operations occur.
We first argue that eventually the tree stops changing.
Since no successful chromatic tree operations occur after $C$, the only steps that can change the tree after $C$ are successful invocations of \sct\ performed by \tryrebalance.
Boyar, Fagerberg and Larsen~\cite{Boyar97amortizationresults} proved that after a bounded number of rebalancing steps, the tree becomes a red-black tree, and then no further rebalancing steps can be applied.
Thus, eventually the tree must stop changing.

This implies that every invocation of \func{Search} (and, hence, \func{Get}) terminates after a finite number of steps, unless the process executing it crashes.
Thus, \func{Search} and \func{Get} are non-blocking.
It follows that no invocation of \func{Search} or \func{Get} occurs after $C$.
To prove progress for the other operations, we consider two cases.

Suppose processes take infinitely many steps in \ins\ or \del\ operations.
Then, processes perform infinitely many invocations of \tryins, \trydel\ or \tryrebalance.
By Lemma~\ref{lem-chromatic}.\ref{claim-chromatic-invariants-follow-template} and Lemma~\ref{lem-chromatic-tryrebalance-follows-template-infinitely-often}, infinitely many of these invocations follow the tree update template.
By Theorem~\ref{thm-dotreeup-progress}, infinitely many of them will succeed, so infinitely many must succeed after $C$, which is a contradiction.
%%% what if they succeed but don't change anything? well, the point is we're saying they don't succeed.
%Only one successful invocations of \tryins\ or \trydel\ by each process can occur after $T$.
%So, there must be infinitely many successful calls to \tryrebalance.
%Boyar, Fagerberg and Larsen proved \cite{Boyar97amortizationresults} 
%proved that after a bounded number of
%rebalancing steps, the tree becomes a RBT, and then no further rebalancing
%steps can be applied, a contradiction.

Now, suppose there is a configuration $C'$ (after $C$) after which no process takes a step in an \ins\ or \del\ operation.
%Then, since all invocations of \sct\ are performed by processes executing \ins\ or \del, no process invokes \sct\ after $C'$.
%Thus, the tree does not change after $C'$, which means that any invocations of \func{Search} and \func{Get} that start after $C'$ will terminate.
%It follows that \func{Search} and \func{Get} are lock-free.
Then, every operation after $C'$ must be an invocation of \func{Predecessor} or \func{Successor}.
%Since \func{Search} represents the \func{SearchPhase} procedure in the template, we learn that the \func{SearchPhase} procedure is lock-free.
%Then, eventually, no process invokes \sct\ (since the queries 
%\func{get}, \func{Successor} and \func{Predecessor} do not
%perform \sct s).
Observe that every invocation of \func{Predecessor} or \func{Successor} is an execution of a \vlt-\func{Query} algorithm (see the definition in Section~\ref{template-progress}).
By progress property P2 of \llt, \sct\ and \vlt\ (see Section~\ref{progress-spec}), infinitely many invocations of \sct\ or \vlt\ must succeed.
Since there are only finitely many invocations of \sct, infinitely many invocations of \vlt\ must succeed, and infinitely many of these must occur after $C'$.
Observe that \vlt\ is performed only by invocations of \func{Predecessor} and \func{Successor}, and an invocation of \func{Predecessor} or \func{Successor} is successful precisely if it performs a successful \vlt.
Therefore, infinitely many successful invocations of \func{Predecessor} and/or \func{Successor} must occur after $C'$, which is a contradiction.
\end{chapscxproof}

%\eric{According to Trevor's email of Feb 8, 5:23 am there might be something to add to the proof of progress.}

\section{Bounding the height of the tree}
\label{height-bound}

We now show that the height of the chromatic search tree at any time is $O(c + \log n)$ where
$n$ is the number of keys stored in the tree and $c$ is the number of \ins\ and \del\ operations currently in progress.
Since we always perform rebalancing steps that satisfy VIOL, if we reach a leaf without finding the violation that an \ins\ or \del\ created, then the violation has been eliminated.
This allows us to prove that the number of violations in the tree at any time is bounded above by $c$. %, the number of insertions and deletions that are currently in progress.
Further, since removing all violations would yield a red-black tree with height $O(\log n)$, and eliminating each violation reduces the height by at most one, the height of the chromatic tree is $O(c + \log n)$.

%We now show that the height of the chromatic search tree is $O(\log n + c)$ where
%$n$ is the number of keys stored in the tree and $c$ is the number of incomplete
%\ins\ and \del\ operations.
%Each \ins\ or \del\ can create one new violation.
%We prove that no \ins\ or \del\ terminates until the violation it created is destroyed.
%Thus, the number of violations in the tree at any time is bounded by $c$, and the required 
%bound follows.

\begin{defn}
Let $x$ be a node that is in the data structure.  We say that $x.w-1$ \textbf{overweight violations occur at $x$} if $x.w >1$.
We say that a \textbf{red-red violation occurs at $x$} if $x$ and its parent in the data structure both have weight 0.
%We say that a \textbf{violation occurs at $x$} 
%if either an overweight or a red-red violation occurs at $x$.
\end{defn}

The following lemma says that red-red violations can never be created at a node, except when the node is first added to the data structure.

\begin{lem}
\label{no-new-red-red}
Let $v$ be a node with weight 0.
Suppose that when $v$ is added to the data structure, its (unique) parent has non-zero weight. 
Then $v$ is never the child of a node with weight 0.
\end{lem}
\begin{chapscxproof}
Node weights are immutable.
It is easy to check by inspection of each transformation in Figure~\ref{fig-chromatic-rotations}
that if $v$ is not a newly created node and it
acquires a new parent in the transformation with weight 0,
then $v$ had a parent of weight 0 prior to the transformation.
\end{chapscxproof}

\begin{defn}
A process $P$ is \textbf{in a cleanup phase for} $k$ if it is executing an \ins$(k)$ or a \del$(k)$ and it has performed a successful \sct\ inside a \tryins\ or \trydel\ that returns $createdViolation=\true$. 
If  $P$ is between line \ref{cleanup-start} and \ref{cleanup-end}, 
$location(P)$ and $parent(P)$ are the values of $P$'s local variables $l$ and $p$; otherwise $location(P)$ is $entry$ and $parent(P)$ is \nil.
\end{defn}

We use the following invariant to show that each violation in the data structure has a
pending update operation that is responsible for removing it before terminating:
either that process is on the way towards the violation, or it will find another violation and
restart from the top of the tree, heading towards the violation.

\begin{lem}
In every configuration, there exists an injective mapping $\rho$ from violations to processes such that, for every violation $x$, 
\begin{compactitem}
\item
{\rm (A)} process $\rho(x)$ is in a cleanup phase for some key $k_x$ and 
\item
{\rm (B)} $x$ is on the search path from $entry$ for $k_x$ and
\item {\rm (C)} either\\ 
{\rm (C1)} the search path for $k_x$ from $location(\rho(x))$ contains the violation $x$, or\\
{\rm (C2)} $location(\rho(x)).w=0$ and $parent(\rho(x)).w=0$, or\\
{\rm (C3)} in the prefix of the search path for $k_x$ from $location(\rho(x))$ up to and including
the first non-finalized node (or the entire search path if all nodes are finalized), there
is a node with weight greater than 1 or two nodes in a row with weight~0.
\end{compactitem}
\end{lem}
\begin{chapscxproof}
In the initial configuration, there are no violations, so the invariant is trivially satisfied.
We show that any step $S$ by any process $P$ preserves the invariant.  
We assume there is a function $\rho$ satisfying the claim for the configuration $C$ immediately before $S$ and show that there is a function $\rho'$ satisfying the claim for the configuration $C'$ immediately after $S$.
The only step that can cause a process to leave its cleanup phase is
the termination of an \ins\ or \del\ that is in its cleanup phase.
The only
steps that can change $location(P)$ and $parent(P)$ are $P$'s execution of line \ref{cleanup-end} or the read of the child pointer on line \ref{move-l-left} or \ref{move-l-right}.  (We think of all
of the updates to local variables in the braces on those lines as happening atomically
with the read of the child pointer.)
The only steps that can change child pointers or finalize nodes  
are successful \sct s.   No other steps $S$ can cause the invariant to become false.

{\bf Case 1} $S$ is the termination of an \ins\ or \del\ that is in its cleanup phase:
We choose $\rho'=\rho$.
$S$ happens when the test in line \ref{cleanup-terminate} is true, meaning that $location(P)$ is a leaf.
Leaves always have weight greater than 0.  The weight of the leaf cannot be greater than 1, because then
the process would have exited the loop in the previous iteration after the test at line  \ref{find-violation} returned true (since weights of nodes never change).
Thus, $location(P)$ is a leaf with weight 1.
So, $P$ cannot be $\rho(x)$ for any violation $x$, so $S$ cannot make the invariant become false.

{\bf Case 2} $S$ is an execution of line \ref{cleanup-end}:
We choose $\rho'=\rho$.
Step $S$ changes $location(P)$ to $entry$.  
If $P\neq \rho(x)$ for any violation $x$, then this step cannot affect the truth of the invariant.  
Now suppose $P=\rho(x_0)$ for some violation $x_0$.
The truth of properties (A) and (B) are not affected by a change in $location(P)$
and property (C) is not affected for any violation $x\neq x_0$.
Since $\rho$ satisfies property (B) for violation $x_0$ before $S$, it will satisfy 
property (C1) for $x_0$ after $S$.

{\bf Case 3} $S$ is a read of the left child pointer on line \ref{move-l-left}:
We choose $\rho'=\rho$.
Step $S$ changes $location(P)$ from some node $v$ to node $v_L$, which is $v$'s left child when $S$ is performed.
If $P\neq \rho(x)$ for any violation $x$, then this step cannot affect the truth of the invariant.  
So, suppose $P=\rho(x_0)$ for some violation $x_0$.
By (A), $P$ is in a cleanup phase for $k_{x_0}$.
The truth of (A) and (B) are not affected by a change in $location(P)$
and property (C) is not affected for any violation $x\neq x_0$.
So it 
remains to prove that (C) is true for violation $x_0$ in $C'$.

First, we prove $v.w\leq 1$, and hence there is never an overweight violation at $v$.
If $v$ is $entry$, then $v.w=1$.
Otherwise, $S$ does not occur during the first iteration of \cleanup's inner loop.
In the previous iteration, $v.w\leq 1$ at line \ref{find-violation} (otherwise, the loop would
have terminated).

Next, we prove that there is no red-red violation at $v$ when $S$ occurs.
If $v$ is $entry$ or is not in the data structure when $S$ occurs, 
then there cannot be a red-red violation at $v$ when $S$ occurs, by definition.
Otherwise, node $v$ was read as the child of some other node $u$ in the previous iteration
of \cleanup's inner loop
and line \ref{find-violation} found that $u.w\neq 0$ or $v.w\neq 0$ (otherwise the loop would have terminated).
So, at some time before $S$ (and when $v$ was in the data structure), 
there was no red-red violation at $v$.
By Lemma \ref{no-new-red-red}, there is no red-red violation at $v$ when $S$ is performed.

Next, we prove that (C2) cannot be true for $x_0$ in configuration $C$.
If $S$ is in the first iteration of \cleanup's inner loop, then $location(P)=entry$, which has weight 1.
If $S$ is not in the first iteration of \cleanup's inner loop, then the previous iteration found
$parent(P).w\neq 0$ or $location(P).w\neq 0$ (otherwise the loop would have terminated).

So we consider two cases, depending on whether (C1) or (C3) is true in configuration $C$.

{\bf Case 3a} (C1) is true in configuration $C$:
Thus, when $S$ is performed, 
the violation $x_0$ is on the search path for $k_{x_0}$ from $v$, but it is not at $v$ (as argued above).
$S$ reads the {\it left} child of $v$, so $k_{x_0} < v.k$ (since the key of node $v$ never changes).
So, $x_0$ must be on the search path for $k_{x_0}$ from $v_L$.  This means (C1) is satisfied for $x_0$ in configuration $C'$.

{\bf Case 3b} (C3) is true in configuration $C$:
We argued above that $v.w \le 1$, so the prefix must contain two nodes in a row with weight 0.
%If the prefix contains a node with weight greater than 1, we argued above that $v.w\neq 1$, so
%after $S$, (C3) is true.
%So suppose the prefix contains two nodes in a row with weight 0.
If they are the first two nodes,
$v$ and $v_L$, then (C2) is true after $S$.  Otherwise, (C3) is still true after $S$.

{\bf Case 4} $S$ is a read of the right child pointer on line \ref{move-l-right}:
The argument is symmetric to Case 3.

{\bf Case 5} $S$ is a successful \sct:
We must define the mapping $\rho'$ for each violation $x$ in configuration $C'$.
By Lemma \ref{no-new-red-red} and the fact that node weights are immutable, no transformation in Figure~\ref{fig-chromatic-rotations} can create a new violation
at a node that was already in the data structure in configuration $C$.
So, if $x$ is at a node that was in the data structure in configuration $C$,
$x$ was a violation in configuration $C$, and $\rho(x)$ is well-defined.
In this case, we let $\rho'(x)=\rho(x)$.

If $x$ is at a node that was added to the data structure by $S$, then we must define
$\rho(x)$ on a case-by-case basis for all transformations described in Figure~\ref{fig-chromatic-rotations}.
(The symmetric operations are handled symmetrically.)

\begin{figure}[h]
%\hspace{-5mm}
%\begin{minipage}{1\textwidth}
\centering
\noindent
\begin{tabular}{|l|l|l|}\hline
Transformation & Red-red violations $x$ created by $S$ & $\rho'(x)$\\\hline
RB1 & none created & --\\\hline
RB2 & none created & --\\\hline
BLK & at $n$ (if $\ux.w=1$ and $u.w=0$) & $\rho$(red-red violation at one of $\uxll,\uxlr,\uxrl,\uxrr$)$\dagger$\\\hline
PUSH & none created & --\\\hline
W1,W2,W3,W4 & none created & -- \\\hline
W5 & at $n$ (if $\ux.w=u.w=0$) & $\rho$(red-red violation at $\ux$) \\\hline
W6 & at $n$ (if $\ux.w=u.w=0$) & $\rho$(red-red violation at $\ux$)\\\hline
W7 & none created & --\\\hline
INSERT1 & at $n$ (if $\ux.w=1$ and $u.w=0$) & process performing the \ins\\\hline
INSERT2 & none created & -- \\\hline
DELETE & at $n$ (if $\ux.w=\uxr.w=u.w=0$)&$\rho$(red-red violation at $\ux$)\\\hline
\end{tabular}
\caption{Description of how $\rho'$ maps red-red violations at newly added nodes to processes responsible for them. %to where $x$ is a red-red violation at a newly added node.
$\dagger$By inspection of the decision tree in Figure \ref{fig-decision-tree}, BLK is only applied if one of $\uxll,\uxlr,\uxrl$ or $\uxrr$ has weight 0, and therefore a red-red violation, in configuration $C$, and this red-red violation is eliminated by the transformation.}
\label{fig-table-violations-redred}
\end{figure}

If $x$ is a red-red violation at a newly added node, 
we define $\rho'(x)$ according to the table in Figure~\ref{fig-table-violations-redred}.
%
%\footnotetext{By inspection of the decision tree in Figure \ref{fig-decision-tree}, BLK is only applied if one of $\uxll,\uxlr,\uxrl$ or $\uxrr$ has weight 0, and therefore a red-red violation, in configuration $C$, and this red-red violation is eliminated by the transformation.}
%%\footnotetext[2]{By inspection of the decision tree in Figure \ref{fig-decision-tree}, W1, W2, W3 and W4 are applied only if $\ux.w>0$.}
%%\eric{Should the second footnote be removed now that we have changed the rebalancing steps to make this a precondition?}
%
For each newly added node that has $k$ overweight violations after $S$,
$\rho'$ maps them to the $k$ distinct processes $\{\rho(q) : q\in Q\}$, where
$Q$ is given by the table in Figure~\ref{fig-table-violations-overweight}.

\begin{figure}[h]
\centering
\noindent
\begin{tabular}{|l|ll|l|}\hline
Transformation & \multicolumn{2}{c|}{Overweight violations created by $S$} & Set $Q$ of overweight violations before $S$ \\\hline
RB1 & $\ux.w-1$ at $n$ &(if $\ux.w>1$) & $\ux.w-1$ at $\ux$\\\hline
RB2 & $\ux.w-1$ at $n$ &(if $\ux.w>1$) & $\ux.w-1$ at $\ux$\\\hline
BLK & $\ux.w-2$ at $n$ &(if $\ux.w>2$) & $\ux.w-2$ of the $\ux.w-1$ at $\ux$\\\hline
PUSH & $\ux.w$ at $n$  &(if $\ux.w>0$)  & $\ux.w-1$ at $\ux$, and 1 at $\uxl$\\
PUSH & $\uxl.w-2$ at $\nL$ &(if $\uxl.w>2$) & $\uxl.w-2$ of the $\uxl.w-1$ at $\uxl$\\\hline
W1 & $\ux.w-1$ at $n$  &(if $\ux.w>1$)  & $\ux.w-1$ at $\ux$\\
W1 & $\uxl.w-2$ at $\nLL$ &(if $\uxl.w>2$) & $\uxl.w-2$ of the $\uxl.w-1$ at $\uxl$\\
W1 & $\uxrl.w-2$ at $\nLR$ &(if $\uxrl.w>2$) & $\uxrl.w-2$ of the $\uxrl.w-1$ at $\uxl$\\\hline
W2 & $\ux.w-1$ at $n$ & (if $\ux.w>1$) & $\ux.w-1$ at $\ux$\\
W2 & $\uxl.w-2$ at $\nLL$ & (if $\uxl.w>2$) & $\uxl.w-2$ of the $\uxl.w-1$ at $\uxl$\\\hline
W3 & $\ux.w-1$ at $n$  &(if $\ux.w>1$)  & $\ux.w-1$ at $\ux$\\
W3 & $\uxl.w-2$ at $\nLLL$ &(if $\uxl.w>2$) & $\uxl.w-2$ of the $\uxl.w-1$ at $\uxl$\\\hline
W4 & $\ux.w-1$ at $n$ &(if $\ux.w>1$) & $\ux.w-1$ at $\ux$\\
W4 & $\uxl.w-2$ at $\nLL$ &(if $\uxl.w>2$) & $\uxl.w-2$ of the $\uxl.w-1$ at $\uxl$\\\hline
W5 & $\ux.w-1$ at $n$ &(if $\ux.w>1$) & $\ux.w-1$ at $\ux$\\
W5 & $\uxl.w-2$ at $\nLL$ &(if $\uxl.w>2$) & $\uxl.w-2$ of the $\uxl.w-1$ at $\uxl$\\\hline
W6 & $\ux.w-1$ at $n$ &(if $\ux.w>1$) & $\ux.w-1$ at $\ux$\\
W6 & $\uxl.w-2$ at $\nLL$ &(if $\uxl.w>2$) & $\uxl.w-2$ of the $\uxl.w-1$ at $\uxl$\\\hline
W7 & $\ux.w$ at $n$  &(if $\ux.w>0$)  & $\ux.w-1$ at $\ux$, and 1 at $\uxl$\\
W7 & $\uxl.w-2$ at $\nL$ &(if $\uxl.w>2$) & $\uxl.w-2$ of the $\uxl.w-1$ at $\uxl$\\
W7 & $\uxr.w-2$ at $\nR$ &(if $\uxr.w>2$) & $\uxr.w-2$ of the $\uxr.w-1$ at $\uxr$\\\hline
INSERT1 & $\ux.w-2$ at $n$ &(if $\ux.w>2$) & $\ux.w-2$ of the $\ux.w-1$ at $\ux$\\\hline
INSERT2 & $\ux.w-1$ at $n$ &(if $\ux.w>1$) & $\ux.w-1$ at $\ux$\\\hline
DELETE & $\ux.w+\uxr.w-1$ at $n$ & (if $\ux.w+\uxr.w>1$) & $\max(0,\ux.w-1)$ at $\ux$ and\\
	&							&					  & $\max(0,\uxr.w-1)$ at $\uxr \dagger$\\
\hline
\end{tabular}
\caption{Description of how $\rho'$ maps overweight violations at newly added nodes to processes responsible for them.
(Note that $Q$ is a set, since a node can have many overweight violations.)
%Definition of mapping $\rho'$ where $Q$ is a set of overweight violations before a transformation $S$.
$\dagger$In this case, the number of violations in $Q$ is one too small if both $\ux.w$ and $\uxr.w$ are greater than 0, so the remaining violation is assigned to the process that performed the \del's \sct.}
\label{fig-table-violations-overweight}
\end{figure}

%\footnotetext[2]{In this case, the number of violations in $Q$ is one too small if both $\ux.w$ and $\uxr.w$ are greater than 0, so the remaining violation is assigned to the process that performed the \del's \sct.}

The function $\rho'$ is injective, since $\rho'$ maps each violation created by $S$ to
a distinct process that $\rho$ assigned to a violation that has been removed by $S$, with only two exceptions:  for red-red violations caused by \func{InsertNew} and one overweight violation
caused by \func{Delete}, $\rho'$ maps 
the red-red violation to the process that has just begun its cleanup phase (and therefore
was not assigned any violation by $\rho$).

Let $x$ be any violation in the tree in configuration $C'$.  We show that $\rho'$ satisfies
properties (A), (B) and (C) for $x$ in configuration $C'$.

{\bf Property (A)}:
Every process in the image of $\rho'$ was either in the image of $\rho$ or a process that just
entered its cleanup phase at step $S$, so every process in the image of $\rho'$ is in its
cleanup phase.  

{\bf Property (B) and (C)}: We consider several subcases.

{\bf Subcase 5a}
Suppose $S$ is an \func{InsertNew}'s \sct, and $x$ is the red-red violation  created by $S$.
Then, $P$ is in its cleanup phase for the inserted key, which is one of the children of the node containing the red-red violation $x$.
Since the tree is a BST, $x$ is on the search path for this key, so (B) holds.

In this subcase, $location(\rho'(x)) = entry$ since $P=\rho'(x)$ has just entered its cleanup phase.
So property (B) implies property (C1).

{\bf Subcase 5b}
Suppose $S$ is a \func{Delete}'s \sct, and $x$ is the overweight violation assigned to $P$ by $\rho'$. 
Then, $P$ is in a cleanup phase for the deleted key, which was in one of the children of $\ux$ before $S$.
Therefore, $x$ (at the root of the newly inserted subtree) is on the search path for this key, so (B) holds.

As in the previous subcase, $location(\rho'(x)) = entry$ since $P=\rho'(x)$ 
has just entered its cleanup phase.
So property (B) implies property (C1).

{\bf Subcase 5c}
If $x$ is at a node that was added to the data structure by $S$ (and is not covered by the above
two cases), then $\rho'(x)$ is
$\rho(y)$ for some violation $y$ that has been removed from the tree by $S$, as described
in the above two tables.  
Let $k$ be the key such that process $\rho(y)=\rho'(x)$ is in the cleanup phase for $k$.
By property (B), $y$ was on the search path for $k$ before $S$.
It is easy to check by inspection of the tables and Figure~\ref{fig-chromatic-rotations} that 
any search path that went through $y$'s node in configuration $C$ goes through $x$'s node in configuration $C'$.
(We designed the tables to have this property.)
Thus, since $y$ was on the search path for $k$ in configuration $C$, 
$x$ is on the search path for $k$ in configuration $C'$, satisfying property (B).

If (C2) is true for violation $y$ in configuration $C$, then (C2) is true for $x$
in configuration $C'$ (since
$S$ does not affect $location()$ or $parent()$ and $\rho(y)=\rho'(x)$).
If (C3) is true for violation $y$ in configuration $C$, then (C3) is true for $x$ in 
configuration $C'$ (since any node that is finalized remains finalized forever, and its child pointers do not change).

So, for the remainder of the proof of subcase 5c, suppose (C1) is true for $y$ in configuration $C$.
Let $l=location(\rho(y))$ in configuration $C$.
Then $y$ is on the search path for $k$ from $l$ in configuration $C$.

First, suppose $S$ removes $l$ from the data structure.
\begin{compactitem}
\item
If $y$ is a red-red violation at node $l$ in configuration $C$, then
the red-red violation was already there when process $\rho(y)$ read $l$ as the child 
of some other node (by Lemma \ref{no-new-red-red}) and (C2) is true for $x$ in configuration $C'$.
\item
If $y$ is an overweight violation at node $l$ in configuration $C$, then it makes (C3) true for $x$ in configuration $C'$.
\item
Otherwise, 
since both $l$ and its descendant, the parent of the node that contains $y$, are removed by $S$,
the entire path between these two nodes is removed from the data structure by $S$.
%\trevor{It this because $y$ is removed, and a node in the data structure cannot point to a node that has been removed?}
%\eric{It's more like the fact that R is a contiguous subgraph}
So, all nodes along this path are finalized by $S$ because Constraint~\ref{constraint-finalized-iff-removed} is satisfied.  Thus, the violation $y$ makes
(C3) true for $x$ in configuration $C'$.
%\trevor{Regarding which lemma to use, all of the nodes on this removed path have to be in the $R$ sequence of $S$, so this should simply follow from Lemma~\ref{lem-dotreeup-constraints-invariants}.\ref{claim-dotreeup-finalized-before-removed} or, alternatively, the fact that Constraint~\ref{constraint-finalized-iff-removed} is satisfied.}
\end{compactitem}

Now, suppose $S$ does not remove $l$ from the data structure.
In  configuration $C$, the search path from $l$ for $k$ contains $y$.
It is easy to check by inspection of the tables defining $\rho'$ and Figure~\ref{fig-chromatic-rotations} that 
any search path from $l$ that went through $y$'s node in configuration $C$ 
goes through $x$'s node in  configuration  $C'$.
So, (C1) is true in configuration $C'$.

{\bf Subcase 5d}
If $x$ is at a node that was in the data structure in configuration $C$, 
then $\rho'(x)=\rho(x)$.
Let $k$ be the key such that this process is in the cleanup phase for $k$.
Since $x$ was on the search path for $k$ in configuration $C$ 
and $S$ did not remove $x$ from the data structure,
$x$ is still on the search path for $k$ in configuration $C'$ (by inspection of Figure~\ref{fig-chromatic-rotations}).
This establishes property (B).

If (C2) or (C3) is true for $x$ in configuration $C$, 
then it is also true for $x$ in configuration $C'$, for
the same reason as in Subcase 5c. 

So, suppose (C1) is true for $x$ in configuration $C$.
Let $l=location(\rho(x))$ in configuration $C$.  
Then, (C1) says that $x$ is on the search path for $k$ from $l$ in configuration $C$. 
If $S$ does not change any of the child pointers on this path between $l$ and $x$, then
$x$ is still on the search path from $location(\rho'(x)) = l$ in configuration $C'$, 
so property (C1) holds for $x$ in $C'$.
So, suppose $S$ does change the child pointer of some node on this path from $old$ to $new$.
Then the search path from $l$ for $k$ in configuration $C$
goes through $old$ to some node $f$ in the Fringe set 
$F$ of $S$ and then onward to the node containing violation $x$.
By inspection of the transformations in Figure~\ref{fig-chromatic-rotations},
the search path for $k$ from $l$ in configuration $C'$ 
goes through $new$ to the same node $f$, and then
onward to the node containing the violation $x$.
Thus, property (C1) is true for $x$ in configuration $C'$.
\end{chapscxproof}

\begin{cor}
\label{violation-bound}
The number of violations in the data structure is bounded by the number of incomplete \ins\ and \del\ operations.
\end{cor}

%\eric{Trevor: I did some editing from here to the end.  I strengthened the following lemma's claim 1, simplified the proof. 
%I also modified the proof of claim 2 because it wasn't clear before which node was $topmost$,
%so I rephrased.
% Remove this comment if you approve.}

In the following discussion, we are discussing ``pure'' chromatic trees, without the dummy
nodes with key $\infty$ that appear at the top of our tree.
The sum of weights on a path from the root to a leaf of a chromatic tree
is called the \textit{path weight} of that leaf.
The {\it height} of a node $v$, denoted $h(v)$ is the maximum number of nodes on a path from $v$ to a leaf descendant of $v$.
We also define the \textit{weighted height} of a node $v$ as follows.
 \begin{displaymath}
   wh(v) = \left\{
     \begin{array}{ll}
       v.w & \mbox{if } v \mbox{ is a leaf} \\
       \max(wh(v.left), wh(v.right))+v.w & \mbox{otherwise}
     \end{array}
   \right.
  \end{displaymath}
%Since the path weights of all leaves of a chromatic tree are equal, we can simplify $wh(v)$.
% \begin{displaymath}
%   wh(v) = \left\{
%     \begin{array}{ll}
%       v.w & \mbox{if }v \mbox{ is a leaf} \\
%       wh(v.left)+v.w &  \mbox{otherwise}
%     \end{array}
%   \right.
%  \end{displaymath}

\begin{lem} \label{lem-chromatic-claims}
Consider a chromatic tree rooted at $root$ that contains $n$ nodes and $c$ violations.
Suppose $T$ is any red black tree rooted at $root_T$ that results from performing a sequence 
of rebalancing steps on the tree rooted at $root$ to eliminate all violations.
Then, the following claims hold.
\begin{enumerate}
\item $h(root) \le 2wh(root) + c$
\label{claim-chromatic-h-wh}
%\item $wh(root)$ changes by at most $c$ as a result of fixing all violations
%\label{claim-chromatic-wh-changes-by-at-most-c}
\item $wh(root) \le wh(root_T) + c$
\label{claim-chromatic-wh-whT}
\item $wh(root_T) \le h(root_T)$
\label{claim-chromatic-whT-hT}
\end{enumerate}
\end{lem}
\begin{chapscxproof}
\textbf{Claim~\ref{claim-chromatic-h-wh}:}
Consider any path from $root$ to a leaf.
It has at most $wh(root)$ non-red nodes.
So, there can be at most $wh(root)$ red nodes that do not have red parents on the path (since $root$ has weight 1).
There are at most $c$ red nodes on the path that have red parents.
So the total number of nodes on the path is at most $2wh(root)+c$.

\textbf{Claim~\ref{claim-chromatic-wh-whT}:}
Consider any rebalancing step that is performed by replacing some node 
$\ux$ by $n$ (using the notation of Figure~\ref{fig-chromatic-rotations}).
If $\ux$ is not the root of the chromatic tree, then $wh(\ux)=wh(n)$,
since the path weights of all leaves in a chromatic tree must be equal. 
(Otherwise, the path weight to a leaf in the subtree rooted at $n$ would become
different from the path weight to a leaf outside this subtree.)

Thus, the only rebalancing steps that can change the weighted height of the root are those
where $\ux$ is the root of the tree.
Recall that the weight of the root is always one.
If $\ux$ and $n$ are supposed to have different weights according to Figure~\ref{fig-chromatic-rotations},
then blindly setting the weight of the 
$n$ to one will have the effect of changing the weighted height of the root.
By inspection of Figure~\ref{fig-chromatic-rotations}, the only transformation that increases
the weighted height of the root is BLK, because it is the only transformation where
the weight of $n$ is supposed to be less than the weight of $\ux$.
%only \textsc{BLK}, \textsc{W7}, and \textsc{Push} where the weights of $\ux$ and $n$ are different. 
%By trivial inspection of the rebalancing steps, only \textsc{red-push1}, \textsc{red-push2}, \textsc{weight-dec3} and \textsc{weight-push} can change the weighted height of the root.
%Furthermore, these rebalancing steps increase or decrease the weighted height of the root by one, and %, in doing so,
Thus, each application of BLK at the root increases the weighted height of the root by one, but also
eliminates at least one red-red violation at a grandchild of the root (without introducing any new violations).  Since none of the rebalancing
steps increases the number of violations in the tree,
%reduce the number of violations in the tree by at least one.
%(Since the topmost node is $root$, none of these rebalancing steps change the weight of $root$.
%With this fact, it is easy to prove that these rebalancing steps do not create any new violations.
%We now describe the violations that these rebalancing steps remove.
%\textsc{BLK} eliminates a red-red violation at a grandchild of $root$, \textsc{W7} eliminates overweight violations at the left and right children of $root$, and \textsc{Push} eliminates an overweight violation at the left child of $root$.)
performing any sequence of steps that eliminates $c$ violations will change the weighted height of the root by at most $c$. %result in the weighted height of the root increasing or decreasing by at most $c$.
The claim then follows from the fact that $T$ is produced by eliminating $c$ violations from the chromatic tree rooted at $root$.

\textbf{Claim~\ref{claim-chromatic-whT-hT}:}
Since $T$ is a RBT, it contains no overweight violations. 
Thus, the weighted height of the tree is a sum of zeros and ones. 
It follows that $wh(root_T) \le h(root_T)$.
\end{chapscxproof}

\begin{cor} \label{cor-chromatic-height-and-violations}
If there are $c$ incomplete \ins\ and \del\ operations and the data structure contains
$n$ keys, then its height is $O(log\ n + c)$.
\end{cor}
\begin{chapscxproof}
Let $root$, $T$, and $root_T$ be defined as in Lemma~\ref{lem-chromatic-claims}.
We immediately obtain $h(root) \le 2h(root_T) + 3c$  from Corollary \ref{violation-bound} and Lemma~\ref{lem-chromatic-claims}.  Since the height of a RBT is $O(\log n)$, it follows that the height of our data structure is $O(\log n + c)$ (including the two dummy nodes at the top of the tree with key $\infty$).
\end{chapscxproof}

%\trevor{Describe what \tryrebalance\ does before calling \func{DoRB2}: some traversal/search (well, that's really in \cleanup), necessary \llt s.}
%%\trevor{Should I rewrite \func{DoRB2} to perform some traversal/search, and all of its necessary \llt s, or should I mention that these things are done by \tryrebalance\ before calling \func{DoRB2}?}
%\trevor{Should add some sort of rudimentary arguments to show that $V$, $R$ and $n$ satisfy the constraints for the tree update template?}%  I guess we should probably discuss why the various constraints are satisfied...}

%\begin{enumerate}
%\item When we rotate
%\item How we choose which rebalancing step to perform (designed decision tree to move violations up on the search path)
%\item Why violations get eliminated
%\item How to implement with \sct
%\item Structure of insert/delete/cleanup (repeatedly: search, try ex, exit condition varies)
%\item Explain why height is good
%\end{enumerate}

%\vspace{-5mm}

\section{Allowing more violations}

Forcing insertions and deletions to rebalance the chromatic tree after creating only a single violation can cause unnecessary rebalancing steps to be performed, for example, because an overweight violation created by a deletion might be eliminated by a subsequent insertion.
In practice, we can reduce the total number of rebalancing steps that occur by modifying our \ins\ and \del\ procedures so that \cleanup\ is invoked only once the number of violations on a path from $entry$ to a leaf exceeds some constant $k$.
The resulting data structure has height $O(k + c + \log n)$.
%\trevor{I think this might actually be $O(k + c + \log n)$...  I think it should be possible to show exactly $2+k+c+2\lceil\log n\rceil$ (but maybe hard to get $k+c$ and not $3(k+c)$).}
Since searches are significantly faster than updates, slightly increasing search costs to reduce update costs yields performance benefits for many workloads.
%Our experiments show that this technique yields moderate performance improvements for workloads that include updates.

\section{Experimental results} \label{sec-chromatic-exp}

\begin{figure*}[tb]
\def\darkness{45}
\def\expscale{0.75}
\def\expleftwidth{0.03\textwidth}
\centering
\hspace{-1cm}
\begin{minipage}{\expleftwidth}
%\hspace{3mm}
%\vspace{3mm}
\end{minipage}
\begin{minipage}{0.333\textwidth}
\small
\centering
\hspace{6mm}
\textbf{50\% \func{Ins}, 50\% \func{Del}, 0\% \func{Get}}
\end{minipage}
\begin{minipage}{0.333\textwidth}
\small
\centering
\hspace{4.5mm}
\textbf{20\% \func{Ins}, 10\% \func{Del}, 70\% \func{Get}}
\end{minipage}
\begin{minipage}{0.333\textwidth}
\small
\centering
\hspace{6mm}
\textbf{0\% \func{Ins}, 0\% \func{Del}, 100\% \func{Get}}
\end{minipage}\\
\hspace{-1cm}
\begin{minipage}{\expleftwidth}
\centering
\vspace{-2mm}
\includegraphics[scale=\expscale]{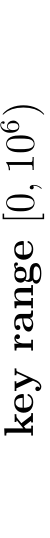}
\vspace{2mm}

\vspace{-3mm}
\includegraphics[scale=\expscale]{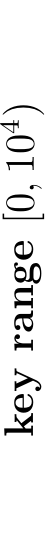}
\vspace{3mm}

\vspace{-6mm}
\includegraphics[scale=\expscale]{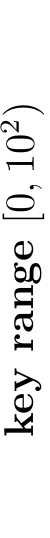}
\vspace{6mm}
\end{minipage}
\begin{minipage}{0.333\textwidth}
\centering
\includegraphics[scale=\expscale]{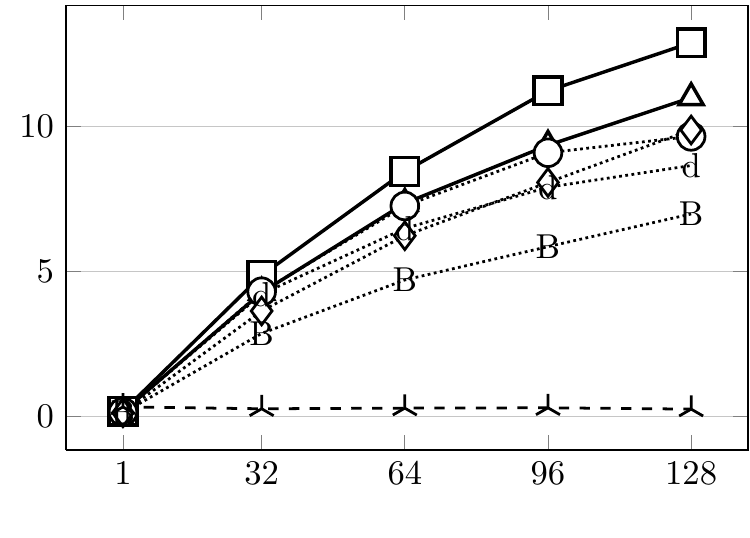}

\vspace{-2mm}
\includegraphics[scale=\expscale]{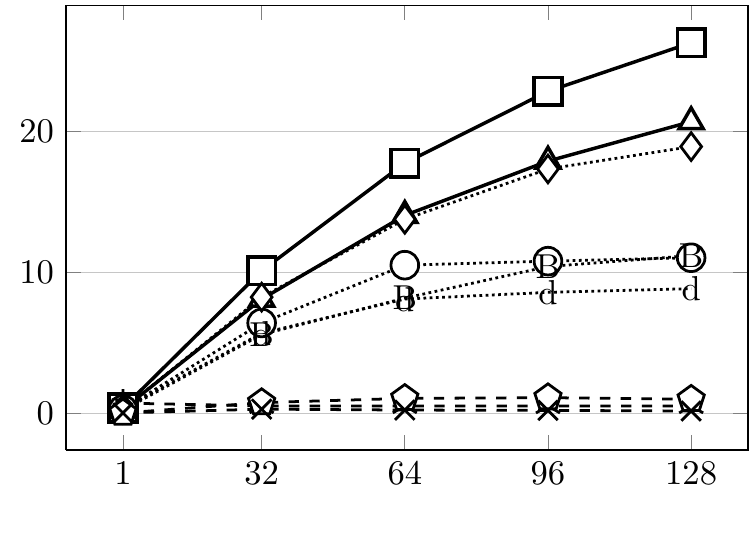}

\vspace{-2mm}
\includegraphics[scale=\expscale]{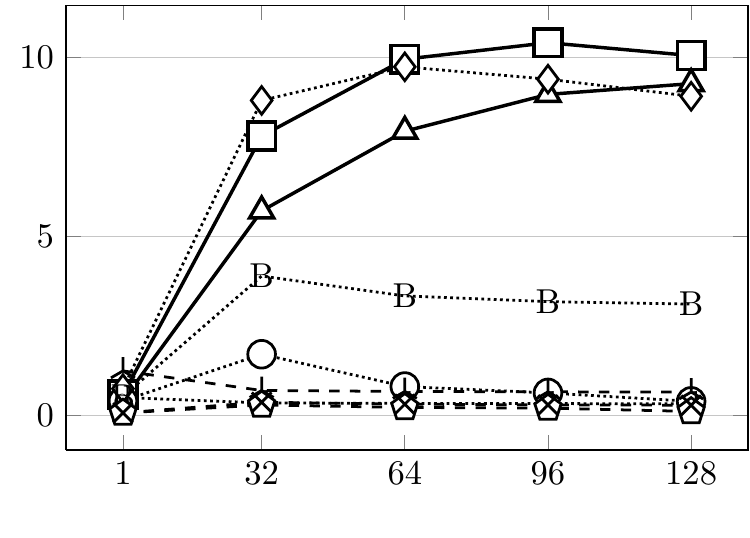}
\end{minipage}
\begin{minipage}{0.333\textwidth}
\centering
\vspace{-1.45mm}
\includegraphics[scale=\expscale]{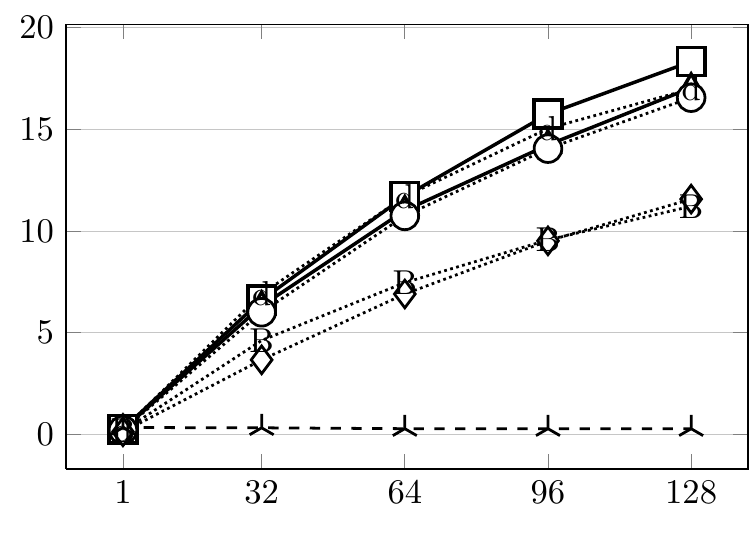}
\vspace{1.45mm}

\vspace{-2mm}
\includegraphics[scale=\expscale]{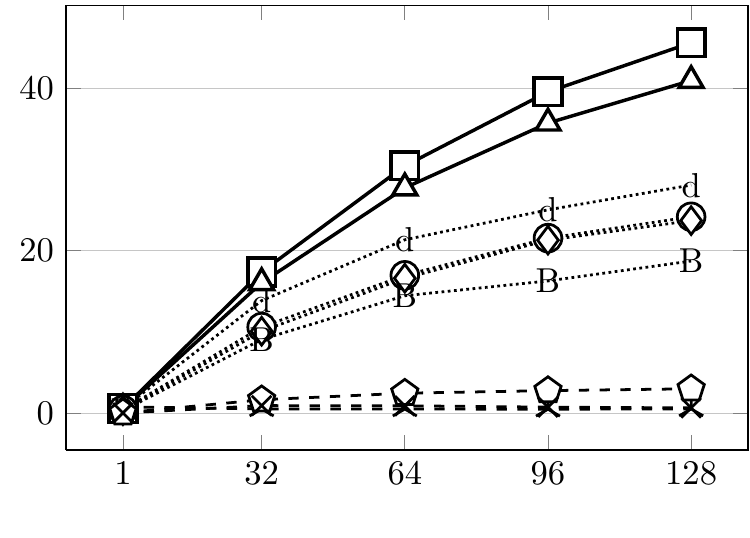}

\vspace{-2mm}
\includegraphics[scale=\expscale]{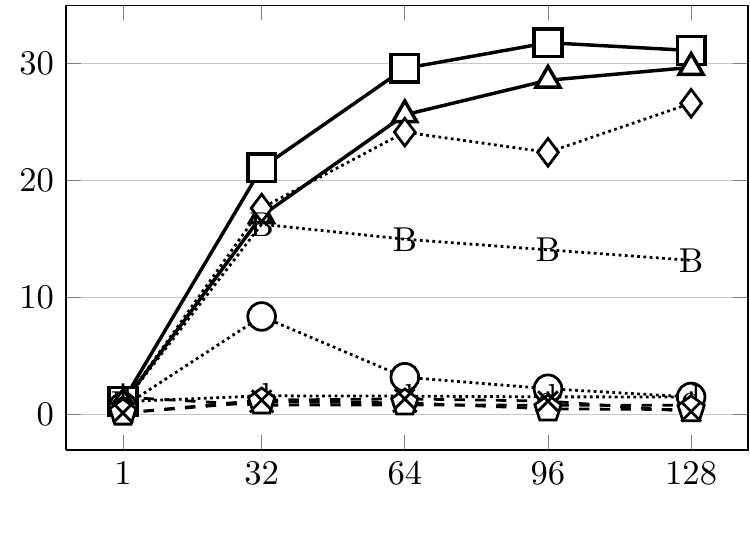}
\end{minipage}
\begin{minipage}{0.333\textwidth}
\centering
\includegraphics[scale=\expscale]{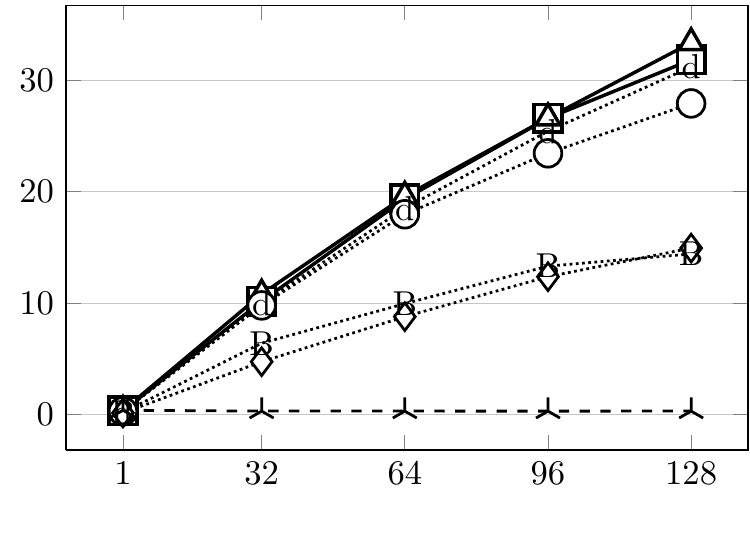}

\vspace{-2mm}
\vspace{-0.88mm}
\includegraphics[scale=\expscale]{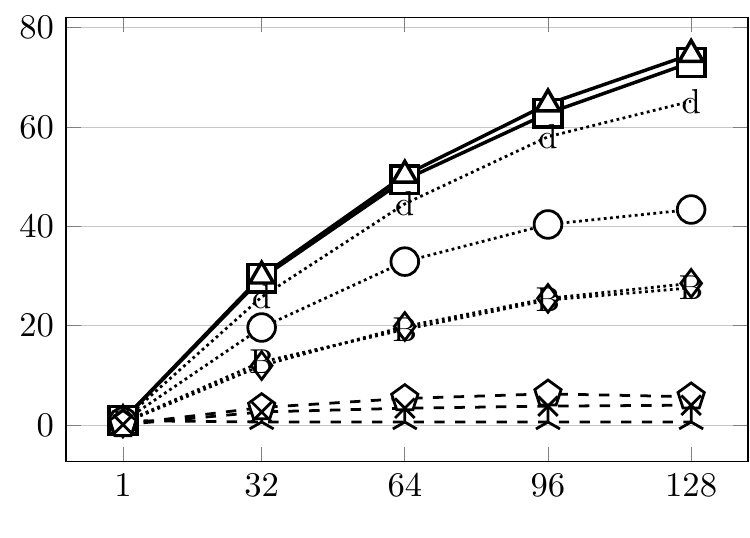}
\vspace{0.88mm}

\vspace{-2mm}
\includegraphics[scale=\expscale]{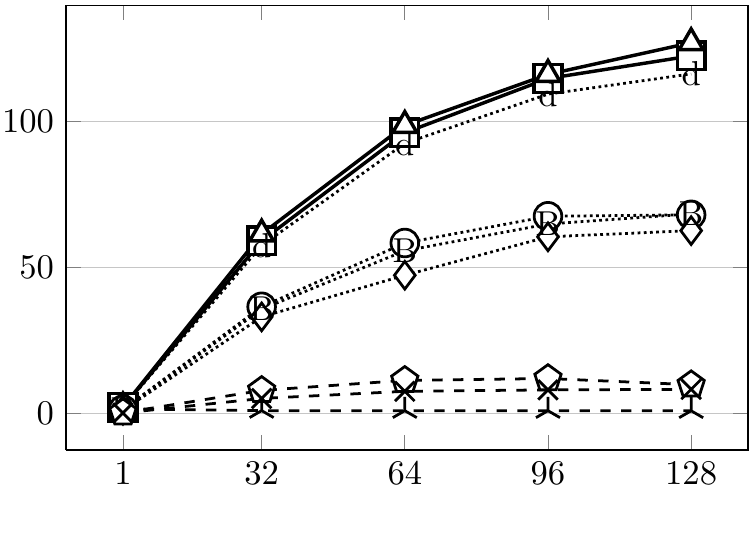}
\end{minipage}\\
\vspace{-4mm}
\includegraphics[scale=0.15]{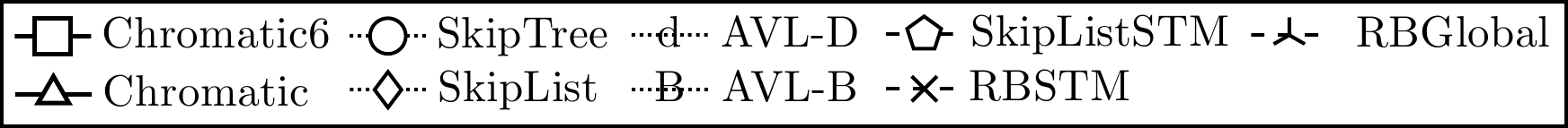}
\vspace{-1mm}
\caption{\textit{Multithreaded} throughput (millions of operations/second) for 2-socket SPARC T2+ (128 hardware threads) on y-axis versus number of threads on x-axis.}
\label{fig-experiments}
\vspace{-1mm}
\end{figure*}

We compared the performance of our chromatic tree (Chromatic), and the variant of our chromatic tree that invokes \cleanup\ only when the number of violations on a path exceeds six (Chromatic6), against several leading data structures that implement ordered dictionaries:
the non-blocking skip-list (SkipList) of the Java Class Library,
the non-blocking multiway search tree (SkipTree) of Spiegel and Reynolds~\cite{SR10},
the lock-based relaxed-balance AVL tree with non-blocking searches (AVL-D) of Drachsler et~al.~\cite{Drachsler2014}, and
the lock-based relaxed-balance AVL tree (AVL-B) of Bronson et~al.~\cite{BCCO10:ppopp}.
Our comparison also includes an STM-based red-black tree optimized by Oracle engineers (RBSTM) \cite{Herlihy2003}, an STM-based skip-list (SkipListSTM), and the highly optimized Java red-black tree, \texttt{java.util.TreeMap}, with operations protected by a global lock (RBGlobal).
The STM data structures are implemented using DeuceSTM 1.3.0, which is one of the fastest STM implementations that does not require modifications to the Java virtual machine.
We used the offline instrumentation capability of DeuceSTM to minimize the overhead of the STM instrumentation. %at running time that might skew our results
All of the implementations used were made publicly available by their respective authors.
For a fair comparison between data structures, we made slight modifications to RBSTM and SkipListSTM to use generics, instead of hardcoding the type of keys as \texttt{int}, and to store values in addition to keys. %implement the dictionary ADT, rather than the set ADT.
%These changes decreased the throughput of these algorithms by approximately 14\%.
\begin{shortver}
Java code for Chromatic and Chromatic6 is available from
\mbox{\url{http://implementations.tbrown.pro}}.
\end{shortver}
%\trevor{Should we create an anonymous URL for the submission?}
%\cite{paper2}.  It's weird for a paper to cite itself.
%\eric{Remember to add this to bib file}

We tested the data structures for three different operation mixes, 0i-0d, 20i-10d and 50i-50d, where $x$i-$y$d denotes $x$\% \ins s, $y$\% \del s, and $(100-x-y)$\% \func{Get}s.
These operation mixes represent workloads where:
all operations are queries, 
a moderate proportion operations are \ins s and \del s, and 
all operations are \ins s and \del s. %, respectively.
%(We also performed experiments for 5i-5d, with similar results to 0i-0d.)
We used three key ranges, $[0,10^2), [0,10^4)$ and $[0,10^6)$, to test different contention levels.
%, since the data structures can only grow roughly as large as the number of keys they contain.
For example, for key range $[0,10^2)$, data structures will be small, so 
%processes performing operations on uniformly random keys are likely to interfere with one another.
updates are likely to affect overlapping parts of the data structure.

For each data structure, each operation mix, each key range, and each thread count in 
\{1, 32, 64, 96, 128\}, we ran five trials which each measured the total throughput (operations per second) of all threads for five seconds.
Each trial began with an untimed prefilling phase, which continued until the data structure was within 5\% of its expected size in the steady state.
For operation mix 50i-50d, the expected size is half of the key range.
This is because, eventually, each key in the key range has been inserted or deleted at least once, and the last operation on any key is equally likely to be an insertion (in which case it is in the data structure) or a deletion (in which case it is not in the data structure).
Similarly, 20i-10d yields an expected size of two thirds of the key range since, eventually, each key has been inserted or deleted and the last
operation
%update
on a particular key is twice as likely to be an insertion as a deletion.
For 0i-0d, we prefilled the data structures so that they contain half of the key range.

%by performing random operations with operation mix 50i-50d, on uniformly random keys drawn from the 
%specified key range.
%\eric{Does it make sense to have the structure half-full for the 20i-10d case? (Why didn't you use 15i-15d in that case anyway?)}
%Each trial then ran for five seconds, and measured the total throughput (operations per second) of all threads, as each thread performed random operations according to the specified operation mix, on uniformly random keys. % in the specified range.

We used a Sun Enterprise T5240 with 32GB of RAM and two UltraSPARC T2+ processors, for a total of 16 $\times$ 1.2GHz cores supporting a total of 128 hardware threads.
%We were interested in measuring true concurrency, so we focused on the performance up to 16 threads.
%We also included results for 32 threads to see how performance changes when threads must share cores.
The Sun 64-bit JVM version 1.7.0\_03 was run in server mode, with 3GB minimum and maximum heap sizes.
Different experiments run within a single instance of a Java virtual machine (JVM) are not statistically independent, so each batch of five trials was run in its own JVM instance.
Prior to running each batch, a fixed set of three trials was run to cause the Java HotSpot compiler to optimize the running code.
Garbage collection was manually triggered before each trial.
The heap size of 3GB was small enough that garbage collection was performed regularly (approximately ten times) in each trial.
%This allowed us to measure the true performance of each algorithm, including the impact of garbage collection in the steady state.
We did not pin threads to cores, since this is unlikely to occur in practice. %threads will typically not be pinned in practice.
%We decided on a heap size of 1GB after performing preliminary experiments to find a heap size that was 
%small enough to regularly trigger garbage collection and large enough to keep standard deviations small.
%Since production software often has memory constraints, we believe allowing regular garbage collection 
%yields more realistic results.
%If the heap size is large enough that garbage collection is not performed, then the performance of 
%Chromatic further increases, relative to the other algorithms.

Figure~\ref{fig-experiments} shows our experimental results.
Our algorithms are drawn with solid lines.
Competing handcrafted implementations are drawn with dotted lines.
Implementations with coarse-grained synchronization are drawn with dashed lines.
Error bars are not drawn because they are mostly too small to see: The standard deviation is less than 2\% of the mean for half of the data points, and less than 10\% of the mean for 95\% of the data points.
The STM data structures are not included in the graphs for key range $[0, 10^6)$, because of the enormous length of time needed just to perform prefilling (more than 120 seconds per five second trial).

Chromatic6 nearly always outperforms Chromatic.
The only exception is for an all query workload, where Chromatic performs slightly better.
Chromatic6 is prefilled with the Chromatic6 insertion and deletion algorithms, so it has a slightly larger average leaf depth than Chromatic; this accounts for the performance difference.
In every graph, Chromatic6 rivals or outperforms the other data structures, even the highly optimized implementations of SkipList and SkipTree which were crafted with the help of Doug Lea and %members of 
the Java Community Process JSR-166 Expert Group.
%The sole exception is under high contention (key range $[0, 10^2)$) with
%%a large number of updates
%only \ins\ and \del\ operations (50i-50d), where SkipList outperforms the others
%% data structures
%by a wide margin, due to its simplicity.
%%In this case, the relative simplicity of the SkipList data structure allows it to perform well.
%Chromatic is the only other data structure that scales well to 32 threads in this case.
%SkipTree likely suffers because the large degree of its nodes decreases possible concurrency, 
%while AVL's lock-based design requires threads to wait. % for locks.
Under high contention (key range $[0,10^2)$), Chromatic6 outperforms every competing data structure except for SkipList in case 50i-50d and AVL-D in case 0i-0d.
In the former case, SkipList %achieves less than half the throughput of Chromatic6 for an all query workload, it
approaches the performance of Chromatic6 when there are many \ins s and \del s, due to the simplicity of its updates.
In the latter case, the non-blocking searches of AVL-D allow it to perform nearly as well as Chromatic6; this is also evident for the other two key ranges.
SkipTree, AVL-D and AVL-B all experience negative scaling beyond 32 threads when there are updates.
For SkipTree, this is because its nodes contain many child pointers, and processes modify a node by replacing it (severely limiting concurrency when the tree is small).
For AVL-D and AVL-B, this is likely because processes waste time waiting for locks to be released when they perform updates.
%AVL-B must additionally acquired locks while the tree is being traversed to find a location to modify.
%
Under moderate contention (key range $[0,10^4)$), in cases 50i-50d and 20i-10d, Chromatic6 significantly outperforms the other data structures. % for most thread counts.
Under low contention, the advantages of a non-blocking approach are less pronounced, but Chromatic6 is still at the top of each graph (likely because of low overhead and searches that ignore updates).
%data structures all scale well, and Chromatic and SkipTree 
%%compares favourably with the other data structures, 
%outperform SkipList and~AVL.
%(The differences in case 50i-50d are not statistically significant.)
%The fact that Chromatic rivals or outperforms SkipList and SkipTree in most cases is particularly impressive, since their highly optimized implementations were crafted with the help of Doug Lea and members of the Java Community Process JSR-166 Expert Group. %, whereas relatively little effort has been spent optimizing Chromatic.

\begin{figure}[tb]
\vspace{-3mm}
\centering
\includegraphics[scale=0.33]{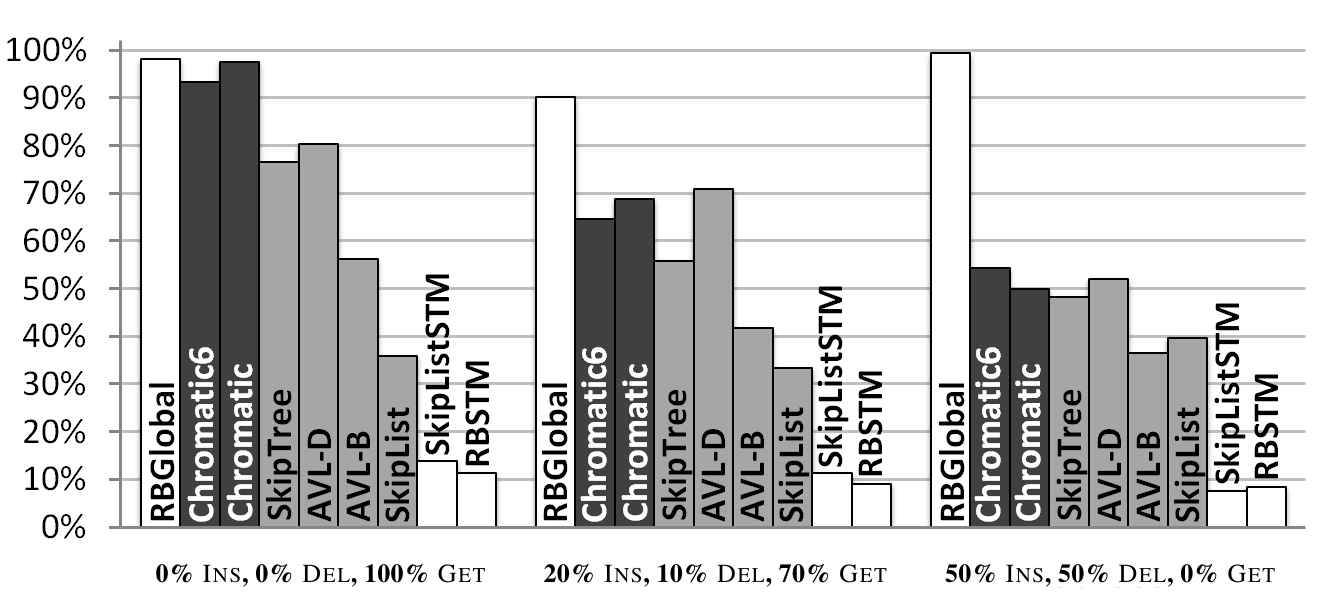}
%\vspace{-5mm}
\caption{\textit{Single threaded} throughput of the data structures relative to Java's sequential RBT for key range $[0,10^6)$. % . \trevor{Add the phrasing ``single threaded'' to the y-axis, and reformat the workload titles on the x-axis to match the other graphs.}
}
\label{fig-single-thread}
%\vspace{-2mm}
\end{figure}

Figure~\ref{fig-single-thread} compares the single-threaded performance of the data structures, relative to that of the sequential RBT, \texttt{java.util.TreeMap}.
This demonstrates that the overhead introduced by our technique is relatively small.
%Over all workloads, Chromatic6 and Chromatic rival or outperform the other hand crafted data structures.
%For the all query and moderate update workloads, Chromatic6 and Chromatic perform almost as well as RBGlobal.
%This makes Chromatic6 and Chromatic an attractive option for low contention applications with few or moderately many updates, which are common in practice.
%\trevor{Should maybe give a better discussion of the various cases, and some attempt to explain why we are faster.}

Although balanced BSTs are designed to
give performance guarantees for worst-case sequences of operations,
the experiments are performed using random sequences.
For such sequences, BSTs \textit{without} rebalancing operations
are balanced with high probability and, hence, will have better performance
because of their lower overhead.
Thus, better experiments are needed to evaluate balanced BSTs.

\begin{figure}[tb]
\centering
\includegraphics[width=1\textwidth]{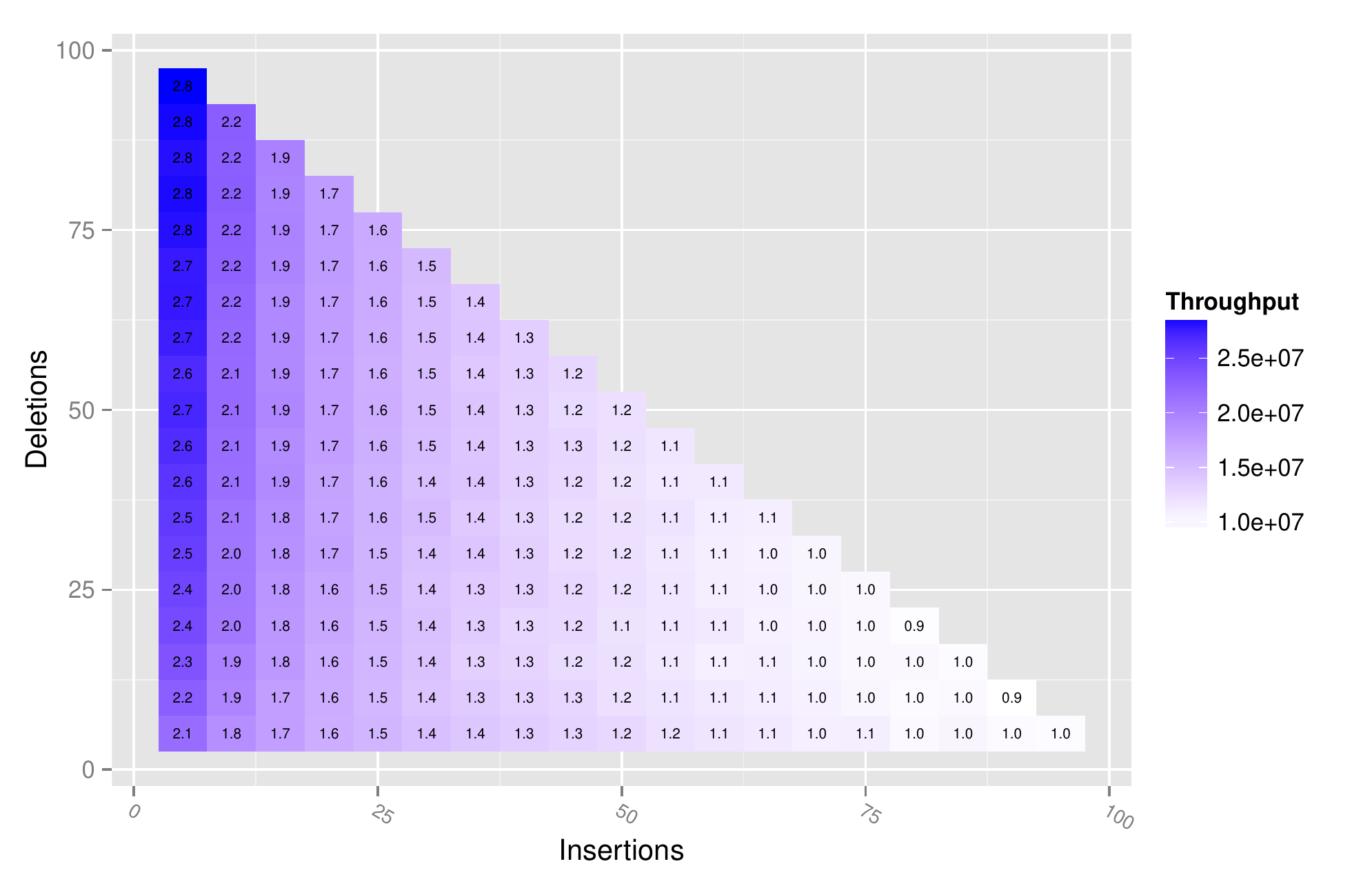}
\caption{Heatmap showing throughput (operations per second) for \textbf{Chromatic6} over a wide variety of operation mixes, with key range $[0,10^6)$ and 128 threads.}
\label{fig-heatmap-chromatic}
\end{figure}

\begin{figure}[tb]
\centering
\includegraphics[width=1\textwidth]{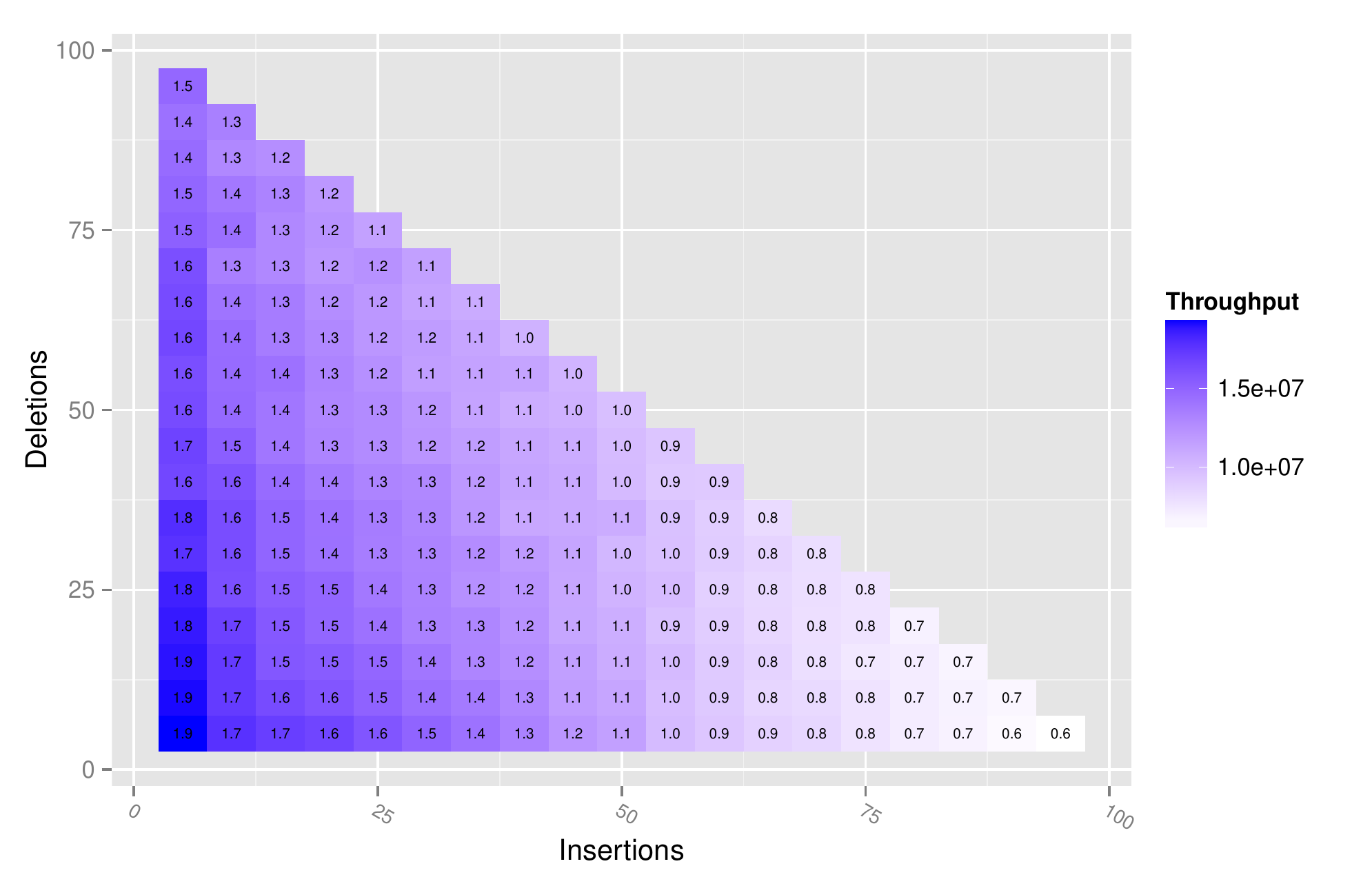}
\caption{Heatmap showing throughput (operations per second) for \textbf{SkipTree} over a wide variety of operation mixes, with key range $[0,10^6)$ and 128 threads.}
\label{fig-heatmap-skiptree}
\end{figure}

\begin{figure}[tb]
\centering
\includegraphics[width=1\textwidth]{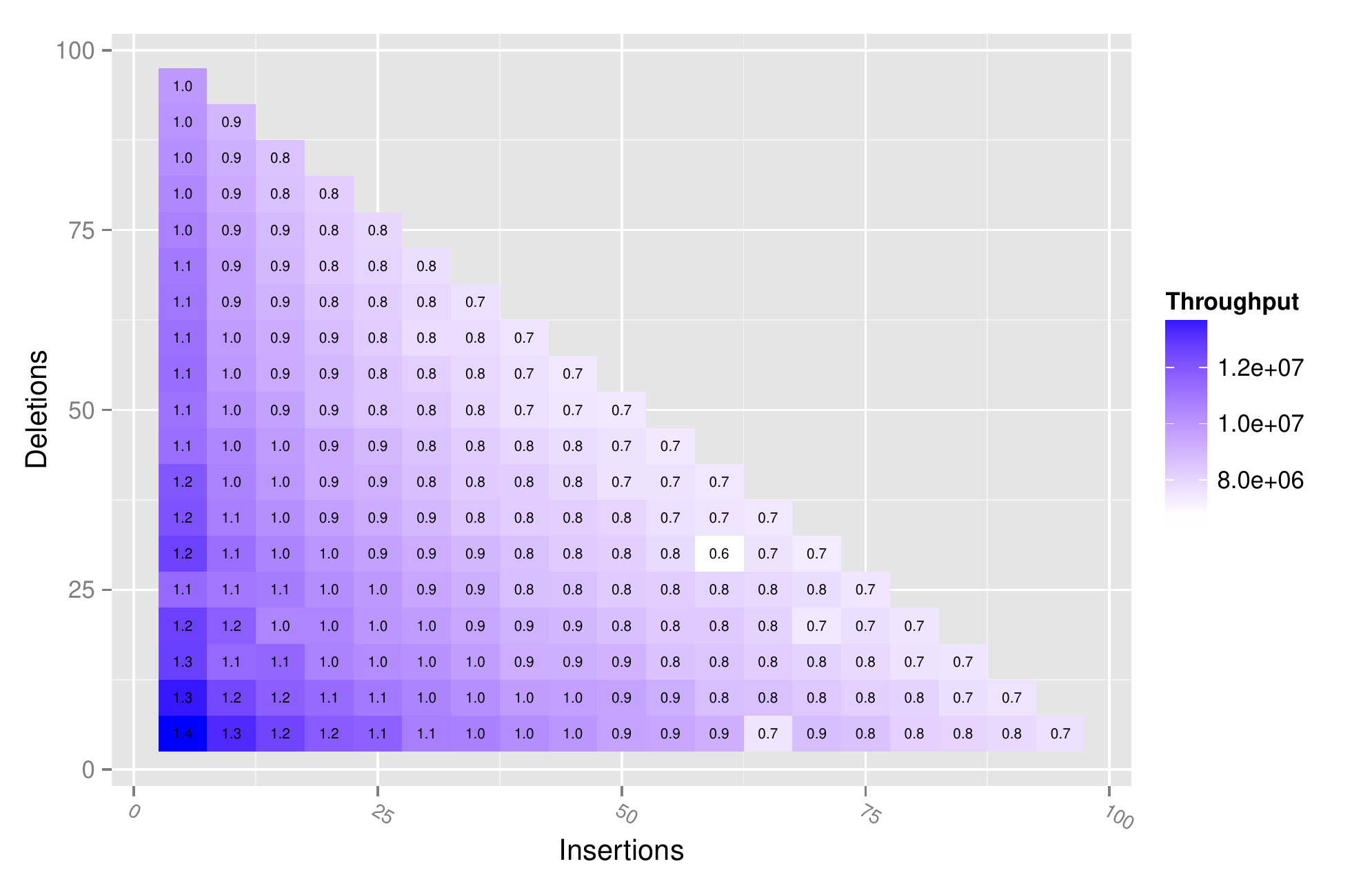}
\caption{Heatmap showing throughput (operations per second) for \textbf{AVL-B} over a wide variety of operation mixes, with key range $[0,10^6)$ and 128 threads.}
\label{fig-heatmap-avlb}
\end{figure}

\paragraph{Exploring additional operation mixes}
The results above explored a fairly limited set of operation mixes.
In the interest of obtaining a more complete picture of the performance of a subset of these algorithms, we also performed some supplementary experiments for a wider variety of operation mixes.
Specifically, for each of the data structures in \{Chromatic6, SkipTree, AVL-B\}, we ran trials of the same sort that we performed above, with key range $[0,10^6)$ and 128 threads, for \textit{all} operation mixes with $x$\% insertions and $y$\% deletions, where $x, y \in \{5, 10, 15, ..., 95\}$ and $x+y \le 100$.
The results for each data structure appear in Figure~\ref{fig-heatmap-chromatic}, %Figure~\ref{fig-heatmap-skiplist},
Figure~\ref{fig-heatmap-skiptree} and Figure~\ref{fig-heatmap-avlb}.

Broadly, the results show that throughput increases as the fraction of operations that are searches increases.
The results additionally show that throughput of Chromatic6 increases as the ratio of deletions to insertions increases (causing the size of the tree to be smaller in the steady state).
This demonstrates that Chromatic6 scales well, even as contention increases.
However, the results for SkipTree and AVL-B do not show this effect. Instead, performance \textit{decreases} as the ratio of deletions to insertions increases.
For SkipTree, this is because nodes contain many keys, so a updates in a small tree can contend on a large fraction of the data structure.
For AVL-B, we believe this is due to increased contention as a result of hand-over-hand locking, which must always start at the root.

%\begin{figure}[tb]
%\centering
%\includegraphics[width=1\textwidth]{chap-template/chromatic/heatmap-nozero-sl.pdf}
%\caption{Heatmap showing throughput (operations per microsecond) for \textbf{SkipList} over a wide variety of operation mixes, with key range $[0,10^6)$ and 128 threads.}
%\label{fig-heatmap-skiplist}
%\end{figure}

\section{Summary} \label{sec-chromatic-conclusion}

In this chapter, we demonstrated the use of our tree update template %that can be used to obtain non-blocking implementations of any data structure based on a down-tree, and demonstrated
%We demonstrated 
%its use
by implementing a non-blocking chromatic tree.
To the authors' knowledge, this is the first provably correct, non-blocking balanced BST with fine-grained synchronization.
Proving the correctness of a direct implementation of a chromatic tree from hardware primitives such as CAS would have been completely intractable.
By developing our template abstraction and the chromatic tree in tandem, we were able to minimize overhead, so the chromatic tree is very efficient.
%We assume safe garbage collection, which is provided by managed languages like Java and C\#, but it should be possible to eliminate this assumption by using, for example, the new efficient memory reclamation scheme of Aghazadeh et~al. \citep{AGW13}.
%Our experiments show that a Java implementation of our chromatic tree outperforms other known, highly tuned concurrent data structures that implement the ordered dictionary ADT.
%
%

We hope that this work sparks interest in developing more relaxed-balance sequential versions of data structures, since it is now easy to obtain efficient concurrent implementations of them using our template.
We also hope this work leads to a greater diversity of advanced lock-free balanced trees.
Towards this end, we present four different lock-free balanced trees in this thesis.
Additionally, since the paper that presented our tree update template and lock-free chromatic tree was first published at PPoPP 2014, the \textit{weak AVL trees} of Haeupler, Sen and Tarjan~\cite{Haeupler15} have been implemented using the tree update template by He and Li~\cite{He:2016}.

\chapter{Relaxed AVL tree implemented with the template} \label{chap-ravl}
% !TEX root = paper.tex

%In this chapter, we describe some other data structures that either have been, or could be, implemented using the template in Chapter~\ref{chap-template}.
%Here, we give only a high level overview of how each could be implemented, with the understanding that the implementation details would be similar to those of the chromatic tree (Chapter~\ref{chap-chromatree}), and the relaxed $(a,b)$-tree (Chapter~\ref{chap-abtree}).
%
%\section{AVL trees with relaxed balance}

Several papers have proposed more concurrency friendly versions of AVL trees that allow rebalancing steps to be decoupled from insertions and deletions~\cite{BGMS98, larsen1994avl, Lar00, nurmi1996relaxed}.
This decoupling allows the tree to temporarily become unbalanced while insertions and deletions are in progress, which allows a greater degree of concurrency.
We consider the relaxation proposed by Larsen~\cite{larsen1994avl}.
However, the other proposals could also be implemented using our template.
Given a copy of~\cite{larsen1994avl} and a description of the tree update template, a first year undergraduate student produced a Java implementation of a relaxed-balance AVL tree in less than a week.
We start by defining AVL trees, and then describe Larsen's relaxation.

\section{AVL trees}
An AVL tree~\cite{adelsonvelskii1963algorithm} is a balanced search tree with stricter balance than a red-black tree (or chromatic tree).
Each \textit{internal} node $r$ in an AVL tree has a \textit{balance factor} $bf(r) = h(r.left) - h(r.right)$, where $h(r)$ is the height of $r$, defined as follows.
 \begin{displaymath}
   h(r) = \left\{
     \begin{array}{ll}
       0 & : r \mbox{ is a leaf} \\
       max\{h(r.left), h(r.right)\}+1 & : \mbox{otherwise}
     \end{array}
   \right.
  \end{displaymath}
Balance is maintained in an AVL tree with the following invariant.

\noindent\textbf{AVL balance invariant:} $bf(r) \in \{-1, 0, 1\}$ for all nodes $r$ in the tree.

\noindent In other words, in an AVL tree, the heights of the left and right subtrees of each node can differ by at most one.
This invariant yields a tree with height at most $\log_{\phi} (\sqrt{5} (n+2)) - 3$ where $\phi$ is the golden ratio, and $n$ is the number of keys in the tree (see pp.460 of \cite{Knuth:1998:ACP:280635}).
This is approximately $1.44 \log_2 (n+2)$, which is somewhat better than the upper bound of $2 \log_n (n+1)$ on the height of a red-black tree.

\section{Relaxed AVL trees}
In a relaxed AVL (RAVL) tree, each node $r$ has an integer \textit{tag} that is zero if $r$ is the root and, otherwise, satisfies $r.tag \ge -1$ for internal nodes and $r.tag \ge 1$ for leaves.
Similar to balance factors in an AVL tree, each \textit{internal} node $r$ in a RAVL tree has a \textit{relaxed balance factor} $rbf(r) = rh(r.left) - rh(r.right)$, where $rh(r)$ is the \textit{relaxed height} of $r$, defined as follows.
 \begin{displaymath}
   rh(r) = \left\{
     \begin{array}{ll}
       r.tag & : r \mbox{ is a leaf} \\
       max\{rh(r.left), rh(r.right)\}+1+r.tag & : \mbox{otherwise}
     \end{array}
   \right.
  \end{displaymath}
Intuitively, positive (resp., negative) tags let a node pretend that it is the root of a taller (resp., shorter) subtree.
Note that a node can pretend to be the root of a \textit{much} taller subtree, but it can only pretend to be the root of a slightly shorter subtree, since $tag \ge -1$.
In analogy to AVL trees, we have the following balance invariant.

\noindent\textbf{RAVL balance invariant:} $rbf(r) \in \{-1, 0, 1\}$ for all nodes $r$ in the tree.

\begin{figure}[tb]
\hspace{-5mm}
\begin{minipage}{0.49\textwidth}
\centering
\includegraphics[width=1\linewidth]{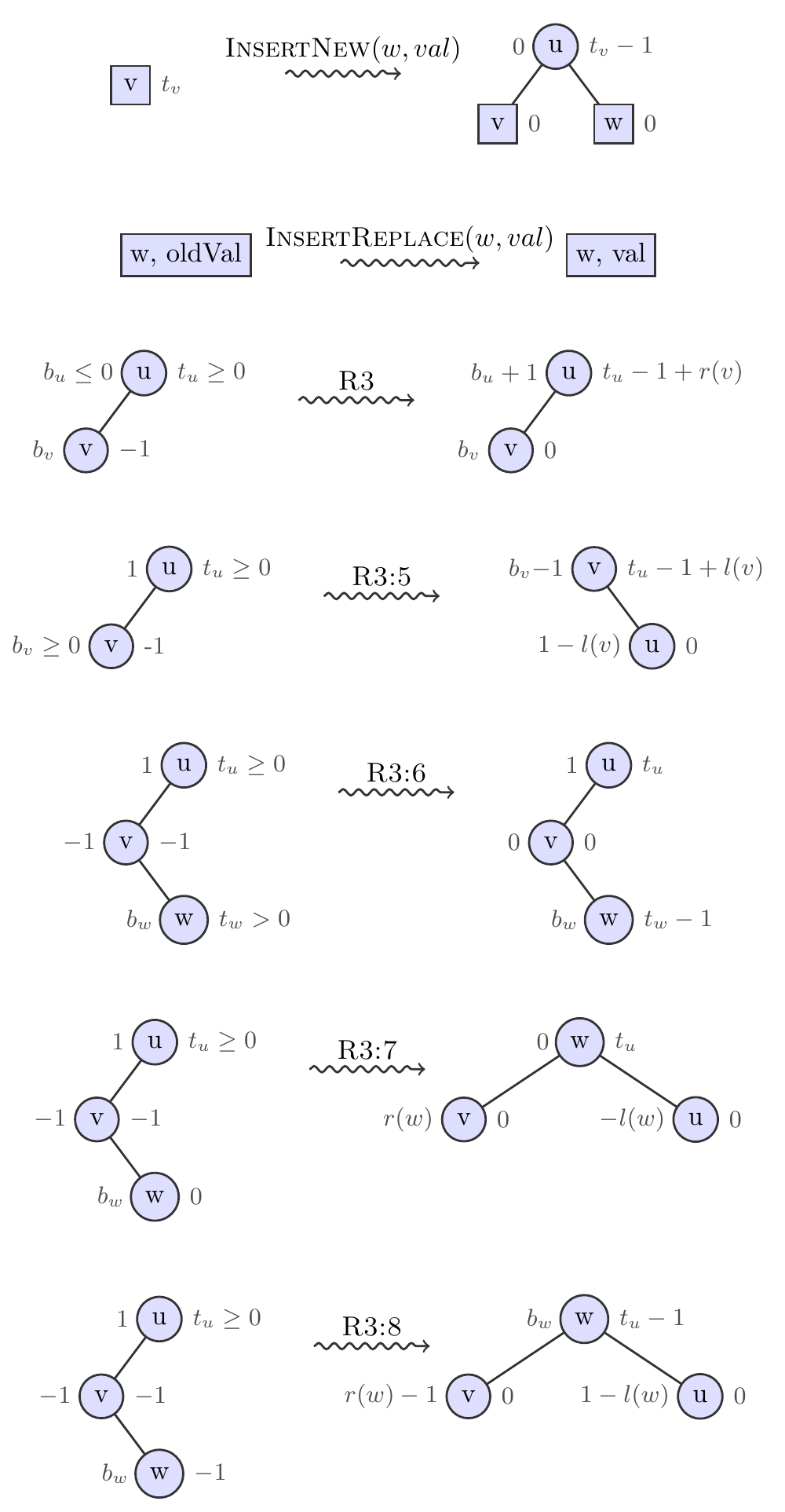}
\end{minipage}
\begin{minipage}{0.02\textwidth}
\end{minipage}
\begin{minipage}{0.49\textwidth}
\centering
\includegraphics[width=1\linewidth]{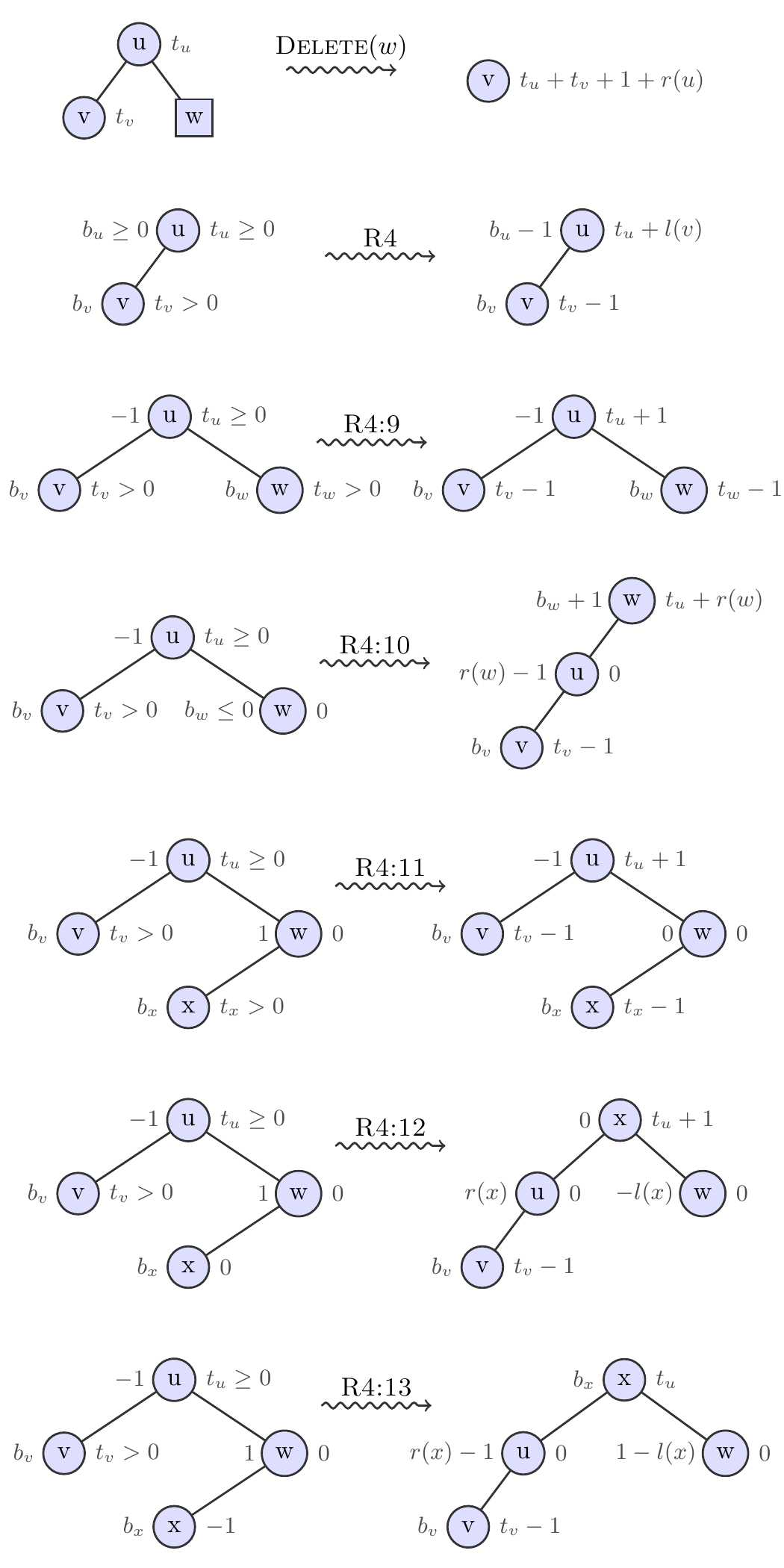}
\end{minipage}
\caption{Updates for the RAVL tree.
Relaxed balance factors appear on the left of nodes, and tag values on the right.}
\label{fig-ravl-updates}
\end{figure}

We now describe the updates to the RAVL tree, which appear in Figure~\ref{fig-ravl-updates}.
Relaxed balance factors appear to the left of nodes, and tag values appear to the right.
To simplify the presentation, we define two functions $l(u)$ and $r(u)$, such that $l(u) = 1$ if $rbf(u) = -1$, and $l(u) = 0$ otherwise, and $r(u) = 1$ if $rbf(u) = 1$, and $r(u) = 0$ otherwise.

We consider a dictionary implemented using a RAVL tree.
Recall that a dictionary represents a set of key-value pairs, and offers a \func{Get} operation to search for the value associated with a key, an \ins\ operation to add a new key-value pair (or replace the value in an existing key-value pair), and a \del\ operation to remove any existing key-value pair for a given key.
The tree is leaf-oriented, which means the key-value pairs in the dictionary are contained entirely in the leaves.
Internal nodes contain routing keys which direct searches to the appropriate leaf.

To insert a key-value pair $\langle k, val \rangle$, a process begins searching for the leaf $l$ where it should appear.
If the leaf contains $k$, then the process performs \func{InsertReplace}, which replaces the leaf with a new copy that contains $val$ instead of the value previously associated with $k$.
Otherwise, the process performs \func{InsertNew}, which replaces the leaf by a subtree consisting of an internal node with two leaves as its children.
The left leaf contains the smaller of $k$ and the key in $l$, and the right leaf contains the larger of $k$ and the key in $l$.
Both leaves have tag value zero and relaxed balance factor zero.
The internal node contains the same key as the right leaf and has relaxed balanced factor zero.
The tag value of the internal node is $l.tag - 1$.

This choice of tag value for the internal node maintains the RAVL balance invariant.
To see why, consider the following.
Let $l$ be the leaf found by the search, $n$ be the internal node after the update, and $n_l$ and $n_l$ be its left and right children.
Let $rbf$ (resp., $rbf'$) be the relaxed balance factor before (resp., after) the update, and $rh$ (resp., $rh'$) be the relaxed height before (resp., after) the update.
Suppose the RAVL balance invariant holds before the update.
We prove it holds after the update.
We first argue that the new internal node $n$ has relaxed balance factor zero.
By definition, $rh'(n_l) = n_l.tag = 0$ and $rh'(n_r) = n_r.tag = 0$, so $rbf'(n) = rh'(n_l) - rh'(n_r) = 0$.
Now, we argue that no other internal node has its relaxed balance factor changed by the update.
Observe that the update can only affect the relaxed balance factors of the ancestors of $l$.
To argue that the update does not change the relaxed balance factor of any ancestor of $l$, it suffices to prove $rh'(n) = rh(l)$.
By definition, $rh'(n) = max\{rh'(n_l), rh'(n_r)\} + 1 + (l.tag - 1)$.
Thus, $rh'(n) = max\{0, 0\} + 1 + (l.tag - 1) = l.tag$.
Since $l$ is a leaf, $rh(l) = l.tag$, so $rh'(n) = rh(l)$.

To delete any key-value pair with key $k$, a process searches for $k$, ending at a leaf $l$.
If the leaf does not contain $k$, then the deletion terminates.
Otherwise, the process performs \func{Delete}, which removes $l$ and its parent, leaving only the sibling $s$ of $l$.
After the update, $s$ has tag value $l.tag + s.tag + 1 + r(u)$.
It is straightforward to verify that this choice of tag value maintains the RAVL balance invariant.

We now describe how rebalancing works in a RAVL tree.
If a node $r$ in a RAVL tree has a negative tag value, then we say a \textit{negative violation} occurs at $r$.
If $r$ has a positive tag value, then $r.tag$ \textit{positive violations} occur at $r$.
Observe that, if a RAVL tree $T$ contains no violations, then for all $r \in T$, $r.tag = 0$, so $rh(r) = h(r)$, so $rbf(r) = bf(r)$, so $T$ is an AVL tree.
Thus, the goal of rebalancing is to perform rebalancing steps to eliminate all violations, while maintaining the RAVL balance invariant, to produce an AVL tree.

The updates starting with \func{R} in Figure~\ref{fig-ravl-updates} are rebalancing steps.
For each rebalancing step, in the left side of the diagram, there is a violation that is either eliminated or moved upwards in the tree by the update.
For example, \func{R3} applies when there is a negative violation at $v$, and \func{R3} either eliminates this violation or moves it to the parent.
Each of the rebalancing steps also has a mirror-image that is obtained by flipping the left and right child pointers, and flipping any relaxed balance factor inequalities.
We append \func{sym} to a rebalancing step's name to indicate that it is the horizontal symmetry of the original rebalancing step.
For example, \func{R3sym} applies if $b_u \ge 0$ and the right child of $u$ has $tag = -1$.

The set of rebalancing steps presented here is slightly different from the set presented in~\cite{larsen1994avl}.
There, the rebalancing steps were introduced simply as \textit{operation 3}, \textit{operation 4}, and so on, through \textit{operation 13}.
Operations 3 and 4 correspond to \func{R3} and \func{R4}, respectively.
However, unlike \func{R3} and \func{R4}, operations 3 and 4 do not require $t_u \ge 0$.
Performing these operations when $t_u < 0$ causes the relaxed balance factor of a node to become -2 or +2, violating the RAVL balance invariant.
Conceptually, the authors of~\cite{larsen1994avl} treated these relaxed balance factors of -2 and +2 as a new type of violation, and provided operations 5 through 13 to eliminate this new type of violation.
Specifically, operations 5 through 8 were designed to be performed after operation 3, and operations 9 through 13 were designed to be performed after operation 4.
Thus, the elimination of a positive or negative violation required performing two rebalancing steps: either operation 3 or 4, followed by one of the operations 5 through 13.
Here, rather than performing two separate rebalancing steps, we simply combine the two steps into one.
Thus, the rebalancing steps presented in Figure~\ref{fig-ravl-updates} consist of \func{R3} and \func{R4}, as well as \textit{composite} rebalancing steps \func{R3:5}, \func{R3:6}, \func{R3:7}, \func{R3:8}, \func{R4:9}, \func{R4:10}, \func{R4:11}, \func{R4:12} and \func{R4:13}.
Each composite rebalancing step \func{R}$x$\func{:}$y$ has the same effect as first performing operation $x$, then performing operation $y$ on the result.

\section{Implementation overview}

Here, we give only a high level overview of how the RAVL tree could be implemented, with the understanding that the implementation details would be similar to those of the chromatic tree in Chapter~\ref{chap-chromatree} and the relaxed $(a,b)$-tree that we describe in Chapter~\ref{chap-abtree} (both of which are described and proved in full detail).

Broadly, the implementation is very similar to our chromatic tree.
%At a high level, the implementation is very similar to the implementations of the chromatic tree and the relaxed $(a,b)$-tree.
\ins\ and \del\ respectively invoke \tryins\ and \trydel\ procedures, which perform a BST search, and return a leaf and its parent (and possibly grandparent).
The updates in Figure~\ref{fig-ravl-updates} are implemented (using the template) just like the updates in the chromatic tree.
Whenever an \func{InsertNew} or \func{Delete} update is performed, a tag violation might be created.
If an \ins\ or \del\ creates a violation, it invokes a \func{Cleanup} procedure.
\func{Cleanup} repeatedly searches for the key that was inserted or deleted, fixing any violations it sees, until it performs a search and sees no violations.

We now explain how \func{Cleanup} decides which rebalancing step to perform when it finds a violation.
If a violation occurs at the left (resp., right) child of a node, we call it a \textit{left violation} (resp., \textit{right violation}).
The rebalancing steps in Figure~\ref{fig-ravl-updates} are applied when a left violation is found, and the symmetric updates are applied when a right violation is found.
Without loss of generality, suppose \func{Cleanup} encounters a left violation.
In the following, we refer to nodes by their names in Figure~\ref{fig-ravl-updates}.
The violation found by \func{Cleanup} always occurs at node $v$.
Observe that $t_u \ge 0$, or else \func{Cleanup} would have stopped at node $u$.
If the violation is a negative violation, then one of \func{R3}, \func{R3:5}, \func{R3:6}, \func{R3:7} or \func{R3:8} applies.
Otherwise, the violation is a positive violation, and there is either a nearby negative violation that can be fixed, or one of \func{R4}, \func{R4:9}, \func{R4:10}, \func{R4:11}, \func{R4:12} or \func{R4:13} applies.

We give the complete decision tree for determining which rebalancing step to perform.
Suppose a negative violation occurs at $v$.
If $rbf(u) \le 0$, then \func{R3} applies.
So, suppose $rbf(u) > 0$.
By the RAVL balance invariant, $rbf(u) = 1$.
If $rbf(v) \ge 0$, then \func{R3:5} applies.
So, suppose $rbf(v) < 0$.
If $w.tag > 0$, then \func{R3:6} applies.
Otherwise, if $w.tag = 0$, then \func{R3:7} applies.
Otherwise, \func{R3:8} applies.

Now, suppose a positive violation occurs at $v$.
If $rbf(u) \ge 0$, then \func{R4} applies.
So, suppose $rbf(u) < 0$.
If $w.tag > 0$, then \func{R4:9} applies.
So, suppose $w.tag \le 0$.
In this case, before attempting to fix the positive violation, we first check if a negative violation occurs at $v$'s sibling $w$.
If so, the decision tree for negative violations is used, and that violation is fixed first.
So, we can assume that $w.tag \ge 0$, which implies that $w.tag = 0$.
If $rbf(w) \le 0$ then \func{R4:10} applies.
So, suppose $rbf(w) > 0$.
By the RAVL balance invariant, $rbf(w) = 1$.
If the left child $x$ of $w$ satisfies $x.tag > 0$, then \func{R4:11} applies.
Otherwise, if $x.tag = 0$, then \func{R4:12} applies.
Otherwise, \func{R4:13} applies.

\paragraph{Correctness and progress}

Since we demonstrate with the chromatic tree and relaxed $(a,b)$-tree how such a proof would proceed, we leave this as an exercise.
A proof of correctness and progress for the RAVL tree would follow the exact same structure as the proof for the chromatic tree.
One can simply copy the proof for the chromatic tree, and work through it, making minor changes.

\paragraph{Java implementation}

Given copies of \cite{larsen1994avl} and \cite{Brown:2014}, a first year undergraduate student produced a Java implementation of the RAVL tree in less than a week.
Its performance was slightly lower than that of Chromatic.
Similar to Chromatic6, we modified \func{Cleanup} so that it only performs rebalancing steps if at least $k$ violations appear in a path.
Doing this improved performance for the RAVL tree, bringing it in line with Chromatic6.
This suggests that the performance bottleneck in each data structure is the search phase of operations.
\begin{shortver}
The code for this Java implementation is available from \mbox{\url{http://implementations.tbrown.pro}}.
\end{shortver}

\section{Towards a height bound}

We claim that an RAVL tree containing $n$ keys has height at most $O(\log n + c)$, where $c$ is the number of unfinished insertions and deletions.
A full proof of this would be similar to the proof of the height bound for the chromatic tree.
Such a proof has two components: a proof that each RAVL tree contains at most $c$ violations, and a proof that an RAVL tree containing $c$ violations has height $O(\log n + c)$.
The first component would follow the same approach as the proof for the chromatic tree, so we leave it as an exercise.
However, the second component is somewhat subtle, so we prove it.

\begin{lem} \label{lem-avl-height-and-violations}
A relaxed AVL tree containing $n$ keys and $c$ violations has height $O(\log n + c)$.
\end{lem}
\begin{chapscxproof}
Suppose a relaxed AVL tree rooted at $root$ contains $n$ keys and $c$ violations.
By Corollary~2 of \cite{larsen1994avl}, if the topmost node in a rebalancing step is not $root$, then the rebalancing step does not change the relaxed height of the topmost node.
In fact, each rebalancing step would leave the relaxed height of the root unchanged if it they were able to change its tag value.
However, the tag value of the root is always zero, so, a rebalancing step that would normally set the tag value of the topmost node to a non-zero tag value (and, in doing so, preserve the relaxed height of the topmost node) will have the effect of changing the root's relaxed height.
Thus, the root is the only node that can have its relaxed height changed by a rebalancing step (and it is clearly the topmost node in any rebalancing step that modifies it).
In order to understand how the relaxed height of the root can be changed by a rebalancing step, we consider how the relaxed height of the topmost node in a rebalancing step changes depending on whether its tag value is forced to zero.

Consider any rebalancing step $S$.
Before $S$, let $u$ be its topmost node (shown in Figure~\ref{fig-ravl-updates}), $u_L$ and $u_R$ be its children, and $t$  be the value of $u.tag$.
After $S$, let $u'$ be the topmost node, $u_L'$ and $u_R'$ be its children, and $t'$ be the value of $u.tag$.
By definition, $rh(u) = max\{rh(u_L), rh(u_R)\} + 1 + t$ and $rh(u') = max\{rh(u_L'), rh(u_R')\} + 1 + t'$.
By Corollary~2 of~\cite{larsen1994avl}, $rh(u) = rh(u')$.
%Thus, $max\{rh(u_L), rh(u_R)\} + 1 + t = max\{rh(u_L'), rh(u_R')\} + 1 + t'$.

Now, suppose that $S$ sets $u.tag$ to 0 instead of $t'$ (which is what happens if $u = root$).
After $S$, let $u''$ be the topmost node.
Observe that the children of $u''$ are simply $u_L'$ and $u_R'$.
We compute $rh(u'') - rh(u)$.
By definition $rh(u'') = max\{rh(u_L'), rh(u_R')\} + 1 + 0$.
Thus, $rh(u'') - rh(u) = max\{rh(u_L'), rh(u_R')\} + 1 - (max\{rh(u_L), rh(u_R)\} + 1 + t) = max\{rh(u_L'), rh(u_R')\} - max\{rh(u_L), rh(u_R)\} - t$.
Since $rh(u) = rh(u')$, we have $max\{rh(u_L), rh(u_R)\} + 1 + t = max\{rh(u_L'), rh(u_R')\} + 1 + t'$, so $max\{rh(u_L), rh(u_R)\} = max\{rh(u_L'), rh(u_R')\} + t' - t$.
Substituting for $max\{rh(u_L), rh(u_R)\}$ in the equation for $rh(u'') - rh(h)$, we have $rh(u'') - rh(u) = max\{rh(u_L'), rh(u_R')\} - (max\{rh(u_L'), rh(u_R')\} + t' - t) - t$.
The right side of this equation simplifies to $-t'$.
Therefore, $S$ changes the relaxed height of $u$ by $-t'$.
By inspection of the rebalancing steps, $t'$ is at least $t-1$ and at most $t+1$, where $t$ is the value of $u.tag$ before $S$.
Thus, when a rebalancing step is applied where $u = root$ (so $t = 0$), $t'$ is at least $-1$ and at most $1$.
It follows that $S$ changes the relaxed height of the root by $-1$, $0$ or $1$.

By inspection of the rebalancing steps, each rebalancing step where $u = root$ reduces the number of violations in the tree by at least one.
Therefore, removing $c$ violations from the tree changes $|rh(root)|$ by at most $c$.
Suppose $T$ is any relaxed AVL tree rooted at $root_T$ that results from removing all $c$ violations from the tree rooted at $root$ by performing rebalancing steps.
Then, $|rh(root) - rh(root_T)| \le c$.
Since $T$ contains no violations, it is an AVL tree.
Hence, $rh(root_T) = h(root_T)$, which implies $|rh(root) - h(root_T)| \le c$.
Since the tree rooted at $root$ contains only $c$ violations, we have $|rh(root) - h(root)| \le c$.
Since $|rh(root) - h(root)| \le c$ and $|rh(root) - h(root_T)| \le c$, we have $|h(root) - h(root_T)| \le 2c$.
Finally, since $h(root_T) \in O(\log n)$, $h(root) \in O(\log n + c)$.
\end{chapscxproof}

\chapter{Relaxed $(a,b)$-tree implemented with the template} \label{chap-abtree}
% !TEX root = paper.tex

\section{$(a,b)$-trees and the relaxation}

We first define $(a,b)$-trees, then explain how they are relaxed.
An $(a,b)$-tree is a balanced, leaf-oriented search tree.
Leaf-oriented B-trees, 2-3 trees and 2-3-4 trees are all special cases of $(a,b)$-trees.%
\footnote{Using $(a,b)$-tree notation, a 2-3 tree is a $(2,3)$-tree, a 2-3-4 tree is a $(2,4)$-tree, and a B-tree is either an $(a,2a)$-tree or an $(a,2a-1)$-tree (since there are two common definitions of a B-tree).}
Excluding the root, each leaf in an $(a,b)$-tree has between $a$ and $b$ keys, and each internal node has between $a$ and $b$ child pointers, where $b \ge 2a - 1$.
If the root is a leaf, then it has between 1 and $b$ keys.
Otherwise, it has between 2 and $b$ child pointers.
Each internal node has one more child pointer than it has keys.
%Each leaf has the same number of pointers as keys, and these pointers point to \textit{data}, rather than other nodes.
The \textit{degree} of a node $u$, denoted $|u|$, is the number of pointers it contains.
%The keys of a node are maintained in order.
%
%\footnote{Some authors specify B-trees by $(a,b) = (a,2a-1)$, but the amortized results in \cite{LF95} do not apply to this variant.}
%
We denote the parent of a node $r$ by $\pi(r)$.
The \textit{level} of $r$ is defined as follows.
 \begin{displaymath}
   l(r) = \left\{
     \begin{array}{ll}
       0 & : r \mbox{ is a leaf} \\
       l(\pi(r)) + 1 & : \mbox{otherwise}
     \end{array}
   \right.
  \end{displaymath} 
As in B-trees, the balance invariant in $(a,b)$-trees is quite strict: All leaves in an $(a,b)$-tree have the same level.
Consequently, insertions and deletions of keys can involve a significant amount of work rebalancing the tree.
In particular, rebalancing is necessary whenever a process tries to insert a key into a leaf that is already full (i.e., contains $b$ keys), or delete a key from a leaf that contains only $a$ keys.
Sometimes rebalancing can be achieved with a small localized update, but other times it affects many nodes (even an entire path from the root to a leaf).
%Sometimes, the necessary rebalancing can affect every node on the path from the root to a leaf.
Synchronizing on many nodes adds algorithmic complexity and limits concurrency.

%\paragraph{Updates}
%We consider a dictionary implemented using $(a,b)$-trees.
%Insertion in a leaf that contains less than $b$ keys is easy: the new key is simply added to the leaf.
%
%
%
%If an insertion would cause a leaf $r$ to have more than $b$ keys, then $r$ gains a new sibling, and the keys of $r$, along with the key being inserted, are evenly distributed between $r$ and its new sibling.
%If
%Since $b \ge 2a$, there are enough keys in $r$ to give both new leaves  at least $a$ keys, so no rebalancing is necessary.
%However, if a deletion would cause a leaf to have less than $a$ keys, it must perform a rebalancing operation to merge with a predecessor or successor, and the resulting leaf may contain more than $b$ keys, necessitating another rebalancing operation.
%Henceforth, we refer to these rebalancing operations as \textit{rotations}.
%These rotations may propagate a problem all the way up the tree, before it finally disappears at the root.
%In order to produce a non-blocking implementation of an $(a,b)$-tree, we must perform entire updates atomically, including any necessary rotations.
%The need to atomically replace a large (non-constant) number of nodes would make updates complex and inefficient, and contention near the root would severely limit concurrency.

%To simplify algorithms and improve concurrency, 
In order to avoid these issues, relaxed $(a,b)$-trees: allow leaves to have different levels, allow nodes to contain fewer keys/pointers, split up monolithic rebalancing operations into small localized updates, and decouple rebalancing from insertion and deletion~\cite{LF95}.
Each leaf in a relaxed $(a,b)$-tree has between zero and $b$ keys, and each internal node has between one and $b$ child pointers, where $b \ge 2a - 1$.
Each node is augmented with a tag bit, which is zero for leaves and the root.
We say a node is \textit{tagged} if its tag bit is set.
The \textit{relaxed level} of a node $r$ is defined as follows.
 \begin{displaymath}
   rl(r) = \left\{
     \begin{array}{ll}
       r.tag & : r \mbox{ is a leaf} \\
       rl(\pi(r)) + (1 - r.tag) & : \mbox{otherwise}
     \end{array}
   \right.
  \end{displaymath} 
The relaxed level of $r$ is just like the level of $r$, except any tagged nodes on the path from $r$ to the root are not counted.
The balance invariant in relaxed $(a,b)$-trees is: All leaves in a relaxed $(a,b)$-tree have the same \textit{relaxed level}.

\begin{figure}[tbph]
\centering
\vspace{-8mm}
\begin{tabular}{ | m{2.5cm} | >{\centering\arraybackslash} m{12.5cm} | }
\hline

%\vspace{7mm}
\func{DeletePair} & \includegraphics[scale=0.85]{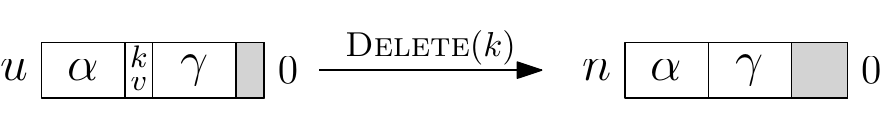} \\
\hline

\vspace{7mm}
\func{ReplacePair} & \includegraphics[scale=0.85]{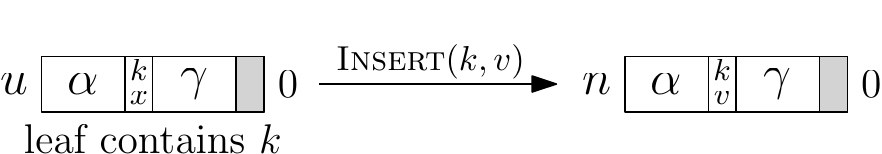} \\
\func{InsertPair} & \includegraphics[scale=0.85]{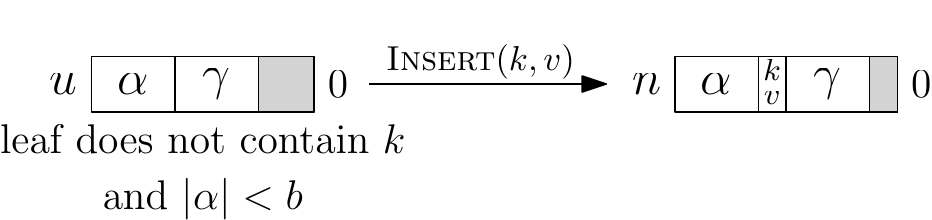} \\
\func{Overflow} & \includegraphics[scale=0.85]{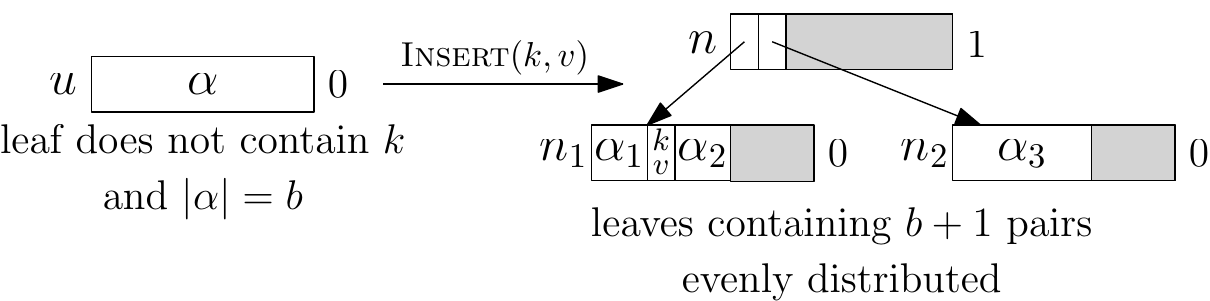} \\
\hline

\vspace{7mm}
\func{RootUntag} & \includegraphics[scale=0.85]{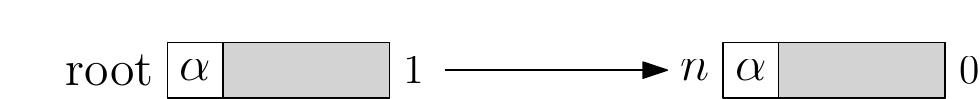} \\
\func{RootAbsorb} & \includegraphics[scale=0.85]{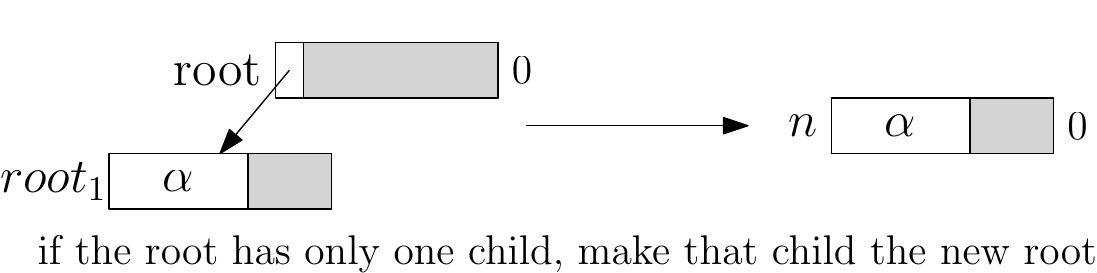} \\
\hline

\vspace{7mm}
\func{AbsorbChild} & \includegraphics[scale=0.85]{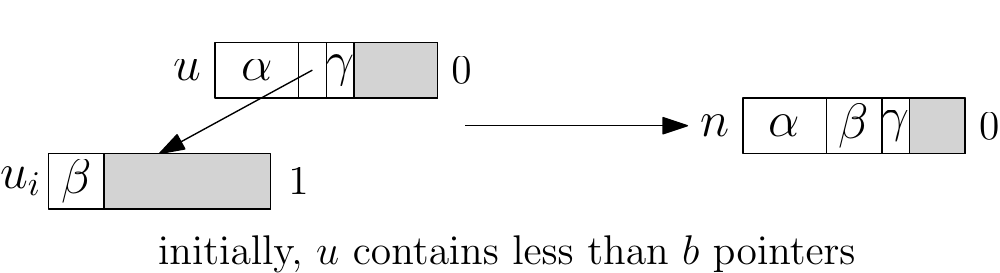} \\
\func{PropagateTag} & \includegraphics[scale=0.85]{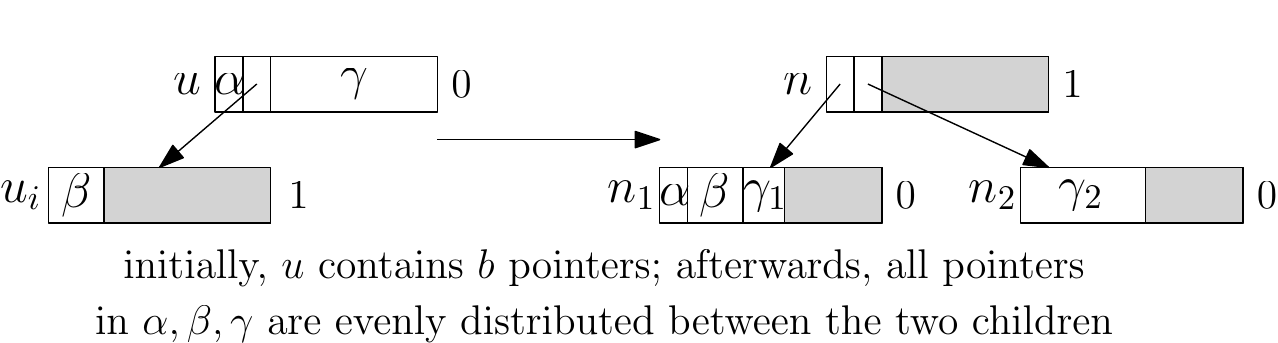} \\
\hline

\vspace{7mm}
\func{AbsorbSibling} & \includegraphics[scale=0.85]{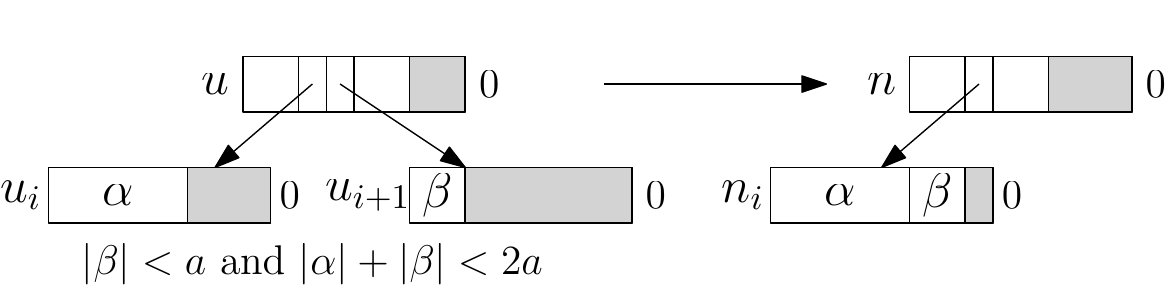} \\
\func{Distribute} & \includegraphics[scale=0.85]{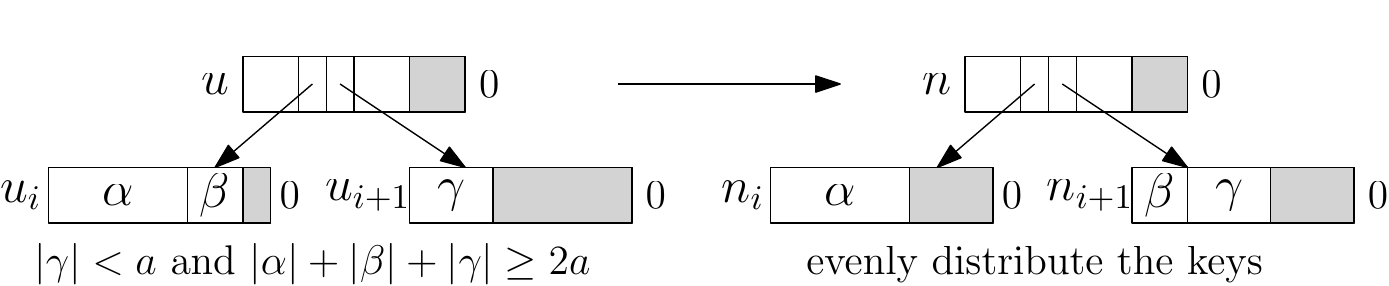} \\
\hline
\end{tabular}
\vspace{-2mm}
%\vspace{-5mm}
\caption{Updates to relaxed $(a,b)$-trees.
In leaves, a single key-value pair is drawn with the key on top, and the value on the bottom.
%Greek letters represent groups of keys and pointers/values.
Tag bits appear to the right of nodes.
Note that tagged nodes have exactly two pointers.
$|\alpha|$ denotes the degree of $\alpha$ (i.e., the number of pointers it contains).
}
\label{fig-abtree-updates}
\end{figure}

We now describe the updates to a relaxed $(a,b)$-tree, which appear in Figure~\ref{fig-abtree-updates}.
There, tag bits appear to the right of nodes.
We consider a dictionary implemented using a relaxed $(a,b)$-tree.
Recall that a dictionary represents a set of key-value pairs, and offers a \func{Get} operation to search for the value associated with a key, an \ins\ operation to add a new key-value pair (or replace the value in an existing key-value pair), and a \del\ operation to remove any existing key-value pair for a given key.
The tree is leaf-oriented, which means the key-value pairs in the dictionary are contained entirely in the leaves.
Internal nodes contain routing keys which direct searches to the appropriate leaf.
Leaves contain keys and pointers to values.
Let $kv(u)$ be the set of key-value pairs in a node $u$.

To insert a key-value pair $\langle k, v \rangle$, a process begins searching for the leaf $l$ where it should appear.
If the leaf already contains some key-value pair $\langle k, x \rangle$, then the process performs \func{ReplacePair}, which replaces that key-value pair with $\langle k, v \rangle$.
Now, suppose the dictionary does not contain any key value pair with key $k$.
If $l$ contains fewer than $b$ keys, then the process performs \func{InsertPair}, which inserts $\langle k, v \rangle$ into $l$.
Otherwise, the process performs \func{Overflow}, which conceptually replaces $l$ with a small subtree consisting of an internal node and two leaves.
The key-value pairs in $kv(l) \cup \{\langle k, v \rangle\}$ are distributed evenly between the leaves.

To delete any key-value pair with key $k$, a process searches for $k$, ending at a leaf $l$.
If the leaf does not contain any key-value pair with key $k$, then the deletion terminates.
Otherwise, the process performs \func{DeletePair}, which removes any key-value pair with key $k$ from $l$.

We now describe how rebalancing works in a relaxed $(a,b)$-tree.
If a node $r$ is tagged, then we say a \textit{tag violation} occurs at $r$.
If $r$ is an untagged node that is not the root, and $r$ has degree less than $a$, then we say a \textit{degree violation} occurs at $r$.
We say a \textit{degree violation} occurs at the root only if it is an internal node with degree \textit{one}.
(Note that tag violations effectively supersede degree violations.
That is, a degree violation does occur at a node if a tag violations occurs at that node.)
Observe that a relaxed $(a,b)$-tree that contains no violations is an $(a,b)$-tree.
Therefore, the objective of rebalancing is to eliminate violations, while maintaining the invariant that the data structure is a relaxed $(a,b)$-tree.

At a high level, violations are always created at leaves.
Tag violations are created by \func{Overflow} and degree violations are created by \func{DeletePair}.
Other updates either remove violations or propagate them up the tree towards the root, where they can always be removed.

We now describe precisely how violations are created, moved and removed by the $(a,b)$-tree updates in Figure~\ref{fig-abtree-updates}.
\func{Overflow} creates a tag violation at the new internal node, and \func{DeletePair} creates a degree violation at a leaf precisely if the leaf contains $a$ keys just before the \func{DeletePair} removes one of them.
\func{InsertPair} removes a degree violation at a leaf precisely if the leaf contains $a$ keys just after the \func{InsertPair} adds a key to it.
\func{RootUntag} removes a tag violation at the root, and will create a degree violation in the process precisely if the root is internal and has degree one.
\func{RootAbsorb} removes a degree violation at the root.
If there is a tag violation at the child of the root before the \func{RootAbsorb}, then it removes that tag violation, as well, unless the child of the root had degree one, in which case the degree violation is conceptually \textit{moved} from the child to the root.
\func{AbsorbChild} removes a tag violation at $u_i$.
If its parent $u$ has degree $a-1$ before the update, and $|\alpha|+|\gamma|+|\beta| \ge a$, then the update also removes a degree violation at $u$.
\func{PropagateTag} moves a tag violation from $u_i$ to $u$.
\func{Distribute} removes a degree violation at $u_i$ or $u_{i+1}$.
(In Figure~\ref{fig-abtree-updates}, a degree violation is depicted at $u_{i+1}$, but the update is also applicable if the degree violation appears at $u_i$, instead. Note that \func{Distribute} does not apply if degree violations simultaneously appear at both $u_i$ and $u_{i+1}$, since it requires $|\alpha|+|\beta|+|\gamma| \ge 2a$.)
\func{AbsorbSibling} is more complex.

Like \func{Distribute}, \func{AbsorbSibling} attempts to fix a degree violation at $u_i$ or $u_{i+1}$.
If a single degree violation appears at either $u_i$ or $u_{i+1}$ (depicted at $u_{i+1}$ in Figure~\ref{fig-abtree-updates}), then \func{AbsorbSibling} removes it.
However, unlike \func{Distribute}, \func{AbsorbSibling} can be applied when degree violations occur at both $u_i$ and $u_{i+1}$ (which occurs precisely when $|\alpha| < a$ and $|\alpha|+|\beta| \ge a$).
In this case, \func{AbsorbSibling} removes both degree violations.
However, since \func{AbsorbSibling} removes the node $u_{i+1}$, if $u$ contains $a$ pointers before the update, then the update will create a degree violation at $u$.
Therefore, depending on the degrees of $u$, $u_i$ and $u_{i+1}$, \func{AbsorbSibling} will either: (1) remove degree violations at $u_i$ and $u_{i+1}$, (2) remove degree violations at $u_i$ and $u_{i+1}$ and create one at $u$, (3) remove a degree violation at either $u_i$ or $u_{i+1}$, or (4) remove a degree violation at either $u_i$ or $u_{i+1}$ and create one at $u$.
In case (2) and (4), we think of the violation created at $u$ as having been \textit{moved} there from $u_i$ or $u_{i+1}$ (so, conceptually, no new violation is created).
Thus, \func{AbsorbSibling} removes up to two degree violations, and possibly moves a degree violation from a node to its parent.

\section{Implementation}

\begin{figure}[tb]
\begin{framed}
%\hspace*{-7mm}
%\begin{minipage}[t]{85mm}
\def\namewidth{18mm}
\preplisting
\begin{lstlisting}[mathescape=true,style=nonumbers]
 type// \node
     //\com User-defined fields
     //\wcnarrow{$tag$}{tag bit (immutable)}
     //\wcnarrow{$k_1, k_2, ..., k_d$}{keys (immutable)}
     //\wcnarrow{$p_1, p_2, ..., p_d$}{pointers (mutable)}
     //\wcnarrow{$d$}{degree of the node (immutable)}
     //\com Fields used by \llt/\sct\ algorithm
     //\wcnarrow{$\info$}{pointer to \op}
     //\wcnarrow{$marked$}{marked bit}
\end{lstlisting}
\end{framed}
%\end{minipage}
	\caption{Data definition for a node in the $(a,b)$-tree.}
	\label{code-abtree-data}
\end{figure}

Let $a$ be the minimum degree of nodes, and $b$ be the maximum degree.
We represent each node by a \rec\ with $b$ mutable pointers, and $b$ immutable keys, as well as immutable fields $d$ and $tag$ which contain the node's degree and tag bit, respectively.
(See Figure~\ref{code-abtree-data}.)
The degree $d$ represents the number of pointers that are used, and the tag bit indicates whether the node is tagged.
Internal nodes have one fewer key than pointers, so $k_d$ is unused in an internal node.
Leaves have exactly as many pointers as keys.

To avoid special cases when the $(a,b)$-tree is empty, we add a sentinel node at the top of the tree. % (see Figure \ref{fig-treetop}).
The sentinel node $entry$ always has one child and no keys.
(Every search that passes through it will simply follow that one child pointer.)
The sole child of this sentinel node is initially an empty leaf (with $d=0$ and no keys or pointers).
The actual $(a,b)$-tree is rooted at the child of the sentinel node.
For convenience, we use $root$ to refer to the current child of the sentinel node.

\begin{figure}[tb]
\begin{framed}
\prepnewlisting
%\vspace{-5mm}
%\hrule
%\vspace{-2mm}
\begin{lstlisting}[mathescape=true]
 //\func{Get}$(key)$
   $\langle -, -, l \rangle := \func{Search}(key)$
   if $l \mbox{ contains } key$ then return $\mbox{ the value associated with } key$
   else return $\nil$ // \\ \vspace{-2mm} \hrule \vspace{1mm} %
      
 //\func{Search}$(key)$
   $gp := \nil; p := entry; l := entry.p_1$
   while $l$// is internal
     $gp := p; p := l$
     $i := 1$
     while $i < l.d$ and $key \ge l.k_i$ do $i := i + 1$ //\medcom Locate appropriate child pointer to follow
     $l := l.p_i$ //\medcom Follow the child pointer
   return $\langle gp, p, l \rangle$
\end{lstlisting}
\end{framed}
	\caption{\func{Get} and \func{Search}.}
	\label{code-abtree-search}
\end{figure}

Detailed pseudocode for \func{Get}, \ins\ and \del\ is given in Figure~\ref{code-abtree-search}, \ref{code-abtree-ins} and~\ref{code-abtree-del}.
\func{Get}, \ins\ and \del\ each execute an auxiliary procedure, \func{Search}($key$), which appears in Figure~\ref{code-abtree-search}.
\func{Search}$(key)$ starts at $entry$ and traverses nodes as in an ordinary B-tree search, reading child pointers until reaching a leaf, which it then returns (along with the leaf's parent and grandparent).
Sometimes, the grandparent returned by an invocation of \func{Search} is not accessed.
It is easy to argue, by inspection of the code, and because of the sentinel node, that the leaf's parent always exists (i.e., is not $\nil$), and its grandparent exists whenever it is accessed.
We define the \textit{search path} for $key$ at any time to be the path that \func{Search}($key$) would follow, if it were done instantaneously.
The \func{Get}($key$) operation simply executes a \func{Search}($key$) and then returns the value found in the leaf if it contains $key$, and $\bot$ otherwise.

\begin{figure}[p]
\begin{framed}
\def\namewidth{18mm}
\preplisting
\begin{lstlisting}[mathescape=true]
 //\ins$(key, value)$
   //\com Identical to \ins\ procedure for the chromatic tree (see Figure~\ref{code-chromatic-ins}). \vspace{2mm}\hrule\vspace{2mm} %
    
 //\tryins$(key, value)$ 
   //\tline{\com Returns $\langle \true, \bot \rangle$ if $key$ was not in the dictionary %
               and inserting it caused a violation,} %
              {$\langle \false, \bot \rangle$ if $key$ was not in the dictionary %
               and inserting it did not cause a violation,} %
              {$\langle \false, oldValue \rangle$ if $\langle key, oldValue \rangle$ was in the dictionary, and %} %
              %{
              \fail\ if we should try again}

   //\com Search for $key$ in the tree
   $\langle -, p, l \rangle := \func{Search}(key)$
   
   //\com Template iteration 0 (parent of leaf)
   $result_p := \llt(p)$
   if $result_p \in \{\fail, \finalized\}$ then return $\fail$
   if $l \notin result$ then return $\fail$// \medcom{\func{Conflict}: verify $p$ still points to $l$}\label{code-abtree-tryins-conflict1}
   //Let $p.p_i$ be the child pointer of $p$ that pointed to $l$ at the previous line

   //\com Template iteration 1 (leaf)
   $result_l := \llt(l)$
   if $result_l \in \{\fail, \finalized\}$ then return $\fail$

   //\com Computing \func{SCX-Arguments} from locally stored values (and immutable fields)
   $V := \langle p, l \rangle$
   $R := \langle l \rangle$
   //$fld :=$ a pointer to $p.p_i$
   if $l \mbox{ contains } key$ then //\hfill\com Replace the value associated with an existing key $ $ 
     //Let $oldValue$ be the value associated with $key$ in $l$
     //$newNode :=$ new copy of $l$ that contains $\langle key, value \rangle$ instead of $\langle key, oldValue \rangle$
     $createdViolation := false$
   else if $l.d < b$ then //\hfill\com Insert a new key-value pair into a full leaf $ $ 
     //$newNode :=$ new copy of $l$ that has the new key-value pair inserted
     $oldValue := \nil$
     $createdViolation := false$
   else //\com $l.d = b$ \hfill\com Insert a new key-value pair into a non-full leaf $ $ 
     //$newNode :=$ pointer to a subtree of three newly created nodes: one internal node and two leaves, configured as in \func{Overflow} in Figure~\ref{fig-abtree-updates} (so that the key-value pairs in $kv(l) \cup \{\langle key, value \rangle\}$ are evenly distributed between the leaves, and the internal node is tagged)
     $oldValue := \nil$
     $createdViolation := true$

   if $\sct(V, R, fld, newNode)$ then return $\langle createdViolation, oldValue \rangle$
   else return $\fail$
\end{lstlisting}
\end{framed}
	\caption{Pseudocode for \func{Insert} and \tryins . Here, $a$ is the minimum degree of nodes, and $b$ is the maximum degree of nodes.}
	\label{code-abtree-ins}
\end{figure}

\subsection{Detailed description of insertion}
Pseudocode for \ins\ and its helper function \tryins\ appear in Figure~\ref{code-abtree-ins}.
\ins\ is identical to the \ins\ procedure for the chromatic tree.
It simply repeatedly invokes \tryins\ until it successfully performs the insertion, and then invokes \cleanup\ if it created a violation.
\tryins\ takes $key$ and $value$ as its arguments, and returns either \fail, or a pair $\langle createdViolation, oldValue \rangle$.
In this pair, $createdViolation$ is a bit that indicates whether the invocation of \tryins\ created a violation, and $oldValue$ is the value that was previously associated with $key$.
If \tryins\ returns \fail, this signals to \ins\ that the update was not successful, and \ins\ should perform another invocation of \tryins\ to try again.

\tryins\ first invokes \func{Search}$(key)$, which returns a leaf $l$ and its parent $p$.
Then, it performs \llt\ on $p$ and $l$.
If either \llt\ returns \fail\ or \finalized, then \tryins\ returns \fail.
So, we assume both \llt s succeed.
After the \llt$(p)$, \tryins\ verifies that $p$ still points to $l$ (since $p$ may have changed since it was visited by \func{Search}).
This is conceptually part of the \func{Conflict} procedure in the template.
After the \llt$(l)$, \tryins\ computes \sct-\func{Arguments}.

The creation of the arguments to \sct\ is straightforward.
If $l$ already contains $key$, then \tryins\ performs \func{ReplacePair} by setting $oldValue$ to the value that was associated with $key$ in $l$, and replacing $l$ with a new node in which the previous association $\langle key, oldValue \rangle$ is replaced with $\langle key, value \rangle$.
This does not create any violation, so $createdViolation$ is set to false.
Otherwise, if $l$ contains fewer than $b$ keys, then \tryins\ performs \func{InsertPair} by replacing $l$ with a new copy that has $\langle key, value \rangle$ inserted.
Since there was no value previously associated with $key$, $oldValue$ is set to $\nil$.
As in the previous case, this does not create any violation, so $createdViolation$ is set to false.
Otherwise, $l$ already contains $b$ keys, so \tryins\ performs \func{Overflow} by replacing $l$ with a subtree of three newly created nodes (configured as shown in Figure~\ref{fig-abtree-updates}).
Since there was no value previously associated with $key$, $oldValue$ is set to $\nil$.
Since \func{Overflow} creates a violation, $createdViolation$ is set to true.

\begin{figure}[p]
\begin{framed}
\def\namewidth{18mm}
\preplisting
\begin{lstlisting}[mathescape=true]
 //\del$(key)$
   //\com Identical to \del\ procedure for the chromatic tree (see Figure~\ref{code-chromatic-del}). \vspace{2mm}\hrule\vspace{2mm} %
    
 //\trydel$(key)$ 
   //\qline{\com Returns $\langle \false, \bot \rangle$ if $key$ was not in the dictionary,} %
              {$\langle \false, oldValue \rangle$ if $\langle key, oldValue \rangle$ was in the dictionary and deleting it did not create a violation,} %
              {$\langle \true, oldValue \rangle$ if $\langle key, oldValue \rangle$ was in the dictionary and deleting it created a violation, and} %
              {\fail\ if we should try again}

   //\com Search for $key$ in the tree
   $\langle -, p, l \rangle := \func{Search}(key)$
   
   //\com Template iteration 0 (parent of leaf)
   $result_p := \llt(p)$
   if $result_p \in \{\fail, \finalized\}$ then return $\fail$
   if $l \notin result$ then return $\fail$// \medcom{\func{Conflict}: verify $p$ still points to $l$}
   //Let $p.p_i$ be the child pointer of $p$ that pointed to $l$ at the previous line

   //\com Template iteration 1 (leaf)
   $result_l := \llt(l)$
   if $result_l \in \{\fail, \finalized\}$ then return $\fail$

   //\com Computing \func{SCX-Arguments} from locally stored values (and immutable fields)
   $V := \langle p, l \rangle$
   $R := \langle l \rangle$
   //$fld :=$ a pointer to $p.p_i$
   if $l \mbox{ does not contain } key$ then //\medcom The tree does not contain $key$ $ $ 
     $oldValue := \nil$
     $createdViolation := false$
     return $\langle createdViolation, oldValue \rangle$
   else
     //Let $oldValue$ be the value associated with $key$ in $l$
     //$newNode :=$ new copy of $l$ that does not contain $\langle key, oldValue \rangle$
     $createdViolation := (l.d = a)$ 
     if $\sct(V, R, fld, newNode)$ then return $\langle createdViolation, oldValue \rangle$
     else return $\fail$
\end{lstlisting}
\end{framed}
	\caption{Pseudocode for \func{Delete} and \trydel . Here, $a$ is the minimum degree of nodes, and $b$ is the maximum degree of nodes.}
	\label{code-abtree-del}
\end{figure}

\subsection{Detailed description of deletion}
Pseudocode for \del\ and its helper function \trydel\ appear in Figure~\ref{code-abtree-del}.
\del\ is identical to the \del\ procedure for the chromatic tree.
It simply repeatedly invokes \trydel\ until it successfully performs the deletion, and then invokes \cleanup\ if it created a violation.
\trydel\ takes $key$ as its argument, and returns either \fail, or a pair $\langle createdViolation, oldValue \rangle$.
If \trydel\ returns \fail, this signals that the update was not successful, and \del\ should perform another invocation of \trydel\ to try again.

\trydel\ is identical to \tryins\ up until it begins computing  \sct-\func{Arguments}, so we jump straight to the description of how it computes \sct-\func{Arguments}.
If $l$ does not contain $key$, then $key$ is not in the tree, so \trydel\ simply returns $\langle \false, oldValue \rangle$.
Otherwise, \trydel\ performs \func{DeletePair} by setting $oldValue$ to the value that was associated with $key$ in $l$, and replacing $l$ with a new copy that does not contain $key$.
The replacement of $l$ creates a (degree) violation if and only if $l$ contains exactly $a$ key-value pairs.
So, $createdViolation$ is set to the result of the expression $l.d = a$.
(Observe that, if $l.d < a$, then the degree violation was actually created by a \textit{previous} deletion at $l$.)

\begin{figure}[p]
\begin{framed}
\preplisting
\begin{lstlisting}[mathescape=true]
 //\cleanup$(key)$
   //\com Ensures the violation created by an \func{Overflow} or \func{DeletePair} (which is, in turn, caused by an \ins$(key, value)$ or \del$(key)$) gets eliminated
   while $\true$ //\medcom Repeatedly search until no violation is found
     //\com Conceptually, \func{SearchPhase} starts here
     $gp := \nil$; $p := entry$; $l := entry.p_1$ //\medcom Save the three last nodes traversed\label{abtree-cleanup-initialize-search}\label{code-abtree-cleanup-start}
     $ix_p = 0$; $ix_l = 1$ //\medcom Also save the indices of the pointers to $p$ and $l$

     //\com Base case: check for violations at the root of the $(a,b)$-tree
     if $l.tag = 1$ then $\func{TryRootUntag}(p, ix_l, l)$//\label{abtree-cleanup-rootuntag}
     else if $l\mbox{ is internal}$ and $l.d = 1$ then $\func{TryRootAbsorb}(p, ix_l, l)$//\label{abtree-cleanup-rootabsorb}
     else //\com Continue to search for $key$, looking for a violation
       loop//\label{abtree-cleanup-search-start}
         if //$l$ is a leaf then \textbf{return}\label{abtree-cleanup-search-return} \medcom Arrived at leaf without finding any violation
         $ix_p := ix_l$; $ix_l := 1$
         while $ix_l < l.d$ and $key \ge l.k_{ix_l}$ //\medcom Locate appropriate child pointer to follow
           $ix_l := ix_l + 1$
         $gp := p$; $p := l$; $l := l.p_{ix_l}$ //\medcom Follow the child pointer\label{code-abtree-cleanup-follow-pointer}
         if $l.tag = 1$ or $l.d < a$ then exit loop //\label{abtree-cleanup-search-exit}\medcom Exit the loop if there is a violation at $l$

     //\com Note: if we got here, we found a violation at $l$ and $gp \neq \nil$
     //\com Try to fix any tag violation at $l$ (recall: tag violations supersede degree violations)
     if $l.tag = 1$ then//\label{abtree-cleanup-l-tagged}
       if $p.d + l.d \le b+1$ then $\func{TryAbsorbChild}(gp, ix_p, p, ix_l, l)$//\label{abtree-cleanup-absorbchild}
       else $\func{TryPropagateTag}(gp, ix_p, p, ix_l, l)$//\label{abtree-cleanup-propagatetag}
     else //\com Try to fix the degree violation at $l$ (assert: $l.d < a$ and $gp \neq \nil$)\label{abtree-cleanup-l-untagged}
       $ix_s := (ix_p > 0\ ?\ ix_p - 1 : ix_p + 1)$ //\medcom Compute the index of a sibling of $l$
       $s := p.p_{ix_s}$ //\medcom Get a pointer to this sibling
       //\com We can only fix the degree violation at $l$ if $s$ is not tagged, so we first check if $s$ is tagged
       if $s.tag = 1$ then//\label{abtree-cleanup-s-tagged}
         //\com Fix the tag violation at $s$
         if $p.d + s.d \le b+1$ then $\func{TryAbsorbChild}(gp, ix_p, p, ix_s, s)$//\label{abtree-cleanup-absorbchild2}
         else $\func{TryPropagateTag}(gp, ix_p, p, ix_s, s)$//\label{abtree-cleanup-propagatetag2}
       else
         //\com Both $l$ and $s$ are untagged, so we try to fix the degree violation at $l$\label{abtree-cleanup-s-untagged}
         if $l.d + s.d < 2a$ then $\func{TryAbsorbSibling}(gp, ix_p, p, ix_l, l, ix_s, s)$//\label{abtree-cleanup-absorbsibling}
         else $\func{TryDistribute}(gp, ix_p, p, ix_l, l, ix_s, s)$//\label{abtree-cleanup-distribute}\label{code-abtree-cleanup-end}
\end{lstlisting}
\end{framed}
	\caption{Pseudocode for \cleanup. Here, $a$ is the minimum degree of nodes, and $b$ is the maximum degree of nodes.}
	\label{code-abtree-cleanup}
\end{figure}

\subsection{The rebalancing algorithm}
As in the chromatic tree, we assign responsibility for a violation to the process that created it.
Whenever a process creates a violation, it executes a \func{Cleanup} procedure that %$(key)$.
%Specifically, if an \ins$(key, value)$ or \del$(key)$ creates a violation, it invokes \func{Cleanup}$(key)$.
%\func{Cleanup}$(key)$ 
repeatedly searches for violations and performs rebalancing steps to eliminate them, terminating when it no longer finds any. %$on the search path for $key$, 
%until it can no longer find any. %it performs a search without encountering any stopping at the first violation it encounters.
%If it finds no violation during a search, then it terminates.
%Otherwise, it performs a rebalancing step.

At a high level, in each iteration of the outer loop, \func{Cleanup}$(key)$ searches for $key$, stopping at the first violation it encounters.
If it does not find a violation, then it terminates.
Otherwise, it determines which rebalancing step it should perform, and then invokes the appropriate procedure in \{\func{TryRootUntag}, \func{TryRootAbsorb}, \func{TryAbsorbChild}, \func{TryPropagateTag}, \func{TryAbsorbSibling}, \func{TryDistribute}\}. 
Conceptually, each of these procedures implements the update phase of an update in Figure~\ref{fig-abtree-updates}, taking the return value $m$ of the search phase as its argument.
(Note that we discuss precisely how these procedures are implemented, below.)
After invoking one of these procedures, \func{Cleanup} moves to the next iteration of the loop.

%%and invokes \sct\ to perform the rebalancing step.
%%a $success$ bit that is \true\ if all of its invocations of \llt\ returned snapshots and \sct  
%\sct-\func{Arguments} for the corresponding rebalancing step.
%The procedure returns \fail\ if one of its invocations of \llt\ returns \fail\ or \finalized, or 
%
%If the invocations of \llt\ performed by the procedure all returns snapshots, and either returns the set of argument for an \sct\ that will perform the rebalancing step, as well as a $success$ bit that indicates whether 

%First, it traverses the tree looking for a violation.
%Second, if it finds a violation, it then determines which rebalancing step it should perform.
%Third, it constructs 
More specifically, a process executing \func{Cleanup} first checks if there is a tag violation at $root$ (line~\ref{abtree-cleanup-rootuntag}).
If so, it invokes \func{TryRootUntag}. %, which performs a sequence of \llt s and computes \sct-\func{Arguments} for the update, and then invokes \sct\ and continues to the next iteration.
Otherwise, the process checks if there is a degree violation at $root$ (line~\ref{abtree-cleanup-rootabsorb}).
If so, it invokes \func{TryRootAbsorb}.
Otherwise, it enters the loop at line~\ref{abtree-cleanup-search-start}, where it searches for a violation on the search path for $key$.
The process exits this loop when it reaches a leaf without finding a violation and returns from \func{Cleanup} (at line~\ref{abtree-cleanup-search-return}), or finds a violation and exits the loop (at line~\ref{abtree-cleanup-search-exit}).
Consequently, at line~\ref{abtree-cleanup-l-tagged}, we know there is a violation at $l$.

If $l$ is tagged, then \func{Cleanup} invokes \func{TryAbsorbChild} or \func{TryPropagateTag}, as appropriate.
Suppose $l$ is not tagged.
Then there is a degree violation at $l$.
Degree violations are fixed using \func{AbsorbSibling} and \func{Distribute}, which manipulate $l$ and its sibling $s$.
Since these rebalancing steps require that $l$ and $s$ are both untagged, \func{Cleanup} must first check whether $s$ is tagged.
If so, \func{Cleanup} invokes \func{TryAbsorbChild} or \func{TryPropagateTag} to fix the tag violation at $s$.
(After fixing a tag violation at $s$, \func{Cleanup} will move to the next iteration of the loop, and search again.)
%If the degree violation at $l$ is not fixed by another process, then \func{Cleanup} will find it again in a subsequent iteration.)
Otherwise, \func{Cleanup} invokes \func{TryAbsorbSibling} or \func{TryDistribute}, as appropriate, to fix the degree violation at $l$.

The argument for why this \func{Cleanup} algorithm yields an upper bound on the height of the tree is very similar to the argument for the chromatic tree.
It relies on the fact that an \ins$(key, value)$ or \del$(key)$ always creates its violation at the terminal leaf on the search path for $key$, and a violation can only be moved from a node to its parent.
Thus, intuitively, a process can always find any violation it created while performing an \ins$(key, value)$ or \del$(key)$ by searching for $key$.
(And, similarly, if a process finds no violation while searching for $key$, then any violation it created has been eliminated.)
We will discuss this further in Section~\ref{sec-abtree-height}.

%the argument relies on a few key facts.
%An \ins$(key, value)$ or \del$(key)$ always creates its violation at the terminal leaf on the search path for $key$.
%Furthermore, whenever a violation at a node $u$ is moved by a rebalancing step, it is simply moved to the parent of $u$.
%Note that this entails replacing $u$ and the parent of $u$ by changing a pointer $p_i$ in the grandparent of $u$.
%If a search reaches the parent of $u$ (by reading $p_i$ before the change), then the search
%
%ince moving a violation from $u$ to its parent $\pi(u)$ entails replacing $u$ and $\pi(u)$, a search that reaches  and its parent, the parent 
%if a violation is moved from $u$ to $\pi(u)$ as a search is about to traverse an edge from $u$ to $\pi(u)$, the search will either reach $u$, 
%
% (so violations cannot be moved in such a way that a search will miss them)., and (3) violations cannot be moved in such a search cannot skip over a violation maintain the invariant that if there is a violation on the search path for $key$, then a search .

\begin{figure}[p]
%\begin{minipage}{0.38\textwidth}
%%\begin{framed}
%\vspace{-1.6cm}
%\hspace{-1cm}
%\includegraphics[scale=0.65]{chap-template/abtree/example-absorbsibling.pdf}
%%\end{framed}
%\end{minipage}
%\begin{minipage}{0.62\textwidth}
\vspace{-5mm}
\begin{framed}
\centering
\vspace{2mm}
\includegraphics[scale=0.85]{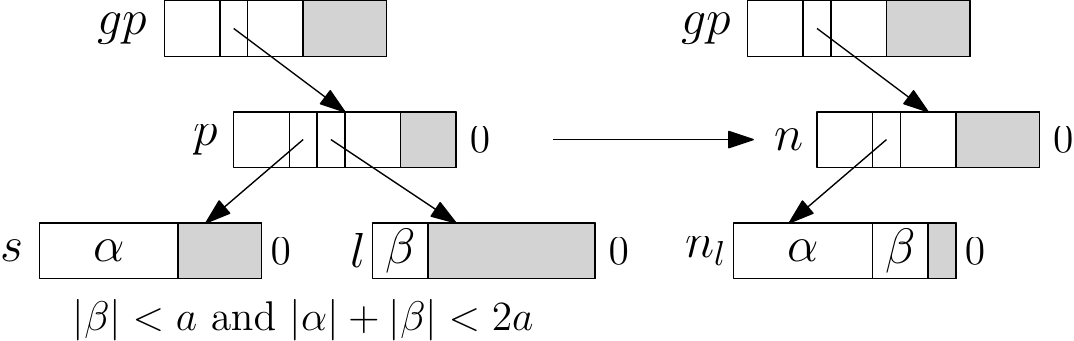} \vspace{2mm}\hrule
\preplistingnonumbers
\begin{lstlisting}[mathescape=true]
 //$\func{TryAbsorbSibling}(gp, ix_p, p, ix_l, l, ix_s, s)$
   //\com Implements the \func{UpdatePhase} for the \func{AbsorbSibling} update. Precondition: there is a degree violation at $l$, and there are no other violations at $p, s$ or $l$.

   //\com Template iteration 0 (grandparent of $l$)
   $result_{gp} := \llt(gp)$
   if $result_{gp} \in \{\fail, \finalized\}$ then return $\fail$
   //Let $ptr_p$ be the value in $result_{gp}$ that was read from $gp.p_{ix_p}$
   if $ptr_p \neq p$ then return $\fail$// \medcom{\func{Conflict}: verify $gp$ still points to $p$}

   //\com Template iteration 1 (parent of $l$)
   $result_p := \llt(p)$
   if $result_p \in \{\fail, \finalized\}$ then return $\fail$
   //Let $ptr_l$ be the value in $result_p$ that was read from $p.p_{ix_l}$
   //Let $ptr_s$ be the value in $result_p$ that was read from $p.p_{ix_s}$
   if $ptr_l \neq l$ then return $\fail$// \medcom{\func{Conflict}: verify $p$ still points to $l$}
   if $ptr_s \neq s$ then return $\fail$// \medcom{\func{Conflict}: verify $p$ still points to $s$}

   //\com Determine whether $s$ is the left or right sibling of $l$ (used to order the nodes in $V$ top-to-bottom, then left-to-right)
   if $ix_s < ix_l$ then $left := s$; $right := l$ //\medcom{$s$ is the left sibling of $l$}
   else $left := l$; $right := s$//\medcom{$s$ is the right sibling of $l$}

   //\com Template iteration 2 (the leftmost of $l$ and $s$)
   $result_{left} := \llt(left)$
   if $result_{left} \in \{\fail, \finalized\}$ then return $\fail$

   //\com Template iteration 3 (the rightmost of $l$ and $s$)
   $result_{right} := \llt(right)$
   if $result_{right} \in \{\fail, \finalized\}$ then return $\fail$

   //\com Compute \sct-\func{Arguments}: Start by creating new nodes according to the right-hand of the diagram above (using the immutable fields of $gp, p, l$ and $s$, and the results of the \llt s)
   //create node $n_l$ with key-value pairs $kv(l) \cup kv(s)$
   //create node $new$ by copying $p$, removing the child pointer to $s$ (and the corresponding key), and changing the pointer to $l$ to point to $n_l$
   $V := \langle gp,p,left,right \rangle$
   $R := \langle p,left,right \rangle$
   $fld := \&gp.p_{ix_p}$

   //\com Perform the \sct\ to change child pointer $ix_p$ of node $gp$
   $\sct(V, R, fld, new)$
\end{lstlisting}
\end{framed}
%\end{minipage}
	\caption{
	Implementing rebalancing step \func{AbsorbSibling}.
	Other rebalancing steps are handled similarly using the diagrams shown in Figure~\ref{fig-abtree-updates}.
    Here, $a$ is the minimum degree of nodes, and $b$ is the maximum degree of nodes.
	}
	\label{code-abtree-absorbsibling}
\end{figure}

\subsection{Implementing a rebalancing step}

We now explain how \func{TryAbsorbSibling} is implemented.
The procedures for the other rebalancing steps are similar.
Pseudocode appears in Figure~\ref{code-abtree-absorbsibling}.
\func{TryAbsorbSibling} takes four nodes $gp, p, l$ and $s$, as well as three integers $ix_p, ix_l$ and $ix_s$, as its arguments.
When \func{TryAbsorbSibling} is invoked by \func{Cleanup}, there is a degree violation at the node $l$, and no other violations at $p, s$ or $l$.
Furthermore, during its most recent search phase, \func{Cleanup} saw $gp.p_{ix_p} = p$, $p.p_{ix_l} = l$ and $p.p_{ix_s} = s$.

\func{TryAbsorbSibling} begins by invoking \llt$(gp)$.
Conceptually, this marks the beginning of iteration zero of the loop in the template.
If any \llt\ performed by \func{TryAbsorbSibling} returns \fail\ or \finalized, then \func{TryAbsorbSibling} immediately returns \fail, which prompts \func{Cleanup} to search again for violations to fix.
%So, suppose this invocation returns a snapshot.
If the \llt$(gp)$ returns a snapshot, then \func{TryAbsorbSibling} verifies that $gp.p_{ix_p}$ still points to $p$, according to the result of the \llt$(gp)$.
If $gp.p_{ix_p}$ no longer points to $p$, then \func{TryAbsorbSibling} immediately returns \fail.
In terms of the template, this verification is considered to be part of the \func{Conflict} procedure.

In iteration one of the loop in the template, \func{TryAbsorbSibling} invokes \llt$(p)$.
%Suppose this invocation returns a snapshot.
If this \llt\ returns a snapshot, then \func{TryAbsorbSibling} verifies that $p.p_{ix_l}$ still points to $l$ and $p.p_{ix_s}$ still points to $s$, according to the result of the \llt$(p)$.
If $p.p_{ix_l}$ no longer points to $l$ or $p.p_{ix_s}$ no longer points to $s$, then \func{TryAbsorbSibling} immediately returns \fail.
(As above, these verification steps are considered to be part of \func{Conflict}.)

Recall that the template places an ordering constraint on the nodes in the $V$ sequence that will be passed to \sct.
Specifically, the sequences $V$ constructed by all updates that take place entirely during a period of time when no \sct s change the tree structure must be ordered consistently according to a fixed tree traversal algorithm (for example, an in-order traversal or a breadth-first traversal).
We use a breadth-first traversal (i.e., top-to-bottom, then left-to-right).
So, \func{TryAbsorbSibling} next determines whether $s$ is the left or right sibling of $l$, and gives the alias $left$ to the leftmost of the two, and the alias $right$ to the rightmost of the two.

In iterations two and three of the loop in the template, \func{TryAbsorbSibling} simply invokes \llt$(left)$ and \llt$(right)$.
Finally, \func{TryAbsorbSibling} computes \sct-\func{Arguments}, and invokes \sct\ to perform the update.
The return value of \sct\ is ignored, because it does not matter whether the \sct\ succeeds or fails.
In either case, \func{Cleanup} will continue to perform rebalancing steps until it performs a search without encountering any violation.
It is straightforward to verify that each iteration of the outer loop in \func{Cleanup} follows the template.

\section{Correctness proof}

In the following, the terms \textbf{search path} and \textbf{range} are defined as they were in the proof of the chromatic tree (in Section~\ref{chromatic-correctness}).

\begin{lem} \label{lem-abtree}
Our implementation of a relaxed $(a,b)$-tree satisfies the following claims.
\begin{compactenum}
\item \tryins\ and \trydel\ follow the tree update template and satisfy all constraints specified by the template.
Each iteration of the outer loop in \func{Cleanup} follows the template.
\label{claim-abtree-invariants-follow-template}
\item The node $entry$ always has no keys and one child pointer. The child pointer points to a node $root \neq entry$.
\label{claim-abtree-invariants-top-of-tree}
\item If a node $v$ is in the data structure in some configuration $C$ and $v$ was on the search path for key $k$ in some earlier configuration $C'$, then $v$ is on the search path for $k$ in $C$. (Equivalently, updates to the tree do not shrink the range of any node in the tree.) \label{lem-abtree-searchpath}
\item If an invocation of \func{Search}$(k)$ reaches a node $v$, then there was some earlier configuration during the search when $v$ was on the search path for $k$. \label{lem-abtree-searches}
\item The \func{Search} procedure used by \tryins\ and \trydel\ satisfies DTP. \label{lem-abtree-dtp}
\item All invocations of \tryins\ and \trydel\ that perform a successful \sct\ are atomic (including their search phases). All invocations of \func{TryRootUntag}, \func{TryRootAbsorb}, \func{TryAbsorbChild}, \func{TryPropagateTag}, \func{TryAbsorbSibling} and \func{TryDistribute} that perform a successful \sct\ are atomic. \label{lem-abtree-atomicity}
\item The tree rooted at the child of $entry$ is always a relaxed $(a,b)$-tree. \label{lem-abtree-searchtree}
\end{compactenum}
\end{lem}
\begin{chapscxproof}
We prove these claims together by induction on the sequence of steps (invocations and responses of procedures, atomic invocations of \func{Search} (for Claim~\ref{lem-abtree-dtp}) reads from shared memory, and \llt s and \sct s) in an execution.
That is, we assume that all of the claims hold before an arbitrary step in the execution, and prove they hold after the step.

\medskip

\noindent\textbf{Claim~\ref{claim-abtree-invariants-follow-template}:}
This claim follows almost immediately from inspection of the code.
The only subtlety is showing that no process invokes $\llt(r)$ where $r = \nil$.
Suppose the inductive hypothesis holds just before an invocation $I$ of $\llt(r)$.

Suppose $I$ occurs in \tryins\ or \trydel.
Then, $r$ must be one of the nodes $p$ or $l$ returned by \func{Search}.
By inspection of the code, $l$ is not \nil.
By inductive Claim~\ref{claim-abtree-invariants-top-of-tree}, $l$ has a parent, so $p$ is not \nil.
Now, suppose $I$ occurs in \func{TryAbsorbSibling}.
Then, $r$ must be one of the nodes $gp$, $p$ or $l$ passed to \func{TryAbsorbSibling}, or a sibling $s$ of $l$ obtained from an \llt$(p)$.
Before \func{Cleanup} invokes \func{TryAbsorbSibling} it must execute one iteration of the loop at line~\ref{abtree-cleanup-search-start}, where it sets $gp := p$.
It is straightforward to argue that $l, p$ and $gp$ are all non-\nil\ when the loop terminates.
Thus, if $r \in \{gp, p, l\}$, then $r \neq \nil$.
It remains to prove $s \neq \nil$.
By inductive Claim~\ref{lem-abtree-searchtree}, when $I$ obtains $s$ from the result of its \llt$(p)$, the tree is a relaxed $(a,b)$-tree, and $p$ has at least two children.
Thus, $s$ exists (and is non-\nil).
The proof for \func{TryAbsorbChild}, \func{TryPropagateTag} and  \func{TryDistribute} is similar.
The proof when $I$ occurs in \func{TryRootUntag} or \func{TryRootAbsorb} is even simpler, since $gp$ is \textit{not} passed as an argument.
Thus, we need only argue that $l$ and $p$ are non-\nil, which follows immediately from line~\ref{abtree-cleanup-initialize-search}.

\medskip

\noindent\textbf{Claim~\ref{claim-abtree-invariants-top-of-tree}:}
The only step that can modify the tree (and, hence, affect this claim) is an invocation $S$ of \sct\ performed by an invocation $I$ of \tryins, \trydel, \func{TryRootUntag}, \func{TryRootAbsorb}, \func{TryAbsorbChild}, \func{TryPropagateTag}, \func{TryAbsorbSibling} or \func{TryDistribute}.
Proving this claim entails arguing that $I$ cannot replace the entry point, or modify it in any way except by changing its single child pointer.

Suppose the inductive hypothesis holds just before $S$.
By inductive Claim~\ref{claim-abtree-invariants-follow-template}, $I$ follows the tree update template up until it performs $S$.
By Lemma~\ref{lem-effective-updatephase-atomic} and Lemma~\ref{lem-dotreeup-constraints-invariants}, the update phase of $I$ atomically performs performs one of the transformations in Figure~\ref{fig-abtree-updates}.
All of these transformations simply change a single child pointer to replace one or more nodes.
However, each node that is replaced has a parent, which $entry$ does not.
Thus, $entry$ cannot be replaced.

\medskip

\noindent\textbf{Claim~\ref{lem-abtree-searchpath}:}
Initially, the claim is trivially true (since the tree only contains $entry$, which is on every search path).
In order for $v$ to change from being on the search path for $k$ in configuration $C'$ to no longer being on the search path for $k$ in configuration $C$, the tree must change between $C'$ and $C$.
Thus, there must be a successful \sct\ $S$ between $C'$ and $C$.
Moreover, this is the only kind of step that can affect this claim.
We show $S$ preserves the property that $v$ is on the search path for $k$.

By inductive Claim~\ref{claim-abtree-invariants-follow-template}, $S$ is performed by a template operation.
Thus, by Lemma~\ref{lem-dotreeup-constraints-invariants}, $S$ changes a pointer of a node from $old$ to $new$, removing a connected set $R$ of nodes (rooted at $old$) from the tree, and inserting a new connected set $N$ of nodes.
If $v$ is not a descendant of $old$ immediately before $S$, then this change cannot remove $v$ from the search path for $k$.
So, suppose $v$ is a descendant of $old$ immediately prior to $S$.

Since $v$ is in the data structure in both $C'$ and $C$, it must be in the data structure at all times between $C'$ and $C$ by Lemma~\ref{lem-dotreeupdate-rec-cannot-be-added-after-removal}.
Therefore, $v$ is a descendant of $old$, but $S$ does not remove $v$ from the tree.
Recall that the fringe $F_R$ is the set of nodes that are children of nodes in $R$, but are not themselves in $R$ (see Figure~\ref{fig-replace-subtree} and Figure~\ref{fig-replace-subtree2}).
By definition, $v$ must be a descendant of a node $f \in F_R$.
Moreover, since $v$ is on the search path for $k$ just before $S$, so is $f$.
We argue, for each possible tree modification in Figure~\ref{fig-abtree-updates}, that if any node in $F_R$ is on the search path for $k$ prior to $S$, then it is still on the search path for $k$ after $S$.
We proceed by cases.

\textit{Case~1:} Suppose $S$ performs a \func{InsertPair}, \func{Overflow} or \func{DeletePair} update.
Since $S$ replaces a leaf with either a new leaf, or a new internal node and two new leaves, the fringe set is empty.
Thus, the claim is vacuously true.

\textit{Case~2:} Suppose $S$ performs a \func{RootUntag} or \func{RootAbsorb} update.
Then, $S$ does not change the range of any node in the fringe set, so the claim holds.

\textit{Case~3:} Suppose $S$ performs a \func{TryAbsorbChild} update. % or \func{TryPropagateTag}.
By inductive Claim~\ref{lem-abtree-searchtree}, the tree is a relaxed $(a,b)$-tree before $S$.
Since leaves are never tagged in a relaxed $(a,b)$-tree, the node $u_i$ in the depiction of \func{AbsorbChild} in Figure~\ref{fig-abtree-updates} must be internal.
Thus, one can think of each of the nodes $u_i$ and $u$ as a sequence of alternating pointers and keys, starting and ending with a pointer.
Consequently, $\alpha$, $\beta$ and $\gamma$ (in Figure~\ref{fig-abtree-updates}) can be thought of as sequences of alternating pointers and keys, where $\alpha$ starts with a pointer and ends with a key, $\beta$ starts and ends with pointers, and $\gamma$ starts with a key and ends with a pointer.
%Consider an inorder traversal of the tree that outputs the sequence of pointers and keys it encounters in the internal nodes it visits.
%
The fringe $F_R$ is the set of nodes pointed to by $\alpha$, $\beta$ and $\gamma$.
Observe that the keys in $u_i$ and $u$ partition the range of $u$, and this partition defines the range of each node in $F_R$.
Specifically, the partition begins with the left endpoint of the range of $u$, then continues with the alternating pointers and keys of $\alpha$, $\beta$ and $\gamma$, and finally ends with the right endpoint of the range of $u$.
%The range of each node in $F_R$ is then defined by the keys that appear immediately to its left and right in the partition.
The update does not change this partition, so the range of each node in $F_R$ is the same before and after the update.
Therefore, if a node in $F_R$ is on the search path to $key$ before the update, it is still on the search path after the update.
The cases for \func{PropagateTag}, \func{AbsorbSibling} and \func{Distribute} follow the exact same reasoning.

\medskip

\noindent\textbf{Claim~\ref{lem-abtree-searches}:}
Consider a read $r$ of a child pointer in an invocation $I$ of \func{Search}$(k)$.
This is the only kind of step that can affect this claim.
Let $v$ be the node pointed to by the value returned by $r$.
We prove that $r$ preserves the claim.
If $r$ is the first read of a child pointer by $I$, then $v$ is $entry$, which is always on the search path for $k$, so $r$ preserves the claim.

Now, suppose there is a previous read $r'$ of a child pointer by $I$.
By the inductive hypothesis, the node $v'$ that was returned by $r'$ was on the search path for $k$ in some configuration $C'$ after the beginning of $I$ and before $r'$.
Our goal is to prove that there is a configuration $C$, after $C'$ and before $r$, when $v$ is on the search path for $k$.
If $v'$ is in the tree when $I$ performs $r$, then $C$ is the configuration just before $r$.
Otherwise, $C$ is the last configuration before $v'$ was removed from the tree.

Without loss of generality, suppose $I$ reaches $v$ by following the $i$th child pointer of $v'$.
By inductive Claim~\ref{lem-abtree-searchtree}, the data structure is a relaxed $(a,b)$-tree just before $r$, so $k \ge v'.k_j$ for $j < i$ and $k < v'.k_j$ for $j \ge i$.
We now prove that $v'.p_i$ points to $v$ in $C$.
Suppose $v'$ is in the tree when $I$ performs $r$ (so $C$ is just before $r$).
This case follows immediately from our assumption that $r$ reads a pointer to $v$ from $v'.p_i$.
Now, suppose $v'$ is not in the tree when $r$ occurs, so $C$ is the last configuration before $v'$ was removed.
Recall that $v'$ is in the tree in $C'$.
By Lemma~\ref{lem-dotreeupdate-rec-cannot-be-added-after-removal}, $v'$ cannot be added back into the data structure after it is removed.
Since $v'$ is in the tree in $C'$, and is not in the tree when $I$ subsequently performs $r$, $v'$ must be removed after $C'$ and before $r$ occurs.
By inductive Claim~\ref{claim-abtree-invariants-follow-template} and template Constraint~\ref{constraint-finalized-iff-removed}, $v'$ becomes finalized precisely when it is removed.
Since $v'$ cannot change after it is finalized, and $v'.p_i$ points to $v$ when $r$ occurs (which is after $v'$ is removed), we can see that $v'.p_i$ must point to $v$ in $C$ (which is just before $v'$ is removed).

Finally, we prove that $v$ is on the search path for $k$ in $C$.
Since $v'$ was on the search path for $k$ in $C'$, and it is in the data structure in $C$, which is after $C'$ but before $r$, inductive Claim~\ref{lem-abtree-searchpath} implies that $v'$ is on the search path for $k$ in $C$.
Since $v = v'.p_i$, $k \ge v'.k_j$ for $j < i$ and $k < v'.k_j$ for $j \ge i$, we can see that $v$ must also be on the search path for $k$ in $C$.

\medskip

\noindent\textbf{Claim~\ref{lem-abtree-dtp}:}
Initially, the claim holds vacuously (since no steps have been taken).
Suppose an invocation $S$ of \func{Search}$(k)$ in \tryins\ terminates and returns $m$.
(The proof for \trydel\ is similar.)
Consider any configuration $C$, after $S$ returns $m$, in which all of the nodes in $m$ are in the tree and their fields agree with the values in $m$.
Suppose the inductive hypothesis up until $C$.
We prove that an invocation $S'$ of \func{Search}$(k)$ in \tryins\ would return $m$ if $S'$ were performed atomically just after configuration $C$.

The value $m = \langle -, p, l \rangle$ returned by $S$ contains a leaf $l$ and its parent $p$.
By inductive Claim~\ref{lem-abtree-searches}, $p$ and $l$ were each on the search path at some point during $S$ (which is before $C$).
Since $p$ and $l$ are in the tree in $C$, inductive Claim~\ref{lem-abtree-searchpath} implies that they are on the search path for $k$ in $C$.
Therefore, $S'$ will visit each of them.
Conceptually, $m$ also encodes the fact that $p$ points to $l$.
This fact is checked at line~\ref{code-abtree-tryins-conflict1} as part of the \func{Conflict} procedure.
By our assumption (that the fields of the nodes in $m$ in configuration $C$ agree with their values in $m$), $p$ is also the parent of $l$ when $S'$ is performed.
Consequently, $S'$ will also return $m = \langle -, p, l \rangle$.

\medskip

\noindent\textbf{Claim~\ref{lem-abtree-atomicity}:}
By inductive Claim~\ref{lem-abtree-dtp} and Theorem~\ref{thm-effectivedtp-atomic}, all invocations of \tryins\ and \trydel\ that perform a successful \sct\ are atomic.
By Lemma~\ref{lem-effective-updatephase-atomic}, all invocations of \func{TryRootUntag}, \func{TryRootAbsorb}, \func{TryAbsorbChild}, \func{TryPropagateTag}, \func{TryAbsorbSibling} and \func{TryDistribute} that perform a successful \sct\ are atomic.

\medskip

\noindent\textbf{Claim~\ref{lem-abtree-searchtree}:}
Only successful invocations of \sct\ can affect this claim.
Successful invocations of \sct\ are performed only in \tryins, \trydel\ and the rebalancing procedures: \func{TryRootUntag}, \func{TryRootAbsorb}, \func{TryAbsorbChild}, 
\func{TryPropagateTag}, \func{TryAbsorbSibling} and \func{TryDistribute}.
The claim holds in the initial state of the tree (which is described in Claim~\ref{claim-abtree-invariants-top-of-tree}).
We show that every successful invocation $S$ of \sct\ preserves the claim.
We proceed by cases.

\textit{Case~1:} $S$ is performed in an invocation $I$ of \tryins$(key, value)$ or \trydel$(key)$.
By Claim~\ref{lem-abtree-atomicity}, the entirety of $I$ is atomic, including its search phase.
Thus, when $I$ occurs, its search procedure returns the unique leaf $l$ on the search path for $key$.
Consequently, $I$ atomically performs one of the transformations \func{DeletePair}, \func{ReplacePair}, \func{InsertPair}, or \func{Overflow} in Figure~\ref{fig-abtree-updates} to replace $l$ (and possibly some of its neighbouring nodes).
Since $I$ is entirely atomic (including its search phase), and it simply performs one of the $(a,b)$-tree updates, it is easy to verify that it preserves the claim.

\textit{Case~2:} $S$ is performed in an invocation $I$ of one of the rebalancing procedures. %\func{TryRootUntag}, \func{TryRootAbsorb}, \func{TryAbsorbChild}, \func{TryPropagateTag}, \func{TryAbsorbSibling} or \func{TryDistribute}.
By Claim~\ref{lem-abtree-atomicity}, the update phase of $I$ is atomic (but the search phase is not necessarily atomic).
Therefore, $I$ atomically performs one of the rebalancing transformations at some location in the tree (but not necessarily the same rebalancing transformation, at the same location, that it would perform if $I$'s search were also part of the atomic update).
All of the rebalancing transformations preserve the claim (regardless of where in the tree they are performed).
\end{chapscxproof}

We define the linearization points for relaxed $(a,b)$-tree operations as follows.
\begin{compactitem}
\item \func{Get}($key$) is linearized at a time during the operation when the leaf reached was on the search path for $key$.
(This time exists, by Lemma~\ref{lem-abtree}.\ref{lem-abtree-searches}.)
\item An \ins\ is linearized at its successful \sct\ inside \tryins\ (if such an \sct\ exists).
(Note: every \ins\ that terminates performs a successful \sct.)
\item A \del\ that returns $\bot$ is linearized at a time during the operation when the leaf returned by its last invocation of \func{Search} was on the search path for $key$.
(This time exists, by Lemma~\ref{lem-abtree}.\ref{lem-abtree-searches}.)
\item A \del\ that does not return $\bot$ is linearized at its successful \sct\ inside \trydel\ (if such an \sct\ exists).
(Note: every \del\ that terminates, but does not return $\bot$, performs a successful \sct.)
\end{compactitem}
It is easy to verify that every operation that terminates is assigned a linearization point during the operation.

\begin{thm}
The relaxed $(a,b)$-tree is a linearizable implementation of a dictionary with the operations \func{Get}, \ins\ and \del.
\end{thm}
\begin{chapscxproof}
Lemma~\ref{lem-abtree}.\ref{lem-abtree-atomicity} proves that the \sct s implement atomic changes to the tree as shown in Figure~\ref{fig-abtree-updates}.
By inspection of these transformations, the set of keys and associated values stored in leaves are not altered by any rebalancing steps.
Moreover, the transformations performed by each linearized \ins\ and \del\ maintain the invariant that the set of keys and associated values stored in leaves of the tree is exactly the set that should be in the dictionary.
When an invocation of \func{Get}$(key)$ is linearized, the search path for $key$ ends at the leaf returned by its invocation of \func{Search}.
If that leaf contains $key$, \func{Get} returns the associated value, which is correct.
If that leaf does not contain $key$, then, by Lemma~\ref{lem-abtree}.\ref{lem-abtree-searchtree}, it is nowhere else in the tree, so \func{Get} is correct to return $\bot$.
\end{chapscxproof}

\section{Progress proof}

Our goal is to prove that, if processes take steps infinitely often, then relaxed $(a,b)$-tree operations succeed infinitely often.
At a high level, this follows from Theorem~\ref{thm-dotreeup-progress} (the final progress result for template operations), and the fact that at most $(i+d)\lfloor \log_a (|T|+i)/2 \rfloor+1$ rebalancing steps can be performed after $i$ insertions and $d$ deletions have been performed on a standard $(a,b)$-tree $T$ (proved in \cite{DBLP:journals/ijfcs/LarsenF96}).
Additionally, note that for $b \ge 2a$, only amortized $O(1)$ rebalancing steps are needed per insertion or deletion to maintain balance.

\begin{thm}
The relaxed $(a,b)$-tree operations are non-blocking.
\end{thm}
\begin{chapscxproof}
To derive a contradiction, suppose there is some configuration $C$ after which some processes continue to take steps but no successful relaxed $(a,b)$-tree operations occur.
We first argue that eventually the tree stops changing.
Since no successful relaxed $(a,b)$-tree operations occur after $C$, the only steps that can change the tree after $C$ are successful invocations of \sct\ performed by invocations of \func{TryRootUntag}, \func{TryRootAbsorb}, \func{TryAbsorbChild}, \func{TryPropagateTag}, \func{TryAbsorbSibling} or \func{TryDistribute}.
Larsen and Fagerberg~\cite{DBLP:journals/ijfcs/LarsenF96} proved that after a bounded number of rebalancing steps, a relaxed $(a,b)$-tree becomes a standard $(a,b)$-tree, and then no further rebalancing steps can be applied.
Thus, eventually the tree must stop changing.

This implies that every invocation of \func{Search} (and, hence, \func{Get}) succeeds after a finite number of steps, unless the process executing it crashes.
%Thus, \func{Search} and \func{Get} are non-blocking.
It follows that no invocation of \func{Search} or \func{Get} occurs after $C$.
Therefore, eventually processes only take steps in \ins\ and/or \del\ operations.
Thus, processes perform infinitely many invocations of \tryins\ and/or \trydel, and/or infinitely many iterations of the outer loop in \func{Cleanup}.
By Lemma~\ref{lem-abtree}.\ref{claim-abtree-invariants-follow-template}, \tryins, \trydel, and iterations of the outer loop in \func{Cleanup}, all follow the template.
Thus, Theorem~\ref{thm-dotreeup-progress} implies that infinitely many of these template updates will succeed.
Since the number of rebalancing steps that can be performed is finite if the number of successful insertions and deletions is finite, there must be infinitely many successful insertions and/or deletions.
Consequently, infinitely many must succeed after $C$, which is a contradiction.
\end{chapscxproof}

\section{Bounding the height of the tree} \label{sec-abtree-height}

We now show that the height of the relaxed $(a,b)$-tree at any time is $O(c + \log_a n)$ where
$n$ is the number of keys stored in the tree and $c$ is the number of \ins\ and \del\ operations \textit{currently} in progress.
(When no process is performing \ins\ or \del, the $c$ term disappears.)
At a high level, the upper bound holds for the following reason.
Since we always perform rebalancing steps that satisfy VIOL, if we reach a leaf without finding the violation that an \ins\ or \del\ created, then the violation has been eliminated.
Since each \ins\ or \del\ creates and most one violation, and eliminates it before terminating, we can prove that the number of violations in the tree at any time is bounded above by $c$.
Further, since removing all violations would yield an $(a,b)$-tree tree with height $O(\log_a n)$, and eliminating each violation reduces the height by \textit{at most} one, the height of the relaxed $(a,b)$-tree is $O(c + \log_a n)$.

Note that this is a very pessimistic upper bound, because many violations do not increase the height of the tree.
In our experiments, the height is typically approximately $1 + \log_a n$.
To the authors' knowledge, a tree of height $c+\log_a n$ is only achieved in a specific pathology where all processes cooperate on the same path to cause $n$ consecutive tag violations (by having one process perform \func{Overflow} at a node, then a second process perform \func{Overflow} at one of the leaves created by the first \func{Overflow}, then a third process perform \func{Overflow} at one of the leaves created by the second \func{Overflow}, and so on).
Observe that, to cause an \func{Overflow} at a leaf that was created by another \func{Overflow}, processes must insert $b-a+1$ keys at the leaf.
Additionally, no operation that performs an \func{Overflow} update can terminate until all processes have finished constructing this pathology, or else the operation will perform rebalancing steps to eliminate the tag violations.

\begin{defn}
Let $x$ be a node that is in the data structure.
We say that a \textit{tag violation occurs at $x$} if $x$ has its \textit{tag} bit set.
Suppose $x$ is not $entry$, and no tag violation occurs at $x$.
We say that a \textbf{degree violation occurs at $x$} if (1) $x$ is $root$ (i.e., $entry.p_1$) and $x$ is an internal node with one pointer, or (2) $x$ is not $root$ and $x$ has fewer than $a$ pointers.
\end{defn}

\begin{defn}
A process $P$ is \textbf{in a cleanup phase for} $k$ if it is executing an \ins$(k, value)$ or a \del$(k)$ and it has performed a successful \sct\ inside a \tryins\ or \trydel\ that returns $createdViolation=\true$. 
If $P$ is between line \ref{code-abtree-cleanup-start} and \ref{code-abtree-cleanup-end}, then
$location(P)$ is the value of $P$'s local variable $l$; %and $parent(P)$ are the values of $P$'s local variables $l$ and $p$; otherwise, $location(P)$ is the $entry$ node and $parent(P)$ is \nil.
otherwise, $location(P)$ is the $entry$ node.
\end{defn}

We use the following invariant to show that each violation in the data structure has a
pending update operation that is responsible for removing it before terminating:
either that process is on the way towards the violation, or it will find another violation and
restart from the top of the tree, heading towards the violation.

%    degree violations at new nodes
%        
%    deletepair: if $u.d \le a$
%    replacepair: no
%    insertpair: no
%    overflow: no
%    rootuntag: at $n$ (if $root$ is internal and $root.d = 1$)
%    rootabsorb: at $n$ (if $root_1$ is internal and $root_1.d = 1$)
%    absorbchild: at $n$ (if $u.d \le a-2$)
%    propagatetag: no
%    absorbsibling:
%        at $n$ (if $u.d \le a$)
%        at $n_i$ (if $u_i.d + u_{i+1}.d < a$)
%    distribute: at $n$ (if $u.d < a$)

\begin{lem}
In every configuration, there exists an injective mapping $\rho$ from violations to processes such that, for every violation $x$, 
\begin{compactitem}
\item
{\rm (A)} process $\rho(x)$ is in a cleanup phase for some key $k_x$ and 
\item
{\rm (B)} $x$ is on the search path for $k_x$ from $root$ (i.e., $entry.p_1$)  and
\item {\rm (C)} either\\ 
{\rm (C1)} the search path for $k_x$ from $location(\rho(x))$ contains the violation $x$, or\\
%{\rm (C2)} $location(\rho(x)).w=0$ and $parent(\rho(x)).w=0$, or\\
{\rm (C2)} in the prefix of the search path for $k_x$ from $location(\rho(x))$ up to and including
the first non-finalized node (or the entire search path if all nodes are finalized), there
is a node where a violation occurs. %with weight greater than 1 or two nodes in a row with weight~0.
\end{compactitem}
\end{lem}
\begin{chapscxproof}
In the initial configuration, there are no violations, so the invariant is trivially satisfied.
We show that any step $S$ by any process $P$ preserves the invariant.  
We assume there is a function $\rho$ satisfying the claim for the configuration $C$ immediately before $S$ and show that there is a function $\rho'$ satisfying the claim for the configuration $C'$ immediately after $S$.
The only step that can cause a process to leave its cleanup phase is
the termination of an \ins\ or \del\ that is in its cleanup phase.
The only
steps that can change $location(P)$ and $parent(P)$ are $P$'s execution of line~\ref{abtree-cleanup-initialize-search} or the read of the child pointer on line~\ref{code-abtree-cleanup-follow-pointer}.
(We think of all of the updates to local variables on those lines as happening atomically with the read of the child pointer.)
The only steps that can change child pointers or finalize nodes are successful \sct s.
No other steps $S$ can cause the invariant to become false.

{\bf Case 1} $S$ is the termination of an \ins\ or \del\ that is in its cleanup phase:
We choose $\rho'=\rho$.
$S$ happens when the test in line~\ref{abtree-cleanup-search-return} is true, meaning that $location(P)$ is a leaf.
Leaves are never tagged.
There cannot be a degree violation at the leaf, because then the process would have exited the loop in the previous iteration after the test at line  \ref{abtree-cleanup-search-exit} returned true (since set of pointers/values in a leaf never changes).
Thus, no violation occurs at $location(P)$.
So, $P$ cannot be $\rho(x)$ for any violation $x$, so $S$ cannot make the invariant become false.

{\bf Case 2} $S$ is an execution of line~\ref{abtree-cleanup-initialize-search}:
We choose $\rho'=\rho$.
Step $S$ changes $location(P)$ to $entry.p_1$ and $parent(P)$ to $entry$.  
If $P \neq \rho(x)$ for any violation $x$, then this step cannot affect the truth of the invariant.  
Now suppose $P=\rho(x_0)$ for some violation $x_0$.
The truth of properties (A) and (B) are not affected by a change in $location(P)$ and property (C) is not affected for any violation $x \neq x_0$.
Since $\rho$ satisfies property (B) for violation $x_0$ before $S$, it will satisfy property (C1) for $x_0$ after $S$.

{\bf Case 3} $S$ is a read of the child pointer on line~\ref{code-abtree-cleanup-follow-pointer}:
We choose $\rho'=\rho$.
Step $S$ changes $location(P)$ from some node $l$ to node $l_i$, which is $l$'s $i$th child when $S$ is performed.
If $P \neq \rho(x)$ for any violation $x$, then this step cannot affect the truth of the invariant.
So, suppose $P=\rho(x_0)$ for some violation $x_0$.
By (A), $P$ is in a cleanup phase for $k_{x_0}$.
The truth of (A) and (B) are not affected by a change in $location(P)$ and property (C) is not affected for any violation $x\neq x_0$.
So it remains to prove that (C) is true for violation $x_0$ in $C'$.

First, we prove there is no violation at $l$.
Suppose $S$ occurs in the first iteration of \cleanup's inner loop.
Then, $l$ was $root$ when it was read from $entry.p_1$ at line~\ref{abtree-cleanup-initialize-search}.
Before entering the inner loop, \cleanup\ saw that $l$ was untagged (so there is no tag violation at $l$), and was not an internal node with one pointer (so there is no degree violation at $l$).
Thus, there is no violation at $l$ in this case.
Now, suppose $S$ does not occur in the first iteration of \cleanup's inner loop.
In the previous iteration, \cleanup\ saw $l.tag = 0$ and $l.d \ge a$ at line~\ref{abtree-cleanup-search-exit}, otherwise the loop would have terminated.
So, there is no violation at $l$.

We consider two cases, depending on whether (C1) or (C2) is true in configuration $C$.

{\bf Case 3a} (C1) is true in configuration $C$:
Thus, when $S$ is performed, the violation $x_0$ is on the search path for $k_{x_0}$ from $l$, but it is not at $l$ (as argued above).
$S$ reads the $i$th child of $l$, so $k \ge l.k_j$ for $j < i$ and $k < l.k_j$ for $j \ge i$ (since the keys of node $l$ never change).
So, $x_0$ must be on the search path for $k_{x_0}$ from $l_i$.
This means (C1) is satisfied for $x_0$ in configuration $C'$.

{\bf Case 3b} (C2) is true in configuration $C$:
This proof is similar to the previous case.

{\bf Case 4} $S$ is a successful \sct:
We must define the mapping $\rho'$ for each violation $x$ in configuration $C'$.
Since tag bits are immutable and the number of pointers in nodes do not change, no transformation in Figure~\ref{fig-abtree-updates} can create a new violation at a node that was already in the data structure in configuration $C$.
So, if $x$ is at a node that was in the data structure in configuration $C$,
$x$ was a violation in configuration $C$, and $\rho(x)$ is well-defined.
In this case, we let $\rho'(x)=\rho(x)$.

\begin{figure}[h]
%%\hspace{-5mm}
%%\begin{minipage}{1\textwidth}
\begin{center}
\centering
\noindent
\begin{tabular}{|l|l|l|}\hline
Transformation & Tag violations $x$ created by $S$ & $\rho'(x)$ \\\hline
\func{DeletePair} & none created & --\\\hline
\func{ReplacePair} & none created & --\\\hline
\func{InsertPair} & none created & --\\\hline
\func{Overflow} & at $n$ & process performing the \ins \\\hline
\func{RootUntag} & none created & -- \\\hline
\func{RootAbsorb} & none created & -- \\\hline
\func{AbsorbChild} & none created & -- \\\hline
\func{PropagateTag} & at $n$ & $\rho$(tag violation at $u_i$) \\\hline
\func{AbsorbSibling} & none created & -- \\\hline
\func{Distribute} & none created & -- \\\hline
\end{tabular}
\end{center}
\caption{Description of how $\rho'$ maps tag violations at newly added nodes to processes responsible for them.}
\label{fig-table-violations-tag}
\end{figure}

\begin{figure}[h]
\begin{center}
\centering
\noindent
\begin{tabular}{|l|l|l|}\hline
Transformation & Degree violations $x$ created by $S$ & $\rho'(x)$ \\\hline
\func{DeletePair} & at $n$ (if $u.d = a$) & process performing the \del \\\hline
\func{DeletePair} & at $n$ (if $u.d < a$) & $\rho$(degree violation at $u$) \\\hline
\func{ReplacePair} & none created & -- \\\hline
\func{InsertPair} & none created & -- \\\hline
\func{Overflow} & none created & -- \\\hline
\func{RootUntag} & at $n$ (if $root$ is internal and $root.d = 1$) & $\rho$(tag violation at $u$) \\\hline
\func{RootAbsorb} & at $n$ (if $root_1$ is internal and $root_1.d = 1$) & $\rho$(degree violation at $root_1$) \\\hline
\func{AbsorbChild} & at $n$ (if $u.d \le a-2$) & $\rho$(degree violation at $u$) \\\hline
\func{PropagateTag} & none created & -- \\\hline
\func{AbsorbSibling} & at $n_i$ (if $u_i.d + u_{i+1}.d < a$) & $\rho$(degree violation at $u_i$) \\\hline
\func{AbsorbSibling} & at $n$ (if $u.d = a$ and $u_i.d + u_{i+1}.d < a$) & $\rho$(degree violation at $u_{i+1}$) \\\hline
\func{AbsorbSibling} & at $n$ (if $u.d = a$ and $u_i.d + u_{i+1}.d \ge a$) & $\rho$(degree violation at $u_i$ or $u_{i+1}$) \\\hline
\func{AbsorbSibling} & at $n$ (if $u.d < a$) & $\rho$(degree violation at $u$) \\\hline
\func{Distribute} & at $n$ (if $u.d < a$) & $\rho$(degree violation at $u$) \\\hline
\end{tabular}
\end{center}
\caption{Description of how $\rho'$ maps degree violations at newly added nodes to processes responsible for them.
Here, $u.d$ denotes the degree of a node $u$.}
\label{fig-table-violations-degree}
\end{figure}

If $x$ is at a node that was added to the data structure by $S$, then we must define $\rho(x)$ on a case-by-case basis for all transformations described in Figure~\ref{fig-abtree-updates}.
If $x$ is a tag violation at a newly added node, we define $\rho'(x)$ according to the table in Figure~\ref{fig-table-violations-tag}.
If $x$ is a degree violation at a newly added node, we define $\rho'(x)$ according to the table in Figure~\ref{fig-table-violations-degree}.

The function $\rho'$ is injective, since $\rho'$ maps each violation created by $S$ to a distinct process that $\rho$ assigned to a violation that has been removed by $S$, with only two exceptions: for tag violations caused by \func{Overflow} and degree violation caused by \func{DeletePair}.
In these exceptional cases, $\rho'$ maps these violations to the process that has just begun its cleanup phase (and therefore was not assigned any violation by $\rho$).

Let $x$ be any violation in the tree in configuration $C'$.
We show that $\rho'$ satisfies
properties (A), (B) and (C) for $x$ in configuration $C'$.

{\bf Property (A)}:
Every process in the image of $\rho'$ was either in the image of $\rho$ or a process that just
entered its cleanup phase at step $S$, so every process in the image of $\rho'$ is in its
cleanup phase.  

{\bf Property (B) and (C)}: We consider several subcases.

{\bf Subcase 4a}
Suppose $S$ is an \func{Overflow}'s \sct, and $x$ is the tag violation  created by $S$.
Then, $P$ is in its cleanup phase for the inserted key, which is one of the children of the node containing the tag violation $x$.
Since the tree is a search tree, $x$ is on the search path for this key, so (B) holds.
In this subcase, $location(\rho'(x)) = entry$ since $P=\rho'(x)$ has just entered its cleanup phase.
So property (B) implies property (C1).

{\bf Subcase 4b}
Suppose $S$ is a \func{DeletePair}'s \sct, and $x$ is the degree violation assigned to $P$ by $\rho'$. 
Then, $P$ is in a cleanup phase for the deleted key, which was in $u$ before $S$.
Therefore, $x$ (at the replacement leaf $n$) is on the search path for this key, so (B) holds.
As in the previous subcase, $location(\rho'(x)) = entry$ since $P=\rho'(x)$ 
has just entered its cleanup phase.
So property (B) implies property (C1).

{\bf Subcase 4c}
If $x$ is at a node that was added to the data structure by $S$ (and is not covered by the above two cases), then $\rho'(x)$ is $\rho(y)$ for some violation $y$ that has been removed from the tree by $S$, as described
in the above two tables.
Let $k$ be the key such that process $\rho(y)=\rho'(x)$ is in the cleanup phase for $k$.
By property (B), $y$ was on the search path for $k$ before $S$.
By inspection of the tables, and Lemma~\ref{lem-abtree}.\ref{lem-abtree-searchpath}, 
%It is easy to check by inspection of the tables and Figure~\ref{fig-abtree-updates} that 
any search path that went through $y$'s node in configuration $C$ goes through $x$'s node in configuration $C'$.
(We designed the tables to have this property.)
Thus, since $y$ was on the search path for $k$ in configuration $C$, 
$x$ is on the search path for $k$ in configuration $C'$, satisfying property (B).

If (C2) is true for violation $y$ in configuration $C$, then (C2) is true for $x$ in configuration $C'$ (since any node that is finalized remains finalized forever, and its child pointers do not change).

So, for the remainder of the proof of subcase 4c, suppose (C1) is true for $y$ in configuration $C$.
Let $l=location(\rho(y))$ in configuration $C$.
Then $y$ is on the search path for $k$ from $l$ in configuration $C$.

First, suppose $S$ removes $l$ from the data structure.
\begin{compactitem}
\item
If $y$ is a violation at node $l$ in configuration $C$, then it makes (C2) true for $x$ in configuration $C'$.
\item
Otherwise, since both $l$ and its descendant, the parent of the node that contains $y$, are removed by $S$, the entire path between these two nodes is removed from the data structure by $S$.
So, all nodes along this path are finalized by $S$ because Constraint~\ref{constraint-finalized-iff-removed} is satisfied.
Thus, the violation $y$ makes (C2) true for $x$ in configuration $C'$.
\end{compactitem}

Now, suppose $S$ does not remove $l$ from the data structure.
In configuration $C$, the search path from $l$ for $k$ contains $y$.
By inspection of the tables defining $\rho'$, and Lemma~\ref{lem-abtree}.\ref{lem-abtree-searchpath}, any search path from $l$ that went through $y$'s node in configuration $C$ goes through $x$'s node in  configuration $C'$.
So, (C1) is true in configuration $C'$.

{\bf Subcase 4d}
If $x$ is at a node that was already in the data structure in configuration $C$, then $\rho'(x)=\rho(x)$.
Let $k$ be the key such that this process is in the cleanup phase for $k$.
Since $x$ was on the search path for $k$ in configuration $C$ 
and $S$ did not remove $x$ from the data structure,
$x$ is still on the search path for $k$ in configuration $C'$ (by Lemma~\ref{lem-abtree}.\ref{lem-abtree-searchpath}).
This establishes property (B).

If (C2) is true for $x$ in configuration $C$, then it also holds for $x$ in $C'$, for the same reason as in Subcase 4c. 
So, suppose (C1) is true for $x$ in configuration $C$.
Let $l=location(\rho(x))$ in configuration $C$.
Then, (C1) says that $x$ is on the search path for $k$ from $l$ in configuration $C$. 
If $S$ does not change any of the child pointers on this path between $l$ and $x$, then $x$ is still on the search path from $location(\rho'(x)) = l$ in configuration $C'$, so property (C1) holds for $x$ in $C'$.
So, suppose $S$ does change the child pointer of some node on this path from $old$ to $new$.
Then the search path for $k$ from $l$ in configuration $C$ goes through $old$ to some node $f$ in the fringe set $F_R$ of $S$ and then onward to the node containing violation $x$.
By Lemma~\ref{lem-abtree}.\ref{lem-abtree-searchpath}, the search path for $k$ from $l$ in configuration $C'$ goes through $new$ to the same node $f$, and then onward to the node containing the violation $x$.
Thus, property (C1) is true for $x$ in configuration $C'$.
\end{chapscxproof}

\begin{cor} \label{lem-abtree-violation-bound}
The number of violations in the data structure is bounded by the number of incomplete \ins\ and \del\ operations.
\end{cor}

%\begin{lem} \label{lem-ab-claims}
%Consider a relaxed $(a,b)$-tree $T$ rooted that contains $n$ nodes and $c$ violations.
%Suppose relaxed $(a,b)$-tree rebalancing steps are performed on $T$ until it no longer contains any violations, and becomes a standard $(a,b)$-tree $T'$.
%The height of $T'$ is at most $c$ greater than the height of $T$.
%\end{lem}
%\begin{proof}
%The rebalancing steps in a relaxed $(a,b)$-tree maintain the invariant that all leaves have the same relaxed level.
%Thus, every pair of leaves $u_1$ and $u_2$ in $T$ satisfy $rl(u_1) = rl(u_2)$.
%The only time the relaxed level of a leaf changes is when a \func{RootUntag} update changes the $tag$ bit of the root from one to zero, uniformly incrementing the relaxed level of every leaf by one.
%\func{RootUntag} also decreases the number of violations in the tree by one.
%Thus, eliminating $c$ violations can increase the relaxed level of every leaf by at most $c$.
%
%Once all violations are eliminated, each node $u$ satisfies $l(u) = rl(u)$, and the height of the tree is simply the relaxed level of any leaf.
%
%The claim then follows from the fact that $T'$ is produced by performing rotations to eliminate $c$ violations.
%
%\end{proof}

\begin{lem} \label{lem-ab-claims}
Consider a relaxed $(a,b)$-tree $T$ rooted that contains $n$ nodes and $c$ violations.
Suppose $T'$ is any $(a,b)$-tree that results from performing sufficient rotations on $T$ to eliminate all violations.
Then, the following claims hold for any leaf $u$ in $T$, and any leaf $u' \in T'$.
\begin{enumerate}
\item $l(u) \le rl(u) + c$
\label{claim-ab-l-rl}
\item $rl(u) \le rl(u') + c$
\label{claim-ab-rl-rlT}
\item $rl(u') = l(u')$
\label{claim-ab-rlT-lT}
\end{enumerate}
\end{lem}
\begin{chapscxproof}
\textbf{Claim~\ref{claim-ab-l-rl}:}
Immediate from the definitions of $l$ and $rl$.

\textbf{Claim~\ref{claim-ab-rl-rlT}:}
The rebalancing steps in a relaxed $(a,b)$-tree maintain the invariant that all leaves have the same relaxed level.
Thus, every pair of leaves $u_1$ and $u_2$ in $T$ satisfy $rl(u_1) = rl(u_2)$.
The only time the relaxed level of a leaf changes is when a \func{RootUntag} update changes the $tag$ bit of the root from one to zero, uniformly incrementing the relaxed level of every leaf by one.
\func{RootUntag} also decreases the number of violations in the tree by one.
Thus, eliminating $c$ violations can increase the relaxed level of every leaf by at most $c$.
The claim then follows from the fact that $T'$ is produced by performing rotations to eliminate $c$ violations.

\textbf{Claim~\ref{claim-ab-rlT-lT}:}
Immediate from definitions of $l$ and $rl$, and the fact that, since $T'$ is an $(a,b)$-tree, no node is tagged.
\end{chapscxproof}

\begin{cor} \label{cor-ab-height-and-violations}
Our implementation of a relaxed $(a,b)$-tree containing $n$ keys has height $O(c+ \log_a n)$, where $c$ is the number of incomplete \ins\ and \del\ operations.
\end{cor}
\begin{chapscxproof}
Let $r$, $T$, $r'$ and $T'$ be defined as in Lemma~\ref{lem-ab-claims}.
By Corollary~\ref{lem-abtree-violation-bound}, $T$ contains at most $c$ violations.
Thus, we immediately obtain $l(r) \le l(r') + 2c$ from Lemma~\ref{lem-ab-claims}.
Since $T'$ is an $(a,b)$-tree, $l(r') \in O(\log_a n)$.
Therefore, $l(r) \in O(c + \log_a n)$.
\end{chapscxproof}

\section{Experiments} \label{sec-abtree-exp}

We implemented the relaxed $(a,b)$-tree in C++, using a fast memory reclamation scheme called DEBRA that is described in Chapter~\ref{chap-debra}.
Since this implementation was done in C++, a direct comparison with the Java implementation of the chromatic tree described earlier would not make sense.
(Any performance differences due to the algorithms would be conflated with the differences between C++ and Java.)
So, for comparison, we also implemented an unbalanced version of the chromatic tree (BST) in C++, which simply does not perform any rebalancing steps.\footnote{We subsequently implemented the full chromatic tree in C++, but we did not have that implementation when these experiments were run. Subsequent experiments with random insertions and deletions of uniformly random keys indicated that the performance of BST is the same, or slightly better than that of the chromatic tree, for the workloads shown here.
This is because random operations on uniform keys yield a fairly balanced tree, so BST enjoys the benefits of balance without paying the overhead of performing rebalancing steps.}
BST used the same memory reclamation scheme as the relaxed $(a,b)$-tree.

We ran a small set of experiments on an Intel E7-4830 v3 with 12 cores and 2 hyperthreads per core, for a total of 24 hardware contexts.
Each core has a private 32KB L1 cache and 256KB L2 cache (which is shared between hyperthreads on a core).
All cores share a 30MB L3 cache.
The machine has 128GB of RAM and runs Ubuntu 14.04 LTS.

All code was compiled with the GNU C++ compiler (G++) 4.8.4 with build target x86\_64-linux-gnu and compilation options \texttt{-std=c++0x -mcx16 -O3}.
Thread support was provided by the POSIX Threads library.
We used the scalable allocator jemalloc 4.2.1~\cite{Evans:2006}, which greatly improved performance for both BST and the relaxed $(a,b)$-tree.

We pinned one thread to each hardware context.
For thread counts up to 12, we pinned at most one thread to each core, so hyperthreading is engaged only for thread counts 13 through 24.

In our experiments, we fixed the parameters for the $(a,b)$-tree at $a=6$ and $b=16$.
With $b = 16$, each node occupies four consecutive cache lines.
Since $(a,b)$-trees require $b \ge 2a-1$, with $b=16$, we must have $a \le 8$.
We chose to make $a$ slightly smaller than 8 in order to exploit a performance tradeoff: a smaller minimum degree slightly increases depth, but decreases the number of degree violations that are created in an execution.
With $a = 6$, after a leaf is created by an \func{Overflow} update (containing $b/2 = 8$ keys), three keys must be deleted from the leaf (with no intervening insertions into the leaf) before a degree violation is created.
(In contrast, with $a=8$, the first deletion from a leaf created by \func{Overflow} causes a degree violation, and triggers rebalancing.)

We compared the performance of BST and the relaxed $(a,b)$-tree using a simple randomized microbenchmark.
For each algorithm $A \in \{$BST, relaxed $(a,b)$-tree$\}$ and update rate $U \in \{100, 10, 0\}$, we ran five timed \textit{trials} for several thread counts $n$.
Each trial proceeded in two phases: \textit{prefilling} and \textit{measuring}.
In the prefilling phase, $n$ concurrent threads performed 50\% \textit{Insert} and 50\% \textit{Delete} operations on keys drawn uniformly randomly from $[0, 10^6)$ until the size of the tree converged to a steady state (containing approximately $10^6/2$ keys).
Next, the trial entered the measuring phase, during which threads began counting how many operations they performed.
(These counts were eventually summed over all threads and reported in our graphs.)
In this phase, each thread performed $(U/2)$\% \textit{Insert}, $(U/2)$\% \textit{Delete} and $(100-U)$\% \textit{Find} operations on keys drawn uniformly from $[0,10^6)$ for ten seconds.

As a way of validating correctness in each trial, each thread maintained a \textit{checksum}.
Each time a thread inserted a new key, it added the key to its checksum.
Each time a thread deleted a key, it subtracted the key from its checksum.
Observe that, at the end of any correct trial, the sum of all thread checksums must be equal to the sum of keys in the tree.

\begin{figure}[tb]
    \centering
    \setlength\tabcolsep{1pt}
    \begin{tabular}{m{0.33\linewidth}m{0.33\linewidth}m{0.33\linewidth}}
        \fcolorbox{black!50}{black!20}{\parbox{\dimexpr \linewidth-2\fboxsep-2\fboxrule}{\centering {0\% updates}}} &
        \fcolorbox{black!50}{black!20}{\parbox{\dimexpr \linewidth-2\fboxsep-2\fboxrule}{\centering {10\% updates}}} &
        \fcolorbox{black!50}{black!20}{\parbox{\dimexpr \linewidth-2\fboxsep-2\fboxrule}{\centering {100\% updates}}}
        \\
        \includegraphics[width=\linewidth]{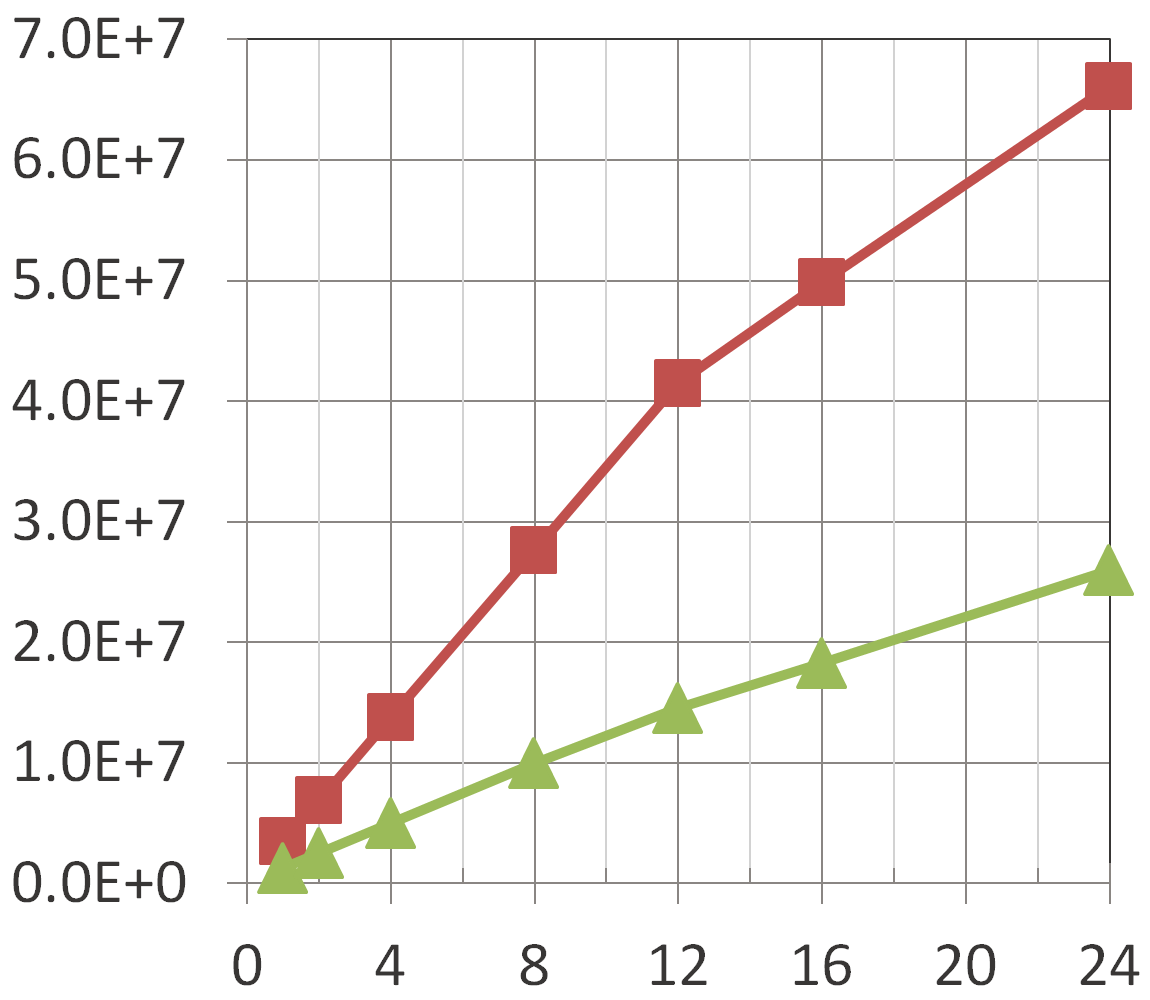} &
        \includegraphics[width=\linewidth]{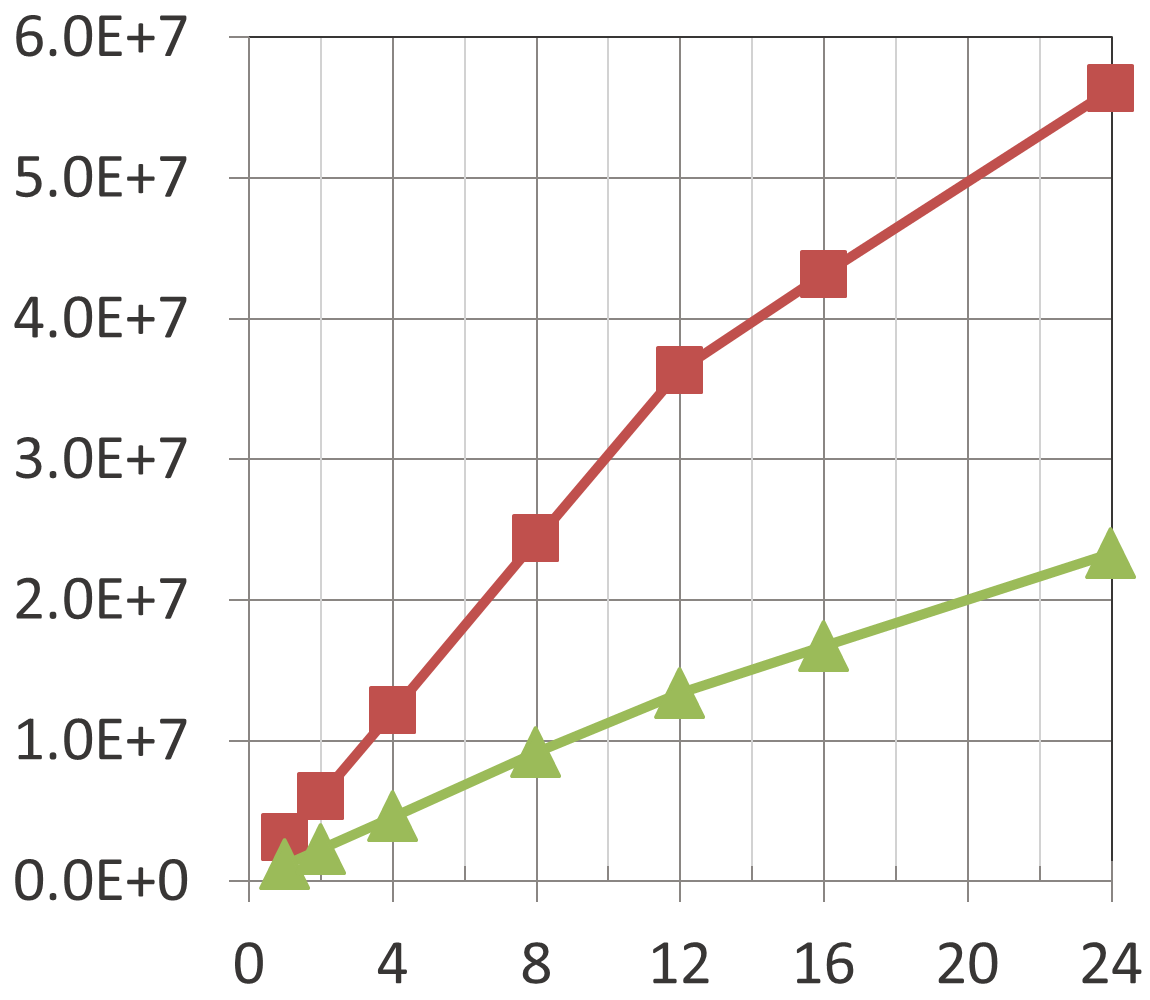} &
        \includegraphics[width=\linewidth]{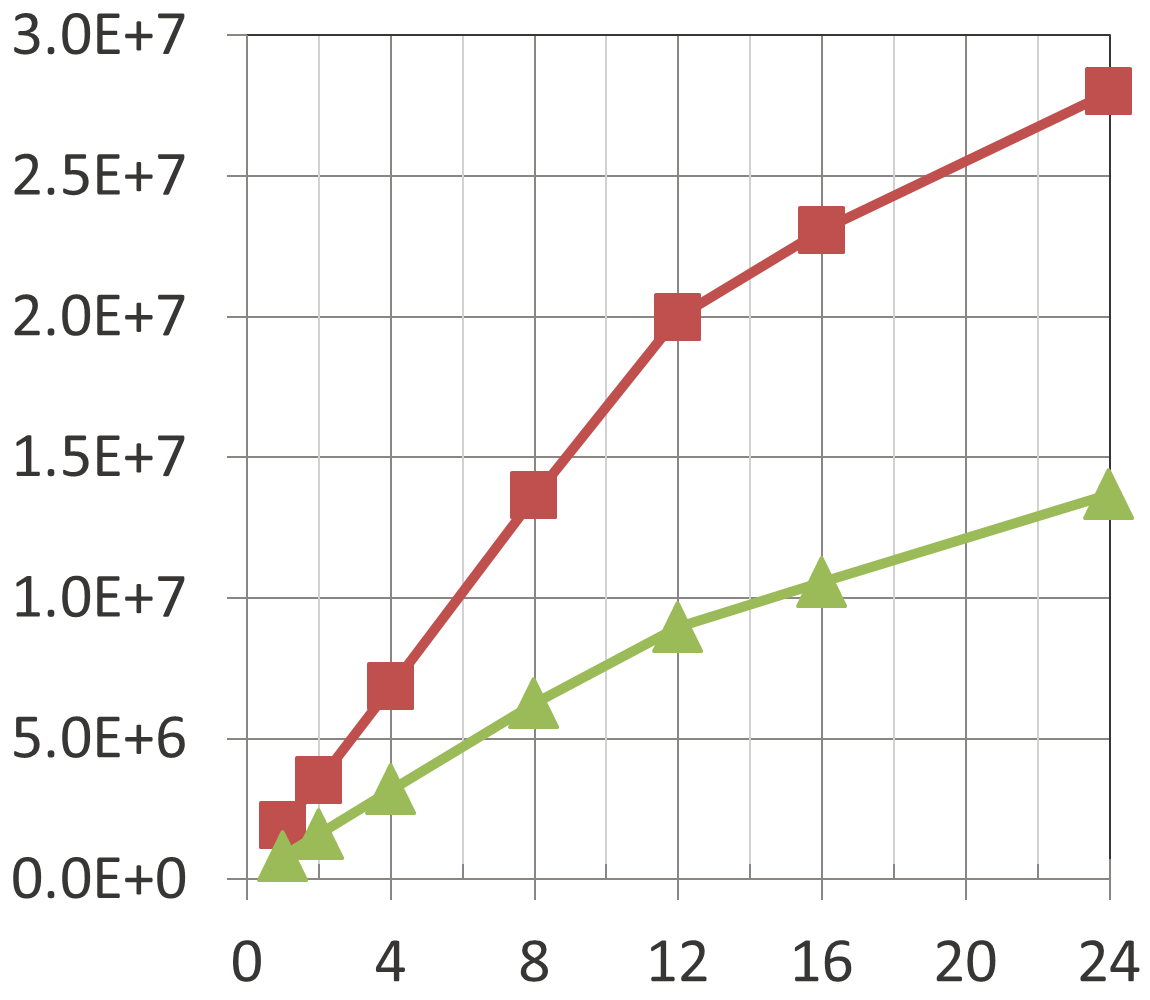}
        \\
    \end{tabular}
	\includegraphics[width=0.4\linewidth]{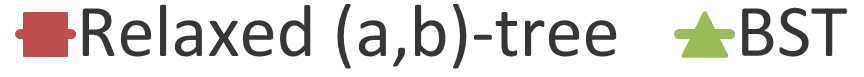}
\caption{Performance results for a microbenchmark comparing BST and the relaxed $(a,b)$-tree.
The x-axis represents the number of concurrent threads.
The y-axis represents operations per microsecond.}
\label{fig-abtree-experiments}
\end{figure}

The results for this benchmark appear in Figure~\ref{fig-abtree-experiments}.
In these experiments, the key range is fairly large, and the relaxed $(a,b)$-tree has a significant performance advantage over BST.
With 100\% updates, it approximately doubles the performance of the BST, and the performance gap grows wider as the proportion of operations that are searches increases.
This performance advantage comes from the smaller depth and improved cache locality offered by the fat nodes in the $(a,b)$-tree.

We briefly explain why cache performance is better in a tree with fat nodes than in a BST.
In a modern system, whenever a process reads a memory address, the cache coherence protocol actually reads the entire cache line containing the address (typically 64 bytes) and stores it in the cache.
In fact, in recent Intel systems, there is a cache \textit{prefetcher} that additionally fetches an \textit{adjacent} cache line, as well as a \textit{streamer} that may fetch even more subsequent, consecutive cache lines (up to an entire page of memory).
Thus, each read from memory loads \textit{at least} two full cache lines from memory (and often three or four).

For example, reading a key in a BST node might also cause the value and child pointers to be loaded into the cache at the same time, if they happen to reside in the same cache line (or the adjacent one).
In this particular experiment, a BST node was 40 bytes, so each time a node was accessed, at least $128-40 = 88$ bytes of unrelated information was also loaded into the cache.
This unrelated information clutters the cache, negatively impacting its utilization.
Furthermore, when a process is searching in the tree, each time a pair of cache lines are retrieved from memory, they likely contain only one relevant key, and the next relevant key cannot be accessed until more cache lines are loaded, which introduces significant latency.
This is sometimes called the \textit{pointer chasing} problem.

In contrast, fat nodes \textit{harness} prefetching behaviour by storing several cache lines worth of relevant information consecutively.
Consequently, cache utilization is improved, and accessing a key has the side effect of loading several other relevant keys and/or pointers from memory, at the same time (reducing the impact of pointer chasing).

Preliminary experiments suggest that there is an even larger performance gap with larger trees, and that the relaxed $(a,b)$-tree remains competitive with the BST even when the key range is two orders of magnitude smaller than in the workload shown here.
In light of these results, future code libraries should consider including concurrent data structures with fat nodes, like the relaxed $(a,b)$-tree.

\chapter{\bslack s: space efficient B-trees} \label{chap-bslack}
% !TEX root = paper.tex

\begin{hide}
B-trees are balanced trees designed for block-based storage media.
Internal nodes contain between $b/2$ and $b$ child pointers, and one less key.
Leaves contain between $b/2$ and $b$ keys.
All leaves have the same depth, so access times are predictable.
If memory can be allocated on a per-byte basis, nodes can simply be allocated the precise amount of space they need to store their data, and no space is wasted.
However, typically, all nodes have the same, fixed capacity, and some of the capacity of nodes is wasted.
As much as 50\% of the capacity of each node is wasted in the worst case.
This is particularly problematic when data structures are being implemented in hardware, since memory allocation and reclamation schemes are often very simplistic, allowing only a single block size to be allocated (to avoid fragmentation).
Furthermore, since hardware devices must include sufficient resources to handle the worst-case, good expected behaviour is not enough to allow hardware developers to reduce the amount of memory included in their devices.
To address this problem, we introduce \textit{\bslack s}, which are a variant of B-trees with substantially better worst-case space complexity.
We also introduce \rbslack s, which are a variant of \bslack s that are more amenable to concurrent implementation.
%A non-blocking implementation of a \bslack\ can be obtained using the tree update template described in \cite{paper2}.
%(Details are %will be
%provided in the full version of the paper.)
%This technique is particularly efficient with small, localized updates.
%The resulting implementation will allow updates to proceed without interfering with searches, and will continue to function correctly even if processes crash.

The development of \bslack s was inspired by a collaboration with a manufacturer of internet routers, who wanted to build a concurrent router based on a tree.
In such embedded devices, memory is limited, so it is important to use it as efficiently as possible.
%Many embedded systems are also concurrent.
A suitable tree would have a simple algorithm for updates, small space complexity, fast searches, and searches that would not be blocked by concurrent updates.
Updates were expected to be
%very
infrequent.
One naive approach is to rebuild the entire tree after each update.
Keeping an old copy of the tree while rebuilding a new copy would allow searches to proceed unhindered, but this would double the space required to store the tree.

%Search trees can be either node-oriented, in which each key is stored in an internal node or leaf, or leaf-oriented, in which each key is stored at a leaf and the keys of internal nodes serve only to direct a search to the appropriate leaf.
%In a node-oriented B-tree, the leaves and internal nodes have different sizes (because internal nodes contain keys and pointers to children, and leaves contain only keys).
%So, if only one block size can be allocated, a significant amount of space is wasted.
%Moreover, deletion in a node-oriented tree sometimes requires stealing a key from a successor (or predecessor), which can be in a different part of the tree.
%This is a problem for concurrent implementation, since the operation can involve a large number of nodes, namely, the nodes on the path between the node and its successor.

\bslack s are leaf-oriented trees with many desirable properties.
The average degree of nodes is high, exceeding $b-2$ for trees of height at least three.
Their space complexity is better than all of their competitors. % (see Appendix~\ref{sec-bslack-space-complexity}). %, even some that do not support deletion.
Consider a dictionary implemented by a leaf-oriented search tree, in which, along with each key, a leaf stores a pointer to associated data.
Suppose that each key and each pointer to a child or to data occupies a single word.
Then, $\frac{2b}{b-3}n$ is an upper bound on the number of words needed to store a \bslack\ with $n > b^3$ keys.
%($\bigstar$ A tighter bound appears in Appendix~\ref{sec-maxslack-to-bslack}. \trevor{also in the analysis section's summary...})
For large $b$, this tends to $2n$, which is optimal.
Section~\ref{sec-analysis} gives a more complex upper bound, which is much better when $b$ is small.
\bslack s have logarithmic height, and the number of rebalancing steps performed after a sequence of $m$ updates to a \bslack\ of size $n$ is amortized $O(\log (n+m))$ per update.
Furthermore, the number of rebalancing steps needed to rebalance the tree can be reduced to amortized \textit{constant} per update at the cost of slightly increased space complexity. %in the full version of the paper. %(see Appendix~\ref{sec-constant-rebalancing}).

The rest of this paper is organized as follows.
Section~\ref{sec-bslack-related} surveys related work.
Section~\ref{sec-bslack} introduces \bslack s and \rbslack s.
Height, average degree, space complexity and rebalancing costs of \rbslack s (and, hence, of \bslack s) are analyzed in Section~\ref{sec-analysis}.
Section~\ref{sec-constant-rebalancing} describes the variant of \bslack s with amortized constant rebalancing. %a small modification to \bslack s that requires only an amortized constant number of rebalancing steps per insertion or deletion to maintain balance.
Section~\ref{sec-bslack-space-complexity} compares the worst-case space complexity of \bslack s to competing data structures.
Single-threaded experimental results appear in Section~\ref{sec-bslack-exp}.
Section~\ref{sec-bslack-cleanup} discusses some implementation issues surrounding rebalancing.
Section~\ref{sec-bslack-concurrent} describes an approach for obtaining a lock-free implementation of a \bslack\ using the template described in Chapter~\ref{chap-template}.
%Finally, we conclude in Section~\ref{sec-bslack-conclusion}.
\end{hide}

\section{Related work} \label{sec-bslack-related}

\begin{hide}
B-trees were initially proposed by Bayer and McCreight in 1970 \cite{bayer1970organization}. Insertion into a full node in a B-tree causes it to split into two nodes, each half full.
Deletion from a half-full node causes it to merge with a neighbour.
Arnow, Tenenbaum and Wu proposed P-trees \cite{arnow1985p}, which enjoy moderate improvements to \textit{average} space complexity over B-trees, but waste 66\% of each node in the worst case.

A number of generalizations of B-trees have been suggested that achieve much less waste if no deletions are performed.
Bayer and McCreight also proposed B*-trees in \cite{bayer1970organization}, which improve upon the worst-case space complexity of B-trees.
At most a third of the capacity of each node in a B*-tree is wasted.
This is achieved by splitting a node only when it and one of its neighbours are both full, replacing these two nodes by three nodes.
K{\"u}spert \cite{kuspert1983} generalized B*-trees to trees where each node contains between $\lfloor \frac{bm}{m+1} \rfloor$ and $b$ pointers or keys, where $m \le b-1$ is a design parameter.
Such a tree behaves just like a B*-tree everywhere except at the leaves.
An insertion into a full leaf causes keys to be shifted among the nearest $m-1$ siblings to make room for the inserted key.
If the $m-1$ nearest siblings are also full, then these $m$ nodes are replaced by $m+1$ nodes which evenly share keys.
Large values of $m$ yield good worst-case space complexity.
However, with large $m$, updates would be complicated and inefficient in a concurrent setting.

Baeza-Yates and Per-\r{a}ke Larson introduced B+trees with partial expansions \cite{baeza1989performance}.
Several node sizes are used, each a multiple of the block size.
An insertion to a full node causes it to expand to the next larger node size.
With three node sizes, at most 33\% of each node can be wasted, and worst-case utilization improves with the number of block sizes used.
However, this technique simply pushes the complexity of the problem onto the memory allocator. Memory allocation is relatively simple for one block size, but it quickly becomes impractical for simple hardware to support larger numbers of block sizes.

Culik, Ottmann and Wood introduced strongly dense multiway trees (SDM-trees) \cite{culik1981dense}.
An SDM-tree is a node-oriented tree in which all leaves have the same depth, and the root contains at least two pointers.
Apart from the root, every internal node $u$ with fewer than $b$ pointers has at least one sibling.
Each sibling of $u$ has $b$ pointers if it is an internal node and $b$ keys if it is a leaf.
Insertion can be done in $O(b^3 + (\log n)^{b-2})$ time.
Deletion is not supported, but the authors mention that the insertion algorithm could be modified to obtain a deletion algorithm, and the time complexity of the resulting algorithm ``would be at most $O(n)$ and at least $O((\log n)^{b-1})$.''
Besides the long running times for each operation (and the lack of better amortized results), the insertion algorithm is very complex and involves many nodes, which makes it poorly suited for hardware implementation.
Furthermore, in a concurrent setting, an extremely large section of the tree would have to be modified atomically, which would severely limit concurrency.

Srinivasan introduced a leaf-oriented B-tree variant called an \textit{Overflow tree} \cite{Srinivasan01011991}.
For each parent of a leaf, its children are divided into one or more groups, and an overflow node is associated with each group.
The tree satisfies the B-tree properties and the additional requirement that each leaf contains at least $b-1-s$ keys, where $s \ge 2$ is a design parameter and $b$ is the maximum degree of nodes.
Inserting a key into a full leaf causes the key to be inserted into the overflow node instead; if the overflow node is full, the entire group is reorganized.
Deleting from a leaf is the same as in a B-tree unless it will cause the leaf to contain too few keys, in which case, a key is taken from the overflow node; if a key cannot be taken from the overflow node, the entire group is reorganized.
Each search must look at an overflow node.
The need to atomically modify and search two places at once makes this data structure poorly suited for concurrent implementation.

Hsuang introduced a class of node-oriented trees called \textit{H-trees} \cite{Huang:1985:HTO:3857.3858}, which are a subclass of B-trees parameterized by $\gamma$ and $\delta$.
These parameters specify a lower bound on the number of grandchildren of each internal node (that has grandchildren), and a lower bound on the number of keys contained in each leaf, respectively.
Larger values of $\delta$ and $\gamma$ yield trees that use memory more efficiently.
When $\delta$ and $\gamma$ are as large as possible, each leaf contains at least $b-3$ keys, and each internal node has zero or at least $\lfloor \frac{b^2+1}{2} \rfloor$ grandchildren.
The paper presents $O(\log n)$ insertion and deletion algorithms for node-oriented H-trees.
The algorithms are very complex and involve many cases.
H-trees have a minimum average degree of approximately $b/\sqrt{2}$ for internal nodes, which is much smaller than the $b-2$ of \bslack s (for trees of height at least three).

Section~\ref{sec-bslack-space-complexity} 
%\textbf{[[[fix this]]]} The full version of the paper %Appendix~\ref{sec-bslack-space-complexity}
describes families of B-trees, H-trees and Overflow trees which require significantly more space than  \bslack s.

Rosenberg and Snyder introduced \textit{compact B-trees} \cite{Rosenberg1979}, which can be constructed from a set of keys using the minimum number of nodes possible.
No compactness preserving insertion or deletion procedures are known.
The authors suggested using regular B-tree updates and periodically compacting a data structure to improve efficiency.
However, experiments in \cite{arnow1984empirical} showed that starting with a compact B-tree and adding only 1.6\% more keys using standard B-tree operations reduced storage utilization from 99\% to 67\%.
Thus, to maintain reasonable space complexity, the expensive tree rebuilding/compaction algorithm would have to be executed prohibitively often.

An impressive paper by Br{\"o}nnimann et~al. \cite{bronnimann2007putting} presented three ways to transform an arbitrary sequential dictionary into a more space efficient one.
One of these ways will be discussed here; of the other two, one is extremely complex and poorly suited for concurrent hardware implementation, and the other pushes the complexity onto the memory allocator.

This transformation takes any sequential tree data structure and modifies it by replacing each key in the sequential data structure with a \textit{chunk}, which is a group of $b-2$, $b-1$ or $b$ keys, where $b$ is the memory block size.
All chunks in the data structure are also kept in a doubly linked list to facilitate iteration and movement of keys between chunks. % over the keys in the data structure.
For instance, a BST would be transformed into a tree in which each node has zero, one or two children, and $b-2$, $b-1$ or $b$ keys.
All keys in chunks in the left subtree of a node $u$ would be smaller than all keys in $u$'s chunk, and all keys in chunks in the right subtree of $u$ would be larger than all keys in $u$'s chunk.
A search for key $k$ behaves the same as in the sequential data structure until it reaches the only chunk that can contain $k$, and searches for $k$ within the chunk.
An insertion first searches for the appropriate chunk, then it inserts the key into this chunk.
Inserting into a full chunk requires shifting the keys of the $b$ closest neighbouring chunks to make room.
If these $b$ chunks are full, then $b$ consecutive keys from these chunks are first placed in a new chunk, and the remaining $b(b-1)$ keys are distributed amongst the original $b$ chunks. %a key is taken from each, and a new node containing $b$ keys is inserted using the sequential data structure's insertion algorithm.
Deletion is similar.
%\textbf{[[[this is not precise; you can't just take one key from each. you have to take $b$ consecutive keys out and put them in a new node, then distribute the remaining $b(b-1)$ keys amongst the original $b$ nodes.]]]}
Each operation in the resulting data structure runs in $O(f(n)+b^2)$ steps, where $f(n)$ is the number of steps taken by the sequential data structure to perform the same operation.

After this transformation, a B-tree with maximum degree $b$ requires $2n+O(n/b)$ words to store $n$ keys and pointers to data.
In the worst-case, each chunk wastes $2/b$ of its space, which is somewhat worse than in \bslack s.
Furthermore, supporting fast searches can introduce significant complexity to the hardware design.
Suppose this transformation is applied to a B-tree.
Then a node in the transformed B-tree contains up to $b-1$ chunks, each of which contains $b-2, b-1$ or $b$ keys, and occupies one block of memory.
Thus in order to determine which child pointer should be followed, a search must load up to $b-1$ blocks.
In order for searches to be fast, hardware must therefore be able to quickly load up to $b-1$ blocks from memory (a total of $\theta(b^2)$ memory words).
This represents a significant challenge for hardware design.
%Therefore, hardware must be able to quickly load up to $b-1$ blocks (up to $b(b-1)$ words) from memory at once, or else deciding which child pointer a search should follow will be slow.

%$\bigstar$
As we saw in Chapter~\ref{chap-abtree}, B-trees with relaxed balance have previously been proposed.
%
%Insertion and deletion in a balanced search tree typically involves performing one or more \textit{rebalancing steps} in order to maintain balance.
%\textit{Relaxed balanced} search trees decouple rebalancing from insertion and deletion, so that rebalancing steps can be delayed or interleaved with insertions and deletions.
%A brief survey of relaxed balanced search trees can be found in \cite{DBLP:journals/acta/Larsen98}.
%Larsen and Fagerberg introduced \textit{relaxed B-trees} \cite{DBLP:journals/ijfcs/LarsenF96}, which are a relaxed balanced version of B-trees.
%Relaxed B-trees have very simple insertion and deletion operations, and rebalancing steps can be performed (in any order) to transform a relaxed B-tree into a B-tree.
%%If rebalancing is delayed and multiple insertions or deletions are performed, then nodes may violate B-tree properties by containing too few keys and pointers.
%Several of the updates to \bslack s are derivative of updates to relaxed B-trees.
%Larsen and Fagerberg also improved the space complexity of relaxed B-trees while the tree is out of balance \cite{JL01abtrees}, but the worst-case space complexity is no better than in B-trees.
%
In particular, several of the updates to \bslack s are derivative of updates to the relaxed $(a,b)$-tree.
However, limited work has been done on improving the space complexity of relaxed balance data structures.
Larsen and Fagerberg improved the space complexity of the relaxed $(a,b)$-tree \textit{while the tree is out of balance}~\cite{JL01abtrees}, but its worst-case space complexity remains unchanged. %no better than in B-trees.
\end{hide}

\section{\bslack s} \label{sec-bslack}

\begin{hide}
A \bslack\ is a variant of a B-tree. %with maximum degree $b > 4$.
Each node stores its keys in sorted order, so binary search can be used to determine which child of an internal node should be visited next by a search, or whether a leaf contains a key.
Let $p_0, p_1, ..., p_m$ be the sequence of pointers contained in an internal node, and $k_1, k_2, ..., k_m$ be its sequence of keys.
For each $1 \le i \le m$, the subtree pointed to by $p_{i-1}$ contains keys strictly smaller than $k_i$, and the subtree pointed to by $p_i$ contains keys greater than \textit{or equal to} $k_i$.
We say that the \textit{degree of an internal node} is the number of non-\nil\ pointers it contains, and the \textit{degree of a leaf} is the number of keys it contains.
This unusual definition of degree simplifies our discussion.
The degree of node $v$ is denoted $deg(v)$.
If the maximum possible degree of a node is $b$, and its degree is $b-x$, then we say it contains $x$ units of \textit{slack} (or simply $x$ \textit{slack}).

\noindent A \bslack\ is a leaf-oriented search tree with maximum degree $b > 4$ in which:
\begin{compactenum}[\hspace{4.1mm}\bfseries P1:]
\item every leaf has the same depth,
\label{prop-bslack-depth}
\item internal nodes contain between 2 and $b$ pointers (and one less key),%
\label{prop-bslack-internal}
\item leaves contain between 0 and $b$ keys, and
\label{prop-bslack-leaf}
\item for each internal node $u$, the total slack contained in the \textit{children} of $u$ is at most $b-1$.
%\item for each internal node $u$ with children $v_1, v_2, ..., v_{|c(u)|}$,\\
%$deg(v_1)+deg(v_2)+...+deg(v_{|c(u)|}) \ge |c(u)|b-b$.
\label{prop-bslack-slack}
\end{compactenum}
P\ref{prop-bslack-slack} is the key property that distinguishes \bslack s from other variants of B-trees. %(and we derive the name Bounded-slack tree, or \bslack, from it).
%For any internal node $u$, the sum of the degrees of its children is at most $|c(u)|b$.
%If an internal node has $|c(u)|$ children, then the sum of the degrees of its children is at most $|c(u)|b$.
%Thus, P\ref{prop-bslack-slack} says that the amount of \textit{slack}, or wasted space, amongst the children of an internal node is at most $b$ (hence the name \bslack).
It limits the aggregate space wasted by a number of nodes, as opposed to limiting the space wasted by each node.
Alternatively, P\ref{prop-bslack-slack} can be thought of as a lower bound on the sum of the degrees of the children of each internal node.
Formally, for each internal node with children $v_1, v_2, ..., v_l$, $deg(v_1)+deg(v_2)+...+deg(v_l) \ge lb-(b-1) = lb-b+1$.
This interpretation is useful to show that all nodes have large subtrees.
For instance, it %, the relation above
implies that a node $u$ with two internal children must have at least $b+1$ grandchildren.
If these grandchildren are also internal nodes, we can conclude that % use the same relation at each of these grandchildren (and some arithmetic) to show that
$u$ must have at least $b^2-b+2$ great grandchildren.

A tree that satisfies P\ref{prop-bslack-depth}, and in which every node has degree $b-1$, is an example of a \bslack.
%Another example can be constructed as follows.
Another example of a \bslack\ is a tree of height two, where $b$ is even, the root has degree two, its two children have degree $b/2$ and $b/2+1$, respectively, and the grandchildren of the root are leaves with degree $b$, except for two, one in the left subtree of the root, and one in the right subtree, that each have degree one.
This tree contains the smallest number of keys of any \bslack\ of height two.
\end{hide}

\subsection{\Rbslack s} \label{sec-rbslack}

\begin{hide}
A relaxed balance search tree decouples updates that rebalance (or reorganize the keys of) the tree from updates that modify the set of keys stored in the tree~\cite{relaxedbalance}.
%We now define a \textit{relaxed balance} variant of \bslack s, called \textit{\rbslack s}.
The advantages of this decoupling are twofold.
First, updates to a relaxed balance version of a search tree are smaller, so a greater degree of concurrency is possible in a multithreaded setting.
Second, for some applications, it may be useful to temporarily disable rebalancing to allow a large number of updates to be performed quickly, and to gradually rebalance the tree afterwards.

A \rbslack\ is a relaxed balance version of a \bslack\ that has weakened the properties.
A \weight\ of zero or one is associated with each node.
These \weight s serve a purpose similar to the colors red and black in a red-black tree.
We define the \textit{relaxed depth} of a node to be one less than the sum of the \weight s on the path from the root to this node.
A \rbslack\ is a leaf-oriented search tree with maximum degree $b > 4$ in which:
%\noindent A \rbslack, where $b > 4$, is a leaf-oriented search tree in which:
\begin{compactenum}[\hspace{4.1mm}\bfseries P1$'$:]
\setcounter{enumi}{-1}
\item every node with \weight\ zero contains exactly two pointers,
\label{prop-rbslack-weight-zero}
\item every leaf has the same relaxed depth,
\label{prop-rbslack-depth}
\item internal nodes contain between 1 and $b$ pointers (and one less key), and%
\label{prop-rbslack-internal}
%\item leaves contain between 0 and $b$ keys
%\label{prop-bslack-leaf}
\end{compactenum}
\begin{compactenum}[\hspace{4.1mm}\bfseries P1\hspace{1.01mm}:]
\setcounter{enumi}{2}
\item leaves contain between 0 and $b$ keys
%\label{prop-rbslack-leaf}
\end{compactenum}

%, by replacing P\ref{prop-bslack-depth} with P\ref{prop-bslack-depth}$'$, replacing P\ref{prop-bslack-internal} with the property that internal nodes have between one and $b$ pointers, and eliminating P\ref{prop-bslack-slack}.
To clarify the difference between \bslack s and \rbslack s, we identify several types of \textit{violations} of the \bslack s properties that can be present in a \rbslack.
We say that a \textit{\weight\ violation} occurs at a node with \weight\ zero, a \textit{slack violation} occurs at a node that violates P\ref{prop-bslack-slack}, and a \textit{degree violation} occurs at an internal node with only one child (violating P\ref{prop-bslack-internal}).
%Note that these are violations of the \bslack\ properties, but not violations of the \rbslack\ properties.
Observe that P1 is satisfied in a \rbslack\ with no \weight\ violations.
Likewise, P2 is satisfied in a \rbslack\ with no degree violations, and P4 is satisfied in a \rbslack\ with no slack violations.
Therefore, a \rbslack\ that contains no violations is a \bslack.
%Although a \rbslack\ satisfies much weaker properties than a \bslack,
Rebalancing steps can be performed to eliminate violations, and gradually transform any \rbslack\ into a \bslack.
%Specifically, we use rebalancing updates to set all \weight s to one, and to ensure that P\ref{prop-bslack-slack} is satisfied.
\end{hide}

\subsection{Updates to \rbslack s}

\begin{figure}[tbph]
\centering
%\vspace{-10mm}
%\renewcommand{\arraystretch}{3}
\begin{tabular}{ | m{2.2cm} | >{\centering\arraybackslash} m{13cm} | }
\hline

%\vspace{7mm}
\textbf{Delete} & \includegraphics[scale=0.85]{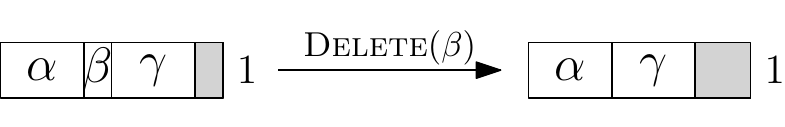} \\
\hline

\vspace{7mm}
%\func{ReplacePair} & \includegraphics[scale=0.85]{chap-template/abtree/ops-replacepair.pdf} \\
%\func{InsertPair} & \includegraphics[scale=0.85]{chap-template/abtree/ops-insertpair.pdf} \\
%\func{Overflow} & \includegraphics[scale=0.85]{chap-template/abtree/ops-overflow.pdf} \\
\textbf{Insert} & \includegraphics[scale=0.85]{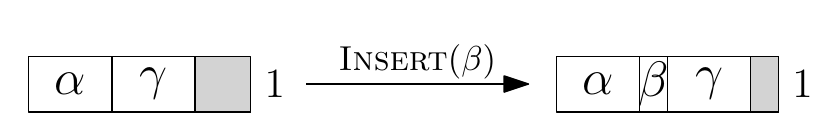} \\
\textbf{Overflow} & \includegraphics[scale=0.85]{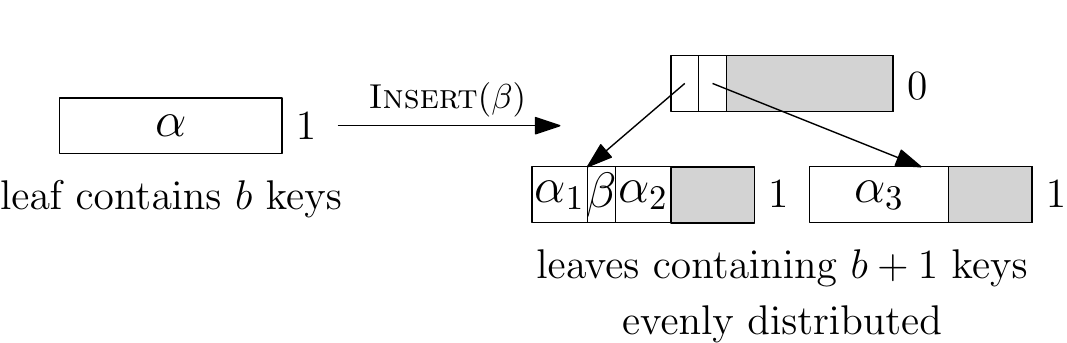} \\
%\textbf{Insert-Distribute} & \includegraphics[scale=0.7]{chap-bslack/figures/ops-insert-distribute.pdf}\\
%\textbf{Insert-Overflow} & \includegraphics[scale=0.7]{chap-bslack/figures/ops-insert-overflow.pdf} \\
\hline

\vspace{7mm}
\textbf{Root-Zero} & \includegraphics[scale=0.85]{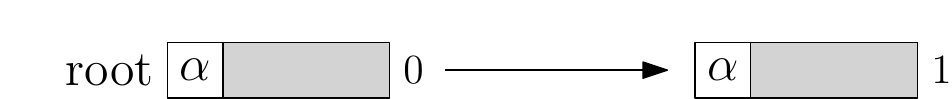} \\
\textbf{Root-Replace} & \includegraphics[scale=0.85]{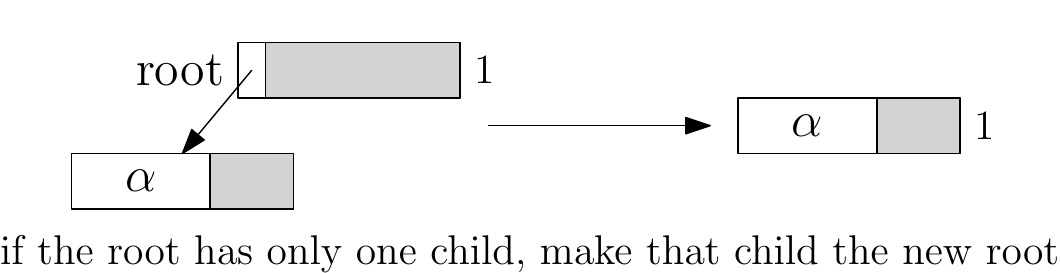} \\
\hline

\vspace{7mm}
\textbf{Absorb} & \includegraphics[scale=0.85]{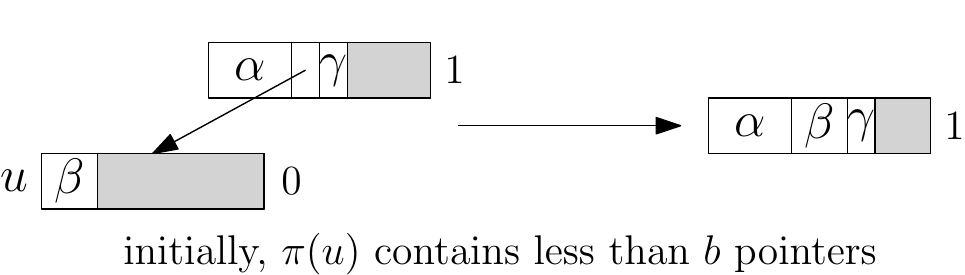} \\
\textbf{Split} & \includegraphics[scale=0.85]{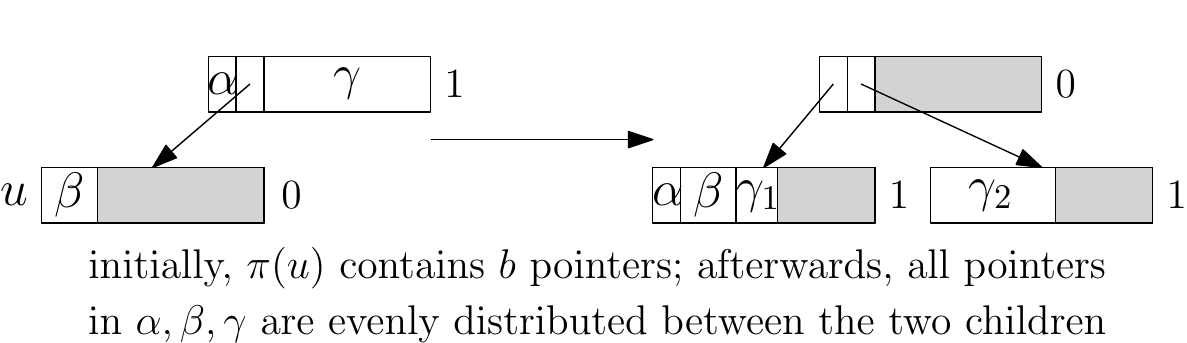} \\
\hline

\vspace{7mm}
\textbf{Compress} & \includegraphics[scale=0.85]{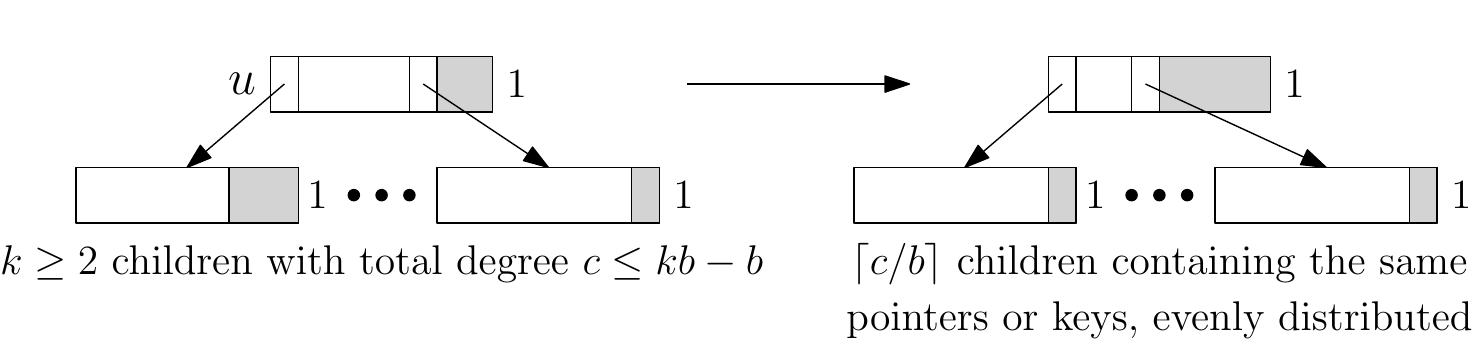} \\
\textbf{One-Child} & \includegraphics[scale=0.85]{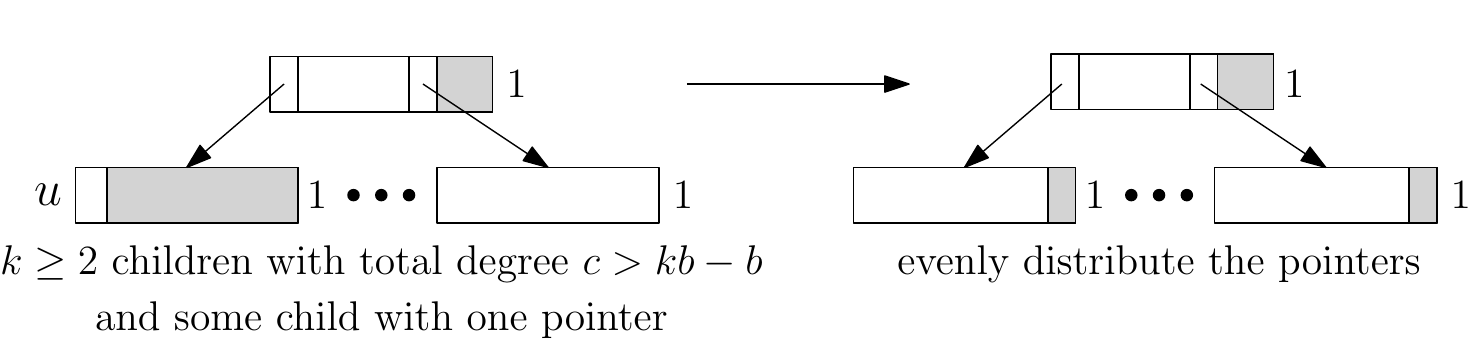} \\
\hline
\end{tabular}
\vspace{-2mm}
%\vspace{-5mm}
\caption{Updates to \bslack s (and \rbslack s).
Nodes with \weight\ zero contain exactly two pointers.
}
\label{fig-bslack-updates}
\end{figure}

We now describe the algorithms for inserting and deleting keys in a \rbslack\ (in a way that maintains P\ref{prop-rbslack-weight-zero}$'$, P\ref{prop-rbslack-depth}$'$, P\ref{prop-rbslack-internal}$'$ and P\ref{prop-bslack-leaf}).
We use the \textbf{Insert}, \textbf{Delete} and \textbf{Overflow} updates introduced by Larsen and Fagerberg in their work on relaxed $(a,b)$-trees~\cite{LF95}.
These updates appear in Figure~\ref{fig-bslack-updates}.
There, \weight s appear to the right of nodes, and shaded regions represent slack.
If $u$ is a node that is not the root, then we let $\pi(u)$ denote the parent of $u$.
The insertion and deletion algorithms always ensure that all leaves have \weight\ one.
We also study how these updates change the amount of slack in nodes, and how they create, move or eliminate violations.

\textbf{Deletion.}
First, a search is performed to find the leaf $u$ where the deletion should occur.
If the leaf does not contain the key to be deleted, then the deletion terminates immediately, and the tree does not change.
If the leaf contains the key to be deleted, then the key is removed from the sequence of keys stored in that leaf. %, and the \textit{cleanup phase} begins.
%Note that deletion maintains the properties of a \rbslack.
%\trevor{can create P4 violation only}
%\trevor{if exists, call cleanup, which fixes this violation}
%In the cleanup phase, if the children of $\pi(u)$ contain more than $b$ slack in total, then a Compress update is performed at $\pi(u)$ to reduce the number of children (and, hence, the total slack shared amongst the children of $\pi(u)$) and evenly distribute keys.
%(The Compress %, Absorb and Split updates are
%update is described in detail in Section~\ref{sec-rebalancing}.)
%As we discuss in Section~\ref{sec-rebalancing}, a Compress update that affects a node $v$ and its children may cause P\ref{prop-bslack-slack} to be violated at $\pi(v)$, and at some of the children of $v$.
%Compress updates are performed until no node violates~P\ref{prop-bslack-slack}, at which point the deletion terminates.
Deleting this key may create a slack violation. % in which case rebalancing steps can be performed to eliminate this violation.

%If, after the deletion, the children of $\pi(u)$ contain more than $b$ slack, then a Compress update must be performed at $\pi(u)$ to reduce the number of children and evenly distribute keys.
%(In actuality, it does not matter whether keys are evenly distributed.
%We choose to evenly distribute them because this will also evenly distribute any slack, and hopefully, make subsequent insertions more likely to reach a node that contains some slack, so that they involve less work.)
%The Compress update will increase the slack of the grandparent of $u$ by one, so another Compress update may be necessary.

\textbf{Insertion.}
To perform an insertion, a search is first performed to find the leaf $u$ where the insertion should occur.
If $u$ contains some slack, then the key is added to the sequence of keys in $u$, and the insertion terminates.
Otherwise, $u$ cannot accommodate the new key, so Overflow is performed.
Overflow replaces $u$ by a subtree of height one consisting of an internal node with \weight\ zero, and two leaves with \weight\ one.
%Doing this naively would violate P\ref{prop-bslack-depth}$'$.
%One way to do this correctly is to replace a leaf of \weight\ one by a subtree whose root has \weight\ zero, and whose leaves have \weight\ one.
The $b$ keys stored in $u$, plus the new key, are evenly distributed between the children of the new internal node.
If $u$ was the root before the insertion, then the new internal node becomes the new root.
Otherwise, $u$'s parent $\pi(u)$ before the insertion is changed to point to the new internal node instead of $u$.
%After the insertion, P\ref{prop-bslack-internal} and P\ref{prop-bslack-leaf} are satisfied.
%Any leaves in the subtree rooted at $\pi(u)$ have the same relaxed depth after the update as they did beforehand.
After Overflow, there is a \weight\ violation at the new internal node.
%since there is a node with \weight\ zero, P\ref{prop-bslack-depth} is violated.
Additionally, since the new internal node contains $b-2$ slack, whereas $u$ contained no slack, there may be a slack violation at $\pi(u)$. %P\ref{prop-bslack-slack} may be violated at $\pi(u)$.
%Finally, rebalancing steps are performed to eliminate any violations.
%
%After this, the cleanup phase begins.
%In the cleanup phase, if the grandparent of $u$ is not the root, then it has \weight\ zero, so Absorb and Split updates are performed until no node contains \weight\ zero.
%As soon as a Absorb is performed, the zero \weight\ is eliminated.
%However, if no Absorb is performed, then Split updates are performed to push the zero \weight\ up in the tree, until it reaches the root, where it will be eliminated.
%Once no node has weight zero, Compress updates must be performed wherever any node violated P\ref{prop-bslack-slack}.
%Absorb updates cannot cause any node to violate P\ref{prop-bslack-slack}, but Split updates can cause P\ref{prop-bslack-slack} to be violated at the two child nodes.
%Since Compress updates may necessitate other Compress updates, they are performed until no node violates~P\ref{prop-bslack-slack}, at which point the insertion terminates.
%Let $p_0, ..., p_m$ and $k_0, ..., k_{m-1}$ be the pointers and keys, respectively, of $\pi(u)$.
%Pointers $p_{\lfloor b/2 \rfloor}, ..., p_m$ and keys $k_{\lfloor b/2 \rfloor+1}, ..., k_{m-1}$ are moved from $\pi(u)$ to a new node, which becomes $\pi(u)$'s right sibling under $gp$.
%Key $k_{\lfloor b/2 \rfloor}$ is then moved from $\pi(u)$ to $gp$.

Delete, Insert and Overflow maintain the properties of a \rbslack.
They will also %The same algorithms can also be used to insert and delete keys in a \bslack, and will
maintain the properties of a \bslack, provided that rebalancing steps are performed to remove any violations that are created. %after each insertion or deletion.

\subsection{Rebalancing steps} \label{sec-rebalancing}

\begin{hide}

The rebalancing steps are also based on the work of Larsen and Fagerberg.
In fact, \textbf{Root-Zero}, \textbf{Root-Replace}, \textbf{Absorb} and \textbf{Split} are the same as in~\cite{LF95}.
However, the \textbf{Compress} and \textbf{One-Child} operations are newly introduced by this work.
These operations ensure that P\ref{prop-bslack-slack} and P\ref{prop-bslack-internal} are satisfied, respectively.

%%Every \rbslack\ can be transformed into a \bslack\ by performing a sequence of rebalancing steps.
%%Observe that Insert-Overflow can increase the number of \weight s with value -1 in the tree (by one), and it is the only update that can do this.
%%The \textit{Split} updates can be used to set a \weight\ value of -1 to zero, or to move it towards the root, where it will be set to zero.
%There are six different rebalancing steps for \rbslack s: Root-Zero, Root-Replace, Absorb, Split, One-Child and Compress.
%\textbf{[[[describe the first four as the same operations as in kim's paper]]]}
If there is a degree violation at the root, then Root-Replace is performed.
If there is no degree violation at the root, but there is a weight violation at the root, then Root-Zero is performed.
If there is a \weight\ violation at an internal node that is not the root, then Absorb or Split is performed.
Suppose there are no \weight\ violations.
If there is a degree violation at a node $u$ and no degree or slack violation at $\pi(u)$, then One-Child is performed.
If there is a slack violation at a node $u$ and no degree violation at $u$, then Compress is performed.
%Suppose there are no degree violations.
%If there is a slack violation at $u$, then Compress is performed.
%The first two can be performed only when there is a violation at the root and the next three can be performed only when there is a violation at a node other than the root.
Figure~\ref{fig-bslack-updates} illustrates these rebalancing steps.
%We say that a \textit{\weight\ violation} occurs at a node with \weight\ zero, a \textit{slack violation} occurs at a node that violates P\ref{prop-bslack-slack}, and a \textit{degree violation} occurs at an internal node with only one child (violating P\ref{prop-bslack-internal}).
%Note that these are violations of the \bslack\ properties, but not violations of the \rbslack\ properties.
%Additionally, a \rbslack\ that contains no violations is a \bslack.
The goal of rebalancing is to eliminate all violations, while maintaining the \rbslack\ properties.

\textbf{Root-Zero.}
Root-Zero changes the \weight\ of the root from zero to one, eliminating a \weight\ violation, and incrementing the relaxed depth of every node.
If P\ref{prop-bslack-depth}$'$ held before Root-Zero, it holds afterwards.
%If every pair of leaves had the same relaxed depth before Root-Zero, this remains true afterwards.

\textbf{Root-Replace.}
Root-Replace replaces the root $r$ by its only child $u$, and sets $u$'s \weight\ to one.
This eliminates a degree violation at $r$, and any \weight\ violation at $u$.
If $u$ had \weight\ zero before Root-Replace, then the relaxed depth of every leaf is the same before and after Root-Replace.
Otherwise, the relaxed depth of every leaf is decremented by Root-Replace.
In both cases, if P\ref{prop-bslack-depth}$'$ held before Root-Replace, it holds afterwards.

\textbf{Absorb.}
Let $u$ be a non-root node with \weight\ zero.
Absorb is performed when $\pi(u)$ contains less than $b$ pointers.
In this case, the two pointers in $u$ are moved into $\pi(u)$, and $u$ is removed from the tree.
Since the pointer from $\pi(u)$ to $u$ is no longer needed once $u$ is removed, $\pi(u)$ now contains at most $b$ pointers.
The only node that was removed is $u$ and, since it had \weight\ zero, the relaxed depth of every leaf remains the same.
Thus, if P\ref{prop-bslack-depth}$'$ held before Absorb, it also holds afterwards.
%Moreover, the depth of every leaf in the subtree rooted at $u$ has now been decremented\trevor{(?include?)}.
%Moving the pointers from $u$ to $\pi(u)$ decrements the level of every leaf that was originally in the subtree rooted at $u$.
%Since $u$ had \weight\ zero, and these pointers move to a node with \weight\ one, the relaxed level of every leaf remains the same after the update. % (so the update maintains P\ref{prop-rbslack-depth}'). %allowing us to set the \weight\ of the new node to zero without violating P\ref{prop-rbslack-depth}'.
Absorb eliminates a \weight\ violation at $u$, but may create a slack violation at $\pi(u)$.

\textbf{Split.}
Let $u$ be a non-root node with \weight\ zero.
Split is performed when $\pi(u)$ contains exactly $b$ pointers.
In this case, there are too many pointers to fit in a single node.
We create a new node $v$ with \weight\ one, and evenly distribute all of the pointers and keys of $u$ and $\pi(u)$ (except for the pointer from $\pi(u)$ to $u$) between $u$ and $v$.
Now $\pi(u)$ has two children, $u$ and $v$.
The \weight\ of $u$ is set to one, and the \weight\ of $\pi(u)$ is set to zero.
As above, this does not change the relaxed depth of any leaf, so P\ref{prop-bslack-depth}$'$ still holds after Split.
Split moves a \weight\ violation from $u$ to $\pi(u)$ (closer to the root, where it can be eliminated by a Root-Zero or Root-Replace), but may create slack violations at $u$ and $v$.

\textbf{Compress.}
%Before a Compress update can be performed, 
Compress is performed when there is a slack violation at an internal node $u$, there is no degree violation at $u$, and there are no \weight\ violations at $u$ or any of its $k \ge 2$ children.
Let $c \le kb-b$ be the number of pointers or keys stored in the children of $u$.
Compress evenly distributes the pointers or keys contained in the children of $u$ amongst the first $\lceil c/b \rceil$ children of $u$, and discards the other children.
This will also eliminate any degree violations at the children of $u$ if $c > 1$.
After the update, $u$ satisfies P\ref{prop-bslack-slack}.
Compress does not change the relaxed depth of any node, so P\ref{prop-bslack-depth}$'$ still holds after.
Compress removes at least one child of $u$, so it increases the slack of $u$ by at least one, possibly creating a slack violation at $\pi(u)$.
(However, it decreases the total amount of slack in the tree by at least $b-1$.)
Thus, after a Compress, it may be necessary to perform another Compress at $\pi(u)$.
Furthermore, as Compress distributes keys and pointers, it may move nodes with different parents together, under the same parent.
Even if two parents initially satisfied P\ref{prop-bslack-slack} (so the children of each parent contain a total of less than $b$ slack), the children of the combined parent may contain $b$ or more slack, creating a slack violation.
Therefore, after a Compress, it may also be necessary to perform Compress at some of the children of $u$.

\textbf{One-Child.}
One-Child is performed when there is a degree violation at an internal node $u$, there are no \weight\ violations at $u$ or any of its siblings, and there is no violation of any kind at $\pi(u)$.
Let $k$ be the degree of $\pi(u)$.
Since there is no slack violation at $\pi(u)$, there are a total of $c > kb-b = b(k-1)$ pointers stored in $u$ and its siblings.
Since $u$ has only one child pointer, each of its other $k-1$ siblings must contain $b$ pointers.
One-Child evenly distributes the keys and pointers of the children of $\pi(u)$. %are evenly distributed.
One-Child does not change the relaxed depth of any node, so P\ref{prop-bslack-depth}$'$ still holds after.
One-Child eliminates a degree violation at $u$, but, like Compress, it may move children with different parents together under the same parent, possibly creating slack violations at some children of $\pi(u)$.
So, it may be necessary to perform Compress at some of the children of $\pi(u)$.

All of these updates maintain P\ref{prop-rbslack-weight-zero}$'$, P\ref{prop-bslack-internal}$'$ and P\ref{prop-bslack-leaf}.
%Since P\ref{prop-rbslack-weight-zero}$'$ holds, $\pi(u)$ contains at most $b-1$ pointers when Absorb is applied, and exactly $b$ pointers when Split is applied.
While rebalancing steps are being performed to eliminate the violation created by an insertion or deletion, there is at most one node with \weight\ zero.

We prove that a rebalancing step can be applied in any \rbslack\ that is not a \bslack. %\ will be transformed into a \bslack\ after a finite number of rebalancing steps have been performed (in any order).

\begin{lem} \label{lem-relaxed-then-rebalance}
Let $T$ be a \rbslack.
If $T$ is not a \bslack, then a rebalancing step can be performed.
\end{lem}
\begin{chapscxproof}
If $T$ is not a \bslack, it contains a \weight\ violation, a slack violation or a degree violation.
If there is \weight\ violation, then Root-Zero, Absorb or Split can be performed.
Suppose there are no \weight\ violations.
Let $u$ be the node at the smallest depth that has a slack or degree violation.
Suppose $u$ has a degree violation.
If $u$ is the root, then Root-Replace can be performed.
Otherwise, $\pi(u)$ has no violation, so One-Child can be performed.
Suppose $u$ does not have a degree violation.
Then, $u$ must have a slack violation, and Compress can be performed.
%
%If there is a slack violation at an internal node, but no degree violation at that node, then Compress can be performed.
%If there is a degree violation at an internal node, but no slack or degree violation at its parent, then One-Child can be performed.
%
%If there is a degree violation at an internal node, and a slack or degree violation at its parent, then consider the internal node $u$ with the smallest depth that has a slack violation.
%There is no slack violation at a smaller depth than $u$.
%Consider the node $v$ at the smallest depth that has a degree violation.
%If $v$ is the root, then Root-Replace can be performed.
%Otherwise, One-Child can be performed at $\pi(v)$, since there is no degree violation at $\pi(v)$.
%\qed
\end{chapscxproof}
\end{hide}

\section{Analysis} \label{sec-analysis}

\begin{hide}
%Due to space constraints, this section merely gives an outline of results proved about \bslack s.
%In the full version of the paper, we provide %In Appendix~\ref{sec-appendix-analysis}, we provide
%a detailed analysis of \bslack s that store $n$ keys, by giving: an upper bound on the height of the tree, a lower bound on the average degree of nodes (and, hence, utilization), and an upper bound on the space complexity.
This section provides a detailed analysis of \bslack s that store $n$ keys, by giving: an upper bound on the height of the tree, a lower bound on the average degree of nodes (and, hence, utilization), and an upper bound on the space complexity.
We first give a brief outline of the proofs and results, and then provide the full details.

Arbitrary \bslack s are difficult to analyze, so we begin by studying a class of trees called $b$-\maxslack s.
A $b$-\maxslack\ has a root with degree two, and satisfies P\ref{prop-bslack-depth}, P\ref{prop-bslack-internal} and P\ref{prop-bslack-leaf}, but instead of P\ref{prop-bslack-slack}, the children of each internal node contain a total of exactly $b$ slack.
Thus, a $b$-\maxslack\ is a \rbslack, but not a \bslack.
%A $(b,k)$-\maxslack\ is a $b$-\maxslack\ with a root of degree $k$.
Consider a $b$-\maxslack\ $T$ of height $h$ that contains $n$ keys.
We prove that the total degree at depth $\delta \le h$ in $T$ is $d(\delta) = 2^{-\delta}(\alpha^{\delta}+\gamma^{\delta})$, where $\alpha = b+\sqrt{b^2-4b}$ and $\gamma = b-\sqrt{b^2-4b}$.
%For $\delta \ge 0$, $d(\delta)$ is an integer.
Since the total degree at the lowest depth is precisely the number of keys in the tree, every $b$-\maxslack\ of height $h$ contains exactly $d(h)$ keys.
%We immediately obtain the upper bound $h \le h_0$, where $h_0$ is the smallest integer that satisfies $d(h_0) \ge n$.
Furthermore, when $h \ge 3$, we also have $(b-2)^{h} < d(h) \le b^{h}$.
Therefore, for $n > b^3$ (which implies height at least three), $h$ satisfies $\lceil \log_b n \rceil \le h \le \lceil \log_{b-2} n\rceil$.
We also prove that the average degree of nodes in $T$ is $\frac{b\cdot d(h-1)-b+2}{b\cdot d(h-2)-b+3}$, which is greater than $b-2$ for $h \ge 3$.

We next prove some connections between \maxslack s and \bslack s.
First, we show that each $b$-\maxslack\ of height $h$ has a smaller total degree of nodes at each depth than any \bslack\ of height $h$.
We do this by starting with an arbitrary \bslack\ of height $h$, and repeatedly removing pointers and keys from the children of each internal node that satisfies P\ref{prop-bslack-slack} (taking care not to violate P1, P2 or P3), until we obtain a $b$-\maxslack.
It follows that each $b$-\maxslack\ of height $h$ contains fewer keys than any \bslack\ of height $h$.
Consequently, every $b$-\maxslack\ with $n$ keys has height at least as large as any \bslack\ with $n$ keys.
We next prove that every $b$-\maxslack\ of height $h$ has a smaller average node degree than any \bslack\ of height $h$.
As above, the proof starts with an arbitrary \bslack\ of height $h$, and removes pointers and keys from nodes until the tree becomes a $b$-\maxslack.
However, in this proof, every time we remove a pointer, we must additionally show that the average degree of nodes in the tree decreases.

%We then show that an \maxslack\ of height $h$ whose root has degree two has fewer keys than any \bslack\ of height $h$ (Corollary~\ref{cor-2maxslack-worst-case-height}) and a smaller average degree than any \bslack\ of height $h$ (Lemma~\ref{lem-D-lower-bound-for-bslack} and \ref{lem-2maxslack-worst-case-Dbar}).
%We use this lower bound on average degree to compute an upper bound on the space complexity of any \bslack\ with $n$ keys.

We then compute the \textit{space complexity} of a \bslack\ containing $n$ keys, which is the number of words needed to store it.
%We then use the lower bound on the average degree of nodes in a \bslack\ $T$ of height $h$ to compute the space complexity of $T$.
Consider a leaf-oriented tree with maximum degree $b$.
For simplicity, we assume that each key and each pointer to a child or data occupies one word in memory.
Thus, a leaf occupies $2b$ words, and an internal node occupies $2b-1$ words.
A memory block size of $2b$ is assumed.
Let $\bar D$ be the average degree of nodes.
Then, $U = \bar D/b$ is the proportion of space that is utilized (which we call the \textit{average space utilization} of the tree), and $1-U$ is the proportion of space that is wasted.
The space complexity is $2bF$, where $F$ is the number of nodes in the tree.
Suppose the tree contains $n$ keys.
%Let us compute its space complexity.
By definition, the sum of the degrees of all nodes is $F - 1 + n$, since each node,
except the root, has a pointer into it and the degree of a leaf is the number of keys it contains.
Additionally, $F \bar D$ is equal to the sum of degrees of all nodes, so $F = (n - 1)/(\bar D - 1)$.
Therefore, the space complexity is $2b(n - 1)/(\bar D - 1)$. % = 2b(n-1)/(bU - 1)$.
In order to compute an upper bound on the space complexity for a \bslack\ of height $h$, we simply need a lower bound on $\bar D$.
%Therefore, $S(n) = 2b(n - 1)/(\bar D - 1)$, which is related to utilization by $S(n) = 2b(n-1)/(bU - 1)$.
Above, we saw that $\bar D > b-2$ for \bslack s of height at least three.
It follows that a \bslack\ with $n > b^3$ keys %(which must have height at least three)
has space complexity at most $2b(n-1)/(b-3) < 2n \frac{b}{b-3}$.
(Recall that $2n$ is optimal under these assumption.)
%\textbf{[[[Be VERY careful with the following... I think it's right, though.]]]}
A slightly tighter upper bound is $2b(n-1) / \big( \frac{b\cdot d(h-1)-b+2}{b\cdot d(h-2)-b+3}-1 \big)$.

Section~\ref{sec-bslack-space-complexity} 
%The full version of the paper 
describes pathological families of B-trees, Overflow trees and H-trees, and compares the space complexity of example trees in these families with the worst-case upper bound on the space complexity of a \bslack.
By studying these families, we obtain lower bounds on the space complexity of these trees that are above the upper bound for \bslack s.

We also study the number of rebalancing steps necessary to maintain balance in a \rbslack.
Consider a \rbslack\ obtained by starting from a \bslack\ containing $n$ keys and performing a sequence of $i$ insertions and $d$ deletions.
We prove that such a \rbslack\ will be transformed back into a \bslack\ after at most $2i(2 + \lfloor \log_{\lfloor\frac{b}{2}\rfloor} (n+i)/2 \rfloor) + d/(b-1)$ rebalancing steps, irrespective of which rebalancing steps are performed, and in which order.
Hence, insertions perform amortized $O(\log(n+i))$ rebalancing steps and deletions perform an amortized constant number of rebalancing steps.
\end{hide}

%\subsection{Analysis of \bslack s} \label{sec-appendix-analysis}
%
%We analyze properties of \bslack s and \maxslack s with maximum degree $b$.
%A $(b,k)$-\maxslack\ is an \maxslack\ in which the root has degree $k$ and each node has maximum degree $b$.

\subsection{Analysis of \maxslack s} \label{sec-maxslack}

\begin{hide}
%We study properties of \bslack s and \maxslack s with maximum degree $b$.
For our analysis, it is helpful to generalize the definition of $b$-\maxslack s to the following.
A $(b,k)$-\maxslack\ is the same as $b$-\maxslack, except that the root has degree $k$, instead of degree two. % in which the root has degree $k$ and each node has maximum degree $b$.

%A tree in which the children of each internal node share exactly $b$ slack %(formally, for each internal node with children $v_1, v_2, ..., v_l$, $deg(v_1)+deg(v_2)+...+deg(v_l) = lb-b$) 
%is called a \textit{\maxslack}.
%A \maxslack\ with a root of degree $k \ge 2$ is called a \textit{$(b,k)$-\maxslack}.
%In this section, we prove several results about $(b,k)$-\maxslack s.
%In particular, we compute the total degree of nodes at each depth, and the average degree of nodes in a $(b,k)$-\maxslack\ of height $h$.
%As we will see, $(b,2)$-\maxslack s are very important for our analysis, because they represent the worst case for \bslack s in several respects.
%In Section~\ref{sec-maxslack-to-bslack}, we prove that a $(b,2)$-\maxslack\ has the greatest height of any \bslack\ containing the same number of keys, and the smallest average degree of nodes of any \bslack\ with the same height.

%We now prove several results about \maxslack s.
We now %Our first task is to 
compute the total degree of nodes at each depth in a $(b,k)$-\maxslack.
As we will see, for each $k$, the total degree of nodes at a given depth $\delta$ is the same in every $(b,k)$-\maxslack\ of height at least $\delta$.
%Recall that the degree of a leaf is the number of keys in the leaf.
%So, the following lemma also computes the number of keys in the leaves of a $(b,k)$-\maxslack.

\begin{lem} \label{lem-N}
%The number of pointers or keys at depth $\delta$ in a $(b,k)$-\maxslack\ is:
The total degree of nodes at depth $\delta$ in a $(b,k)$-\maxslack\ of height $h$ is:
 \begin{displaymath}
   d(\delta, k) = \left\{
     \begin{array}{ll}
       k & \mbox{if } \delta = 0 \\
       kb-b & \mbox{if } \delta = 1 \\
       b(d(\delta-1, k) - d(\delta-2, k))& \mbox{if } 1 < \delta \le h
     \end{array}
   \right.
  \end{displaymath}
\end{lem}
\begin{chapscxproof}
In the following, we use $c(u)$ to denote the set of children of node $u$.
Let %$k \ge 2$, and 
$T$ be a $(b,k)$-\maxslack.
The proof is by induction on $\delta$.
The base cases $\delta = 0$ and $\delta = 1$ are immediate from the definition of a $(b,k)$-\maxslack.
Consider $\delta > 1$.
Let $T_\delta$ be the set of nodes at depth $\delta$ in $T$.
Then, %$$|T_{\delta+1}| = 
$$d(\delta,k) = \sum_{v \in T_{\delta}} deg(v) = \sum_{u \in T_{\delta-1}} \sum_{v \in c(u)} deg(v).$$
Since $T$ is an \maxslack, $$\sum_{v \in c(u)} deg(v) = deg(u)b-b = b(deg(u)-1), \ \mbox{so}$$ $$d(\delta, k) = \sum_{u \in T_{\delta-1}} b(deg(u)-1) = b(d(\delta-1,k) - |T_{\delta-1}|) = b(d(\delta-1,k) - d(\delta-2,k)).$$
%Since the number of nodes at depth $\delta-1$ is equal to the total degree of nodes at depth $\delta-1$, we have $d(\delta,k) = b(d(\delta-1,k)-d(\delta-2,k)).$
%\qed
\end{chapscxproof}

Since $d(\delta, k)$ is a linear homogeneous recurrence relation with constant coefficients, we use the technique described in Section~2.1.1(a) of \cite{greene1982mathematics} to obtain the following closed form solution.

\begin{lem} \label{lem-N-closed-form}
$d(\delta, k) = 2^{-\delta} (k_1 (b+\sqrt{b^2-4b})^{\delta} + k_2(b-\sqrt{b^2-4b})^{\delta}),$ where $k_1 = \frac{bk-2b}{2\sqrt{b^2-4b}} + \frac{k}{2}$ and $k_2 = k - k_1$.
\end{lem}
%\begin{chapscxproof}
\details{
We obtain a closed form solution for the recurrence %of Lemma~\ref{lem-nodes-at-depth-recurrence} 
using a technique described in Section~2.1.1(a) of \cite{greene1982mathematics}.
A linear homogeneous recurrence relation with constant coefficients is an equation of the form $a_n = c_1 a_{n-1} + c_2 a_{n-2} + ... + c_d a_{n-d}$.
The coefficients $c_1, ..., c_d$ yield the characteristic polynomial $p(t) = t^d - c_1 t^{d-1} - c_2 t^{d-2} - ... - c_d$.
If the roots $r_1, ..., r_d$ of $p(t)$ are unique, then $a_n = k_1 r_1^n + k_2 r_2^n + ... + k_d r_d^n$, where the coefficients $k_1, ..., k_d$ are chosen to satisfy the base cases for the recurrence.

Observe that $d(\delta) = b(d(\delta-1) - d(\delta-2))$ can be rewritten as $a_n = b a_{n-1} + (-b) a_{n-2}$, which yields the characteristic polynomial $p(t) = t^2 - bt + b$.
The quadratic equation yields roots $\frac{b \pm \sqrt{b^2 - 4b}}{2}$, which are unique for $b > 4$.
Thus, we have $a_n = k_1 \big(\frac{b + \sqrt{b^2 - 4b}}{2}\big)^n + k_2 \big(\frac{b - \sqrt{b^2 - 4b}}{2}\big)^n$.
Plugging in the base cases, we have $a_0 = k = k_1 + k_2$ and $a_1 = kb-b = k_1 \frac{b + \sqrt{b^2 - 4b}}{2} + k_2 \frac{b - \sqrt{b^2 - 4b}}{2}$.
The first equation yields $k_2 = k - k_1$.
After some basic algebra, %the second yields
$k_1 = \frac{kb - 2b}{2\sqrt{b^2-4b}} + \frac{k}{2}$.
%Clearly, $k_1 = k_2 = 1$ is a solution to both of these equations.
%\qed
}
%\end{chapscxproof}

\begin{cor} \label{cor-N-closed-form-two-maxslack}
$d(\delta, 2) = 2^{-\delta} ((b+\sqrt{b^2-4b})^{\delta} + (b-\sqrt{b^2-4b})^{\delta})$.
\end{cor}
%\begin{chapscxproof}
\details{
When $k=2$, $k_1 = \frac{bk-2b}{2\sqrt{b^2-4b}} + \frac{k}{2} = 1$ and $k_2 = k - k_1 = 1$.
}
%\qed
%\end{chapscxproof}

Since $b+\sqrt{b^2-4b}$ asymptotically approaches $2b$ as $b$ increases, and $b-\sqrt{b^2-4b}$ approaches zero, $d(\delta, 2)$ approximately grows like $b^{\delta}$ for large $b$.
%
%THE FOLLOWING IS WRONG. THIS NEEDS TO BE COMPUTED USING Dbar, NOT d!
%Simple algebra establishes the following bounds on $d(\delta, 2)$.
%\begin{cor} \label{cor-N-bounds}
%For $h \ge 3$, $\big(\frac{b}{2}\big)^h < (b-1.4)^h < (b-\frac{\gamma}{2})^h \le d(\delta, 2).$
%\end{cor}
Simple algebra establishes the following bounds on $d(\delta, 2)$.
\begin{cor} \label{cor-N-bounds}
For $h \ge 3$, $\big(\frac{b}{2}\big)^h < d(h, 2) \le b^h.$
\end{cor}

For the following lemma, we
%We
used symbolic mathematics software to obtain the partial derivatives of $d(\delta, k)$ with respect to $\delta$ and $k$, %.
%By proving that these derivatives are positive, we established the following lemmas.
and prove that they are positive.

\begin{lem} \label{lem-N-increasing-with-delta}
$d(\delta, k)$ is an increasing function of $\delta$ and $k$.
\end{lem}

\begin{lem} \label{lem}
The total degree of nodes in every $(b,k)$-\maxslack\ of height $h$ is
\begin{displaymath}
   D(h,k) = \left\{
     \begin{array}{ll}
       k & : h = 0 \\
       k + b(d(h-1,k)-1) & : h > 0
     \end{array}
   \right.
  \end{displaymath}
Moreover, $D(h, k)$ is increasing in $h$ and $k$.
\end{lem}
\begin{chapscxproof}
By Lemma~\ref{lem-N}, $D(h,k) = \sum_{\delta=0}^h d(\delta,k) = k + (kb-b) + \sum_{\delta=2}^h b(d(\delta-1,k)-d(\delta-2,k))$.
This telescoping sum reduces to $D(h,k) = k + (kb-b) + b(d(h-1,k)-d(0,k)) = k + b(d(h-1,k)-1)$.
Lemma~\ref{lem-N-increasing-with-delta} %and Lemma~\ref{lem-N-increasing-with-k}
implies that $D(h, k)$ is increasing in $h$ and $k$.
%The result follows from basic algebra. %Basic algebra yields $D(h,k) = b(d(h-1,k)-1)+3$.
%\qed
\end{chapscxproof}

We now consider the average degree of nodes at each depth in a $(b,2)$-\maxslack.
In any $(b,2)$-\maxslack, the two children of the root must share exactly $2b-b = b$ pointers or keys.
Let us build intuition with an example.
Consider a $(b,2)$-\maxslack\ in which the children of the root evenly share $b$ pointers or keys.
The grandchildren of the root must share a total of exactly $((b/2)b-b) + ((b/2)b-b) = b^2-2b$ pointers or keys.
Thus, the average degree of nodes at depth zero is two, at depth one is $b/2$, and at depth two is $b-2$.
We prove that every $(b,2)$-\maxslack\ $T$ of height $h$ has the smallest average node degree of any $(b,k)$-\maxslack\ of height $h$.
We first derive expressions for the average degree of nodes in an \maxslack, and the average degree at each depth.

Since the total degree of nodes at depth $\delta$ is $d(\delta,k)$, and the total number of nodes at depth $\delta$ is $d(\delta-1,k)$, we obtain the following.

\begin{lem} \label{lem-d}
The average degree of nodes at depth $\delta$ in any $(b,k)$-\maxslack\ of height $h$ is:
\begin{displaymath}
   \bar d(\delta,k) = \left\{
     \begin{array}{ll}
       k & : \delta = 0 \\
       \frac{d(\delta,k)}{d(\delta-1,k)}& : 0 < \delta \le h
     \end{array}
   \right.
  \end{displaymath}
\end{lem}

We used our mathematics software to obtain the partial derivatives of $\bar d(\delta, k)$ with respect to $\delta$ and $k$.
We proved the following two lemmas by showing that $\frac{\partial}{\partial \delta} \bar d(\delta,k) > 0$ for $2 \le k \le b-2$, $\frac{\partial}{\partial \delta} \bar d(\delta,k) < 0$ for $k > b-2$, and $\frac{\partial}{\partial k} \bar d(\delta,k) > 0$.

\begin{lem} \label{lem-d-change-with-delta}
$\bar d(\delta,k)$ is an increasing function of $\delta$ for $2 \le k \le b-2$, and is a decreasing function of $\delta$ for $k \ge b-1$.
\end{lem}
\details%
{
\begin{chapscxproof}
We prove this lemma by demonstrating that the partial derivative of $\bar d(\delta,k)$ with respect to $\delta$ is positive for $2 \le k \le b-2$, and negative for $k > b-2$.
%This derivative would be extremely difficult to compute by hand, so we used a symbolic mathematics package.
%The result was a very complicated rational function that can be simplified to 
Using our mathematics software, we obtained a very complicated rational function that we simplified to 
$$\frac{\partial}{\partial \delta} \bar d(\delta,k) = \frac{-(k^2-bk+b)(4b)^{\delta} \sqrt{b^2-4b}\big(\alpha \gamma \log_e \frac{\alpha}{\gamma}\big)}{((2k-\gamma)(\alpha^{\delta}+\gamma^{\delta}))^2},$$
where $\alpha = b + \sqrt{b^2-4b}$ and $\gamma = b - \sqrt{b^2-4b}$.
Observe that $1 < \gamma < b < \alpha < 2b$.
Thus, the denominator is positive except when $2k - \gamma = 0$.
\trevor{Proving $2k-\gamma > 0$ is MUCH easier than this... Just use $\gamma < 2$.}
This is the case precisely when $k^2 - kb + b = 0$, which occurs when $k = \frac{\gamma}{2}$ or $k = \frac{\alpha}{2}$.
Since $b > 4$, $\frac{\gamma}{2} < 2$ and $b-2 < \frac{\alpha}{2} < b-1$.
\trevor{Where the hell did I get $b-2$ and $b-1$ from? Something seems wrong.}
Therefore, the denominator is always positive when $k \ge 2$. % is an integer.
Since $b \ge 5$, $4b > 0$ and $\sqrt{b^2-4b} > 0.$
Since $1 < \gamma < \alpha$, $\alpha\gamma\log_e \frac{\alpha}{\gamma} > 0$.
Therefore, $\frac{\partial}{\partial \delta} \bar d(\delta,k) > 0$ is equivalent to $k^2-bk+b < 0$.
As above, $k^2-bk+b = 0$ when $k=\frac{\gamma}{2}$ or $k=\frac{\alpha}{2}$.
For all $k \in (\frac{\gamma}{2}, \frac{\alpha}{2})$, $k^2-bk+b < 0$, so $\frac{\partial}{\partial \delta} \bar d(\delta,k) > 0$ for $2 \le k \le b-2$.
\trevor{Look at ``demonstrating that depthwise average degree is increasing with depth.nb'' to fix this proof.}
%\qed
\end{chapscxproof}
}

\begin{lem} \label{lem-d-increasing-with-k}
$\bar d(\delta,k)$ is an increasing function of $k$.
\end{lem}
%%\begin{chapscxproof}
%\details{
%Using our symbolic mathematics package, we obtained %a rational function that simplified to
%$$\frac{\partial}{\partial k} \bar d(\delta,k) = \frac{4^{\delta+1}b^\delta (b-4)}{((2k-\gamma)(\alpha^{\delta}+\gamma^{\delta}))^2},$$
%where $\alpha = b + \sqrt{b^2-4b}$ and $\gamma = b - \sqrt{b^2-4b}$.
%As we argued above, the denominator is always positive when $k \ge 2$ is an integer.
%Thus, we need only show $4^{\delta+1}b^\delta (b-4) > 0$, which is true because $b \ge 5$.
%}
%%\qed
%%\end{chapscxproof}
%
%\begin{lem} \label{lem-d-lower-bound}
%$\bar d(\delta,k) \ge b-2$ for $2 \le \delta \le h$.
%\end{lem}
%\begin{chapscxproof}
%$\bar d(2,2) = \frac{d(2,2)}{d(1,2)} = \frac{b(2b-b - 2)}{2b-b} = b-2$.
%%$\bar d(2,k) = \frac{d(2,k)}{d(1,k)} = \frac{b(kb-b - b)}{kb-b} = \frac{b(k-1)-k}{k-1} = \frac{b(k-1)-(k-1)-1}{k-1} = \frac{(k-1)(b-1)}{k-1} = b-1-\frac{1}{k-1} \ge b-2$.
%By Lemma~\ref{lem-d-change-with-delta}, $\bar d(\delta,2) \ge b-2$ for $\delta \ge 2$.
%Finally, Lemma~\ref{lem-d-increasing-with-k} implies that $\bar d(\delta,k) \ge b-2$ for all $\delta,k \ge 2$.
%\qed
%\end{chapscxproof}

Let $T$ be a \bslack\ of height $h$.
Since every node in $T$ except for the root is pointed to by exactly one child pointer, the number of nodes in $T$ is the total degree of all nodes at depths zero through $h-1$, plus one for the root, which is exactly $D(h-1,k)+1$.
Thus, we obtain the following.

\begin{lem} \label{lem-D-closed-form}
The average degree of nodes in any $(b,k)$-\maxslack\ of height $h$ is:
 \begin{displaymath}
   \bar D(h,k) = \left\{
     \begin{array}{ll}
       k & : h = 0 \\
       \frac{D(h,k)}{D(h-1,k)+1}& : h > 0
     \end{array}
   \right.
  \end{displaymath}
\end{lem}
%\begin{chapscxproof}
%By definition, the average degree of nodes in $T$ is the total degree of nodes in $T$ divided by the number of nodes. %total space reserved for pointers in internal nodes and keys in leaves.
%The total degree of nodes in $T$ is exactly $D(h,k)$.
%Since every node except for the root is pointed to by exactly one child pointer, the number of nodes in $T$ is the total degree of all nodes at depths zero through $h-1$, plus one for the root.
%This is exactly $D(h-1,k)+1$.
%\qed
%\end{chapscxproof}

\begin{lem} \label{lem-D-change-with-h}
$\bar D(h,k)$ is an increasing function of $h$ for $2 \le k \le b-2$, and is a decreasing function of $h$ for $k \ge b-1$.
\end{lem}
\begin{chapscxproof}
%Fix $k$.
%Without loss of generality, suppose $k \le b-2$.
%(The argument is symmetric for $k > b-2$; simply negate every inequality and swap max and min.)
%The average degree of the nodes at depths zero through $h$ is the same in every $(b,k)$-\maxslack\ of height at least $h$.
%By definition, $\bar D(h,k)$ must be a weighted average of the terms $\bar d(0,k), \bar d(1,k), ..., \bar d(h,k)$.
%Thus, $\bar D(h+1,k)$ is a weighted average of $\bar D(h,k)$ and $\bar d(h+1,k)$
%By Lemma~\ref{lem-d-change-with-delta}, $\bar d(h+1,k) > \max_{i = 0...h} \bar d(i,k)$.
%Therefore, $\bar d(h+1,k) > \bar D(h,k)$, which implies that $\bar D(h+1,k) > \bar D(h,k)$.
%
The average node degree at depth $\delta \in \{0,1,...,h\}$ is the same in every $(b,k)$-\maxslack\ of height at least $h$.
By definition, $\bar D(h,k)$ must be a weighted average of the terms $\bar d(0,k), \bar d(1,k), ..., \bar d(h,k)$.
Thus, $\bar D(h+1,k)$ is a weighted average of $\bar D(h,k)$ and $\bar d(h+1,k)$
Suppose $k \le b-2$.
By Lemma~\ref{lem-d-change-with-delta}, $\bar d(h+1,k) > \max_{i \le h} \bar d(i,k)$.
Therefore, $\bar d(h+1,k) > \bar D(h,k)$, which implies that $\bar D(h+1,k) > \bar D(h,k)$.
Now, suppose $k \ge b-1$.
By Lemma~\ref{lem-d-change-with-delta}, $\bar d(h+1,k) < \min_{i \le h} \bar d(i,k)$.
Therefore, $\bar d(h+1,k) < \bar D(h,k)$, which implies that $\bar D(h+1,k) < \bar D(h,k)$.
%\qed
\end{chapscxproof}

We used our mathematics software to obtain the partial derivative of $\bar D(h,k)$ with respect to $k$, and proved that it is positive, yielding the following result.

\begin{lem} \label{lem-D-increasing-with-k}
$\bar D(h,k)$ is an increasing function of $k$.
\end{lem}
%\begin{chapscxproof}
\details{
We prove $\frac{\partial}{\partial k} \bar D(h,k) > 0$.
As above, we used our mathematics software to compute the derivative, yielding a function that we simplified to
%This derivative would be extremely difficult to compute by hand, so we used a symbolic mathematics package.
%The result is a (very complicated) rational function.
$$\frac{\partial}{\partial k} \bar D(h, k) = \frac{2^h b(b-4)(2^{h+1} (1+b^{h+1}) - \alpha^{h+1} - \gamma^{h+1})}{((\gamma^h - \alpha^h)(\gamma(b+k) + 2b) + 2^{h+1}\sqrt{b^2-4b}(1+k-b))^2},$$
where $\alpha = b + \sqrt{b^2-4b}$ and $\gamma = b - \sqrt{b^2-4b}$.

We first prove that the numerator is positive.
Since $b > 4$, $2^h b(b-4) > 0$, so we need only prove %$2^{h+1}(1+b^{h+1}) > b \gamma^h - \gamma^h \sqrt{b^2-4b} + b \alpha^h + \alpha^h \sqrt{b^2-4b}$ or, equivalently, $2^{h+1}(1+b^{h+1}) > \gamma^h (b - \sqrt{b^2-4b}) + \alpha^h (b + \sqrt{b^2-4b})$.
%Substituting $\alpha$ and $\gamma$, we obtain
$2^{h+1}(1+b^{h+1}) > \alpha^{h+1} + \gamma^{h+1}$.
By the binomial theorem, $(\alpha + \gamma)^{h+1} = \sum_{k=0}^{h+1} \binom{h+1}{k} \alpha^{(h+1)-k} \gamma^k = \alpha^{h+1} + \gamma^{h+1} + \sum_{k=1}^h \binom{h}{k} \alpha^{h-k} \gamma^k$.
Since $\alpha + \gamma = 2b$, $(2b)^{h+1} = \alpha^{h+1} + \gamma^{h+1} + \sum_{k=1}^h \binom{h}{k} \alpha^{h-k} \gamma^k \ge \alpha^{h+1} + \gamma^{h+1}$.
Since $2^{h+1}(1+b^{h+1}) > (2b)^{h+1}$, we have established $2^{h+1}(1+b^{h+1}) > \alpha^{h+1} + \gamma^{h+1}$. %, which proves that $\frac{\partial}{\partial k} \bar D(h,k) > 0$.

We now prove that the denominator is positive.
Since the denominator is of the form $(f(k))^2$, it suffices to prove $f(k) = (\gamma^h - \alpha^h)(\gamma(b+k) + 2b) + 2^{h+1}\sqrt{b^2-4b}(1+k-b) \neq 0$.
We first try to show $f(k) < 0$.
Since $\gamma < \alpha$ and $\gamma, b, k > 0$, $(\gamma^h - \alpha^h)(\gamma(b+k) + 2b) < 0$.
The expression $2^{h+1}\sqrt{b^2-4b}$ is positive, so $f(k) < 0$ will be true if $k \le b-1$.
Suppose $k = b$.
Then, $f(k) = f(b) = \alpha^h - \gamma^h - 2^{h-1} \sqrt{1-4/b}$.
We prove $f(b) > 0$.
Since $b > 4$, $f(b) > \alpha^h - \gamma^h - 2^{h-1}$.
By the binomial theorem, $2^{h-1} + \gamma^h < 2^h + \gamma^h < (2+\gamma)^h$.
Observe that $\alpha > \gamma+2$.
Thus, $2^{h-1} + \gamma^h < (2+\gamma)^h < \alpha^h$, so $f(b) > 0$.
Therefore, in all cases, $f(k) \neq 0$.
}
%\qed
%\end{chapscxproof}

We proved a simple lower bound on $\bar D(h,k)$ using our mathematics software.

\begin{lem} \label{lem-D-lower-bound}
%$\bar D(h,k) \ge b-4$ for $h = 2$, and
$\bar D(h,k) > b-2$ for $h \ge 3$ (and $b \ge 5$).
\end{lem}
\end{hide}

\subsection{Relating \bslack s to \maxslack s} \label{sec-maxslack-to-bslack}

\begin{hide}
In this section, we first prove that a $(b,2)$-\maxslack\ has the greatest height of any \bslack\ containing the same number of keys, and a smaller average degree of nodes than any \bslack\ with the same height.
Then, we compute upper and lower bounds on the space used to store a \bslack\ with $n$ keys.

%%In the following, we use $d(\delta)$ as shorthand for $d(\delta, 2)$.
%We now prove that any $(b,2)$-\maxslack s has the minimum total degree, at each depth, of any \bslack\ of the same height.
%Since the degree at the lowest depth of the tree is simply the number of keys in the tree, this yields an immediate height bound for \bslack s.

\begin{prop} \label{lem-internal-with-slack-has-degree-three-child}
Let $u$ be an internal node in a \bslack. %\ with children $v_1, v_2, ..., v_l$.
If the total slack contained in the children of $u$ is less than $b$, then some child of $u$ has degree at least three.
%If $deg(v_1)+deg(v_2)+...+deg(v_l) > lb-b$, then some $v_i$ has degree at least three.
\end{prop}
\begin{chapscxproof}
Suppose the total slack contained in the children $v_1, v_2, ..., v_l$ of $u$ is less than $b$.
Then, $deg(v_1)+deg(v_2)+...+deg(v_l) > lb-b$.
By P\ref{prop-bslack-internal}, $l \ge 2$.
Since $b > 4$, it follows that $lb-b = b(l-1) > 4(l-1) \ge 2l$.
Therefore, the average degree of the children of $u$ is $(deg(v_1)+deg(v_2)+...+deg(v_l))/l > \frac{lb-b}{l} > 2$.
%\qed
\end{chapscxproof}

\begin{lem} \label{lem-N-worst-in-two-maxslack}
Every $(b,2)$-\maxslack\ of height $h$ has a smaller total degree of nodes, at each depth, than any \bslack\ of height $h$.
%Every $(b,2)$-\maxslack\ of height $h$ contains the minimum number of keys of any \bslack\ of height $h$. % Every \bslack\ $T$ of height $h$ that contains the minimum number of keys of any tree of height $h$ is a $(b,2)$-\maxslack.
\end{lem}
\begin{chapscxproof}
Let $T$ be a \bslack\ of height $h$. % that is not a $(b,2)$-\maxslack.
We can transform $T$ into a $(b,2)$-\maxslack\ by removing keys and pointers.
Removing a key, or a pointer from a node that has at least three pointers, does not affect P\ref{prop-bslack-depth}, P\ref{prop-bslack-internal} or P\ref{prop-bslack-leaf}.
%Since $T$ is not a $(b,2)$-\maxslack, for some internal node $u$, the total slack contained in the children of $u$ is less than $b$.
Let $u$ be any internal node whose children share a total of less than $b$ slack.
We arbitrarily remove a key or pointer from the child, $v$, of $u$ with the largest degree.
By Proposition~\ref{lem-internal-with-slack-has-degree-three-child}, $v$ must have degree at least three.
%Removing a key or pointer from $v$ increases the total slack contained in the children of $u$ by one, so it does not violate P\ref{prop-bslack-slack}.
%Therefore, $T$ is still a \bslack.
We can repeat this process until $T$ is a $(b,2)$-\maxslack.
%\qed
\end{chapscxproof}

\begin{cor} \label{cor-2maxslack-worst-case-height}
Every $(b,2)$-\maxslack\ with $n$ keys has a larger height than any \bslack\ with $n$ keys.
\end{cor}

\begin{cor} \label{cor-2maxslack-fewest-keys}
Any \bslack\ of height $h$ contains more keys than every $(b,2)$-\maxslack\ of height $h$, and, hence, more than $d(h,2)$ keys.
\end{cor}

\begin{lem} \label{lem-D-lower-bound-for-bslack}
Every $(b,2)$-\maxslack\ of height $h$ has a smaller average node degree than any \bslack\ of height $h$.
\end{lem}
\begin{chapscxproof}
We first describe how to transform a \bslack\ into an \maxslack\ of the same height while decreasing the average node degree.
Observe that, since an \maxslack\ satisfies P\ref{prop-bslack-depth}, P\ref{prop-bslack-internal} and P\ref{prop-bslack-leaf}, any \bslack\ will become an \maxslack\ if pointers and keys are removed until, for each internal node $u$, the children of $u$ share a total of $b$ slack.
%The high-level idea 
%%of the transformation 
%is to start at the parents of leaves, and work up towards the root, level by level.
%At each step, we remove some keys or pointers to transform an internal node of degree $k$ whose children are either leaves, or the roots of \maxslack s, into the root of a $k$-maxslack tree.
%Every time we remove a pointer, we must ensure that we do not increase the average node degree.
%
The proof is by induction on the height of the tree.
Let $u$ be the root of a \bslack\ $T$. % and $v_1, v_2, ..., v_l$ be its children.

In the base case, the children of $u$ are leaves.
Arbitrarily removing keys from the children of $u$ until the children contain a total of exactly $b$ slack will transform $T$ into an \maxslack\ while decreasing the average degree of nodes.

Now, suppose the children of $u$ are internal.
By the inductive hypothesis, we can transform each subtree rooted at a child of $u$ into an \maxslack.
After these transformations, for every internal node in every subtree rooted at a child of $u$, the children of this internal node contain a total of $b$ slack.
If the children of $u$ contain a total of $b$ slack, then $T$ is an \maxslack.
Otherwise, we would like to remove some grandchild of $u$, to increase this slack.
%(If we increase this slack to $b$, then $T$ will be an \maxslack.)
As we argued in the proof of Lemma~\ref{lem-N-worst-in-two-maxslack}, removing a pointer from a node that has the largest degree amongst its siblings yields a \bslack. %does not affect P\ref{prop-bslack-depth}, P\ref{prop-bslack-internal}, P\ref{prop-bslack-leaf} or P\ref{prop-bslack-slack}.
However, we must carefully choose which grandchild to remove so that we decrease the average degree of nodes.
%
%More specifically, we removing the grandchild that is the root of the tree with the largest average node degree, we can avoid increasing the average node degree.
%To avoid increasing the average node degree, we must remove the grandchild that is the root of the tree with the largest average node degree.
%However, removing a node could increase the average degree of nodes, so we must carefully choose which grandchild to remove.
Let $v$ be the child of $u$ that is the root of the tree with the largest average degree.
By Lemma~\ref{lem-D-increasing-with-k}, $v$ has the largest degree amongst its siblings.
By Lemma~\ref{lem-internal-with-slack-has-degree-three-child}, $v$ must have at least three pointers, so removing one of its children does not violate P\ref{prop-bslack-internal}.
It is easy to verify that removing one of $v$'s children will not violate P\ref{prop-bslack-depth} or P\ref{prop-bslack-leaf}.
We remove the child of $v$ that is the root of the tree with the largest average degree.
Since this tree 
%the tree rooted at $w$ 
has the largest average degree of any tree rooted at a child of $v$, removing it decreases the average degree of $T$. %the tree rooted at $v$.
(This is because every other subtree rooted at a child of $v$ has the same or smaller average degree, and removing this child of $v$ decreases the degree of $v$.)
%Since the average degree of the tree rooted at $v$ does not increase, neither does the average degree of $T$.
%Each child of $r$ is still the root of an \maxslack, because removing a pointer does not decrease the slack of any node.
%
%It is easy to verify that each child of $r$ is still the root of an \maxslack.
We can repeatedly apply this transformation until the children of $u$ contain a total of $b$ slack, at which point $T$ is an \maxslack.

We now prove the main result.
Given a \bslack\ $T$, we first transform it into an \maxslack.
Then, if the root of $T$ has more than two children, we transform $T$ into a $(b,2)$-\maxslack\ by keeping the two children %of the root 
that are the roots of the trees with the largest average degrees, and throwing away the rest.
%\qed
\end{chapscxproof}

Since the average degree of nodes represents the fraction of space that is utilized, this %result also 
implies that every $(b,2)$-\maxslack\ of height $h$ wastes a larger proportion of space than any \bslack\ of the same height.
We can also obtain a lower bound on the average degree (and, hence, the fraction of space that is utilized) for any \bslack\ containing $n$ keys.

\begin{lem} \label{lem-2maxslack-worst-case-Dbar}
A \bslack\ with $n \ge 2$ keys has average degree greater than $\bar D(\lceil \log_b n \rceil - 1, 2)$.
\end{lem}
\begin{chapscxproof}
Let $T$ be a \bslack\ of height $h$ containing $n$ keys.
By Lemma~\ref{lem-D-lower-bound-for-bslack}, $T$ has average degree greater than $\bar D(h,2)$.
Lemma~\ref{lem-D-change-with-h} implies that $\bar D(h,2)$ increases with $h$, so it suffices to find a lower bound on $h$.
Since $b$ is the maximum possible degree for any node in $T$, $h$ is at least $\lceil \log_b n \rceil - 1$.
%If every node in $T$ has the maximum possible degree, $b$, then $n = b^{h+1}$, so $h = \lceil \log_b n \rceil - 1$.
%It follows that the height of $T$ is at least $\lceil \log_b n \rceil - 1$.
%\qed
\end{chapscxproof}

We can now compute bounds on $\func{S}(n)$, the space complexity of a \bslack\ containing $n$ keys.
%$\func{S}(n)$ is the total number of words of memory occupied by all nodes.
%To simplify this computation, we assume keys and pointers each occupy a single word in memory.
%The number of words occupied by a node is then $2b$.
%(Technically, internal nodes use up to $2b-1$ words, but we are assuming only one block size can be allocated.)
%%The number of words occupied by a node is then $2b-1$ for internal nodes %($b$ pointers and $b-1$ keys) 
%%and $2b$ for leaves. %($b$ pointers and $k$ keys).
%Therefore, $\func{S}(n) = 2bF$, where $F$ is the total number of nodes in the tree.
Recall from Section~\ref{sec-bslack} that $\func{S}(n) = 2b(n-1)/(\bar D - 1)$.
%We can obtain an upper bound on $F$ by using the lower bound on average degree that was derived above.
%By definition, the total degree of the tree is $F - 1 + n$, since each node has a pointer into it, except for the root, and the degree of a leaf is the number of keys it contains.
%Let $D^* \le b$ be the average degree of the tree.
%Then, $F D^*$ is equal to the total degree, $F - 1 + n$, so $F = \frac{n - 1}{D^* - 1}$.
%%Therefore, $F$ is maximized when average degree is minimized.
By Lemma~\ref{lem-2maxslack-worst-case-Dbar}, $D^* \ge \bar D(\lceil \log_b n \rceil - 1, 2)$.
Let $s = \bar D(\lceil \log_b n \rceil - 1, 2)$.
Then, $$\frac{n-1}{b-1} \le F \le \frac{n-1}{s - 1} \ \mbox{ and } \ \frac{2b(n-1)}{b-1} \le \func{S}(n) \le \frac{2b(n-1)}{s - 1}.$$
%We can also obtain an upper bound that is looser, but somewhat easier to understand.
%When $n > b^3$, $\lceil \log_b n \rceil - 1 \ge 3$, so Lemma~\ref{lem-D-lower-bound} implies $s > b-1.4$.
%Therefore, $\func{S}(n) < \frac{2b(n-1)}{b-2.4} < \frac{2b}{b-2.4}n$.
%
\\

\end{hide}

\subsection{Amortized logarithmic rebalancing} \label{sec-log-rebalancing}

\begin{hide}
%%
%%In practice, 
%In the worst-case, $Br(T,\delta) \le b-1$, since a tree $T$ in which every node has $b-1$ children is a \bslack.
%Beyond depth two, $Br(T,\delta)$ quickly closes the gap between $b-2$ and its worst-case upper bound of $b-1$.
%For instance, in a $16$-slack tree $T$, where $Br(T,\delta) \le 15$ in the worst-case, $Br(T,3) \approx 14.86$ and $Br(T,4) \approx 14.92$.
%Note that %Since the children of each internal node can share $b$ slack, 
%$Br(T, \delta) \le b-1$ in the worst-case, since the children of each internal node can share $b$ slack.
%Interestingly, the average node degree in every \bslack increases quite quickly with depth, exceeding $b-2$ at depth three.
% 
%that, as the depth of a node goes to $\infty$, the number of pointers or keys shared \textit{among its children} goes to $b^2-b$ (which implies that the average number of pointers or keys per node goes to $b-1$).

%We now prove that every \rbslack\ can be transformed into a \bslack\ by performing a finite number of rebalancing steps.  (We prove amortized logarithmic rebalancing in Section~\ref{sec-analysis}.)
%%, and that at most an amortized logarithmic number of rebalancing steps is needed to do this transformation.
%Note that any results proved for \rbslack s also hold for \bslack s.

In the following, we assume that the tree is initially a \bslack.
After a sequence of insertions and deletions, the tree is a \rbslack.
The goal of this section is to establish an upper bound on the number of rebalancing steps needed to transform this \rbslack\ back into a \bslack.

We assume that the updates shown in Figure~\ref{fig-bslack-updates} are performed sequentially.
In a concurrent setting, locks or lock-free methodologies such as the template in Chapter~\ref{chap-template} can be used to ensure that updates appear to atomically operate on mutually exclusive sets of nodes (so that the effect will be the same as if the updates were performed sequentially in some order).

Our analysis follows the approach taken in \cite{LF95}.
Consider any arbitrary \bslack\ $T$.
Initially, we associate every key in the tree with the leaf that contains it.
When a key is inserted into a leaf $u$, we associate the key with $u$.
After a key is deleted from $u$, the key is still associated with $u$.
If the node $u$ is deleted, then all keys associated with $u$ are instead associated with another node.
Two cases arise.
If $u$ is deleted by a Root-Replace, then all keys associated with $u$ are instead associated with the only child of $u$.
Otherwise, $u$ is deleted by Absorb or Compress, and all keys associated with $u$ are instead associated with the node that was the parent of $u$ before the Absorb or Compress.

Let $\sigma$ be a sequence of updates to $T$, and $w$ be an internal node in the tree after the updates in $\sigma$ have been performed.
We define the \textit{\iset} of $w$ to be the multiset of all keys associated with nodes in the subtree rooted at $w$.
Therefore, the \iset\ of the root contains $n+i$ keys, where $i$ is the number of insertions in $\sigma$ and $n$ is the size of $T$.

We also define the \textit{relaxed height} of a node $u$ in a \rbslack.
Suppose we formed a \textit{new} \rbslack\ $T'$ by detaching the subtree rooted at $u$ from the \rbslack\ that contains it. %Consider the \rbslack\ $T$ that would be formed by detaching the subtree rooted $u$ from the \rbslack\ that contains it.
The relaxed height of $u$, denoted $rh(u)$, is then the relaxed depth of the leaves in $T'$.

The following lemma relates the relaxed height of a node to the number of keys in its \iset.

\begin{lem}
Consider a \bslack\ $T$ containing at least two keys, and a sequence of updates to it.
Then, let $u$ be any node in the resulting tree.
If $u$ is the root, then its \iset\ contains at least $2 \big\lfloor \frac b 2 \big\rfloor^{rh(u)-weight(u)}$ keys.
Otherwise, its \iset\ contains at least $\big\lfloor \frac b 2 \big\rfloor^{rh(u)}$ keys.
\end{lem}
\begin{chapscxproof}
The proof is by induction on the sequence of updates performed on $T$.

\textbf{Base case.}
Let $u$ be any node in $T$, $T_u$ be the tree rooted at $u$, and $h_u$ be the height of $T_u$.
Any subtree of a \bslack\ is a \bslack, so $T_u$ is a \bslack.
By Corollary~\ref{cor-2maxslack-fewest-keys}, $T_u$ must contain at least $d(h_u,2)$ keys.
By Corollary~\ref{cor-N-bounds}, $d(h_u,2) > (\frac b 2)^{h_u}$.
Since $T_u$ is a \bslack, every node has \weight\ one, so the relaxed height of each node is equal to its height and $h_u = rh(u)$.
Therefore, $d(h_u,2) > (\frac b 2)^{rh(u)} \ge \lfloor \frac b 2 \rfloor^{rh(u)} \ge 2\lfloor \frac b 2 \rfloor^{rh(u)-1} = 2 \lfloor \frac b 2 \rfloor^{rh(u)-weight(u)}$.

\textbf{Inductive step.}
Suppose the claim holds before an update $U$.
We prove it holds after $U$.
Let $rh'(u)$ be the relaxed height of a node $u$ after $U$.

Suppose $U$ is Delete.
Then, each \iset\ remains the same, and every node has the same relaxed height before and after $U$.

Suppose $U$ is Insert.
Then, each \iset\ either gains one new key, or remains the same, and every node has the same relaxed height before and after $U$.

Suppose $U$ is Root-Replace.
Let $p$ be the old root, and $u$ be its only child.
If $u$ has \weight\ one before $U$, then $rh'(u) = rh(p)-1$.
Otherwise, $u$ has \weight\ zero before $U$, so $rh'(u) = rh(p)$.
Any keys associated with $p$ before $U$ are associated with $u$ after $U$, so the \iset\ of the root is the same before and after $U$.

Suppose $U$ is Root-Zero.
Let $r$ be the relaxed height of the root before $U$.
By the inductive hypothesis, the \iset\ of the root contains at least $2 \big\lfloor \frac b 2 \big\rfloor^{r}$ keys before $U$.
Moreover, $U$ does not change any \iset.
After $U$, the relaxed height of the root increases to $r+1$ because the \weight\ of the root changes from zero to one, so the \iset\ of the root contains at least $2 \big\lfloor \frac b 2 \big\rfloor^r = 2 \big\lfloor \frac b 2 \big\rfloor^{(r+1)-1} = 2 \big\lfloor \frac b 2 \big\rfloor^{rh(root)-weight(root)}$.

Suppose $U$ is Absorb.
Let $u$ be the child before $U$ and $p$ be its parent.
In this case, $u$ is removed by $U$, and all of its associated keys are instead associated with $p$.
The \iset, \weight\ and relaxed height of $p$ are all the same before and after $U$.

Suppose $U$ is Split.
Let $u$ be the child before $U$ and $p$ be its parent.
In this case, $U$ creates a new child, $v$, of $p$ and moves all of $p$'s pointers (except for its pointers to $u$ and $v$) into $u$ and $v$, so that $u$ and $v$ each contain at least $\lfloor \frac{b+1}{2} \rfloor \ge \frac b 2$ pointers.
Observe that $rh'(u) = rh'(v) = rh'(p) = rh(p)$.
Each pointer in $u$ or $v$ after $U$ points to a node with relaxed height $rh(p)-1$.
By the inductive hypothesis, the \iset\ of every such node contains at least $\lfloor \frac b 2 \rfloor^{rh(p)-1}$ keys.
Therefore, the \iset s of $u$ and $v$ each contain at least $\frac b 2 \lfloor \frac b 2 \rfloor^{rh(p)-1} \ge \lfloor \frac b 2 \rfloor^{rh(p)} = \lfloor \frac b 2 \rfloor^{rh'(u)} = \lfloor \frac b 2 \rfloor^{rh'(v)}$ keys, and the \iset\ of $p$ contains at least $\lfloor \frac b 2 \rfloor^{rh'(u)} + \lfloor \frac b 2 \rfloor^{rh'(v)} = 2 \lfloor \frac b 2 \rfloor^{rh'(p)}$ keys.

Suppose $U$ is Overflow.
Let $u$ be the leaf that is full.
In this case, $U$ creates a new leaf $r$ and an internal node $p$ with \weight\ zero and pointers to $u$ and $v$, and moves half of the keys from $u$ into $v$.
After $U$, the \iset\ of $p$ contains at least $b+1$ keys.
Since $u$ and $v$ are leaves, $rh(p) = 1$, so $b+1 \ge 2 \lfloor \frac b 2 \rfloor^{rh(p)}$.
The \iset s of $u$ and $v$ each contain at least $\frac b 2$ keys.
Since $rh(u) = rh(v) = 1$, $\frac b 2 \ge \lfloor \frac b 2 \rfloor^{rh(u)} = \lfloor \frac b 2 \rfloor^{rh(v)}$.

Suppose $U$ is Compress or One-Child.
Let $p$ be the upper node and $k$ be its degree.
Observe that $U$ does not change the weight or relaxed height of any node, and does not remove any key from the \iset\ of $p$.
After $U$, $p$ has $\lceil \frac c b \rceil$ children that evenly share $c$ pointers or keys.
Thus, each child contains at least $\lfloor \frac{c}{\lceil c/b \rceil} \rfloor \ge \lfloor \frac{c}{c/b+1} \rfloor = \lfloor \frac{cb}{c+b} \rfloor = \lfloor \frac{b}{1+b/c} \rfloor$ pointers or keys.
If $U$ is One-Child, then $c > kb-b$ and $k \ge 2$, so $c > b$ and $\lfloor \frac{b}{1+b/c} \rfloor > \lfloor \frac b 2 \rfloor$.
If $U$ is Compress, then two cases arise.
If $p$ has at least two children after $U$, then $\lceil \frac c b \rceil \ge 2$, so $c/b+1 \ge 2$ and $b/c \le 1$.
Therefore, each child of $p$ contains at least $\lfloor \frac{b}{1+b/c} \rfloor \ge \lfloor \frac b 2 \rfloor$ pointers or keys after $U$.
Otherwise, after $U$, the single child of $p$ contains all of the pointers and keys of the children that were removed, so its \iset\ is at least as large as it was before $U$.
The claim then follows immediately from the inductive hypothesis.
%\qed
\end{chapscxproof}

\begin{cor}
Consider a \rbslack\ that results from performing a sequence of operations, $i$ of which are insertions, on a \bslack\ containing $n$ keys.
%Consider a \rbslack\ that results from performing a sequence of operations on a \bslack\ $T$, $i$ of which are insertions.
%Let $T$ be a \bslack\ that is transformed into a \rbslack\ by a sequence of operations. %Suppose a sequence of operations are performed on a \bslack\ $T$.
The relaxed height of the root, and, hence, any node in this \rbslack\ is at most $\big\lfloor \log_{\lfloor\frac{b}{2}\rfloor} \frac{n+i}{2} \big\rfloor + 1$.
\end{cor}

\begin{lem} \label{lem-upper-bound-on-splits}
After a sequence of operations, $i$ of which are insertions, on a \bslack\ containing $n$ keys, the total number of Absorb and Root-Zero updates that can be performed is at most $i$, and the number of Split updates that can be performed is at most $i (1+\big\lfloor \log_{\lfloor\frac{b}{2}\rfloor} \frac{n+i}{2} \big\rfloor)$.
\end{lem}
\begin{chapscxproof}
Absorb or Split is performed when a node has \weight\ zero.
Overflow is the only update that increases the number of zero \weight s in the tree, and at most $i$ Overflows occur, so there are at most $i$ nodes with \weight\ zero in the tree.
Absorb and Root-Zero each decrease the number of zero \weight s in the tree by one, so at most $i$ of these updates can be performed.
Root-Zero, Root-Replace, Absorb and Split are the only updates that can change the relaxed height of a node with \weight\ zero.
Root-Zero, Root-Replace and Absorb each change a zero \weight\ to one, and Split moves a zero \weight\ from a node with relaxed height $r$ to a node with relaxed height $r+1$.
%Split moves a zero \weight\ from a node with relaxed height $r$ to a node with relaxed height $r+1$.
%Split is the only update that moves a zero \weight, and no other update changes the relaxed height of a node with \weight\ zero.
Therefore, each zero \weight\ will remain at a node with the same relaxed height until it is moved by Split or changed to one by Root-Zero, Root-Replace or Absorb.
Since the relaxed height of any node in the tree is at most $\big\lfloor \log_{\lfloor\frac{b}{2}\rfloor} \frac{n+i}{2} \big\rfloor + 1$, each of the $i$ zero \weight s in the tree can be moved by Split at most $\big\lfloor \log_{\lfloor\frac{b}{2}\rfloor} \frac{n+i}{2} \big\rfloor + 1$ times.
%\qed
\end{chapscxproof}

\begin{lem} \label{lem-amortized-rebalancing}
Let $T$ be a \rbslack\ of height $h$ that is obtained by performing any sequence of $i$ insertions and $d$ deletions on an initially empty \rbslack.
At most $2i(4+\frac 3 2 \big\lfloor \log_{\lfloor\frac{b}{2}\rfloor} \frac{n+i}{2} \big\rfloor) + 2d/(b-1)$ rebalancing steps can be applied to~$T$.
\end{lem}
\begin{chapscxproof}
Let $c$ be the total degree of the children of a parent where Compress or One-Child is performed.
We first bound the number of One-Child updates that can be performed.
If a node has exactly one pointer, we say a \textit{pointer violation} occurs at that node.
One-Child is performed only when a pointer violation occurs at a child of the parent and $c > kb-b$.
Since $c > kb-b$ and $k \ge 2$, $c \ge b+1$, so the parent will have at least two children after One-Child.
Furthermore, each child of the parent will have degree at least $\lfloor b/2 \rfloor$.
Thus, One-Child removes every pointer violation at a child of the parent, and does not create any pointer violation.
Root-Replace removes a pointer violation at the root, decreasing the number of pointer violations in the tree by one.
However, Compress can \textit{increase} the number of pointer violations in the tree by one if $c \le b$.
No other update changes the number of pointer violations in the tree.
Therefore, the total number of One-Child and Root-Replace updates that can be performed is bounded above by the number of Compress updates.

We bound the number of Compress updates by studying the change in the total amount of slack in the tree that is caused by each type of update.
It is convenient to ignore the slack in any node with a zero \weight\ value, since Compress cannot affect any such node.
Compress redistributes a total of $c < kb-b$ pointers or keys from $k$ nodes to $\lceil c/b \rceil$ nodes.
Since $c \le kb-b = b(k-1)$, $\lceil c/b \rceil \le k-1$, Compress will remove at least one node from the tree.
Removing this node removes $b$ slack, and increases slack at the parent by one.
Thus, Compress reduces the total amount of slack in the tree by at least $b-1$ (and by even more, if more than one node is removed).
Split is applied precisely when the parent of a node with \weight\ value zero contains exactly $b$ pointers (and no slack).
Since the node with \weight\ value zero contains exactly two pointers, $b+1$ pointers are moved into the nodes with \weight\ value one in Figure~\ref{fig-bslack-updates}, so Split increases the total slack in the tree by exactly $b-1$.
By Lemma~\ref{lem-upper-bound-on-splits}, Split can be performed at most $i (1+\big\lfloor \log_{\lfloor\frac{b}{2}\rfloor} \frac{n+i}{2} \big\rfloor)$ times, so the total amount of slack created by Split is at most $i(1+\big\lfloor \log_{\lfloor\frac{b}{2}\rfloor} \frac{n+i}{2} \big\rfloor)(b-1)$.
It is easy to verify that Insert, Insert-Distribute and Absorb each decrease the total slack by one, and that Delete and Insert-Overflow increase the total slack by one and $2(b-1)$, respectively.
Thus, the total slack created by Delete and Insert-Overflow updates is $d + 2i(b-1)$, so the total slack in $T$ is at most $i(1+\big\lfloor \log_{\lfloor\frac{b}{2}\rfloor} \frac{n+i}{2} \big\rfloor)(b-1) + d + 2i(b-1) = i(b-1)(3+\big\lfloor \log_{\lfloor\frac{b}{2}\rfloor} \frac{n+i}{2} \big\rfloor) + d$.
Therefore, at most $i(3+\big\lfloor \log_{\lfloor\frac{b}{2}\rfloor} \frac{n+i}{2} \big\rfloor) + d/(b-1)$ Compress updates can occur.

By Lemma~\ref{lem-upper-bound-on-splits} at most $i$ Absorb and Root-Zero updates and $i (1+\big\lfloor \log_{\lfloor\frac{b}{2}\rfloor} \frac{n+i}{2} \big\rfloor)$ Split updates can occur.
Since the total number of Root-Replace and One-Child updates is at most the number of Compress updates, the number of Root-Replace, One-Child and Compress updates that can occur is at most $2i(3+\big\lfloor \log_{\lfloor\frac{b}{2}\rfloor} \frac{n+i}{2} \big\rfloor) + 2d/(b-1)$.
Therefore, at most $2i(3+\big\lfloor \log_{\lfloor\frac{b}{2}\rfloor} \frac{n+i}{2} \big\rfloor) + 2d/(b-1) + i + i (1+\big\lfloor \log_{\lfloor\frac{b}{2}\rfloor} \frac{n+i}{2} \big\rfloor) = 2i(4+\frac 3 2 \big\lfloor \log_{\lfloor\frac{b}{2}\rfloor} \frac{n+i}{2} \big\rfloor) + 2d/(b-1)$ rebalancing steps can be applied to $T$.
%\qed
\end{chapscxproof}

This result implies that the number of rebalancing steps needed to rebalance the tree after a sequence of deletions is amortized constant, and after a sequence of insertions is amortized logarithmic in: the size of the tree the last time it was a \bslack\ plus the number of insertions that have occurred since then. %, where $h$ is the height of the tree after all insertions.
%In Section~\ref{sec-analysis}, we prove that the height is logarithmic.
Section~\ref{sec-constant-rebalancing} explains how \bslack s can be modified to obtain amortized constant rebalancing by slightly increasing the amount of slack shared amongst the children of an internal node.

%\trevor{note to self: we should be able to get $O(b \cdot h(T))$ rebalancing updates from the previous lemma, and plug in the $h(T)$ later.}
%
%\trevor{Mention: by making a small change to the compress operation, we get amortized constant rebalancing. Explain in appendix.}
\end{hide}

\section{\bslack s with amortized constant rebalancing} \label{sec-constant-rebalancing}

\begin{hide}
The main challenge in achieving amortized constant rebalancing is ensuring that long sequences of Split and Compress operations occur infrequently.
Split can necessitate other Splits higher in the tree and many Compresses.
Compress can necessitate many other Compresses.
This makes Split and Compress particularly problematic.
%The other rebalancing steps do not lead to long sequences of rebalancing steps.

Split occurs only when an internal node is full.
If a Compress at an internal node leaves some slack in each of its children, 
then Splits will not immediately occur at the children.
With this in mind, we make some small modifications.
%to P\ref{prop-bslack-slack}, the definition of a slack violation and Compress.
P\ref{prop-bslack-slack} is replaced with P\ref{prop-bslack-slack}$'$, which says that, for each internal node $u$ of degree $k$, the total slack contained in the children of $u$ is at most $b+k-1$ (so the worst-case slack per node is only one greater than in a standard \bslack).
A slack violation then occurs at any internal node that violates P\ref{prop-bslack-slack}$'$.
The children of an internal node of degree $k$ where a slack violation occurs will have total degree less than $kb-(b+k-1)=(k-1)(b-1)$.
Thus, Compress is performed only at internal nodes whose children have total degree $c \le (k-1)(b-1)$, and One-Child is performed only at internal nodes whose children have total degree $c > (k-1)(b-1)$.
This threshold is chosen so that Compress is only performed when it can remove one node and still leave each child with one slack (so that each child can accommodate one more key before necessitating a Split).
We then change Compress so that it evenly distributes the $c$ pointers or keys of its children amongst $\lceil \frac{c}{b-1} \rceil$ nodes, instead of $\lceil \frac{c}{b} \rceil$.
This way, each child is guaranteed to have at least one slack afterwards.

We prove that the number of rebalancing steps is amortized constant using the potential method. 
The potential of a node $u$, denoted $\phi(u)$, captures the intuition that
a node is bad if it contains too much slack, is full, or has \weight\ zero.
%it is bad when a node contains too much slack, and it is bad when a node is full or has \weight\ zero.
\begin{displaymath}
   \phi(u) = \left\{
     \begin{array}{ll}
       b-deg(u) & \mbox{if $deg(u) < b$ and $u$ has \weight\ one} \\
       b & \mbox{otherwise (i.e.,\ $deg(u) = b$ or $u$ has \weight\ zero)}
     \end{array}
   \right.
\end{displaymath}
The potential of a tree $T$, denoted $\Phi(T)$, is the sum of potentials of its nodes.

We now study how $\Phi(T)$ is changed by deletion, insertion, and each rebalancing step.
Let $u$ be a node with degree $k$. Recall that leaves never have \weight\ zero.

\textbf{Delete.}
If $u$ is full, then $\phi(u)$ changes from $b$ to $1$.
Otherwise, $\phi(u)$ changes from $b-k$ to $b-(k-1)$.
So, $\phi(u)$ increases by at most one.

\textbf{Insert.}
If the insertion fills $u$, then $\phi(u)$ changes from $1$ to $b$.
Otherwise, $\phi(u)$ changes from $b-k$ to $b-(k+1)$.
So, $\phi(u)$ increases by at most $b-1$.

\textbf{Overflow.}
A full node with potential $b$ turns into a node with \weight\ zero, which has potential $b$, and two nodes with \weight\ one that share a total of $b-1$ slack.
Thus, $b$ potential is replaced by $b+b-1=2b-1$ potential, which is an increase of $b-1$.

\textbf{Absorb.}
Let $u$ be the node with \weight\ zero. Its parent $\pi(u)$ has \weight\ one.
Beforehand, $u$ has degree two and $\pi(u)$ contains $j \leq b-1$ pointers, so is it not full.
%Let $v$ be the upper node and $u$ be the lower node.
%Since $u$ has \weight\ zero, it has degree two.
%Thus, $\pi(u)$ is not full before the Absorb.
Absorb decreases potential by $b$ by eliminating $u$.
It also moves a pointer from $u$ to $\pi(u)$.
%Let $j$ be the degree of $v$ before the Absorb.
If $j=b-1$, then $\phi(\pi(u))$ changes from 1 to $b$, increasing potential by $b-1$, for a net decrease of one.
Otherwise, $\phi(\pi(u))$ changes from $b-j$ to $b-(j-1)$, for a net decrease in potential of $b-1$.

\textbf{Split.}
Let $u$ be the node with \weight\ zero. Its parent $\pi(u)$ has \weight\ one, but it is full, so
it has potential $b$ behorehand. Afterwards, it has \weight\ zero, so its potential does not change.
%Let $v$ be the upper node and $u$ be the lower node.
Before the Split, $u$ has potential $b$, and it is split into two nodes, each of \weight\ one, that share a total of $b-1$ slack.
After the Split, the sum of their potentials is $b-1$.
Thus, the potential of the tree is decreased by one.

\textbf{Compress.}
%We would like to show that Compress decreases the total potential in the tree.
Let $u$ be a node with $k \geq 2$ children that have total degree at most $(k-1)(b-1)$.
%the upper node.
%Compress removes at least one node and distributes its keys to the remaining siblings.
%THIS IS NOT QUITE ACCURRATE
The modified version of Compress will leave at least one slack at each child of $u$, so the total potential of $u$'s children will be the total amount of slack they contain, which decreases by at least $b$, since
at least one of the children is removed.
Removing a child of $u$ also increases the slack at $u$ by one, which increases $\phi(u)$ by one (unless $u$ was full, in which case it decreases $\phi(u)$).
For each child of $u$ that is removed by Compress, $b$ slack is eliminated at at the children of $u$, and one slack is added at the parent.
Therefore, the total potential of the tree decreases by $b-1$ for each child removed by Compress.
Since at least one child is removed, the total potential of the tree decreases by at least $b-1$.

\textbf{One-Child.}
Since One-Child evenly distributes keys, it cannot create any more full nodes than existed beforehand.
It does not affect \weight s, and it does not remove any key or pointer.
So, One-Child does not affect the total potential of the tree.

Since $\Phi(T)$ is increased by one for Delete and $b-1$ for Insert (and Overflow), and no other operation increases it, after $i$ insertions and $d$ deletions, $\Phi(T) \le (b-1)i+d$.
We can use $\Phi(T)$ to bound the number of rebalancing steps that can be performed on $T$.
Let $C$, $A$, $S$, $R_0$, $R_r$ be the number of Compresses, Absorbs, Splits, Root-Zeros and Root-Replaces, respectively.
We immediately obtain $(b-1)C + A + S + 2(R_0 + R_r) \le \Phi(T) \le (b-1)i+d$.
Astute readers will notice that One-Child has not yet made an appearance.
By the same argument as in Lemma~\ref{lem-amortized-rebalancing}, the number of One-Childs is at most $C$.
Therefore, the number of rebalancing steps is constant per update.

In fact, we can achieve tighter bounds if we are more careful.
By the same argument as in Lemma~\ref{lem-upper-bound-on-splits}, $A + R_0 \le i$.
Additionally, the same argument used to show that the number of One-Childs is at most $C$ applies to Root-Replace, so $R_r \le C$.
Therefore, on average, there is at most one Absorb or Root-Zero per insertion, at most one Compress (and One-Child) and Root-Replace per insertion, and at most one Split per insertion.
Similarly, on average, there is at most one Absorb, Split, Root-Replace or Root-Zero per deletion, and at most one Compress (and One-Child) per $b-1$ deletions.

The increase in space complexity associated with these changes is very small.
Since the worst-case slack per node is only one greater than in a \bslack, the minimum average degree of this modified \bslack\ is at most one less than in a \bslack.
So, worst-case lower bound on utilization changes from $\bar D/b$ to $(\bar D - 1)/b$, and the space complexity upper bound changes from $2b(n-1)/(\bar D - 1) < 2b(n-1)/(b-3)$ to $2b(n-1)/((\bar D - 1) - 1) < 2b(n-1)/(b-4)$.
\end{hide}

\section{Space complexity of competing trees} \label{sec-bslack-space-complexity}

\begin{hide}
In this section, we study the space complexity of some pathological families of B-trees, Overflow trees and H-trees.
The maximum degree of nodes is $b$, the block size is $2b$, and all trees are leaf-oriented.

\begin{itemize}
\item \textbf{B-tree.} The root has degree two, and all other nodes have degree $b/2$.
\item \textbf{Overflow tree.} The root has degree two, the internal nodes have degree $b/2$, and the leaves have degree $b-3$.
Overflow groups are chosen to be as large as possible, to minimize wasted space.
Specifically, for each parent $u$ of a leaf, $u$'s children are all in a single group, with one shared overflow node.
Thus, each overflow node is shared by $b/2$ leaves (which contain a total of $(b-3)(b/2) = b^2/2-3b/2$ keys).
\item \textbf{H-tree.} Parameters $\gamma$ and $\delta$ are chosen to be as large as possible, to minimize wasted space.
The root has degree two, the internal nodes have degree $\lceil b/\sqrt{2} \rceil$, and the leaves have degree $b-2$.
(H-trees are node-oriented, which would significantly inflate their space complexity on a system with only one block size.
%So, the H-trees considered here are leaf-oriented.
The space complexity bounds shown in Figure~\ref{fig-spk} ignore this, and are thus quite charitable.
To actually achieve such good space complexity bounds for H-trees, one would have to completely redesign the data structure to be leaf-oriented.) % A node-oriented H-tree would have significantly worse space complexity on a system with only one block size.)
\end{itemize}

We assume that a key and a pointer each occupy a single word in memory.
For each family, and each choice of maximum degree in $\{8,16,32\}$, we consider the minimum height tree from the family containing at least $10^6$ keys, and computed its space complexity.
%Each tree contains between $10^6$ and $2 \times 10^6$ keys.
%We computed the space complexity of each of these trees.
Observe that the resulting space complexity values are \textit{lower bounds} on the worst-case space complexity for these data structures.
These space complexity values appear in Figure~\ref{fig-spk}, along with \textit{very pessimistic upper bounds} on the space complexity for any \bslack\ ($b \in \{8,16,32\}$) containing at least $50,000$ keys.
(The aforementioned upper bounds actually apply to \maxslack s, which allow the slack shared amongst the children of a node to be one greater than in a \bslack.
Additionally, despite the fact that the space complexity of \bslack s \textit{improves as the number of keys grows}, these upper bounds only assume the tree contains at least 50,000 keys, in contrast to the other data structures, which contain at least $10^6$ keys.)
Therefore, these results are actually quite charitable to the other data structures.

Nevertheless, the advantage of \bslack s is clear.
The optimal space to store $n$ keys and pointers to associated data is $2n$.
H-trees, the closest competitor to \bslack s, use more than double the space beyond what is optimal.
If these trees were modified to implement a set instead of a dictionary (by eliminating data and allowing leaves to contain up to $2b$ keys), then the optimal space would become $n$, and it is expected that relative differences in space complexity between the trees would increase further.

\renewcommand{\tabcolsep}{2mm}
\begin{figure}[tb]
\centering
\begin{tabular}{|c|ccccc|}
\hline
Max degree & B-tree & Overflow tree & H-tree & \bslack & Optimal \\
\hline
8  & $\ge 5.333n$ & $\ge 5.066n$ & $\ge 3.840n$ & $< 2.789n$ & $2.000n$ \\
16 & $\ge 4.571n$ & $\ge 3.120n$ & $\ge 2.685n$ & $< 2.301n$ & $2.000n$ \\
32 & $\ge 4.266n$ & $\ge 2.492n$ & $\ge 2.307n$ & $< 2.145n$ & $2.000n$ \\
\hline
\end{tabular}
%\begin{tabular}{|c|ccccc|}
%\hline
%Operation & B-tree & Overflow tree & H-tree & SDM-tree & \bslack \\
%\hline
%Insert & $\Theta(\log n)$ & $\Theta(\log n)$ & $\Theta(\log n)$ & $O((\log n)^{m-2})$ & amortized $O(\log n)$ \\
%Delete & $\Theta(\log n)$ & $\Theta(\log n)$ & $\Theta(\log n)$ & $O(n)$ & amortized $O(\log n)$ \\
%\hline
%\end{tabular}
\caption{
    Space complexity of example trees, and worst-case bound for \bslack s.
}
\label{fig-spk}
\end{figure}
\end{hide}

\section{Sequential experiments} \label{sec-bslack-exp}

\begin{hide}
\paragraph{Java implementation}
The \bslack\ was implemented as a sequential data structure in Java.
Each update $U$ shown in Figure~\ref{fig-bslack-updates} was implemented as follows.
A process performing $U$ creates new a node for each node on the right hand side of the depiction of $U$ in Figure~\ref{fig-bslack-updates}, and replaces the nodes on the left hand side by these new nodes.
The nodes that were replaced by $U$ are eventually reclaimed by Java's automatic garbage collection.
If a Delete, Insert or Overflow creates a violation, then the process invokes a \textbf{Cleanup} procedure to perform rebalancing steps until the tree no longer contains any violations.

For simplicity, we implemented Cleanup as a recursive procedure.
Cleanup takes the node $u$ where a violation occurs as its argument, and attempts to perform a rebalancing step to fix the violation at $u$.
If this rebalancing step creates any new violations, or replaces any nodes with existing violations and moves the violations to new nodes, then recursive invocations of Cleanup are performed to eliminate these violations.
If an invocation of Cleanup sees that the node whose violation it was supposed to fix has already been replaced, then it knows the invocation of Cleanup that replaced it will make a recursive call to fix the violation, wherever it was moved.
Thus, this invocation of Cleanup can simply return.
The downside of a recursive implementation of Cleanup is that stack overflow may occur if rebalancing steps create a large number of violations.
Other ways to implement the Cleanup procedure are discussed in Section~\ref{sec-bslack-cleanup}.

Java is not an ideal language for implementing the \bslack, since it gives very little control over memory layout.
%This implementation simply serves as a proof of concept.
%Java is not an ideal language in which to implement \bslack s, since it gives very little control over memory layout.
The purpose of our Java implementation is simply to serve as a guide for any implementers who are interested in porting the \bslack\ tree to other languages.
Each node is implemented as an array of keys, and an array of child pointers.
Unlike in C/C++, in Java, arrays cannot be embedded directly in a node.
Instead, nodes contain pointers to arrays, which are located elsewhere in memory.
Thus, even after a process has loaded (a cache line that contains) a node, accessing a key or child pointer of that node still requires performing additional loads from %loading additional cache lines from 
potentially distant locations in main memory.
This makes the implementation somewhat inefficient.
Furthermore, the implementation does not directly satisfy the space complexity upper bounds computed in Section~\ref{sec-analysis}.
%However, we still expect the space complexity of the Java implementation to be fairly good, since the allocator allows multiple block pointers to arrays simply add a few extra words of overhead to each node, 
However, if we ignore these complications and pretend that keys and pointers are embedded directly inside nodes, then we can still use this implementation to study structural properties of \bslack s in practice.

\paragraph{Methodology}
%We performed several different experiments using our Java implementation.
We performed randomized experimental trials for several $b$ values, simulated workloads, and tree sizes.
Each trial was divided into two phases.
In the first phase, a \bslack\ was created and initialized by inserting and deleting (with 50\% probability each) keys drawn uniformly randomly from $[0,size)$, until the tree stabilized, containing approximately $size/2$ keys.
In the second phase, one million random insertions and deletions were performed with some specified probabilities, and each rebalancing step was recorded.
We considered probabilities: 50\% insertion and 50\% deletion (50i-50d), 90\% insertion and 10\% deletion (90i-10d), and 10\% insertion and 90\% deletion (10i-90d).
At the end of each trial, average degree and space complexity were computed (under the assumption that keys and pointers were actually embedded directly in nodes).

\paragraph{Results}
We discuss a small selection of the results.
For 50i-50d, $b=16$ and $size=2^{20}=$ 1,048,576, there were approximately 1.2 rebalancing steps per successful update, the average degree was approximately 15.5, and the space complexity was less than $2.209n$.
In fact, even with a rather small $size$ of $2^{12}=$ 4,096, the average degree was approximately 15.4, and the space complexity was less than $2.226n$, which is substantially better than the theoretical average degree lower bound of 12.7 and space complexity upper bound of $2.726n$.
This suggests that the performance of \bslack s is much better in practice than what is suggested by our theoretical results.
For 50i-50d, $b=32$, and $size=2^{20}$, there were approximately 1.1 rebalancing steps per successful update, the average degree was approximately 31.5, and the space complexity was less than $2.097n$.
For 10i-90d, $b=16$ and $size=2^{20}$, there was less than one rebalancing step per successful update.
Even with 90\% of updates being deletion, the average degree remained the same as in the 50i-50d case, and the space complexity was less than $2.213n$.
For 90i-10d, there were approximately 1.2 rebalancing steps per successful update.
These results suggest that little rebalancing is required for random updates on uniform keys.

\begin{figure}[t]
\centering
\includegraphics[width=0.66\linewidth]{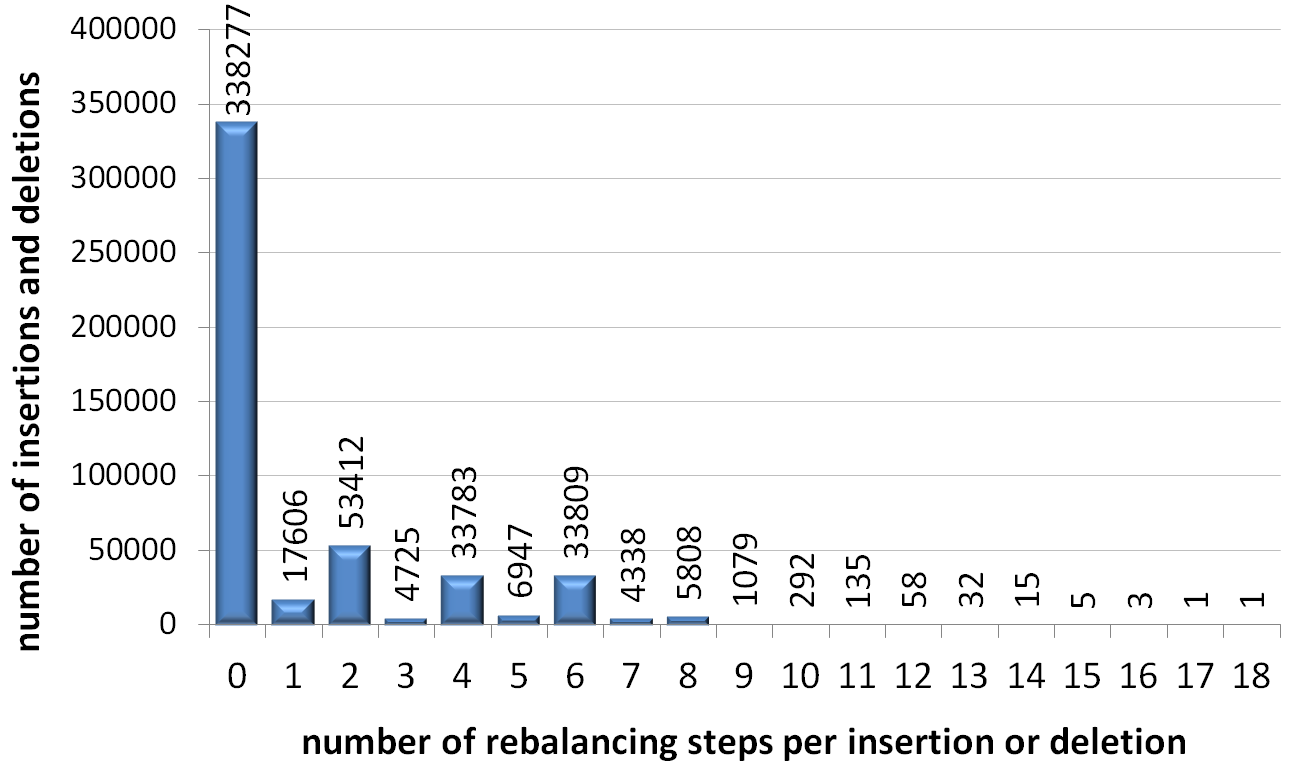}
\caption{Histogram showing the frequency of updates that perform a certain number of rebalancing steps in a randomized trial.}
\label{fig-bslack-histogram}
\end{figure}

\paragraph{Distribution of rebalancing}
It is also interesting to understand how many rebalancing steps are necessitated by each insertion or deletion. % the longest running updates.
So, we plotted a histogram for one trial with parameters 50i-50d, $b=16$ and $size=2^{20}=$ 1,048,576 in Figure~\ref{fig-bslack-histogram}.
(Results for other trials, values of $b$ and simulated workloads are similar.)
The $x$-axis shows the number of rebalancing steps that were performed by a single insertion or deletion, and the $y$-axis shows how many insertions or deletions performed $x$ rebalancing steps.
The total number of successful insertions and deletions was 500,326, and the height of the tree was four when the trial finished.
The most rebalancing steps performed by an insertion or deletion was 18, 67.6\% of successful insertions and deletions performed no rebalancing, 97.6\% performed six or less rebalancing steps, and 99.9\% performed less than ten rebalancing steps.
\end{hide}

\section{Implemention issues for rebalancing} \label{sec-bslack-cleanup}

\begin{hide}
One way to implement rebalancing is to explicitly maintain a collection of pointers to internal nodes where rebalancing steps must be performed.
After an update creates a violation, a rebalancing step is performed to fix that violation.
Every time a rebalancing step creates a violation, a pointer to the node where the violation occurs (and possibly also a pointer to its parent) is added to the collection.
An update does not terminate until it has emptied the queue and performed rebalancing steps to fix all violations in the tree.

Observe that violations can only occur at internal nodes.
The number of internal nodes is quite small compared to $n$ (close to $n/(b-2)^2$ in \bslack s of height at least three).
So, even if a collection contains \textit{every} internal node, the worst-case collection size may be reasonable for some applications.
Since the amortized number of rebalancing steps per update is small, most updates will result in a small collection.
We recommend using a small, fixed-size queue % implemented with a circular array, 
and switching to a more computationally expensive algorithm if the queue becomes full.
For instance, when a process tries to enqueue a pointer and the queue is full, it can simply discard that pointer, and continue the algorithm, recording the fact that the a pointer was discarded.
Eventually, after enough rebalancing steps are performed, the queue becomes empty, and the process can traverse the tree to find any outstanding violations, repopulate the queue, and continue the algorithm.
Although this approach is very expensive once the queue becomes full, it will not significantly increase the average running time of updates if the queue rarely becomes full.
The experimental results in Section~\ref{sec-bslack-exp} indicate that this approach could be practical, even with a very small bound on queue size.
\end{hide}

\chapter{\Rbslack s implemented with the template} \label{chap-lfbslack}
% !TEX root = paper.tex

\section{Implementation} \label{sec-bslack-concurrent}

\begin{figure}[tb]
\begin{framed}
%\hspace*{-7mm}
%\begin{minipage}[t]{85mm}
\def\namewidth{18mm}
\preplisting
\begin{lstlisting}[mathescape=true,style=nonumbers]
 type// \node
     //\com Fields used by \llt/\sct\ algorithm
     //\wcnarrow{$\info$}{pointer to \op}
     //\wcnarrow{$marked$}{marked bit}
     //\com User-defined fields
     //\wcnarrow{$weight$}{weight bit (immutable)}
     //\wcnarrow{$leaf$}{leaf bit (immutable)}
     //\wcnarrow{$searchKey$}{an auxiliary key for rebalancing (immutable)}
     //\wcnarrow{$d$}{degree of the node (immutable)}
     //\wcnarrow{$k_1, k_2, ..., k_d$}{keys (immutable)}
     //\wcnarrow{$p_1, p_2, ..., p_d$}{pointers (mutable)}
\end{lstlisting}
\end{framed}
%\end{minipage}
	\caption{Data definition for a node in a \rbslack.}
	\label{code-bslack-data}
\end{figure}

We present an implementation of a \rbslack\ using the template described in Chapter~\ref{chap-template}.
This implementation is very similar to the implementation of the $(a,b)$-tree in Chapter~\ref{chap-abtree}, except for the way that rebalancing is performed.

Let $b$ be the maximum degree of nodes.
We represent each node by a \rec\ with $b$ mutable pointers, and $b$ immutable keys, as well as immutable fields $d$, $weight$, $leaf$ and $searchKey$. % which respectively contain the node's degree, weight bit, a bit indicating whether the node is a leaf, and an auxiliary key used for rebalancing.
(See Figure~\ref{code-bslack-data}.)
The field $d$ contains the number of pointers that are used.
The $weight$ field contains the node's weight.
The bit $leaf$ indicates whether the node is a leaf.
The field $searchKey$ contains an auxiliary key that can be used to locate the node during rebalancing.
If a node contains at least one key, then $searchKey$ is just the first key, $k_1$, which might seem redundant.
However, when a node contains \textit{no} keys, $searchKey$ provides a process with a key that it can use to search for the node (which would otherwise be difficult to locate).
Internal nodes have one fewer key than pointers, so the key $k_d$ is unused in an internal node.
Leaves have exactly as many pointers as keys.

To avoid special cases when the tree is empty, we add a sentinel node at the top of the tree.
The sentinel node $entry$ always has one child and no keys.
(Every search that passes through it will simply follow that one child pointer.)
The sole child of this sentinel node is initially an empty leaf (with $d=0$ and no keys or pointers).
The actual \rbslack\ is rooted at the child of the sentinel node.
For convenience, we use $root$ to refer to the current child of the sentinel node.

%\begin{figure}[tb]
%\begin{framed}
%\prepnewlisting
%%\vspace{-5mm}
%%\hrule
%%\vspace{-2mm}
%\begin{lstlisting}[mathescape=true]
% //\func{Get}$(key)$
%   $\langle -, -, l \rangle := \func{Search}(key)$
%   if $l \mbox{ contains } key$ then return $\mbox{ the value associated with } key$
%   else return $\nil$ // \\ \vspace{-2mm} \hrule \vspace{1mm} %
%      
% //\func{Search}$(key)$
%   $gp := \nil; p := entry; l := entry.p_1$
%   while $l$// is internal
%     $gp := p; p := l$
%     $i := 1$
%     while $i < l.d$ and $key \ge l.k_i$ do $i := i + 1$ //\medcom Locate appropriate child pointer to follow
%     $l := l.p_i$ //\medcom Follow the child pointer
%   return $\langle gp, p, l \rangle$
%\end{lstlisting}
%\end{framed}
%	\caption{\func{Get} and \func{Search}.}
%	\label{code-bslack-search}
%\end{figure}

Detailed pseudocode for \func{Get} (and its auxiliary subroutine \func{Search}), \ins\ and \del\ is given in Figure~\ref{code-abtree-search}, \ref{code-bslack-ins} and~\ref{code-bslack-del}.
The implementations of \func{Get} and \func{Search} are identical to the same procedures in the relaxed $(a,b)$-tree implementation.
%\func{Get}, \ins\ and \del\ each execute an auxiliary procedure, \func{Search}($key$), which appears in Figure~\ref{code-abtree-search}.
The implementations of \ins\ and \del\ are also very similar to those of the relaxed $(a,b)$-tree.
The difference lies in the way that these procedures trigger rebalancing.
In our discussion of \ins\ and \del, we focus on the differences between the \ins\ and \del\ procedures for the \rbslack\ and relaxed $(a,b)$-tree.
The way we perform rebalancing in the \rbslack\ is quite different from rebalancing in the relaxed $(a,b)$-tree, and we carefully discuss the \rbslack\ rebalancing algorithm.

\begin{figure}[ph]
\begin{framed}
\def\namewidth{18mm}
\prepnewlisting
\begin{lstlisting}[mathescape=true]
 //\ins$(key, value)$
   //\com Returns $\bot$ if $key$ was not in the dictionary. Otherwise, this returns the value previously associated with $key$.
   loop
     //\com \textbf{Start of operation attempt}
     //\com Search for $key$ in the tree
     $\langle -, p, l \rangle := \func{Search}(key)$
   
     //\com Template iteration 0 (parent of leaf)
     $result_p := \llt(p)$
     if $result_p \in \{\fail, \finalized\}$ then continue //\medcom{goto next iteration (to retry)}
     if $l \notin result$ then // \medcom{\func{Conflict}: verify $p$ still points to $l$}\label{code-bslack-tryins-conflict1}
       continue //\medcom{goto next iteration (to retry)}
     //Let $p.p_i$ be the child pointer of $p$ that pointed to $l$ at the previous line

     //\com Template iteration 1 (leaf)
     $result_l := \llt(l)$
     if $result_l \in \{\fail, \finalized\}$ then continue //\medcom{goto next iteration (to retry)}

     //\com Computing \func{SCX-Arguments} from locally stored values (and immutable fields)
     $V := \langle p, l \rangle$
     $R := \langle l \rangle$
     //$fld :=$ a pointer to $p.p_i$
     if $l \mbox{ contains } key$ then //\medcom{Replace the value associated with an existing key}
       //Let $oldValue$ be the value associated with $key$ in $l$
       //$n :=$ new copy of $l$ that contains $\langle key, value \rangle$ instead of $\langle key, oldValue \rangle$
       $overflow := \false$
     else if $l.d < b$ then //\medcom{Insert a new key-value pair into a full leaf}
       //$n :=$ new copy of $l$ that has the new key-value pair inserted
       $oldValue := \bot$
       $overflow := \false$
     else //\com $l.d = b$ \medcom{Insert a new key-value pair into a non-full leaf}
       //$n :=$ pointer to a subtree of three newly created nodes: one internal node and two leaves, configured as in \textbf{Overflow} in Figure~\ref{fig-bslack-updates} (so that the key-value pairs in $kv(l) \cup \{\langle key, value \rangle\}$ are evenly distributed between the leaves, and the internal node has weight zero). The $searchKey$ of each new node is its first key.
       $oldValue := \bot$
       $overflow := \true$
       //\com If $n$ will replace $root$, then we also perform \textbf{Root-Zero} as part of the same atomic update
       if $p = entry$ then $n.weight = 1$ //\label{lfbslack-insert-rootzero}

     $success := \sct(V, R, fld, n)$ //\label{lfbslack-ins-scx}
     //\com \textbf{End of operation attempt}
     if $success$ then
       if $overflow$ then //\medcom{After an \textbf{Overflow}, we may need to rebalance}
         $\func{FixWeight}(n)$ //\medcom{Check for weight violation at $n$}
         $\func{FixSlack}(p)$ //\medcom{Check for slack violation at $p$}\label{lfbslack-ins-fixslack}
       return $oldValue$
     else continue //\medcom{goto next iteration (to retry)}
\end{lstlisting}
\end{framed}
	\caption{Pseudocode for \func{Insert}. Here, $b$ is the maximum degree of nodes.}
	\label{code-bslack-ins}
\end{figure}

\subsection{Insertion}
Pseudocode for \ins\ appears in Figure~\ref{code-bslack-ins}.
As in the relaxed $(a,b)$-tree, \ins\ takes two arguments: $key$ and $value$.
If $key$ is not in the dictionary, then \ins\ inserts the key-value pair $\langle key, value \rangle$, and returns $\bot$.
Otherwise, \ins\ replaced the existing key-value pair $\langle key, oldValue \rangle$ in the dictionary with $\langle key, value \rangle$ and returns $oldValue$.

\ins\ differs from the relaxed $(a,b)$-tree's \ins\ procedure in three ways.
First, instead of having a helper function \tryins\ that is invoked repeatedly in a loop, the body of the \tryins\ function is inlined directly in \ins.
Second, when new nodes are created for an Overflow update, we set the $searchKey$ field of each new node to the node's first key.
Note that each new node contains at least one key, and can be located by searching for its first key (using the \func{SearchNode} procedure in Figure~\ref{code-bslack-optimistic-nofixneeded}, which is described below).
Third, after performing a successful \sct, instead of invoking \cleanup\ to perform any needed rebalancing, two rebalancing procedures called \func{FixWeight} and \func{FixSlack} are invoked.
These procedures are discussed in Section~\ref{bslack-conc-rebalancing}.
%Each invocation is supplied with a pointer to a specific node at which a violation occurs (or is suspected to occur).

\begin{figure}[ph]
\begin{framed}
\def\namewidth{18mm}
\preplisting
\begin{lstlisting}[mathescape=true]
 //\del$(key)$
   //\com Returns $\bot$ if $key$ was not in the dictionary. Otherwise, this returns the value associated with $key$.
   loop
     //\com \textbf{Start of operation attempt}
     //\com Search for $key$ in the tree
     $\langle -, p, l \rangle := \func{Search}(key)$
   
     //\com Template iteration 0 (parent of leaf)
     $result_p := \llt(p)$
     if $result_p \in \{\fail, \finalized\}$ then continue //\medcom{goto next iteration (to retry)}
     if $l \notin result$ then // \medcom{\func{Conflict}: verify $p$ still points to $l$}
       continue //\medcom{goto next iteration (to retry)}
     //Let $p.p_i$ be the child pointer of $p$ that pointed to $l$ at the previous line

     //\com Template iteration 1 (leaf)
     $result_l := \llt(l)$
     if $result_l \in \{\fail, \finalized\}$ then continue //\medcom{goto next iteration (to retry)}

     //\com Computing \func{SCX-Arguments} from locally stored values (and immutable fields)
     $V := \langle p, l \rangle$
     $R := \langle l \rangle$
     //$fld :=$ a pointer to $p.p_i$
     if $l \mbox{ does not contain } key$ then //\medcom The tree does not contain $key$ $ $ 
       return $\bot$
     else
       //Let $oldValue$ be the value associated with $key$ in $l$
       //$n :=$ new copy of $l$ that does not contain $\langle key, oldValue \rangle$
       $success := \sct(V, R, fld, n)$ //\label{lfbslack-del-scx}
       //\com \textbf{End of operation attempt}
       if $success$ then
         $\func{FixSlack}(p)$ //\medcom Check for slack violation at $p$\label{lfbslack-del-fixslack}
         return $oldValue$
       else continue //\medcom{goto next iteration (to retry)}
\end{lstlisting}
\end{framed}
	\caption{Pseudocode for \func{Delete}.}
	\label{code-bslack-del}
\end{figure}

\subsection{Deletion}
Pseudocode for \del\ appears in Figure~\ref{code-bslack-del}.
As in the relaxed $(a,b)$-tree, \del\ takes one argument, $key$.
If $key$ is not in the dictionary, then \del\ simply returns $\bot$.
Otherwise, \del\ removes the existing key-value pair $\langle key, oldValue \rangle$ from the dictionary and returns $oldValue$.

\del\ differs from the relaxed $(a,b)$-tree's \del\ procedure in three ways.
First, instead of having a helper function \trydel\ that is invoked repeatedly in a loop, the body of the \trydel\ function is inlined directly in \del.
Second, when the new node is created for a \del, we set its $searchKey$ field to the $searchKey$ of the node $l$ that it is replacing.
Note that, since $l$ could be reached by searching for $l.searchKey$ (using the \func{SearchNode} procedure) before the \del, so can the new node after the \del.
Third, after performing a successful \sct, instead of invoking \cleanup\ to perform any needed rebalancing, the rebalancing procedure \func{FixSlack} is invoked.
%This invocation is supplied with a pointer to a specific node at which a violation occurs (or is suspected to occur).

\begin{figure}[h]
\centering
\begin{tabular}{|l|ll|}
\hline
\textbf{Update $U$ by $P$} & \multicolumn{2}{l|}{\textbf{Violations that $P$ becomes responsible for after $U$}} \\\hline
Delete          & possible $\langle slack, parent \rangle$ & \\\hline %if $Cslack(parent) \ge b-1$ before $U$\\\hline
Insert          & none possible & \\\hline
Overflow        & $\langle weight, n \rangle$, and & \\
                & possible $\langle slack, parent \rangle$ & \\\hline %if $Cslack(parent) \ge 2$ before $U$ \\\hline
Root-Zero       & $\langle degree, n \rangle$, and & if $\langle degree, root \rangle \in \beta$ \\
                & $\langle slack, n \rangle$ & if $\langle slack, root \rangle \in \beta$ \\\hline
Root-Replace    & $\langle degree, n \rangle$, and & if $\langle degree,$ the child of $root \rangle \in \beta$ \\
                & $\langle slack, n \rangle$ & if $\langle slack,$ the child of $root \rangle \in \beta$ \\\hline
Absorb          & possible $\langle slack, n \rangle$ & \\\hline%if $\langle slack, \pi(u) \rangle \in \beta$ \\\hline
Split           & $\langle weight, n \rangle$, and & \\
                & possible $\langle slack, n \rangle$, and & \\
                & possible $\langle slack, n.p_1 \rangle$, and & \\
                & possible $\langle slack, n.p_2 \rangle$, and & \\
                & possible $\langle slack, parent \rangle$ & \\\hline
Compress        & possible $\langle slack,$ any child of $n \rangle$, and & \\
                & possible $\langle slack, parent \rangle$, and & \\
                & $\langle degree, n \rangle$ & if the children of $n$ have total degree $c \le b$ \\\hline
One-Child       & possible $\langle slack,$ any child of $n \rangle$ & \\\hline
\end{tabular}
\caption{
%Characterization of $\beta' \setminus \beta$ for each update $U$ to a \rbslack.
%Here, %$Cslack(u)$ is the total slack shared amongst the children of $u$, 
%$n$ is the topmost new node created by the update, $u$ corresponds to the node labeled as such in Figure~\ref{fig-bslack-updates}, $\pi(u)$ is the parent of $u$, $parent$ is the node whose child pointer is changed by the update (which is the parent of $n$ after the update), and $root$ refers to the root of the \rbslack\ (not the sentinel node $entry$).}
Violations that a process $P$ becomes responsible for after performing an update $U$.
Here, $n$ is the topmost node shown on the right-hand side of $U$'s diagram in Figure~\ref{fig-bslack-updates}, $parent$ is the node whose child pointer is changed by $U$ (i.e., the parent of $n$ after $U$), and $root$ is the root of the \rbslack\ (not the sentinel node $entry$).}
\label{fig-bslack-violation-movement}
\end{figure}

\subsection{The rebalancing algorithm} \label{bslack-conc-rebalancing}
Our algorithm for rebalancing a \rbslack\ is considerably different from the algorithm for rebalancing the relaxed $(a,b)$-tree.
In the relaxed $(a,b)$-tree, each insertion or deletion can increase the number of violations in the tree by at most one.
Whenever a process creates a new violation, it takes responsibility for that violation, and performs rebalancing steps to fix it.
Each rebalancing step that the process performs will either fix the violation (eliminating it and decreasing the number of violations in the tree), or move the violation from a node to its parent.
Since a violation can move only from a node to its parent, a process can always find the violation it is responsible for by searching for the key it inserted or deleted to create the violation.
Thus, a process performs rebalancing by repeatedly searching for a key, and performing rebalancing steps to fix any violations it sees, until it performs a search and sees no violations.
%In the relaxed $(a,b)$-tree, each time a process created a violation $x$, it remembers the $key$ that it inserted or deleted to create $x$.
%The process then performs rebalancing by repeatedly searching from the root for $key$, fixing any violations it encounters, until it no longer sees any violations.
%This works because relaxed $(a,b)$-trees satisfy the following property: whenever the violation $x$ is moved from one node to another by an update, it remains on the search path to $key$.
%
However, in a \rbslack, a rebalancing step can create several new violations, which are not necessarily on the search path to any one key.
Therefore, processes cannot simply search for a single key, fixing any violations they see.

We think of a violation as a pair $\langle type, node \rangle$, where $type$ is \textit{slack}, \textit{degree} or \textit{weight}, and $node$ is a node \textit{in the tree} where the violation occurs.
Consider any update $U$ to a \rbslack\ (see Figure~\ref{fig-bslack-updates}).
Observe that $U$ changes a child pointer of a node $parent$, removing a connected set $R$ of nodes from the tree and inserting a connected set $N$ of nodes into the tree.
Let $\beta$ be the set of violations in the tree before $U$, and $\beta'$ be the set of violations in the tree after $U$.
The process that performs $U$ takes responsibility for all violations in $\beta' \setminus \beta$.
All of the violations at nodes in $R$ are no longer in the tree after $U$, and any violations at nodes in $N$ are in the tree after $U$, but were not in the tree before $U$.
Additionally, $U$ may create a slack violation at $parent$.
Furthermore, observe that $U$ can create violations only at nodes in $N \cup \{parent\}$, and can create only a slack violation at $parent$.
Therefore, $\beta' \setminus \beta$ is the set of violations at nodes in $N$, as well as a possible slack violation at $parent$.
(Technically, in our implementation, a process if $P$ performs an update that \textit{could} create a slack violation at $parent$, then it will take responsibility for any slack violation at $parent$, even if that violation was in $\beta$. So, $P$ \textit{actually} takes responsibility for $(\beta' \setminus \beta) \cup \{\mbox{any slack violation at $parent$ just after $U$}\}$.)
In Figure~\ref{fig-bslack-violation-movement}, we precisely characterize the set of violations that a process takes responsibility for after each update to a \rbslack.
The information in this figure was obtained in a straightforward way by inspecting the updates to \rbslack s in Figure~\ref{fig-bslack-updates}.

%Each time a process performs an update that removes a connected set $R$ of nodes from the tree and inserts a connected set $N$ of nodes into the tree, it takes responsibility for \textit{all} violations that occur at nodes in $N$, as well as any new slack violation that it creates at the parent of the topmost node in $N$.
Note that, after $U$ removes the nodes in $R$ from the tree, %Consequently, 
any processes that were previously responsible for violations occurring at nodes in $R$ are no longer responsible for them. %once a node is removed from the tree, any processes that were responsible for violations occurring at that node are no longer responsible for them.
This approach turns out to be fairly practical.
If a process is trying to fix a violation at a node $u$, and it fails to perform a rebalancing step because $u$ has been removed from the tree, then it can simply proceed as if it had successfully fixed the violation (since the violation has either been eliminated, or another process is now responsible for it).

Consistent with Figure~\ref{fig-bslack-violation-movement}, after a process performs an Overflow update in \ins, it invokes \func{FixWeight} to fix the weight violation at the topmost node $n$ in $N$, and then invokes \func{FixSlack} to fix any possible slack violation at the parent $p$ of $n$.
Similarly, after a process performs a successful \del, it invokes \func{FixSlack} to fix any possible slack violation at $p$.

\subsection{Fixing weight violations}

\begin{figure}[ph]
\vspace{-3mm}
\begin{framed}
\preplisting
\begin{lstlisting}[mathescape=true]
 //\func{FixWeight}$(node)$
   if $node.weight = 1$ then return //\label{lfbslack-fixweight-noweightviol}
   //\com Assert: node is internal (because leaves have weight one), and is not entry or root (because both have weight one)
   //\com Optimistically check to see if $node$ was already removed (so we are no longer responsible for its violations)
   if $\llt(node) = \finalized$ then return //\label{lfbslack-fixweight-finalized}
   loop
     //\com \textbf{Start of operation attempt}
     //\com Search for $node$ and fix any weight violation at $node$
     $result := \func{SearchNode}(node)$
     //\com If $result = \fail$, then another update removed $node$, so we are no longer responsible for its violations.
     if $result = \fail$ then return else $\langle gp, ix_p, p, ix_l, l \rangle := result$ //\label{lfbslack-fixweight-searchnode-fail}

     //\com We cannot fix a weight violation at $l$ if there is a weight violation at $p$, so we first check $p$
     if $p.weight = 0$ then
       //\com \textbf{End of operation attempt}
       $\func{FixWeight}(p)$ //\label{lfbslack-fixweight-exception1}
       continue //\medcom go to next iteration (to retry)
        
     //\com Template iteration 0 (grandparent of $node$)
     $result_{gp} := \llt(gp)$ //\label{lfbslack-fixweight-templateafterhere}
     if $result_{gp} \in \{\fail, \finalized\}$ then continue //\medcom goto next iteration (to retry)
     if $result_{gp}.p_{ix_p} \neq p$ then continue //\medcom{\func{Conflict}: verify $gp$ still points to $p$}
     //\com Template iteration 1 (parent of $node$)
     $result_p := \llt(p)$
     if $result_p \in \{\fail, \finalized\}$ then continue //\medcom goto next iteration (to retry)
     if $result_p.p_{ix_l} \neq l$ then continue //\medcom{\func{Conflict}: verify $p$ still points to $l$}
     //\com Template iteration 2 ($node$)
     $result_l := \llt(l)$
     if $result_l \in \{\fail, \finalized\}$ then continue //\medcom goto next iteration (to retry)

     //\com Computing \sct-\func{Arguments} from locally stored values (and immutable fields)
     $V := \langle gp, p, l \rangle$; $R := \langle p, l \rangle$; $fld := \mbox{ a pointer to } gp.p_{ix_p}$
     if $p.d + l.d - 1 \le b$ //\medcom{If the contents of $l$ and $p$ fit in a single node}
       //$n :=$ pointer to a new internal node configured as in Absorb in Figure~\ref{fig-bslack-updates}. The $searchKey$ of the new node is its first key.
       $success := \sct(V, R, fld, n)$ //\label{lfbslack-fixweight-absorb-scx}
       //\com \textbf{End of operation attempt}
       if $success$ then
         $\func{FixSlack}(n)$
         return //\label{lfbslack-fixweight-did-absorb}
     else //\medcom{The contents of $l$ and $p$ cannot fit in a single node}
       //$n :=$ pointer to a subtree of three newly created nodes, configured as in Split in Figure~\ref{fig-bslack-updates}. The $searchKey$ of each new node is its first key.
       //\com If $n$ will replace $root$, then we also perform Root-Zero as part of the same atomic update
       if $gp = entry$ then $n.weight = 1$ //\label{lfbslack-fixweight-rootzero}
       $success := \sct(V, R, fld, n)$ //\label{lfbslack-fixweight-split-scx}
       //\com \textbf{End of operation attempt}
       if $success$ then
         if $gp \neq entry$ then $\func{FixWeight}(n)$ //\label{lfbslack-fixweight-fixweight1}
         for each $x \in \{n, left, right, gp\}$ do $\func{FixSlack}(x)$//\label{lfbslack-fixweight-fixslack-parent}
         return //\label{lfbslack-fixweight-did-split}
\end{lstlisting}
\end{framed}
\vspace{-5mm}
	\caption{Pseudocode for \func{FixWeight}. Here, $b$ is the maximum degree of nodes.}
	\label{code-bslack-fixweight}
\end{figure}

Pseudocode for \func{FixWeight} appears in Figure~\ref{code-bslack-fixweight}.
The procedure takes a single argument, $node$, which points to a node that is known (or suspected) to contain a weight violation.
At a high level, \func{FixWeight} repeatedly: locates $node$ in the tree, determines whether it contains a weight violation, and attempts to fix the weight violation.
It continues to do this until it either successfully fixes the weight violation, or $node$ is removed by another process (so this process is no longer responsible for fixing any violations at $node$).

We now describe \func{FixWeight} in detail.
An invocation $I$ of \func{FixWeight} begins by checking whether $node$ has weight zero.
If not, then there is no weight violation at $node$, so $I$ simply returns.
So, suppose $node$ has weight zero (meaning there is a weight violation at $node$).

The next step is an (optional) optimization that avoids searching for $node$ if it has already been removed from the tree. %$node$ checks whether $node$ has already been removed from the tree by another process (so that we are no longer responsible for fixing any violations at $node$).
This optimization entails invoking $\llt(node)$ and checking whether $node$ is \finalized.
It is easy to verify that our lock-free \rbslack\ implementation satisfies the following property: a node is finalized precisely when it is removed from the tree.
(This is equivalent to saying that every node removed from the tree by an invocation of \sct$(V,R,fld,new)$ appears in $R$.)
%Consequently, if a process performs an invocation of $\llt(node)$ that returns \finalized, then $node$ has been removed from the tree, and the process is no longer responsible for fixing any violations at $node$.
So, if $\llt(node)$ returns \finalized, then $I$ can simply return.

Suppose the invocation of $\llt(node)$ does not return \finalized.
Then, $I$ enters a loop wherein it will repeatedly attempt to locate and fix a violation at $node$ until either it succeeds, or $node$ is removed from the tree by another process.
Each iteration of this loop follows the tree update template (described in Chapter~\ref{chap-template}) until just before it invokes \func{FixWeight} or \func{FixSlack} (described in Section~\ref{sec-fixdegreeorslack}).

The first step in the loop is to search for $node$ by invoking a procedure called \func{SearchNode}, which appears in Figure~\ref{code-bslack-optimistic-nofixneeded}.
\func{SearchNode} is similar to \func{Search}, except that (1) it uses the $searchKey$ of $node$ to locate it in the tree, (2) the search stops as soon as it encounters $node$, and (3) it returns \fail\ if it does not encounter $node$.
\func{SearchNode} either returns \fail, or $\langle gp, ix_p, p, ix_{node}, node \rangle$ such that, during the search, $p$ was read from $gp.p_{ix_p}$ and $node$ was subsequently read from $p.p_{ix_{node}}$.
It is straightforward to prove the following using the techniques in Chapter~\ref{chap-abtree}.
If \func{SearchNode} returns \fail, then $node$ was removed from the tree at some point before \func{SearchNode} terminated.
%Otherwise, $node$ was in the tree at some point during the search.
Thus, if $I$'s invocation of \func{SearchNode}$(node)$ returns \fail, then the process executing $I$ is no longer responsible for any violations at $node$, so $I$ can simply return.
So, suppose $I$'s invocation of \func{SearchNode}$(node)$ does not return \fail.

\fakeparagraph{Fixing any weight violation at $p$ first}
Then, $I$ will attempt to fix the weight violation at $node$ by performing Absorb or Split.
However, neither of these updates can be performed if the parent $p$ of $node$ has weight zero.
So, if $p$ has weight zero, $I$ invokes \func{FixWeight}$(p)$ to fix it and then skips to the next iteration of the loop (to retry fixing the weight violation at $node$).

Note that $I$ \textit{does not follow the template} if it performs this invocation of \func{FixWeight}$(p)$.
We briefly explain why this is the case.
$I$ does not invoke \sct, so if it were to follow the template, its final \func{UpdateNotNeeded} procedure would have to return \true.
If \func{UpdateNotNeeded} were to return \true, then the template operation would have to be successful and return \func{Result}.
However, the template operation will continue to the next iteration of the loop and perform another attempt after invoking \func{FixWeight} (which conceptually corresponds to the template operation returning \fail\ and trying again).
In contrast, when $I$ invokes \func{FixWeight} at line~\ref{lfbslack-fixweight-fixweight1}, it does so after performing a successful \sct, and it subsequently returns successfully.
(So, in that case, we can think of the template operation as having terminated successfully before invoking \func{FixWeight}.)
As we will see, deviating from the template adds an additional step to the progress proof.

\fakeparagraph{Performing \llt s}
Suppose $p$ has weight one.
Then, $I$ invokes \llt$(gp)$.
Conceptually, this step represents the beginning of iteration 0 of the tree update template.
After invoking \llt$(gp)$, $I$ verifies that the $gp$ still points to $p$.
This step conceptually represents the \func{Conflict} procedure in the template.
If the \llt\ returns \fail\ or \finalized, or if $gp$ does not point to $p$, then $I$ skips to the next iteration of the loop (to retry).
So, suppose $I$ does not skip to the next iteration.

Then, $I$ invokes \llt$(p)$.
This represents the beginning of iteration 1 of the template.
Next, $I$ verifies that $p$ still points to $node$ (as part of the template's \func{Conflict} procedure).
If the \llt\ returns \fail\ or \finalized, or if $p$ does not point to $l$, then $I$ skips to the next iteration of the loop.
So, suppose $I$ does not skip to the next iteration.

Finally, $I$ invokes \llt$(l)$ (where $l = node$).
This represents iteration 2 of the template.
If the \llt\ returns \fail\ or \finalized\, then $I$ skips to the next iteration of the loop.
So, suppose it does not.

\fakeparagraph{Computing \sct-\func{Arguments}, performing \sct\ and rebalancing}
Next, $I$ computes the arguments $V$, $R$, $fld$ and $n$ for an invocation of \sct$(V, R, fld, n)$ by using locally stored values and immutable fields of nodes.
Both Absorb and Split change a pointer of $gp$ (at index $ix_p$) from $p$ to point to some new node $n$, removing $p$ and $l$ from the tree.
Since $p$ and $l$ are removed, $R = \langle p, l \rangle$.
Since $gp$ is changed, it must be in $V$.
Since $R$ is a subsequence of $V$, $V = \langle gp, p, l \rangle$.
In order to compute $n$, $I$ must decide whether it should perform Absorb or Split.
So, it checks whether the contents of $l$ and $p$ can fit in a single node.

Suppose the contents of $l$ and $p$ can fit in a single node.
Then, $I$ attempts to perform Absorb by creating a single new internal node $n$ (configured as shown in Absorb in Figure~\ref{fig-bslack-updates}), and invoking \sct.
(The node $l$ must be internal, since there was a weight violation at $l$, and weight violations cannot occur at leaves.
Thus, the new node created by this update must be internal.)
The $searchKey$ of $n$ is simply its first key.
If the invocation of \sct\ returns \true, then, in accordance with Figure~\ref{fig-bslack-violation-movement}, $I$ invokes \func{FixSlack} to fix any slack violation at $n$.

Now, suppose the contents of $l$ and $p$ cannot fit in a single node.
Then, $I$ attempts to perform Split by creating a subtree of three new nodes (configured as shown in Split in Figure~\ref{fig-bslack-updates}) rooted at $n$, and invoking \sct.
As we argued in the previous case, $l$ must be internal, so the new nodes must all be internal, as well.
The $searchKey$ of each new node is its first key.
If $gp = entry$ (and the \sct\ is successful), then the \sct\ will replace the $root$ of the \rbslack\ with $n$.
In this case, $I$ also performs Root-Zero as part of the same atomic update (by setting $n.weight = 1$ before invoking \sct).
If the invocation of \sct\ returns \true, then, in accordance with Figure~\ref{fig-bslack-violation-movement}, $I$ invokes \func{FixWeight} to fix a weight violation at $n$ (unless it eliminated the violation by performing Root-Zero as part of the \sct), and then invokes \func{FixSlack} to fix possible slack violations at $gp$ and each new node.
%(Although the row for Root-Zero in Figure~\ref{fig-bslack-violation-movement} suggests that $I$ should also try to fix a possible degree violation at $n$ if it performed Root-Zero, there can be a degree violation at $n$ only if there was previously a degree violation at $root$.
%As we discuss in the next section, we implement \func{FixSlack} in such a way that degree violations can never occur at $root$.
%Thus, $I$ need not worry about degree violations at $n$.)

\subsection{Fixing degree and slack violations} \label{sec-fixdegreeorslack}

Pseudocode for \func{FixSlack} appears in Figure~\ref{code-bslack-fixdegreeorslack}.
The procedure takes a single argument, $node$, which points to a node that is suspected to contain a degree or slack violation.
\func{FixSlack} is quite similar to \func{FixWeight}.
It repeatedly: locates $node$ in the tree, determines whether it contains a degree or slack violation, and attempts to fix the violation.
It continues to do this until it either successfully fixes the violation, or $node$ is removed by another process.

We give a detailed description of \func{FixSlack}.
%% warning: manual word break
An invocation $I$ of \func{FixSlack} begins by invoking a procedure called \func{NoFixNeeded}, which appears in Figure~\ref{code-bslack-optimistic-nofixneeded}.
If \func{NoFixNeeded} returns \true, then $I$ simply returns.
This is an (optional) optimization that avoids searching for $node$ if it can determine there is no violation to fix.
\func{NoFixNeeded} is described in detail in Section~\ref{sec-nofixneeded}.
Suppose \func{NoFixNeeded} returns \false.

Then, $I$ enters a loop wherein it will repeatedly attempt to locate and fix a degree or slack violation at $node$ (using Compress or One-Child) until either it succeeds, or $node$ is removed from the tree by another process.
Each iteration of this loop follows the tree update template (described in Chapter~\ref{chap-template}) until just before it invokes \func{FixWeight} or \func{FixSlack}.

As in \func{FixWeight}, the first step in the loop is to search for $node$ by invoking \func{SearchNode}.
Recall that \func{SearchNode} either returns \fail, or $\langle gp, ix_p, p, ix_{node}, node \rangle$ such that, during the search, $p$ was read from $gp.p_{ix_p}$ and $node$ was subsequently read from $p.p_{ix_{node}}$.
If it returns \fail, then the process executing $I$ is no longer responsible for any violations at $node$, so $I$ can simply return.
So, suppose \func{SearchNode} does not return \fail.
Then, the local variable $l$ is the same as $node$.

\begin{figure}[tbh]
\begin{framed}
\preplisting
\begin{lstlisting}[mathescape=true]
 //\func{FixSlack}$(node)$
   if $\func{NoFixNeeded}(node)$ then return //\medcom Optimistically check if this procedure is needed \label{lfbslack-fixslack-nofixneeded}
   loop
     //\com \textbf{Start of operation attempt}
     //\com Search for $node$ and fix any degree or weight violation at $node$
     $result := \func{SearchNode}(node)$
     //\com If $result = \fail$, then another update removed $node$, so we are no longer responsible for its violations.
     if $result = \fail$ then return else $\langle gp, ix_p, p, ix_l, l \rangle := result$ //\label{lfbslack-fixslack-searchnode-fail}

     //\com Determine whether Root-Replace should be performed
     if $gp = \nil$ and $node.d = 1$ then //\medcom Degree violation at $node = root$
       $result := \func{DoRootReplace}(p, ix_l, l)$
       //\com \textbf{End of operation attempt}
       if $result = \func{Retry}$ then continue //\medcom goto next iteration (to retry)
       for each $x \in result$ do $\func{FixSlack}(x)$
       return //\label{lfbslack-fixslack-did-rootreplace}

     if $node.d > 1$ //\com If there is no degree violation at $node$, then we are trying to perform Compress \label{lfbslack-fixslack-nodegree}
       //Take one extra step in the search, so that $p$ becomes the topmost node in the Compress diagram in Figure~\ref{fig-bslack-updates} \label{lfbslack-fixslack-extrastep}

     //\com Note: If there is a degree violation at $node$, then $l = node$. Otherwise, $p = node$.
     $result := \func{DoCompress}(gp, ix_p, p, ix_l, l)$
     //\com \textbf{End of operation attempt}
     if $result = \func{Retry}$ then continue //\medcom goto next iteration (to retry)
     for each $x \in result$ do $\func{FixSlack}(x)$ //\label{lfbslack-fixslack-fixdegreeorslack2}
     return //\label{lfbslack-fixslack-did-compress}
\end{lstlisting}
\end{framed}
\vspace{-5mm}
	\caption{Pseudocode for \func{FixSlack}.}
	\label{code-bslack-fixdegreeorslack}
\end{figure}

\begin{figure}[ph]
\vspace{-5mm}
\begin{framed}
\preplisting
\begin{lstlisting}[mathescape=true]
 //\func{DoRootReplace}$(p, ix_l, l)$
   //\com Template iteration 0 (parent of $node$)
   $result_p := \llt(p)$
   if $result_p \in \{\fail, \finalized\}$ then return $\func{Retry}$
   if $result_p.p_{ix_l} \neq l$ then return $\func{Retry}$ //\medcom{\func{Conflict}: verify $p$ still points to $l$}

   //\com Template iteration 1 ($node$)
   $result_l := \llt(l)$
   if $result_l \in \{\fail, \finalized\}$ then return $\func{Retry}$
   //Let $c$ be the single child pointer in $result_l$

   //\com Computing \sct-\func{Arguments} from locally stored values (and immutable fields)
   $V := \langle p, l \rangle$; $R := \langle l \rangle$; $fld := p.p_{ix_l}$
   //$n :=$ a pointer to a newly created copy of $c$ with $weight$ one
   //\com Note: if $c.weight = 0$ then this update also eliminates a weight violation
   if $\sct(V, R, fld, n)$ then return $n$ //\label{lfbslack-dorootreplace-scx}
   return $\func{Retry}$

 //\func{DoCompress}$(gp, ix_p, p, ix_l, l)$
   //\com Template iteration 0 (grandparent of $node$) %    //\com Note: $gp \neq \bot$, since \func{FixSlack}$(node)$ is not invoked when $node$ is a child of $entry$

   $result_{gp} := \llt(gp)$
   if $result_{gp} \in \{\fail, \finalized\}$ then return $\func{Retry}$
   if $result_{gp}.p_{ix_p} \neq p$ then return $\func{Retry}$ //\medcom{\func{Conflict}: verify $gp$ still points to $p$}

   //\com Template iteration 1 (parent of $node$)
   $result_p := \llt(p)$
   if $result_p \in \{\fail, \finalized\}$ then return $\func{Retry}$
   if $result_p.p_{ix_l} \neq l$ then return $\func{Retry}$ //\medcom{\func{Conflict}: verify $p$ still points to $l$}

   //\com Perform $\llt$s on the nodes in $result_p$
   $failed := \false$
   for $i = 1..p.d$
     //\com Template iteration $i+1$ (child $i$ of $p$)
     if $\llt(result_p.p_i) \in \{\fail, \finalized\}$ then return $\func{Retry}$

   //\com Before fixing a degree or slack violation at $node$, we must fix any weight violations at $p$ or its children
   if $\func{FixAllWeightViolations}(p, result_p)$ then return $\func{Retry}$ //\com Retry if we fixed a weight violation //\label{lfbslack-fixslack-exception1}

   //\com Determine whether there is a slack violation at $node = p$, or a degree violation at $node = l$
   $pGrandDegree := 0$
   for $i = 1..p.d$ do $pGrandDegree := pGrandDegree + result_p.p_i.d$
   $slack = p.d * b - pGrandDegree$ //\medcom Total slack shared amongst the nodes of $result_p$
   if $slack < b$ and $node.d > 1$ then return $\emptyset$ //\medcom{\func{UpdateNotNeeded}: no violation at $node$} \label{lfbslack-fixslack-change}

   //\com Computing \sct-\func{Arguments} from locally stored values (and immutable fields)
   $V := \langle gp, p, result_p.p_1, result_p.p_2, ... \rangle$; $R := \langle p, result_p.p_1, result_p.p_2, ... \rangle$; $fld := gp.p_{ix_p}$
   $\langle doRootReplace, n, nChildren \rangle := \func{CreateCompressedNodes}(gp, p, result_p, pGrandDegree)$
   if $\sct(V, R, fld, n)$ then //\label{lfbslack-docompress-scx}
     if $doRootReplace$ then return $\{n\}$ //\medcom Did Compress/One-Child \textbf{and} Root-Replace
     else return $\{gp, n\} \cup nChildren$ //\medcom Did Compress/One-Child\label{lfbslack-docompress-return-parent}
   return $\func{Retry}$
\end{lstlisting}
\end{framed}
\vspace{-5mm}
	\caption{Pseudocode for \func{DoCompress} and \func{DoRootReplace}. Here, $b$ is the maximum degree of nodes.}
	\label{code-bslack-docompress-and-dorootreplace}
\end{figure}

\begin{figure}[ph]
\begin{framed}
\preplisting
\begin{lstlisting}[mathescape=true]
 //\func{SearchNode}$(node)$
   //\com If $node$ is in the tree, this returns $\langle gp, ix_p, p, ix_{node}, node \rangle$ such that, during the search, $p$ was read from $gp.p_{ix_p}$ and $node$ was subsequently read from $p.p_{ix_{node}}$. Otherwise, this returns \fail.
   $key := node.searchKey$ //\medcom We locate $node$ using $node.searchKey$
   $gp := \bot$; $p := entry$; $l := p.p_1$ //\medcom Save the last three nodes encountered \label{lfbslack-searchnode-start}
   $ix_p := 0$; $ix_l := 1$ //\medcom{Invariant: $l$ was read from $p.p_{ix_l}$}
   while $l \neq node$ and not $l.leaf$ //\label{lfbslack-searchnode-exitloop}
     $ix_p := ix_l$
     while $ix_l < l.d$ and $key \ge l.k_{ix_l}$ //\medcom Locate appropriate child pointer to follow
       $ix_l := ix_l + 1$
     $gp := p$; $p := l$; $l := l.p_{ix_l}$ //\medcom Follow the child pointer \label{lfbslack-searchnode-end}

   //\com Check whether we reached $node$
   if $l \neq node$ then
     return $\fail$ //\com Another update removed $l$, so we are no longer responsible for any violation at $l$.
   return $\langle gp, ix_p, p, ix_l, l \rangle$ //\vspace{2mm}\hrule\vspace{2mm} %

 //\func{NoFixNeeded}$(node)$
   if $node.leaf$ return $\true$ //\medcom Degree and slack violations cannot occur at leaves \label{lfbslack-nofixneeded-leaf}

   //\com Optimistically check to see if $node$ was already removed (so we are no longer responsible for its violations)
   $result := \llt(node)$
   if $result = \finalized$ then return $\true$ //\label{lfbslack-nofixneeded-finalized}
   if $result = \fail$ then return $\false$ //\medcom Cannot determine whether there is a violation
   if $node.weight = 0$ then return $\false$ //\medcom Weight violations must be fixed before others
   if $node.d = 1$ then return $\false$ //\medcom Found a degree violation at $node$

   //\com Use the snapshot to (try to) determine whether there is a slack violation at $node$
   $slack := 0$; $numLeaves := 0$
   for $i = 1 .. node.d$
     $slack := slack + (b - result.p_i.d)$
     if $result.p_i.weight = 0$ then return $\false$ //\medcom Weight violations must be fixed before others
     if $slack \ge b$ then return $\false$ //\medcom Possible slack violation at $node$ \label{lfbslack-nofixneeded-change}
   return $\true$ //\medcom{No violation to fix} \label{lfbslack-nofixneeded-nodegreeorslack} \vspace{2mm}\hrule\vspace{2mm} %

 //\func{FixAllWeightViolations}$(p, result_p)$
   for $i = 1 .. p.d$
     if $result_p.p_i.weight = 0$ then
       //\com \textbf{End of operation attempt}
       $foundWeightViolation := \true$
       $\func{FixWeight}(result_p.p_i)$ //\label{lfbslack-fixallweight-fixweight1}
   if $p.weight = 0$ then
     //\com \textbf{End of operation attempt}
     $foundWeightViolation := \true$
     $\func{FixWeight}(p)$ //\label{lfbslack-fixallweight-fixweight2}
   return $foundWeightViolation$ //\medcom If we fixed a weight violation, retry the search
\end{lstlisting}
\end{framed}
\vspace{-5mm}
	\caption{Pseudocode for \func{SearchNode}, \func{NoFixNeeded} and \func{FixAllWeightViolations}. Here, $b$ is the maximum degree of nodes.}
	\label{code-bslack-optimistic-nofixneeded}
\end{figure}

\begin{figure}[tbh]
\begin{framed}
\preplisting
\begin{lstlisting}[mathescape=true]
 //\func{CreateCompressedNodes}$(gp, p, result_p, pGrandDegree)$
   //\com Determine how to divide keys and values as evenly as possible, into as few new nodes as possible
   $numNewChildren := \lceil pGrandDegree / b \rceil$ //\label{lfbslack-createcompressednodes-change}
   $ceilNodes := pGrandDegree\mbox{ \textit{mod} }numNewChildren$
   $floorNodes := numNewChildren - ceilNodes$
   $ceilDegree := \lceil pGrandDegree / numNewChildren \rceil$
   $floorDegree := \lfloor pGrandDegree / numNewChildren \rfloor$
   
   //\com Note: since there are no weight violations at any nodes in $result_p$, they are either all leaves or all internal
   $pChildrenAreLeaves := (result_p.p_1.leaf)$ //\com Determine which is the case

   //\com Create new node(s)
   if $gp = entry$ and $numNewChildren = 1$ then //\label{lfbslack-createcompressednodes-rootreplace-start}
     //\com If we would replace $root$ by a node with degree one, then we perform Root-Replace as well
     //$n :=$ pointer to a newly created node that contains all of the keys and pointers of the nodes in $result_p$. This node is a leaf if $pChildrenAreLeaves = \true$, and is internal otherwise. The $searchKey$ of $n$ is the same as the $searchKey$ of $p$ (since $n$ will replace $p$). \label{lfbslack-createcompressednodes-rootreplace-createnodes}
     return $\langle \true, n, \emptyset \rangle$ //\label{lfbslack-createcompressednodes-rootreplace-end}
   else
     //$n :=$ pointer to a subtree consisting of a newly created parent and $numNewChildren$ newly created children, configured as in Compress in Figure~\ref{fig-bslack-updates}. The keys and pointers of the nodes in $result_p$ are divided such that $ceilNodes$ of the new children have degree $ceilDegree$, and $floorNodes$ of the new children have degree $floorDegree$. These children are leaves if $pChildrenAreLeaves = \true$, and are internal otherwise. If $numNewChildren = 1$, then the $searchKey$ of each new node is the same as the $searchKey$ of $p$, otherwise the $searchKey$ of each node is its first key. \label{lfbslack-createcompressednodes-compress-createnodes}
     return $\langle \false, n, \{\mbox{children of }n\} \rangle$
\end{lstlisting}
\end{framed}
	\caption{Pseudocode for \func{CreateCompressedNodes}. Here, $b$ is the maximum degree of nodes.}
	\label{code-bslack-createcompressnodes}
\end{figure}

\subsubsection{Using Root-Replace to fix a degree violation at the root}

Next, $I$ checks whether $gp = \nil$ and $node.d = 1$.
If so, then $p = entry$ and $node$ is the $root$.
So, $I$ will attempt to perform Root-Replace to eliminate the degree violation.
It does this by performing an invocation $I'$ of \func{DoRootReplace}$(p, ix_l, l)$.
The pseudocode for \func{DoRootReplace} appears in Figure~\ref{code-bslack-docompress-and-dorootreplace}.

\fakeparagraph{Performing \llt s}
$I'$ begins by invoking \llt$(p)$.
Conceptually, this step represents the beginning of iteration 0 of the tree update template.
After invoking \llt$(p)$, $I'$ verifies that the $p$ still points to $l$.
This step conceptually represents the \func{Conflict} procedure in the template.
If the \llt\ returns \fail\ or \finalized, or if $p$ does not point to $l$, then $I'$ returns \func{Retry}, which causes $I$ to skip to the next iteration of loop (to retry).
So, suppose $I'$ does not return \func{Retry}.

Then, $I'$ invokes \llt$(l)$.
This represents the beginning of iteration 1 of the template.
If the \llt\ returns \fail\ or \finalized, then $I'$ return \func{Retry}.
So, suppose $I'$ does not return \func{Retry}.

\fakeparagraph{Computing \sct-\func{Arguments}}
Next, $I'$ computes the arguments $V$, $R$, $fld$ and $n$ for an invocation of \sct$(V, R, fld, n)$ by using locally stored values and immutable fields of nodes.
Root-Replace changes a pointer of $p$ (at index $ix_l$) from $l$ (which is $root$) to point to some new node $n$, removing $l$ from the tree.
Since $l$ is removed, $R = \langle l \rangle$.
Since $p$ is changed, it must be in $V$.
Since $R$ is a subsequence of $V$, $V = \langle p, l \rangle$.
Recall that the degree of $l = node$ was seen to be one before \func{DoRootReplace} was invoked by $I$.
Since the degree of a node never changes, $l$ still has degree one.
Let $c$ be the single child of $l$ that was returned by the invocation of $\llt(l)$ performed by $I'$.
$I'$ creates $n$ by copying $c$ and setting the $weight$ of the new copy to one.
If there was a weight violation at $c$, then this update eliminates it.

\fakeparagraph{Performing \sct\ and subsequent rebalancing}
Next, $I'$ invokes \sct$(V, R, fld, n)$ to replace $l$ with $n$.
If the invocation of \sct\ returns \false, then $I'$ simply returns \func{Retry}.
So, suppose the \sct\ returns \true.
Then, according to Figure~\ref{fig-bslack-violation-movement}, the process performing $I'$ may now be responsible for a degree or slack violation at $n$.
So, $I'$ returns $n$ to $I$.
$I$ then invokes \func{FixSlack}$(n)$ to fix any degree or slack violation at $n$.

\subsubsection{Using Compress to fix both degree and slack violations}

Now, suppose $gp \neq \nil$ or $node.d \neq 1$, so $I$ will not perform Root-Replace.
Instead, it will attempt to perform Compress or One-Child.
Although we conceptually view Compress and One-Child as different updates, we can actually use Compress instead of One-Child to fix degree violations.
According to Figure~\ref{fig-bslack-updates}, Compress applies when $u$ has $k \ge 2$ children with total degree $c \le kb-b$ (so there is a slack violation at $u$).
Suppose we follow the tree update template to implement Compress by replacing $u$ and its children with a new internal node and $\lceil c/b \rceil$ new children (where the keys and pointers in the children of $u$ before the update are evenly distributed amongst the new children after the update).
According to Figure~\ref{fig-bslack-updates}, One-Child applies when $\pi(u)$ has $k \ge 2$ children with total degree $c > kb-b$ (so there is a degree violation at $u$, but no slack violation at $\pi(u)$).
Each of the $k$ children of $\pi(u)$ can have degree at most $b$, so $kb \ge c$.
Since $kb \ge c > kb-b$, we have $\lceil c/b \rceil = k$.
So, whenever the preconditions for One-Child are satisfied, if we simply pretend there is a slack violation at $\pi(u)$, and perform Compress (as if we were going to fix that violation), then we will create a new internal node and $\lceil c/b \rceil = k$ new children (which evenly share the keys and pointers that were contained in the children of $\pi(u)$ before the update).
This is exactly what One-Child does.

Observe that a Compress update where $\pi(u)$ is the topmost node replaced by the update will fix either a slack violation at $\pi(u)$ or a degree violation at $u$.
So, if the violation to be fixed at $node$ is a degree violation, then we fix it by performing Compress where the \textit{parent} of $node$ is the topmost node replaced by the update.
In the pseudocode, $l = node$, and $p$ is its parent.
So, we fix the violation by performing Compress where $p$ is the topmost node replaced by the update.
However, if there is no degree violation at $node$, then the violation we are interested in fixing at $node$ must be a slack violation.
In this case, we fix the violation by performing Compress where $node$ (instead of its parent) is the topmost node replaced by the update.
In order to facilitate the use of the same code for both cases, if the violation is a slack violation, then $I$ takes one more step in the search.
After this extra step, the variable $p$ points to $node$, and $l$ points to some child of $node$.
So, as in the previous case, we fix the violation by %So, as in the case of a degree violation,
performing Compress where $p$ is the topmost node replaced by the update.

$I$ performs the necessary Compress update by performing an invocation $I'$ of \func{DoCompress}$(gp, ix_p, p, ix_l, l)$.
Pseudocode for \func{DoCompress} appears in Figure~\ref{code-bslack-docompress-and-dorootreplace}.

\fakeparagraph{Performing \llt s}
$I'$ begins by invoking \llt$(gp)$.
Conceptually, this step represents the beginning of iteration 0 of the tree update template.
After invoking \llt$(gp)$, $I'$ verifies that the $gp$ still points to $p$.
This step conceptually represents the \func{Conflict} procedure in the template.
If the \llt\ returns \fail\ or \finalized, or if $gp$ does not point to $p$, then $I'$ returns \func{Retry} (which causes $I$ to skips to the next iteration of its loop and retry).
So, suppose $I'$ does not return \func{Retry}.

Then, $I'$ invokes \llt$(p)$.
This represents the beginning of iteration 1 of the template.
Next, $I'$ verifies that $p$ still points to $l$ (as part of the template's \func{Conflict} procedure).
If the \llt\ returns \fail\ or \finalized, or if $p$ does not point to $l$, then $I'$ returns \func{Retry}.
So, suppose $I'$ does not return \func{Retry}.

$I'$ then performs \llt s on each node pointed to by the snapshot that was returned by its \llt$(p)$.
These steps correspond to template iterations 2 through $p.d+1$, where $p.d$ is the degree of $p$.
If any of these invocations of \llt\ return \fail\ or \finalized, then $I'$ returns \func{Retry}.
So, suppose $I'$ does not return \func{Retry}.

\fakeparagraph{Fixing nearby weight violations first}
Compress cannot be performed if there are any weight violations at $p$ or any of its children.
So, $I'$ invokes \func{FixAllWeightViolations}$(p, result_p)$ to fix any weight violations at these nodes before proceeding (see Figure~\ref{code-bslack-optimistic-nofixneeded}).
\func{FixAllWeightViolations} checks for weight violations at $p$ or the nodes in $result_p$, and invokes \func{FixWeight} to fix each violation it finds.
\func{FixAllWeightViolations} returns \true\ if it performs any invocation of \func{FixWeight} and \false\ otherwise.

Note that $I$ \textit{does not follow the template} if this invocation of \func{FixAllWeightViolations} performs an invocation of \func{FixWeight}.
As we will see, deviating from the template adds an additional step to the progress proof.
If the invocation of \func{FixAllWeightViolations}$(p, result_p)$ returns \true, then $I'$ returns \func{Retry}.
So, suppose $I'$ does not return \func{Retry}.

\fakeparagraph{Computing \func{UpdateNotNeeded}}
Now that $I'$ has performed \llt$(gp)$ and \llt$(p)$, and confirmed that there are no weight violations at $p$ or any of its children, it checks whether a Compress update is necessary.
Conceptually, this is part of the \func{UpdateNotNeeded} procedure in the template.
To do this, $I'$ uses the snapshot of $p$'s children that was returned by its invocation of \llt$(p)$ to compute the total slack shared amongst the children of $p$.
$I'$ then checks whether there is a slack violation at $p$, or degree violation at $l$.
(Observe that a slack violation at $p$ and a degree violation at $l$ can each be fixed by a Compress where $p$ is the topmost node replaced by the update.)
%By the semantics of \llt\ and \sct, $I'$ can now check whether there is a slack violation at $p$, or degree violation at $l$ (both of which would be fixed by a Compress where $p$ is the topmost node replaced by the update), with the following knowledge.
%If $I'$ finds a slack or degree violation, and $I'$ subsequently performs a successful \sct$(V, R, fld, new)$ where $p, l \in V$, then the slack or degree violation 
%If there is no slack or degree 
If there is no slack or degree violation, then $I'$ simply returns $\emptyset$ (which will cause $I$ to return, successfully, without performing any additional rebalancing).
So, suppose $I'$ finds a slack or degree violation.

%Now that $I'$ has performed \llt$(gp)$ and \llt$(p)$, and confirmed that there are no weight violations at $p$ or any of its children, it must once again check whether there is a slack violation at $node$.
%This is because any mutable fields of $gp$ or $p$ that $I'$ read might have changed between when it read them, and when $I'$ performed the corresponding \llt.
%(Checking for degree or slack violations conceptually falls into the \func{Conflict} procedure in the tree update template.)
%%The semantics of \llt\ and \sct\ guarantee that, if $I'$ performs a successful \sct$(V, R, fld, new)$ with $p \in V$, then any violations $I'$ sees at $node$ \textit{after} its \llt$(p)$.
%$I'$ uses the snapshot of $p$'s children that was returned by its invocation of \llt$(p)$ to compute the total slack shared amongst the children of $p$.
%$I'$ then checks whether there is a degree violation at $node$, or a slack violation at $p$.
%Recall that, if there is a degree violation at $node$, then we do not take an extra step in the search, so $node = l$.
%Otherwise, we take an extra step in the search, so $p = node$.
%Thus, in each case, $I'$ is actually checking whether there is a violation at $node$.
%If there is no slack or degree violation at $node$, then $I'$ returns (since there is no degree or slack violation to fix).

\fakeparagraph{Computing \sct-\func{Arguments}}
Next, $I'$ computes the arguments $V$, $R$, $fld$ and $n$ for an invocation of \sct$(V, R, fld, n)$ by using locally stored values and immutable fields of nodes.
Compress changes a pointer of $gp$ (at index $ix_p$) from $p$ to point to some new node $n$, removing $p$ and its children from the tree.
Since $p$ and its children are removed, $R = \langle p, result_p.p_1, result_p.p_2, ... \rangle$.
Since $gp$ is changed, it must be in $V$.
Since $R$ is a subsequence of $V$, $V = \langle gp, p, result_p.p_1, result_p.p_2, ... \rangle$.
In order to create the new nodes needed for the Compress update, $I'$ invokes a procedure called \func{CreateCompressedNodes}, which appears in Figure~\ref{code-bslack-createcompressnodes}.

\fakeparagraph{Creating the new node(s)}
At a high level, \func{CreateCompressedNodes} creates a small subtree of new nodes that would result by applying Compress (and possibly also Root-Replace) to $p$ and its children.
These new nodes will be used by \func{DoCompress} to replace $p$ and its children in the tree.

More specifically, \func{CreateCompressedNodes} takes four arguments: a pointer $gp$, a pointer $p$, a set of pointers $result_p$ and a natural number $pGrandDegree$.
These arguments correspond to the variables with the same names in \func{FixSlack}.
An invocation $I''$ of \func{CreateCompressedNodes}$(gp, p, result_p, pGrandDegree)$ begins by determining how to divide the $pGrandDegree$ keys and values in the nodes of $result_p$ as evenly as possible into $numNewChildren = \lceil pGrandDegree/b \rceil$ new nodes (where $b$ is the maximum degree of nodes).
Specifically, it divides them into $ceilNodes$ nodes with degree $ceilDegree$, and $floorNodes$ with degree $floorDegree$.

Next, $I''$ determines whether the nodes in $result_p$ are leaves or internal nodes.
Observe that, since there are no weight violations at any of the nodes in $result_p$, property P1$'$ of \rbslack s implies that these nodes are either all leaves or all internal nodes.
Therefore, $I''$ checks whether the first node in $result_p$ is a leaf, and stores the result in a variable $pChildrenAreLeaves$.
If so, all nodes in $result_p$ are leaves.
Otherwise, all nodes in $result_p$ are internal nodes.

If $gp = entry$ and $numNewChildren = 1$, then Compress would replace the $root$ of the tree by a new node with only one child $n$, necessitating a subsequent Root-Replace update.
In this case, $I''$ will simply cause Root-Replace to be performed at the same time as the Compress update by returning the new child $n$ as the replacement for $p$.
This new node $n$ contains all of the keys and pointers of the nodes in $result_p$.
It is a leaf if the nodes in $result_p$ are leaves, and an internal node otherwise.
The $searchKey$ of $n$ is the same as the $searchKey$ of $p$.
Since $p$ was reachable by searching for $p.searchKey$ before this update, and $n$ replaces node $p$ in the tree (and none of the ancestors of $p$ are replaced or changed by the update, except to replace $p$ by $n$), $n$ will also be reachable by searching for $p.searchKey$ after this update.
Finally, $I''$ returns $\langle \true, n, \emptyset \rangle$, which indicates that the newly created node $n$ is the result of performing both Compress and Root-Replace.

Now, suppose $gp \neq entry$ or $numNewChildren \neq 1$.
In this case, $I''$ creates a new parent for the $numNewChildren$ newly created children, and returns points to this parent, and the new children.
The new children are leaves if the nodes in $result_p$ are leaves, and are internal nodes otherwise.
Two subcases arise.

First, suppose $numNewChildren = 1$.
Then, the $searchKey$ fields of the single new child and its new parent $n$ are both the same as the $searchKey$ of $p$.
By the same argument as in the previous case, $n$ is reachable by searching for $p.searchKey$ after this update.
Moreover, since $n$ has only one child, clearly that child is reachable by searching for the same $searchKey$.

Second, suppose $numNewChildren \neq 1$.
Then, the $searchKey$ of each new node is its first key.
We briefly argue that each new node contains at least one key.
Since $numNewChildren \neq 1$, and P2$'$ says that each internal node has at least one pointer, we obtain $numNewChildren > 1$.
Thus, the new parent node $n$ contains at least one key.
Since $I''$ divides the key and pointers as evenly as possible, and into as few new child nodes as possible, each new child has degree at least $\lfloor b/2 \rfloor$.
So, each new child contains at least $\lfloor b/2 \rfloor - 1$ keys.
Since $b \ge 5$, $\lfloor b/2 \rfloor - 1 \ge 2$.

Finally, $I''$ returns $\langle \false, n, \{\mbox{children of }n\} \rangle$, which indicates that $n$ is the result of performing Compress.

\fakeparagraph{Performing \sct\ and subsequent rebalancing}
After the invocation of \func{CreateCompressedNodes} by $I'$ returns $\langle doRootReplace, n, nChildren \rangle$, $I'$ invokes \sct$(V, R, fld, n)$ to replace the subtree rooted at $p$ with the subtree rooted at $n$.
If the invocation of \sct\ returns \false, then $I'$ simply returns \func{Retry}.
So, suppose the \sct\ returns \true.
Then, $I'$ checks whether $doRootReplace = \true$.
If so, $I'$ atomically performed both Compress and Root-Replace, so $n$ is the only new node created by the update.
Consequently, by Figure~\ref{fig-bslack-violation-movement}, $I$ need only check for degree or slack violations at $n$.
Thus, $I'$ returns $\{n\}$, which will cause $I$ to invoke \func{FixSlack}$(n)$.
However, if $doRootReplace = \false$, then $I$ must check for a degree violation at $n$, and slack violations at $gp$ and all new children of $n$.
Thus, $I'$ returns $\{gp, n\} \cup nChildren$.
$I$ will then invoke \func{FixSlack}$(x)$ for each $x \in \{gp, n\} \cup nChildren$.

%%\trevor{note: the following is not true. there can be degree violations at $root$. i believe it's true that there cannot be weight violations, though.}
%As a minor point, as we described above, whenever a rebalancing update would necessitate a subsequent Root-Replace or Root-Zero update, we opportunistically perform Root-Replace or Root-Zero (as appropriate) as part of the same atomic update.
%Consequently, in our implementation, there is \textit{never} a degree or weight violation at the $root$.

As a minor point, as we described above, whenever a rebalancing update would necessitate a subsequent Root-Zero update, we always perform Root-Zero as part of the same atomic update.
Consequently, in our implementation, there is \textit{never} a weight violation at the $root$.

%\textbf{for describing which violations could exist after a combination of compress/one-child and root-replace.}
% in this case, the compress creates a node n with exactly
% one child. this child may have a slack violation, and
% n may have a degree violation. additionally, p may have
% a slack violation.
% however, after we also perform root-replace, n is removed
% altogether, so there are no violations at n.
% note that the n in the root-replace comment above refers
% to the single child of the node n referred to by the
% compress comment.
% thus, the only possible violations after the root-replace
% are a slack violation at the child, a degree violation
% at the child, and a slack violation at p.
% we check for (and attempt to fix) each of these.
%
% note: it is impossible for there to be a weight violation at childrenNewP or p, since these nodes must have weight=$\true$ for the compress/one-child+root-replace operation to be applicable, and we consequently CREATE childrenNewP[0] and p with weight=$\true$ above

\subsection{An optimization to avoid unnecessary searches while rebalancing} \label{sec-nofixneeded}

An invocation $I'$ of \func{NoFixNeeded} begins by checking whether $node$ is a leaf.
If it is, then $I'$ returns \true, since degree and slack violations cannot occur at leaves.
So, suppose $node$ is internal.
Then, $I'$ performs \llt$(node)$ and checks whether it returns \finalized.
If so, $node$ has been removed from the tree by another process (and we are no longer responsible for any violations at $node$), so $I'$ returns \true.
Otherwise, if the \llt\ returns \fail, then $I'$ cannot determine whether there is a violation at $node$, so it returns \false.
So, suppose the \llt\ does not return \finalized\ or \fail\ (which means it returns a snapshot of the pointers of $node$).

Next, $I'$ checks if $node$ has weight zero.
If so, there is a weight violation at $node$.
Since a degree or slack violation at $node$ cannot even be fixed if there is a weight violation at $node$, the weight violation must be fixed before proceeding, so $I'$ returns \false.
So, suppose there is no weight violation at $node$.
Next, $I'$ checks if there is a degree violation at $node$, and returns \false\ if there is.
So, suppose there is no degree violation at $node$.

$I'$ uses the snapshot of $node$'s children that was returned by the \llt$(node)$ to check for a slack violation at $node$.
Specifically, it uses the degree of each of these children to compute the total slack shared amongst these children, and it also checks whether there are weight violations at any of these children.
If $I'$ finds a weight violation, it must be fixed before proceeding (for the same reasons we gave in the case where there is a weight violation at $node$), so $I'$ returns \false.
Suppose $I'$ does not find any weight violations.
Then, if the total slack is less than $b$, there was no degree violation at $node$ when the \llt$(node)$ occurred, so $I'$ returns \true.
Otherwise, there may be a slack violation at $node$, so $I'$ returns \false.

\section{Correctness proof}

For convenience, we define a few terms.
Consider an iteration of the outer loop in \ins, \del, \func{FixWeight} or \func{FixSlack}.
Within this iteration, we call the sequence of steps between the comments ``Start of operation attempt'' and ``End of operation attempt'' an \textbf{operation attempt}.
More precisely, an operation attempt begins when a process performs a step just after a ``Start of operation attempt'' comment, and ends when it reaches an ``End of operation attempt'' comment, exits the loop, goes to the next iteration, or starts a new operation attempt.
Each invocation of \ins\ or \del\ performs a (non-empty) sequence of operation attempts.
Each invocation of \func{FixWeight} or \func{FixSlack} either returns before entering the loop, or performs a sequence of operation attempts.
As we will see below, some operation attempts follow the template, and some do not.
The terms \textbf{search path} and \textbf{range} are defined as they were in the proof of the chromatic tree (in Section~\ref{chromatic-correctness}).

\begin{lem} \label{lem-bslack}
Our implementation of a \rbslack\ satisfies the following claims.
\begin{compactenum}
\item Each operation attempt $A$ %performed by \ins\ or \del\
%, \func{FixWeight} or \func{FixSlack}
follows the tree update template and satisfies all constraints specified by the template, unless:
(1) $A$ occurs in \func{FixWeight}, and it invokes \func{FixWeight} at line~\ref{lfbslack-fixweight-exception1}, or (2) $A$ occurs in \func{FixSlack}, and it performs an invocation of \func{FixWeight} at line~\ref{lfbslack-fixallweight-fixweight1} or line~\ref{lfbslack-fixallweight-fixweight2}.
\label{claim-bslack-invariants-follow-template}
\item The node $entry$ has weight one, no keys, and a single child pointer to a node $root \neq entry$ with weight one.
\label{claim-bslack-invariants-top-of-tree}
\item If a node $v$ is in the data structure in some configuration $C$ and $v$ was on the search path for key $k$ in some earlier configuration $C'$, then $v$ is on the search path for $k$ in $C$. (Equivalently, updates to the tree do not shrink the range of any node in the tree.) \label{lem-bslack-searchpath}
\item If an invocation of \func{Search}$(k)$ reaches a node $v$, then there was some earlier configuration during the search when $v$ was on the search path for $k$. \label{lem-bslack-searches}
\item The \func{Search} procedure used by \ins\ and \del\ satisfies DTP. \label{lem-bslack-dtp}
\item All operation attempts in \ins\ and \del\ that perform a successful \sct\ are atomic (including their search phases).
All operation attempts in \func{FixWeight} and \func{FixSlack} that perform a successful \sct\ have atomic update phases.
\label{lem-bslack-atomicity}
\item The tree rooted at the child of $entry$ is always a \rbslack. \label{lem-bslack-searchtree}
\end{compactenum}
\end{lem}
\begin{chapscxproof}
We prove these claims together by induction on the sequence of steps %(invocations and responses of procedures, atomic invocations of \func{Search} (for Claim~\ref{lem-bslack-dtp}) reads from shared memory, and \llt s and \sct s) 
in an execution.
That is, we assume that all of the claims hold before an arbitrary step in the execution, and prove they hold after the step.

\medskip

\noindent\textbf{Claim~\ref{claim-bslack-invariants-follow-template}:}
This claim follows almost immediately from inspection of the code.
The only subtlety is showing that no process invokes $\llt(r)$, \func{FixWeight}$(r)$ or \func{FixSlack}$(r)$ where $r = \nil$.
Suppose the inductive hypothesis holds just before an invocation $I$ of $\llt(r)$, \func{FixWeight}$(r)$ or \func{FixSlack}$(r)$.

We first observe that, if an invocation of \func{Search} or \func{SearchNode} that returns $\langle gp, ix_p, p, ix_l, l \rangle$, then $p, l \neq \nil$.
This follows by inductive Claim~\ref{claim-bslack-invariants-top-of-tree} and inspection of the code.

Suppose $I$ is an invocation of \func{FixWeight}$(r)$.
If $I$ occurs in \ins, then $r$ is a newly created node.
If $I$ occurs in \func{FixWeight}, then $r = p$ or $r = n$.
In the first case, $r \neq \nil$ since it was returned by \func{SearchNode}.
In the second case, $r$ is a newly created node.
If $I$ occurs in \func{FixAllWeightViolations}, then $r = p$ or $r = result_p.p_i$, where $result_p$ is a result of an \llt$(p)$.
In the first case, $r \neq \nil$ since it was returned by \func{SearchNode}.
In the second case, $r$ was returned by an \llt$(p)$ where $p \neq \nil$, so $r \neq \nil$ by inductive Claim~\ref{lem-bslack-searchtree}.

Suppose $I$ is an invocation of \func{FixSlack}$(r)$.
If $I$ occurs in \ins\ or \del, then $r = p$, where $p$ was returned by \func{Search} (and hence is non-\nil).

If $I$ occurs in \func{FixWeight}, then $r$ is either $gp$ or a newly created node.
We argue that $gp \neq \nil$.
The pointer $gp$ was returned by \func{SearchNode}$(node)$, which also returned $l$ and $p$.
Since this invocation of \func{SearchNode} does not return \fail, $l = node$.
Also observe that $node = l$ was seen to have weight zero before \func{SearchNode} was invoked.
The $weight$ field of a node cannot change, so $l$ always has weight zero.
By Claim~\ref{claim-bslack-invariants-top-of-tree}, $entry$ and its child always have weight one.
So, $l$ cannot be $entry$ or its child.
It follows that $gp \neq \nil$.

If $I$ occurs in \func{FixSlack}, then $r$ is either $gp$ or a newly created node.
We argue that $gp \neq \nil$.
By inspection of the pseudocode, if $r = gp$, then $I$ was invoked at line~\ref{lfbslack-fixslack-fixdegreeorslack2}.
Thus, $r$ is either the first or second pointer returned by \func{SearchNode}, depending on whether line~\ref{lfbslack-fixslack-extrastep} is executed.
Two subcases arise.

\textit{Case 1:} suppose $node.d > 1$ at line~\ref{lfbslack-fixslack-nodegree}, so line~\ref{lfbslack-fixslack-extrastep} is executed.
By Claim~\ref{claim-bslack-invariants-top-of-tree}, $entry.d = 1$, so $node \neq entry$.
Since \func{SearchNode} does not return \fail, the variable $l$ returned by \func{SearchNode} is the same as $node$.
Thus, after line~\ref{lfbslack-fixslack-extrastep}, $p = node \neq \nil$ and $l \neq \nil$.
Since $p \neq entry$, we have $gp \neq \nil$.

\textit{Case 2:} suppose $node.d \le 1$ at line~\ref{lfbslack-fixslack-nodegree}, so line~\ref{lfbslack-fixslack-extrastep} is not executed.
By Claim~\ref{lem-bslack-searchtree}, P\ref{prop-rbslack-internal}$'$ is satisfied, so $node.d = 1$.
Since we only execute this case if we previously saw that $gp \neq \nil$ or $node.d \neq 1$, and we know that $node.d = 1$, we obtain $gp \neq \nil$.

Now, suppose $I$ is an invocation of \llt$(r)$.
Then, $I$ can occur in \ins, \del, \func{FixWeight}, \func{NoFixNeeded}, \func{DoRootReplace} or \func{DoCompress}.

If $I$ occurs in \ins\ or \del, then $r$ must be one of the nodes $p$ or $l$ returned by \func{Search}.
As we argued above, $l, p \neq \nil$.

If $I$ occurs in \func{FixWeight}$(node)$, then $r$ must either be $node$, or one of the nodes $gp$, $p$ or $l$ returned by \func{SearchNode}$(node)$.
We have already argued above that the argument $node$ is non-\nil.
If $r$ is $p$ or $l$, then $r \neq \nil$ (as argued above).
The argument for $gp \neq \nil$ is identical to the argument above for the case where $I$ is an invocation of \func{FixWeight} that occurs in \func{FixWeight}.

%We argue that $gp \neq \nil$.
%Observe that $node$ is seen to have weight zero before \func{SearchNode}$(node)$ is invoked.
%The $weight$ field of a node cannot change, so $node$ always has weight zero.
%By Claim~\ref{claim-bslack-invariants-top-of-tree}, $entry$ and $root$ have weight one.
%So, $node$ must be a descendant of $root$.
%Moreover, since \func{SearchNode} does not return \fail, $l = node$.
%Thus, $l$ has a parent and a grandparent (by Claim~\ref{lem-bslack-searchtree}).
%It follows that $gp$ and $p$ are non-\nil.

If $I$ occurs in \func{NoFixNeeded}$(node)$, then $r = node$.
By inspection of the pseudocode, \func{NoFixNeeded}$(node)$ is invoked only by \func{FixSlack}$(node)$, and we have argued above that the argument $node$ to \func{FixSlack} is non-\nil.

If $I$ occurs in \func{DoRootReplace}$(p, ix_l, l)$, then $r = p$ or $r = l$.
Observe that \func{DoRootReplace} is invoked only by \func{FixSlack}$(node)$, and the arguments $p$ and $l$ to \func{DoRootReplace} were returned by an invocation of \func{SearchNode}$(node)$ (along with another pointer $gp$).
As we argued above, $p, l \neq \nil$.

If $I$ occurs in \func{DoCompress}$(gp, ix_p, p, ix_l, l)$, then $r = gp$ or $r = p$ or $r = l$ or $r$ is one of the pointers returned by an invocation of \llt$(p)$ performed by the \func{DoCompress}.
The argument that $l, p \neq \nil$ is identical to the argument in the previous case.
If $r$ was returned by an \llt$(p)$, then $r \neq \nil$ by inductive Claim~\ref{lem-bslack-searchtree}.
The argument that $gp \neq \nil$ is identical to the case where $I$ is an invocation of \func{FixSlack} that occurs in \func{FixSlack}.

%Two subcases arise.
%
%\textit{Case 1:} suppose $node.d > 1$ at line~\ref{lfbslack-fixslack-nodegree}, so the body of the if-statement is executed.
%By Claim~\ref{claim-bslack-invariants-top-of-tree}, $entry.d = 1$, so $node \neq entry$.
%Therefore, $node$ is a descendant of $entry$.
%Before $I$, \func{FixSlack} performs an invocation of \func{NoFixNeeded} that sees $node$ is an internal node.
%Since \func{SearchNode} does not return \fail, the variable $l$ returned by \func{SearchNode} is the same as $node$.
%Thus, after line~\ref{lfbslack-fixslack-extrastep}, $p = node \neq \nil$ and $l \neq \nil$.
%Since $p$ is a descendant of $entry$, we have $gp \neq \nil$.
%Furthermore, since $p$ is internal, Claim~\ref{lem-bslack-searchtree} implies that all nodes in $result_p$ are non-\nil.
%
%\textit{Case 2:} suppose $node.d \le 1$ at line~\ref{lfbslack-fixslack-nodegree}, so the body of the if-statement is not executed.
%Since \func{SearchNode} does not return \fail, $l = node$.
%By inspection of \func{SearchNode}, an invocation of \func{SearchNode}$(entry)$ will return \fail, so $node \neq entry$.
%By Claim~\ref{lem-bslack-searchtree}, P\ref{prop-rbslack-internal}$'$ is satisfied, so $node.d = 1$.
%By Claim~\ref{claim-bslack-invariants-top-of-tree}, $node \neq root$.
%Therefore, Claim~\ref{lem-bslack-searchtree} implies that $l = node$ is a descendant is of $root$.
%It follows that $p$ and $gp$ are non-\nil.
%The argument that the nodes in $result_p$ are non-\nil\ is the same as in the last case.

\medskip

\noindent\textbf{Claim~\ref{claim-bslack-invariants-top-of-tree}:}
The only step that can modify the tree (and, hence, affect this claim) is an invocation $S$ of \sct\ performed by an invocation $I$ of \ins, \del, \func{FixWeight} or \func{FixSlack}.
Proving this claim entails (1) arguing that $I$ cannot replace the entry point, or modify it in any way except by changing its single child pointer and (2) if $I$ replaces $root$, then it replaces it with a new node that has weight one.

Suppose the inductive hypothesis holds just before $S$.
By inductive Claim~\ref{claim-bslack-invariants-follow-template}, $I$ follows the tree update template up until it performs $S$.
By Lemma~\ref{lem-dotreeup-constraints-invariants} and Lemma~\ref{lem-effective-updatephase-atomic}, the update phase of $I$ is performed atomically.
Thus, by inspection of the pseudocode, $I$ atomically performs Delete, Insert, Overflow, Root-Replace, Absorb, Split or Compress, or a combination of Root-Zero and Overflow or Split, or a combination of Root-Replace and Compress.

We first argue that $entry$ cannot be replaced.
All of these transformations simply change a single child pointer to replace one or more nodes.
However, each node that is replaced has a parent, which $entry$ does not.

Now, suppose $I$ replaces $root$.
The only updates that could replace $root$ by a new node with weight zero are Overflow and Split.
However, whenever an Overflow or Split would do so, Root-Zero is also performed as part of the same atomic update (see line~\ref{lfbslack-insert-rootzero} and line~\ref{lfbslack-fixweight-rootzero}) so that the new $root$ has weight one.
%The only update that could replace $root$ by a new internal node with degree less than two is Compress.
%However, whenever a Compress would do so, Root-Replace is also performed as part of the same atomic update (see line~\ref{lfbslack-createcompressednodes-rootreplace-start} through line~\ref{lfbslack-createcompressednodes-rootreplace-end}).
%\trevor{HOWEVER, it seems like you really can have a degree violation at the root...}

% performs performs one of the transformations in Figure~\ref{fig-bslack-updates}.
%All of these transformations simply change a single child pointer to replace one or more nodes.
%However, each node that is replaced has a parent, which $entry$ does not.
%Thus, $entry$ cannot be replaced.

\medskip

\noindent\textbf{Claim~\ref{lem-bslack-searchpath}:}
The proof of this claim is very similar to the proof of Lemma~\ref{lem-abtree}.\ref{lem-abtree-searchpath}.
Initially, the claim is trivially true (since the tree only contains $entry$, which is on every search path).
In order for $v$ to change from being on the search path for $k$ in configuration $C'$ to no longer being on the search path for $k$ in configuration $C$, the tree must change between $C'$ and $C$.
Thus, there must be a successful \sct\ $S$ between $C'$ and $C$.
Moreover, this is the only kind of step that can affect this claim.
We show $S$ preserves the property that $v$ is on the search path for $k$.

By inductive Claim~\ref{claim-bslack-invariants-follow-template} and inspection of the pseudocode, $S$ is performed by a template operation.
Thus, by Lemma~\ref{lem-dotreeup-constraints-invariants}, $S$ changes a pointer of a node from $old$ to $new$, removing a connected set $R$ of nodes (rooted at $old$) from the tree, and inserting a new connected set $N$ of nodes.
If $v$ is not a descendant of $old$ immediately before $S$, then this change cannot remove $v$ from the search path for $k$.
So, suppose $v$ is a descendant of $old$ immediately prior to $S$.

Since $v$ is in the data structure in both $C'$ and $C$, it must be in the data structure at all times between $C'$ and $C$ by Lemma~\ref{lem-dotreeupdate-rec-cannot-be-added-after-removal}.
Therefore, $v$ is a descendant of $old$, but $S$ does not remove $v$ from the tree.
Recall that the fringe $F_R$ is the set of nodes that are children of nodes in $R$, but are not themselves in $R$ (see Figure~\ref{fig-replace-subtree} and Figure~\ref{fig-replace-subtree2}).
By definition, $v$ must be a descendant of a node $f \in F_R$.
Moreover, since $v$ is on the search path for $k$ just before $S$, so is $f$.
We argue, for each possible tree modification in Figure~\ref{fig-bslack-updates}, that if any node in $F_R$ is on the search path for $k$ prior to $S$, then it is still on the search path for $k$ after $S$.
We proceed by cases.

\textit{Case~1:} Suppose $S$ performs Insert, Overflow or Delete.
Since $S$ replaces a leaf with either a new leaf, or a new internal node and two new leaves, the fringe set is empty.
Thus, the claim is vacuously true.

\textit{Case~2:} Suppose $S$ performs Root-Replace or Root-Zero update.
Then, $S$ does not change the range of any node in the fringe set, so the claim holds.

\textit{Case~3:} Suppose $S$ performs Absorb.
By inductive Claim~\ref{lem-bslack-searchtree}, the tree is a \rbslack\ before $S$.
Since leaves always have weight one in a \rbslack, the node $u$ in the depiction of Absorb in Figure~\ref{fig-bslack-updates} must be internal.
Thus, one can think of each of the nodes $u$ and $\pi(u)$ as a sequence of alternating pointers and keys, starting and ending with a pointer.
Consequently, $\alpha$, $\beta$ and $\gamma$ (in Figure~\ref{fig-bslack-updates}) can be thought of as sequences of alternating pointers and keys, where $\alpha$ starts with a pointer and ends with a key, $\beta$ starts and ends with pointers, and $\gamma$ starts with a key and ends with a pointer.
%Consider an inorder traversal of the tree that outputs the sequence of pointers and keys it encounters in the internal nodes it visits.
%
The fringe $F_R$ is the set of nodes pointed to by $\alpha$, $\beta$ and $\gamma$.
Observe that the keys in $u$ and $\pi(u)$ partition the range of $\pi(u)$, and this partition defines the range of each node in $F_R$.
Specifically, the partition begins with the left endpoint of the range of $\pi(u)$, then continues with the alternating pointers and keys of $\alpha$, $\beta$ and $\gamma$, and finally ends with the right endpoint of the range of $\pi(u)$.
The update does not change this partition, so the range of each node in $F_R$ is the same before and after the update.
Therefore, if a node in $F_R$ is on the search path to $key$ before the update, it is still on the search path after the update.
The cases for Split, Compress and One-Child follow the exact same reasoning.

\medskip

\noindent\textbf{Claim~\ref{lem-bslack-searches}:}
The proof of this claim is identical to the proof of Lemma~\ref{lem-abtree}.\ref{lem-abtree-searches} (except for the text substitution ``relaxed $(a,b)$-tree'' $\rightarrow$ ``\rbslack'').

\medskip

\noindent\textbf{Claim~\ref{lem-bslack-dtp}:}
The proof of this claim is very similar to the proof of Lemma~\ref{lem-abtree}.\ref{lem-abtree-dtp}.
Initially, the claim holds vacuously (since no steps have been taken).
Suppose an invocation $S$ of \func{Search}$(k)$ in \ins\ terminates and returns $m$.
(The proof for \del\ is similar.)
Consider any configuration $C$, after $S$ returns $m$, in which all of the nodes in $m$ are in the tree and their fields agree with the values in $m$.
Suppose the inductive hypothesis up until $C$.
We prove that an invocation $S'$ of \func{Search}$(k)$ in \ins\ would return $m$ if $S'$ were performed atomically just after configuration $C$.

The value $m = \langle -, p, l \rangle$ returned by $S$ contains a leaf $l$ and its parent $p$.
By inductive Claim~\ref{lem-bslack-searches}, $p$ and $l$ were each on the search path at some point during $S$ (which is before $C$).
Since $p$ and $l$ are in the tree in $C$, inductive Claim~\ref{lem-bslack-searchpath} implies that they are on the search path for $k$ in $C$.
Therefore, $S'$ will visit each of them.
Conceptually, $m$ also encodes the fact that $p$ points to $l$.
This fact is checked at line~\ref{code-bslack-tryins-conflict1} as part of the \func{Conflict} procedure.
By our assumption (that the fields of the nodes in $m$ in configuration $C$ agree with their values in $m$), $p$ is also the parent of $l$ when $S'$ is performed.
Consequently, $S'$ will also return $m = \langle -, p, l \rangle$.

\medskip

\noindent\textbf{Claim~\ref{lem-bslack-atomicity}:}
By inductive Claim~\ref{lem-bslack-dtp} and Theorem~\ref{thm-effectivedtp-atomic}, all operation attempts that occur in \ins\ and \del\ and execute a successful \sct\ are atomic.
By Lemma~\ref{lem-effective-updatephase-atomic}, all operation attempts that occur in \func{FixWeight} and \func{FixSlack} and  execute a successful \sct\ have atomic update phases.

\medskip

\noindent\textbf{Claim~\ref{lem-bslack-searchtree}:}
Only successful invocations of \sct\ can affect this claim.
Successful invocations of \sct\ are performed only in \ins, \del\ and the rebalancing procedures: \func{FixWeight} and \func{FixSlack}.
The claim holds in the initial state of the tree (which is described in Claim~\ref{claim-bslack-invariants-top-of-tree}).
We show that every successful invocation $S$ of \sct\ preserves the claim.
We proceed by cases.

\textit{Case~1:} $S$ is in an operation attempt $A$ that occurs in an invocation of \ins$(key, value)$ or \del$(key)$.
By Claim~\ref{lem-bslack-atomicity}, the entire operation attempt $A$ is atomic (including the search phase).
Thus, the invocation of \func{Search} by $A$ returns the unique leaf $l$ on the search path for $key$.
Consequently, $A$ atomically performs one of the transformations \func{Insert}, \func{Overflow} or \func{Delete} to replace $l$ (and possibly some of its neighbouring nodes).
Since $A$ is entirely atomic, and it simply performs one of the \rbslack\ updates, it is easy to verify that it preserves the claim.

\textit{Case~2:} $S$ is in an operation attempt $A$ that occurs in an invocation of one \func{FixWeight} or \func{FixSlack}. %of the rebalancing procedures. %\func{TryRootUntag}, \func{TryRootAbsorb}, \func{TryAbsorbChild}, \func{TryPropagateTag}, \func{TryAbsorbSibling} or \func{TryDistribute}.
By Claim~\ref{lem-bslack-atomicity}, the update phase of $A$ is atomic (but the search phase is not necessarily atomic).
Therefore, $A$ atomically performs one or two rebalancing transformations at some location in the tree (but not necessarily the same rebalancing transformations, at the same location, that it would perform if $A$'s search were also part of the atomic update).
All of the rebalancing transformations preserve the claim (regardless of where in the tree they are performed).
\end{chapscxproof}

We define the linearization points for \rbslack\ operations as follows.
\begin{compactitem}
\item \func{Get}($key$) is linearized at a time during the operation when the leaf reached was on the search path for $key$.
(This time exists, by Lemma~\ref{lem-bslack}.\ref{lem-bslack-searches}.)
\item An \ins\ is linearized at its successful \sct\ (if such an \sct\ exists).
(Note: every \ins\ that terminates performs a successful \sct.)
\item A \del\ that returns $\bot$ is linearized at a time during the operation when the leaf returned by its last invocation of \func{Search} was on the search path for $key$.
(This time exists, by Lemma~\ref{lem-bslack}.\ref{lem-bslack-searches}.)
\item A \del\ that does not return $\bot$ is linearized at its successful \sct\ (if such an \sct\ exists).
(Note: every \del\ that terminates, but does not return $\bot$, performs a successful \sct.)
\end{compactitem}
It is easy to verify that every operation that terminates is linearized, and that each linearized operation has a linearization point that is during the operation. %linearization point falls inside the operation it belongs to.

\begin{thm}
The \rbslack\ is a linearizable implementation of a dictionary with the operations \func{Get}, \ins\ and \del.
\end{thm}
\begin{chapscxproof}
Lemma~\ref{lem-bslack}.\ref{lem-bslack-atomicity} proves that each successful \sct\ atomically performs one or two of transformations shown in Figure~\ref{fig-bslack-updates}.
By inspection of these transformations, the set of keys and associated values stored in leaves are not altered by any rebalancing steps.
Moreover, the transformations performed by each linearized \ins\ and \del\ maintain the invariant that the set of keys and associated values stored in leaves of the tree is exactly the set that should be in the dictionary.
When an invocation of \func{Get}$(key)$ is linearized, the search path for $key$ ends at the leaf returned by its invocation of \func{Search}.
If that leaf contains $key$, \func{Get} returns the associated value, which is correct.
If that leaf does not contain $key$, then, by Lemma~\ref{lem-bslack}.\ref{lem-bslack-searchtree}, it is nowhere else in the tree, so \func{Get} is correct to return $\bot$.
\end{chapscxproof}

\section{Progress proof}

Our goal is to prove that, if processes take steps infinitely often, then \rbslack\ operations succeed infinitely often.
At a high level, this follows from Theorem~\ref{thm-dotreeup-progress} (the final progress result for template operations), and Lemma~\ref{lem-amortized-rebalancing}, which states that at most $2i(4+\frac 3 2 \big\lfloor \log_{\lfloor\frac{b}{2}\rfloor} \frac{n+i}{2} \big\rfloor) + 2d/(b-1)$ rebalancing steps can be performed after $i$ insertions and $d$ deletions have been performed on an empty \bslack.

Theorem~\ref{thm-dotreeup-progress} applies only if processes perform infinitely many template operations, so we must prove that processes will perform infinitely many template operations if they take steps infinitely often.
The subtlety is that some operation attempts % of \func{FixWeight} and \func{FixSlack}
might not follow the template.
Specifically, as we saw in Lemma~\ref{lem-bslack}.\ref{claim-bslack-invariants-follow-template}, an operation attempt $A$ does not follow the template if: (1) $A$ occurs in \func{FixWeight}, and it invokes \func{FixWeight} at line~\ref{lfbslack-fixweight-exception1}, or (2) $A$ occurs in \func{FixSlack}, and it performs an invocation of \func{FixWeight} at line~\ref{lfbslack-fixallweight-fixweight1} or line~\ref{lfbslack-fixallweight-fixweight2}.
One can imagine a pathology in which non-blocking progress is violated, because processes perform only finitely many template operations, but take infinitely many steps in \func{FixWeight} and/or \func{FixSlack}.
%One can imagine a pathology in which non-blocking progress is violated, because processes perform only finitely many operation attempts that follow the template, but perform infinitely many operation attempts that do \textit{not} follow the template.
We first prove that this does not happen.
Then, we prove the main result.

%\begin{lem}
%If processes perform only finitely many template operations, then only finitely many invocations of 
%\end{lem}
%\begin{chapscxproof}
%However, this can happen only finitely many times, since (1) nodes are finalized when they are removed from the tree and are never reinserted into the tree after being removed, (2) only finitely many nodes are ever inserted into the tree since there are only finitely many template operations, and (3) whenever an operation attempt invokes \func{FixWeight}$(node)$, it previously saw (during the operation attempt) that $node$ was in the tree.
%(Regarding (3), the operation attempt performs an invocation of \func{SearchNode}, and either this invocation returns $node$, or it returns the parent $p$ of $node$ and the operation attempt then performs an invocation of \llt$(p)$ that returns a pointer to $node$.)
%\end{chapscxproof}

%\begin{lem} \label{lem-bslack-searches-terminate-if-tree-stops-changing}
%If only finitely many template operations are performed, then all invocations of \func{Search}, \func{SearchNode} and \func{Get} perform a finite number of steps.
%\end{lem}
%\begin{chapscxproof}
%Since the tree is changed only by successful invocations of \sct, and \sct\ is invoked only by template operations, the tree eventually stops changing.
%Consequently, every invocation of \func{Search} (and, hence, \func{Get}) terminates after a finite number of steps, unless the process executing it crashes.
%\end{chapscxproof}

\begin{obs} \label{obs-lfbslack-finalized-when-removed}
Nodes are finalized precisely when they are removed from the data structure (and are never reinserted into the data structure).
\end{obs}
%\begin{chapscxproof}
%By inspection of the pseudocode (and the semantics of \sct).
%\end{chapscxproof}

\begin{lem} \label{lem-bslack-fix-follows-template-infinitely-often}
If processes take infinitely many steps in \func{FixWeight} and/or \func{FixSlack}, then infinitely many template operations are performed.
\end{lem}
\begin{chapscxproof}
Suppose, to derive a contradiction, that processes take infinitely many steps in \func{FixWeight} and/or \func{FixSlack}, but only finitely many template operations are performed.
Since the tree is changed only by successful invocations of \sct, and \sct\ is invoked only by template operations, the tree eventually stops changing.
Consequently, every invocation of \func{Search} or \func{SearchNode} %(and, hence, \func{Get}) 
terminates after a finite number of steps, unless the process executing it crashes.
%By Lemma~\ref{lem-bslack-searches-terminate-if-tree-stops-changing}, all invocations of \func{Search} and \func{SearchNode} perform a finite number of steps.

%\trevor{instead of the following paragraph, prove: (1) if processes take infinitely many steps in fixweight, there are infinitely many operation attempts in fixweight, and (2) if processes take infinitely many steps in fixdegreeorslack, there are infinitely many operation attempts in fixdegreeorslack.}
%
%\textbf{Claim:} If processes take infinitely many steps in \func{FixWeight} (resp., \func{FixSlack}), then there are infinitely many operation attempts in \func{FixWeight} (resp., \func{FixSlack}).
%
%Suppose processes take infinitely many steps in \func{FixWeight}, but there is some configuration $C'$ after which there are no operation attempts in \func{FixWeight}.
%Then, every invocation of \func{FixWeight} after $C'$ returns at line~\ref{lfbslack-fixweight-noweightviol} or line~\ref{lfbslack-fixweight-finalized}.
%By inspection of the code, these invocations of \func{FixWeight} are wait-free (since our implementation of \llt\ is wait-free).
%Therefore, processes must perform infinitely many invocations of \func{FixWeight}.
%Since there are no operation attempts in \func{FixWeight} after $C'$, \trevor{i think this is not true. reverting to other proof, below.}

Suppose there is some configuration $C'$ after which there are no operation attempts in \func{FixWeight} or \func{FixSlack}.
Then, every invocation of \func{FixWeight} after $C'$ returns at line~\ref{lfbslack-fixweight-noweightviol} or line~\ref{lfbslack-fixweight-finalized}, and every invocation of \func{FixSlack} after $C'$ returns at line~\ref{lfbslack-fixslack-nofixneeded}.
By inspection of the code, these invocations of \func{FixWeight} and \func{FixSlack} are wait-free (since their loops are bounded, and our implementation of \llt\ is wait-free).
Therefore, processes must perform infinitely many invocations of \func{FixWeight} and/or \func{FixSlack}.
Since there are no operation attempts in \func{FixWeight} or \func{FixSlack} after $C'$, these invocations must be performed by operation attempts in \ins\ and/or \del.
Each operation attempt in \ins\ or \del\ performs a finite number of invocations of \func{FixWeight} and/or \func{FixSlack}.
Thus, there must be infinitely many operation attempts in \ins\ and/or \del.
However, by Lemma~\ref{lem-bslack}.\ref{claim-bslack-invariants-follow-template}, these operation attempts are in fact template operations.
Thus, infinitely many template operations are performed.
Since this contradicts our assumption, there must be infinitely many operation attempts in \func{FixWeight} and/or \func{FixSlack}.

%Now, suppose there are infinitely many operation attempts in \func{FixWeight} and/or \func{FixSlack}.
Since only finitely many template operations are performed, there must be some configuration $C'$ after which no operation attempt is a template operation.
Thus, by Lemma~\ref{lem-bslack}.\ref{claim-bslack-invariants-follow-template}, after $C'$, every operation attempt in \func{FixWeight} invokes \func{FixWeight} at line~\ref{lfbslack-fixweight-exception1}, and every operation attempt in \func{FixSlack} invokes \func{FixWeight} at line~\ref{lfbslack-fixallweight-fixweight1} or line~\ref{lfbslack-fixallweight-fixweight2}.
We consider two subcases.

\textbf{Case 1:} suppose processes begin infinitely many operation attempts in \func{FixWeight}.
Let $q$ be any process that takes infinitely many steps in \func{FixWeight}.
Consider the first invocation $I$ of \func{FixWeight}$(node)$ by $q$ that begins after $C'$.
Since $I$ is after $C'$, it performs another invocation $I'$ of \func{FixWeight}$(\pi(node))$ at line~\ref{lfbslack-fixweight-exception1} where $\pi(node)$ was read from $node.left$ or $node.right$ in \func{SearchNode} and $\pi(node).weight$ was seen to be zero just before $I$ started $I'$.
Since there are no template operations after $C'$, the tree does not change after $C'$, so $\pi(node)$ is the parent of $node$.
The invocation $I'$ also performs another invocation of \func{FixWeight}$(\pi(\pi(node)))$ where $\pi(\pi(node))$ is the parent of $\pi(node)$ and $\pi(\pi(node)).weight = 0$, and so on.
By Lemma~\ref{lem-bslack}.\ref{claim-bslack-invariants-top-of-tree} and Lemma~\ref{lem-bslack}.\ref{lem-bslack-searchtree}, $node$ is a descendent of $entry$, so $q$ will eventually invoke \func{FixWeight}$(entry)$.
However, $entry.weight \neq 0$, so the invocation of \func{FixWeight}$(entry)$ will not invoke \func{FixWeight} at line~\ref{lfbslack-fixweight-exception1}, which is a contradiction.

%\trevor{high level idea: eventually an invocation of fixweightviolation will not invoke fixweight at line 94, because of $p=entry$...
%%
%more specifically, after $C'$, any invocation of fixweight by q will invoke fixweight at line 94. consider the first invocation I of fixweight(node) by q that begins after $C'$. this invocation perform another invocation I' of fixweight$(\pi(node))$ where $\pi(node)$ was read from node.left or node.right in searchnode, and $\pi(node).weight = 0$. since the tree does not change after $C'$, $\pi(node)$ is the parent of node. the invocation I' then performs another invocation I'' of fixweight$(\pi(\pi(node)))$, where $\pi(\pi(node))$ is the grandparent of node, and $\pi(\pi(node)).weight = 0$, and so on. by lemma~(bslack).(top of tree), node is a descendent of entry, so q will eventually invoke fixweight(entry). however, entry.$weight = 0$, so that invocation of fixweight will NOT invoke fixweight, which is a contradiction.}

\textbf{Case 2:} suppose processes begin infinitely many operation attempts in \func{FixSlack}.
We argue that processes also begin infinitely many operation attempts in \func{FixWeight}.
Suppose not.
Since there are no template operation after $C'$, every operation attempt that occurs in \func{FixSlack} and starts after $C'$ must invoke \func{FixWeight} (in \func{FixAllWeightViolations}).
Furthermore, since we have assumed there are only finitely many operation attempts in \func{FixWeight}, eventually, each invocation of \func{FixWeight} must return before taking any step in the loop.
Consequently, eventually, every invocation of \func{FixWeight}$(node)$ must see $node.weight = 0$ or \llt$(node) = \finalized$.
When a process executing \func{FixAllWeightViolations} invokes \func{FixWeight}$(node)$, it does so after seeing $node.weight = 1$.
Thus, eventually, every invocation of \func{FixWeight}$(node)$ must see \llt$(node) = \finalized$.

However, as we now argue, invocations of \func{FixWeight} can perform only finitely many invocations of \llt\ that return \finalized.
Only finitely many invocations of \llt\ can be performed by operation attempts that begin before $C'$.
Consider any invocation of \func{FixWeight} performed by an operation attempt $A$ that begins after $C'$.
We prove that no invocation of \llt\ performed by $A$ returns \finalized.
Before invoking \func{FixWeight}$(node)$, $A$ performs an invocation of \func{SearchNode}.
This invocation either returns $node$, or it returns the parent $p$ of $node$ and the operation attempt then performs an invocation of \llt$(p)$ that returns a pointer to $node$.
Since the tree does not change after $C'$, in each case, $node$ is in the tree during $A$.
%By inspection of the pseudocode (and the semantics of \sct), nodes are finalized precisely when they are removed from the data structure (and are never reinserted into the data structure).
Consequently, $node$ cannot be finalized.
However, this contradicts our argument above that, eventually, every invocation of \func{FixWeight}$(node)$ must see \llt$(node) = \finalized$.
Thus, our assumption that processes only begin finitely many operation attempts in \func{FixWeight} must be invalid.
Therefore, the proof of this case follows from the proof of Case 1.

Since both cases lead to a contradiction, our original assumption must be incorrect.
Thus, infinitely many template operations must be performed.
\end{chapscxproof}

\begin{thm}
The \rbslack\ operations are non-blocking.
\end{thm}
\begin{chapscxproof}
To derive a contradiction, suppose there is some configuration $C$ after which some processes continue to take steps but no successful \rbslack\ operations occur.
We first argue that eventually the tree stops changing.
Since no successful \rbslack\ operations occur after $C$, the only steps that can change the tree after $C$ are successful invocations of \sct\ performed by invocations of \func{FixWeight} or \func{FixSlack}.
By Lemma~\ref{lem-amortized-rebalancing}, at most $2i(4+\frac 3 2 \big\lfloor \log_{\lfloor\frac{b}{2}\rfloor} \frac{n+i}{2} \big\rfloor) + 2d/(b-1)$ rebalancing steps can be performed after $i$ insertions and $d$ deletions have been performed on an empty \bslack\ (and then no further rebalancing steps can be applied).
Thus, eventually, the tree must stop changing.

This implies that every invocation of \func{Search}, \func{SearchNode} or \func{Get} terminates after a finite number of steps, unless the process executing it crashes.
Consequently, any invocation of \func{Get} performed after $C$ will be successful (unless the process executing it crashes).
Thus, eventually, no process takes a step in \func{Get}. %by a process that does not crash will terminate successfully It follows that no invocation of \func{Get} occurs after $C$.

Suppose processes take infinitely many steps in \func{FixWeight} and/or \func{FixSlack} after $C$.
Then, by Lemma~\ref{lem-bslack-fix-follows-template-infinitely-often}, infinitely many template operations are performed, 
Thus, by Theorem~\ref{thm-dotreeup-progress}, infinitely many of these template updates will succeed.
However, this contradicts the above argument that the tree eventually stops changing.
%Consequently, some template update will succeed after $C$, which will many must succeed after $C$, which is a contradiction.

Thus, eventually, processes take steps only in \ins\ and/or \del. %no processes take steps in \func{FixWeight} or \func{FixSlack}.
Consequently, infinitely many operation attempts are performed in \ins\ and \del.
By Lemma~\ref{lem-bslack}.\ref{claim-bslack-invariants-follow-template}, infinitely many template operations are performed.
Moreover, by Theorem~\ref{thm-dotreeup-progress}, infinitely many of these template updates will succeed, which contradicts our argument that the tree eventually stops changing.
%
%
%Thus, processes perform infinitely many invocations of \tryins\ and/or \trydel, and/or infinitely many iterations of the outer loop in \func{Cleanup}.
%By Lemma~\ref{lem-bslack}.\ref{claim-bslack-invariants-follow-template}, \tryins, \trydel, and iterations of the outer loop in \func{Cleanup}, all follow the template.
%Thus, Theorem~\ref{thm-dotreeup-progress} implies that infinitely many of these template updates will succeed.
%Since the number of rebalancing steps that can be performed is finite if the number of successful insertions and deletions is finite, there must be infinitely many successful insertions and/or deletions.
%Consequently, infinitely many must succeed after $C$, which is a contradiction.
\end{chapscxproof}

\section{Balance proof}

\begin{lem} \label{lem-bslack-u-on-searchpath-to-its-searchkey}
In all configurations, for each node $u$ in the tree, $u$ is on the search path to $u.searchKey$.
\end{lem}
\begin{chapscxproof}
In the initial configuration, there are only two nodes, $entry$ and $root$, and both are on the search path to \textit{every} key, so the invariant holds.
We show that any step $S$ by any process $P$ preserves the invariant.
Since the $searchKey$ field of a node is immutable, the only steps that can affect this invariant are steps that insert nodes into the tree.
The only step that can insert a node into the tree is a successful \sct s $S$ (at line~\ref{lfbslack-ins-scx}, line~\ref{lfbslack-del-scx}, line~\ref{lfbslack-fixweight-absorb-scx}, line~\ref{lfbslack-fixweight-split-scx}, line~\ref{lfbslack-dorootreplace-scx} or line~\ref{lfbslack-docompress-scx}).
$S$ atomically performs one or two of the \rbslack\ updates in Figure~\ref{fig-bslack-updates}.

It is straightforward to verify that, if $S$ occurs at line~\ref{lfbslack-ins-scx}, line~\ref{lfbslack-fixweight-absorb-scx} or line~\ref{lfbslack-fixweight-split-scx}, then each node $u$ inserted by $S$ contains at least one key, and the $searchKey$ of $u$ is its first key.
Thus, in each of these cases, Lemma~\ref{lem-bslack}.\ref{lem-bslack-searchtree} implies that $u$ is on the search path to $u.searchKey$.
If $S$ occurs at line~\ref{lfbslack-del-scx}, then $S$ replaces a leaf $l$ with a new node $n$ that has the same $searchKey$.
Since we have assumed that $l$ is on the search path to $l.searchKey$ before $S$, $n$ is on the search path to $n.searchKey = l.searchKey$ after $S$.
The argument is similar for the case where $S$ occurs at line~\ref{lfbslack-dorootreplace-scx}.
If $S$ occurs at line~\ref{lfbslack-docompress-scx}, three cases arise.

\textbf{Case 1:} $S$ replaces an internal node $p$ and its children with a single new node $n$ (created at line~\ref{lfbslack-createcompressednodes-rootreplace-createnodes}) that has the same $searchKey$ as $p$.
Since we have assumed that $p$ is on the search path to $p.searchKey$ before $S$, $n$ is on the search path to $n.searchKey = p.searchKey$ after $S$.

\textbf{Case 2:} $S$ replaces $p$ and its children with a new node $n$ that has a single child $n_c$ (both created at line~\ref{lfbslack-createcompressednodes-compress-createnodes} when $numNewChildren = 1$), where $n$ and $n_c$ both have the same $searchKey$ as $p$.
Since we have assumed that $p$ is on the search path to $p.searchKey$ before $S$, $n$ is on the search path to $n.searchKey = p.searchKey$ after $S$.
Since $n_c$ is the only child of $n$, it is also on the search path to $n_c.searchKey = n.searchKey$ after $S$.

\textbf{Case 3:} for each node $u$ inserted by $S$ (created at line~\ref{lfbslack-createcompressednodes-compress-createnodes} when $numNewChildren > 1$), the $searchKey$ of $u$ is its first key.
It is straightforward to verify that $u$ contains at least one key (since $numNewChildren > 1$ and there is at most $b-1$ slack shared amongst the new children).
By Lemma~\ref{lem-bslack}.\ref{lem-bslack-searchtree}, $u$ is on the search path to $u.searchKey$.
\end{chapscxproof}

\begin{defn}
If a process $P$ is between line~\ref{lfbslack-searchnode-start} and line~\ref{lfbslack-searchnode-end}, then $location(P)$ is the value of $P$'s local variable $l$ and $target(P)$ is the value of $P$'s local variable $node$.
Otherwise, $location(P) = target(P) = entry$.
\end{defn}

\begin{lem} \label{lem-bslack-target-on-searchpath-from-location}
In all configurations, for each process $P$, if $target(P)$ is not finalized, then $target(P)$ is on the search path to $target(P).searchKey$ starting from $location(P)$.
\end{lem}
\begin{chapscxproof}
In the initial configuration, $location(P) = target(P) = entry$ for each process $P$, so the invariant trivially holds. %there are only two nodes, $entry$ and $root$, and both are on the search path to \textit{every} key, so the invariant holds.
We show that any step $S$ by any process $P$ preserves the invariant.
Since the $searchKey$ field of a node is immutable, the only steps that can affect this invariant are steps that modify child pointers in the tree, and steps that change $target(P)$ or $location(P)$.
The only step $S$ that can modify a child pointer in the tree is a successful \sct s at line~\ref{lfbslack-ins-scx}, line~\ref{lfbslack-del-scx}, line~\ref{lfbslack-fixweight-absorb-scx}, line~\ref{lfbslack-fixweight-split-scx}, line~\ref{lfbslack-dorootreplace-scx} or line~\ref{lfbslack-docompress-scx}.
The only step $S$ that can change $target(P)$ or $location(P)$ is a read of a child pointer at line~\ref{lfbslack-searchnode-start} or line~\ref{lfbslack-searchnode-end}.

\textbf{Case 1:} $S$ occurs at line~\ref{lfbslack-searchnode-start} in an invocation of \func{SearchNode}$(node)$.
Step $S$ changes $location(P)$ to $entry.p_1$ and $target(P)$ to $node$.
By Lemma~\ref{lem-bslack-u-on-searchpath-to-its-searchkey}, $node$ is on the search path to $node.searchKey$ starting from $entry$ (and, hence, starting from $entry.p_1 = location(P)$).

\textbf{Case 2:} $S$ occurs at line~\ref{lfbslack-searchnode-end}.
We use $l(P)$ to denote the value of $location(P)$ before $S$, and $l'(P)$ to denote the value of $location(P)$ after $S$.
Observe that $l(P) \neq node$, since $P$ would have exit the loop at line~\ref{lfbslack-searchnode-exitloop} otherwise.
Since (1) we have assumed that $target(P) = node$ was on the search path to $node.searchKey$ from $l(P)$ before $S$, (2) $l(P) \neq node$, and (3) $l'(P)$ is a child of $l(P)$ that was chosen using $node.searchKey$, $node$ is still on the search path to $node.searchKey$ from $l'(P)$ after $S$.

\textbf{Case 3:} $S$ is a successful \sct.
%
%\trevor{EDIT FROM HERE. idea: to affect the claim, you have to replace a node that is an ancestor of target(P) and a descendent of location(P). (could it be location(P) itself?) argue that any such replacement leaves target(P) on the search path to target(P).searchKey starting from location(P).}
$S$ atomically performs one or two of the \rbslack\ updates in Figure~\ref{fig-bslack-updates}, removing a connected set $R$ of nodes rooted at $top$ from the tree, and replacing them with a set $N$ of newly created nodes rooted at $n$.
In order to affect the invariant, $S$ must remove and replace at least one node on the path from $location(P)$ to $target(P)$.
Three subcases arise.
%First, suppose $S$ removes $location(P)$ and $target(P)$.
%Then, $S$ removes the entire path between the two, and each node on that path is finalized by $S$, so nothing on that path changes after it is removed.
%Thus, since we assumed $target(P)$ was on the search path to $target(P).searchKey$ starting from $location(P)$ before $S$, this is still true after $S$.

\textbf{Subcase 1}: $target(P) \in R$.
In this case, $target(P)$ is finalized after $S$, so the invariant trivially holds.

\textbf{Subcase 2}: $target(P) \notin R$ and $location(P) \in R$.
Let $F_R$ be the fringe of $R$, that is, the set of nodes that are not in $R$, but are pointed to by a node in $R$.
Before $S$, the search path to $target(P).searchKey$ starting from $location(P)$ enters $R$ by passing through $top$, and then exits $R$ by passing through some node $u \in F_R$ (eventually passing through $target(P)$, which is either $u$ or a descendant of $U$).
Since $location(P)$ and the parent of $u$ are both in $R$, and $R$ is a connected set of nodes, all of the nodes on the path from $location(P)$ to the parent of $u$ are removed from the tree and finalized by $S$.
Consequently, they do not change after they are removed.
Thus, the search path to $target(P).searchKey$ starting from $location(P)$ still passes through $u$ after $S$.
Since $S$ does not modify any node on the path from $u$ to $target(P)$, the invariant holds.

\textbf{Subcase 3}: $target(P) \notin R$ and $location(P) \notin R$.
%Then, $S$ removes some other node $u \neq target(P)$ on the path from $location(P)$ to $target(P)$.
%After $S$, the search to $target(P).searchKey$ passes through 
%Let $F_R$ be the fringe of $R$, that is, the set of nodes that are not in $R$, but are pointed to by a node in $R$.
Since $location(P), target(P) \notin R$, and $R$ is a connected set of nodes, one of which we have assumed is on the path from $location(P)$ to $target(P)$, $location(P)$ is a proper ancestor of all nodes in $R$, and some node $u \in F_R$ is an ancestor of $target(P)$.
Before $S$, the search path to $target(P).searchKey$ starting from $location(P)$ enters $R$ by passing through $top$, and then exits $R$ by passing through $u$.
It is straightforward to verify that, regardless of which transformations in Figure~\ref{fig-bslack-updates} are performed by $S$, after $S$, the search path to $target(P).searchKey$ starting from $location(P)$ enters $N$ by passing through $n$, and then exits $N$ by passing through $u$.
(The transformations were designed to satisfy this property.)
%(In other words, if a node in $F_R$ was on the search path to some key before a transformation in Figure~\ref{fig-bslack-updates}, it remains on this search path after the transformation.)
Since $S$ does not modify any node on the path from $u$ to $target(P)$, the invariant holds. %$target(P)$ is on the search path to $target(P).searchKey$ starting from $location(P)$ after $S$.
\end{chapscxproof}

\begin{cor} \label{cor-bslack-searchnode-fail-implies-finalized}
If an invocation of \func{SearchNode}$(node)$ returns \fail, then $node$ is finalized.
\end{cor}

\begin{defn}
Consider an execution in which processes perform invocations of \ins, \del\ and \func{Get} (and these procedures invoke other procedures).
%Invocations of \func{Get} invoke \func{Search}.
%Invocations of \ins\ and \del\ invoke \func{Search}, \func{FixWeight}, \func{FixSlack}, \llt\ and \sct.
%Invocations of \func{FixWeight} and \func{FixSlack} invoke \func{SearchNode}, \func{FixWeight}, \func{FixSlack}, \llt\ and \sct.
%Invocations of \func{FixSlack} additionally invoke \func{NoFixNeeded}
In any given configuration, each process is executing a particular procedure, which is either \ins, \del\ or \func{Get}, or was invoked by another procedure.
Each process has a \textbf{call chain}, which consists of a sequence of invocations $I_1, I_2, ..., I_k$ where $I_{i+1}$ was invoked by $I_i$ for all $i$.
Observe that, for each process, $I_1 \in \{\ins, \del, \func{Get}\}$, and for each $i > 1$, $I_i \notin \{\ins, \del, \func{Get}\}$.
\end{defn}

\begin{defn}
A process $P$ is \textbf{performing weight cleanup for} $node$ in configuration $C$ if its call chain includes \func{FixWeight}$(node)$.
Similarly, $P$ is \textbf{performing slack cleanup for} $node$ in configuration $C$ if its call chain includes \func{FixSlack}$(node)$.
\end{defn}

We now % the following invariant to 
show that each violation in the data structure has a process that is responsible for removing it.
Recall that a violation is defined as an ordered pair containing a type and a node.
%Above, we defined violations as an ordered pair containing a type and a node.
%In contrast, in the following, we think of each violation as an \textit{abstract entity} that is created by an update, and then moves from node to node in the tree until it is eliminated by another update.

\begin{lem} \label{lem-bslack-mapping}
Consider any execution $C_0 \cdot S_1 \cdot C_1 \cdot S_2 \cdot C_2 ...$, where $\mathcal{C} = \{C_0, C_1, ...\}$ is the set of configurations, $\mathcal{S} = \{S_1, S_2, ...\}$ is the set of steps, and $\mathcal{P}$ is the set of processes that take steps.
Let $\mathcal{V}$ and $\mathcal{N}$ be the sets of violations and nodes, respectively, that are ever in the tree.
There exists a \textbf{responsibility} function $\rho : \mathcal{C} \times \mathcal{V} \rightarrow \mathcal{P} \times \mathcal{N}$ such that, for every configuration $C_i$, and every weight (resp., slack or degree) violation $x = \langle type, loc \rangle$ in the tree in configuration $C_i$, the following holds.
Let $\rho(C_i, x) = \langle P, node \rangle$. \\
(1) $node = loc$, and \\
(2) $P$ is performing \ins\ or \del\ in configuration $C_i$, and is also performing weight (resp., slack) cleanup for $node.searchKey$, or will do so before its \ins\ or \del\ terminates, and \\
(3) If $x$ is a violation in configuration $C_{i-1}$, and $\rho(C_{i-1}, x) = \langle Q, - \rangle$ and $\rho(C_i, x) = \langle P, - \rangle$, where $P \neq Q$, then $S_i$ must be a successful \sct\ by $P$.
\end{lem}
\begin{chapscxproof}
In the initial configuration, there are no violations, so the invariant is trivially satisfied.
We show that any step $S_i$ by any process $P$ preserves the invariant.  
We assume that $\rho$ satisfies the claim for configurations $C_0, C_1, ..., C_{i-1}$ and show that it satisfies the claim for the configuration $C_i$ just after $S_i$.
The only step that can cause $P$ to stop performing cleanup for $node.searchKey$ is the termination of an invocation of \func{FixWeight}$(node)$ or \func{FixSlack}$(node)$ by $P$.
The only steps that can add and remove nodes and violations are successful \sct s.
No other steps $S_i$ can cause the invariant to become false.

{\bf Case 1} $S_i$ is the termination of an invocation of \func{FixWeight}$(node)$ by $P$.
We let $\rho(C_i, x) = \rho(C_{i-1}, x)$ for each violation $x$ in configuration $C_i$.
Part (3) of the invariant is immediate.
Since $S_i$ is not a successful \sct, it does not add or remove nodes or violations, so the same violations exist, at the same nodes, in $C_{i-1}$ and $C_i$.
Thus, part (1) of the invariant holds.

By definition, $P$ is performing weight cleanup for $node.searchKey$.
Recall that the call chain for each process starts with a \func{Get}, \ins\ or \del.
If $P$'s call chain starts with \func{Get}, then $P$ cannot be executing \func{FixWeight}, so $P$ must be performing \ins\ or \del.
We show that $\rho$ does not map any weight violation to $node$ in configuration $C_{i-1}$ (and, hence, in configuration $C_i$), so $S_i$ cannot make part (2) of the invariant become false.
$S_i$ is a return statement at line~\ref{lfbslack-fixweight-noweightviol}, line~\ref{lfbslack-fixweight-finalized}, line~\ref{lfbslack-fixweight-searchnode-fail}, line~\ref{lfbslack-fixweight-did-absorb} or line~\ref{lfbslack-fixweight-did-split}.
Suppose $S_i$ occurs at line~\ref{lfbslack-fixweight-noweightviol}
Then, $node.weight = 1$ in configuration $C_{i-1}$, so $\rho$ does not map any weight violation to $node$ in $C_{i-1}$.
Now, suppose $S_i$ occurs at line~\ref{lfbslack-fixweight-finalized}.
Then, $node$ is finalized before $S_i$ (and it remains finalized thereafter, so it is finalized in configuration $C_{i-1}$), so $\rho$ does not map any violation to $node$ in $C_{i-1}$.
Now, suppose $S_i$ occurs at line~\ref{lfbslack-fixweight-searchnode-fail}.
Then, \func{SearchNode} searched for $node.searchKey$ and returned \fail, so Corollary~\ref{cor-bslack-searchnode-fail-implies-finalized} implies that $node$ is finalized before $C_{i-1}$ (so it is finalized in $C_{i-1}$).
Consequently, $\rho$ does not map any violation to $node$ in $C_{i-1}$.
Now, suppose $S_i$ occurs at line~\ref{lfbslack-fixweight-did-absorb} or line~\ref{lfbslack-fixweight-did-split}.
Then, $node$ was finalized by the preceding \sct, so $\rho$ does not map any violation to $node$ in $C_{i-1}$.

{\bf Case 2} $S_i$ is the termination of an invocation of \func{FixSlack}$(node)$ by $P$.
We let $\rho(C_i, x) = \rho(C_{i-1}, x)$ for each violation $x$ in configuration $C_i$.
Part (3) of the invariant is immediate.
Since $S_i$ is not a successful \sct, it does not add or remove nodes or violations, so the same violations exist, at the same nodes, in $C_{i-1}$ and $C_i$.
Thus, part (1) of the invariant holds.

By definition, $P$ is performing slack cleanup for $node.searchKey$.
Recall that the call chain for each process starts with a \func{Get}, \ins\ or \del.
If $P$'s call chain starts with \func{Get}, then $P$ cannot be executing \func{FixSlack}, so $P$ must be performing \ins\ or \del.
We show that $\rho$ does not map any slack or degree violation to $node$ in configuration $C_{i-1}$ (and, hence, in configuration $C_i$), so $S_i$ cannot make part (2) of the invariant become false.
$S_i$ is a return statement at line~\ref{lfbslack-fixslack-nofixneeded}, line~\ref{lfbslack-fixslack-searchnode-fail}, line~\ref{lfbslack-fixslack-did-rootreplace} or line~\ref{lfbslack-fixslack-did-compress}.
We start with the easy cases.
If $S_i$ occurs at line~\ref{lfbslack-fixslack-searchnode-fail}, then \func{SearchNode} searched for $node.searchKey$ and returned \fail\ before $C_{i-1}$, so Corollary~\ref{cor-bslack-searchnode-fail-implies-finalized} implies that $node$ is finalized before $C_{i-1}$.
Consequently, in configuration $C_{i-1}$, $node$ is finalized, so $\rho$ does not map any violation to $node$.
If $S_i$ occurs at line~\ref{lfbslack-fixslack-did-rootreplace} or line~\ref{lfbslack-fixslack-did-compress}, then $node$ was finalized by the preceding \sct, so $\rho$ does not map any violation to $node$ in configuration $C_{i-1}$.

Now, suppose $S_i$ occurs at line~\ref{lfbslack-fixslack-nofixneeded}.
Then, it immediately follows an invocation of \func{NoFixNeeded} that returns \true\ at line~\ref{lfbslack-nofixneeded-leaf}, line~\ref{lfbslack-nofixneeded-finalized} or line~\ref{lfbslack-nofixneeded-nodegreeorslack}.
If it returns \true\ at line~\ref{lfbslack-nofixneeded-leaf}, then $node$ is a leaf, so there cannot be any violation at $node$ (so $\rho$ cannot map any violation to $node$ in $C_{i-1}$).
If it returns \true\ at line~\ref{lfbslack-nofixneeded-finalized}, then $node$ is finalized before $C_{i-1}$ (so $\rho$ cannot map any violation to $node$ in $C_{i-1}$).
If it returns \true\ at line~\ref{lfbslack-nofixneeded-nodegreeorslack}, then \func{NoFixNeeded} performs \llt$(node)$ and obtains a snapshot of the children of $node$, and sums the (immutable) degree fields of these children, then determines that there was no degree or slack violation at $node$ in the configuration $C_j$ ($j < i-1$) just before the \llt\ was performed.
Since there was no degree or slack violation at $node$ in $C_j$, $\rho$ does not map any degree or slack violation to $node$ in $C_j$.
By part (3) of the invariant, prior to $S_i$, a process can take responsibility for violations \textit{only} by performing a successful \sct, and cannot cause other processes to become responsible for violations.
Consequently, since $\rho$ does not map any degree or slack violation to $node$ in configuration $C_j$, and there is no successful \sct\ by $P$ after $C_j$ and before $S_i$ (by inspection of the code), $\rho$ cannot map any degree or slack violation to $\langle P, node \rangle$ in configuration $C_{i-1}$ (although $\rho$ \textit{might} map a degree or slack violation to $\langle Q, node \rangle$, where $Q \neq P$).

{\bf Case 3} $S_i$ is a successful \sct.
%We define $\rho(C_i, x)$ and show that , case-by-case, after each update in Figure~\ref{fig-bslack-updates}.
Let $parent$ be the node whose pointer is changed by $S_i$.
Recall that $x$ is a violation of type $type$ at $loc$ in configuration $C_i$ (just after $S_i$).
We consider four subcases depending on the value of $x = \langle type, loc \rangle$.

%\trevor{idea: each violation (1) is still in the tree at the same node as before and $\rho$ has not changed for it, or (2) is eliminated, or (3) is one of the ones we list in the table, that we take responsibility for.}

%\trevor{show part (1)}
%We show part (1) of the invariant holds for each violation $x$ that exists in configuration $C_i$.
%
%\trevor{show part (2)}
%Recall that the call chain for each process starts with a \func{Get}, \ins\ or \del.
%If $P$'s call chain starts with \func{Get}, then $P$ cannot perform \sct\ (since \func{Get} does not perform \sct, and does not invoke any procedure that performs \sct).
%Therefore, $P$ is performing \ins\ or \del.
%
%\trevor{show part (3)}
%
%We show that $\rho(C_i, x)$ is well defined.
%\trevor{[note] this case seems quite a lot simpler than in the other trees, since $P$ simply takes responsibility for every violation at a new node, and any new slack violation at $parent$, and there are no other changes to the violations in the tree, so $\rho'=\rho$ for those violations.}

\textit{Subcase 1}: $type$ is weight or degree and $loc$ was in the tree in configuration $C_{i-1}$.
In this case, we let $\rho(C_i, x) = \rho(C_{i-1}, x)$ (preserving part (3) of the invariant).
Since tag bits are immutable and the number of pointers in nodes do not change, no transformation in Figure~\ref{fig-bslack-updates} can create a new weight or degree violation at a node that was already in the data structure.
Thus, $x$ was a violation (that occurred at $loc$) in configuration $C_{i-1}$, so $\rho(C_{i-1}, x)$ is well-defined.
Since $\rho(C_i, x) = \rho(C_{i-1}, x)$ and $loc$ is not changed or removed by $S_i$, parts (1) and (2) of the invariant are preserved.

\textit{Subcase 2}: $type$ is slack and $loc \neq parent$ was in the tree in configuration $C_{i-1}$.
In this case, we let $\rho(C_i, x) = \rho(C_{i-1}, x)$ (preserving part (3) of the invariant).
As Figure~\ref{fig-bslack-violation-movement} shows, only \func{Delete}, \func{Overflow}, \func{Split} and \func{Compress} can create a slack violation at $parent$.
Furthermore, $S_i$ cannot create a new slack violation at any other node $loc \neq parent$ that was in the tree in configuration $C_{i-1}$.
Thus, $x$ was a violation (that occurred at $loc$) in configuration $C_{i-1}$, so $\rho(C_{i-1}, x)$ is well-defined.
Since $\rho(C_i, x) = \rho(C_{i-1}, x)$ and $loc$ is not changed or removed by $S_i$, parts (1) and (2) of the invariant are preserved.

\textit{Subcase 3}: $type$ is slack and $loc = parent$ was in the tree in configuration $C_{i-1}$.
In this case, we let $\rho(C_i, x) = \langle P, parent \rangle$ (satisfying part (3) of the invariant).
By Observation~\ref{obs-lfbslack-finalized-when-removed} and the semantics of \sct, $S_i$ does not remove $parent$ from the tree, so $parent$ is in the tree just after $S_i$.
Since $x$ occurs at $loc$ in configuration $C_i$ and $loc = parent$, and $\rho(C_i, x) = \langle P, parent \rangle$, part (1) of the invariant holds.

Recall that the call chain for each process starts with a \func{Get}, \ins\ or \del.
If $P$'s call chain starts with \func{Get}, then $P$ cannot perform \sct, so $P$ must be performing \ins\ or \del.
We argue that $P$ performs an invocation $I$ of \func{FixSlack}$(parent)$ before its \ins\ or \del\ terminates.
By Figure~\ref{fig-bslack-violation-movement}, in order for there to be a slack violation at $parent$ after $S_i$, $S_i$ must perform Compress (but not Root-Replace), Delete, Overflow or Split.
If $S_i$ performs Compress (but not Root-Replace) at line~\ref{lfbslack-docompress-scx}, then $doRootReplace = \false$, so \func{DoCompress} returns $gp = parent$.
Consequently, $P$ performs $I$ at line~\ref{lfbslack-fixslack-fixdegreeorslack2}.
If $S_i$ performs Delete (at line~\ref{lfbslack-del-scx}), then $P$ performs $I$ at line~\ref{lfbslack-del-fixslack}.
If $S_i$ performs Overflow (at line~\ref{lfbslack-ins-scx}), then $P$ performs $I$ at line~\ref{lfbslack-ins-fixslack}.
If $S_i$ performs Split (at line~\ref{lfbslack-fixweight-split-scx}), then $P$ performs $I$ at line~\ref{lfbslack-fixweight-fixslack-parent}.
Thus, part (2) of the invariant holds.

\textit{Subcase 4}: $loc$ was not in the tree in configuration $C_{i-1}$.
In this case, $\rho(C_i, x) = \langle P, loc \rangle$ (satisfying parts (1) and (3) of the invariant).
Recall that the call chain for each process starts with a \func{Get}, \ins\ or \del.
If $P$'s call chain starts with \func{Get}, then $P$ cannot perform \sct, so $P$ must be performing \ins\ or \del.
It remains to argue that, if $x$ is a weight (resp., degree or slack) violation, then $P$ performs an invocation $I$ of \func{FixWeight}$(loc)$ (resp., \func{FixSlack}$(loc)$) before its \ins\ or \del\ terminates.
Since $loc$ is not in the tree before $S_i$, $S_i$ must have performed an update that created and inserted $loc$ into the tree.
Consequently, in the terminology of Figure~\ref{fig-bslack-violation-movement}, $loc$ is $n$ or a child of $n$.
(We know $loc \neq parent$, because $parent$ is not newly created by the update.)
Figure~\ref{fig-bslack-violation-movement} shows precisely where $x$ could be after each type of update.

For example, if $S_i$ performs Split, then $x$ is either a weight violation at $loc = n$ or a slack violation at $loc \in \{n, n.p_1, n.p_2\}$.
(We know $loc \neq parent$, because $parent$ is not a newly created node.)
In this case, $S_i$ occurs at line~\ref{lfbslack-fixweight-split-scx}.
$P$ will invoke \func{FixSlack} for $n$ and its two children at line~\ref{lfbslack-fixweight-fixslack-parent} before its invocation of \ins\ or \del\ terminates.
If $gp = entry$, then $n$ is the new root of the \rbslack, and its weight was set to one at line~\ref{lfbslack-fixweight-rootzero}, so there is no weight violation at $n$.
Otherwise, $P$ will also invoke \func{FixWeight}$(n)$ at line~\ref{lfbslack-fixweight-fixweight1} before its invocation of \ins\ or \del\ terminates.
Consequently, part (2) of the invariant holds in this case.
The argument for the other updates is very similar.
\end{chapscxproof}

\begin{cor}
The lock-free \rbslack\ is a \bslack\ whenever no process is executing \ins\ or \del.
\end{cor}
\begin{chapscxproof}
If no process is executing \ins\ or \del, then Lemma~\ref{lem-bslack-mapping}.2 implies that there are no violations in the tree.
A \rbslack\ with no violations is a \bslack.
\end{chapscxproof}

Curiously, the proof of Case 2 in the preceding lemma was significantly more complex than the corresponding cases, in the corresponding lemmas, for the Chromatic tree, AVL tree and relaxed $(a,b)$-tree.
We briefly explain why.
In the other trees, violations can only be created at newly inserted nodes, because whether a node has a violation depends only on the node's immutable fields.
However, in a \rbslack, whether a node has a violation depends not only on the node's immutable fields, but also on the immutable fields of its \textit{children}.
Since the children of a node can change, it is possible for a node to be in the tree and have no violation in one configuration, and, in a subsequent configuration, be in the tree and \textit{have} a violation.
This possibility makes the proof of Case 2 quite subtle. %Addressing this possibility adds some complexity to this proof.

\section{Modifications for amortized constant rebalancing} \label{sec-lfbslack-constant-rebalancing-mod}

We briefly explain how to modify our pseudocode to obtain amortized constant rebalancing, as described in Section~\ref{sec-constant-rebalancing}.
This requires two changes.
First, the definition of a slack violation is modified slightly, so that a slack violation occurs at a node $u$ whenever the total slack shared amongst the children of $u$ is less than $b+u.d$ (instead of %whenever the total slack is less than 
$b$).
Second, the Compress update changes so that the total degree $c$ shared amongst the children before the update are evenly distributed amongst $\lceil c / (b-1) \rceil$ new nodes after the update (instead of $\lceil c / b \rceil$ new nodes).
These changes do not affect the proof.

The specific changes to the pseudocode follow.
At line~\ref{lfbslack-fixslack-change} of \func{FixSlack}, the condition $slack < b$ becomes $slack < b + p.d$.
Similarly, at line~\ref{lfbslack-nofixneeded-change} of \func{NoFixNeeded}, the condition $slack \ge b$ becomes $slack \ge b + node.d$.
Finally, at line~\ref{lfbslack-createcompressednodes-change} of \func{CreateCompressedNodes}, $numNewChildren$ is set to $\lceil pGrandDegree / (b-1) \rceil$.

\subsection{Adding a range query operation} \label{sec-lfbslack-rq}

A \func{RangeQuery} operation takes, as its arguments, two keys $low$ and $high$, and returns all key-value pairs present in the dictionary whose keys are in $[low, high)$.
These operations are commonly used in databases.
Consequently, a \func{RangeQuery} operation is a common addition to the (ordered) dictionary ADT.

We give a simple implementation of a lock-free \func{RangeQuery} operation for the \rbslack\ that uses \llt\ and \vlt\ to obtain a snapshot of the desired range.
Recall that a \vlt$(V)$ by process $p$ returns \true\ if no node $u \in V$ has changed since $p$ last performed \llt$(u)$, and \false\ otherwise.

\begin{figure}[p]
\begin{framed}
\preplisting
\begin{lstlisting}[mathescape=true]
 //\func{RangeQuery}$(low, high)$
 //\textbf{retry:}
   //$result :=$ empty sequence
   //$V :=$ empty sequence
   //$q :=$ empty queue
  
   //\com Depth first traversal (of relevant subtrees)
   $q.enqueue(entry)$
   while not $q.isEmpty()$
     $node := q.dequeue()$
     if $\llt(node) \in \{\fail, \finalized\}$ then goto //\textbf{retry}
     //Let $c_1, c_2, ..., c_n$ be the child pointers returned by \llt$(node)$
     $nkeys := n - 1$

     //\com Visit $node$
     //Add $node$ to $V$
     if $node.leaf$ //\hfill\com $node$ is a leaf, so we record its relevant keys
       //Add the keys in $node \cap [low, high)$ to $result$
     else //\hfill\com $node$ is internal, so we explore its children
       //\com Find right-most sub-tree that could contain a key in $[low, high)$
       $r := nkeys$
       while $r > 0$ and $high < node.k_{r-1}$ //\hfill\com Subtree rooted at $c_r$ contains only keys greater than $high$
         $r := r - 1$ //\hfill\com Skip child $c_r$ of $node$
       //\com Find left-most sub-tree that could contain a key in $[low, high)$
       $l := 0$ //\hfill\com Index of smallest key in $node$
       while $l < nkeys$ and $low \ge k_l$ //\hfill\com Subtree rooted at $c_l$ contains only keys less than $low$
         $l := l + 1$ //\hfill\com Skip child $c_l$ of $node$

       //\com Enqueue relevant children to continue the BFS
       for $i = l..r$ do $q.enqueue(c_i)$
      
   //\com Validation
   if not $\vlt(V)$ then goto //\textbf{retry}
   return $result$
\end{lstlisting}
\end{framed}
\vspace{-5mm}
	\caption{Pseudocode for \func{RangeQuery}.}
	\label{code-lfbslack-rq}
\end{figure}

The implementation of \func{RangeQuery}$(low, high)$ appears in Figure~\ref{code-lfbslack-rq}.
It first performs a breadth-first traversal of the tree, pruning any subtrees that it determines cannot intersect $[low, high)$.
The traversal visits the children of a node from left to right, and performs \llt\ on each node it visits.
If any \llt\ returns \fail\ or \finalized, then a node visited by the traversal has changed, so the \func{RangeQuery} restarts its traversal.
So, suppose all invocations of \llt\ return snapshots.
Then, the \func{RangeQuery} invoked \vlt$(V)$, where $V$ is the sequence of nodes visited by the traversal.
If the \vlt\ returns \false, then a node visited by the traversal has changed (or replaced), so the \func{RangeQuery} restarts its traversal.
Otherwise, the nodes visited by the traversal form a snapshot, and the \func{RangeQuery} returns $K \cap [low, high)$, where $K$ is the set of keys in the \textit{leaves} in $V$.

Each \func{RangeQuery} that performs an invocation of \vlt$(V)$ which returns \true\ is linearized at this invocation of \vlt.
We briefly explain why this algorithm is correct.
Consider a \func{RangeQuery} $R$ that returns $K \cap [low, high)$.
Since (1) all child pointers followed by the traversal are obtained from snapshots returned by \llt, (2) the \vlt\ succeeds only if none of these nodes have changed, and (3) a node in the tree is excluded from $V$ only if its subtree cannot intersect $[low, high)$, $K$ must contain all keys that are in $[low, high)$ when the \vlt\ occurs.
Thus, $K \cap [low, high)$ contains precisely the keys in the tree that are in $[low, high)$ when $R$ is linearized.

Now, we briefly explain why the algorithm satisfies lock-freedom.
With this algorithm, individual \func{RangeQuery} operations can be susceptible to starvation is updates are frequent, but they are guaranteed to succeed in a finite number of steps if there are no updates.
Lock-freedom is satisfied as long as updates continue to succeed, and once updates stop succeeding, any ongoing \func{RangeQueries} will eventually succeed (satisfying lock-freedom), unless the processes performing them crash.

\section{Experiments}

We implemented the lock-free \rbslack\ in C++, using a fast memory reclamation scheme called DEBRA that is described in Chapter~\ref{chap-debra}.
This implementation includes the optimization for amortized constant rebalancing that is described in Section~\ref{sec-lfbslack-constant-rebalancing-mod}.
We then performed a series of experiments to compare the lock-free \rbslack\ with the unbalanced BST and relaxed $(a,b)$-tree implementations described in Section~\ref{sec-abtree-exp}.
(These implementations also reclaim memory using DEBRA.)

For the relaxed $(a,b)$-tree, we set $a = 6$ and $b = 16$ (for the same reason that we detailed in Section~\ref{sec-abtree-exp}).
Similarly, for the \rbslack, we set the maximum degree $b$ to 16.
A single node size of 224 bytes was used for the \rbslack\ and the relaxed $(a,b)$-tree (just large enough to hold 16 4-byte keys and 8-byte pointers, as well as the meta-data for the tree algorithm and for \llt\ and \sct).
A single \op\ size of 320 bytes was used for the \rbslack\ and the relaxed $(a,b)$-tree.
%Thus, each node occupies nearly four cache lines.
In the BST, each node occupies 40 bytes (enough to hold a 4-byte key, 4-byte value, two 8-byte child pointers, and meta-data for \llt\ and \sct), and each \op\ occupies 120 bytes.
Nodes and \op s were not padded. % in any implementation.

\paragraph{Experimental system}

%We performed experiments on two different experimental systems.
%Each of these systems has a non-uniform memory architecture (NUMA) in which threads have significantly different access costs to different parts of memory depending on which processor they are currently executing on.
%
%The first is a 2-socket Intel E7-4830 v3 with 12 cores per socket and 2 hyperthreads (HTs) per core, for a total of 48 threads.
%Each core has a private 32KB L1 cache and 256KB L2 cache (which is shared between HTs on a core).
%All cores on a socket share a 30MB L3 cache.
%
%The second is a 4-socket AMD Opteron 6380 with 8 cores per socket and 2 HTs per core, for a total of 64 threads.
%Each core has a private 16KB L1 data cache and 2MB L2 cache (which is shared between HTs on a core).
%All cores on a socket share a 6MB L3 cache.
%
%Both machines have 128GB of RAM, and run Ubuntu 14.04 LTS.
%All code was compiled with the GNU C++ compiler (G++) 4.8.4 with build target x86\_64-linux-gnu and compilation options \texttt{-std=c++0x -mcx16 -O3}.
%Thread support was provided by the POSIX Threads library.
%We used the scalable allocator jemalloc 4.2.1~\cite{Evans:2006}, which greatly improved performance for all algorithms.
%
%We pin threads to cores on the different sockets in a round-robin fashion.
%On the Intel machine, hyperthreading is engaged for thread counts 25-48.
%On the AMD machine, hyperthreading is engaged for thread counts 33-64.
%%(Consequently, the effect of hyperthreading can be seen on our graphs.)

We performed experiments on a 4-socket AMD Opteron 6272 with 16 cores per socket, for a total of 64 threads.
Each core has a private 64KB L1 data cache, and 2MB L2 cache that is shared with one other core.
All cores on a socket share a 16MB L3 cache.
The size of a cacheline is 64 bytes.
This system has a non-uniform memory architecture (NUMA) in which threads have significantly different access costs to different parts of memory depending on which processor they are currently executing on.
The machine has 128GB of RAM, and runs Ubuntu 14.04 LTS.

All code was compiled with the GNU C++ compiler (G++) 4.8.4 with build target x86\_64-linux-gnu and compilation options \texttt{-std=c++0x -mcx16 -O3}.
Thread support was provided by the POSIX Threads library.
We used the scalable allocator jemalloc 4.2.1~\cite{Evans:2006}, which greatly improved performance for all algorithms.
We used the Performance Application Programming Interface (PAPI) library~\cite{Browne:2000} to collect statistics from hardware performance counters to determine cache miss rates, stall times, instructions retired, and so on.
We pin threads to cores on the different sockets in a round-robin fashion (such that at most one thread is pinned to each core).

\subsection{Steady-state performance} \label{sec-lfbslack-exp1}

In this section, we describe a simple randomized benchmark for different workloads consisting of basic dictionary operations (\func{Get}, \ins\ and \del) on uniformly random keys drawn from fixed key ranges.
The goal is to study the performance of these basic dictionary operations for each of the three tree algorithms in the steady state (wherein the tree contains approximately half of the keys in the key range, and is neither growing nor shrinking).
Since the three tree algorithms are all leaf-oriented, updates always perform modifications tree close to a leaf.
Consequently, the internal structure of the tree is somewhat calcified in the steady state (especially near the top of the tree).
%Consequently, in the steady state, the internal nodes in the tree are modified less frequently, the higher they appear in the tree.
%
Since \rbslack s are intended to be used in workloads with many \func{Get}s and few updates, we study workloads containing up to 5\% \ins\ and 5\% \del\ operations.

\begin{figure}[t]
    \centering
    \setlength\tabcolsep{0pt}
    \begin{tabular}{m{0.04\linewidth}m{0.31\linewidth}m{0.31\linewidth}m{0.31\linewidth}}
        &
        \fcolorbox{black!50}{black!20}{\parbox{\dimexpr \linewidth-2\fboxsep-2\fboxrule}{\centering {\large 0i-0d}}} &
        \fcolorbox{black!50}{black!20}{\parbox{\dimexpr \linewidth-2\fboxsep-2\fboxrule}{\centering {\large 1i-1d}}} &
        \fcolorbox{black!50}{black!20}{\parbox{\dimexpr \linewidth-2\fboxsep-2\fboxrule}{\centering {\large 5i-5d}}}
        \\
        \vspace{-3mm}\rotatebox{90}{\normalsize Key range $[0, 10^7)$} &
        \includegraphics[width=\linewidth]{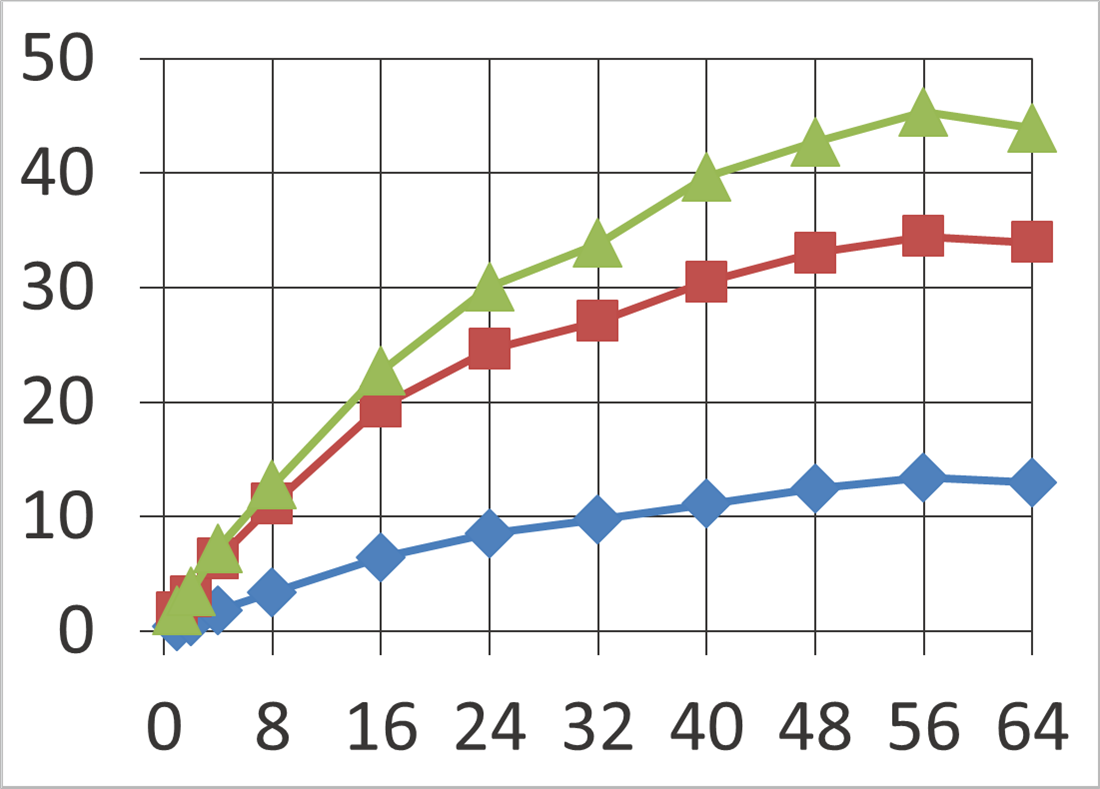} &
        \includegraphics[width=\linewidth]{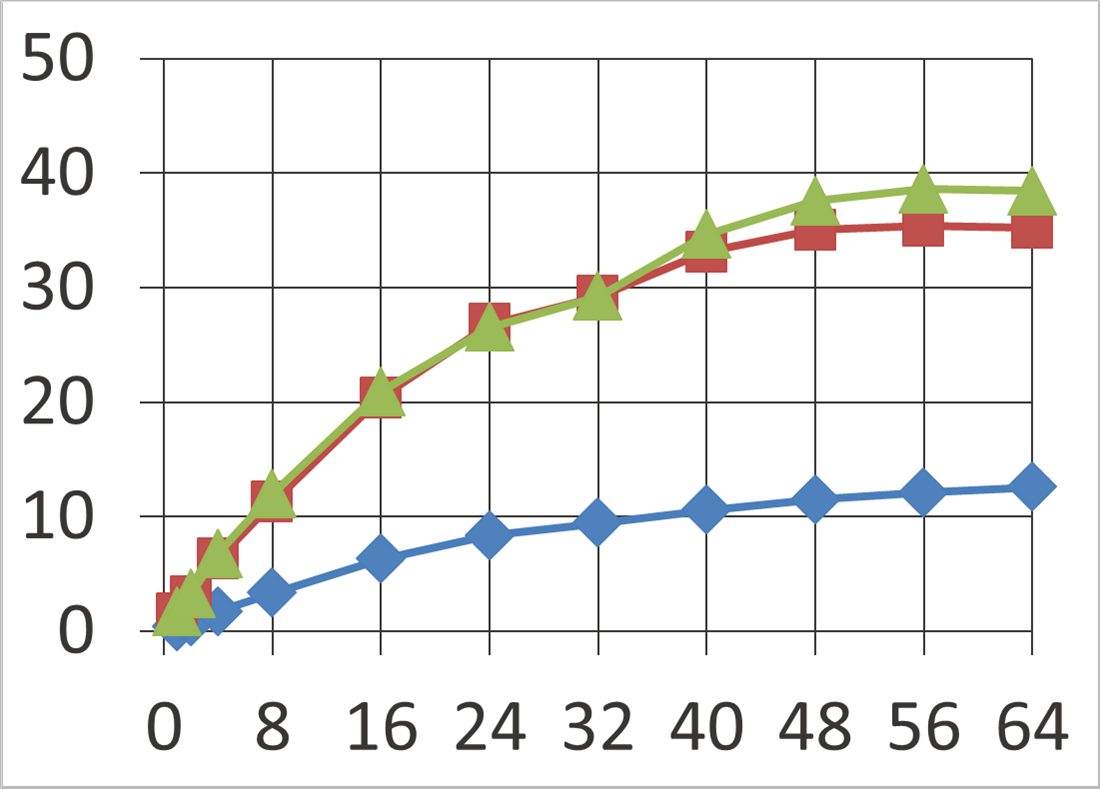} &
        \includegraphics[width=\linewidth]{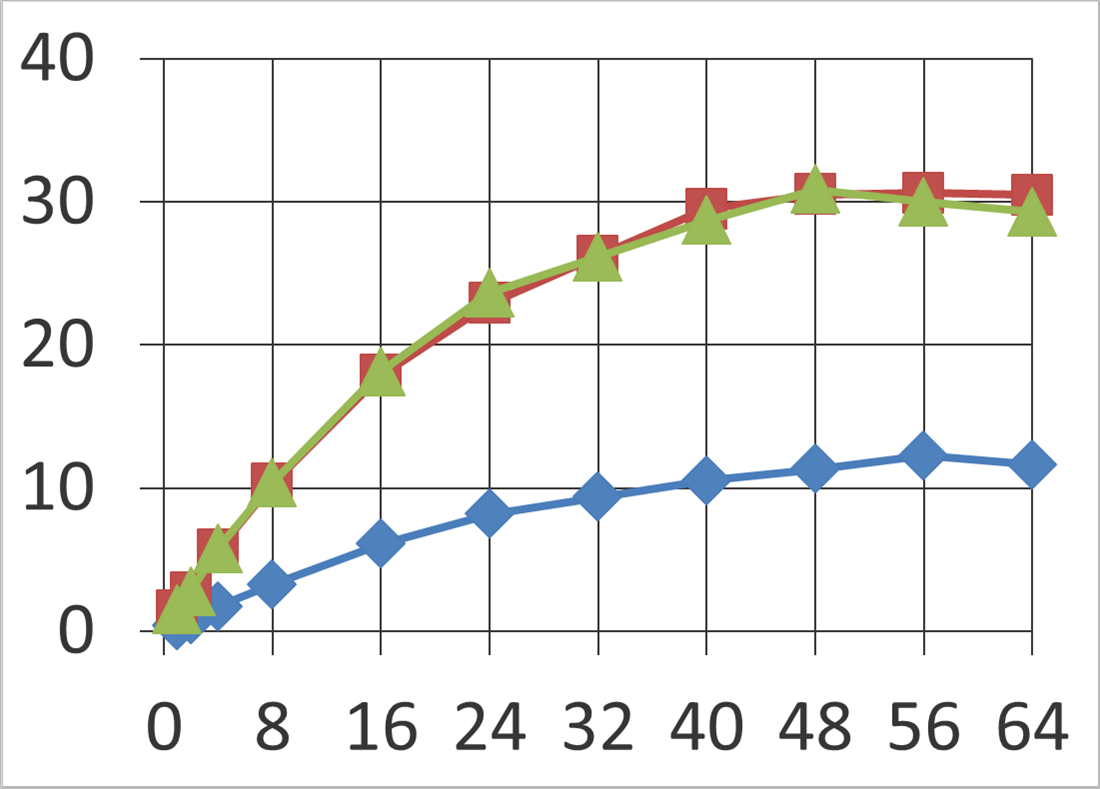}
        \\
        \vspace{-9mm}\rotatebox{90}{\normalsize Key range $[0, 10^6)$} &
        \vspace{-6mm}\includegraphics[width=\linewidth]{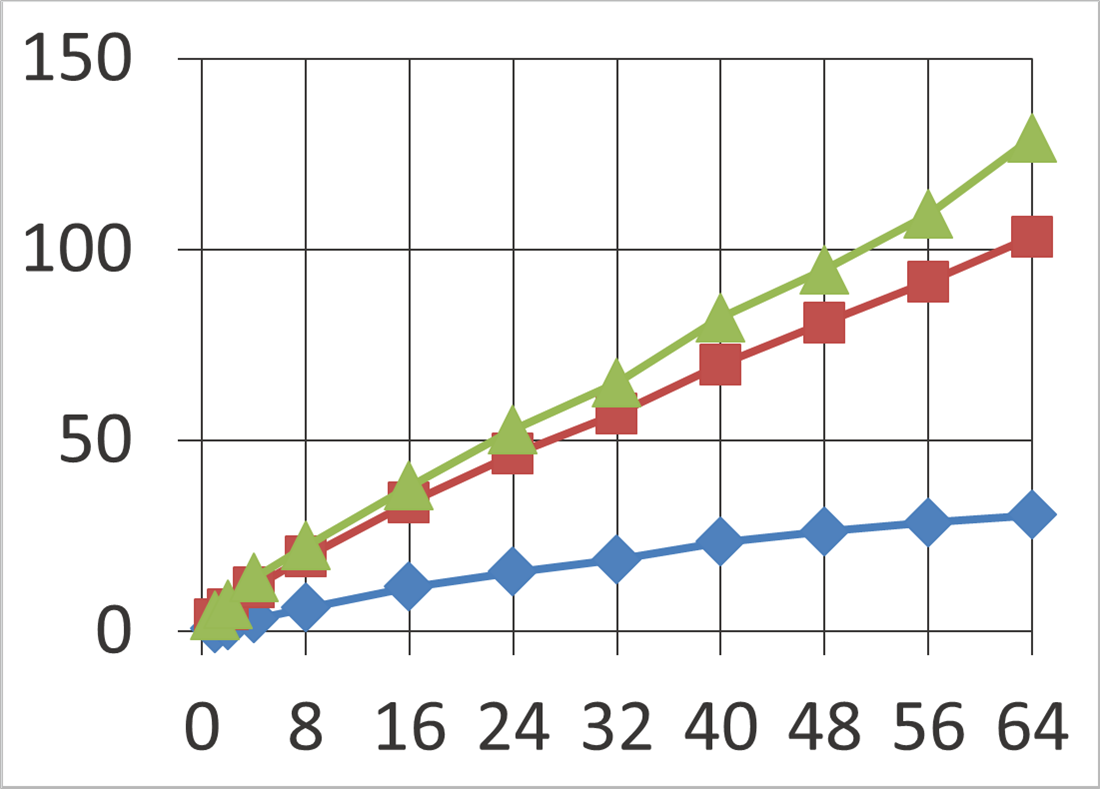} &
        \vspace{-6mm}\includegraphics[width=\linewidth]{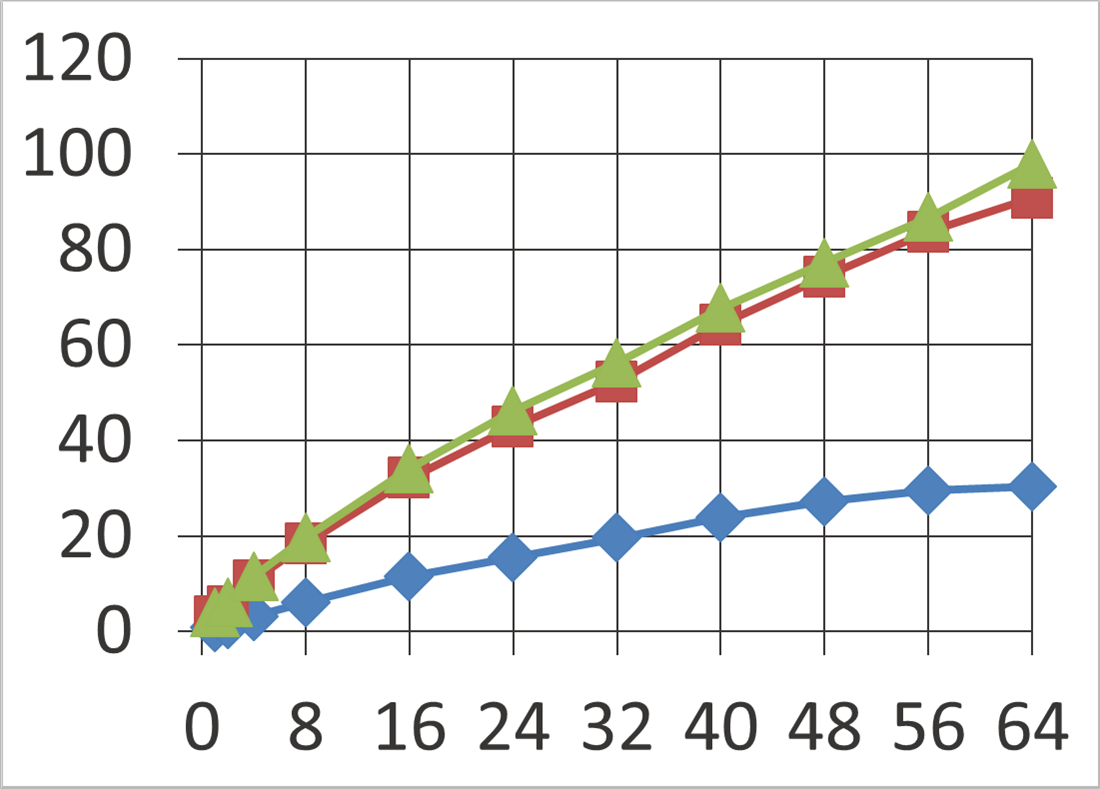} &
        \vspace{-6mm}\includegraphics[width=\linewidth]{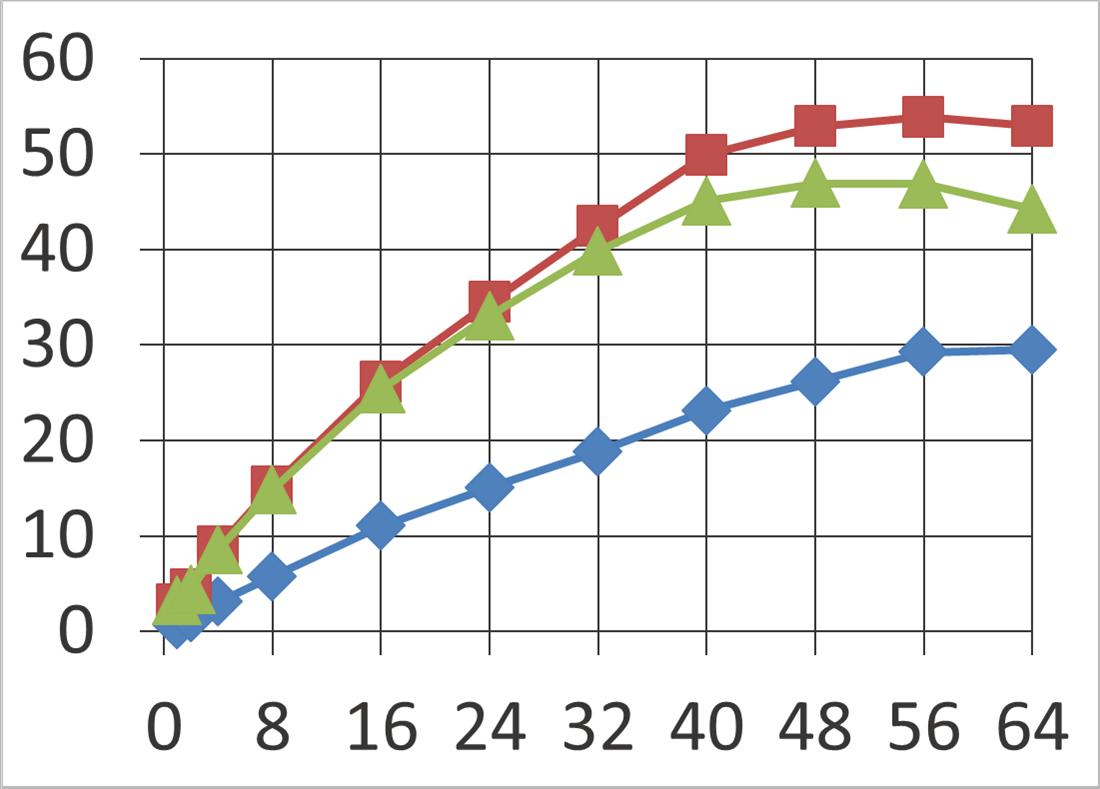}
        \\
        \vspace{-9mm}\rotatebox{90}{\normalsize Key range $[0, 10^5)$} &
        \vspace{-6mm}\includegraphics[width=\linewidth]{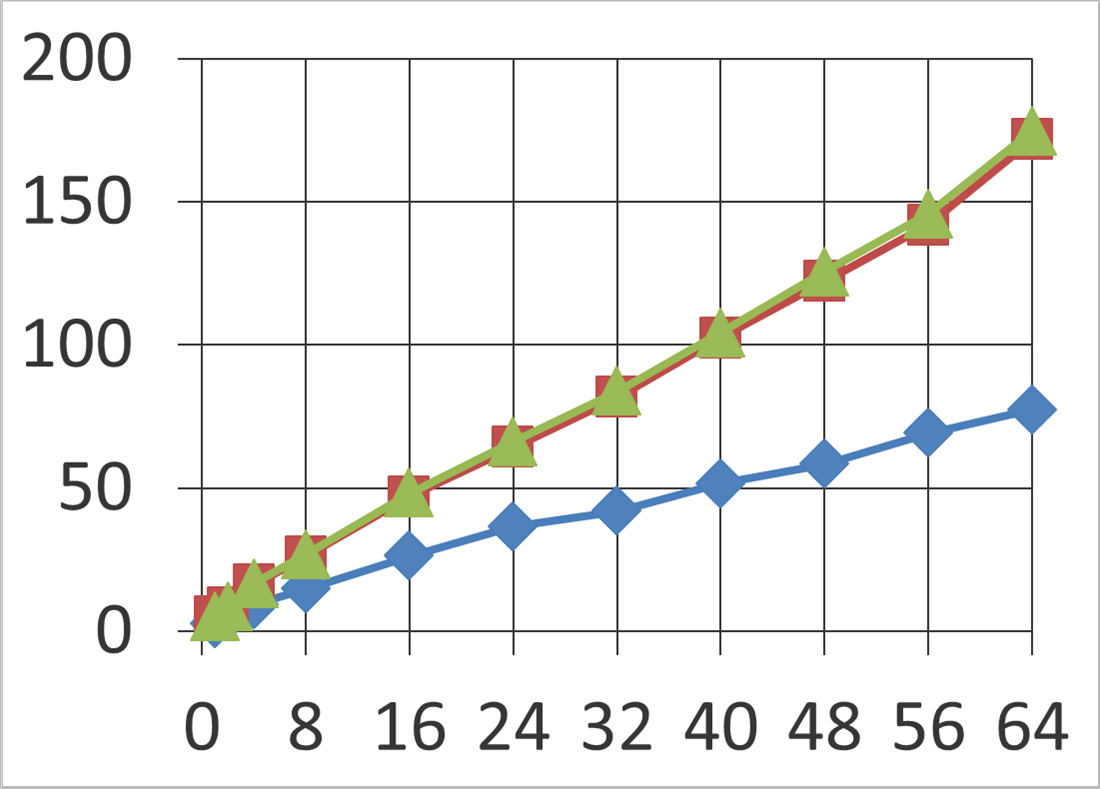} &
        \vspace{-6mm}\includegraphics[width=\linewidth]{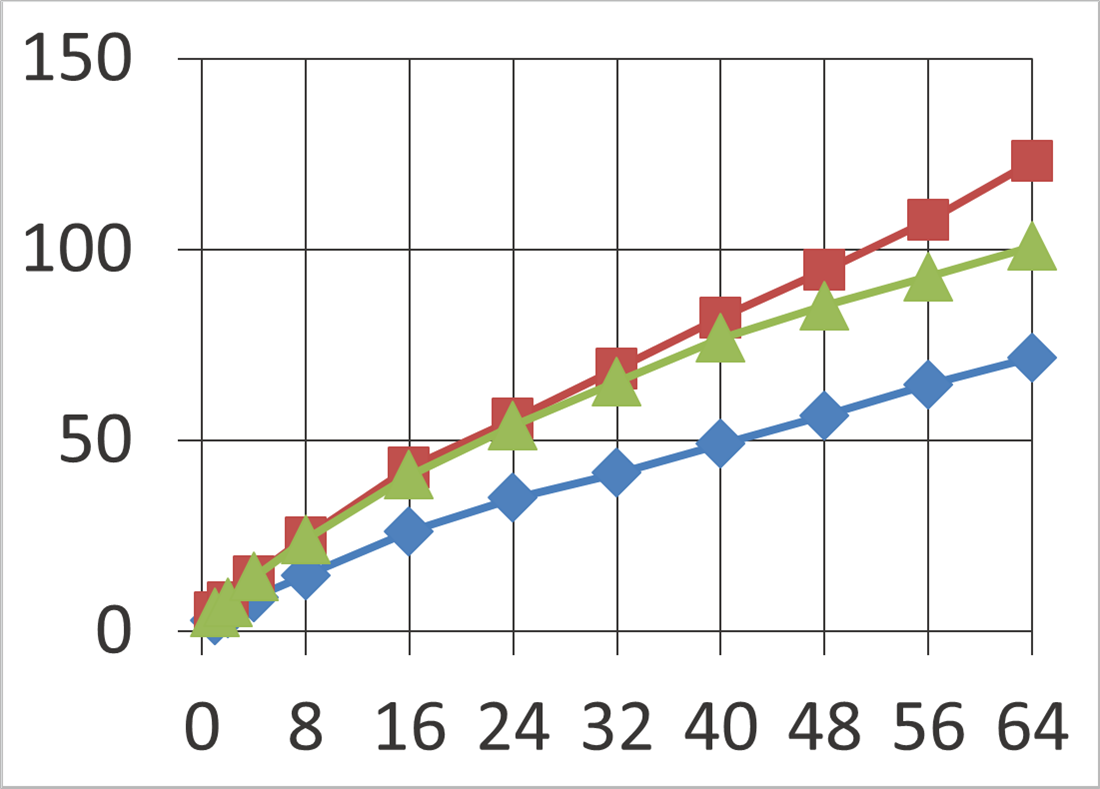} &
        \vspace{-6mm}\includegraphics[width=\linewidth]{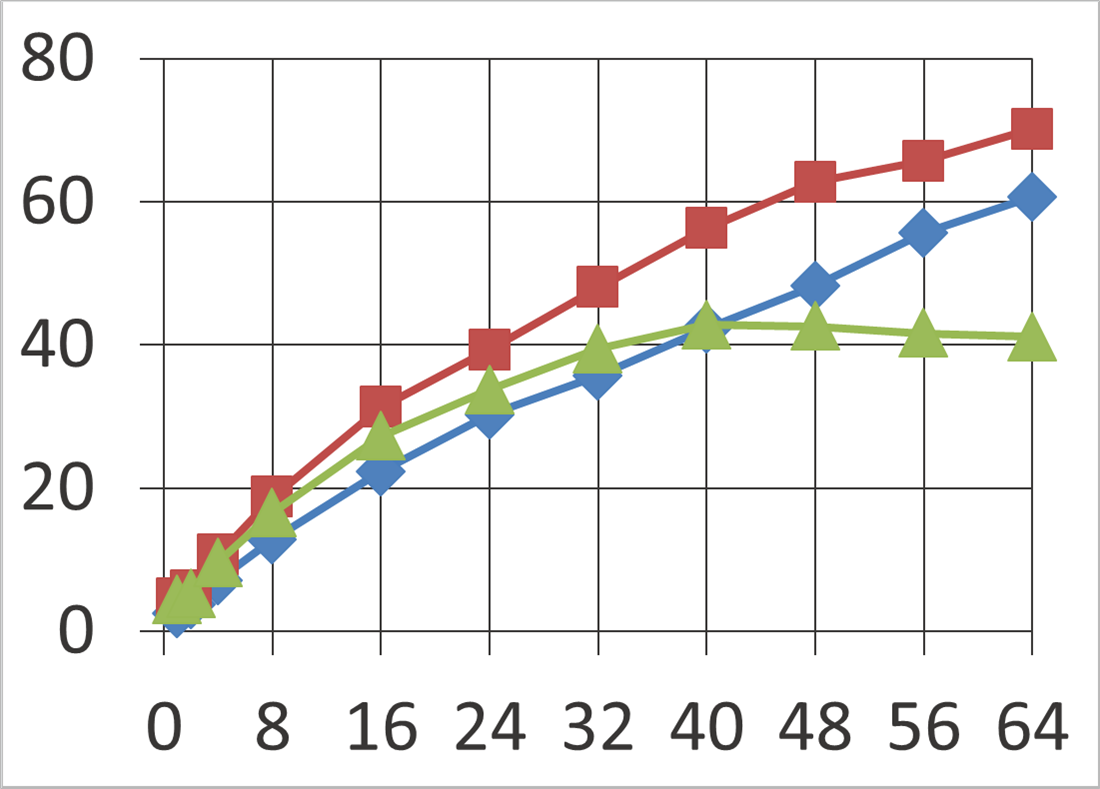}
        \\
    \end{tabular}
    \vspace{-2mm}
%    \hspace{0.07\linewidth}
	\includegraphics[width=0.8\linewidth]{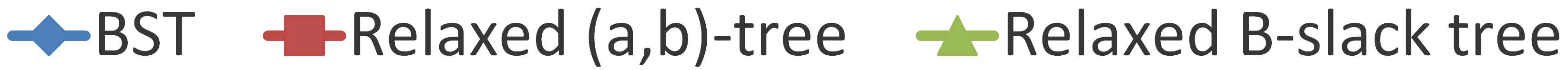}
\caption{Experimental results.
The x-axis represents the number of concurrent threads.
The y-axis represents operations per microsecond.}
\label{fig-bslack-exp}
\end{figure}

%We ran a simple randomized microbenchmark to study the performance of our implementations.
Our microbenchmark was implemented as follows.
For each algorithm $A \in \{$\Rbslack, Relaxed $(a,b)$-tree, BST$\}$ workload $W \in \{$0i-0d, 1i-1d, 5i-5d$\}$ (where $x$i-$y$d represents $x$\% \ins s, $y$\% \del s and $(100-x-y)$\% \func{Get}s) and key range size $S \in \{10^5, 10^6, 10^7\}$, we ran five timed \textit{trials} for several thread counts $n$.
Each trial proceeded in two phases: \textit{prefilling} and \textit{measuring}.
In the prefilling phase, $n$ concurrent threads performed 50\% \textit{Insert} and 50\% \textit{Delete} operations on keys drawn uniformly randomly from $[0, S)$ until the size of the tree converged to a steady state. % (containing approximately $S/2$ keys).
Note that a steady state is achieved when the tree contains $S/2$ keys, since an insertion or deletion of a key $k$ drawn uniformly from $[0, S)$ is then equally likely to succeed or fail.
Next, the trial entered the measuring phase, during which threads began counting how many operations they performed.
%(These counts were eventually summed over all threads and reported in our graphs.)
In this phase, each thread performed
random operations according to the workload $W$
% $x$\% \textit{Insert}, $y$\% \textit{Delete} and $(100-x-y)$\% \textit{Get} operations (where $W = x$i-$y$d)
on keys drawn uniformly from $[0,S)$ for three seconds.

As a way of validating correctness in each trial, each thread maintains a \textit{checksum}.
Each time a thread inserts (resp., deletes) a key, it adds the key to (resp., subtracts from) its checksum.
At the end of the trial, the sum of all thread checksums must be equal to the sum of keys in the tree.

\paragraph{Results}

Results appear in Figure~\ref{fig-bslack-exp}.
As the first column shows, searches are much faster in the trees with large nodes than in the BST.
(Note that the BST has been found to perform approximately as well as the Chromatic tree in these types of workloads, since insertions and deletions of uniform random keys yield approximately balanced trees.)

For the two larger key ranges, the \rbslack\ outperforms the relaxed $(a,b)$-tree.
We argue that this is due to the slightly smaller height and better cache utilization of \rbslack\ nodes, which results from their higher average node degree.
(Naturally, cache utilization only matters once the tree can no longer entirely fit in cache, which is only the case in the key ranges $[0, 10^6)$ and $[0, 10^7)$.)
As an example, with 64 threads and key range $[0,10^6)$, the \rbslack\ has average node degree 14.54 and height 4, and the relaxed $(a,b)$-tree has average node degree 9.82 and height 5.
Moreover, we used PAPI to measure the number of cycles during which a processor is stalled while waiting for a resource (e.g., a load from main memory), and found that operations experienced approximately 78\% more stalled cycles in the relaxed $(a,b)$-tree.
In our analysis, we prefer to use stalled cycles as a metric to explain performance, rather than cache misses, because cache misses are not all equally costly.
Whereas measuring the number of cache misses only captures \textit{how many times} bad things happen, counting stalled cycles captures \textit{how much impact} they have on the execution (in the aggregate).

\begin{figure}[t]
\centering
\includegraphics[width=0.4\linewidth]{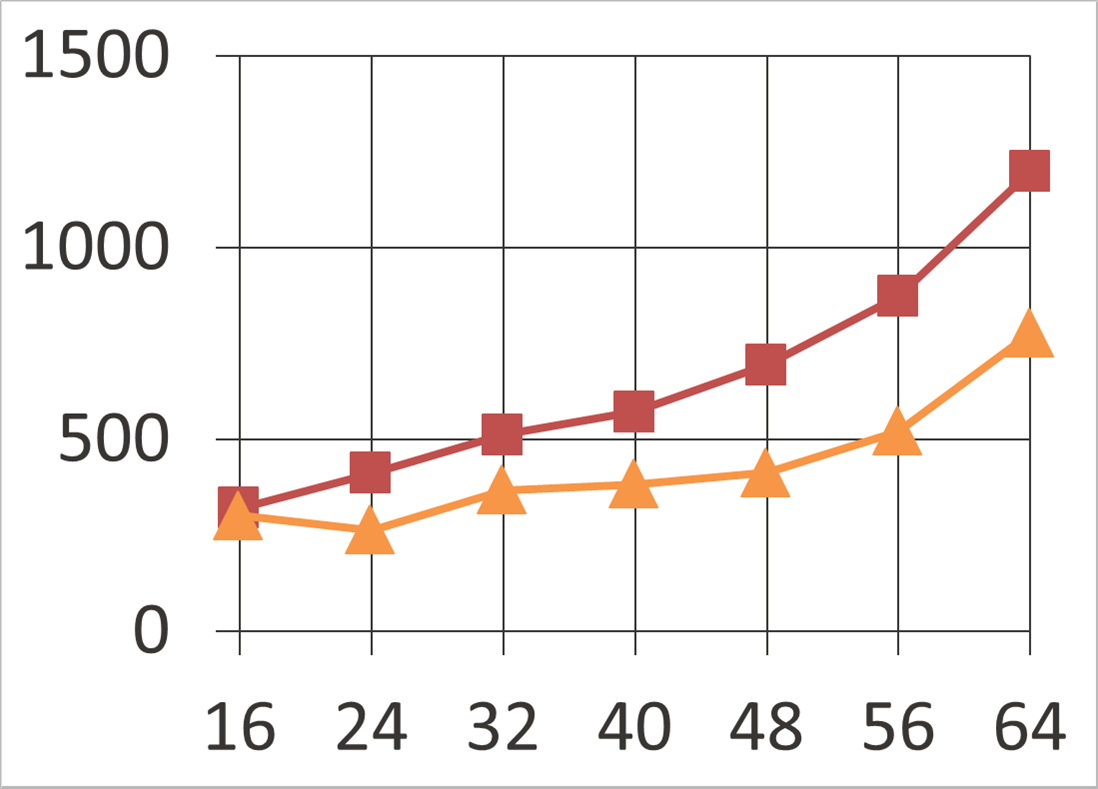}
\caption{Stalled cycle measurements for 0i-0d with key range $[0, 10^7)$.
Only the \rbslack\ and the relaxed $(a,b)$-tree are depicted.
The y-axis shows the average number of stalled cycles per operation.
The x-axis shows the number of concurrent threads.}
\label{fig-bslack-exp-0i0d10m-stalls}
\end{figure}

Note that, for key range $[0, 10^7)$, the failure to scale at high thread counts is due to cache effects, rather than contention.
Figure~\ref{fig-bslack-exp-0i0d10m-stalls} shows the average number of stalled cycles per operation versus the number of concurrent threads in case 0i-0d with key range $[0, 10^7)$, for the \rbslack\ and the relaxed $(a,b)$-tree.
(The BST is omitted, since it experiences more than ten times as many stalled cycles than the other trees due to its poor cache utilization, so it would dominate the graph.)
Since the L3 cache is shared between all threads on a socket, as the number of threads grows, the effective per-thread L3 cache size decreases.
Consequently, threads experience more cache misses, and, hence, more stalled cycles per operation.
The \rbslack\ utilizes the cache more effectively, because of its higher node degree, so it experiences fewer stalled cycles.
%\trevor{talk about the cause of the decrease in performance at high thread counts in the top left graph. it's not concurrency / scaling, it's cache efficiency dropping as hyperthreading reduces the effective cache size per thread.}

As the number of update operations increases, the performance of the \rbslack\ decreases relative to the other two algorithms.
We believe this is due to the overhead of the \rbslack's stricter rebalancing.
However, the \rbslack\ still manages to match or outperform the other algorithms in half of the workloads that contain updates.
These results suggest that the \rbslack\ is a good alternative to a relaxed $(a,b)$-tree when the tree is expected to be very large (which is the primary use case for trees with large node degree) and few updates occur.

\subsection{Tree building performance}

In this section, we study the performance of the three tree algorithms in the \textit{prefilling phase} (as they work to \textit{reach} a steady state).
Whereas the internal structure of the tree undergoes relatively few changes in the steady state, the tree structure is completely built from nothing in the prefilling phase.
Thus, by measuring the time needed to prefill each data structure, we can get some indication of the cost of building (and balancing) the internal structure of the tree.

\begin{figure}[t]
\centering
\includegraphics[width=0.5\linewidth]{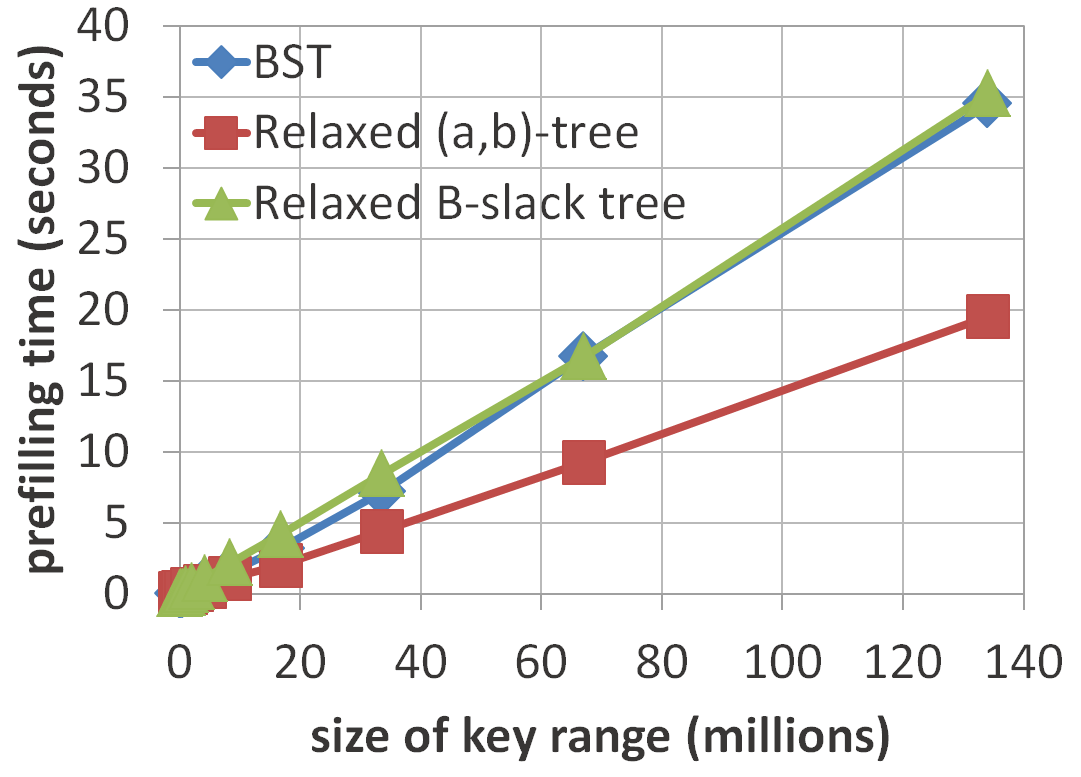}
\caption{Experiment showing the time needed to prefill a data structure with 64 threads, for a variety of key ranges.} %Tree building performance for key ranges $[0, 2^{17})$, $[0, 2^{18})$, ..., $[0, 2^{27})$.}
\label{fig-bslack-exp-prefilling-time}
\end{figure}

We performed \textit{trials} to measure the time needed to prefill each data structure using 64 threads, for range sizes: $2^{17}$, $2^{18}$, $2^{19}$, ..., $2^{27}$ (from 131,072 to approximately 134 million).
Each trial was implemented as a sequence of \textit{prefilling intervals}.
At the start of each interval, the main thread creates 64 new threads, each of which waits on a barrier until all threads have been created.
Then, all threads perform 50\% insertion and 50\% deletion operations on keys drawn uniformly from the key range for 100 milliseconds.
After 100 milliseconds, the main thread destroys these 64 threads, and computes the size of the tree.
If the size is within 3\% of the expected size of the tree in the steady state, then prefilling terminates, and the main thread outputs the number of prefilling intervals that have been performed.
%This allows us to measure the time needed to prefill the tree %to within 3\% of the expected steady state size 
%with a maximum error of approximately 100 milliseconds.
This allows us to approximate the time needed to prefill the tree (with a granularity of 100 milliseconds).

The results appear in Figure~\ref{fig-bslack-exp-prefilling-time}.
The relaxed $(a,b)$-tree is a clear winner.
Its prefilling takes slightly more than \textit{half} of the time needed for the BST and \rbslack.
Interestingly, the relaxed $(a,b)$-tree shows approximately the same performance advantage over the BST and \rbslack, although it outperforms them for very different reasons.
The relaxed $(a,b)$-tree outperforms the BST because of the latter's poor cache utilization.
However, it outperforms the \rbslack\ (\textit{in spite of} the \rbslack's superior cache utilization) because of the high overhead of rebalancing in the \rbslack.

Note that prefilling represents a workload in which all operations are updates, which is not an intended use case for the \rbslack.
In fact, this workload serves as a worst-case for the \rbslack.

\subsection{Memory usage of the final trees}

In this section, we study the amount of memory used by a static \rbslack\ that is constructed by many concurrent threads. %occupies is used to represent the final trees (at the end of each trial).
For each key range, and algorithm, we ran a trial in which 64 threads prefilled the data structure, and then the trial terminated.
We then computed the total amount of memory used by nodes in the final data structure.

\begin{figure}[t]
\centering
\includegraphics[width=0.66\linewidth]{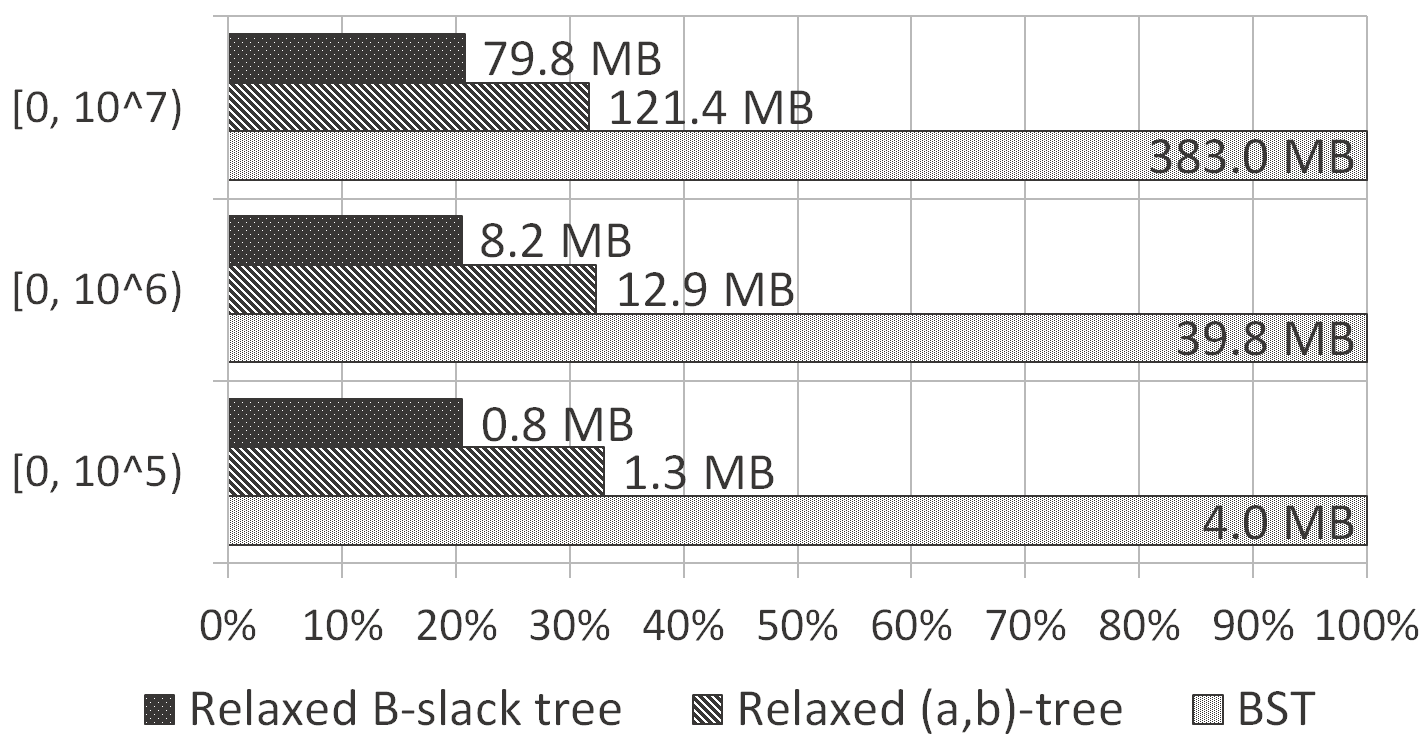}
\caption{Experiment showing the amount of memory occupied by nodes in a static data structure constructed by 64 concurrent threads, for three different key ranges.}
\label{fig-bslack-exp-memory-usage}
\end{figure}

The results appear in Figure~\ref{fig-bslack-exp-memory-usage}.
There, the memory usage for each data structure is expressed as a percentage value, relative to the memory usage of the BST (which is always 100\%).
Each data point for the BST is further annotated with its absolute memory usage in megabytes.
Consistently, the relaxed $(a,b)$-tree uses approximately a third of the memory used by the BST, and the \rbslack\ uses approximately one fifth.
The \rbslack\ achieves significant space savings over the relaxed $(a,b)$-tree, which uses between 52\% and 60\% more memory on average.

A more sophisticated analysis of memory consumption would go beyond the static tree at the end of each trial and study the usage of memory by threads in the process of building the trees. %compute the total amount of memory used by threads in the process of building the trees.
We leave this for future work.

%\trevor{experiment 3: comparing memory usage for the lock-free \rbslack\ with memory usage for the other trees. (might need debra+ to get a good result. this will involve using qprotect to preserve a reference to viol in each rebalancing procedure before entering a Q state.)}

%\trevor{experiment 4: start from a b-slack tree, then run with 48 threads for k seconds, then stop all threads without having them terminate normally, and measure the density of nodes and height? (to understand if there's a gap in practice between the result in corollary 10.13 [which can only be applied if all ins/del ops terminate] and the tree properties in practice)}

\subsection{Workloads with range queries}

In this section, we study workloads containing \func{RangeQuery} operations.
Recall that a \func{RangeQuery} takes, as its arguments, two keys $low$ and $high$, and returns all key-value pairs present in the dictionary whose keys are in $[low, high)$.
%This is a common operation in databases.
\func{RangeQuery} operations can be significantly more efficient in data structures where nodes contain many keys, since they do not need to visit as many nodes to cover the range $[low, high)$.

This experiment was performed the same way as the experiment described in Section~\ref{sec-lfbslack-exp1}, but with different workloads: $W \in$ \{0i-0d-10rq, 1i-1d-10rq, 5i-5d-10rq\}, where $x$i-$y$d-$z$rq represents $x$\% \ins, $y$\% \del, $z$\% \func{RangeQuery} and $(100-x-y-z)$\% \func{Get} operations.
For each operation \func{RangeQuery}$(low, high)$, we set $low$ to be a key drawn uniformly randomly from the key range, and we set $high = low+1000$.
(Thus, in the steady state, when a tree is expected to contain half of the keys in a fixed key range, a \func{RangeQuery} is expected to return 500 keys.)
%For each algorithm $A \in$ \{\Rbslack, Relaxed $(a,b)$-tree, BST\}, workload $W \in$ \{0i-0d-10rq, 1i-1d-10rq, 5i-5d-10rq\} (where $x$i-$y$d-$z$rq represents $x$\% \ins, $y$\% \del, $z$\% \func{RangeQuery} and $(100-x-y-z)$\% \func{Get} operations), and key range size $S \in \{10^5, 10^6, 10^7\}$, we ran five timed \textit{trials} for several thread counts $n$.
%In each trial, the data structure was first \textit{prefilled}, as described in Section~\ref{sec-lfbslack-exp1}.
%Then, the trial entered the \textit{measuring phase}, during which each thread performed operations according to the workload $W$ on keys drawn uniformly from $[0, S)$ for three seconds.

\begin{figure}[tb]
    \centering
    \setlength\tabcolsep{0pt}
    \begin{tabular}{m{0.04\linewidth}m{0.31\linewidth}m{0.31\linewidth}m{0.31\linewidth}}
        &
        \fcolorbox{black!50}{black!20}{\parbox{\dimexpr \linewidth-2\fboxsep-2\fboxrule}{\centering {\large 0i-0d-10rq}}} &
        \fcolorbox{black!50}{black!20}{\parbox{\dimexpr \linewidth-2\fboxsep-2\fboxrule}{\centering {\large 1i-1d-10rq}}} &
        \fcolorbox{black!50}{black!20}{\parbox{\dimexpr \linewidth-2\fboxsep-2\fboxrule}{\centering {\large 5i-5d-10rq}}}
        \\
        \vspace{-3mm}\rotatebox{90}{\normalsize Key range $[0, 10^7)$} &
        \includegraphics[width=\linewidth]{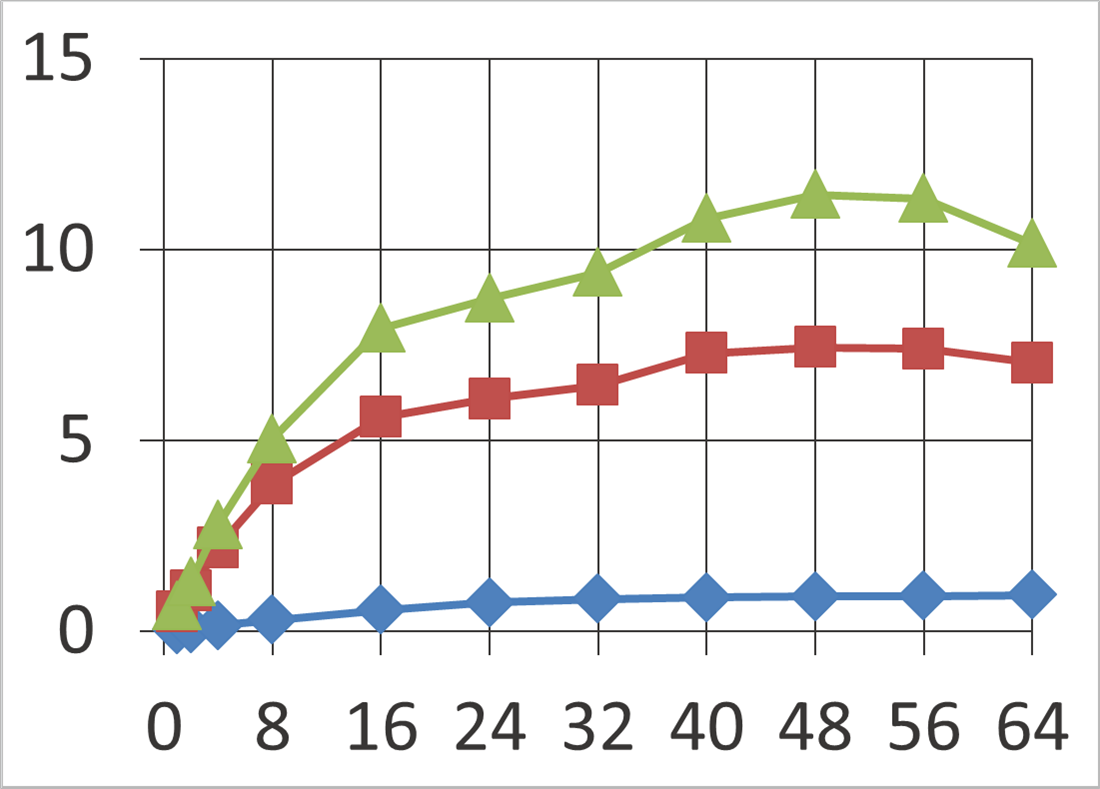} &
        \includegraphics[width=\linewidth]{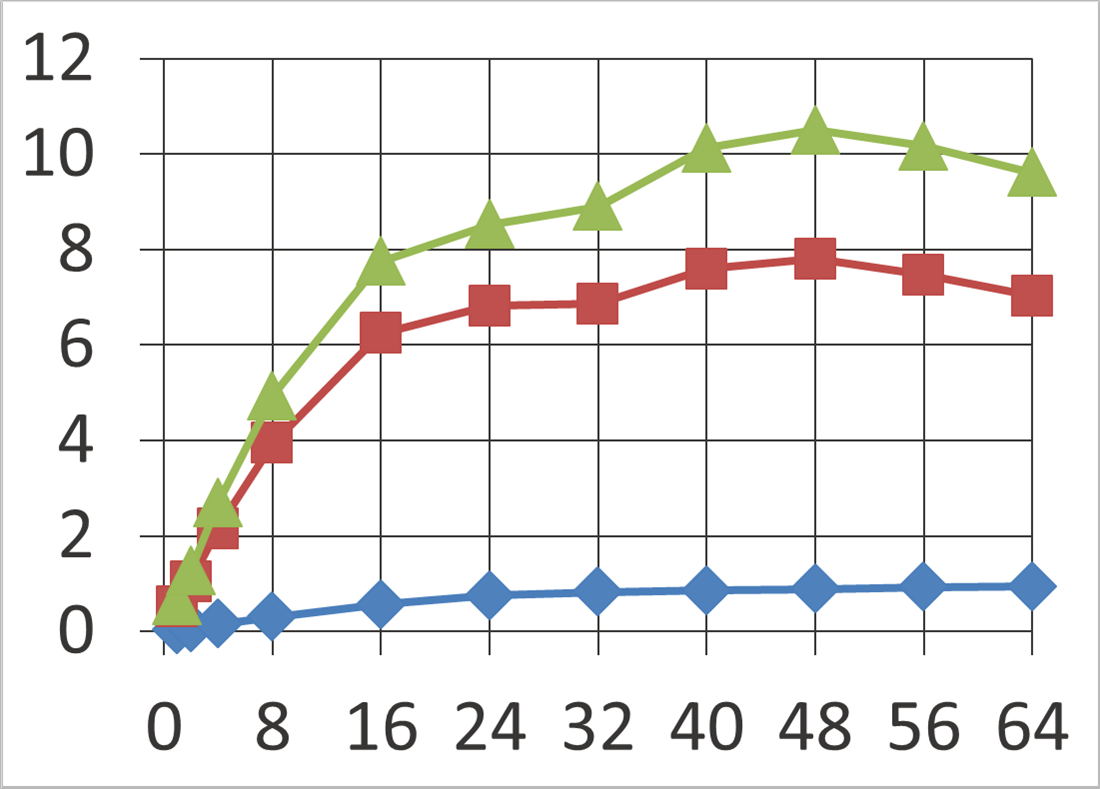} &
        \includegraphics[width=\linewidth]{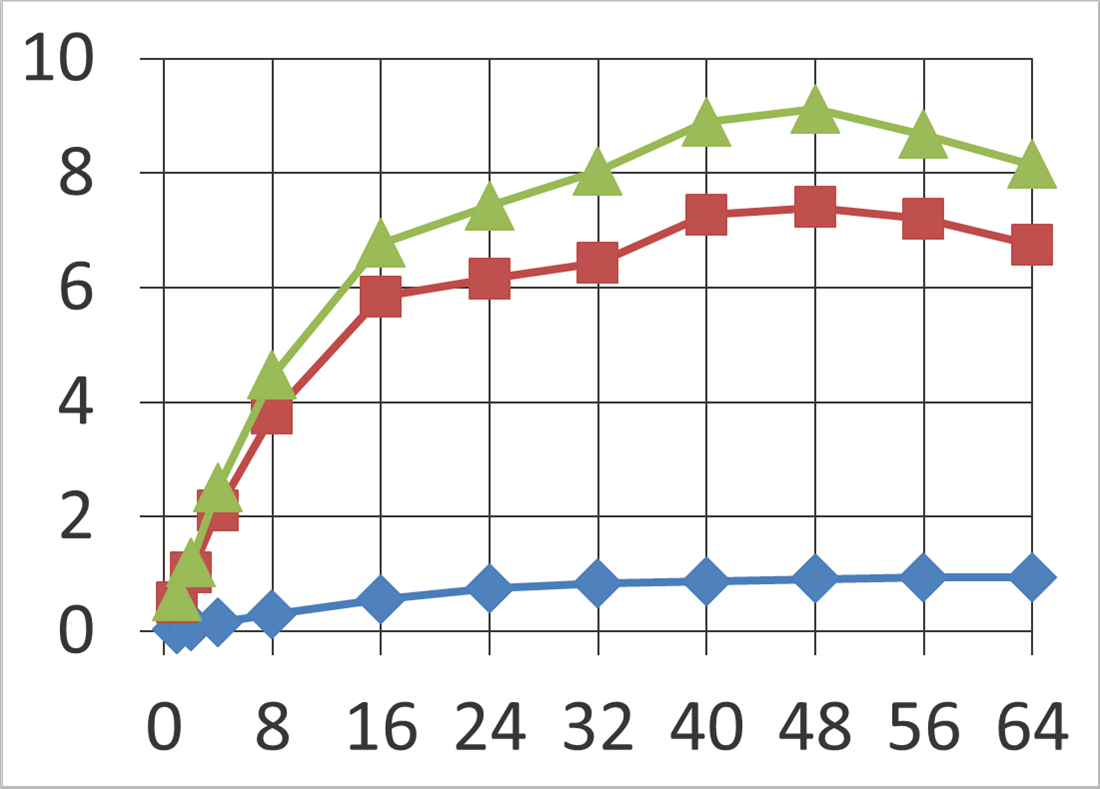}
        \\
        \vspace{-9mm}\rotatebox{90}{\normalsize Key range $[0, 10^6)$} &
        \vspace{-6mm}\includegraphics[width=\linewidth]{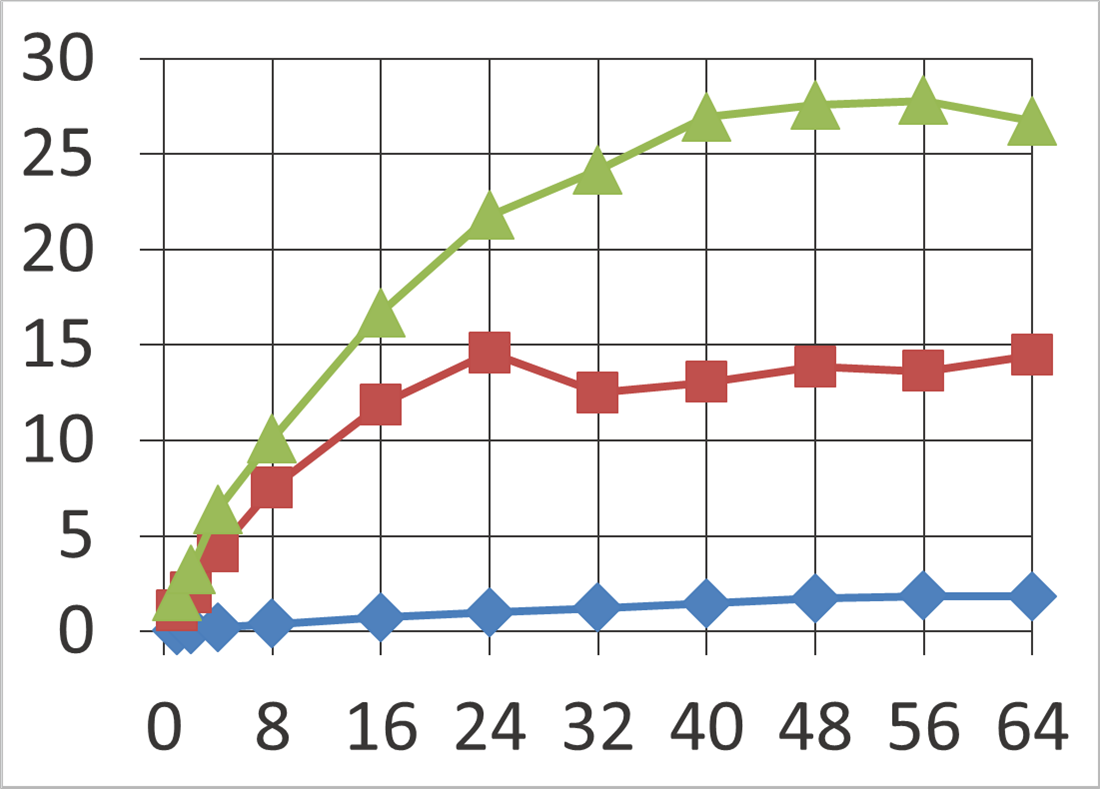} &
        \vspace{-6mm}\includegraphics[width=\linewidth]{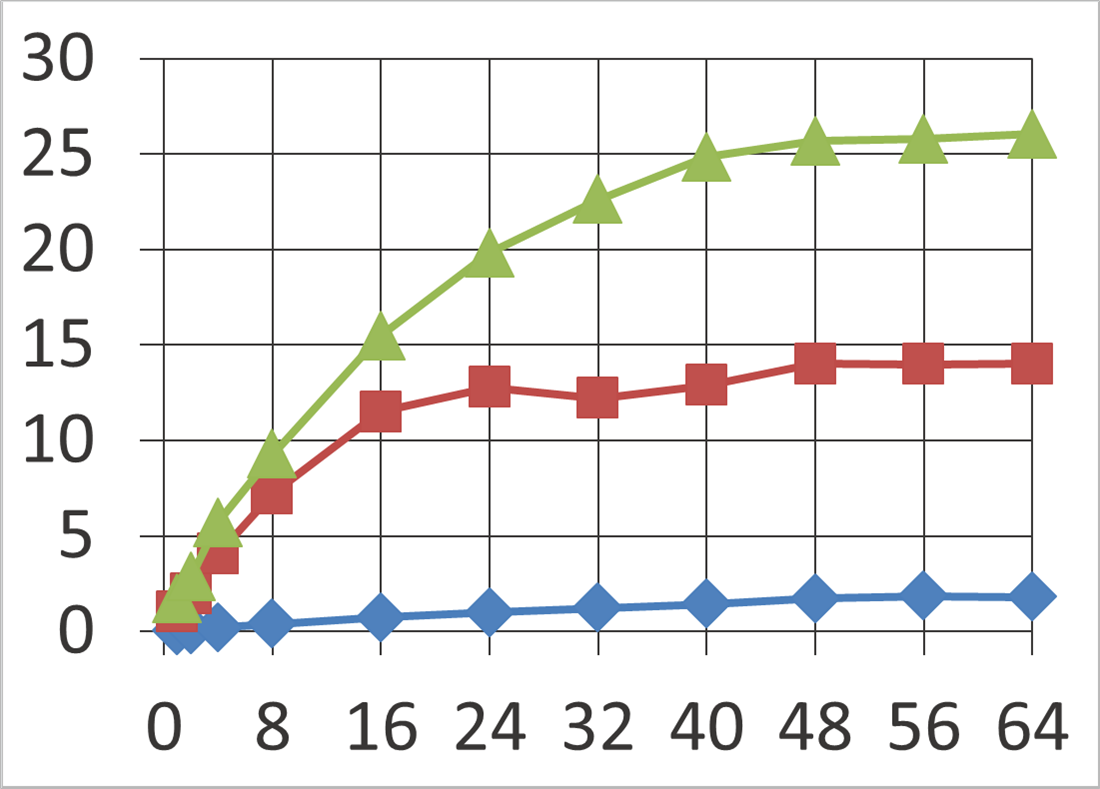} &
        \vspace{-6mm}\includegraphics[width=\linewidth]{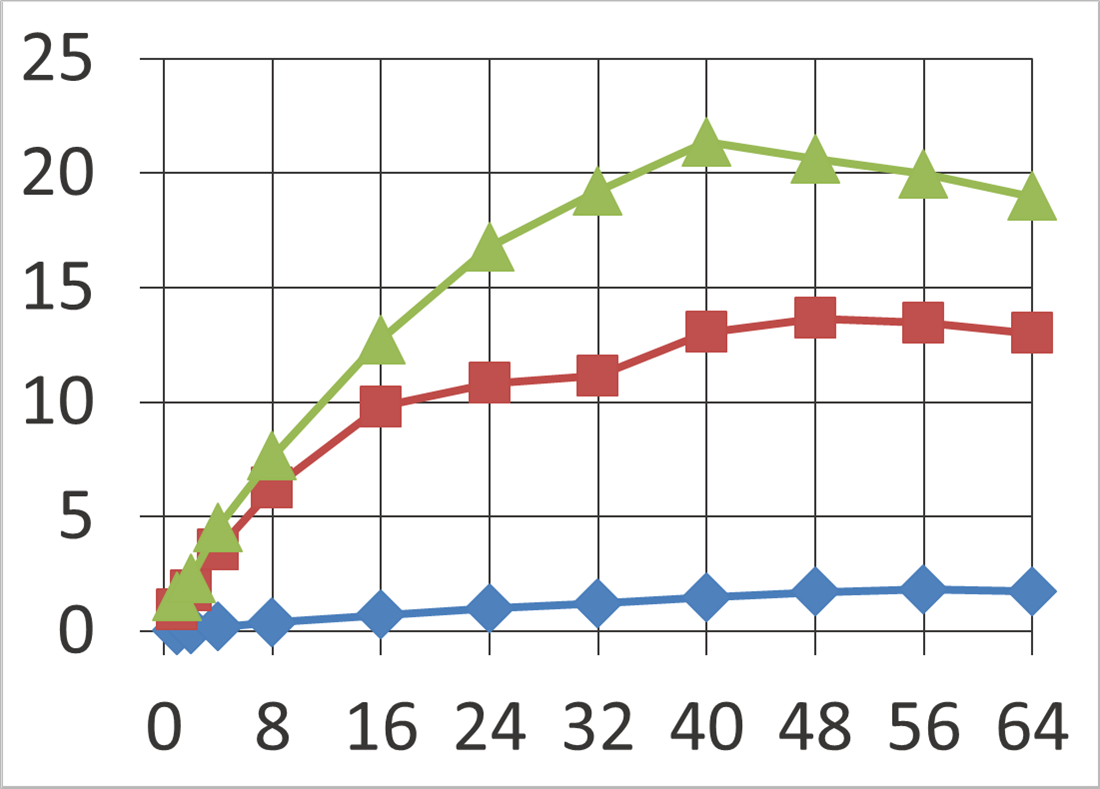}
        \\
        \vspace{-9mm}\rotatebox{90}{\normalsize Key range $[0, 10^5)$} &
        \vspace{-6mm}\includegraphics[width=\linewidth]{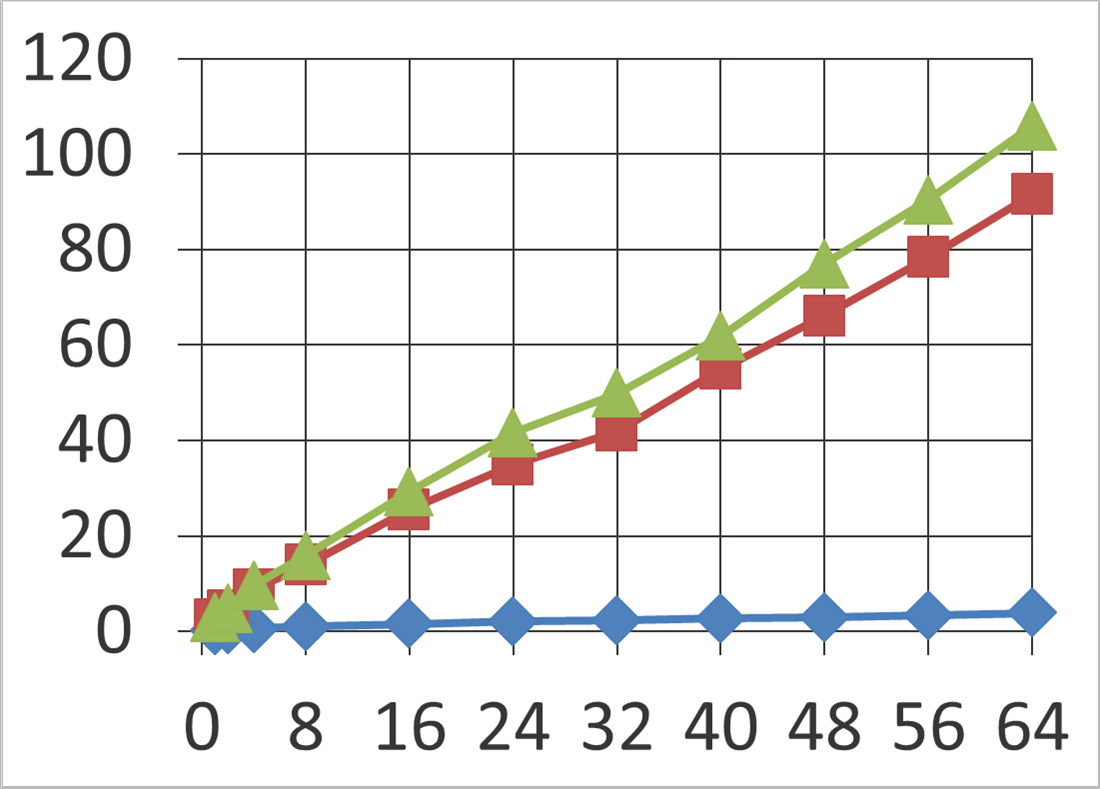} &
        \vspace{-6mm}\includegraphics[width=\linewidth]{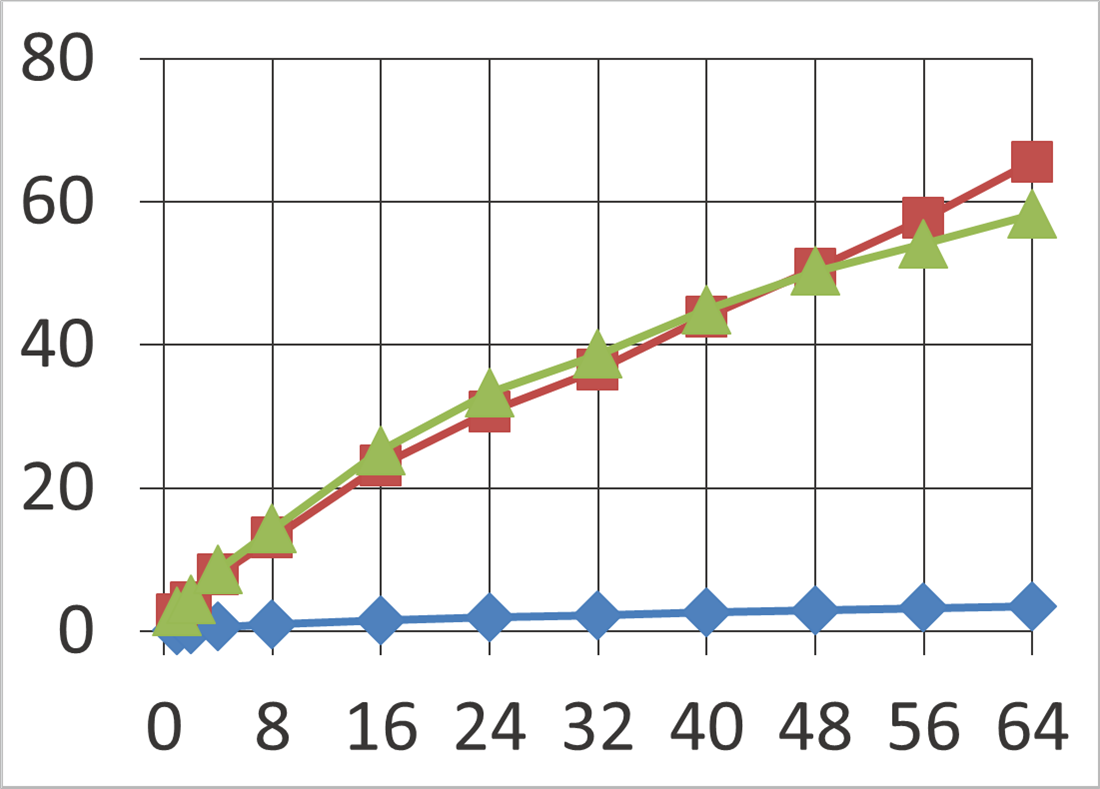} &
        \vspace{-6mm}\includegraphics[width=\linewidth]{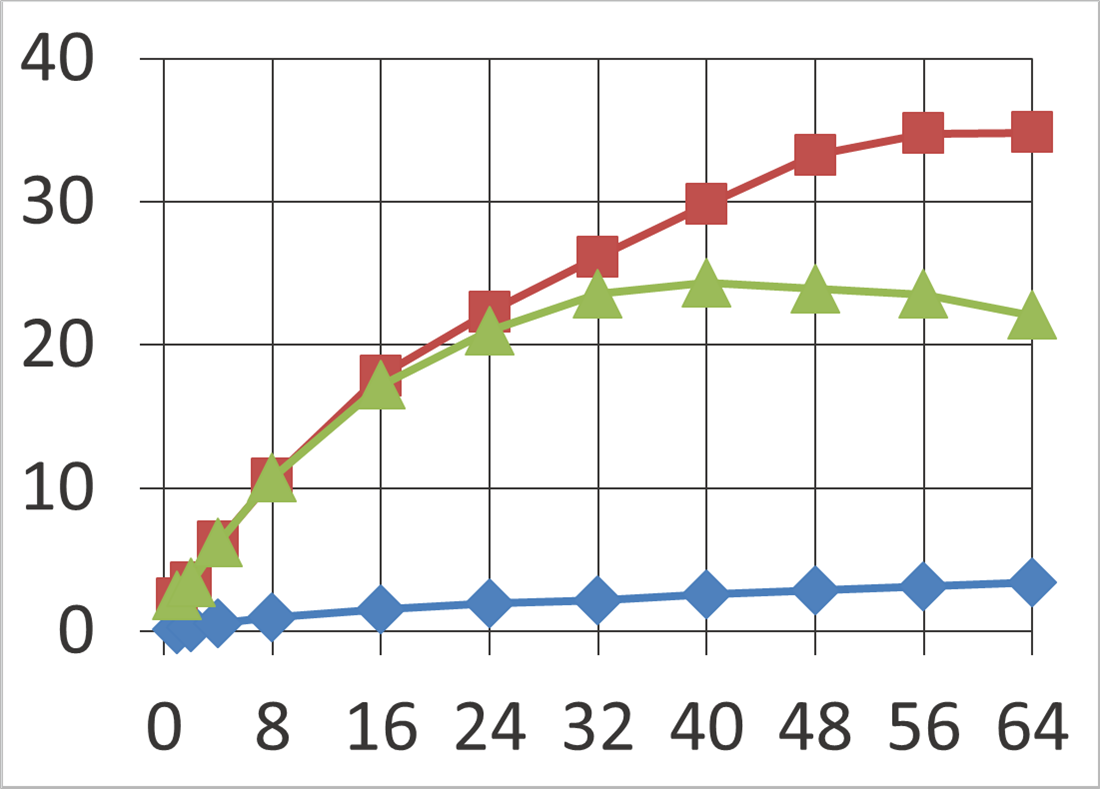}
        \\
    \end{tabular}
    \vspace{-2mm}
%    \hspace{0.07\linewidth}
	\includegraphics[width=0.8\linewidth]{chap-bslack/figures/graphs_pomela1/legend.png}
\caption{Experimental results for workloads including range queries.
The x-axis represents the number of concurrent threads.
The y-axis represents operations per microsecond.}
\label{fig-bslack-exp-rq}
\end{figure}

Results appear in Figure~\ref{fig-bslack-exp-rq}.
Broadly, the BST performs extremely poorly, and the \rbslack\ performs significantly better than the relaxed $(a,b)$-tree, except in the smallest key range.

In the BST, \func{RangeQuery} operations are inefficient, since they must visit many more nodes than in the other trees.
Additionally, since the BST uses so much more memory than the other data structures, only a small part of the BST can fit in cache in any workload shown
Thus, the cache performance for the BST is quite poor.

%Observe that \func{RangeQuery} operations load a much larger number of memory locations into the cache than \ins, \del\ or \func{Get} operations.
In the \rbslack, \func{RangeQuery} operations visit slightly fewer nodes than in the relaxed $(a,b)$-tree (because of the \rbslack's higher average node degree).
Moreover, when compared to the other trees, a larger proportion of the \rbslack\ fits in the cache.

We now discuss the performance differences between the \rbslack\ and the relaxed $(a,b)$-tree.

In the smallest key range, $[0, 10^5)$, both trees easily fit in cache (occupying less than one tenth of it).
In workload 0i-0d-10rq, the \rbslack\ performs slightly better, because its \func{RangeQueries} visit approximately 32\% fewer nodes, on average.
In workloads 1i-1d-10rq and 5i-5d-10rq, the performance of the \rbslack\ decreases relative to the relaxed $(a,b)$-tree, because of the higher cost of updates in the \rbslack.
These updates are more costly primarily because more rebalancing is needed to maintain the stricter balance property.
Consequently, if one performs the same sequence of \ins\ and \del\ operations in a \rbslack\ and a relaxed $(a,b)$-tree, many more nodes will be modified (or replaced) on average in the \rbslack.
Each time a node is modified (or replaced) by a process on one socket, last level cache invalidations occur on all other sockets.
Hence, the next time a process on another socket accesses the node, it will incur an expensive last level cache miss.
Note that this effect is \textit{amplified} in the smallest key range, because the static tree easily fits in cache, so these cross-socket cache invalidations are the primary source of cache misses.

\begin{figure}[tb]
\centering
\begin{tabular}{cc}
{\large (a) Key range $[0, 10^5)$} &
{\large (b) Key range $[0, 10^6)$} \\
\includegraphics[width=0.4\linewidth]{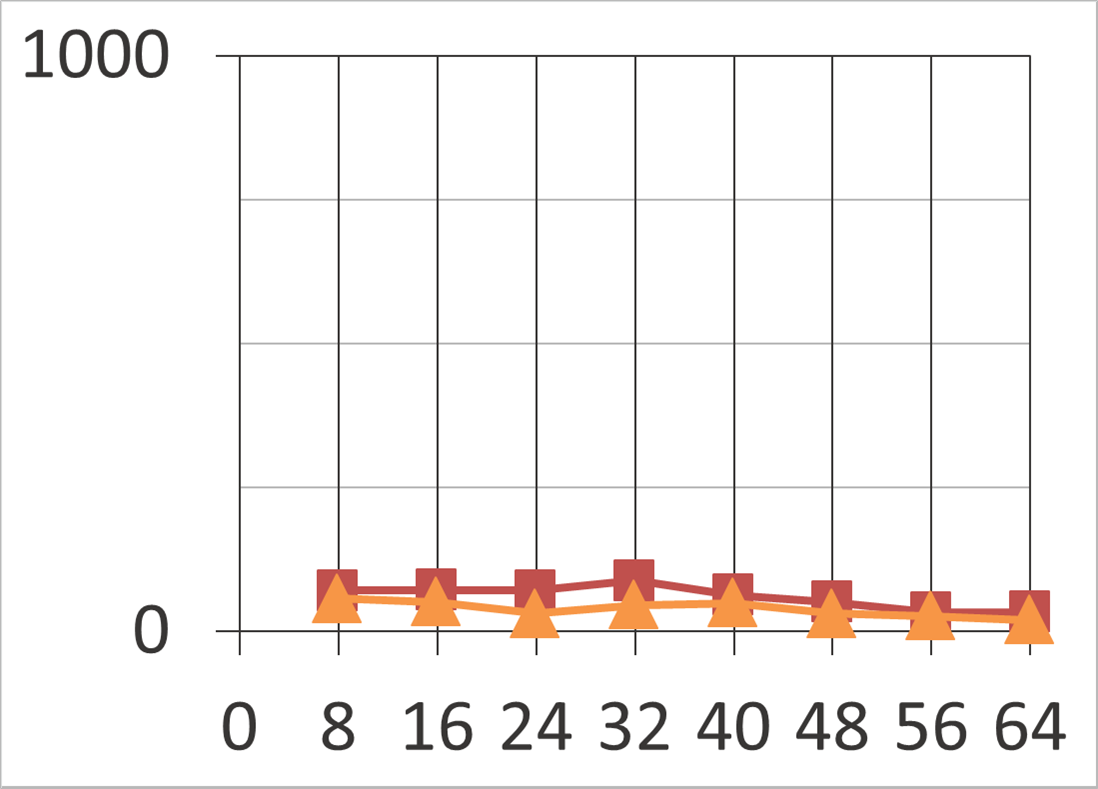} &
\includegraphics[width=0.4\linewidth]{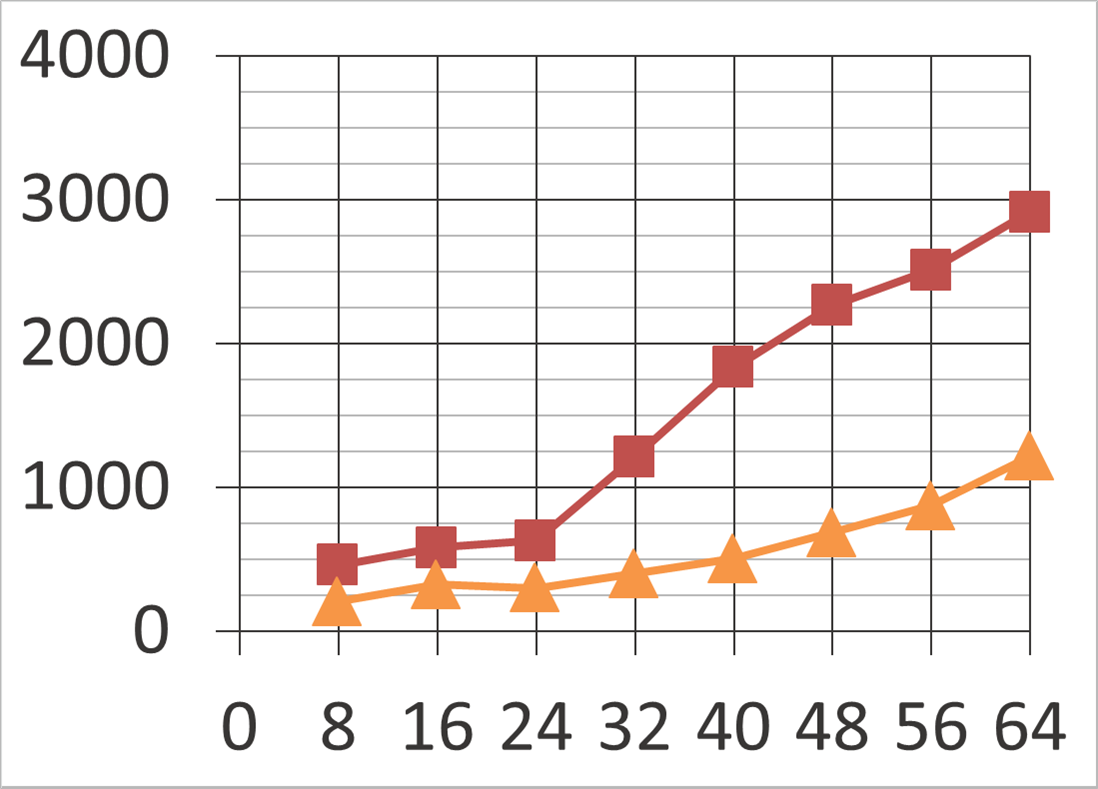}
\end{tabular}
\caption{Stalled cycle measurements for workload 0i-0d-10rq.
Only the \rbslack\ and the relaxed $(a,b)$-tree are depicted.
The y-axis shows the average number of stalled cycles per operation.
The x-axis shows the number of concurrent threads.}
\label{fig-bslack-exp-0i0d10rq1m-stalls}
\end{figure}

In the middle key range, $[0, 10^6)$, both static trees would fit in entirely the cache (with the \rbslack\ occupying approximately half of the last level cache, and the relaxed $(a,b)$-tree occupying nearly the entire last level cache), if there were nothing else stored in the cache.
However, this was not the case.
Consider Figure~\ref{fig-bslack-exp-0i0d10rq1m-stalls}(b), which shows that the number of stalled cycles per operation in the relaxed $(a,b)$-tree increases significantly after 24 threads.
This coincides with the flattening of the relaxed $(a,b)$-tree after 24 threads in the 0i-0d-10rq graph in Figure~\ref{fig-bslack-exp-rq}.
Compare Figure~\ref{fig-bslack-exp-0i0d10rq1m-stalls}(b) with Figure~\ref{fig-bslack-exp-0i0d10rq1m-stalls}(a), which shows the graph for key range $[0, 10^5)$, wherein both trees are small enough that they easily fit in the last level cache.
If the trees in key range $[0, 10^6)$ were actually contained in the last level cache during our trials, then we would expect these two graphs to look similar.
The number of stalled cycles per operation also increases with the number of threads for the \rbslack\ (in Figure~\ref{fig-bslack-exp-0i0d10rq1m-stalls}(b)), but it increases more slowly, and is much smaller at all thread counts (since the \rbslack\ uses less space).

We briefly consider what else might occupy the last level cache (so that the trees cannot fit).
Because of the implementation of \llt\ and \sct, nodes in the tree point to \op s created by the invocations of \sct\ that last modified them.
These \op s are accessed by invocations of \llt, which are performed by \ins, \del\ and \func{RangeQuery}, so they were also part of certain threads' working sets (and, hence, sometimes appeared in the cache).
%(Of course, threads also perform other cached accesses for, e.g., reclamation of nodes and \op s.)
Threads also use a certain amount of stack space as they perform operations, and the amount used varies throughout the execution (and can grow relatively large, since we implemented rebalancing recursively in both trees).
Recent accesses by a thread to its stack memory are cached. %, so they are also part of a thread's working set.
Moreover, %Ultimately, a non-trivial fraction of the cache is used by threads to cache accesses to memory that does not contain nodes in the tree, and the fraction of 
cache space devoted to these memory accesses grows with the number of threads.

The graphs for the largest key range, $[0, 10^7)$, look similar to the graphs for key range $[0, 10^6)$, except that the performance advantage of the \rbslack\ over the relaxed $(a,b)$-tree is smaller in $[0, 10^7)$.
As we discussed above, part of the performance advantage of the \rbslack\ over the relaxed $(a,b)$-tree in $[0, 10^6)$ was due to the fact that a significantly larger portion of the \rbslack\ fit in the last level cache.
However, in $[0, 10^7)$, only a small part of either tree fits in the cache, so caching effects have less impact on performance.
Thus, in these graphs, the performance advantage of the \rbslack\ is primarily due to its higher average node degree, and correspondingly smaller height.

%However, as threads perform range queries, they use a significant amount of memory to store the queue that is used to perform the breadth-first search (and their accesses to this queue are, of course, cached).
%The memory used to store the queue significantly decreases the available cache space for the static trees, especially as the number of threads grows.

\chapter{Reclaiming memory} \label{chap-debra}
% !TEX root = paper.tex

\begin{thesisnot}
In concurrent data structures that use locks, it is typically straightforward to free memory to the operating system after a node is removed from the data structure.
For example, consider a singly-linked list implementation of the set abstract data type using hand-over-hand locking.
Hand-over-hand locking allows a process to lock a node (other than the head node) only if it holds a lock on the node's predecessor.
To traverse from a locked node $u$ to its successor $v$, the process first locks $v$, then it unlocks $u$.
To delete a node $u$, a process first locks the head node, then performs hand-over-hand locking until it reaches $u$'s predecessor and locks it.
The process then locks $u$ (to ensure no other process holds a lock on $u$), removes it from the list, frees it to the operating system, and finally unlocks $u$'s predecessor.
It is easy to argue that no other process has a pointer to $u$ when $u$ is freed.

In contrast, memory reclamation is one of the most challenging aspects of lock-free data structure design.
Lock-free algorithms (also called non-blocking algorithms) guarantee that as long as some process continues to take steps, eventually some process will complete an operation.
%Wait-free algorithms provide the stronger guarantee that any process that takes steps will eventually complete its operation.
The main difficulty in performing memory reclamation for a lock-free data structure is that a process can be sleeping while holding a pointer to an object that is about to be freed.
Thus, carelessly freeing an object can cause a sleeping process to access freed memory when it wakes up, crashing the program or producing subtle errors.
Since nodes are not locked, processes must coordinate to let each other know which nodes are safe to reclaim, and which might still be accessed.
(This is also a problem for data structures with lock-based updates and lock-free queries.
The solutions presented hererin apply equally well to lock-based data structures with lock-free queries.)
%
%Techniques for determining which nodes are safe to reclaim can be divided into \textit{automatic techniques}, which do not require a programmer to specify when an object has been removed from the data structure, and \textit{non-automatic techniques}, which do. %the programmer to do this and possibly more. %involve programmers in memory management.
%It is conceptually useful to divide the work of reclaiming an object into two parts: determining when an object is removed from the data structure, and determining when this object can be freed to the operating system.
%The former is the responsibility of either an \textit{automatic} memory reclamation scheme, or a lock-free data structure.
%The latter is always the responsibility of the memory reclamation scheme.
%Automatic techniques offer a significantly simpler programming environment, but can be inefficient. %less efficient than more specialized techniques.
%Non-automatic techniques are particularly useful when developing data structures for software libraries, since optimization can pay large dividends when code is reused.
%Additionally, non-automatic techniques can be used to build data structures that serve as building blocks to construct new automated techniques.
\end{thesisnot}

\begin{thesisonly}%
In this chapter, we study the problem of performing \textit{safe memory reclamation} for lock-free data structures.
\end{thesisonly}
\begin{wrapfigure}{r}{0.5\textwidth}
\includegraphics[width=\linewidth]{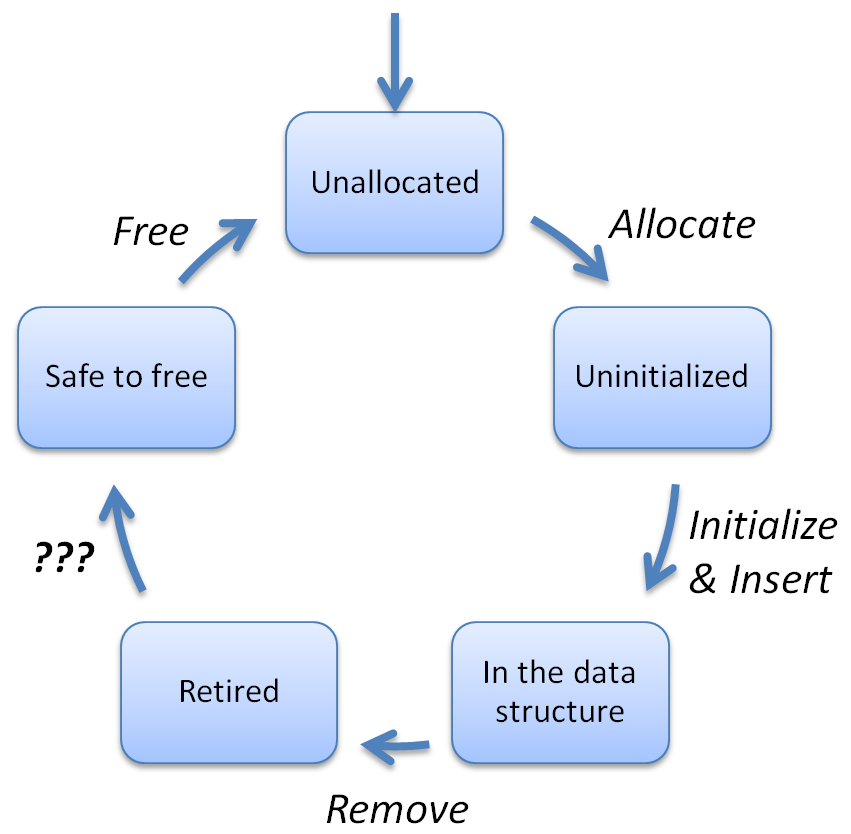}
\caption{The lifecycle of a record.}
\label{fig-lifecycle-of-record}
\end{wrapfigure}
%
%The \textit{safe memory reclamation} problem is defined as follows.
Consider a linked data structure that contains \textit{records}, which point to one another.
Figure~\ref{fig-lifecycle-of-record} shows the lifecycle of a record.
Initially all records are \textit{unallocated}.
A process can \textit{allocate} a record, after which we say the record is \textit{uninitialized}.
The process then initializes and inserts the record into the data structure.
Eventually, a process may remove the record from the data structure, after which we say the record is \textit{retired}.
The goal of safe memory reclamation is to determine when it is safe to \textit{free} a retired record, returning it to an unallocated state.
Once it is safe to free a record, one can either free it, or immediately reuse it.
In either case, we say that the record has been \textit{reclaimed}.

It is conceptually useful to divide the work of reclaiming an object into two parts: determining when a record is retired, and determining when it is safe to free (because no process can reach it by following pointers).
We call a memory reclamation scheme \textit{automatic} if it performs all of the work of reclaiming a record, and \textit{non-automatic} if it requires a data structure operation to invoke a procedure whenever it removes a record from the data structure.
Non-automatic memory reclamation schemes may also require a data structure operation to invoke procedures when other events occur, for instance, when the operation begins, or when it accesses a new record.
A lock-free algorithm can invoke these procedures only if they are lock-free or wait-free.
(Otherwise, they will break the lock-free progress property.)

%Many memory reclamation schemes also provide guarantees on 
To specify progress for a memory reclamation scheme, it is also necessary to provide some guarantee that records will eventually be reclaimed.
Ideally, one would guarantee that \textit{every} retired record is eventually reclaimed.
However, it is impossible to guarantee such a strong property when processes can crash.
Since it is not safe to free a record while a process has a pointer to it, and one cannot distinguish between a crashed process and a very slow one, a crashed process can prevent some records from ever being reclaimed.
%
%Note that, in many memory reclamation schemes, a process which crashes while  these guarantees typically allow for a bounded number of records that are never reclaimed.
%This is necessary to allow memory reclamation schemes in which processes
We call a memory reclamation scheme \textit{fault-tolerant} if crashed processes can only prevent a bounded number of records from being reclaimed.

%The former is the responsibility of either an \textit{automatic} memory reclamation scheme, or a lock-free data structure.
%The latter is always the responsibility of the memory reclamation scheme.
%Techniques for determining which nodes are safe to reclaim can be divided into \textit{automatic techniques}, which do not require a programmer to specify when an object has been removed from the data structure, and \textit{non-automatic techniques}, which do. %the programmer to do this and possibly more. %involve programmers in memory management.
Automatic memory reclamation offers a simple programming environment, but can be inefficient, and is typically not fault-tolerant.
Non-automatic memory reclamation schemes are particularly useful when developing data structures for software libraries, since optimizations can pay large dividends when code is reused.
Additionally, non-automatic techniques can be used to build data structures that serve as building blocks to construct new automated techniques.

Garbage collectors comprise the largest class of automatic memory reclamation schemes. % can be called garbage collectors. %can all be called garbage collection algorithms, with the possible exception .
%and reference counting are both automated techniques (with the caveat that the latter cannot always be used).
The literature on garbage collection is vast, and lies outside the scope of this thesis.
Garbage collection schemes are surveyed in \cite{Jones1996, Schoeberl2010}.
Reference counting is another memory reclamation scheme that can be automatic.
Limited forms of reference counting are also used to construct non-automatic memory reclamation schemes.

Non-automatic techniques can broadly be grouped into five categories: \textit{unsafe reclamation}, \textit{reference counting}, \textit{hazard pointers}, \textit{epoch based reclamation} and \textit{transactional memory assisted reclamation}.
Unsafe reclamation algorithms do not implement safe memory reclamation.
Instead, they immediately reclaim records without waiting until they can safely be freed, which can cause numerous problems.
%Thus, a process may free or reuse a records while another process can still reach it by following pointers, which can cause numerous problems.
For example, suppose a process $p$ reads a field $f$ of a record $r$ and sees A, then performs a
\begin{thesisnot}
compare-and-swap (CAS)
\end{thesisnot}
\begin{thesisonly}
CAS
\end{thesisonly}
instruction to change $f$ from A to B. %, and interprets the success of this CAS to mean that $f$ contained A at all times between the read and CAS.
%As long as $r$ is not reclaimed, 
If $r$ is reclaimed before it can safely be freed, then $p$ can still have a pointer to $r$ \textit{after} $r$ has been reclaimed.
Thus, $p$'s CAS can be performed after $r$ has been reclaimed.
And, if $f$ happens to contain A when $p$'s CAS is performed, then the CAS will erroneously succeed, effectively changing a different record than the one $p$ intended to change.
Unsafe reclamation algorithms must ensure that such problems do not occur. %, for example, by attaching version numbers to fields.
Algorithms in the other categories implement safe memory reclamation, and cannot experience such problems.
%\trevor{update this list. possibly object pools -> unsafe or immediate reclamation, and addition of something? oa?}
%\trevor{safe memory reclamation---some schemes don't do this. if you don't impl. safe memory reclamation, then you need to deal with aba problem etc. (maybe sidestep talking about aba problem until in the model, or maybe talk about aba problem in intro ``if you try to use x when it's not safe to free yet, this is what can happen''.)}
As will be discussed in Section~\ref{sec-debra-related}, %and Appendix~\ref{appendix-related}, 
existing techniques either do not work for, or are inefficient for, many natural lock-free data structures. %, %a large and interesting class of data structure implementations, 
%including a lock-free singly-linked list that can atomically remove two consecutive nodes, and a lock-free tree that supports rotations.

In %traditional 
epoch based reclamation (EBR), the execution is divided into epochs.
Each record removed from the data structure in epoch $e$ is placed into a shared \textit{limbo bag} for epoch $e$.
Limbo bags are maintained for the last three epochs. %A limbo bag is maintained for each of the last three epochs.Three bags are maintained: one for the current epoch, and one for each of the previous two epochs.
Each time a process starts an operation, it reads and announces the current epoch, and checks the announcements of other processes.
If all processes have announced the current epoch, then a new epoch begins, and the contents of the oldest limbo bag can be reclaimed.
%Synchronizing on shared limbo bags and scanning all announcements for each operation is costly.
%Additionally,
If a process sleeps or crashes during an operation, then no memory can be reclaimed, so EBR is not fault tolerant.

The first contribution of this work is DEBRA, a distributed variant of EBR with numerous advantages over classical EBR.
DEBRA supports \textit{partial fault tolerance} by allowing reclamation to continue after a process crashes, as long as that process was not performing an operation on the data structure.
%This also reduces delays in reclaiming records when some processes run slowly, or perform other work besides operating on the data structure.
DEBRA also significantly improves the performance of EBR by: amortizing the cost of checking processes' announced epochs over many operations, eliminating the shared limbo bags in favour of private limbo bags for each process, and optimizing for good cache performance (even on NUMA systems).
DEBRA performs O(1) steps at the beginning and end of each data structure operation and O(1) steps each time an record is removed from the data structure.

% want a way to advance the epoch without waiting for someone who may be very slow or has crashed
% 

Our main contribution is DEBRA+, the first fault tolerant epoch based reclamation scheme.
The primary challenge was to find a way to allow a process to advance the current epoch without waiting for another process, which may have crashed.
In order to advance the epoch, we must ensure that such a process will not access any retired record if it takes another step.
The technique we introduce for adding fault tolerance to DEBRA uses signals, an interprocess communication mechanism supported by many operating systems (e.g., Linux and UNIX).
%For operating systems with signaling functionality (e.g., Linux and Unix), we propose and implement a new mechanism for adding fault tolerance to DEBRA,
%completely eliminates the main disadvantage of DEBRA. %and providing high performance, even with a very large number of threads.
%We call the resulting technique DEBRA+.
With DEBRA+, at most $O(mn^2)$ records are in the limbo bags, waiting to be freed, where $n$ is the number of processes and $m$ is the largest number of records removed from the data structure by one operation. %, and $c$ is a threshold typically chosen to be $\Omega(nr)$ for better performance.
%Experiments show that this mechanism improves the performance of DEBRA by up to 97\% for large thread counts.

A major problem arises when auditioning non-automatic techniques to see which performs best %how each performs 
for a given lock-free data structure. %one encounters when programming lock-free data structures and
%algorithms is that the algorithm must be programmed specifically to be compatible
%with the memory reclamation scheme.
The set of operations exposed to the programmer by non-automatic techniques varies widely.
This is undesirable for several reasons.
%In practice, this means memory reclamation cannot
%be treated as a black box, and it is difficult to change the different memory
%reclamation scheme for an existing algorithm.
%This is undesirable for There are many reasons that this is
%suboptimal. First,
First, from a theoretical standpoint, there is no reason that a data
structure should have to be aware of how its records are allocated and reclaimed.
Second, it is difficult to interchange memory reclamation schemes to determine
which performs best in a given situation.
Presently, doing this entails writing many
different versions of the data structure, each version tailored to a specific memory
reclamation scheme.
Third, as new advancements in memory reclamation appear, existing data structures have to be reimplemented (and their correctness painstakingly re-examined) before they can reap the benefits.
Updating lock-free data structures in this way is non-trivial.
Fourth, moving code to a machine with a different memory consistency model can require changes to the memory reclamation scheme, which currently requires changing the code for the lock-free data structure as well.
Fifth, if several instances of a data structure are used for very different purposes (e.g., many small trees with strict memory footprint requirements and one large tree with no such requirement), then it may be appropriate to use different memory reclamation schemes for the different instances.
Currently, %If memory reclamation code is intertwined with data structure code, 
this requires writing and maintaining code for different versions of the data structure.

These issues were all considered in the context of sequential data structures when
the C++ standard template libraries were implemented.
Their solution was to introduce an \textit{Allocator} abstraction, which allows a data structure to perform \textit{allocate} and \textit{deallocate} operations. %on an Allocator whose implementation is \textit{not} known to the data structure.
With this abstraction, the memory allocation scheme for a data structure
can easily be changed without modifying %or compiling (or even having access to) 
the data structure code at all.
Unfortunately, the Allocator abstraction cannot be applied directly to lock-free programming, since its operations do not align well with the operations of lock-free memory reclamation schemes.
(For example, it requires the data structure to know when it is safe to free a record.)
%Lock-free programming begs for a
%generalization of the Allocator abstraction, so that lock-free data structures can
%benefit from the same decoupling of memory management and data structure
%implementation.
The main challenge in developing an appropriate generalization of the Allocator abstraction is to find the right set of operations to expose to the data structure implementation, so that the result is simultaneously highly efficient, versatile and easy to program.
In Section~\ref{sec-abstraction}, we present our third contribution, the first generalization of the Allocator abstraction for lock-free programming.

Experiments on C++ implementations of DEBRA and DEBRA+ show that overhead is very low.
Compared with performing \textit{no reclamation} at all, DEBRA is on average 4\% slower, and at worst 21\% slower, over a wide variety of thread counts and workloads.
In some experiments, DEBRA actually \textit{improves} performance by as much as 20\%.
Although it seems impossible to achieve better performance when computation is spent reclaiming records, DEBRA reduces the memory footprint of the data structure, which improves memory locality and cache performance.
%Over a wide variety of thread counts and workloads, DEBRA introduces only 10\% overhead, on average. %, than performing no memory reclamation. %performance by an average of 6.5\%.
Adding fault tolerance to DEBRA adds 2.5\% overhead, on average.
%This overhead could be almost completely eliminated by an improved implementation of a GCC compiler feature. % that is officially provided by GCC. % a feature that some compilers currently claim to support. %asynchronous unwind tables.
%DEBRA+ averages 10\% overhead when compared to no reclamation.
Section~\ref{sec-debra-exp} presents extensive experiments comparing DEBRA and DEBRA+ with other leading memory reclamation schemes.
For example, DEBRA+ also outperforms a highly efficient implementation of hazard pointers by an average of 75\%. %between 30\% and 177\%, averaging a 75\% improvement.

%\trevor{bulleted contributions? (including some snippet about changing memory reclamation scheme with 1 line of code, from the abstract?)}

\begin{thesisnot}
\section{Model} \label{sec-prelim}

We consider an asynchronous shared memory system with $n$ processes.
Each process has local memory that is not accessible by any other process, and there is a shared memory accessible by all processes.
Memory is divided into primitive objects, which have atomic operations that are provided directly by the hardware.
Examples include read/write registers, compare-and-swap (CAS) objects, and double-wide compare-and-swap (DWCAS) objects.
A \textit{\record} is a collection of primitive objects, which we refer to as \textit{fields}.
A data structure consists of a fixed set of \textit{entry points}, which are pointers to \record s, and the \record s that are reachable by following one or more pointers from an entry point.
%A \record\ is \textit{retired from a data structure} when it changes from being in the data structure to not being in the data structure.
%A \record\ is \textit{inserted into a data structure} when it changes from being not in the data structure to being in the data structure.
%The system is equipped with a \textit{memory allocator}.
%Initially, no \record\ is accessible to any process.
%\textit{Allocating} a \record\ makes it accessible and provides the process that requested it with a pointer to the \record.
%A \record\ can also be \textit{freed}, which makes it inaccessible.
%Accessing a freed \record\ results in the failure of the entire system.
The system has an memory \textit{allocator} that provides operations to \textit{allocate} and \textit{free} \record s.
Initially, all \record s in shared memory are \textit{unallocated}.
Accessing an unallocated record will cause program failure.
Allocating a \record\ provides the process that requested it with a pointer to it, and makes it accessible by any process that has a pointer to it.
A \record\ can also be \textit{freed}, which returns it to the \textit{unallocated} state.

\trevor{describe htm}

\trevor{cache coherency}

\trevor{numa}
\end{thesisnot}

%\newpage
\section{Related work} \label{sec-debra-related}

There are many existing techniques for reclaiming memory in lock-free data structures.
%We discuss the most relevant ones here, and discuss further related work in Appendix~\ref{appendix-related}.
%However, all of these techniques have some disadvantages.
We give a detailed survey of the literature, and identify significant problems with some of the most widely used memory reclamation algorithms, and some of the most recent ones.
These problems are poorly understood, and make these reclamation algorithms unsuitable for use with a large class of lock-free algorithms.

\paragraph{Reference Counting (RC).}
RC augments each \record\ $o$ with a counter that records the number of pointers that processes, entry points and \record s have to $r$.
A \record\ can safely be freed once its reference count becomes zero.
Reference counts are updated every time a pointer to a \record\ is created or destroyed.
Naturally, a process must first read a pointer to reach a \record\ before it can increment the \record's reference counter.
This window between when a \record\ is reached and when its reference counter is updated reveals the main challenge in designing a RC scheme: the reference count of a freed \record\ must not be accessed. % (because it is not accessible).
However, the reference count of a retired \record\ can be accessed.

Detlefs et~al.~\cite{Detlefs2002} introduced lock-free reference counting (LFRC), which is applicable to arbitrary lock-free data structures.
LFRC uses the double compare-and-swap (DCAS) synchronization primitive, which atomically modifies two arbitrary words in memory, to change the reference count of a \record\ only if a certain pointer still points to it.
DCAS is not natively available in modern hardware, but it can be implemented from CAS~\cite{Harris:2002}. %with hardware transactions \cite{Dice2009} and from CAS \cite{Attiya2008}.
Herlihy et~al.~\cite{Herlihy2005} subsequently improved LFRC to single-word lock-free reference counting (SLFRC), which uses single-word CAS instead of DCAS.
To prevent the reference count of a freed \record\ from being accessed, the implementation of SLFRC uses a variant of Hazard Pointers (described below) to prevent \record s from being freed until any pending accesses have finished.
Lee \cite{Lee2010} developed a distributed reference counting scheme from fetch-and-increment and swap.
%It may be possible to extend it to other data structures.
Each node contains several limited reference counts, and %which respectively count: incoming pointers from objects, incoming pointers from processes
the true reference count for a node is distributed between itself and its parent.
The scheme was developed for in-trees (in which each node only has a pointer to its parent), but it may be generalizable. % to other data structures.

RC requires extra memory for \textit{each \record} and it cannot reclaim \record s whose pointers form a cycle (since their reference counts will never drop to zero).
Manually breaking cycles to allow reclamation requires knowledge of the data structure and adds more overhead.
RC has high overhead, since following a pointer involves incrementing, and later decrementing, its reference count, which is expensive.
Experiments confirm that RC is less efficient than other techniques \cite{Hart2007}.

\paragraph{Hazard Pointers (HPs).}
Michael introduced HPs \cite{Michael2004}, and provided a wait-free implementation from atomic read/write registers.
(Herlihy et~al.~\cite{Herlihy2005} independently developed another version of HPs called Pass-the-Buck (PTB), providing a lock-free implementation from CAS, and a wait-free implementation from double-wide CAS.) %\trevor{they guarantee that every record passed to their retire operation will eventually be freed. [check this]})
%Whereas RC tracks the number of incoming pointers to each \record, 
HPs track which \record s might be accessed by each process.
Before a process can access a field of a \record\ $r$, or use a pointer to $r$ as the expected value for a CAS, it must first acquire a hazard pointer to $r$. %by writing a pointer to $r$ in a shared \textit{announce} array with one part for each process. %and ensure that the hazard pointer was acquired before $r$ is retired.
%This announces to all processes that $r$ might be accessed and cannot safely be freed.
To correctly use HPs, one must satisfy the following constraint. %can be applied to any lock-free data structure that satisfies the following constraint.
%In order to use hazard pointers, a lock-free algorithm must satisfy the following constraint.
%
%\begin{compactenum}[{\bf Constraint~1:}]
%	\item 
Suppose a \record\ $r$ is retired at time $t_r$ and later accessed by a process $p$ at time $t_a$.
%	If $r$ remains retired between $t_r$ and $t_a$, then 
Then,
one of $p$'s HPs must continuously point to $r$ from before $t_r$ until after $t_a$.
%\end{compactenum}
%
This constraint implies that, after a \record\ $r$ is retired and is not pointed to by any HP, no process can acquire a HP to $r$ until it is freed, allocated again, and inserted back into the data structure.
%no process can obtain a hazard pointer to a \record\ after a time when it is retired and is not pointed to by any hazard pointer and before the next time it is inserted into the data structure. %will not be accessed by any process until it is next inserted into the data structure.
Therefore, a process can safely free a retired \record\ after scanning all HPs and seeing that none of them point to $r$.

To acquire a HP to $r$, a process first announces the HP by writing a pointer to $r$ in a shared memory location only it can write to. %with one part for each process.
This announces to all processes that $r$ might be accessed and cannot safely be freed.
Then, the process verifies that $r$ is in the data structure.
If $r$ is not in the data structure, then the process can behave as if its operation had failed due to contention (typically by %aborting and 
restarting its operation), without threatening the progress guarantees of the data structure.
%
%In some data structures, processes cannot easily determine whether records are in the data structure.
%If a process restarts its operation without being able to determine that 
%\trevor{talk about next sentence later. i have one issue with it. faith just wants me to fix up the singular-plural discrepancies.}
As we will discuss below, for many data structures, a process cannot easily tell with certainty whether a record is in the data structure.
Such data structures are modified in ad-hoc ways so that operations restart whenever they cannot tell whether a record is in the data structure.
This requires reproving the data structures' progress guarantees (a subtlety that has been missed by many).

%%Typically, there are configurations in which one can easily determine that a record is \textit{definitely} in the data structure, or that a record \textit{might} not be in the data structure, but it is very difficult to determine that a record is \textit{definitely} not in the data structure.
%Such data structures are modified in ad-hoc ways so that operations restart whenever they cannot tell whether a record is retired.
%This requires reproving the data structures' progress guarantees.
%%If the \record\ is retired, then the process behaves as if its operation had failed due to contention; this typically means restarting the operation.
%%The constraint implies that, %Since a process must acquire a hazard pointer \textit{before} the \record\ it points to is retired,
%%a \record\ that is retired and is not pointed to by any hazard pointer will not be accessed by any process until it is next inserted into the data structure.

On modern Intel and AMD systems, a \textit{memory barrier} must be issued immediately after a HP is announced to ensure that the announcement is immediately flushed to main memory, where it can be seen by other processes.
Otherwise, %this memory barrier is \textit{not} issued, then 
a HP announcement might be delayed so that a process performing reclamation will miss it, and erroneously free the \record\ it protects.
Memory barriers are costly, and this introduces significant overhead.

Many lock-free algorithms require only a small constant number $k$ of HPs per process, since a HP can be released once a process will no longer access the \record\ to which it points during an operation.
Scanning the HPs of all processes takes $\Theta(nk)$ steps, so doing this each time a \record\ is retired would be costly.
However, with a small modification, the expected amortized cost to retire an object is $O(1)$. % can be retired in $O(1)$ expected amortized time. %the amortized cost of retiring a \record\ can be improved.
Each process maintains a local collection of \record s that it has removed from the data structure.
(Note that each record is only removed by one process.)
When a collection contains $nk+\Omega(nk)$ objects, the process creates a hash table $T$ containing every HP, and, for each object $o$ in its collection, checks whether $o$ is in $T$.
If not, $o$ is freed. % and, for each object $o \in C_p$, $p$ frees $o$ if $o \notin T$. %checks if the object is in $T$, and frees any object that is not. %checks if each object is in $T$to see which objects can be freed.
%Suppose each process $p$ maintains a local collection $C_p$ of retired \record s.
%When $C_p$ contains $nk+\Omega(nk)$ objects, $p$ creates a hash table $T$ containing every HP, and, for each object $o \in C_p$, $p$ frees $o$ if $o \notin T$. %checks if the object is in $T$, and frees any object that is not. %checks if each object is in $T$to see which objects can be freed.
Since there are at most $nk$ \record s in $T$,
$\Omega(nk)$ \record s can be freed. %a large enough batch.
%Specifically, when a collection contains $nk+\Omega(nk)$ elements, the process invokes a procedure \textit{ScanAndFree}, which scans all hazard pointers and frees any \record s in the collection that are not pointed to by any hazard pointer.
%(A collection must contain $nk+\Omega(nk)$ elements when a process scans hazard pointers so that $\Omega(nk)$ \record s can be freed, even if $nk$ \record s in the collection are pointed to by hazard pointers.)
%If each collection is implemented using a hash table, then ScanAndFree takes expected amortized constant time per freed \record.
%If each collection is implemented using a balanced tree, then ScanAndFree takes amortized $O(\log nk)$ time per freed \record.
%In each case, 
%The algorithm is wait-free, 
Thus, the number of \record s waiting to be freed is $O(kn^2)$. % ($O(kn^2)$ if processes maintain collections of $nk+\Omega(nk)$ elements).
%Space overhead is $O(nk)$ words to store HPs
%HPs have slightly less overhead than RC, since only a write and a memory barrier must be performed before a pointer can be followed.
To obtain good performance, one typically chooses a fairly large constant for the $\Omega(kn^2)$ term.

Aghazadeh et~al.~\cite{Aghazadeh2014} introduced an improved version of HPs with a worst case constant time procedure for scanning HPs each time a \record\ is retired.
Their algorithm maintains two queues of length $nk$ for each process.
These queues are used to incrementally scan HPs as \record s are retired.
The algorithm adds a limited type of reference count to each \record\ that tracks the number of incoming references from one of the queues.
%The queues occupy $O(kn^2)$ words, and 
Note that $\Theta(\log(nk))$ bits are reserved in each \record\ for the reference count.

Recall that, with traditional HPs, a process must issue a costly memory barrier immediately after announcing a HP.
Dice et~al.~\cite{Dice2016} recently introduced three techniques for implementing HPs more efficiently by eliminating these frequent barriers. % and replacing them with another mechanism (for ensuring that each process can see all HPs that were announced before it started performing reclamation).
The first technique harnesses the \textit{memory protection} mechanism of modern operating systems.
Whenever a process is about to scan all hazard pointers to reclaim memory, it first \textit{write-protects} the memory pages that contain all hazard pointers.
Enabling write-protection on a page acts like a global memory barrier that causes all processes to flush any pending writes to the page before it is write-protected.
Thus, from the perspective of the process performing memory reclamation, it is as if all processes had been issuing memory barriers immediately after their HP announcements.
Unfortunately, this technique is \textit{not} lock-free, since a process that crashes during reclamation will cause all processes to block.
Nevertheless, it could be useful for lock-based algorithms that perform searches without acquiring locks.
The second technique exploits an idiosyncrasy of certain x86 architectures to achieve a non-blocking variant of the first technique.
However, the authors stress that this is not a portable solution. %do not recommend it as a portable or reliable option.
The third technique is a hardware-assisted mechanism that relies on an extension to current processor architectures suggested by the authors.

\paragraph{Problems with Hazard Pointers.}

A major problem with HPs is that they cannot be used with many lock-free data structures.
Recall that in order to acquire a HP to a record $r$, one must first announce a HP to $r$, then verify that $r$ is not retired.
If one can determine that $r$ is \textit{definitely not retired}, then a HP to $r$ has successfully been acquired.
On the other hand, if one can determine that $r$ is \textit{definitely retired}, then the operation can simply behave as if it failed due to contention (typically restarting).
However, in many data structures, it is not clear how an operation can determine whether a node is \textit{definitely} retired or not retired.
And, if the operation cannot tell for sure which is the case, then behaving as if it failed due to contention can cause the data structure to lose its progress guarantee, as in the following example.

Many data structures with lock-free operations use \textit{marking} to prevent processes from erroneously performing modifications to records just before, or after, they are removed from the data structure.
Specifically, before a record is removed from the data structure, it is marked, and no process is allowed to change a marked node.
Search operations can often traverse marked nodes, and even leave the data structure to traverse some retired nodes, and still succeed (see, e.g.,~\cite{arbel2014concurrent,Brown:2014,BH11,Drachsler2014,Ellen:2010,Ellen2014,Heller2005,Howley:2012,Natarajan:2014,Ramachandran2015,Shafiei:2013}).

As an example, consider a lock-free singly-linked list in which nodes are marked before they are retired, and operations can traverse retired nodes.
Suppose that, while searching this list, a process $p$ has acquired a HP to node $u$ and needs to acquire a HP to the next node, $u'$.
To acquire the HP to $u'$, $p$ reads and announces the pointer to $u'$, and must then \textit{verify that $u'$ is in the list}.

One might initially think of checking whether $u'$ is marked to determine whether it is in the list.
Since nodes are marked before they are retired, a node that is not marked is definitely not retired.
Unfortunately, there are two problems with this approach.
First, in order to check whether $u'$ is marked, $p$ would already need to have a HP to $u'$.
Second, if $p$ were to see that $u'$ \textit{is} marked, then it would learn nothing about whether $u'$ is actually in the list.
(Since nodes are marked \textit{before} they are retired, a process may have marked $u'$ but not yet retired it.)

We can get around the first problem by having $p$ check whether the \textit{previous} node $u$ is marked, instead of checking whether $u'$ is marked.
(Recall that $p$ already has a HP to $u$.)
If $u$ points to $u'$ and is not marked, then $u$ and $u'$ are definitely both in the list.
However, this does not resolve the second problem: if $u$ is marked, we learn nothing about whether $u$ is in the list (and, hence, we learn nothing about whether $u'$ is in the list).
All we have done is reduced the problem of determining whether $u'$ is in the list to the problem of determining whether $u$ is in the list.

To resolve the second problem, one might think of having $p$ continue to move backwards in the list until it reaches a node that is not marked.
(Since nodes do not change after they are marked, if $p$ were to find a node that is not marked, and points to the chain of marked nodes $p$ traversed, then $p$ would know that all of these nodes are in the list.)
However, unless $p$ holds HPs to \textit{every} node it traverses while moving backwards in the list, it will run into the first problem again.
In the worst case, $p$ will have to hold HPs to every node it visits during its search.
Thus, the number of HPs needed by each process can be \textit{arbitrarily large}.
This can introduce significant space and time overhead, and also enables a crashed process to prevent an arbitrarily large number of nodes from being reclaimed.
Additionally, since the list is singly-linked, $p$ will have to remember all of the previous nodes that it visited, which may require changes to the algorithm.

Another possible way to verify that $u'$ is in the list is to recursively start a new search looking for $u'$.
Of course, this new search can encounter the exact same problem at a different node on the way to $u'$.
Additionally, $p$ would require additional HPs to perform this search (without releasing the HPs it currently holds on $u$).
Clearly, this approach would be extremely complex and inefficient.
Moreover, for some data structures, the amortized cost of operations depends on the number of marked nodes they traverse.
Searching again from an entry point can provably lead to an asymptotic increase in the amortized cost of operations~\cite{Ellen2014}.

Given that lots of data structures use HPs in practice, one might ask how they deal with these problems.
In practice, it is common for data structures that use HPs to simply restart an operation whenever a marked node is encountered.
This is usually done without any concern for whether restarting operations will preserve the data structure's progress guarantees.
We argue that this will not always preserve lock-freedom.
(And we suspect that this will \textit{almost always violate} lock-freedom.)
In our example, suppose $p$ sees that $u$ is marked, and restarts its search without being certain that $u$ is actually retired.
If the process that marked $u$ crashed before actually retiring $u$, then after $p$ restarts the search, it will simply encounter $u$ again, see that it is marked, and restart again.
Thus, $p$ will restart its search forever, never making progress.
Other processes can get into similar infinite loops just as easily, so this can cause all processes to block, violating lock-freedom.
It is straightforward to see that restarting in this way will violate lock-freedom for search procedures in many data structures (including~\cite{arbel2014concurrent,Brown:2014,BH11,Drachsler2014,Ellen:2010,Ellen2014,Heller2005,Howley:2012,Natarajan:2014,Ramachandran2015,Shafiei:2013}).

Additionally, HPs introduce complications in data structures with \textit{helping}. % mechanisms of many lock-free data structures.
In many lock-free data structures, whenever a process $p$ is prevented from making progress by another operation $O$ (perhaps because $O$ has marked a node that $p$ would like to modify), $p$ \textit{helps} $O$ by performing the steps necessary to complete $O$ (removing any marked nodes from the data structure), on behalf of the process that started $O$.
This involves accessing several nodes, and modifying some of them.
Of course, before $p$ can access these nodes, it must acquire HPs to them.
As we explained above, in order for $p$ to acquire HPs to these nodes, it must be able to determine that they are definitely not retired.
Consequently, one cannot use helping to resolve the problems described above.

Note that, more generally, these problems occur whenever a process can traverse a pointer from a retired node to another retired node (regardless of whether nodes are marked).
%This is the case in many lock-free data structures (including \cite{Brown:2014,BH11,Drachsler2014,Ellen:2010,Natarajan:2014,Shafiei:2013}).
It appears this scenario can occur in any data structure where retired records can point to records that are still in the data structure (which can later be retired), and searches do not help other operations (which means they cannot restart without violating lock-free progress).
%We conjecture this is possible in any lock-free data structure where searches do not help other operations, and records are retired without first 

Considering the exposition above, it seems likely that precisely determining whether nodes are retired after announcing HPs is approximately as difficult as maintaining a reference count in each node that counts the number of incoming pointers from other nodes in the data structure.

\paragraph{Beware \& Cleanup (B\&C).}
B\&C
%A technique called \textit{Beware \& Cleanup} (B\&C) 
was introduced by Gidenstam et~al.~\cite{Gidenstam2009} to allow processes to acquire hazard pointers to \record s that have been retired but not reclaimed.
A limited form of RC is added to HPs to ensure that a retired \record\ is freed only when no other %retired 
\record\ points to it.
%Broadly speaking, HPs are used to guarantee the safety of processes' local pointers to \record s, and RC is used to guarantee the safety of retired \record s' pointers.
%In B\&C, 
The reference count of a \record\ counts incoming pointers from other \record s, but does not count incoming pointers from processes' local memories.
Before reclaiming a \record, a process must verify that its reference count is zero, and that no HP points to it.
Consequently, %A retired \record\ cannot be reclaimed while its reference count is non-zero, so 
after announcing a HP to a \record\ $r$, to determine whether $r$ is retired, it suffices to check whether its reference count is nonzero. %there is still a pointer to the \record\ from another \record.
%A retired \record\ cannot be reclaimed while its reference count is non-zero or any HP points to it.
%It follows that, when an operations encounters a record, it do not have to restart when they reach a retired \record.

Unfortunately, if a data structure allows retired records to point to other retired records, then a retired record's reference count may never be zero.
To address this issue, the authors make the following assumption: ``each [pointer] in a [retired record] that references another [retired record] can be replaced with a reference to [a record that is not retired], with retained semantics for all of the involved threads.''
To use B\&C, one must implement a procedure that takes a retired record and changes it so that it no longer points to any retired record.
Designing such a procedure and proving that it ``retains semantics for all of the involved threads'' is non-trivial.
Additionally, B\&C's algorithm for retiring \record s is extremely complicated, and the technique has significantly higher overhead than HPs.

The authors did not mention the problems with HPs described above.
They stated that operations using HPs would need to restart whenever they encountered a retired \record, but did not consider how such operations would actually determine whether a \record\ is retired.
The goal of the work was simply to \textit{improve performance} for operations that frequently restart.
Nevertheless, their technique does solve the problems with HPs described above.
Regrettably, it does not appear to be practical.

%Unfortunately, B\&C's algorithm for retiring \record s is extremely complicated, and the technique has significantly higher overhead than HPs.
%Additionally, B\&C can only be applied to data structures that satisfy the following property: ``each [pointer] in a [retired record] that references another [retired record] can be replaced with a reference to [a record that is not retired], with retained semantics for all of the involved threads.''
%To use B\&C, one must implement a procedure that takes a retired record and modifies it so that it no longer points to any retired record.
%Designing such a procedure and proving that it ``retains semantics for all of the involved threads'' is non-trivial.

\paragraph{ThreadScan (TS).}
TS is a variant of hazard pointers that avoids making an expensive announcement for each record accessed by treating the private memory of each process as if every pointer it contains is an announcement~\cite{Alistarh:2015}. %uses operating system signals to enable threads obtain a progress guarantee.
TS was developed independently, at the same time as DEBRA+.
Like DEBRA+, TS uses uses operating system signals to enable threads obtain a progress guarantee.

Each process maintains a single-writer multi-reader \textit{delete buffer} in shared memory that contains the \record s it has removed from the data structure, but has not yet freed.
When process $p$'s buffer becomes sufficiently large, $p$ starts \textit{reclamation} by acquiring a global lock (which prevents other processes from starting reclamation).
It then collects the records in the buffers of all processes, and sends a signal to every other process.
%\trevor{write about aggregating all processes' delete buffers.}
Whenever a process $q$ receives a signal, it scans through its own private memory (stack and registers) %to determine whether it has any pointers to \record s, 
and \textit{marks} each \record\ that it has a pointer to.
Then, $q$ sends an acknowledgment back to $p$, indicating that it has finished marking \record s.
Process $p$ waits until it receives acknowledgments from all processes, and then frees any \record s in its buffer that are not marked.

The authors claim that TS offers strong progress guarantees under the assumption that: ``the operating system does not starve threads.''
However, this assumption is extremely strong.
It implies that processes cannot fail, which means that TS cannot be used by a lock-free algorithm.
There are two reasons why TS needs this assumption.
First, TS uses a global lock to ensure that only one process is performing reclamation at a time.
%(Moreover, it seems difficult to eliminate this lock in favour of a lock-free approach, since the process holding the lock also aggregates all processes' delete buffers while the lock is held.)
Second, a process performing reclamation must wait to receive acknowledgments from all other processes.

Although TS is not fault-tolerant, it is a very attractive option in practice, since it is easy to use, and it appears to be quite efficient.
In order to use TS, one simply invokes a procedure whenever a process has just removed a \record\ from the data structure.
However, TS can only be applied to algorithms that satisfy a set of assumptions.
One of these assumptions is particularly problematic: ``nodes in [a process'] delete buffer have already been removed [from the data structure], and cannot be accessed through shared references, or through references present on the heap.''
Restated using our terminology, this assumption says that records in a process' delete buffer are retired, and cannot be accessed by following pointers from other records (even other retired records).
%\trevor{use square brackets and transliterate their terminology to that used in this paper, or give some translation after.}
The authors claim that this assumption ``follows from the definition of concurrent memory reclamation~\cite{Michael2004,Herlihy2005}.'' %seemingly implying that one cannot perform concurrent lock-free memory reclamation for data structures that do not satisfy it.
However, it is not clear that this assumption follows from the definitions in those papers. %this is not true, and 
%it is not clear that this assumption follows from the definitions for HPs~\cite{Michael2004} and PTB~\cite{Herlihy2005}.

Consider the relevant part of the definition in~\cite{Michael2004}:
A hazard pointer to a \record\ must be acquired before the \record\ is retired.
%Once a \record\ is removed from a shared data structure, it cannot be reached by following pointers in shared memory starting from (one of) the root(s) of the data structure.
This admits the possibility that a process can hold hazard pointers to (and, hence, safely access) potentially many retired \record s, as long as the necessary hazard pointers were acquired before these \record s were retired.
Thus, it is theoretically possible for a process to follow pointers from retired \record s to other retired \record s (although there are some problems in practice with traversing pointers from retired \record s, as we discussed above). % is reachable by following pointers in shared memory from other \record s that have been removed from the data structure (provided that 
Similarly, in the pass the buck algorithm of Herlihy et~al.~\cite{Herlihy2005}, it is safe to access any \record\ that is \textit{injail}, which means it has been allocated since it was last freed.
Since \record s are retired before being freed, any \record\ that is retired but has not yet been freed is still \textit{injail}.
%\trevor{explain more. faith was confused at how next sentence follows from prev. she didn't see that you retire before freeing, so retired-but-not-freed things are still injail, so you can safely access them.}
Thus, it is perfectly fine to follow pointers in shared memory from retired \record s to other retired \record s.
Therefore, it would appear that TS's assumption above is strictly stronger than each of these.

\paragraph{Applicability of TS.}
TS's assumption prevents it from being used with algorithms where a process can traverse a pointer in shared memory from a retired \record\ to another retired \record .
This includes all of the algorithms listed where we discussed the problem with HPs.
We briefly consider what happens if such an algorithm is used with TS.

Suppose a process $p$ has a pointer in its private memory to a node $u$ that another process $q$ has removed from the data structure, and is about to read a pointer in $u$ that points to another node $u'$ which $q$ has also been removed from the data structure.
%Let $q$ be the process that removed $u$ and $u'$ from the data structure (so $u$ and $u'$ are in $q$'s buffer).
Then, suppose that, before $p$ reads the pointer in $u$ that it would follow to reach $u'$, $q$ begins reclamation, and signals all processes, including $p$.
This causes all processes to stop what they are doing and help $q$ perform reclamation, by scanning their private memories for any pointers to nodes in $q$'s buffer (including $u$ and $u'$).
If no process has a pointer to $u'$ in its memory, then no process will mark node $u'$, and process $q$ will free $u'$.
This can occur because $p$ has a pointer to $u$ in its private memory, but does have a direct pointer to $u'$.
However, since $u$ points to $u'$, after $p$ finishes helping $q$ perform its reclamation, and resumes its regular algorithm, it will follow the pointer from $u$ to $u'$, performing an illegal access to a freed node.

\paragraph{Dynamic Collect}

Dragojevi{\'c} et~al. \cite{Dragojevic2011} explored how hardware transactional memory (HTM) can be used to easily produce several implementations of a \textit{dynamic collect} object.
A dynamic collect object has four operations: \textit{Register}, \textit{DeRegister}, \textit{Update} and \textit{Collect}.
Intuitively, one can think of a dynamic collect object as a collection of hazard pointers that can increase and decrease in size. %solving the same problem as hazard pointers. %They then used a dynamic collect object to maintain a collection of hazard pointers that can increase and decrease in size.
Register adds a new HP to the dynamic collect object, and DeRegister removes one.
Update sets a HP. % to  point to an object.
An invocation of Collect returns the set of HPs that were set before the Collect began (and were not removed during the Collect), and possibly some others. % hazard pointers. %that were set to point to an object during the Collect.
%Several straightforward implementations using arrays and lists  are presented.
Similarly to HPs, it is not clear how one could use dynamic collect to reclaim memory for a lock-free data structure wherein processes can traverse pointers from retired records to other retired records.

One interesting observation made by the paper is that, in RC schemes, if a process traverses several records inside a transaction, then some increments and decrements of reference counts can be elided.
We explain with an example.
Let $r_1$, $r_2$ and $r_3$ be records in a data structure.
Suppose a process executing in a transaction follows a pointer from $r_1$ to $r_2$, and a pointer from $r_2$ to $r_3$.
Observe that it is not necessary to increment or decrement $r_2$'s reference count, since it would be incremented when the process follows the pointer from $r_1$ to $r_2$, and decremented when the process follows the pointer from $r_2$ to $r_3$ (and the atomicity of transactions guarantees that neither change will be visible on its own).
In general, if a process follows a chain of pointers inside a transaction, only the reference counts of the first and last records must be updated. %from $r_1$ to $r_2,$ to $r_3$, and so on, all the way to $r_m$, then the reference counts of $r_2$ through $r_{m-1}$ need not be updated.
%This can significantly reduce the burden placed on the processor's cache coherence protocol by memory reclamation.
The authors described how to efficiently traverse large linked lists by splitting a traversal into many small transactions, and using this technique to eliminate most of the overhead of reference counting.

\paragraph{StackTrack (ST).}
%\trevor{contributions: (1) dynamically divides the operation into smaller transactions, using shorter transactions if longer ones typically abort, and vice versa. (2) at the end of each transaction, a process automatically publishes HPs to everything in its private memory.}
Alistarh et~al.~\cite{Alistarh2014} introduced an algorithm called StackTrack, which relies on the cache coherence protocol of the processor to detect conflicts between a process that frees a \record\ and a process that accesses the \record.
The key idea is to execute each operation of the lock-free data structure in a transaction, and use the implementation of HTM to automatically monitor all pointers stored in the private memory of processes without having to explicitly announce pointers to them before they are accessed.
%This is possible because,
If a transaction accesses a \record\ which is freed during the transaction, then the HTM system will simply abort the transaction, instead of causing the failure of the entire system.
%Unfortunately, transactions are not guaranteed to commit in hardware.
%Thus, if each data structure operation were always executed in a single transaction, then one could simply free records as soon as they are retired.
%Unfortunately, current implementations of HTM offer no guarantees that transactions will ever commit, and larger

To decrease the probability of aborting transactions, each operation is split into many small transactions called \textit{segments}.
This takes advantage of the fact that lock-free algorithms do \textit{not} depend on transactions for atomicity.
Segments are executed in order, with each segment being repeatedly attempted until it either commits or has aborted $k$ times.
The size of segments is adjusted dynamically based on how frequently segments commit or abort.
If segments frequently abort, then they are made smaller (which increases overhead, but makes them more likely to commit), and vice versa.
If a segment aborts $k$ times, then a non-transactional fallback code path for the segment is executed instead.
This fallback path uses HPs.
%
%Each time a segment fails to commit, it is adjusted so that it contains fewer instructions, which has the effect of making it more likely to commit the next time it is retried.
A process $p$ executing a segment on the fallback path may need to access some records that it obtained pointers to while executing its previous segment.
In order to access these records, $p$ must know that they have not been freed.
Thus, $p$ must already have HPs to these records when it begins executing on the fallback path.
Consequently, at the end of each segment executed \textit{in a transaction} by a process $p$, all pointers in $p$'s private memory are announced as HPs.
%\trevor{at the end of each segment, all pointers in the process' private memory are announced as HPs. (the reason why they have to do this is: [explain what happens if you don't do this.]))}
%Splitting operations into small transactions introduces overhead, and also causes some problems whose solutions introduce further overhead.
%%Although a segment will abort if it accesses a \record\ that was concurrently freed, a problem can arise if the \record\ was freed just before the segment began. %when a process has just finished one segment, but has not yet started the next.
%%The additional mechanisms introduced to deal with this problem inflate the write sets of segments, and add overhead when \record s are freed.
%\trevor{kill following after rewrite}
%For example, at the end of each transaction, any pointers that will be used by the next transaction are announced, so that they are not freed before the next transaction begins.

Any time a process wants to free a \record\ $r$, it must first verify that no HP points to $r$.
%%\trevor{perhaps say something about each txn announcing any of its pointers that will be used in the next txn, and keeping retired \record s in per-process lists until a list reaches a certain size, then, for each retired \record\ $r$ in the list, scanning all announcements and processes' local memories to see if there are any pointers to $r$, and freeing $r$ if there are no pointers to it.}
Each time a process removes a record $r$ from the data structure, it places $r$ in a local list.
If the size of the list exceeds a predefined threshold, then the process invokes a procedure called \textit{ScanAndFree}, which iterates over each record $r$ in the list, and frees $r$ if it is not announced.
%
%%Since currently available implementations of HTM do not offer progress guarantees, a \textit{fallback path} that does not use HTM must be provided to make the algorithm lock-free.
%ST falls back to HPs when transactions cannot succeed in hardware. %, but less efficient.
%%ST encounters the same problem as HPs when a data structure allows a searching process to traverse multiple consecutive retired \record s.
%%Like HPs, ST requires operations to restart when they encounter a retired \record.
%%\textbf{[this seems imprecise. when running in htm, accessing a concurrently freed \record\ is not an issue. accessing a \record\ that is freed before the txn begins is not an issue because of announcing at the end of each txn, and scanning announcements by freeing threads. however, on the fallback path, where you use something like HPs, i think it suffers from the same problem as HPs. does this kind of thing also happen when processes are all on the htm path, and someone is scanning? need to reread the paper...]}
%The paper does not give an upper bound on the number of retired \record s that are not freed.
%Currently,
ST requires a programmer to insert code before and after each operation, whenever a record is retired, and after every few lines of lock-free data structure code.
%(although, it should be possible to write a compiler to automate much of this).
%\trevor{work this next line in. do so by properly crediting ideas currently listed under stacktrack to these guys. splitting txns and "announcing" with refcounts. stacktrack just uses hps.}
%Dragojevi{\'c} et~al.~\cite{Dragojevic2011} explored simple schemes for memory reclamation using HTM.

\paragraph{A Problem with StackTrack.}
%Recall that each data structure operation is split into many small transactions, and at the end of each transaction, any pointers that will be used later on are announced, so that they are not freed before the next transaction begins.
%%In StackTrack, each data structure operation is split into many small transactions.
%%At the end of each transaction, any pointers that will be used later on are announced, so that they are not freed before the next transaction begins.
%Any time a process $p$ wants to free a \record\ $r$, it must first verify that $r$ is not announced.
%For efficiency, each time a process removes a record $r$ from the data structure, it places $r$ in a local list.
%If the size of the list exceeds a predefined threshold, then the process invokes a procedure called \textit{ScanAndFree}, which iterates over each record $r$ in the list, and frees $r$ if it is not announced.

Although no such constraint is stated in the paper, ST cannot be applied to any data structure in which an operation traverses a pointer from a retired \record\ to another retired \record %that atomically removes two or more \record s from the data structure 
~\cite{Matveev2014}.
This includes all of the data structures mentioned above where we discussed the problems with HPs.

%A major problem can occur when StackTrack is used to reclaim memory for a data structure in which one removed record can point to another removed record.
We briefly explain what happens when ST is used to reclaim memory for such a data structure.
Consider a simple list consisting three nodes, $A$, $B$ and $C$, where $A$ points to $B$ and $B$ points to $C$.
For simplicity, we assume the list supports an operation that atomically deletes two consecutive nodes. % (but this assumption can be removed).
%Initially, $A$, $B$ and $C$ are all reachable.
%Then, $A$'s pointer to $B$ is changed to NULL, so that $B$ and $C$ are no longer in the data structure.
%$B$ and $C$ are now unreachable.
%Consider a system with two processes $p$ and $q$.
Suppose process $p$ is searching the list for $C.key$, and process $q$ is concurrently removing $B$ and $C$.
Process $p$ starts a transaction, obtains a pointer to $B$ by reading $A.next$, announces $B$, and commits the transaction.
Then, process $q$ starts a transaction, changes $A.next$ to NULL (to remove $B$ and $C$ from the list), commits the transaction, and adds $B$ and $C$ to its list of removed records.
Next, $q$ invokes \textit{ScanAndFree} and sees that $p$ has announced $B$, so $q$ does not free $B$.
However, $p$ has not announced $C$, so $q$ frees $C$.
Then, $p$ starts a new transaction and reads $B.next$, obtaining a pointer to $C$, and then follows that pointer, which causes the transaction to abort.
This transaction will abort every time it is retried.
Consequently, the data structure operation cannot make progress on the fast path.
Furthermore, as we described above, the HP fallback path cannot accommodate data structures where operations traverse pointers from a retired \record\ to another retired \record. %this problem occurs while process $p$ is executing on the fallback path, then it might follow the pointer to $C$ while not executing a transaction, which will cause the system to crash.

\paragraph{Epochs.}
A process is in a \textit{quiescent state} whenever it does not have a pointer to any \record\ in the data structure.
A grace period is any time interval during which every process has a point when it is in a quiescent state.
%Quiescent state-based reclamation (QSBR)~\cite{McKenney:1998} uses the fact that a \record\ retired by a process can safely be freed after any subsequent grace period.
%%As an example, in a non-preemptive OS kernel environment (where a process will not undergo a context switch until it voluntarily chooses to), if processes are not permitted to hold pointers to \record s in the data structure when they context switch, then a while a process is switched out, it can be considered to be in a quiescent state.
%%A process can then force a grace period to elapse by requesting that the OS schedule it to run on each physical processor, in sequence.
%Applying QSBR to a data structure requires manually identifying quiescent states, which may be impractical for some data structures.
%Additionally, QSBR is not fault-tolerant, because a stalled process can indefinitely prevent reclamation of any memory.
%
Fraser \cite{Fraser2004} described epoch based reclamation (EBR), which %is similar to QSBR, but
assumes that a process is in a quiescent state between its successive data structure operations.
%restricts its attention to data structures for which each process will enter a quiescent state at the end of each data structure operation.
More specifically, EBR can be applied only if processes cannot save pointers read during an operation and access them during a later operation.
\begin{fullver}
We expand on the brief description of EBR given at the beginning of the chapter.
\end{fullver}
\begin{shortver}
We expand on the brief description of EBR given above.
\end{shortver} 
EBR uses a single global counter, which records the current \textit{epoch}, and an announce array.
Each data structure operation first reads and announces the current epoch $\epsilon$, and then checks whether all processes have announced the current epoch.
If so, it increments the current epoch using CAS.
%%At the beginning of an operation by process $p$, a value $v$ is read from the global counter and announced in $p$'s slot of the \textit{announce} array.
%Then, $p$ reads each entry of the \textit{announce} array to determine whether each process has announced the current epoch (value of the global counter). 
%If so, $p$ uses CAS to change the global counter from $v$ to $v+1$.
%Note that whenever the global counter contains  value $v$,  each value in the  \textit{announce} array  is either $v-1$ or $v$.
%Suppose the global counter contains value $v$.
The key observation is that the period of time starting from when the epoch was changed from $\epsilon-2$ to $\epsilon-1$ until it was changed from $\epsilon-1$ to $\epsilon$ is a grace period (since each process announced a new value, and, hence, started a new operation).
So, any \record s retired in epoch $\epsilon-2$ can safely be freed in epoch $\epsilon$.
%In other words, it is safe to free any \record s that were retired two epochs ago.
Whenever a \record\ is retired in epoch $\epsilon$, it is appended to a limbo bag for that epoch.
It is sufficient to maintain three limbo bags (for epochs $\epsilon$, $\epsilon-1$ and $\epsilon-2$, respectively).
%To efficiently free retired \record s, they are kept in three lists, one for the current epoch, one for the last epoch, and one for the epoch before that.
Whenever the epoch changes, every \record\ in the oldest limbo bag is freed, and that limbo bag becomes the limbo bag for the current epoch.

Since EBR only introduces a small amount of overhead at the beginning of each \textit{operation}, it is significantly more efficient than HPs, which requires costly synchronization \textit{each time a new \record\ is accessed}.
%EBR is significantly more efficient than HPs efficient, since it adds at most $n$ reads, one write and one CAS per \textit{operation} (plus any shared memory operations to add retired \record\ to the shared lists).
%This is significantly less overhead than hazard pointers or reference counting, which require costly synchronization for each pointer read.
The penalty for writing to memory only at the beginning of each operation, rather than each time a new \record\ is accessed, is that processes have little information about which \record s might be accessed by a process that is suspected to have crashed.
%In contrast, hazard pointers and reference counting both give complete information about which \record s might still be accessed.
Consequently, EBR is not fault tolerant.

%Note that synchronizing on shared limbo bags makes it significantly less efficient than DEBRA.

Quiescent state-based reclamation (QSBR)~\cite{McKenney:1998} is a generalization of EBR that can be used with data structures where processes can save pointers read during an operation and access them during a later operation.
However, to use QSBR, one must manually identify times when individual processes are quiescent.

\paragraph{Drop the Anchor (DTA).}
Braginsky, Kogan and Petrank \cite{Braginsky2013} introduced DTA, a specialized technique for singly-linked lists, which explores a middle ground between HPs and EBR.
Instead of acquiring a HP each time a pointer to a node is read, a HP is acquired only once for every $c$ pointers read.
When a HP to a node $u$ is acquired, it prevents other processes from reclaiming $u$ and the next $c-1$ nodes currently in the list.
(It also prevents other processes from reclaiming any nodes that are inserted amongst these nodes.)
Suppose a process $q$ performs $s$ operations without seeing any progress by another process $p$.
Then, $q$ will cut all nodes that $p$ might access out of the list, replacing them with new copies
It will also mark the old nodes so that $p$ can tell what has happened.
If $p$ has crashed, then the nodes that $q$ cuts out of the list can never be freed.
However, if $p$ has not crashed, then it will eventually see what has happened, and attempt to free these marked nodes.
Observe that memory reclamation can continue in the list regardless of whether $p$ has crashed.
Consequently, DTA is fault tolerant.
%This allows memory reclamation to continue, even if processes crash.

DTA has been shown to be efficient \cite{Alistarh2014,Braginsky2013}.
However, it is not clear how it could be extended to work for other data structures.
Additionally, DTA needs to be integrated with the mechanism for synchronizing updates to the linked list, because a sequence of nodes can be cut out of the list concurrently with other updates.

In the worst case, the number of retired nodes that cannot be freed is $\Omega(scn^2)$.
To see why, consider the following.
Let $p$ be a process with a HP at a node $u_1$ that allows it to access nodes $u_1, ..., u_c$.
Suppose each process other inserts $s$ nodes between $u_1$ and $u_c$ before $p$ performs another step. %and then a process $q$ sees that $p$ has not made progress.
Observe that $p$'s next step might access any of the $\Omega(scn)$ nodes starting at $u_1$ and ending at $u_c$.
Thus, a process $q$ that suspects $p$ has crashed will cut all of these nodes out of the list.
None of these nodes can be freed if $p$ crashes.
If this is repeated for each process $p \neq q$, then there will be $\Omega(scn^2)$ nodes that cannot be reclaimed.

\medskip
\noindent\textbf{QSense (QS).}
%\trevor{point out this and threadscan are looking at restricted models. these are not asynchronous. QS is incorrect on true async system, but TS just fails to make progress.}
%
Recently, Balmau et~al.~\cite{Balmau2016} introduced QS, another algorithm that combines HPs and EBR.
Like DTA, QS adds fault-tolerance to EBR by using HPs.
Like the accelerated implementations of HPs in~\cite{Dice2016}, the performance benefit of QS over HPs comes from a reduction in the overhead of memory barriers issued to ensure that HP announcements are visible to all threads.
At a high level, QS uses two execution paths: an EBR-based fast path and a HP-based slow path.
As long as processes continue to make progress, they continue to use the fast path.
If a process has not made progress for a sufficiently long time, then all processes switch to the slow path.
To guarantee that nodes are not erroneously freed when the algorithm switches from the fast path to the slow path, the fast path and slow path both acquire HPs.
On the fast path, EBR is effectively used to reduce the cost of reclamation by eliminating the need to scan HPs to determine whether nodes can be freed.

In order to reduce the overhead of issuing memory barriers for the HPs acquired in QS, the authors make the following observation: In a modern operating system, running on an x86/64 architecture, whenever a process experiences a context switch, the kernel code for performing a context switch %executes a strong synchronization primitive that 
issues at least one memory barrier.
Suppose a process $p$ announces a HP at time $t$, and does \textit{not} perform a memory barrier after announcing the HP.
Additionally, suppose another process $q$ begins scanning HPs (to perform reclamation) at time $t' > t$.
If $p$ experiences a context switch between $t$ and $t'$, then $q$ will see $p$'s HP announcement, as if $p$ had issued a memory barrier.
(This is somewhat similar to the HP schemes of Dice et al.~\cite{Dice2016} discussed above, which also harness operating system and/or hardware primitives to eliminate memory barriers.)

The authors introduce \textit{rooster processes} to trigger context switches for all processes at regular intervals.
Each processor has a rooster process pinned to it that sleeps for some fixed length of time $T$, then wakes up (forcing a context switch), then immediately sleeps again.
Whenever a process wants to reclaim a node $u$ that was retired at time $t$, it waits until time $t+T+\epsilon$ (for some small $\epsilon > 0$), by which point a rooster process should have woken up, forcing a context switch and guaranteeing that the reclaiming process can see any HPs announced before the node was removed from the data structure.
The $\epsilon$ term above is necessary because, in real systems, when a process requests to sleep for $T$ time units, it may sleep longer.
If a rooster process sleeps for more than $t+T+\epsilon$ time, then its failure to trigger a timely context switch might cause a reclaiming process to miss a HP and erroneously free a node that is still in use.
Thus, \textit{QS only works under the following assumption}: a bound on $\epsilon$ must be known and rooster processes never fail.
Consequently, QS does not work in a fully asynchronous system.
(In comparison, the TS algorithm also makes timing assumptions, but it only loses its \textit{progress} guarantee in a fully asynchronous system.)

%QS is lock-free only under the assumption that rooster processes never fail.
%Additionally, 
QS switches from the slow path to the fast path only when every process has completed an operation since it last switched to the slow path.
Consequently, if a process crashes either before or while processes are executing on the slow path, then all processes remain on the slow path forever (where they cannot use EBR to reclaim memory).
This is not true for DEBRA+, which allows EBR to continue, even if a process has crashed.
Furthermore, since QS uses HPs, it cannot be used with data structures where operations traverse pointers from a retired record to another retired record (as we described above).

\medskip
\noindent\textbf{Optimistic Access (OA).}
Cohen and Petrank~\cite{Cohen2015} introduced an approach where algorithms optimistically access parts of memory that might have already been reclaimed, and check after each read whether this was the case.
(Their approach was developed independently, at the same time as DEBRA+.)
Crucially, OA relies on processes being able to access reclaimed memory without causing the system to crash.
Consequently, an algorithm that uses OA must either (a) never release memory to the operating system, or (b) trap segmentation fault and bus fault signals and ignore them.
%
%To allow processes to optimistically access reclaimed memory without crashing the program, an algorithm that uses OA must either (a) never release memory to the operating system, or (b) trap segmentation fault and bus fault signals and ignore them.
If option (a) is used, then OA has the same downsides as OPs.
On the other hand, if option (b) is used, then one must cope with the downsides of trapping segmentation and bus faults.
In particular, if an algorithm contains bugs that cause segmentation or bus faults, then option (b) makes it significantly more difficult to identify these bugs, because one cannot easily distinguish between faults caused by a program bug and faults caused by accesses to reclaimed memory.
Such algorithms cannot be used in software that already traps segmentation or bus faults, including many common debugging tools.
(Incidentally, %in modern operating systems, the kernel code for sending and receiving signals acquires locks, so, 
like DEBRA+, option (b) is lock-free only if the operating system's signaling mechanism is lock-free.)

OA can be used only with algorithms that appear in a \textit{normalized} form.
At a high level, an operation in a normalized algorithm proceeds in three phases: CAS generation, CAS execution and wrap-up.
In the CAS generation phase, the operation reads shared memory and constructs a sequence of CAS steps.
These CAS steps are performed in the CAS execution phase.
Any additional processing is performed in the wrap-up phase.
In the CAS execution and wrap-up phases, the operation can be helped by other processes.
As we will see, the class of algorithms in lock-free normalized form is quite similar to the class of algorithms that can use DEBRA+ in a straightforward way.

At a high-level, OA works as follows.
Each process $p$ has a \textit{warning} bit that is cleared whenever $p$ starts a new operation, and is set whenever any process begins reclaiming memory.
After $p$ performs a read from shared memory, it checks whether its warning bit is set, and, if so, jumps back to a safe checkpoint earlier in the operation (intuitively restarting the operation).
Unlike reads, CAS operations cannot be optimistically performed on reclaimed memory, since this could cause data corruption.
Thus, \textit{before} performing CAS on any \record, a process first announces a HP to ensure that another process does not reclaim the \record\ (and possibly does the same for the old value and new value for the CAS), and then checks whether its warning bit is set.
If so, it releases its HP and jumps back to a safe checkpoint.
Whenever $p$ retires a \record, it places it in a local buffer.
If this local buffer becomes sufficiently large, then $p$ performs reclamation.
To perform reclamation, $p$ simply sets the warning bits of all processes, and then frees every \record\ in its local buffer that is not pointed to by a HP.

%Observe that normalized lock-free algorithms must use CAS to perform all changes to shared memory.
%Thus, to obtain a normalized version of a lock-free algorithm that performs \textit{writes} to shared memory (e.g.,~\cite{Brown:2013}), one must replace each write to shared memory with a CAS operation, adding overhead.
%(This is not true for DEBRA+, which supports lock-free algorithms that perform writes.)
Before one can use OA to reclaim memory for a lock-free algorithm, one must first transform the algorithm into normalized lock-free form, then replace each shared memory read or CAS with a small procedure.
%When using OA, a lock-free algorithms is modified as follows. %, reads, writes and CAS operations on shared addresses in the original lock-free algorithm are instrumented as follows.
Each shared memory read is replaced with: two reads, a branch, a possible write, and a possible jump.
Each shared memory write or CAS is replaced with: four to seven writes, %(plus one to six more writes later in the code), 
one read, one branch, a CAS, a memory fence, three possible bit-masking operations, and a possible jump.

Conceptually, OA is somewhat simpler than DEBRA+.
However, in practice, OA would likely be less efficient, and more time consuming to apply to complex data structures, since it requires code modifications for each read, write and CAS on shared memory, and for each \record\ retired.
It is also quite likely to be less efficient, for this reason.
In a subsequent paper, Cohen and Petrank presented an extended version of OA called automatic optimistic access (AOA).
AOA uses garbage collection techniques to eliminate the need for a procedure to be invoked whenever a record is retired~\cite{Cohen2015AOA}.

\begin{figure*}[tb]
    \setlength\tabcolsep{4pt}
	\vspace{-2mm}
	%\hspace{-15mm}
    \small
	\centering
	\begin{tabular}{|l|c|c|c|c|c|c|c|c|c|c|c|c|}
	\hline
	Necessary code modifications 			   & RC & HP & B\&C & TS & ST & EBR & DTA & QS & OA & DEBRA & DEBRA+ \\
	\hline
	\hspace{2mm} per accessed \record\ 	& \cmark & \cmark & \cmark & & \cmark & & \cmark & \cmark & \cmark & & \\
	\hspace{2mm} per operation 	 & & & & \cmark & & \cmark & \cmark & \cmark & & \cmark & \cmark \\
	\hspace{2mm} per retired \record\ 	& & \cmark & \cmark & \cmark & \cmark & \cmark & \cmark & \cmark & \cmark & \cmark & \cmark \\
	\hspace{2mm} other 			& a & b & a & & b, c & & d & b & e & & f \\
	\hline
%	\begin{tabular}{@{}l@{}}Can traverse pointer from\\retired \record\ to retired \record\end{tabular} & \cmark & \xmark & \cmark & \xmark & \xmark & \cmark & \cmark & \xmark & \cmark & \cmark & \cmark \\
    Special timing assumptions & & & & For progress & & & & For correctness & & & \\
    \hline
    Fault tolerant & \cmark & \cmark & \cmark & & \cmark & & \cmark & \cmark & \cmark & & \cmark \\
    \hline
    \begin{tabular}{@{}l@{}}Termination of memory\\reclamation procedures\end{tabular} & L & W & L & Blocking & L & L/W & L & L$_{rooster}$ & L & W & W$_{sig}$ \\
    \hline
	\begin{tabular}{@{}l@{}}Can traverse pointer from\\retired \record\ to retired \record\end{tabular} & \cmark & & \cmark & & & \cmark & \cmark & & \cmark & \cmark & \cmark \\
	%Unfreeable retired \record s & $\infty$ & $\Theta(mn)$ & $\Theta(kn^2)$ & $O(n(k+d))$ & ??? & $\Theta(kn^2)$ & $\infty$ & $\Omega(scn^2)$ & $\infty$ & $\Theta(mn^2)$ \\
	%%\hspace{2mm} steps per freed \record\             & $\infty$ & - & e.a. $O(1)$ & - & - & $\infty$ & - & a. $O(1)$ \\
	\hline
	\end{tabular}
	\vspace{-2mm}
	\caption{
%\trevor{timing assumptions / model of computation, whether they need htm, only support some d.s., progress}
	Summary of reclamation schemes.
	\textbf{Other code modifications:} (a) break cycles in pointers;
	(b) write recovery code for when a process fails to acquire a HP;
	(c) insert transaction \textit{checkpoints} after every few lines of code;
	(d) integrate crash recovery with the lock-free data structure's synchronization mechanisms (and only works for lists);
	(e) transform lock-free algorithm into normalized form, and then instrument every read, write and CAS on shared memory;
	(f) write crash recovery code (which is trivial for many data structures).
	%Variables: 
	%$n$ is the number of processes, $k$ is the largest number of hazard pointers used by any process,
	%$c$ is the number of \record s protected by an anchor, $s$ is the number of operations a process performs before it suspects another process has crashed, $m$ is the maximum number of \record s removed from the data structure by a single operation, and $d$ is the maximum, over all executions and retired \record s $r$, of the total number of pointers that have ever pointed to $r$, plus the outdegree of $r$.
    \textbf{Termination of memory reclamation procedures:} (L) lock-free; (L$_{rooster}$) lock-free if rooster processes cannot crash; (W) wait-free; (W$_{sig}$) wait-free if the operating system's signaling mechanism is wait-free
	}
	\label{fig-related}
	\vspace{-2mm}
\end{figure*}

\medskip
\noindent\textbf{Applying each technique.}
The effort required to apply these techniques varies widely.
See Figure~\ref{fig-related} for a summary.

%RC can be applied nearly automatically by a compiler, but any cycles in pointers must be manually broken (a task that requires knowledge of the data structure implementation).
%All others require a procedure to be invoked whenever a \record\ has been retired.
%In addition, HPs, B\&C and DTA each require a programmer to study a data structure implementation to determine which pointers need to be protected, and insert code to obtain the necessary protection.
%EBR requires a procedure to be invoked at the beginning of each operation.
%Currently, ST also requires a programmer to insert code at the beginning and end of each operation, and after every few lines of lock-free data structure code (although, it should be possible to write a compiler to automate much of this).

\section{DEBRA: Distributed Epoch Based Reclamation} %Technique without fault tolerance}
\label{sec-technique}

\begin{figure*}
\lstset{escapechar=@,style=customc}
\centering
\begin{minipage}{0.44\textwidth}
\centering
(Applying DEBRA)
\begin{lstlisting}[frame=single]
Value search(Key key) {
+   leaveQstate();

    Node *node = root;
    
    while (!node.isLeaf()) {

        if (key < node->key)) {
            node = node->left;
        } else {
            node = node->right;
        }






    }
    if (key == node->key) {
+       Value result = node->value;
+       enterQstate();
        return result;
    }
+   enterQstate();
    return NO_VALUE;
}
\end{lstlisting}
\end{minipage}
\hspace{3mm}
\begin{minipage}{0.48\textwidth}
\centering
(Applying hazard pointers)
\begin{lstlisting}[frame=single]
Value search(Key key) {
+   announce root
+   if (root is retired) restart search
    Node *node = root;
+   Node *prev = NULL;
    while (!node.isLeaf()) {
+       prev = node;
        if (key < node->key)) {
            node = node->left;
        } else {
            node = node->right;
        }
+       announce node
+       if (node is retired) {
+           release hazard pointer to prev
+           restart search
+       }
+       release hazard pointer to prev
    }
    if (key == node->key) {
+       Value result = node->value;
+       release hazard pointer to node
        return result;
    }
+   release hazard pointer to node
    return NO_VALUE;
}
\end{lstlisting}
\end{minipage}
\caption{Applying DEBRA and HPs to a search in a binary search tree. (+) denotes a new line. %Note that determining whether ``node is \textit{retired}'' (and not just \textit{marked}) can be extremely difficult, as we see in Section~\ref{sec-debra-related}.
}
\label{fig-using-debra-vs-hp}
\end{figure*}

In this section, we present DEBRA, a distributed version of EBR with numerous advantages over classical EBR.

First, DEBRA supports a sort of \textit{partial fault tolerance}.
Consider an execution of a data structure that uses EBR.
Observe that a process that sleeps for a long time will delay reclamation for all other processes, \textit{even if it is not currently executing an operation on the data structure}.
In DEBRA, a process can prevent other operations from reclaiming memory \textbf{only} if it is currently executing an operation on the data structure.
Thus, if a process sleeps for a long time or crashes while it is not executing an operation on the data structure, other processes will continue to reclaim memory as usual.
This can have a significant impact in real applications, where operations on a data structure may represent a small part of the execution.
It also makes it possible to terminate some of the processes operating on a data structure, or reassign them to different tasks, without permanently halting reclamation for all other processes.

Second, recall that in EBR, each time a process begins a new operation, it reads the epoch announcements of all processes.
This can be expensive, especially on NUMA systems, where reading the epoch announcements of processes on other sockets is likely to incur extremely expensive last-level cache misses.
In DEBRA, we read the epoch announcements of all processes \textit{incrementally} over many operations.
This can slightly delay the reclamation of some records, but it dramatically reduces the overall cost of reading epoch announcements.

Third, instead of having all processes synchronize on shared epoch bags, each process has its own local epoch bags, and reclamation proceeds independently for each process.
Additionally, these bags are carefully optimized for good cache performance and extremely low overhead.

\paragraph{Using DEBRA}

%Our new memory reclamation scheme, 
DEBRA provides four operations:
%We define a DEBRA object, which is used to reclaim records.
%A DEBRA object provides four operations: 
\leaveq$()$, \enterq$()$, \retire$(r)$ and \isq$()$, where $r$ is a record.
Each of these operations takes $O(1)$ steps in the worst-case.
Let $T$ be a lock-free data structure.
%Recall that a process is in a \textit{quiescent state} whenever it does not have a pointer to any \record\ in the data structure.
To use DEBRA, $T$ simply invokes \leaveq\ at the beginning of each operation, \enterq\ at the end of each operation, and \retire$(r)$ each time a record $r$ is retired (i.e., removed from $T$).
Like EBR, DEBRA assumes that a process does not hold a pointer to any record between successive operations on $T$.
%Another constraint is that \retire$(r)$ can be invoked only once each time $r$ is retired.
%Each time a record $r$ is retired, \retire$(r)$ can be invoked only once.
Each process alternates invocations of \leaveq\ and \enterq, beginning with an invocation of \leaveq. %can invoke \leaveq\ only if it has invoked \enterq\ since its last invocation of \leaveq, or it has never invoked \leaveq.
%It is permissible for a process to invoke \enterq\ repeatedly.
Each process is said to be \textit{quiescent} initially and after invoking \enterq, and is said to be \textit{non-quiescent} after invoking \leaveq.
An invocation of \isq\ by a process returns true if it is quiescent, and false otherwise. %$p$ invoked \enterq\ since its last invocation of \leaveq, or if $p$ has never invoked \leaveq, and false otherwise.
Figure~\ref{fig-using-debra-vs-hp} is an example of DEBRA applied to code for searching a lock-free binary search tree.
For comparison, it also shows how HPs could be applied \textit{\textbf{if}} one could determine whether a node is retired.
Note that determining whether node is \textit{retired} (and not just \textit{marked}) can be extremely difficult, as discussed in Section~\ref{sec-debra-related} (where we discussed problems with HPs).
%which we have argued can be extremely difficult (in Section~\ref{sec-debra-related}, where we discussed problems with HPs).

\paragraph{Implementation}

\begin{figure*}
\lstset{escapechar=@,style=customc}
\begin{lstlisting}[frame=single]
process local variables:
    long pid;                           // process id
    long checkNext;                     // the next process whose announcement should be checked
    blockbag *bags[0..2];               // limbo bags for the last three epochs
    blockbag *currentBag;               // pointer to the limbo bag for the current epoch
    long index;                         // index of currentBag in bags[0..2]
    long opsSinceCheck;                 // # ops performed since checking another process' announcement
shared variables:
    long epoch;                         // current epoch
    long announce[n];                   // per-process announced epoch and quiescent bit
    objectpool *pool;                   // pointer to object pool

bool getQuiescentBit(long otherPid)      { return announce[otherPid] & 1; }
void setQuiescentBitTrue(long otherPid)  { announce[otherPid] = announce[otherPid] | 1; }
void setQuiescentBitFalse(long otherPid) { announce[otherPid] = announce[otherPid] & ~1; }
bool isEqual(long readEpoch, long announcement) {
    return readEpoch == (announcement & ~1); // compare read epoch to epoch-bits from announcement
}

void retire(record *p) { currentBag->add(p); }
bool isQuiescent()     { return getQuiescentBit(pid); }
void enterQstate()     { setQuiescentBitTrue(pid); }
bool leaveQstate() {
    bool result = false;
    long readEpoch = epoch;
    if (!isEqual(readEpoch, announce[pid])) { // our announcement differs from the current epoch
        opsSinceCheck = checkNext = 0;    // we are now scanning announcements for a new epoch
        rotateAndReclaim();
        result = true;                    // result indicates that we changed our announcement
    }
    // incrementally scan all announcements
    if (++opsSinceCheck >= CHECK_THRESH) {
        opsSinceCheck = 0;
        long other = checkNext % n;
        if (isEqual(readEpoch, announce[other]) || getQuiescentBit(other)) {
            long c = ++checkNext;
            if (c >= n && c >= INCR_THRESH) { // if we scanned every announcement
                CAS(&epoch, readEpoch, readEpoch+2);
    }   }   }
    announce[pid] = readEpoch;          // announce new epoch with quiescent bit = false
    return result;
}
void rotateAndReclaim() { // rotate limbo bags and reclaim records retired two epochs ago
    index = (index+1) % 3;              // compute index of oldest limbo bag
    currentBag = bags[index];           // reuse the oldest libmo bag as the new currentBag
    pool->moveFullBlocks(currentBag);   // move all full blocks to the pool
}
\end{lstlisting}
\caption{C++ style pseudocode for DEBRA, where $n$ is the number of processes, INCR\_THRESH is the minimum number of times a process must invoke \leaveq\ before it can increment the epoch, and CHECK\_THRESH is the number of times a process must invoke \leaveq\ before it will check the epoch announcement of another process.}
\label{fig-debra}
\end{figure*}

C++ style psuedocode for DEBRA appears in Figure~\ref{fig-debra}.
Each process $p$ has three limbo bags, denoted $bag_0$, $bag_1$ and $bag_2$, which contain records that it removed from the data structure.
%Specifically, $bag_i$ contains records the process removed from the data structure  in the current epoch, the previous epoch, .
%At any point, one of these bags 
%These bags are called \textit{currBag}, \textit{prevBag} and \textit{pprevBag}.
At any point, one of these bags is designated as $p$'s limbo bag for the current epoch, and is pointed to by a local variable \textit{currentBag}.
Whenever $p$ removes a record from the data structure, it simply adds it to \textit{currentBag}.
%A variable \textit{index} records the index of \textit{currentBag} in $\langle bag_0, bag_1, bag_2 \rangle$.
Each process has a \textit{quiescent bit}, which indicates whether the process is currently quiescent.
The only thing $p$ does when it enters a quiescent state is set its quiescent bit.
Whenever $p$ \textit{leaves} a quiescent state, it reads the current epoch $e$ and announces it in \textit{announce}$_p$.
If this changes the value of \textit{announce}$_p$, then the contents of the oldest limbo bag can be reused or freed.
In this case, $p$ changes \textit{currentBag} to point to the oldest limbo bag, and then moves the contents of \textit{currentBag} to an object pool. %, and the limbo bags are renamed so that $bag_3$ becomes the new $bag_0$, $bag_2$ becomes the new $bag_3$, $bag_1$ becomes the new $bag_2$ and $bag_0$ becomes the new $bag_1$. %\textit{currBag} becomes the new \textit{prevBag}, \textit{prevBag} becomes the new \textit{pprevBag}, and \textit{pprevBag} becomes the new \textit{currBag}.
Next, $p$ attempts to determine whether the epoch can be advanced, which is the case if each process is either quiescent or has announced $e$.
%To facilitate checking this condition, each process has a \textit{quiescent bit}, which indicates whether the process is currently quiescent.
To do this efficiently, $p$ checks the announcements and quiescent bits of all processes \textit{incrementally}, reading one announcement and one quiescent bit in each \leaveq\ operation.
Process $p$ repeatedly checks the announcement and quiescent bit of the same process $q$ in each of its \leaveq\ operations, until $q$ either announces the current epoch or becomes quiescent, or until the epoch changes.
A local variable \textit{checkNext} keeps track of the next process whose announcement should be checked.
Once \textit{checkNext} is $n$, $p$ performs a CAS to increment the current epoch.
% (specifically, the announcement of process \textit{checkNext} modulo $n$).
% it resets a local variable, \textit{checkNext}, which records the number of announcements it has scanned since it last announced a new epoch.
%The implementations of \isq$()$, \enterq$()$ and \retire$(r)$ are trivial: \isq\ returns the process' quiescent bit, \enterq\ sets the quiescent bit, and \retire\ simply adds $r$ to the object pool.
%The only thing that $p$ does when it enters a quiescent state is set its quiescent bit. %Whenever $p$ \textit{enters} a quiescent state, it simply sets its quiescent bit.
%\trevor{give reference to pseudocode. (a bit later?) move the figure up here (or later).}

% after r is put in a bag...
% r put into current bag, and after, the currentBag pointer must change 3 times (before the bag's contents are moved to the pool and the bag is reused).
% each of 3 changes happens in invocations of leaveq that announces a new epoch.
% between any pair of these invocations of leaveq, the epoch must change.
% so, there are 2 epoch changes between t1 and t2.
% between these 2 epoch changes, for each process q, there is a time when q was quiescent or announced a new epoch (and q can only announce a new epoch when it is quiescent).
% so, q was quiescent at some point in [t1, t2], and could not hold a pointer to r at that time.
% it follows that q must obtain a pointer to r by following pointers from an entry point starting at some time after t1.
% but, r is removed from the data structure before t1, so this is a contradiction.

\paragraph{Correctness}

%It is fairly straightforward to show that 
DEBRA reclaims a record only when no process has a pointer to it:
Suppose $p$ places a record $r$ in limbo bag $b$ at time $t_1$, and moves $r$ from $b$ to the pool at time $t_2$.
Assume, to obtain a contradiction, that a process $q$ has a pointer to $r$ at time $t_2$.
At time $t_1$, $b$ is $p$'s current limbo bag, and just before time $t_2$, $b$ is changed from being $p$'s oldest limbo bag to being $p$'s current limbo bag, again.
Thus, \textit{currentBag} must be changed at least three times between $t_1$ and $t_2$.
Since $p$ changes \textit{currentBag} only in an invocation of \leaveq\ that changes \textit{announce}$_p$, $p$ must perform at least three such invocations between $t_1$ and $t_2$.
The current epoch must change between any pair of invocations of \leaveq\ that change \textit{announce}$_p$, so the current epoch must change at least twice between $t_1$ and $t_2$.
Consider two consecutive changes of the current epoch, from $e$ to $e'$ and from $e'$ to $e''$.
At some point between these two changes, $q$ must either be quiescent or have announced $e'$.
%\trevor{the following few sentences are confusing...}
Process $q$ must announce $e'$ after reading the current epoch and seeing $e'$, before the current epoch changes from $e'$ to $e''$.
Thus, $q$ must announce $e'$ after the current epoch changes from $e$ to $e'$, and before it changes from $e'$ to $e''$.
Since $q$ can only announce an epoch when it is in a quiescent state, $q$ must therefore be quiescent at some point between $t_1$ and $t_2$.
This means $q$ must obtain its pointer to $r$ by following pointers from an entry point after it was quiescent, which is after $t_1$.
However, $r$ is removed from the data structure before $t_1$, and, hence, it is no longer reachable by following pointers from an entry point.
This is a contradiction.

\paragraph{Object pool}

The object pool shared by all processes is implemented as a collection of $n$ \textit{pool bags}, one per process, and one shared bag.
Whenever a process moves a record to the pool, it places the record in its pool bag.
If its pool bag is too large, it moves some records to the shared bag.
Whenever a process wants to allocate a record, it first attempts to remove one from its pool bag.
If its pool bag is empty, it attempts to take some records from the shared bag.
If the process fails to take any record from the shared bag, then it will allocate some new records and place them in its pool bag. %\trevor{now, this works differently. we allocate a whole block of records}.

\paragraph{Block bags}

For efficiency, each pool bag and limbo bag is implemented as a \textit{blockbag}, which is a singly-linked list of \textit{blocks}.
Each block contains a \textit{next} pointer and up to $B$ records.
(In our experiments, $B = 256$.)
The head block in a blockbag always contains fewer than $B$ records, and every subsequent block contains exactly $B$ records.
With this invariant, it is straightforward to design constant time operations to add and remove records in a blockbag, and to move all full blocks from one blockbag to another.
This allows a process to move all full blocks of records in its oldest limbo bag to the pool highly efficiently after announcing a new epoch.
However, if the process only moves \textit{full} blocks to the pool, then this limbo bag may be left with some records in its head block.
These records \textit{could} be moved to the pool immediately, but it is more efficient to leave them in the bag, and simply move them to the pool later, once the block that contains them is full.
One consequence of not moving these records to the pool is that each limbo bag can contain at most $B-1$ records that were retired two or more epochs ago.
This does not affect correctness.
The shared bag is implemented as a lock-free singly-linked list of blocks with operations to add and remove a full block. %\textit{add} and \textit{remove} operations.
Moving entire blocks to and from the shared bag greatly reduces
%the frequency with which processes must 
synchronization costs. %on the shared bag.

%\textbf{using blocks is very efficient.}
%\textbf{however, it means we have to allocate and deallocate blocks.}
Operating on blocks instead of individual records significantly reduces overhead. %, and improves memory locality and cache performance.
However, it also requires a process to allocate and deallocate blocks.
To reduce the number of blocks that are allocated and deallocated during an execution, each process has a bounded \textit{block pool} that is used by all of its local blockbags.
Instead of deallocating a block, a process returns the block to its block bool.
If the block pool is already full, then the block is freed.
Experiments show that allowing each process to keep up to 16 blocks in its block pool reduces the number of blocks allocated by more than 99.9\%.
No blocks are allocated for the shared bag, since blocks are simply moved between pool bags and the shared bag.

\paragraph{Minor optimizations}

%C++ style psuedocode for DEBRA appears in Figure~\ref{fig-debra}.
We make two additional optimizations in our implementation.
First, the least significant bit of \textit{announce}$_p$ is used as $p$'s quiescent bit.
This allows both values to be read and written atomically, which reduces the number of reads and writes to shared memory.
Second, a process attempts to increment the current epoch only after invoking \leaveq\ at least INCR\_THRESH times, where INCR\_THRESH is a constant (100 in our experiments).
This is especially helpful when the number of processes is small.
For example, in a single process system, without this optimization, the process will advance the epoch and try to move records to the pool at the beginning of every single operation, introducing unnecessary overhead.

\paragraph{Optimizing for NUMA systems}

%%On a system with a uniform memory architecture, two different processors have the same cost to access any given part of memory.
%\trevor{define this stuff clearly in the model. introduce cache hierarchies and cache protocols: shared mode, exclusive mode, invalidations and misses. [maybe fix definitions like this in the abtree section.]}
%On a NUMA system, different processors can have widely differing costs to access different parts of memory.
%Memory on a NUMA system is conceptually divided into different \textit{NUMA zones}, and certain processors have affinities for certain NUMA zones.
%Alternatively, we say that memory in certain NUMA zones is \textit{local} to a given processor, or group of processors.
%For example, in a multi-socket system, processors on one socket can often access shared caches on that socket significantly more efficiently than they can access caches on another socket.
%In this case, we would say that caches on the same socket are local, and caches on another socket are remote.
%\trevor{not all of these definitions are necessary. we may be able to get away with talking about processes that each have their own private cache, and groups of processes that share a cache.}
%%In other words, different processors can access memory in different NUMA zones (significantly) more efficiently than others.
%%Typically, single-socket systems have uniform architectures, and multi-socket systems have some degree of non-uniformity.

Memory layout and access pattern can have a significant impact on the performance of an algorithm on a NUMA system.
%As an example, consider a two-socket system with a two level cache hierarchy, consisting of a private L1 cache for each processor, and per-socket L2 caches which are shared between processors on the same socket.
%A processor can load data extremely quickly from its L1 cache, quickly from its L2 cache, slowly from the other socket's L2 cache, and extremely slowly from main memory.
%Suppose $n$ threads are executing an algorithm in which one process $p$ repeatedly increments a counter $C$, and all other processes repeatedly read $C$.
%Let $r_1$ and $r_2$ be any two consecutive reads of $C$ by a process $q$.
%If $C$ does not change between $r_1$ and $r_2$, then $q$ can read $C$'s value directly from its L1 cache, which is extremely fast.
%Otherwise, $q$ must read $C$'s value either from $p$'s L2 cache, or from main memory.
%If $p$ and $q$ are on the same socket, and share the same L2 cache, then $q$ can read $C$'s value directly from its L2 cache, which is fast.
%However, if $p$ and $q$ are on different sockets, then $q$ must read $C$'s value from the L2 cache on the other socket, which is slow.
%In this case, we say that $q$ incurs a \textit{remote cache miss} when it performs $r_2$.
%Every time $p$ increments $C$, it causes \textit{all processes on the other socket} to incur expensive remote cache misses.
%
Recall that each invocation of \leaveq\ by a process $q$ reads one announcement \textit{announce}$_p$ (which is periodically changed by process $p$).
If $p$ is on the same socket as $q$ (so $p$ and $q$ share the last-level cache), then this read will usually be quite fast, since a write by $p$ to \textit{announce}$_p$ will not invalidate $q$'s cached copy of \textit{announce}$_p$.
However, if they are on different sockets, then a write by $p$ to \textit{announce}$_p$ will invalidate $q$'s cached copy, and will cause a subsequent read of \textit{announce}$_p$ by $q$ to incur a last-level cache miss.
The cost of these last-level cache misses is not noticeable in all workloads, but we \textit{did} notice it in experimental workloads where all processes performed short read-only operations on small data structures that fit in the last-level cache (so the only last-level cache misses were caused by these reads).
To reduce the impact of these cache misses, we suggest reading an announcement only once every CHECK\_THRESH invocations of \leaveq\ (where CHECK\_THRESH is a small constant).
Of course, checking announcements less frequently can delay reclamation for some records.

\section{Adding fault tolerance} \label{sec-debraplus}

The primary disadvantage of DEBRA is that a crashed or slow process can stay in a non-quiescent state for an arbitrarily long time.
This prevents any other processes from freeing memory.
Although we did not observe this pathological behaviour in our experiments, many applications require a bound on the number of records waiting to be freed.

%• Explain very carefully that the recovery code consists of a few steps that are not a problem if they are interrupted and not executed, followed by the lock-free algorithm's help procedure, which is never a problem if you crash while executing it (because lock-free algorithms let other processes clean up the mess left by a process). Lock-free algorithms are designed so that if a process crashes while it is in the middle of an operation and leaves the d.s. in an inconsistent state, other processes can repair that state. The onus is on a process that wants to access part of a data structure to restore that part of the data structure to a consistent state before using it.
%\trevor{good place to talk about how lock-free algorithms are ripe for adding crash recovery mechanisms to, since they are always designed so that any mess left by a crashed process will be cleaned up by other processes.}

Lock-free data structures are ideal for building fault tolerant systems, because they are designed to be provably fault tolerant.
If a process crashes while it is in the middle of any lock-free operation, and it leaves the data structure in an inconsistent state, other processes can always repair that state.
The onus is on a process that wants to access part of a data structure to restore that part of the data structure to a consistent state before using it.
Consequently, a lock-free data structure always provides procedures to repair and access parts of the data structure that are damaged (by a process crash) or undergoing changes.
(Furthermore, these procedures are necessarily designed so that processes can crash while executing them, and other processes can still repair the data structure and continue to make progress.)
We use these procedures to design a mechanism that allows a process to \textit{neutralize} another process that is preventing it from advancing the epoch.
%In this section, we describe how to implement fault tolerance for DEBRA by allowing a process to \textit{neutralize} another process that is preventing it from advancing the epoch.

A novel aspect of DEBRA+ is our %We make
use of two features offered by Unix, Linux and other POSIX-compliant operating systems. %Unix based operatiof uses a feature of Unix based operating systems (e.g., Linux and Solaris)
The first is \textit{signaling}, an inter-process communication mechanism.
Signals can be sent to a process by the operating system, and by other processes.
When a process receives a signal, the code it was executing is interrupted, and the process begins executing a \textit{signal handler}, instead.
When the process returns from the signal handler, it resumes executing from where it was interrupted.
A process can specify what action it will take when it receives a particular signal by registering a function as its signal handler. % that should be executed when it receives a signal. by \textit{registering} a function as its signal handler.

The second feature is \textit{non-local goto}, which allows a process to begin executing from a different instruction, \textit{outside} of the current function, by using two procedures: \textit{sigsetjmp} and \textit{siglongjmp}.
A process first invokes \textit{sigsetjmp}, which saves its local state and returns false.
Later, the process can invoke \textit{siglongjmp}, which restores the state saved by \textit{sigsetjmp} immediately prior to its return, but causes it to return true instead of false.
%Non-local goto is implemented with a pair of procedures called \textit{sigsetjmp} and \textit{siglongjmp}.
%\textit{sigsetjmp}
%%
%Invoking \textit{sigsetjmp} saves the local state of the process, and the process can subsequently use \textit{siglongjmp} to restore that state.
%%Invoking \textit{siglongjmp} restores the local state of the process to one that was previously saved by a call to a procedure called \textit{sigsetjmp}.
The standard way to use these primitives is with the idiom: %to invoke it in the antecedent of an if-statement (e.g.,
``if (sigsetjmp(...)) alternate(); else usual();''. %\{ X; \} else \{ Y; \}''. %).
A process that executes this code will save its state and execute usual().
Then, if the process later invokes \textit{siglongjmp}, it will restore the state saved by \textit{sigsetjmp} and immediately begin executing alternate().
%The \textit{sigsetjmp} procedure returns zero when a process initially invokes it.
%Later, if the process performs \textit{siglongjmp} and returns to an earlier invocation of \textit{sigsetjmp}, the 

At the beginning of each operation by a process $q$, $q$ invokes \textit{sigsetjmp}, following the idiom described above, and then proceeds with its operation.
Another process $p$ can interrupt $q$ by sending a signal to $q$. %using a procedure called \textit{pthread\_kill} to send a signal to $q$.
We design $q$'s signal handler so that, if $q$ was interrupted while it was in a quiescent state, then $q$ will simply return from the signal handler and resume its regular execution from wherever it was interrupted.
However, if $q$ was interrupted in a non-quiescent state, then it is neutralized: it will enter a quiescent state and perform a \textit{siglongjmp}.
Then, $q$ will execute special \textit{recovery code}, which allows it to clean up any mess it left because it was neutralized.
Since $q$'s signal handler performs \textit{siglongjmp} only if $q$ was interrupted in a non-quiescent state, $q$ will not perform \textit{siglongjmp} while it is executing recovery code.
Hence, if $q$ receives a signal while it is executing recovery code, it will simply return from the signal handler and resume executing %the recovery code
wherever it left off.

%To avoid any problems that might occur if processes are interrupted or neutralized while executing recovery code, 
%It consists of a small number of steps that can be interrupted and not executed
%Explain very carefully that the recovery code consists of a few steps that are not a problem if they are interrupted and not executed, followed by the lock-free algorithm's help procedure, which is never a problem if you crash while executing it (because lock-free algorithms let other processes clean up the mess left by a process).

Our technique requires the operating system to guarantee that, after a process $p$ sends $q$ a signal, %uses \textit{pthread\_kill} to send $q$ a signal, 
the next time process $q$ takes a step, it will execute its signal handler.
(This requirement is satisfied by the Linux kernel~\cite{kerrisk2010linux}. %version 4.10.1 of the Linux kernel, and has been satisfied by many previous versions, as far as we have gone back.
It can also be weakened with small modifications to DEBRA+, which are discussed at the end of this section.)
With this guarantee, after $p$ has sent a signal to $q$, it knows that $q$ will not access any retired record until $q$ has executed its recovery code and subsequently executed \leaveq.
%This allows $p$ to proceed as if $q$ is quiescent, as soon as $p$ has sent $q$ a signal.
Thus, as soon as $p$ has sent $q$ a signal, $p$ can immediately proceed as if $q$ is quiescent.

\paragraph{Operations for which recovery is simple}

\begin{wrapfigure}{r}{0.4\textwidth}
\vspace{-5mm}
\includegraphics[width=\linewidth]{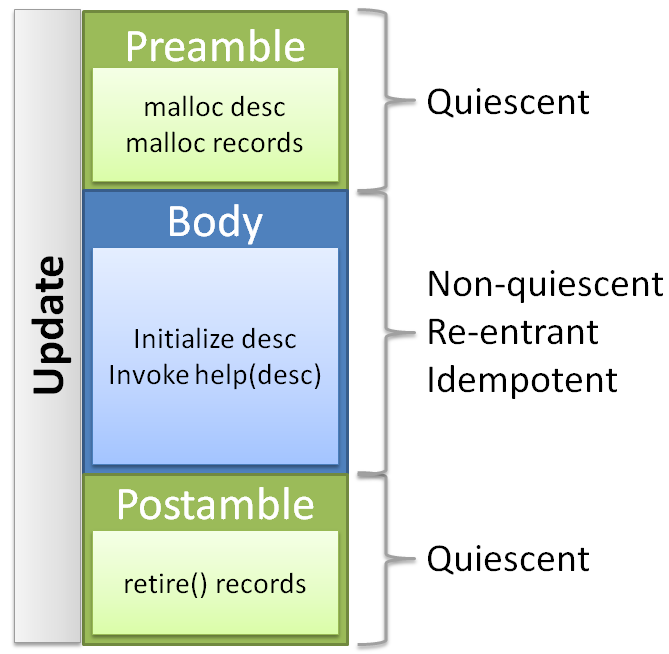}
%\caption{Operations for which recovery is simple.}
%\label{fig-debraplus-recoverable-data-structures}
\vspace{-5mm}
\end{wrapfigure}
The main difficulty is using DEBRA+ is designing recovery code.
Although recovery must be tailored to the data structure, it is straightforward for lock-free operations of the following form.
Each operation is divided into three parts: a quiescent bookkeeping \textit{preamble}, a non-quiescent \textit{body}, and a quiescent bookkeeping \textit{postamble}.
Processes can be neutralized while executing in the body, but cannot be neutralized while executing in the preamble or postamble (because a process will not call \textit{siglongjmp} while it is quiescent).
Consequently, processes should not do anything in the body that will corrupt the data structure if they are neutralized part way through. % and then start executing it again from the beginning.
%In other words, the body should be reentrant.
Allocation, deallocation, manipulation of process-local data structures that persist between operations, and other non-reentrant actions should occur only in the preamble and postamble.

\paragraph{Applying DEBRA+}

Figure~\ref{fig-applying-debraplus} shows how to apply DEBRA+ to the type of operations described above.
The remainder of this section explains the steps shown there.
Consider an operation $O$.
In the quiescent preamble, $O$ allocates a special record called a \textit{descriptor}. %, which will be used to describe how to perform the operation.
%The descriptor functions like a road-map that explains can be used by other processes to help $O$ 
A pointer to this descriptor is stored in a process local variable called $desc$.
Any other records that might be needed by the body are also allocated in the preamble.

\begin{figure}
%\hspace{-8mm}
\begin{minipage}{\linewidth}
%\begin{minipage}{0.296\textwidth}
%\centering
%(No reclamation)
%\lstset{escapechar=@,style=customc}
%\begin{lst}[frame=single]
%
%
%
%int doOperationXYZ(args) {
%   int result;
%   // allocate records
%   while (!finished) {
%    
%    
%    
%    
%
%    
%        
%      do search phase
%      create descriptor
%      assign result
%
%
%
%
%
%      help(descriptor)
%
%        
%   }
%   // deallocate any unused
%
%   return result;
%}
%\end{lstlisting}
%\end{minipage}
%\hspace{2mm}
%\begin{minipage}{0.296\textwidth}
%\centering
%(With DEBRA)
%\lstset{escapechar=@,style=customc}
%\begin{lstlisting}[frame=single]
%
%
%
%int doOperationXYZ(args) {
%   int result;
%   // allocate records
%   while (!finished) {
%    
%    
%    
%    
%
%    
%+     leaveQstate();
%      do search phase
%      create descriptor
%      assign result
%
%
%
%
%
%      help(descriptor)
%+     enterQstate();
%        
%   }
%   // deallocate any unused
%+  perform retire() calls
%   return result;
%}
%\end{lstlisting}
%\end{minipage}
%\hspace{2mm}
%\begin{minipage}{0.385\textwidth}
%\centering
%(With DEBRA+)
\lstset{escapechar=@,style=customc}
\begin{lstlisting}[name=applying,aboveskip=0pt,belowskip=0pt]
process local variables:
   descriptor *desc; @\vspace{1.5mm}@
\end{lstlisting}
\begin{lstlisting}[name=applying,aboveskip=0pt,belowskip=0pt,backgroundcolor=\color{gray!20}]
void signalHandler(args):
  if (isQuiescent()) then
    enterQstate();
    siglongjmp(...);        // jump to recovery code@\vspace{1.5mm}@
\end{lstlisting}
\begin{lstlisting}[name=applying,aboveskip=0pt,belowskip=0pt]
int doOperationXYZ(args):
  ...                       // quiescent preamble
  while (!done)
\end{lstlisting}
\begin{lstlisting}[name=applying,aboveskip=0pt,belowskip=0pt,backgroundcolor=\color{gray!20}]
    if (sigsetjmp(...))     // begin recovery code
      if (isRProtected(desc)) done = help(desc);
      RUnprotectAll();
    else                    // end recovery code
\end{lstlisting}
\begin{lstlisting}[name=applying,aboveskip=0pt,belowskip=0pt,backgroundcolor=\color{gray!0}]
      leaveQstate();        // begin body
\end{lstlisting}
\begin{lstlisting}[name=applying,aboveskip=0pt,belowskip=0pt]
      do search phase
      initialize *desc
\end{lstlisting}
\begin{lstlisting}[name=applying,aboveskip=0pt,belowskip=0pt,backgroundcolor=\color{gray!20}]
      RProtect each record that will be accessed, or used as the old value of a CAS, by help(desc)
      RProtect(desc);
\end{lstlisting}
\begin{lstlisting}[name=applying,aboveskip=0pt,belowskip=0pt]
      done = help(desc);
\end{lstlisting}
\begin{lstlisting}[name=applying,aboveskip=0pt,belowskip=0pt,backgroundcolor=\color{gray!0}]
      enterQstate();        // end body
\end{lstlisting}
\begin{lstlisting}[name=applying,aboveskip=0pt,belowskip=0pt,backgroundcolor=\color{gray!20}]
      RUnprotectAll();
\end{lstlisting}
\begin{lstlisting}[name=applying,aboveskip=0pt,belowskip=0pt]
  ...                       // quiescent postamble
  perform retire() calls
\end{lstlisting}
\end{minipage}
%\end{minipage}
\vspace{-2mm}
\caption{
	Applying DEBRA+ to a typical lock-free operation.
	Lines 14 and 20 are necessary for both DEBRA and DEBRA+.
	Gray lines are necessary for fault tolerance (DEBRA+).
} % + denotes a line changed to support DEBRA+.}
\label{fig-applying-debraplus}
%\vspace{-5mm}
\end{figure}

The body first reads some records, and then initializes the descriptor $desc$.
Intuitively, this descriptor contains a description of all the steps the operation $O$ will perform.
The body then executes a \textit{help} procedure, which uses the information in the descriptor to perform the operation.
We assume that the descriptor includes all pointers that the \textit{help} procedure will follow (or use as the expected value of a CAS). %describes how another process can help the operation to complete.
The \textit{help} procedure can also be used by other processes to help the operation complete.
In a system where processes may crash, a process whose progress is blocked by another operation cannot simply wait for the operation to complete, so helping is necessary.
The \textit{help} procedure for any lock-free algorithm is typically reentrant and idempotent, because, at any time, one process can pause, and another process can begin helping.
%(More generally, every lock-free algorithm is necessarily designed to function correctly if a process is interrupted at some step in the body and stops executing there.)
The end of the body is marked with an invocation of \enterq\ (which is, itself, reentrant and idempotent).
The quiescent postamble then invokes \textit{retire} for each record that was removed from the data structure by the operation.

%In a system where processes might crash, a process cannot simply wait for another operation to complete.
%Thus, a typical lock-free operation $O$ proceeds by first reading records, then creating and announcing a special record called a \textit{descriptor}, which describes how another process can help the operation to complete, and finally invoking an idempotent \textit{help} procedure that uses the information in the descriptor to perform the operation.
%This \textit{help} procedure can also be used by other processes to help $O$ complete.
%The \textit{help} procedure for any lock-free algorithm is always reentrant and idempotent, because, at any time, one process can pause, and another process can begin helping.
%We restrict our attention to operations with a body that follows this sequence of steps. %first creates a descriptor and then invokes \textit{help} to perform the operation.
%%Any operation that can be split into a quiescent preamble, a reentrant body that creates a descriptor and invokes an idempotent \textit{help} procedure to perform the operation, and a quiescent postamble, is easy to recover from.
%\trevor{maybe mention that this class includes every data structure that follows the tree update template?}

\paragraph{Recovery} % for such operations}
We now describe recovery for an operation $O$ performed by a process $p$.
Suppose $p$ receives a signal, enters a quiescent state, and then performs \textit{siglongjmp} to begin executing its recovery code.
Although there are many places where $p$ might have been executing when it was neutralized, it is fairly simply to determine what action it should take next.
The main challenge is determining whether another process already performed $O$ on $p$'s behalf.
To do this, $p$ checks whether it announced a descriptor for $O$ before it was neutralized.
If it did, then some other process might have seen this descriptor and started helping $O$.
So, $p$ invokes \textit{help} (which is safe even if another process already helped $O$, since \textit{help} is idempotent).
Otherwise, %no other process is aware of $O$, so 
$p$ can simply restart the body of $O$.

%In the recovery code, $p$ must determine whether it should retry $O$.
%Another process may have already performed $O$ on $p$'s behalf, so $p$
%First, $p$ checks whether it announced a descriptor for $O$ before it was interrupted.
%If not, then no other process is aware of $O$, and $p$ can simply restart the body of $O$.
%Otherwise, $p$ invokes \textit{help}.
%%If not, then no other process is aware of $O$, and $p$ can simply restart the body of $O$.
%%However, if $p$ announced a descriptor for $O$, then some other process may have already looked at the descriptor and completed the operation.
%%In this case, $p$ can simply invoke the \textit{help} procedure to complete $O$.
%Since \textit{help} is idempotent, it is safe for $p$ to invoke it even if another process already helped $O$.

DEBRA allows a non-quiescent process executing an operation to safely access any record that it reached by following pointers from an entry point during the operation.
However, DEBRA does \textit{not} allow quiescent processes to safely access any records. %accesses by quiescent processes are \textit{not} protected by DEBRA.
In DEBRA+, once a process $p$ has been sent a signal, other processes treat $p$ as if were quiescent.
Furthermore, $p$ enters a quiescent state before executing recovery code, and it remains in a quiescent state throughout the recovery code.
%Consequently, after $p$ starts executing its signal handler, it cannot safely access any records it has pointers to, since they might already have been freed. %it can no longer safely access any records it previously had pointers to
%Moreover, $p$ enters a quiescent state
%has  so $p$ can not safely access any records. As soon as you execute your signal handler, you enter a quiescent state and remain in a quiescent state throughout the recovery code.
However, the help procedure in $p$'s recovery code must access the descriptor record, and possibly some of the records to which it points. % (\textit{even though $p$ is quiescent}).
%Since these records are not protected by DEBRA, they might already have been freed.
%To prevent these records from being freed before the recovery code can runhelp procedure can use them, we use HPs in a very limited way to protect records so they can be accessed by recovery code. [introduce RProtect]
Thus, we need an additional mechanism in DEBRA+ to allow $p$ to access this limited set of \record s even though it is quiescent.
We use HPs in a very limited way to prevent these records from being freed by other processes before $p$ has finished %either its operation or 
its recovery code.
This lets $p$ safely run its recovery code in a quiescent state, so that other processes can continue to advance the current epoch and reclaim memory. %which does not preventing the epoch from being changed. %way, they can be accessed by recovery code.

We now describe how HPs are used.
Let $S$ be the set of records that will be accessed, or used as the old value of a CAS, by \textit{help}$(desc)$.
In the body of an operation by $p$, %before invoking \textit{help}$(desc)$, 
$p$ announces HPs to all records in $S$ by invoking \textit{RProtect}$(r)$ for each $r \in S$. %in the set $S$ of records that will be accessed, or used as the old value of a CAS, by \textit{help}$(desc)$.
%It does this by invoking \textit{RProtect} for each $r \in S$.
Then, $p$ invokes \textit{RProtect}$(desc)$ to announce a HP to the descriptor, and invokes \textit{help}$(desc)$.
After performing \textit{help}$(desc)$ and invoking \enterq, $p$ invokes \textit{RUnprotectAll} to release all of its HPs.
Note that, since \textit{RProtect} is performed in the body, while $p$ is non-quiescent, $p$ might be neutralized while executing \textit{RProtect}.
Hence, \textit{RProtect} must be reentrant and idempotent.

When executing recovery code, $p$ first invokes \textit{isRProtected}$(desc)$ (which returns true if some HP points to $desc$ and false otherwise) to determine whether it announced a HP to $desc$.
Suppose it did.
Since $p$ announces a HP to the descriptor $d$ \textit{after} announcing HPs to all records in $S$, when $p$ performs recovery, if it sees that it announced a HP to $d$, then it knows it already announced HPs to all records in $S$.
Thus, its HPs will prevent everything it will access during its recovery from being reclaimed until it has finished using them.
So, $p$ can safely execute \textit{help}$(desc)$.
%If so, then each $r$ in $S$ is also protected, and $p$ is ready to safely execute \textit{help}$(desc)$.
Now, suppose $p$ did not announce a HP to $desc$.
Since $p$ announces a HP to $desc$ \textit{before} invoking \textit{help}$(desc)$, this means $p$ has not yet invoked \textit{help}$(desc)$, so no other process is aware of $p$'s operation.
Therefore, $p$ can simply terminate its recovery code and restart its operation.
At the end of the recovery code, $p$ invokes \textit{RUnprotectAll} to release all of its HPs.

Recall that, in DEBRA, whenever a process $p$ announces a new epoch, it can immediately reclaim all records in its oldest limbo bag and move them to its pool.
In DEBRA+, some of the records in $p$'s oldest limbo bag might be pointed to by HPs, so $p$ cannot simply move all of these to its pool.
Before $p$ can move a record $r$ to the pool, it must first verify that no HP points to it. % by invoking \textit{isRProtected}$(r)$, which returns true if some HP points to $r$, and false otherwise.
We discuss how this can be done efficiently, below.

\paragraph{Complexity}

Thanks to our new \textit{neutralizing} mechanism, we can bound the number of \record s waiting to be freed.
Each time a process $p$ performing \leaveq\ encounters a process $q$ that is not quiescent and has not announced epoch $e$, $p$ checks whether the size of its own current limbo bag exceeds some constant $c$.
If so, $p$ neutralizes $q$.
After $p$'s current limbo bag contains at least $c$ elements, and $p$ performs $n$ more data structure operations, it will have performed \leaveq\ $n$ times, and each non-quiescent process will either have announced the current epoch or been neutralized by $p$.
Consequently, $p$ will advance the current epoch, and, the next time it performs \leaveq, it will announce the new epoch and reclaim records.
%Thus, once $p$'s current limbo bag contains at least $c$ elements, $p$ will perform at most $n$ more operations before announcing a new epoch and reclaiming memory.
It follows that $p$'s current limbo bag can contain at most $c+O(nm)$ elements, where $m$ is the largest number of \record s that can be removed from the data structure by a high-level operation.
Therefore, the total number of \record s waiting to be freed is $O(n(nm+c))$.

In DEBRA, all full blocks in a limbo bag are moved to the pool in constant time.
In DEBRA+, \record s can be moved to the pool only if no HP points to them, so this is no longer possible.
One way to move \record s from a limbo bag $b$ to the pool is to iterate over each \record\ $r$ in $b$, and check if a HP points to $r$.
To make this more efficient, we move records from $b$ to the pool only when $b$ contains $nk+\Omega(nk)+B$ \record s, where $k$ is the number of HPs needed per process, and $B$ is the maximum number of \record s in a block.
%Then, by inserting every HP into a hash table 
Before we begin iterating over \record s in $b$, we create a hash table containing every HP.
Then, we can check whether a HP points to $r$ in O(1) expected time.
We can further optimize by rearranging \record s in $b$ so that we can still move full blocks to the pool, instead of individual records.
To do this, we iterate over the \record s in $b$, and move the ones pointed to by HPs to the beginning of the blockbag.
All full blocks that do not contain a \record\ pointed to by a HP are then moved in O(1) time.
%As we iterate over the \record s in $b$, we move each \record\ that is pointed to by a HP to the beginning of $b$.
%This can be done in a single pass by simultaneously iterating over $b$ with two pointers, one that keeps track of how far we have iterated,
%Then, we move all full blocks after the last \record\ pointed to by a HP to the pool in constant time.
Since there are at most $nk$ HPs, and we scan only when $b$ contains at least $nk+B+\Omega(nk)$ \record s, we will be able to move at least $max\{B, \Omega(nk)\}$ \record s to the pool.
Thus, the expected amortized cost to move a record to the pool (or free it) is O(1).

\begin{figure}[ph]
%\vspace{-3mm}
\lstset{escapechar=@,style=customc}
\begin{lstlisting}[frame=single]
process local variables:
    hashtable scanning;                 // hash table used to collect all RProtected records
shared variables:
    arraystack RProtected[n];           // array of RProtected record* for each process

bool isRProtected(record *r) { return RProtected[pid].contains(r); }
bool RProtect(record *r)     { RProtected[pid].add(r);             } // O(1) time
void RUnprotectAll()         { RProtected[pid].clear();            } // O(1) time
bool leaveQstate() {
    bool result = false;
    long readEpoch = epoch;
    if (!isEqual(readEpoch, announce[other])) { // our announcement differs from the current epoch
        checkNext = 0;                    // we are now scanning announcements for a new epoch
        rotateAndReclaim();
        result = true;                  // result indicates that we changed our announcement
    }
    // incrementally scan all announcements
    if (++opsSinceCheck >= CHECK_THRESH) {
        opsSinceCheck = 0;
        long other = checkNext % n;
        if (isEqual(readEpoch, announce[other]) || isQuiescent(other) || suspectNeutralized(other)) {
            long c = ++checkNext;
            if (c >= n && c >= INCR_THRESH) { // if we have scanned every announcement
                CAS(&epoch, readEpoch, readEpoch+1);
    }   }   }
    announce[pid] = readEpoch;          // announce new epoch with quiescent bit = false
    return result;
}
void rotateAndReclaim() {
    index = (current+1) % 3;      // compute index of oldest limbo bag
    currentBag = bags[index];     // reuse the oldest limbo bag as the new currentBag
    // if currentBag contains sufficiently many records to get amortized O(1) time per record
    if (currentBag->getSizeInBlocks() >= scanThreshold) {
        // hash all announcements
        scanning.clear();
        for (int other=0; other < n; ++other) {
            int sz = RProtected[other].size();
            for (int i=0; i<sz; ++i) {
                record *hp = RProtected[other].get(i);
                if (hp != NULL) {
                    scanning.insert(hp);
        }   }   }
        // if any records in currentBag are RProtected, swap them to the front
        blockbag_iterator it1 = currentBag->begin();
        blockbag_iterator it2 = currentBag->begin();
        while (it1 != currentBag->end()) {
            if (scanning.contains(*it1)) {  // record pointed to by it1 is RProtected
                swap(it1, it2);             // swap records pointed to by it1 and it2
                it2++;                      // advance iterator it2
            }
            it1++;                          // advance iterator it1
        }
        // now, every record after it2 can be freed, so we reclaim all full blocks after it2
        pool->moveFullBlocks(it2);          // O(1) time
}   }
bool suspectNeutralized(long other) {
    return (currentBag->getSizeInBlocks() >= SUSPECT_THRESHOLD_IN_BLOCKS)
        && (!pthread_kill(getPthreadID(other), SIGQUIT)); // successfully send signal to other
}
\end{lstlisting}
\vspace{-2mm}
%\begin{figure} [tb]
\lstset{escapechar=@,style=customc}
\begin{lstlisting}[frame=single]
void signalhandler(int signum, siginfo_t *info, void *uctx) {
    // if the process is not in a quiescent state, it jumps to a different instruction
    // and cleans up after itself, instead of continuing its current operation.
    if (!isQuiescent()) {
        enterQstate();
        siglongjmp(...);
}   } // otherwise, the process simply continues its operation as if nothing had happened.
\end{lstlisting}
%\caption{C++ code for the signal handler a process executes whenever it has been suspected and neutralized.}
%\label{fig-sighandler}
%\end{figure}
\caption{C++ style pseudocode for data and procedures \textbf{added to DEBRA} to obtain DEBRA+, where $n$ is the number of processes, INCR\_THRESH is the minimum number of times a process must invoke \leaveq\ before it can increment the epoch, and CHECK\_THRESH is the number of times a process must invoke \leaveq\ before it will check the epoch announcement of another process.}
\label{fig-debraplus}
\end{figure}

\paragraph{Implementation}

C++ style pseudocode for DEBRA+ appears in Figure~\ref{fig-debraplus}.
There, only the procedures that are different from DEBRA are shown.
%The implementation follows the description above, including the optimization to move records from a limbo bag to a pool in expected amortized O(1) steps.
There are three main differences from DEBRA.
First, in an invocation of \leaveq\ by process $p$, if $p$ encounters a process $q$ that has not announced the current epoch, and is not quiescent, $p$ invokes a procedure called \textit{suspectNeutralized}.
This procedure checks whether $p$'s current limbo bag contains more than a certain number of records, and, if so, neutralizes $q$.
Recall that $p$ neutralizes $q$ by sending a signal to $q$.
The signal handler that is executed by $q$ when it receives a signal is called \textit{signalhandler}.
Second, a limited version of HPs is provided by procedures \textit{isRProtected}, \textit{RProtect} and \textit{RUnprotectAll}.
Third, the procedure \textit{rotateAndReclaim} implements the algorithm described above efficiently moving records from a limbo bag to a pool (only if the records are not pointed to by HPs) in expected amortized O(1) steps per record.

%Figure~\ref{fig-applying-debraplus} summarizes how to apply DEBRA+ to an operation of the type described in this section. %, and, for comparison, also shows how DEBRA can be applied.
One problem with DEBRA+ is that it seems difficult to apply to lock-based data structures, because it is dangerous to interrupt a process and force it to restart while it holds a lock.
Of course, there is little reason to use DEBRA+ for a lock-based data structure, since locks can cause deadlock if processes crash. % are inherently not fault tolerant.) %its progress guarantees are meaningless in the context of locking, which is inherently not fault tolerant.)
For lock-based data structures, DEBRA can be used, instead.

%%\section{Object pool implementation}
%\section{Alternative implementation options} \label{appendix-debraplus}
\paragraph{Alternative implementation options}

Above, we specified
%In Section~\ref{sec-debraplus}, we specify
the guarantee that the operating system signaling mechanism must provide: after a process $p$ sends process $q$ a signal, the next time $q$ takes a step, it will execute its signal handler.
It is possible to modify DEBRA+ to work with a weaker guarantee.
Suppose the operating system instead guaranteed that, after a process $p$ sends process $q$ a signal, $q$ is guaranteed to begin executing its signal handler when it next experiences a context switch.
Then, after $p$ sends a signal to $q$, it can either wait or defer reclamation until $q$ is next context switched or is not running.
For most operating system schedulers, every process is context switched out after a bounded length of time (a scheduling quantum).
Many operating systems also provide mechanisms to determine whether a process is currently running, how many context switches it has undergone, how much time remains until it will be context switched, and so on.
For example, Linux provides access to this information through a virtual file system (rooted at "/proc").

%\trevor{explain how to get thread id}

%\trevor{other alternative to sigsetjmp, siglongjmp: try/catch/throw/async-unwind-tables...}
%
%\trevor{for the psuedocode, it remains to explain the thresholds (scanThreshold, SUSPECT\_THRESHOLD\_IN\_BLOCKS), blockbag (moveFullBlocks, getSizeInBlocks), blockbag\_iterator ($++$, $*$, swap), objectpool (moveFullBlocks), arraystack (get, add, size, clear), hashtable (clear, insert, get), getPthreadID, pthread\_kill}

%\section{Ways to get process ID in a signal handler}

%\section{Blockbag pseudocode? Blockbag iterator pseudocode? Block pseudocode? Hashtable pseudocode? Block pool pseudocode?}

\section{A lock-free memory management abstraction} \label{sec-abstraction}

There are many compelling reasons to separate memory allocation and reclamation from data structure code.
Although the steps that a programmer must take to apply a non-automatic technique to a data structure vary widely, it is possible to find a small, natural set of operations that allows %design an abstraction that allows 
a programmer to write data structure code once, and easily plug in many popular memory reclamation schemes.
%The key challenge is to find the correct set of memory management operations to expose to the programmer.
In this section, we describe a record management abstraction, called a Record Manager, that is easy to use, and provides sufficient flexibility to support: HPs (all versions), B\&C, TS, EBR, DTA, QS, DEBRA and DEBRA+.
(ST is not supported because it requires a programmer to insert transactions throughout the code, and to annotate the beginning and end of each stack frame.
OA is not supported because it requires each read and CAS to be instrumented.)

%\trevor{diagram of record manager and its components, and the operations they offer?}

A Record Manager has three components: an Allocator, a Reclaimer and a Pool. %(see Figure~\ref{fig-debraplus-recoverable-data-structures}).
The Allocator determines how records will be allocated (e.g., by individual calls to \textit{malloc} or by handing out records from a large range of memory) and freed.
The Reclaimer is given records after they are removed from the data structure, and determines when they can be safely handed off to the Pool.
The Pool determines when records are handed to the Allocator to be freed, and whether a process actually uses the Allocator to allocate a new record.

We implement data structures, Allocators, Reclaimers and Pools in a modular way, so that they can be combined easily.
This clean separation into interchangeable components allows, e.g., the same Pool implementation to be used with both a HP Reclaimer and a DEBRA Reclaimer.
Modularity is typically achieved with inheritance, but inheritance introduces significant runtime overhead.
For example, when the precompiled data structure invokes \textit{retire}, it does not know which of the precompiled versions of \textit{retire} it should use, so it must perform work at runtime to choose the correct implementation.
In C++, this can be done more efficiently with \textit{template parameters}, which allow a compiler to reach into %compiled 
code and replace placeholder calls with calls to the correct implementations.
Unlike inheritance, templates introduce no overhead, since the correct implementation is compiled into the code.
Furthermore, if the correct implementation is a small function, the compiler can simply insert its code directly into the calling function (eliminating the function call altogether).

\begin{wrapfigure}{r}{0.5\textwidth}
\vspace{-5mm}
\includegraphics[width=\linewidth]{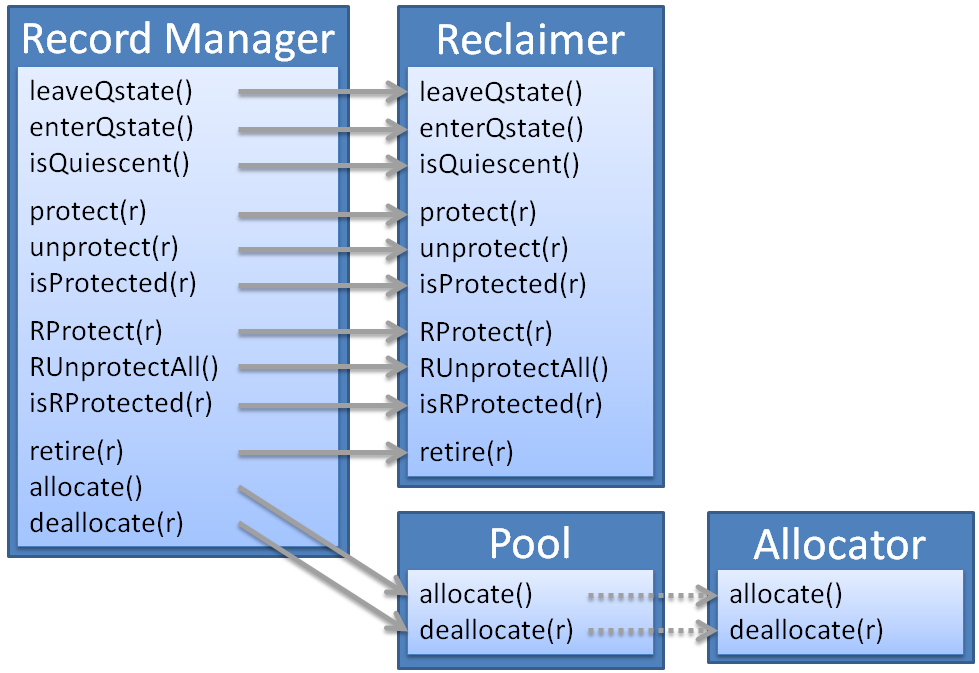}
\caption{Operations provided by a Record Manager, Reclaimer, Pool and Allocator. Solid arrows (resp., dashed arrows) indicate that an operation on one object invokes (resp., may invoke) an operation on another object.}
%\label{fig-debraplus-recoverable-data-structures}
%\vspace{-3mm}
\end{wrapfigure}
A programmer interacts with the Record Manager, which exposes the operations of the Pool and Reclaimer.
A Pool provides \textit{allocate} and \textit{deallocate} operations.
A Reclaimer provides operations for the basic events that memory reclamation schemes are interested in: starting and finishing data structure operations (\leaveq\ and \enterq), reaching a new pointer and disposing of it (\textit{protect} and \textit{unprotect}), and retiring a record (\textit{retire}).
It also provides operations to check whether a process is quiescent (\isq) and whether a pointer can be followed safely (\textit{isProtected}).
Finally, it provides operations for making information available to recovery code (\textit{RProtect}, \textit{RUnprotectAll}, \textit{isRProtected}).
%A Reclaimer provides ten operations: \enterq, \leaveq, \isq, \textit{retire}, \textit{RProtect}, \textit{RUnprotectAll}, \textit{isRProtected}, \textit{protect}, \textit{unprotect} and \textit{isProtected}.
%This might seem excessive, but a successful abstraction must instrument the basic events that memory reclamation schemes are interested in, such as starting and finishing operations (\enterq\ and \leaveq), and reaching a new pointer and disposing of it (\textit{protect} and \textit{•}.
%We also have to provide a mechanism for making information available to recovery code (RProtect).
%
%cover events that memory reclamation schemes might be interested in: reach a new pointer, start a new operation, finished with a pointer; we also have to provide mechanisms for making information available after a process crashes: RProtect

Most of these are described in Section~\ref{sec-technique} and Section~\ref{sec-debraplus}.
We describe the rest here.
\textit{protect}, which must be invoked on a record $r$ before accessing any field of $r$, returns true if the process successfully protects $r$ (and, hence, is permitted to access its fields), and returns false otherwise.
%The second and third arguments of \textit{protect} are: a pointer to a function, and a sequence of arguments to be passed to that function (discussed below).
Once a process has successfully protected $r$, it remains protected until the process invokes \textit{unprotect}$(r)$ or becomes quiescent.
\textit{isProtected}$(r)$ returns true if $r$ is currently protected by the process.

%We now discuss how Reclaimers can be implemented for several non-automatic techniques.
Reclaimers for DEBRA and DEBRA+ are effectively described in Section~\ref{sec-technique} and Section~\ref{sec-debraplus}. % (with Reclaimer pseudocode in Figure~\ref{fig-debra} and Figure~\ref{fig-debraplus}).
For these techniques, \textit{unprotect} does nothing, and \textit{protect} and \textit{isProtected} simply return true.
(Consequently, these calls are optimized out of the code by the compiler.)
For HPs, \leaveq, \textit{RProtect} and \textit{RUnprotectAll} all do nothing, and \isq\ and \textit{isRProtected} simply return false.
\textit{unprotect}$(r)$ releases a HP to $r$, and \enterq\ clears all announced HPs.
\textit{protect} announces a HP to a record and executes a function, which determines whether that record is in the data structure.
%(Thus, the data structure implementation must provide \textit{protect} with a pointer to a function and a sequence of arguments that can be used to determine whether $r$ is in the data structure.)
%If $f(a)$ returns true, then \textit{protect} returns true.
%Otherwise, it returns false.
\textit{retire}$(r)$ places $r$ in a bag, and, if the bag contains sufficiently many records, it constructs a hash table $T$ containing all HPs, and moves all records not in $T$ to the Pool (as described in Section~\ref{sec-debra-related}). % in O(1) expected amortized time (as described in Section~\ref{sec-debra-related}).

%For efficiency, 
A predicate called \textit{supportsCrashRecovery} %and \textit{supportsHelping}, 
is added to Reclaimers to allow a programmer to add crash recovery to a data structure without imposing overhead for Reclaimers that do not support crash recovery.
For example, a programmer can check whether \textit{supportsCrashRecovery} is true before invoking \textit{RProtect}.
%This predicate allows a programmer to, e.g., invoke \textit{RProtect} on records that are needed for crash recovery \textit{only} for Reclaimers that support crash recovery. %, or disable helping for techniques like HPs where helping is problematic.
The code statement ``if (\textit{supportsCrashRecovery})'' %and ``if (supportsHelping)'' 
statically evaluates to ``if (true)'' or ``if (false)'' at compile time, once the Reclaimer template has been filled in. %has replaced the dummy Reclaimer type with an actual Reclaimer.
Consequently, these if-statements are completely eliminated by the compiler. %do not appear in running code.
In our experiments, this predicate is used to invoke \textit{sigsetjmp} %(and their overhead) 
only for DEBRA+ (eliminating overhead for the other techniques). %for all techniques except DEBRA+.
%they let you do useful things with zero overhead. for example, you can RProtect some nodes that will be needed for crash recovery only if crash recovery is supported, or disable helping if HPs are plugged into a data structure. they are used to eliminate sigsetjmp calls (and their overhead) for all techniques except for DEBRA+.}

%\trevor{flow of objects?}

\section{Experiments} \label{sec-debra-exp}

Our primary experimental system was an Intel i7 4770 machine with 4 cores, 8 hardware threads and 16GB of memory, running Ubuntu 14.04 LTS.
%(When the paper~\cite{Brown:2015} this work is based on was published, there was no system with more than 4 cores that provided HTM and, hence, supported ST.)
All code was compiled with GCC 4.9.1-3 and the highest optimization level (-O3).
Google's high performance Thread Caching malloc (tcmalloc-2.4) was used.
%For each data structure, we run six 10 second \textit{trials}, each of which randomly performs insertions, deletions and searches, according to some probability distribution, on uniformly random keys drawn from a fixed key range.
%For each data structure, and each combination of 
%we run six 10 second \textit{trials} for every combination of reclamation techniques, thread counts, operation mixes and key ranges.

\begin{figure}[t]
	\newcommand\graphscale{0.25}
	%\hspace{-17mm}
	\centering
	\hspace{-7mm}
	\begin{tabular}{llll}
	&
	\hfill \small\textbf{50\% ins, 50\% del} \hfill \hfill &
	\hfill \small\textbf{25\% ins, 25\% del, 50\% search} \hfill \hfill &
	\hfill \small\textbf{100\% search} \hfill \hfill
	\\
	\includegraphics[scale=\graphscale]{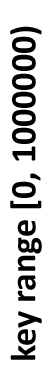} &
	\includegraphics[scale=\graphscale]{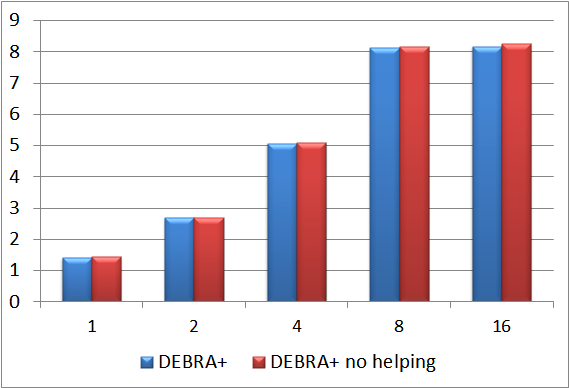} &
	\hspace{-3.5mm}\includegraphics[scale=\graphscale]{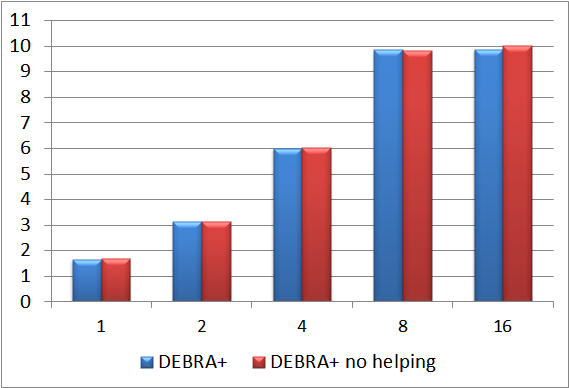} &
	\hspace{-3.5mm}\includegraphics[scale=\graphscale]{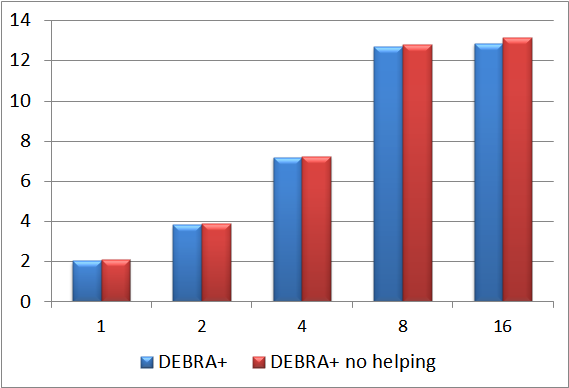}
	\vspace{-0.5mm}
	\\
	\includegraphics[scale=\graphscale]{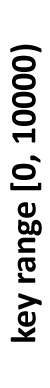} &
	\includegraphics[scale=\graphscale]{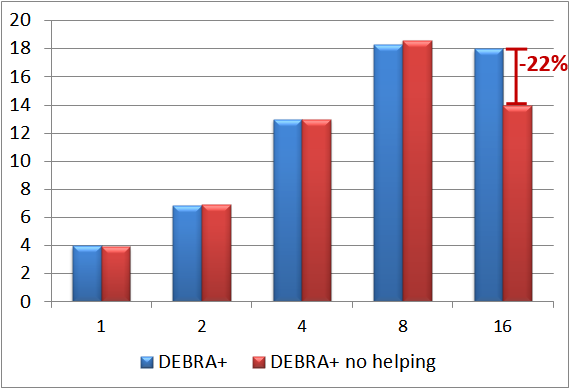} &
	\hspace{-3.5mm}\includegraphics[scale=\graphscale]{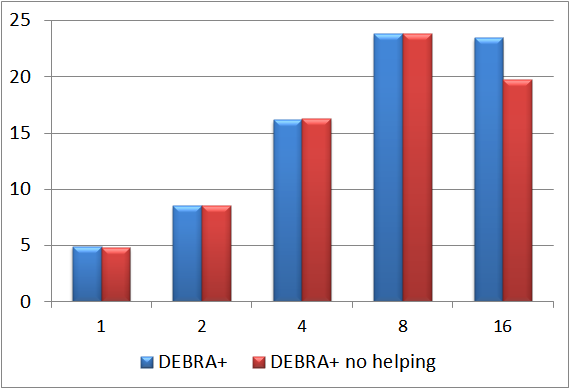} &
	\hspace{-3.5mm}\includegraphics[scale=\graphscale]{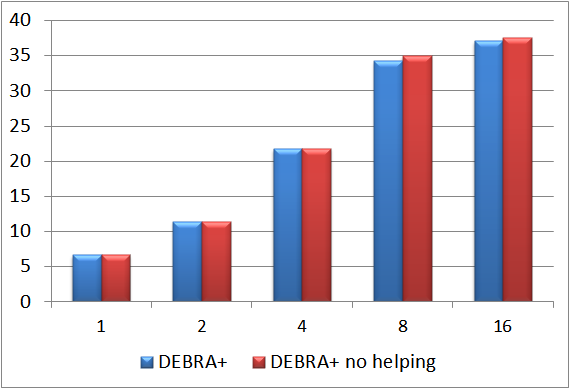}
	\\
	\end{tabular}
    \vspace{-2mm}
	\caption{Overhead introduced by restarting BST operations instead of helping whenever a marked node is reached.}
	\label{fig-debra-nohelp}
\end{figure}

%\trevor{mention it would be cool to compare with OA and QS, but that is left for future work.}
%To see how DEBRA+ and DEBRA perform on a real system, w
We ran experiments to compare the performance of various Reclaimers: DEBRA, DEBRA+, HP, ST and no reclamation (None).
We used the Record Manager abstraction to perform allocation and reclamation for a lock-free balanced binary search tree (BST)~\cite{Brown:2014}.
Searches in this BST can traverse pointers from retired nodes to other retired nodes, so ST cannot be used, and we must confront the problems described in Section~\ref{sec-debra-related} to apply HP.
Properly dealing with HP's problems would be highly complex and inefficient, so we simply restart any operation that suspects a node is retired.
Consequently, applying HP causes the BST to lose its lock-free progress guarantee.
To determine whether this significantly affects the performance of HP, we added the same restarting behaviour to DEBRA, and observed that its impact on performance was small.
(See Figure~\ref{fig-debra-nohelp}.)
Note that the HP scheme was tuned for high performance (instead of space efficiency) by allowing processes to accumulate large buffers of retired nodes before attempting to reclaim memory.

Code for ST was graciously provided by its authors.
They used a lock-based skip list to compare None, HP and ST.
We modified their code to use a Record Manager for allocating and pooling nodes, and used it %the resulting code 
to compare None, DEBRA, HP and ST.
The actual reclamation code for HP and ST is due to the authors of ST.
%We modify a lock-based skip list implemented by the authors of ST to use a Record Manager, and used it to compare None, DEBRA, HP and ST.
%(The algorithm was modified to use the Record Manager's \textit{Allocator} to allocate nodes, and reclaimed nodes were returned to the Record Manager's \textit{Pool}, to be reused. However, for HP and ST, reclamation was performed using code written by the authors of ST.)
Since the skip list uses locks, it cannot use DEBRA+. %we cannot use DEBRA+ to reclaim memory for it.

\paragraph{Experiment 1}
Our first experiment compared the overhead of performing reclamation for the various Reclaimers.
In this experiment, each Reclaimer performed all the work necessary to reclaim nodes, but nodes were not actually reclaimed (and, hence, were not reused).
The Record Manager used a \textit{Bump Allocator}: each process requests a large region of memory from the operating system at the beginning of an execution, and then divides that region into nodes, which it allocates in sequence.
Since nodes were not actually reclaimed, we eliminated the Pool component of the Record Manager.
In this experiment, a data structure suffers the overhead of reclamation, but does \textit{not} enjoy its benefits (namely, a smaller memory footprint and fewer cache misses).

For the balanced BST, we ran eight \textit{trials} for each combination of Reclaimers in \{None, DEBRA, DEBRA+, HP\}, thread counts (in \{1, 2, ..., 16\}), operation mixes in \{25i-25d, 50i-50d\} (where $x$i-$y$d means $x$\% insertions, $y$\% deletions and ($100-x-y$)\% searches) and key ranges in \{[0, 10000), [0, 1000000)\}.
For the skip list, the thread counts and operation mixes were the same, but ST was used instead of DEBRA+, and there was only one key range, [0, 200000).
In each trial, the data structure was first prefilled to half of the key range, then the appropriate number of threads performed random operations (according to the operation mix) on uniformly random keys from the key range for two seconds.
The average of each set of eight trials became a data point in a graph.
(Unfortunately, the system quickly runs out of memory when nodes are not reclaimed, so it is not possible to run all trials for longer than two seconds.
However, we ran long trials for many cases to verify that the results do not change.)

\begin{figure}[t]
    \begin{minipage}{\textwidth}
    \setlength\tabcolsep{0pt}
    \centering
    \begin{tabular}{m{0.02\linewidth}m{0.3\linewidth}m{0.3\linewidth}}
        &
        \multicolumn{2}{c}{\fcolorbox{black!80}{black!40}{\parbox{\dimexpr 0.6\linewidth-2\fboxsep-2\fboxrule}{\centering\textbf{Experiment 1 on 8-thread Intel i7-4770}}}}
        \\
        &
        \fcolorbox{black!50}{black!20}{\parbox{\dimexpr \linewidth-2\fboxsep-2\fboxrule}{\centering {\footnotesize 50\% ins, 50\% del}}} &
        \fcolorbox{black!50}{black!20}{\parbox{\dimexpr \linewidth-2\fboxsep-2\fboxrule}{\centering {\footnotesize 25\% ins, 25\% del, 50\% search}}}
        \\
        \rotatebox{90}{{\footnotesize \textbf{BST range} $[0, 10^6)$}} &
        \includegraphics[width=\linewidth]{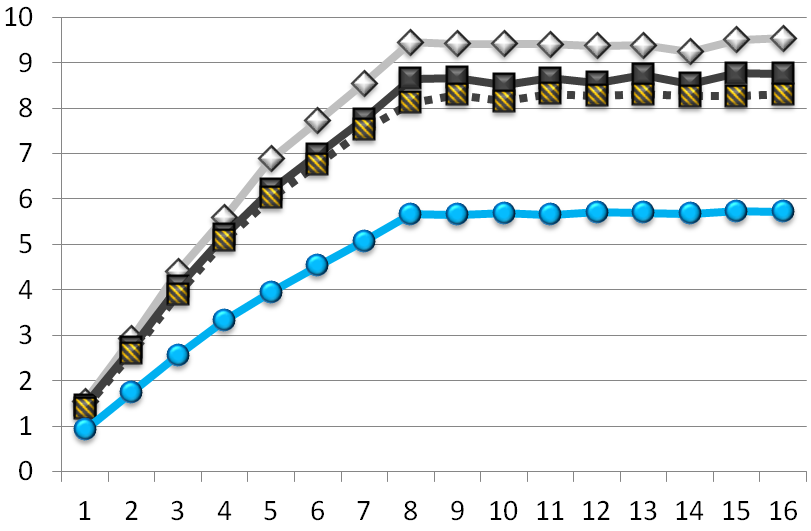} &
        \includegraphics[width=\linewidth]{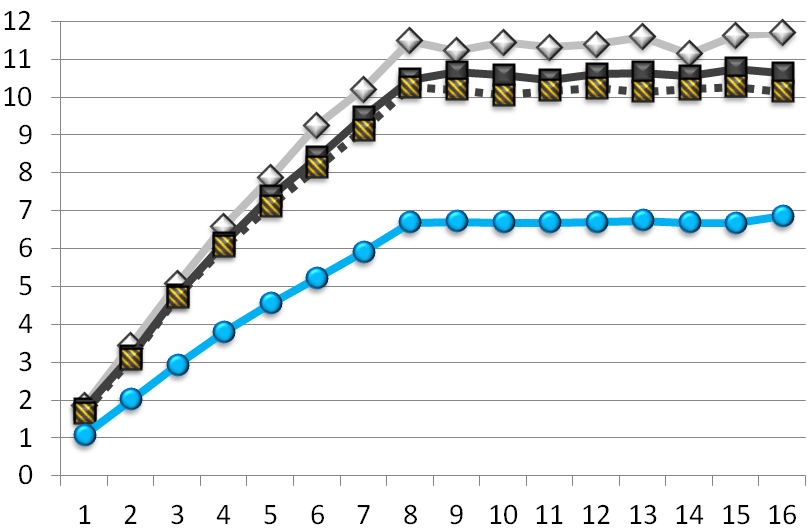}
        \\
        \rotatebox{90}{{\footnotesize \textbf{BST range} $[0, 10^4)$}} &
        \includegraphics[width=\linewidth]{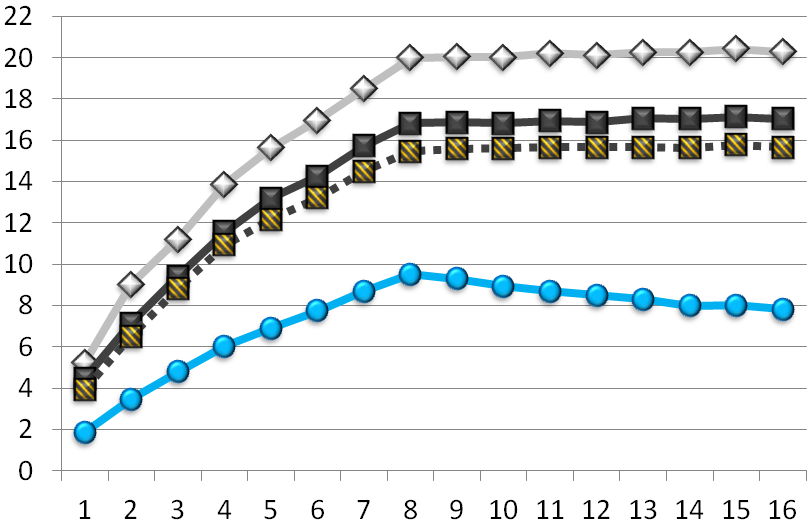} &
        \includegraphics[width=\linewidth]{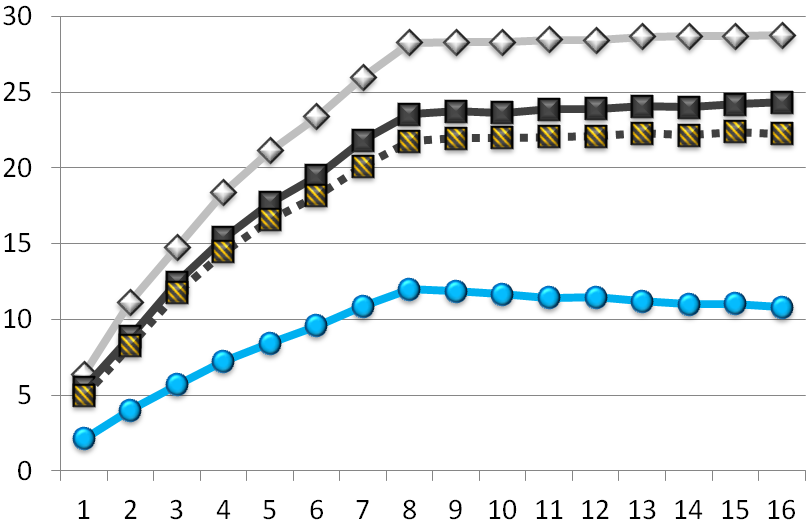}
        \\
        \rotatebox{90}{{\footnotesize \textbf{Skiplist range} $[0, 2 \cdot 10^5)$}} &
        \includegraphics[width=\linewidth]{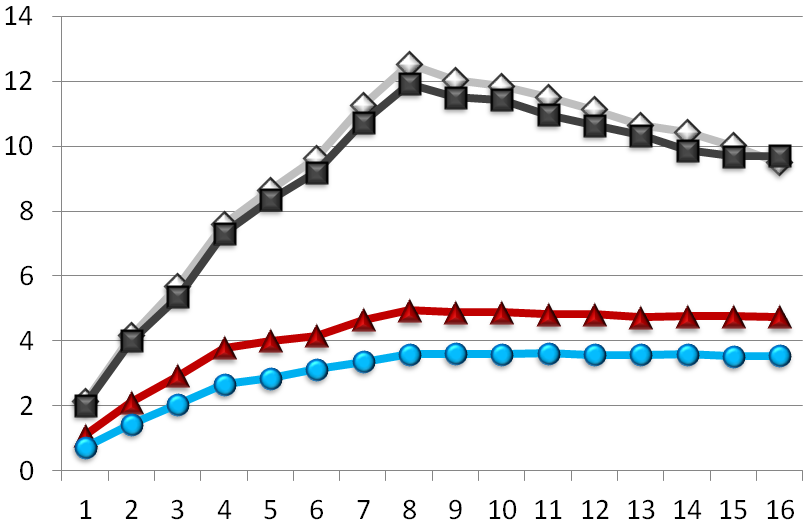} &
        \includegraphics[width=\linewidth]{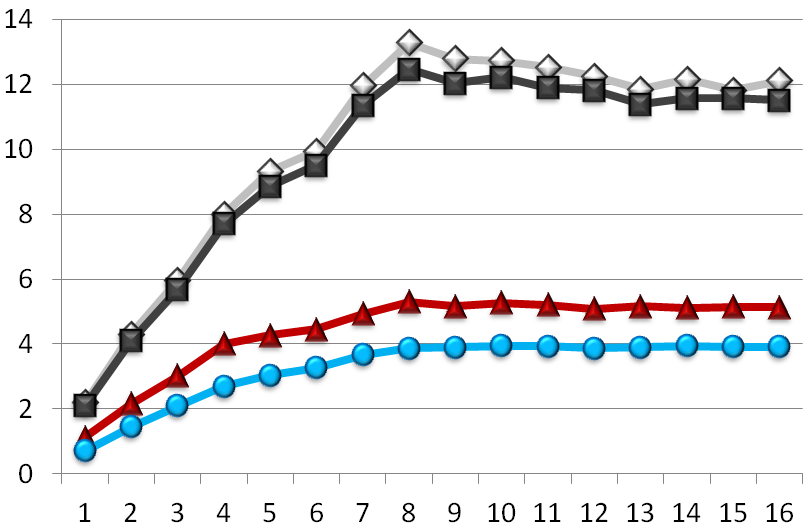}
        \\
        &
        \multicolumn{2}{c}{\includegraphics[width=0.6\textwidth]{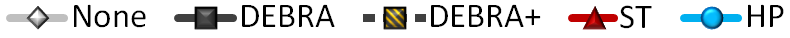}}
        \\
    \end{tabular}
    \end{minipage}
    \vspace{-2mm}
	\caption{
		Results for Experiment 1 (Overhead of reclamation).
		The x-axis shows the number of processes.
		The y-axis shows throughput, in millions of operations per second.
	}
	\label{fig-exp1}
\end{figure}

The results in Figure~\ref{fig-exp1} show that DEBRA and DEBRA+ have extremely low overhead.
In the BST, DEBRA has between 5\% and 22\% overhead (averaging 12\%), and DEBRA+ has between 7\% and 28\% overhead (averaging 17\%).
Compared to HP, on average, DEBRA performs 94\% more operations and DEBRA+ performs 83\% more.
This is largely because DEBRA and DEBRA+ synchronize once \textit{per operation}, whereas HP synchronizes each time a process reaches a new node.
In the skip list, DEBRA has up to 6\% overhead (averaging 4\%), outperforms HP by an average of 200\%, and also outperforms ST by between 93\% and 168\% (averaging 133\%).
ST has significant overhead.
For instance, on average, it starts almost four transactions per operation (each of which announces one or more pointers), and runtime checks are frequently performed to determine if a new transaction should be started.

\paragraph{Experiment 2}

In our second experiment, nodes were actually reclaimed.
The Reclaimers were each paired with the same Pool as DEBRA.
The only exception was None, which does not use a Pool.
(Thus, None is the same as in the first experiment.) %So, None does not have to check whether a Pool is empty before fetching a new node from the Allocator.) % is also eliminated.)
%ST was not included in this experiment, because the code from the authors currently does not work properly when nodes are actually reclaimed.
%The ST authors are currently revising the code.
%Presumably, it will be fixed soon.

\begin{figure}[t]
    \begin{minipage}{\textwidth}
    \setlength\tabcolsep{0pt}
    \centering
    \begin{tabular}{m{0.02\linewidth}m{0.3\linewidth}m{0.3\linewidth}}
        &
        \multicolumn{2}{c}{\fcolorbox{black!80}{black!40}{\parbox{\dimexpr 0.6\linewidth-2\fboxsep-2\fboxrule}{\centering\textbf{Experiment 2 on 8-thread Intel i7-4770}}}}
        \\
        &
        \fcolorbox{black!50}{black!20}{\parbox{\dimexpr \linewidth-2\fboxsep-2\fboxrule}{\centering {\footnotesize 50\% ins, 50\% del}}} &
        \fcolorbox{black!50}{black!20}{\parbox{\dimexpr \linewidth-2\fboxsep-2\fboxrule}{\centering {\footnotesize 25\% ins, 25\% del, 50\% search}}}
        \\
        \rotatebox{90}{{\footnotesize \textbf{BST range} $[0, 10^6)$}} &
        \includegraphics[width=\linewidth]{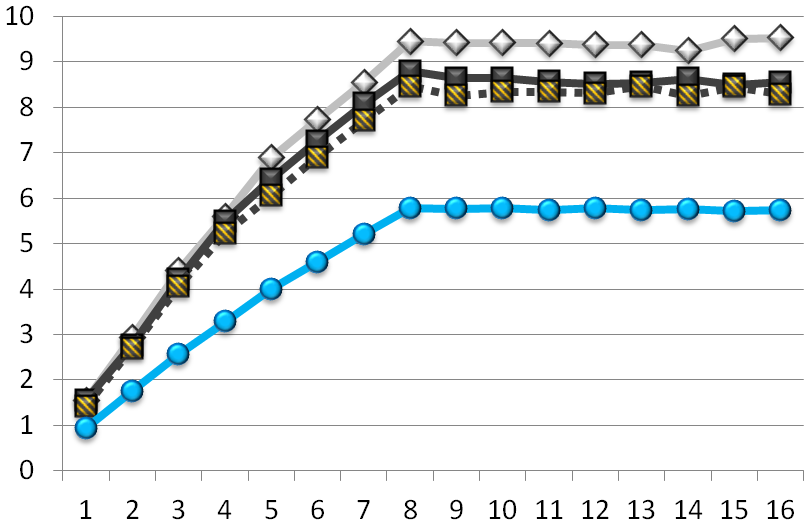} &
        \includegraphics[width=\linewidth]{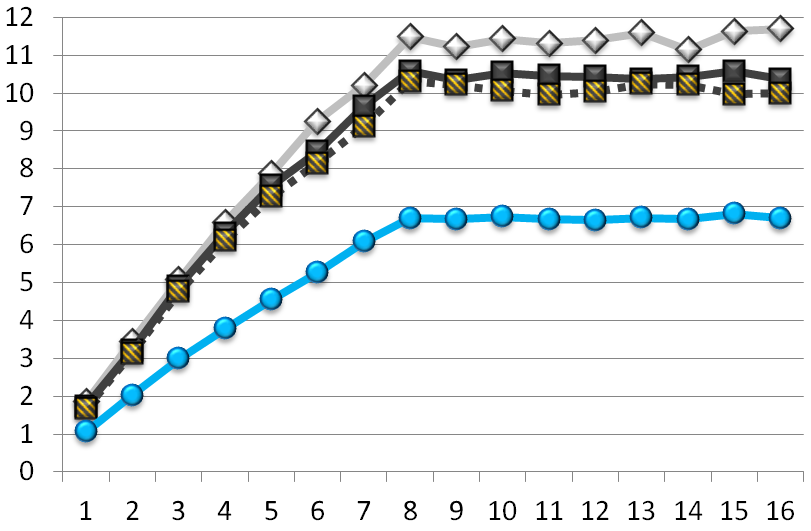}
        \\
        \rotatebox{90}{{\footnotesize \textbf{BST range} $[0, 10^4)$}} &
        \includegraphics[width=\linewidth]{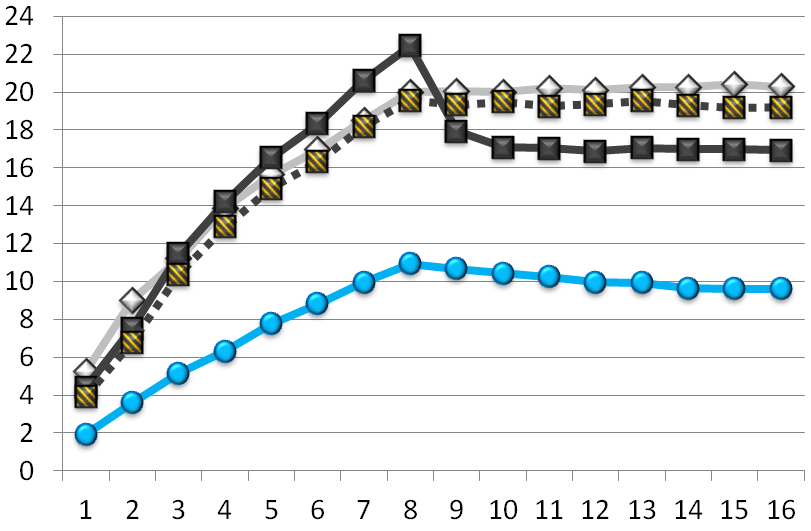} &
        \includegraphics[width=\linewidth]{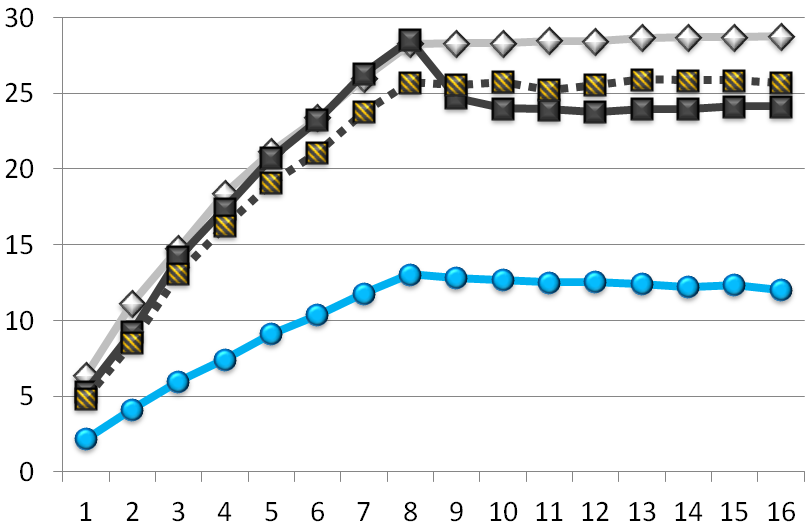}
        \\
        \rotatebox{90}{{\footnotesize \textbf{Skiplist range} $[0, 2 \cdot 10^5)$}} &
        \includegraphics[width=\linewidth]{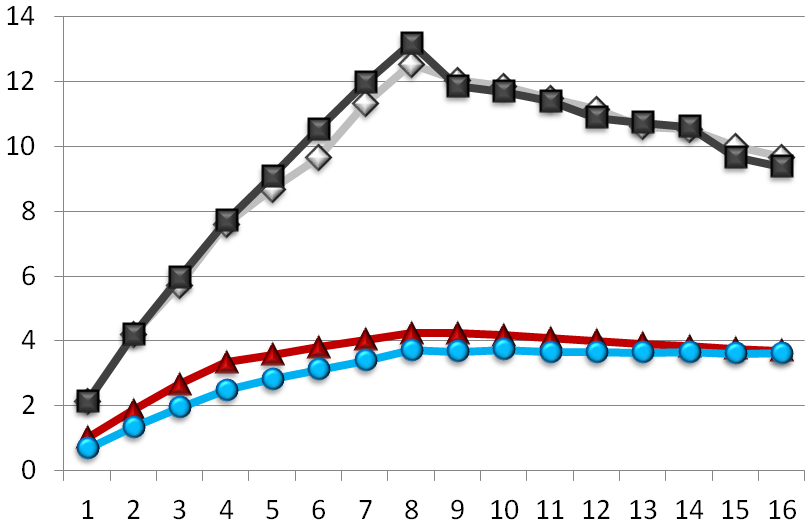} &
        \includegraphics[width=\linewidth]{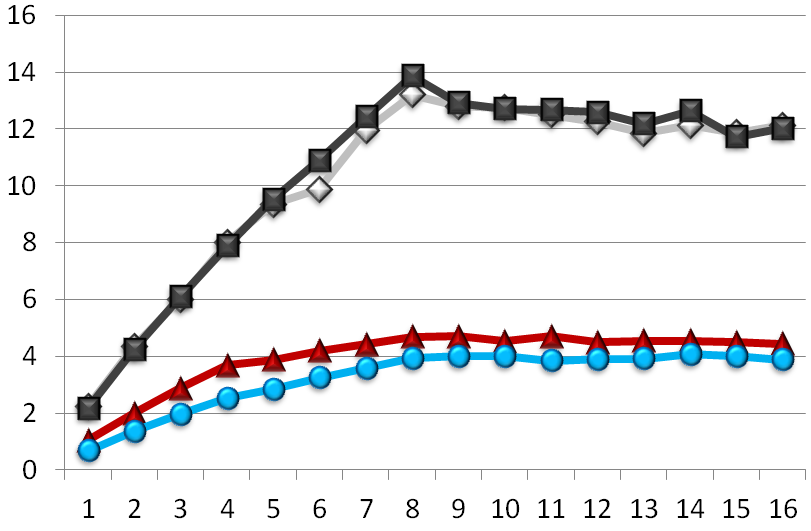}
        \\
        &
        \multicolumn{2}{c}{\includegraphics[width=0.6\textwidth]{chap-debra/graphs/legend.png}}
        \\
    \end{tabular}
    \end{minipage}
    \vspace{-2mm}
	\caption{
		Experiment 2 (Using a Bump Allocator and a Pool).
	}
	\label{fig-exp2}
\end{figure}

%\begin{figure}[t]
%	\newcommand\graphscale{0.14}
%	\vspace{-2mm}
%	\hspace{-3mm}
%	\begin{tabular}{lll}
%	&
%	\hfill 50\% ins, 50\% del \hfill \hfill &
%	\hfill 25\% ins, 25\% del, 50\% search \hfill \hfill %&
%%	\hfill \textbf{Skip list range [0, 200000)} \hfill \hfill
%	\\
%	\includegraphics[scale=\graphscale]{chap-debra/images/row-header-bst1mv2.png} &
%	\hspace{-4mm}
%	\includegraphics[scale=\graphscale]{chap-debra/graphs/exp2bst50i50d1m.png} &
%	\hspace{-4mm}
%	\includegraphics[scale=\graphscale]{chap-debra/graphs/exp2bst25i25d1m.png}
%	\vspace{-3.5mm}
%	\\
%	\includegraphics[scale=\graphscale]{chap-debra/images/row-header-bst10kv2.png} &
%	\hspace{-4mm}
%	\includegraphics[scale=\graphscale]{chap-debra/graphs/exp2bst50i50d10k.png} &
%	\hspace{-4mm}
%	\includegraphics[scale=\graphscale]{chap-debra/graphs/exp2bst25i25d10k.png}
%	\vspace{-3.5mm}
%	\\
%	\includegraphics[scale=\graphscale]{chap-debra/images/row-header-sl200kv2.png} &
%	\hspace{-4mm}
%	\includegraphics[scale=\graphscale]{chap-debra/graphs/exp2sl50i50d200k-v2.png} &
%	\hspace{-4mm}
%	\includegraphics[scale=\graphscale]{chap-debra/graphs/exp2sl25i25d200k-v2.png}
%	\end{tabular}
%	\vspace{-3.5mm}
%	\begin{center}
%		\includegraphics[scale=0.3]{chap-debra/graphs/legend.png}
%	\end{center}
%	\vspace{-6mm}
%	\caption{
%		Experiment 2 (Using a Bump Allocator and a Pool).
%%		Results for Experiment 2 (Using a Bump Allocator and a Pool).
%%		The x-axis shows the number of processes.
%%		The y-axis shows throughput, in millions of operations per second.
%	}
%%	\vspace{-2mm}
%	\label{fig-exp2}
%\end{figure}

\begin{figure}[t]
    \begin{minipage}{\textwidth}
    \setlength\tabcolsep{0pt}
    \centering
    \begin{tabular}{m{0.02\linewidth}m{0.3\linewidth}m{0.3\linewidth}}
        &
        \multicolumn{2}{c}{\fcolorbox{black!80}{black!40}{\parbox{\dimexpr 0.6\linewidth-2\fboxsep-2\fboxrule}{\centering\textbf{Experiment 2 on 64-thread Oracle T4-1}}}}
        \\
        &
        \fcolorbox{black!50}{black!20}{\parbox{\dimexpr \linewidth-2\fboxsep-2\fboxrule}{\centering {\footnotesize 50\% ins, 50\% del}}} &
        \fcolorbox{black!50}{black!20}{\parbox{\dimexpr \linewidth-2\fboxsep-2\fboxrule}{\centering {\footnotesize 25\% ins, 25\% del, 50\% search}}}
        \\
        \rotatebox{90}{{\footnotesize \textbf{BST range} $[0, 10^6)$}} &
        \includegraphics[width=\linewidth]{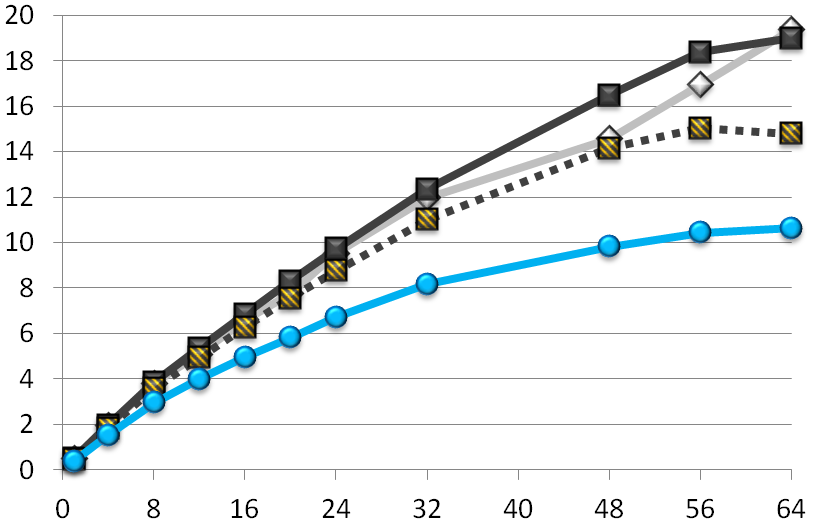} &
        \includegraphics[width=\linewidth]{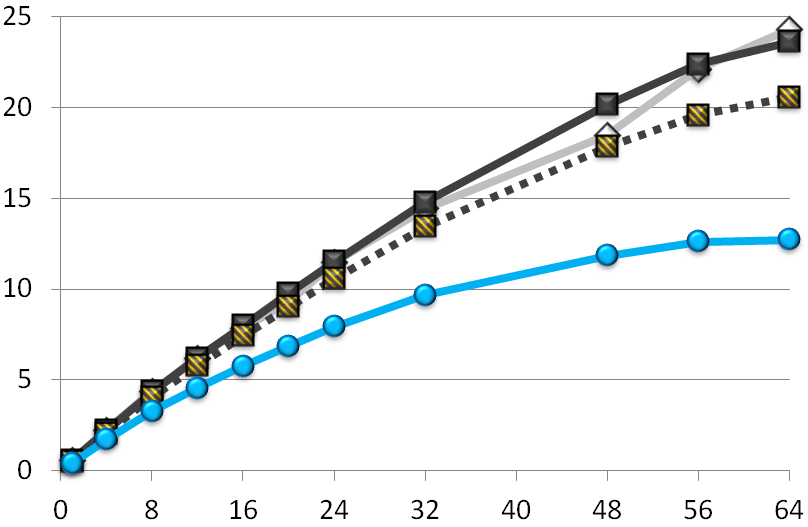}
        \\
%        \rotatebox{90}{{\footnotesize \textbf{BST range} $[0, 10^4)$}} &
%        \includegraphics[width=\linewidth]{chap-debra/graphs/exp2bst50i50d10k.png} &
%        \includegraphics[width=\linewidth]{chap-debra/graphs/exp2bst25i25d10k.png}
%        \\
%        \rotatebox{90}{{\footnotesize \textbf{Skiplist range} $[0, 2 \cdot 10^5)$}} &
%        \includegraphics[width=\linewidth]{chap-debra/graphs/exp2sl50i50d200k-v2.png} &
%        \includegraphics[width=\linewidth]{chap-debra/graphs/exp2sl25i25d200k-v2.png}
%        \\
        &
        \multicolumn{2}{c}{\includegraphics[width=0.6\textwidth]{chap-debra/graphs/legend.png}}
        \\
    \end{tabular}
    \end{minipage}
    \vspace{-2mm}
	\caption{
		Extra results for Experiment 2 on Oracle T4-1.
(ST could not be run, since the Oracle T4-1 does not support hardware transactional memory.)
	}
	\label{fig-oracle-exp2}
\end{figure}

%\begin{figure}[t]
%	\newcommand\graphscale{0.14}
%	\vspace{-2mm}
%	\hspace{-3mm}
%	\begin{tabular}{lll}
%	&
%	\hfill 50\% ins, 50\% del \hfill \hfill &
%	\hfill 25\% ins, 25\% del, 50\% search \hfill \hfill %&
%%	\hfill 100\% search \hfill \hfill %&
%%	\hfill \textbf{Skip list range [0, 200000)} \hfill \hfill
%	\\
%	\includegraphics[scale=\graphscale]{chap-debra/images/row-header-bst1mv2.png} &
%	\hspace{-4mm}
%	\includegraphics[scale=\graphscale]{chap-debra/graphs/oracle-exp2bst50i50d1m.png} &
%	\hspace{-4mm}
%	\includegraphics[scale=\graphscale]{chap-debra/graphs/oracle-exp2bst25i25d1m.png}
%	%\includegraphics[scale=\graphscale]{chap-debra/graphs/oracle-exp2bst0i0d1m.png}
%	\vspace{-3.5mm}
%	\end{tabular}
%	\caption{
%		Extra results for Experiment 2 on Oracle T4-1.
%	}
%	\vspace{-3mm}
%	\label{fig-oracle-exp2}
%\end{figure}

The results appear in Figure~\ref{fig-exp2}.
In the BST, DEBRA is only 8\% slower than None on average, and, for some data points, DEBRA actually improves performance by up to 12\%.
This is possible because DEBRA reduces the memory footprint of the data structure, which allows a larger fraction of the allocated nodes to fit in cache and, hence, reduces the number of cache misses.
DEBRA+ is between 2\% and 25\% slower than None, averaging 10\%.
Compared to HP, DEBRA is between 48\% and 145\% faster, averaging 80\%, and DEBRA+ is between 43\% and 123\% faster, averaging 76\%.
In the skip list, DEBRA performs \textit{as well as None}.
%Furthermore, DEBRA outperforms ST by between 87\% and 218\%, averaging 162\%.
DEBRA also \textit{outperforms} ST by between 108\% and 211\%, averaging 160\%.
%\trevor{WHY DOES DEBRA TAKE A PERFORMANCE HIT AT 9-16 AND DEBRA+ IS OK?}

\begin{figure}[t]
	\vspace{-2mm}
	\centering
	\includegraphics[width=0.5\linewidth]{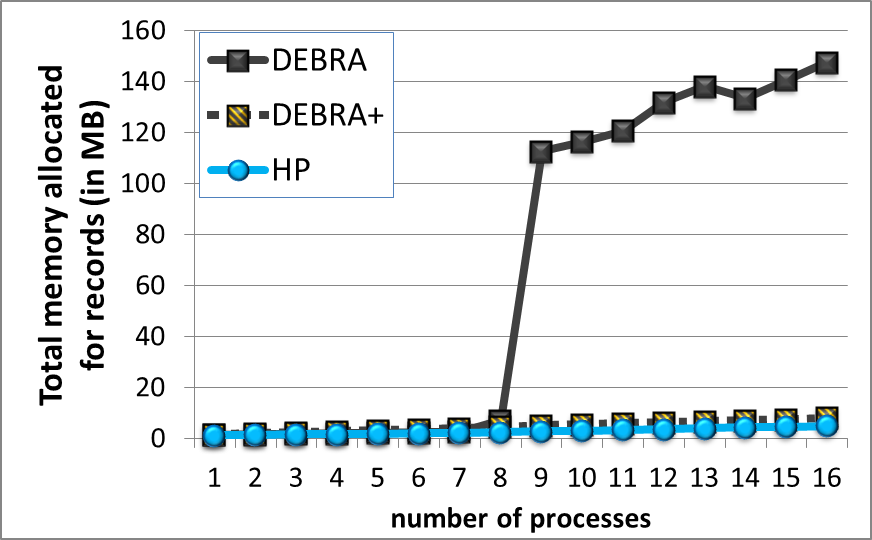}
	\vspace{-2mm}
	\caption{
		Memory allocated for records in Experiment 2 in the BST with keyrange 10,000 and workload 50i-50d.
	}
	\label{fig-allocated}
	\vspace{-2mm}
\end{figure}

To measure the benefit of neutralizing slow processes, we tracked the total amount of memory allocated for records in each trial.
Since we used bump allocation, this simply required determining how far each bump allocator's pointer had moved during the execution.
Thus, we were able to compute the total amount of memory allocated \textit{after} each trial had finished (without having any impact on the trial while it was executing).
Figure~\ref{fig-allocated} shows the total amount of memory allocated for records in the second experiment in the BST with key range 10,000 and workload 50i-50d.
(The other cases were similar.)
DEBRA, DEBRA+ and HP all perform similarly up to eight processes.
However, for more than eight processes, some processes are always context switched out, and they often prevent DEBRA from advancing the epoch in a timely manner.
DEBRA+ fixes this issue.
With 16 processes, DEBRA+ neutralizes processes an average of 935 times per trial, reducing memory usage by an average of 94\% over DEBRA.

We also ran the second experiment on a NUMA Oracle T4-1 system with 8 cores and 64 hardware contexts.
Figure~\ref{fig-oracle-exp2} shows a representative sample of the results.
Note that ST could not be run on this machine, since it does not support HTM.

\paragraph{Experiment 3}
Our third experiment is like the second, except we used a different Allocator, which does not preallocate memory.
The Allocator's \textit{allocate} operation simply invokes \textit{malloc} to request memory from the operation system (and its \textit{deallocate} operation invokes \textit{free}).

\begin{figure}[t]
    \begin{minipage}{\textwidth}
    \setlength\tabcolsep{0pt}
    \centering
    \begin{tabular}{m{0.02\linewidth}m{0.3\linewidth}m{0.3\linewidth}}
        &
        \multicolumn{2}{c}{\fcolorbox{black!80}{black!40}{\parbox{\dimexpr 0.6\linewidth-2\fboxsep-2\fboxrule}{\centering\textbf{Experiment 3 on 8-thread Intel i7-4770}}}}
        \\
        &
        \fcolorbox{black!50}{black!20}{\parbox{\dimexpr \linewidth-2\fboxsep-2\fboxrule}{\centering {\footnotesize 50\% ins, 50\% del}}} &
        \fcolorbox{black!50}{black!20}{\parbox{\dimexpr \linewidth-2\fboxsep-2\fboxrule}{\centering {\footnotesize 25\% ins, 25\% del, 50\% search}}}
        \\
        \rotatebox{90}{{\footnotesize \textbf{BST range} $[0, 10^6)$}} &
        \includegraphics[width=\linewidth]{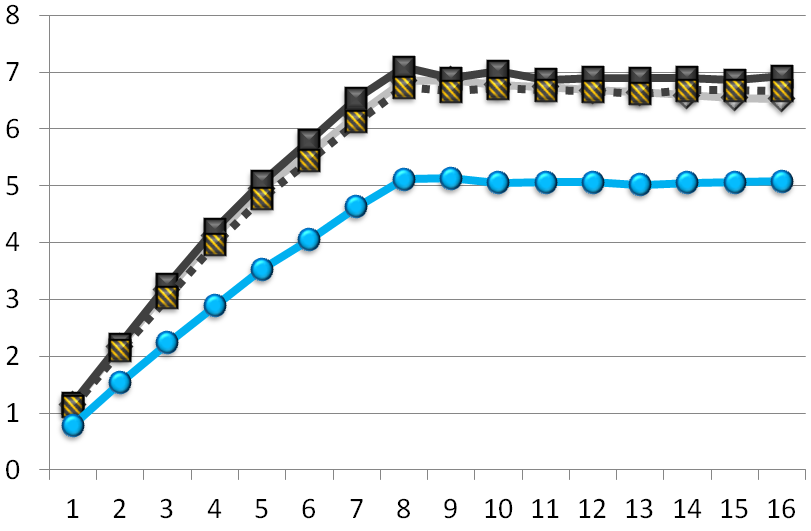} &
        \includegraphics[width=\linewidth]{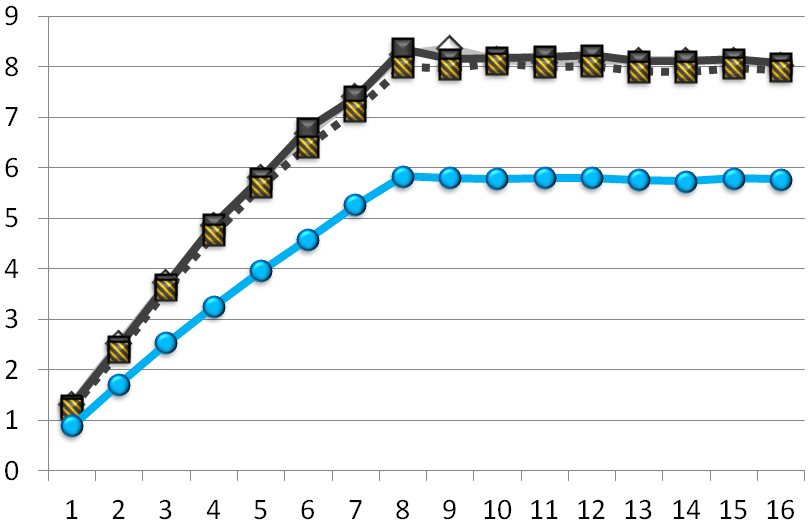}
        \\
        \rotatebox{90}{{\footnotesize \textbf{BST range} $[0, 10^4)$}} &
        \includegraphics[width=\linewidth]{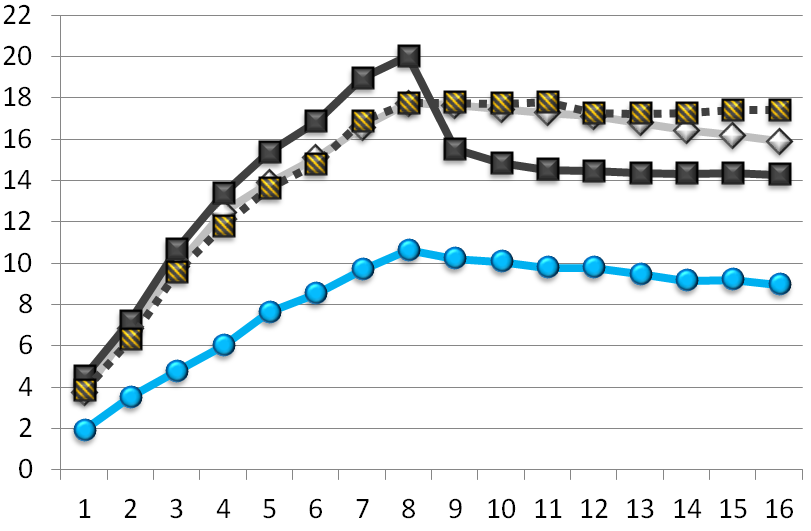} &
        \includegraphics[width=\linewidth]{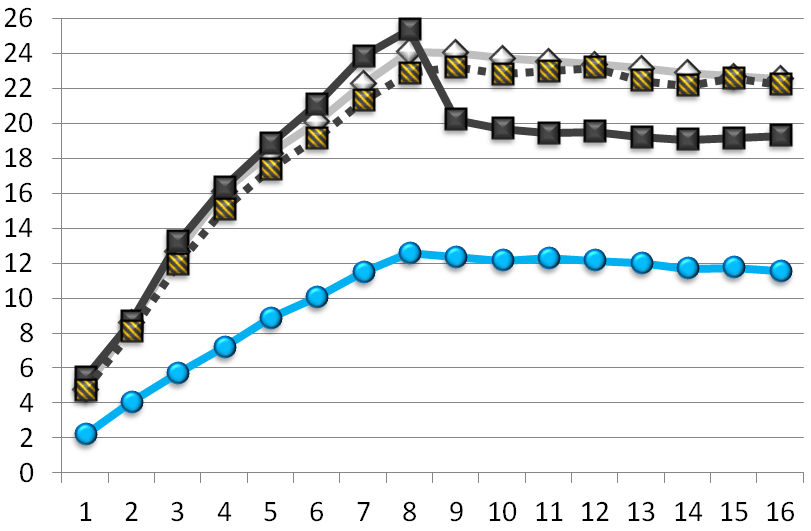}
        \\
        \rotatebox{90}{{\footnotesize \textbf{Skiplist range} $[0, 2 \cdot 10^5)$}} &
        \includegraphics[width=\linewidth]{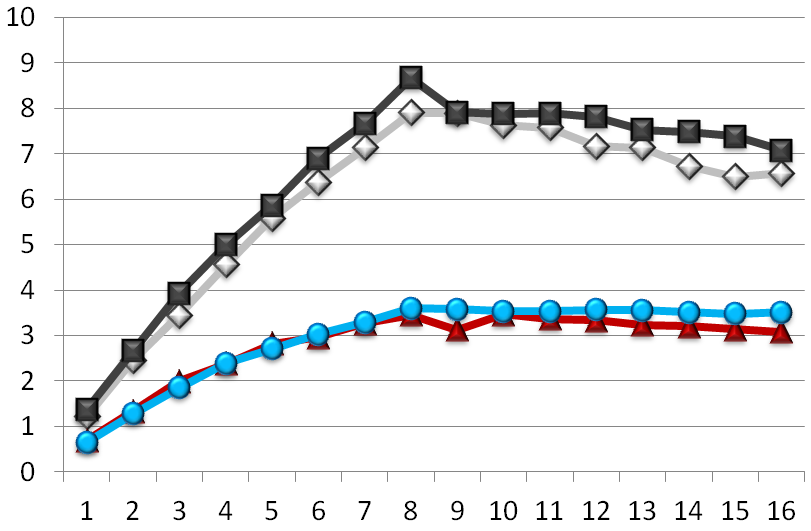} &
        \includegraphics[width=\linewidth]{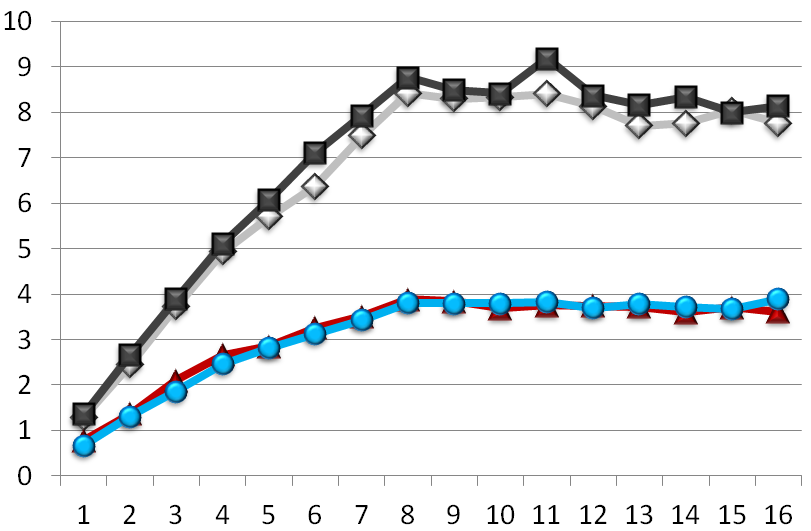}
        \\
        &
        \multicolumn{2}{c}{\includegraphics[width=0.6\textwidth]{chap-debra/graphs/legend.png}}
        \\
    \end{tabular}
    \end{minipage}
    \vspace{-2mm}
	\caption{
		Experiment 3 (Using malloc and a Pool).
	}
	\label{fig-exp3}
\end{figure}

%\begin{figure}[t]
%	\newcommand\graphscale{0.14}
%	\vspace{-2mm}
%	\hspace{-3mm}
%	\begin{tabular}{lll}
%	&
%	\hfill 50\% ins, 50\% del \hfill \hfill &
%	\hfill 25\% ins, 25\% del, 50\% search \hfill \hfill
%	\\
%	\includegraphics[scale=\graphscale]{chap-debra/images/row-header-bst1mv2.png} &
%	\hspace{-4mm}
%	\includegraphics[scale=\graphscale]{chap-debra/graphs/exp3bst50i50d1m.png} &
%	\hspace{-4mm}
%	\includegraphics[scale=\graphscale]{chap-debra/graphs/exp3bst25i25d1m.png}
%	\vspace{-3.5mm}
%	\\
%	\includegraphics[scale=\graphscale]{chap-debra/images/row-header-bst10kv2.png} &
%	\hspace{-4mm}
%	\includegraphics[scale=\graphscale]{chap-debra/graphs/exp3bst50i50d10k.png} &
%	\hspace{-4mm}
%	\includegraphics[scale=\graphscale]{chap-debra/graphs/exp3bst25i25d10k.png}
%	\vspace{-3.5mm}
%	\\
%	\includegraphics[scale=\graphscale]{chap-debra/images/row-header-sl200kv2.png} &
%	\hspace{-4mm}
%	\includegraphics[scale=\graphscale]{chap-debra/graphs/exp3sl50i50d200k-v2.png} &
%	\hspace{-4mm}
%	\includegraphics[scale=\graphscale]{chap-debra/graphs/exp3sl25i25d200k-v2.png}
%	\\
%	\end{tabular}
%	\vspace{-3.5mm}
%	\begin{center}
%		\includegraphics[scale=0.3]{chap-debra/graphs/legend.png}
%	\end{center}
%	\vspace{-6mm}
%	\caption{
%		Experiment 3: (Using malloc and a Pool).
%%		Results for Experiment 3: (Using malloc and a Pool).
%%		The x-axis shows the number of processes.
%%		The y-axis shows throughput, in millions of operations per second.
%	}
%	\label{fig-exp3}
%\end{figure}

The results (which appear in Figure~\ref{fig-exp3}) are similar to the results for the second experiment.
However, the absolute throughput is significantly smaller than in the previous experiments, because of the overhead of invoking \textit{malloc}.
Although HP and ST are negatively affected, proportionally, they slow down less than None, DEBRA and DEBRA+.
This illustrates an important experimental principle: \textit{overhead should be minimized, because uniformly adding overhead to an experiment disproportionately impacts low-overhead algorithms, and obscures their advantage}.

\section{Summary}

In this work, we presented a distributed variant of EBR, called DEBRA.
Compared to EBR, DEBRA significantly reduces synchronization overhead and offers high performance even with many more processes than physical cores. %replaces EBR's shared limbo bags with per-process limbo bags, significantly reducing synchronization overhead.
%Experiments over a wide variety of thread counts, workloads and contention levels show that overhead is extremely low.
Our experiments show that, compared with performing no reclamation at all, DEBRA is 4\% slower on average, 21\% slower at worst, and up to 20\% \textit{faster} in some cases.
Moreover, DEBRA outperforms StackTrack by an average of 138\%.
DEBRA is easy to use, and only adds O(1) steps per data structure operation and O(1) steps per retired record. % removed from the data structure.

We also presented DEBRA+, the first epoch based reclamation scheme that allows processes to continue reclaiming memory after a process has crashed.
In an $n$ process system, 
%With DEBRA+, 
the number of objects waiting to be freed is $O(mn^2)$, where $m$ is the largest number of objects retired by one data structure operation.
The cost to reuse or free a record is O(1) expected amortized time. %(and could be improved to O(1) worst case time with a small modification).
In our experiments, DEBRA+ reduced memory consumption over DEBRA by 94\%.
Compared with performing no reclamation, DEBRA+ is only 10\% slower on average.
DEBRA+ also outperforms a highly efficient implementation of hazard pointers by an average of 70\%. %between 30\% and 134\%, averaging 70\%.

We introduced the \textit{Record Manager}, the first generalization of the C++ \textit{Allocator} abstraction that is suitable for lock-free programming.
A Record Manager separates memory reclamation code from lock-free data structure code, which allows a dynamic data structure to be implemented without knowing how its records will be allocated, reclaimed and freed.
This abstraction adds virtually no overhead.
It is highly flexible, allowing a programmer to interchange techniques for reclamation, object pooling, allocation and deallocation by changing one line of code.
%
%%\trevor{although i think it might be important to have a conclusion, to summarize the contributions (since they're not bulleted in the intro), and to leave the reader thinking the right things, this might be too similar to the abstract to warrant appearing here...}

Besides DEBRA and DEBRA+, the neutralizing technique introduced in this work is of independent interest.
%Recall that neutralizing processes less than 1000 times (out of tens of millions of operations) improved memory consumption by 94\% in our experiments.
%Perhaps neutralized processes were locked in some sort of pathological behaviour, and neutralizing them cured them of their pathology.
%If so, neutralizing might prove useful for eliminating other pathological behaviours.
%the memory reclamation aspect of this work is 
%There are many opportunities for i
%Although this work represents an important step towards fast, simple, fault tolerant memory reclamation for lock-free algorithms, there are still many opportunities for improvement.
%It may be possible to extend the neutralizing technique to other operating systems (particularly, Windows). % (such as Windows), or whether additional operating system support is necessary.
It would be useful to find different ways to neutralize processes, so, for example, the neutralizing technique could be used with different operating systems.
%Perhaps there is different way to neutralize processes. %but this requires further research.
%More broadly, it is unclear what kind of operating system support is necessary to enable effective fault tolerant techniques such as neutralizing. %should be provided to support fault tolerant applications.
There may also be opportunities to apply neutralizing in other contexts, such as garbage collection. %, where algorithms often use locking to stop program threads while the collector runs.
Finally, it would also be interesting to understand whether these ideas can be extended to lock-based algorithms (even for a restricted class, such as reentrant and idempotent algorithms).

\chapter{Reusing descriptors} \label{chap-descriptors}
\renewcommand{\func}{\textit}
\begin{thesisonly}
\renewcommand{\llt}{LLX}
\renewcommand{\sct}{SCX}
\end{thesisonly}

\lstset{mathescape=true, gobble=1, numbers=left, numberstyle=\tiny, basicstyle=\footnotesize, numberblanklines=false, escapeinside={//}{\^^M}, breaklines=true, keywordstyle=\bfseries, otherkeywords={for,each,in,true,false}, morekeywords={type,subtype,break,continue,if,else,end,loop,while,do,done,exit, when,then,return,read,and,or,not,for,each,boolean,procedure,invoke,next,iteration,until,goto}}

%!TEX root = paper.tex

\begin{thesisnot}
\section{Introduction}

\begin{fullver}
As core counts continue to rise in modern processors, it is increasingly important for applications to be scalable.
Designing scalable concurrent software is notoriously difficult (even for experts), and programmers must rely on efficient concurrent library code to be effective.
Concurrent data structures represent some of the most fundamental building blocks in libraries, and are important in both theory and practice.
\end{fullver}

Many concurrent data structures use locks, but locks have downsides, such as susceptibility to convoying, deadlock and priority inversion.
Lock-free data structures avoid these downsides, and can be quite efficient.
They guarantee that some process will always makes progress, even if some processes halt unexpectedly.
%\end{fullver}
%Lock-freedom guarantees that some process always makes progress, even if some processes halt unexpectedly.
This guarantee is typically achieved with \textit{helping}, which allows a process to harness any time that it would otherwise
%The goal of helping is to harness any time that a process would otherwise 
spend waiting for another operation to complete. % in order to further the progress of the other operation.
Specifically, whenever a process $p$ is prevented from making progress by another operation, it attempts to perform some (or all) of the work of the other operation, on behalf of the process that started it.
This way, even if the other process has crashed, its operation can be completed, so that it no longer blocks $p$. %removing any obstructions that it created. %and any shared resources that are reserved for use exclusively by that operation can be released. %that would be is to have a process that is blocked by an operation performed by another process, it helps the other process to make progress before continuing its own operation.

\end{thesisnot}

In simple lock-free data structures (e.g., \cite{Valois:1995,Harris:2001,Michael:2002,Natarajan:2014,Lea}), a process can determine how to help an operation that blocks it by inspecting a small part of the %some local neighbourhood in the 
data structure.
\begin{thesisonly}
In more complex lock-free data structures (such as~\cite{Ellen:2010,Howley:2012,Shafiei:2013,Brown:2014}, and those implemented with \llt\ and \sct), processes publish \textit{descriptors} for their operations, and helpers look at these descriptors to determine how to help.
\end{thesisonly}
\begin{thesisnot}
In more complex lock-free data structures~\cite{Ellen:2010,Howley:2012,Shafiei:2013,Brown:2014}, processes publish \textit{descriptors} for their operations, and helpers look at these descriptors to determine how to help.
\end{thesisnot}
\xspace
A descriptor typically encodes a sequence of steps that a process should follow in order to complete the operation that created it.

Since lock-free algorithms cannot use mutual exclusion, many helpers can simultaneously help an operation, potentially long after the operation has terminated.
\begin{fullver}
Thus, to avoid situations where helpers read inconsistent data in a descriptor and corrupt the data structure, or try to access a descriptor that has been freed to the operating system and crash, each descriptor must remain consistent and accessible until it can be determined that no helper will ever access it again.
\end{fullver}
\begin{shortver}
Thus, to avoid situations where helpers read inconsistent data in a descriptor and corrupt the data structure, each descriptor must remain consistent and accessible until no helper will ever access it again.
\end{shortver}
This leads to \textit{wasteful algorithms} which allocate a new descriptor for each operation. %, and eventually reclaim descriptors using lock-free memory reclamation algorithms.
%%Naturally, the descriptors used for helping must eventually be freed to the operating system, or reused.
%We call these \textit{wasteful algorithms}. % (or \textit{wasteful implementations}).

\begin{thesisnot}
In this work, 
\end{thesisnot}
\begin{thesisonly}
In this chapter, 
\end{thesisonly}
we introduce two simple abstract data types (ADTs) that capture the way descriptors are used by wasteful algorithms.
The \textit{immutable descriptor} ADT appears in Section~\ref{sec-adt-immutable}.
It provides two operations, \func{CreateNew} and \func{ReadField}, which respectively create and initialize a new descriptor, and read one of its fields.
The \textit{mutable descriptor} ADT, which appears in Section~\ref{sec-adt-mutable}, extends the immutable descriptor ADT by adding two operations: \func{WriteField} and \func{CASField}.
These allow a helper to modify fields of the descriptor (e.g., to indicate that the operation has been partially or fully completed).
We also give examples of wasteful algorithms whose usage of descriptors is captured by these ADTs.

The natural way to implement the immutable and mutable descriptor ADTs is to have \func{CreateNew} allocate memory and initialize it, and to have \func{ReadField}, \func{WriteField} and \func{CASField} perform a read, write and CAS, respectively.
%Any lock-free algorithm that uses this natural implementation is, of course, a wasteful algorithm.
%%If the ADT is implemented so that a descriptor is created by allocating new memory, %the creation of a descriptor is implemented by %these operations are implemented so that a descriptor is created by 
%%%allocating new memory, 
%%then lock-free algorithms that use this implementation are wasteful descriptor algorithms.
%
%Our transformation places no requirements on the way that an algorithm reclaims memory (including the memory allocated for descriptors).
%However, we note that 
Every implementation of one of these ADTs must eventually reclaim the descriptors it allocates.
\begin{fullver}
Otherwise, the algorithm would eventually exhaust memory, and either cause the system to crash, or block while waiting for more memory to become available, violating lock-free progress.
%Indeed, one alternative to our technique is to use a lock-free memory reclamation algorithm so that descriptors can be reclaimed and reused.
Usually, a lock-free memory reclamation algorithm is used for the reclamation of descriptors.
We briefly explain why reclaiming descriptors this way is expensive.
\end{fullver}
\begin{shortver}
Otherwise, the algorithm would eventually exhaust memory.
We briefly explain why reclaiming descriptors is expensive.
\end{shortver}

It is non-trivial to determine when a descriptor is safe to free.
In order to safely free a descriptor, a process must know that the descriptor is no longer \textit{reachable}.
This means no other process can reach the descriptor by following pointers in shared memory \textit{or} in its private memory.
State of the art lock-free memory reclamation
\begin{thesisonly}
algorithms~(e.g., DEBRA and Hazard Pointers)
\end{thesisonly}
\begin{thesisnot}
algorithms~\cite{Michael:2004,Brown:2015}
\end{thesisnot}
can determine when no process has a pointer to an object in its \textit{private} memory, but typically require the underlying algorithm to identify a time after which the object is no longer reachable from \textit{shared} memory (and then invoke a \textit{Retire} function).
%Thus, for each descriptor, the algorithm must identify a point when the descriptor is no longer \textit{reachable}.
%At some point after an operation attempt $o$ completes, its descriptor is no longer \textit{reachable}.

Thus, for each high-level operation attempt $O$, an algorithm must identify a time $t$ such that no operation attempt started after $t$ can encounter a pointer in shared memory to $O$'s descriptor. %descriptor reference in shared memory for $O$'s descriptor.
In an algorithm where each operation attempt removes all pointers to its descriptor from shared memory before it terminates, $t$ is when $O$ completes.
However, in some algorithms (such as the implementation of \llt\ and \sct\ in Chapter~\ref{chap-scx}), pointers to descriptors are ``lazily'' cleaned up by subsequent operation attempts.
In such an algorithm, $t$ may be long after $O$ completes (and, consequently, $t$ may be quite difficult to identify).
The overhead of reclaiming descriptors comes both from identifying $t$, and from actually running a lock-free memory reclamation algorithm.

Additionally, in some applications, such as embedded systems, it is important to have a small, predictable number of descriptors in the system. % (e.g., to avoid exhausting memory).
In such cases, one must use memory reclamation algorithms that prioritize having a small \textit{descriptor footprint}, i.e., the largest number of descriptors in the system at one time.
Such algorithms incur high overhead.
For example, \textit{hazard pointers}~\cite{Michael:2004} can be used to achieve a small descriptor footprint, but it must perform costly memory fences \textit{every} time a process tries to access a new
\begin{thesisonly}
descriptor (as we described in Chapter~\ref{chap-debra}).
\end{thesisonly}
\begin{thesisnot}
descriptor.
\end{thesisnot}

%Unfortunately, most state of the art lock-free memory reclamation algorithms either introduce high overhead, or allow the descriptor footprint to become very large. %fail to satisfy one of our two goals. % described above.
%For example, the popular \textit{hazard pointer} algorithm could be used to guarantee a very tight upper bound on the descriptor footprint, but it would achieve this upper bound by performing costly memory barriers each time a process tries to access a descriptor (incurring extremely high overhead).
%At the other end of the spectrum, \textit{epoch based reclamation} could be used to reclaim descriptors with very low overhead, but it would offer no bound on the descriptor footprint.
%In general, if a memory reclamation algorithm is tuned to aggressively reclaim descriptors to minimize the descriptor footprint, then reclamation will introduce significant runtime overhead.
%(And, if not, it will allow the descriptor footprint to become large.)
%% in order to maintain a small, predictable number of descriptors in the system, a memory reclamation algorithm must aggressively reclaim descriptors.
%state of the art algorithms that do this (e.g., hazard pointers) introduce significant runtime overhead.
%hazard pointers~\cite{??} is a state of the art example of such an algorithm.
%unfortunately, doing the cost of aggressively reclaiming is such algorithms incur significant runtime overhead for doing so.
%moreover, algorithms that 

To circumvent the aforementioned problems, we introduce a \textit{weak descriptor} ADT (in Section~\ref{sec-weak-descriptors}) that has slightly \textit{weaker semantics} than the mutable descriptor ADT, but can be implemented \textit{without memory reclamation}. %significantly more efficiently.
The crucial difference %between the descriptor ADT and the weak descriptor ADT 
is that each time a process invokes \func{CreateNew} to create a new descriptor, it \textit{invalidates} all of its previous descriptors.
An invocation of \func{ReadField} on an invalid descriptor \textit{fails} and returns a special value $\bot$.
Invocations of \func{WriteField} and \func{CASField} on invalid descriptors have no effect.
%This behaviour is sensible because a descriptor is used only to help the operation that created it, and helping is no longer needed if the operation has terminated.
We believe the weak descriptor ADT can be useful in designing new lock-free algorithms, since an invocation of \func{ReadField} that returns $\bot$ can be used to inform a helper that it no longer needs to continue helping (making further accesses to the descriptor unnecessary).
%We provide an efficient implementation of the weak descriptor ADT with drastically smaller space complexity than the natural implementation of the descriptor ADT.

We also identify a class of lock-free algorithms that use the descriptor ADT, and which can be \textit{transformed} to use the weak descriptor ADT (in Section~\ref{sec-weak-transformation}). %, greatly improving their performance.
At a high level, these are algorithms in which (1) each operation attempt creates a descriptor and invokes a \func{Help} function on it, and (2) \func{ReadField}, \func{WriteField} and \func{CASField} operations occur only inside invocations of \func{Help}.
Intuitively, the fact that these operations occur only in \func{Help} makes it easy to determine how the transformed algorithm should proceed when it performs an invalid operation: the operation being helped must have already terminated, so it no longer needs help.
%Thus, one can simply return from \func{Help}.
We prove correctness for our transformation, and demonstrate
%We also demonstrate 
its use by transforming a wasteful implementation of a double-compare-single-swap (DCSS) primitive~\cite{Harris:2002}.

We then present an extension to our weak descriptor ADT, and show how an even larger class of lock-free algorithms can be transformed to use this extension (in Section~\ref{sec-adt-extended}).
In particular, the algorithms in this class can also perform \func{ReadField} operations \textit{outside} of \func{Help}.\xspace
\begin{thesisnot}
We prove correctness for the transformation, and demonstrate its use by transforming wasteful implementations of a $k$-compare-and-swap ($k$-CAS) primitive~\cite{Harris:2002} and \llt\ and \sct.
These primitives can be used to implement a wide variety of advanced lock-free data structures.
For example, \llt\ and \sct\ have been used to implement lists, chromatic trees, relaxed AVL trees, relaxed $(a,b)$-trees, relaxed $b$-slack trees and weak AVL trees~\cite{Brown:2014,BrownPhD,He:2016}.
\end{thesisnot}
\begin{thesisonly}
We prove correctness and progress for the transformation, and demonstrate its use by transforming a wasteful implementation of a $k$-compare-and-swap ($k$-CAS) primitive~\cite{Harris:2002}, as well as the \llt\ and \sct\ implementation in Chapter~\ref{chap-scx}.
\end{thesisonly}

We used mostly known techniques to produce an efficient, provably correct implementation of our extended weak descriptor ADT in Section~\ref{sec-extended-impl}.
With this implementation, the transformed algorithms for $k$-CAS, and LLX and SCX, have some desirable properties.
In the original $k$-CAS algorithm, \textit{each operation attempt} allocates at least $k+1$ new descriptors.
In contrast, the transformed algorithm allocates only two descriptors \textit{per process, once, at the beginning of the execution}, and these descriptors are reused.
Similarly, in the original algorithm for LLX and SCX, each SCX operation creates a new descriptor, but the transformed algorithm allocates only one descriptor per process, at the beginning of the execution.
%
%If an algorithm in this class allocated a new descriptor \textit{for each operation attempt}, and no operation attempt invokes another operation that creates a descriptor, then the transformed algorithm will allocate only \textit{one descriptor per process}, at the start of the execution (and each process will simply reuse its descriptor).
This entirely eliminates dynamic allocation \textit{and} memory reclamation for descriptors (significantly reducing overhead), and results in an extremely small descriptor footprint.

We present extensive experiments on a 64-thread AMD system and a 48-thread Intel system (in Section~\ref{sec-exp}).
These experiments use a variety of workloads to compare our transformed implementations with wasteful implementations that use state of the art memory reclamation algorithms.
Our results show that our transformed implementations always perform at least as well as their wasteful counterparts, and \textit{significantly} outperform them in some workloads.
In a $k$-CAS microbenchmark, our implementation outperformed wasteful implementations using fast distributed epoch-based reclamation~\cite{Brown:2015}, hazard pointers~\cite{Michael:2004} and read-copy-update (RCU)~\cite{Desnoyers:2012} by up to 2.3x, 3.3x and 5.0x, respectively.
In a microbenchmark using a binary search tree (BST) implemented with LLX and SCX, our transformed implementation is up to 57\% faster than the next best wasteful implementation.

%The remainder of this paper is structured as follows.
%\trevor{write paper roadmap.}
%%We start by defining the model in Section~\ref{sec-model}.
%%Then, in Section~\ref{sec-transformation}, we describe our transformation and the class of lock-free algorithms to which it applies. 
%%\begin{shortver}
%%Next, we give two example applications of our transformation in Section~\ref{sec-example} to demonstrate its use. 
%%\end{shortver}
%%\begin{fullver}
%%Next, we give three example applications of our transformation in Section~\ref{sec-example} to demonstrate its use. 
%%\end{fullver} 
%%%Specifically, we transform a double-compare single-swap (DCSS) implementation in Section~\ref{sec-dcss} as a simple example.
%%%We also demonstrate two more difficult transformations of a $k$-compare single-swap ($k$-CAS) algorithm (Section~\ref{sec-kcas}) and a binary search tree (BST) algorithm (Section~\ref{sec-bst}.
%%~In Section~\ref{sec-exp}, we perform an experimental study of different reusable and wasteful descriptor implementations of several algorithms over a variety of workloads.
%%Section~\ref{sec-exp-wraparound} contains our experimental study of errors caused by sequence number wraparound.
%%Related work is discussed in Section~\ref{sec-related}.
%%Finally, we conclude in Section~\ref{sec-desc-conclusion}.

The crucial observation in this work is that, in algorithms where descriptors are used only to facilitate helping, a descriptor is no longer needed once the operation that created it has terminated.
This allows a process to reuse a descriptor as soon as its operation attempt finishes, instead of allocating a new descriptor for each operation attempt, and waiting considerably longer (and incurring much higher overhead) to reclaim it using standard memory reclamation techniques.
The challenge in this work is to characterize the set of algorithms that can benefit from this observation, and to design and prove the correctness of a transformation that takes such algorithms and produces new algorithms that simply reuse a small number of descriptors.
As a result of developing this transformation, we also produce significantly faster implementations of %strong synchronization primitives, such as 
$k$-CAS, and \llt\ and \sct.
%!TEX root = paper.tex

%\vspace{-2mm}
\begin{thesisnot}
\section{Model} \label{sec-model}

%\trevor{eliminate or move this into the model chapter. valid/invalid descriptors might be something we want to move elsewhere within this chapter, though?}

We study a shared memory system with $n$ processes numbered $1..n$.
Each process has a private memory, and there is a shared memory accessible by all processes.
Shared memory consists of base objects (such as read/write registers, compare-and-swap objects, or descriptors), each of which offers a set of operations.
%
%All shared memory locations are initially \textit{unallocated}, and accessing them will cause the entire system to crash.
%The system has an \textit{allocator} that provides two operations: \textit{allocate} and \textit{free}. %processes can use to \textit{allocate} regions of memory.
%The allocate operation takes a \textit{size} argument, expressed in bytes, and returns a pointer $ptr$ to a newly allocated region of memory of the requested size.
%Once a memory region is \textit{allocated}, processes can freely access it. % (without causing a program failure).
%A process can subsequently invoke \textit{free}$(ptr)$ to return that memory to the unallocated state.
%%(This is also called \textit{reclaiming} memory.)

%shared memory 
%multi process 
%unmanaged evironment (can be extended to managed ones to reduce GC work)
%things to write about: 
%$n$ is the number of processes. 
%CAS 
%allocation reclamation 

A \textit{descriptor} object is \textit{valid} when it is first created, and becomes \textit{invalid} when it is invalidated by a \func{CreateNew} operation.
All other base objects are \textit{always valid}.
Any operation on a valid object (resp., an invalid object) is a valid operation (resp., an invalid operation).
In the following, the terms \textit{operation} and \textit{valid operation} refer to low-level atomic operations on base objects, and the term \textit{high-level operation} refers to an ADT operation that is being implemented \textit{using} base objects as atomic building blocks.

The remainder of the model is relevant mainly when we prove the correctness of our transformations.

A \emph{configuration} describes the state of all processes and base objects.
%\maya{maybe change to: A \emph{configuration} describes the state of all the processes and all shared variables.}
The \textit{state} of a process consists of \textit{private memory}, which is accessible only to the process, and a \textit{program counter}, which is a pointer to the step of the algorithm that the process will execute when it next takes a step.
In the \emph{initial} configuration, each process and base object is in its initial state.
An \emph{execution} $\pi$ is an alternating sequence of configurations
and steps, $C_0 \cdot s_1 \cdots s_i \cdot C_i\cdots$,
where $C_0$ is the initial configuration,
and each configuration $C_i$ is the result of executing step $s_i$
in configuration $C_{i-1}$.
%A \emph{prefix} $\sigma$ of $\pi$ is a sub-sequence of $\pi$ starting in $C_0$ and ending with a configuration.
%An \emph{interval} of $\pi$ is a sub-sequence that starts with a step and ends with a configuration.

%We also define what it means for two configurations $C$ and $C'$ to be indistinguishable to a process $p$, which we denote by $C \similar{p} C'$.
%Typically, indistinguishability is defined so that, if $C \similar{p} C'$, then the states of $p$ and all base objects are the same in $C$ and $C'$.
%However, in this work, we only care about the states of $p$ and all \textit{valid base objects}.
%So, we define $C \similar{p} C'$ to hold if: every object that is valid in both configurations has the same state in $C$ and $C'$, and each process $p \in P$ has the same state in $C$ and $C'$. 
%Note that $C \similar{p} C'$ is well defined when $C$ and $C'$ are configurations in executions of different algorithms.
%For convenience we also define $C \similar{P} C'$, where $P \neq \emptyset$ is a set of processes, to mean $C \similar{p} C'$ for each $p \in P$.
%(As a special case, we define $C \similar{\emptyset} C'$ to mean: every object that is valid in both configurations has the same state in $C$ and $C'$.)

An execution is \emph{linearizable}~\cite{Herlihy:1990} if it is possible to identify, for each high-level operation, a \emph{linearization point} that occurs during the high-level operation, so that the response of each high-level operation is the same as if it were performed atomically at its linearization point.
An algorithm is \emph{linearizable} if all possible executions are linearizable.
\end{thesisnot}
%!TEX root = paper.tex

\section{Wasteful Algorithms} \label{sec-wasteful}
In this section, we describe increasingly complex classes of lock-free wasteful algorithms, and progressively build up a descriptor ADT to capture their behaviour.
First, we consider algorithms in which descriptors are not changed after they are initialized.
Such descriptors are called \textit{immutable}.
We then discuss algorithms in which descriptors are modified by helpers.
We call such descriptors \textit{mutable}.

%We start by describing one common way to implement operations in a lock-free wasteful algorithm.
For the sake of illustration, we start by describing one common way that lock-free wasteful algorithms are implemented. %to implement \textit{high-level} operations in a lock-free wasteful algorithm.
Consider a lock-free algorithm that implements a set of \textit{high-level} operations.
%For the sake of illustration, consider a lock-free operation that consists of one or more \textit{attempts}, which either succeed, or fail due to contention.
Each high-level operation consists of one or more \textit{attempts}, which either succeed, or fail due to contention.
Each high-level operation attempt accesses a set of objects (e.g., individual memory locations or nodes of a tree).
Conceptually, a high-level operation attempt locks a subset of these objects and then possibly modifies some of them. %before modifying any of them. %modifying the data structure.
%Each operation attempt conceptually locks each object it wants to modify.
These locks are special: instead of providing exclusive access to a \textit{process}, they provide exclusive access to a \textit{high-level operation attempt}.
Whenever a high-level operation attempt by a process $p$ is unable to lock an object because it is already locked by another high-level operation attempt $O$, $p$ first \textit{helps} $O$ to complete, before continuing its own attempt or starting a new one.
By helping $O$ complete, $p$ effectively removes the locks that prevent it from making progress.
Note that $p$ is able to access objects locked for a different high-level operation attempt (which is not possible in traditional lock-based algorithms), but only for the purpose of helping the other high-level operation attempt complete.

We now discuss how helping is implemented.
Each high-level operation or operation attempt allocates a new \textit{descriptor} object, and fills it with information that describes any modifications it will perform.
This information will be used by any processes that help the high-level operation attempt.
For example, if the lock-free algorithm performs its modifications with a sequence of CAS steps, then the descriptor might contain the addresses, expected values and new values for the CAS steps.

A high-level operation attempt locks each object it would like to access by publishing pointers to its descriptor, typically using CAS.
Each pointer may be published in a dedicated field for descriptor pointers, or in a memory location that is also used to store application values.
%An operation attempt then locks the objects that it wishes to access , typically by using CAS to publish a pointer to its descriptor in the each object. 
%An operation attempt then locks an object by publishing a pointer to its descriptor in the object, typically using CAS.
%%Thus, whenever a process fails to lock an object, it can obtain a pointer to the descriptor for the operation that blocked it.
%The pointer to the descriptor may be announced in a dedicated field for descriptor pointers, or it may be announced in a memory location that is also used to store application data.
For example, in the BST of Ellen~et~al., nodes have a separate field for descriptor pointers~\cite{Ellen:2010}, but in Harris' implementation of multi-word CAS from single-word CAS, high-level operations temporarily replace application values with pointers to descriptors~\cite{Harris:2002}.
%Our transformation applies to both of these algorithms.
%In order to help an operation $o$, a process uses the information stored in $o$'s descriptor to determine how to proceed.
%When the operation attempt is finished, it removes any pointers to its descriptor to indicate that it no longer needs help.

%Whenever a process encounters a pointer to a descriptor (for an operation attempt that is not its own), it invokes a decision procedure \textit{IsActive}$(ptr)$, where $ptr$ is a pointer to the operation attempt's descriptor.
%\textit{IsActive} returns false \textit{only} if the operation attempt was completed before \textit{IsActive} terminated.
%%If \textit{IsActive}$(ptr)$ terminates after the operation attempt has been completed, then \textit{IsActive} returns false.
%Otherwise \textit{IsActive} returns true.
%(Consequently, if \textit{IsActive} returns false, the operation attempt no longer needs to be helped.)
%If \textit{IsActive} returns true, then the process \textit{may} decide to help the operation attempt by invoking a function \textit{Help}($ptr$).
When a process encounters a pointer $ptr$ to a descriptor (for a high-level operation attempt that is not its own), it may decide to help the other high-level operation attempt by invoking a function \textit{Help}$(ptr)$.
Typically, \textit{Help}$(ptr)$ is also invoked by the process that started the high-level operation.
That is, the mechanism used to help is the same one used by a process to perform its own high-level operation attempt.
%Note that \textit{Help}$(ptr)$ can be invoked by several helping processes in addition to the process that started the operation.
%The only shared memory locations that \textit{Help}$(ptr)$ can access are those pointed to by fields of the descriptor. %accesses only the shared memory locations accesses only stack variables, the contents of the descriptor, and any memory locations pointed to by the descriptor.
%We say that the operation attempt has \textit{completed} once an invocation of \textit{Help}$(ptr)$ (by any process) has terminated.
%If all accesses to fields of a descriptor occur inside \textit{Help}, then it is straightforward to apply our transformation.
%Otherwise, some algorithm specific knowledge may be necessary, as we discuss below.
%%All accesses to fields of a descriptor occur inside \textit{IsActive} or \textit{Help}.

Wasteful algorithms typically assume that, whenever an operation attempt allocates a new descriptor, it uses fresh memory that has never previously been allocated.
If this assumption is violated, then an \textit{ABA problem} may occur.
Suppose a process $p$ reads an address $x$ and sees $A$, then performs a CAS to change $x$ from $A$ to $C$, and interprets the success of the CAS to mean that $x$ contained $A$ at all times between the read and CAS.
If another process changes $x$ from $A$ to $B$ and back to $A$ between $p$'s read and CAS, then $p$'s interpretation is invalid, and an ABA problem has occurred.
%Of course, if memory used to store a descriptor is reclaimed, and later reallocated and used to store another descriptor, this will violate the assumption that each descriptor is allocated new memory which has never previously been allocated.
Note that safe memory reclamation algorithms will reclaim a descriptor only if no process has, or can obtain, a pointer to it.
Thus, no process can tell whether a descriptor is allocated fresh or reclaimed memory. %or memory that was previously reclaimed, 
So, safe memory reclamation will not introduce ABA problems.

\subsection{Immutable descriptors} \label{sec-adt-immutable}

%For simplicity, we first assume that helpers never modify the contents of a descriptor.
%
%\paragraph{Immutable descriptor ADT}
%
%%\begin{figure}
%%	\lstinputlisting[frame=single,tabsize=2,firstnumber=auto,name=adt,style=customc]{immutable-ADT.c}	
%%	\caption{Immutable descriptors ADT\label{code-immutable-adt}}
%%\end{figure} 

We give a trivial \textit{immutable descriptor} ADT that captures the way that descriptors are used by the class of wasteful algorithms we just described.
%The ADT for immutable descriptors appears in Figure~\ref{code-immutable-adt}.
%\trevor{edit adt again. define descriptor types as sets of fields. talk about fields and values of fields.}
%A descriptor \textit{type} is defined as a set of fields.
A \textit{descriptor} has a set of fields, and each field contains a value. %associates each field $f \in T$ with a value.
The ADT offers two operations: \func{CreateNew} and \func{ReadField}. %, which have the following semantics.
%\begin{description}
%	\item[\func{CreateNew}] 
\func{CreateNew} takes, as its arguments, a descriptor type and a sequence of values, one for each field of the descriptor. %set $pairs$ that contains exactly one pair $(f, value)$ for each $f \in T$.
It returns a unique descriptor pointer $des$ that has never previously been returned by \func{CreateNew}.
%    (Most often, the descriptor reference is a pointer.)
Every descriptor pointer returned by \func{CreateNew} represents a new immutable descriptor object.
%    The descriptor referenced by $des$ associates each field $f$ with its (unique) value in $pairs$.
%	\item[\func{ReadField}]
\func{ReadField} takes, as its arguments, a descriptor pointer $des$ and a field $f$, and returns the value of $f$ in $des$.
%    If $des$ was returned by a previous \func{CreateNew}$(T, pairs)$ operation with $(f, value) \in pairs$, then \func{ReadField} returns $value$.
%    Otherwise, \func{ReadField} is undefined.
%It returns the value of $f$ in $des$. %for $f$ that was passed to the invocation of \func{CreateNew} that returned $des$.
%%    (I.e., if $des$ was returned by \func{CreateNew}$(T, pairs)$, then \func{ReadField} returns $v$, where $(f, v) \in pairs$.)
%\end{description}
%Every descriptor reference returned by \func{CreateNew} represents a new immutable descriptor object.

\begin{shortver}
We require the immutable descriptor ADT operations to be lock-free, so they can be used to implement lock-free data structures.
(We discuss this further in Appendix~\ref{appendix-desc-imm-progress}.)
\end{shortver}

\begin{fullver}
In wasteful algorithms, whenever a process wants to create a new descriptor, it simply invokes \func{CreateNew}.
Whenever a helper wants to access a descriptor, it invokes \func{ReadField}.
\end{fullver}
\begin{fullver}
\paragraph{Progress}
If the immutable descriptor ADT is implemented so that \func{CreateNew} allocates and initializes a new descriptor, and \func{ReadField} reads and returns a field of a descriptor, then its operations will be \textit{wait-free} (i.e., each operation will terminate after a finite number of its own steps).
However, wait-free descriptor operations are not necessary to guarantee lock-freedom for high-level operations that use descriptors. %if we start with a lock-free algorithm that uses wait-free createnew and readfield, we can replace the wait-free impl of createnew and readfield with a lock-free impl of createnew and readfield, and the alg will still be lock-free.
Instead, we simply require descriptor operations to be lock-free.
We now explain why this is sufficient to implement lock-free data structures.

Consider a lock-free algorithm that %implements a set of \textit{high-level} operations, and 
uses a wait-free implementation of the immutable descriptor ADT.
Suppose we transform this algorithm by replacing the wait-free implementation of the descriptor ADT with a \textit{lock-free} implementation.
We argue that the transformed algorithm remains lock-free.
In other words, we show that, if processes take infinitely many steps in the transformed algorithm, then infinitely many high-level operations complete.
In the original algorithm, if processes take infinitely many steps, then infinitely many high-level operations will complete.
The only steps we change to obtain the transformed algorithm are invocations of \func{CreateNew} and \func{ReadField}, some of which might no longer terminate.
%In the transformed algorithm, some of these invocations might be non-terminating.
Therefore, the only way the transformed algorithm can \textit{fail} to satisfy lock-freedom is if, eventually, all processes take steps only in non-terminating invocations of \func{CreateNew} and \func{ReadField}.
(Otherwise, processes take infinitely many steps of the original algorithm, so infinitely many high-level operations will succeed.)
In this case, only finitely many invocations of \func{CreateNew} and \func{ReadField} will terminate.
However, since \func{CreateNew} and \func{ReadField} are lock-free, infinitely many invocations of \func{CreateNew} and/or \func{ReadField} must terminate. %, so this is impossible \trevor{the ``this'' qualifier is hard to tie to its reference two sentences back}.
Thus, a lock-free implementation of the immutable descriptor ADT is sufficient to implement lock-free algorithms.
\end{fullver}

\paragraph{Example Algorithm: DCSS}

We use the double-compare single-swap (\func{DCSS}) algorithm of Harris et al.~\cite{Harris:2002} as an example of a lock-free algorithm that fits the preceding description.
Its usage of descriptors is easily captured by the immutable descriptor ADT.
A \func{DCSS}($a_1,e_1,a_2,e_2,n_2$) operation does the following \textit{atomically}.
It checks whether the values in addresses $a_1$ and $a_2$ are equal to a pair of expected values, $e_1$ and $e_2$.
If so, it stores the value $n_2$ in $a_2$ and returns $e_2$.
Otherwise it returns the current value of $a_2$.

\begin{figure}[tb]
%	\noindent
%\begin{minipage}{0.4825\textwidth}
%\hspace{2mm}
%\begin{minipage}{\linewidth}
\begin{lstlisting}[name=dcss,frame=single]
 type $\func{DCSSdes}: \{\vaddr_1, \vexp_1, \vaddr_2, \vexp_2, \vnew_2\}$

 //\com \textbf{DCSS ADT operations}
 $\func{DCSS}(a_1, e_1, a_2, e_2, n_2):$
   $des := \func{CreateNew}(\func{DCSSdes}, a_1, e_1, a_2, e_2, n_2)$//\label{code-dcss-throw-allocate}
   $fdes := flag(des)$// \label{code-dcss-throw-flag}
   loop //\label{code-dcss-throw-do}
     $r := \func{CAS}(a_2, e_2, fdes)$ //\label{code-dcss-throw-publish-cas}
     if $r\ \mbox{is flagged}$ then $\func{DCSSHelp}(r)$ //\label{code-dcss-throw-dcss-help}
     else exit loop //\label{code-dcss-throw-while}
   if $r = e_2$ then $\func{DCSSHelp}(fdes)$ //\label{code-dcss-throw-dcss-finish}
   return $r$ //\label{code-dcss-throw-dcss-return}

 $\func{DCSSRead}(addr):$
   loop
     $r := *addr$ //\label{code-dcss-throw-read-read}
     if $r\ \mbox{is flagged}$ then $\func{DCSSHelp}(r)$ //\label{code-dcss-throw-read-help}
     else exit loop
   return $r$ //\label{code-dcss-throw-read-return}

 //\com \textbf{Private procedures}
 $\func{DCSSHelp}(fdes):$
   $des := unflag(fdes)$
   $addr_1 := \func{ReadField}(des, \vaddr_1)$ //\label{code-dcss-throw-help-read-a1}
   $addr_2 := \func{ReadField}(des, \vaddr_2)$
   $exp_1 := \func{ReadField}(des, \vexp_1)$
   if $*addr_1 = exp_1$ then //\label{code-dcss-throw-help-compare-a1}
     $new_2 := \func{ReadField}(des, \vnew_2)$
     $\func{CAS}(addr_2, fdes, new_2)$ //\label{code-dcss-throw-help-cas-new}
   else
     $exp_2 := \func{ReadField}(des, \vexp_2)$
     $\func{CAS}(addr_2, fdes, exp_2)$ //\label{code-dcss-throw-help-cas-old}
\end{lstlisting}
    \vspace{-4.5mm}
	\caption{Code for the \func{DCSS} algorithm of Harris et~al.~\cite{Harris:2002} using the \textit{immutable descriptor} ADT.}
    \label{code-throw-DCSS}
%    The code on the right is transformed to use the weak descriptor ADT.
%    Changed lines in the transformed code are marked with an asterisk (*).
%    Newly added lines are marked with a plus sign (+).}
\end{figure}

Pseudocode for the \func{DCSS} algorithm appears in %on the left-hand side of 
Figure~\ref{code-throw-DCSS}.
At a high level, \func{DCSS} creates a descriptor, and then attempts to lock $a_2$ by using CAS to replace the value in $a_2$ with a pointer to its descriptor.
%%This descriptor will be used by each operation attempt.
%%Each iteration of the loop at lines~\ref{code-dcss-throw-do}-\ref{code-dcss-throw-while} performs an operation attempt.
%%%At a high level, 
%%Each operation attempt 
%In each iteration, it uses CAS to attempt to install a pointer to its descriptor directly in $a_2$ (replacing an value). %, and, if it succeeds, it reads $a_1$ and checks whether it matches the expected value.
Since the \func{DCSS} algorithm replaces values with descriptor pointers, it needs a way to distinguish between values and descriptor pointers (in order to determine when helping is needed).
So, it steals a bit from each memory location and uses this bit to \textit{flag} descriptor pointers. % as such.
%Thus, \func{DCSS} actually locks $a_2$ by replacing the value in $a_2$ with a \textit{flagged} descriptor pointer (which we simply call a \textit{flagged pointer}).

We now give a more detailed description.
\func{DCSS} starts by creating and initializing a new descriptor $des$ at line~\ref{code-dcss-throw-allocate}.
It then flags $des$ %The descriptor pointer for the \func{DCSS} operation's descriptor is flagged 
at line~\ref{code-dcss-throw-flag}.
We call the result \textit{fdes} a \textit{flagged pointer}.
\func{DCSS} then attempts to lock $a_2$ in the loop at lines~\ref{code-dcss-throw-do}-\ref{code-dcss-throw-while}.
In each iteration, it tries to store its flagged pointer in $a_2$ using CAS. % at line~\ref{code-dcss-throw-publish-cas}.
If the CAS is successful, then the operation attempt invokes \func{DCSSHelp} to complete the operation (at line~\ref{code-dcss-throw-dcss-finish}).
Now, suppose the CAS fails.
Then, the \func{DCSS} checks whether its CAS failed because $a_2$ contained another \func{DCSS} operation's flagged pointer (at line~\ref{code-dcss-throw-dcss-help}).
If so, it invokes \func{DCSSHelp} to help the other \func{DCSS} complete, and then retries its CAS.
%Otherwise, the CAS failed because $a_2$ contained a value different from $e_2$.
\func{DCSS} repeatedly performs its CAS (and helping) until the \func{DCSS} either succeeds, or fails because $a_2$ did not contain~$e_2$.

\func{DCSSHelp} takes a flagged pointer $fdes$ as its argument, and begins by unflagging $fdes$ (to obtain the actual descriptor pointer for the operation).
Then, it reads $a_1$ and checks whether it contains $e_1$ (at line~\ref{code-dcss-throw-help-compare-a1}).
If so, it uses CAS to change $a_2$ from $fdes$ to $n_2$, completing the \func{DCSS} (at line~\ref{code-dcss-throw-help-cas-new}).
Otherwise, it uses CAS to change $a_2$ from $fdes$ to $e_2$, effectively aborting the \func{DCSS} (at line~\ref{code-dcss-throw-help-cas-old}).
Note that this code is executed by the process that created the descriptor, and also possibly by several helpers.
Some of these helpers may perform a CAS at line~\ref{code-dcss-throw-help-compare-a1} and some may perform a CAS at line~\ref{code-dcss-throw-help-cas-new}, but only the first of these CAS steps can succeed.
%%\begin{fullver}
%	The intuition behind why helpers do not perform conflicting changes is that only one of these CAS steps can succeed, and it can only succeed the first time it is performed by any process helping a given operation.
%%\end{fullver}
%%\begin{shortver}
%%	The intuition behind why helping works is that only one of these CAS steps can succeed, and it can only succeed the first time it is performed by any process helping a given operation.
%%\end{shortver}

When a program uses DCSS, some addresses can contain either values or descriptor pointers. %in any algorithm that uses \func{DCSS}, 
So, each read of such an address must be replaced with an invocation of a function called \func{DCSSRead}.
\func{DCSSRead} takes an address $addr$ as its argument, and begins by reading $addr$ (at line~\ref{code-dcss-throw-read-read}).
It then checks whether it read a descriptor pointer (at line~\ref{code-dcss-throw-read-help}) and, if so, invokes \func{DCSSHelp} to help that \func{DCSS} complete.
\func{DCSSRead} repeatedly reads and performs helping until it sees a value, which it returns (at line~\ref{code-dcss-throw-read-return}).

\subsection{Mutable descriptors} \label{sec-adt-mutable}
In some more advanced lock-free algorithms, each descriptor also contains information about the \textit{status} of its high-level operation attempt, and this status information is used to coordinate helping efforts between processes.
Intuitively, the status information gives helpers some idea of what work has already been done, and what work remains to be done.
%To coordinate helping efforts between processes, the descriptor may contain some information about the \textit{state} of the operation attempt.
Helpers use this information to direct their efforts, and update it as they make progress.
As a trivial example, the state information might simply be a bit that is set (by the process that started the high-level operation, or a helper) once the high-level operation succeeds.

As another example, in an algorithm where high-level operation attempts proceed in several phases, the descriptor might store the current phase, which would be updated by helpers as they successfully complete phases.
Observe that, since lock-free algorithms cannot use mutual exclusion, helpers often use CAS to avoid making conflicting changes to status information, which is quite expensive.
Updating status information may introduce contention.
Even when there is no contention, it adds overhead.
Lock-free algorithms typically try to minimize updates to status information.
Moreover, status information is usually simplistic, and is encoded using a small number of bits.

Status information might be represented as a single field in a descriptor, or it might be distributed across several fields.
Any fields of a descriptor that contain status information are said to be \textit{mutable}.
All other fields are called \textit{immutable}, because they do not change during an operation.

\paragraph{Mutable descriptor ADT}

%\begin{figure}
%	\lstinputlisting[frame=single,tabsize=2,firstnumber=auto,name=adt,style=customc]{mutable-ADT.c}	
%	\caption{Mutable descriptors ADT\label{code-mutable-adt}}
%\end{figure} 

We now extend the immutable descriptor ADT to provide operations for changing (mutable) fields of descriptors.
The \textit{mutable descriptor} ADT offers four operations: \func{CreateNew}, \func{WriteField}, \func{CASField} and \func{ReadField}.
The semantics for \func{CreateNew} and \func{ReadField} are the same as in the immutable descriptor ADT.
%We now give the semantics for the other two operations.
%%First, it is helpful to define the \textit{current value} of a field $f$ of a descriptor of type $T$ with pointer $des$ in a configuration $C$.
%
%\begin{description}
%	\item[\func{WriteField}]
\func{WriteField} takes, as its arguments, a descriptor pointer $des$, a field $f$ and a value $v$.
It stores $v$ in field $f$ of $des$. %sets the value associated with $f$ in $des$ to $v$.
%\item[\func{CASField}] 
\func{CASField} takes, as its arguments, a descriptor pointer $des$, a field $f$, an expected value $exp$ and a new value $v$.
    Let $v_f$ be the value of $f$ in $des$ just before the \func{CASField}.
    If $v_f = exp$, then \func{CASField} stores $v$ in $f$.
    \func{CASField} returns $v_f$.
%%    Let $v'$ be the value passed to the last (before the \func{CASField}) \func{CASField}$(T, des, f, exp', v')$ that returns $exp'$, or \func{WriteField}$(T, des, f, v')$.
%%    If there is no such operation before the \func{CASField}, then let $v'$ be the value for $f$ that was passed to the \func{CreateNew} operation that returned $des$.
%%    \func{CASField} returns $v'$.
%	\item[\func{ReadField}] takes, as its arguments, a descriptor type $T$, a descriptor pointer $des$ (for a descriptor of type $T$), and a field $f \in T$.
%    \func{ReadField} returns the value that is associated with $f$ in $des$ (just before the \func{ReadField}).
%%    It returns the value $v$ passed to the last (before the \func{ReadField}) \func{CASField}$(T, des, f, exp, v)$ that returns $exp$, or \func{WriteField}$(T, des, f, v)$.
%%    If there is no such operation before the \func{ReadField}, then it returns the value for $f$ that was passed to the \func{CreateNew} operation that returned $des$.
%\end{description}

As in the immutable descriptor ADT, we require the operations of the mutable descriptor ADT to be lock-free.

\begin{figure}[th!]
\begin{lstlisting}[name=kcas,frame=single]
 type $\func{k-CASdes}: \{\vstate, \vaddr_1, \vexp_1, \vnew_1,$ $\vaddr_2, \vexp_2, \vnew_2, \ldots, \vaddr_k, \vexp_k, \vnew_k \}$
  
 //\com \textbf{k-CAS ADT operations}
 $\func{k-CAS}(a_1, e_1, n_1, a_2, e_2, n_2, \ldots, a_k, e_k, n_k):$
   $des := \func{CreateNew}(\func{k-CASdes}, Undecided, a_1, e_1, n_1,\ldots )$ //\label{line-kcasthrow-kcas-createnew}
   $fdes :=$// flagged version of $des$\label{line-kcasthrow-kcas-flag}
   return $\func{k-CASHelp}(fdes)$//\label{line-kcasthrow-kcas-help}
  
 $\func{k-CASRead}(addr):$
   loop
     $r := \func{DCSSRead}(addr)$//\label{line-kcasthrow-kcasread-dcssread}
     if $r \mbox{ is flagged}$ then $\func{k-CASHelp}(r)$ //\label{line-kcasthrow-kcasread-help}
     else exit loop
   return $r$ 
 
 //\com \textbf{Private procedures}
 $\func{k-CASHelp}(fdes):$
   $des :=$// remove the flag from $fdes$ \label{line-kcasthrow-kcashelp-unflag}
   //\com Use DCSS to store $fdes$ in each of $a_1, a_2, \ldots, a_k$ 
   //\com \textit{only} if $des$ has $\vstate$ $\textit{Undecided}$ and $a_i = e_i$ for all $i$
   if $\func{ReadField}(des,\vstate) = \textit{Undecided}$ then//\label{line-kcasthrow-kcashelp-phase1-ifundecided}
     $state := Succeeded$//\label{line-kcasthrow-kcashelp-phase1-setstatesucceeded}
     for $i=1 ... k$ do//\label{line-kcasthrow-kcashelp-phase1-start}
 $retry\_entry$: //\label{line-kcasthrow-kcashelp-phase1-labelretry}
       $a_1 := \func{ReadField}(des,\vstate)$ //\label{line-kcasthrow-kcashelp-phase1-readstate}
       $a_2 := \func{ReadField}(des,\vaddr_i)$ //\label{line-kcasthrow-kcashelp-phase1-readaddr}
       $e_2 := \func{ReadField}(des,\vexp_i)$ //\label{line-kcasthrow-kcashelp-phase1-readexp}
       $val := \func{DCSS}(\langle des,\vstate \rangle,\textit{Undecided},a_2,e_2,fdes)$ //\label{line-kcasthrow-kcashelp-phase1-dcss}
       if $val \mbox{ is flagged}$ then //\label{line-kcasthrow-kcashelp-phase1-iskcas}
         if $val \neq fdes$ then //\label{line-kcasthrow-kcashelp-phase1-ifdifferentkcas}
           $\func{k-CASHelp}(val)$//\label{line-kcasthrow-kcashelp-phase1-help}
           goto $retry\_entry$//\label{line-kcasthrow-kcashelp-phase1-gotoretry}
       else
         if $val \neq e_2$ then //\label{line-kcasthrow-kcashelp-phase1-ifnotexpected}
           $state := Failed$ //\label{line-kcasthrow-kcashelp-phase1-setstatefailed}
           break //\label{line-kcasthrow-kcashelp-phase1-break}
     $\func{CASField}(des,\vstate,\textit{Undecided},state)$ //\label{line-kcasthrow-kcashelp-phase1-casstate}\label{line-kcasthrow-kcashelp-phase1-end}

   //\com Replace $fdes$ in $a_1, ..., a_k$ with $n_1, ..., n_k$ or $e_1, ..., e_k$
   $state : = \func{ReadField}(des,\vstate)$ //\label{line-kcasthrow-kcashelp-phase2-readstate}
   for $i=1 ... k$ do //\label{line-kcasthrow-kcashelp-phase2-start}
     $a = \func{ReadField}(des,\vaddr_i)$ //\label{line-kcasthrow-kcashelp-phase2-readaddr}
     if $state = Succeeded$ then //\label{line-kcasthrow-kcashelp-ifsucceeded}
       $new := \func{ReadField}(des,\vnew_i)$ //\label{line-kcasthrow-kcashelp-phase2-readnew}
     else
       $new := \func{ReadField}(des,\vexp_i)$ //\label{line-kcasthrow-kcashelp-phase2-readexp}
     $\func{CAS}(a,fdes,new)$ //\label{line-kcasthrow-kcashelp-phase2-casnew}\label{line-kcasthrow-kcashelp-phase2-end}
   return $(state = Succeeded)$ //\label{line-kcasthrow-kcashelp-return}
\end{lstlisting}
% $\func{IsDCSS}(val):$
% $\langle -, -, flagbit \rangle := val$
% return $flagbit$
% $\func{DCSSFlag}(des):$
% return  $\langle des, false, true \rangle$
\vspace{-4.5mm}
\caption{Code for the \func{k-CAS} algorithm of Harris et~al.~\cite{Harris:2002} using the \textit{mutable descriptor} ADT.}
\label{code-throw-kCAS}
\vspace{-3mm}
\end{figure}

\paragraph{Example Algorithm: $k$-CAS}

%We first give the semantics for $k$-CAS.
A \textit{$k$-CAS}($a_1, ..., a_k,$ $e_1, ..., e_k,$ $n_1, ..., n_k$) operation atomically does the following.
First, it checks if each address $a_i$ contains its expected value $e_i$.
If so, it writes a new value $n_i$ to $a_i$ for all $i$ and returns true.
Otherwise it returns false.

The $k$-CAS algorithm of Harris~et~al.~\cite{Harris:2002} is an example of a lock-free algorithm that has descriptors with mutable fields.
At a high level, %We start with a high-level sketch of the algorithm.
a $k$-CAS operation $O$ in this algorithm starts by creating a descriptor that contains its arguments. %: $a_1, a_2, ..., a_k, e_1, e_2, ..., e_k, n_1, n_2, ..., n_k$.
It then tries to lock each location $a_i$ \textit{for the operation} $O$ by changing the contents of $a_i$ from $e_i$ to $des$, where $des$ is a pointer to $O$'s descriptor.
If it successfully locks each location $a_i$, then it changes each $a_i$ from $des$ to $n_i$, and returns true.
%However, if it fails to lock a location, two cases arise.
If it fails because $a_i$ is locked for another operation, then it helps the other operation to complete (and unlock its addresses), and then tries again.
If it fails because $a_i$ contains an application value different from $e_i$, then the $k$-CAS fails, and unlocks each location $a_j$ that it locked by changing it from $des$ back to $e_j$, and returns false.
(The same thing happens if $O$ fails to lock $a_i$ because the operation has already terminated.)

We now give a more detailed description of the algorithm.
Pseudocode appears in Figure~\ref{code-throw-kCAS}.
A $k$-CAS operation creates its descriptor at line~\ref{line-kcasthrow-kcas-createnew}, and then %that contains all arguments to the operation ($a_i, e_i$ and $n_i$, for all $i$).
%Then, it 
invokes a function \func{k-CASHelp} to complete the operation.
%\func{HelpKCAS} performs the two phases of the $k$-CAS operation described above.
%The first phase consists of lines~\ref{line-kcasthrow-kcashelp-phase1-start}-\ref{line-kcasthrow-kcashelp-phase1-end}, and the second phase consists of lines~\ref{line-kcasthrow-kcashelp-phase2-start}-\ref{line-kcasthrow-kcashelp-phase2-end}.
%
In addition to the arguments to its $k$-CAS operation, a $k$-CAS descriptor contains a 2-bit \textit{state} field that initially contains \textit{Undecided} and is changed to \textit{Succeeded} or \textit{Failed} depending on how the operation progresses.
This \textit{state} field is used to coordinate helpers.

Let $p$ be a process performing (or helping) a $k$-CAS operation $O$ that created a descriptor $d$.
If $p$ fails to lock some address $a_i$ in $d$, then $p$ attempts to change the \textit{state} of $d$ using CAS from \textit{Undecided} to \textit{Failed}.
On the other hand, if $p$ successfully locks each address in $d$, then $p$ attempts to change the \textit{state} of $d$ using CAS from \textit{Undecided} to \textit{Succeeded}.
Since the \textit{state} field changes only from \textit{Undecided} to either \textit{Failed} or \textit{Succeeded}, only the first CAS on the state field of $d$ will succeed.
The $k$-CAS implementation then uses a lock-free DCSS primitive (the one presented in Section~\ref{sec-adt-immutable}) to ensure that $p$ can lock addresses for $O$ \textit{only} while $d$'s \textit{state} is \textit{Undecided}.
%This is achieved by replacing every CAS that changes $a_i$ from $e_i$ to $d$ with a DCSS that does the same thing, but succeeds only if \textit{state} is \textit{Undecided}.
%Intuitively, 
This prevents helpers from erroneously performing successful CAS steps after the $k$-CAS operation is already over.

Recall that the DCSS algorithm allocates a descriptor for each DCSS operation.
A $k$-CAS operation performs potentially \textit{many} DCSS operations (at least $k$ for a successful $k$-CAS), and also allocates its own $k$-CAS descriptor.
The $k$-CAS algorithm need not be aware of DCSS descriptors (or of the bit reserved in each memory location by the DCSS algorithm to flag values as DCSS descriptor pointers), since it can simply use the \func{DCSSRead} procedure described above whenever it accesses a memory location that might contain a DCSS descriptor.
However, the converse is \textit{not true}, since the $k$-CAS algorithm performs DCSS on the \textit{state} field of a $k$-CAS descriptor.
Of course, the \textit{state} field must be accessed using the $k$-CAS descriptor's \func{ReadField} operation.
So, to allow DCSS to access the \textit{state} field, we must modify DCSS slightly.
First, instead of passing an address $a_1$ to DCSS, we pass a pointer to the $k$-CAS descriptor and the name of the \textit{state} field (at line~\ref{line-kcasthrow-kcashelp-phase1-dcss} of Figure~\ref{code-throw-kCAS}).
Second, we replace the read of $addr_1$ in DCSS (at line~\ref{code-dcss-throw-help-compare-a1} of Figure~\ref{code-throw-DCSS}) with an invocation of \func{ReadField}.

Since $k$-CAS descriptor pointers are temporarily stored in memory locations that normally contain application values, the $k$-CAS algorithm needs a way to determine whether a value in a memory location is an application value or a $k$-CAS descriptor pointer.
In the DCSS algorithm, the solution was to reserve a bit in each memory location, and use this bit to \textit{flag} the value contained in the location as a pointer to a DCSS descriptor.
Similarly, the $k$-CAS algorithm reserves a bit in each memory location to flag a value as a $k$-CAS descriptor pointer.
The $k$-CAS and DCSS algorithms need not be aware of each other's reserved bits, but they should not reserve the same bit (or else, e.g., a DCSS operation could encounter a $k$-CAS descriptor pointer, and interpret it as a DCSS descriptor pointer).

When the $k$-CAS algorithm is used, some memory addresses may contain either values or descriptor pointers, so reads of such addresses must be replaced by a \func{k-CASRead} operation.
This operation reads an address, and checks whether it contains a $k$-CAS descriptor pointer.
If so, it helps the $k$-CAS operation to complete, and tries again.
Otherwise, it returns the value it read.
For further details, refer to~\cite{Harris:2002}.

\section{Weak descriptors} \label{sec-weak-descriptors}
In this section we present a \textit{weak descriptor} ADT that has weaker semantics than the mutable descriptor ADT, but can be implemented more efficiently (in particular, without requiring any memory reclamation for descriptors).
We identify a class of algorithms that use the mutable descriptor ADT, and which can be transformed to use the weak descriptor ADT, instead.

We first discuss a restricted case where operation attempts only create a single descriptor, and we give an ADT, transformation and proof for that restricted case.
(In the next section, we describe how the ADT and transformation can be modified slightly to support operation attempts that create multiple descriptors.)

%\begin{figure}
%	\lstinputlisting[frame=single,tabsize=2,firstnumber=auto,name=adt,style=customc]{weak-ADT.c}	
%	\caption{Weak descriptors ADT\label{code-weak-adt}}
%\end{figure} 

\subsection{Weak descriptor ADT} \label{sec-adt-weak}
The weak descriptor ADT is a variant of the mutable descriptor ADT that allows some operations to \textit{fail}. % (if they are accessing a descriptor of type $T$ created by a process that has since created a new descriptor of type $T$).
%
%The following definitions helps to simplify the explanation of how and when operations fail. % in the semantics for the weak descriptor ADT.
To ease the discussion, we introduce the concept of descriptor validity.
Let $des$ be a pointer returned by a \func{CreateNew} operation $O$ by a process $p$, and $d$ be the descriptor pointed to by $des$.
%Let $T$ be a descriptor type, and $des$ be a descriptor pointer (for a descriptor of type $T$) returned by a \func{CreateNew} operation $O$ by process $p$.
%The pointer $des$ represents a unique descriptor object $d$.
In each configuration, $d$ is either \textbf{valid} or \textbf{invalid}.
Initially, $d$ is valid.
If $p$ performs another \func{CreateNew} operation $O'$ \textit{after} $O$, then $d$ becomes invalid immediately after $O'$ (and will never be valid again).
%(We refer interchangeably to descriptors and descriptor pointers when stating whether a descriptor is valid or invalid.)

We say that a \func{ReadField}$(des, ...)$, \func{WriteField}$(des, ...)$ or \func{CASField}$(des, ...)$ operation is performed \textbf{on a descriptor} $d$, where $des$ is a pointer to $d$.
An operation on a valid (resp., invalid) descriptor is said to be valid (resp., invalid).
Invalid operations have no effect on any base object, and return a special value $\bot$ (which is never contained in a field of any descriptor) instead of their usual return value.
We say that a \func{CreateNew}$(T, ...)$ operation $O$ is performed \textbf{on a descriptor} $d$ if $O$ returns a pointer to $d$.
Observe that a \func{CreateNew} operation is always valid.

The semantics for \func{CreateNew} are the same as in the mutable descriptor ADT.
The semantics for the other three operations are the same as in the mutable descriptor ADT, except that they can be invalid.
%Invalid operations have no effect on the descriptor's state, and return $\bot$ (except for \func{WriteField}, which does not return a value).

As in the other descriptor ADTs, we require the operations of the weak descriptor ADT to be lock-free.

\subsection{Transforming a class of algorithms to use the weak descriptor ADT} \label{sec-weak-transformation}

We now formally define a class of lock-free algorithms that use the mutable descriptor ADT, and can easily be transformed so that they use the weak descriptor ADT, instead.
We say that a process $p$ \textbf{owns} a descriptor $d$ if it performed a \func{CreateNew} operation that returned a pointer $des$ to $d$.
Note that, in the following, we abuse notation slightly be referring interchangeably to a descriptor and a pointer to it.

We say that a step $s$ of an execution is \textit{nontrivial} if it changes the state of an object $o$ in shared memory, and \textit{trivial} otherwise.
In particular, all invalid operations are trivial, and an unsuccessful CAS or a CAS whose expected and new values are the same are both trivial.

%More formally, let $O$ be any operation on $o$.
%If $O$ would return a different value depending on whether it is performed before or after $s$, then $s$ is \textit{nontrivial}.
%Otherwise, if each operation on $o$ would return the same value regardless of whether it is executed before or after $s$, then $s$ is \textit{trivial}.
%For example, suppose $o$ is a weak descriptor.
%If $s$ is a valid \func{CASField}$(T, des, f, exp, v)$ that returns $exp$ where $v \neq exp$, then $s$ is nontrivial (since a \func{ReadField}$(T, des, f)$ operation will return a different value before and after $s$).
%Similarly, if $s$ is a valid \func{WriteField}$(T, des, f, v)$ and the value associated with $f$ in $des$ just before $s$ is different from $v$, then $s$ is nontrivial.
%Otherwise, $s$ is trivial.

%\trevor{define what it means to change an object in shared memory. intuitively, a step $s$ changes an object in shared memory if $s$ could cause a subsequent operation on the object to return a different response (than it would before $s$).}

\begin{definition} \label{def:WCA}
	Weak-compatible algorithms (WCA) are lock-free wasteful algorithms that use the mutable descriptor ADT, and have the following properties:  
	
	\begin{enumerate}
		\item Each high-level operation attempt $O$ by a process $p$ may create (and initialize) a single descriptor $d$.
        Inside $O$, $p$ may perform at most one invocation of a function \func{Help}$(d)$ (and $p$ may not invoke \func{Help}$(d)$ outside of $O$).
        \label{def:WCA:prop:alg-steps}
        \item A process may help any operation attempt $O'$ by another process by invoking \func{Help}$(d')$ where $d'$ is the descriptor that was created by $O'$.
        \label{def:WCA:prop:helping}
		\item If $O$ terminates at time $t$, %If an invocation of \func{Help}$(d)$ terminates at time $t$, 
        then any steps taken in an invocation of \func{Help}$(d)$ after time $t$ are \textit{trivial} (i.e., do not \textbf{change} the state of \textbf{any} shared object, incl. $d$).\label{def:WCA:prop:not-change-mem}
		\item While a process $q \neq p$ is performing \func{Help}$(d)$, $q$ cannot change any variables in its private memory that are still defined once \func{Help}$(d)$ terminates (i.e.,~variables that are local to the process $q$, but are not local to \func{Help}).
%        \item If a process $q \neq p$ invokes \func{Help}$(d)$, then $q$ \textbf{cannot} change any variables that are still defined once \func{Help}$(d)$ terminates (i.e.,~variables that are local to the process, but are not local to \func{Help}). %$q$'s process state immediately after \func{Help}$(d)$ is the same as its state before \func{Help}$(d)$, except that its program counter is now immediately after \func{Help}.
        \label{def:WCA:prop:return-help}
		\item All accesses (read, write or CAS) to a field of $d$ occur inside either \func{Help}$(d)$ or $O$.
        \label{def:WCA:prop:desc-ops-only-in-help}
	\end{enumerate}
\end{definition}

At a high level, properties~\ref{def:WCA:prop:alg-steps} and~\ref{def:WCA:prop:helping} of WCA describe how descriptors are created and helped.
Property~\ref{def:WCA:prop:return-help} intuitively states that, whenever a process $q$ finishes helping another process perform its operation attempt, $q$ knows only that it finished helping, and does not remember anything about what it did while helping the other process.
In particular, this means that $q$ cannot pay attention to the return value of \func{Help}.
We explain why this behaviour makes sense.
If $q$ creates a descriptor $d$ as part of a high-level operation attempt $O$ and invokes \func{Help}$(d)$, then $q$ might care about the return value of \func{Help}, since it needs to compute the response of $O$.
However, if $q$ is just helping another process $p$'s high-level operation attempt $O$, then it does not care about the response of \func{Help}, since it does not need to compute the response of $O$.
The remaining properties, \ref{def:WCA:prop:not-change-mem} and~\ref{def:WCA:prop:desc-ops-only-in-help}, allow us to argue that the contents of a descriptor are no longer needed once the operation that created it has terminated (and, hence, it makes sense for the descriptor to become invalid).
%
%%\trevor{re. (3): all i know when i finish helping is that i finished helping.}
%%
%%\trevor{clarify (3): if you are helping someone else, then you don't pay attention to the return value of help. if you created the descriptor, then you might care about what help returns, because you need to compute the response of the operation. on the other hand, if you are just helping someone else's operation, then you don't care about the response of help, because you don't need to compute the response of the operation that you helped.}
%
In Section~\ref{sec-adt-extended}, we will study a larger class of algorithms with a weaker version of property~\ref{def:WCA:prop:desc-ops-only-in-help}. % requirement than the one in Definition~\ref{def:WCA}.\ref{def:WCA:prop:desc-ops-only-in-help}.)

\paragraph{The transformation}

Each algorithm in WCA can be transformed in a straightforward way into an algorithm that uses the weak descriptor ADT as follows.
%First, we replace the implementation of the mutable descriptor ADT operations with an implementation of the corresponding weak descriptor ADT operations.
Consider any \func{ReadField} or \func{CASField} operation $op$ performed by a high-level operation attempt $O$ in an invocation of \func{Help}$(d)$, where $d$ was created by a \textit{different} high-level operation attempt $O'$.
Note that $op$ is performed while $O$ is \textit{helping} $O'$.
After $op$, a check is added to determine whether $op$ was invalid, in which case $p$ returns from \func{Help} immediately.
%After each \func{ReadField} or \func{CASField} operation performed by a high-level operation attempt $O$ in an invocation of \func{Help}$(d)$, where $d$ was created by a \textit{different} high-level operation attempt, a check is added to determine whether the operation was invalid, in which case $p$ returns from \func{Help} immediately.
(In this case, \func{Help} does not need to continue, since $op$ will be invalid only if $O'$ has already been completed by the process that owns $d$ or a helper.)

\paragraph{Example Algorithm: DCSS}

\begin{figure}[t]
\begin{lstlisting}[name=applyingdcss,frame=single,mathescape=true]
 $\func{DCSSHelp}(fdes):$
   $des := \func{Unflag}(fdes)$
   $addr_1 := \func{ReadField}(des, \vaddr_1)$ //\label{code-dcss-weak-ugly-help-read-a1}
   if $addr_1 = \bot$ then return //\label{code-dcss-weak-ugly-help-checkreturn-v}
   $addr_2 := \func{ReadField}(des, \vaddr_2)$
   if $addr_2 = \bot$ then return //\label{code-dcss-weak-ugly-help-checkreturn-addr2}
   $exp_1 := \func{ReadField}(des, \vexp_1)$ //\label{code-dcss-weak-ugly-help-read-exp1}
   if $exp_1 = \bot$ then return //\label{code-dcss-weak-ugly-help-checkreturn-exp1}
   if $*addr_1 = exp_1$ then //\label{code-dcss-weak-ugly-help-compare-a1}
     $new_2 := \func{ReadField}(des, \vnew_2)$ //\label{code-dcss-weak-ugly-help-read-new2}
     if $new_2 = \bot$ then return //\label{code-dcss-weak-ugly-help-checkreturn-new2}
     $\func{CAS}(addr_2, fdes, new_2)$ //\label{code-dcss-weak-ugly-help-cas-new}
   else
     $exp_2 := \func{ReadField}(des, \vexp_2)$ //\label{code-dcss-weak-ugly-help-read-exp2}
     if $exp_2 = \bot$ then return //\label{code-dcss-weak-ugly-help-checkreturn-exp2}
     $\func{CAS}(addr_2, fdes, exp_2)$ //\label{code-dcss-weak-ugly-help-cas-old}
\end{lstlisting}
\vspace{-4.5mm}
\caption{Applying the transformation to \func{DCSS}.} %Code for the \func{DCSS} algorithm of Harris et~al.~\cite{Harris:2002} that has been \textit{transformed} to use the \textit{weak descriptor} ADT.
%We include only procedures that differ from Figure~\ref{code-throw-DCSS}.}
\label{code-weak-ugly-DCSS}
\end{figure}

Figure~\ref{code-weak-ugly-DCSS} shows code for the \func{DCSS} algorithm in Figure~\ref{code-throw-DCSS} %of Harris et~al.~\cite{Harris:2002} 
that has been \textit{transformed} to use the weak descriptor ADT. %\textit{weak descriptor} ADT.
There, we include only the \func{DCSSHelp} procedure, since it is the only one that differs from Figure~\ref{code-throw-DCSS}.
The transformation adds lines~\ref{code-dcss-weak-ugly-help-checkreturn-v}, \ref{code-dcss-weak-ugly-help-checkreturn-addr2}, \ref{code-dcss-weak-ugly-help-checkreturn-exp1}, \ref{code-dcss-weak-ugly-help-checkreturn-new2} and \ref{code-dcss-weak-ugly-help-checkreturn-exp2} to \textit{check} whether the preceding invocations of \func{ReadField} are invalid.

\paragraph{Correctness}

%\trevor{note: if i remove this proof in favour of just keeping the later one, i must edit the chapter / section description earlier that says this proof exists.}

We argue that our transformation takes a linearizable algorithm $\mathcal{A} \in$ WCA that uses mutable descriptors and produces a linearizable algorithm $\mathcal{A}'$ that uses weak descriptors.
Consider any execution $e'$ of the transformed algorithm $\mathcal{A'}$.
We prove there exists an execution $e$ of the original algorithm $\mathcal{A}$ that performs the \textit{same} high-level operations, in the same order, and with the same responses, as in $e'$.
We explain how this helps.
Since $e$ is a correct execution of the original algorithm $\mathcal{A}$, the high-level operations performed in $e$ must respect the sequential specification(s) of the object(s) implemented in $\mathcal{A}$.
Furthermore, since $e'$ performs the same high-level operations, in the same order, and with the same responses, the high-level operations in $e'$ must also respect the sequential specification(s) of the same object(s).
Therefore, the transformed algorithm $\mathcal{A'}$ is correct.

We construct %the execution 
$e$ %of $\mathcal{A}$ 
as follows.
By Property~\ref{def:WCA:prop:desc-ops-only-in-help} of WCA, all \func{ReadField}, \func{WriteField} and \func{CASField} operations occur in \func{Help}.
Whenever a check by a process $p$ follows a \func{ReadField} or \func{CASField} in $e'$ that returns $\bot$ (because the operation attempt $O$ being helped by $p$ has already terminated), we replace that check by a consecutive sequence of steps in which $p$ finishes its invocation of \func{Help}.
All other checks immediately following \func{ReadField} or \func{CASField} are simply removed.

By Property~\ref{def:WCA:prop:not-change-mem} of WCA, none of the steps added to $e$ change the state of any shared object.
So, these steps will not change the behaviour of any other process.
We also argue that none of these steps make any changes to $p$'s private memory that persist after $p$ finishes its invocation of \func{Help}.
(I.e., any changes these steps make to $p$'s private memory are \textit{reverted} by the time $p$ finishes its invocation of \func{Help}, so $p$'s private memory is the same just after the invocation of \func{Help} as it was just before the invocation of \func{Help}.)
So, these steps will not change the behaviour of $p$ after it finishes its invocation of \func{Help}.
%$p$'s private memory has the same contents after it finishes its invocation of \func{Help} as just before it started performing \func{Help}.
Observe that, whenever a process performs a \func{ReadField} or \func{CASField} operation on a descriptor that it created, this operation will return a value different from $\bot$.
This is due to Property~\ref{def:WCA:prop:alg-steps} of WCA, and the definition of the weak descriptor ADT, which states that $d$ becomes invalid only after $O$ has terminated.
Since $p$'s invocation of \func{ReadField} or \func{CASField} returns $\bot$, $p$ must therefore be performing \func{Help}$(d)$ where $d$ was created by a \textit{different} process.
Thus, Property~\ref{def:WCA:prop:return-help} of WCA implies that, after $p$ performs the sequence of steps to finish its invocation of \func{Help}$(d)$, its private memory has the same state as it did just before it invoked \func{Help}. %, except that its program counter points to the step immediately after \func{Help}.

%Note that the same high-level operations are performed in $e$ and $e'$, in the same order, and they return the same responses in both executions.

\paragraph{Reading immutable fields efficiently}
\begin{shortver}
If an invocation of \func{Help}$(des)$ accesses many immutable fields of a descriptor, then we can optimize it by replacing many \func{ReadField} operations with something more efficient.
This optimization is described in Appendix~\ref{appendix-desc-weak-readimmutables}.
%, we add another operation, \func{ReadImmutables}, which atomically returns \textit{all} of a descriptor's immutable fields.
%There, we show how \func{ReadImmutables} can be used in the transformed DCSS code in Figure~\ref{code-weak-ugly-DCSS} (yielding much simpler code, as a side benefit).
\end{shortver}
\begin{fullver}
If an invocation of \func{Help}$(des)$ accesses many immutable fields of a descriptor, then we can optimize it by replacing many \func{ReadField} operations with something more efficient.
To this end, we can add a new operation, \func{ReadImmutables}.
%At a high level, 
This operation reads and returns \textit{all} of a descriptor's immutable fields, unless the descriptor is invalid, in which case it returns $\bot$.
%Its semantics follow.
%This operation could easily be implemented by performing a sequence of \func{ReadField} operations (on immutable fields), but it can be implemented more efficiently as a single operation.
%
%or \func{ReadImmutables}$(T, des)$ 

%\begin{description}
%    \item[\func{ReadImmutables}] takes, as its argument, a descriptor pointer $des$.
%    If a \func{ReadImmutables} operation $O$ is invalid, then it returns $\bot$.
%    Otherwise, $O$ returns the value associated with each field $f \in T$ in $des$ (just before $O$).
%\end{description}

To use \func{ReadImmutables} in \func{Help}$(des)$, one can simply perform, at the beginning of \func{Help}, a \func{ReadImmutables} operation, followed by an \textit{if}-statement that checks whether it the operation invalid, and, if so, returns immediately.
Then, in the body of \func{Help}$(des)$, each invocation of \func{ReadField}$(des, f)$, where $f$ is immutable, is replaced with a direct read from the set of values returned by \func{ReadImmutables}.
\begin{figure}[tb]
\begin{lstlisting}[name=dcssreadimmutables,frame=single]
 $\func{DCSSHelp}(fdes):$
   $des := \func{Unflag}(fdes)$
   $values := \func{ReadImmutables}(des)$ //\label{code-weak-nice-DCSS-readimmutables}
   if $values = \bot$ then return
   $\langle addr_1, exp_1, addr_2, exp_2, new_2 \rangle := values$

   if $*addr_1 = exp_1$ then
     $\func{CAS}(addr_2, fdes, new_2)$
   else
     $\func{CAS}(addr_2, fdes, exp_2)$
\end{lstlisting}
%\begin{lstlisting}[frame=single,mathescape=true]
% $\func{DCSSHelp}(fdes):$
%   $des := \func{Unflag}(fdes)$
%   $values := \func{ReadImmutables}(\func{DCSSdes}, des)$ //\label{code-weak-nice-DCSS-readimmutables}
%   if $values = \bot$ then return
%   $addr_1 := values.\vaddr_1$
%   $addr_2 := values.\vaddr_2$
%   $exp_1 := values.\vexp_1$
%   if $*addr_1 = exp_1$ then
%     $new_2 := values.\vnew_2$
%     $\func{CAS}(addr_2, fdes, new_2)$
%   else
%     $exp_2 := values.\vexp_2$
%     $\func{CAS}(addr_2, fdes, exp_2)$
%\end{lstlisting}
\vspace{-4.5mm}
\caption{Using \func{ReadImmutables} to optimize and streamline the transformed DCSS algorithm.}
\label{code-weak-nice-DCSS}
\end{figure}
We demonstrate this approach on the transformed pseudocode for DCSS in Figure~\ref{code-weak-ugly-DCSS}.
Figure~\ref{code-weak-nice-DCSS} shows the result. %how \func{ReadImmutables} is used with our approach.
Since all fields of a DCSS descriptor are immutable, \textit{every} invocation of \func{ReadField} can be replaced with a direct read from the result of the \func{ReadImmutables} operation performed at line~\ref{code-weak-nice-DCSS-readimmutables}.
(This will not be the case in an algorithm where the \func{Help} procedure reads mutable fields.)
%When a \func{Help} procedure performs several \func{ReadField} operations, 
Since \func{ReadImmutables} replaces several invocations of \func{ReadField}, it has the added benefit of making code simpler and shorter.
\end{fullver}

\section{Extended Weak Descriptors} \label{sec-adt-extended}

In this section, we describe an extended version of the weak descriptor ADT, and an extended version of the transformation in Section~\ref{sec-weak-transformation}.
This extended transformation weakens property~\ref{def:WCA:prop:desc-ops-only-in-help} of WCA so that \func{ReadField} operations on a descriptor $d$ can also be performed \textit{outside} of \func{Help}$(d)$.
%Thus, the extended transformation can be applied to a larger class of algorithms.

%\trevor{give example of \llt\ and \sct\ early on in this section, to motivate this?}

\paragraph{Extended weak descriptor ADT}
%We first describe an \textit{extended weak descriptor} ADT.
This ADT is the same as the weak descriptor ADT, except that \func{ReadField} is extended to take, as an additional argument, a default value $dv$ that is returned instead of $\bot$ when the operation is invalid.
Observe that the weak descriptor ADT is a special case of the extended weak descriptor ADT where each argument $dv$ to an invocation of \func{ReadField} is $\bot$.

%\trevor{what about extending \func{ReadImmutables}?}

\paragraph{The extended transformation}
We now describe how the \textit{weak transformation} in Section~\ref{sec-weak-transformation} is extended.
%(We defer a definition of the class of algorithms that can be transformed until this description.)
%As in the weak transformation, after each \func{ReadField} or \func{CASField} operation on a descriptor $d$ by a process $p$ in an invocation of \func{Help}$(d)$, where $d$ was created by another process, a check is added to determine whether the operation was invalid, in which case $p$ returns from \func{Help} immediately.
%Additionally, 
The transformation is similar to the one in Section~\ref{sec-weak-transformation}.
For \func{CASField} and \func{WriteField} operations, the transformation is the same.
However, an invocation of \func{ReadField}$(des, f)$ is handled differently depending on whether it occurs inside an invocation of \func{Help}$(des)$.
If it is, it is replaced with an invocation of \func{ReadField}$(des, f, \bot)$ followed by the check, as in the WCA transformation.
If not, it is replaced with an invocation of \func{ReadField}$(des, f, dv)$, where the choice of $dv$ is specific to the algorithm being transformed.

Let $\mathcal{A}$ be any algorithm that uses mutable descriptors, and satisfies properties \ref{def:WCA:prop:alg-steps}-\ref{def:WCA:prop:return-help} of WCA algorithms (see Definition~\ref{def:WCA}), as well as a weaker version of property \ref{def:WCA:prop:desc-ops-only-in-help} which states: every write or CAS to a field of a descriptor $d$ must occur in an invocation of \func{Help}$(d)$.
Consider an extended transformation of $\mathcal{A}$.
Let $e$ be an execution of $\mathcal{A}$ and let $e'$ be an execution that is the same as $e$, except that one (arbitrary) descriptor $d$ becomes invalid at some point $t$ after the high-level operation attempt $O$ that created $d$ terminates.
(When we say that $d$ becomes invalid at time $t$, we mean that after $t$, each invocation of \func{ReadField}$(d, f, dv)$ that is performed outside of \func{Help}$(d)$ returns its default value $dv$.)

Let $O'$ be any high-level operation attempt in $e'$ which, after $t$, performs \func{ReadField} on $d$ outside of \func{Help}$(d)$.
We say that the extended transformation is \textit{correct for} $\mathcal{A}$ if, for all choices of $e$, $e'$, $d$, $t$, and $O'$, the exact same changes are performed by $O'$ in $e$ and $e'$ to any variables that are still defined once $O'$ terminates (i.e.,~variables that are local to the process performing $O'$, but are not local to $O'$, and variables in shared memory), and $O'$ returns the same response in both executions.
An algorithm $\mathcal{A}$ is an \textit{extended weak-compatible algorithm} (and is in the class \textit{EWCA}) if there is an extended transformation that is correct for $\mathcal{A}$.

\begin{shortver}
\fakeparagraph{Correctness}
Due to a lack of space, the proof %for the extended transformation
appears in Appendix~\ref{appendix-desc-extended-weak-proof}.
\end{shortver}

\begin{fullver}
\paragraph{Correctness}
%
%!TEX root = paper.tex

Consider any extended transformation which is correct for a linearizable algorithm $\mathcal{A}$ that uses mutable descriptors.
We prove the result of applying this transformation to $\mathcal{A}$ is a linearizable algorithm $\mathcal{A}'$ that uses extended weak descriptors.
Specifically, let $e'$ be any execution of $\mathcal{A}'$.
We prove there is an execution $e$ of $\mathcal{A}$ that performs the same high-level operations, in the same order, with the same responses, as in $e'$.

%Consider a high-level operation attempt $O$ that performs a \func{ReadField} or \func{CASField} operation on the descriptor $d$ created by $O$.
%This operation will be valid, because, by definition of the extended weak descriptor ADT, $d$ becomes invalid only after $O$ has terminated.
%
%By Property~\ref{def:WCA:prop:desc-ops-only-in-help} of WCA, all \func{WriteField} and \func{CASField} operations on a descriptor $d$ occur in \func{Help}$(d)$.
%However, unlike in Section~\ref{sec-weak-transformation}, \func{ReadField} operations can appear elsewhere.

First, we define an execution $e_0$.
Whenever a check in $e'$ by a process $p$ in \func{Help}$(d)$ determines that the preceding \func{ReadField} or \func{CASField} on a descriptor $d$ is \textit{invalid} (which means that the operation attempt being helped by $p$ has already terminated), we replace that check by a consecutive sequence of steps in which $p$ finishes its invocation of \func{Help}$(d)$.
By Property~\ref{def:WCA:prop:not-change-mem} of WCA, none of these added steps change the state of any shared variable.
Moreover, by Property~\ref{def:WCA:prop:return-help} of WCA, $p$ does not change any variable that is still defined after its invocation of \func{Help}, so $p$ has the same local state after \func{Help} in $e_0$ and $e'$.
Whenever such a check determines that the preceding \func{ReadField} or \func{CASField} is \textit{valid}, we simply remove this check.
Observe that each invalid operation in $e_0$ is an invalid \func{ReadField} operation on some descriptor $d$ performed outside of \func{Help}$(d)$.

Let $d_1, d_2, ...$ be the sequence of descriptors created in $e_0$.
We inductively construct a sequence $e_1, e_2, ...$ of executions such that $e_i$ differs from $e_{i-1}$ only in that descriptor $d_i$ never becomes invalid in $e_i$.
Specifically, for each high-level operation attempt $O'$ that performs an invalid \func{ReadField} operation on descriptor $d_i$ outside of \func{Help}$(d_i)$, consider the first such \func{ReadField} operation $R$.
All of the steps of $O'$ prior to $R$ are the same in $e_i$ as in $e_{i-1}$.
After $R$, $O'$ continues to take steps in $e_i$, but each \func{ReadField} operation that $O'$ performs on a field $f$ of $d_i$ returns the contents of $f$ (instead of a default value).
This may result in $O'$ executing completely different code paths in $e_{i-1}$ and $e_i$.
However, by the definition of an extended transformation that is correct for $\mathcal{A}$, $O'$ returns the same response in $e_i$ and $e_{i-1}$ and performs the \textit{exact same changes} to any variables that are still defined once $O'$ terminates.
Thus, for each variable $v$ that is still defined once $O'$ terminates, we can schedule the sequence of changes to $v$ in the exact same way in $e_i$ and $e_{i-1}$ (which implies that any reads in $e_{i-1}$ which see these changes can be scheduled appropriately in $e_i$). %, and $O'$ will return the same response in both executions.

Since the claim holds for all $i$, there is an execution $e$ in which no descriptor becomes invalid (so $e$ is an execution of $\mathcal{A}$), and the same high-level operation attempts are performed, in the same order, and with the same responses.
\end{fullver}

\paragraph{Multiple descriptors per operation attempt}
%\trevor{edit for extended adt}
\begin{shortver}
\vspace{-3.6mm}
In some lock-free algorithms, an operation can create several different descriptors, and potentially invoke a different \func{Help} procedure for each descriptor.
In Appendix~\ref{appendix-desc-weak-multiple}, we describe how to adjust the definitions and proof above to support these kinds of algorithms.
\end{shortver}
\begin{fullver}
In some lock-free algorithms, a high-level operation attempt can create several different descriptors, and potentially invoke a different \func{Help} procedure for each descriptor.
We describe how to adjust the definitions above to support these kinds of algorithms.
For simplicity, we think of there being a single \func{Help} procedure that checks the type of the descriptor passed to it, and behaves differently for different types.

In order to allow a high-level operation attempt to create multiple descriptors without simply invalidating the ones it previously created, we update the definition of valid and invalid descriptors.
Let $des$ be a pointer to a descriptor $d$ of type $T$ returned by a \func{CreateNew} operation $C$ performed by process $p$.
Initially, $d$ is valid.
If $p$ performs another \func{CreateNew} operation $C'$ with the \textit{same descriptor type} $T$ after $C$, then $d$ becomes invalid immediately after $C'$ (and will never be valid again).

With this definition of valid and invalid descriptors, it might initially seem like an operation cannot create multiple descriptors of the same type $T$.
However, this turns out not to be a problem.
If an operation should create multiple descriptors of type $T$, we can simply imagine creating multiple \textit{clone} types $T_1, T_2, ...$ that have the exact same fields as $T$.
To create $k$ descriptors of type $T$, one would then create $k$ clone types, and have an operation invoke \func{CreateNew} once for each clone type.
(However, we are unaware of any algorithms in which a high-level operation attempt creates multiple descriptors of the same type.)

We also slightly modify Property~\ref{def:WCA:prop:alg-steps} of (extended) weak-compatible algorithms, as follows, to accommodate the use of multiple descriptors.
Each high-level operation attempt $O$ by a process $p$ may create (and initialize) a sequence $D$ of descriptors, each with a \textbf{unique type}.
Inside $O$, $p$ may perform at most one invocation of a function \func{Help}$(d)$ for each $d \in D$ (and $p$ may not invoke \func{Help}$(d)$ outside of $O$).
Note that the proof for the extended weak transformation goes through unchanged.

\end{fullver}

\subsection{Example Algorithm: k-CAS}
%\paragraph{Example Algorithm - k-CAS}

In this section, we explain how the extended transformation is applied to the $k$-CAS algorithm presented in Section~\ref{sec-adt-mutable}.
There is only one place in the algorithm where an invocation $I$ of \func{ReadField} on a $k$-CAS descriptor $des$ is performed \textit{outside} of \func{Help}$(des)$ (the \func{Help} procedure for $k$-CAS).
(Note that no invocations of \func{ReadField} on a DCSS descriptor $des'$ are performed outside of \func{HelpDCSS}$(des')$.)
Specifically, $I$ reads the \textit{state} field of a $k$-CAS descriptor inside the modified version of 
\func{HelpDCSS}.
\begin{fullver}
Recall that the $k$-CAS algorithm passes a $k$-CAS descriptor pointer and the name of the \textit{state} field as the first argument to DCSS at line~\ref{line-kcasthrow-kcashelp-phase1-dcss} of Figure~\ref{code-throw-kCAS}, and the DCSS algorithm is modified to use \func{ReadField} at line~\ref{code-dcss-throw-help-compare-a1} of Figure~\ref{code-throw-DCSS} to read this \textit{state} field.
\end{fullver}
\begin{shortver}
Recall that the $k$-CAS algorithm passes a $k$-CAS descriptor pointer and the name of the \textit{state} field as the first argument to DCSS, and the DCSS algorithm is modified to use \func{ReadField} (at line~\ref{code-dcss-throw-help-compare-a1} of Figure~\ref{code-throw-DCSS}) to read this \textit{state} field.
\end{shortver}
We choose the default value $dv = \textit{Succeeded}$ for this invocation of \func{ReadField}.
%The default value $dv$ for $I$ is \textit{Succeeded}.
We explain why this extended transformation of the $k$-CAS algorithm is correct.

When $I$ is performed at line~\ref{code-dcss-throw-help-compare-a1} of Figure~\ref{code-throw-DCSS}, its response is compared with $exp_1$, which contains \textit{Undecided}.
If $I$ returns \textit{Undecided}, then the CAS at line~\ref{code-dcss-throw-help-cas-new} is performed, and the process $p$ performing $I$ returns from \func{HelpDCSS}.
Otherwise, the CAS at line~\ref{code-dcss-throw-help-cas-old} is performed, and $p$ returns from \func{HelpDCSS}.

Suppose $I$ is invalid.
Then, we know the $k$-CAS operation attempt that created $des$ has been completed.
We use the following algorithm specific knowledge.
After a $k$-CAS operation attempt has completed, its $k$-CAS descriptor has \textit{state} \textit{Succeeded} or \textit{Failed} (and is never changed back to \textit{Undecided}).
(This can be determined by inspection of the code.)
Thus, if $I$ were valid, its response would \textit{not} be \textit{Undecided}, and $p$ would perform the CAS at line~\ref{code-dcss-throw-help-cas-old} and return from \func{HelpDCSS}.
Since $dv = Succeeded$, $p$ does exactly the same thing when $I$ is invalid.
%the descriptor $des$ still existed, then its \textit{state} would not be \textit{Undecided}, and the branch would \textit{not} be taken.
(Note that the exact value of \textit{state} is unimportant.
It is only important that it is not \textit{Undecided}.)

\subsection{Example Algorithm: LLX and SCX}
%\paragraph{Example Algorithm - LLX and SCX}

\begin{shortver}
We also applied our transformation to the lock-free wasteful implementation of the LLX and SCX primitives presented by Brown~et~al.~\cite{Brown:2013}.
%The arguments for its correctness are similar to those for the $k$-CAS algorithm.
%Details will appear in the full version of this paper.
%However, we present experimental results for this algorithm, below.
Full details appear in Appendix~\ref{appendix-extended-example-bst}.
\end{shortver}
\begin{fullver}
\afterpage{\clearpage}
\begin{figure}[p]
\begin{framed}
\vspace{-2mm}
\def\pwidth{4cm}
\def\namewidth{30mm}
\begin{lstlisting}[name=llxscxdata,style=nonumbers]
 type $\func{\sct des}$
   //\com Immutable descriptor fields
   //\wcnarrow{$\vnfreeze$}{number of \rec s to be frozen}
   //\wcnarrow{$\vnfinalize$}{number of \rec s to be finalized}
   //\wcnarrow{$\vV_1, \vV_2, ...$}{\rec s to be frozen}
   //\wcnarrow{$\vR_1, \vR_2, ...$}{\rec s to be finalized (must be a subsequence of $\langle \vV_1, \vV_2, ... \rangle$)}
   //\wcnarrow{$\vdes_1, \vdes_2, ...$}{descriptor pointers read from $\vV_1.\info, \vV_2.\info, ...$}
   //\wcnarrow{$\vfld$}{pointer to a field of some $\vV_i$}
   //\wcnarrow{$\vnew$}{value to be written into the field $\vfld$}
   //\wcnarrow{$\vold$}{value previously read from the field $\vfld$} 
   //\com Mutable descriptor fields
   //\wcnarrow{$\vstate$}{one of \{\freezing, \done, \retry\}}
   //\wcnarrow{$\vallfrozen$}{Boolean} \vspace{1mm} \hrule \vspace{-3mm}
\end{lstlisting}
\begin{thesisnot}
\begin{lstlisting}[name=llxscxdata,style=nonumbers]
 type $\rec$
   //\com User-defined fields (e.g., for a node in a tree)
   //\wcnarrow{$m_1, \ldots, m_y$}{mutable fields}
   //\wcnarrow{$i_1, \ldots, i_z$}{immutable fields}
   //\com Fields used by \llt /\sct\ algorithm
   //\wcnarrow{$\info$}{descriptor pointer}
   //\wcnarrow{$marked$}{Boolean} \vspace{1mm} \hrule \vspace{-3mm}
\end{lstlisting}
\end{thesisnot}
\begin{lstlisting}[name=llxscx]
 //\llt$(r)$ by process $p$
   $marked_1 := r.marked$ // \label{desc-ll-read-marked1}
   $r\info := r.\info$ // \label{desc-ll-read} 
   $state := \func{ReadField}(r\info,\vstate)$ // \label{desc-ll-read-state}
   $marked_2 := r.marked$ // \label{desc-ll-read-marked2}
   if $state = \retry$ or $(state = \done$ and not $marked_2)$ then //  \label{desc-ll-check-frozen} \sidecom{if $r$ was not frozen at line~\ref{desc-ll-read-state}}
     read $r.m_1,...,r.m_y$ //and record the values in local variables $m_1,...,m_y$%
        \label{desc-ll-collect}
     if $r.\info = r\info$ then//\label{desc-ll-reread}\sidecom{if $r.\info$ contains the same} 
       //store $\langle r, r\info, \langle m_1, ..., m_y \rangle \rangle$ in $p$'s local table \sidecom{descriptor as on line~\ref{desc-ll-read}}\label{desc-ll-store}
       return $\langle m_1, ..., m_y \rangle$ //\label{desc-ll-return} \vspace{2mm}%

   if $state = \freezing$ then $\func{Help}(r\info)$ //\label{desc-ll-help}
   if $marked_1$ then return $\finalized$ // \label{desc-ll-return-finalized}
   else return $\fail$ // \label{desc-ll-return-fail} \vspace{1mm} \hrule %
\vspace{1mm}%

 //\sct$(V = \langle V_1, V_2, ..., V_k \rangle, R = \langle R_1, R_2, ..., R_l \rangle, fld, new)$ by process $p$
 //\tline{\com Preconditions: (\presctlinked) for each $r$ in $V$, $p$ has performed an invocation $I_r$ of \llt$(r)$ linked to this \sct}%
         {\hspace{19.5mm}(\presctabainit) $new$ is not the initial value of $fld$}%
         {\hspace{19.5mm}(\presctaba) for each $r$ in $V$, no $\sct(V', R', fld, new)$ was linearized before $I_r$ was linearized}
   //Let $des_1, des_2, ..., des_k$ be the descriptor pointers for $V_1, V_2, ..., V_k$ in $p$'s local table of \llt\ results \label{desc-sct-create-llresults}
   //Let $old$ be the value for $fld$ stored in $p$'s local table of \llt\ results\label{desc-sct-create-old}
   $des := \func{CreateNew}(\func{\func{\sct des}}, \{(\vnfreeze, k), (\vnfinalize, l), (\vV_1,V_1), (\vV_2,V_2), ..., (\vV_k,V_k),$ $(\vR_1,R_1), (\vR_2,R_2), ..., (\vR_l,R_l), (\vdes_1,des_1), (\vdes_2,des_2), ..., (\vdes_k,des_k),$ $(\vfld,fld), (\vnew,new), (\vold,old), (\vstate,\freezing), (\vallfrozen,\false)\})$//\label{desc-sct-create-op}
   return $\help(des)$ // \label{desc-sct-call-help} \vspace{1mm} \hrule %
\vspace{1mm}%

 //\help$(des)$ 
   //\com \mbox{Freeze all \rec s in $des.\vV$ to protect their mutable fields from being changed by other \sct s}
   $\langle \textit{nfreeze}, \textit{nfinal}, V_1, V_2, ..., V_{\textit{nfreeze}}, R_1, R_2, ..., R_{\textit{nfinal}}, des_1, des_2, ..., des_{\textit{nfreeze}}, fld, new, old \rangle := \func{ReadImmutables}(des)$
   for $i = 1 .. \textit{nfreeze}$ do//\label{desc-help-fcas-loop-begin}
     if not $\cas(V_i.\info,des_i,des)$ then //\sidecom{\textbf{\fcas}}\label{desc-help-fcas}
       if $V_i.\info \neq des$ then //\label{desc-help-check-frozen}
         //\com \mbox{Could not freeze $V_i$ because it is frozen for another \sct}
         if $\func{ReadField}(des, \vallfrozen) = \true$ then//\sidecom{\textbf{\fcstep}}\label{desc-help-fcstep}
           //\com the \sct\ has already completed successfully 
           return $\true$ // \label{desc-help-return-true-loop} 
         else
           //\com Atomically unfreeze all \rec s frozen for this \sct 
           $\func{WriteField}(des, \vstate, \retry)$ //\sidecom{\textbf{\astep}}\label{desc-help-astep}
           return $\false$ // \label{desc-help-return-false} \vspace{2mm}
   //\com Finished freezing \rec s (Assert: $state \in \{\freezing, \done\}$) 
   $\func{WriteField}(des, \vallfrozen, \true)$//\sidecom{\textbf{\fstep}}\label{desc-help-fstep}
   for $i = 1 .. \textit{nfinal}$ do
     $R_i.marked := \true$ //\sidecom{\textbf{\markstep}}\label{desc-help-markstep}
   //$\cas(fld, old, new)$ \sidecom{\textbf{\upcas}}\label{desc-help-upcas}
   //\com Finalize all $R_i$ in $R$, and unfreeze all $V_i$ in $V$ that are not in $R$
   $\func{WriteField}(des, \vstate, \done)$//\sidecom{\textbf{\cstep}}\label{desc-help-cstep}
   return $\true$ // \label{desc-help-return-true}
\end{lstlisting}
\vspace{-2mm}
\end{framed}
    \vspace{-6mm}
    \caption{Code for the \llt\ and \sct\ %algorithm of Brown et~al.~\cite{Brown:2013} 
        using the \textit{mutable descriptor} ADT.}
    \label{code-desc-scx}
\end{figure}

\begin{thesisnot}
In this section, we explain how the extended transformation is applied to the multiword synchronization primitives load-linked-extended (\llt) and store-conditional-extended (\sct) of Brown et~al.~\cite{Brown:2013}.
Note that Brown et~al.~\cite{Brown:2014} also used these primitives to design a tree update template that can be followed to produce a fast lock-free implementation of any data structure based on a down-tree (a directed acyclic graph where each node has indegree one).
Thus, by optimizing \llt\ and \sct, we also optimize the tree update template, and all of the data structures that have been implemented with it.
\end{thesisnot}
\begin{thesisonly}
In this section, we explain how the extended transformation is applied to the \llt\ and \sct\ implementation in Chapter~\ref{chap-scx}.
\end{thesisonly}
Pseudocode for \llt\ and \sct\ using mutable descriptors is presented in Figure~\ref{code-desc-scx}.
\begin{thesisonly}
Recall that each \sct\ creates a new descriptor called an \op, which has two mutable fields: a 2-bit \textit{state} field and an \textit{allFrozen} bit.
The \textit{state} field contains one of three values: \textit{InProgress}, \textit{Committed} and \textit{Aborted}.
\end{thesisonly}

\begin{thesisnot}
\llt\ and \sct\ operate on multi-field \textit{data records}, which can be used to represent, e.g., nodes in a tree, or records in a table.
Like descriptors, \textit{data records} contain mutable and immutable fields.
However, whereas descriptors are used only to facilitate helping, and are not part of a sequential data structure, data records are.

\llt$(r)$ attempts to take a snapshot of the mutable fields of a \rec\ $r$.
If it is concurrent with an \sct\ involving~$r$, it may return \fail, instead.
Individual fields of a \rec\ can also be read directly.
An \sct$(V,R,fld,new)$ takes as its arguments a sequence $V$ of \rec s, a subsequence $R$ of $V$, a pointer $fld$ to a mutable field of one \rec\ in~$V$, and a new value $new$ for that field.
The \sct\ tries to atomically: store the value $new$ in the field that $fld$ points to and {\it finalize} each \rec\ in $R$.
Once a \rec\ is finalized, its mutable fields cannot be changed by any subsequent \sct, and any \llt\ of the \rec\ will return \finalized\ instead of a snapshot.

Before a process invokes \sct, it must perform an \llt$(r)$ on each \rec\ $r$ in $V$.
The last such \llt\ by the process is said to be {\it linked} to the \sct, and the linked \llt\ must return a snapshot of $r$ (not \fail\ or \finalized).
An \sct($V, R, fld, new$) by a process modifies the data structure and returns \true\ only if no \rec\ $r$ in $V$ has changed since its linked \llt($r$); otherwise the \sct\ fails and returns \false.
Although \llt\ and \sct\ can fail, their failures are limited in such a way that they can be used to build data structures with lock-free progress.
See \cite{Brown:2013} for a more formal specification of these primitives.

Each \sct\ operation creates a new descriptor called an \op.
\llt\ and \sct\ requires each \rec\ $r$ to have a dedicated field $r.\info$ that stores a pointer to an \op, and this field is only ever accessed by \llt\ and \sct\ operations.
Each \rec\ also has a \textit{marked} bit which is accessed only by \llt\ and \sct.
This field is used by \sct\ to finalize \rec s.
We say that a \rec\ is \textit{marked} if its \textit{marked} bit is set.
\op s have two mutable fields: a 2-bit \textit{state} field and an \textit{allFrozen} bit.
The \textit{state} field contains one of three values: \textit{InProgress}, \textit{Committed} and \textit{Aborted}.
%These fields are embedded in the same word as the sequence number in SCX records.
%The original algorithm does not steal any bits from pointers, since there is a dedicated field in each node for descriptor pointers.
\end{thesisnot}

%We use the following algorithm specific knowledge.
The following properties of the \llt\ and \sct\ algorithm are relevant for our purposes.
\begin{compactenum}[P1.]
	\item Before the first invocation of \func{Help}$(des)$ for an \sct\ $O$ (performed by $O$ or a helper) has been completed, the \op\ $des$ created by $O$ has its \textit{state} field set to \textit{Committed} or \textit{Aborted}, and, after this, the \textit{state} field of $des$ is never changed again.
    \item A marked \rec\ remains marked forever.
    \item A marked \rec\ cannot point to an \op\ with $\textit{state} = \textit{Aborted}$.
    \item Each time the $\info$ field of a \rec\ changes, it changes to a new value that has never previously been stored there (to avoid the ABA problem).
\end{compactenum}

There is only one place in the code where an invocation $I$ of $\func{ReadField}(\op, d,$ $f, dv)$ can occur outside of \func{Help}$(des)$: at line~\ref{desc-ll-read-state} of \llt\ in Figure~\ref{code-desc-scx}.
$I$ reads the \textit{state} field of $d$.
We choose the default value $dv = \textit{Committed}$ for $I$.
We give a rigorous, but straightforward, proof that this extended transformation of \llt\ and \sct\ is correct.
%(We stress that this is a straightforward proof.)
%(We stress that this is a straightforward proof, and the following is somewhat pedantic. We give this level of detail simply to demonstrate what a through, rigorous proof would look like, and to show that there is nothing difficult about these arguments.)
%\trevor{stress this is an easy and pedantic proof. we just include it to give an example with full details, and to show that there is nothing difficult about it.}

Let $e$ be an execution of the original \llt\ and \sct\ algorithm $\mathcal{A}$, and let $e'$ be an execution that is the same as $e$, except that one arbitrary \op\ $d$ becomes invalid at some point $t$ after the \sct\ operation attempt $O$ that created $d$ terminates.
Let $O'$ be any \llt\ in $e$ which, after $t$, performs an invocation $I$ of \func{ReadField} on $d$ outside of \func{Help}$(d)$.
We must prove that $O'$ performs the exact same changes in $e$ and $e'$ to any variables that are still defined after $O'$ terminates, and returns the same response in both executions.

%$I$ is valid in $e$ and invalid in $e'$.
Since $I$ is invalid in $e'$, by definition, the \sct\ $O$ that created $d$ must have terminated before $I$.
Thus, by P1, $I$ must return \textit{Committed} or \textit{Aborted} in $e$.
If $I$ returns \textit{Committed} in $e$, then $I$ returns the same response in $e$ and $e'$, so $O'$ is exactly the same in both executions.
Now, suppose $I$ returns \textit{Aborted} in $e$.
We consider three cases, depending on where $O'$ returns in $e$.

\textit{Case 1:} $O'$ returns at line~\ref{desc-ll-return} in $e$.
If $marked_2 = \false$, then $O'$ behaves exactly the same way in $e$ and $e'$.
So, suppose $marked_2 = \true$.
Then, $O'$ will enter the if-statement at line~\ref{desc-ll-check-frozen} in $e$, but not in $e'$.
In this case, $O'$ saw that the \rec\ $r$ pointed to an \op\ with $\textit{state} = \textit{Aborted}$ when it performed line~\ref{desc-ll-read}, and that $r$ was marked when it performed line~\ref{desc-ll-read-marked2}.
By P3, $r$ cannot simultaneously be marked and point to an \op\ with $\textit{state} = \textit{Aborted}$, so $r.\info$ must change between these two lines.
By P4, it must change to a value different from $r\info$, so the if-block at line~\ref{desc-ll-reread} will not be executed in $e$.
However, this contradicts our assumption that $O'$ returns at line~\ref{desc-ll-return}.

\textit{Case 2:} $O'$ returns \finalized\ at line~\ref{desc-ll-return-finalized} in $e$.
Observe that $O'$ does not execute line~\ref{desc-ll-store} in $e$ (since it would then return at the following line). %, so $O'$ does not change any variable that is still defined after it terminates.
We first prove that $O'$ does not execute line~\ref{desc-ll-store} in $e'$.
Since $O'$ sees $marked_1 = \true$ just before returning at line~\ref{desc-ll-return-finalized} in $e$, P2 implies that $marked_2 = \true$ (in both $e$ and $e'$).
Since $I$ returns \textit{Committed} in $e'$, $O'$ will not enter the if-block at line~\ref{desc-ll-check-frozen} in $e'$.
Thus, $O'$ reaches line~\ref{desc-ll-help} in both $e$ and $e'$.

Since $I$ returns \textit{Committed} in $e'$, and we have assumed $I$ returns \textit{Aborted} in $e$, $O'$ will not invoke \func{Help} at line~\ref{desc-ll-help} in $e$ or $e'$.
Therefore, $O'$ does not change any variable that is still defined after it terminates.
So, it suffices to prove that $O'$ returns \finalized\ (at line~\ref{desc-ll-return-finalized}) in $e'$.
%Since $O'$ returns at line~\ref{desc-ll-return-finalized} in $e$ after seeing $marked_1 = \true$, we know $marked_1 = \true$ in $e'$.
However, this is immediate from the fact that $marked_1 = \true$ in $O'$ in $e$ (and, hence, in $e'$).

\textit{Case 3:} $O'$ returns \fail\ at line~\ref{desc-ll-return-fail} in $e$.
The proof is similar to the previous case, except $marked_1 = \false$ in $O'$ in $e$, so when $O'$ reaches line~\ref{desc-ll-return-finalized}, it will enter the \textit{else}-block and return \fail\ in both $e$ and $e'$.

\end{fullver}

%note to self: we COULD simplify llx to the following, and doing so would be more in line with the C++ code. but, we should still be able to argue correctness without this change.
%
%    bool marked1 = node->marked
%    SCXRecord<K,V> *scx1 = node->scxRecord
%    int state = scx1->state
%    bool marked2 = node->marked
%    
%    if state == ABORTED or (state == COMMITTED and not marked2) then
%        read mutables
%        SCXRecord<K,V> *scx2 = node->scxRecord
%        if scx1 == scx2 then
%            return <scx1, mutables>    // success
%
%    help(scx1)
%    if marked1 then
%        return finalized
%    else
%        return fail

\begin{thesisnot}
\section{Implementing the extended weak descriptor ADT} \label{sec-extended-impl}
\end{thesisnot}
\begin{thesisonly}
\section{Implementing the extended weak descriptor ADT} \label{sec-extended-impl}
\end{thesisonly}

%\begin{figure}[t]
%	%\vspace{-3mm}
%	\includegraphics[width=\linewidth]{chap-desc/figures/seqptr.pdf}
%	%\vspace{-7mm}
%	\caption{Bit-representation of a \textit{descriptor pointer} for the extended weak descriptor ADT implementation.} % to a reusable descriptor.}
%	\label{fig-seqptr}
%	%\vspace{-4mm}
%\end{figure}

In this section, we give an efficient implementation of the extended weak descriptor ADT. %(with support for operation attempts that create multiple descriptors).
Note, however, that this implementation uses largely known techniques (similar to~\cite{Marathe:2008}), and is not the main contribution of this work.
\begin{shortver}
	Due to lack of space, we defer the details of the implementation and the proof %of correctness and progress 
    to Appendix~\ref{appendix-detailed-implementation}.
Here, we give only a high-level overview.
\end{shortver}

\begin{fullver}
At a high level, each
\end{fullver}
\begin{shortver}
Each
\end{shortver}
process $p$ uses a \textit{single} descriptor object $D_{T,p}$ in shared memory to represent \textit{all} descriptors of type $T$ that it ever creates.
The descriptor object $D_{T,p}$ conceptually represents $p$'s \textit{current} descriptor of type $T$.
At different times in an execution, $D_{T,p}$ represents different \textit{abstract descriptors} created by $p$.
We store a sequence number in $D_{T,p}$ that is incremented every time $p$ performs \func{CreateNew}$(T, -)$. % to create a new descriptor of type $T$.
%\trevor{more carefully explain descriptor pointers, here.}
Instead of using traditional descriptor pointers, we represent each descriptor pointer as a pair of bit fields stored in a single word.
These bit fields contain the name of the process who owns the descriptor, and a sequence number that indicates which invocation of \func{CreateNew} conceptually created this descriptor.
%$des$ returned from an invocation of \func{CreateNew}$(T, -)$ by $p$ as a pair of bit fields
%Each descriptor pointer $des$ returned from an invocation of \func{CreateNew}$(T, -)$ by $p$ encodes %the descriptor type $T$, 
%the process name $p$ and the sequence number that represents the newly created abstract descriptor.
When a descriptor pointer is passed to an operation $O$ on the abstract descriptor, $O$ compares the sequence number in $des$ with the current sequence number in $D_{T,p}$ to determine whether the operation is valid or invalid.
Thus, incrementing the sequence number in $D_{T,p}$ effectively makes all abstract descriptors of type $T$ that were previously created by $p$ \textit{invalid}.

%Conceptually, each process has only a \textit{single} descriptor of each type, but there are different \textit{versions} of each descriptor.
%Whenever a process $p$ performs \func{CreateNew}$(T, pairs)$, it conceptually \textit{creates a new version} of its descriptor $D_{T,p}$ and \textit{destroys all previous versions}.
%The idea of destroying previous versions of descriptors is analogous to the idea of \textit{invalidating} descriptors in Section~\ref{sec-weak-transformation} (where an invocation of \func{CreateNew}$(T, pairs)$ by $p$ causes all descriptors of type $T$ that were returned by $p$'s previous invocations of \func{CreateNew}$(T, -)$ to become \textit{invalid}).
%
%Conceptually, the high level idea is to have each process use only a \textit{single} descriptor of each type (used by the application) for each process, but to .

\begin{fullver}
%!TEX root = paper.tex

\begin{figure}[ph!]
\begin{lstlisting}[name=extendedweak,frame=single]
 //\com \textbf{Data types}
 //Descriptor of type $T:$
   $mutables = \langle seq, mut_1, mut_2, ... \rangle$ //\hfill\com Mutable fields
   $imm_1, imm_2, ... $ //\hfill\com Immutable fields

 //\com \textbf{Shared variables}
   $D_{T,p}$ //for each descriptor type $T$ and process $p$

 //\com \textbf{ADT operations}
 //$\func{CreateNew}(T, v_1, v_2, ...)$ by process $p:$
   $oldseq$ := $D_{T,p}.mutables.seq$
   $D_{T,p}.mutables.seq$ := $oldseq+1$ //\label{code-impl-create-new-lin}
   for each field $f$ in $D_{T,p}$
     let $value$ //be the corresponding value in $\{v_1, v_2, ...\}$
     if $f \mbox{ is immutable}$ then
       $D_{T,p}.f := value$
     else 
       $D_{T,p}.mutables.f := value$
   $D_{T,p}.mutables.seq$ := $oldseq+2$ //\label{code-extended-weak-before-fence}
   return $\langle p, oldseq+2 \rangle$ //\label{code-extended-weak-after-fence}\com Descriptor identifier
   
 //$\func{ReadField}(des, f, dv):$
   $\langle q, seq \rangle := des$
   if $f \mbox{ is immutable}$ then
     $result := D_{T,q}.f$ //\label{code-impl-read-field-immutable-lin}
   else
     $result := D_{T,q}.mutables.f$ //\label{code-impl-read-field-mutable-lin}
   if $seq \neq D_{T,q}.mutables.seq$ then return $dv$ //\label{code-impl-read-field-invalid-lin}
   return $result$ //\label{code-impl-read-field-valid-return}

 //$\func{ReadImmutables}(des):$
   $\langle q, seq \rangle := des$
   for each $f$ in $des$
     if $f \mbox{ is immutable}$ then //add $D_{T,q}.f$ to $result$
   if $seq \neq D_{T,q}.mutables.seq$ then return $\bot$ //\label{code-impl-read-immutables-lin}
   return $result$//\label{code-impl-read-immutables-valid-return}

 //$\func{WriteField}(des, f, value):$
   $\langle q, seq \rangle := des$
   loop
     $exp := D_{T,q}.mutables$ //\label{code-impl-write-field-read}
     if $exp.seq \neq seq$ then return //\label{code-impl-write-field-invalid-lin}
     $new := exp$
     $new.f := value$
     if $\func{CAS}(\&D_{T,q}.mutables, exp, new)$ then return//\label{code-impl-write-field-success-lin}

 //$\func{CASField}(des, f, \textit{fexp}, \textit{fnew}):$
   $\langle q, seq \rangle := des$
   loop
     $exp := D_{T,q}.mutables$ //\label{code-impl-cas-field-read}
     if $exp.seq \neq seq$ then return $\bot$ //\label{code-impl-cas-field-invalid-lin}
     if $exp.f \neq \textit{fexp}$ then return $exp.f$ //\label{code-impl-cas-field-failed-lin}
     $new := exp$
     $new.f := \textit{fnew}$
     if $\func{CAS}(\&D_{T,q}.mutables, exp, new)$ then //\label{code-impl-cas-field-success-lin}
       return $fnew$ //\label{code-impl-cas-field-valid-return}
\end{lstlisting}
\vspace{-3mm}
\caption{Pseudocode for the \textit{extended weak descriptor} ADT implementation.}
\label{code-reuse-impl}
\end{figure}

%We now give a more detailed description.
\begin{fullver}
\paragraph{Detailed description}
Complete
\end{fullver}
\begin{shortver}
Complete
\end{shortver}
pseudocode %for the implementation 
appears in Figure~\ref{code-reuse-impl}.
We start by describing the data types and shared variables.
%Multiple types of descriptors can be defined. % by the application that uses this implementation.
%The number of fields in each descriptor type, and the names of the fields, are determined by the application.
%For each descriptor type $T$ and each process $p$, there is a descriptor $D_{T,p}$ of type $T$ in shared memory.
Each descriptor contains zero or more immutable fields, and zero or more mutable fields (which are determined by the \textit{descriptor type}), as well as a sequence number field \textit{seq}.
Recall that $D_{T,p}$ represents different abstract descriptors at different times. 
Note that the immutable fields of $D_{T,p}$ are only immutable for as long as $D_{T,p}$ represents the same abstract descriptor.
When $D_{T,p}$ is reused, so that it represents a different abstract descriptor, its immutable fields can be reinitialized.
Usually very few bits are required for the mutable fields, since they exist solely to capture the state of an ongoing operation (and it is inefficient to frequently change the state of a descriptor).
(Every lock-free algorithm we are aware of uses at most a small constant number of bits for its mutable fields.)
%Thus, it is realistic to expect that a sequence number and any mutable fields can be packed into a single word.
%In the following, we assume that $mutables$ can fit in a single word.
Consequently, we think of the sequence field and the mutable fields of a descriptor $d$ as being packed together in a single word $mutables$ of $d$ (with subfields for the sequence field and each mutable field).
(Note that, if more space is needed for mutable fields in some future algorithm, we can eliminate this assumption about the size of $mutables$, as we explain below.)
We use $d.f$ to denote an immutable field $f$ of $d$, $d.mutables$ to denote the field $mutables$ of $d$, and $d.mutables.f$ to denote a mutable field $f$ of $d$.

%We assume that the field $mutables$ can be modified atomically using CAS.
Since $mutables$ fits in a single word, it can be modified atomically using CAS.
By having CAS atomically operate on a mutable field and the sequence number, %on a field of a descriptor includes the sequence number in its \textit{expected value}, so 
we can ensure that a descriptor changes only if its sequence number has not changed.

We now describe the operations.
An invocation of \func{CreateNew}$(T, ...)$ by process $p$ first increments the sequence number of $D_{T,p}$, then initializes all of its fields, then increments the sequence number again and returns a new descriptor pointer (with the up-to-date sequence number).
Observe that the descriptor pointers returned by \func{CreateNew} always have even sequence numbers, and the sequence number of a descriptor is odd while it is being initialized by \func{CreateNew}.
Consequently, while a descriptor is being initialized, its sequence number does not match any descriptor pointer in the system, so no process can read or modify the descriptor's fields.

Note that this approach of incrementing a sequence number twice has been used in different contexts such as in transactional memory, where the least significant bit represents whether the sequence number is locked or unlocked.
Here, the idea is slightly different, since the least significant bit represents whether the descriptor is currently being reused and initialized, or is safe to access.
(Nevertheless, in some sense, one can think of the bit indicating whether the descriptor is currently being initialized as a sort of lock.
It does not prevent other processes from making progress (since operations on the descriptor will terminate, but will simply be invalid), but it does prevent them from accessing fields of the descriptor as they are being changed.)

An invocation of \func{ReadField}$(des, f, \textit{default})$ by $p$ %extracts the sequence number $seq$ from $des$, 
reads the value $v$ of the mutable or immutable field $f$ from $D_{T,p}$ followed by its sequence number $s$.
If $s$ matches the sequence number in the descriptor pointer $des$, then $v$ is returned.
Otherwise, \textit{default} is returned.

\func{ReadImmutables} is similar to \func{ReadField}, except it reads all immutable fields, instead of a single field, and it returns $\bot$ instead of \textit{default}.

An invocation $I$ of \func{WriteField}$(des, f, value)$ by $p$ performs a sequence of one or more \textit{attempts}.
In each attempt, it reads the contents $old$ of $mutables$, including the sequence number $s$, from $D_{T,p}$, then checks whether $s$ matches the sequence number in the descriptor pointer $des$.
If the sequence numbers do not match, then the abstract descriptor represented by $des$ is invalid, so $I$ returns without changing $f$.
Otherwise, $I$ uses CAS to try to change $D_{T,p}.mutables$ from $old$ to $new$, which is a copy of $old$ in which the contents of field $f$ have been changed (locally) to contain $value$.
Observe that this CAS will succeed only if the sequence number in $D_{T,p}.mutables$ matches the sequence number in $des$.
If the CAS succeeds, then $I$ returns.
Otherwise, $I$ performs another attempt. %(since it cannot tell whether the CAS failed because another process changed the sequence number or another mutable field).

Note that \func{WriteField} is less efficient than performing a direct write to memory.
However, since mutable fields are used merely to encode the status of an ongoing operation, there are usually very few changes to a descriptor.
%(Additionally, our experiments suggest that one can obtain significant performance improvements over traditional wasteful algorithms with this implementation.)

%\trevor{perhaps mention this is less efficient than writes, but seems from our experiments to be fast, still. [maybe mention that there is a possibility for the thread that creates a descriptor to use real writes instead of cas]}

%\trevor{possibly combine this with writefield, since the only difference is the return values and the check of whether old.f is equal to fexp.}
\func{CASField} is quite similar to \func{WriteField}. %, except that (1) it also takes an expected value \textit{fexp} for $f$ as an argument and it does not change $f$ unless $f$ contains \textit{fexp}, and (2) it returns a value.
%We describe the differences.
The only differences are (1) \func{CASField} has different return values and, (2) in each attempt, it performs an additional check to determine whether $old.f$ is equal to $\textit{fexp}$, and, if not, returns $old.f$.
%An invocation $I$ of \func{CASField}$(T, des, f, \textit{fexp}, \textit{fnew})$ by $p$ performs a sequence of one or more \textit{attempts}.
%In each attempt, it reads the contents $old$ of $mutables$, including the sequence number $s$, from $D_{T,p}$, then checks whether $s$ matches the sequence number in the descriptor pointer $des$.
%If the sequence numbers do not match, then the abstract descriptor represented by $des$ is invalid, so $I$ returns $\bot$ without changing $f$.
%So, suppose the sequence numbers match.
%Then, $I$ checks whether $old.f$ contains \textit{fexp}.
%If not, $I$ returns $old.f$. %then $f$ does not contain \textit{fexp}, so $I$ returns false.
%Otherwise, $I$ uses CAS to try to change $D_{T,p}.mutables$ from $old$ to a copy of $old$ in which $f$ contains $value$.
%%Observe that this CAS will succeed only if the sequence number in $D_{T,p}.mutables$ matches the sequence number in $des$.
%If the CAS succeeds, then $I$ returns $old.f$.
%Otherwise, it performs another attempt.

\paragraph{Practical considerations}

One might wonder, in an algorithm with multiple types of descriptors, why the type of a descriptor is not also encoded in descriptor pointers.
%\trevor{perhaps mention somewhere that algorithms that use multiple descriptors already know what kind of descriptor they are accessing?}
In algorithms that use multiple descriptor types, any time the original algorithm accesses a field of a descriptor, it typically must know what kind of descriptor it is accessing (if, for no other reason, to compute the address of the desired field within the descriptor).
In such algorithms, it would not be necessary for descriptor pointers to carry this extra information.
For algorithms that access descriptors without knowing their exact types, one can include the descriptor type in descriptor pointers.
%Thus, the algorithm already knows which global array it should use to locate a descriptor, and it is not necessary for our transformation to explicitly encode the descriptor type in its pointers.

%\begin{fullver}
%	Some non-blocking applications ``steal'' the two (on a 32-bit system) or three (on a 64-bit system) lowest-order bits from pointers to encode additional information.~
%\end{fullver}
%\begin{shortver}
	Some lock-free algorithms ``steal'' up to three bits from pointers to encode additional information, typically to distinguish between application values and (potentially, various types of) descriptors.~
%\end{shortver}
To accommodate such algorithms, one can slightly shrink the sequence number in our descriptor pointers, and reserve the three lowest-order bits for use by other algorithms. %our transformation reserves the three lowest-order bits of our pointers as \textit{user-defined} bits to be used by the underlying algorithm. % concatenate them (as lower order bits) to our pointer values.

One obvious way to store the descriptors for each thread is to create an array for each descriptor type, with a slot containing a descriptor for each process.
In this kind of implementation, it is extremely important to pad each slot to avoid false sharing~\cite{scott1993false}.
We suggest allocating at least two cache lines for each descriptor (128 bytes on modern Intel and AMD machines).

%Researchers in concurrent algorithms typically write pseudocode assuming a sequentially consistent memory model.
To improve efficiency, modern Intel and AMD processors implement a relaxed memory model called total store order (TSO) that allows certain steps in a program to be executed out of order.
Specifically, a read that occurs after a write in a program can actually be executed \textit{before} the write, as long as the read and write are not accessing the same address.
This can render a concurrent algorithm incorrect if it requires a write by a process $p$ to be visible to other processes \textit{before} $p$ performs a subsequent read.
One can prevent this reordering by placing a memory fence (or barrier) between the write and read.
CAS instructions also act as memory fences.
Our implementation does not require any memory fences (beyond those implied by CAS instructions).
This is an attractive property, since memory fences incur high overhead.

%\paragraph{ABA problems}

Our implementation uses unbounded sequence numbers.
However, in practice, sequence numbers are bounded, and they may wrap around.
If wraparound occurs, then two invocations of \func{CreateNew} might return the same descriptor pointer.
This can cause an \textit{ABA problem} if the high-level algorithm that uses descriptors relies on the uniqueness of descriptor pointers returned by \func{CreateNew}.
%Suppose a process $p$ reads an address $x$ and sees $A$, then performs a CAS that changes $x$ from $A$ to $C$, and interprets the success of this CAS to mean that $x$ contained $A$ at all times between the read and CAS.
%If other processes change $x$ from $A$ to $B$ and from $B$ back to $A$ between $p$'s read and CAS, then $p$'s interpretation is invalid, and we say an ABA problem has occurred.

We argue that the sequence number can be made sufficiently large on modern systems for this to be a non-issue.
Consider a system with a 64-bit word size.
Recall that a sequence number appears both in each descriptor pointer, and also in the $mutables$ field of each descriptor.
A descriptor pointer contains only a process name and a sequence number, so if $n$ bits are reserved for the process name, then $64-n$ bits remain for the sequence number.
The $mutables$ field contains the descriptor's mutable fields and a sequence number, so if $m$ bits are reserved for mutable fields, then $64-m$ bits remain for the sequence number.
Thus, if we use 14-bit process names (as the Linux kernel does), and the mutable fields of each descriptor fit in at most 14 bits, then 50 bits remain for the sequence number.
We are unaware of any algorithm that requires more than three bits for mutable fields in its descriptors, so this is realistic.
In this case, a single process must perform $2^{50}$ operations to trigger even a single wraparound.
If we assume that a single process can perform one million operations per second, this will take 35 years of continuous execution.
If this is still a concern, then one can use double-wide CAS (DWCAS), which is implemented on modern Intel and AMD systems, instead of CAS, to atomically operate on two adjacent words (containing a much larger sequence number).

Although we are unaware of any current lock-free algorithms that use more than three bits for mutable fields in descriptors, some future algorithm may use more.
If the mutable fields of a descriptor cannot fit in the same word as a sequence number, then our approach must be modified.
If the mutable fields and a sequence number can fit in two adjacent words, then one can simply use DWCAS instead of CAS.
Otherwise, one can store mutable fields in their own separate words, and \textit{replicate} the sequence number, storing a copy in the word adjacent to each mutable field.
To change a mutable field, one would then perform DWCAS on the word containing the mutable field, and its adjacent sequence number.
When the descriptor is reused, instead of incrementing a single sequence number, one would increment all sequence numbers.

%\trevor{write about replicating the sequence number for large mutable fields. the write-up from the ppopp rebuttal is commented out here.}
%%One would modify a mutable field by performing double-wide CAS (available on modern Intel/AMD processors) on the field and its adjacent sequence number. When the descriptor is reused, instead of incrementing a single sequence number, one would increment all sequence numbers. However, we would stress that it is a reasonable assumption for all mutable fields to fit in one word with the sequence number. Having studied dozens of lock-free algorithms, we are not aware of a single one that violates this assumption. Recall that mutable fields are used only to record the status of an ongoing operation. Hence, they rarely exceed one or two bits.
%%
%%
%%% OLD DESCRIPTION:
%%If the mutable fields cannot be co-located with the sequence number, then a double-wide CAS can be used.
%%If a double-wide CAS is still not sufficient, then the sequence number can be replicated and attached to each mutable field (details omitted due to lack of space).

In order to choose how many bits should be devoted to the process name in descriptor pointers, one must know an upper bound on the number of processes.
We stress that this is not an onerous constraint, because the upper bound does not need to be tight.
%For instance, one can devote 16 bits to the process name, and support up to 65,536 concurrent processes, which is sufficient for the foreseeable future.
Note that one need not initially allocate descriptors for all processes that \textit{could} be running in the system.
It is straightforward to allocate a descriptor for a process the first time it invokes \func{CreateNew} (potentially even in batches, to amortize the cost and improve control over memory layout).

\medskip

\fakeparagraph{Correctness}
We now prove that our implementation %of extended weak descriptors 
is linearizable. %provides linearizable executions that follow the semantics of the extended weak descriptor ADT presented in Section~\ref{sec-adt-extended}. 
We first give the linearization points for all operations. % are as follows.
\begin{itemize}
\item Each invocation of \func{CreateNew} is linearized at the increment of the sequence number at line~\ref{code-impl-create-new-lin}.
\item If an invocation $I$ of \func{ReadField}$(des, f, dv)$ returns at line~\ref{code-impl-read-field-invalid-lin}, then it is linearized at the read of the sequence number at the same line.
If $I$ returns at line~\ref{code-impl-read-field-valid-return}, then it is linearized at the preceding read of the field $f$: for immutable fields this is line~\ref{code-impl-read-field-immutable-lin}, and for mutable fields this is line~\ref{code-impl-read-field-mutable-lin}.
\item Each invocation of \func{ReadImmutables} is linearized at the read of the sequence number at line~\ref{code-impl-read-immutables-lin}.
\item If an invocation $I$ of \func{WriteField}$(des, f, value)$ returns at line~\ref{code-impl-write-field-invalid-lin}, then it is linearized at the last read of the sequence number at the same line.
If $I$ returns at line~\ref{code-impl-write-field-success-lin}, then it
is linearized at the successful CAS at the same line.
\item If an invocation $I$ of \func{CASField}$(des, f, \textit{fexp}, \textit{fnew})$ returns at line~\ref{code-impl-cas-field-invalid-lin}, then it is linearized at the last read of the sequence number at the same line.
If $I$ returns at line~\ref{code-impl-cas-field-valid-return}, then it is linearized at the successful CAS at the previous line.
If $I$ returns at line~\ref{code-impl-cas-field-failed-lin}, then it
is linearized at the last read at the same line.
\end{itemize}

\begin{Observation}
	%Let $D_{T,p}$ be a descriptor of type $T$ owned by process p. 
	The sequence number of $D_{T,p}$ (also denoted $D_{T,p}.mutables.seq$) is written only by $p$ in invocations of \func{CreateNew}($T,-$).\label{impl-observ-write-to-seq}
\end{Observation}
%\begin{chapscxproof}
%	There are only two other writes to $D_{T,p}.mutables$, one at line~\ref{code-impl-write-field-success-lin} of \func{WriteField}($T,des,f,-$) and the other  line~\ref{code-impl-cas-field-success-lin} of \func{CASField}($T,des,f,-,-$), where $des = \langle p,- \rangle$. Since $f$ cannot represent the sequence field (it is private to the implementation), in both cases $exp.seq$ is equal to $new.seq$, thus, the CAS operation does not change the sequence field. 
%\end{chapscxproof}
%\trevor{previous lemma is trivial; immediate from the code. create an "observation" newtheroem type and make that and the following lemmas both observations.}

\begin{Observation}
	Every descriptor pointer returned by \func{CreateNew} has an even sequence number, and the linearization point of \func{CreateNew} always changes the sequence number of the descriptor to an odd number. \label{impl-observ-even-vs-odd}
\end{Observation}

\begin{Observation}
	The sequence number returned by a \\
	\func{CreateNew}($T,-$) operation by p is $2+v$ where $v$ is the sequence number returned by $p$'s previous \func{CreateNew}($T,-$) operation, or $v = 0$ if $p$ has not performed \func{CreateNew}($T,-$).\label{impl-observ-unique-des}
	%\maya{TODO - fix overflow}
\end{Observation}

We now prove that the above linearization points are correct. 
Let $e$ be an execution of our implementation of extended weak descriptors.
%To prove the correctness of our implementation assume $e'$ includes the low level atomic steps that are executed inside the the extended weak descriptor ADT implementation. 
Let $O_1,O_2\cdots O_k$ be the extended weak descriptor operations executed in $e$ in the order they are linearized. %, and let $\ell_1,\ell_2\cdots \ell_k$ be the respective linearization points of these operations.
%%
%%Let $e_l$ be the serial execution of the same ADT operations according to the linearization order. 
%%It is easy to show that the serial execution $e_l$ follows the semantics presented in section~\ref{sec-adt-extended}.
%%We show that each $O_i$ in $e_l$ has the same return value as $O_i$ in $e$.
%%We show that the return values of $O_1,O_2\cdots O_k$ respect the semantics presented in section~\ref{sec-adt-extended}.
%%
%In the following, we write $s <_e s'$ to mean that step $s$ occurs before step $s'$ in execution $e$. %the happens before relation $<_{e}$ to denote the order between low level steps of an execution $e$.
%% as well as the order of high level operations in a serial execution $e$.   
%%
%%Note: we prove correctness assuming unbounded sequence numbers.
\begin{shortver}
Note that we prove correctness assuming unbounded sequence numbers.
The implications of bounded sequence numbers were considered above. %are considered in Appendix~\ref{appendix-impl-practical}.
\end{shortver}

\begin{theorem}
	The responses of $O_1,O_2\cdots O_k$ respect the semantics of the extended weak descriptor ADT.
\end{theorem}  
  
\begin{chapscxproof}
By strong induction on the sequence of extended weak descriptor operations that terminate in $e$.
Base case: the claim vacuously holds when no operations have returned.

%according to definition~\ref{} \maya{need extended version of definition 1} the first linearization point $s_{1}$ has to be the linearization point of \func{CreateNew} operation by some process $p$.
%Since this is the first operation by the process, any return value is unique and therefore respects the semantics. 

%Since $p$ is the only process that can write to $D_{T,p}.mutables.seq$ (note that both CAS operations at lines~\ref{code-impl-write-field-success-lin} and~\ref{code-impl-cas-field-success-lin} do not change $D_{T,q}.mutables.seq$), $O_1$ reads the initial zero value of $oldseq$ and returns $\langle p, 2 \rangle$ as required. 

Induction step: assume the return values of $O_{1}, O_{2}\cdots O_{i-1}$ follow the semantics of the extended weak descriptor ADT. 
%Consider $O_i$.
Let $p$ be the process that performs $O_i$, and $T$ be the type of descriptor on which $O_i$ is performed.

Suppose $O_i$ is a \func{CreateNew}($T,-$) operation.
By Observation~\ref{impl-observ-unique-des}, %since the sequence number is combined with $p$, the return value has never been returned by a previous invocation of \func{CreateNew}($T,-$).
$O_i$ returns a unique descriptor pointer.

%\textbf{\func{CreateNew}}. By lemma~\ref{impl-proof-write-to-seq} \func{CreateNew} is the only operation that changes the sequence number. Since writes to the sequence number are monotonically increasing, if $p$ read value $oldseq$ then the return value $\langle p, oldseq + 2 \rangle$ is unique 
%\trevor{has never been returned by a previous invocation of createnew by this process; note that this is a place where you use the inductive hypothesis}. 
%\trevor{replace the preceding proof chunk with an observation that says the sequence number returned by a createnew operation by p returns $2+v$ where $v$ is the sequence number returned by $p$'s previous createnew operation (of course, $v = 0$ is $p$ has not performed createnew). then we can use the observation to say that of course createnew returns a unique value, since the sequence number is combined with p.}
%%$p$ reads the same value of $oldseq$ in both $e_l$ and $e$. Thus, $O_i$ return the same value $ \langle p,oldseq+2\rangle$ in both $e_l$ and $e$. 

In each of the following cases, $O_i$ takes a descriptor pointer $des$ as one of its arguments.
Let $q$ and $seq$ be the process name sequence number in $des$, respectively.
%For the remaining operations, assume that $des = \langle q,seq \rangle$ for some process $q$ (if the operation is valid, $q$ might be equal to $p$). 
Let $O_{init}$ be the \func{CreateNew}($T,-$) by $q$ that returned $des$. %, and let $\ell_{init}$ be the linearization point of $O_{init}$.
Since $des$ is returned by $O_{init}$ before it is passed to any operation, $O_{init}$ is linearized before $O_i$. %we have $\ell_{init} <_e \ell_i$.

Suppose $O_i$ is a \func{ReadField} that returns the default value at line~\ref{code-impl-read-field-invalid-lin}, a \func{CASField} that returns $\bot$ at line~\ref{code-impl-cas-field-invalid-lin} or a \func{ReadImmutables} that returns $\bot$ at line~\ref{code-impl-read-immutables-lin}.
We argue that $des$ is invalid when $O_i$ is linearized.
In each case, $O_i$ returns after seeing that the sequence number of $D_{T,q}$ no longer contains $seq$.
Thus, this sequence number must change after $des$ is returned by $O_{init}$, and before $O_i$ is linearized. %after it was last read by $O_i$ at line~\ref{code-impl-read-field-read}, \ref{code-impl-cas-field-read} or \ref{code-impl-read-immutables-read}, respectively, and before $O_i$ is linearized.
By Observation~\ref{impl-observ-write-to-seq}, this change to the sequence number of $D_{T,q}$ must be performed by a \func{CreateNew}$(T, -)$ operation $O_{change}$ by $q$ (which occurs after $O_{init}$, and before $O_i$ is linearized).
$O_{change}$ changes the sequence number twice, and is linearized at the first change.
Thus, $O_{change}$ is linearized after $O_{init}$ and before $O_i$, which means that $des$ is \textit{invalid} when $O_i$ is linearized.
%By inspection of the code, $O_i$ returns the default value if it is a \func{ReadField}, and $\bot$, otherwise.
%If $O_i$ returns at is an invocation of \func{ReadField}, then $O_i$ returns $default$ as required. Otherwise, $O_i$ is an invocation of \func{CASField} or \func{ReadImmutables} and $O_i$ returns $\bot$ as required.
%
%%from $seq$ to some different value $seq'$ since it was last read by $O_i$.
%Let $s_{change}$ be the step that performed this change.
%By Observations~\ref{impl-observ-write-to-seq}-\ref{impl-observ-unique-des}, $s_{change}$ changes the sequence number from $seq$ to $seq+1$ at line~\ref{code-impl-create-new-lin} of \func{CreateNew}($T,-$) operation by $q$ and $\ell_{init} <_e s_{change} <_e \ell_i$.
%Since line~\ref{code-impl-create-new-lin} is the linearization point of \func{CreateNew} and above claims, $O_i$ is an invalid operation. 
%
%If $O_i$ is an invocation of \func{ReadField}, then $O_i$ returns $default$ as required. Otherwise, $O_i$ is an invocation of \func{CASField} or \func{ReadImmutables} and $O_i$ returns $\bot$ as required

Now suppose $O_i$ is a \func{ReadField} that returns at line~\ref{code-impl-read-field-valid-return}, a \func{CASField} that returns at line~\ref{code-impl-cas-field-valid-return}, or a \func{ReadImmutables} that returns at line~\ref{code-impl-read-immutables-valid-return}.
We first argue that $des$ is valid when $O_i$ is linearized.
In each case, $O_i$ sees that the sequence number of $D_{T,q}$ is $seq$ at some time $t$, which is either when $O_i$ is linearized, or is after $O_i$ is linearized.
By Observation~\ref{impl-observ-write-to-seq} and Observation~\ref{impl-observ-unique-des}, whenever the sequence number of $D_{T,q}$ is changed, it is changed to a new value that it never previously contained.
Thus, since the sequence number of $D_{T,q}$ contains $seq$ when $O_{init}$ terminates, and it contains $seq$ at time $t$, it contains $seq$ at all times after $O_{init}$ terminates and before $t$.
Hence, the sequence number of $D_{T,q}$ contains $seq$ when $O_i$ is linearized.
By Observation~\ref{impl-observ-write-to-seq}, $q$ does not perform any \func{CreateNew}$(T, -)$ after $O_{init}$ and before $t$, so $des$ is \textit{valid} when $O_i$ is linearized.

We now argue that the response of $O_i$ is correct if it is a \func{ReadField}$(des, f, dv)$.
The proof is similar when $O_i$ is a \func{CASField} or \func{ReadImmutables}.

%\trevor{edit here}
%
%\trevor{where does observation 2 come into this?}

%Case I: $O_i$ is a \func{ReadField}$(des, f, dv)$.
If $f$ is immutable, then it is changed only by \func{CreateNew}$(T, -)$ operations by $q$.
Since $des$ is valid when $O_i$ is linearized, $O_{init}$ performs the last change to $f$ before $O_i$ is linearized.
Recall that $O_i$ start after $O_{init}$ terminates.
Thus, the write of $f$ in $O_{init}$ happens before the invocation of $O_i$, and $O_i$ will return the value written to $f$ by $O_{init}$. %the value returned by $O_i$ read when $O_i$ is linearized is the value written to $f$ by $O_{init}$ as required. 

If $f$ is mutable, then let $O_{change}$ be the operation that performs the last change to $f$ before $O_i$ is linearized, and $v$ be the value that it stores in $f$.
Observe that $O_i$ returns $v$.
We show that $O_{change}$ is the last operation that changes $f$ and is linearized before $O_i$.
If $O_{change}$ is the same as $O_{init}$, then we are done.
Otherwise, since we have argued that $q$ does not perform any \func{CreateNew}$(T, -)$ after $O_{init}$ and before $t$, $O_{change}$ must be a \func{WriteField} or \func{CASField}.
In each case, $O_{change}$ can change $f$ only once, with a successful CAS (at line~\ref{code-impl-cas-field-success-lin} or line~\ref{code-impl-write-field-success-lin}).
Since $O_{change}$ is linearized at this CAS, it is linearized before $O_i$.
Moreover, since we have assumed that $O_{change}$ is the last operation to change $f$ before $O_i$, no other operation that changes $f$ linearized after $O_{change}$ and before $O_i$.
\end{chapscxproof} 

\paragraph{Progress}

The proof of lock-free progress is straightforward.
Suppose, to obtain a contradiction, that there is an execution in which processes take infinitely many steps, but only finitely many (extended weak descriptor) operations terminate.
Then, after some time $t$, no operation terminates, which means there is at least one operation $O$ in which a process takes infinitely many steps.
By inspection of Figure~\ref{code-reuse-impl}, $O$ must be a \func{WriteField} or \func{CASField} operation.
Suppose $O$ is a \func{WriteField} operation.
Then, each time $O$ executes line~\ref{code-impl-write-field-invalid-lin}, it sees $old.seq = seq$, and each time it executes line~\ref{code-impl-write-field-success-lin}, its CAS fails and returns $old$ without changing $D_{T,q}.mutables$.
Observe that the CAS will fail only if $D_{T,q}.mutables$ changes after it is read at line~\ref{code-impl-write-field-read} and before the CAS at line~\ref{code-impl-write-field-success-lin}.
Thus, $D_{T,q}.mutables$ changes infinitely many times in the execution.
Since $D_{T,q}.mutables$ can be changed only by \func{WriteField} or \func{CASField} operations, and any operation that changes $D_{T,q}.mutables$ immediately terminates, there must be infinitely many operations of \func{WriteField} or \func{CASField} that terminate, which is a contradiction.
The proof is similar when $O$ is a \func{CASField} operation.

\end{fullver}

%!TEX root = paper.tex

%\vspace{-2mm}
\section{Experiments} \label{sec-exp}

%\subsection{System configurations}

Our experiments were run on two large-scale systems.
The first is a 2-socket Intel E7-4830 v3 with 12 cores per socket and 2 hyperthreads (HTs) per core, for a total of 48 threads.
Each core has a private 32KB L1 cache and 256KB L2 cache (which is shared between HTs on a core).
All cores on a socket share a 30MB L3 cache.
The second is a 4-socket AMD Opteron 6380 with 8 cores per socket and 2 HTs per core, for a total of 64 threads.
Each core has a private 16KB L1 data cache and 2MB L2 cache (which is shared between HTs on a core).
All cores on a socket share a 6MB L3 cache.

Since both machines have multiple sockets and a non-uniform memory architecture (NUMA), in all of our experiments, we pinned threads to cores so that the first socket is filled first, then the second socket is filled, and so on.
Furthermore, within each socket, each core has one thread pinned to it before hyperthreading is engaged.
Consequently, our graphs clearly show the effects of hyperthreading and NUMA.

\begin{fullver}
For example, on the Intel machine, from thread counts 1 to 12 all threads are running on a single socket and at most one thread is pinned to each core. \textbf{(socket 1: no HTs; socket 2: empty)}.
From 13 to 24, all threads are running on a single socket and cores either have one or two threads pinned to them \textbf{(socket 1: HTs; socket 2: empty)}.
From 25 to 36, each core on the first socket has two threads pinned to it, and the remaining threads are each pinned to unique cores on the second socket \textbf{(socket 1: HTs; socket 2: no HTs)}.
Finally, from 37 to 48, each core on the first socket has two threads pinned to it, and cores on the second socket have one or two threads pinned to them \textbf{(socket 1: HTs; socket 2: HTs)}.
\end{fullver}

Both machines have 128GB of RAM.
Each runs Ubuntu 14.04 LTS.
All code was compiled with the GNU C++ compiler (G++) 4.8.4 with build target x86\_64-linux-gnu and compilation options \texttt{-std=c++0x -mcx16 -O3}.
Thread support was provided by the POSIX Threads library.
\begin{fullver}
~We used the Performance Application Programming Interface (PAPI) library~\cite{Browne:2000} to collect statistics from hardware performance counters to determine cache miss rates, stall times, instructions retired, and so on.
\end{fullver}
\begin{shortver}
~We used the Performance Application Programming Interface (PAPI) library~\cite{Browne:2000} to collect statistics from hardware counters to determine cache miss rates, stall times, etc.
\end{shortver}
\begin{shortver}
~We used the scalable allocator jemalloc 4.2.1~\cite{Evans:2006}, which greatly improved performance for all algorithms.
\end{shortver}

\begin{fullver}
The system (glibc) allocator was found to have poor scaling and overall performance.
Instead, we used jemalloc 4.2.1, a fast user-space allocator designed to minimize contention and improve scalability~\cite{Evans:2006}.
The library was dynamically linked with \texttt{LD\_PRELOAD}, which is the recommended method.
This allocator was found to yield vastly superior performance for all algorithms, in all benchmarks.
We also tried the tcmalloc allocator from Google's Perftools library, which is another common choice for concurrency-friendly allocation.
However, performance with tcmalloc was substantially worse for all algorithms than with jemalloc.

On the AMD machine, transparent huge-pages were disabled manually in the jemalloc implementation by changing the default allocation chunk size from $2^{21}$ to $2^{19}$ using the environment parameter setting \texttt{MALLOC\_CONF=lg\_chunk:19}.
This maintained or improved the performance for all algorithms in all workloads, and did not change the performance relationship between any pair of algorithms.
The same change did not improve performance on the Intel machine (for any algorithm or workload), so the original chunk size was used.

For read-heavy workloads, it was necessary to force distribution of pages across NUMA nodes to get consistently high performance.
To achieve this, we used \texttt{numactl --interleave=all} for all workloads.
(Doing this did not negatively impact the performance of any workload, but its benefit was less noticeable for write-heavy workloads.)

%As a sidenote, on the AMD machine, it took a significant amount of experimentation with different configurations to achieve stable, reproducible measurements with small variance.
%Machines based on this line of processors are extremely difficult to get trustworthy results from (primarily owing to their high degree of NUMA-ness).
%Thread pinning, scalable allocation and page interleaving across NUMA nodes were all necessary components for obtaining trustworthy results.
%Our experience suggests that results gathered on similar machines in future must minimally address these concerns to have a hope of being valid.
\end{fullver}

\begin{shortver}
\paragraph{$k$-CAS microbenchmark}
In
\end{shortver}
\begin{fullver}
\subsection{$k$-CAS microbenchmark} \label{sec-exp-kcas}
In
\end{fullver}
order to compare our reusable descriptor technique with algorithms that reclaim descriptors, we implemented $k$-CAS with several memory reclamation schemes.
Specifically, we implemented a lock-free memory reclamation scheme that aggressively frees memory called \textit{hazard pointers}~\cite{Michael:2004}, a (blocking) epoch-based reclamation scheme called \textit{DEBRA}~\cite{Brown:2015}, and reclamation using the read-copy-update (RCU) primitives~\cite{Desnoyers:2012} (also blocking).
We use \textit{Reuse} as shorthand for our reusable descriptor based algorithm, and \textit{DEBRA}, \textit{HP} and \textit{RCU} to denote the 
\begin{shortver}
other algorithms.
\end{shortver}
\begin{fullver}
algorithms that use DEBRA, hazard pointers and RCU, respectively.
\end{fullver}

The paper by Harris~et~al. also describes an optimization to reduce the number of DCSS descriptors that are allocated by embedding them in the $k$-CAS descriptor.
We applied this optimization, and found that it did not significantly improve performance.
Furthermore, it complicated reclamation with hazard pointers.
Thus, we did not use this optimization.

\fakeparagraph{Methodology}
We compared our implementations of $k$-CAS using a simple array-based microbenchmark.
For each algorithm $A \in \{$\textit{Reuse}, \textit{DEBRA}, \textit{HP}, \textit{RCU}$\}$, array size $S \in \{2^{14}, 2^{20}, 2^{26}\}$ and $k$-CAS parameter $k \in \{2, 16\}$, we run ten timed \textit{trials} for several thread counts $n$.
In each trial, an array of a fixed size $S$ is allocated and each entry is initialized to zero.
Then, $n$ concurrent threads run for one second, during which each thread repeatedly chooses $k$ uniformly random locations in the array, reads those locations, and then performs a $k$-CAS (using algorithm $A$) to increment each location by one.

As a way of validating correctness in each trial, each thread keeps track of how many successful $k$-CAS operations it performs.
At the end of the trial, the sum of entries in the array must be $k$ times the total number of successful $k$-CAS operations over all threads.

\begin{shortver}
\begin{figure}[t]
    \centering
    \setlength\tabcolsep{0pt}
    \begin{minipage}{0.49\linewidth}
    \begin{tabular}{m{0.05\linewidth}m{0.465\linewidth}m{0.465\linewidth}}
        &
        \multicolumn{2}{c}{
            %\dimexpr \linewidth-2\fboxsep-2\fboxrule
            \fcolorbox{black!80}{black!40}{\parbox{\dimexpr 0.93\linewidth-2\fboxsep-2\fboxrule}{\centering\textbf{2x 24-thread Intel E7-4830 v3}}}
        }
        \\
        &
        \fcolorbox{black!50}{black!20}{\parbox{\dimexpr \linewidth-2\fboxsep-2\fboxrule}{\centering {\footnotesize 2-CAS}}} &
        \fcolorbox{black!50}{black!20}{\parbox{\dimexpr \linewidth-2\fboxsep-2\fboxrule}{\centering {\footnotesize 16-CAS}}}
        \\
        \rotatebox{90}{Array size $2^{26}$} &
        \includegraphics[width=\linewidth]{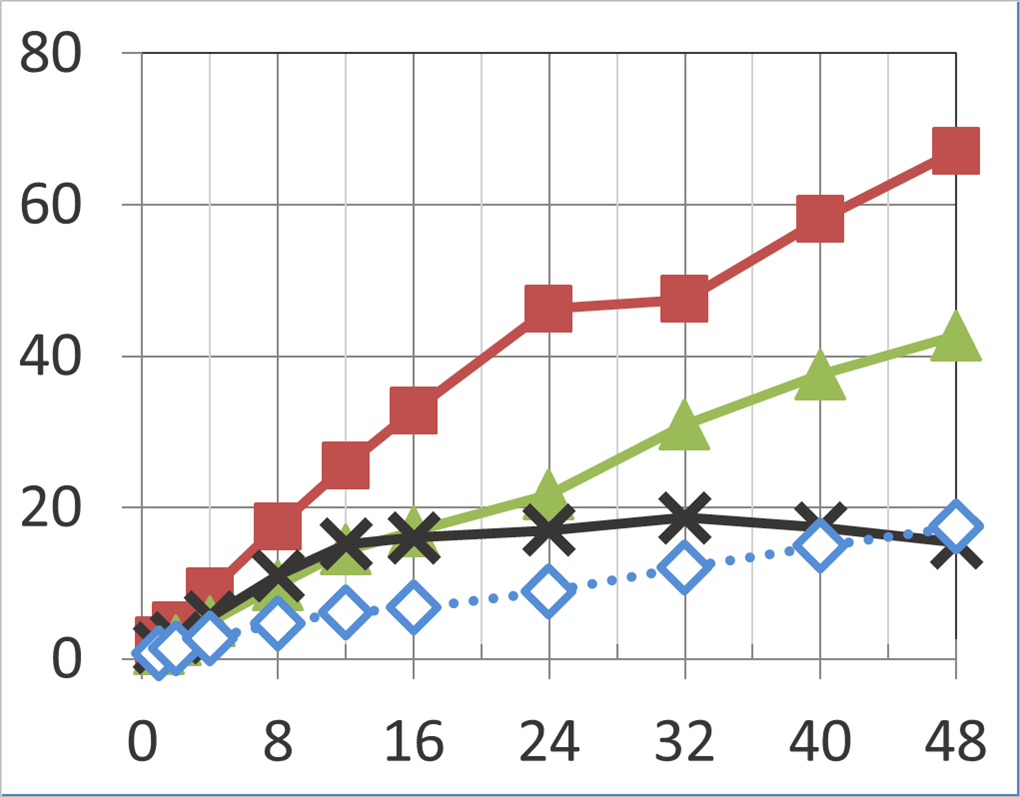} &
        \includegraphics[width=\linewidth]{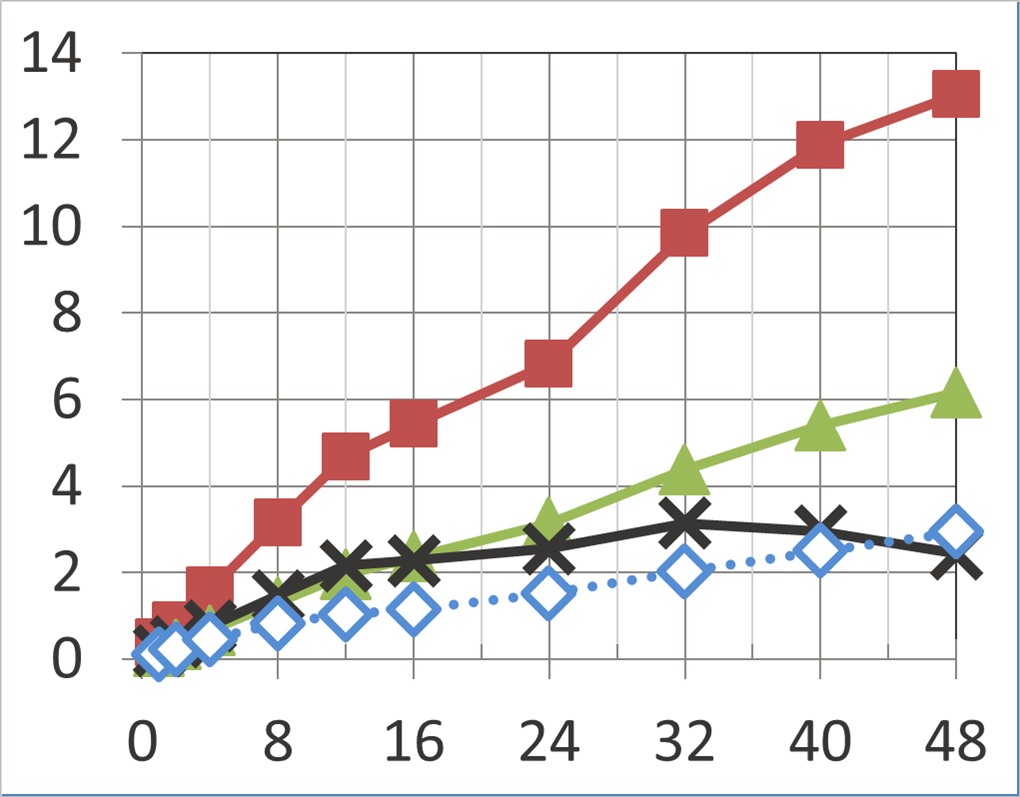}
        \\
        \vspace{-1.5mm}\rotatebox{90}{Array size $2^{20}$} &
        \vspace{-1.5mm}\includegraphics[width=\linewidth]{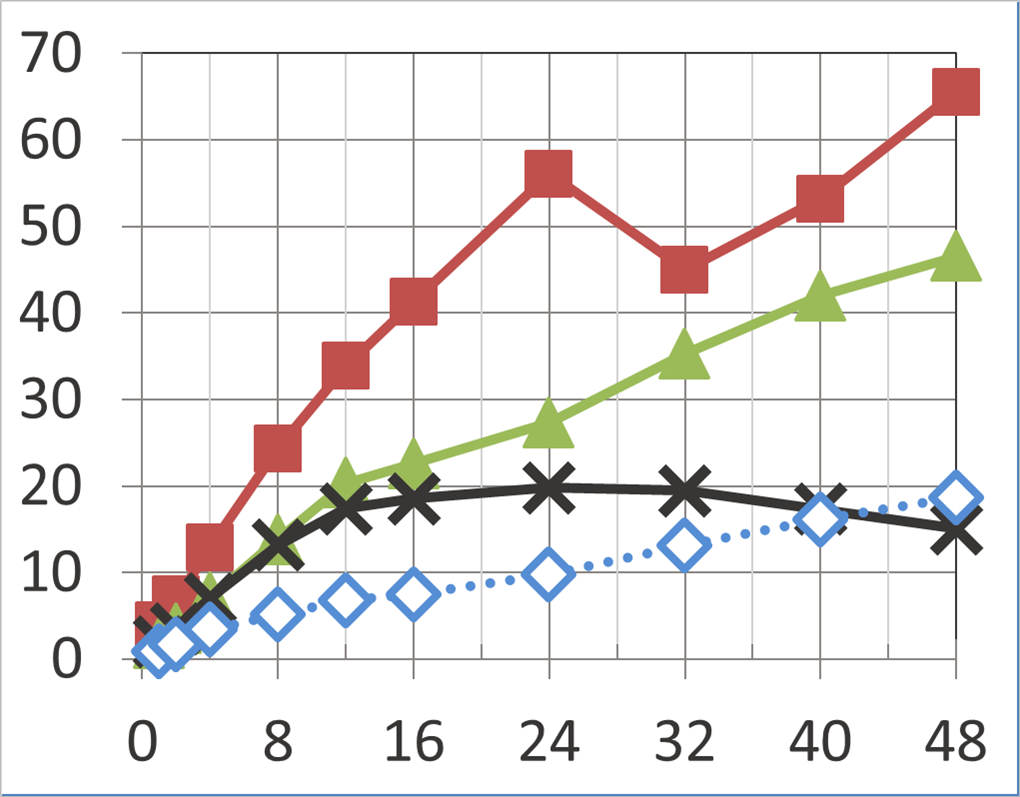} &
        \vspace{-1.5mm}\includegraphics[width=\linewidth]{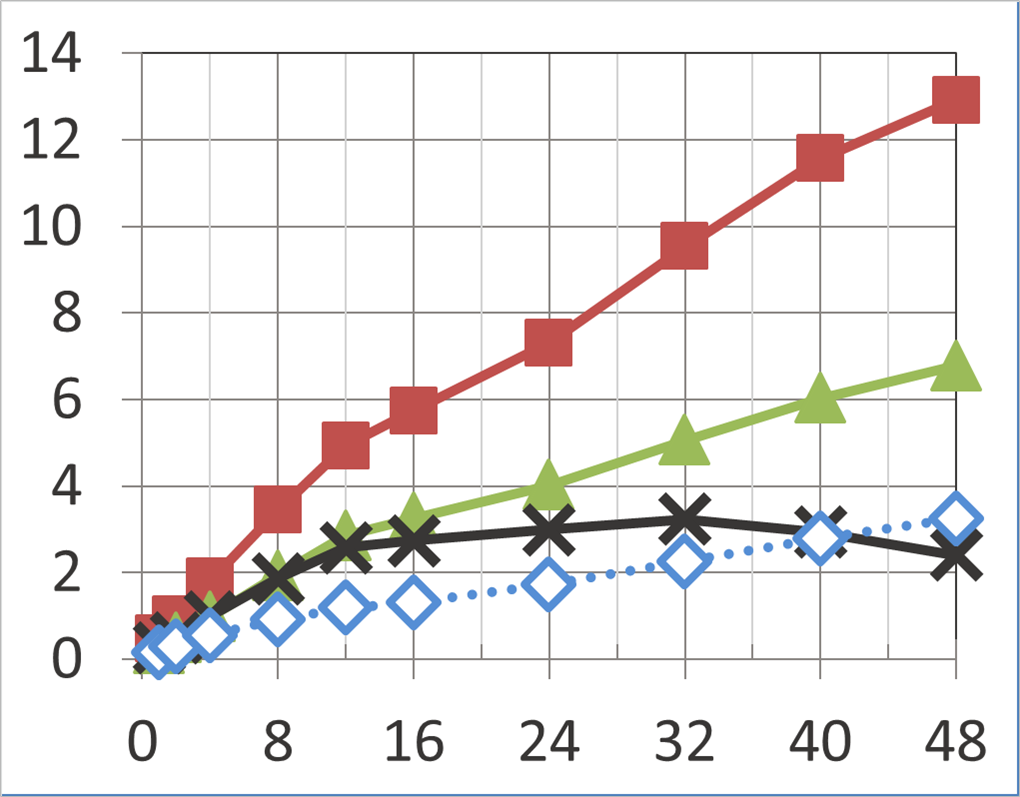}
        \\
        \vspace{-1.5mm}\rotatebox{90}{Array size $2^{14}$} &
        \vspace{-1.5mm}\includegraphics[width=\linewidth]{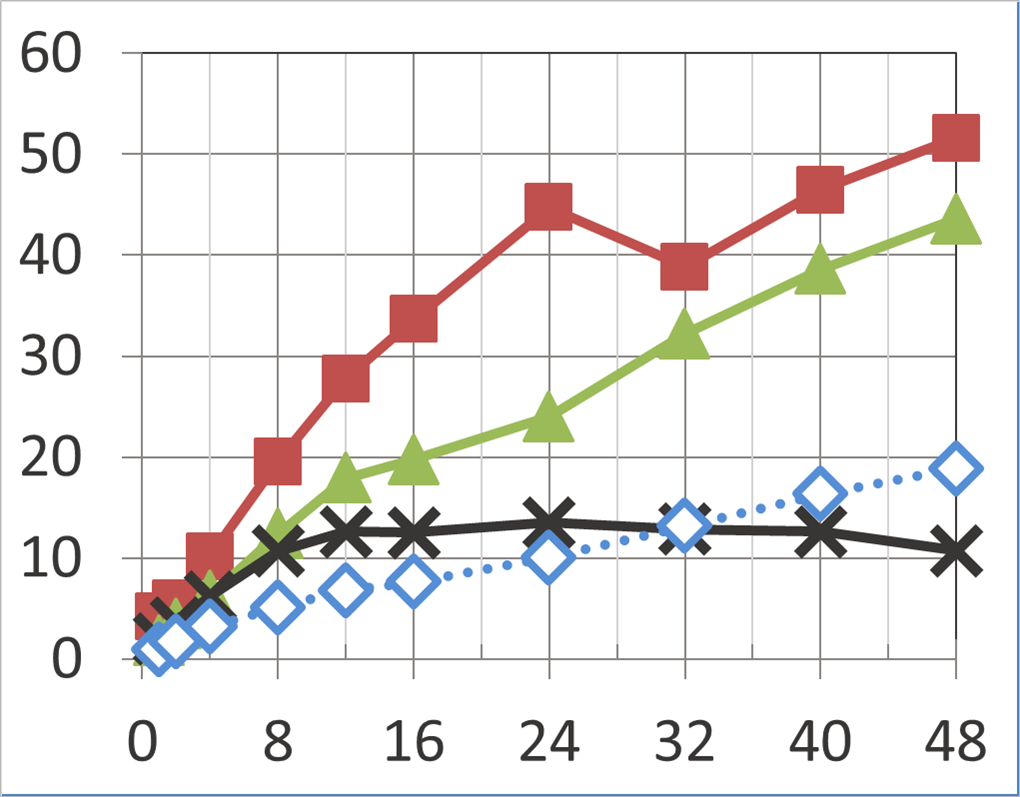} &
        \vspace{-1.5mm}\includegraphics[width=\linewidth]{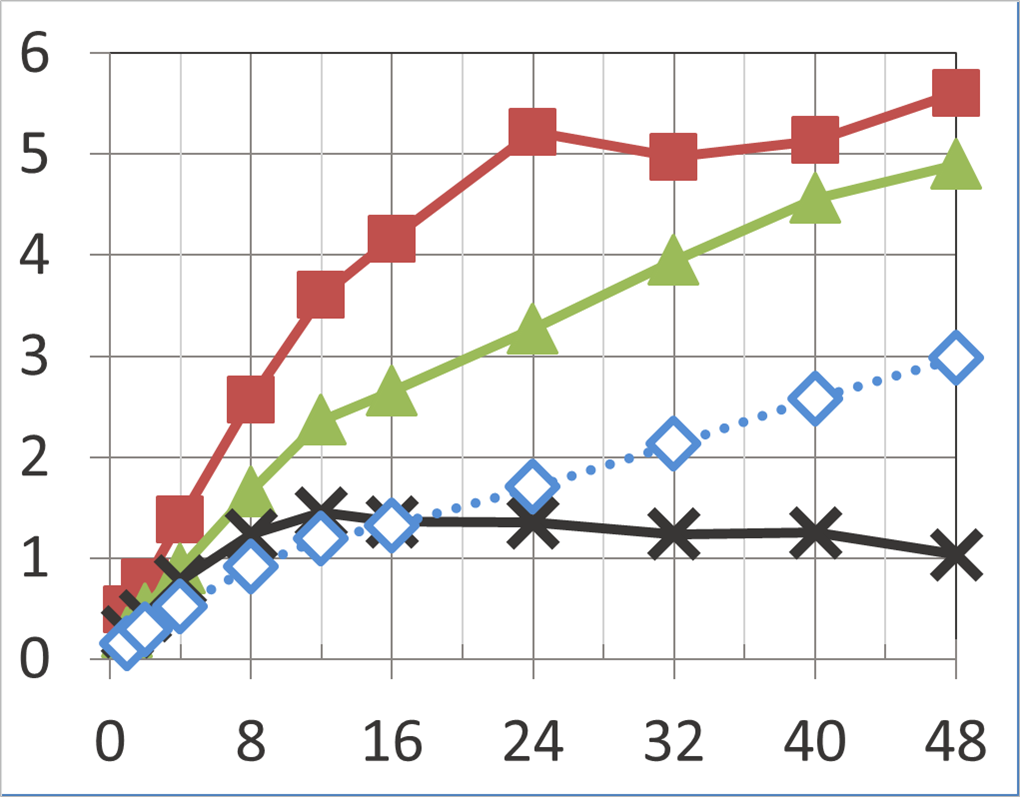}
        \\
    \end{tabular}
    \end{minipage}
    \begin{minipage}{0.49\linewidth}
    \begin{tabular}{m{0.05\linewidth}m{0.465\linewidth}m{0.465\linewidth}}
        &
        \multicolumn{2}{c}{
            %\dimexpr \linewidth-2\fboxsep-2\fboxrule
            \fcolorbox{black!80}{black!40}{\parbox{\dimexpr 0.93\linewidth-2\fboxsep-2\fboxrule}{\centering\textbf{4x 16-thread AMD Opteron 6380}}}
        }
        \\
        &
        \fcolorbox{black!50}{black!20}{\parbox{\dimexpr \linewidth-2\fboxsep-2\fboxrule}{\centering {\footnotesize 2-CAS}}} &
        \fcolorbox{black!50}{black!20}{\parbox{\dimexpr \linewidth-2\fboxsep-2\fboxrule}{\centering {\footnotesize 16-CAS}}}
        \\
        \rotatebox{90}{Array size $2^{26}$} &
        \includegraphics[width=\linewidth]{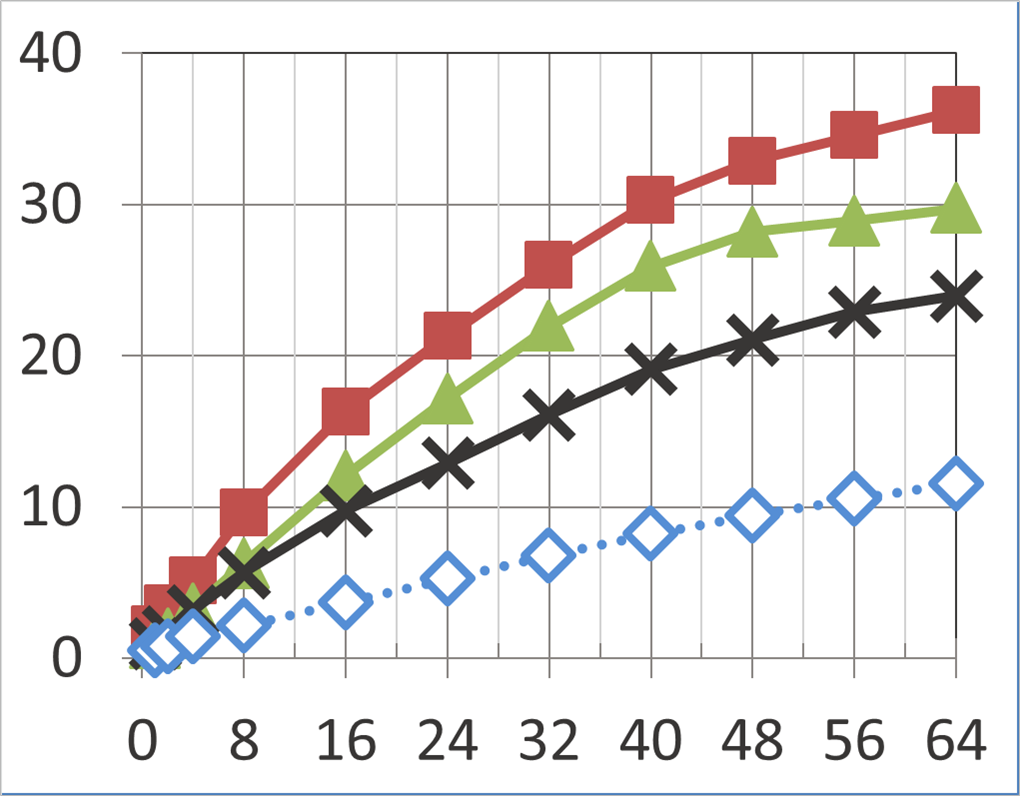} &
        \includegraphics[width=\linewidth]{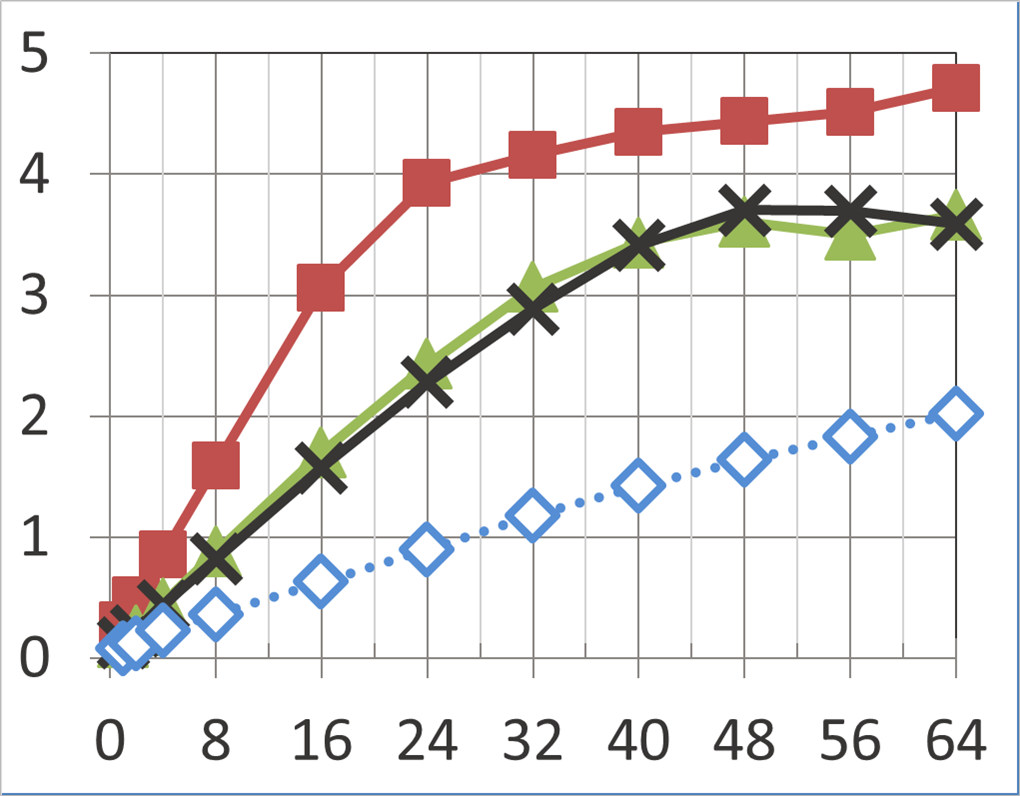}
        \\
        \vspace{-1.5mm}\rotatebox{90}{Array size $2^{20}$} &
        \vspace{-1.5mm}\includegraphics[width=\linewidth]{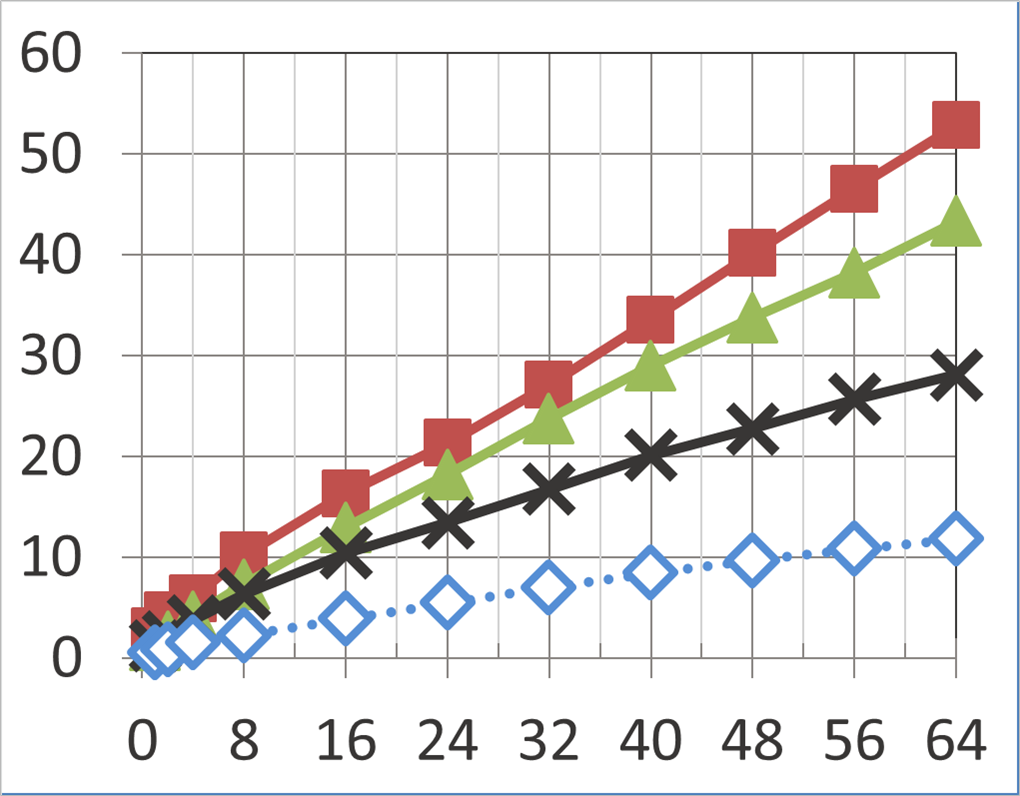} &
        \vspace{-1.5mm}\includegraphics[width=\linewidth]{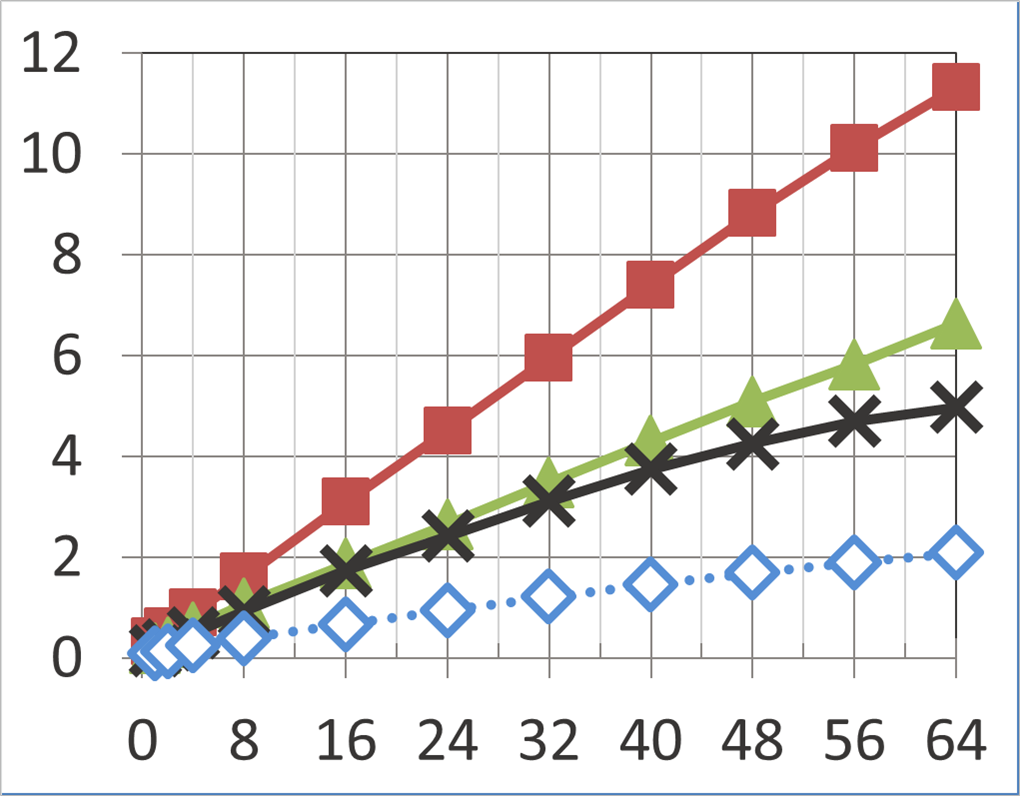}
        \\
        \vspace{-1.5mm}\rotatebox{90}{Array size $2^{14}$} &
        \vspace{-1.5mm}\includegraphics[width=\linewidth]{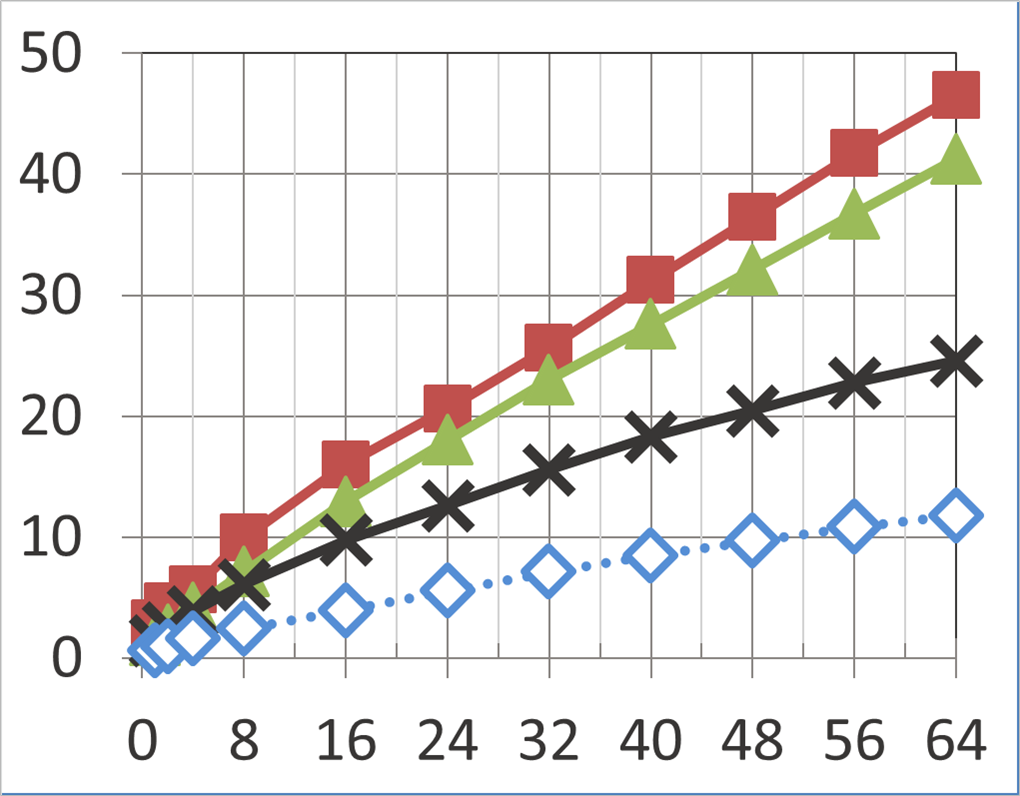} &
        \vspace{-1.5mm}\includegraphics[width=\linewidth]{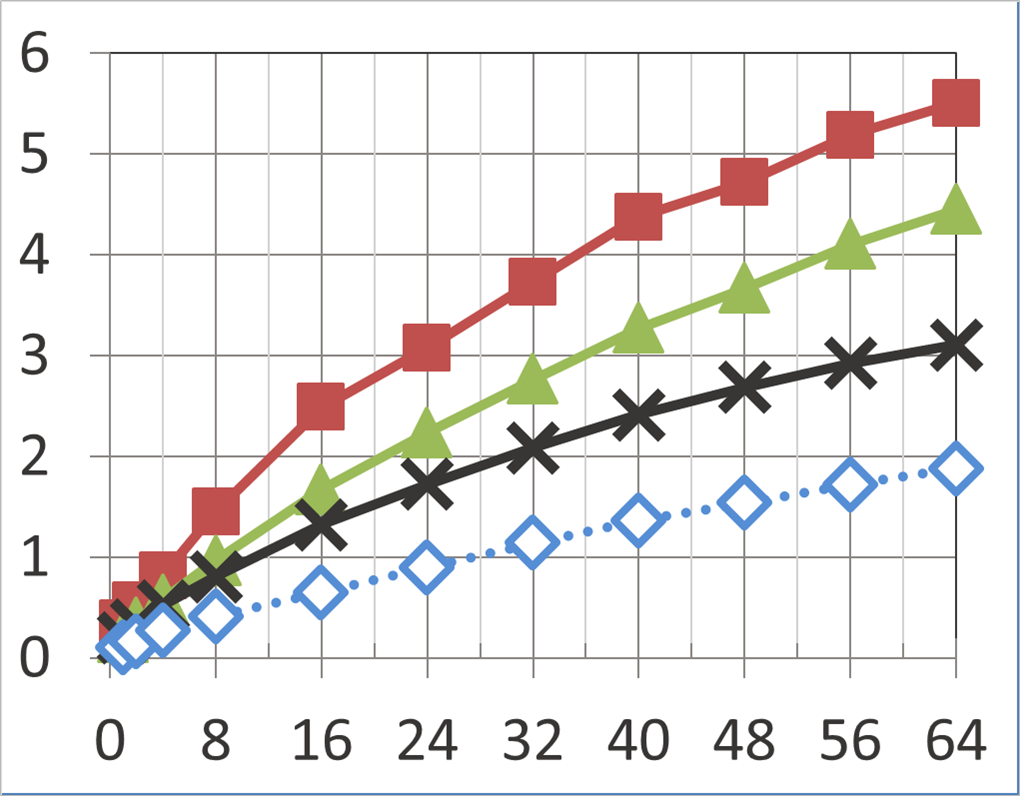}
        \\
    \end{tabular}
    \end{minipage}
%\vspace{-3mm}
\includegraphics[width=0.5\linewidth]{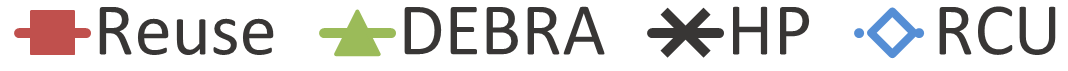}
\vspace{-2mm}
\caption{Results for a \textbf{$k$-CAS microbenchmark}.
The x-axis represents the number of concurrent threads.
The y-axis represents operations per microsecond.}
\label{fig-exp-kcas}
%\vspace{-4mm}
\end{figure}
\end{shortver}

\begin{fullver}
\begin{figure}[p]
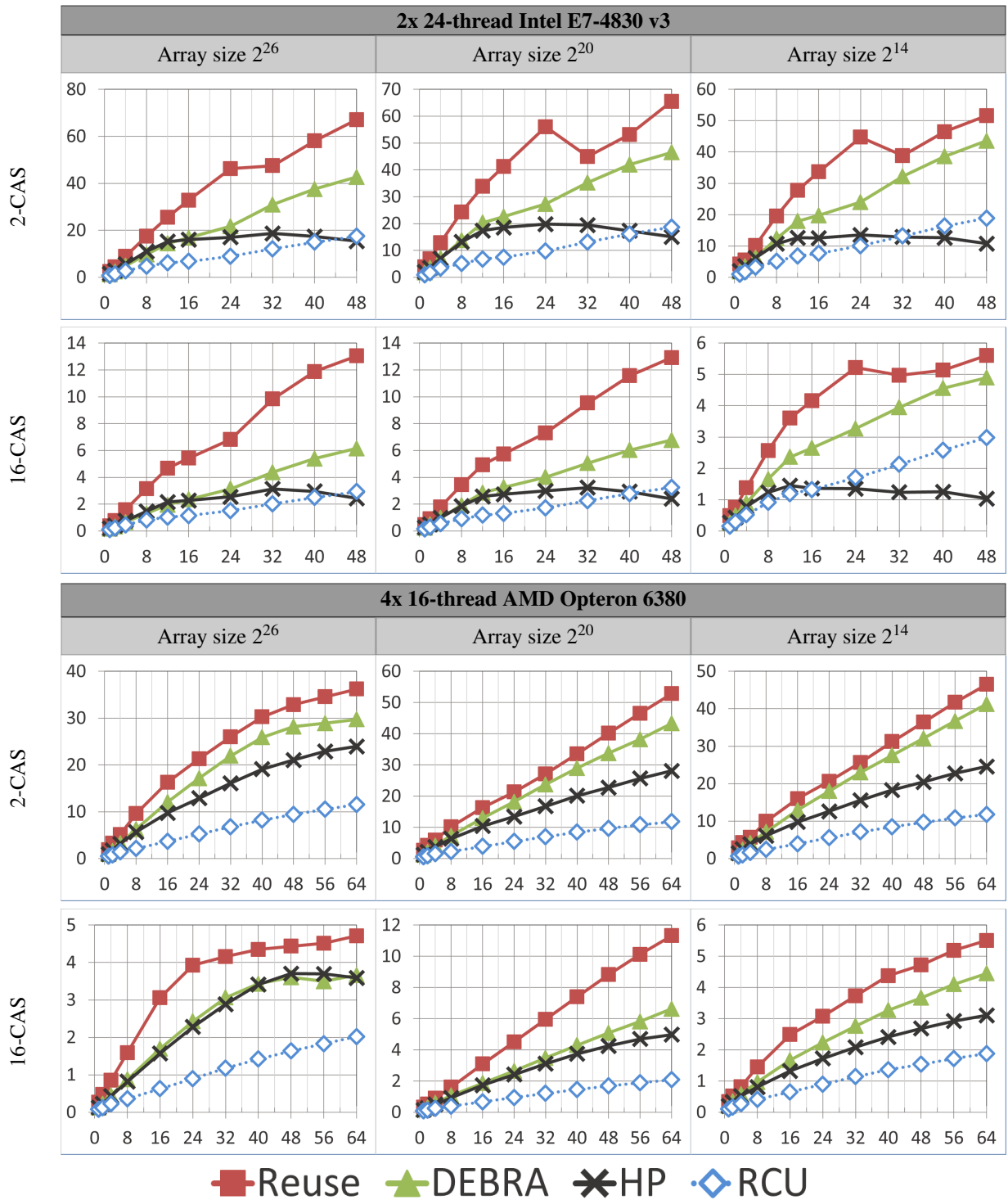

%    \vspace{-3mm}
    \centering
    \setlength\tabcolsep{0pt}
    \begin{tabular}{m{0.05\linewidth}m{0.31\linewidth}m{0.31\linewidth}m{0.31\linewidth}}
        &
        \multicolumn{3}{c}{
            %\dimexpr \linewidth-2\fboxsep-2\fboxrule
            \fcolorbox{black!80}{black!40}{\parbox{\dimexpr 0.93\linewidth-2\fboxsep-2\fboxrule}{\centering\textbf{2x 24-thread Intel E7-4830 v3}}}
        }
        \\
        &
        \fcolorbox{black!50}{black!20}{\parbox{\dimexpr \linewidth-2\fboxsep-2\fboxrule}{\centering {\normalsize Array size $2^{26}$}}} &
        \fcolorbox{black!50}{black!20}{\parbox{\dimexpr \linewidth-2\fboxsep-2\fboxrule}{\centering {\normalsize Array size $2^{20}$}}} &
        \fcolorbox{black!50}{black!20}{\parbox{\dimexpr \linewidth-2\fboxsep-2\fboxrule}{\centering {\normalsize Array size $2^{14}$}}}
        \\
        \rotatebox{90}{2-CAS} &
        \includegraphics[width=\linewidth]{chap-desc/figures/graphs/kcas-reuse-vs-throw8_48_tapuz40_lgchunk21_67108864-k2.png} &
        \includegraphics[width=\linewidth]{chap-desc/figures/graphs/kcas-reuse-vs-throw8_48_tapuz40_lgchunk21_1048576-k2.png} &
        \includegraphics[width=\linewidth]{chap-desc/figures/graphs/kcas-reuse-vs-throw8_48_tapuz40_lgchunk21_16384-k2.png}
        \\
        \rotatebox{90}{16-CAS} &
        \includegraphics[width=\linewidth]{chap-desc/figures/graphs/kcas-reuse-vs-throw8_48_tapuz40_lgchunk21_67108864-k16.png} &
        \includegraphics[width=\linewidth]{chap-desc/figures/graphs/kcas-reuse-vs-throw8_48_tapuz40_lgchunk21_1048576-k16.png} &
        \includegraphics[width=\linewidth]{chap-desc/figures/graphs/kcas-reuse-vs-throw8_48_tapuz40_lgchunk21_16384-k16.png}
        \\
    \end{tabular}
    \begin{tabular}{m{0.05\linewidth}m{0.31\linewidth}m{0.31\linewidth}m{0.31\linewidth}}
        &
        \multicolumn{3}{c}{
            %\dimexpr \linewidth-2\fboxsep-2\fboxrule
            \fcolorbox{black!80}{black!40}{\parbox{\dimexpr 0.93\linewidth-2\fboxsep-2\fboxrule}{\centering\textbf{4x 16-thread AMD Opteron 6380}}}
        }
        \\
        &
        \fcolorbox{black!50}{black!20}{\parbox{\dimexpr \linewidth-2\fboxsep-2\fboxrule}{\centering {\normalsize Array size $2^{26}$}}} &
        \fcolorbox{black!50}{black!20}{\parbox{\dimexpr \linewidth-2\fboxsep-2\fboxrule}{\centering {\normalsize Array size $2^{20}$}}} &
        \fcolorbox{black!50}{black!20}{\parbox{\dimexpr \linewidth-2\fboxsep-2\fboxrule}{\centering {\normalsize Array size $2^{14}$}}}
        \\
        \rotatebox{90}{2-CAS} &
        \includegraphics[width=\linewidth]{chap-desc/figures/graphs/kcas-reuse-vs-throw8_64_csl-pomela6_lgchunk19_67108864-k2.png} &
        \includegraphics[width=\linewidth]{chap-desc/figures/graphs/kcas-reuse-vs-throw8_64_csl-pomela6_lgchunk19_1048576-k2.png} &
        \includegraphics[width=\linewidth]{chap-desc/figures/graphs/kcas-reuse-vs-throw8_64_csl-pomela6_lgchunk19_16384-k2.png}
        \\
        \rotatebox{90}{16-CAS} &
        \includegraphics[width=\linewidth]{chap-desc/figures/graphs/kcas-reuse-vs-throw8_64_csl-pomela6_lgchunk19_67108864-k16.png} &
        \includegraphics[width=\linewidth]{chap-desc/figures/graphs/kcas-reuse-vs-throw8_64_csl-pomela6_lgchunk19_1048576-k16.png} &
        \includegraphics[width=\linewidth]{chap-desc/figures/graphs/kcas-reuse-vs-throw8_64_csl-pomela6_lgchunk19_16384-k16.png}
        \\
    \end{tabular}
	\includegraphics[width=0.6\linewidth]{chap-desc/figures/graphs/legend-small.png}
    %\vspace{-3mm}
\caption{Results for a \textbf{$k$-CAS microbenchmark}.
The x-axis represents the number of concurrent threads.
The y-axis represents operations per microsecond.}
\label{fig-exp-kcas}
%\vspace{-4mm}
\end{figure}
\end{fullver}

\fakeparagraph{Results}
The results for this benchmark appear in Figure~\ref{fig-exp-kcas}.
Error bars are not drawn on the graphs, since more than 97\% of the data points have a standard deviation that is less than 5\% of the mean (making them essentially too small to see).

Overall, \textit{Reuse} outperforms every other algorithm, in every workload, on both machines.
Notably, on the Intel machine, its throughput is \textit{2.2 times} that of the next best algorithm at 48 threads with $k=16$ and array size $2^{26}$.
On the AMD machine, its throughput is 1.7 times that of the next best algorithm at 64 threads with $k=16$ and array size $2^{20}$.

% 24 threads
%PAPI_L1_DCM=0.632535
%PAPI_L2_TCM=0.843646
%PAPI_L3_TCM=0.711768
%PAPI_RES_STL=277.819
%PAPI_TOT_CYC=-1
%PAPI_TOT_ISR=231.838

% 25 threads
%PAPI_L1_DCM=1.91134
%PAPI_L2_TCM=2.09996
%PAPI_L3_TCM=1.59849
%PAPI_RES_STL=415.654
%PAPI_TOT_CYC=-1
%PAPI_TOT_ISR=511.316

% 32 threads
%PAPI_L1_DCM=1.10443
%PAPI_L2_TCM=1.34277
%PAPI_L3_TCM=0.950947
%PAPI_RES_STL=162.592
%PAPI_TOT_CYC=-1
%PAPI_TOT_ISR=228.895

On the Intel machine, with $k=2$, NUMA effects are quite noticeable for \textit{Reuse} in the jump from 24 to 32 threads, as threads begin running on the second socket.
According the statistics we collected with PAPI, this decrease in performance corresponds to an increase in cache misses.
For example, with $k=2$ and an array of size $2^{26}$ in the Intel machine, jumping from 24 threads to 25 increases the number of L3 cache misses per operation from 0.7 to 1.6 (with similar increases in L1 and L2 cache misses and pipeline stalls).
We believe this is due to cross-socket cache invalidations.
%The resulting cache misses must be served from the second socket's L3 cache, or from main memory, instead of being served from the shared L3 cache on the first socket.

From the three graphs for $k=2$ on Intel, we can see that the effect is more severe with larger absolute throughput (since the additive overhead of a cache miss is more significant).
Consequently, the effect is masked by the much smaller throughput of the slower algorithms, and by the substantially lower throughputs in the $k=16$ case, except when the array is of size $2^{14}$.
In the array of size $2^{14}$, contention is extremely high, since each of the 48 threads are accessing 16 $k$-CAS addresses, each of which causes contention on the entire cache line of 8 words, for a total of 6144 array entries contended at any given time.
Thus, cache misses become a dominating factor in the performance on two sockets.
\begin{fullver}
These effects were not observed on the AMD machine.
The number of cache misses is not significantly different when crossing socket boundaries, which suggests an architectural robustness to NUMA effects that is not seen on the Intel machine.
\end{fullver}
\begin{shortver}
The number of cache misses is not significantly different when crossing socket boundaries, which suggests a robustness to NUMA effects that is not seen on the Intel machine.
\end{shortver}

Interestingly, absolute throughputs on the AMD machine are larger with array size $2^{20}$ than with sizes $2^{14}$ and $2^{26}$.
This is because the $2^{20}$ array size represents a sweet spot with less contention than the $2^{14}$ size and better cache utilization than the $2^{26}$ size.
For example, with 64 threads and $k=16$, \textit{Reuse} incurred approximately 50\% more cache misses with size $2^{26}$ than with size $2^{20}$, and approximately 50\% of operations helped one another with size $2^{14}$, whereas less than 1\% of operations helped one another with size $2^{20}$.

Note, however, that this is not true on the Intel machine.
There, $2^{26}$ is almost always as fast as $2^{20}$, because of the very large shared L3 cache (which is 5x larger than on the AMD machine).
This is reflected in the increased number of cycles where the processor is stalled (e.g., waiting for cache misses to be served) when moving from size $2^{20}$ to $2^{26}$.
On the Intel machine, stalled cycles increase by $85\%$ per operation, whereas on the AMD machine they increase by a whopping $450\%$ per operation.

\begin{shortver}
\fakeparagraph{Additional experiments}
Appendix~\ref{appendix-exp} presents three additional experiments.
The first is a study of memory usage in the $k$-CAS microbenchmark.
The second is a microbenchmark evaluating the performance of a lock-free binary search tree implemented using a transformed version of the LLX and SCX primitives.
The third is an empirical study of the likelihood of ABA problems with sequence numbers of varying bit-widths.
\end{shortver}
\begin{fullver}
\begin{figure}[t]
	%    \vspace{-3mm}
	\centering
	\setlength\tabcolsep{0pt}
	\begin{tabular}{m{0.05\linewidth}m{0.46\linewidth}} %m{0.46\linewidth}}
		&
		\multicolumn{1}{c}{%2}{c}{
			%\dimexpr \linewidth-2\fboxsep-2\fboxrule
			\fcolorbox{black!80}{black!40}{\parbox{\dimexpr 0.46\linewidth-2\fboxsep-2\fboxrule}{\centering\textbf{2x 24-thread Intel E7-4830 v3}}}
		}
		\\
		&
%		\fcolorbox{black!50}{black!20}{\parbox{\dimexpr \linewidth-2\fboxsep-2\fboxrule}{\centering {\footnotesize Descriptor allocations (in MB)}}} &
		\fcolorbox{black!50}{black!20}{\parbox{\dimexpr \linewidth-2\fboxsep-2\fboxrule}{\centering {\footnotesize Descriptor footprint (in bytes)}}}
		\\
		\rotatebox{90}{$S=2^{26}, k=16$} &
		\includegraphics[width=\linewidth]{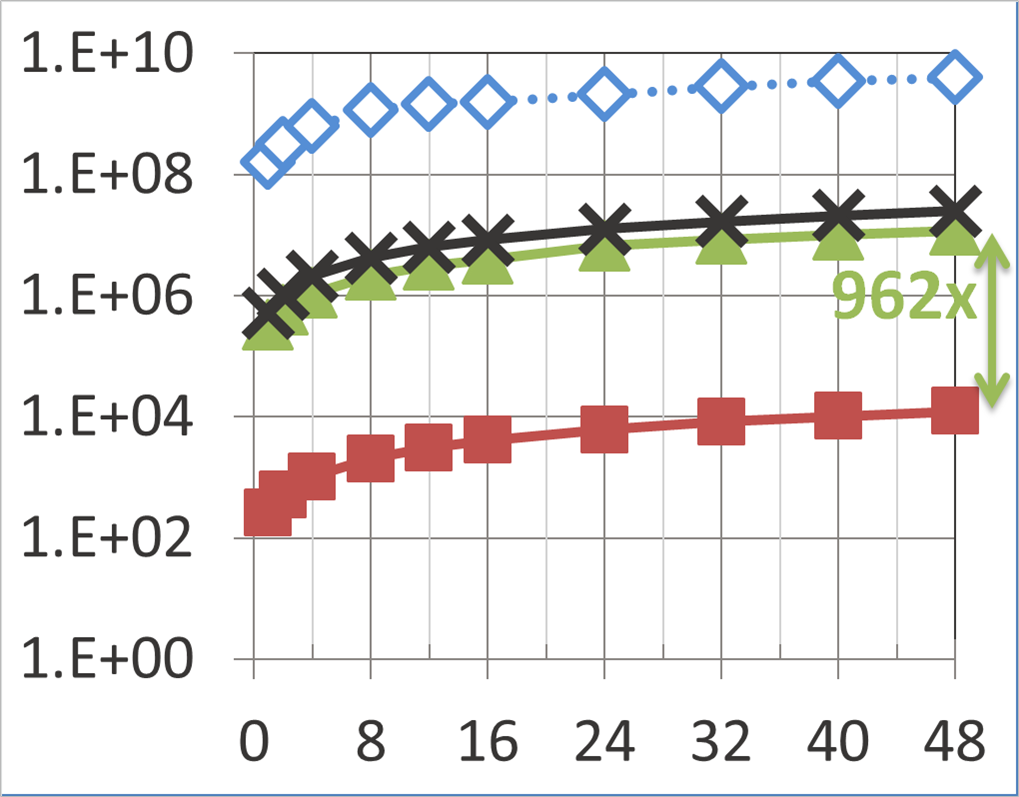}
		\\
	\end{tabular}
	\includegraphics[width=0.5\linewidth]{chap-desc/figures/graphs/legend-small.png}
	%\vspace{-3mm}
	\caption{Memory usage for the \textbf{$k$-CAS microbenchmark}.
		The x-axis represents the number of concurrent threads. \textit{Note the logarithmic scale.}} % in the graph on the right.}}
	\label{fig-exp-memory}
	%\vspace{-4mm}
\end{figure}

\begin{fullver}
\paragraph{Memory usage}
We
\end{fullver}
\begin{shortver}
\subsection{Memory usage in the $k$-CAS benchmark}
We
\end{shortver}
studied memory usage for all algorithms, in all workloads, on both systems, but we only show results for array size $2^{26}$ and $k=16$, because the other graphs are very similar.
These results appear in Figure~\ref{fig-exp-memory}.
%
%The left graph shows the total number of descriptor allocations performed throughout an execution.
%These results were obtained by having each thread increment a private counter on each call to \texttt{malloc}.
%\textit{Reuse} performs almost no allocations, since it only allocates a fixed number of descriptors at the start of an execution.
%
In particular, we are interested in the descriptor footprint, i.e., the maximum amount of memory ever occupied by descriptors in an execution.
Unfortunately, computing the descriptor footprint exactly would require excessive synchronization between threads.
Thus, we approximate the descriptor footprint by computing the descriptor footprint \textit{for each thread}, and then summing those individual footprints.
(This is only an approximation, since different threads may hit their peak memory usage for descriptors at different times.)
The graph in Figure~\ref{fig-exp-memory} contains the results of this approximation.

\begin{fullver}
These results were obtained as follows.
Each thread used three private variables: \textit{totalFree}, \textit{totalMalloc} and \textit{maxFootprint}.
Each time a thread invoked \texttt{free}, it incremented \textit{totalFree} by the size of the descriptor being freed.
Each time a thread invoked \texttt{malloc}, it incremented \textit{totalMalloc} by the size of the descriptor being allocated, and then set $\mbox{\textit{maxFootprint}} = max\{\mbox{\textit{maxFootprint}}, \mbox{\textit{totalMalloc}} - \mbox{\textit{totalFree}}\}$.
The per-thread \textit{maxFootprint}s are then summed to obtain the data points in the graph.
\end{fullver}

Note that the $y$-axis is a logarithmic scale.
The results show that \textit{DEBRA} and \textit{HPs} use almost \textbf{three orders of magnitude} more memory than \textit{Reuse} at their peaks, and \textit{RCU} uses nearly three orders of magnitude more memory than \textit{DEBRA} and \textit{HPs}.
\textit{RCU}'s memory usage is significantly higher because reclamation is delayed significantly longer than in the other algorithms. %, so the working set size is larger.

\subsection{BST microbenchmark}

Unlike in the $k$-CAS algorithm, where memory reclamation was only needed for descriptors, in the BST, memory reclamation is always needed for nodes.
To compare our technique with different memory reclamation options, we implemented four variants of the BST algorithm: \textit{DEBRA/DEBRA}, \textit{DEBRA/Reuse}, \textit{RCU/RCU} and \textit{RCU/Reuse}.
Here, an algorithm named \textit{X/Y} uses \textit{X} to reclaim nodes and \textit{Y} for descriptors.
For example, \textit{DEBRA/Reuse} uses DEBRA to reclaim nodes and has reusable descriptors.

\begin{thesisnot}
Unfortunately, we could not create a variant of the BST using hazard pointers.
As part of the \textit{finalizing} mechanism, this BST implementation \textit{marks} nodes before deleting them.
Furthermore searches are allowed to traverse marked nodes, regardless of whether they have been deleted, and subsequently succeed.
These algorithmic properties make it infeasible to use hazard pointers~\cite{Brown:2015}.
\end{thesisnot}

\medskip

\paragraph{Methodology}
We compared our BST variants using a simple randomized microbenchmark.
For each algorithm $A \in \{$\textit{DEBRA/DEBRA}, \textit{DEBRA/Reuse}, \textit{RCU/RCU}, \textit{RCU/Reuse}$\}$, key range size $K \in \{10^5, 10^6\}$ and update rate
\begin{fullver}
$U \in \{100, 10, 0\}$,
\end{fullver}
\begin{shortver}
$U \in \{100, 0\}$,
\end{shortver}
we run ten timed \textit{trials} for several thread counts $n$.
Each trial proceeds in two phases: \textit{prefilling} and \textit{measuring}.
In the prefilling phase, $n$ concurrent threads perform 50\% \textit{Insert} and 50\% \textit{Delete} operations on keys drawn uniformly randomly from $[0, K)$ until the size of the tree converges to a steady state (containing approximately $K/2$ keys).
Next, the trial enters the measuring phase, during which threads begin counting how many operations they perform.
(These counts are eventually summed over all threads and reported in our graphs.)
In this phase, each thread instead performs $(U/2)$\% \textit{Insert}, $(U/2)$\% \textit{Delete} and $(100-U)$\% \textit{Find} operations on keys drawn uniformly from $[0,K)$ for one second.

As a way of validating correctness in each trial, each thread maintains a \textit{checksum}.
Each time a thread inserts a new key, it adds the key to its checksum.
Each time a thread deletes a key, it subtracts the key from its checksum.
At the end of the trial, the sum of all thread checksums must be equal to the sum of keys in the tree.

\medskip

\paragraph{Results}
\begin{shortver}
\begin{figure}[t]
    \centering
    \setlength\tabcolsep{0pt}
    \begin{minipage}{0.49\linewidth}
    \begin{tabular}{m{0.05\linewidth}m{0.465\linewidth}m{0.465\linewidth}}
        &
        \multicolumn{2}{c}{
            %\dimexpr \linewidth-2\fboxsep-2\fboxrule
            \fcolorbox{black!80}{black!40}{\parbox{\dimexpr 0.93\linewidth-2\fboxsep-2\fboxrule}{\centering\textbf{2x 24-thread Intel E7-4830 v3}}}
        }
        \\
        &
        \fcolorbox{black!50}{black!20}{\parbox{\dimexpr \linewidth-2\fboxsep-2\fboxrule}{\centering {\footnotesize range $[0, 10^5)$}}} &
        \fcolorbox{black!50}{black!20}{\parbox{\dimexpr \linewidth-2\fboxsep-2\fboxrule}{\centering {\footnotesize range $[0, 10^6)$}}}
        \\
        \rotatebox{90}{100\% updates} &
        \includegraphics[width=\linewidth]{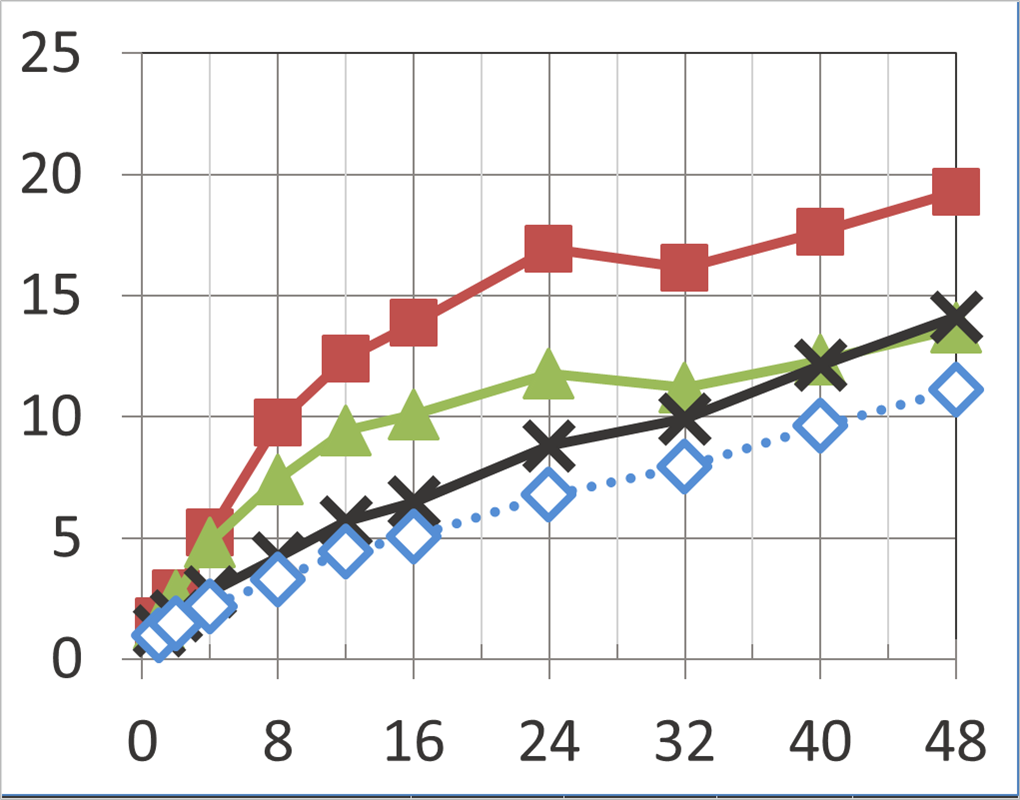} &
        \includegraphics[width=\linewidth]{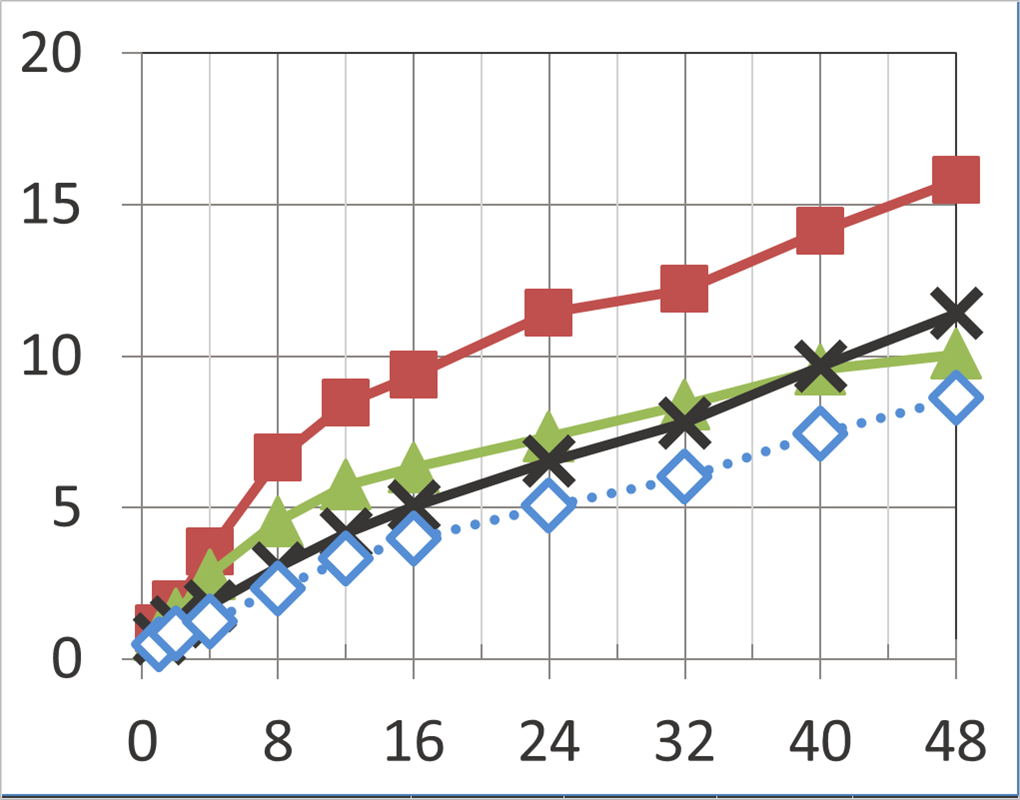}
        \\
        \rotatebox{90}{0\% updates} &
        \includegraphics[width=\linewidth]{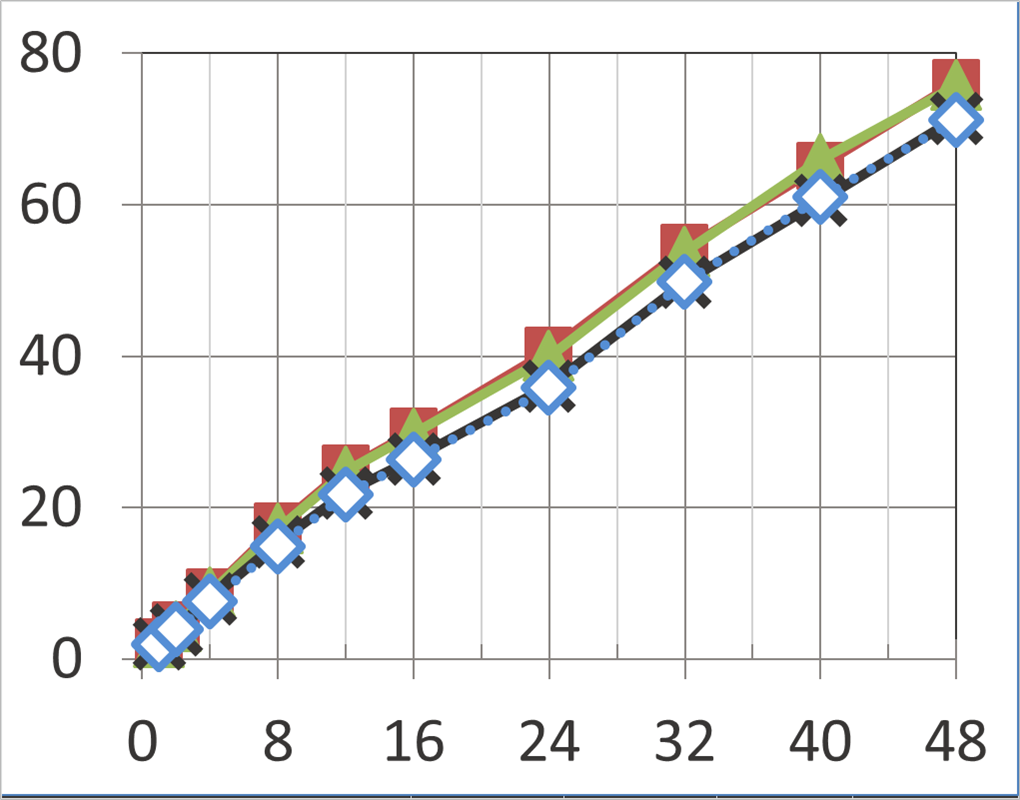} &
        \includegraphics[width=\linewidth]{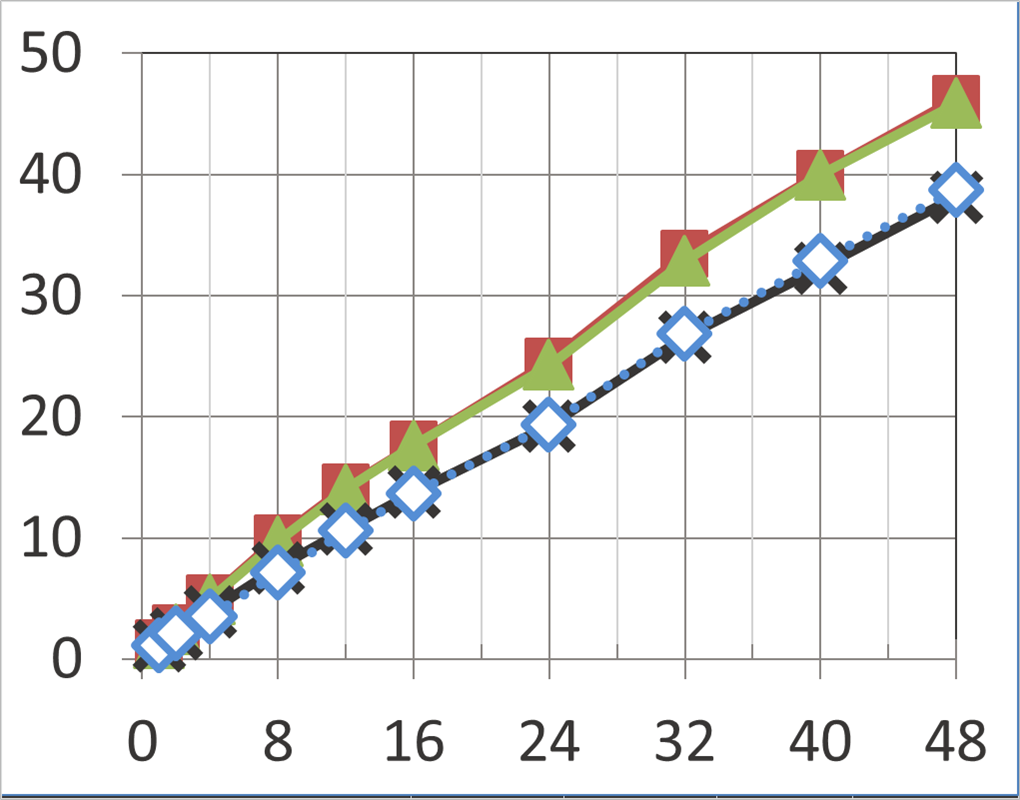}
        \\
    \end{tabular}
    \end{minipage}
    \begin{minipage}{0.49\linewidth}
    \begin{tabular}{m{0.05\linewidth}m{0.465\linewidth}m{0.465\linewidth}}
        &
        \multicolumn{2}{c}{
            %\dimexpr \linewidth-2\fboxsep-2\fboxrule
            \fcolorbox{black!80}{black!40}{\parbox{\dimexpr 0.93\linewidth-2\fboxsep-2\fboxrule}{\centering\textbf{4x 16-thread AMD Opteron 6380}}}
        }
        \\
        &
        \fcolorbox{black!50}{black!20}{\parbox{\dimexpr \linewidth-2\fboxsep-2\fboxrule}{\centering {\footnotesize range $[0, 10^5)$}}} &
        \fcolorbox{black!50}{black!20}{\parbox{\dimexpr \linewidth-2\fboxsep-2\fboxrule}{\centering {\footnotesize range $[0, 10^6)$}}}
        \\
        \rotatebox{90}{100\% updates} &
        \includegraphics[width=\linewidth]{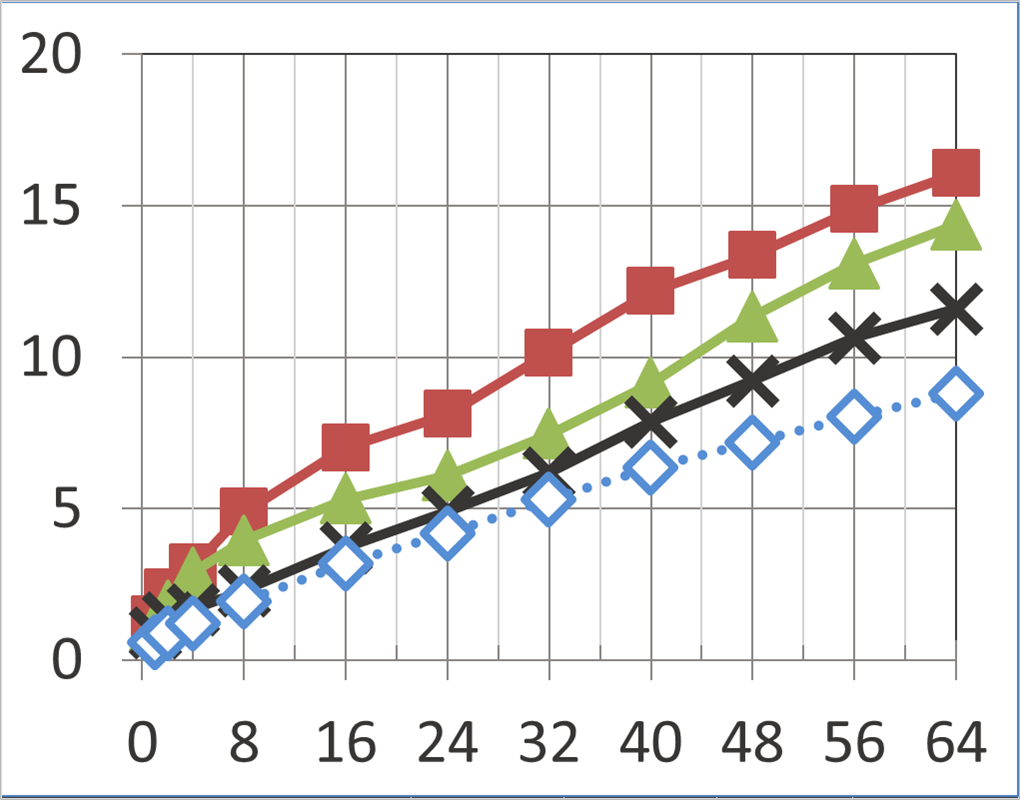} &
        \includegraphics[width=\linewidth]{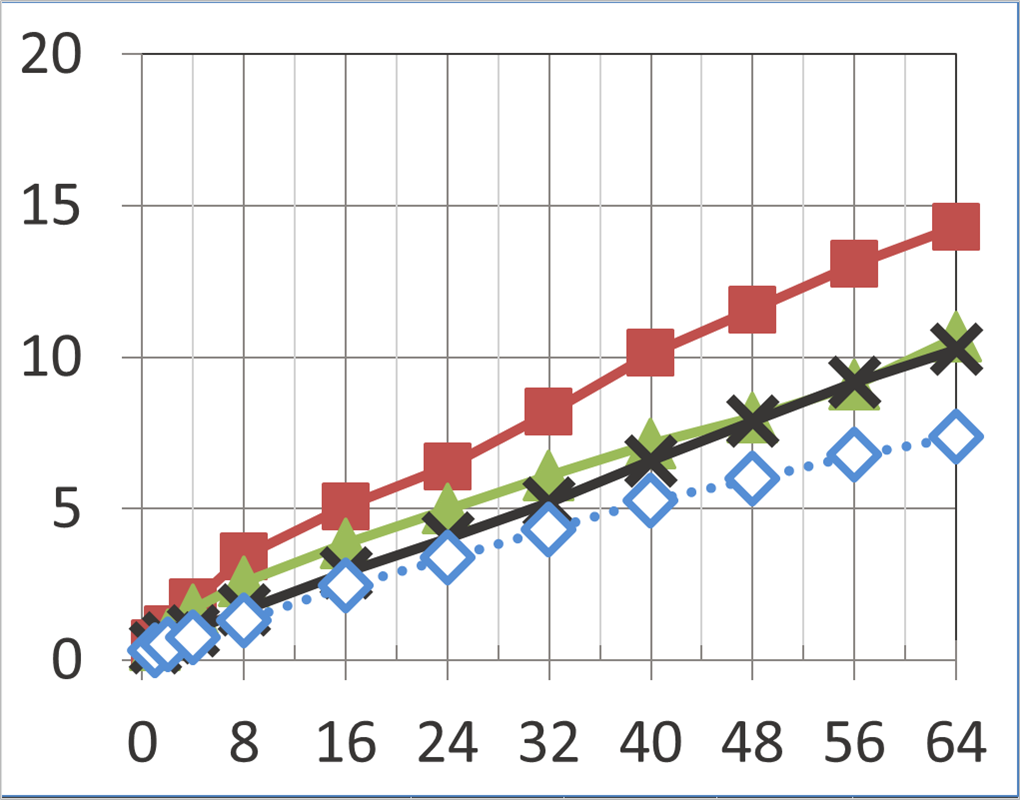}
        \\
        \rotatebox{90}{0\% updates} &
        \includegraphics[width=\linewidth]{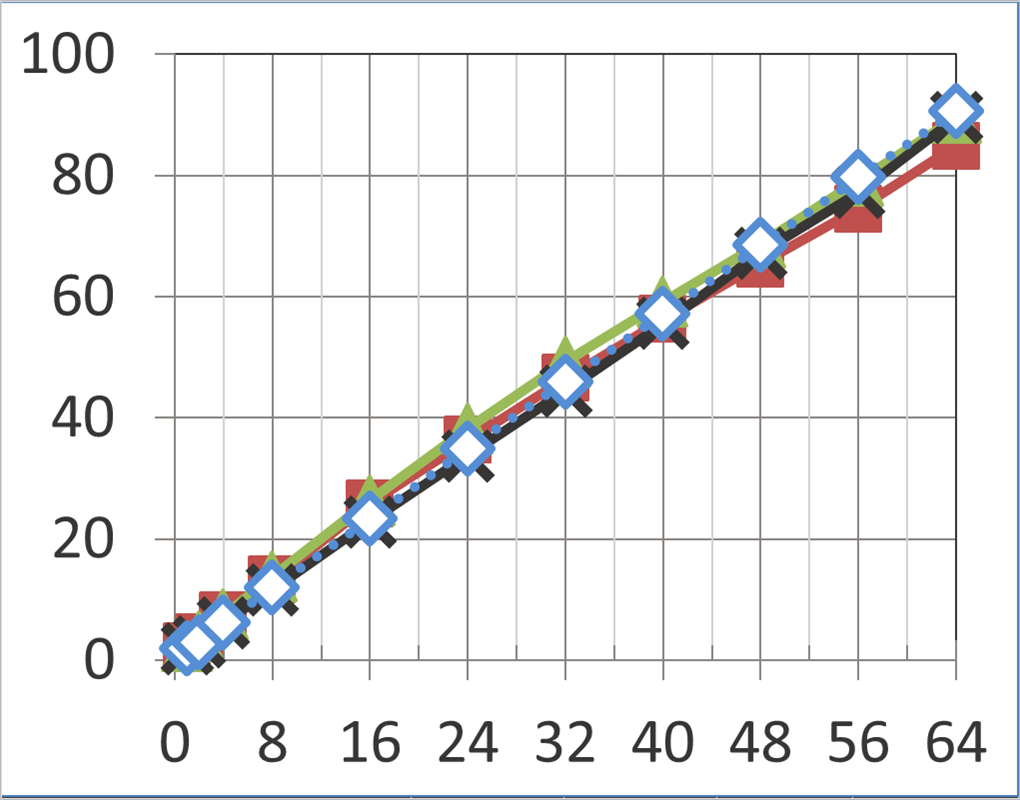} &
        \includegraphics[width=\linewidth]{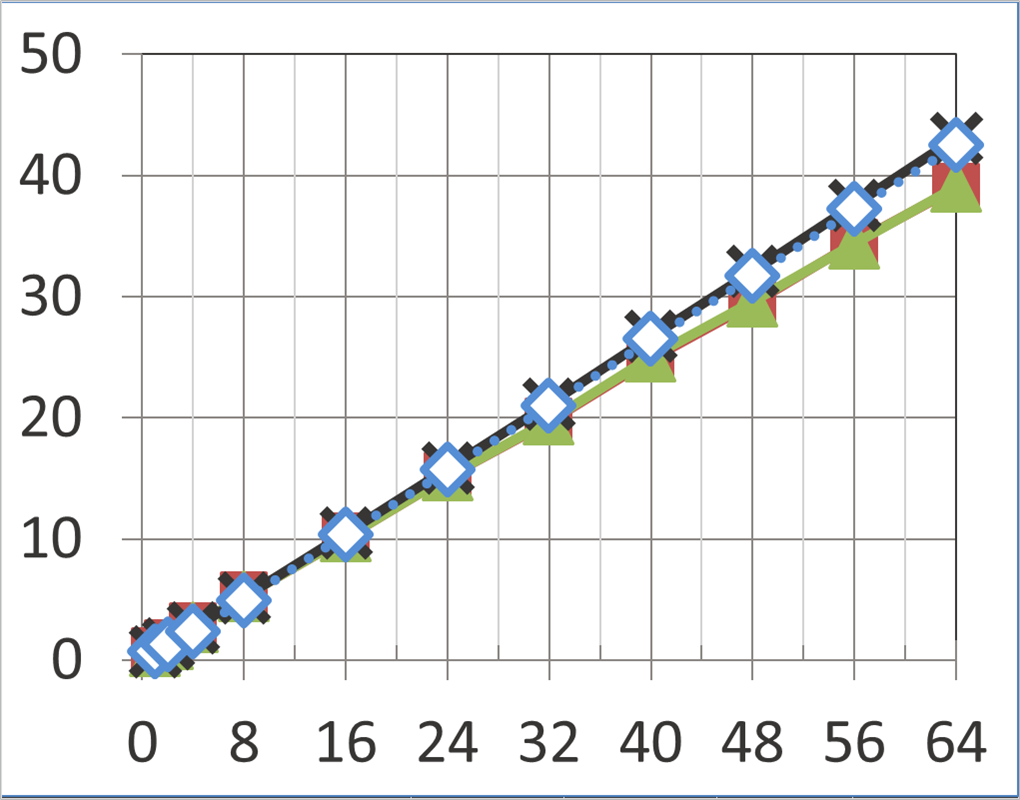}
        \\
    \end{tabular}
    \end{minipage}
	\includegraphics[width=\linewidth]{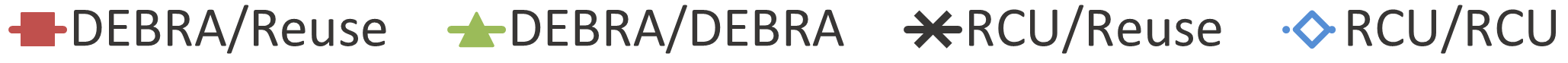}
   % \vspace{-5mm}
\caption{Results for a \textbf{BST microbenchmark}.
The x-axis represents the number of concurrent threads.
The y-axis represents operations per microsecond.}
\label{fig-exp-bst}
\end{figure}
\end{shortver}
\begin{fullver}
\begin{figure}[p]
    \centering
    \setlength\tabcolsep{0pt}
    \begin{tabular}{m{0.05\linewidth}m{0.31\linewidth}m{0.31\linewidth}m{0.31\linewidth}}
        &
        \multicolumn{3}{c}{
            %\dimexpr \linewidth-2\fboxsep-2\fboxrule
            \fcolorbox{black!80}{black!40}{\parbox{\dimexpr 0.93\linewidth-2\fboxsep-2\fboxrule}{\centering\textbf{2x 24-thread Intel E7-4830 v3}}}
        }
        \\
        &
        \fcolorbox{black!50}{black!20}{\parbox{\dimexpr \linewidth-2\fboxsep-2\fboxrule}{\centering {\normalsize 100\% updates}}} &
        \fcolorbox{black!50}{black!20}{\parbox{\dimexpr \linewidth-2\fboxsep-2\fboxrule}{\centering {\normalsize 10\% updates}}} &
        \fcolorbox{black!50}{black!20}{\parbox{\dimexpr \linewidth-2\fboxsep-2\fboxrule}{\centering {\normalsize 0\% updates}}}
        \\
        \rotatebox{90}{\normalsize range $[0, 10^6)$} &
        \includegraphics[width=\linewidth]{chap-desc/figures/graphs/reuse-vs-throw8_48_tapuz40_lgchunk21_1000000k-50i-50d.png} &
        \includegraphics[width=\linewidth]{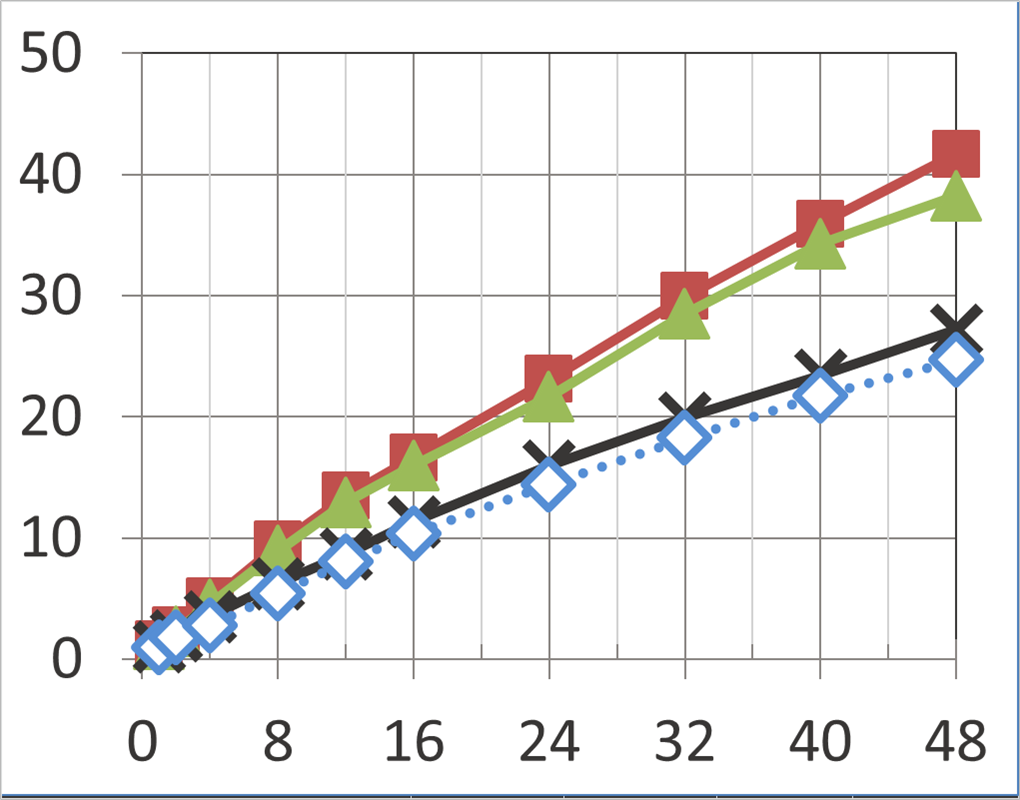} &
        \includegraphics[width=\linewidth]{chap-desc/figures/graphs/reuse-vs-throw8_48_tapuz40_lgchunk21_1000000k-0i-0d.png}
        \\
        \rotatebox{90}{\normalsize range $[0, 10^5)$} &
        \includegraphics[width=\linewidth]{chap-desc/figures/graphs/reuse-vs-throw8_48_tapuz40_lgchunk21_100000k-50i-50d.png} &
        \includegraphics[width=\linewidth]{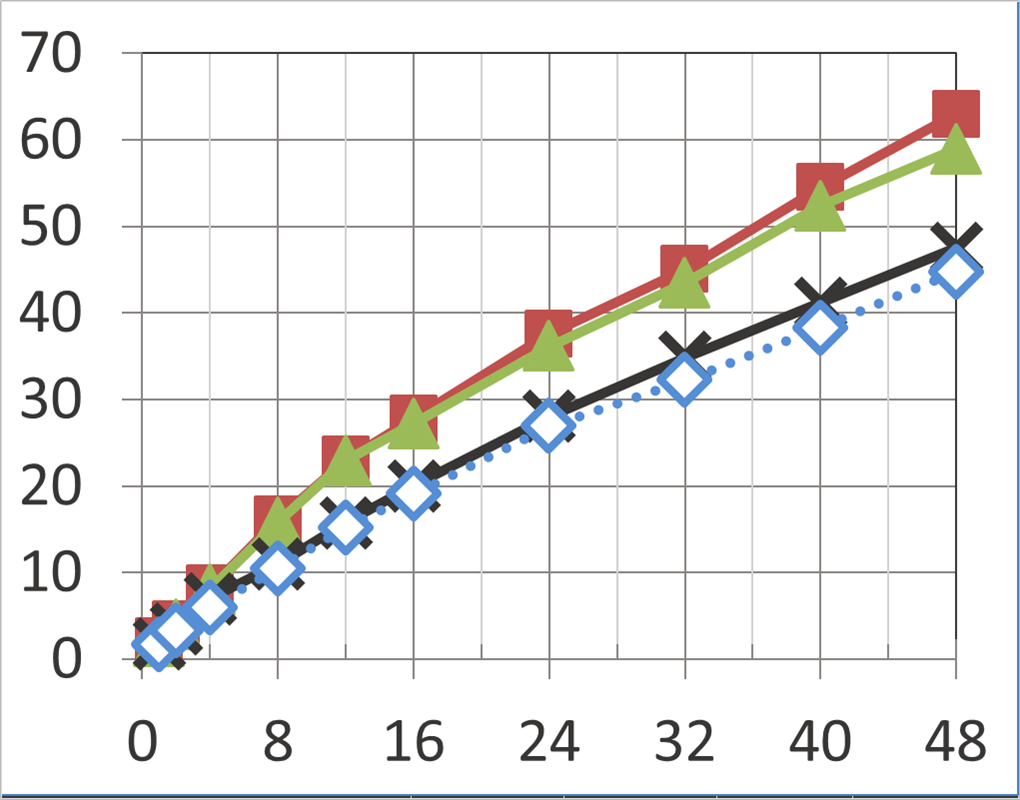} &
        \includegraphics[width=\linewidth]{chap-desc/figures/graphs/reuse-vs-throw8_48_tapuz40_lgchunk21_100000k-0i-0d.png}
        \\
    \end{tabular}

    \begin{tabular}{m{0.05\linewidth}m{0.31\linewidth}m{0.31\linewidth}m{0.31\linewidth}}
        &
        \multicolumn{3}{c}{
            %\dimexpr \linewidth-2\fboxsep-2\fboxrule
            \fcolorbox{black!80}{black!40}{\parbox{\dimexpr 0.93\linewidth-2\fboxsep-2\fboxrule}{\centering\textbf{4x 16-thread AMD Opteron 6380}}}
        }
        \\
        &
        \fcolorbox{black!50}{black!20}{\parbox{\dimexpr \linewidth-2\fboxsep-2\fboxrule}{\centering {\normalsize 100\% updates}}} &
        \fcolorbox{black!50}{black!20}{\parbox{\dimexpr \linewidth-2\fboxsep-2\fboxrule}{\centering {\normalsize 10\% updates}}} &
        \fcolorbox{black!50}{black!20}{\parbox{\dimexpr \linewidth-2\fboxsep-2\fboxrule}{\centering {\normalsize 0\% updates}}}
        \\
        \rotatebox{90}{\normalsize range $[0, 10^6)$} &
        \includegraphics[width=\linewidth]{chap-desc/figures/graphs/reuse-vs-throw8_64_csl-pomela6_lgchunk19_1000000k-50i-50d.png} &
        \includegraphics[width=\linewidth]{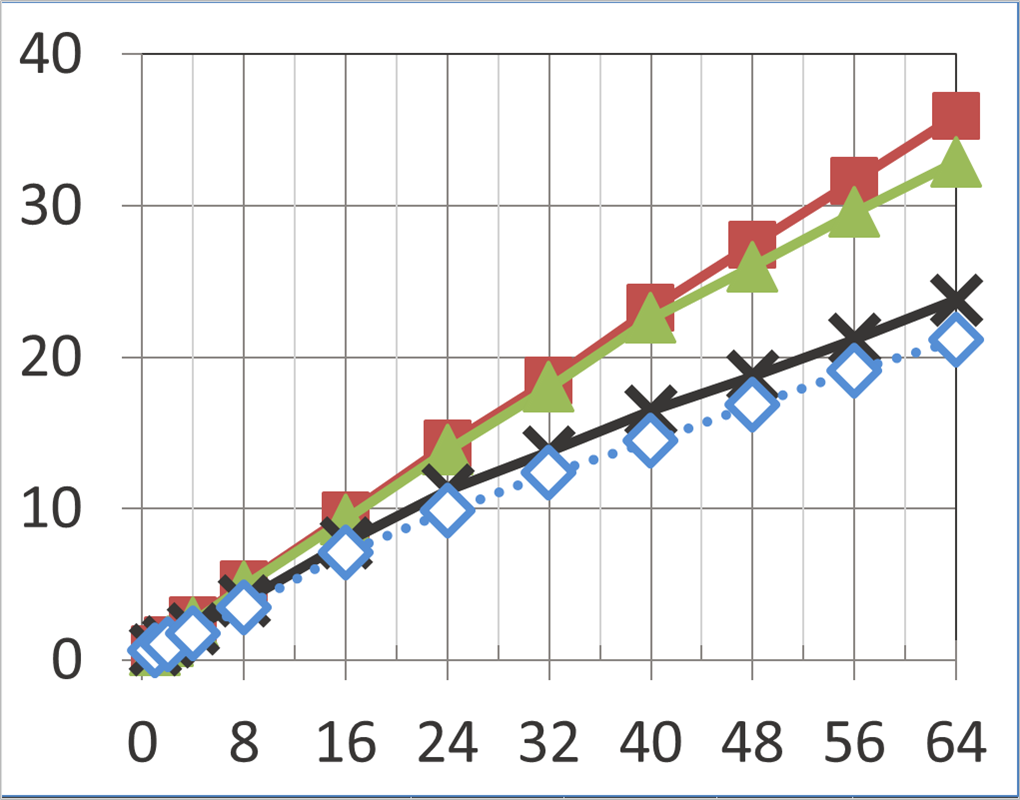} &
        \includegraphics[width=\linewidth]{chap-desc/figures/graphs/reuse-vs-throw8_64_csl-pomela6_lgchunk19_1000000k-0i-0d.png}
        \\
        \rotatebox{90}{\normalsize range $[0, 10^5)$} &
        \includegraphics[width=\linewidth]{chap-desc/figures/graphs/reuse-vs-throw8_64_csl-pomela6_lgchunk19_100000k-50i-50d.png} &
        \includegraphics[width=\linewidth]{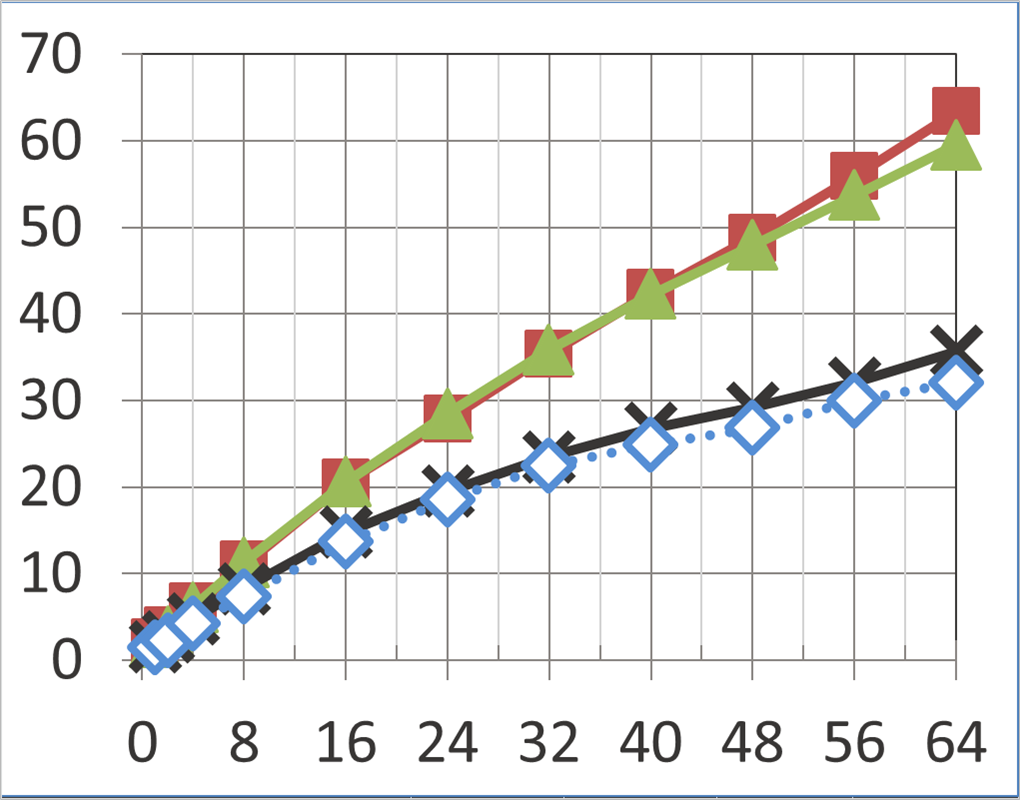} &
        \includegraphics[width=\linewidth]{chap-desc/figures/graphs/reuse-vs-throw8_64_csl-pomela6_lgchunk19_100000k-0i-0d.png}
        \\
    \end{tabular}
	\includegraphics[width=0.8\linewidth]{chap-desc/figures/graphs/legend-bst.png}
   % \vspace{-5mm}
\caption{Results for a \textbf{BST microbenchmark}.
The x-axis represents the number of concurrent threads.
The y-axis represents operations per microsecond.}
\label{fig-exp-bst}
\end{figure}
\end{fullver}
The results for this benchmark appear in Figure~\ref{fig-exp-bst}.
The \textit{Reuse} variants perform at least as well as the pure reclamation variants in every case, and significantly outperform the reclamation variants in the 100\% update workload.
Most notably, on the Intel machine with key range $[0, 10^6]$ and 48 threads, \textit{DEBRA/Reuse} outperforms \textit{DEBRA/DEBRA} by 57\%, and \textit{RCU/Reuse} outperforms \textit{RCU/RCU} by 33\%.
As expected, \textit{Reuse} does not perform significantly faster than the reclamation variants in the workloads with no updates.
This is because searches do not create descriptors.
However, crucially, our transformation does not impose any overhead on searches, either. %, as the 0\% updates workload shows.

\subsection{Studying sequence number wraparound} \label{sec-exp-wraparound}

\begin{figure}[th]
   % \vspace{-2mm}
    \centering
    \begin{tabular}{m{0.5\linewidth}m{0.25\linewidth}}
        \includegraphics[width=\linewidth]{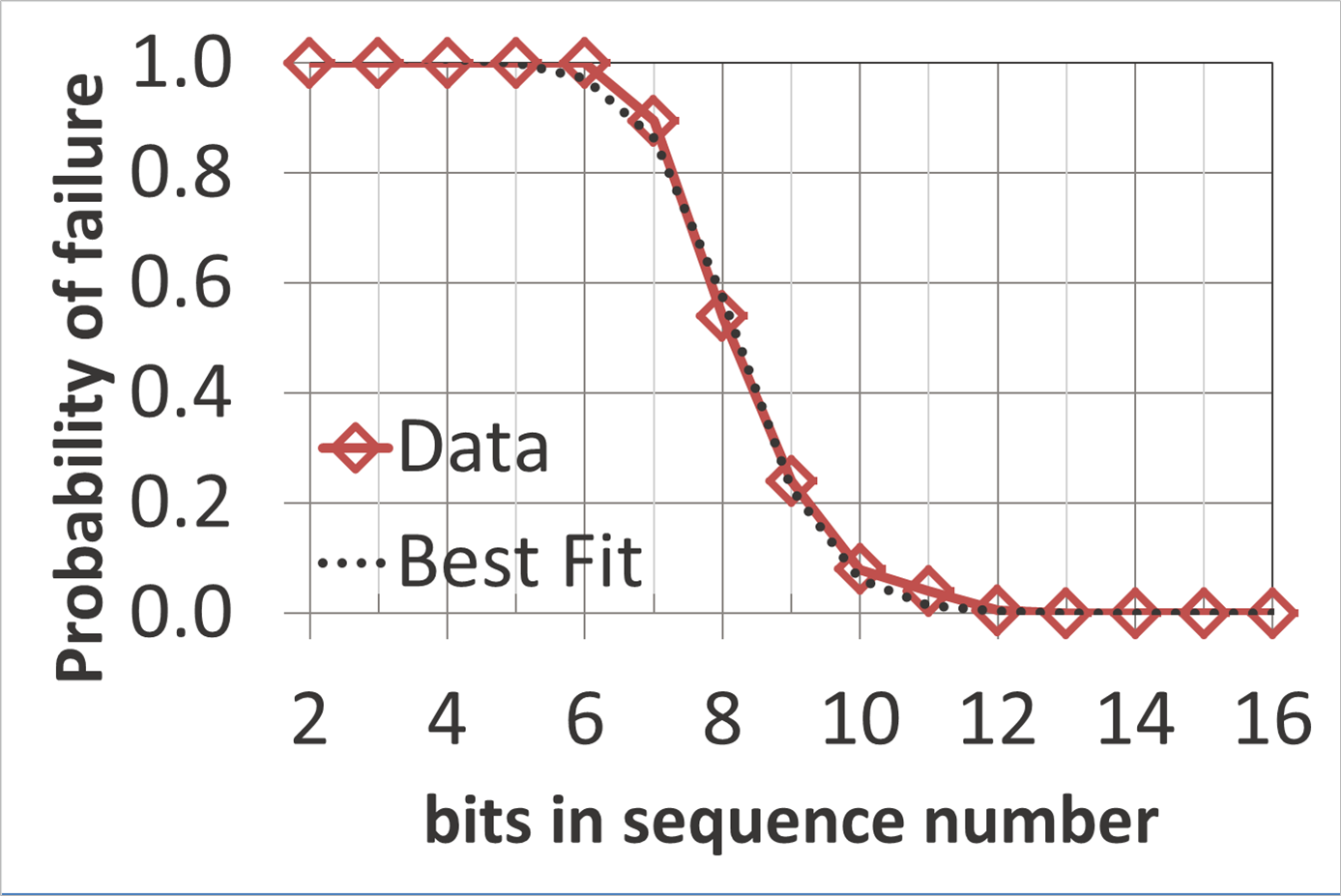} &
        \begin{tabular}{|r|l|}
            \hline
            Bits & E[time until error] \\
            \hline
            16 & 4.5 hours \\
            24 & 116 years \\
            32 & 26078192 years \\
            48 & $> 10^{18}$ years \\
            \hline
        \end{tabular}
    \end{tabular}
   % \vspace{-4mm}
    \caption{
Experiment studying sequence number wraparound. 
%Left: graph showing often program errors occur with sequence numbers of different bit-widths in a BST workload with 100\% updates.
%Right: table showing the expected time until an error occurs in this workload (using data extrapolated from a best-fit curve).
}
    \label{fig-exp-wraparound}
\vspace{-4mm}
\end{figure}

\begin{shortver}
We performed experiments on the larger AMD machine to study how frequently errors occur in a complex algorithm when sequence numbers of varying bit-widths experience wraparound.
For this experiment, we implemented a lock-free binary search tree (BST) using the technique of Brown~et~al.~\cite{Brown:2014}, and applied our transformation to it.
For each bit-width $B \in \{2,3,4,...,48\}$, we performed 200 trials in which 64 threads run concurrently. %for 100 milliseconds before terminating.
In each trial, all threads performed 50\% insertions and 50\% deletions on uniform random keys drawn from $[0, 10^5)$ until the tree contained approximately 50,000 keys.
Then, each thread began recording how many operations it performed, and ran for 100 milliseconds before terminating.
To help determine whether a trial was correct, each thread maintained a \textit{checksum} by adding each new key it successfully inserted, and subtracting each key it successfully deleted.
(At the end of a trial, the sum of these checksums must match the sum of keys in the tree.)
\end{shortver}
\begin{fullver}
We performed experiments on the larger AMD machine to study how frequently errors occur when sequence numbers of varying bit-widths experience wraparound.
For each bit-width $B \in \{2,3,4,...,48\}$, we performed 200 trials in which 64 threads run for 100 milliseconds before terminating.
Each trial was the same as a trial in our BST experiments with 100\% updates and key range $[0, 10^5)$.
\end{fullver}

We identified three different types of errors in these trials.
First, at the end of a trial, the sum of the checksums maintained by all threads would fail to match the sum of keys in the tree.
Second, threads would enter infinite loops due to the tree structure being corrupted, e.g., because a cycle was introduced.
(We identified this type of error by waiting until some thread had run twice as long as it should have.)
Third, an invalid memory access would cause immediate program failure (e.g., due to segmentation fault or bus error).

For each $B$ value, we divided the number of failed runs by 200 to estimate the probability of a trial failing.
A graph showing the resulting estimated probability distribution appears in Figure~\ref{fig-exp-wraparound}.
For small $B$ values, trials frequently experienced errors.
However, for $B \ge 13$, we did not observe a single error in 200 trials (despite the fact that wraparound consistently occurred in every trial).
For $B \ge 16$, trials were not sufficiently long for wraparound to consistently occur.
The results appear in Figure~\ref{fig-exp-wraparound}.

As is common in physics when studying unknown functions, we make an educated guess that the distribution is sigmoidal, 
%The measured distribution appears to be sigmoidal, 
which means it is of the form $f(x) = a/(1+e^{-b(x-c)})$ for constants $a, b$ and $c$.
We determined a sigmoidal curve of best fit from the data, obtaining the function $f(x) = 1/(1+e^{1.53969(x-8.199181)})$, which
%We used the statistical software package $R$ to fit a sigmoid function to our data.
%In order to achieve horizontal asymptotes of 0 and 1, it was necessary to fix $a = 1$.
%The resulting function, $f(x) = 1/(1+e^{1.53969(x-8.199181)})$,
is plotted as the \textit{Best Fit} curve on the graph in Figure~\ref{fig-exp-wraparound}.
As the graph shows, the error between the best fit curve and the measured data is extremely small.
Although we do not have a justification for the shape of this distribution, we think it is worthwhile to put forth a hypothesis and study its consequences. %properties under the assumption that the hypothesis is correct.

We used $f(x)$ to extrapolate on the data to estimate the expected time until an error occurs in this workload for several bit-widths that would be impractical to test experimentally.
These extrapolations appear in the table on the right of Figure~\ref{fig-exp-wraparound}.
They should be taken with a grain of salt, since the error in our estimation likely grows quickly with $B$.
However, the extrapolations suggest that even $B = 32$ would be quite safe for this workload.
To our knowledge, this kind of experimental exploration of the practicality of unbounded sequence numbers has not previously been done.

%Left: graph showing often program errors occur with sequence numbers of different bit-widths in a BST workload with 100\% updates.
%Right: table showing the expected time until an error occurs in this workload (using data extrapolated from a best-fit curve).

\end{fullver}
%!TEX root = paper.tex

\vspace{-2mm}
\section{Related Work} \label{sec-related}

Several papers have presented universal constructions or strong primitives for non-blocking algorithms in which operations create descriptors~\cite{Israeli:1994, Anderson:1995, Afek:1995, Moir:1997, Harris:2002, LMS09, JP05:opodis, Marathe:2008, Attiya:2011, Brown:2013}. % and help one another
A subset of these algorithms employ ad-hoc techniques for reusing descriptors~\cite{Israeli:1994, Anderson:1995, Afek:1995, Moir:1997, Marathe:2008, LMS09, JP05:opodis}.
The rest assume descriptors will be allocated for each operation and eventually reclaimed.

Most of the ad-hoc techniques for reusing descriptors have significant downsides.
Some are complex and tightly integrated into the underlying algorithm, or rely on highly specific algorithmic properties (e.g., that descriptors contain only a single word). %and it is not clear how they could be applied to other algorithms.
%Unfortunately, these algorithms are very complex, and it is difficult to see how their ad-hoc techniques for avoiding descriptor allocation could be applied to other implementations.  
Others use synchronization primitives that atomically operate on large words, which are not available on modern systems, and are inefficient when implemented in software.
Yet others introduce high space overhead (e.g., by attaching a sequence number to \textit{every} memory word).
Some techniques also incur significant runtime overhead (e.g., by invoking expensive synchronization primitives just to \textit{read fields} of a descriptor).
%Additionally, these techniques use large LL/SC objects, which are not available on many systems, and are inefficient when implemented in software.
%[could we say: we conjecture that simpler universal constructions could be designed if one did not need to worry about limiting the allocation of descriptors, or reclaiming them?]
Furthermore, these techniques give, at best, a vague idea of how one might reuse descriptors for arbitrary algorithms, and it would be difficult %a significant undertaking 
to determine how to use them in practice.
Our work avoids all of these downsides, and provides a concrete approach for transforming a large class of algorithms.

%\textbf{talk about kcss; effectively has single-word descriptors that it reuses using a fast obstruction free impl of ll/sc. not clear how it could be generalized to descriptors with more than 1 word}

Barnes~\cite{Barnes:1993} introduced a technique for producing non-blocking algorithms that can be more efficient (and sometimes simpler) than the universal constructions described above.
%Unlike these universal constructions,
With Barnes' technique, each operation creates a new descriptor.
Creating a new descriptor for each operation allows his technique to avoid the ABA problem while remaining conceptually simple.
Each operation conceptually locks each location it will modify by installing a pointer to its descriptor, and then performs it modifications and unlocks each location.
Barnes' technique is the inspiration for the class WCA. %of lock-free algorithms to which our transformation can be applied.
%Many algorithms have since been introduced that use variants of the 
Many algorithms have since been introduced using variants of this technique~\cite{Harris:2002, Ellen:2010, Attiya:2011, Howley:2012, Shafiei:2013, Brown:2013, Brown:2014}.
Several of these algorithms are quite efficient in practice despite the overhead of creating and reclaiming descriptors.
Our technique can significantly improve the space and time overhead of such algorithms.

\trevorlater{CITE MY PODC2017 PAPER HERE, ALSO}
%%%%%
Recent work has identified ways to use hardware transactional memory (HTM) to reduce descriptor allocation~\cite{Liu2015}.
Currently, HTM is supported only on recent Intel and IBM processors.
Other architectures, such as AMD, SPARC and ARM have not yet developed HTM support.
Thus, it is important to provide solutions for systems with no HTM support.
Additionally, even with HTM support, our approach is useful.
Current (and likely future) implementations of HTM offer no progress guarantees, so one must provide a lock-free fallback path to guarantee lock-free progress.
The techniques in~\cite{Liu2015} accelerate the HTM-based \textit{fast path}, but do nothing to reduce descriptor allocations on the fallback path.
In some workloads, many operations run on the fallback path, so it is important for it to be efficient.
Our work provides a way to accelerate the fallback path, and is orthogonal to work that optimizes the fast path.

The \textit{long-lived renaming} (LLR) problem is related to our work (see~\cite{Brodsky2011} for a survey), but its solutions do not solve our problem.
LLR provides processes with operations to \textit{acquire} one unique resource from a pool of resources, and subsequently \textit{release} it.
One could imagine a scheme in which processes use LLR to reuse a small set of descriptors by invoking \textit{acquire} instead of allocating a new descriptor, and eventually invoking \textit{release}.
Note, however, that a descriptor can safely be released only once it can no longer be accessed by any other process.
Determining when it is safe to release a descriptor is as hard as performing general memory reclamation, and would also require delaying the release (and subsequent acquisition) of a descriptor (which would increase the number of descriptors needed).
In contrast, our weak descriptors eliminate the need for memory reclamation, and allow immediate reuse.

%!TEX root = paper.tex

\begin{thesisonly}
\section{Summary}
\end{thesisonly}
\begin{thesisnot}
\section{Conclusion} 
\end{thesisnot}
\label{sec-desc-conclusion}

We presented a novel technique for transforming algorithms that throw away descriptors into algorithms that reuse descriptors.
Our experiments show that our transformation yields significant performance improvements for a lock-free $k$-CAS algorithm. % for a lock-free $k$-CAS algorithm, over a variety of workloads, when compared to implementations that perform memory reclamation on descriptors.
Furthermore, our transformation reduces peak memory usage %of the $k$-CAS algorithm 
by nearly three orders of magnitude over the next best implementation.
\begin{shortver}
We also applied our transformation to a lock-free binary search tree implemented using LLX and SCX, and saw significant performance improvements in update-heavy workloads. %a wide variety of experiments.
%Due to a lack of space, they appear in Appendix~\ref{appendix-exp}.
%The results show that our transformation improves performance by up to 57\% over the next best implementation.
\end{shortver}

\begin{fullver}
We also applied our transformation to a lock-free implementation of \llt\ and \sct, and studied its performance by doing rigorous experiments on a lock-free binary search tree that uses \llt\ and \sct. %, and performed rigorous experiments to study its performance.
These experiments demonstrated a significant performance advantage for our transformed algorithm in workloads that perform many updates.
Our transformed \llt\ and \sct\ algorithm has the potential to accelerate many algorithms that use \llt\ and \sct.
%The results show that our transformation can significantly improve the performance of algorithms implemented using \llt\ and \sct.
%For example, on the Intel system, in a microbenchmark where threads perform insertions and deletions in a tree containing approximately half a million keys, our transformed algorithm performed 57\% more operations per second than the next best implementation that uses state of the art memory reclamation to recycle descriptors.
\end{fullver}

We believe our transformation can be used to improve the performance and memory usage of many other algorithms that throw away descriptors.
Moreover, we hope that our extended weak descriptor ADT will aid in the design of more efficient, complex algorithms, by allowing algorithm designers to benefit from the conceptual simplicity of throwing away descriptors without paying the practical costs of doing so.

\begin{thesisnot}
\begin{fullver}
\section*{Acknowledgments}

We thank Faith Ellen for her gracious help in developing proofs for our transformations, and her insightful comments.
The paper is greatly improved as a result.
This work was supported by the Israel Science Foundation (grant 1749/14), the Yad-HaNadiv foundation, the Natural Sciences and Engineering Research Council of Canada, and Global Affairs Canada.
Maya Arbel-Raviv is supported in part by the Technion Hasso Platner Institute Research School.
\end{fullver}
\end{thesisnot}

\chapter{Accelerating the template with HTM} \label{chap-3path}
\begin{thesisonly}
\renewcommand{\llt}{LLX}
\renewcommand{\sct}{SCX}
\end{thesisonly}

\begin{thesisnot}
\section{Introduction}

Concurrent data structures are crucial building blocks in multi-threaded software.
There are many concurrent data structures implemented using locks, but locks can be inefficient, and are not fault tolerant (since a process that crashes while holding a lock can prevent all other processes from making progress).
%If a data structure is protected by a single global lock, it does not scale.
%
Thus, it is often preferable to use hardware synchronization primitives like compare-and-swap (\textsc{CAS}) instead of locks.
This enables the development of \textit{lock-free} (or \textit{non-blocking}) data structures, which guarantee that at least one process will always continue to make progress, even if some processes crash.
However, it is notoriously difficult to implement lock-free data structures from CAS, and this has inhibited the development of advanced lock-free data structures.

One way of simplifying this task is to use a higher level synchronization primitive that can atomically access multiple locations. %~\cite{Barnes:1993,Israeli:1994,ST97,Harris:2002}.
For example, consider a $k$-word compare-and-swap ($k$-CAS), which atomically: reads $k$ locations, checks if they contain $k$ expected values, and, if so, writes $k$ new values.
$k$-CAS is highly expressive, and it can be used in a straightforward way to implement \textit{any} atomic operation.
Moreover, it can be implemented from CAS and registers~\cite{Harris:2002}.
However, since $k$-CAS is so expressive, it is difficult to implement efficiently. %, and is thought by many to be impractical.

Brown et~al.~\cite{Brown:2013} developed a set of new primitives called \llt\ and \sct\ that are less expressive than $k$-CAS, but can still be used in a natural way to implement many advanced data structures.
%These primitives, \textit{load-link-extended} (\llt), \textit{validate-extended} (\vlt) and \textit{store-conditional-extended} (\sct), are generalizations of the well known single-word primitives \textit{load-link} (\textsc{LL}), \textit{validate} (\textsc{VL}) and \textit{store-conditional} (\textsc{SC}).
These primitives can be implemented much more efficiently than $k$-CAS.
At a high level, \llt\ returns a snapshot of a node in a data structure, and after performing \llt s on one or more nodes, one can perform an \sct\ to atomically: change a field of one of these nodes, and \textit{finalize} a subset of them, \textit{only if} none of these nodes have changed since the process performed \llt s on them.
Finalizing a node prevents any further changes to it, which is useful to stop processes from erroneously modifying deleted parts of the data structure.
%checks whether any node in a given set $V$ has changed since the process last performed \llt\ on it, and, if not, writes a new value to a field of a node in $V$, and \textit{finalizes} a connected subset of these nodes, so they can never be changed again.
%Brown et~al. presented an efficient implementation of these primitives from CAS and a full proof.
In a subsequent paper, Brown et~al. used \llt\ and \sct\ to design a \textit{tree update template} that can be followed to produce lock-free implementations of down-trees (trees in which all nodes except the root have in-degree one) with any kinds of update operations~\cite{Brown:2014}.
They demonstrated the use of the template by implementing a chromatic tree, which is an advanced variant of a red-black tree (a type of balanced binary search tree) that offers better scalability. %is more concurrency friendly. %suitable for a concurrent settings. %complex balanced binary search tree in which rebalancing can be performed in small atomic steps that can be freely interleaved with other updates.
The template has also been used to implement many other advanced data structures, including lists, relaxed AVL trees, relaxed ($a,b$)-trees, relaxed $b$-slack trees and weak AVL trees~\cite{Brown:2014,BrownPhD,He:2016}.
Some of these data structures are highly efficient, and would be well suited for inclusion in data structure libraries.

%\trevor{motivate the need for faster implementations of llx/scx/vlx and the template.}
In this work, we study how the new hardware transactional memory (HTM) capabilities found in recent processors (e.g., by Intel and IBM) can be used to produce significantly faster implementations of the tree update template. %\llt\ and \sct\ primitives, and the tree update template. %, that are significantly more efficient than those found in the literature.
\end{thesisnot}
\begin{thesisonly}
In this chapter, we study how the new hardware transactional memory (HTM) capabilities found in recent processors (e.g., by Intel and IBM) can be used to produce significantly faster implementations of the tree update template.
\end{thesisonly}
By accelerating %\llt, \sct\ and 
the tree update template, we also provide a way to accelerate all of the data structures that have been implemented with it.
Since library data structures are reused many times, even minor performance improvements confer a large benefit.

%Our implementations make use of new hardware transactional memory (HTM) capabilities that can be found in recent processors by, e.g., Intel and IBM.
%Herein, we consider Intel's implementation of HTM.
%
%Recently, processor manufacturers such as Intel and IBM have introduced hardware transactional memory (HTM) in their processors.
%\trevor{mention other manufacturers, and say we consider intel's impl.?}
HTM allows a programmer to run blocks of code in transactions, which either commit and take effect atomically, or abort and have no effect on shared memory.
Although transactional memory was originally intended to \textit{simplify} concurrent programming, researchers have since realized that HTM can also be used effectively to \textit{improve the performance of existing concurrent code}~\cite{Liu2015,timnat2015practical,makreshanski2015lock}:
Hardware transactions typically have very little overhead, so they can often be used to replace other, more expensive synchronization mechanisms.
For example, instead of performing a sequence of CAS primitives, it may be faster to perform reads, if-statements and writes inside a transaction. % can sometimes be accelerated by replacing it with a transaction that performs reads, if-statements and writes.
%
%This work is targeted at researchers and developers of library code, who design complex handcrafted data structures, and are always interested in advanced optimization techniques.
%Our goal is to show them a new way to use HTM as a tool to further accelerate their handcrafted data structures.
Note that this represents a \textit{non-standard use of HTM}: we are \textit{not} interested in its ease of use, but, rather, in its ability to reduce synchronization costs.

Although hardware transactions are fast, it is surprisingly difficult to obtain the full performance benefit of HTM.
Here, we consider Intel's HTM, which is a \textit{best-effort} implementation.
This means it offers \textit{no guarantee} that transactions will ever commit.
Even in a single threaded system, a transaction can repeatedly abort because of internal buffer overflows, page faults, interrupts, and many other events.
So, to guarantee progress, any code that uses HTM must also provide a \textit{software fallback path} to be executed if a transaction fails.
%As we will see, 
The design of the fallback path %decision of whether to allow operations on the fallback path to run concurrently with hardware transactions 
profoundly impacts the performance of HTM-based algorithms.

\fakeparagraph{Allowing concurrency between two paths}
Consider an operation $O$ that is implemented using the tree update template.
One natural way to use HTM to accelerate $O$ is to use the original operation as a fallback path, and then obtain an HTM-based fast path by wrapping $O$ in a transaction, and performing optimizations to improve performance~\cite{Liu2015}.
We call this the \textbf{2-path concurrent} algorithm (\textit{2-path con}).
Since the fast path is just an optimized version of the fallback path, transactions on the fast path and fallback path can safely run concurrently.
If a transaction aborts, it can either be retried on the fast path, or be executed on the fallback path.
%However, \textit{supporting concurrency between the fast path and fallback path adds significant overhead}.
Unfortunately, supporting concurrency between the fast path and fallback path can add significant overhead on the fast path.

The first source of overhead is \textit{instrumentation} on the fast path that manipulates the \textit{meta-data} used by the fallback path to synchronize processes.
For example, lock-free algorithms often create a \textit{descriptor} for each update operation (so that processes can determine how to help one another make progress), and store pointers to these descriptors in \rec s, where they act as locks.
The fast path must also manipulate these descriptors and pointers so that the fallback path can detect changes made by the fast path.
%(For lock-based algorithms, the fast path typically must \textit{read} locks, but can avoid actually acquiring them.)

The second source of overhead comes from constraints imposed by algorithmic assumptions made on the fallback path. % on the fast path. % can operate
%Additionally, if the fast path and fallback path run concurrently, then %In addition to the overhead of manipulating synchronization meta-data on the fast path, %Since the algorithms for the fast path and fallback path run concurrently,
%the algorithm for the fallback path imposes constraints on the way that the fast path algorithm can function, and these constraints can introduce overhead.
The tree update template implementation in~\cite{Brown:2014} assumes that only child pointers can change, and all other fields of nodes, such as keys and values, are never changed (after a node is created).
%suppose we used the lock-free, unbalanced binary search tree (BST) of Ellen~at~al.~\cite{Ellen:2010} as a fallback path.
%In this algorithm, processes can directly modify the child pointers of a node, but cannot directly change other fields (such as its key or value).
Changes to these other \textit{immutable} fields must be made by replacing a node with a new copy that reflects the desired change.
%Processes communicate with one another using \textit{flagging} (which indicates that a node will have one of its child pointers changed), and \textit{marking} (which indicates that a node will be deleted).
%
Because of this assumption on the fallback path, transactions on the fast path \textit{cannot} directly change any field of a node other than its child pointers. % (if they run concurrently with transactions on the fallback path).
This is because the fallback path has no mechanism to detect such a change (and may, for example, erroneously delete a node that is concurrently being modified by the fast path).
%Thus, the fast path must create new nodes to create immutable fields, which makes it much less efficient.
%
%Any fast path that is paired with this algorithm as its fallback path cannot directly change any field of a node other than its child pointers, because the fallback path has no mechanism to detect such a change (and may, e.g., erroneously delete a node that is concurrently being modified by the fast path).
Thus, just like the fallback path, the fast path must replace a node with a new copy to change one of its immutable fields, which can be much less efficient than changing the field directly.
%This problem is not unique to the BST of Ellen~et~al.
%In fact, the exact same constraint would be also imposed by many other lock-free tree algorithms~\cite{BH11,Natarajan:2014,PBO11,Shafiei:2013}.

\fakeparagraph{Disallowing concurrency between two paths}
To avoid the overheads described above, concurrency is often \textit{disallowed} between the fast path and fallback path.
The simplest example of this approach is a technique called \textbf{transactional lock elision} (TLE) \cite{rajwar2001speculative,rajwar2002transactional}.
TLE is used to implement an operation by wrapping its sequential code in a transaction, and falling back to acquire a global lock after a certain number of transactional attempts.
At the beginning of each transaction, a process reads the state of the global lock and aborts the transaction if the lock is held (to prevent inconsistencies that might arise because the fallback path is not atomic).
Once a process begins executing on the fallback path, all concurrent transactions abort, and processes wait until the fallback path is empty before retrying their transactions.

If transactions never abort, then \textit{TLE represents the best performance we can hope to achieve}, because the fallback path introduces almost no overhead and synchronization is performed entirely by hardware.
Note, however, that TLE is not lock-free.
%If transactions abort infrequently, then processes rarely execute on the fallback path, and a high degree of concurrency can be achieved.
%In practice, TLE performs exceptionally well in workloads that experience few aborts~\cite{afek2014reduced,diegues2014virtues,yoo2013performance}.
%In fact, TLE is so effective that straightforward implementations can achieve significantly higher performance in some workloads than state of the art lock-free algorithms.
Additionally, in workloads where operations periodically run on the fallback path, performance can be very poor.

As a toy example, consider a TLE implementation of a binary search tree, with a workload consisting of insertions, deletions and \textit{range queries}.
A range query returns all of the keys in a range [$lo, hi$).
Range queries access many memory locations, and cause frequent transactional aborts due to internal processor buffer overflows (capacity limits).
Thus, range queries periodically run on the fallback path, where they can lead to numerous performance problems.
Since the fallback path is sequential, range queries (or any other long-running operations) cause a severe \textit{concurrency bottleneck}, because they prevent transactions from running on the fast path while they slowly complete, serially.

One way to mitigate this bottleneck is to replace the sequential fallback path in TLE with a lock-free algorithm, and replace the global lock with a fetch-and-increment object $F$ that counts how many operations are running on the fallback path. % (or, alternatively, use a scalable non-zero indicator~\cite{ellen2007snzi}).
Instead of aborting if the lock is held, transactions on the fast path abort if $F$ is non-zero.
We call this the \textbf{2-path non-concurrent} algorithm (\textit{2-path $\overline{con}$}).
In this algorithm, if transactions on the fast path retry only a few times before moving to the fallback path, or do not wait between retries for the fallback path to become empty, then the lemming effect~\cite{Dice2009} can occur. (The lemming effect occurs when processes on the fast path rapidly fail and move to the fallback path, simply because other processes are on the fallback path.)
This can cause the algorithm to run only as fast as the (much slower) fallback path.
However, if transactions avoid the lemming effect by retrying many times before moving to the fallback path, and waiting between retries for the fallback path to become empty, then processes can spend most of their time \textit{waiting}.
%
%The only way to avoid all of these performance pathologies is to allow transactions to continue to execute on the fast path \textit{while} transactions are executing on the fallback path.
%However, as we discussed above, this can involve high overhead from instrumentation on the fast path.
\begin{fullver}
The performance problems discussed up to this point are summarized on the left side of Figure~\ref{fig-problem}.
\end{fullver}

\fakeparagraph{The problem with two paths}
In this paper, we study two different types of workloads: \textbf{light workloads}, in which transactions rarely run on the fallback path, and \textbf{heavy workloads}, in which transactions more frequently run on the fallback path.
In light workloads, algorithms that allow concurrency between paths perform very poorly (due to high overhead) in comparison to algorithms that disallow concurrency.
However, in heavy workloads, algorithms that disallow concurrency perform very poorly (since transactions on the fallback path prevent transactions from running on the fast path) in comparison to algorithms that allow concurrency between paths.
Consequently, all two path algorithms have workloads that yield poor performance.
Our experiments confirm this, showing surprisingly poor performance for two path algorithms in many cases.
%
%are precisely the workloads where data structures can benefit from HTM \textit{and} the interaction between the fast path and fallback path matters.

\begin{fullver}
\begin{figure}
%    \vspace{-7mm}
    \centering
    \begin{minipage}{0.5\textwidth}
        \centering
        \includegraphics[width=\linewidth]{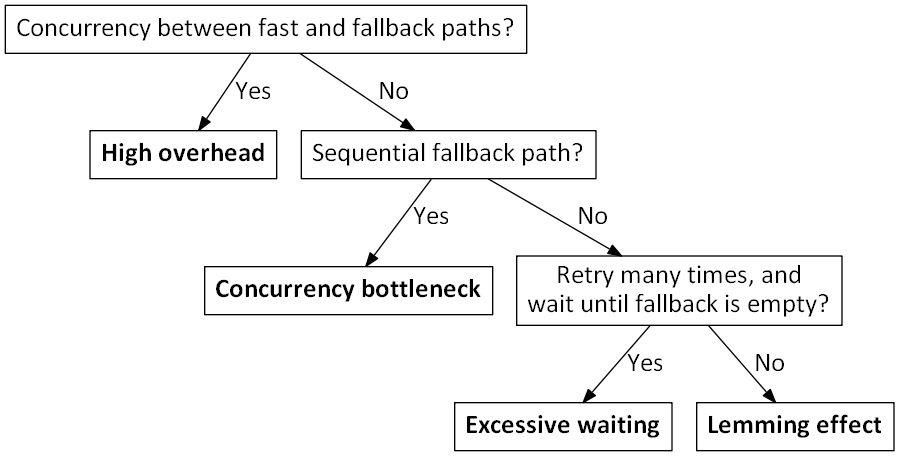}
    \end{minipage}
    \hspace{10mm}
    \begin{minipage}{0.4\textwidth}
        \centering
        \includegraphics[width=\linewidth]{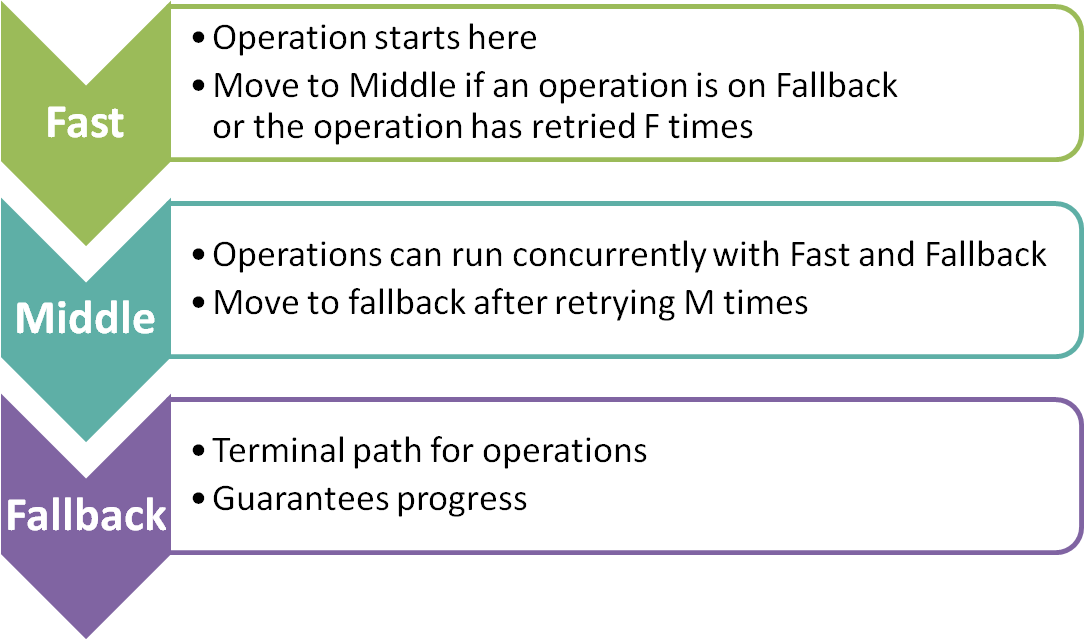}
    \end{minipage}
\caption{(Left) Performance problems affecting two-path algorithms. (Right) Using three execution paths.}
%\caption{Using three execution paths.}
\label{fig-problem}
\end{figure}
\end{fullver}

\fakeparagraph{Using three paths}
We introduce a technique that simultaneously achieves high performance for both light and heavy workloads by using three paths: an HTM fast path, an HTM middle path and a non-transactional fallback path.
\begin{fullver}
(See the illustration on the right side of Figure~\ref{fig-problem}.)
\end{fullver}
Each operation begins on the fast path, and moves to the middle path after it retries $F$ times.
\begin{shortver}
\begin{wrapfigure}{r}{0.35\textwidth}
    \vspace{-4mm}
    \hspace{-2mm}
    \includegraphics[width=\linewidth]{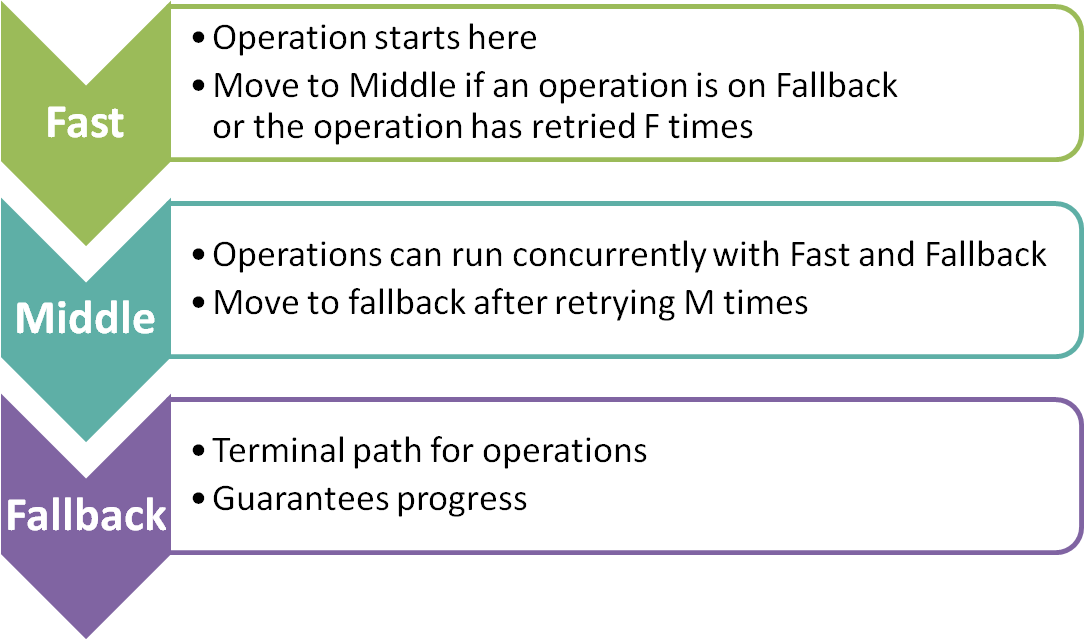}
    \vspace{-4mm}
\end{wrapfigure}%
\end{shortver}
An operation on the middle path moves to the fallback path after retrying $M$ times on the middle path.
The fast path does not manipulate any synchronization meta-data used by the fallback path, so operations on the fast path and fallback path cannot run concurrently. 
Thus, whenever an operation is on the fallback path, all operations on the fast path move to the middle path.
The middle path manipulates the synchronization meta-data used by the fallback path, so operations on the middle path and fallback path can run concurrently. 
Operations on the middle path can also run concurrently with operations on the fast path (since conflicts are resolved by the HTM system).
%Once all operations on the fallback path finish, new operation attempts start on the fast path.
We call this the \textbf{3-path} algorithm (\textit{3-path}).

We briefly discuss why this approach avoids the performance problems described above. %satisfies our goals.
Since transactions on the fast path do not run concurrently with transactions on the fallback path, transactions on the fast path run with \textit{no instrumentation overhead}.
%the fallback path does not impose any overhead on the fast path.
When a transaction is on the fallback path, transactions can freely execute on the middle path, \textit{without waiting}.
The \textit{lemming effect} does not occur, since transactions do not have to move to the fallback path simply because a transaction is on the fallback path.
Furthermore, we enable a high degree of concurrency, because the fast and middle paths can run concurrently, and the middle and fallback paths can run concurrently.

We performed experiments to evaluate our new template algorithms by comparing them with the original template algorithm.
In order to compare the different template algorithms, we used each algorithm to implement two data structures: a binary search tree (BST) and a relaxed ($a,b$)-tree.
We then ran microbenchmarks to compare the performance (operations per second) of the different implementations in both light and heavy workloads.
The results show that our new template algorithms offer significant performance improvements.
For example, on an Intel system with 72 concurrent processes, our best implementation of the relaxed ($a,b$)-tree outperformed the implementation using the original template algorithm by an average of 410\% over all workloads.

\medskip

\noindent\textbf{Contributions}
\begin{compactitem}
%    \item We study heavy workloads, an often ignored class of workloads that we believe are important for assessing the performance of algorithms that use HTM.
    \item We present four accelerated implementations of the tree update template
\begin{thesisnot}
    of Brown~et~al.
\end{thesisnot}
    that explore the design space for HTM-based implementations: \textit{2-path con}, \textit{TLE}, \textit{2-path $\overline{con}$}, and \textit{3-path}.
%    \item We identify a number of performance issues affecting the HTM-based implementations that use two paths, and explain how a third execution path can be used to circumvent these problems.
    \item We highlight the importance of studying both light and heavy workloads in the HTM setting. Each serves a distinct role in evaluating algorithms: light workloads demonstrate the potential of HTM to improve performance by reducing overhead, and heavy workloads capture the performance impact of interactions between different execution paths.
    \item We demonstrate the effectiveness of our approach by accelerating two different lock-free data structures: an unbalanced BST, and a relaxed ($a,b$)-tree.
    Experimental results show a significant performance advantage for our accelerated implementations.
%    \item We present experimental results that show a significant performance advantage for our accelerated implementations over their 
%
%their instrumented and uninstrumented two path counterparts, and over the original algorithms. %For example, on a 72-thread Intel system, our three path ($a,b$)-tree implementation outperforms a two path implementation by up to 251\% (averaging a 127\% improvement), and outperforms a handcrafted lock-free implementation by up to 421\% (averaging a 216\% improvement).
    %neelam [USE THIS ONE]
    %Row Labels	Max of 3path/2path	Average of 3path/2path	Max of 3path/tle	Average of 3path/tle	Max of 3path/nonhtm	Average of 3path/nonhtm
    %abtree	251.0%	127.3%	133.3%	37.0%	420.8%	215.5%
\end{compactitem}
\medskip

\begin{thesisonly}
The remainder of this chapter is structured as follows.
\end{thesisonly}
\begin{thesisnot}
The remainder of the paper is structured as follows.
The model is introduced in Section~\ref{sec-3path-model}.
Section~\ref{sec-3path-background} describes \llt\ and \sct, and the tree update template.
\end{thesisnot}
\begin{fullver}
We describe an HTM-based implementation of \llt\ and \sct\ in Section~\ref{sec-htmscx}. %, which is used by our first accelerated template implementation.
\end{fullver}
In Section~\ref{sec-3path-algs}, we describe our four template implementations, and argue correctness and progress.
%In Section~\ref{sec-3path-alg2}, we describe the TLE-based algorithm, and the 2-path non-concurrent algorithm.
%In Section~\ref{sec-3path-alg3}, we describe the 3-path algorithm.
\begin{fullver}
In Section~\ref{sec-3path-ds}, we describe two data structures that we use in our experiments. %as examples to demonstrate the effectiveness of our accelerated template implementations.
\end{fullver}
Experimental results %from three different systems
are presented in Section~\ref{sec-3path-exp}.
\begin{fullver}
In Section~\ref{sec-3path-memrecl}, we describe a way to reclaim memory more efficiently for \textit{3-path} algorithms.
\end{fullver}
Related work is surveyed in Section~\ref{sec-3path-related}.
\begin{fullver}
%In Section~\ref{sec-3path-hybridnorec}, we compare the performance of our approach with that of hybrid transactional memory. %one of our accelerated implementations with an implementation describe how one of the data structures in Section~\ref{sec-3path-ds} could be implemented using a fast hybrid transactional memory algorithm, and present experiments that compare the performance of such an implementation to our accelerated implementations.
%In Section~\ref{sec-3path-other-uses}, we describe two other potential uses for our \textit{3-path} algorithms: % by showing how one could 
%accelerating data structures that use the read-copy-update (RCU) or $k$-compare-and-swap primitives.
In Section~\ref{sec-3path-other-uses}, we describe how our approach could be used to accelerate data structures that use the read-copy-update (RCU) or $k$-compare-and-swap primitives.
%In Section~\ref{sec-3path-search-nontxn}, we present an optimization to \textit{3-path} template algorithms that allows an operation to traverse the tree outside of a transaction. %searches phase of an operation to be performed outside of its transaction.
\end{fullver}
\begin{thesisnot}
Finally, we conclude in Section~\ref{sec-3path-conclusion}.
\end{thesisnot}
\begin{thesisonly}
Finally, we summarize in Section~\ref{sec-3path-conclusion}.
\end{thesisonly}

\begin{thesisnot}
%\vspace{-2mm}
\section{Model} \label{sec-3path-model}

We consider an asynchronous shared memory system with $n$ processes, and Intel's implementation of HTM.
Arbitrary blocks of code can be executed as transactions, which either commit (and appear to take place instantaneously) or abort (and have no effect on the contents of shared memory).
A transaction is started by invoking \textit{txBegin}, is committed by invoking \textit{txEnd}, and can be aborted by invoking \textit{txAbort}.
Intel's implementation of HTM is best-effort, which means that the system can force transactions to abort at any time, and no transactions are ever guaranteed to commit.
Each time a transaction aborts, the hardware provides a reason why the abort occurred.
Two reasons are of particular interest.
\textit{Conflict} aborts occur when two processes contend on the same cache-line.
Since a cache-line contains multiple machine words, \textit{conflict} aborts can occur even if two processes never contend on the same memory location.
\textit{Capacity} aborts occur when a transaction exhausts some shared resource within the HTM system.
Intuitively, this occurs when a transaction accesses too many memory locations.
(In reality, \textit{capacity} aborts also occur for a variety of complex reasons that make it difficult to predict when they will occur.)
%Some common data structure operations, such as range queries, are highly likely to experience \textit{capacity} aborts on the fast path.
%Aborts can also be manually triggered within a transaction by invoking \textit{Abort}. %a primitive called \textit{Abort}. %(c), where $c$ is a reason for the abort (that is provided by the programmer).
%Whenever one of the fast path algorithms we discuss in this paper aborts because it detects that a process is on the fallback path, the abort is caused explicitly by our code, and we specify \textit{fallback} as the reason.

\section{Background} \label{sec-3path-background}

\fakeparagraph{The \llt\ and \sct\ primitives}
The load-link extended (\llt) and store-conditional extended (\sct) primitives are multi-word generalizations of the well-known load-link (LL) and store-conditional (SC), and they have been implemented from single-word \cas\ \cite{Brown:2013}.
\llt\ and \sct\ operate on \rec s, each of which consists of a fixed number of mutable fields (which can change), and a fixed number of immutable fields (which cannot). 

\llt($r$) attempts to take a snapshot of the mutable fields of a \rec\ $r$.
If it is concurrent with an \sct\ involving~$r$, it may return \fail, instead.
Individual fields of a \rec\ can also be read directly.
An \sct($V,$ $R,$ $fld,$ $new$) takes as its arguments a sequence $V$ of \rec s, a subsequence $R$ of $V$, a pointer $fld$ to a mutable field of one \rec\ in~$V$, and a new value $new$ for that field.
The \sct\ tries to atomically store the value $new$ in the field that $fld$ points to and {\it finalize} each \rec\ in $R$.
Once a \rec\ is finalized, its mutable fields cannot be changed by any subsequent \sct, and any \llt\ of the \rec\ will return \finalized\ instead of a snapshot.

Before a process $p$ invokes \sct, it must perform an \llt($r$) on each \rec\ $r$ in $V$.
For each $r \in V$, the last \llt($r$) performed by $p$ prior to the \sct\ is said to be {\it linked} to the \sct, and this linked \llt\ must return a snapshot of $r$ (not \fail\ or \finalized).
An \sct($V, R, fld, new$) by a process modifies the data structure and returns \true\ (in which case we say it \textit{succeeds}) only if no \rec\ $r$ in $V$ has changed since its linked \llt($r$); otherwise the \sct\ fails and returns \false.
Although \llt\ and \sct\ can fail, their failures are limited in such a way that they can be used to build data structures with lock-free progress.
See \cite{Brown:2013} for a more formal specification. % of these primitives.

\begin{shortver}
We briefly describe the implementation of \sct.
Each \rec\ is augmented with two fields: $marked$ and $\info$.
Each \sct($V,$ $R,$ $fld,$ $new$) starts by creating an \op\ $D$, which contains all of the information necessary to perform the \sct, and then completes the \sct\ by invoking a function called \func{Help}, and passing $D$ as its argument.
%The \op\ also contains a $state$ field, which initially contains the value InProgress.
%When the \sct\ finishes, the $state$ field will contain either Committed or Aborted, depending on whether the \sct\ succeeded.
%
Invocations of \sct\ synchronize with one another by taking a special kind of lock on each \rec\ in $V$.
These locks grant exclusive access to an \textit{\sct\ operation}, rather than to a \textit{process}.
%Henceforth, we use the term \textit{freezing} (resp., unfreezing), instead of locking (resp., unlocking), to differentiate this kind of locking from typical mutual exclusion.
%A \rec\ $u$ is \textit{frozen} for an \sct\ $S$ with \op\ $U$ if $u.\info$ points to $U$, and either $U.state =$ Committed and $u.marked = true$ (in which case we say $u$ is \textit{finalized}), or $U.state =$ InProgress.
%
An \sct\ $S$ locks a \rec\ $u$ by using CAS to store a pointer to its \op\ in $u.\info$. %(at the \fcas\ step in Figure~\ref{code-3path-scxo}).
Suppose $S$ successfully locks all \rec s in its $V$ sequence.
Then, $S$ finalizes each \rec\ $u \in R$ by setting $u.marked := \true$, and releases its locks on all \textit{other} \rec s (leaving finalized \rec s permanently locked). %a \textit{marked} bit in $u$ (at the \markstep\ in Figure~\ref{code-3path-scxo}).
Whenever a process encounters a (non-finalized) node locked for $S$, it invokes \func{Help}($D$) to help $S$ complete and release its locks.
\end{shortver}
Observe that \sct\ can only change a single value in a \rec\ (and finalize a sequence of \rec s) atomically.
Thus, to implement an \textit{operation} that changes multiple fields, one must create \textit{new} \rec s that contain the desired changes, and use \sct\ to change \textit{one} pointer to replace the old \rec s. % with the new ones.
%(One also uses the same approach for operations that change immutable fields.)

%
%Under these constraints, the implementation of \llt,  \sct, and \vlt\ in \cite{Brown:2013} guarantees that there is a linearization of all \sct s that modify the data structure (which may include \sct s that do not terminate because a process crashed, but \textit{not} any \sct s that fail), and all \llt s that return, but do not fail.

%\textbf{Semantics of LLX and SCX.}
%LLX($u$) returns a snapshot of all the fields of node $u$.
%SCX($V,$ $R,$ $fld,$ $new$) takes, as its arguments, a sequence $V$ of nodes, a subsequence $R$ of $V$, a field $fld$ of a node in $V$, and a value $new$.
%A process performs LLX on a sequence of nodes in $V$, then performs SCX($V,$ $R,$ $fld,$ $new$) which succeeds only if none of the nodes in $V$ have changed since the process last performed LLX on them.
%If the SCX succeeds, it atomically changes $fld$ to $new$ and \textit{finalizes} all of the nodes in $R$ (which prevents them from ever changing again).
%Finalizing makes it easy to ensure that a node is not erroneously changed by another operation after a \textit{Delete} removes it from the data structure.
%These primitives are implemented from single word CAS, so they can be implemented on most modern systems.

\begin{fullver}
We now describe the implementation of \llt\ and \sct.
%\begin{figure}[tb]
%\begin{framed}
%\hspace{1mm}
%\begin{minipage}{0.54\linewidth}
%\def\namewidth{17mm}
%\small
%\preplisting
%\begin{lstlisting}[mathescape=true,style=nonumbers]
% type $\op$
%   //\wcnarrow{$V$}{sequence of \rec s}
%   //\wcnarrow{$R$}{subsequence of $V$ to be finalized}
%   //\wcnarrow{$fld$}{pointer to a field of a \rec\ in $V$}
%   //\wcnarrow{$new$}{value to be written into the field $fld$}
%   //\wcnarrow{$old$}{value previously read from the field $fld$} 
%   //\wcnarrow{$state$}{one of \{\freezing, \done, \retry\}}
%   //\wcnarrow{$\freezingdone$}{Boolean}
%   //\wcnarrow{$\llresults$}{sequence of pointers, one read from}
%   //\wcnarrow{\mbox{ }}{the \info\ field of each element of $V$}
%\end{lstlisting}
%\end{minipage}
%\begin{minipage}{0.44\linewidth}
%\def\namewidth{18mm}
%\small
%\preplisting
%\begin{lstlisting}[mathescape=true,style=nonumbers]
% type $\rec$
%   //\com User-defined fields 
%   //\wcnarrow{$m_1, \ldots, m_y$}{mutable fields}
%   //\wcnarrow{$i_1, \ldots, i_z$}{immutable fields}
%   //\com Fields used by  \llt /\sct\ algorithm
%   //\wcnarrow{$\info$}{pointer to an \op}
%   //\wcnarrow{$marked$}{Boolean}
%   //\vspace{8mm}
%\end{lstlisting}
%\end{minipage}
%\end{framed}
%    \vspace{-5mm}
%	\caption{Type definitions for shared objects used to implement \llt, \sct, and \vlt.}
%	\label{code-3path-scxo-data}
%\end{figure}
%
\begin{figure*}[h!]
%\vspace{-3mm}
\small
\def\pwidth{4cm}
\prepnewlisting
%\hrule
\begin{framed}
\vspace{-1.5mm}
\hspace{5mm}
\begin{minipage}{0.57\linewidth}
\def\namewidth{17mm}
\preplisting
\footnotesize
\begin{lstlisting}[mathescape=true,style=nonumbers]
 type $\op$
   //\wcnarrow{$V$}{sequence of \rec s}
   //\wcnarrow{$R$}{subsequence of $V$ to be finalized}
   //\wcnarrow{$fld$}{pointer to a field of a \rec\ in $V$}
   //\wcnarrow{$new$}{value to be written into the field $fld$}
   //\wcnarrow{$old$}{value previously read from the field $fld$} 
   //\wcnarrow{$state$}{one of \{\freezing, \done, \retry\}}
   //\wcnarrow{$\llresults$}{sequence of pointers read from $r.\info$ for each $r \in V$}
   //\wcnarrow{$\freezingdone$}{Boolean}
\end{lstlisting}
\end{minipage}
\begin{minipage}{0.38\linewidth}
\def\namewidth{18mm}
\preplisting
\footnotesize
\begin{lstlisting}[mathescape=true,style=nonumbers]
 type $\rec$
   //\com User-defined fields 
   //\wcnarrow{$m_1, \ldots, m_y$}{mutable fields}
   //\wcnarrow{$i_1, \ldots, i_z$}{immutable fields}
   //\com Fields used by  \llt /\sct\ algorithm
   //\wcnarrow{$\info$}{pointer to an \op}
   //\wcnarrow{$marked$}{Boolean}
   //\vspace{4.5mm}
\end{lstlisting}
\end{minipage}
\vspace{-1.5mm} \hrule \vspace{-1mm}
\footnotesize
\begin{lstlisting}[mathescape=true]
  //\llt$_O(r)$ by process $p$
    $marked_1 := r.marked$ // \label{ll-read-marked1}
    $r\info := r.\info$ // \label{ll-read} 
    $state := r\info.state$ // \label{ll-read-state}
    $marked_2 := r.marked$ // \label{ll-read-marked2}
    if $state = \retry$ or $(state = \done$ and not $marked_2)$ then //  \label{ll-check-frozen} \sidecom{if $r$ was not frozen at line~\ref{ll-read-state}}
      read $r.m_1,...,r.m_y$ //and record the values in local variables $m_1,...,m_y$%
      \label{ll-collect}
      if $r.\info = r\info$ then//\label{ll-reread}\sidecom{if $r.\info$ points to the same} 
        //store $\langle r, r\info, \langle m_1, ..., m_y \rangle \rangle$ in $p$'s local table %
\sidecom{\op\ as on line~\ref{ll-read}}\label{ll-store}
        return $\langle m_1, ..., m_y \rangle$ // \label{ll-return}  \vspace{2mm}
    if ($r\info.state = \done$ or ($r\info.state = \freezing$ and $\help(r\info)))$ and $marked_1$ then// \label{ll-check-finalized}
      return $\finalized$ // \label{ll-return-finalized}
    else
      if $r.\info.state = \freezing$ then $\help(r.\info)$ // \label{ll-help-fail} 
      return $\fail$ // \label{ll-return-fail} \vspace{1mm} \hrule %
\vspace{1mm}
  //\sct$_O(V, R, fld, new)$ by process $p$
  //\tline{\com Preconditions: (\presctlinked) for each $r$ in $V$, $p$ has performed an invocation $I_r$ of \llt$(r)$ linked to this \sct}%
          {\hspace{19.5mm}(\presctabainit) $new$ is not the initial value of $fld$}%
          {\hspace{19.5mm}(\presctaba) for each $r$ in $V$, no $\sct(V', R', fld, new)$ was linearized before $I_r$ was linearized}
    //\dline{Let $\llresults$ be a pointer to a table in $p$'s private memory containing,}%
            {for each $r$ in $V$, the value of $r.\info$ read by $p$'s last \llt$(r)$}
    //Let $old$ be the value for $fld$ returned by $p$'s last \llt$(r)$\vspace{1.5mm}%
    return $\help(\mbox{pointer to new \op} (V, R, fld, new, old, \freezing,  \false, \llresults ))$ // \label{sct-create-op}\label{sct-call-help} \vspace{1mm} \hrule %
\vspace{1mm}
  //\help$(scxPtr)$ 
    //\com \mbox{Freeze all \rec s in $scxPtr.V$ to protect their mutable fields from being changed by other \sct s}
    for each $r$ in $scxPtr.V \mbox{ enumerated in order}$ do//\label{help-fcas-loop-begin}
      //Let $r$\info\ be the pointer indexed by $r$ in $scxPtr.\llresults$ \label{help-rinfo}
      if not $\cas(r.\info,r\info,scxPtr)$ then //\sidecom{\textbf{\fcas}}\label{help-fcas}
        if $r.\info \neq scxPtr$ then // \label{help-check-frozen} 
          //\com \mbox{Could not freeze $r$ because it is frozen for another \sct}
          if $scxPtr.\freezingdone = \true$ then//\sidecom{\textbf{\fcstep}}\label{help-fcstep}
            //\com the \sct\ has already completed successfully 
            return $\true$ // \label{help-return-true-loop} 
          else
            //\com Atomically unfreeze all \rec s frozen for this \sct 
            $scxPtr.state := \retry$ //\sidecom{\textbf{\astep}}\label{help-astep}
            return $\false$ // \label{help-return-false} \vspace{2mm}
    //\com Finished freezing \rec s (Assert: $state \in \{\freezing, \done\}$) 
    $scxPtr.\freezingdone := \true$//\sidecom{\textbf{\fstep}}\label{help-fstep}
    for each $r \in scxPtr.R$ do $r.marked := \true$ //\sidecom{\textbf{\markstep}}\label{help-markstep}
    //$\cas(scxPtr.fld, scxPtr.old, scxPtr.new)$ \sidecom{\textbf{\upcas}}\label{help-upcas} \vspace{2mm}
    //\com Finalize all $r$ in $R$, and unfreeze all $r$ in $V$ that are not in $R$ 
    $scxPtr.state := \done$//\sidecom{\textbf{\cstep}}\label{help-cstep}
    return $\true$ // \label{help-return-true}
\end{lstlisting}
\vspace{-1mm}
\end{framed}
    \vspace{-6mm}
	\caption{Data types and pseudocode for the original \llt\ and \sct\ algorithm.}
	\label{code-3path-scxo}
\end{figure*}
%
%%\noindent
%\textbf{Implementation from CAS.}
%We give a high level description of how SCX is implemented from CAS to facilitate an explanation of how we optimized it. %since it will be useful when we are discussing optimizations on the middle path.
%This description will also be useful when we describe certain optimizations that are possible on the fast path are \textit{not} possible on the middle path.
\begin{thesisnot}
Pseudocode appears %for the original, CAS-based implementation of \llt\ and \sct\ appears 
in Figure~\ref{code-3path-scxo}.
\end{thesisnot}
Each invocation $S$ of \sct$_O(V, R, fld, new)$ starts by creating an \op\ $D$, which contains all of the information necessary to perform $S$, and then invokes \func{Help}$(D)$ to perform it.
The \op\ also contains a $state$ field, which initially contains the value \freezing.
When $S$ finishes, the $state$ field of $D$ will contain either \done\ or \retry\, depending on whether the $S$ succeeded.

$S$ synchronizes with other invocations of \sct$_O$ by taking a special kind of lock on each \rec\ in $V$.
These locks grant exclusive access to an \textit{operation}, rather than to a \textit{process}.
Henceforth, we use the term \textit{freezing} (resp., unfreezing), instead of locking (resp., unlocking), to differentiate this kind of locking from typical mutual exclusion.
A \rec\ $u$ is \textit{frozen} for $S$ if $u.\info$ points to $D$, and either $D.state =$ \done\ and $u.marked = true$ (in which case we say $u$ is \textit{finalized}), or $D.state = \freezing$.
%Note that $u$ is additionally \textit{finalized} if $D.state =$ \done\ and $u.marked = true$.

So, $S$ freezes a \rec\ $u$ by using CAS to store a pointer to $D$ in $u.\info$ (at the \fcas\ step in Figure~\ref{code-3path-scxo}).
Suppose $S$ successfully freezes all \rec s in its $V$ sequence.
Then, $S$ prepares to finalize each \rec\ $u \in R$ by setting a \textit{marked} bit in $u$ (at the \markstep\ in Figure~\ref{code-3path-scxo}).
Finally, $S$ changes $fld$ to $new$, and atomically releases all locks by setting $D.state$ to \done\ (at the \cstep\ in Figure~\ref{code-3path-scxo}). %a bit that indicates the \sct$_O$ has finished (so pointers to its \op\ no longer represent acquired locks).
Observe that setting $D.state$ to \done\ has the effect of atomically finalizing all \rec s in $R$ and unfreezing all \rec s in $V \setminus R$.

Now, suppose $S$ was prevented from freezing some \rec\ $u$ because another invocation $S'$ of \sct$_O$ had already frozen $u$ (i.e., the \fcas\ step by $S$ failed, and it saw $r.\info \neq scxPtr$ at the following line).
Then, $S$ aborts by setting the $D.state$ to \retry\ (at the \astep\ in Figure~\ref{code-3path-scxo}).
This has the effect of atomically unfreezing any \rec s $S$ had frozen.
Note that, before the process that performed $S$ can perform another invocation of \sct$_O$ with $u$ in its $V$-sequence, it must perform \llt$(u)$.
If $u$ is still frozen for $S'$ when this \llt$(u)$ is performed, then the \llt\ will use the information stored in the \op\ at $u$ to \textit{help} $S'$ complete and unfreeze $u$.
(\op s also contain another field $\freezingdone$ that is used to coordinate any processes helping the \sct, ensuring that they do not make conflicting changes to the $state$ field.)

The correctness argument for this algorithm is subtle, and we leave the details to~\cite{Brown:2013}, but one crucial property of this algorithm is relevant to this work:
%This implementation of \llt\ and \sct$_O$ satisfies a particular property (that it depends on for its correctness).

\noindent\textbf{P1.} between any two changes to (the user-defined fields of) a \rec\ $u$, a pointer to a new \op\ (that has never before been contained in $u.\info$) is stored in $u.\info$.

\noindent
This property is used to determine whether a \rec\ has changed between the last \llt$_O$ on it and a subsequent invocation of \sct$_O$.
Consider an invocation $S$ of \sct$_O(V,R,fld,new)$ by a process $p$.
Let $u$ be any \rec in $V$, and $L$ be the last invocation of \llt$_O(u)$ by $p$.
$L$ reads $u.\info$ and sees some value $ptr$.
$S$ subsequently performs a \fcas\ step to change $u.\info$ from $ptr$ to point to its \op, freezing $u$.
If this CAS succeeds, then $S$ infers that $u$ has not changed between the read of $u.\info$ in the \llt\ and the \fcas\ step.

%\trevor{possibly talk about the high level description of \llt\ and \sct, and mention descriptor pointers and marked bits. possibly do this in the section where we implement htm-based \llt\ and \sct, instead.}

\fakeparagraph{Progress properties}
\begin{thesisonly}
We briefly recall the progress property for \llt\ and \sct.
\end{thesisonly}
Specifying a progress guarantee for \llt\ and \sct\ operations is subtle, because if processes repeatedly perform \llt\ on \rec s that have been finalized, or repeatedly perform failed \llt s, then they may never be able to invoke \sct.
In particular, it is not sufficient to simply prove that \llt s return snapshots infinitely often, since \textit{all} of the \llt s in a sequence must return snapshots before a process can invoke \sct.
To simplify the progress guarantee for \llt\ and \sct, we make a definition.
An \sct-\func{Update} algorithm is one that performs \llt s on a sequence $V$ of \rec s and invokes \sct$(V, R, fld, new)$ if they all return snapshots.
\begin{thesisnot}
The progress guarantee in~\cite{Brown:2013} is then stated as follows.
\end{thesisnot}
\begin{thesisonly}
The progress guarantee in Chapter~\ref{chap-scx} was then stated as follows.
\end{thesisonly}

\begin{compactenum}[\hspace{3.4mm}{\bf PROG}:]
\item Suppose that
    (a) there is always some non-finalized \rec\ reachable by following pointers from an entry point, 
    (b) for each \rec\ $r$, each process performs finitely many invocations of \llt$(r)$ that return \finalized, and
    (c) processes perform infinitely many executions of \sct-\func{Update} algorithms.
    Then, infinitely many invocations of \sct\ succeed.
\end{compactenum}
\end{fullver}

\fakeparagraph{The tree update template}
The tree update template implements lock-free updates that atomically replace an old connected subgraph $R$ of a down-tree by a new connected subgraph $N$.
Such an update can implement any change to the tree, such as an insertion into a BST or a rotation in a balanced tree.
\begin{wrapfigure}{r}{0.5\linewidth}
	\input{chap-template/tree-fig-1.pdf_t}
	\caption{Example of the tree update template.
%			$R$ is the set of nodes to be removed,
%			$N$ is a tree of new nodes that have never before appeared in the tree, and
%			$F_N$ is the set of children of $N$ (and of $R$).
%			Nodes in $F_N$ may have children.  
%			The shaded nodes (and possibly others) are in the sequence $V$  of the \sct\ that performs the update.
%			The darkly shaded nodes are finalized by the \sct.
			}
	\label{fig-3path-replace-subtree}
\end{wrapfigure}
The old subgraph includes all nodes with a field (including a child pointer) to be modified.
The new subgraph may have pointers to nodes in the old tree.
Since every node in a down-tree has indegree one, the update can be performed by changing a single child pointer of some node $parent$.
However, problems could arise if a concurrent operation changes the part of the tree being updated.
For example, nodes in the old subgraph, or even $parent$, could be removed from the tree before $parent$'s child pointer is changed.
The template takes care of the process coordination required to prevent such problems.

%\begin{figure}[tb]
%	\centering
%	\input{chap-template/tree-fig-2.pdf_t}
%	\caption{Examples of two special cases of the tree update template when no nodes are removed from the tree.  
%	(a) Replacing a \nil\ child pointer: In this case, $R=F_N=\emptyset$.  
%	(b) Inserting new nodes in the middle of the tree: In this case, $R=\emptyset$ and $F_N$ consists of a single node.}
%	\label{fig-3path-replace-subtree2}
%\end{figure}

Each tree node is represented by a \rec\ with a fixed number of child pointers as its mutable fields. % (but different nodes may have different numbers of child fields).  
Each child pointer either points to a \rec\ or contains \nil\ (denoted by $\multimap$ in our figures).
Any other data in the node is stored in immutable fields.
Thus, if an update must change some of this data, it makes a new copy of the node with the updated data.
There is a \rec\ $entry$ which acts as the entry point to the data structure and is never deleted.
%%(This \rec\ points to the root of a down-tree.)
%An empty tree can be represented by, e.g., having $entry$ point to an empty \rec\ (containing no fields).
%A node is {\em in the tree} if it can be reached by following pointers from $entry$.

At a high level, an update that follows the template proceeds in two phases: the \textit{search phase} and the \textit{update phase}.
In the search phase, the update searches for a location where it should occur.
Then, in the update phase, the update performs \llt s on a connected subgraph of nodes in the tree, including $parent$ and the set $R$ of nodes to be removed from the tree.
Next, it decides whether the tree should be modified, and, if so, creates a new subgraph of nodes and performs an \sct\ that atomically changes a child pointer, as shown in Figure~\ref{fig-3path-replace-subtree}, and finalizes any nodes in $R$.
%Figure~\ref{fig-3path-replace-subtree2} shows two special cases where $R$ is empty.
%(As a minor note, search operations that do not perform any update can also be said to follow the template.)
See~\cite{Brown:2014} for further details.
\end{thesisnot}

\begin{fullver}
\section{HTM-based \llt\ and \sct} \label{sec-htmscx}

In this section, we describe an HTM-based implementation of \llt\ and \sct.
This implementation is used by our first accelerated template implementation, \textit{2-path con}, which is described in Section~\ref{sec-3path-algs}.

\begin{figure*}[t]
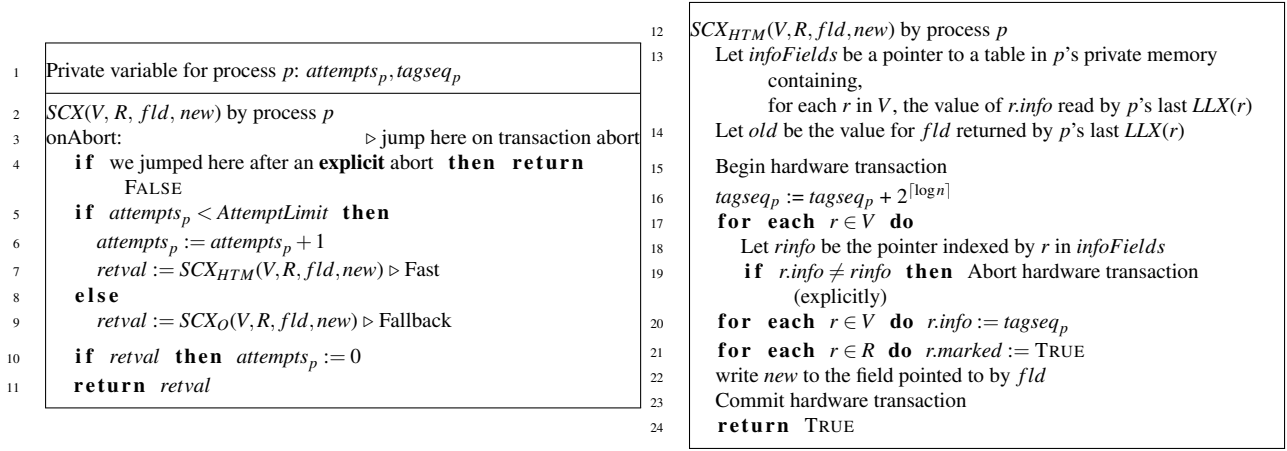

%\vspace{-2mm}
\footnotesize
\def\pwidth{4cm}
%\hrule
%\vspace{-2mm}
\begin{minipage}{0.48\linewidth}
\begin{framed}
\prepnewlisting
\begin{lstlisting}[mathescape=true]
 //Private variable for process $p$: $\textit{attempts}_p, \textit{tagseq}_p$ \vspace{1mm}\hrule\vspace{1mm}
 //\sct($V,$ $R,$ $fld,$ $new$) by process $p$
 //onAbort: \hfill\com jump here on transaction abort
   if $\mbox{we jumped here after an \textbf{explicit} abort}$ then return $\false$ //\label{newscx-scx-explicit-return}
   if $\textit{attempts}_p < \textit{AttemptLimit}$ then
     $\textit{attempts}_p := \textit{attempts}_p + 1$
     //$retval := \sct_{HTM}$($V,R,fld,new$) \com Fast
   else
     //$retval := \sct_O$($V,R,fld,new$) \com Fallback \vspace{1.5mm}
   if $retval$ then $\textit{attempts}_p := 0$
   return $retval$ //\vspace{-1mm}
\end{lstlisting}
\end{framed}
\end{minipage}
\hspace{0.03\linewidth}
\begin{minipage}{0.48\linewidth}
\begin{framed}
\preplisting
\begin{lstlisting}[mathescape=true]
 //\sct$_{HTM}$($V, R, fld, new$) by process $p$
   //\dline{Let $\llresults$ be a pointer to a table in $p$'s private memory containing,}%
           {for each $r$ in $V$, the value of $r.\info$ read by $p$'s last \llt($r$)}
   //Let $old$ be the value for $fld$ returned by $p$'s last \llt($r$)\vspace{1.5mm}%

   //Begin hardware transaction \label{newscx-scxhtm-xbegin}
   //\textit{tagseq}$_p$ := \textit{tagseq}$_p$ + $2^{\lceil \log n \rceil}$
   for each $r \in V$ do //\label{newscx-scxhtm-freezing-loop-start}
     //Let $r$\info\ be the pointer indexed by $r$ in $\llresults$
     if $r.\info \neq r\info$ then //Abort hardware transaction (explicitly) \label{newscx-scxhtm-abort} \label{newscx-scxhtm-freezing-loop-end}
   for each $r \in V$ do $r.\info := \textit{tagseq}_p$
   for each $r \in R$ do $r.marked := \true$
   //write $new$ to the field pointed to by $fld$
   //Commit hardware transaction \label{newscx-scxhtm-commit}
   return $\true$ //\vspace{-1mm}
\end{lstlisting}
\end{framed}
\end{minipage}
\begin{thesisnot}
    \vspace{-5mm}
\end{thesisnot}
	\caption{HTM-based implementation of \sct.}
	\label{code-3path-newscx}
\end{figure*}

\begin{shortver}
\fakeparagraph{HTM-based \llt\ and \sct}
\end{shortver}
In the following, we use \sct$_O$ and \llt$_O$ to refer to the original lock-free implementation of \llt\ and \sct.
We give an implementation of \sct\ that uses an HTM-based fast path called \sct$_{HTM}$, and \sct$_O$ as its fallback path.
Hardware transactions are instrumented so they can run concurrently with processes executing \sct$_O$.
This algorithm guarantees lock-freedom and achieves a high degree of concurrency.
Pseudocode appears in Figure~\ref{code-3path-newscx}.
%To use the algorithm, one invokes \llt\ and \sct\ as usual.
%
At a high level, an \sct$_{HTM}$ by a process $p$ starts a transaction, then attempts to perform a highly optimized version of \sct$_O$.
Each time a transaction executed by $p$ aborts, control jumps to the onAbort label, at the beginning of the \sct\ procedure.
If a process \textit{explicitly} aborts a transaction at line~\ref{newscx-scxhtm-abort}, % executes a step that explicitly aborts the transaction, 
then \sct\ returns \false\ at line~\ref{newscx-scx-explicit-return}.
Each process has a budget \textit{AttemptLimit} that specifies how many times it will attempt hardware transactions before it will fall back to executing \sct$_O$.

In \sct$_O$, \op s are used (1) to facilitate helping, and (2) to lock \rec s and detect changes to them.
In particular, \sct$_O$ guarantees the following property.
\textbf{P1}: between any two changes to (the user-defined fields of) a \rec\ $u$, a new \op\ pointer is stored in $u.\info$.
However, \sct$_{HTM}$ does not create \op s.
In a transactional setting, helping causes unnecessary aborts, since executing a transaction that performs the same work as a running transaction will cause at least one (and probably both) to abort.
Helping in transactions is also \textit{not necessary} to guarantee progress, since progress is guaranteed by the fallback path.
So, to preserve property P1, we give each process $p$ a \textit{tagged sequence number} \textit{tseq}$_p$ that contains the process name, a sequence number, and a \textit{tag} bit.
The \textit{tag} bit is the least significant bit.
On modern systems where pointers are word aligned, the least significant bit in a pointer is always zero.
Thus, the tag bit allows a process to distinguish between a tagged sequence number and a pointer.
In \sct$_{HTM}$, instead of having $p$ create a new \op\ and store pointers to it in \rec s to lock them, $p$ increments its sequence number in $\textit{tseq}_p$ and stores $\textit{tseq}_p$ in \rec s.
Since no writes performed by a transaction $T$ can be seen until it commits, it never actually needs to hold any locks.
Thus, every value of $\textit{tseq}_p$ stored in a \rec\ represents an unlocked value, and writing $\textit{tseq}_p$ represents $p$ locking and immediately unlocking a node.

After storing $\textit{tseq}_p$ in each $r \in V$, \sct$_{HTM}$ finalizes each $r \in R$ by setting $r.marked := \true$ (mimicking the behaviour of \sct$_O$).
Then, it stores $new$ in the field pointed to by $fld$, and commits.
Note that eliminating the creation of \op s on the fast path also eliminates the need to \textit{reclaim} any created \op s, which further reduces overhead.

The \sct$_{HTM}$ algorithm also necessitates a small change to \llt$_O$, to handle tagged sequence numbers.
An invocation of \llt$_O$($r$) reads a pointer $r\info$ to an \op, follows $r\info$ to read one of its fields, and uses the value it reads to determine whether $r$ is locked.
However, $r\info$ may now contains a tagged sequence number, instead of a pointer to an \op.
So, in our modified algorithm, which we call \llt$_{HTM}$, before a process tries to follow $r\info$, it first checks whether $r\info$ is a tagged sequence number, and, if so, behaves as if $r$ is unlocked.
\begin{shortver}
Due to lack of space, we defer a more detailed description, and a proof of correctness and progress, to Appendix~\ref{appendix-correctness}.
\end{shortver}
\begin{fullver}
The code for \llt$_{HTM}$ appears in Figure~\ref{code-3path-htmsct-transformation3}.

\subsection{Correctness and Progress}
In this section, we prove correctness and progress for our HTM-based implementation of \llt\ and \sct.
We do this by showing that one can start with \llt$_O$ and \sct$_O$, and obtain our HTM-based implementation by applying a sequence of transformations.
Intuitively, these transformations preserve the semantics of \sct\ and maintain backwards compatibility with \sct$_O$ so that the transformed versions can be run concurrently with invocations of \sct$_O$.
More formally, for each execution of a transformed algorithm, there is an execution of the original algorithm in which: the same operations are performed, they are linearized in the same order, and they return the same results.
For each transformation, we sketch the correctness and progress argument, since the transformations are simple and a formal proof would be overly pedantic. %with the understanding that a formal proof would .

\begin{shortver}
\subsection{Original lock-free implementation of \llt\ and \sct}

We start by giving a more detailed description of \llt$_O$ and \sct$_O$.

\end{shortver}

\begin{shortver}
\subsection{Transforming \llt$_O$ and \sct$_O$ to obtain our HTM-based implementation}
\end{shortver}

\paragraph{Adding transactions}

\begin{figure*}[th]
%\vspace{-2mm}
\footnotesize
\def\pwidth{4cm}
\prepnewlisting
%\hrule
\vspace{-2mm}
\begin{framed}
\begin{lstlisting}[mathescape=true]
  //\sct$_1(V, R, fld, new)$ by process $p$
    //\dline{Let $\llresults$ be a pointer to a table in $p$'s private memory containing,}%
            {for each $r$ in $V$, the value of $r.\info$ read by $p$'s last \llt$(r)$}
    //Let $old$ be the value for $fld$ returned by $p$'s last \llt$(r)$\vspace{1.5mm}%

    //Begin hardware transaction
    $scxPtr := \mbox{pointer to new \op}(V,R,fld,new,old,\freezing,\false,\llresults)$ //\label{htmsct-transformation1-new-op}
    //\com \mbox{Freeze all \rec s in $scxPtr.V$ to protect their mutable fields from being changed by other \sct s}
    for each $r$ in $scxPtr.V \mbox{ enumerated in order}$ do//\label{htmsct-transformation1-fcas-loop-begin}
      //Let $r$\info\ be the pointer indexed by $r$ in $scxPtr.\llresults$ \label{htmsct-transformation1-rinfo}
      if not $\cas(r.\info,r\info,scxPtr)$ then //\sidecom{\textbf{\fcas}}\label{htmsct-transformation1-fcas}
        if $r.\info \neq scxPtr$ then // \label{htmsct-transformation1-check-frozen} 
          //\com \mbox{Could not freeze $r$ because it is frozen for another \sct}
          if $scxPtr.\freezingdone = \true$ then//\sidecom{\textbf{\fcstep}}\label{htmsct-transformation1-fcstep}
            //\com the \sct\ has already completed successfully 
            //Commit hardware transaction\label{htmsct-transformation1-commit1}
            return $\true$ // \label{htmsct-transformation1-return-true-loop} 
          else //Abort hardware transaction (explicitly) \label{htmsct-transformation1-commit2} \vspace{1.5mm}%

    //\com Finished freezing \rec s (Assert: $state \in \{\freezing, \done\}$) 
    $scxPtr.\freezingdone := \true$//\sidecom{\textbf{\fstep}}\label{htmsct-transformation1-fstep}
    for each $r \in scxPtr.R$ do $r.marked := \true$ //\sidecom{\textbf{\markstep}}\label{htmsct-transformation1-markstep}
    //$\cas(scxPtr.fld, scxPtr.old, scxPtr.new)$ \sidecom{\textbf{\upcas}}\label{htmsct-transformation1-upcas} \vspace{2mm}
    //\com Finalize all $r$ in $R$, and unfreeze all $r$ in $V$ that are not in $R$ 
    $scxPtr.state := \done$//\sidecom{\textbf{\cstep}}\label{htmsct-transformation1-cstep}
    //Commit hardware transaction\label{htmsct-transformation1-commit3}
    return $\true$ // \label{htmsct-transformation1-return-true} \vspace{-1mm}
\end{lstlisting}
\end{framed}
    \vspace{-5mm}
	\caption{HTM-based \sct: after adding transactions.}
	\label{code-3path-htmsct-transformation1}
\end{figure*}

For the first transformation, we replaced the invocation of \help\ in \sct$_O$ with the body of the \help\ function, and wrapped the code in a transaction.
Since the fast path simply executes the fallback path algorithm in a transaction, the correctness of the resulting algorithm is immediate from the correctness of the original \llt\ and \sct\ algorithm.
%(We argue that lock-free progress is satisfied only after the final transformation.)

We also observe that it is not necessary to commit a transaction that sets the $state$ of its \op\ to \retry\ and returns \false.
The only effect that committing such a transaction would have on shared memory is changing some of the $\info$ fields of \rec s in its $V$ sequence to point to its \op.
In \sct$_O$, $\info$ fields serve two purposes.
First, they provide pointers to an \op\ while its \sct\ is in progress (so it can be helped).
Second, they act as locks that grant exclusive access to an \sct$_O$, and allow an invocation of \sct$_O$ to determine whether any user-defined fields of a \rec\ $r$ have changed since its linked \llt$(r)$ (using property P1).
However, since the effects of a transaction are not visible until it has already committed, a %n \sct\ operation being performed by a 
transaction no longer needs help by the time it modified any $\info$ field.
And, since an \sct$_O$ that sets the $state$ of its \op\ to \retry\ does not change any user-defined field of a \rec, these changes to $\info$ fields are not needed to preserve property P1.
The only consequence of changing these $\info$ fields is that other invocations of \sct$_O$ might needlessly fail and return \false, as well.
So, instead of setting $state = \retry$ and committing, we \textit{explicitly abort} the transaction and return \false.
%Of course, this makes it unnecessary to set $state = \retry$ in the transaction.
Figure~\ref{code-3path-htmsct-transformation1} shows the result of this transformation: \sct$_1$.
(Note that aborting transactions does not affect correctness---only progress.)

\begin{figure*}[th]
%\vspace{-2mm}
\footnotesize
\def\pwidth{4cm}
\prepnewlisting
%\hrule
\vspace{-2mm}
\begin{framed}
\begin{lstlisting}[mathescape=true]
  //Private variable for process $p$: $\textit{attempts}_p$\vspace{1mm}\hrule\vspace{1mm}
  //\sct$(V, R, fld, new)$ by process $p$
  //onAbort: \hfill\com jump here on transaction abort \label{htmsct-usage-onabort}
    if $\mbox{we jumped here after an explicit abort in the code}$ then return $\false$
    if $\textit{attempts}_p < \textit{AttemptLimit}$ then //\label{htmsct-usage-check-attempts}
      $\textit{attempts}_p := \textit{attempts}_p + 1$
      $retval := \sct_1(V,R,fld,new)$ //\hfill\com invoke HTM-based \sct
    else
      $retval := \sct_O(V,R,fld,new)$ //\hfill\com fall back to original \sct\vspace{1.5mm}
    if $retval$ then $\textit{attempts}_p := 0$//\hfill\com reset $p$'s attempt counter before returning \true\label{htmsct-usage-return-true}
    return $retval$ //\vspace{-1mm}
\end{lstlisting}
\end{framed}
    \vspace{-5mm}
	\caption{How the HTM-based \sct$_1$ is used to provide lock-free \sct.}
	\label{code-3path-htmsct-usage}
\end{figure*}

Of course, we must provide a fallback code path in order to guarantee progress.
Figure~\ref{code-3path-htmsct-usage} shows how \sct$_1$ (the fast path) and \sct$_O$ (the fallback path) are used together to implement lock-free \sct.
%(We defer a progress proof until the transformations have all been described.)
In order to decide when each code path should be executed, we give each process $p$ a private variable $\textit{attempts}_p$ that contains the number of times $p$ has attempted a hardware transaction since it last performed an \sct$_1$ or \sct$_O$ that succeeded (i.e., returned \true).
The \sct\ procedure checks whether $\textit{attempts}_p$ is less than a (positive) threshold \textit{AttemptLimit}.
If so, $p$ increments $\textit{attempts}_p$ and invokes \sct$_1$ to execute a transaction on the fast path.
If not, $p$ invokes \sct$_O$ (to guarantee progress).
Whenever $p$ returns \true\ from an invocation of \sct$_1$ or \sct$_O$, it resets its budget $\textit{attempts}_p$ to zero, so it will execute on the fast path in its next \sct.
Each time a transaction executed by $p$ aborts, control jumps to the onAbort label, at the beginning of the \sct\ procedure.
If a process explicitly aborts a transaction it is executing (at line~\ref{htmsct-transformation1-commit2} in $\sct_1$), % executes a step that explicitly aborts the transaction, 
then control jumps to the onAbort label, \textit{and} the \sct\ returns \false\ at the next line.

\paragraph{Progress}

\begin{thesisonly}
\begin{thesisonly}
We briefly recall the progress property for \llt\ and \sct.
\end{thesisonly}
Specifying a progress guarantee for \llt\ and \sct\ operations is subtle, because if processes repeatedly perform \llt\ on \rec s that have been finalized, or repeatedly perform failed \llt s, then they may never be able to invoke \sct.
In particular, it is not sufficient to simply prove that \llt s return snapshots infinitely often, since \textit{all} of the \llt s in a sequence must return snapshots before a process can invoke \sct.
To simplify the progress guarantee for \llt\ and \sct, we make a definition.
An \sct-\func{Update} algorithm is one that performs \llt s on a sequence $V$ of \rec s and invokes \sct$(V, R, fld, new)$ if they all return snapshots.
\begin{thesisnot}
The progress guarantee in~\cite{Brown:2013} is then stated as follows.
\end{thesisnot}
\begin{thesisonly}
The progress guarantee in Chapter~\ref{chap-scx} was then stated as follows.
\end{thesisonly}

\begin{compactenum}[\hspace{3.4mm}{\bf PROG}:]
\item Suppose that
    (a) there is always some non-finalized \rec\ reachable by following pointers from an entry point, 
    (b) for each \rec\ $r$, each process performs finitely many invocations of \llt$(r)$ that return \finalized, and
    (c) processes perform infinitely many executions of \sct-\func{Update} algorithms.
    Then, infinitely many invocations of \sct\ succeed.
\end{compactenum}
\end{thesisonly}

%We now prove progress. %sketch the progress proof.
%Every time a transaction commits, an \sct\ returns \true.
%Thus, as long as transactions continue to commit, PROG is satisfied.
%If transactions stop committing, then eventually all transactions abort.
%Since each process 
%
\begin{thesisnot}
It is proved in~\cite{Brown:2013} that PROG is satisfied by \llt$_O$ and \sct$_O$.
\end{thesisnot}
\begin{thesisonly}
\llt$_O$ and \sct$_O$ satisfy PROG.
\end{thesisonly}
We argue that PROG is satisfied by the implementation of \llt\ and \sct\ in Figure~\ref{code-3path-htmsct-usage}. %when \llt\ is implemented as $\llt_O$ and \sct\ is implemented as \sct$_1$. %We argue that \llt$_O$ and \sct$_1$ provide the same progress guarantee as \llt$_O$ and \sct$_O$. %The progress argument is fairly straightforward.
To obtain a contradiction, suppose the antecedent of PROG holds, but only finitely many invocations of \sct\ return \true.
Then, after some time $t$, no invocation of \sct\ returns \true.

\textit{Case 1:} Suppose processes take infinitely many steps in transactions.
By inspection of the code, each transaction is wait-free, and \sct\ returns \true\ immediately after a transaction commits.
Since no transaction commits after $t$, there must be infinitely many aborts.
However, each process can perform at most \textit{AttemptLimit} aborts since the last time it performed an invocation of \sct\ that returned \true.
So, only finitely many aborts can occur after $t$---a contradiction.

\textit{Case 2:} Suppose processes take only finitely many steps in transactions.
Then, processes take only finitely many steps in \sct$_1$.
It follows that, after some time $t'$, no process takes a step in \sct$_1$.
Therefore, in the suffix of the execution after $t'$, processes only take steps in \sct$_O$ and \llt$_O$.
%If we consider any tagged sequence numbers that \llt$_{HTM}$ encounters to simply be pointers to \op s with $state =$ \done\, then each \llt$_{HTM}$ behaves the exact same way as \llt$_O$.
However, since \llt$_O$ and \sct$_O$ satisfy PROG, infinitely many invocations of \sct\ must succeed after $t'$, which is a contradiction.

\paragraph{Eliminating \textit{most} accesses to fields of \op s created on the fast path}

\begin{figure*}[th]
%\vspace{-2mm}
\footnotesize
\def\pwidth{4cm}
\prepnewlisting
%\hrule
\vspace{-2mm}
\begin{framed}
\begin{lstlisting}[mathescape=true]
  //Private variable for process $p$: $\textit{attempts}_p$\vspace{1mm}\hrule\vspace{1mm}
  //\sct$_2(V, R, fld, new)$ by process $p$
    //\dline{Let $\llresults$ be a pointer to a table in $p$'s private memory containing,}%
            {for each $r$ in $V$, the value of $r.\info$ read by $p$'s last \llt$(r)$}
    //Let $old$ be the value for $fld$ returned by $p$'s last \llt$(r)$\vspace{1.5mm}%

    //Begin hardware transaction
    $scxPtr := \mbox{pointer to new \op}(-,-,-,-,-,\freezing,-,-)$ //\label{htmsct-transformation2-new-op}
    //\com \mbox{Freeze all \rec s in $V$ to protect their mutable fields from being changed by other \sct s}
    for each $r$ in $V \mbox{ enumerated in order}$ do//\label{htmsct-transformation2-fcas-loop-begin}
      //Let $r$\info\ be the pointer indexed by $r$ in $\llresults$ \label{htmsct-transformation2-rinfo}
      if not $\cas(r.\info,r\info,scxPtr)$ then //\sidecom{\textbf{\fcas}}\label{htmsct-transformation2-fcas}
        if $r.\info \neq scxPtr$ then //Abort hardware transaction (explicitly)\vspace{1.5mm}%

    //\com Finished freezing \rec s
    for each $r \in R$ do $r.marked := \true$ //\hspace{4mm}\com Finalize each $r \in R$ \sidecom{\textbf{\markstep}}\label{htmsct-transformation2-markstep}
    //$\cas(fld, old, new)$ \sidecom{\textbf{\upcas}}\label{htmsct-transformation2-upcas}
    $scxPtr.state := \done$//\sidecom{\textbf{\cstep}}\label{htmsct-transformation2-cstep}
    //Commit hardware transaction
    return $\true$ // \label{htmsct-transformation2-return-true}\vspace{-1mm}
\end{lstlisting}
\end{framed}
    \vspace{-5mm}
	\caption{HTM-based \sct: after eliminating \textbf{most} accesses to fields of \op s created on the fast path.}
	\label{code-3path-htmsct-transformation2}
\end{figure*}

In \llt$_O$ and \sct$_O$, helping is needed to guarantee progress, because otherwise, %\llt$_O$ operations help \sct$_O$ operations in order to guarantee progress.
%This helping is needed because, otherwise, 
an invocation of \sct$_O$ that crashes while one or more \rec s are frozen for it could cause every invocation of \llt$_O$ to return \fail\ (which, in turn, could prevent processes from performing the necessary linked invocations of \llt$_O$ to invoke \sct$_O$).
%However, when the \sct$_O$ algorithm is performed inside a transaction, it can \textit{never} cause an invocation of \llt$_O$ to return \fail .
%We briefly explain why.
%If an \llt$_O$ accesses a memory location that is in the data set of a concurrent transaction $T$, then the transaction will simply abort (to preserve atomicity).
%Let $D$ be the \op\ created by $T$.
%Since \llt$_O$ cannot see any effects of $T$ until $T$ has committed, if \llt$_O$ can see $D$, then $D.state$ is \done\ or \retry.
%If $D.state =$ \retry, then \trevor{THIS IS THE WRONG EXPLANATION. WE DON'T EVEN KEEP SCX RECORDS, SO WE CAN'T DISTINGUISH BETWEEN ABORTED AND COMMITTED.}
However, as we mentioned above, since transactions are atomic, a process cannot see any of their writes (including the contents of any \op\ they create and publish pointers to) until they have committed, at which point they no longer need help.
%Thus, a process executing in \llt$_O$ or \sct$_O$ can never see that a \rec\ is \textit{currently} frozen for an invocation of \sct$_1$ (they can only see that it was frozen at some point in the past). [NOT TRUE BECAUSE OF FINALIZED]
%In contrast, since transactions are atomic, a process cannot see any data written by a transaction in \sct$_1$ (including the contents of any \op\ created by a transaction) until the transaction has committed, at which point the operation it was performing no longer needs help.
Thus, it is not necessary to help transactions in \sct$_1$.\footnote{In fact, helping transactions would be \textit{actively harmful}, since performing the same modifications to shared memory as an in-flight transaction \textit{will cause it to abort}. This leads to very poor performance, in practice.}
%Additionally, note that progress is guaranteed by the fallback path, not the fast path.
%(This is because each invocation of \sct$_O$ is wait-free, and a bounded number of transactional attempts are made before 

In fact, it is easy to see that processes will not help any \op\ created by a transaction in \sct$_1$.
Observe that each transaction in $\sct_1$ sets the $state$ of its \op\ to \done\ before committing.
Consequently, if an invocation of \llt$_O$ reads $r.\info$ and obtains a pointer $r\info$ to an \op\ created by a transaction in $\sct_1$, then $r\info$ has $state$ \done.
Therefore, by inspection of the code, \llt$_O$ will not invoke \help$(r\info)$.
%That is, processes will not help any \op\ created by a transaction in $\sct_1$.
%(As we will explain below, this does not violate lock-freedom, since progress is guaranteed by the fallback path, not the fast path.)

Since \llt$_O$ never invokes \help$(r\info)$ for any $r\info$ created by a transaction in \sct$_1$, most fields of an \op\ created by a transaction are accessed only by the process that created the \op.
The only field that is accessed by other processes is the $state$ field (which is accessed in \llt$_O$).
%Furthermore, the $state$ field is only ever set to \done\ by a transaction, so we know precisely what it must contain if it was written by a transaction.
Therefore, it suffices for a transaction in \sct$_1$ to initialize only the $state$ field of its \op.
As we will see, any accesses to the other fields can simply be eliminated or replaced with locally available information.

Using this knowledge, we transform $\sct_1$ in Figure~\ref{code-3path-htmsct-transformation1} into a new procedure called $\sct_2$ in Figure~\ref{code-3path-htmsct-transformation2}.
First, instead of initializing the entire \op\ when we create a new \op\ at line~\ref{htmsct-transformation1-new-op} in $\sct_1$, we initialize only the $state$ field.
We then change any steps that read fields of the \op\ (lines~\ref{htmsct-transformation1-fcas-loop-begin}, \ref{htmsct-transformation1-rinfo}, \ref{htmsct-transformation1-fcstep}, \ref{htmsct-transformation1-markstep} and~\ref{htmsct-transformation1-upcas} in $\sct_1$) to use locally available information, instead.

Next, we eliminate the \textit{\fstep} at line~\ref{htmsct-transformation1-fstep} in $\sct_1$, which changes the $\freezingdone$ field of the \op.
Recall that $\freezingdone$ is used by \sct$_O$ to prevent helpers from making conflicting changes to the $state$ field of its \op.
When a \textit{\fcas} fails in an invocation $S$ of \sct$_O$ (at line~\ref{help-fcas} of \func{Help} in Figure~\ref{code-main}), it indicates that either $S$ will fail due to contention, or another process had already helped $S$ to complete successfully.
The $\freezingdone$ bit allows a process to distinguish between these two cases.
Specifically, it is proved in~\cite{Brown:2013} that a process will see $\freezingdone = \true$ at line~\ref{help-fcas} of \func{Help} if and only if another process already helped $S$ complete and set $\freezingdone := \true$.
However, since we have argued that processes never help transactions (and, in fact, no other process can even \textit{access} the \op\ until the transaction that created it has committed), $\freezingdone$ is always \false\ at the corresponding step (line~\ref{htmsct-transformation1-fcstep}) in $\sct_1$.
This observation allows us to eliminate the entire \textit{if} branch at line~\ref{htmsct-transformation1-fcstep} in \sct$_1$.

Clearly, this transformation preserves PROG.
%\trevor{clearly progress is preserved.}
%
Note that \sct$_2$ (and each of the subsequent transformed variants) is used in the same way as \sct$_1$: Simply replace \sct$_1$ in Figure~\ref{code-3path-htmsct-usage} with \sct$_2$.

\paragraph{Completely eliminating accesses to fields of \op s created on the fast path}

\begin{figure*}[th]
%\vspace{-2mm}
\footnotesize
\def\pwidth{4cm}
\prepnewlisting
%\hrule
\vspace{-2mm}
\begin{framed}
\begin{lstlisting}[mathescape=true]
  //Private variable for process $p$: $\textit{attempts}_p$ \vspace{1mm}\hrule\vspace{1mm}
  //\llt$_{HTM}(r)$ by process $p$ \com Precondition: $r \neq \nil$.
    $marked_1 := r.marked$ // \hfill\com{order of lines~\ref{ll-read-marked1}--\ref{ll-read-marked2} matters} \label{htmllt-read-marked1}
    $r\info := r.\info$ // \label{htmllt-read} 
 *  $state := (r\info\ \&\ 1)\ ?\ \mbox{\done}\ : r\info.state$ //\hfill\com if \textit{rinfo} is tagged, take \textit{state} to be \done\ \label{htmllt-read-state}
    $marked_2 := r.marked$ // \label{htmllt-read-marked2}
    if $state = \retry$ or $(state = \done$ and not $marked_2)$ then //\hfill\com{if $r$ was not frozen at line~\ref{ll-read-state}} \label{htmllt-check-frozen}
      read $r.m_1,...,r.m_y$ //and record the values in local variables $m_1,...,m_y$%
      \label{htmllt-collect}
      if $r.\info = r\info$ then//\hfill\com{if $r.\info$ points to the same \op\ as on line~\ref{ll-read}} \label{htmllt-reread}
        //store $\langle r, r\info, \langle m_1, ..., m_y \rangle \rangle$ in $p$'s local table \label{htmllt-store}
        return $\langle m_1, ..., m_y \rangle$ // \label{htmllt-return}  \vspace{2mm}
    if $state = \freezing$ then $\help(r\info)$
    if $marked_1$ then//%
    \label{htmllt-check-finalized}
      return $\finalized$ // \label{htmllt-return-finalized}
    else
      return $\fail$ // \label{htmllt-return-fail} \vspace{1.5mm} \hrule \vspace{1.5mm}%

  //\sct$_3(V, R, fld, new)$ by process $p$
    //\dline{Let $\llresults$ be a pointer to a table in $p$'s private memory containing,}%
            {for each $r$ in $V$, the value of $r.\info$ read by $p$'s last \llt$(r)$}
    //Let $old$ be the value for $fld$ returned by $p$'s last \llt$(r)$\vspace{1.5mm}%

    //Begin hardware transaction
    $scxPtr := \mbox{pointer to new \op}(-,-,-,-,-,-,-,-)$ //\label{htmsct-transformation3-new-op}
    //\com \mbox{Freeze all \rec s in $V$ to protect their mutable fields from being changed by other \sct s}
    for each $r$ in $V \mbox{ enumerated in order}$ do//\label{htmsct-transformation3-fcas-loop-begin}
      //Let $r$\info\ be the pointer indexed by $r$ in $\llresults$ \label{htmsct-transformation3-rinfo}
      if not $\cas(r.\info,r\info, (scxPtr\ \&\ 1))$ then //\sidecom{\textbf{\fcas}}\label{htmsct-transformation3-fcas}
        if $r.\info \neq (scxPtr\ \&\ 1)$ then //Abort hardware transaction (explicitly)\label{htmsct-transformation3-abort}\vspace{1.5mm}%

    //\com Finished freezing \rec s
    for each $r \in R$ do $r.marked := \true$ //\hspace{4mm}\com Finalize each $r \in R$ \sidecom{\textbf{\markstep}}\label{htmsct-transformation3-markstep}
    //$\cas(fld, old, new)$ \sidecom{\textbf{\upcas}}\label{htmsct-transformation3-upcas}
    //Commit hardware transaction\label{htmsct-transformation3-commit}
    return $\true$ // \label{htmsct-transformation3-return-true}\vspace{-1mm}
\end{lstlisting}
\end{framed}
    \vspace{-5mm}
	\caption{HTM-based \sct: after completely eliminating accesses to fields of \op s created on the fast path.}
	\label{code-3path-htmsct-transformation3}
\end{figure*}

We now describe a transformation that completely eliminates all accesses to the $state$ fields of \op s created by transactions in $\sct_2$ (i.e., the last remaining accesses by transactions to fields of \op s).

We transform \sct$_2$ into a new procedure \sct$_3$, which appears in Figure~\ref{code-3path-htmsct-transformation3}.
First, the \cstep\ in $\sct_2$ is eliminated.
Whereas in $\sct_2$, we stored a pointer to the \op\ in $r.\info$ for each $r \in V$ at line~\ref{htmsct-transformation2-fcas}, we store a \textit{tagged pointer} to the \op\ at line~\ref{htmsct-transformation3-fcas} in $\sct_3$.
A tagged pointer is simply a pointer that has its least significant bit set to one.
Note that, on modern systems where pointers are word aligned, the least significant bit in a pointer to an \op\ will be zero.
Thus, the least significant bit in a tagged pointer allows processes to distinguish between a tagged pointer (which is stored in $r.\info$ by a transaction) from a regular pointer (which is stored in $r.\info$ by an invocation of $\sct_O$).
Line~\ref{htmsct-transformation3-abort} in $\sct_3$ is also updated to check for a tagged pointer in $r.\info$.

In order to deal with tagged pointers, we transform \llt$_O$ into new procedure called \llt$_{HTM}$, that is used instead of \llt$_O$ from here on.
Any time an invocation of \llt$_O$ would follow a pointer that was read from an $\info$ field $r.\info$, \llt$_{HTM}$ first checks whether the value $r\info$ read from the $\info$ field is a pointer or a tagged pointer.
If it is a pointer, then \llt$_{HTM}$ proceeds exactly as in \llt$_O$.
However, if $r\info$ is a tagged pointer, then \llt$_{HTM}$ proceeds as if it had seen an \op\ with \textit{state} \done\ (i.e., whose \sct\ has already returned \true).
We explain why this is correct.
If $r\info$ contains a tagged pointer, then it was written by a transaction $T$ that committed (since it changed shared memory) at line~\ref{htmsct-transformation3-commit} in \sct$_3$, just before returning \true.
Observe that, in \sct$_2$, the $state$ of the \op\ is set to \done\ just before \true\ is returned.
In other words, if not for this transformation, $T$ would have set the $state$ of its \op\ to \done.
So, clearly it is correct to treat $r\info$ as if it were an \op\ with $state = \done$.

Since this transformation simply changes the \textit{representation} of an \op\ $D$ with $state =$ \done\ that is created by a transaction (and does not change how the algorithm behaves when it encounters $D$), it preserves PROG.

%\trevor{idea: progress is preserved because \llt$_{HTM}$ returns the same thing as \llt$_O$ would, and makes no changes to shared memory, except in \help, which it invokes if and only if \llt$_O$ would. (bit fuzzy what "\llt$_O$ would" means, since the steps/configurations have to be interleaved, and a mapping between configurations has to be there, and you really need a whole different execution where \llt$_O$ is the correct function to use, because all \sct\ operations were implemented solely with \sct$_O$, etc...)}

%\trevor{promising idea: can we simply say progress is preserved, because we're just changing the representation of some committed scx records (and not changing whether or how they are used by the llx/scx algorithm)? [use this observation to give an intuitive, simplified explanation of the correctness argument for the transformation, also?]}

\paragraph{Eliminating the creation of \op s on the fast path}

Since transactions in $\sct_3$ are not helped, we would like to eliminate the \textit{creation} of \op s in transactions, altogether. %there is little reason for them to use \op s.
However, since \op s are used as part of the \textit{freezing} mechanism in \sct$_O$ on the fallback path, we cannot simply eliminate the steps that freeze \rec s, or else transactions on the fast path will not synchronize with \sct$_O$ operations on the fallback path.
Consider an invocation $S$ of \sct$_O$ by a process $p$ that creates an \op\ $D$, and an invocation $L$ of \llt$(r)$ linked to $S$.
When $S$ uses CAS to freeze $r$ (by changing $r.\info$ from the value seen by $L$ to $D$), it interprets the success of the CAS to mean that $r$ has not changed since $L$ (relying on property P1).
If a transaction in \sct$_3$ changes $r$ without changing $r.\info$ (to a new value that has never before appeared in $r.\info$), then it would violate P1, rendering this interpretation invalid.
Thus, transactions in \sct$_3(V, R, fld, new)$ must change $r.\info$ to a new value, for each $r \in V$.

%\begin{figure}[th]
%\centering
%\includegraphics[width=0.5\linewidth]{chap-scx/tagged-seq-num.pdf}
%\caption{The bit fields of a tagged sequence number. Here, $n$ is an \textbf{upper bound} on the total number of processes.}
%\label{fig-scx-tagged-seq-num}
%\end{figure}

We transform \sct$_3$ into a new procedure \sct$_r$, which appears in Figure~\ref{code-3path-htmsct-transformation4}.
We now explain what a transaction $T$ in an invocation $S$ of \sct$_4$ by a process $p$ does instead of creating an \op\ and using it to freeze \rec s.
We give each process $p$ a \textit{tagged sequence number} \textit{tseq}$_p$, which consists of three bit fields: a tag-bit, a process name, and a sequence number. % (illustrated in Figure~\ref{fig-scx-tagged-seq-num}).
The tag-bit, which is the least significant bit, is always one.
This tag-bit distinguishes tagged sequence numbers from pointers to \op s (similar to tagged pointers, above).
The process name field of \textit{tseq}$_p$ contains $p$.
The sequence number is a non-negative integer that is initially zero.
Instead of creating a new \op\ (at line~\ref{htmsct-transformation3-new-op} in $\sct_3$), $S$ increments the sequence number field of \textit{tseq}$_p$.
Then, instead of storing a pointer to an \op\ in $r.\info$ for each $r \in V$ (at line~\ref{htmsct-transformation3-fcas} in \sct$_3$), $T$ stores $\textit{tseq}_p$.
(Line~\ref{htmsct-transformation3-abort} is also changed accordingly.)
The combination of the process name and sequence number bit fields ensure that whenever $T$ stores $\textit{tseq}_p$ in an $\info$ field, it is storing a value that has never previously been contained in that field.\footnote{Technically, with a finite word size it is possible for a sequence number to overflow and wrap around, potentially causing P1 to be violated. %in which case $p$ may sometimes write values to an $\info$ field that have previously been contained in the $\info$ field.
On modern systems with a 64-bit word size, we suggest representing a tagged sequence number using 1 tag-bit, 15 bits for the process name (allowing up to 32,768 concurrent processes) and 48 bits for the sequence number.
In order for a sequence number to experience wraparound, a \textit{single process} must then perform $2^{48}$ operations.
According to experimental measurements for several common data structures on high performance systems, this would take at least a decade of continuous updates.
Moreover, if wraparound is still a concern, one can replace the \fcas\ steps in \sct$_O$ with double-wide CAS instructions (available on all modern systems) which atomically operate on 128-bits, making wraparound virtually impossible.}

\begin{figure*}[th]
%\vspace{-2mm}
\footnotesize
\def\pwidth{4cm}
\prepnewlisting
%\hrule
\vspace{-2mm}
\begin{framed}
\begin{lstlisting}[mathescape=true]
  //Private variable for process $p$: $\textit{tseq}_p$ \vspace{1mm}\hrule\vspace{1mm}
  //\sct$_4(V, R, fld, new)$ by process $p$
    //\dline{Let $\llresults$ be a pointer to a table in $p$'s private memory containing,}%
            {for each $r$ in $V$, the value of $r.\info$ read by $p$'s last \llt$(r)$}
    //Let $old$ be the value for $fld$ returned by $p$'s last \llt$(r)$\vspace{1.5mm}%

    //Begin hardware transaction
    //\textit{tseq}$_p$ := \textit{tseq}$_p$ + $2^{\lceil \log n \rceil}$ \hfill \com{increment $p$'s tagged sequence number}
    //\com \mbox{Freeze all \rec s in $V$ to protect their mutable fields from being changed by other \sct s}
    for each $r$ in $V \mbox{ enumerated in order}$ do
      //Let $r$\info\ be the pointer indexed by $r$ in $\llresults$
      if not $\cas(r.\info,r\info,\textit{tseq}_p)$ then //\sidecom{\textbf{\fcas}}\label{htmsct-transformation4-fcas}
        if $r.\info \neq \textit{tseq}_p$ then //Abort hardware transaction (explicitly)\vspace{1.5mm}%

    //\com Finished freezing \rec s
    //\com Finalize each $r \in R$, update $fld$, and unfreeze all $r \in (V \setminus R)$
    for each $r \in R$ do $r.marked := \true$ //\sidecom{\textbf{\markstep}}
    //$\cas(fld, old, new)$ \sidecom{\textbf{\upcas}}\label{htmsct-transformation4-upcas}
    //Commit hardware transaction \label{htmsct-transformation4-commit2}
    return $\true$ //\vspace{-1mm}
\end{lstlisting}
\end{framed}
    \vspace{-5mm}
	\caption{HTM-based \sct: after eliminating \op\ creation on the fast path.}
	\label{code-3path-htmsct-transformation4}
\end{figure*}

%Using this knowledge, we transform $\sct_3$ in Figure~\ref{code-3path-htmsct-transformation3} into a new implementation called $\sct_4$ in Figure~\ref{code-3path-htmsct-transformation4}.
%First we eliminate the creation of the new \op\ at line~\ref{htmsct-transformation3-new-op} in $\sct_3$.
%In its place, we increment the process' tagged sequence number $\textit{tseq}_p$.
%We then replace the references to the \op\ pointer at lines~\ref{htmsct-transformation3-fcas} and~\ref{htmsct-transformation3-commit} with $\textit{tseq}_p$.

Observe that $\llt_{HTM}$ does not require any further modification to work with tagged sequence numbers, since it distinguishes between tagged sequence numbers and \op s using the tag-bit (the exact same way it distinguished between tagged pointers and pointers to \op s).
Moreover, it remains correct to treat tagged sequence numbers as if they are \op s with $state$ \done\ (for the same reason it was correct to treat tagged pointers that way).
Progress is preserved for the same reason as it was in the previous transformation: we are simply changing the \textit{representation} of \op s with $state =$ \done\ that are created by transactions.
%\trevor{preserves progress, since we are just changing the representation of  "empty" \op s for completed \sct s with unique addresses.}

Note that this transformation eliminates not only the \textit{creation} of \op s, but also the need to \textit{reclaim} those \op s.
Thus, it can lead to significant performance improvements.

\paragraph{Simple optimizations}

\begin{figure*}[th]
%\vspace{-2mm}
\footnotesize
\def\pwidth{4cm}
\prepnewlisting
%\hrule
\vspace{-2mm}
\begin{framed}
\begin{lstlisting}[mathescape=true]
  //Private variable for process $p$: $\textit{tseq}_p$ \vspace{1mm}\hrule\vspace{1mm}
  //\sct$_5(V, R, fld, new)$ by process $p$
    //\dline{Let $\llresults$ be a pointer to a table in $p$'s private memory containing,}%
            {for each $r$ in $V$, the value of $r.\info$ read by $p$'s last \llt$(r)$}
    //Let $old$ be the value for $fld$ returned by $p$'s last \llt$(r)$\vspace{1.5mm}%

    //Begin hardware transaction
    //\textit{tseq}$_p$ := \textit{tseq}$_p$ + $2^{\lceil \log n \rceil}$ \hfill \com{increment $p$'s tagged sequence number}
    //\com \mbox{Freeze all \rec s in $V$ to protect their mutable fields from being changed by other \sct s}
    for each $r$ in $V \mbox{ enumerated in order}$ do
      //Let $r$\info\ be the pointer indexed by $r$ in $\llresults$
      if $r.\info = r\info$ then $r.\info := \textit{tseq}_p$ //\label{htmsct-transformation5-freezing}
      else //Abort hardware transaction (explicitly) \vspace{1.5mm}%

    //\com Finished freezing \rec s
    //\com Finalize each $r \in R$, update $fld$, and unfreeze all $r \in (V \setminus R)$
    for each $r \in R$ do $r.marked := \true$ //\sidecom{\textbf{\markstep}}
    if $fld = old$ then $fld := new$ //\label{htmsct-transformation5-update}
    //Commit hardware transaction
    return $\true$ //\vspace{-1mm}
\end{lstlisting}
\end{framed}
    \vspace{-5mm}
	\caption{HTM-based \sct: after replacing CAS with sequential code and optimizing.}
	\label{code-3path-htmsct-transformation5}
\end{figure*}

Since any code executed inside a transaction is atomic, we are free to replace atomic synchronization primitives inside a transaction with sequential code, and reorder the transaction's steps in any way that does not change its sequential behaviour. % so long as doing so does not change the sequential behaviour, and , without changing the transaction's behaviour. 
We now describe how to transform \sct$_4$ by performing two simple optimizations.

For the first optimization, we replace each invocation of CAS$(x, o, n)$ with sequential code: \textbf{if} $x = 0$ \textbf{then} $x := n, result := \true$ \textbf{else} $result := \false$.
If the CAS is part of a condition for an if-statement, then we execute this code just before the if-statement, and replace the invocation of CAS with $result$.
We then eliminate any \textit{dead code} that cannot be executed.
Figure~\ref{code-3path-htmsct-transformation5} shows the transformed procedure, \sct$_5$.

More concretely, in place of the CAS at line~\ref{htmsct-transformation4-fcas} in \sct$_4$, we do the following.
First, we check whether $r.\info = r\info$.
If so, we set $r.\info := \textit{tseq}_p$ and continue to the next iteration of the loop.
Suppose not.
If we were naively transforming the code, then the next step would be to check whether $r.\info$ contains $\textit{tseq}_p$.
However, %since there are no other processes helping this \sct , 
$p$ is the only process that can write $\textit{tseq}_p$, and it only writes $\textit{tseq}_p$ just before continuing to the next iteration.
Thus, $r.\info$ cannot possibly contain $\textit{tseq}_p$ in this case, which makes it unnecessary to check whether $r.\info = \textit{tseq}_p$.
Therefore, we execute the \textit{else}-case, and explicitly abort the transaction.
Observe that, if \sct$_5$ is used to replace \sct$_1$ in Figure~\ref{code-3path-htmsct-usage}, then this explicit abort will cause \sct\ to return \false\ (right after it jumps to the onAbort label).
In place of the CAS at line~\ref{htmsct-transformation4-upcas} in \sct$_4$, we can simply check whether $fld$ contains $old$ and, if so, write $new$ into $fld$.

In fact, it is not necessary to check whether $fld$ contains $old$, because the transaction will have aborted if $fld$ was changed after $old$ was read from it.
We explain why.
Let $S$ be an invocation of \sct$_5$ (in Figure~\ref{code-3path-htmsct-transformation5}) by a process $p$, and let $r$ be the \rec\ that contains $fld$.
Suppose $S$ executes line~\ref{htmsct-transformation5-update} in \sct$_5$, where it checks whether $fld = old$.
Before invoking $S$, $p$ performs an invocation $L$ of \llt$(r)$ linked to $S$.
Subsequently, $p$ reads $old$ while performing $S$.
After that, $p$ freezes $r$ while performing $S$.
If $r$ changes after $L$, and before $p$ executes line~\ref{htmsct-transformation5-freezing}, then $p$ will see $r.\info \neq r\info$ when it executes line~\ref{htmsct-transformation5-freezing} (by property P1, which has been preserved by our transformations).
Consequently, $p$ will fail to freeze $r$, and $S$ will perform an explicit abort and return \false, so it will \textit{not} reach line~\ref{htmsct-transformation5-update}, which contradicts our assumption (so this case is impossible).
On the other hand, if $r$ changes after $p$ executes line~\ref{htmsct-transformation5-freezing}, and before $p$ executes line~\ref{htmsct-transformation5-update}, then the transaction will abort due to a data conflict (detected by the HTM system).
Therefore, when $p$ executes line~\ref{htmsct-transformation5-update}, $fld$ must contain $old$.
%Since $S$ checks whether $fld$ contains $old$ after freezing $r$, and there are no helpers, $r$ is still frozen for $S$ when $S$ checks whether $fld$ contains $old$.
%Recall that a \rec\ can be changed by an \sct\ only while the \rec\ is frozen for the \sct.

\begin{shortver}
\begin{figure*}[th]
%\vspace{-2mm}
\footnotesize
\def\pwidth{4cm}
\prepnewlisting
%\hrule
%\vspace{-2mm}
\begin{framed}
\begin{lstlisting}[mathescape=true]
  //Private variable for process $p$: $\textit{tseq}_p$ \vspace{1mm}\hrule\vspace{1mm}
  //\sct$_{HTM}(V, R, fld, new)$ by process $p$
    //\dline{Let $\llresults$ be a pointer to a table in $p$'s private memory containing,}%
            {for each $r$ in $V$, the value of $r.\info$ read by $p$'s last \llt$(r)$}
    //Let $old$ be the value for $fld$ returned by $p$'s last \llt$(r)$\vspace{1.5mm}%

    //Begin hardware transaction
    //\textit{tseq}$_p$ := \textit{tseq}$_p$ + $2^{\lceil \log n \rceil}$ \hfill \com{increment $p$'s tagged sequence number} \label{htmsct-tseq-increment}
    for each $r \in V$ do //\hfill\com abort if any $r \in V$ has changed since the linked \llt$(r)$ \label{htmsct-abort}
      //Let $r$\info\ be the pointer indexed by $r$ in $\llresults$ \label{htmsct-rinfo}
      if $r.\info \neq r\info$ then //Abort hardware transaction (explicitly)
    for each $r \in V$ do $r.\info := \textit{tseq}_p$ //\hfill\com change $r.\info$ to a new value, for each $r \in V$ \label{htmsct-change-info} \label{htmsct-fcas}
    for each $r \in R$ do $r.marked := \true$//\hfill\com mark each $r \in R$ (so it will be finalized) \label{htmsct-mark}
    //write $new$ to the field pointed to by $fld$ \hfill\com perform the update \label{htmsct-write-fld}
    //Commit hardware transaction
    return $\true$ //\vspace{-1mm}
\end{lstlisting}
\end{framed}
    \vspace{-5mm}
	\caption{Final implementation of \sct$_{HTM}$.}
	\label{code-3path-htmsct}
\end{figure*}
\end{shortver}

For the second optimization, we split the loop in Figure~\ref{code-3path-htmsct-transformation5} into two.
The first loop contains all of the steps that check whether $r.\info = r\info$, and the second loop contains all of the steps that set $r.\info := \textit{tseq}_p$.
This way, all of the writes to $r.\info$ occur after all of the reads and if-statements.
The advantage of delaying writes for as long as possible in a transaction is that it reduces the probability of the transaction %aborting or 
causing other transactions to abort. %(because a write performed inside a transaction can cause another transaction to abort immediately, whereas reads are less likely to cause other transactions to abort).
As a minor point, whereas the loop in \sct$_O$ iterated over the elements of the sequence $V$ in a particular order to guarantee progress, it is not necessary to do so here, since progress is guaranteed by the fallback path, not the fast path.
%In some data structures, this may eliminate a step where the sequence $V$ is sorted (although in many data structures such sorting is unnecessary).
Clearly, this transformation does not affect correctness or progress.
\begin{shortver}
The final result, % of this final transformation, 
\sct$_{HTM}$, appears in Figure~\ref{code-3path-htmsct}.
\end{shortver}

\end{fullver}
\end{fullver}

\section{Accelerated template implementations} \label{sec-3path-algs}

\begin{shortver}

\end{shortver}

\begin{thesisonly}
\subsection{The \textit{2-path con} algorithm}
\end{thesisonly}
\begin{thesisnot}
\fakeparagraph{The \textit{2-path con} algorithm}
\end{thesisnot}
We now use our HTM-based \llt\ and \sct\ to obtain an HTM-based implementation of a template operation $O$.
The fallback path for $O$ is simply a lock-free implementation of $O$ using \llt$_O$ and \sct$_O$.
The fast path for $O$ starts a transaction, then performs the same code as the fallback path, except that it uses the HTM-based \llt\ and \sct. %\llt$_{HTM}$ and \sct$_{HTM}$ in place of \llt$_O$ and \sct$_O$.
Since the \textit{entire operation} is performed inside a transaction, we can optimize the invocations of \sct$_{HTM}$ that are performed by $O$ as follows.
Lines~\ref{newscx-scxhtm-xbegin} and~\ref{newscx-scxhtm-commit} can be eliminated, since \sct$_{HTM}$ is already running inside a large transaction. 
Additionally, lines~\ref{newscx-scxhtm-freezing-loop-start}-\ref{newscx-scxhtm-freezing-loop-end} can be eliminated, since the transaction will abort due to a data conflict if $r.\info$ changes after it is read in the (preceding) linked invocation of \llt($r$), and before the transaction commits.
%Finally, since we eliminated line~\ref{newscx-scxhtm-freezing-loop-end}, which is the only explicit abort, line~\ref{newscx-scx-explicit-return} can also be eliminated.
The proof of correctness and progress for \textit{2-path con} follows immediately from the proof of the original template and the proof of the HTM-based \llt\ and \sct\ implementation.

Note that it is not necessary to perform the entire operation in a single transaction.
In Section~\ref{sec-3path-search-outside}, we describe a modification that allows a read-only \textit{searching} prefix of the operation to be performed before the transaction begins. %outside of the transaction. %non-transactionally. %, search phase of an operation to be performed \textit{before} the transaction.

\begin{thesisonly}
\subsection{The \textit{TLE} algorithm}
\end{thesisonly}
\begin{thesisnot}
\fakeparagraph{The \textit{TLE} algorithm}
\end{thesisnot}
%
%This algorithm is straightforward.
To obtain a TLE implementation of an operation $O$, we simply take \textit{sequential code} for $O$ and wrap it in a transaction on the fast path.
The fallback path acquires and releases a global lock instead of starting and committing a transaction, but otherwise executes the same code as the fast path.
To prevent the fast path and fallback path from running concurrently, transactions on the fast path start by reading the lock state and aborting if it is held.
An operation attempts to run on the fast path up to \textit{AttemptLimit} times (waiting for the lock to be free before each attempt) before resorting to the fallback path.
The correctness of TLE is trivial.
Note, however, that TLE only satisfies deadlock-freedom (not lock-freedom).

%Although operations implemented using the tree update template can only change a single pointer (and accomplish other changes by creating new nodes), a TLE-based implementation of $O$ need not respect this requirement.
%(
%
%Since there is no concurrency between the fast path and fallback path, the fallback path does not impose any restrictions on the implementation of $O$ on the fast path.

\begin{thesisonly}
\subsection{The \textit{2-path $\overline{con}$} algorithm}
\end{thesisonly}
\begin{thesisnot}
\fakeparagraph{The \textit{2-path $\overline{con}$} algorithm}
\end{thesisnot}
We can improve concurrency on the fallback path and guarantee lock-freedom by using a lock-free algorithm on the fallback path, and a global fetch-and-increment object $F$ instead of a global lock.
%This algorithm is somewhat similar to TLE, but with a lock-free fallback path, and a global fetch-and-increment object $F$ instead of a global lock.
Consider an operation $O$ implemented with the tree update template.
We describe a \textit{2-path $\overline{con}$} implementation of $O$.
The fallback path increments $F$, then executes the lock-free tree update template implementation of $O$, and finally decrements $F$.
The fast path executes \textit{sequential code} for $O$ in a transaction.
To prevent the fast path and fallback path from running concurrently, transactions on the fast path start by reading $F$ and aborting if it is nonzero.
An operation attempts to run on the fast path up to \textit{AttemptLimit} times (waiting for $F$ to become zero before each attempt) before resorting to the fallback path.
\begin{fullver}
(Note that $F$ can actually be a \textit{counter} object, instead of a fetch-and-increment object. %it actually suffices for $F$ to be a \textit{counter} object, which is somewhat weaker than a fetch-and-increment object, since the former can be implemented using only registers.
The former is somewhat weaker, and can be implemented using only registers.)
\end{fullver}

Recall that operations implemented using the tree update template can only change a single pointer atomically (and can perform multiple changes atomically only by creating a connected set of new nodes that reflect the desired changes).
Thus, each operation on the fallback path simply creates new nodes and changes a single pointer (and assumes that all other operations also behave this way).
However, since the fast path and fallback path do not run concurrently, the fallback path does \textit{not} impose this requirement on the fast path.
Consequently, the fast path can make (multiple) direct changes to nodes. %(which can be considerably more efficient).
Unfortunately, as we described above, this algorithm can still suffer from concurrency bottlenecks.

\begin{thesisonly}
\subsection{The \textit{3-path} algorithm}
\end{thesisonly}
\begin{thesisnot}
\fakeparagraph{The \textit{3-path} algorithm}
\end{thesisnot}
One can think of the \textit{3-path} algorithm as a kind of hybrid between the \textit{2-path con} and \textit{2-path $\overline{con}$} algorithms that obtains their benefits while avoiding their downsides.
Consider an operation $O$ implemented with the tree update template.
We describe a 3-path implementation of $O$.
As in \textit{2-path $\overline{con}$}, there is a global fetch-and-increment object $F$, and the fast path executes \textit{sequential code} for $O$ in a transaction.
The middle path and fallback path behave like the fast path and fallback path in the \textit{2-path con} algorithm, respectively. % (\textit{with} the optimization described above for running each fast path operation in a single large transaction).
Each time an operation begins (resp., stops) executing on the fallback path, it increments (resp., decrements) $F$.
(If the scalability of fetch-and-increment is of concern, then a \textit{scalable non-zero indicator} object~\cite{ellen2007snzi} can be used, instead.)
This prevents the fast and fallback paths from running concurrently.
As we described above, operations begin on the fast path, and move to the middle path after \textit{FastLimit} attempts, or if they see $F \neq 0$.
Operations move from the middle path to the fallback path after \textit{MiddleLimit} attempts.
Note that an operation never waits for the fallback path to become empty---it simply moves to the middle path.

Since the fast path and fallback path do not run concurrently, the fallback path does not impose any overhead on the fast path, except checking if $F = 0$ (offering low overhead for light workloads).
Additionally, when there are operations running on the fallback path, hardware transactions can continue to run on the middle path (offering high concurrency for heavy workloads).

\begin{shortver}
Due to lack of space, correctness and progress arguments are deferred to Appendix~\ref{appendix-3path-correctness}.
\end{shortver}
\begin{fullver}
\paragraph{Correctness}
The correctness argument for \textit{3-path} is straightforward.
The goal is to prove that all template operations are linearizable, regardless of which path they execute on.
%the correctness of interactions between the fallback path and middle path, and interactions between the middle path and fast path.
\begin{shortver}
Recall that the fallback path and middle path behave like the fast path and fallback path in \textit{2-path con}, and the correctness of \textit{2-path con} was proved in Appendix~\ref{appendix-correctness}.
\end{shortver}
\begin{fullver}
Recall that the fallback path and middle path behave like the fast path and fallback path in \textit{2-path con}, and the correctness of \textit{2-path con} was proved above.
\end{fullver}
It follows that, if there are no operations on the fast path, then the correctness of operations on the middle path and fallback path is immediate from the correctness of \textit{2-path con}.
Of course, whenever there is an operation executing on the fallback path, no operation can run on the fast path.
Since operations on the fast path and middle path run in transactions, they are atomic, and any conflicts between the fast path and middle path are handled automatically by the HTM system.
Therefore, all template operations are linearizable.

\paragraph{Progress}
The progress argument for \textit{3-path} relies on a three simple assumptions.
\begin{compactenum}[\bf {A}1.]
\item The sequential code for an operation executed on the fast path must terminate after a finite number of steps if it is run on a static tree (which does not change during the operation).
\item In an operation executed on the middle path or fallback path, the search phase must terminate after a finite number of steps if it is run on a static tree.
%Note that these assumptions are preconditions of lock-freedom.
\item In an operation executed on the middle path or fallback path, the update phase can modify only a finite number of nodes.
\end{compactenum}

\smallskip

We give a simple proof that \textit{3-path} satisfies lock-freedom. % if the sequential code for each operation terminates. %progress property $\mathcal{P} \in \{$deadlock- and livelock-freedom, lock-freedom, wait-freedom$\}$ if the fallback path satisfies $\mathcal{P}$ and each \textbf{transaction} terminates.
%We present the argument for $\mathcal{P} = $ lock-freedom.
%(The other arguments are similar.)
%\trevor{for example, for wait-freedom... others are similar.}
To obtain a contradiction, suppose there is an execution in which after some time $t$, some process takes infinitely many steps, but no operation terminates.
Thus, the tree does not change after $t$.
We first argue that no process takes infinitely many steps in a transaction $T$.
If $T$ occurs on the fast path, then A1 guarantees it will terminate.
If $T$ occurs on the middle path, then A2 and A3 guarantee that it will terminate.
%Suppose $T$ occurs on the fast path.
%Then, $T$ performs sequential code for an operation on the data structure.
%We assume this sequential code is written in such a way that it will terminate after a finite number of steps if it is run on a static tree.
%Since the tree does not change after $t$, this operation will eventually terminate.
%%Since its update phase modifies a finite set of nodes, the operation will eventually terminate.
%Now suppose $T$ occurs on the middle path.
%Then, $T$ performs a template operation, which consists of a search phase followed by an update phase.
%Since the tree does not change after $t$, the search phase will eventually terminate.
%Since the update phase can modify only finitely many nodes, it will eventually terminate.
%Thus, $T$ eventually terminates.
Therefore, eventually, processes only take steps on the fallback path.
Progress then follows from the fact that the original tree update template implementation (our fallback path) is lock-free.
%
%Each operation that takes infinitely many steps eventually executes only on the fallback path, since the number of transactional attempts on the fast path and middle path are bounded, and each transaction terminates.
%After this point, the execution is essentially an execution of the fallback path algorithm, so $\mathcal{P}$ is satisfied.
\end{fullver}

\section{Example data structures} \label{sec-3path-ds}

\begin{thesisonly}
We use two data structures as examples to help us study our accelerated template algorithms: an unbalanced BST (similar to the Chromatic tree in Chapter~\ref{chap-chromatree}, but with no rebalancing), and the relaxed $(a,b)$-tree described in Chapter~\ref{chap-abtree}.
\end{thesisonly}
\begin{thesisnot}
We use two data structures as examples to help us study our accelerated template algorithms: an unbalanced BST, and a relaxed $(a,b)$-tree.
\end{thesisnot}
In this section, we briefly describe these algorithms, and %description the unbalanced BST and the relaxed $(a,b)$-tree.
%We also 
give additional details on their \textit{3-path} implementations. % of these data structures.

Each data structure implements the ordered dictionary ADT, which stores a set of keys, and associates each key with a value.
An ordered dictionary offers four operations: \textsc{Insert}$(key, value)$, \textsc{Delete}$(key)$, \textsc{Search}$(key)$ and \textsc{RangeQuery}$(lo, hi)$.
\begin{thesisonly}
Recall that both
\end{thesisonly}
\begin{thesisnot}
Both
\end{thesisnot}
data structures are \textit{leaf-oriented} (also called \textit{external}), which means that all of the keys in the dictionary are stored in the leaves of the tree, and internal nodes contain \textit{routing} keys which simply direct searches to the appropriate leaf.
This is in contrast to \textit{node-oriented} or \textit{internal} trees, in which internal nodes also contain keys in the set.

\subsection{Unbalanced BST}

%%In this section, we describe a lock-free \textit{leaf-oriented} unbalanced BST that implements an ordered dictionary (which stores a set of keys, and associates key with a value) and offers four operations: \textsc{Insert}$(key, value)$, \textsc{Delete}$(key)$, \textsc{Search}$(key)$ and \textsc{RangeQuery}$(lo, hi)$.
%In a \textit{leaf-oriented} (also called \textit{external}) tree, all of the keys in the set are stored in the leaves of the tree, and internal nodes contain \textit{routing} keys which simply direct searches to the appropriate leaf.
%(This is in contrast to \textit{node-oriented} or \textit{internal} trees, in which internal nodes also contain keys in the set.)
%%\textit{Lock-free} (also called non-blocking) algorithms guarantee that the system as a whole will always make progress (but individual operations may starve).
%We now describe the \textit{3-path} implementation of the unbalanced BST.

\begin{figure}
    %    \vspace{-7mm}
    \centering
    \includegraphics[width=\linewidth]{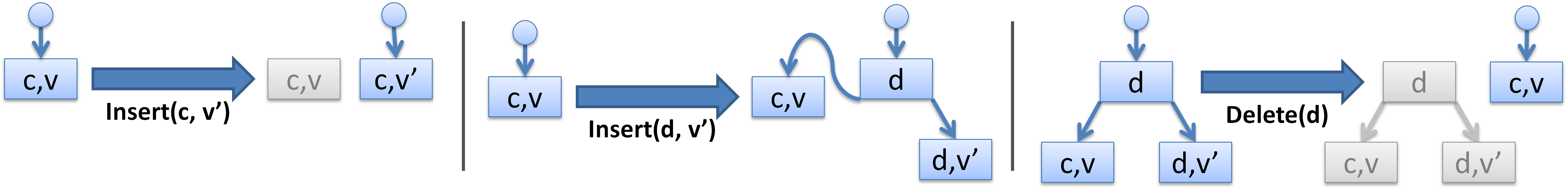}
    \caption{Fallback path operations for the unbalanced BST.}
    \label{fig-lfbst-ops}
\end{figure}

\textbf{Fallback path.}
The fallback path consists of a lock-free implementation of the operations in Figure~\ref{fig-lfbst-ops} using the (original) tree update template. %handcrafted implementation obtained by using the methodology for lock-free trees of Brown~et~al.~\cite{Brown:2014}.
\begin{thesisonly}
Note that this implementation is similar to the Chromatic tree in Chapter~\ref{chap-chromatree}, except it does not perform any rebalancing.
\end{thesisonly}
%Figure~\ref{fig-lfbst-ops} shows the insertion and deletion operations for the tree.
As required by the template, these operations change child pointers, but do not change the key or value fields of nodes directly.
Instead, to replace a node's key or value, the node is replaced by a new copy.
If $key$ is not already in the tree, then \textit{Insert}$(key, value)$ inserts a new leaf and internal node.
Otherwise, \textit{Insert}$(key, value)$ replaces the leaf containing $key$ with a new leaf that contains the updated value.
\textit{Delete}$(key)$ replaces the leaf $l$ being deleted and its parent with a new copy of the sibling of $l$.
%We now briefly explain why these operations sometimes replace nodes with copies, rather than modifying them directly.
%Although this algorithm is a state of the art concurrent data structure, it has several significant inefficiencies, which we briefly discuss.
%We briefly discuss
%We discuss three of these inefficiencies, here. (Liu~et~al.~\cite{Liu2015} also discuss numerous inefficiencies these )

%%\textbf{Copying nodes to avoid race conditions.}
%One might wonder why a node should be replaced instead of having its fields changed directly, which would be significantly more efficient.
%It turns out that it is very difficult to avoid race conditions in lock-free algorithms if fields other than pointers are changed directly.
%For example, in a BST, if an \textit{Insert} changes the value associated with a key at some node $u$, and a \textit{Delete} simultaneously removes the node $u$ from the tree, then the effect of the \textit{Insert} is lost.
%To avoid this problem, the \textit{Delete} of $u$ should succeed only if $u$ is still pointed to by its parent \textit{and} $u$'s still contains the value we expect.
%However, this is difficult to ensure when using classical atomic synchronization primitives like single word compare and swap (CAS) or load-linked/store-conditional (LL/SC), since they operate only on a single word.
%
%\textbf{Copying nodes to avoid the ABA problem.}
It may seem strange that \textit{Delete} creates a new copy of the deleted leaf's sibling, instead of simply reusing the existing sibling (which is not changed by the deletion).
\begin{thesisonly}
Recall that this comes from a requirement of the tree update template: each invocation of \sct$(V, R, fld, new)$ must change the field $fld$ to a value that it has \textit{never previously contained} (to avoid the ABA problem).
\end{thesisonly}
\begin{thesisnot}
This comes from a requirement of the tree update template: each invocation of \sct$(V, R, fld, new)$ must change the field $fld$ to a value that it has \textit{never previously contained}.
%This requirement is satisfied by ensuring that each operation uses \sct\ to change a pointer to point to a newly created node.
%
This requirement is motivated by a particularly tricky aspect of lock-free programming: avoiding the \textit{ABA problem}.
The ABA problem occurs when a process $p$ reads a memory location $x$ and sees value $A$, then performs a CAS on $x$ to change it from $A$ to $C$, and \textit{interprets} the success of this CAS to mean that $x$ has not changed between when $p$ read $x$ and performed the CAS on it.
In reality, after $p$ read $x$ and before it performed the CAS, another process $q$ may have changed $x$ to $B$, and then back to $A$, rendering $p$'s interpretation %of the successful CAS 
invalid.
In practice, the ABA problem can result in data structure operations being applied multiple times, or lost altogether.
%One way to ensure that the ABA problem does not occur is to perform CAS only on pointers to freshly allocated memory (that no other process has a pointer to).
The ABA problem cannot occur if each successful CAS on a field stores a value that has never previously been contained in the field (since, then, $q$ cannot change $x$ from $B$ back to $A$).
So, in the template, the ABA problem is avoided by having each operation use \sct\ to store \textit{a pointer to a newly created node} (which cannot have previously been contained in any field).
%One way to avoid the ABA problem is to tag each word that will be changed by a CAS with a \textit{version number}, and use a double-wide CAS (DWCAS) that can atomically operate on both the word and its tag.
\end{thesisnot}

\textbf{Middle path.}
The middle path is the same as the fallback path, except that each operation is performed in a large transaction, and the HTM-based implementation of \llt\ and \sct\ is used instead of the original implementation.

\textbf{Fast path.}
\begin{figure}
%    \vspace{-9mm}
    \centering
    \includegraphics[width=\linewidth]{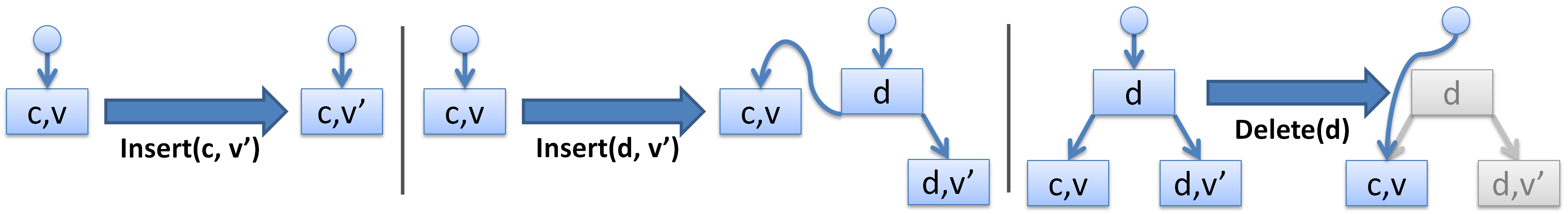}
    \caption{
        Fast path operations for the unbalanced BST.
        (\textit{Insert}$(d, v')$ is the same as on the fallback path.)
    }
    \label{fig-lfbst-ops-fastpath}
\end{figure}
The fast path is a sequential implementation of the BST, where each operation is executed in a transaction.
Figure~\ref{fig-lfbst-ops-fastpath} shows the insertion and deletion operations on the fast path.
%The first step in each transaction is to read the global fetch-and-add object $F$ and abort if it is non-zero (to prevent concurrency between the fast path and fallback path).
Unlike on the fallback path, operations on the fast path directly modify the keys and values of nodes, and, hence, can avoid creating nodes in some situations.
If $key$ is already in the tree, then \textit{Insert}$(key, value)$ directly changes the value of the leaf that contains $key$.
Otherwise, \textit{Insert}$(key, value)$ creates a new leaf and internal node and attaches them to the tree.
\textit{Delete}$(key)$ changes a pointer to remove the leaf containing $key$ and its parent from the tree.

%\noindent
\textbf{How the fast path improves performance.}
The first major performance improvement on the fast path comes from a reduction in node creation.
Each invocation of \textit{Insert}$(key, value')$ that sees $key$ in the tree can avoid creating a new node by writing $value'$ directly into the node that already contains $key$.
In contrast, a new node had to be created on the middle path, since the middle path runs concurrently with the fallback path, which assumes that the keys and values of nodes do not change.
Additionally, each invocation of \textit{Delete} that sees $key$ in the tree can avoid creating a new copy of the sibling of the deleted leaf.
This optimization was not possible on the middle path, because the fallback path assumes that each successful operation writes a pointer to a newly created node.
The second major improvement comes from the fact that reads and writes suffice where invocations of LLX and SCX were needed on the other paths.

\subsection{Relaxed $(a,b)$-tree}
%
%\subsection{Example: lock-free (a, b)-tree}
%
\begin{thesisonly}
Recall that a relaxed $(a,b)$-tree is an advanced balanced tree in which nodes contain at most $b$ keys, and, if there are no ongoing updates (insertions and deletions), nodes contain at least $a$ keys, and all leaves have the same depth.
\end{thesisonly}
\begin{thesisnot}
The relaxed $(a,b)$-tree~\cite{JL01abtrees} is a generalization of a B-tree (a balanced tree in which each node contains many keys).
%In this section, we describe a three path algorithm for an external $(a,b)$-tree~\cite{JL01abtrees} (a generalization of a B-tree) that implements an ordered dictionary.
Larsen introduced the relaxed $(a,b)$-tree as a sequential data structure that was well suited to fine-grained locking.
Internal nodes contain up to $b-1$ \textit{routing} keys, and have one more child pointer than the number of keys.
Leaves contain up to $b$ key-value pairs (which are in the dictionary).
Values may be pointers to large data objects.
The \textit{degree} of an internal node (resp. leaf) is the number of pointers (resp. keys) it contains.
When there are no ongoing updates (insertions and deletions) in a relaxed $(a,b)$-tree, all leaves have the same depth, and nodes have \textit{degree} at least $a$ and at most $b$, where $b \ge 2a-1$.
Maintaining this balance condition requires rebalancing steps similar to the \textit{splits} and \textit{joins} of B-trees.
(See~\cite{JL01abtrees} for further details on the rebalancing steps.)
%For simplicity, we do not describe these rebalancing steps.
\end{thesisnot}

\textbf{Fallback path.}
\begin{thesisonly}
The fallback path is the lock-free relaxed $(a,b)$-tree in Chapter~\ref{chap-abtree}, which uses the (original) tree update template.
For convenience, we briefly recall how its operations work.
For simplicity, we omit a description of the rebalancing steps here, and only describe the insertion and deletion operations.
\end{thesisonly}
\begin{thesisnot}
The fallback path consists of a lock-free implementation of the relaxed $(a,b)$-tree operations using the (original) tree update template.
\end{thesisnot}
%We implemented an $(a,b)$-tree using the methodology (for applying LLX and SCX) of Brown et~al.~\cite{Brown:2014}.
%To our knowledge, prior to this work, no lock-free implementation of $(a,b)$-trees has appeared.
%Similar to the lock-free BST, keys and values are not changed directly.
%Instead, keys and values are changed by replacing nodes with copies.
If $key$ is in the tree, then \textit{Insert}$(key, value)$ replaces the leaf containing $key$ with a new copy that contains $(key, value)$.
Suppose $key$ is not in the tree.
Then, \textit{Insert} finds the leaf $u$ where the key should be inserted.
If $u$ is not full (has degree less than $b$), then it is replaced with a new copy that contains $(key, value)$.
Otherwise, $u$ is replaced by a subtree of three new nodes: one parent and two children.
The two new children evenly share the key-value pairs of $u$ and $(key, value)$.
The new parent $p$ contains only a single routing key and two pointers (to the two new children), and is \textit{tagged}, which indicates that the subtree rooted at $p$ is too tall, and rebalancing should be performed to shrink its height.
\textit{Delete}$(key)$ replaces the leaf containing $key$ with a new copy $new$ that has $key$ deleted.
If the degree of $new$ is smaller than $a$, then rebalancing must be performed.

\textbf{Middle path.}
This path is obtained from the fallback path the same way as in the unbalanced BST.
%As in the unbalanced BST, the middle path  %The middle path was obtained from the fallback path the same way as for the external BST. %the same way as the middle path for the external BST.

\textbf{Fast path.}
The fast path is a sequential implementation of a relaxed $(a,b)$-tree whose operations are executed inside transactions.
Like the external BST, the major performance improvement over the middle path comes from the facts that (1) operations create fewer nodes, and (2) reads and writes suffice where LLX and SCX were needed on the other paths.
In particular, \textit{Insert}$(key, value)$ and \textit{Delete}$(key)$ simply directly modify the keys and values of leaves, instead of creating new nodes, except in the case of an \textit{Insert} into a full node $u$.
In that case, two new nodes are created: a parent and a sibling for $u$.
%The key-value pairs of $u$ and $(key, value)$ are then evenly distributed between $u$ and its new sibling.
(Recall that this case resulted in the creation of three new nodes on the fallback path and middle path.)
Note that reducing node creation is more impactful for the relaxed $(a,b)$-tree than for the unbalanced BST, since nodes are much larger.
As a minor point, we found that it was faster in practice to perform \textit{rebalancing steps} by creating new nodes, and simply replacing the old nodes with the new nodes that reflect the desired change (instead of rebalancing by directly changing the keys, values and pointers of nodes).

\begin{thesisnot}
\vspace{-2.5mm}
\end{thesisnot}
\section{Experimental results} \label{sec-3path-exp}
\begin{thesisnot}
\vspace{-1.5mm}
\end{thesisnot}
We used two different Intel systems for our experiments: %an i7-4770 with eight hardware threads, 
a dual-socket 12-core E7-4830 v3 with hyperthreading for a total of 48 hardware threads (running Ubuntu 14.04LTS), and a dual-socket 18-core E5-2699 v3 with hyperthreading for a total of 72 hardware threads (running Ubuntu 15.04).
Each machine had 128GB of RAM.
We used the scalable thread-caching allocator (tcmalloc) from the Google perftools library.
All code was compiled on GCC 4.8+ with arguments \texttt{-std=c++0x -O2 -mcx16}.
(Using the higher optimization level \texttt{-O3} did not significantly improve performance for any algorithm, and decreased performance for some algorithms.)
%All systems supported Intel's implementation of HTM.
On both machines, we \textit{pinned} threads such that we saturate one socket before scheduling any threads on the other.

\begin{shortver}
\textbf{Data structures implemented with the template.}
We used two data structures to study the performance of our accelerated template implementations: an unbalanced BST, and a relaxed $(a,b)$-tree.
The relaxed $(a,b)$-tree is an advanced balanced tree in which nodes contain up to $b$ keys, and, when there are no ongoing updates, they contain at least $a$ keys (where $b \ge 2a-1$).
\end{shortver}
\begin{fullver}
\textbf{Data structure parameters.}
Recall that nodes in the relaxed $(a,b)$-tree contain up to $b$ keys, and, when there are no ongoing updates, they contain at least $a$ keys (where $b \ge 2a-1$).
\end{fullver}
In our experiments, we fix $a=6$ and $b=16$.
With $b = 16$, each node occupies four consecutive cache lines.
Since $b \ge 2a-1$, with $b=16$, we must have $a \le 8$.
We chose to make $a$ slightly smaller than 8 in order to exploit a performance tradeoff: a smaller minimum degree may slightly increase depth, but decreases the number of rebalancing steps that are needed to maintain balance.
\begin{shortver}
Further details appear in Appendix~\ref{appendix-ds}.
\end{shortver}

\textbf{Template implementations studied.}
We implemented each of the data structure with four different template implementations: \textit{3-path}, \textit{2-path con}, \textit{TLE} and the original template implementation, which we call \textit{Non-HTM}.
(\textit{2-path $\overline{con}$} is omitted, since it performed similarly to TLE, and cluttered the graphs.)
%These correspond to the three path algorithms we described, two path algorithms using the middle and fallback paths we described (which can execute concurrently), algorithms obtained by applying TLE to the sequential code for the fast paths we described, and algorithms consisting of the fallback paths we described.
%The \textit{3 path} variations are the three path algorithms we described.
%The \textit{2 path} variations are two path algorithms using the middle and fallback paths we described. (Recall that these paths can execute concurrently.)
%The \textit{TLE} variations were obtained by applying transactional lock elision to the sequential code for the fast paths we described.
%(The \textit{Non-HTM} variations use the original tree update template.)
%
The \textit{2-path con} and \textit{TLE} implementations perform up to 20 attempts on the fast path before resorting to the fallback path.
\textit{3-path} performs up to 10 attempts (each) on the fast path and middle path.
We implemented memory reclamation using DEBRA~\cite{Brown:2015}, an epoch based reclamation scheme.
\begin{shortver}
A more efficient way to reclaim memory for \textit{3-path} is proposed in Appendix~\ref{appendix-memrecl}.
\end{shortver}
\begin{fullver}
A more efficient way to reclaim memory for \textit{3-path} is proposed in Section~\ref{sec-3path-memrecl}
\end{fullver}
%RCU primitives for the internal BST were provided by the Userspace RCU library~\cite{Desnoyers:2012}.
%We were unable to install this library on the E5-2699 v3 system, so we could not use the internal BST on that system.

%A PODC reviewer suggested comparing with a hybrid TM such as PhTM.
%Unfortunately, none of the hybrid TMs described above have publicly available code, and we have not yet had time to pursue this, but we will investigate further.
%We expect that the three path algorithms will significantly outperform any hybrid TM based versions, since they are handcrafted implementations, and hybrid TM typically involves considerable overhead.
%%said that I should compare with … code not publicly available, and I have not had time to do this, but I will investigate further.
%%	Must say something about competitors: they won't be competitive, not publicly available

\subsection{Light vs. Heavy workloads}

\fakeparagraph{Methodology}
%For the external BST, ($a,b$)-tree and internal BST,
We study two workloads: in \textbf{light}, $n$ processes perform updates (50\% insertion and 50\% deletion), and in \textbf{heavy}, $n-1$ processes perform updates, and one thread performs 100\% range queries (RQs).
For each workload and data structure implementation, and a variety of thread counts, we perform a set of five randomized trials.
In each trial, $n$ processes perform either updates or RQs (as appropriate for the workload) for one second, and counted the number of completed operations.
Updates are performed on keys
%\begin{wrapfigure}{r}{0.52\linewidth}
\begin{figure}[tb]
\centering
    \setlength\tabcolsep{0pt}
    \begin{tabular}{m{0.02\textwidth}m{0.3\textwidth}m{0.3\textwidth}}%m{0.24\textwidth}}
        &
        \multicolumn{2}{c}{%4}{c}{
            %\dimexpr \textwidth-2\fboxsep-2\fboxrule
            \fcolorbox{black!80}{black!40}{\parbox{\dimexpr 0.6\textwidth-2\fboxsep-2\fboxrule}{\centering\textbf{2x 24-thread Intel E7-4830 v3}}}
        }
        \\
        &
        \fcolorbox{black!50}{black!20}{\parbox{\dimexpr \linewidth-2\fboxsep-2\fboxrule}{\centering {\footnotesize Unbalanced BST (LLX/SCX)\\ Updates w/key range [$0,10^4$)}}} &
%        \fcolorbox{black!50}{black!20}{\parbox{\dimexpr \linewidth-2\fboxsep-2\fboxrule}{\centering {\footnotesize External BST (LLX/SCX)\\ Updates w/key range [$0,10^6$)}}} &
        \fcolorbox{black!50}{black!20}{\parbox{\dimexpr \linewidth-2\fboxsep-2\fboxrule}{\centering {\footnotesize (a, b)-tree (LLX/SCX)\\ Updates w/key range [$0,10^6$)}}}
        \\
        \vspace{-3mm}\rotatebox{90}{{\footnotesize\textbf{Light} workload}} &
        \includegraphics[width=\linewidth]{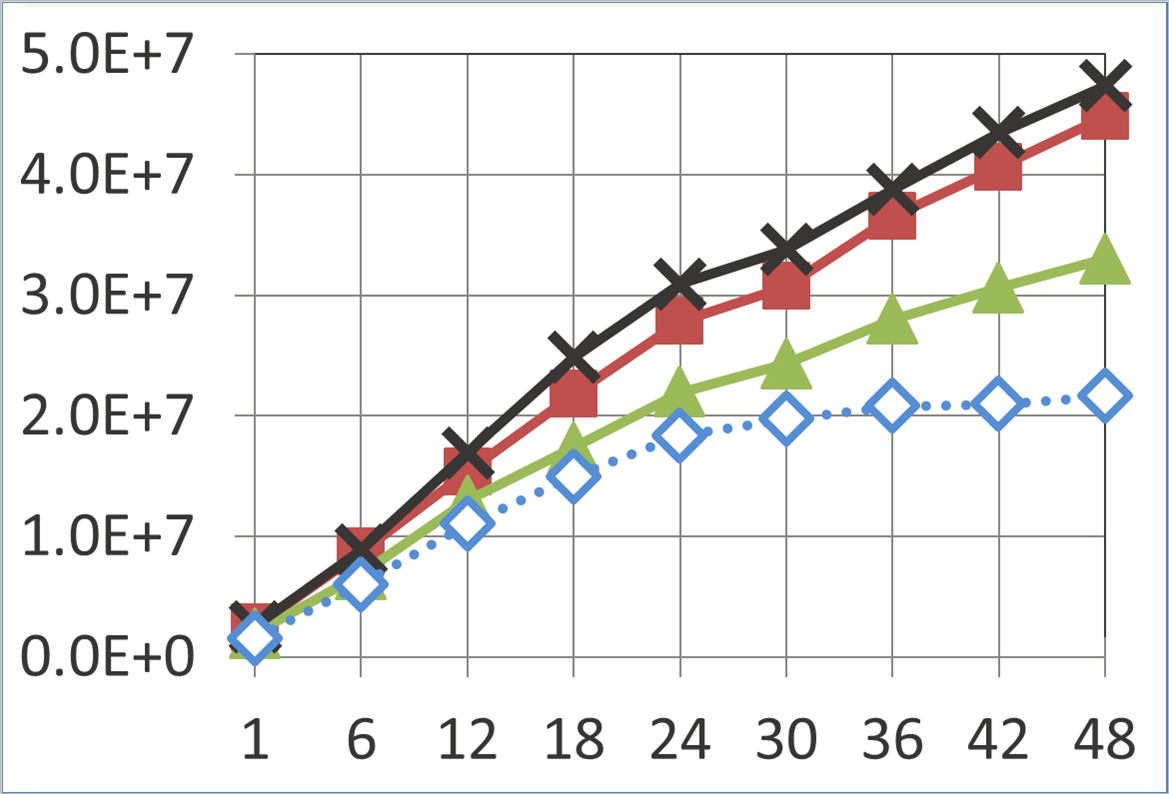} &
        \includegraphics[width=\linewidth]{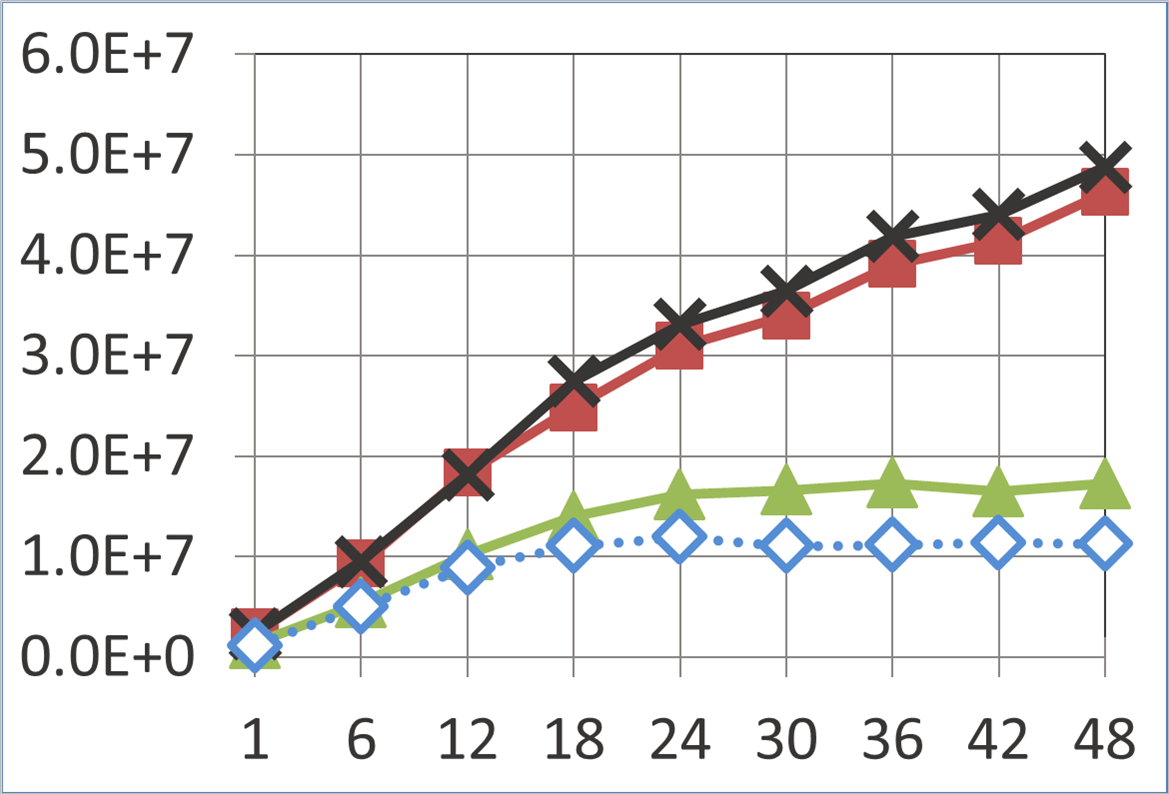}
        \\
        \vspace{-4.5mm}\rotatebox{90}{{\footnotesize\textbf{Heavy} workload}} &
        \vspace{-4.5mm}\includegraphics[width=\linewidth]{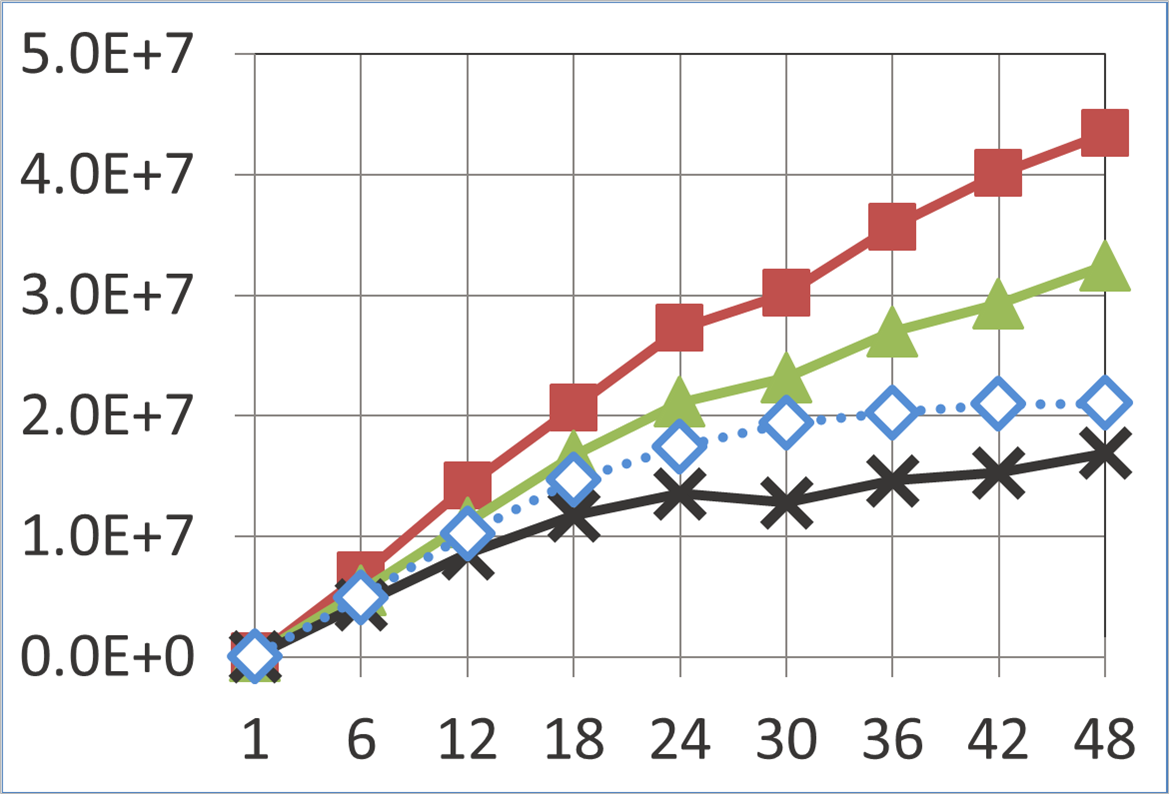} &
        \vspace{-4.5mm}\includegraphics[width=\linewidth]{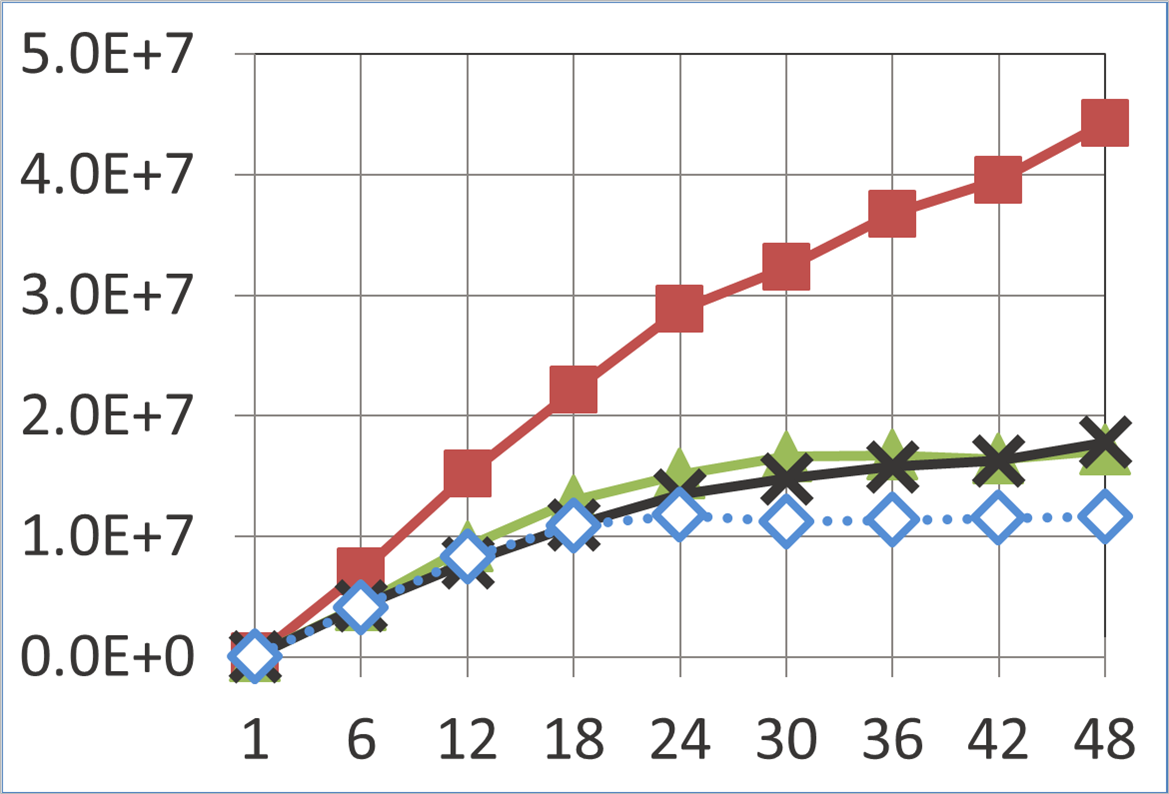}
        \\
    \end{tabular}

    \vspace{-1mm}
    \hspace{0.02\textwidth}\includegraphics[width=0.6\linewidth]{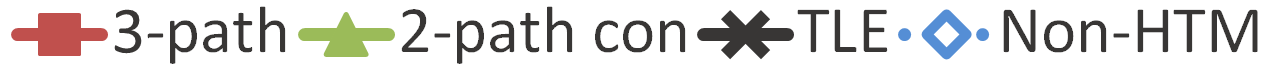}
%    \vspace{-2mm}
    \caption{Results (48-thread system) showing throughput (operations per second) versus the number of concurrent processes.}
    \label{fig-graphs}
    \vspace{-3mm}
\end{figure}
drawn uniformly randomly from a fixed key range [$0, K$).
RQs are performed on ranges [$lo, lo+s$) where $lo$ is uniformly random in [$0, K$) and $s$ is chosen, according to a probability distribution described below, from [$1, 1000$] for the BST and [$1, 10000$] for the ($a,b$)-tree.
%\trevor{explain distribution!!!!}
(We found that nodes in the ($a,b$)-tree contained approximately 10 keys, on average, so the respective maximum values of $s$ for the BST and ($a,b$)-tree resulted in range queries returning keys from approximately the same number of nodes in both data structures.)
To ensure that we are measuring steady-state performance, at the start of each trial, the data structure is prefilled by having threads perform 50\% insertions and 50\% deletions on uniform keys until the data structure contains approximately half of the keys in [$0, K$).

We verified the correctness of each data structure after each trial by %the results of each trial by 
computing \textit{key-sum hashes}.
Each thread maintains the sum of all keys it successfully inserts, minus the sum of all keys it successfully deletes.
At the end of the trial, the total of these sums over all threads must match the sum of keys in the tree. % (and we verified this).

\fakeparagraph{Probability distribution of $s$}
We chose the probability distribution of $s$ to produce many small RQs, and a smaller number of very large ones.
%To achieve this, we generated a random value of $s$ as follows.
%First, draw a real number $x$ uniformly from [$0, 1$).
%Then, square it, biasing the distribution towards zero.
%Finally, multiply the result by 1000 for the BST, or 10000 for the ($a,b$)-tree.
To achieve this, we chose $s$ to be $\lfloor x^2 S \rfloor + 1$, where $x$ is a uniform real number in [$0, 1$), and $S = 1000$ for the BST and $S = 10000$ for the ($a,b$)-tree.
By squaring $x$, we bias the uniform distribution towards zero, creating a larger number of small RQs. %The idea is to square $x$ to bias the distribution towards zero

\fakeparagraph{Results}
We briefly discuss the results from the 48 thread machine, which appear in Figure~\ref{fig-graphs}.
The BST and the relaxed ($a,b$)-tree behave fairly similarly.
Since the ($a,b$)-tree has large nodes, it benefits much more from a low-overhead fast path (in \textit{TLE} or \textit{3-path}) which can avoid creating new nodes during updates.
In the light workloads, \textit{3-path} performs significantly better than \textit{2-path con} (which has more overhead) and approximately as well as \textit{TLE}.
On average, the \textit{3-path} algorithms completed 2.1x as many operations as their \textit{non-HTM} counterparts (and with 48 concurrent processes, this increases to 3.0x, on average).
%Since contention is low and updates access a relatively small number of nodes (compared to RQs), almost all operations succeed in hardware, so \textit{TLE} essentially represents the best performance we can achieve in this case. %, since operations almost never run Less than 0.25\% of operations executed on the fallback path, so TLE represents the best performance we can achieve in this case. %when operations almost never run on the fallback path.)
In the heavy workloads, \textit{3-path} significantly outperforms \textit{TLE} (completing 2.0x as many operations, on average), which suffers from \textit{excessive waiting}. %but the other algorithms are mostly unaffected.
Interestingly, \textit{3-path} is also significantly faster than \textit{2-path con} in the heavy workloads.
This is because, even though RQs are always being performed, some RQs can succeed on the fast path, so many update operations can still run on the fast path in \textit{3-path}, where they incur much less overhead (than they would in \textit{2-path con}).

\begin{shortver}
Experimental results from the 72-thread machine appear in Appendix~\ref{appendix-exp}.
%There, \textit{3-path} shows an even larger performance advantage over \textit{Non-HTM}.
\end{shortver}
\begin{fullver}
\begin{figure}[tb]
\begin{minipage}{\textwidth}
    \centering
    \setlength\tabcolsep{0pt}
    \begin{tabular}{m{0.02\textwidth}m{0.3\textwidth}m{0.3\textwidth}}%m{0.24\textwidth}}
        &
        \multicolumn{2}{c}{%4}{c}{
            %\dimexpr \textwidth-2\fboxsep-2\fboxrule
            \fcolorbox{black!80}{black!40}{\parbox{\dimexpr 0.6\textwidth-2\fboxsep-2\fboxrule}{\centering\textbf{2x 36-thread Intel E5-2699 v3}}}
        }
        \\
        &
        \fcolorbox{black!50}{black!20}{\parbox{\dimexpr \linewidth-2\fboxsep-2\fboxrule}{\centering {\footnotesize Unbalanced BST (LLX/SCX)\\ Updates w/key range [$0,10^4$)}}} &
        %        \fcolorbox{black!50}{black!20}{\parbox{\dimexpr \linewidth-2\fboxsep-2\fboxrule}{\centering {\footnotesize External BST (LLX/SCX)\\ Updates w/key range [$0,10^6$)}}} &
        \fcolorbox{black!50}{black!20}{\parbox{\dimexpr \linewidth-2\fboxsep-2\fboxrule}{\centering {\footnotesize (a, b)-tree (LLX/SCX)\\ Updates w/key range [$0,10^6$)}}}
        \\
        \vspace{-3mm}\rotatebox{90}{{\footnotesize\textbf{Light} workload}} &
        \includegraphics[width=\linewidth]{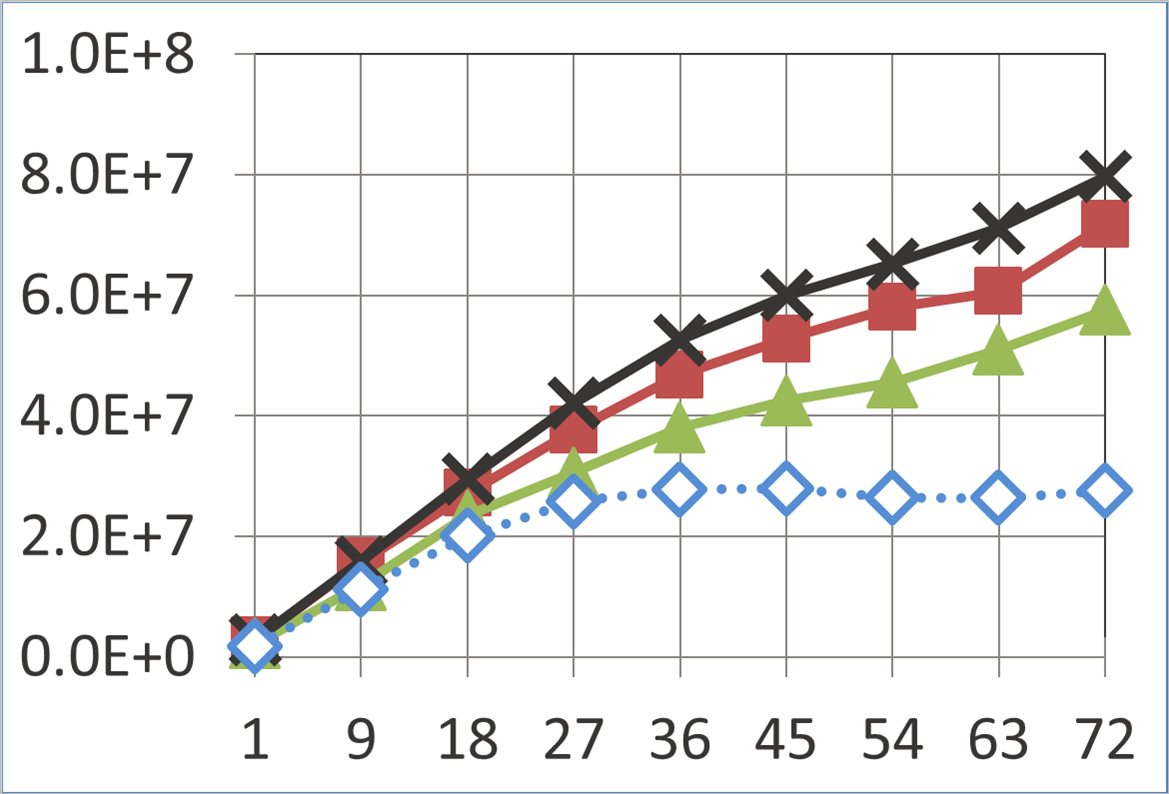} &
        \includegraphics[width=\linewidth]{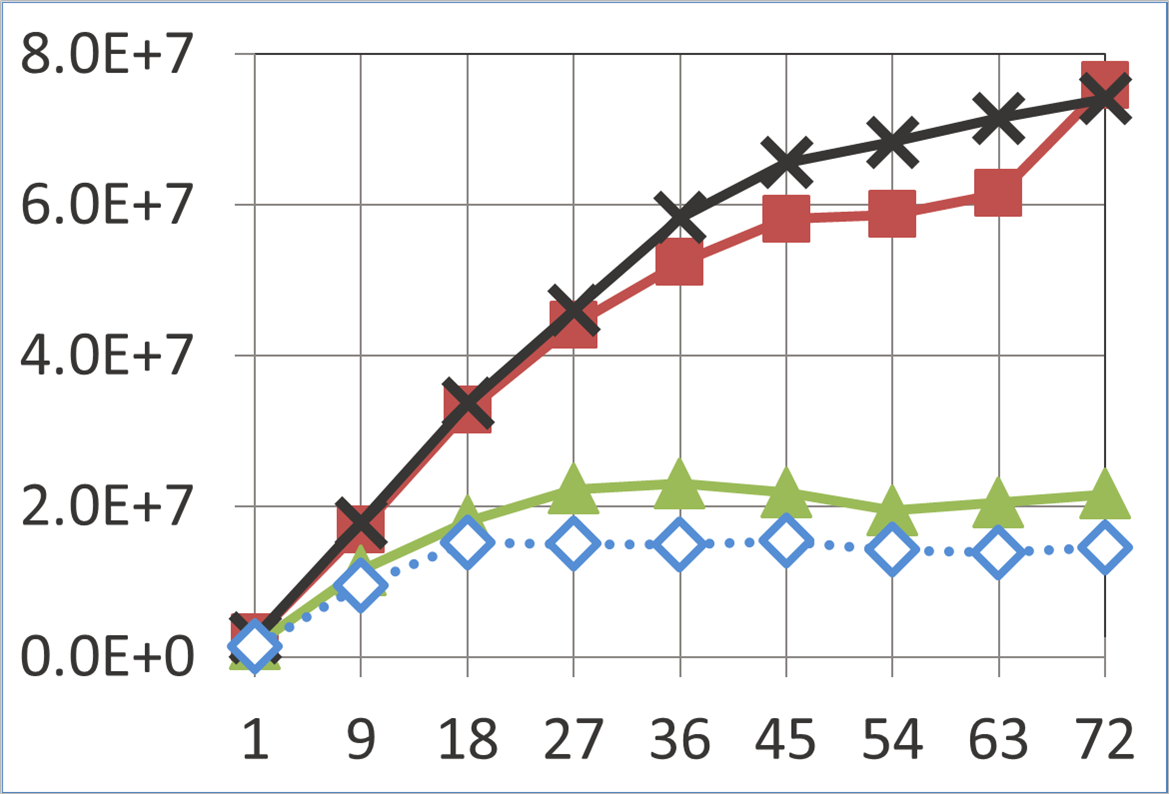}
        \\
        \vspace{-4.5mm}\rotatebox{90}{{\footnotesize\textbf{Heavy} workload}} &
        \vspace{-4.5mm}\includegraphics[width=\linewidth]{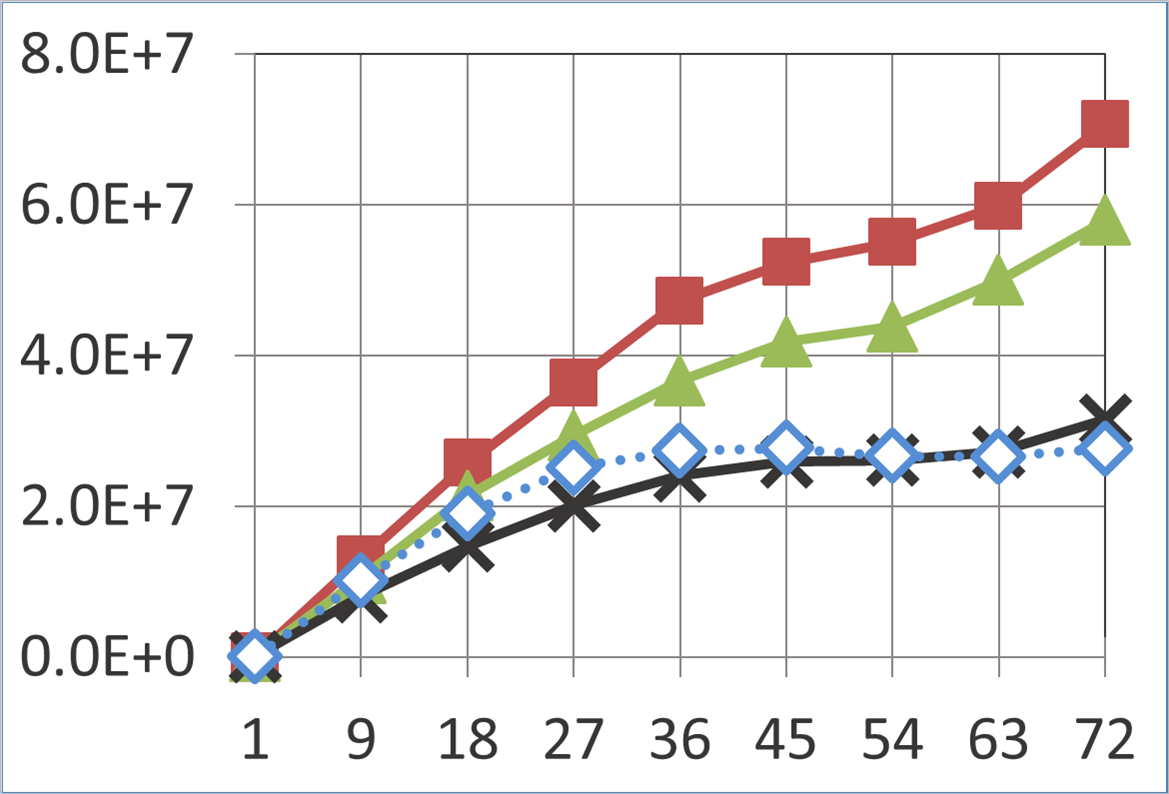} &
        \vspace{-4.5mm}\includegraphics[width=\linewidth]{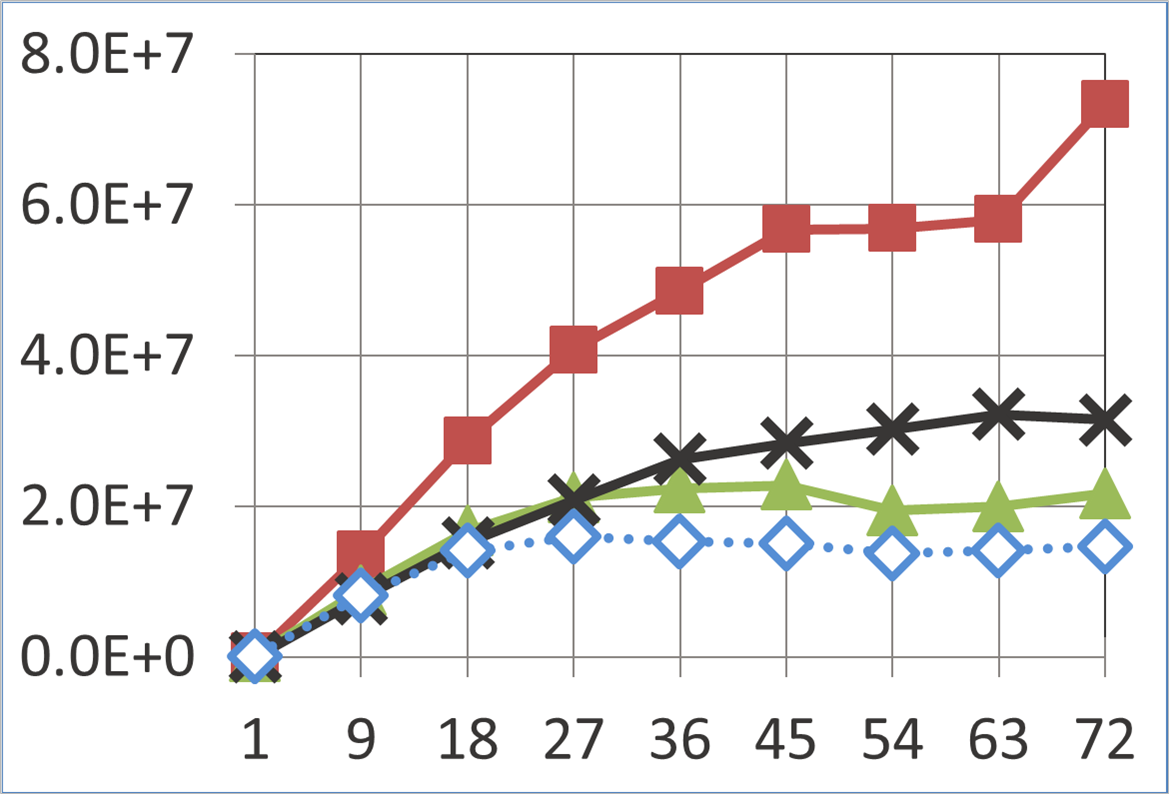}
        \\
    \end{tabular}
    \vspace{1mm}
    
    \includegraphics[width=0.6\textwidth]{chap-3path/figures/perfgraphs/legendv4.png}
\end{minipage}
\caption{Supplementary results from the 2x36 thread Intel E5-2699 v3.}
\label{fig-3path-exp-neelam}
\end{figure}
Supplementary experimental results from the 72-thread machine appear in Figure~\ref{fig-3path-exp-neelam}
\end{fullver}
\subsection{Code path usage and abort rates}
To gain further insight into the behaviour of our accelerated template implementations, we gathered some additional metrics about the experiments described above.
Here, we only describe results from the 48-thread Intel machine.
(Results from the 72-thread Intel machine were similar.)

\fakeparagraph{Operations completed on each path}
We started by measuring how often operations completed successfully on each execution path.
This revealed that operations almost always completed on the fast path.
Broadly, over all thread counts, the minimum number of operations completed on the fast path in any trial was 86\%, and the average over all trials was 97\%.

In each trial that we performed with 48 concurrent threads, at least 96\% of operations completed on the fast path, \textit{even in the workloads with RQs}.
Recall that RQs are the operations most likely to run on the fallback path, and they are only performed by a single thread, so they make up a relatively small fraction of the total operations performed in a trial.
In fact, our measurements showed that the number of operations which completed on the fallback path was never more than a fraction of one percent in our trials with 48 concurrent threads.

In light of this, it might be somewhat surprising that the performance of TLE was so much worse in heavy workloads than light ones.
However, the cost of serializing threads is high, and this cost is compounded by the fact that the operations which complete on the fallback path are often long-running.
Of course, in workloads where more operations run on the fallback path, the advantage of improving concurrency between paths would be even greater.

%It is also instructive to consider a simple workload in which a single-threaded performs 100\% RQs. %heavy workloads, since 
%%Recall that one thread always performs 100\% RQs in our heavy workloads.
%%So, in trials with small thread counts, a much larger proportion of the operations performed were RQs. % (and, in a heavy workload with a single thread, \textit{all} operations are RQs).
%%Consequently, a larger fraction of operations could not complete on the fast path. %, and complete on the fallback path (or middle path).
%%Although in our trials with six concurrent threads, at least 98\% of all operations completed on the fast path, with \textit{one} thread,
%We ran such a workload to study the basic behaviour of RQs in each of the algorithms.
%In \textit{2-path con}, essentially all RQs completed on the (instrumented) fast path.
%This is expected, since 
%In \textit{TLE}, approximately 99\% of RQs completed on the fast path for the BST, but less than 1\% of RQs completed on the fast path for the ($a,b$)-tree.
%This is likely

\begin{figure}[t]
    \centering
    \includegraphics[width=0.6\linewidth]{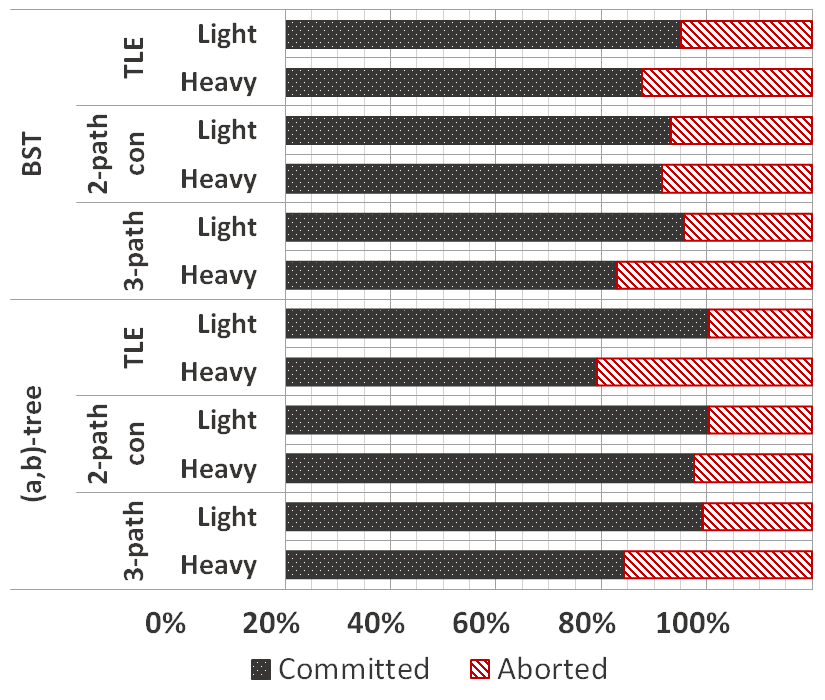} %&
    \caption{Summary of how many transactions commit vs. how many abort in our experiments on the 48-thread Intel machine.}
    \label{fig-abortrate}
\end{figure}

\fakeparagraph{Commit/abort rates}
We also measured how many transactions committed and how many aborted, on each execution path, in each of our trials.
Figure~\ref{fig-abortrate} summarizes the average commit/abort rates for each data structure, template implementation and workload.
Since nearly all operations completed on the fast path, we decided not to distinguish between the commit/abort rate on the fast path and the commit/abort rate on the middle path.

\subsection{Comparing with hybrid transactional memory}
\textit{Hybrid transactional memory} (hybrid TM) combines hardware and software transactions to hide the limitations of HTM and guarantee progress.
This offers an alternative way of using HTM to implement concurrent data structures.
Note, however, that state of the art hybrid TMs use locks.
So, \textbf{they cannot be used to implement lock-free data structures.}
Regardless, to get an idea of how such implementations would perform, relative to our accelerated template implementations, we implemented the unbalanced BST using Hybrid NOrec, which is arguably the fastest hybrid TM implementation with readily available code~\cite{hynorecriegel}.

If we were to use a precompiled library implementation of Hybrid NOrec, then the unbalanced BST algorithm would have to perform a library function call for \textit{each read and write to shared memory}, which would incur significant overhead.
%So, following the work of Riegel et~al.~\cite{hynorecriegel}, we produced an optimized implementation of Hybrid NOrec that has improved concurrency.
So, we directly compiled the code for Hybrid NOrec into the code for the BST, allowing the compiler to inline the Hybrid NOrec functions for reading and writing from shared memory into our BST code, eliminating this overhead.
Of course, if one intended to use hybrid TM in practice (and not in a research prototype), one would use a precompiled library, with all of the requisite overhead.
Thus, the following results are quite charitable towards hybrid TMs.

\begin{figure}[t]
    \centering
    \begin{minipage}{\linewidth}
        \centering
        \setlength\tabcolsep{0pt}
        %        \hspace{-0.1\linewidth}
        \begin{tabular}{m{0.03\linewidth}m{0.6\linewidth}}%m{0.25\textwidth}}%m{0.24\textwidth}}
            &
            %        \multicolumn{2}{c}{%4}{c}{
            %\dimexpr \textwidth-2\fboxsep-2\fboxrule
            \fcolorbox{black!80}{black!40}{\parbox{\dimexpr \linewidth-2\fboxsep-2\fboxrule}{\centering\textbf{2x 24-thread Intel E7-4830 v3}}}
            %        }
            \\
            &
            \fcolorbox{black!50}{black!20}{\parbox{\dimexpr \linewidth-2\fboxsep-2\fboxrule}{\centering {Unbalanced BST (LLX/SCX)\\ Updates w/key range $[0,10^4$)}}} %&
            %        \fcolorbox{black!50}{black!20}{\parbox{\dimexpr \linewidth-2\fboxsep-2\fboxrule}{\centering {\footnotesize (a, b)-tree (LLX/SCX)\\ Updates w/key range $[0,10^6$)}}}
            \\
            \vspace{-3mm}\rotatebox{90}{{\textbf{Light} workload}} &
            \includegraphics[width=\linewidth]{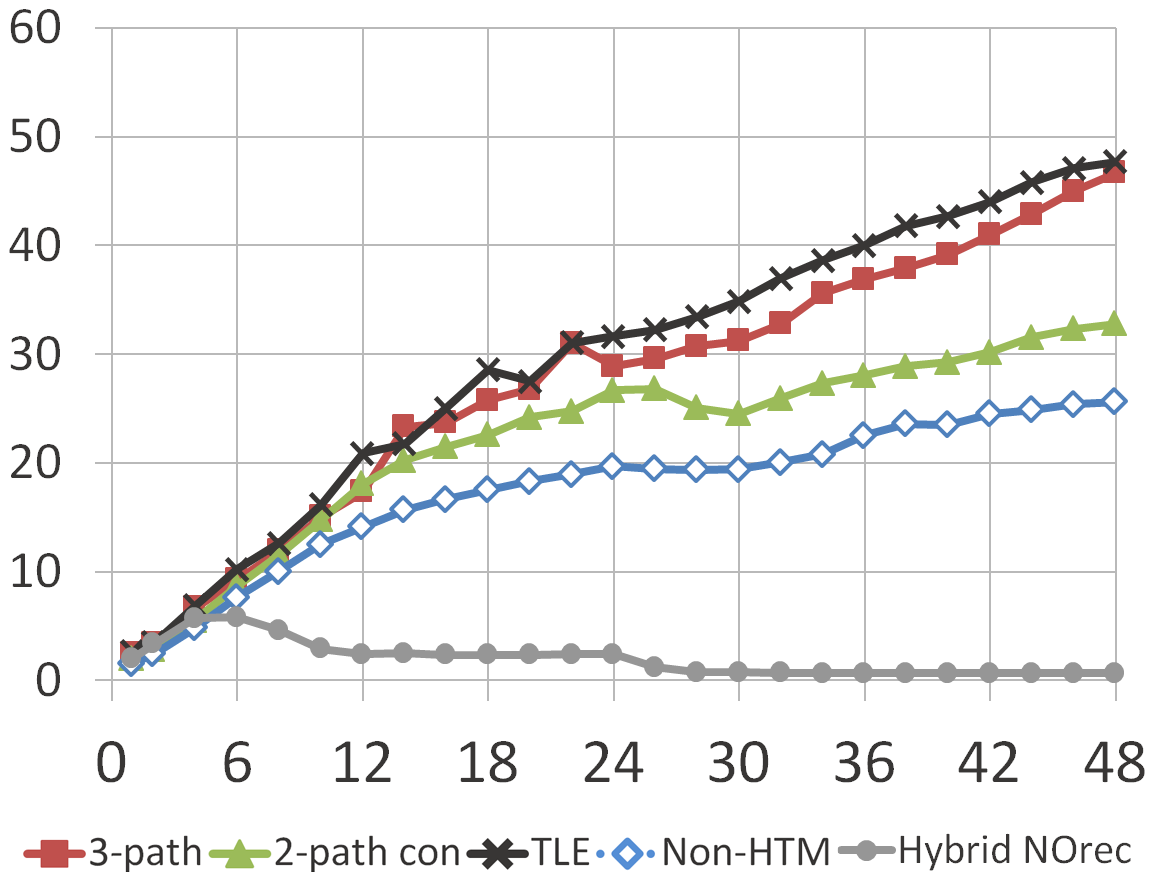} %&
            \\
        \end{tabular}
        %    \vspace{-2mm}
        \caption{Results showing throughput (operations per second) versus number of processes for an unbalanced BST implemented with different tree update template algorithms, and with the hybrid TM algorithm \textit{Hybrid NOrec}.}
        \label{fig-graphs-hybridnorec}
        %    \vspace{-3mm}
    \end{minipage}
\end{figure}

We implemented the BST using Hybrid NOrec by wrapping sequential code for the BST operations in transactions, and manually replacing each read from (resp., write to) shared memory with a read (resp., write) operation provided by Hybrid NOrec.
Figure~\ref{fig-graphs-hybridnorec}
%The following graph 
compares the performance of the resulting implementation to the other BST implementations discussed in Section~\ref{sec-3path-exp}.
%The workload in Figure~\ref{fig-graphs-hybridnorec} is 
%It depicts a light workload, where processes perform 50\% insertions and 50\% deletions on keys drawn uniformly from $[0, 10^4$).
%The x-axis shows the number of concurrent processes.
%The y-axis shows completed operations per microsecond.

The BST implemented with Hybrid NOrec performs relatively well with up to six processes.
However, beyond six processes, it experiences severe negative scaling.
The negative scaling occurs because Hybrid NOrec increments a global counter in each updating transaction (i.e., each transaction that performs at least one write).
This global contention hotspot in updating transactions causes many transactions to abort, simply because they contend on the global counter (and not because they conflict on any data in the tree).
%Clearly, the BST implemented with Hybrid NOrec is not a serious competitor for our accelerated implementations.
However, even without this bottleneck, Hybrid NOrec would still perform poorly in heavy workloads, since it incurs very high instrumentation overhead for software transactions (which must acquire locks, perform repeated validation of read-sets, maintain numerous auxiliary data structures for read-sets and write-sets, and so on).
Note that this problem is not unique to Hybrid NOrec, as every hybrid TM must use a software TM as its fallback path in order to guarantee progress. %, and STMs are known to yield data structures that are significantly slower than hand-crafted ones.
In contrast, in our template implementations, the software-only fallback path is a fast lock-free algorithm.

\section{Modifications for performing searches outside of transactions} \label{sec-3path-search-outside}

In this section, we describe how the \textit{3-path} implementations of the unbalanced BST and relaxed ($a,b$)-tree can be modified so that each operation attempt on the fast path or middle path performs its search phase \textit{before} starting a transaction (and only performs its update phase in a transaction).
(The same technique also applies to the \textit{2-path con} implementations.)
First, note that the lock-free search procedure for each of these data structures is actually a standard, sequential search procedure.
Consequently, a simple sequential search procedure will return the correct result, regardless of whether it is performed inside a transaction.
(Generally, whenever we produce a \textit{3-path} implementation starting from a lock-free fallback path, we will have access to a correct non-transactional search procedure.)

The difficulty is that, when an operation starts a transaction and performs its update phase, it may be working on a part of the tree that was deleted by another operation.
One can imagine an operation $O_d$ that deletes an entire subtree, and an operation $O_i$ that inserts a node into that subtree.
If the search phase of $O_i$ is performed, then $O_d$ is performed, then the update phase of $O_i$ is performed, then $O_i$ may erroneously insert a node into the deleted subtree.

We fix this problem as follows.
Whenever an operation $O$ on the fast path or middle path removes a node from the tree, it sets a $marked$ bit in the node (just like operations on the fallback path do).
Whenever $O$ first accesses a node $u$ in its transaction, it checks whether $u$ has its $marked$ bit set, and, if so, aborts immediately.
This way, $O$'s transaction will commit only if every node that it accessed is in the tree.

We found that this modification yielded small performance improvements (on the order of 5-10\%) in our experiments.
The reason this improves performance is that fewer memory locations are tracked by the HTM system, which results in fewer capacity aborts.
We briefly discuss why the performance benefit is small in our experiments.
The relaxed ($a,b$)-tree has a very small height, because it is balanced, and its nodes contain many keys.
The BST also has a fairly small height (although it is considerably taller than the relaxed ($a,b$)-tree), because processes in our experiments perform insertions and deletions on uniformly random keys, which leads to trees of logarithmic height with high probability.
So, in each case, the sequence of nodes visited by searches is relatively small, and is fairly unlikely to cause capacity aborts.

The performance benefit associated with this modification will be greater for data structures, operations or workloads in which an operation's search phase will access a large number of nodes.
Additionally, IBM's HTM implementation in their POWER8 processors is far more prone to capacity aborts than Intel's implementation, since a transaction \textit{will} abort if it accesses more than 64 different cache lines~\cite{nguyen2015investigation}.
(In contrast, in Intel's implementation, a transaction can potentially commit after accessing tens of thousands of cache lines.)
Thus, this modification could lead to significantly better performance on POWER8 processors.

\section{Reclaiming memory more efficiently} \label{sec-3path-memrecl}

\begin{thesisnot}
    \begin{shortver} % when this appears in the appendix
    For the data structures presented in this paper, we implemented memory reclamation using an epoch based reclamation scheme called DEBRA~\cite{Brown:2015}.
    \end{shortver}
    \begin{fullver} % when this appears as a section
    For the data structures presented in this paper, we implemented memory reclamation using DEBRA.
    \end{fullver}
    This reclamation scheme is designed to reclaim memory for lock-free data structures, which are notoriously difficult to reclaim memory for. %present a particularly difficult problem for reclaiming memory. %This is a particularly difficult type of memory reclamation problem.
    Since processes do not lock nodes before accessing them, one cannot simply invoke \texttt{free()} to release a node's memory back to the operating system as soon as the node is removed from the data structure.
    This is because a process can always be poised to access the node just after it is freed.
\end{thesisnot}
\begin{thesisonly}
    In Chapter~\ref{chap-debra}, we explained that lock-free memory reclamation is hard because a process can always be poised to access a node just after it is freed.
\end{thesisonly}
The penalty for accessing a freed node can be data corruption or a program crash (e.g., due to a segmentation fault).
Once a process has removed a node from a tree, it knows that no other process can reach the node by following pointers starting from the root of the tree.
However, a process may still be able to reach the removed node by starting from a pointer in its \textit{private memory}.
Thus, lock-free memory reclamation schemes must implement special mechanisms to determine when it is safe to free a node (because the node cannot no longer be reached by any process). %that has been removed from the data structure.

However, advanced memory reclamation schemes are unnecessary if (1) \textit{all} accesses to nodes are performed inside transactions, and (2) at the end of each operation on the data structure, the process performing the operation discards all pointers in its private memory that point to nodes.
%With Intel's HTM, accessing freed memory inside a transaction cannot cause a segmentation fault and crash the program.
%Instead, the transaction simply aborts. %either causes the transaction to abort (if the access caused a segmentation fault), or does nothing.
%%Any segmentation fault is hidden by the transaction, and the program does not crash.
%(Note, however, that this is not true for IBM's transactional memory implementation in their POWER8 processors.)
Consider an implementation of a tree that satisfies (1) and (2).
Let $O$ be an operation that removes a node $u$ from the tree, and then frees $u$ immediately thereafter. %, and $O'$ be any operation that accesses $u$ and is concurrent with $O$.
We argue that, if an operation $O'$ accesses $u$, and $O$ subsequently removes $u$ from the tree before $O'$ has terminated, then $O'$ will abort.
(Note that, if $O'$ terminates before $u$ is removed from the tree, then there is no problem.)
By (2), $O'$ must obtain a pointer to $u$ by traversing the tree, starting at the root.
Consider the sequence of pointers followed by $O'$ as it traverses the tree to reach $u$.
At least one of these pointers must change after $O'$ reads it during its traversal, and before $O$ frees $u$ (or else $u$ would still be in the tree after $O$ removed it---a contradiction).
Therefore, $O'$ will abort due to a data conflict.
%Therefore, the first access to $u$ by $O'$ must occur before $O$ has removed $u$ from the tree (since $O'$ cannot reach $u$ by following pointers from the root \textit{after} $u$ has been removed).
%So, $O'$ must access $u$ before $O$ has removed $u$ from the tree.
%Observe that, when $O$ removes $u$ from the tree, it changes a child pointer on the path from the root to $u$.
%If $O'$ read this child pointer during its traversal, then $O'$ will abort due to a data conflict when $O$ changes it.
%
%If $O'$ accesses $u$ at any time \textit{before} $O$ removes it from the tree, then $O'$ will abort.
%This is because, in order for $u$ to be removed from the tree, some pointer that was traversed by $O'$ to reach $u$ must be changed either when $O$ removes the node $u$ from the tree, or before then.
%Now, suppose $O$ only accesses $u$ \textit{after} $O$ removes it from the tree and \textit{before} $O$ frees $u$.
%Then, the contents of $u$ are unchanged since it was in the tree, 
%Now, suppose $O'$ first accesses $u$ \textit{after} $O$ removes it from the tree (e.g., using a pointer stored in the private memory of the process performing $O'$).
%If $O'$ first accesses $u$ \textit{before $O$ frees}
%
%In such an implementation, deleting and immediately freeing a node will simply cause any concurrent transaction that accesses the node (after it is freed) to abort.
%This is because removing the node will change a pointer that was traversed during any concurrent search that reached the node.
Consequently, in such a data structure, reclaiming memory is as easy as invoking \texttt{free()} immediately after a node is removed.

In our three path algorithms, the fast path can only run concurrently with the middle path (but not the fallback path), and the fast path and middle path both satisfy (1) and (2).
Thus, %if every operation on the fast path or middle path runs entirely inside a transaction, then 
memory can be reclaimed on the fast path simply by using \texttt{free()} immediately after removing a node inside a transaction. % (or immediately after committing the transaction).
%This optimization is applicable to the external BST, the $(a,b)$-tree, and the internal BST (but not the linked-list, since the search phase of operations is not performed inside a transaction).
Our performance experiments did \textit{not} implement this optimization, but doing so would likely further improve the performance of the three path algorithms. % relative to the other algorithms.
%\trevor{possibly do a single graph testing this optimization for one data structure---the bst}

%----------- you can free right away
%
%why?
%
%
%mention 2 options for reclamation on the fast path: directly freeing, or using something like DEBRA. direct freeing makes sense only if transactions will not automatically abort on free().

\section{Related work} \label{sec-3path-related}

Hybrid TMs share some similarities to our work, since they all feature multiple execution paths.
The first hybrid TM algorithms allowed HTM and STM transactions to run concurrently~\cite{Kumar2006,Damron2006}.
%An alternative way of obtaining data structure implementations is to use a general technique like \textit{hybrid transactional memory} (hybrid TM).
%%The use of multiple execution paths to improve performance is common in the literature on \textit{hybrid transactional memory} (hybrid TM).
%Hybrid TM combines hardware and software transactions to hide the limitations of an HTM system and guarantee progress.
%%This way, a programmer can write transactions without worrying about whether the hardware will be able to execute them.
%Hybrid TM was simultaneously introduced by Kumar et~al.~\cite{Kumar2006} and Damron et~al.~\cite{Damron2006}.
Hybrid NOrec~\cite{Dalessandro2011} and Reduced hardware NOrec~\cite{Matveev2014} are hybrid TMs that both use global locks on the fallback path, eliminating any concurrency.
We discuss two additional hybrid TMs, Phased TM~\cite{Lev2007} (PhTM) and Invyswell~\cite{Calciu2014}, in more detail.
%%Their algorithms allow HTM and STM transactions to run concurrently.
%Hybrid TMs have also been introduced by Lev et~al.~\cite{Lev2007} (PhTM), Dalessandro et~al.~\cite{Dalessandro2011} (Hybrid NOrec), Matveev and Shavit~\cite{Matveev2014} (Reduced hardware NOrec), and Calciu et~al.~\cite{Calciu2014} (Invyswell).
%Hybrid NOrec and reduced hardware NOrec both use global locks on the fallback path that eliminate any concurrency.
%%(Calciu et~al. also introduced a hybrid TM called Invyswell~\cite{Calciu2014}, but it uses lazy subscription and, consequently, is not safe~\cite{dice2014pitfalls}.) % Therefore, we omit it from our discussion.)
%We discuss PhTM and Invyswell in more detail.

PhTM alternates between five \textit{phases}: HTM-only, STM-only, concurrent HTM/STM, %one that has HTM and STM paths that can run concurrently,
and two global locking phases.
Roughly speaking, PhTM's HTM-only phase corresponds to our uninstrumented fast path, and its concurrent HTM/STM phase corresponds to our middle HTM and fallback paths.
However, their STM-only phase (which allows no concurrent hardware transactions) and global locking phases (which allow no concurrency) have no analogue in our approach.
%PhTM largely leaves it to the implementer to decide when and how to switch between phases.
In heavy workloads, %where most operations succeed in hardware but long-running operations periodically run in software, 
PhTM must oscillate between its HTM-only and concurrent HTM/STM phases to maximize the performance benefit it gets from HTM.
%This oscillation between phases is similar to the way that transactions will alternate between running concurrently on the fast/middle paths, and on the middle/slow paths.
When changing phases, PhTM typically waits until all in-progress transactions complete before allowing transactions to begin in the new mode.
Thus, after a phase change has begun, and before the next phase has begun, there is a window during which new transactions must wait (reducing performance).
One can also think of our three path approach as proceeding in two phases: one with concurrent fast/middle transactions and one with concurrent middle/fallback transactions. %where the fast and middle paths run concurrently, and where the middle and fallback paths run concurrently.
However, in our approach, ``phase changes'' do not introduce any waiting, and there is always concurrency between two execution paths.

%to obtain the performance benefit of HTM for PhTM may must either alternate between its HTM-only and concurrent HTM/STM phases, 

% with the fast path.
%Hybrid NOrec has two paths: an STM path that acquires a global lock at commit-time, and a fast path that can run concurrently with the STM path except when the lock is held.
%
%Reduced hardware NOrec has three paths: a middle path that is similar to the STM path of hybrid NOrec, except it uses a transaction instead of a global lock at commit time, a fast path that can run concurrently with the middle path, and a fallback path that takes a global lock (preventing any concurrency).

%Like PhTM, Invyswell has HTM-only, concurrent HTM/STM and STM-only modes, and two global locking modes.
%However, it imposes numerous additional restrictions 
%(Invyswell actually imposes numerous additional restrictions on when transactions can run concurrently. For simplicity, we have omitted these details. Our three path methodology does not have these restrictions.)

Invyswell is closest to our three path approach.
At a high level, it features an HTM middle path and STM slow path that can run concurrently (sometimes), and an HTM fast path that can run concurrently with the middle path (sometimes) but not the slow path, and two global locking fallback paths (that prevent any concurrency).
Invyswell is more complicated than our approach, and has numerous restrictions on when transactions can run concurrently.
%For simplicity, we have omitted these details.
Our three path methodology does not have these restrictions.
The HTM fast path also uses an optimization called lazy subscription.
It has been shown that lazy subscription can cause opacity to be violated, which can lead to data corruption or program crashes~\cite{dice2014pitfalls}.

Hybrid TM is very general, and it pays for its generality with high overhead.
Consequently, data structure designers can extract far better performance for library code by using more specialized techniques.
Additionally, we stress that state of the art hybrid TMs use locks, so they cannot be used to produce lock-free data structures.
%In the following subsection, we compare the performance of a fast hybrid TM algorithm to our template implementations. %our compare the performance of our BST implementations with an implementation obtained using Hybrid NOrec, the fastest hybrid TM with publicly available code.
%The results show that the BST implementation using Hybrid NOrec is dramatically slower than even the original lock-free BST algorithm.

Different options for concurrency have recently begun to be explored in the context of TLE.
Refined TLE~\cite{dicerefined} and Amalgamated TLE~\cite{afek2015amalgamated} both improve the concurrency of TLE when a process is on the fallback path by allowing HTM transactions to run concurrently with a \textit{single process} on the fallback path.
Both of these approaches still serialize processes on the fallback path.
They also use locks, so they cannot be used to produce lock-free data structures.

%It is important to note that TLE uses locks, and so do the state of the art hybrid TMs.
%Thus, these techniques cannot be used to produce lock-free data structures.

Timnat, Herlihy and Petrank~\cite{timnat2015practical} proposed using a strong synchronization primitive called \textit{multiword compare-and-swap} ($k$-CAS) to obtain fast HTM algorithms.
They showed how to take an algorithm implemented using $k$-CAS and produce a two-path implementation that allows concurrency between the fast and fallback paths.
One of their approaches used a lock-free implementation of $k$-CAS on the fallback path, and an HTM-based implementation of $k$-CAS on the fast path.
They also experimented with two-path implementations that do not allow concurrency between paths, and found that allowing concurrency between the fast path and fallback path introduced significant overhead.
Makreshanski, Levandoski and Stutsman~\cite{makreshanski2015lock} also independently proposed using HTM-based $k$-CAS %to simplify and accelerate %lock-free data structures 
in the context of databases.
%\trevor{make stronger comparison between our approach and the k-cas approaches. maybe just cite brown2013/14 and say llx/scx has significant advantages over kcas for advanced lock-free d.s. such as balanced trees. or, explain in the appendix...}

Liu, Zhou and Spear~\cite{Liu2015} proposed a methodology for accelerating concurrent data structures using HTM, and demonstrated it on several lock-free data structures.
Their methodology uses an HTM-based fast path and a non-transactional fallback path. %that can run concurrently.
The fast path implementation of an operation is obtained by encapsulating part (or all) of the operation in a transaction, and then applying sequential optimizations to the transactional code to improve performance.
Since the optimizations do not change the code's logic, the resulting fast path implements the same logic as the fallback path, so both paths can run concurrently.
Consequently, the fallback path imposes overhead on the fast path. %the fact that transactions on these paths can run concurrently imposes limitations on how the fast path can be designed to improve performance (because transactions on the fast path must remain compatible with those on the fallback path).

Some of the optimizations presented in that paper are similar to some optimizations in our HTM-based implementation of \llt\ and \sct.
For instance, when they applied their methodology to the lock-free unbalanced BST of Ellen et~al.~\cite{Ellen:2010}, they observed that helping can be avoided on the fast path, and that the descriptors which are normally created to facilitate helping can be replaced by a small number of statically allocated descriptors.
However, they did not give details on exactly how these optimizations work, and did not give correctness arguments for them.
In contrast, our optimizations are applied to a more complex algorithm, and are proved correct.

%In the same paper, Liu~et~al. also identified numerous algorithm specific optimizations (some of which can be generalized to other algorithms) that reduced the overhead imposed by the slow path on the fast path.
%The optimizations they describe could be applied to the middle path in our methodology.
%However, some of their optimizations involved carefully modifying the handcrafted fallback path algorithm to eliminate constraints on the fast path.
%This requires a deep understanding of the fallback algorithm, and is not always possible.
%However, in our three path methodology, since the fast path and fallback path do not run concurrently, these optimizations can be applied on the fast path with almost no knowledge of the fallback algorithm.

Multiversion concurrency control (MVCC) is another way to implement range queries efficiently \cite{bernstein1983multiversion, attiya2012single}.
At a high level, it involves maintaining multiple copies of data to allow read-only transactions to see a consistent view of memory and serialize even in the presence of concurrent modifications. %, and garbage collecting  old data only once no transaction can access it.
However, our approach could also be applied to operations that \textit{modify} a range of keys, so it is more general than MVCC. %, since it is not limited to read-only transactions.
%For instance, our approach could be applied to a workload with operations that \textit{modify} a range of keys.

%Studying these problems and developing solutions in the context of data structures, rather than hybrid TM, offers a different perspective, and different opportunities for optimization.
%For example, whereas a hybrid TM must perform the same update requested by the programmer on every path, a three-path BST algorithm has the freedom to change the update being performed, depending on the execution path.

%\section{Comparison with hybrid transactional memory} %\label{sec-3path-hybridnorec}
%
%\input{chap-3path/appendix-hybridnorec}

\section{Other uses for the \textit{3-path} approach}
\label{sec-3path-other-uses}
\subsection{Accelerating data structures that use read-copy-update (RCU)}

In this section, we sketch a \textit{3-path} algorithm for an ordered dictionary implemented with a node-oriented unbalanced BST that uses the RCU synchronization primitives.
%(Recall that, in a node-oriented tree, there are no routing keys, and each key in a node is in the dictionary.)
The intention is for this to serve as an example of how one might use the \textit{3-path} approach to accelerate a data structure that uses RCU.

RCU is both a programming paradigm and a set of synchronization primitives.
The paradigm organizes operations into a search/reader phase and an (optional) update phase.
In the update phase, all modifications are made on a \textit{new copy} of the data, and the old data is atomically replaced with the new copy.
In this work, we are interested in the \textit{RCU primitives} (rather than the paradigm).

\textbf{Semantics of RCU primitives and their uses.}
The basic RCU synchronization primitives are \textit{rcu\_begin}, \textit{rcu\_end} and \textit{rcu\_wait}~\cite{Desnoyers:2012}.
Operations invoke \textit{rcu\_begin} and \textit{rcu\_end} at the beginning and end of the search phase, respectively.
The interval between an invocation of \textit{rcu\_begin} and the next invocation of \textit{rcu\_end} by the same operation is called a \textit{read-side critical section}.
An invocation of \textit{rcu\_wait} blocks until all read-side critical sections that started before the invocation of \textit{rcu\_wait} have ended.
%
%\textbf{Uses of RCU.}
One common use of \textit{rcu\_wait} is to wait, after a node has been deleted, until no readers can have a pointer to it, so that it can safely be freed.
It is possible to use RCU as the sole synchronization mechanism for an algorithm, if one is satisfied with allowing many concurrent readers, but only a single updater at a time.
If multiple concurrent updaters are required, then another synchronization mechanism, such as fine-grained locks, must also be used.
However, one must be careful when using locks with RCU, since locks cannot be acquired inside a read-side critical section without risking deadlock.

\textbf{The CITRUS data structure.}
We consider how one might accelerate a node-oriented BST called CITRUS~\cite{arbel2014concurrent}, which uses the RCU primitives, and fine-grained locking, to synchronize between threads.
First, we briefly describe the implementation of CITRUS.
%however, RCU can also be used in conjunction with fine-grained locking as a secondary synchronization mechanism.
%CITRUS uses the RCU primitives in conjunction with fine-grained locking.
At a high level, RCU is used to allow operations to search without locking, and fine-grained locking is used to allow multiple updaters to proceed concurrently.

The main challenge in the implementation of CITRUS is to prevent race conditions between searches (which do not acquire locks) and deletions.
When an internal node $u$ with two children is deleted in an internal BST, its key is replaced by its successor's key, and the successor (which is a leaf) is then deleted. %is is replaced by its successor.
This case must be handled carefully, or else the following can happen.
%Consider a \textit{Delete}$(key)$, and $S$ be a search for $key'$, and 
Consider concurrent invocations $D$ of \textit{Delete}$(key)$ and $S$ of \textit{Search}$(key')$, where $key'$ is the successor of $key$.
Suppose $S$ traverses past the node $u$ containing $key$, and then $D$ replaces $u$'s key by $key'$, and deletes the node containing $key'$.
The search will then be unable to find $key'$, even though it has been in the tree throughout the entire search.
To avoid this problem in CITRUS, rather than changing the key of $u$ directly, $D$ replaces $u$ with a new copy that contains $key'$.
After replacing $u$, $D$ invokes \textit{rcu\_wait} to wait for any ongoing searches to finish, before finally deleting the leaf containing $key'$.
The primary sources of overhead in this algorithm are invocations of \textit{rcu\_wait}, and lock acquisition costs.

\textbf{Fallback path.}
The fallback path uses the implementation of CITRUS in~\cite{arbel2014concurrent} (additionally incrementing and decrementing the global fetch-and-add object $F$, as described in Section~\ref{sec-3path-algs}).
%consists of an implementation of an node-oriented BST called CITRUS~\cite{arbel2014concurrent} that uses the RCU primitives, and fine-grained locking, to synchronize between threads.

\textbf{Middle path.}
The middle path is obtained from the fallback path by wrapping each fallback path operation in a transaction and optimizing the resulting code.
The most significant optimization comes from an observation that the invocation of \textit{rcu\_wait} in \textit{Delete} is unnecessary since transactions make the operation atomic.
Invocations of \textit{rcu\_wait} are the dominating performance bottleneck in CITRUS, so this optimization greatly improves performance.
A smaller improvement comes from the fact that transactions can avoid acquiring locks.
Transactions on the middle path must ensure that all objects they access are not locked by other operations (on the fallback path), or else they might modify objects locked by operations on the fallback path. %observe the data structure in an inconsistent state.
However, it is not necessary for transaction to actually \textit{acquire} locks.
Instead, it suffices for a transaction to simply \textit{read} the lock state for all objects it accesses (before accessing them) and ensure that they are not held by another process.
This is because transactions \textit{subscribe} to each memory location they access, and, if the value of the location (in this case, the lock state) changes, then the transaction will abort.

\textbf{Fast path.}
The fast path is a sequential implementation of a node-oriented BST whose operations are executed in transactions.
As in the other 3-path algorithms, each transaction starts by reading $F$, and aborts if it is nonzero.
This prevents operations on the fast path and fallback path from running concurrently.
There are two main differences between fast path and the middle path.
First, the fast path does not invoke \textit{rcu\_begin} and \textit{rcu\_end}.
These invocations are unnecessary, because operations on the fast path can run concurrently \textit{only} with other operations on the fast path or middle path, and \textit{neither} path depends on RCU for its correctness.
(However, the middle path \textit{must} invoke these operations, because it runs concurrently with the fallback path, which relies on RCU.)
The second difference is that the fast path does not need to read the lock state for any objects.
Any conflicts between operations on the fast and middle path are resolved directly by the HTM system. %, without the use of locks.

\subsection{Accelerating data structures that use $k$-CAS}

In this section, we sketch how one might produce a \textit{3-path} implementation of a lock-free algorithm that uses the $k$-CAS synchronization primitive.

A $k$-CAS operation takes, as its arguments, $k$ memory locations, expected values and new values, and atomically: reads the memory locations and, if they contain their expected values, writes new values into each of them.
\begin{thesisnot}
$k$-CAS has been implemented from single-word CAS~\cite{Harris:2002}.
We briefly describe this $k$-CAS implementation.
\end{thesisnot}
\begin{thesisonly}
We consider the implementation of $k$-CAS in Chapter~\ref{chap-descriptors}.
For convenience, we give a brief overview of the implementation, here.
\end{thesisonly}
At a high level, a $k$-CAS creates a descriptor object that describes the $k$-CAS operation, then uses CAS to store a pointer to this descriptor in each memory location that it operates on.
Then, it uses CAS to change each memory location to contain its new value.
While a $k$-CAS is in progress, some fields may contain pointers to descriptor objects, instead of their regular values.
Consequently, reading a memory location becomes more complicated: it requires reading the location, then testing whether it contains a pointer to a descriptor object, and, if so, helping the $k$-CAS operation that it represents, before finally returning a value.

\textbf{Fallback path.}
An operation on the fallback path simply increments a global fetch-and-increment object $F$, performs the appropriate operation of the lock-free algorithm, then decrements $F$.
%The fallback path consists of the lock-free singly-linked list in~\cite{timnat2015practical}.
%At a high level, each operation on the fallback path consists of a search phase, optionally followed by an update phase, which is performed using $k$-CAS.

\textbf{Middle path.}
The middle path is the similar to the fallback path, except that the CAS-based implementation of $k$-CAS is replaced with a straightforward implementation using HTM (following the approach in~\cite{timnat2015practical}).
%
%Since the search phase in a linked list can be extremely long, and is likely to cause a transaction to abort (due to capacity limitations), the middle path was obtained by wrapping \textit{only the update phase} of each fallback path operation in a transaction, and optimizing the resulting code.
%The main optimization on the middle path comes from replacing the software implementation of $k$-CAS with straightforward implementation from HTM (using the approach in~\cite{timnat2015practical}).
This HTM-based implementation performs the entire $k$-CAS atomically, and does not need to create a descriptor.
However, to facilitate concurrency with the lock-free implementation of $k$-CAS, the HTM-based $k$-CAS \textit{does} check whether each address it reads contains pointer to a descriptor, and, if so, helps the operation that created the descriptor before trying again.
%\trevor{concurrency????? if no, is it really like 3path? if yes, how can we get away without descriptors?????}
%\trevor{more detail on the htm-based kcas from shahar's work? also check what i do in the code?}
%This approach was introduced by Petrank and Timnat~\cite{timnat2015practical}.

\textbf{Fast path.}
The fast path is a sequential implementation wrapped in a transaction. %in which the \textit{update phase} of each operation is wrapped in a transaction.
The main optimization on the fast path comes from the fact that, since there are no concurrent operations on the fallback path, there are no pointers to $k$-CAS descriptors in shared memory.
Consequently, operations on the fast path do not need to check whether any values they read from shared memory are actually pointers to $k$-CAS descriptors, which significantly reduces overhead.

%\textbf{Preventing fast/fallback concurrency.}
%Observe that our fast path optimization (to avoid checking whether any values that are read are actually pointers to $k$-CAS descriptors) is correct only if the search phase in the fast path does not run concurrently with the update phase of any operation on the fallback path.
%For each of the other data structures we described, each operation runs entirely inside a single transaction.
%Thus, for these data structures, it suffices to verify that the global fetch-and-add object $F$ is zero at the beginning of each transaction to guarantee that operations on the fast path do not run concurrently with operations on the fallback path.
%However, this is not sufficient for the list, since only the update phase of each operation executes inside a transaction.
%So, we implement a fast form of group mutual exclusion that allows many operations on the fast path, or many operations on the fallback path, but not both.
%Consequently, although the fallback path is lock-free, the three path algorithm is only deadlock- and livelock-free (but not lock-free). % because of this mutual exclusion.

%\section{Modifications for performing searches outside of transactions}
%\label{sec-3path-search-nontxn}
%\input{chap-3path/appendix-search-nontxn}

\begin{thesisonly}
\section{Summary} \label{sec-3path-conclusion}
\end{thesisonly}
\begin{thesisnot}
\section{Concluding remarks} \label{sec-3path-conclusion}
\end{thesisnot}

\begin{thesisonly}
In this chapter, 
\end{thesisonly}
\begin{thesisnot}
In this work, 
\end{thesisnot}
we explored the design space for HTM-based implementations of the tree update template of Brown et~al. and presented four accelerated implementations.
We discussed performance issues affecting HTM-based algorithms with two execution paths, and developed an approach that avoids them by using three paths.
We used our template implementations to accelerate two different lock-free data structures, and performed experiments that showed significant performance improvements over several different workloads. %(especially for the \textit{3-path} implementations).
This makes our implementations an attractive option for producing fast concurrent data structures for inclusion in libraries, where performance is critical.

Our accelerated data structures each perform an entire operation inside a single transaction (except on the fallback code path, where no transactions are used).
\begin{fullver}
We discussed how one can improve efficiency by performing the read-only \textit{searching} part of an operation non-transactionally, and simply using a transaction to perform any modifications to the data structure.
\end{fullver}
\begin{shortver}
One can improve efficiency by performing the read-only \textit{searching} part of an operation non-transactionally, and simply using a transaction to perform any modifications to the data structure.
We discuss this optimization in detail in Appendix~\ref{appendix-search-nontxn}.
\end{shortver}
Our \textit{3-path} approach may also have other uses.
\begin{fullver}
As an example, we sketched 
\end{fullver}
\begin{shortver}
As an example, in Appendix~\ref{appendix-other-uses}, we sketch
\end{shortver}
an accelerated \textit{3-path} implementation of a node-oriented BST that uses the read-copy-update (RCU) synchronization primitives.
We suspect that a similar approach could be used to accelerate other data structures that use RCU.
Additionally, we describe how one might produce a \textit{3-path} implementation of a lock-free algorithm that uses the $k$-CAS synchronization primitive.

\addcontentsline{toc}{chapter}{Bibliography} %% This adds a line for the Bibliography in TOC
\bibliographystyle{plain}
\bibliography{bibliography}

\end{document}